# Astrophysical parameters of M dwarfs with exoplanets

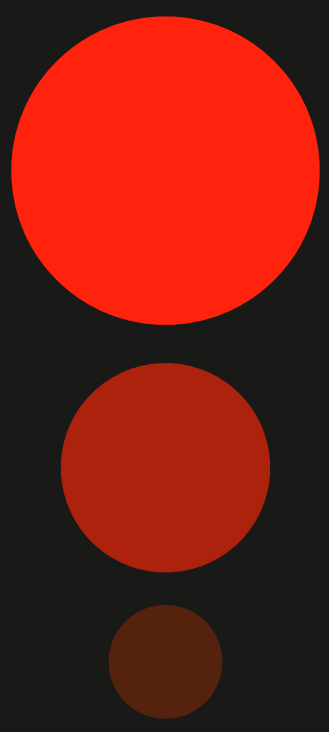

Carlos Cifuentes San Román

DOCTORAL THESIS
2023

Universidad Complutense de Madrid

Facultad de Ciencias Físicas

Tesis doctoral

# *Parámetros astrofísicos de enanas M con exoplanetas*

# *Astrophysical parameters of M dwarfs with exoplanets*

Memoria para optar al grado de doctor presentada por:

Carlos Cifuentes San Román

Supervisado por:

Dr. José Antonio Caballero
Dr. Jorge Sanz Forcada

# Astrophysical parameters *of* M dwarfs *with* exoplanets

CARLOS CIFUENTES SAN ROMÁN

*with external refereeing by*

Prof. Adam J. Burgasser
*and*
Prof. Andrei A. Tokovinin

Madrid, March 2023

*A mi madre*

*“Platero es pequeño, peludo, suave;*
*tan blando por fuera,*
*que se diría todo de algodón,*
*que no lleva huesos.”*

— Juan Ramón Jiménez, *Platero y yo*

# Agradecimientos

*Nuestras madres son nuestro primer firmamento, nuestra primera luz, nuestro universo entero. Este libro trata sobre estrellas: sus colores, sus brillos, su lugar. Hay estrellas que aunque nacieron juntas ahora no están cerca, pero mantienen un vínculo que permanece en el espacio y en el tiempo. Yo tengo la fortuna de tener una estrella que me mira y me acompaña: este libro es para mi madre.*

Mis padres me dieron la vida y lucharon por aquel futuro que *ahora* es — no puedo imaginar un regalo más valioso. Me mostraron siempre el camino del esfuerzo y la empatía. De mi hermana Judith y mi cuñado Iván he aprendido que lo que importa es la sensibilidad y el amor sencillo y sin adornos. A ellos la vida les sigue: Óliver y Mateo, os deseo lo mejor en vuestro propio camino, que es un libro que se acaba de abrir. Mis tías, Eloína y Loli, cada día han sido cariño puro, amor sincero, un hogar para el alma. Sois mi familia y os debo mucho más de lo que podría expresar en palabras.

José Caballero y Jorge Sanz me han ofrecido todo lo que se puede pedir: experiencia, paciencia y cercanía. Ellos me brindaron la libertad de poder descubrir por mi mismo. José ha sido infinitamente generoso con su tiempo, y ha tenido para mi siempre sus puertas abiertas. Una vez él y yo fuimos *The Astrophysical Brothers*, y hubo tantas otras ocasiones en las que la música nos vio unidos. Yo guardo respeto y admiración por él, le estoy profundamente agradecido. Valoro el entusiasmo y la cercanía que María Rosa Zapatero Osorio ha significado. En nuestras reuniones de los viernes yo me noto madurar como científico. En el MPIA Trifon Trifonov me enseñó muchas cosas útiles e importantes. No olvidaré el color de Königstuhl en el otoño de Heidelberg. A Félix un día le regalé miel y manzanas y él me devolvió mucho más. Con Michel Mayor conversé sobre vida en Marte, sobre Suiza, sobre chocolate. Cuando veo su nombre siempre evoco su inspiradora humildad. Me siento afortunado de haber compartido esta etapa con vosotros y vosotras, deseo poder volver a hacerlo.

Dice la RAE que la amistad es "afecto personal, puro y desinteresado, compartido con otra persona, que nace y se fortalece con el trato". Adrián, Bea, Diego, Héctor, Pepe, Sara — comparto con vosotros el placer de la amistad: somos amigos a cambio de nada. Me conocen desde esa edad en la que yo ignoraba hasta cuál era la edad del Universo. A mis colegas en el Centro de Astrobiología: ojalá que la ciencia nos una de nuevo, he sido tan feliz de haber podido compartir este tiempo juntos. Mientras escribo estas líneas, a mi lado está Olga. Ella es mi compañera en el viaje, una luz en el camino, reposo para el corazón.

Esta página no puede ser una lista exhaustiva de quienes traigo a mi memoria al querer agradecer por tanto. Aunque no leas tu nombre, yo habré pensado en ti. Puede incluso que no sepa tu nombre y algo te deba agradecer.

Es vuestro este trabajo, sois todo lo bueno que hay en mi.
Sin vuestra luz no se ven las estrellas.

Carlos, en marzo de 2023.

# Contents

## Abstract

M dwarfs are the most abundant stars in the Universe. They are hosts of a rich diversity of planetary companions, which are much more noticeable in these smaller stars. Advancements in technology continue to push the boundaries of what is possible to be detected on them, even when they exist as faint signals. In many cases, planets orbiting M dwarfs can also be described in remarkable detail. What makes the difference is how deeply we can characterise the host star. This includes to properly model their atmospheres, their abundance of metals, and their activity processes. If they are well described individually, these numerous stars have the potential for providing statistically robust conclusions when combined into larger samples. Carmencita is the input catalogue of nearby M dwarfs for the CARMENES project, which aims to search for potentially habitable Earth-sized planets orbiting them. It contains more than two thousand M dwarfs that are scrutinized by the consortium members from multiple angles and with a high degree of attention to detail.

This thesis contributes to the description of each one of these M dwarfs, including astrometry, photometry, activity, kinematics, and multiplicity, but also to the study of the sample as a whole. We carry out a comprehensive process that involves multi-band photometric and astrometric data analyses. With a minucious examination of every single star and their potential physical companions, we put special effort on the individualised inspection rather than on bulk searches, and in hand-picked data rather than in fully automatised crossmatches. We homogeneously derive stellar luminosities that are eminently empirical, temperatures, masses, and radii, as well as many intermediate products such as bolometric corrections, absolute magnitudes, and colours. From these, we obtain averaged values and empirical relations between fundamental parameters and observables.

Additionally, we address the topic of multiplicity of M dwarfs, describing in detail the existing multiple systems and their components, also proposing the existence of many pairs discovered for the first time. We put the focus on the unresolved binary systems that go unnoticed without spectroscopic studies, but that lead to an incomplete picture of stellar genesis, and to miscalculations of stellar parameters, which negatively impacts the planetary descriptions. The empirical observations presented in this study provide an important benchmark for testing and improving theoretical predictions, because any model of stellar formation and evolution should be able to explain the observed characteristics of these stars.

By taking a careful, individualized approach to the study of M dwarfs, we not only contribute to the study of the Universe's physical processes, but we also pave the way for future discoveries of the potential for life beyond our own planet. Overall, the findings of this study underscore the importance of continued research into the most numerous stars and their planetary systems. We expect that the wealth of data gathered in this thesis will serve as a valuable resource for astronomers and researchers in related fields, and that it will inspire further investigations and new insights into the processes that shape the Universe.

// Resumen

Las enanas M son las estrellas más abundantes en el Universo. Albergan una rica diversidad de compañeros planetarios, que son mucho más evidentes en estas estrellas más pequeñas. Los avances tecnológicos continúan ampliando los límites de lo que es posible detectar en ellas, incluso cuando se manifiestan como señales débiles. En muchos casos, los planetas que orbitan enanas M también pueden describirse con notable detalle. Lo que marca la diferencia es la completitud con el que podemos caracterizar la estrella anfitriona. Esto incluye modelar adecuadamente sus atmósferas, su abundancia en metales y sus procesos de actividad. Si se describen adecuadamente de manera individualizada, estas numerosas estrellas tienen el potencial de proporcionar conclusiones estadísticamente robustas cuando se combinan en muestras más grandes. Carmencita es el catálogo de entrada de enanas M cercanas para el proyecto CARMENES, que tiene como objetivo buscar planetas del tamaño de la Tierra potencialmente habitables orbitando alrededor de ellas. Contiene más de dos mil enanas M que son examinadas por los miembros del consorcio desde múltiples ángulos y con un alto grado de atención al detalle.

Esta tesis contribuye a la descripción de cada una de estas enanas M, incluyendo la astrometría, fotometría, actividad, cinemática y multiplicidad, pero también al estudio de la muestra en su conjunto. Llevamos a cabo un proceso exhaustivo que implica el análisis de datos fotométricos en múltiples bandas, y astrométricos. Con un examen minucioso de cada estrella y sus posibles compañeros físicos, hemos puesto especial énfasis en la inspección individualizada en lugar de en las búsquedas masivas, y en la selección manual de datos en lugar de cruces completamente automatizados con bases de datos. Derivamos homogéneamente luminosidades estelares que son eminentemente empíricas, temperaturas, masas y radios, así como muchos productos intermedios como correcciones bolométricas, magnitudes absolutas y colores. A partir de estos, obtenemos valores promediados y relaciones empíricas entre los parámetros fundamentales y las observables.

Además, abordamos el tema de la multiplicidad de las enanas M, describiendo detalladamente los sistemas múltiples existentes y sus componentes, proponiendo también la existencia de muchas parejas descubiertas por primera vez. Ponemos el foco en los sistemas binarios no resueltos que pasan desapercibidos sin estudios espectroscópicos, pero que conducen a una imagen incompleta de la génesis estelar y a cálculos erróneos de los parámetros estelares, lo que afecta negativamente a las descripciones planetarias. Las observaciones empíricas presentadas en este estudio proporcionan un importante punto de referencia para poner a prueba y mejorar las predicciones teóricas, porque cualquier modelo de formación y evolución estelar debería ser capaz de explicar las características observadas de estas estrellas.

Al tomar un enfoque cuidadoso e individualizado para el estudio de las enanas M, no solo contribuimos al estudio de los procesos físicos del Universo, sino que también abrimos el camino para futuros descubrimientos del potencial de vida más allá de nuestro propio planeta. En general, los hallazgos de este estudio destacan la importancia de continuar la investigación en las estrellas más numerosas y sus sistemas planetarios. Esperamos que la gran cantidad de datos recopilados en esta tesis sirva como un recurso valioso para astrónomos e investigadores en campos relacionados, y que inspire nuevas investigaciones y nuevas perspectivas sobre los procesos que dan forma al Universo.

# Chapter 1

## Introduction

M dwarfs are made of many lives. Their evolution is so slow that from their birth to their final light, many generations of stars like our Sun can go by. This is because M dwarfs make of hydrogen fusion an extremely efficient process. They do not have much mass to convert into energy, but their interiors do a good job at mixing the very few available. They are the first hydrogen-fusing objects that enjoy fully convective interiors, which means that hydrogen is constantly delivered from the surface to the core, making its burning steady and slow. Taking the most of the matter available in this way, however faintly, M dwarfs can glow from dozens to hundreds of billions of years. Meanwhile, the Sun will die one hundred times before the last of their least massive neighbours disappears. In other words, the Universe still needs to be older to see the first red dwarf shut down and die.

M dwarfs are the most abundant stars in the Galaxy, by far exceeding in number stars like our Sun. Planets orbiting M dwarfs are extremely ubiquitous, certainly exceeding the number of stars. With an uncountable amount of possibilities, the number of planets that are suitable for life as we know it could be exceedingly vast. These realisations can safely be extrapolated to other galaxies in the observable Universe. We could say, then, that M dwarfs might be made of many *lifes*. Today we can learn about the diversity and peculiarities of planets orbiting M dwarfs with an unprecedented detail. The problem of how hospitable they can be for life is far from being closed. The future envisions further investigation of the emergence and survival of atmospheres, made possible by the development of extraordinarily well-tailored instrumentation. Disentangling the effects resulting from a plethora of involved factors demands a highly multidisciplinary approach. Among these factors is the behaviour of the host star, which is determined by its fundamental physical parameters, such as its mass. To understand how planets form, evolve, and survive to host life, characterising the host star is the first safe step. The ways are many, and are open — it promises to be a fruitful endeavour.

## 1.1 Know thy star, know thy planet

Over the past two decades, there has been a surge of interest in M dwarfs, as novel horizons previously restricted by technological, instrumental, and computational limitations have been revealed. The current outlook on these stars is fresh and optimistic, particularly in the search for life-hosting exoplanet candidates. Visually discriminating an exoplanet is almost always not a possibility, let alone actually using robots to probe their atmospheres or surfaces directly and bring a sample back to Earth, and so indirect methods to discover planets must be the way. Every discovery of a planet around an M dwarf contributes to our statistical understanding of the existence and occurrence of planets, and the greater the diversity of detected planets, the more comprehensive our knowledge becomes. Bulk studies such as all-sky exoplanet surveys provide many samples to withdraw from, but there are not shortcuts to the meticulous studies of light or radial-velocity curves for individual stars. Therefore, nothing is more important than describing the stars, by empirical observation or by modelling otherwise, as perfectly as possible.

Exoplanet exploration is just one topic of research that will be indisputably benefited by a better characterisation of these cool stars, but many others exist. If we improve our understanding of the individual stars from the most common type in the Universe, we will necessarily gain knowledge about pretty much any global process in the Galaxy. Our understanding of how stars form and evolve to host planets has been achieved by studying individual stars in great detail. With M dwarfs now more accessible than ever before, the possibilities are limitless. Our understanding of the Universe and its processes fundamentally relies on the study of starlight. In order to describe any planet, one must first describe the object that it orbits. Although orphaned, free-floating planets exist and may outnumber those that orbit objects of any type, their description is inherently incomplete and subject to significant uncertainties In any case, it is highly unlikely that life could emerge or survive on these objects.

Planets are composed from the same material as their mother stars (Gonzalez, 1997; Santos et al., 2001; Fischer & Valenti, 2005). As in a family portrait, we should aim at picturing the mother in order to understand the child. Being able to describe the composition, origin, and vital status of a star is the way to go in order to properly describe her planets. In this sense, planets can be known only as good as their stars are, and the planetary diversity found is a reflection of the miscellany of stars. The mass of a planet can be considered its most fundamental property because, when combined with a radius measured from transits, it determines the bulk density. Some planets are prone to an extraordinary characterisation, for instance, if they transit the star that they orbit (e.g. Caballero et al., 2022). More precise determinations of mass and radius translates into better defined constraints on ice mass fraction and size of rocky interior of the orbiting planets, but has little effect on the composition of the gas layer (Dorn et al., 2017). Many efforts have been undertaken in the determination of stellar parameters, including empirical determination of masses, radii, and their relation to luminosity (Veeder, 1974; Henry & McCarthy, 1993; Chabrier & Baraffe, 1997; Delfosse et al., 2000; Bonfils et al., 2005a; Mann et al., 2015, 2019; Terrien et al., 2015; Benedict et al., 2016; Schweitzer et al., 2019), effective temperature, surface gravity, and metallicity (Casagrande et al., 2008; Rojas-Ayala et al., 2013; Montes et al., 2018; Passegger et al., 2018, 2019; Rajpurohit et al., 2018a), or activity and rotation periods (Stauffer & Hartmann, 1986; Reid et al., 1995; Hawley et al., 1996; Morales et al., 2008; Hawley et al., 2014; Newton et al., 2015; Jeffers et al., 2018; Díez Alonso et al., 2019; Schöfer et al., 2019; Lafarga et al., 2023).

Luckily, the parameters of star and planet are fundamentally tied during their detection. For instance, the semi-amplitude of the radial-velocity (RV) wobble that the planet with mass $\mathcal{M}_p$ imprints in the star with mass $\mathcal{M}_\star$ when orbiting with a period $P$ is:

$$K_\star = \frac{1}{\sqrt{1-e^2}} \frac{\mathcal{M}_p \sin i}{(\mathcal{M}_p + \mathcal{M}_\star)^{2/3}} \left(\frac{2\pi G}{P}\right)^{1/3}, \tag{1.1}$$

where $e$ is the eccentricity of the orbit and $i$ is the inclination of the orbital plane. Therefore, the mass of the planet is necessarily a fraction of that of the host star. This is even more evident in the case of the planetary radius. The surface flux $F$ (or amount of light per unit area) that is blocked when the star of radius $\mathcal{R}_\star$ is eclipsed by the planet of radius $\mathcal{R}_p$ is directly proportional to the area that it blocks. A simplified form ignoring the limb darkening can be put as:

$$\frac{\Delta F}{F} = \frac{\mathcal{R}_p^2}{\mathcal{R}_\star^2}. \tag{1.2}$$

Large planets around relatively small stars maximise the observed semiamplitude, $K_\star$, of the reflex velocities, and the surface flux dip, $\Delta F/F$, of the partial occultation of the star, because the effects are much more notable. In the first case is the gravitational tug, in the second case is the proportion of the star that the planet eclipses. The domain of small, potentially habitable rocky planets is most easily approachable in the case of the least massive stars (Anglada-Escudé et al., 2016; Gillon et al., 2017; Zechmeister et al., 2019; Dreizler et al., 2020; Trifonov et al., 2021), and still poses a challenge in the case of Sun-like mass stars. It is important to mention that the very detectability of a planet is reigned by the nature of the star, because the amount of stellar activity or jittery fundamentally limits the precision to derive planetary characteristics.

Another essential parameter that is integral to the definition of a planet is the bolometric luminosity, $\mathcal{L}$. It corresponds to the output stellar energy per unit of time, and it is an intrinsic property directly related to its mass. The Stefann-Boltzmann's law relates the luminosity to the temperature and radius of the star, given that the energy profile as a function of the wavelength can be approximated by that of a black body: $F = \frac{\mathcal{L}}{4\pi\mathcal{R}^2} = \sigma T_{\rm eff}^4$. The luminosity determines the equilibrium temperature of the planets that orbit her, depending on the orbital separation. The equilibrium temperature of the planet, $T_{\rm eq}$, is decisive for the habitability, as it dictates, for example, the fate of an atmosphere on it. If the planet orbits a star of luminosity $\mathcal{L}$ at a separation $a$, the general expression for the equilibrium temperature is:

$$T_{\rm eq} = \left(\frac{\mathcal{L}(1-A_B)}{16\sigma\pi a^2}\right)^{1/4}, \tag{1.3}$$

where $\sigma$ is the Stefan-Boltzmann constant. This formula approximates both star and planet to a black body, and considers a planetary surface that partially reflects a fraction of the incident stellar flux, measured by the Bond albedo, $A_B$. Because of the interdependence of all of these properties, quantifying stellar parameters turns out to be extremely of uttermost importance when characterising the planet (e.g. von Braun et al., 2014a). This is also true for the astrometry of the star: even modest uncertainties in the distance can result in substantial uncertainties in its mass or the equilibrium temperature of the planet, which are dependent on the semi-major axis. Luckily, the *Gaia* mission[1] measured trigonometric parallaxes with unprecedented precision, providing remarkably small relative uncertainties (less than 1 % for sources with $G \lesssim 15$ mag, and 1–10 % for sources with $G \simeq$15–18 mag).

[1]*Gaia* is a space observatory of the European Space Agency (ESA), launched in 2013, which is expected to be operative at least until 2025. This is the first reference to this mission, but it will be recurrent throughout this work given the fundamental role of its data. The name was originally meant an acronym ("Global Astrometric Interferometer for Astrophysics"), but now it remains as a proper noun.

The equilibrium temperature determines the circumstellar habitable zone (CHZ), which is classically defined as the range of orbital separations from a star for which a planetary surface can support liquid water, which translates into $T_{\mathrm{eq}} \simeq 273\text{–}373\,\mathrm{K}$ (Dole, 1964; Hart, 1978; Kasting et al., 1993; Kopparapu et al., 2017, and see Tarter et al. 2007 for the particular M-dwarf case). This approach provides a solid starting point since it allows for more accurate predictions of the expected observational signatures. However, this concept is constantly evolving as new findings emerge from one of the most active areas of astrobiological research, which deals with the origins of life (Ramirez, 2018). Among these findings are: the existence of extraterrestrial liquid water is possible under a range of conditions outside the habitable zone limits (e.g. Murray et al., 2012; Kopparapu et al., 2014; Glein et al., 2015; Choblet et al., 2017; Vance et al., 2018; Kite & Ford, 2018), the possibility of other energy sources besides the parent star to influence positively on biotic processes (Lindegren et al., 2018, and see again Tarter et al. 2007), the fact that life can prosper in extreme environments that do not qualify as habitable in the strict sense (Edwards et al., 2012), the hypothesis of alternative biogenic chemistry (Kan et al., 2016), or the presence and survival of an atmosphere (Lingam & Loeb, 2017). To define what habitable means, and what makes a planet capable of developing and sustaining life, one should questions what is *essential* for giving way to any form of life (see the review by Shahar et al., 2019).

To summarise, our understanding of how planets form and evolve rests fundamentally on the characterisation of their host stars. This includes a clear, updated picture of their nature of the star (Chapter 2), a precise determination of stellar parameters and how their uncertainties propagate (Chapter 3), and a detailed knowledge of the physically bound companions (Chapter 4). In the quest for exoplanets, in particular with expectations for life to prosper, knowing the host star is not only an option, but a necessity. As if it was written in stone on the Temple of Apollo: *Know thy star, know thy planet.*

## 1.2 Brave new worlds

Since the early discoverers of spectrum analysis in astronomy, such as Robert Bunsen and Gustav Kirchhoff, our capability to read between the lines has shifted our perspective about what can be learnt from light alone. William Huggins carried out the first spectroscopic observations in the mid-nineteenth century (Huggins & Miller, 1864), to discover that the chemistry found in the Earth and the Sun was also present in distant stars, comets, and nebulae. His major innovation occurred in 1868, when he was the first to measure the radial velocity of a star in the Doppler shift of its spectral lines (see Huggins & Huggins, 1890). In the spectrum of a white dwarf, van Maanen (1917) discovered elements heavier than helium, which might be indicative of accreted planetary debris (Zuckerman et al., 2007; Farihi, 2016), and so van Maanen's spectrum might be recognised the first early evidence of an exoplanet. Campbell et al. (1988) offered tentative detection of a planet around a star similar to our Sun. The authors considered that the role of the stellar magnetic fields in the observed periodicities was not of trivial interpretation, and therefore a solid confirmation was not possible — the existence of a planet of jovian mass would be later confirmed by Hatzes et al. (2003). Latham et al. (1989) published the first detection of a substellar object candidate outside the Solar System, which presented as 'a probable brown dwarf' in a 84-day orbit around a solar-like star.

It was epoch for many innovative works regarding exoplanetary discovery: Wolszczan & Frail (1992) and Wolszczan (1994) detected and confirmed the first planet outside the Solar System that orbits a pulsar, Hatzes & Cochran (1993) studied radial velocity variations of giant stars, and the notorious discovery of the first solid evidence of an exoplanet orbiting a main sequence, solar-like star by Mayor & Queloz (1995), which was validated later that year (Marcy & Butler, 1995). A similar discovery by Marcy & Butler (1996) and Butler & Marcy (1996) would soon follow. To the date of writing this work, 5250 detections are confirmed exoplanets[2] (Fig. 1.1). Given that three decades have gone by since the first pioneering investigations, this implies that a new planet has been discovered approximately every two days.

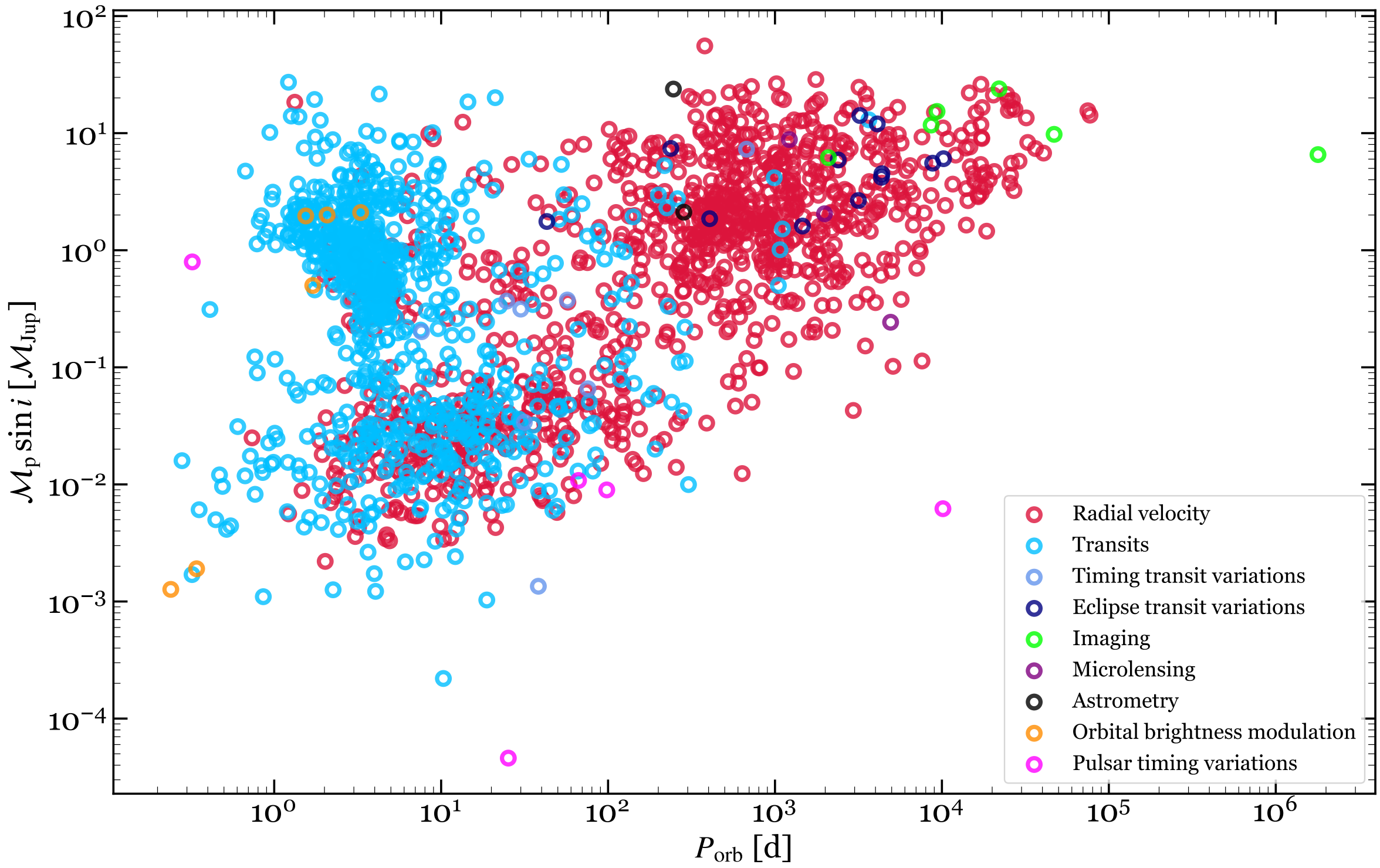

Figure 1.1: Orbital period as a function of planetary mass in jovian units for all exoplanets confirmed to date in the NASA Exoplanet Archive, coloured to differentiate by the discovery method.

Except for the few dozen cases for which there is a direct imaging of the actual planet (e.g. Fig. 1.2), almost all confirmed exoplanets are detected by indirect means. Two techniques account for 95% of the discoveries (Fig. 1.3). The radial-velocity (RV) method has produced 1027 confirmed detections, but it is *eclipsed* by the transit method, with 3945 exoplanets found. In the RV method, because both star and planet revolve around the centre of mass of the system, the orbit of the planet is inferred from the star's orbit, which is measured as a back-and-forth movement in the component of the motion as seen from Earth. In the other case, a planet is said to transit when its orbital path crosses the host star as seen from our perspective. With sufficient accuracy, it is possible to measure a periodic dimming of the light from the star. The effect is equivalent to a solar eclipse, but in another star. As in the case of our Sun, these can be predicted accurately years into the future. The time between transits also tells important information regarding additional planets or other causes that disturb the regularity. Even when the transits method is only sensitive to planets with orbits that are coplanar (or near to) to our line of sight, their discoveries are much more fruitful in number because they can be carried out in all-sky surveys. The majority of the planet detections using transits are attributed to only two of these missions: *Kepler* (Borucki et al., 2010; Howell et al., 2014) and *TESS* (Ricker et al., 2015) have contributed with more than 90% of the transiting planets, with potential candidates already detected in enough number to double the size of the current pool[3]. Other methods can be potentially used during the transits: transit timing variations, eclipse tim-ing variations, or orbital brightness modulation. Other are more sophisticated and demanding from the observational perspective: polarimetry, or microlensing. The latter exploits a fascinating phenomenon: the gravitational lensing effect predicted by Albert Einstein's theory of general relativity. A massive body such a star can bend the light coming from a more distant star. If a planet happens to pass in front of the distant star, it can cause a brief increase in brightness that can be measured.

[2]In this section, all the statistics and figures are made with data from the NASA Exoplanet Archive `https://exoplanets.nasa.gov`, updated as of February 20, 2023.

[3]There are 6153 *TESS*, 2054 *Kepler* and 978 K2 (*Kepler*'s extended mission) candidates to be confirmed. These are transit-like events that appear to be astrophysical in origin.

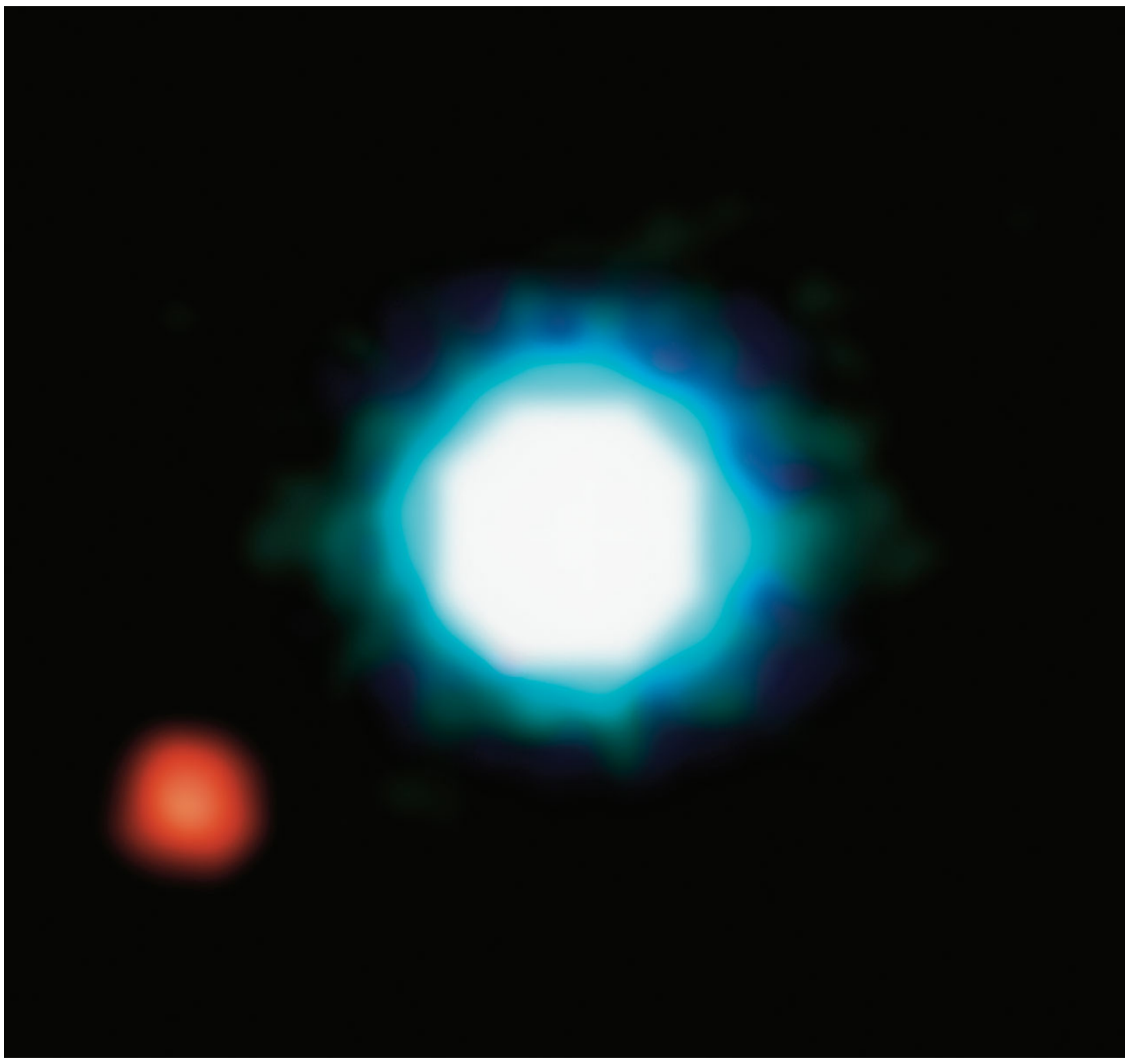

Figure 1.2: 2M1207 b (Chauvin et al., 2004; Mamajek, 2005) is a giant planet ($\sim 5\,\mathcal{M}_{\rm Jup}$, shown in red) around a brown dwarf (2M1207, shown in blue) that constitutes a beautiful example of a directly imaged exoplanet. It is the first planetary-mass companion of a brown dwarf discovered, and the first to be directly imaged beyond the Solar System. The image is composed by three near-infrared exposures (*H*, *K* and *L* passbands) with the NACO adaptive-optics facility at the 8.2-m VLT Yepun telescope at the ESO Paranal Observatory (Credit: ESO).

The technological effort put into the discovery of exoplanets has also produced many other ground- and space-based instruments. Some of the ground based spectrographs with the greatest resolving power are summarised in Table 1.1. In this context, the future depends to a great extent on the precision that can be achieved with Doppler measurements, with a prospect of extraordinary 10-centimetre per second values (Fischer et al., 2016, and compare with the 3 metre-per-second precision of Butler et al. 1996).

One of the most exciting possibilities, for which many observational efforts have already being carried out, is the study of atmospheres of distant worlds (e.g. Ehrenreich et al., 2020). Small, Earth-sized planets are more interesting than large, gaseous worlds, because they could potentially be better hosts for life. The prospect of being able to isolate the spectral features of the atmospheres of these planets is a promising perspective for the future (Tal-Or et al., 2019). Upcoming generations of extremely large telescopes might be able to detect molecules such as $O_2$ in the atmosphere rocky planets in the CHZ of M dwarfs. They could combine high-contrast imaging and high-dispersion spectroscopy techniques to peer into the nearest neighbour planets (Snellen et al., 2015).

Indeed, the proximity of some confirmed exoplanets that also transit, many of them orbiting M dwarfs,

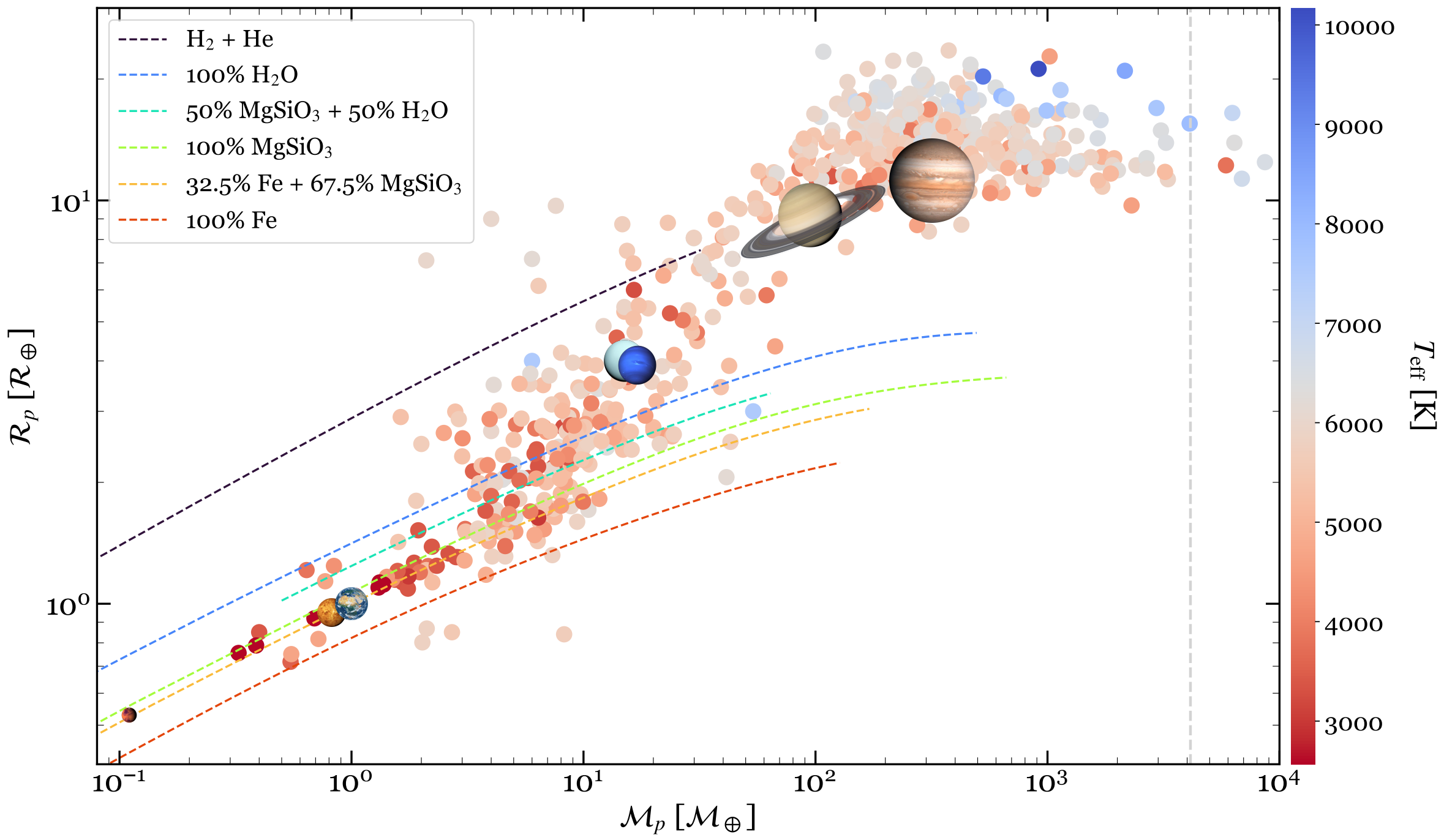

Figure 1.3: Mass-radius diagram for all the confirmed transiting exoplanets with mass determination (from RV or transit time variations), colour-coded by effective temperature of the host star, along with the planets of the Solar System. The dashed coloured curved are theoretical composition models by Zeng et al. (2019) described in the legend, and the dashed grey line represents the limiting mass for the deuterium burning (13 $\mathcal{M}_{\rm Jup}$). Figure adapted from Caballero et al. (2022).

offers a favourable opportunity to probe their atmospheres (Seager & Deming, 2010; Wunderlich et al., 2019). There is already a sizeable body of the literature devoted to these explorations. Some atomic and molecular species investigated include H$\alpha$, He I, Li I, Na I doublet, Mg I, K I, and Ca II infrared tripled (IRT) (Nortmann et al., 2018; Salz et al., 2018; Alonso-Floriano et al., 2019b; Casasayas-Barris et al., 2020; Khalafinejad et al., 2021; Casasayas-Barris et al., 2021; Czesla et al., 2022; Tabernero et al., 2022b), Si (Cont et al., 2022b), Ti and V (Cont et al., 2022a), Fe (Yan et al., 2022), TiO (Cont et al., 2021), and also $H_2O$ (Alonso-Floriano et al., 2019a; Sánchez-López et al., 2019).

Transmission spectroscopy is the method to access to this information observationally, when the light from the star pierces the planet gaseous envelope, if this exists. It is, in some way, as seeing a sunrise (of another 'Sun') in another planet. Nowadays, to observe the atmosphere of an Earth-like planet *from* Earth falls on the edge of the technological capabilities. But the prospects are good: we have the technology, the knowledge, and the targets. Small, rocky planets in the habitable zone of red dwarfs are reachable with today's technology.

## 1.3 M dwarfs

Low-mass, cool stars are remarkably numerous and long-lived objects in the Galaxy. Among them, M dwarfs are by far the most common type of star in the solar neighbourhood, vastly outnumbering their more massive counterparts (Henry et al., 1994, 2006; Reid et al., 2004; Bochanski et al., 2010; Winn & Fabrycky, 2015; Reylé et al., 2021). In their mainly convective interiors, the fusion process is slow and, therefore, the lifespan is long, as they remain on the main sequence for tens of billions of years (Adams & Laughlin, 1997; Baraffe et al., 1998).

Table 1.1: Some of the highest-resolution ground-based spectrographs for exoplanet surveys.

| Instrument[a] | First light | Telescope/ Observatory[b] | Resolution | $\lambda$ coverage [nm] |
|---|---|---|---|---|
| HIRES | 1993 | 10-m WMKO | 85 000 | 300–1000 |
| CORALIE[c] | 1998 | 1.2-m LSO | 60 000 | 390–680 |
| HARPS | 2003 | 3.6-m LSO | 120 000 | 380–690 |
| CRIRES | 2006 | 8.2-m VLT | 100 000 | 950–5200 |
| HARPS-N | 2012 | 3.6-m TNG | 120 000 | 378–691 |
| CARMENES | 2015 | 3.6-m CAHA | 94 600 (VIS) | 520–960 |
| | | | 86 400 (nIR) | 960–1710 |
| ESPRESSO | 2016 | 8.2-m VLT | 200 000 | 380–686 |
| EXPRES | 2018 | 4.3-m DCT | 150 000 | 380–780 |
| SPIRou | 2018 | 3.6-m CFHT | 75 000 | 950–2350 |
| IRD | 2018 | 8.2-m ST | 70 000 | 950–2450 |
| HPF | 2018 | 9.2-m HET | 50 000 | 800–1300 |
| MAROON-X | 2019 | 8.1-m GNT | 85 000 | 500–920 |
| CRIRES+ | 2021 | 8.2-m VLT | 100 000 | 1000–5000 |
| NIRPS | 2022 | 3.6-m LSO | 100 000 | 950–1800 |
| KPF | 2022 | 10-m WMKO | 98 000 | 440–850 |
| ANDES[d] | [2026] | 39-m ELT[e] | 100 000 | 400–1800 |

[a] HIRES: HIgh Resolution Echelle Spectrometer (Vogt et al., 1994); HARPS / HARPS-N: High-Accuracy Radial velocity Planetary Searcher (Mayor et al., 2003); CRIRES / CRIRES+: CRyogenic InfraRed Echelle Spectrograph / ~ Upgrade (Kaeufl et al., 2004); CARMENES: Calar Alto high-Resolution search for M dwarfs with Exoearths with Near-infrared and optical Echelle Spectrographs (see Sect. 2.1); ESPRESSO: Echelle SPectrograph for Rocky Exoplanets and Stable Spectroscopic Observations (Pepe et al., 2010); EXPRES: EXtreme PREcision Spectrograph (Jurgenson et al., 2016); SPIRou: SpectroPolarimètre Infra-Rouge (Artigau et al., 2014); IRD: InfraRed Doppler (Tsujimoto et al., 2018); HPF: Habitable Zone Planet Finder (Mahadevan et al., 2012); MAROON-X: M-dwarf Advanced Radial velocity Observer Of Neighboring eXoplanets (Seifahrt et al., 2018); NIRPS: Near-InfraRed Planet Searcher (Bouchy et al., 2017); KPF: Keck Planet Finder (Gibson et al., 2016); ANDES: ArmazoNes high Dispersion Echelle Spectrograph (Marconi et al., 2022).

[b] CAHA: Centro Astronómico Hispano en Andalucía; CFHT: Canada France Hawaii Telescope; DCT: Discovery Channel Telescope; ELT: Extremely Large Telescope; GNT: Gemini North Telescope; HET: Hobby-Eberly Telescope; LSO: La Silla Observatory; ST: Subaru Telescope; TNG: Telescopio Nazionale Galileo; VLT: Very Large Telescope; WMKO: W. M. Keck Observatory.

[c] CORALIE is not an actual acronym, but the name of the baby daughter of an engineer from the Observatoire de Haute-Provence in France.

[d] Formerly known as HIRES, but was renamed due to the coincidence with Keck's HIRES, which is in operations since the 1990s.

[e] Currently under construction (first light expected in 2028).

Such abundance and prevalence make low-mass stars very attractive targets for multiple areas of astrophysical research. Collectively, M dwarfs are excellent probes for the examination of the Galactic structure (Bahcall & Soneira, 1980; Scalo, 1986; Reid et al., 1997; Chabrier, 2003a; Pirzkal et al., 2005; Caballero et al., 2008; Ferguson et al., 2017), and are also very convenient tracers of Galactic kinematics and evolution (Reid et al., 1995; Gizis et al., 2002; West et al., 2006; Bochanski et al., 2007). Individually, M dwarfs have proven to be interesting targets for the discovery of low-mass exoplanets, and a consid-erable amount of the current literature pays special attention to them (e.g. Boss, 2006; Tarter et al., 2007; Zechmeister et al., 2009; Bonfils et al., 2013a; Mann et al., 2013b; Clanton & Gaudi, 2014; Dressing & Charbonneau, 2015; Fischer et al., 2016; Kopparapu et al., 2017; Reiners et al., 2018b). In particular, low-mass, small-sized stars are specially suited to the search for close-in terrestrial planets because their detection becomes easier with decreasing stellar size and planetary orbital period (Anglada-Escudé et al., 2016; Gillon et al., 2017; Zechmeister et al., 2019).

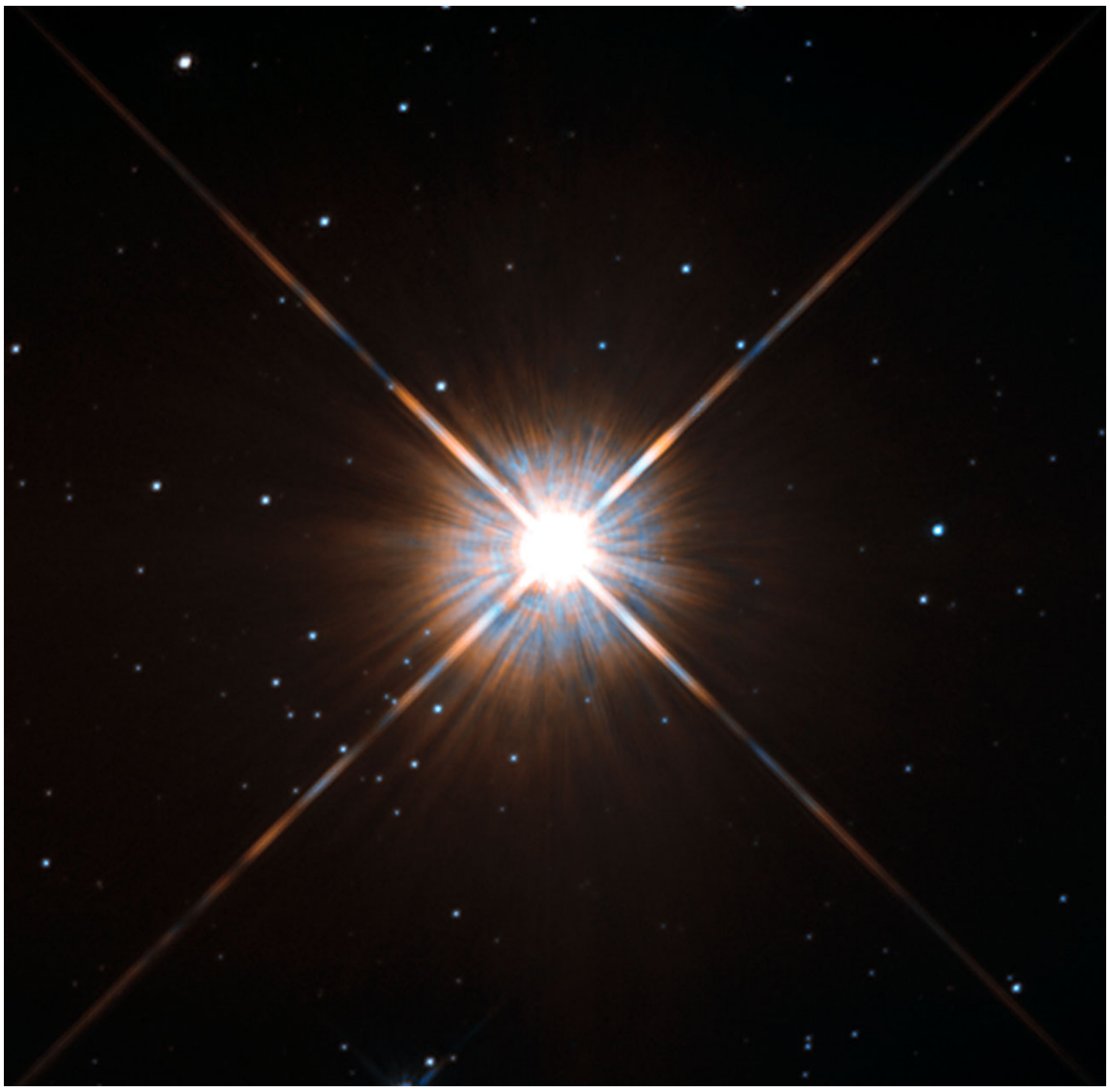

Figure 1.4: Direct image of Proxima Centauri taken using Hubble's Wide Field and Planetary Camera 2 (WFPC2). Credits: ESA/Hubble & NASA.

But to say that M dwarfs are simply *common* would be an understatement. In order to appreciate their ubiquity, we propose an interstellar travel[4]. Departure: our Sun — a G2 V star burning hydrogen in a reasonable peace in its main sequence existence. The nearest stellar neighbour to our Sun is a triple system that contains an M dwarf (Proxima Centauri, M5.5 V — see Fig. 1.4). More than a hundred years passed since the discovery of the star (Innes, 1915; Voûte, 1917)[5] for an Earth-like planet to be detected (Anglada-Escudé et al., 2016)[6]. The second nearest neighbour, which carries the name of its discoverer (E. E. Barnard), is an M dwarf as well (Barnard, 1916), with a controversial planet around (Ribas et al., 2018, but see Lubin et al. 2021). If we ignore the ultracool, little-or-no-glowing objects in Luhman 16 (an L8+T1 binary) and the planetary-mass object WISE 0855-0714 (Y4), the next two stars are M dwarfs, too: Wolf 359 or CN Leo (M6.0 V), and Lalande 21185 (M2.0 V). At 10 pc away from home, we will have spotted 422 stellar and substellar objects, including a few white dwarfs, from which 249 (or 59 out of 100) are spectroscopically classified as M dwarfs (Reylé et al., 2021).

[4] A super-luminous speed is assumed, so distances can be ignored.

[5] Gill (1899) noted the importance of a fair assignation of the prime discoverer.

[6] Two additional, smaller planets, in even closer orbits were found thereafter (Faria et al., 2022), one of them confirmed, and a third, 5.2-year period candidate (Damasso et al., 2020) that awaits confirmation.

None of the stars that are being studied in this research can be seen with the naked eye from Earth. This is because M dwarfs shine primarily in the infrared (the typical peak of emission for M dwarfs lies between ~0.7 and 1.5 $\mu$m), where the human eye is not sensitive. On top of this, they also shine much dimly (less than 10 %) than our Sun. They lie in the lower tail of the main sequence (see Fig. 1.5). Because of their intrinsic faintness, low-mass stars and substellar objects have been widely under-represented until relatively recent times.

With such an abundance, it is not surprising that red dwarfs contribute notably to the luminosity function and constitute an important weight in the mass function baryonic, ordinary matter of the Milky Way (Scalo, 1986; Hawkins & Bessell, 1988; Kroupa et al., 1993; Kirkpatrick et al., 1994; Gould et al., 1996, 1997; Kroupa, 2001; Zheng et al., 2001; Chabrier, 2003b; Covey et al., 2008; Bochanski et al., 2010; Kalirai et al., 2013). Both in the initial mass function (IMF) and the stellar luminosity function (SLF) the existence of unresolved binaries must have an impact that must not be neglected (Kroupa et al., 1991; Piskunov & Mal'Kov, 1991; Chabrier, 2003a; Kroupa & Jerabkova, 2018).

In dwarfs with masses of less than 0.35 $\mathcal{M}_\odot$ (i.e. M3.5-M4.0 type and onwards), convection takes over as the principal mechanism for energy transportation, and fully convective interiors dominate (Delfosse et al., 1998b; Mullan & MacDonald, 2001; Reiners & Basri, 2009). This translates into the fact that a large fraction of mid-M dwarfs to early-L objects are magnetically active (Hawley et al., 1996; Gizis et al., 2000b; West et al., 2004, 2008). In particular, young, low mass stars show very strong chromospheric activity, which intensifies the ultraviolet emission with respect to older populations (Rodriguez et al., 2011). This makes radial-velocity detections more difficult, because some signatures imprinted by active stars can adopt the appearance of the gravitational interaction with one or more planets (France et al., 2013).

If an M dwarf was selected at random, there is a good chance that it would host at least one planet about the size of Earth (Cassan et al., 2012; Bonfils et al., 2013a; Gaidos et al., 2016; Hardegree-Ullman et al., 2019; Mulders et al., 2021; Sabotta et al., 2021). Clanton & Gaudi (2014) estimated the number of planets of *any* kind per M dwarf to be of 1.9 ± 0.5. Using the full four-year of *Kepler* data, Dressing & Charbonneau (2015) estimated the cumulative occurrence rate of planets with 1–4 $\mathcal{R}_\oplus$ and orbital periods less than 200 days to be 2.5 ± 0.2 planets per M dwarf, with $16^{+17}_{-7}$ % and $12^{+10}_{-5}$ % of Earth-size and super-Earth planets per M dwarf habitable zone, respectively. For minimum masses between 1 and 10 $M_\oplus$, Sabotta et al. (2021) estimated in $1.32^{+0.33}_{-0.31}$ planets per M dwarf. More recently, Ribas et al. (2023) reported 1.44 ± 0.20 planets with $\mathcal{M} \sin i < 1000\, \mathcal{M}_\odot$ with periods of less than 1000 days. The common results that emerges from these studies is that each M dwarfs could harbour, *at least*, one planet. These statistics come to reality in the solar neighbourhood, beyond Barnard's star: Lalande 21285 (2 planets, Díaz et al., 2019; Rosenthal et al., 2021), Lacaille 9352 (2 planets, Jeffers et al., 2020), Ross 128 (1 planet, Bonfils et al., 2018b), and so on[7]. Some of them also have additional candidates, but unconfirmed. The evidence collected in the first two decades of the 21st century leaves no trace of doubt: M dwarfs are fertile ground for planetary systems.

We have learned that these planets exist in a rich diversity, which includes Jovian planets (Bayliss et al., 2018; Morales et al., 2019), Saturnian and sub-Saturnian (Hartman et al., 2009; Quirrenbach et al., 2022; Sedaghati et al., 2022), Neptunian and sub-Neptunian (Bonfils et al., 2005b; Reiners et al., 2018a; Luque et al., 2019a, 2022b; Espinoza et al., 2022), super-Earths (Udry & Santos, 2007; Charbonneau et al., 2009; Dittmann et al., 2017; Suárez Mascareño et al., 2017b; Díez Alonso et al., 2018b; Günther et al., 2019; Bluhm et al., 2021; Soto et al., 2021; Toledo-Padrón et al., 2021; Luque et al., 2022a; Damasso et al., 2022; Chaturvedi et al., 2022), and Earth-like planets (Berta-Thompson et al., 2015; Anglada-Escudé et al., 2016; Gillon et al., 2017; Dittmann et al., 2017; Luque et al., 2019b; Kemmer et al., 2020; Amado et al., 2021; Trifonov et al., 2021; Kemmer et al., 2022; Luque et al., 2022a; Caballero et al., 2022).

[7]For CN Leo Tuomi et al. (2019) reported two planets, although this was a very controversial result.

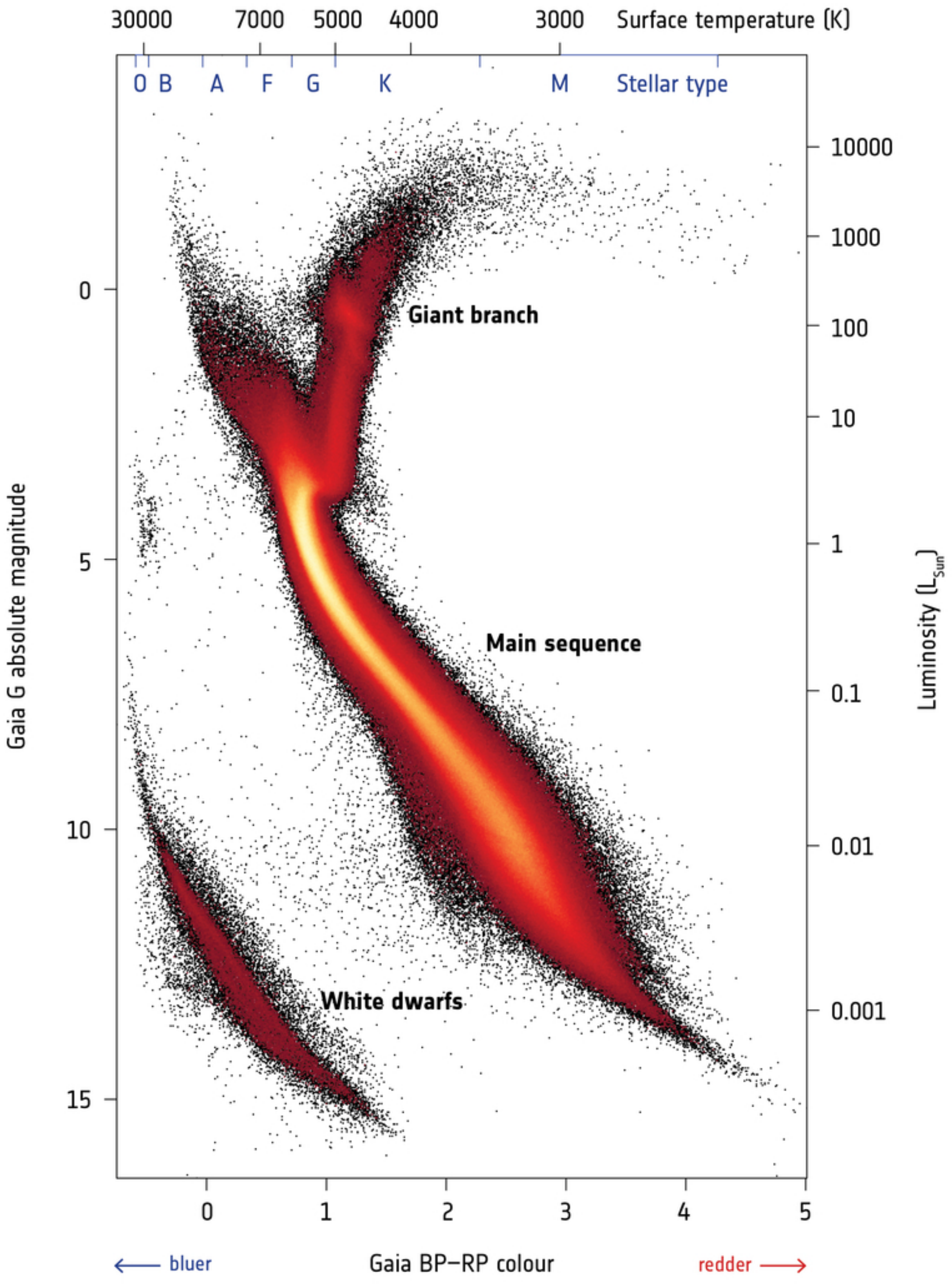

Figure 1.5: Hertzsprung-Russell diagram containing more than four million stars within 1500 parsec from the Sun with *Gaia* DR2 data. Credit: ESA/Gaia/DPAC.

In the case of solar-like stars, a well-studied population, it exists a strong correlation between metallicity and planet occurrence (e.g. Santos et al., 2001). It has been very well studied in the case of gas giants (Gonzalez, 1997; Santos et al., 2000; Laws et al., 2003; Santos et al., 2004; Fischer & Valenti, 2005; Johnson et al., 2010a; Mortier et al., 2012), but remains unclear for super-Earths and Neptune-mass planets (Sousa et al., 2008, 2011; Mayor et al., 2011). Observations suggest that terrestrial planets emerge without any preference on the metal content of the star (Buchhave et al., 2012).

It seems that M dwarfs are less likely to harbour giant, gaseous planets (Endl et al., 2006; Johnson et al., 2007; Mann et al., 2013a; Gan et al., 2022). However, there are counterexamples, such as GJ 3512 ('A giant exoplanet orbiting a very-low-mass star challenges planet formation models', Morales et al., 2019), or the four-planet system in the M4 dwarf IL Aqr, which contains the first jovian planet detected in such a star (Marcy et al., 1998, 2001; Rivera et al., 2005, 2010). Nevertheless, metallicity is a complicated subject in the case of cool stars (see Passegger et al., 2022). Until recent years the systematic study of correlation between fundamental properties and planet occurrence in M dwarfs have been limited due to the small size of confirmed M dwarfs with planets, specially the small-sized. Contrary to FGK stars, there was initially evidence that the planet-hosting M dwarfs had sub-solar metallicities (Bonfils et al., 2005a; Bean et al., 2006), but recent studies point towards the opposite direction (Johnson & Apps, 2009; Rojas-Ayala et al., 2010; Terrien et al., 2012; Hobson et al., 2018). On top of this, the number of low-metallicity red dwarfs does not seem to match the model predictions (Woolf & West, 2012). It has been suggested that the weakened frequency of giant planets found in the red dwarfs can be explained by the lower masses of the host stars, rather than by the effect of metallicity (Johnson & Apps, 2009). Still, the correlation between metallicity and planet occurrence is seen as strong supporting evidence of the core-accretion model of planet formation (Mizuno, 1980; Pollack et al., 1996; Lambrechts & Johansen, 2012). Under this mechanism, planet formation modelling results in a higher occurrence of lower-mass planets (Laughlin et al., 2004; Alibert & Benz, 2017).

The present thesis is all about these common, small, faint, planet-host main sequence stars called M dwarfs. We investigate in detail about 2200 nearby stars of this kind. The sample is the input catalogue of stars for the CARMENES project. The homonymous instrument looks for exo-Earths on a subsample of 350 stars. By meticulously characterising each one of them, we set the foundation for deriving solid and robust global properties for these fascinating objects.

# Chapter 2

## Carmencita

A catalogue should feel like a home for the stars. At home, the stars should be able to find who they are, where is their place on the sky, and how big and luminous they are. And the fingerprint of their spectra, how much *metal* is in their cores. Also, how fast they rotate, how magnetically active they are, how much they apparently shine. Chances are that some of them have a companion that moves along, very close or very wide, or both — they should know that, too. This is what *Carmencita* is about: home for many hundreds of cool stars, each one with so much to tell if one listens carefully. From its very foundation, dozens of parameters have been compiled or measured for every one of them. This has been the effort of many before me, which now embody a respectable amount of the bibliographic content of this work. In this chapter I will describe my own contribution to this catalogue, which includes a meticulous compilation and continuous update, so *Carmencita* feels like home. If I were a red dwarf, I would prefer to be in Carmencita.

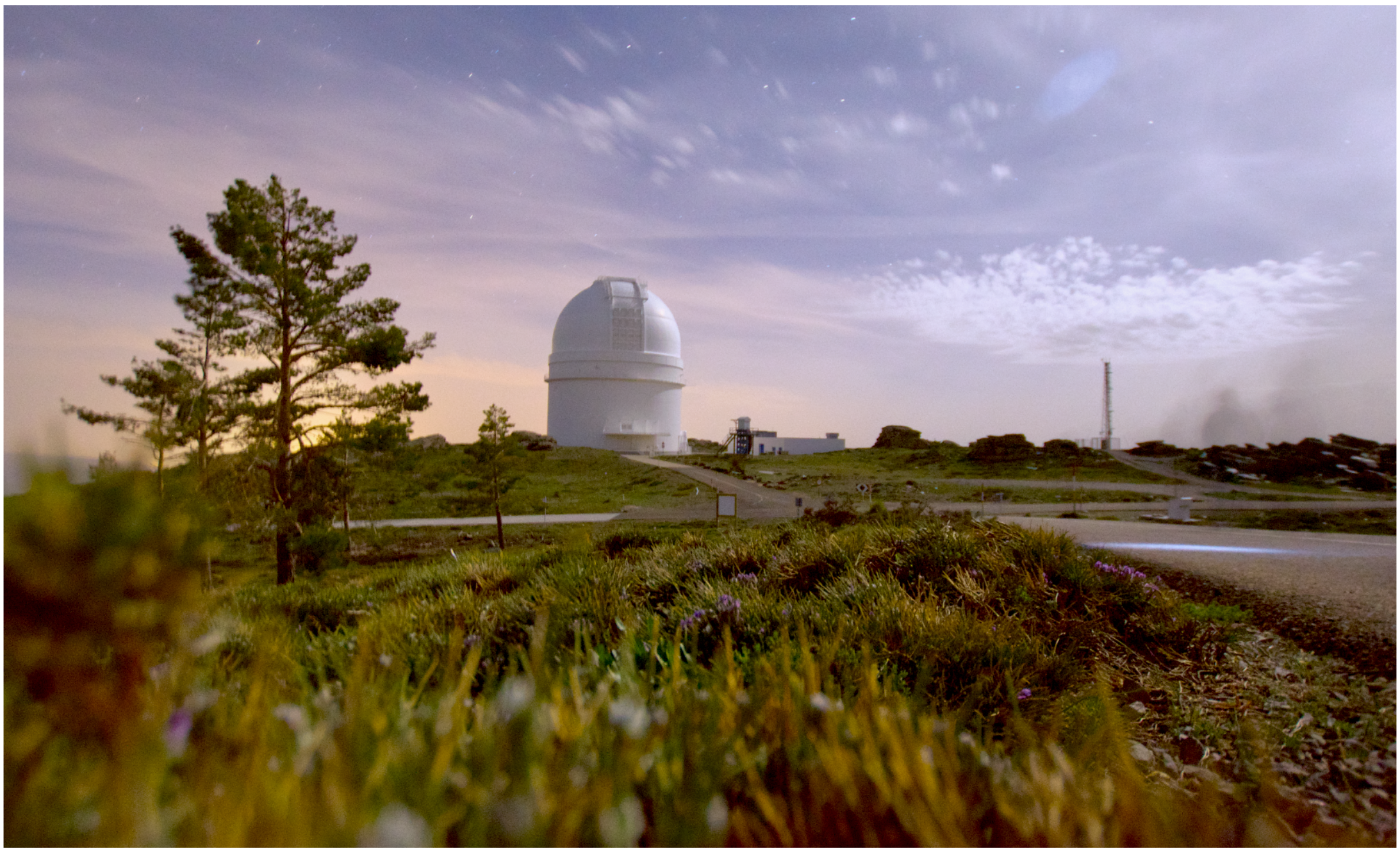

Figure 2.1: Long-exposure moon-washed photography of the dome of the 3.5-m telescope at the Calar Alto Observatory (taken May, 2017).

## 2.1 CARMENES

Calar Alto high-Resolution search for M dwarfs with Exoearths with Near-infrared and optical Echelle Spectrographs (CARMENES) ['kar-men-es] (Caballero et al., 2013b; Quirrenbach et al., 2014) represents an instrument, a science project, and a consortium. Since prospective exercises on exoplanet search identified near-infrared as an area of potential development, CARMENES was designed with enough sensitivity to reach the scientific goal of the project: to detect and characterise terrestrial planets in the habitable zones of late-type main-sequence stars. In particular, the instrument aimed at detecting 2 $\mathcal{M}_\oplus$ planets in the habitable zone of M5 V stars. This translates into a metre-per-second precision in the measurement of radial-velocity variations in extremely faint objects and with long-term stability, which constitutes a technological challenge. CARMENES was designed in such a way that its efficiency is maximum around 1 $\mu$m, which is the emission peak of mid- to late-M dwarfs. The project was conceived as a single large survey targeted at about 300 M-type stars in the solar neighbourhood during guaranteed time observations (GTO) nights at the 3.5-m telescope at the Calar Alto Observatory[1] (Centro Astronómico Hispano en Andalucía (CAHA, Almería, Spain) — see Fig. 2.1 and Fig. 2.2, upper image.

There, the instrument CARMENES is mounted. It consists of a double channel covering two consecutive ranges in the optical (from 520 to 960 nm — VIS) and the near-infrared (from 960 to 1710 nm — NIR) with a very high spatial resolution of R = 80 400 and R = 94 600 in those respective ranges. The spectrum is distributed in 55 and 28 orders for the VIS and NIR channels, respectively (Fig. 2.3). Each channel is an échelle spectrograph, housed in a thermally stabilised vacuum vessel (at a pressure of about $10^{-5}$ mbar) placed in the coudé room of the dome, where exceptional levels of mechanical and thermal stabilisation are ensured. This diminishes the effects that compromise the instrumental stability, mainly pressure variations and possible convection effects. They are mounted on benches and equipped with a temperature stabilisation system that can maintain a temperature constant within $10^{-3}$ K. The VIS spectrograph operates at around room temperature (285 K) with a night stability of about 1 mK, while the NIR spectrograph is cooled down to about 140 K with a night stability of about 3 mK. Each one is simultaneously fed by two octagonal cross-section fibres: one is for calibration, and the other carries the light of the target from the Cassegrain focus of the telescope. Choosing non-circular fibres allows for a higher overall throughput. Using a 100 $\mu$m fibre, their projected size on the sky is 1.5 arcsec[2]. A dichroic beam splitter placed at 960 nm leads the light into these fibres. The calibration is performed with an emission-line lamp and with a Fabry-Pérot etalon, individually for each channel (Th-Ne for VIS, and U-Ne for NIR). The coating of the reflective optics is silver and gold for the VIS and NIR channels, respectively. The detectors are different for each channel: a 4k × 4k pixel CCD for the VIS, and two 2k × 2k pixel CMOS for the NIR. The optical design of the instrument is based on the FEROS design[3] (Kaufer et al., 1997). The interface of the instrument to the telescope or front-end is attached to the Cassegrain focus of the telescope (Fig. 2.2, lower image). With this, CARMENES delivers an exceptional performance in the radial-velocity measurements with an internal precision (or distribution of the formal uncertainties of the RV measurements) of 1.2 m s$^{-1}$ for early and intermediate M-dwarf types, and 5.4 m s$^{-1}$ for the late ones, with the maximum of the distribution (mode) at 0.91 m s$^{-1}$ (Ribas et al., 2023). For more details on the technical side and the performance of the instrument, we also refer to Seifert et al. (2012), Quirrenbach et al. (2014), and Quirrenbach et al. (2016).

[1] CAHA, https://www.caha.es/es/.

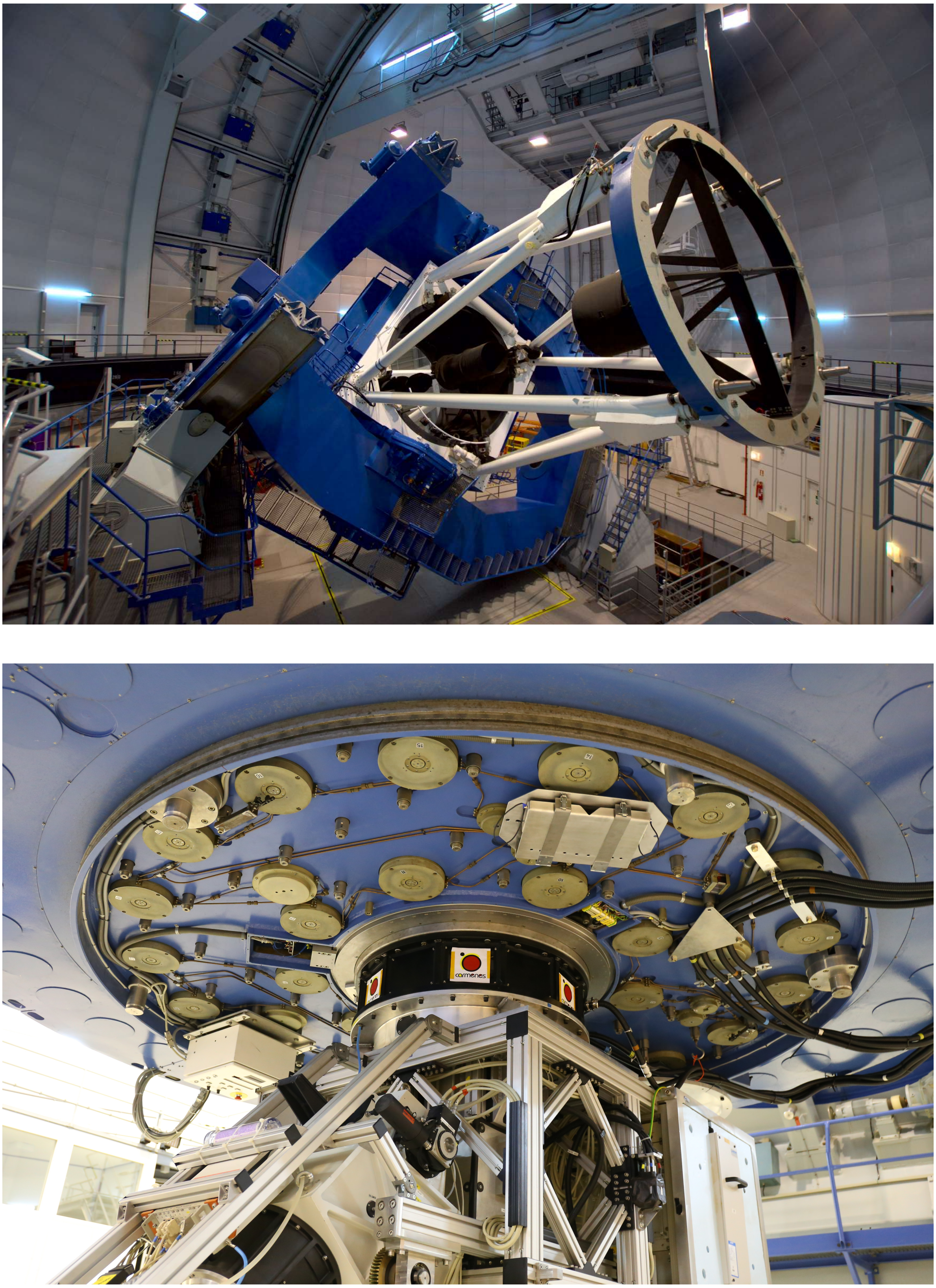

Figure 2.2: Slightly upper view of the 3.5-m telescope at the Calar Alto Observatory with open petals that expose the main mirror (*top*), and a closer view of the Cassegrain focus, which has the attachment of the instrument front-end at the telescope (*bottom*).

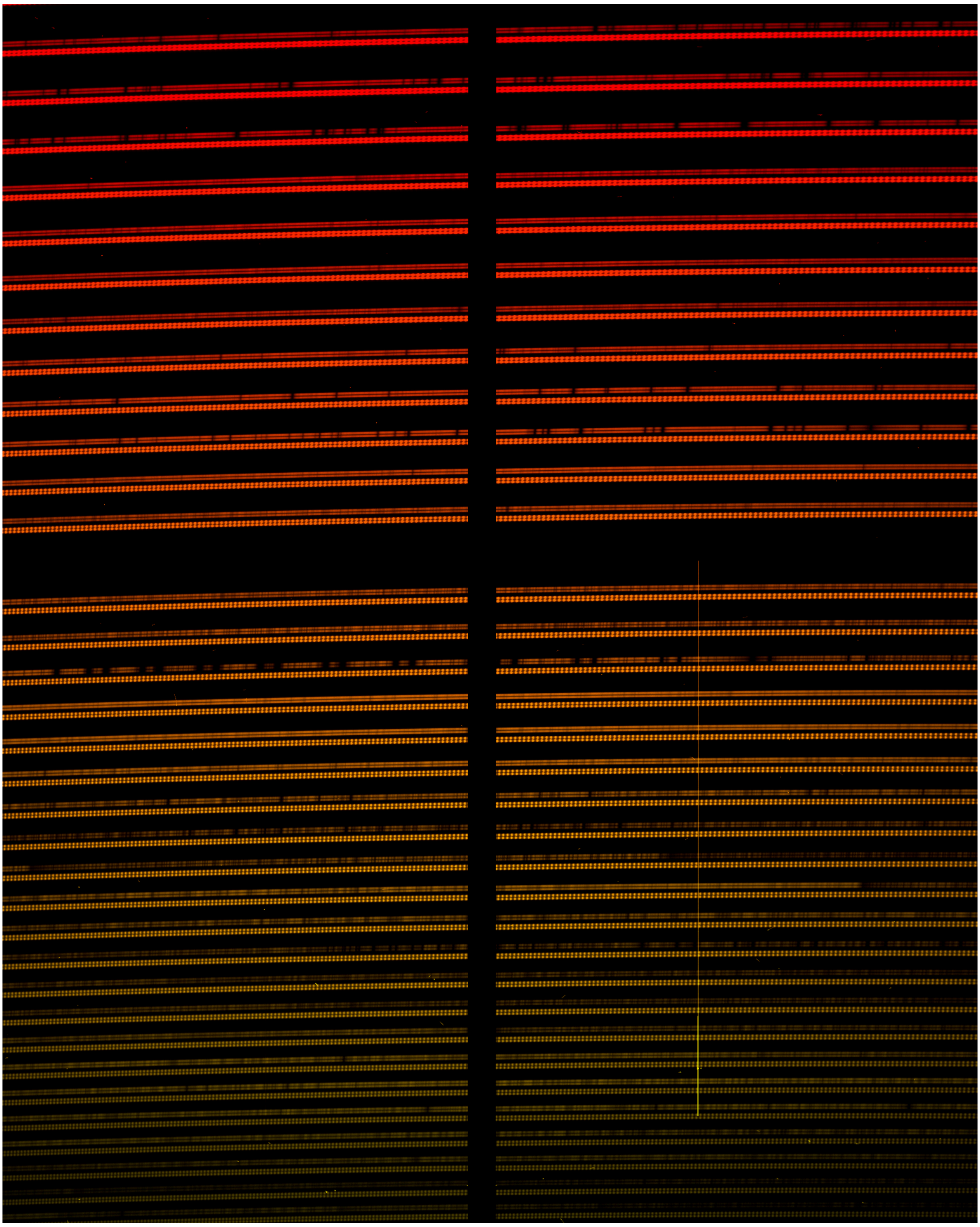

Figure 2.3: Fragment of one of the first CARMENES VIS spectra taken, corresponding to Luyten's Star, along with the Fabry-Pérot for simultaneous calibration. Absorption lines are easily visible.

The consortium was established by more than 200 scientists and engineers from Spanish and German institutions that have contributed to the design, construction and science exploitation of the instrument. The founding members[4] of the CARMENES consortium were chosen on the basis of parity: five in Spain, five in Germany and the formerly Spanish-German Centro Astronómico Hispano en Andalucía (CAHA)[5], in cooperation by the MPG and the CSIC until 2018. The science coordination team includes representatives of these 11 institutions. Even the CARMENES logo was designed to consciously represent a cultural mixture of both Germany and Spain (Fig. 2.4). CARMENES was funded by the MPG, CSIC, Ministerio de Economía y Competitividad (MINECO), and European Regional Development Fund (ERDF), among others. The call for letters of intent for the construction of the next generation instrument for CAHA 3.5-m telescope dates back to March 2008. The first light of the whole instrument was obtained on November 11, 2015, which was prior to the instrument's formal commissioning, on December 14, 2015. The instrument began GTO operations on January 1, 2016, and operated until December 31, 2020. Since January 2021, an upgraded version of the instrument (CARMENES+), which included an enhancement of the cooling system of the NIR channel, carries out 250 additional nights as a legacy project. Subsequent improvements of the instrumentation will incorporate a new wavelength calibration system. While CARMENES has performed radial-velocity *TESS* and K2 (*Kepler*'s extended mission) follow-up (e.g. Palle et al., 2019; Luque et al., 2019a; Nowak et al., 2020; Bluhm et al., 2021; Chaturvedi et al., 2022; Caballero et al., 2022; Radica et al., 2022; Luque et al., 2022a; Trifonov et al., 2023), CARMENES+ will do so for ESA missions *Gaia* and specially, after 2026, PLAnetary Transits and Oscillations of stars (PLATO).

[2]This fact motivates the definition of a boundary between 'close' and 'wide' companions (Chapters 3 and 4), which aims to avoid contamination in the spectra from nearby sources.

[3]This design is a grism cross-dispersed, white-pupil, échelle spectrograph working in quasi-Littrow mode using a two-beam, two-slice, image slicer (Seifert et al., 2012).

[4]Centro de Astrobiología (CAB, CSIC-INTA; Madrid, Spain), Consejo Superior de Investigaciones Científicas (CSIC), Hamburger Sternwarte (HS; Hamburg, Germany), Instituto de Astrofísica de Andalucía (IAA; Granada, Spain), Instituto de Astrofísica de Canarias (IAC; Tenerife, Spain), Institut für Astrophysik Göttingen (IAG; Göttingen, Germany), Institut de Ciències de l'Espai (ICE; Barcelona, Spain), Landessternwarte Königstuhl (LSW; Heidelberg, Germany), Max-Planck-Gesellschaft (MPG), Max-Planck-Institut für Astronomie (MPIA; Heidelberg, Germany), REsearch Consortium On Nearby Stars (RECONS), Thüringer Landessternwarte Tautenburg (TLS; Tautenburg, Germany), and Universidad Complutense de Madrid (UCM; Madrid, Spain).

[5]In 2005, an agreement was signed by the MPIA and IAA to operate jointly the observatory, in a 50-50 proportion, under the denomination Centro Astronómico Hispano-Alemán de Calar Alto. From 2019 to date, the Junta de Andalucía took over the German institution, and joined CSIC to conduct the observatory, making a 100 % Spanish operation, thus acquiring the new denomination.

Figure 2.4: The CARMENES logo is a 'Sol de Miró' (Spain) à la Bauhaus (Germany).

The first data release (DR1) of CARMENES includes all the spectra collected in four years of GTO operations, is soon to be publicly accessible (Ribas et al., 2023). While the main objective of CARMENES is the detection of exoplanets, there is potentially much more to be drawn from high quality spectra. For example, exploiting the diagnostic capabilities of many spectral features: He I infrared triplet (Fuhrmeister et al., 2019b,a; Marfil et al., 2021), K I (Fuhrmeister et al., 2022), Ti and V (Shan et al., 2021), Rb (Abia et al., 2020), or Na, Mg, Si, K, Ca, Ti, V, Cr, Mn, Fe, and Sr (Ishikawa et al., 2022a). These are proven very useful to determine stellar fundamental and atmospheric properties (e.g. Passegger et al., 2018, 2019, 2020, 2022; Schweitzer et al., 2019; Marfil et al., 2021), and also to characterise the stellar activity (e.g. Tal-Or et al., 2018; Fuhrmeister et al., 2018, 2019b; Schöfer et al., 2019; Hintz et al., 2019, 2020; Lafarga et al., 2021), and magnetic fields (Shulyak et al., 2019; Reiners et al., 2022).

## 2.2 Carmencita

When CARMENES saw the first light, the catalogue of M dwarfs to be observed in the guaranteed time was ready. It was baptised 'Carmencita', or the CARMENES Cool dwarf Information and daTa Archive (Caballero et al., 2013a, 2016). To the M dwarfs in Carmencita only two simple criteria were required: to be observable from Calar Alto, Almería (i.e., declination northern of $\delta > -23$ deg), and to be the brightest stars in each spectral subtype, based on the Two Micron All-Sky Survey (2MASS) $J$ magnitude (see Table 1 in Alonso-Floriano et al., 2015b). A variety of sources in the literature provided the known M dwarfs with these characteristics. More than 90 % of the stars in Carmencita come from six works: "The Palomar/MSU nearby star spectroscopic survey" (PMSU, Reid et al., 1995, 2002), "A spectroscopic catalogue of the brightest ($J < 9$) M dwarfs in the northern sky" (Lépine et al., 2013), "G. P. Kuipers spectral classifications of proper-motion stars" (Bidelman, 1985), "An all-sky catalogue of bright M dwarfs" (Lépine & Gaidos, 2011), "Spectral types of M dwarf stars" (Joy & Abt, 1974), and "Spectral classification of high-proper-motion stars" (Lee, 1984).

2MASS is an all-sky coverage survey that provides a photometric solution for 1.6 million sources. The first equatorial coordinates (in the epoch 2000.0 and equinox J2000) and apparent magnitudes (in the $JHK_S$ passbands – centred in 1.24, 1.66, and 2.15 $\mu$m, respectively) were incorporated. The current version of Carmencita (v105) contains 2215 M dwarfs with spectral types from M0.0 V to M9.5 V, and three late-K dwarfs in the boundary K7/M0 V. Among these are the 378 stars intensely monitored by the consortium during GTO (Quirrenbach et al., 2018; Reiners et al., 2018b). The data products derived from the 4-year monitorisation comprise 19 633 spectra for a sample of 362 targets (see Fig. 2.5), and were publicly released on February 2023 as the first data release or DR1 (Ribas et al., 2023).

CARMENES is an extremely sensitive instrument. When looking for planets with the RV method, nearby stellar or substellar companions, physically bound or not, must be extremely well acknowledged, because they may induce real or artificial radial-velocity variations that compromise the measurements. A first classification for suitability was done, avoiding the stars with close ($\rho$ <5 arcsec) companions, either physical or visual, that could also affect the photometric data (Alonso-Floriano et al., 2015b).

GTO dwarfs comprise 362 K7 V to M9 V single stars, which show no evidence of youth or multiplicity. This subsample will be of use throughout this work, especially in the empirical models derived in Chapter 3, because their astrometry and photometry are well-behaved, reliably outlining the main sequence. Alonso-Floriano et al. (2015b) performed a preliminary low-resolution spectral analysis using the Calar Alto Faint Object Spectrograph (CAFOS) at the 2.2-m telescope in Calar Alto, for 753 of the best suited targets.

Carmencita contains equatorial and galactic coordinates, spectral indices, astrophysical parameters, parallactic distances, proper motions, rotational and radial velocities, H$\alpha$ equivalent widths, X-ray count rates and hardness ratios, broadband multi-wavelength photometry from the ultraviolet to the mid-infrared, close and wide multiplicity information, Galactocentric space velocities, and identification in three catalogues (2MASS, *Gaia* DR3, AllWISE). All these parameters are properly referenced and uncertainties are always included, if available. The technical details behind these parameters are omitted here for simplicity, but are included in the descriptions of the tables produced in this work.

Over the years, CARMENES has built Carmencita. Many of the values for different parameters have been measured or calculated by the consortium members. In Table 2.1 we acknowledge this effort by enumerating these contributions. It is important to note that older references, are superseded by newer references, if these achieve more precise determinations based on solid approaches. This is the main potential of a dynamical catalogue like Carmencita: to be able to provide up-to-date, homogeneous, and accessible information for all the members of the consortium. The mentioned table also includes in **boldface** the contributions derived during the process of this thesis[6]. If Carmencita was a house, then (with 2218 rows and 177 columns) it is certainly made of 392 586 individual bricks. A major value of this catalogue is the individual handling that every of these bricks receives, avoiding automatic-only procedures when possible and aiming for a quality-over-quantity approach. When looked as a unity, Carmencita is a statistically robust sample of well-characterised M dwarfs in the neighbourhood of our Sun, and so global properties can be safely withdrawn from its study as a whole.

[6]Additionally, since late 2020 (version 98), I (C. Cifuentes) am in charge of the update and maintenance of Carmencita, the CARMENES input catalogue.

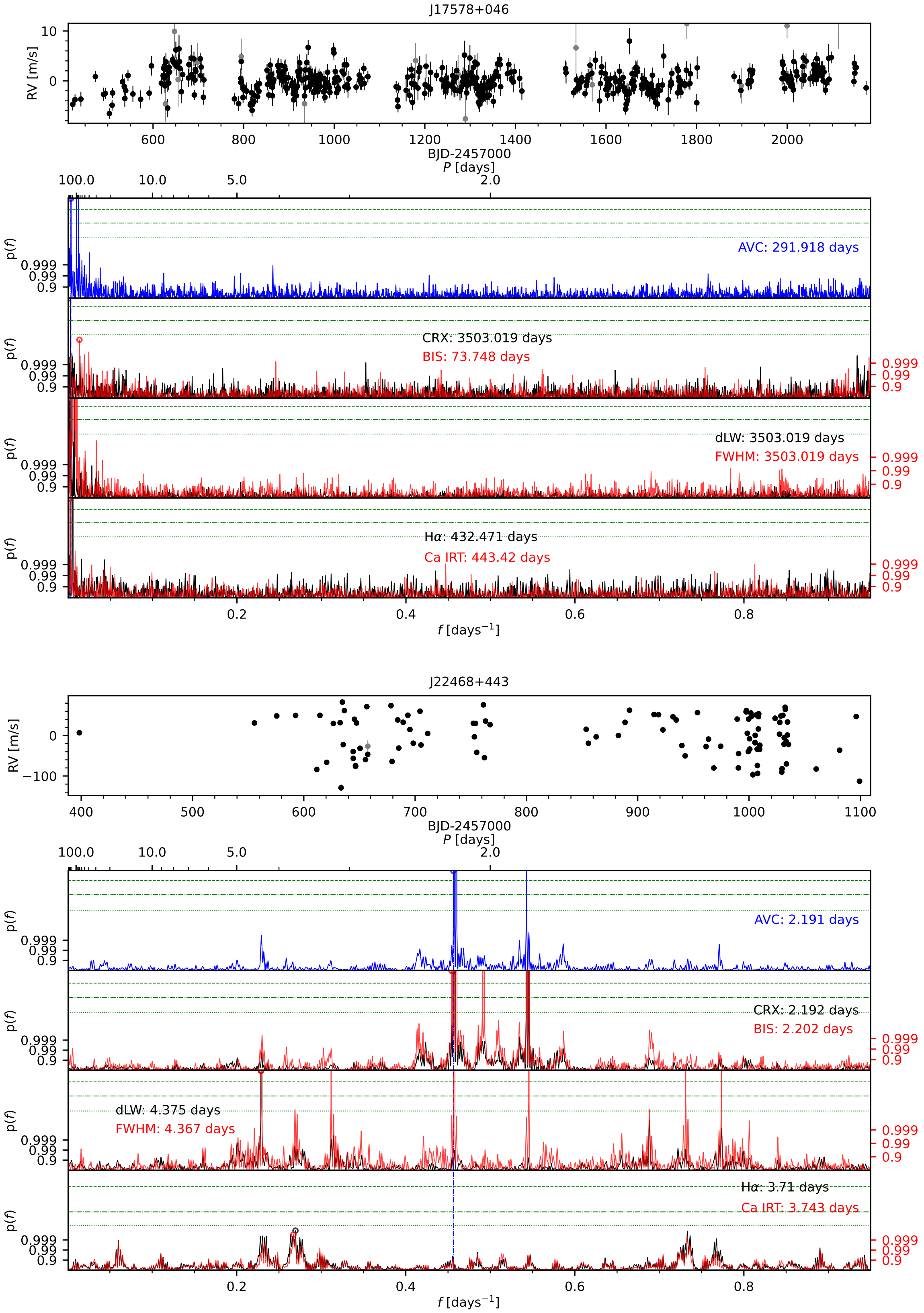

Figure 2.5: Periodograms of the quiet Barnard's Star (J17578+046, *top*) and the highly active EV Lac (J22468+443, *bottom*) obtained from the first data release (DR1) of CARMENES.

Table 2.1: Contribution to the astrophysical parameters of the stars in Carmencita from published (or close to be published) referenced works of the consortium members.

| Parameter[a] | Description | Source(s)[b] |
|---|---|---|
| `SpT` | Spectral type | Alo15a |
| `Teff_K, logg, [Fe/H]` | Effective temperature, surface gravity, iron abundance | Marf21, Pas18,19,20, Cif20, Schw19 |
| `L_Lsol, M_Msol, R_Rsol` | Bolometric luminosity, stellar mass and radius | Cif20*, Schw19, This work (**Chapter 3**) |
| `d_pc` | Heliocentric distance (photometric) | Cor17b, Cif20*, This work (**Chapter 3**) |
| `Vr_kms-1` | Radial velocity | Jef18, Laf20, Bar18,21 |
| `U_kms-1, V_kms-1, W_kms-1,` | Galactocentric velocities | Cor23 (see Section 2.4) |
| `SKG` | Stellar kinematic group | Alo15b, Sha23, this work (see Section 2.4) |
| `P_d` | Rotational period | Die19 |
| `pEWHalpha_A` | $H\alpha$ pseudo-equivalent width | Alo15a, Fuh23, Scho19 |
| `Activity` | Flaring/chromospheric activity indicators | Tal18 |
| `Photometry` | Apparent magnitudes in up to 20 passbands | This work (**Chapter 3**) |
| `Multiplicity` | Multiplicity characteristics | Bar18,21, Jeff18, Cor17a, This work (**Chapter 4**) |

[a] The uncertainties are included in all cases.

[b] Alo15a: Alonso-Floriano et al. (2015b); Alo15b: Alonso-Floriano et al. (2015a); Bar18: Baroch et al. (2018); Bar21: Baroch et al. (2021); Cif20: Cifuentes et al. (2020); Cor17a: Cortés-Contreras et al. (2017a); Cor17b: Cortés-Contreras (2017); Die19: Díez Alonso et al. (2019); Fuh20: Fuhrmeister et al. (2020); Jef18: Jeffers et al. (2018); Laf20: Lafarga et al. (2020); Marf21: Marfil et al. (2021); Pas18: Passegger et al. (2018); Pas19: Passegger et al. (2019); Pas20: Passegger et al. (2020); Scho19: Schöfer et al. (2019); Schw19: Schweitzer et al. (2019); Sha23: Shan et al., in prep. Tal18: Tal-Or et al. (2018). (*): from this reference or computed using the results in it.

## 2.3 Astrometry

*Gaia* has been a game changer, and this is specially true when it comes to astrometric characterisation. For instance, the trigonometric distances to the stars in Carmencita were originally adopted from the General Catalogue of Trigonometric Stellar Parallaxes (van Altena et al., 1995), the new Hipparcos astrometric catalogue (van Leeuwen, 2007), and the MEarth survey (Dittmann et al., 2014). These sources provided data for around 60 % of the stars, while for the remaining 40 %, the use of photometric distances was a must, even though they suffered from a much lower precision than the geometrical measurements from parallaxes. In the case of proper motions, 96 % of the original values in Carmencita came from three catalogues: Lépine and Shara Northern Stars Proper Motion (LSPM-North Lépine & Shara, 2005), the USNO-B1.0 and 2MASS combination (PPMXL; Roeser et al., 2010), and *Hipparcos* (van Leeuwen, 2007). In the present version of Carmencita, *Gaia* DR3 (or DR2) provide proper motions and parallaxes for ~93 % of the stars. For a few stars, parallactic distances are not published, and may be necessary to adopt photometric estimations. It is important to determine potential unresolved binarity, because in those cases the photometric distance would be underestimated.

By representing a colour as a function of the absolute magnitude in one of the two passbands, it is possible to determine a photometric distance. In this work (see Chapter 3, Sect. 3.2.3) we determine a $M_J$ vs. $r - J$ model, using only well-behaved, single stars from GTO, that can serve to approximate the distances photometrically. Using the definition of absolute magnitude in the $r$ passband, $M_r = r - 5\log d_{\rm phot} + 5$, a photometric distance is derived as $d_{\rm phot} = 10^{\frac{5+r-M_r}{5}}$, where $M_r = M_r(r, J)$ is our model derived from polynomial fitting using stars with actual parallactic distances. This is a convenient choice of magnitudes, since they are almost always available for many relatively bright systems.

Table 2.2: Characteristic velocity dispersions in the thin disk, thick disk, and stellar halo, and asymmetric drift (extracted from Bensby et al. 2003).

| | $\chi_{ns}$ | $\sigma_u$ [km s$^{-1}$] | $\sigma_v$ [km s$^{-1}$] | $\sigma_w$ [km s$^{-1}$] | $V_{\rm asym}$ [km s$^{-1}$] |
|---|---|---|---|---|---|
| Thin Disk (D) | 0.94 | 35 | 20 | 16 | –15 |
| Thick Disk (TD) | 0.06 | 67 | 38 | 35 | –46 |
| Halo (H) | 0.0015 | 160 | 90 | 90 | –220 |

## 2.4 Kinematics

Some Carmencita stars belong to moving groups and associations, and some of them considered young. In broad terms, and for an easier first classification, we broadly classified as 'young' those system with an age of less than 1 Ga. These are hereafter denominated stellar kinematic group (SKG), and the column `SKG` in Carmencita contains information about them. In this work we perform a search in a number catalogues in order to update and complete this column, assigning a confirmed or probable membership to any or some kinematic groups. The relation of these SKGs found and the sources for their assignation can be found in Chapter 3 (Sect. 3.2.2). On top of this search, we performed a complementary assignation to SKGs and stellar populations using the `SteParKin` code (see Appendix E). It evaluates the membership of stars to young (< 1 Ga) kinematic moving groups and associations and assigns their stellar populations as proposed by Bensby et al. (2003, 2005), based on their positions and their Galactocentric velocities ($U$, $V$, $W$). These can be derived from the equatorial coordinates ($\alpha$, $\delta$), parallaxes ($\varpi$), proper motions ($\mu_\alpha \cos\delta$, $\mu_\alpha \cos\delta$), and radial velocities ($V_r$), with their corresponding uncertainties:

$$\begin{bmatrix} U \\ V \\ W \end{bmatrix} = \tilde{B} \begin{bmatrix} V_r \\ \frac{k\mu_\alpha \cos\delta}{\varpi} \\ \frac{k\mu_\delta}{\varpi} \end{bmatrix}, \tag{2.1}$$

where $k = 4.74057$ km s$^{-1}$ and $\tilde{B}$ is a 3×3 matrix, result of the converting the equatorial coordinates to the Galactic system, following the approach by Soderblom & Clements (1987). With this, the assignation of a star to a given moving group or association requires that their Galactocentric velocities are within the 3D ellipsoid that defines it.

The stellar population, thin disk (D), thick disk (TD), thin-thick transition disk (TD-D), or halo (H), is also assigned in this step. As Bensby et al. (2003) noted, there is no obvious predetermined way to define a sample of purely thick disk stars in the solar neighbourhood. From the two main methods of finding local thick or thin disk stars, there is the pure kinematical approach, or the combination of kinematics, metallicities, and stellar ages (e.g. Fuhrmann, 1998). The authors adopt the first approach: a population is roughly defined by the observed fraction in the solar neighbourhood ($\chi_{ns}$), the characteristic velocity dispersions ($\sigma_u$, $\sigma_v$, $\sigma_w$), and the asymmetric drift ($v_{\rm asym}$), assuming that the Galactic space velocities ($U$, $V$, $W$) of the stars in these populations have Gaussian distribution:

$$f(U, V, W) = k \exp\left(-\frac{U_{\rm LSR}^2}{2\sigma_U^2} - \frac{(V_{\rm LSR} - V_{\rm asym})^2}{2\sigma_V^2} - \frac{W_{\rm LSR}^2}{2\sigma_W^2}\right), \tag{2.2}$$

where LSR stands for the local standard of rest, and $k$ is a normalisation parameter, defined as:

$$k = \frac{1}{(2\pi)^{3/2}\sigma_U\sigma_V\sigma_W}. \tag{2.3}$$

The values that the authors adopt for the three populations are given in Table 2.2. One caveat worth mentioning in the `SteParKin` approach is the rather optimistic definition of the $UVW$ space of the stellar associations and moving groups as compared to other codes available in the literature. It also does not resolve by itself the overlapping between assignations, but it is labeled as member for both. The controversies found in the overlapping cases were resolved in all cases by analysing individually the consensus in the literature. For those stars with groups already assigned in any of the catalogues, there is coincidence or equivalence in the assignation. For this reason, `SteParKin` has been used as a *complement* to the literature agreement.

Two additional codes of recognition in this topic, extensively used for the probabilistic assignation to kinematic groups are `LACEwING`[7] (Riedel et al., 2017) and `BANYAN` Σ[8] (Gagné et al., 2018b). The assignations given by these tools are incorporated implicitly in the compilation of the catalogues provided by the respective authors (see Table 4.9 in Chapter 4, Sect. 4.4.5) .

The results are showcased in two figures produced by `SteParkin`: a Θ-M (or Toomre) and a Böttlinger diagrams (Figs. 2.6 and 2.7, repectively), using the derived galactocentric velocities. Only one star, namely LP 651-007 (Karmn J02462-049) is found to belong to the Galactic halo.

## 2.5 Fundamental astrophysical parameters

Chapter 3 is devoted to the derivation of stellar parameters and the details. Stellar parameters include bolometric luminosities ($\mathcal{L}$), masses ($\mathcal{M}$), and radii ($\mathcal{R}$). Luminosities are essentially empirically derived, because the intervention of models in the final values is greatly minimised, in favour of multi-wavelength, broadband photometry, and trigonometric distances. Radii are directly derived from the luminosities, given that the effective temperature is known, assuming a black-body approximation for the emitting behaviour of the star. The Stefan-Boltzmann law relates the power radiated from a blackbody with its (effective) temperature, $T_{\rm eff}$. In words, the total energy radiated by a blackbody per unit surface area across all wavelengths per unit time is directly proportional to the fourth power of the blackbody's thermodynamic temperature. In terms of the area of the blackbody of radius $\mathcal{R}$, it becomes: $\mathcal{L} = 4\pi\mathcal{R}^2\sigma T_{\rm eff}^4$, where $\sigma$ is the Stefan-Boltzmann constant. A full derivation process can be found in, e.g., Rybicki & Lightman (1979).

Masses are a parameter of great significance, because they convey information about the past formation, and the future evolution of the star. Luckily, many roads lead to stellar masses, as we explain in Schweitzer et al. (2019) (see Sect. 3.4.1). The effective temperature ($T_{\rm eff}$), surface gravity ($g$), and metallicity (referred as the iron abundance, [Fe/H]), are stellar characteristics that are imprint in the spectra of the stars. The cooler the star, the more complex it becomes to reproduce its spectrum.

The luminosity of a star varies very little for the most part of its lifetime, except for the very beginning and the end. Once in the main sequence, one equation dominates, implying that the more massive a star, the more luminous it is. For instance, in Schweitzer et al. (2019) we found that the relation $\mathcal{L} \propto \mathcal{M}^{2.22\pm0.02}$ holds true for $0.1\,\mathcal{M}_\odot < \mathcal{M} < 0.5\,\mathcal{M}_\odot$ stars, or M1–M8 V. In other words, the main sequence is a sequence of masses. Before settling in this sequence, stars appear more luminous when contracting. This effect is not exclusive of pre-main sequence stars: unresolved binaries are disguised as single, but the measured flux (or apparent magnitude) is actually the contribution of two (in some cases even more) stars, as we investigate in Chapter 4. This makes luminosities, and the derived masses, to be artificially larger than they should be. It must be noted that not only the overluminous stars may be problematic during the mass determination. For instance, it is possible that young stars with spectral types later than ~M4-M5 do not present overluminosity, but the mass determinations derived from main sequence stars may not be not valid. Fortunately, these are very few and can be handled individually[9]. For this reason we did not make any preliminary assumption on the youth of the stars for the calculation of luminosities (see Chapter 3), but addressed each case separately. In this way, the existence of stars that were off the main sequence track in our Hertzsprung-Russell (HR) diagrams came by surprise and hinted towards possible multiplicity that had not been resolved yet.

[7] https://github.com/ariedel/lacewing.

[8] https://github.com/jgagneastro/banyan_sigma.

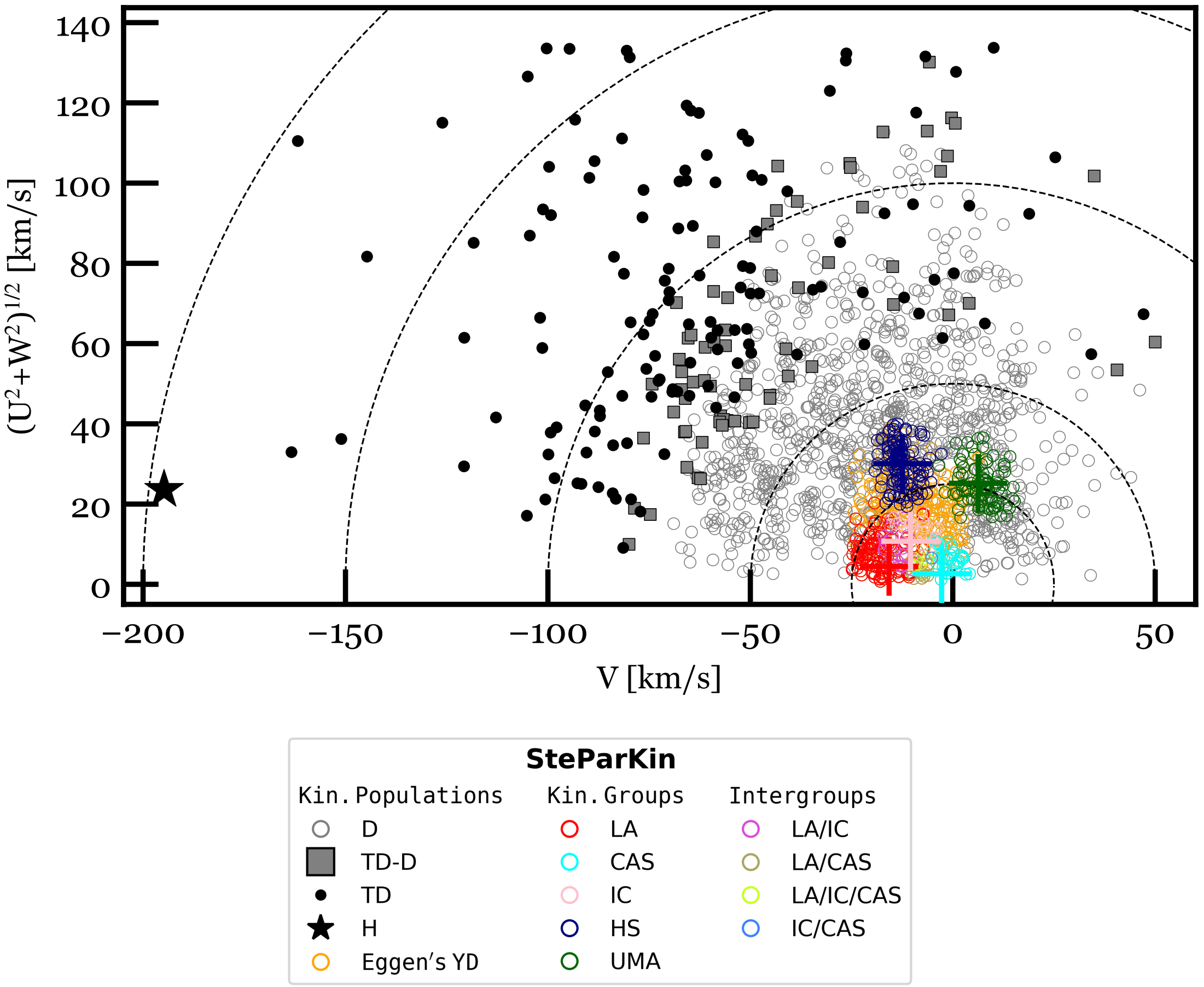

Figure 2.6: Θ-M or Toomre diagram for the Galactocentric velocities of the Carmencita stars.

For the most extreme cases of overluminous stars almost all of them were reported to belong to a young moving groups (YMG). For those, we first compared the membership in the literature with ours, and assigned an age interval and a median value to each of those SKGs. From the several possibilities for modelling the masses of these substellar objects based on their assigned age[10], we preferred the results obtained with the PAdova and TRieste Stellar Evolution Code (PARSEC; Bressan et al., 2012, and see Nguyen et al. 2022 for the V2.0)[11]. Among the possible routes to obtain stellar masses (colour-absolute magnitude diagrams, eclipsing binaries in young open clusters, or luminosity-mass relations), we chose the relation with luminosities, as we found it to be the most consistent one, and also because it benefited from our very precise determination of them (see Chapter 3). In this line, Cardona Guillén et al. (2021)

[9]One notable example is LP 944-20, a field (6.43 pc) ultracool dwarf at the brown dwarf-stellar boundary M9.5 V, which is member of the Castor moving group (see Ribas, 2003b). In Cifuentes et al. (2020) we derived luminosity, radius (from VOSA's effective temperature) and mass that is in very good agreement with its spectral classification ($T_{\rm eff}$ = 2400 K; $\mathcal{L}$ = 2.264 ± 0.022 $10^{-4}\,\mathcal{L}_\odot$; $\mathcal{M}$ = 0.0760 ± 0.0089 $\mathcal{M}_\odot$).

[10]BHAC15 (Baraffe et al., 2015), Mesa Isochrones and Stellar Tracks (MIST; Paxton et al., 2011; Choi et al., 2016), Stellar Parameters of Tracks with Starspots (SPOTS; Somers et al., 2020), PISA (Dell'Omodarme et al., 2012), FRANEC (from the Pisa and

Table 2.3: Overluminous stars identified in HR diagram for which fundamental parameters have been derived using PARSEC isochrone models.

| Karmn | Name | SKG[a] | $\mathcal{L}$ $[10^{-3}\mathcal{L}_\odot]$ | $\mathcal{M}$ $[\mathcal{M}_\odot]$ | $\mathcal{R}$ $[\mathcal{R}_\odot]$ |
|---|---|---|---|---|---|
| J01352–072[†] | Barta 161 12 | $\beta$ Pic | 47.75 ± 0.37 | $0.376^{+0.017}_{-0.021}$ | $0.718^{+0.018}_{-0.013}$ |
| J02088+494 | G 173-039 | AB Dor | 17.82 ± 0.08 | $0.370^{+0.078}_{-0.022}$ | $0.417^{+0.045}_{-0.018}$ |
| J02519+224 | RBS 365 | $\beta$ Pic | 26.89 ± 0.17 | $0.254^{+0.014}_{-0.011}$ | $0.611^{+0.016}_{-0.012}$ |
| J03473-019 | G 80-021 | AB Dor | 31.22 ± 0.22 | $0.469^{+0.052}_{-0.005}$ | $0.494^{+0.055}_{-0.021}$ |
| J04472+206 | RX J0447.2+2038 | IC 2391 | 12.51 ± 0.08 | $0.208^{+0.036}_{-0.031}$ | $0.430^{+0.038}_{-0.024}$ |
| J05019+011 | 1RXJ050156.7+010845 | $\beta$ Pic | 33.83 ± 0.20 | $0.297^{+0.017}_{-0.013}$ | $0.652^{+0.017}_{-0.012}$ |
| J05062+046 | RX J0506.2+0439 | $\beta$ Pic | 28.24 ± 0.19 | $0.262^{+0.014}_{-0.011}$ | $0.619^{+0.017}_{-0.012}$ |
| J05084–210 | 2MJ05082729-2101444 | $\beta$ Pic | 38.99 ± 0.40 | $0.327^{+0.019}_{-0.015}$ | $0.678^{+0.019}_{-0.014}$ |
| J06318+414 | LP 205-044 | Hya | 13.38 ± 0.06 | $0.347^{+0.205}_{-0.017}$ | $0.364^{+0.165}_{-0.010}$ |
| J06574+740 | 2MJ06572616+7405265 | Cas | 7.03 ± 0.63 | $0.269^{+0.009}_{-0.005}$ | $0.284^{+0.017}_{-0.015}$ |
| J07319+362N | BL Lyn | Cas | 18.64 ± 0.08 | $0.413^{+0.007}_{-0.004}$ | $0.397^{+0.004}_{-0.001}$ |
| J07446+035 | YZ CMi[b] | IC 2391 | 11.16 ± 0.09 | $0.192^{+0.033}_{-0.028}$ | $0.342^{+0.161}_{-0.010}$ |
| J09133+688 | G 234-057 | AB Dor | 71.91 ± 1.32 | $0.576^{+0.025}_{-0.023}$ | $0.587^{+0.097}_{-0.024}$ |
| J09449–123 | G 161-071[b] | Arg | 8.97 ± 0.05 | $0.167^{+0.029}_{-0.024}$ | $0.318^{+0.153}_{-0.010}$ |
| J10196+198 | AD Leo | Cas | 23.59 ± 0.11 | $0.446^{+0.006}_{-0.004}$ | $0.429^{+0.002}_{-0.000}$ |
| J11201–104 | LP 733-099 | Cas | 41.65 ± 0.32 | $0.527^{+0.006}_{-0.001}$ | $0.507^{+0.001}_{-0.001}$ |
| J11474+667 | 1RXJ114728.8+664405 | Cas | 8.06 ± 0.08 | $0.289^{+0.011}_{-0.004}$ | $0.297^{+0.006}_{-0.002}$ |
| J12156+526 | StKM 2-809 | UMa | 33.02 ± 0.43 | $0.494^{+0.007}_{-0.002}$ | $0.476^{+0.001}_{-0.001}$ |
| J15218+209 | OT Ser | LA | 45.75 ± 0.22 | $0.525^{+0.200}_{-0.016}$ | $0.519^{+0.223}_{-0.001}$ |
| J16102–193 | K2-33 | USco | 102.62 ± 1.09 | $0.514^{+0.066}_{-0.052}$ | $0.942^{+0.052}_{-0.039}$ |
| J17338+169 | 1RXJ173353.5+165515 | LA | 9.54 ± 0.20 | $0.292^{+0.179}_{-0.022}$ | $0.325^{+0.159}_{-0.013}$ |
| J18174+483 | TYC 3529-1437-1 | LA | 44.88 ± 0.25 | $0.522^{+0.201}_{-0.016}$ | $0.518^{+0.221}_{-0.002}$ |
| J20451–313 | AU Mic | $\beta$ Pic | 98.75 ± 0.86 | $0.606^{+0.025}_{-0.016}$ | $0.862^{+0.022}_{-0.018}$ |
| J22518+317 | GT Peg | IC 2391 | 26.39 ± 0.13 | $0.346^{+0.063}_{-0.044}$ | $0.537^{+0.043}_{-0.026}$ |

[a] AB Dor: AB Doradus, 100 ± 50 Ma (Luhman et al., 2005; Bell et al., 2015; Rodríguez et al., 2018); $\beta$ Pic: $\beta$ Pictoris, $18.5^{+2.0}_{-2.4}$ Ma (Miret-Roig et al., 2020a); Cas: Castor, 300 ± 100 Ma (Ribas, 2003b); IC 2391 SC: Omicron Velorum Supercluster, 40 ± 15 Ma (Barrado y Navascués et al., 1999, 2004); LA: Local Association, $150^{+150}_{-136}$ Ma (López-Santiago et al., 2006); UMa: Ursa Majoris, 300 ± 100 Ma (King et al., 2003); USco: Upper Scorpius, 11 ± 3 Ma (Pecaut et al., 2012).

[b] YZ Cmi and G 161-071 ages were further reviewed, as they appeared as mass outliers in a first estimation. For YZ CMi (Alonso-Floriano et al., 2015a), flagged as a $\beta$ Pictoris possible member that need confirmation, posterior investigations do not mention again this membership (e.g. Loyd et al., 2018). G 161-071 is assigned to Argus/IC 2391 SC based on findings in the literature (e.g. Bell et al., 2015).
(†) Barta 161 12 was identified as a double-lined spectroscopic binary (SB2) by Malo et al. (2014a), but the detection was a false positive.

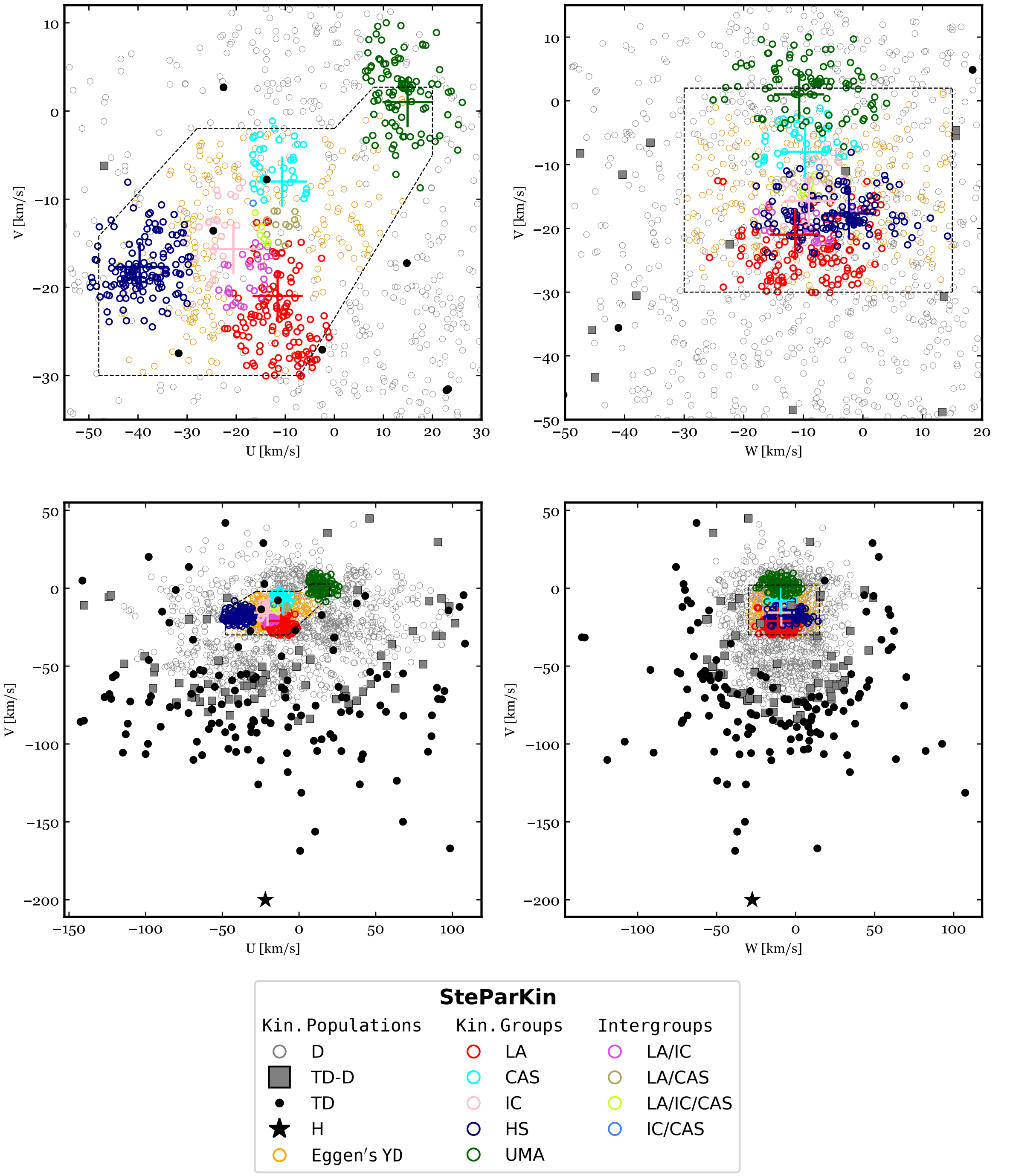

Figure 2.7: Böttlinger diagram for the Galactocentric velocities of the Carmencita stars, representing the $(U, V)$ and $(V, W)$ Galactocentric planes.

Naples groups, Degl'Innocenti et al., 2008), or the Bag of Stellar Tracks and Isochrones (BaSTI; Pietrinferni et al., 2004; Hidalgo et al., 2018), to name a few.

[11] http://stev.oapd.inaf.it/cgi-bin/cmd.

Table 2.4: Late-type stars for which fundamental parameters have been derived using DUSTY00 isochrone models.

| Karmn | Name | Spectral type | SpT ref[a] | $\mathcal{L}$ [$10^{-4}\mathcal{L}_\odot$] | $T_{\rm eff}$[b] [K] | $\mathcal{M}$ [$\mathcal{M}_\odot$] | $\mathcal{R}$ [$\mathcal{R}_\odot$] |
|---|---|---|---|---|---|---|---|
| J02465+164 | LP 411-006 | M6.0 V | PMSU | 9.42 ± 0.05 | 2900 ± 50(1) | 0.103 ± 0.010 | 0.127 ± 0.013 |
| J02530+168 | Teegarden's Star | M7.0 V | Alo15a | 7.22 ± 0.05 | 3034 ± 45(2) | 0.097 ± 0.010 | 0.120 ± 0.012 |
| J03142+286 | LP 299-036 | M6.0 V | PMSU | 10.68 ± 0.06 | 2900 ± 50(1) | 0.113 ± 0.009 | 0.129 ± 0.004 |
| J04198+425 | LSR J0419+4233 | M8.5 V | Lep03 | 4.90 ± 0.03 | 2400 ± 38(1) | 0.089 ± 0.009 | 0.112 ± 0.011 |
| J05394+406 | LSR J0539+4038 | M8.0 V | Lep03 | 5.68 ± 0.03 | 2600 ± 50(1) | 0.092 ± 0.009 | 0.115 ± 0.011 |
| J07403-174 | LP 783-002 | M6.0 V | PMSU | 8.39 ± 0.04 | 3031 ± 26(2) | 0.100 ± 0.010 | 0.124 ± 0.012 |
| J08298+267 | DX Cnc | M6.5 V | Alo15a | 8.36 ± 0.03 | 2997 ± 49(2) | 0.100 ± 0.010 | 0.124 ± 0.012 |
| J08536-034 | LP 666-009 | M9.0 V | Jen09 | 2.81 ± 0.02 | 2400 ± 38(1) | 0.081 ± 0.008 | 0.102 ± 0.010 |
| J09003+218 | LP 368-128 | M6.5 V | Alo15a | 7.94 ± 0.04 | 3061 ± 43(2) | 0.099 ± 0.010 | 0.123 ± 0.012 |
| J09033+056 | NLTT 20861 | M7.0 V | New14 | 18.08 ± 11.03 | 3027 ± 19(2) | 0.139 ± 0.051 | 0.155 ± 0.047 |
| J10482-113 | LP 731-058 | M6.5 V | Alo15a | 6.75 ± 0.04 | 3029 ± 25(2) | 0.095 ± 0.010 | 0.119 ± 0.012 |
| J10564+070 | CN Leo | M6.0 V | Alo15a | 10.12 ± 0.07 | 3071 ± 24(2) | 0.095 ± 0.008 | 0.112 ± 0.002 |
| J14321+081 | LP 560-035 | M6.0 V | New14 | 18.07 ± 0.10 | 3161 ± 86(2) | 0.126 ± 0.011 | 0.142 ± 0.008 |
| J16555-083 | vB 8 | M7.0 V | Alo15a | 5.86 ± 0.03 | 3005 ± 21(2) | 0.092 ± 0.009 | 0.115 ± 0.012 |
| J18356+329 | LSR J1835+3259 | M8.5 V | Schm07 | 2.62 ± 0.01 | 2500 ± 50(1) | 0.081 ± 0.008 | 0.101 ± 0.010 |
| J19169+051S | V1298 Aql | M8.0 V | Alo15a | 4.33 ± 0.03 | 2600 ± 50(1) | 0.087 ± 0.009 | 0.109 ± 0.011 |
| J19255+096 | LSPM J1925+0938 | M8.0 V | New14 | 10.77 ± 0.12 | 2500 ± 50(1) | 0.161 ± 0.011 | 0.175 ± 0.007 |
| J23064-050 | 2MUCD 12171 | M8.0 V | Schm07 | 4.95 ± 0.03 | 2600 ± 50(1) | 0.089 ± 0.009 | 0.112 ± 0.011 |

[a] Alo15a: Alonso-Floriano et al. (2015b); Jen09: Jenkins et al. (2009); Lep03: Lépine et al. (2003); New14: Newton et al. (2014); PMSU: Reid et al. (1995); Schm07: Schmidt et al. (2007).
[b] (1): Obtained from DUSTY00 model fitting; (2) Obtained from spectral synthesis by Marfil et al. (2021).

also suggest PARSEC as the most appropriate grid of synthetic isochrones for mass estimation from luminosities. Therefore, we used our homogeneously derived bolometric luminosities using *Gaia*'s third data release (DR3) latest astrometry and photometry. First, we assigned a kinematic membership to a young moving group for each star (see again Sect. 2.4) and obtain minimum and maximum ages from the literature (see footnote of Table ). This translates, however indirectly, into lower and upper limit of the mass. For each corresponding age, in a $\mathcal{L}$-$\mathcal{M}$ relationship from the PARSEC synthetic models, we interpolated the values of our luminosities to obtain the corresponding masses using a quadratic polynomial least-squares fit. It must be noted that, the determination of masses from these models is only as good as the models are (this 'theoretical gap' applies to all areas of astrophysics). The masses should not be interpreted as the real masses, but the ones *expected* for a given age. The uncertainty would not be limited by our knowledge (or ignorance) about the age of the SKG, but also should incorporate the uncertainty on the assignation to the kinematic group in question, based on their kinematics. Of course, luminosities do contribute with their own uncertainties. Nevertheless, *smaller* uncertainties (magnitudes, parallaxes, effective temperature, etc.) are massively eclipsed by the uncertainty in age.

In Table 2.5 we show the overluminous stars found, for which we have estimated masses from the evolutionary models, while radii are computed using these and the Stefan-Boltzmann relation, with effective temperatures coming from DUSTY model fitting or, preferentially if available, from spectral synthesis by Marfil et al. (2021)[12]. We tabulate their luminosities, the derived masses using PARSEC or DUSTY00 isochrones, the adopted SKG, along with the age range and the reference. While there still are bona fide stars in young SKGs in Carmencita that are *not* overluminous, we do not considered the necessity to redetermine masses for them, as the $\mathcal{M}$-$\mathcal{R}$ relation from Schweitzer et al. (2019) applies nevertheless.

With an analogous procedure, we derived masses for the late-type GTO stars ($\mathcal{L} < 0.1\,\mathcal{L}_\odot$, spectral type later or equal than M6.0 V), using isochrone models (Table 2.4). For these, we found incongruences in all representations that involved masses computed using the mass-radius relation by Schweitzer et al. (2019), because it does not hold true for the least massive stars. We used the Lyon group's DUSTY00 (Chabrier et al., 2000)[13] public code, assuming constant ages of 1, 5, 10 Ga to determine new masses and radii for them. Out of the 18 stars susceptible of having their masses/radii recalculated, two of them (J14321+081 and J09033+056) cannot be characterised with DUSTY00 models, as their luminosity is out of their validity range, and extrapolation beyond the model limits would be needed. These updated values are included in the first data release CARMENES DR1.

[12]The authors employed SteParSyn, a Bayesian spectral synthesis implementation specially suited for late-type stars, following a Markov chain Monte Carlo (MCMC) approach (Tabernero et al., 2022a).

[13] http://perso.ens-lyon.fr/isabelle.baraffe/.

# Chapter 3

# Luminosities

The content of this chapter has been adapted from the article *CARMENES input catalogue of M dwarfs. V. Luminosities, colours, and spectral energy distributions*, published in *Astronomy & Astrophysics* (Cifuentes et al. 2020, A&A, 642, A115).

M dwarfs have gained much relevance in the search for potentially habitable Earth-sized planets in the last years. In our on-going effort to comprehensively and accurately characterise confirmed and potential planet-hosting M dwarfs, in particular for the CARMENES survey, we have carried out an exhaustive multi-band photometric analysis involving spectral energy distributions, luminosities, absolute magnitudes, colours, and spectral types, from which we have derived basic astrophysical parameters. We have carefully compiled photometry in 20 passbands from the ultraviolet to the mid-infrared for a sample of 2479 K5 V to L8 stars and ultracool dwarfs, including 2210 nearby, bright M dwarfs. We combined this information with the latest parallactic distances and close-multiplicity information available, mostly from *Gaia* DR2. For this task we made extensive use of Virtual Observatory tools. We have homogeneously computed accurate bolometric luminosities and effective temperatures of 1843 single stars, derived their radii and masses, studied the impact of metallicity, and compared our findings with the literature. As a result, over 40 000 individually inspected magnitudes, together with the basic data and derived parameters of the stars, individual and averaged by spectral type, have been made public to the astronomical community. In addition, we have reported 40 new close multiple systems and candidates ($\rho < 3.3$ arcsec) and 36 overluminous stars that are assigned to young Galactic populations. In the new era of exoplanet searches around M dwarfs via transit (e.g. *TESS*, *PLATO*) and radial velocity (e.g. CARMENES, NIRPS+HARPS), this work is of fundamental importance for stellar and therefore planetary parameter determination.

## 3.1 Introduction

This work is part of a series of papers devoted to describing the CARMENES input catalogue of M dwarfs. Here we continue the work started by Alonso-Floriano et al. (2015b) on spectral typing from low-resolution spectroscopy of M dwarfs (I), and followed up by Cortés-Contreras et al. (2017b) on multiplicity from high-resolution lucky imaging (II), Jeffers et al. (2018) on activity from high-resolution spectroscopy (III), Díez Alonso et al. (2019) on rotation periods from photometric time series (IV).

In this fifth item of the series, we focus on the analysis of multi-wavelength photometry, from the far ultraviolet to the mid infrared, of a large sample of nearby, bright M dwarfs, including those monitored by CARMENES, as well as some late K dwarfs and early and mid L dwarfs. We derive accurate bolometric luminosities, identify new close binaries, members in young stellar kinematic groups, and other outliers in colour-colour, colour-magnitude, and colour-spectral type diagrams. We also explore different relationships between colours, absolute magnitudes, spectral types, luminosities, masses, and radii. For that, we make extensive use of the second data release of *Gaia* astrometry and photometry (*Gaia* DR2; Gaia Collaboration et al., 2018b), numerous public all-sky surveys from the ground and space, and Virtual Observatory tools such as the `Aladin` interactive sky atlas (Bonnarel et al., 2000), the Tool for OPerations on Catalogues And Tables (`TOPCAT`; Taylor, 2005), and the Virtual Observatory Spectral energy distribution Analyser (`VOSA`; Bayo et al., 2008).

Our work is also connected to that of Schweitzer et al. (2019), who derived masses and radii from effective temperatures (determined from spectral synthesis) and luminosities (measured exactly as in the present work) for 293 M dwarfs monitored by CARMENES. As a result, here we complement the description of the calculation of stellar masses and radii of all planet hosts detected by CARMENES (e.g. Reiners et al., 2018a; Ribas et al., 2018; Trifonov et al., 2018; Zechmeister et al., 2019; Luque et al., 2019b; Morales et al., 2019, to cite a few).

### 3.1.1 Sample

In this Section we describe the building process of our sample, as well as the compilation of their photometric and astrometric data from public catalogues. Our sample is based mainly on Carmencita, the CARMENES input catalogue (Sect. 2.2). In the version used in this Chapter, Carmencita contains 2191 M dwarfs and 3 K dwarfs[1], namely J04167–120 (LP 714–47), J11110+304E (HD 97101 A), and J18198–019 (HD 168442), which satisfied simple selection criteria based on $J$-band magnitude and spectral type regardless of multiplicity, age, or metallicity (cf. Alonso-Floriano et al., 2015b). Except for the three K dwarfs, Carmencita includes M dwarfs visible from the Calar Alto Observatory in Southern Spain ($\delta \gtrsim -23$ deg) with spectral types from M0.0 V to M9.0 V and near-infrared brightnesses between $J = 4.2$ mag and 11.5 mag. The spectral classifications of 2028 M dwarfs (92.5 %) were taken from only three references: Hawley et al. (2002), Lépine et al. (2013), and Alonso-Floriano et al. (2015b), which are equivalent among them according to the latter authors. Of the remaining 163 M dwarfs, most spectral types also came from reliable, equivalent sources (e.g. Gray et al., 2003; Scholz et al., 2005; Riaz et al., 2006), which assures a relative homogeneity in our sample.

As described in the references above, Carmencita is unbiased except for the fact that it may include overluminous and lack underluminous stars in the $J$ band at a fixed spectral type. This fact probably translates into a larger fraction of (overluminous) close multiples and young active stars, and a lower fraction of (underluminous) very low metallicity M-type dwarfs (subdwarfs and extreme subdwarfs; Gizis, 1997; Lépine et al., 2007). From the distribution of the $\zeta$ index, a metallicity proxy measured in low-resolution spectra of a large number of Carmencita stars (cf. Alonso-Floriano et al., 2015b), we extrapolated that most of our M dwarfs have solar-like metallicities, but that there could be a significant number of them with [Fe/H] < −1.0. Nevertheless, volume-completeness samples can be safely drawn from Carmencita, as explained in Chapter 4.

[1] Carmencita is a catalogue in constant development and refinement. For instance, the version used in Chapter 4 includes around 20 more M dwarfs, which are part of the CARMENES follow-up of some targets from the Transiting Exoplanet Survey Satellite (*TESS*) (Ricker et al., 2015) program.

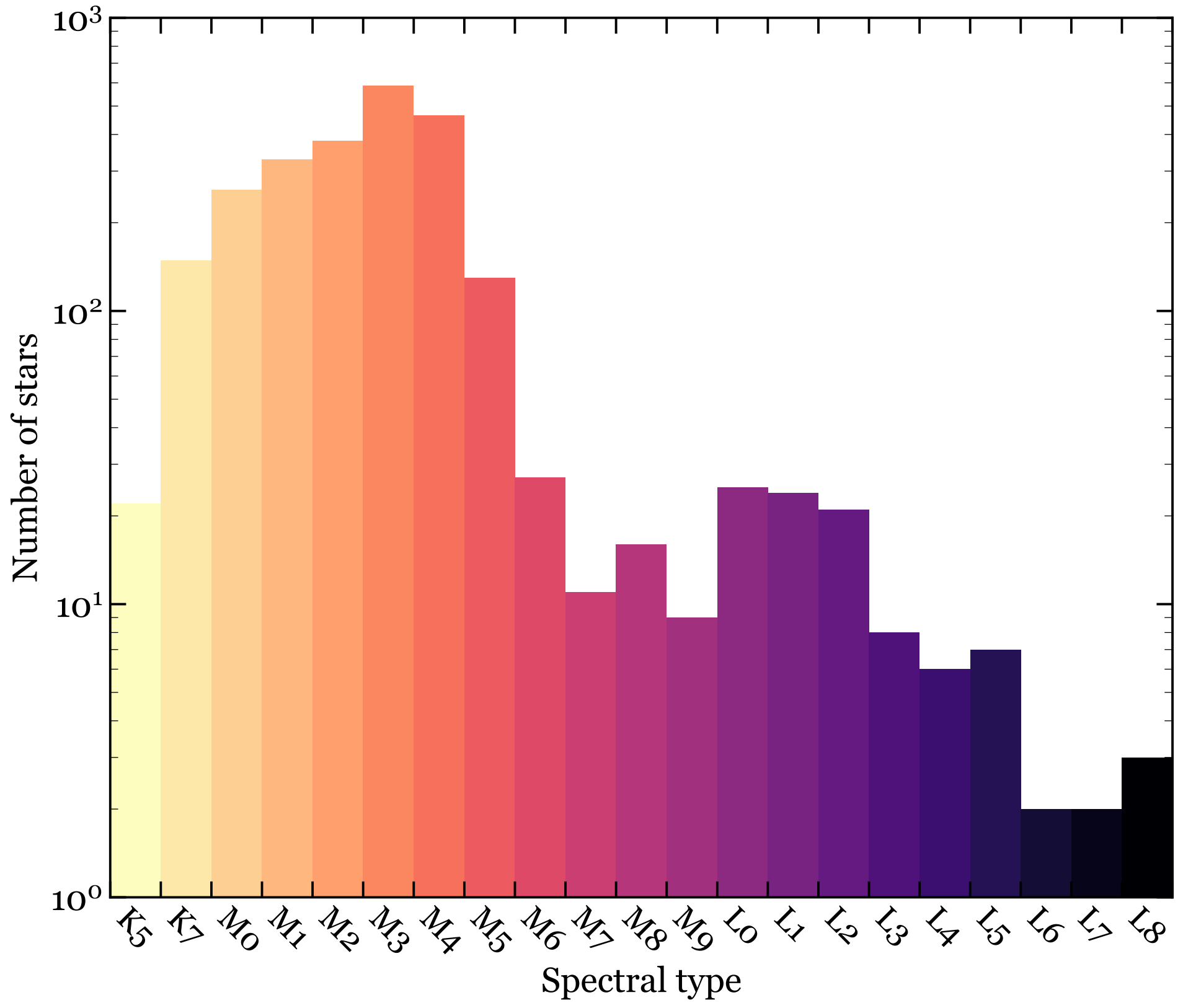

Figure 3.1: Distribution of spectral types in our sample.

In order to extend the photometric sample to a wider spectral range and to avoid any boundary value problem, we complemented Carmencita with additional stars earlier than M0.0 V, and with stars and brown dwarfs later than M6.5 V. The eventual distribution of spectral types is displayed in Fig. 3.1. On the warm side, we included 168 bright stars with spectral types between K5 V and K7 V from Kirkpatrick et al. (1991), Lépine et al. (2013), and Alonso-Floriano et al. (2015b), and the RECONS list of the 100 nearest stars[2] (Henry et al., 2006). We did not include the very bright K stars $\eta$ Cas B, 36 Oph C, BD+01 3942 A, $\xi$ Cap B, 61 Cyg A, and 61 Cyg B, whose photometry is strongly affected by saturation or blending due to close multiplicity.

On the cool side, we first included seven M5.0–9.0 V stars from RECONS with $\delta < -23$ deg. Next, we added 110 ultracool dwarfs from Smart et al. (2017) with a Two Micron All-Sky Survey (2MASS) near-infrared counterpart (Skrutskie et al., 2006) and relative error in *Gaia* DR2 parallaxes ($\delta\varpi/\varpi$) less than 1 %. That addition made 12 M8.0–9.5 V and 98 L0.0–8.0 ultracool dwarfs. We did not include four T-type brown dwarfs (SIMP J013656.57+093347.3, ULAS J141623.94+134836.30, 2MASS 15031961+2525196, and WISE J203042.79+074934.7) and one L dwarf, HD 16270 B, because of their poor 2MASS photo-

[2] http://www.recons.org/TOP100.posted.html.

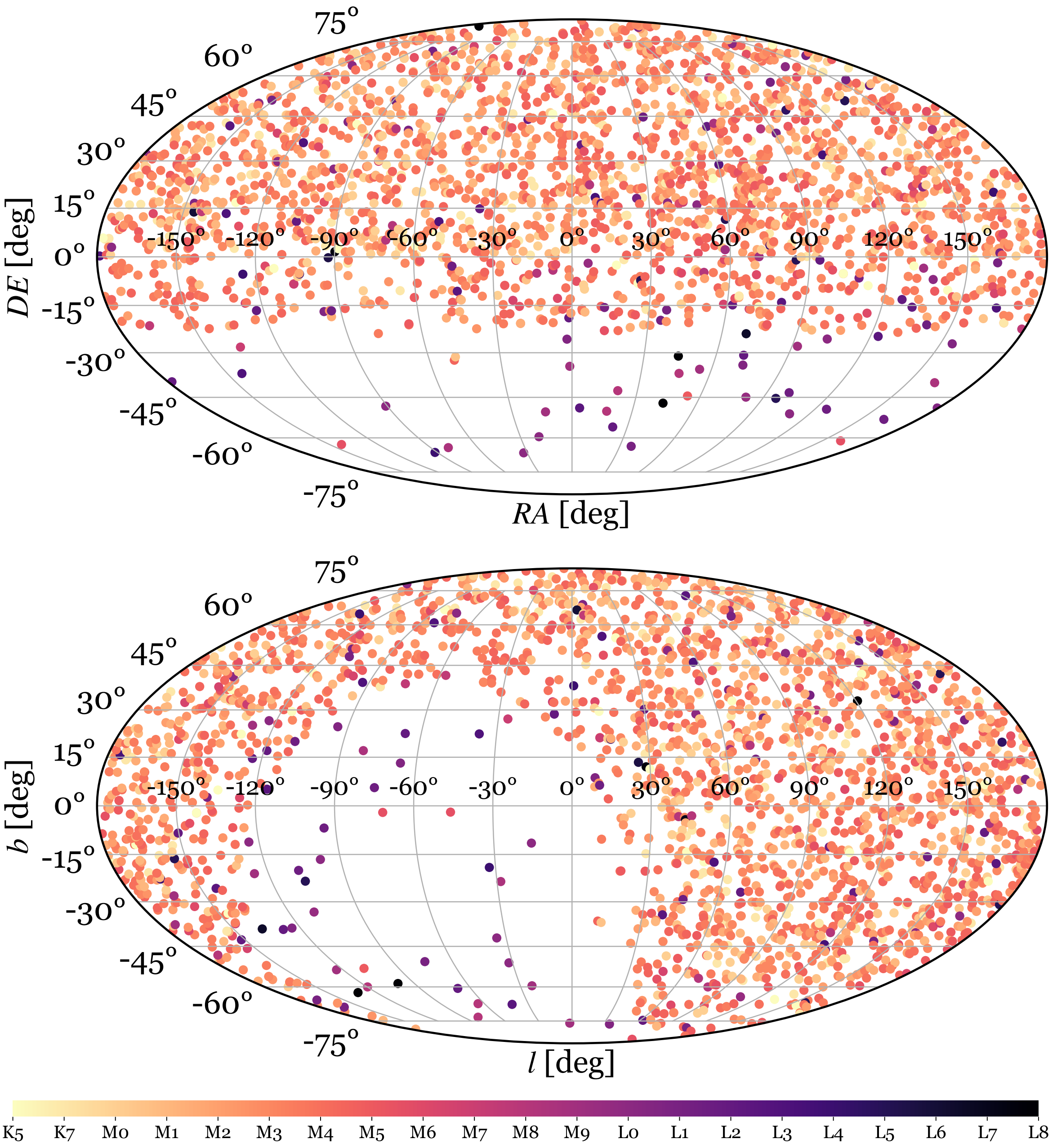

Figure 3.2: Location in the Mollweide-projection sky of the 2479 targets in our sample, which contains Carmencita stars plus the late-K and early-L stars added, colour-coded by spectral type, in equatorial (*top*) and Galactic coordinates (*bottom*). We note the absence of Carmencita M dwarfs with declinations lower than $\delta = -23$ deg.

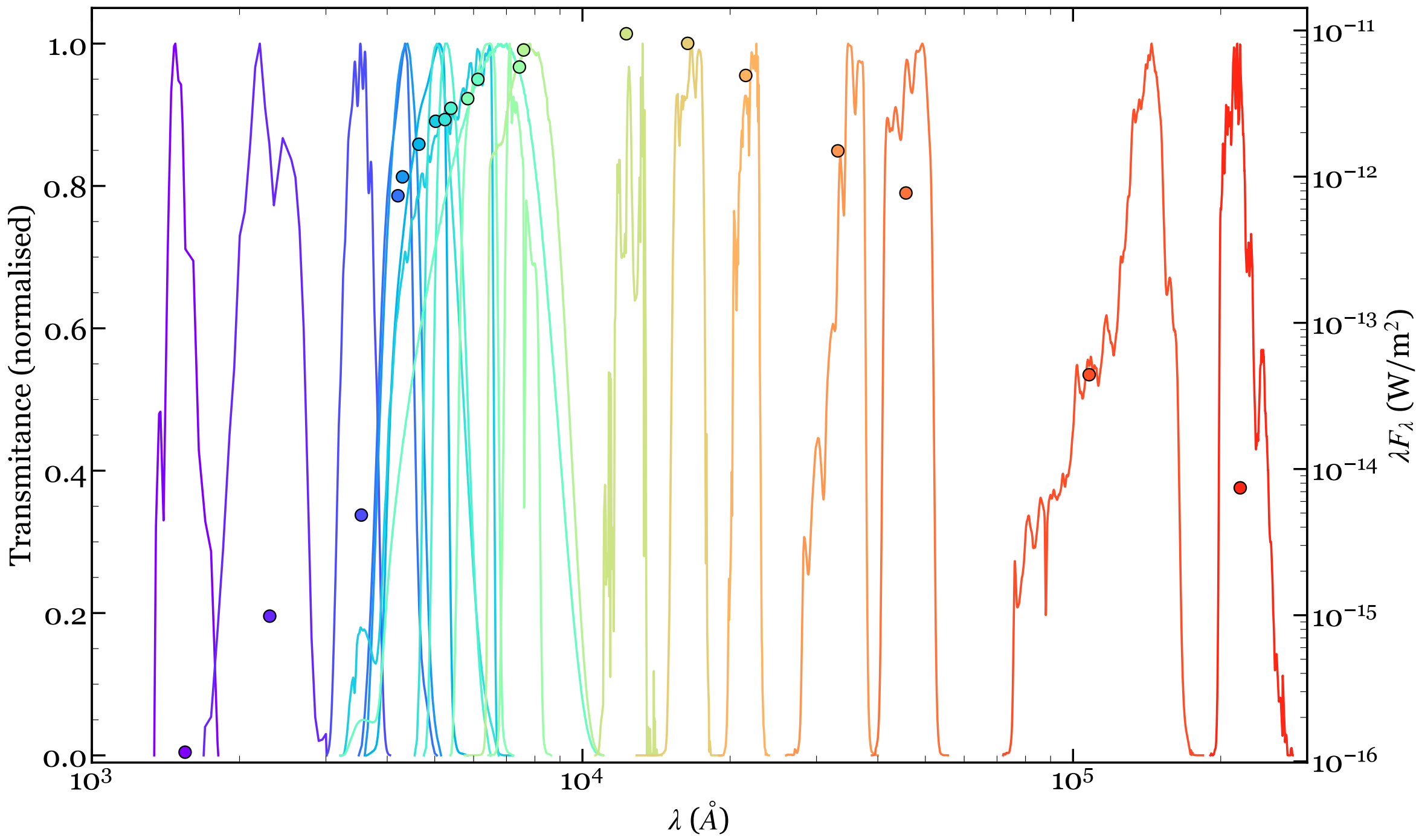

Figure 3.3: Normalised transmission curves of the 20 passbands employed for the compilation of photometry, taken from the SVO Filter Profile Service. For comparison, coloured filled circles depict the spectral energy distribution of DS Leo (Karmn J11026+219, M1.0 V).

metric quality (see Sect. 3.1.2).

As a result, the joint K-M-L spectro-photometric sample contained 2479 targets consisting on 171 late-K dwarfs, 2210 M dwarfs, and 98 L dwarfs. For all targets in the sample we employed and tabulated equatorial coordinates from *Gaia* DR2 except for the 58 stars (five K, 53 M) that were not catalogued by the ESA space mission. For all 58 stars, we used the positions at the epoch of 2MASS projected to the epoch 2015.5 with proper motions from van Leeuwen (2007) and Zacharias et al. (2012), as compiled by Caballero et al. (2016).

The spatial distribution of the 2479 targets is illustrated in Fig. 3.2. For the sake of simplicity, we will use hereafter the term "stars" for the 2479 objects in our sample, including the stellar and substellar objects later than M7 V, also known as ultracool dwarfs (Kirkpatrick et al., 1997).

### 3.1.2 Photometry

For every star in the sample, we compiled multiwavelength broadband photometry covering a wide spectral range from the ultraviolet to the mid-infrared, as illustrated in Fig. 3.3. First of all, with `Aladin` we manually retrieved the 2MASS equatorial coordinates, $JHK_s$ magnitudes and uncertainties, and photometric quality flags of all 2479 stars (we had done this previously for the Carmencita stars; Caballero et al. 2016). Next, we added photometric data from different public catalogues. We started by adding *Gaia* DR2 $G$, $G_{BP}$, and $G_{RP}$ magnitudes, obtained with the query form available in the *Gaia* Archive[3]. We followed by adding magnitudes and uncertainties of the Galaxy Evolution Explorer (*GALEX*) $FUV$ and $NUV$, the Ninth Sloan Digital Sky Survey Data Release (SDSS9) $u'g'r'i'$, Tycho-2 $B_T$ and $V_T$, the AAVSO Photometric All-Sky Survey Data Release 9 (APASS9) $B$ and $V$, the Fourth US Naval Obser-vatory CCD Astrograph Catalog (UCAC4) $BVg'r'i'$, the Carlsberg Meridian Catalogue 15 (CMC15) $r'$, and of the Wide-field Infrared Survey Explorer (AllWISE and WISE) $W1W2W3W4$ (and their quality flags when available). For that, we used the `TOPCAT` automatic positional cross-match tool `CDS X-match` with a search radius of 5 arcsec and the "All" find option. For a few high proper-motion stars, we enlarged the search radius to 10 arcsec. Next we used `Aladin` to: (*i*) visually inspect the automatic cross-matches of all sources (and correct them, especially in mismatched cases of high proper motion and close binary sources), and (*ii*) compile, by hand, the most reliable photometry of Pan-STARRS1 DR1 only for the stars for which $g'$, $r'$, or $i'$ magnitudes were not available in other catalogues (PS1 DR1 delivered up to 60 multi-epoch observations for every star over three years in the five PS1 passbands). Although the inspection must be done individually, the process of loading catalogs and images in `Aladin` can be sped up by using the macro tool option (see Appendix E).

[3]http://gea.esac.esa.int/archive/.

Table 3.1: Passbands employed for the compilation of photometry.

| Band | $\lambda_{\rm eff}$ [Å] | $W_{\rm eff}$ [Å] | $F^0_\lambda$ [W m$^{-2}$ Å$^{-1}$] | Survey[a] | Description |
|---|---|---|---|---|---|
| *FUV* | 1549.0 | 265.6 | $6.491 \times 10^{-12}$ | *GALEX* | GALEX *FUV* |
| *NUV* | 2304.7 | 768.3 | $4.450 \times 10^{-12}$ | *GALEX* | GALEX *NUV* |
| $u'$ | 3594.9 | 558.4 | $3.639 \times 10^{-12}$ | SDSS9 | SDSS $u'$ full transmission |
| $B_T$ | 4206.4 | 708.4 | $6.598 \times 10^{-12}$ | Tycho-2 | Tycho *B* |
| *B* | 4297.2 | 843.1 | $6.491 \times 10^{-12}$ | UCAC4[b] | UCAC4 *B* filter |
| | 4297.2 | 843.1 | $6.491 \times 10^{-12}$ | APASS9[b] | APASS *B* filter |
| $g'$ | 4640.4 | 1158.4 | $5.521 \times 10^{-12}$ | UCAC4 | UCAC4 $g'$ filter |
| | 4640.4 | 1158.4 | $5.521 \times 10^{-12}$ | SDSS9 | SDSS $g'$ full transmission |
| | 4640.4 | 1158.4 | $5.521 \times 10^{-12}$ | APASS9 | APASS $g'$ filter |
| | 4810.8 | 1053.1 | $5.043 \times 10^{-12}$ | PS1 DR1 | PS1 $g'$ filter |
| $G_{B_P}$ | 5020.9 | 2279.5 | $4.035 \times 10^{-12}$ | *Gaia* DR2 | Gaia $G_{BP}$ filter, DR2 |
| | 5035.8 | 2157.5 | $4.079 \times 10^{-12}$ | *Gaia* DR3 | Gaia $G_{BP}$ filter, DR3 |
| $V_T$ | 5243.9 | 1005.7 | $3.984 \times 10^{-12}$ | Tycho-2 | Tycho *V* |
| *V* | 5394.3 | 870.6 | $3.734 \times 10^{-12}$ | UCAC4 | UCAC4 *V* filter |
| | 5394.3 | 870.6 | $3.734 \times 10^{-12}$ | APASS9 | APASS *V* filter |
| $r'$ | 6122.3 | 1111.2 | $2.529 \times 10^{-12}$ | UCAC4 | UCAC4 $r'$ filter |
| | 6122.3 | 1111.2 | $2.529 \times 10^{-12}$ | SDSS9 | SDSS $r'$ full transmission |
| | 6122.3 | 1111.2 | $2.529 \times 10^{-12}$ | APASS9 | APASS $r'$ filter |
| | 6122.3 | 1318.1 | $2.529 \times 10^{-12}$ | CMC15 | SDSS $r'$ full transmission |
| | 6156.4 | 1252.4 | $2.480 \times 10^{-12}$ | PS1 DR1 | PS1 $r'$ filter |
| *G* | 5836.3 | 4358.4 | $2.495 \times 10^{-12}$ | *Gaia* DR2 | Gaia *G* filter, DR2 |
| | 5822.4 | 4052.9 | $2.816 \times 10^{-12}$ | *Gaia* DR3 | Gaia *G* filter, DR3 |
| $i'$ | 7439.5 | 1044.6 | $1.409 \times 10^{-12}$ | UCAC4 | UCAC4 $i'$ filter |
| | 7439.5 | 1044.6 | $1.409 \times 10^{-12}$ | SDSS9 | SDSS $i'$ full transmission |
| | 7439.5 | 1044.6 | $1.409 \times 10^{-12}$ | APASS9 | APASS $i'$ filter |
| | 7503.7 | 1206.6 | $1.372 \times 10^{-12}$ | PS1 DR1 | PS1 $i'$ filter |
| $G_{R_P}$ | 7588.8 | 2943.7 | $1.294 \times 10^{-12}$ | Gaia DR2 | *Gaia* $G_{RP}$ filter, DR2 |
| | 7619.9 | 2924.4 | $1.269 \times 10^{-12}$ | Gaia DR3 | *Gaia* $G_{RP}$ filter, DR3 |
| *J* | 12285.4 | 1624.2 | $3.143 \times 10^{-13}$ | 2MASS | 2MASS *J* |
| *H* | 16386.1 | 2509.4 | $1.144 \times 10^{-13}$ | 2MASS | 2MASS *H* |
| $K_s$ | 21521.6 | 2618.9 | $4.306 \times 10^{-14}$ | 2MASS | 2MASS $K_s$ |
| *W*1 | 33156.6 | 6626.4 | $8.238 \times 10^{-15}$ | (All)WISE | WISE *W*1 filter |
| *W*2 | 45644.9 | 10422.7 | $2.431 \times 10^{-15}$ | (All)WISE | WISE *W*2 filter |
| *W*3 | 107868.4 | 55055.7 | $6.570 \times 10^{-17}$ | (All)WISE | WISE *W*3 filter |
| *W*4 | 219149.6 | 41016.8 | $4.995 \times 10^{-18}$ | (All)WISE | WISE *W*4 filter |

[a] *GALEX* DR5: Galaxy Evolution Explorer, Bianchi et al. (2011); SDSS DR9: Sloan Digital Sky Survey, Ahn et al. (2012); UCAC4: The fourth U.S. Naval Observatory CCD Astrograph Catalog, Zacharias et al. (2012); Pan-STARRS1: Panoramic Survey Telescope and Rapid Response System, Kaiser et al. (2010), Tonry et al. (2012), and Chambers et al. (2016); CMC15: Carlsberg Meridian Catalogue, Niels Bohr Institute et al. (2014); APASS9: The AAVSO Photometric All-Sky Survey, Henden et al. (2016); Tycho-2: Høg et al. (2000); 2MASS: Two Micron All-Sky Survey, Skrutskie et al. (2006); *Gaia* DR2: Gaia Collaboration et al. (2016) with the revised response curves of Maíz Apellániz & Weiler (2018); *Gaia* DR3: Gaia Collaboration et al. (2022b); WISE: Wide-field Infrared Survey Explorer, Cutri & et al. (2012); AllWISE: Cutri & et al. (2014).

[b] UCAC4 and APASS9 *BVgri* are defined as identical to `GCPD/Johnson.B_Landolt`, `GCPD/Johnson.V_Landolt`, `MISC/APASS.sdss_g` / `SLOAN/SDSS.g`, `MISC/APASS.sdss_r` / `SLOAN/SDSS.r`, and `MISC/APASS.sdss_i` / `SLOAN/SDSS.i`, respectively.

The passband name, effective wavelength $\lambda_{\rm eff}$, effective width $W_{\rm eff}$, zero point flux $F^0_\lambda$, survey acronym, and corresponding references of the 20 compiled passbands are listed in Table 3.1. The passband parameters were calculated by `VOSA` with the latest filter transmission curves available at the Filter Profile Service[4] of the Spanish Virtual Observatory. When there were several surveys providing photometric data in the same passband (e.g. $r'$ in UCAC4, SDSS9, APASS9, and PS1 DR1), we prioritised the surveys with the highest spatial resolution, sensitivity, and accuracy. PanSTARRS1 DR1 has slightly different passband parameters from those of the other $g'r'i'$ surveys. Virtually all our K and M dwarfs saturated or were in the non-linear regime in SDSS9 $z'$ and PS1 DR1 $z'y'$, so we did not compile data in these passbands.

*Gaia* $G$, $G_{B_P}$, and $G_{R_P}$ magnitude uncertainties were derived from the uncertainties in the fluxes, while UCAC4 *BVg'r'i'* magnitude uncertainties were collected from an additional `TOPCAT` table access protocol query. However, we chose APASS9 *BV* over UCAC4 *BV* when the UCAC4 uncertainties were 0.00 mag, 0.99 mag, or missing. In the case of poor photometric quality in AllWISE *W1* to *W4* (`Qflag` $\neq$ A,B), we chose the data available in WISE when it improved the quality of AllWISE data. We also identified possible flux excesses in the *Gaia* DR2 $G_{BP}$ and $G_{RP}$ photometric data with the keyword `phot_bp_rp_excess_factor`, following the guidelines of Evans et al. (2018) and Arenou et al. (2018) to separate well-behaved single sources from spurious ones.

$J$ band magnitudes are available for all the stars in the sample, and the completeness in passbands $g'$, $G_{BP}$, $G$, $r'$, $i'$, $G_{RP}$, $H$, $K_s$, $W1$, $W2$, $W3$, and $W4$ is greater than 97 %. For Johnson $B$ and $V$ the completeness is around 86 %, whereas for Tycho-2 $B_T$ and $V_T$ it is only 25 %. At the blue end, $u'$ is complete for 50 % of the sample, and the ultraviolet passbands $FUV$ and $NUV$ are available for 39 % and 14 %, respectively. This is graphically summarised in Fig. 3.4.

In total, we collected 40 094 individual magnitudes. Of them, 39 896 have magnitude uncertainties and 33 594 have good quality photometry, defined as: 2MASS `Qflag` = A (with signal-to-noise ratio ≥10), WISE `Qflag` = A,B, $G_{BP} < 19.5$ mag (see Sect. 3.2.3), and no flux excess in *Gaia* $G_{BP}$ and $G_{RP}$. Figure 3.5 shows the distribution of magnitudes for each band, ordered by increasing $\lambda_{\rm mean}$. The distributions of the bluest bands are broader than the reddest ones, while those of the most complete bands (e.g. $g'$, $r'$, $G$, $J$, $W1$) exhibit small secondary peaks at fainter magnitudes, which correspond to late M and early L dwarfs.

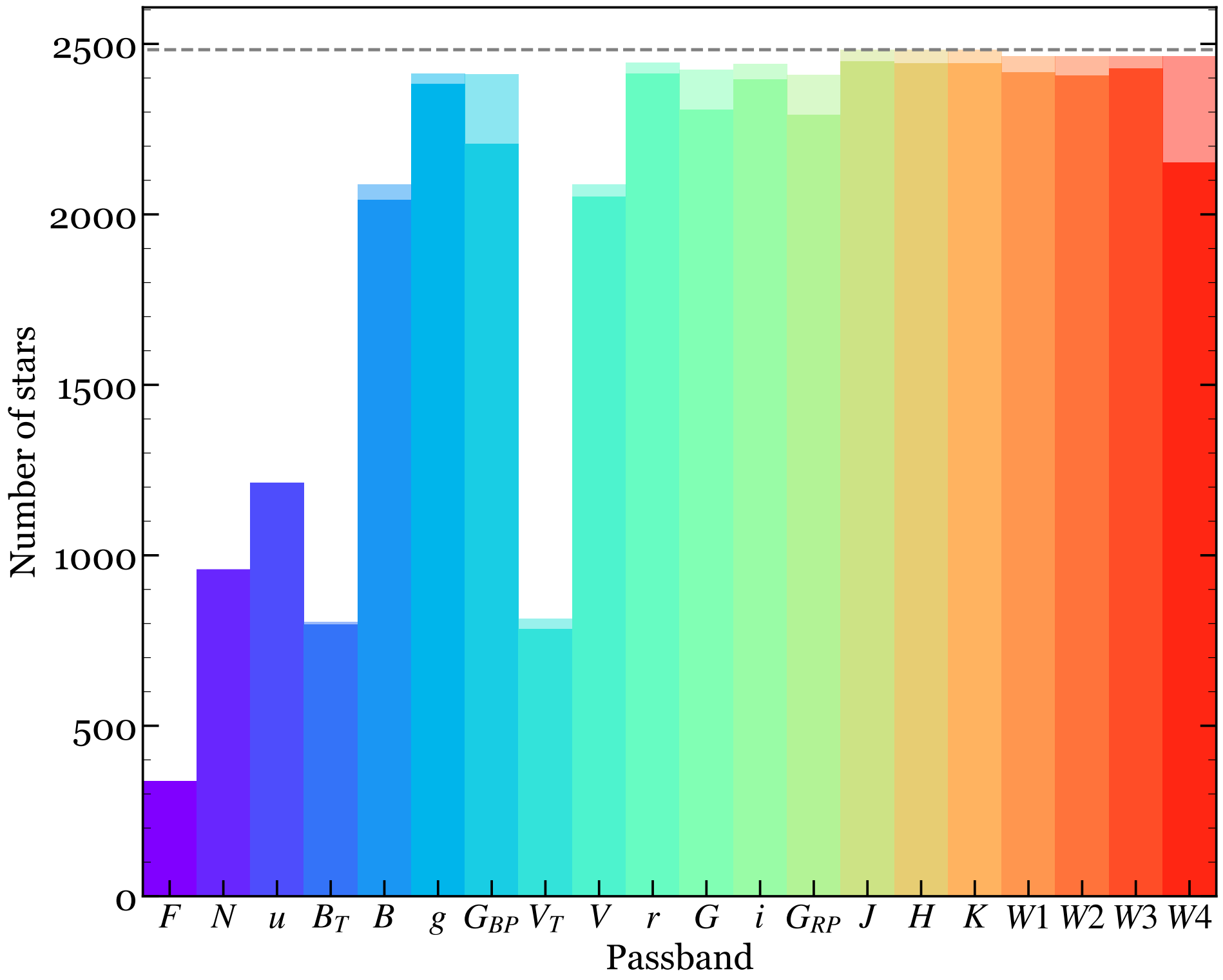

Figure 3.4: Completeness in every passband. Light shaded regions account for measurements with poor quality flags. The dashed horizontal line indicates the total number of stars in the sample.

Table 3.2: Reference of the 2425 parallactic distances in the sample.

| Acronym[a] | Number of stars | Reference |
|---|---|---|
| Gaia2 | 2306 | Gaia Collaboration et al. 2018b |
| HIP2 | 41 | van Leeuwen 2007 |
| Dit14 | 34 | Dittmann et al. 2014 |
| vAl95 | 16 | van Altena et al. 1995 |
| FZ16 | 14 | Finch & Zacharias 2016 |
| Galli18 | 2 | Galli et al. 2018 |
| Hen06 | 2 | Henry et al. 2006 |
| Jao05 | 2 | Jao et al. 2005 |
| Wein16 | 2 | Weinberger et al. 2016 |
| Dahn17 | 1 | Dahn et al. 2017 |
| GC09 | 1 | Gatewood & Coban 2009 |
| Jen52 | 1 | Jenkins 1952 |
| Lep09 | 1 | Lépine et al. 2009 |
| Ried10 | 1 | Riedel et al. 2010 |
| TGAS | 1 | Gaia Collaboration et al. 2016 |

[a] Achronyms used in the on-line table.

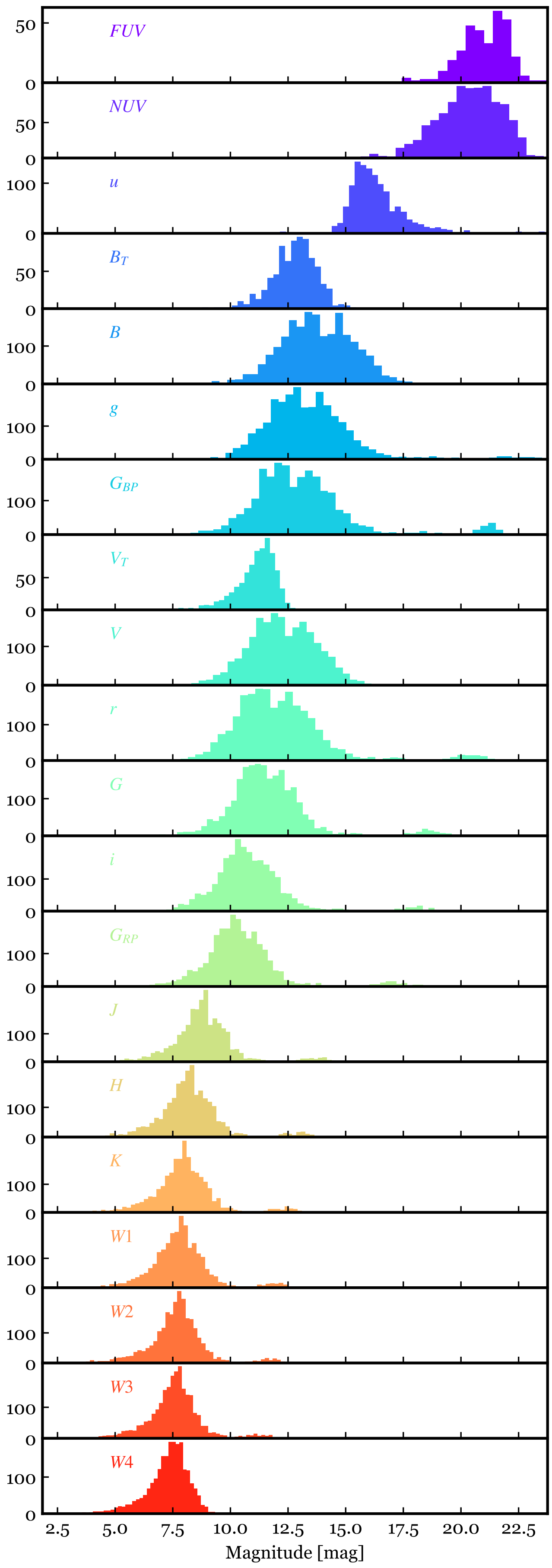

Figure 3.5: Distribution of compiled magnitudes in every passband. The width of the bins follows the Freedman-Diaconis rule.

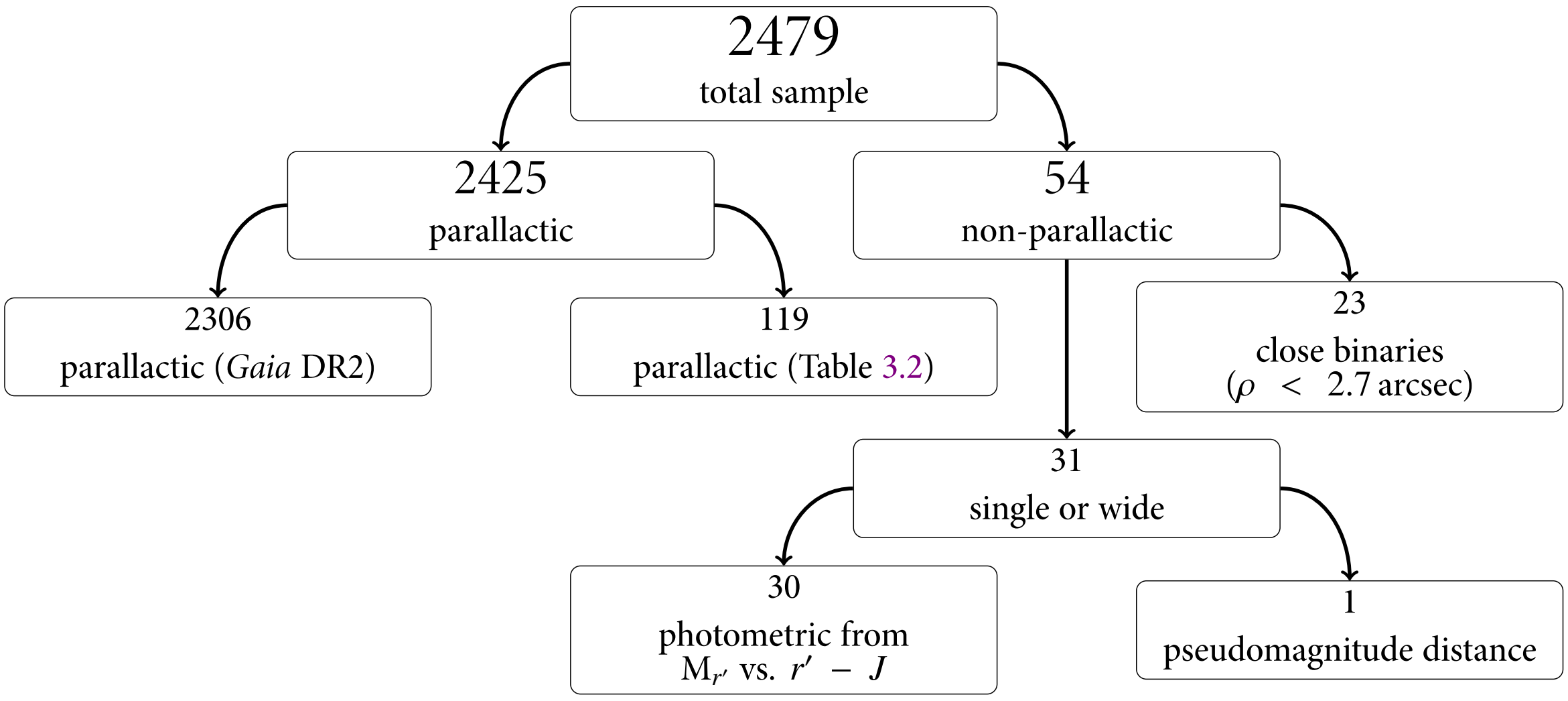

Figure 3.6: Schematic diagram of sources of heliocentric distances.

### 3.1.3 Distances

We compiled equatorial coordinates, proper motions, parallaxes, and astrometric quality indicators from *Gaia* DR2. Of the 2479 stars in our sample, 2425 (97.8 %) had parallactic distances. Of them, 2306 parallaxes came from *Gaia* DR2 (93.0 %) and 119 from a number of references, as detailed in Table 3.2. For 16 stars with unavailable parallactic distances, we used the trigonometric distances of their confirmed proper motion companions from *Gaia* DR2 (ten cases) and van Leeuwen (2007, six cases). As a result, there were 54 stars without parallactic distance, of which 23 are close binaries: four spectroscopic binaries from Reipurth & Mikkola (2012a) and Jeffers et al. (2018), and 19 resolved binaries (16 with $\rho \lesssim$ 0.8 arcsec, and three at $\rho$ = 1.1–2.7 arcsec; see Sect. 3.1.4). The remaining 31 stars are single or have wide companions at angular separations of $\rho$ > 16 arcsec. For 30 of them, we derived photometric distances from $r' - J$ colours following the prescription in Sect. 3.2.3. For the remaining star, a Pleiades member with an $r' - J$ colour outside the validity range, we adopted the "pseudomagnitude" distance to the open cluster of Chelli et al. (2016). As a result, we compiled or derived distances for 2456 stars (i.e. all but the 23 close binaries without parallax). Figure 3.6 shows a schematic summary of the origin of all compiled distances.

Our sample spans a distance range from 1.30 pc (Proxima Centauri) to 171 pc (Haro 6–36). However, ignoring late K dwarfs, overluminous young M dwarfs (in Taurus, Upper Scorpius, and the $\beta$ Pictoris moving group; Sects. 3.2.2 and 4.4.5), and one star with a large parallax uncertainty ($\delta\varpi/\varpi \sim 8$ %), the most distant "regular" M dwarf is LP 415–17, at 73.0 pc (Díez Alonso et al., 2018b; Hirano et al., 2018). Actually, 92 % of the stars are at less than 40 pc, with only half a dozen objects further than 100 pc. The left panel in Fig. 3.7 shows the distance distribution of our K, M, and L sub-samples.

*Gaia* DR2 provides statistical parameters to assess the quality of the astrometric data for each source. The a posteriori mean error of unit weight (`uwe`) is a goodness-of-fit indicator that is implicit in the *Gaia* DR2 solution. Because of its strong dependence on colour and magnitude, a re-normalised `uwe`, or `ruwe`, is a more convenient indicator of well-behaved astrometric solutions (Arenou et al., 2018; Lindegren et al., 2018). The latter authors set a threshold on `ruwe` at 1.4, based on the empirical distribution of a large sample of stars, under which they retained 70 % of their sources. We derived the `ruwe` values for all stars with *Gaia* DR2 measurements in our sample (2421; there are 125 *Gaia* DR2 stars without parallax), and display the corresponding `ruwe` histogram in the right panel of Fig. 3.7. In our case, by retaining 70 % of our sources we re-defined a cut in `ruwe` = 1.41, which is equivalent to the 1.4 value.

[4] http://svo2.cab.inta-csic.es/theory/fps/.

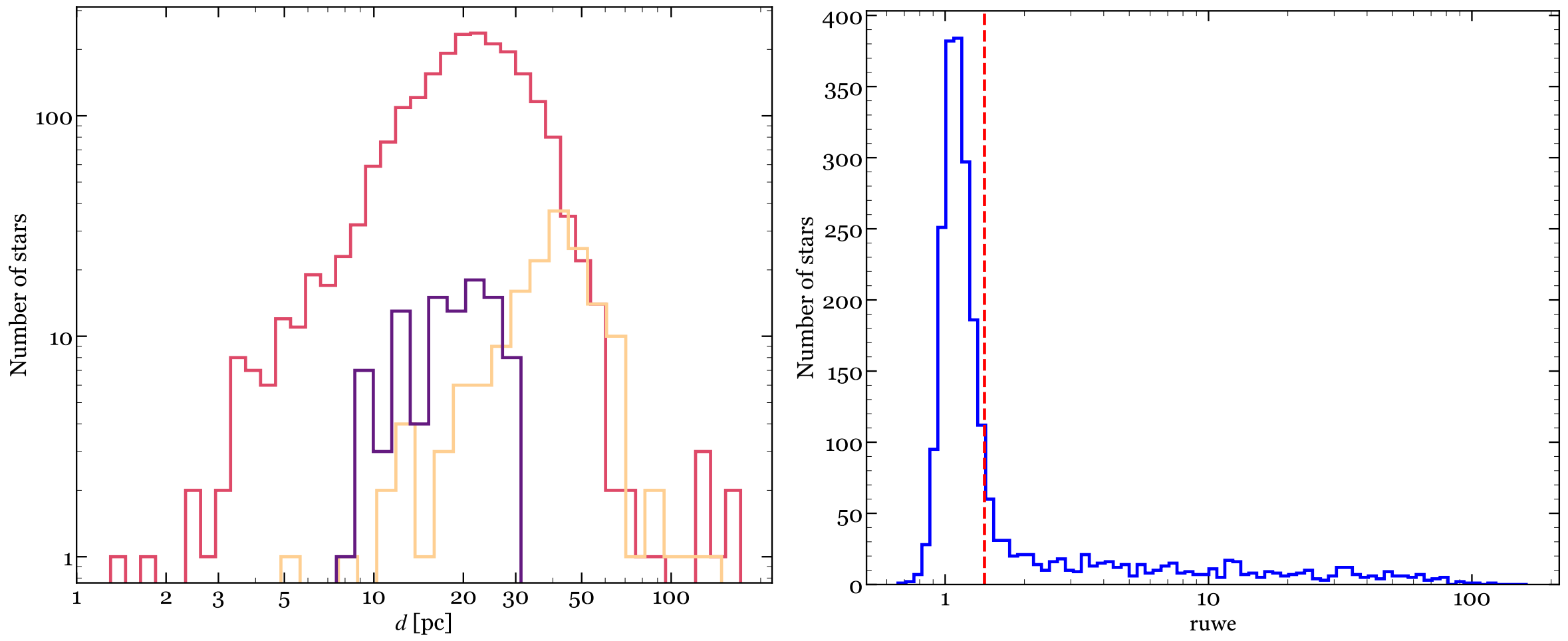

Figure 3.7: Histogram of distances for all stars in the sample, for K (yellow), M (red), and L (violet) dwarfs (*left*), and `ruwe` values for the stars identified in the *Gaia* DR2 catalogue (*right*). The vertical red dashed line in the bottom panel sets the threshold for well-behaved astrometric solutions at `ruwe` = 1.41.

Our sample is not volume limited. First, its basis, the Carmencita catalogue, is not complete. Carmencita contains all known M dwarfs in the solar neighbourhood that are further north than $\delta = -23$ deg with published "spectroscopic" (i.e. non-photometric) spectral types that are brighter than the completeness magnitudes shown in Alonso-Floriano et al. (2015b), meaning they are magnitude limited by spectral subtype: M0.0–0.5 V with $J < 7.3$ mag, M1.0–1.5 V with $J < 7.8$ mag, M2.0–2.5 V with $J < 8.3$ mag, and so on. We refer the reader to the consequences of these selection criteria on the metallicity properties of the sample described in Sect. 3.1.1. Next, the K dwarf and ultracool dwarf additions are not complete either, because, for example, we discarded known K and L dwarf binaries. However, from the distribution of distances, our sample in the Calar Alto sky is complete for M0.0 V, M4.0 V, and M6.0 V stars at approximate distances of 25 pc, 15 pc, and 5 pc, respectively.

### 3.1.4 Close multiplicity

In order to avoid photometric disturbances caused by close sources, we searched for additional *Gaia* DR2 sources within 5 arcsec of our target stars at epoch 2015.5 using the ADQL[5] query form in the *Gaia* Archive. According to Gaia Collaboration et al. (2018b) and, especially, Arenou et al. (2018), *Gaia* can resolve equal-brightness sources separated by down to 0.4 arcsec, which were not resolved in most previous all-sky surveys, such as 2MASS or AllWISE (see Caballero et al. 2019 for a practical example of close binaries resolved for the first time by *Gaia*). For the 2421 stars in our sample that were catalogued by *Gaia*, the search provided 388 additional sources around 353 stars at $\rho < 5$ arcsec. Of them, 324 stars had only one additional source, 24 stars had two sources, 4 stars had three sources, and 1 star had four sources. Besides, for the 58 stars in our sample not tabulated in the *Gaia* catalogue, we used the projected positions as explained in Sect. 3.1.1, which resulted in 11 additional sources around 6 stars. The cases of three or more additional sources corresponded to stars in crowded regions at low Galactic latitudes.

Of the 359 stars with close *Gaia* companion candidates, 166 were already tabulated as members in known physical pairs in the Washington Double Star catalogue (Mason et al., 2001), 4 in Ansdell et al. (2015), and 1 in Heintz (1987). Next, we analysed in detail the remaining 188 systems. Of these, we classified 148 faint sources as background stars and point-like galaxies based on astrometric and photometric criteria: 96 sources have parallaxes $\varpi < 2$ mas and so are located at more than 0.5 kpc; four sources have parallaxes 2 mas $< \varpi <$ 7 mas and turned to be unrelated sources at 47–225 pc (Bayesian distances computed by Bailer-Jones et al. 2018); one source with a parallax of 21.3 mas is located twice as far as the main source; and 47 sources do not have measured parallaxes, proper motions, or 2MASS near-infrared counterparts. In spite of being more than 5 mag fainter than the primary in $G$ band, all 47 sources are visible in digitisations of blue photographic plates of the 1950s (Digitised Sky Survey I), implying that they are background sources much bluer than the stellar primaries[6].

[5] `http://www.ivoa.net/documents/ADQL/`.

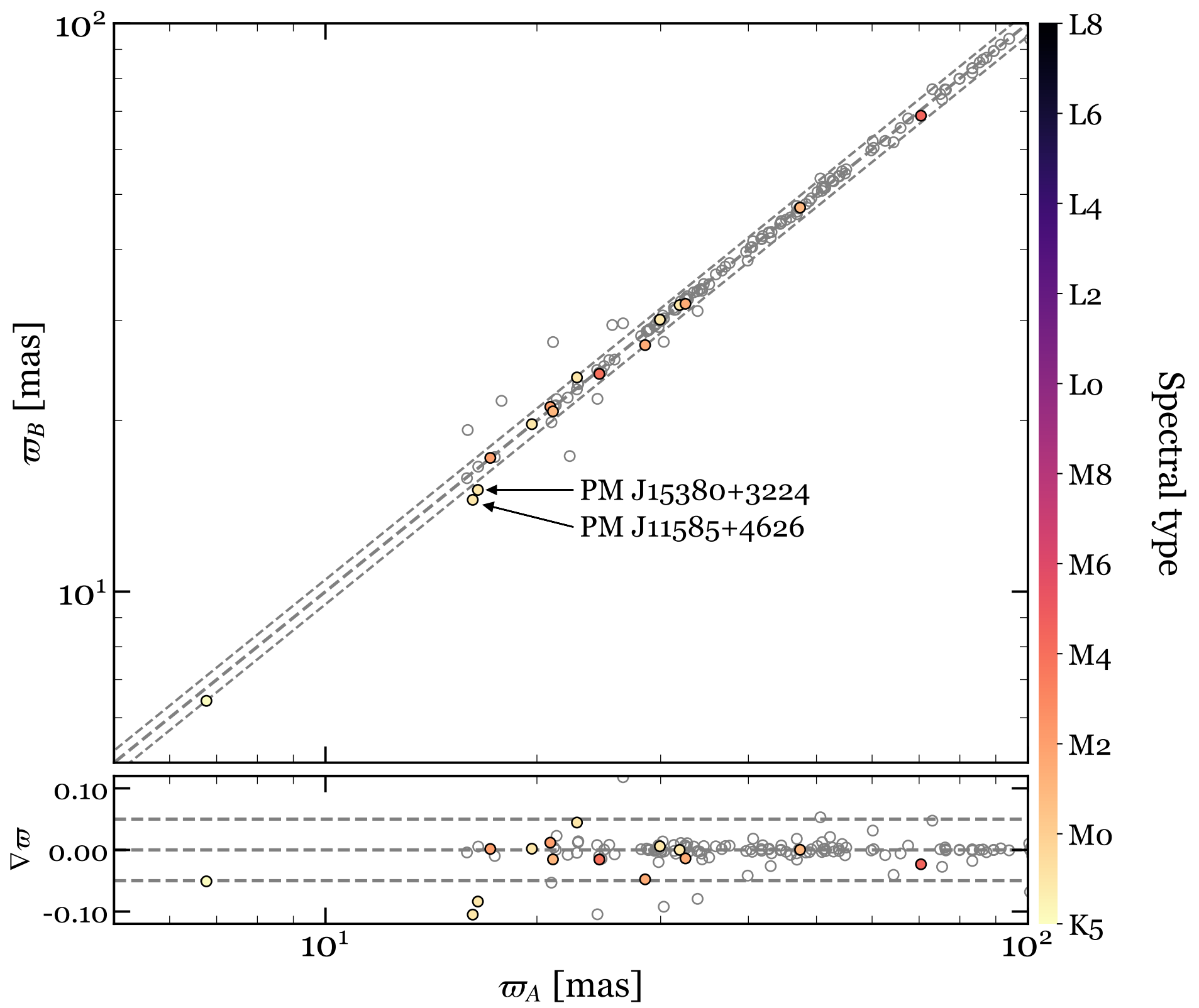

Figure 3.8: Parallax diagram of the primary (A) and secondary (B) components of the 15 new binary systems in Table C.1 with parallactic information in both components, colour-coded by spectral type. The bottom panel shows the normalised difference between both parallaxes, i.e. $\nabla\varpi = (\varpi_B - \varpi_A)/\varpi_A$. Grey empty circles are the 134 previously known pairs in our sample with parallactic information for both components and angular separation of $\rho < 5$ arcsec. The thick and thin dashed lines mark the 1:1 and 1:1 ± 0.05 (i.e. 5 % difference), respectively. Two slight outliers from our list of binary candidates are labelled with their common names.

We investigated the remaining 40 sources not included in the two previous groups. Of them, 15 are in physically bound systems with *Gaia* parallaxes for both components that agree within 1$\sigma$ errors except for two cases, marked in Fig. 3.8. The two systems are bona fide high proper motion pairs, for which we see that the tangential component of the orbital motion and the *Gaia* astrometric solution has not yet taken the close binarity into account. All remaining 25 candidate companions are not visible in the Digitised Sky Survey I and satisfy $\Delta G \lesssim 5$ mag ($\Delta G \sim 0.3$ mag in three cases with $G_{BP}$, $G$, and $G_{RP}$ photometry; see below). In Table C.1 we list the *Gaia* DR2 equatorial coordinates, proper motions, parallaxes, $G$ magnitudes, angular separations $\rho$, and position angles $\theta$ of the 40 new binary systems and candidates. Among them, there are three triple systems consisting of a spectroscopic binary and a fainter companion (see Table C.1 notes). All systems are separated by 3.3 arcsec at most, which explains why other surveys, such as 2MASS, were not able to resolve them.

[6]However, there are certain systems that deserve a high-resolution imaging follow-up, such as J02033–212 (G 272–145), J04429+189 (HD 285968), J05466+441 (Wolf 237), and J11311–149 (LP 732–035).

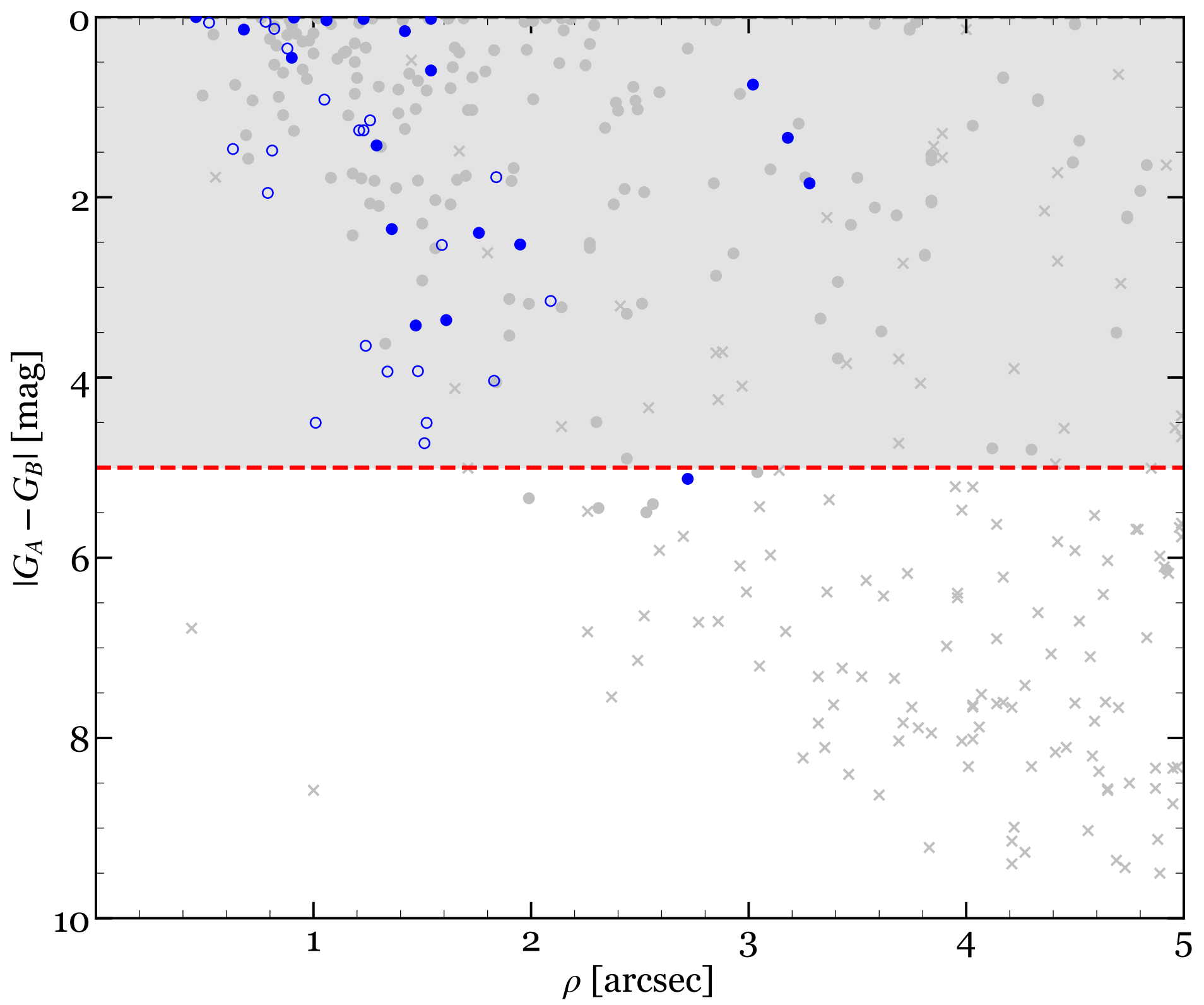

Figure 3.9: Difference in the *Gaia G* magnitude values for the 359 stars with their closest companion within 5 arcsec as a function of angular separation at epoch 2015.5. Known binaries and background stars are depicted with grey filled circles and grey crosses, respectively. New binaries are represented with blue circles, filled if they are confirmed by common parallactic distance, and open if only one component has a measured parallax. The red dashed line marks the boundary at $\Delta G = 5$ mag for contaminated sources.

We consider the photometry of a star as contaminated if the $G$ flux of *any* companion at $\rho < 5$ arcsec, regardless of physical binding, is more than 1 % of its flux, that is if $\Delta G < -2.5 \log(F_{G,\mathrm{B}}/F_{G,\mathrm{A}}) = 5$ mag, where $F_{G,\mathrm{A}}$ and $F_{G,\mathrm{B}}$ represent the fluxes of the primary and secondary components in the $G$ band, respectively. In Fig. 3.9 we plot $\Delta G$ versus $\rho$ of the 359 pairs in our sample with $\rho < 5$ arcsec. Of them, 238 meet the criteria above, and their photometry is therefore flagged as potentially contaminated. To those 238 stars we added other 372 stars from Caballero et al. (2016) that are known to be very close physical systems unresolved by *Gaia* (but resolved with micrometers, speckle, lucky imaging, or adaptive optics systems) and spectroscopic binaries. The 610 "close binaries" are plotted as a reference in most figures afterwards with grey dots, but will not be considered in the following analysis.

The main objective of this section is to investigate the close companions, whether they are physically associated or not, with the purpose of preventing potential sources of photometric contamination. The presence of a nearby companion can adversely affect photometric measurements of a star, particularly when their brightness is similar. Such contamination can have a negative impact on the derived or observed parameters such as luminosity, distance, or colours. It is worth noting that a more comprehensive investigation of multiplicity, encompassing all separations, is the subject of Chapter 4.

Table 3.3: Set of constraints for the spectral energy distribution modelling in VOSA.

| Spectral types | $T_{\rm eff}$ [K] | log $g$ [dex] |
|---|---|---|
| K5 V to M2.0 V | 3300–4600 | 4.5–5.0 |
| M2.5 V to M5.0 V | 2800–3700 | 4.5–5.5 |
| M5.5 V to L8.0 | 1200–3200 | 5.0–5.5 |

**Note:** Iron abundance set to zero ([Fe/H] = 0.0).

## 3.2 Analysis and results

In this section we present the main products of the exploitation of the astrometric and photometric data in the sample, including luminosities, masses, radii, colours, and bolometric corrections.

### 3.2.1 Luminosities

After discarding the 610 close binaries ($\rho < 5$ arcsec), we kept 1843 stars with parallax and whose photometry was not affected by close multiciplicity (however, many of the latter are members of wide multiple systems, as explored in Chapter 4. We used VOSA to compute their basic stellar parameters: bolometric luminosity, $\mathcal{L}$, effective temperature, $T_{\rm eff}$, and surface gravity, log $g$. Among the theoretical model grids available in VOSA for reproducing the observed spectral energy distribution (SED) of each target star, we used the latest `BT-Settl CIFIST` grid from the Lyon group (Husser et al., 2013; Baraffe et al., 2015). We conservatively constrained the possible values of $T_{\rm eff}$ and log $g$ as a function of spectral type as discussed by Pecaut & Mamajek (2013) and Passegger et al. (2018), respectively, and summarised in Table 3.3. We fixed the metallicity to solar (`BT-Settl CIFIST` models are provided for [Fe/H] = 0.0 only) and visual extinction to zero ($A_V = 0$ mag, in view of the closeness of the overall sample; see Sect. 3.1.3). For each star, the VOSA input was the compiled photometry in the passbands in Table 3.1, parallactic distance, and their uncertainties.

In the fitting process, we included the observed fluxes of up to 17 passbands, from optical Tycho-2 $B_T$ to mid-infrared AllWISE $W4$. Since we were only interested in the photospheric emission, we excluded from the fit the other three passbands (i.e. *GALEX* $FUV$ and $NUV$ and SDSS9 $u'$) because the chromospheric emission dominates in the bluest spectral range, especially in late-M dwarfs (Reipurth & Mikkola, 2012a; Stelzer et al., 2013). At wavelengths bluewards of $B_T$ ($\lambda < 4280$ Å) and redwards of $W4$ ($\lambda > 220883$ Å) we followed the VOSA best-fit model (see example in Fig. 3.10). The uncertainty in this assumption was very small, as the estimated fraction of photospheric energy in `BT-Settl CIFIST` spectra bluewards of $B_T$ (in the Wien domain) ranges from 0.46 % to 0.0002 % for M0 V and M8 V, respectively, and redwards of $W4$ (in the Rayleigh-Jeans domain) ranges from 0.0036 % to 0.0087 % for M0 V and M8 V, respectively.

For the best fit, VOSA uses a $\chi^2$ metric, where each photometric point is weighted with its uncertainty. If this uncertainty is blank or artificially set to zero, VOSA assumes a large value instead, which depends on the largest relative error on the SED, and assigns to the point a low weight[7]. The theoretical uncertainties of $T_{\rm eff}$ and log $g$ are determined by the `BT-Settl CIFIST` model grid, which provides synthetic models in steps of 100 K (50 K for spectra cooler than 2400 K) and 0.5 dex, respectively. VOSA estimates the error in the output parameters as half the grid step around the best-fit value.

[7] See http://svo2.cab.inta-csic.es/theory/vosa/.

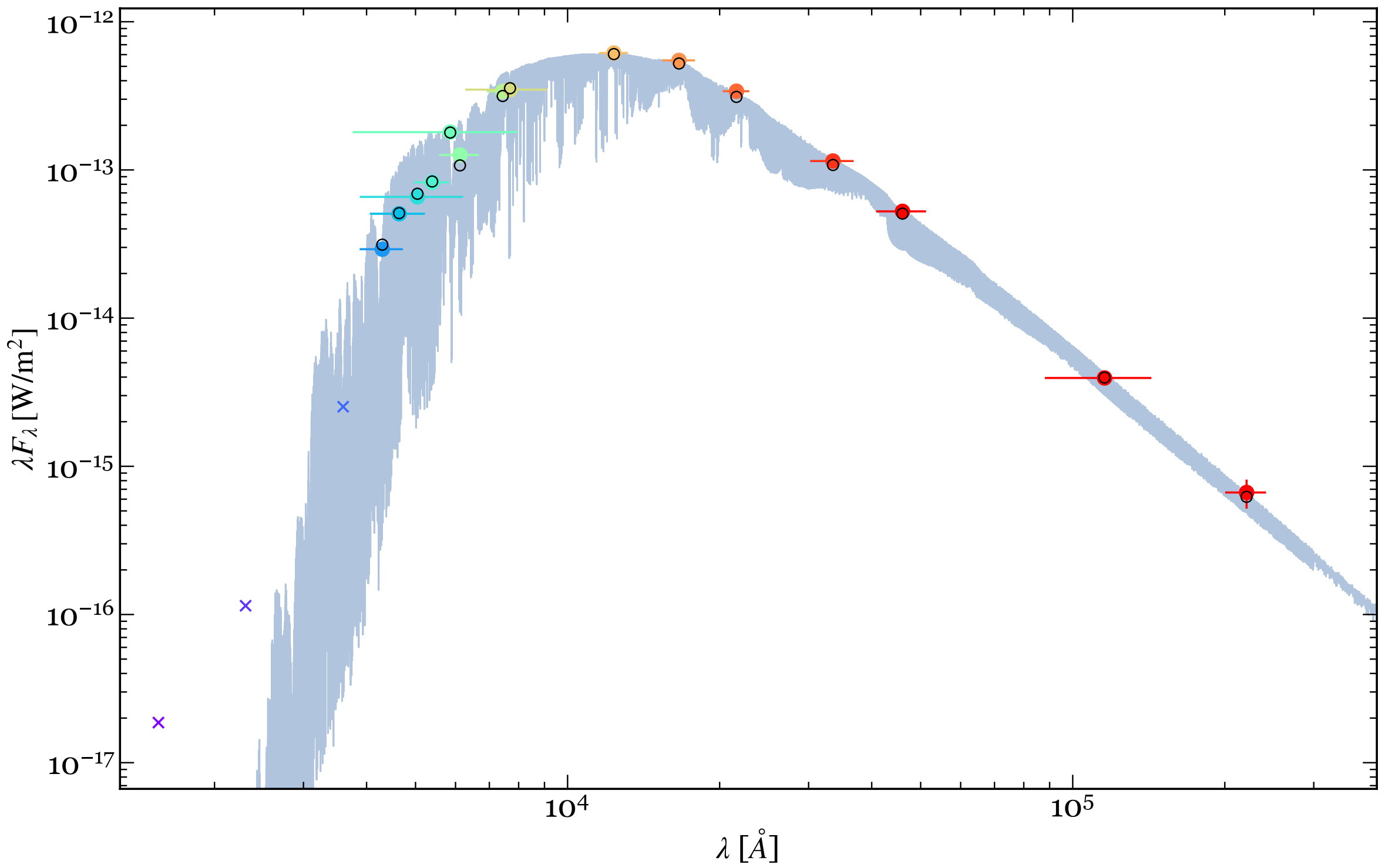

Figure 3.10: Spectral energy distribution of LP 167–071 (J10384+485, M3.0 V). The empirical fluxes (coloured empty circles, following the same colour scheme as in Fig. 3.5) are overimposed on the best-fitting `BT-Settl CIFIST` spectrum (grey; $T_{\rm eff}$ = 3300 K and log $g$ = 5.5). The modelled fluxes are depicted as grey empty circles. Photometric data in the ultraviolet are shown as crosses, and are not considered in the modelling. Horizontal bars represent the effective widths of the bandpasses, while vertical bars (visible only for relatively large values) represent the flux uncertainty derived from the magnitude and parallax errors.

Complementing the `VOSA` automatic identification of photometric outliers in the SED, we inspected all the 1843 individual SEDs and marked 7.1 % of all data points as '`Bad`', as they had bad quality flags (Sect. 3.1.2) or clearly deviated from the SED trend in the optical and, therefore, were not included in the model fitting. After a careful inspection, we also ignored the possible infrared excesses automatically detected by `VOSA`, even for the two single, very young stars in the Taurus-Auriga association (see Sect. 3.2.2).

In Fig. 3.11 we show the distributions of luminosities, effective temperatures, and surface gravities stacked by spectral type. We derived luminosity values ranging from 1.54 $10^{-5}\,\mathcal{L}_\odot$ for the nearby L8 dwarf DENIS-P J0255-4700, to 0.3276 $\mathcal{L}_\odot$ for the K7 V dwarf HD 196795, except for a very young early M member of the $\beta$ Pictoris moving group, namely StKM 1–1155, which has an exceptional luminosity of 1.8817 $\mathcal{L}_\odot$. Although very similar, our luminosities supersede those tabulated by Schweitzer et al. (2019) for the M dwarfs in the CARMENES GTO survey, as we updated some parallactic distances and APASS9 and PS1 DR1 optical magnitudes.

### 3.2.2 Young star candidates

In the two panels of Fig. 3.12 we display two related plots: a Hertzsprung-Russell (HR)[8] diagram with luminosities and effective temperatures from our `VOSA` analysis, and a colour-absolute magnitude diagram with *Gaia* and 2MASS data. After discarding stars with poor astro-photometric data or very close companions, we identified overluminous stars that departed from the main sequence defined by "regular" single stars in the $M_G$ versus $G - J$ diagram, as in the case of StKM 1–1155. We searched the literature for information on their membership in known young kinematic groups (i.e. younger than or of the age of the Hyades, $\tau \lesssim 0.6$ Ga – Perryman et al., 1998; Montes et al., 2001; Zuckerman & Song, 2004). The 36 identified overluminous stars include members of very young associations and moving groups (Taurus-Auriga, Upper Scorpius, $\beta$ Pictoris), moderately young groups (Argus, Tucana-Horologium, Columba, IC 2391 supercluster), middle-aged open clusters and groups (Pleiades, AB Doradus, Hyades), and a miscellanea classification including one star of about 100 Ma (Cruz et al., 2009), an active one that kinematically belongs to the young Galactic disc (Jeffers et al., 2018), and an ultra-fast-rotating, H$\alpha$-variable, X-ray-emitting, young star candidate (IZ Boo – Stephenson, 1986; Fleming, 1998; Mochnacki et al., 2002; Jeffers et al., 2018). The 36 stars and their respective references are listed in Table 3.4. As expected, these sources are also overluminous in the Hertzsprung-Russell diagram. Besides, there are a dozen stars neither tabulated by us nor classified as young star candidates in the literature that are also overluminous, which will deserve attention in forthcoming works. Many of these stars will be actually proven to be two or even more stars disguised as one. In Sect. 4.3.2, and specially in Sect. 4.4.2, we explore in more detail these cases.

[8]This important relation of profound impact on the understanding of stars traces back to the investigations of Hertzsprung (1923), Russell et al. (1923), and Russell (1928).

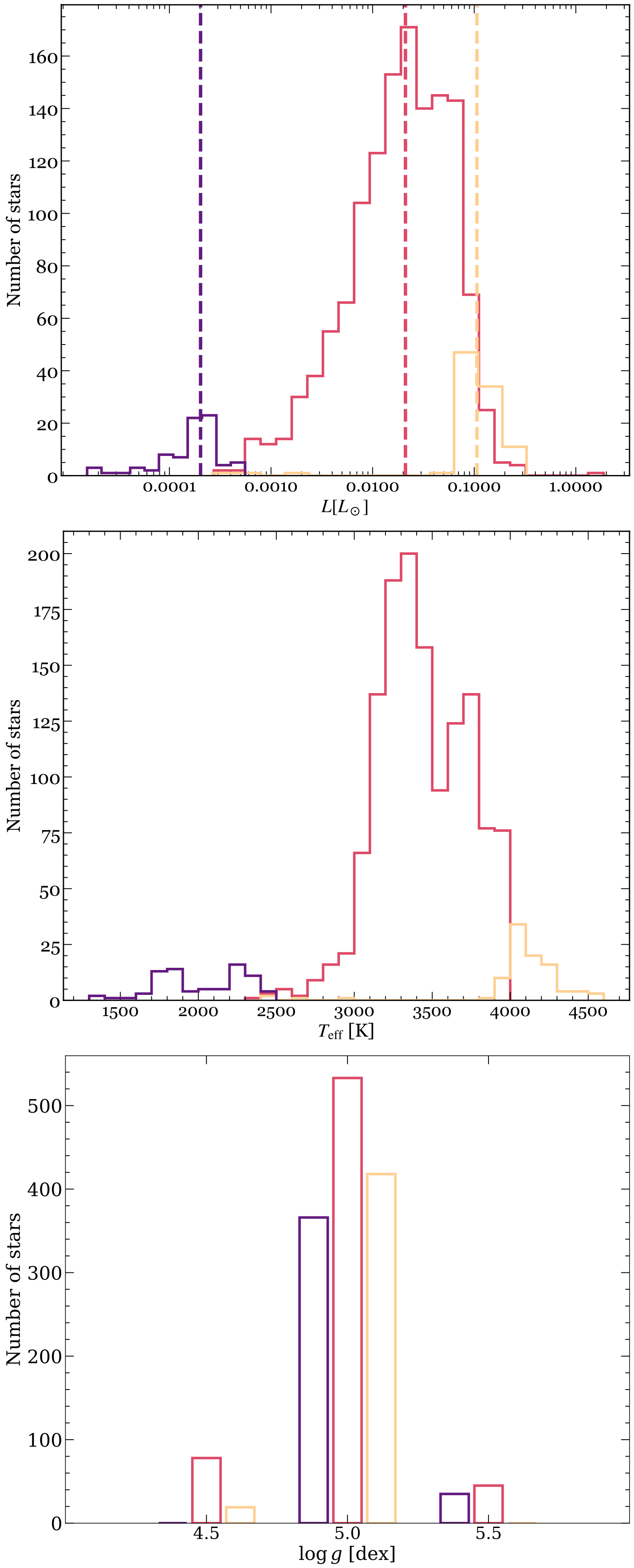

Figure 3.11: Distribution of bolometric luminosities (*top*), effective temperatures (*middle*), and surface gravities (*bottom*) for K (yellow), M (red), and L (violet) dwarfs. Dashed vertical lines mark the median values of bolometric luminosities.

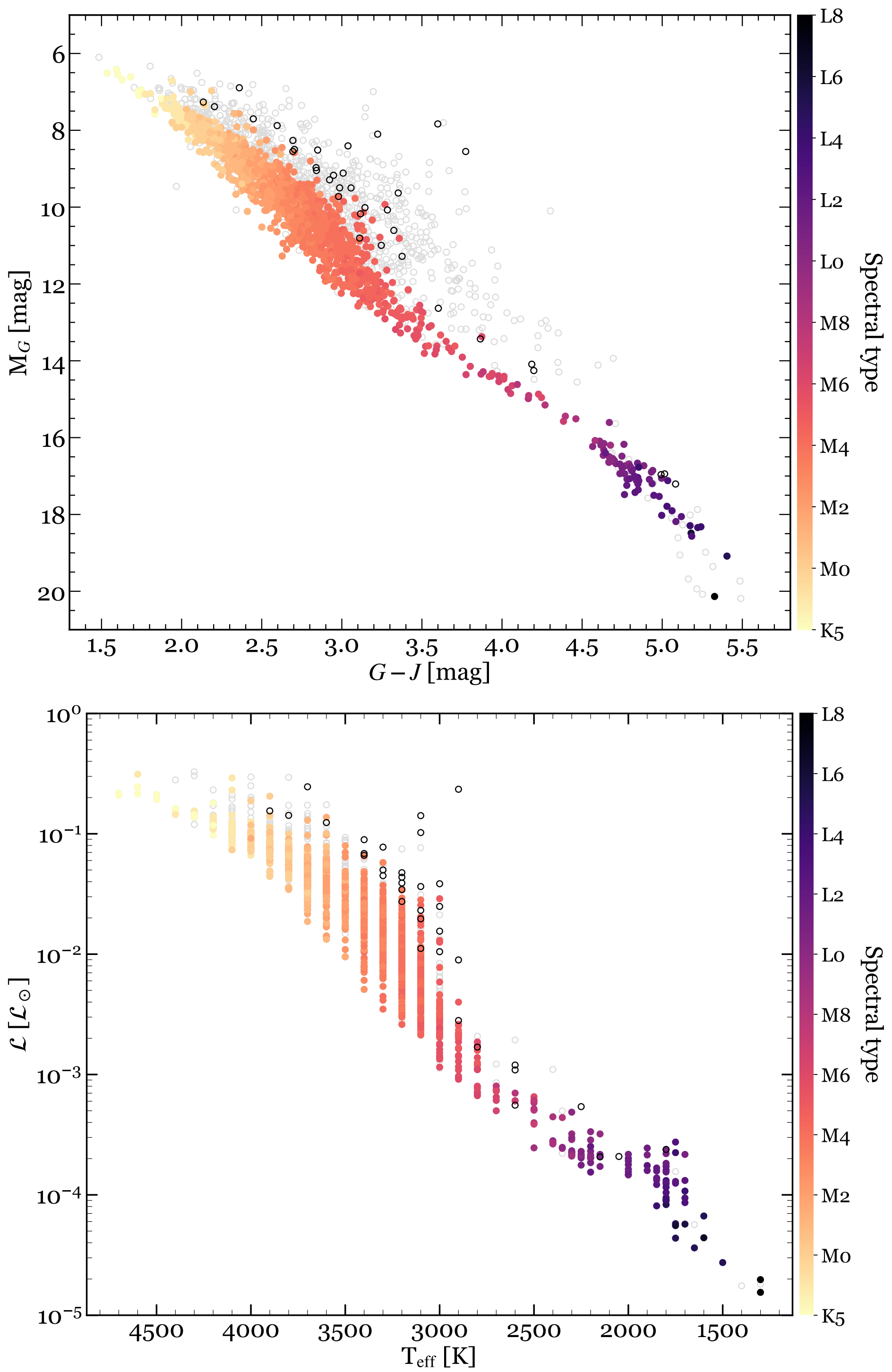

Figure 3.12: Absolute magnitude $M_G$ against $G - J$ colour (*top*), and bolometric luminosity against effective temperature from VOSA (*bottom*). In the top panel, empty grey circles represent stars with poor photometric quality data in $G$, $J$, or both passbands (Sect. 3.1.2), poor astrometric quality data (ruwe >1.41) or non-parallactic distances (Sect. 3.1.3), and close binary stars (Sect. 3.1.4). In the bottom panel, empty grey circles represent stars with poor astrometric quality data or non-parallactic distances, and close binary stars. In both panels, black empty circles are the 36 known young overluminous stars identified in our sample. The remaining "regular" stars are colour-coded by spectral type.

Table 3.4: Overluminous young stars identified in our sample.

| Karmn | Name | Stellar kinematic group | Reference |
|---|---|---|---|
| J0045+1634[a] | 2MUCD 20037 | Argus | Gagné et al. 2014 |
| J01352-072 | Barta 161 12 | $\beta$ Pictoris | Alonso-Floriano et al. 2015a |
| J02443+109W | MCC 401 | $\beta$ Pictoris | Janson et al. 2017 |
| J03510+142 | 2MASS J03510078+1413398 | $\beta$ Pictoris | Gagné et al. 2015 |
| J03548+163 | LP 413–108 | Hyades | Crain et al. 1986 |
| J03565+319 | HAT 214–02089 | Hyades? | Röser et al. 2011 |
| J04206+272 | XEST 16–045 | Taurus | Scelsi et al. 2007 |
| J04238+092 | LP 535–073 | Hyades | Weis et al. 1979 |
| J04238+149 | IN Tau | Hyades | van Rhijn & Raimond 1934 |
| J04252+172 | V805 Tau | Hyades | van Altena 1966 |
| J04369-162 | 2MASS J04365738–1613065 | Tuc-Hor | Malo et al. 2014d |
| J04414+132 | TYC 694–1183–1 | Hyades | Johnson et al. 1962 |
| J0443+0002[a] | 2MUCD 10320 | $\beta$ Pictoris | Alonso-Floriano et al. 2015a |
| J04433+296 | Haro 6-36 | Taurus | Haro et al. 1953 |
| J04595+017 | V1005 Ori | $\beta$ Pictoris | Alonso-Floriano et al. 2015a |
| J05019+011 | 1RXS J050156.7+010845 | $\beta$ Pictoris | Alonso-Floriano et al. 2015a |
| J05084-210 | 2MASS J05082729–2101444 | $\beta$ Pictoris | Alonso-Floriano et al. 2015a |
| J0608-2753[a] | 2MASS 06085283-2753583 | $\beta$ Pictoris | Alonso-Floriano et al. 2015a |
| J07310+460 | 1RXS J073101.9+460030 | Columba | Malo et al. 2013 |
| J07446+035 | YZ CMi | $\beta$ Pictoris | Alonso-Floriano et al. 2015a |
| J09449-123 | G 161-071 | Argus | Bartlett et al. 2017 |
| J11519+075 | RX J1151.9+0731 | $\beta$ Pictoris | Alonso-Floriano et al. 2015a |
| J12508-213 | DENIS J125052.6–212113 | Pleiades? | Clarke et al. 2010 |
| J14200+390 | IZ Boo | Young? | Mochnacki et al. 2002 |
| J14259+142 | StKM 1–1155 | $\beta$ Pictoris | Alonso-Floriano et al. 2015a |
| J15079+762 | HD 135363 B | IC 2391 | Montes et al. 2001; Lépine & Bongiorno 2007 |
| J15166+391 | LP 222–065 | Young disc | Jeffers et al. 2018 |
| J1552+2948[a] | 2MASS J15525906+2948485 | ~100 Ma | Cruz et al. 2009 |
| J15597+440 | RX J1559.7+4403 | AB Dor | Binks & Jeffries 2016 |
| J16102-193 | K2–33 | USco | Preibisch et al. 2001 |
| J17572+707 | LP 044–162 | Argus? | Gagné et al. 2015 |
| J21100-193 | BPS CS 22898-0065 | $\beta$ Pictoris | Alonso-Floriano et al. 2015a |
| J22088+117 | 2MASS J22085034+1144131 | $\beta$ Pictoris | Shkolnik et al. 2017 |
| J23228+787 | NLTT 56725 | Columba | Makarov et al. 2007; Montes et al. 2018 |
| J23301-026 | 2MASS J23301129–0237227 | $\beta$ Pictoris | Alonso-Floriano et al. 2015a |
| J23317-027 | AF Psc | $\beta$ Pictoris | Alonso-Floriano et al. 2015a |

[a] Ultra-cool dwarfs from Smart et al. (2017) not in the CARMENES catalogue of M dwarfs.

### 3.2.3 Diagrams

We present and discuss several diagrams involving colours, absolute magnitudes, and bolometric corrections.

#### Colour-spectral type

We computed 20 average colour indices for adjacent filters and their standard deviation for late-K to late-L dwarfs, using only the good quality photometric data. We list them in Table C.2. The size of the sample for each colour index and spectral type is shown in parentheses. Colour indices computed from samples with less than four elements are included for completeness, albeit with a word of caution. As expected, the amount of data available in the ultraviolet and optical blue passbands decreases for later spectral types (see again Fig. 3.5). In particular, for spectral types M4 V and earlier we have all possible colour combinations, and for spectral types later than M4 V and up to L5 we have all possible colour combinations only between $G$ and $W3$. This colour compilation complements, and most of the time supersedes, previous determinations (Bessell et al., 1998; Dahn et al., 2002; Hawley et al., 2002; Knapp et al., 2004; West et al., 2005; Covey et al., 2007; Zhang et al., 2009; Bochanski et al., 2010; Lépine et al., 2013; Pecaut & Mamajek, 2013; Rajpurohit et al., 2013; Davenport et al., 2014; Filippazzo et al., 2015; Mann et al., 2015; Best et al., 2017).

From all the possible combinations, the *Gaia* DR2-2MASS colour $G - J$ provides one of the most solid estimators of spectral type from late-K to mid-L dwarfs. This is illustrated in Fig. 3.13. Firstly, $G - J$ covers a wide range in colour of about 3.6 mag between K5 V and L8, with a slight flattening restricted to the late L objects. Secondly, it exhibits one of the smallest dispersions in late-M and L dwarfs among all analysed colours, with a median deviation of 0.08 mag. Thirdly, the $G$ and $J$ passbands offer a high availability in this spectral type range, with 97.7 % and 100 % completeness in $G$ and $J$, respectively. Also, faint objects benefit from the reliability of 2MASS and *Gaia* DR2 photometry. This colour index is superior to previous colour indices used to discriminate late spectral types, such as $i' - J$ (Reid et al., 2001; Hawley et al., 2002; West et al., 2005; Covey et al., 2008), and finds a compromise between completeness, photometric data quality (*Gaia* and 2MASS), scatter of the data, and colour interval spanned by the sequence. On the contrary, the use of *Gaia*-only and 2MASS-only colours for spectral typing presents some serious caveats, from the degeneracy of $G_{B_P} - G_{R_P}$ for spectral types M8 V and later, to the narrow interval of 1 mag of $G - G_{R_P}$ from late-K to late-M dwarfs and its pronounced flattening from late-M to mid-L dwarfs, to the blueing of $J - H$ in the M-dwarf domain. In words, it is not possible to differentiate a mid-M dwarf from an L object using the $G_{BP} - G_{RP}$ colour alone, given its degeneracy.

In Fig. 3.14, we plot six additional colour-spectral type diagrams that show the behaviour of other passbands from the near-ultraviolet to mid-infrared, and their adequacy for spectral type estimation. In all cases, data with poor photometric quality are included as empty grey circles, but not considered for any calculation. Firstly, the optical-mid-infrared $G - W3$ colour serves as a useful complement for the $G - J$ colour, especially in the late-M and early-L regime. The $G - W3$ colour also exhibits a monotonic, low-scatter, steady increase from K5 V to L8, although the median of the dispersion is 0.17 mag, twice the value obtained with the $G - J$ index. Additionally, it benefits from the widest interval in colour of all the diagrams, with approximately 7 mag separating K5 V and L8.

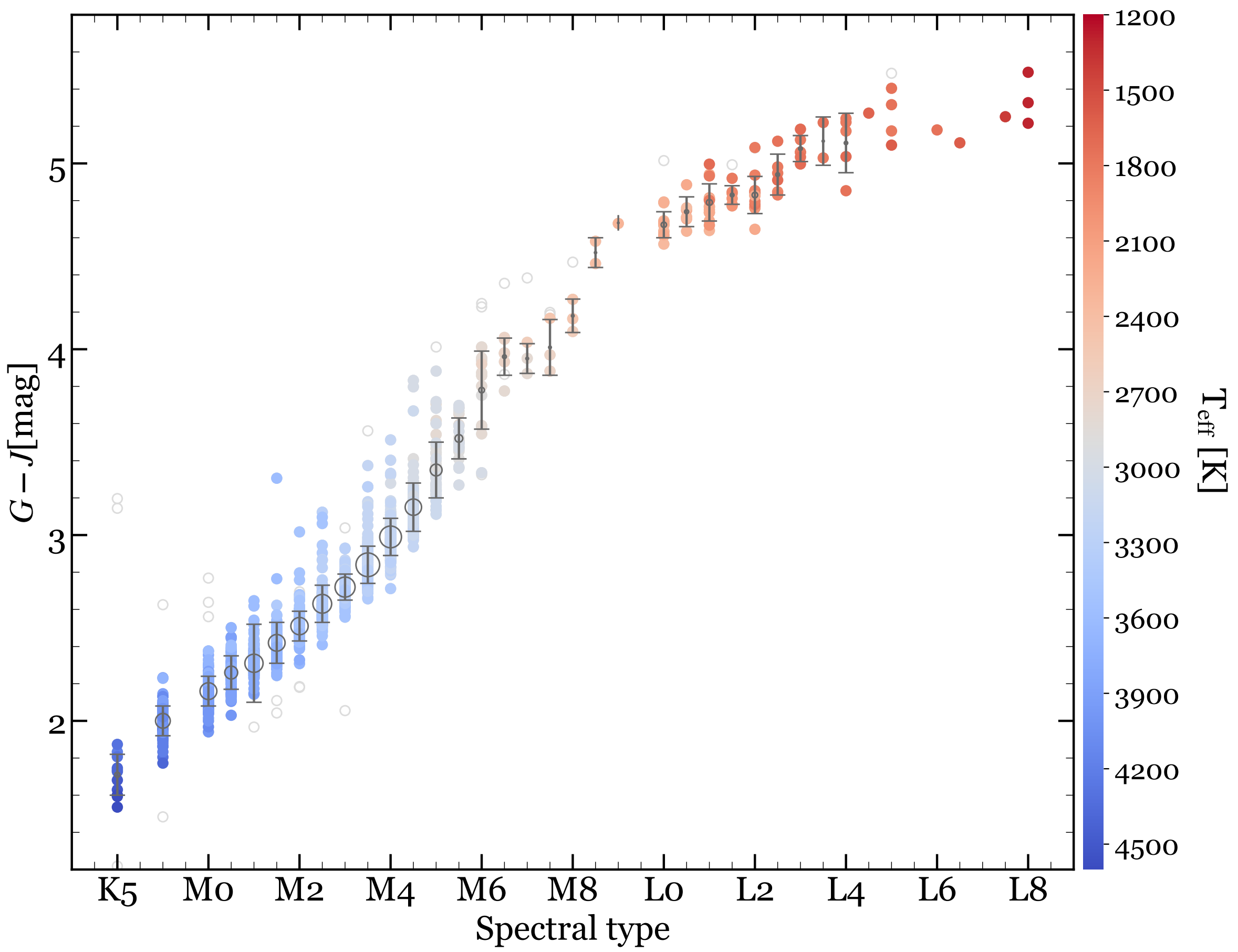

Figure 3.13: $G - J$ colour against spectral type. Grey empty circles mark the average colour for each spectral type with a size proportional to the number of stars and vertical bars accounting for their standard deviation in spectral types with more than one valid colour value. Empty light grey circles on the background depict bad photometric data, as explained in Sect. 3.1.2, and their values are not considered in the calculations of the average colours.

The purely optical colour $r' - i'$, extensively used in the literature, can help to determine spectral types of late-K to late-M dwarfs, but it fails to discriminate the types for cooler objects. It peaks at about 2.8 mag (around M7–8 V), and becomes bluer beyond this point, as shown by, e.g. Hawley et al. (2002) and Liebert & Gizis (2006).

The purely infrared colour $J - W2$ exhibits a remarkably low dispersion from M0 V to M8 V (less than 0.06 mag), but it covers a colour interval of only 0.5 mag. The colour $G_{R_P} - W1$ offers an adequate alternative, with a dispersion slightly larger in the same range (0.09 mag), but spanning five times the colour interval. Furthermore, colours including the $W4$ passband suffer from poor quality data for spectral types M8 V and later.

The $NUV - G_{RP}$ colour is sensitive to both spectral type and ultraviolet flux excesses, which may be caused by chromospheric activity and/or interaction between close binaries. The first case includes "regular" stars later than M3–4 V at the boundary of stellar full convection. The second case comprises, according to Ansdell et al. (2015), young stars (including all our overluminous young stars except one Hyades member) and unidentified binaries, which include unresolved background ultraviolet sources, unresolved old binaries with white dwarf companions, and short-period ($P < 10$ d) tidally interacting binaries that induce ongoing activity on each other. These phenomena give rise to a distinguishable

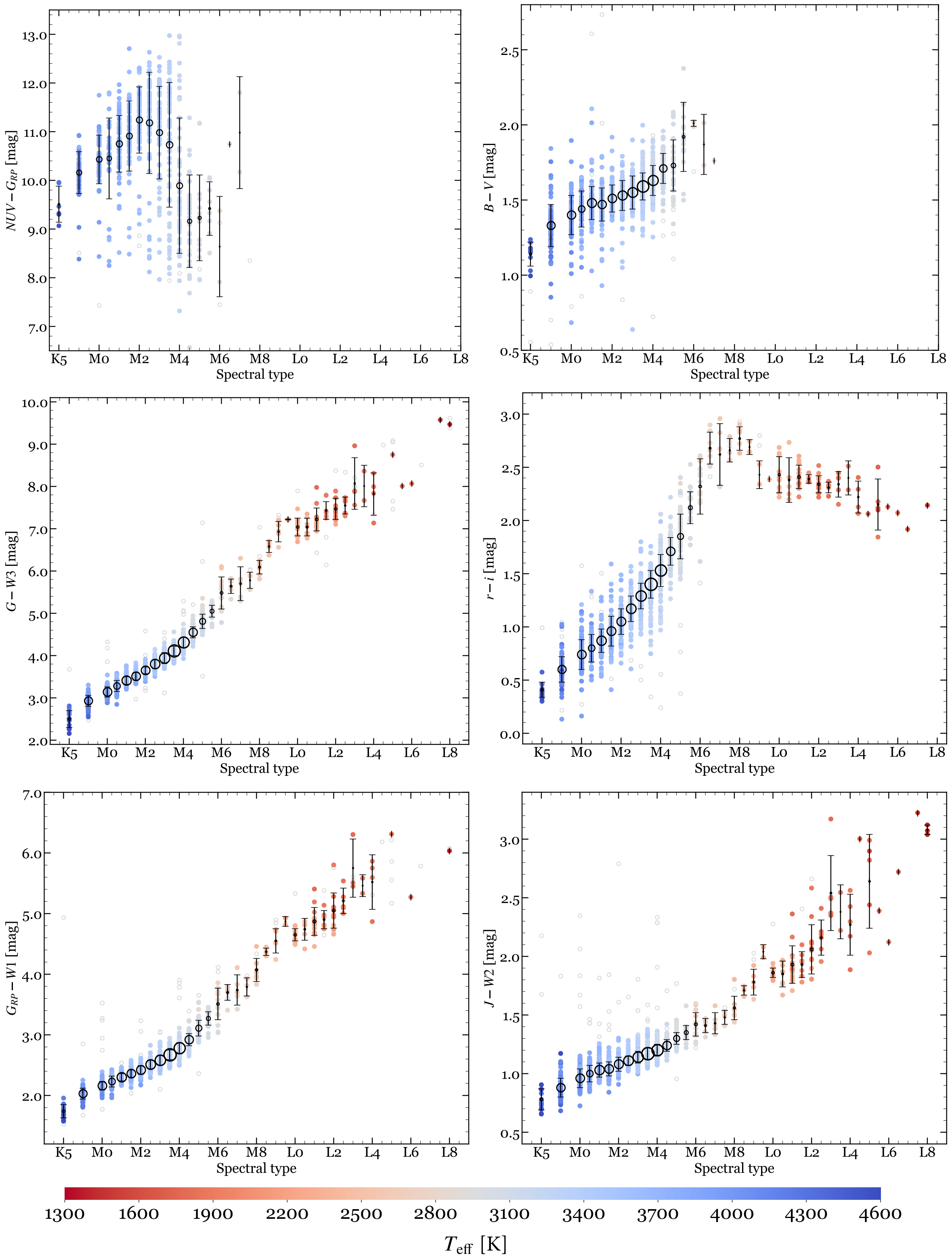

Figure 3.14: Six representative colour-spectral type diagrams, colour-coded by effective temperature.

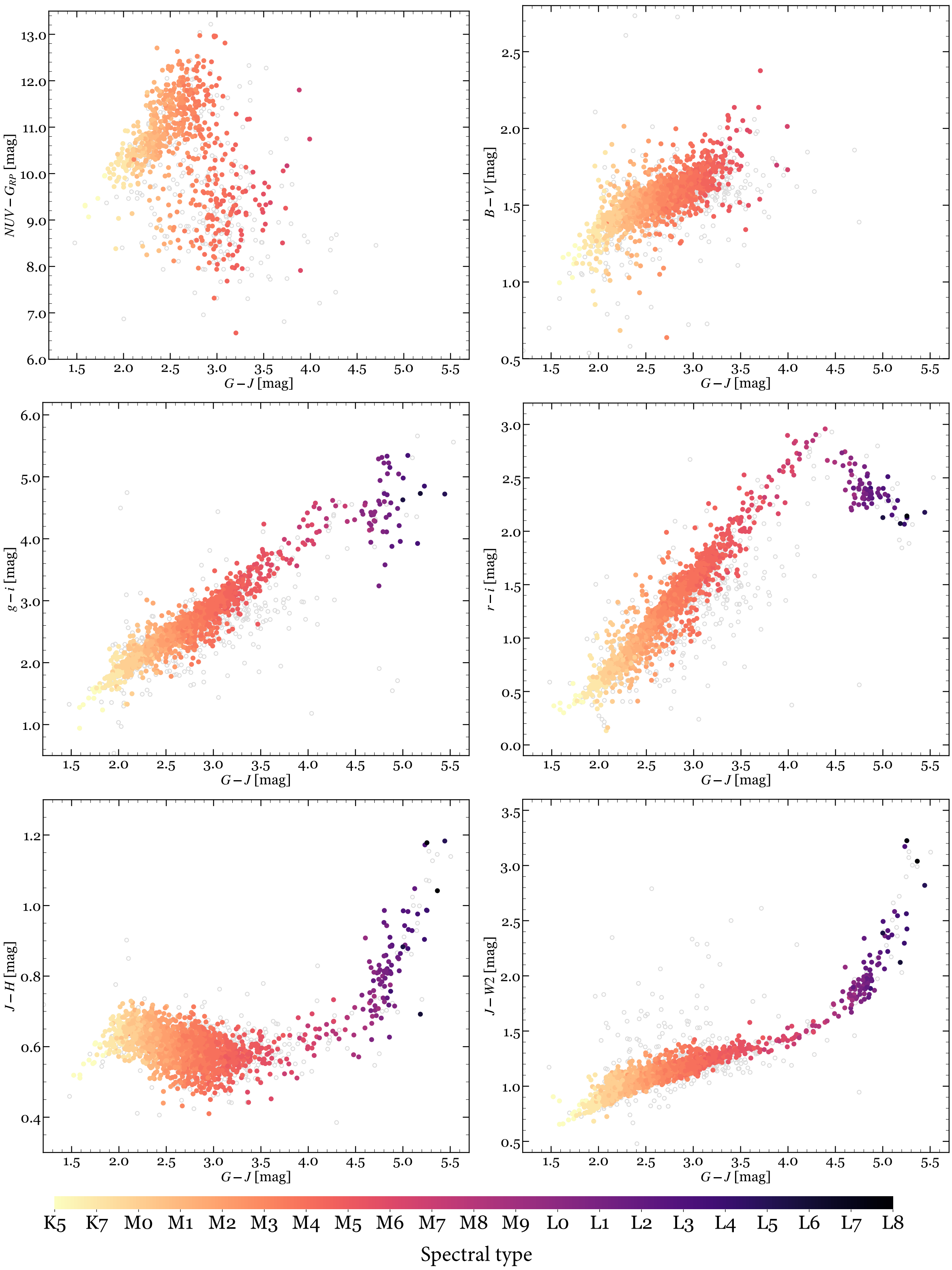

Figure 3.15: Six representative colour-colour diagrams, colour-coded by spectral type.

population appended to the main sequence.

Finally, the optical $B-V$ colour became a commonly used index in the literature, including Bessell et al. (1998), Ramírez & Meléndez (2005), Casagrande et al. (2008), Smith (2018), Sun et al. (2018), del Burgo & Allende Prieto (2018), or Cochrane & Smith (2019), just to name a few. However, the $B-V$ colour has some disadvantages in the M-dwarf domain:

- Both $B$ and $V$ lack the completeness in the optical range that other passbands, such as $G_{BP}$, $r'$, $i'$, or $G_{RP}$, deliver.
- $B-V$ fails to produce a photometric sample statistically that is consistent beyond M5 V, while the *Gaia* DR2, 2MASS, or AllWISE passbands succeed.
- $B-V$ does not correlate with spectral type beyond M5 V.
- The width of the colour interval from late K to mid M is 1 mag, only a few times the scatter of the main sequence (0.12 mag), with a striking flattening between M0 V and M3 V.
- The mean uncertainties of $B$ and $V$ in our sample are 0.056 mag and 0.048 mag, respectively. For comparison, the same parameters for $G$ and $J$ are 0.0012 mag and 0.029 mag, respectively.

Therefore, we discourage the use of $B-V$ as an estimator of spectral type for stars cooler than K5 V. This is especially applicable when the *Gaia* DR2 (and 2MASS or AllWISE) magnitudes are available. The same reasoning above also applies to the $B_T$ and $V_T$ Tycho-2 passbands, which are even less complete.

### Colour-colour

Contrary to apparent magnitudes, observed stellar colours do not depend on distances and is possible to learn relevant information about stars from their colour indices alone. If coming from accurate photometric data, colours can serve as reliable estimators of spectral types. For instance, Strauss et al. (1999) found an extremely red object using Sloan Digital Sky Survey (SDSS) $i'-z'$ colour, which came to be the first T-type object (i.e. methane dwarf) discovered. In the same month, and even in the same journal volume but a few pages later, Burgasser et al. (1999) reported four field-methane dwarf, discovered using 2MASS $J-H$ and $H-K_s$ colours[9]. Tsvetanov et al. (2000) added a second object to the list of field Methane dwarfs, also using colours derived from 2MASS and SDSS. Soon after, Leggett et al. (2000) discovered three more similar objects using colour indices.

Diagrams that compare colours turn out to be also an useful diagnostic of the presence of binary stars, young stellar objects, white dwarfs, giant stars, or even quasars (Hawley et al., 2002; Covey et al., 2007; Casagrande et al., 2008; Davenport et al., 2014; Mann et al., 2015). These sources appear as outliers from a well-defined stellar locus, and they can be isolated and studied individually. Stars significantly brighter in a given colour exhibit an excess that may suggest the existence of relevant physical features, such as circumstellar disks associated to the early stages of evolution, which are the cause of a reddening effect (see for example White & Ghez, 2001).

As in the colour-spectral type diagrams, main sequence stars occupy a well-defined locus in colour-colour diagrams. In spite of the degeneracy beyond M8 V, the narrowest main sequence is observed in the 2MASS-*Gaia* $G_{BP}-G_{RP}$ versus $G-J$ colour-colour diagram shown in the top panel of Fig. 3.16.

[9] The authors identified extremely red objects with $J-H$ <0.2 mag and $H-K_s$<0.2 mag. Figure 3.16 (Chapter 3), can help to contextualise a Methane object as compared with an M dwarf.

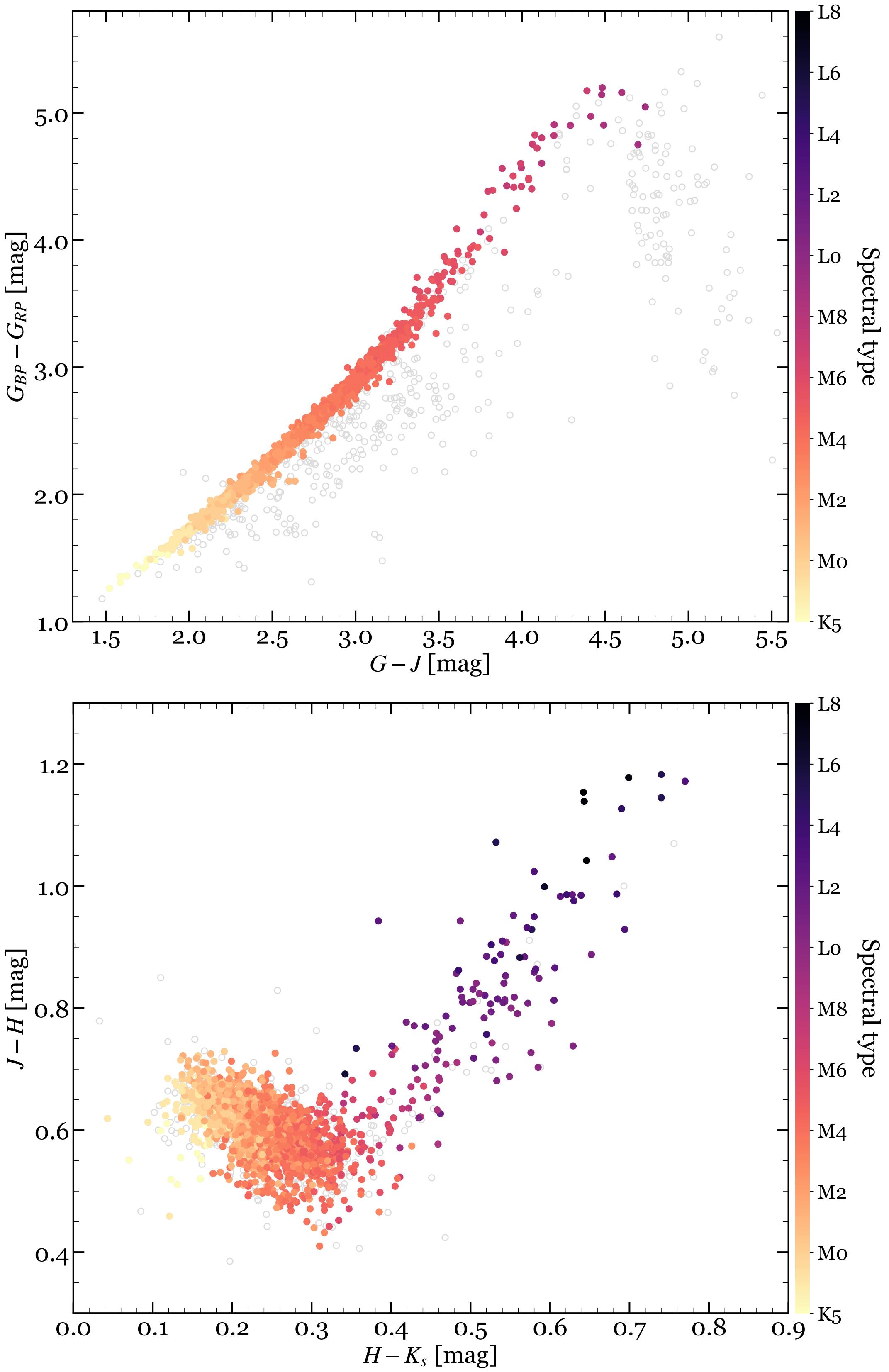

Figure 3.16: Colour-colour diagrams representing $G_{BP} - G_{RP}$ vs. $G - J$ (*top*) and $J - H$ vs. $H - K_s$ (*bottom*). In both panels, empty grey circles represent stars with poor photometric quality data in any of the involved passbands, close binaries, or young stars. The remaining "regular" stars are colour-coded by spectral type.

Outliers in the diagram are mostly unresolved binaries and young stars for colours bluer than $G - J \sim$ 4.5 mag (spectral types earlier than M8 V), albeit other possibilities also exist. For example, G 78–3 (J02455+449), at 57.8 pc (Gaia Collaboration et al., 2018b), is an M5 V star (Hawley et al., 1996), which exhibits a $G - J$ colour typical of an early-M dwarf. For colours redder than $G - J \sim 4.5$ mag, we confirm the findings of Smart et al. (2019), who reported an unreliability in *Gaia* blue-band photometry of very late objects with $G_{BP} > 19.5$ mag due to background underestimation by the *Gaia* automatic pipeline (Gaia Collaboration et al., 2018a; Evans et al., 2018; Smart et al., 2019). As a comparison, in Fig. 3.16 we also show a widely used, 2MASS-only, colour-colour diagram (Kirkpatrick et al., 1999; Knapp et al., 2004; Lépine & Shara, 2005; Hewett et al., 2006; Covey et al., 2007). There, late-K to late-M dwarfs occupy a compact region that ranges from $H - K_s \sim 0.15$ mag, $J - H \sim 0.65$ mag to $H - K_s \sim 0.30$ mag, $J - H \sim$ 0.55 mag, while later stars and brown dwarfs become redder (Kirkpatrick et al., 1999, and references above). We did not notice any near-infrared flux excess, as found in young T Tauri M-type stars and brown dwarfs with warm circumstellar discs (Carpenter, 2001; Caballero et al., 2004; Hernández et al., 2008).

In Fig. 3.15 we display a selection of six additional colour-colour diagrams. In all cases we plot far-ultraviolet to mid infrared-colours against $G - J$. Apart from the stars with poor photometric quality, we also discarded the 2MASS magnitudes of the extraordinarily red 2MUCD 20171 (J03552+113; Faherty et al., 2013) and blue SDSS J141624.08+134826.7 (J1416+1348A; Burgasser et al., 2010) ultracool dwarfs, which were clear outliers in many colour-colour diagrams involving 2MASS magnitudes. As in the colour-spectral type diagrams, the two colour-colour diagrams involving the bluest colours illustrate the two populations of ultraviolet active and inactive sources ($NUV - G_{RP}$) and the poor spectral sequence based on $B - V$ colour. The two diagrams involving UCAC4/SDSS9/APASS9/CMC15/PS1 DR1 $g'r'i'$ passbands, which will also be used at the Vera C. Rubin Observatory for the Legacy Survey of Space and Time (LSST), show a slightly larger spread than *Gaia* data and the double slope of the $r' - i'$ colour also found by Hawley et al. (2002) and Liebert & Gizis (2006). Interestingly, $g' - i'$ has a smaller dispersion in the late K and M dwarf domain than $r' - i'$, but a much larger dispersion at $G - J \gtrsim$ 4.5 mag. This extra scatter at the reddest colours is more likely due to the intrinsic spectral variations at the M/L boundary (à la Hawley et al. 2002; e.g. metallicity) than due to data analysis systematics or Poissonian error at the survey magnitude limits (à la Smart et al. 2019; e.g. background). Finally, the colour-colour diagrams with near-infrared 2MASS and AllWISE data (specially $W3$ and $W4$) are very sensitive to $T_{\rm eff}$ variations at the L spectral types, but quite insensitive in the late-K and M dwarf domain. However, their sensitivity to metallicity must be investigated in detail with, for example, resolved photometry of M-dwarf wide common proper motion companions to FGK-type stars with well-determined stellar astrophysical parameters (Montes et al., 2018).

### Absolute magnitude-colour

In Fig. 3.12 we show the $M_G$ versus $G - J$ diagram. In Fig. 3.17 we show a similar diagram (see more examples in e.g. Dupuy & Liu, 2012), but for $r'$ instead of $G$, and we overplot a quadratic polynomial fit to 278 CARMENES GTO target stars with spectral types ranging from K7 V to M9 V (Reiners et al., 2018b). All of them have well-behaved *Gaia* astrometric solutions (i.e. `ruwe` < 1.41; Fig. 3.7) and do not have close companions (Cortés-Contreras et al., 2017b; Baroch et al., 2018), extreme values of metallicity (Alonso-Floriano et al., 2015b; Passegger et al., 2018, 2019, 2020), young ages (Tal-Or et al., 2018), or large-amplitude photometric variability (Díez Alonso et al., 2019). We also fitted another quadratic polynomial to the $M_G$ versus $G - J$ data of the GTO stars. Thus, with the parameter fits in Table 3.5 and only $r'$ or $G$ and $J$ magnitudes, one can estimate a stellar distance with a median accuracy of 36 % for stars in the colour ranges listed in the column $\Delta X$. From our knowledge of the CARMENES GTO stars, the most important contributor to the fit uncertainty is not the parallax or magnitude error, stellar variability, or unresolved multiplicity, but the intrinsic scatter of the M-dwarf colour sequence due to different metallicity.

Table 3.5: Fit parameters for several empirical relations.

| $Y^a$ [mag] | $X$ [mag] | $a$ [mag] | $b$ [mag$^{-1}$] | $c$ [mag$^{-2}$] | $d$ [mag$^{-3}$] | $e$ [mag$^{-4}$] | $R^2$ | $\Delta X$ [mag] |
|---|---|---|---|---|---|---|---|---|
| $M_{r'}$ | $r' - J$ | +8.38 ± 2.68 | −2.74 ± 2.36 | +1.47 ± 0.68 | −0.132 ± 0.063 | 0 | 0.9398 | [2.0, 5.1] |
| $M_G$ | $G - J$ | +16.24 ± 4.57 | −13.04 ± 4.80 | +5.64 ± 1.66 | −0.622 ± 0.188 | 0 | 0.9308 | [2.0, 4.0] |
| $\log \mathcal{L}/\mathcal{L}_\odot$ | $M_J$ | +2.051 ± 0.075 | −0.662 ± 0.030 | +0.0267 ± 0.0039 | −0.00102 ± 0.00016 | 0 | 0.9923 | [4.4, 11.2) |
| | | −3.906 ± 0.998 | +0.334 ± 0.156 | −0.0263 ± 0.0061 | 0 | 0 | 0.9477 | [11.2, 14.8] |
| $\log \mathcal{L}/\mathcal{L}_\odot$ | $M_G$ | +0.145 ± 0.201 | +0.074 ± 0.060 | −0.0382 ± 0.0060 | +0.00119 ± 0.00019 | 0 | 0.9901 | [6.4, 14.0) |
| | | −2.329 ± 0.687 | +0.092 ± 0.084 | −0.0103 ± 0.0025 | 0 | 0 | 0.9782 | [14.0, 20.2] |
| $BC_G$ | $G - J$ | +0.404 ± 0.187 | +0.161 ± 0.239 | −0.465 ± 0.112 | +0.1159 ± 0.0225 | −0.0115 ± 0.0017 | 0.9960 | (1.5, 5.4] |
| $BC_{r'}$ | $r' - J$ | +0.557 ± 0.085 | −0.036 ± 0.091 | −0.318 ± 0.035 | +0.0552 ± 0.0056 | −0.0037 ± 0.0003 | 0.9983 | (1.5, 7.5] |
| $BC_J$ | $G - J$ | +0.576 ± 0.094 | +0.735 ± 0.104 | −0.132 ± 0.038 | +0.0115 ± 0.0045 | 0 | 0.9547 | (1.5, 4.0] |
| $BC_{W3}$ | $G - J$ | −2.592 ± 0.667 | +5.845 ± 1.005 | −2.611 ± 0.559 | +0.586 ± 0.136 | −0.0496 ± 0.0122 | 0.9727 | (1.5, 4.0] |

[a] In all cases, the polynomial fits follow the form $Y = a + bX + cX^2 + dX^3 + eX^4$ and are applicable in the range $\Delta X$. In all cases, $R^2$ is the correlation coefficient from the Pearson product-moment matrix. These relations should be applied to solar-metallicity stars only (Sect. 3.3).

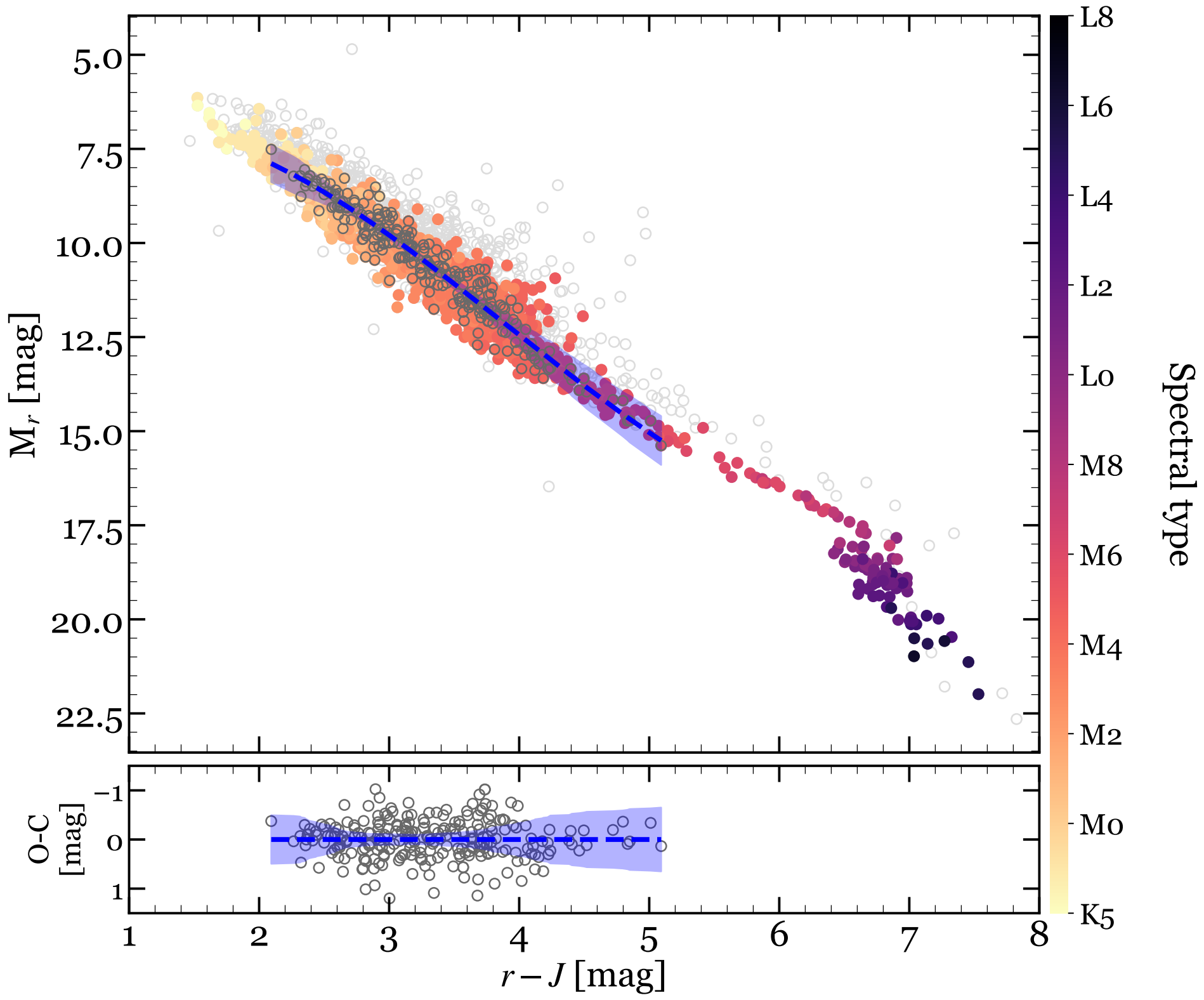

Figure 3.17: Same as Fig. 3.12 but for $M_{r'}$ vs. $r' - J$. The GTO stars in the sample are shown in dark grey. The blue dashed line represents the polynomial fit given in Table 3.5, with a blue shaded region for the 1-$\sigma$ uncertainty region. The fit residuals are shown in the small bottom panel.

The $M_G$ versus $G - J$ relation is particularly helpful because, although there are about 420 million sources with known *Gaia* DR2 and 2MASS magnitudes (Marrese et al., 2019), there are several million near-infrared sources that lack a parallax determination. However, for the 31 single stars in our sample without published trigonometric parallaxes, we estimated photometric distances homogeneously from the $M_{r'}$ versus $r' - J$ relation assuming null extinction. Because of the relatively large uncertainty in the estimates, we did not use these photometric distances throughout our work, but only tabulated them in the on-line summary table described below. In general, we only recommend the use of these relations for estimating photometric distances for stars with solar-like metallicity, as well as good photometric quality (e.g. Bochanski et al., 2007, and see Sect. 3.3).

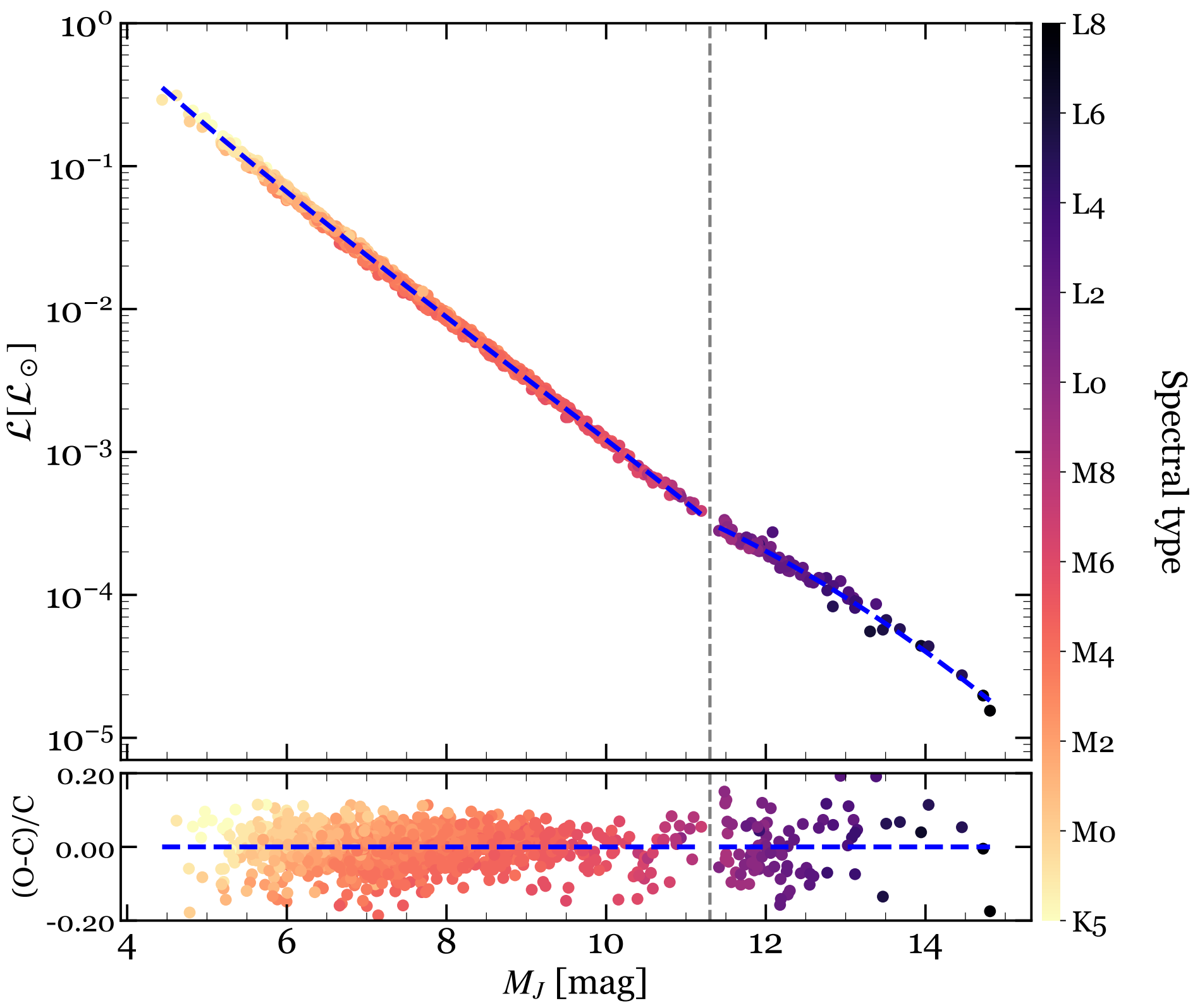

Figure 3.18: Same as Fig. 3.12, but for $\mathcal{L}_{\rm VOSA}$ vs. $M_J$ and normalised fit residuals in the small bottom panel. The vertical dashed line separates the three-degree (late-K and M dwarfs) and two-degree (L dwarfs) fit ranges.

### 3.2.4 Absolute magnitudes and bolometric corrections

The absolute magnitude of a star is directly related to its bolometric luminosity. In our sample, we found that the $J$-band absolute magnitude, $M_J$, provides the correlation with VOSA luminosity that is most complete and that has the smallest scatter. Figure 3.18 shows $\mathcal{L}_{\rm VOSA}$ (in solar units) versus $M_J$ fitted in the late-K- to late-M- and L-dwarf domains, with three-degree and two-degree polynomials, respectively. Although in Table C.2 we list the fit parameters for both luminosities from 2MASS $J$ and *Gaia* $G$, we preferred $J$ over $G$ because the larger effective width of the broad *Gaia* passband introduces more dispersion in the data, quantified by $R^2$. With these relationships in the M-dwarf domain, it is possible to estimate bolometric luminosities from absolute magnitudes $M_J$ and $M_G$ with a relative precision of 4.2 % and 4.5 %, respectively.

We calculated bolometric corrections, $BC_\lambda = M_{\rm bol} - M_\lambda$, for each investigated passband and plot them in Fig. 3.19. For the calculation, we followed the sign criterion of Böhm-Vitense (1989) and the definition of the absolute bolometric magnitude $M_{\rm bol}$ by IAU Resolution B2 (Mamajek et al., 2015), which is independent of the solar luminosity,

$$M_{\rm bol} = -2.5 \log_{10} \frac{\mathcal{L}_\star}{\mathcal{L}_0} = -2.5 \log_{10} \mathcal{L}_\star + M_{\rm bol,0}, \quad (3.1)$$

where $\mathcal{L}_\star$ and $\mathcal{L}_0$ are the luminosity of the star and the zero point of the absolute bolometric magnitude scale, respectively, and $M_{\rm bol,0} \equiv 71.197425$ mag.

From the sample of 2479 stars, for the following analysis we discarded: (*i*) stars with poor photometric or astrometric behaviour based on quality indicators (Sects. 3.1.2 and 3.1.3), (*ii*) close binaries and stars with photometry contaminated by bright nearby companions ($\rho$ < 5 arcsec; Sect. 3.1.4), (*iii*) overluminous objects known to belong to young associations and moving groups (Sect. 3.2.2), and (*iv*) stars with extraordinarily anomalous colours or absolute magnitudes.

Of the different $BC_\lambda$ versus $G - J$ combinations in Fig. 3.19, the narrowest sequence is that of $BC_G$. However, as illustrated by Fig. 3.20, the $BC_{r'}$ versus $r' - J$ sequence is even less scattered and spans wider ranges in X ((1.5, 7.5] mag in $r' - J$ versus (1.5, 5.4] mag in $G - J$) and Y ([−5.8, 0.0] mag in $BC_{r'}$ versus [−3.8, −0.1] mag in $BC_G$), probably due again to the broad $G$ effective width. We fitted polynomials to the relations $BC_G$ versus $G - J$, $BC_{r'}$ versus $r' - J$, $BC_J$ versus $G - J$, and $BC_{W2}$ versus $G - J$, and provide the corresponding parameters and correlation coefficients in Table 3.5. All in all, these relationships are complementary and can help to estimate relatively precise luminosities of M dwarfs with only a handful of widely available data ($G$ and $\varpi$ from *Gaia*, $J$ from 2MASS, $r'$ from a number of surveys including the forthcoming LSST).

### 3.2.5 Masses and radii

Finally, we derived radii $\mathcal{R}$ and masses $\mathcal{M}$ of the well-behaved stars. For $\mathcal{R}$, we used the Stefan-Boltzmann law $L = 4\pi\mathcal{R}^2\sigma T_{\rm eff}^4$ and $\mathcal{L}$ and $T_{\rm eff}$ from `VOSA`. For $\mathcal{M}$ we used the $\mathcal{M}$-$\mathcal{R}$ relation in Eq. 6 of Schweitzer et al. (2019), which came from a compilation of detached, double-lined, double-eclipsing, main-sequence, M-dwarf binaries from the literature[10]. This relation is applicable in a wide range of metallicities for M dwarfs older than a few hundred million years. `VOSA` also computes two stellar radii, one from a model dependent dilution factor and $d$, the other using the Stefan-Boltzmann law, but we did not use them.

## 3.3 Discussion

Here we compare our $\mathcal{L}$, $T_{\rm eff}$, $\mathcal{R}$, $\mathcal{M}$, and photometric data with those in the literature. Tables 3.6 and 3.7 and Figs. 3.21 to 3.28 illustrate the discussion. In particular, in Table 3.6 we show average values of $BC_G$, $BC_J$, $\mathcal{L}$, $T_{\rm eff}$, $\mathcal{M}$, and $\mathcal{R}$ for single, main-sequence stars with spectral types from K5 V to L2.0. The last column, $N$, indicates the number per spectral type bin of well-behaved stars (i.e. with no companions at $\rho$ < 5 arcsec, no overluminosity due to extreme youth, and of good *Gaia* DR2 astrometric and photometric quality). After applying a $3 - \sigma$ clipping, we calculated three-point rolling means and standard deviations between M0.0 V and L2.0 (e.g. tabulated values for M4.0 V stars are the mean and standard deviation of all individual $BC_G$ values of stars with spectral types M3.5, M4.0, and M4.5 V), and simple means and standard deviations for K5 V and K7 V stars. With these rolling means, we conservatively smoothed potential inter-type variability due to the small number of stars per bin at the latest spectral types and the typical uncertainty in M-dwarf spectral type determination, of 0.5 dex (Hawley et al., 2002;

[10] $\mathcal{M} = \alpha + \beta\mathcal{R}$, with $\alpha = -0.0240 \pm 0.0076\,\mathcal{M}_\odot$, $\beta = 1.055 \pm 0.017\,\mathcal{M}_\odot/\mathcal{R}_\odot$, and $\mathcal{M}$ and $\mathcal{R}$ in solar units.

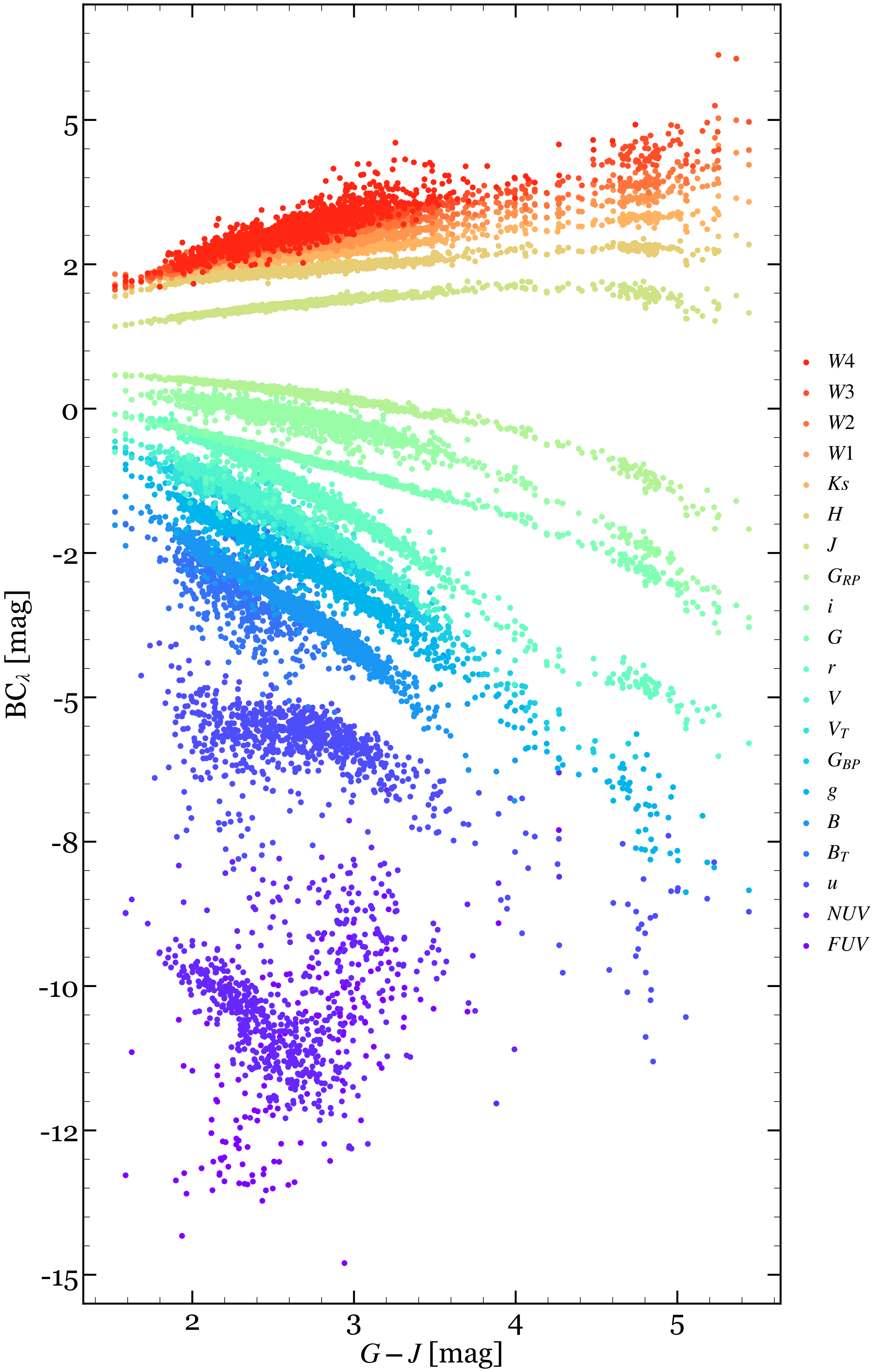

Figure 3.19: Bolometric corrections for every star and passband vs. $G - J$ colour. The coloured sets contain all stars with available photometry in $G$, $J$, and the respective passband.

Table 3.6: Average astrophysical parameters for K5 V to L5 objects.

| Spectral type | $BC_G$ [mag] | $BC_J$ [mag] | $\mathcal{L}$ [$10^{-4}L_\odot$] | $T_{\rm eff}$ [K] | $\mathcal{R}$ [$\mathcal{R}_\odot$] | $\mathcal{M}$ [$\mathcal{M}_\odot$] | $N$ |
|---|---|---|---|---|---|---|---|
| K5 V | −0.202 ± 0.065 | 1.492 ± 0.047 | 1760 ± 424 | 4430 ± 177 | 0.703 ± 0.054 | 0.718 ± 0.057 | 13 |
| K7 V | −0.385 ± 0.047 | 1.613 ± 0.029 | 1000 ± 217 | 4057 ± 100 | 0.634 ± 0.047 | 0.645 ± 0.049 | 74 |
| M0.0 V | −0.477 ± 0.082 | 1.654 ± 0.044 | 786 ± 256 | 3900 ± 146 | 0.600 ± 0.068 | 0.609 ± 0.072 | 104 |
| M0.5 V | −0.551 ± 0.070 | 1.689 ± 0.039 | 606 ± 209 | 3778 ± 112 | 0.564 ± 0.080 | 0.571 ± 0.084 | 60 |
| M1.0 V | −0.615 ± 0.060 | 1.717 ± 0.036 | 505 ± 171 | 3693 ± 85 | 0.541 ± 0.079 | 0.547 ± 0.084 | 112 |
| M1.5 V | −0.664 ± 0.067 | 1.738 ± 0.039 | 424 ± 161 | 3633 ± 93 | 0.509 ± 0.089 | 0.513 ± 0.094 | 95 |
| M2.0 V | −0.740 ± 0.077 | 1.768 ± 0.038 | 335 ± 145 | 3528 ± 111 | 0.477 ± 0.091 | 0.479 ± 0.096 | 98 |
| M2.5 V | −0.824 ± 0.083 | 1.798 ± 0.035 | 242 ± 104 | 3429 ± 101 | 0.432 ± 0.087 | 0.431 ± 0.092 | 118 |
| M3.0 V | −0.910 ± 0.085 | 1.830 ± 0.041 | 183 ± 87 | 3334 ± 88 | 0.393 ± 0.087 | 0.390 ± 0.092 | 142 |
| M3.5 V | −0.992 ± 0.096 | 1.864 ± 0.045 | 133 ± 70 | 3260 ± 92 | 0.350 ± 0.086 | 0.345 ± 0.091 | 192 |
| M4.0 V | −1.063 ± 0.103 | 1.892 ± 0.045 | 99 ± 54 | 3199 ± 88 | 0.315 ± 0.084 | 0.309 ± 0.088 | 169 |
| M4.5 V | −1.163 ± 0.123 | 1.924 ± 0.042 | 67 ± 40 | 3123 ± 88 | 0.269 ± 0.075 | 0.260 ± 0.079 | 85 |
| M5.0 V | −1.296 ± 0.140 | 1.963 ± 0.048 | 35 ± 18 | 3034 ± 91 | 0.211 ± 0.052 | 0.198 ± 0.055 | 47 |
| M5.5 V | −1.445 ± 0.143 | 2.001 ± 0.057 | 20.4 ± 9.3 | 2941 ± 94 | 0.169 ± 0.032 | 0.155 ± 0.034 | 22 |
| M6.0 V | −1.623 ± 0.217 | 2.067 ± 0.066 | 12.4 ± 5.2 | 2850 ± 131 | 0.140 ± 0.022 | 0.124 ± 0.023 | 14 |
| M6.5 V | −1.829 ± 0.223 | 2.101 ± 0.065 | 8.5 ± 3.6 | 2733 ± 124 | 0.127 ± 0.019 | 0.110 ± 0.020 | 6 |
| M7.0 V | −2.018 ± 0.208 | 2.094 ± 0.036 | 5.7 ± 1.1 | 2625 ± 108 | 0.115 ± 0.007 | 0.097 ± 0.007 | 6 |
| M8.0 V | −2.189 ± 0.097 | 2.077 ± 0.048 | 5.22 ± 0.92 | 2492 ± 67 | 0.122 ± 0.007 | 0.105 ± 0.008 | 11 |
| M9.0 V | −2.495 ± 0.223 | 2.067 ± 0.066 | 3.4 ± 1.4 | 2352 ± 113 | 0.107 ± 0.015 | 0.089 ± 0.016 | 7 |
| L0 | −2.729 ± 0.170 | 2.014 ± 0.087 | 2.22 ± 0.29 | 2195 ± 172 | 0.107 ± 0.021 | 0.089 ± 0.023 | 18 |
| L1 | −2.812 ± 0.154 | 1.964 ± 0.093 | 2.00 ± 0.46 | 2076 ± 203 | 0.112 ± 0.020 | 0.094 ± 0.021 | 22 |
| L2 | −2.924 ± 0.187 | 1.900 ± 0.130 | 1.71 ± 0.45 | 1961 ± 193 | 0.114 ± 0.020 | 0.096 ± 0.021 | 20 |
| L3 | −3.109 ± 0.285 | 1.845 ± 0.137 | 1.34 ± 0.42 | 1852 ± 130 | 0.111 ± 0.013 | 0.093 ± 0.014 | 7 |
| L4 | −3.497 ± 0.184 | 1.740 ± 0.136 | 0.78 ± 0.26 | 1733 ± 90 | 0.096 ± 0.014 | 0.077 ± 0.015 | 4 |
| L5 | −3.460 ± 0.219 | 1.801 ± 0.165 | 0.67 ± 0.24 | 1722 ± 98 | 0.090 ± 0.014 | 0.071 ± 0.015 | 7 |

Table 3.7: Average absolute magnitudes for K5 V to L5 objects.

| Spectral type | $M_B$ [mag] | $M_{g'}$ [mag] | $M_{G_{BP}}$ [mag] | $M_V$ [mag] | $M_{r'}$ [mag] | $M_G$ [mag] | $M_{i'}$ [mag] |
|---|---|---|---|---|---|---|---|
| K5 V | 8.32 ± 0.37 | 7.81 ± 0.35 | 7.48 ± 0.39 | 7.23 ± 0.28 | 6.90 ± 0.31 | 6.85 ± 0.35 | 6.51 ± 0.31 |
| K7 V | 9.57 ± 0.44 | 9.00 ± 0.34 | 8.48 ± 0.34 | 8.27 ± 0.35 | 7.70 ± 0.34 | 7.65 ± 0.28 | 7.13 ± 0.26 |
| M0.0 V | 10.11 ± 0.59 | 9.49 ± 0.53 | 8.95 ± 0.52 | 8.73 ± 0.52 | 8.18 ± 0.53 | 8.01 ± 0.43 | 7.46 ± 0.43 |
| M0.5 V | 10.60 ± 0.54 | 9.92 ± 0.52 | 9.39 ± 0.52 | 9.16 ± 0.53 | 8.59 ± 0.52 | 8.38 ± 0.45 | 7.77 ± 0.42 |
| M1.0 V | 10.92 ± 0.48 | 10.24 ± 0.48 | 9.72 ± 0.45 | 9.47 ± 0.47 | 8.90 ± 0.47 | 8.63 ± 0.41 | 7.99 ± 0.41 |
| M1.5 V | 11.26 ± 0.57 | 10.54 ± 0.55 | 10.04 ± 0.54 | 9.78 ± 0.53 | 9.21 ± 0.54 | 8.91 ± 0.48 | 8.23 ± 0.49 |
| M2.0 V | 11.73 ± 0.65 | 10.98 ± 0.62 | 10.49 ± 0.60 | 10.21 ± 0.60 | 9.66 ± 0.61 | 9.27 ± 0.54 | 8.58 ± 0.55 |
| M2.5 V | 12.26 ± 0.70 | 11.50 ± 0.66 | 11.02 ± 0.64 | 10.73 ± 0.65 | 10.18 ± 0.64 | 9.70 ± 0.57 | 8.99 ± 0.58 |
| M3.0 V | 12.81 ± 0.72 | 12.00 ± 0.69 | 11.53 ± 0.69 | 11.26 ± 0.70 | 10.69 ± 0.68 | 10.11 ± 0.61 | 9.38 ± 0.59 |
| M3.5 V | 13.36 ± 0.83 | 12.55 ± 0.80 | 12.09 ± 0.79 | 11.79 ± 0.78 | 11.24 ± 0.77 | 10.57 ± 0.68 | 9.82 ± 0.69 |
| M4.0 V | 13.86 ± 0.89 | 13.03 ± 0.86 | 12.58 ± 0.84 | 12.25 ± 0.83 | 11.70 ± 0.81 | 10.96 ± 0.72 | 10.20 ± 0.72 |
| M4.5 V | 14.53 ± 0.95 | 13.70 ± 0.92 | 13.25 ± 0.91 | 12.88 ± 0.87 | 12.35 ± 0.88 | 11.51 ± 0.78 | 10.74 ± 0.79 |
| M5.0 V | 15.41 ± 1.05 | 14.62 ± 1.00 | 14.16 ± 0.99 | 13.68 ± 0.94 | 13.23 ± 0.96 | 12.20 ± 0.80 | 11.43 ± 0.81 |
| M5.5 V | 16.44 ± 0.92 | 15.73 ± 1.05 | 15.25 ± 1.01 | 14.61 ± 0.78 | 14.31 ± 0.99 | 13.01 ± 0.73 | 12.27 ± 0.68 |
| M6.0 V | 17.58 ± 0.85 | 16.82 ± 0.99 | 16.29 ± 0.96 | 15.63 ± 0.79 | 15.33 ± 0.93 | 13.72 ± 0.64 | 12.98 ± 0.64 |
| M6.5 V | 18.57 ± 0.78 | 17.72 ± 0.91 | 17.26 ± 0.89 | 16.67 ± 0.81 | 16.22 ± 0.83 | 14.34 ± 0.58 | 13.66 ± 0.58 |
| M7.0 V | … | 18.58 ± 0.71 | 18.17 ± 0.61 | … | 17.09 ± 0.60 | 14.89 ± 0.44 | 14.39 ± 0.41 |
| M8.0 V | … | 18.90 ± 0.28 | 18.21 ± 0.35 | … | 17.23 ± 0.31 | 15.15 ± 0.26 | 14.41 ± 0.24 |
| M9.0 V | … | 19.88 ± 0.66 | 18.78 ± 0.77 | … | 18.07 ± 0.54 | 16.01 ± 0.61 | 15.54 ± 0.69 |
| L0 | … | 20.62 ± 0.68 | 19.64 ± 0.25 | … | 18.59 ± 0.29 | 16.57 ± 0.31 | 16.15 ± 0.34 |
| L1 | … | 20.83 ± 0.71 | … | … | 18.74 ± 0.32 | 16.81 ± 0.34 | 16.38 ± 0.32 |
| L2 | … | 21.45 ± 0.95 | … | … | 19.09 ± 0.48 | 17.07 ± 0.42 | 16.71 ± 0.51 |
| L3 | … | 22.15 ± 0.89 | … | … | 19.45 ± 0.56 | 17.56 ± 0.59 | 17.14 ± 0.61 |
| L4 | … | 23.40 ± 0.76 | … | … | 20.49 ± 0.77 | 18.39 ± 0.36 | 18.27 ± 0.75 |
| L5 | … | 23.55 ± 0.79 | … | … | 20.69 ± 0.80 | 18.53 ± 0.30 | 18.54 ± 0.72 |
| **Spectral type** | **$M_{G_{RP}}$ [mag]** | **$M_J$ [mag]** | **$M_H$ [mag]** | **$M_{K_s}$ [mag]** | **$M_{W1}$ [mag]** | **$M_{W2}$ [mag]** | **$M_{W3}$ [mag]** |
| K5 V | 6.10 ± 0.31 | 5.16 ± 0.26 | 4.60 ± 0.24 | 4.46 ± 0.22 | 4.38 ± 0.22 | 4.32 ± 0.32 | 4.29 ± 0.35 |
| K7 V | 6.77 ± 0.25 | 5.65 ± 0.23 | 5.02 ± 0.22 | 4.85 ± 0.20 | 4.75 ± 0.20 | 4.78 ± 0.19 | 4.73 ± 0.17 |
| M0.0 V | 7.08 ± 0.38 | 5.89 ± 0.33 | 5.25 ± 0.34 | 5.07 ± 0.33 | 4.95 ± 0.30 | 4.93 ± 0.25 | 4.90 ± 0.26 |
| M0.5 V | 7.41 ± 0.41 | 6.14 ± 0.37 | 5.51 ± 0.39 | 5.31 ± 0.37 | 5.18 ± 0.36 | 5.13 ± 0.31 | 5.10 ± 0.32 |
| M1.0 V | 7.63 ± 0.39 | 6.30 ± 0.36 | 5.68 ± 0.39 | 5.47 ± 0.37 | 5.33 ± 0.36 | 5.26 ± 0.31 | 5.22 ± 0.32 |
| M1.5 V | 7.87 ± 0.46 | 6.50 ± 0.42 | 5.88 ± 0.46 | 5.66 ± 0.44 | 5.52 ± 0.44 | 5.44 ± 0.39 | 5.39 ± 0.40 |
| M2.0 V | 8.20 ± 0.52 | 6.75 ± 0.47 | 6.15 ± 0.50 | 5.91 ± 0.49 | 5.76 ± 0.48 | 5.67 ± 0.44 | 5.60 ± 0.44 |
| M2.5 V | 8.58 ± 0.55 | 7.06 ± 0.50 | 6.47 ± 0.53 | 6.22 ± 0.52 | 6.07 ± 0.51 | 5.95 ± 0.48 | 5.87 ± 0.47 |
| M3.0 V | 8.95 ± 0.56 | 7.36 ± 0.53 | 6.78 ± 0.55 | 6.52 ± 0.54 | 6.35 ± 0.53 | 6.22 ± 0.50 | 6.12 ± 0.49 |
| M3.5 V | 9.38 ± 0.65 | 7.70 ± 0.60 | 7.13 ± 0.62 | 6.86 ± 0.61 | 6.68 ± 0.59 | 6.53 ± 0.56 | 6.42 ± 0.55 |
| M4.0 V | 9.74 ± 0.69 | 7.99 ± 0.63 | 7.42 ± 0.65 | 7.15 ± 0.63 | 6.96 ± 0.62 | 6.79 ± 0.60 | 6.67 ± 0.59 |
| M4.5 V | 10.25 ± 0.74 | 8.41 ± 0.66 | 7.84 ± 0.67 | 7.56 ± 0.66 | 7.36 ± 0.64 | 7.19 ± 0.63 | 7.05 ± 0.61 |
| M5.0 V | 10.89 ± 0.76 | 8.94 ± 0.66 | 8.37 ± 0.67 | 8.06 ± 0.65 | 7.86 ± 0.64 | 7.67 ± 0.63 | 7.51 ± 0.62 |
| M5.5 V | 11.62 ± 0.58 | 9.56 ± 0.50 | 8.98 ± 0.49 | 8.65 ± 0.47 | 8.43 ± 0.46 | 8.23 ± 0.45 | 8.05 ± 0.44 |
| M6.0 V | 12.24 ± 0.54 | 10.03 ± 0.42 | 9.45 ± 0.41 | 9.10 ± 0.40 | 8.86 ± 0.39 | 8.65 ± 0.40 | 8.47 ± 0.37 |
| M6.5 V | 12.80 ± 0.52 | 10.40 ± 0.37 | 9.80 ± 0.35 | 9.43 ± 0.35 | 9.18 ± 0.34 | 8.98 ± 0.34 | 8.76 ± 0.31 |
| M7.0 V | 13.35 ± 0.39 | 10.78 ± 0.26 | 10.18 ± 0.22 | 9.81 ± 0.18 | 9.53 ± 0.22 | 9.32 ± 0.20 | 9.09 ± 0.12 |
| M8.0 V | 13.58 ± 0.24 | 10.88 ± 0.18 | 10.23 ± 0.17 | 9.83 ± 0.15 | 9.59 ± 0.15 | 9.36 ± 0.14 | 9.08 ± 0.13 |
| M9.0 V | 14.38 ± 0.59 | 11.45 ± 0.43 | 10.74 ± 0.39 | 10.27 ± 0.35 | 9.97 ± 0.32 | 9.70 ± 0.31 | 9.27 ± 0.24 |
| L0 | 14.94 ± 0.31 | 11.86 ± 0.19 | 11.08 ± 0.16 | 10.58 ± 0.16 | 10.23 ± 0.15 | 9.98 ± 0.17 | 9.49 ± 0.20 |
| L1 | 15.15 ± 0.38 | 12.03 ± 0.27 | 11.22 ± 0.23 | 10.72 ± 0.24 | 10.36 ± 0.21 | 10.10 ± 0.21 | 9.60 ± 0.25 |
| L2 | 15.35 ± 0.32 | 12.30 ± 0.42 | 11.44 ± 0.34 | 10.90 ± 0.30 | 10.49 ± 0.23 | 10.22 ± 0.22 | 9.73 ± 0.29 |
| L3 | 15.92 ± 0.60 | 12.62 ± 0.43 | 11.72 ± 0.35 | 11.15 ± 0.29 | 10.68 ± 0.21 | 10.40 ± 0.21 | 9.93 ± 0.34 |
| L4 | 16.79 ± 0.35 | 13.29 ± 0.47 | 12.30 ± 0.40 | 11.69 ± 0.40 | 11.05 ± 0.33 | 10.71 ± 0.34 | 10.28 ± 0.40 |
| L5 | 16.86 ± 0.29 | 13.39 ± 0.51 | 12.44 ± 0.43 | 11.88 ± 0.43 | 11.18 ± 0.35 | 10.85 ± 0.36 | 10.38 ± 0.41 |

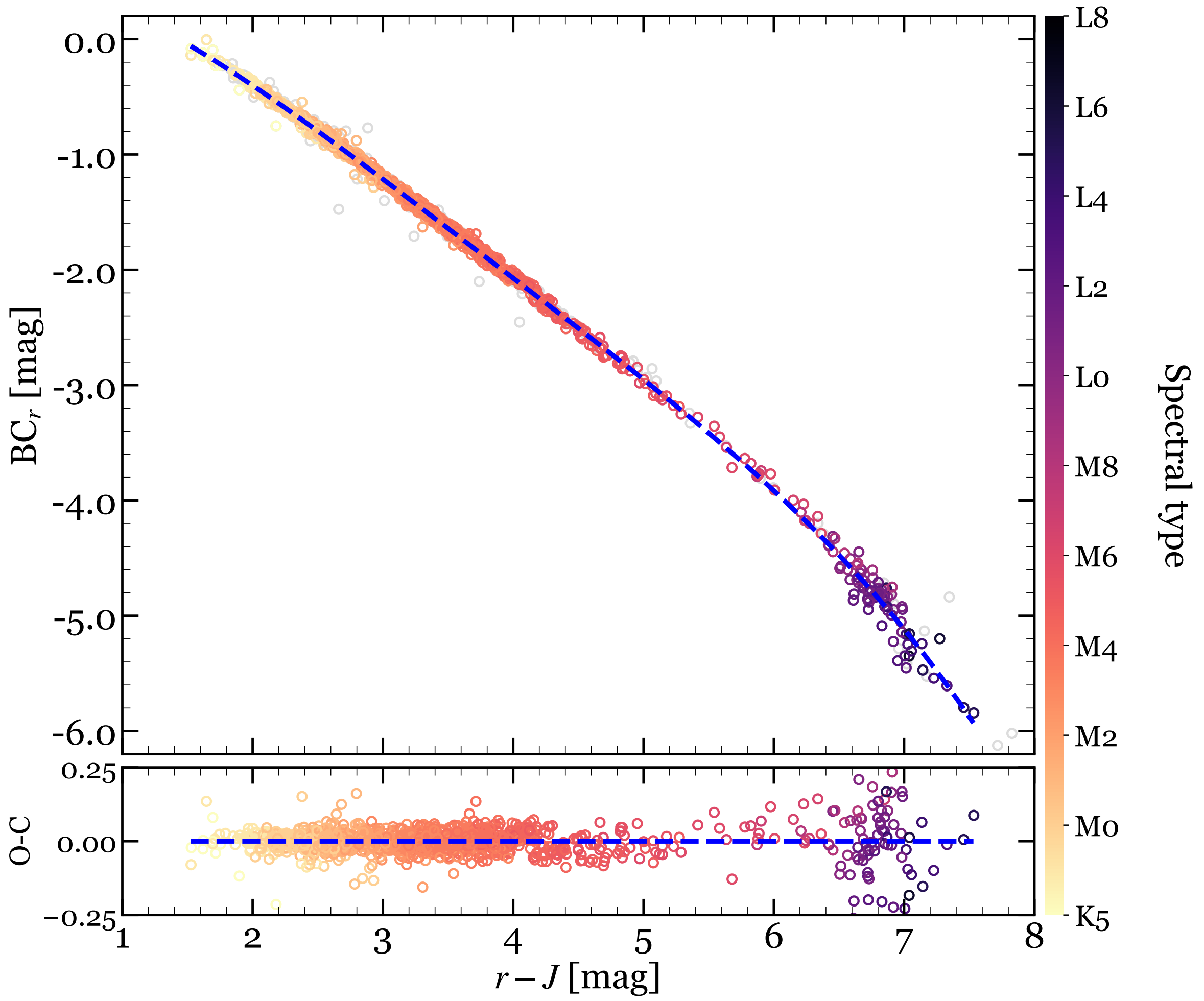

Figure 3.20: Same as Fig. 3.12 but for BC$_{r'}$ vs. $r' - J$. A polynomial fit is shown as a blue dashed line.

Lépine et al., 2013; Alonso-Floriano et al., 2015b). The correspondingly large standard deviations denote the large natural scatter of the main sequence at the earliest spectral types and the difficulty in determining precise parameters at the latest ones. The boundary values for K5 V type were not smoothed and, therefore, must be handled with care. Besides, it is not trivial to distinguish between very late K dwarfs and very early M dwarfs from low-resolution spectroscopy (see again Kirkpatrick et al., 1991; Lépine et al., 2013; Alonso-Floriano et al., 2015b), while the $T_{\rm eff}$ determination does not always help because of different temperature scales in the literature. However, the tabulated value of $\mathcal{L} \approx 0.1\,\mathcal{L}_\odot$ is a more accessible and reliable observational boundary in the K7 V–M0.0 V frontier. This is extensible into the M dominion, even the coolest tail. In this sense, bolometric luminosities could effectively constrain the spectral class of a main sequence star with more reliance that, for instance, the surface temperature. On the other hand, Table 3.7 complements Table C.2 and lists the average absolute magnitudes of K5 V to L2.0 objects in the 14 most representative bands (i.e. all except for *GALEX FUV* and *NUV*, SDSS9 $u'$, Tycho-2 $B_T$ and $V_T$, and *WISE W4*). We applied the same rolling mean and $3 - \sigma$ clipping as in Table 3.6. For each spectral type K5–M7.0 V, a total of 6.227 $10^9$ different colours can be determined from the tabulated absolute magnitudes (e.g. $G - J = M_G - M_J$). For spectral types L0.0–2.0, the number of possible colours is 3 628 800.

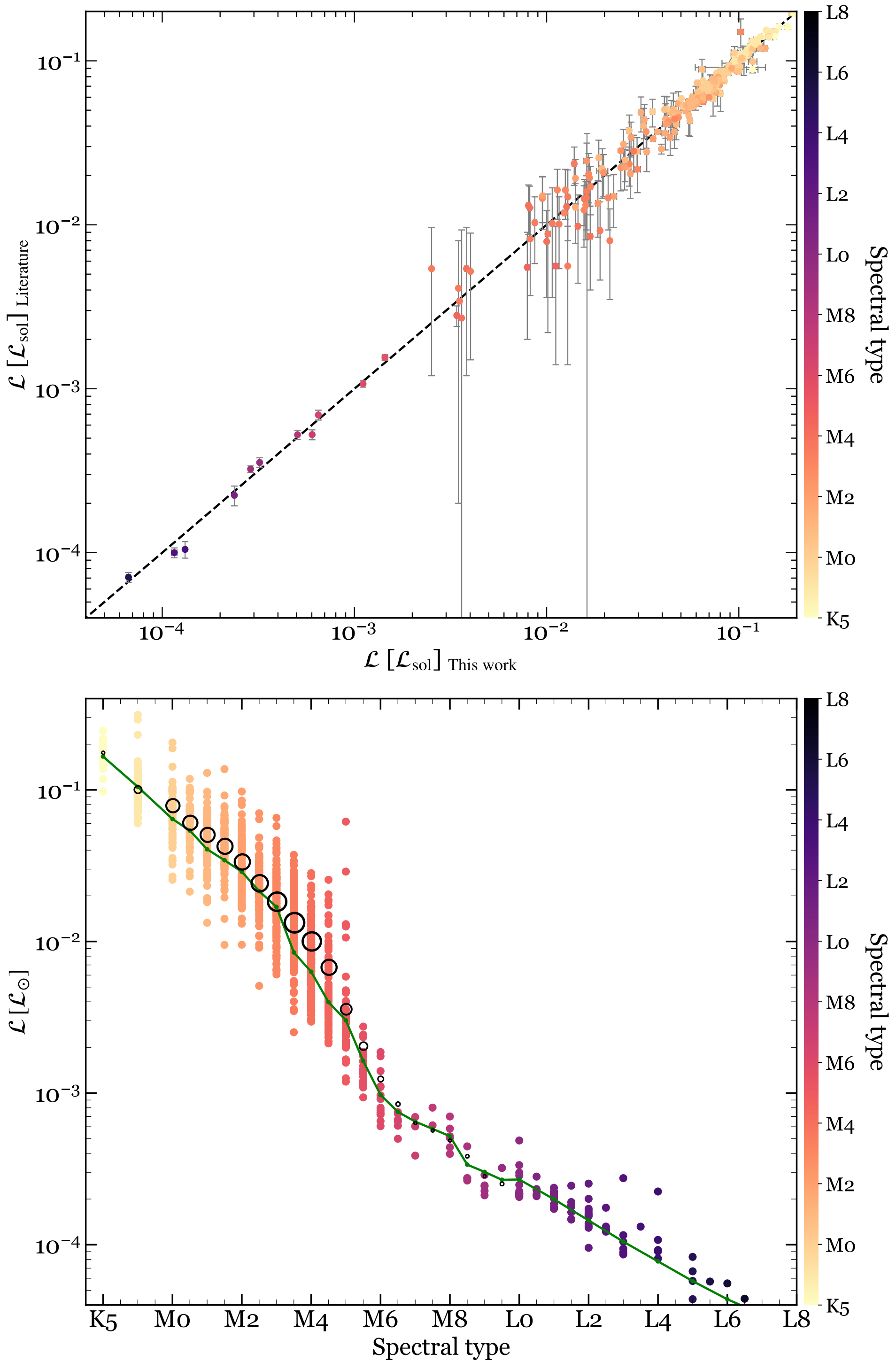

Figure 3.21: Comparison of $\mathcal{L}$ from VOSA and from the literature (*top*) and individual (coloured points) and mean $\mathcal{L}$ (black circles) as a function of spectral type as in Table 3.6 (*bottom*). In the bottom panel, the green line outlines the empirical $\mathcal{L}$-spectral type sequence of Pecaut & Mamajek (2013, with updated values for M0.0 V to M9.5 V; E. E. Mamajek, priv. comm.) and the size of the black points is proportional to the number of stars per spectral type.

**Luminosities** (Fig. 3.21). First, we compared our $\mathcal{L}$ computed with VOSA with those from a number of works in the literature (top panel – Golimowski et al., 2004; Vrba et al., 2004; Howard et al., 2010; Kundurthy et al., 2011; Bonfils et al., 2012; Mann et al., 2013b; Gaidos & Mann, 2014; Affer et al., 2016, 2019; Tuomi et al., 2014; Newton et al., 2015; Anglada-Escudé et al., 2016; Astudillo-Defru et al., 2017b; Dittmann et al., 2017; Gillon et al., 2017; Maldonado et al., 2017; Suárez Mascareño et al., 2017a,b; Gaia Collaboration et al., 2018b; Hirano et al., 2018; Hobson et al., 2018). In spite of (or due to) the relatively large published $\mathcal{L}$ uncertainties of a few ultracool dwarfs, the agreement is in general excellent, especially in the case of Gaia Collaboration et al. (2018b). Our median $\mathcal{L}$ values per spectral type also match those of Pecaut & Mamajek (2013, right panel). When integrated from a well-calibrated, multi-band spectral energy distribution in a wide wavelength coverage and calculated with the latest *Gaia* parallaxes as in this work, $\mathcal{L}$ can become the most reliable "observable" of low-mass stars, instead of the widely used temperature, which is inferred through colours, spectral classification, or expensive, model-dependent, spectral synthesis.

**Effective temperatures** (Fig. 3.22). Next, we compared our $T_{\rm eff}$ from VOSA with the values from the works referred to in the previous paragraph, except from Gaia Collaboration et al. (2018b), plus from Passegger et al. (2019), who in turn compared their $T_{\rm eff}$ with those from Rojas-Ayala et al. (2012), Gaidos & Mann (2014), Maldonado et al. (2015), Mann et al. (2015), Rajpurohit et al. (2018a), and Schweitzer et al. (2019). From the top left panel in Fig. 3.22, our $T_{\rm eff}$ are cooler than those of the literature by $-86 \pm 82$ K. This systematic difference is within the grid step size of the theoretical models used by VOSA, of 100 K or 50 K, but appreciable in the whole $T_{\rm eff}$ = 3000–4000 K range. That VOSA does not interpolate between grid points may partly explain this systematic difference. In the empirical $T_{\rm eff}$-spectral type relation shown in the top right panel, $T_{\rm eff}$ from Rajpurohit et al. (2018a) and Passegger et al. (2019) are, again, slightly warmer than ours in the late- and early-M domains, respectively. However, the agreement with the relation of Pecaut & Mamajek (2013) is exquisite. The K/M and M/L boundaries occur at about 3900 K and 2300 K, respectively, in line with the standard values (e.g. Habets & Heintze, 1981; Kirkpatrick, 2005, see also Table 3.6). In the Hertzsprung-Russell diagram in the bottom left panel, as expected, our targets are significantly less luminous than the very young stars and brown dwarfs of the same $T_{\rm eff}$ tabulated by Pecaut & Mamajek (2013) and Farihi (2016), but our main sequence (excluding young targets) matches that of Newton et al. (2015). The most convincing plot is perhaps the $M_J$ versus $T_{\rm eff}$ diagram in the bottom right panel, where our M-dwarf main sequence perfectly overlaps with those defined by Lépine et al. (2013) and Gaidos & Mann (2014) and extrapolates reasonably well into the ultracool dwarf sequence of Dahn et al. (2002). The absolute magnitude in the vertical axis does not depend on models, Virtual Observatory tools, spectral synthesis, or multi-band photometry, but only on reliable 2MASS $J$-band magnitude and *Gaia* parallaxes.

**Metallicity** (Figs. 3.23 to 3.25). The role of metallicity in the empirical relations between physical parameters of M dwarfs has been the subject of investigation by many teams (e.g. Bonfils et al., 2005a; Woolf & Wallerstein, 2005; Casagrande et al., 2008; Rojas-Ayala et al., 2012; Boyajian et al., 2012; von Boetticher et al., 2019, see Sect. 4.3 in Alonso-Floriano et al. 2015a for a short review). Of them, Mann et al. (2015) showed that empirical relations such as absolute magnitude-radius, radius-temperature, or colour-temperature could benefit from incorporating an additional term that accounts for metallicity. However, mainly because of the limitations of the BT-Settl CIFIST grid of theoretical models stored in the VOSA database, in our work we computed $\mathcal{L}$ and $T_{\rm eff}$ assuming a solar metallicity ([Fe/H] = 0)[11].

[11] Actually, BT-Settl CIFIST models are defined for solar metal abundance, [M/H], but here we used solar iron abundance, [Fe/H], for simplicity.

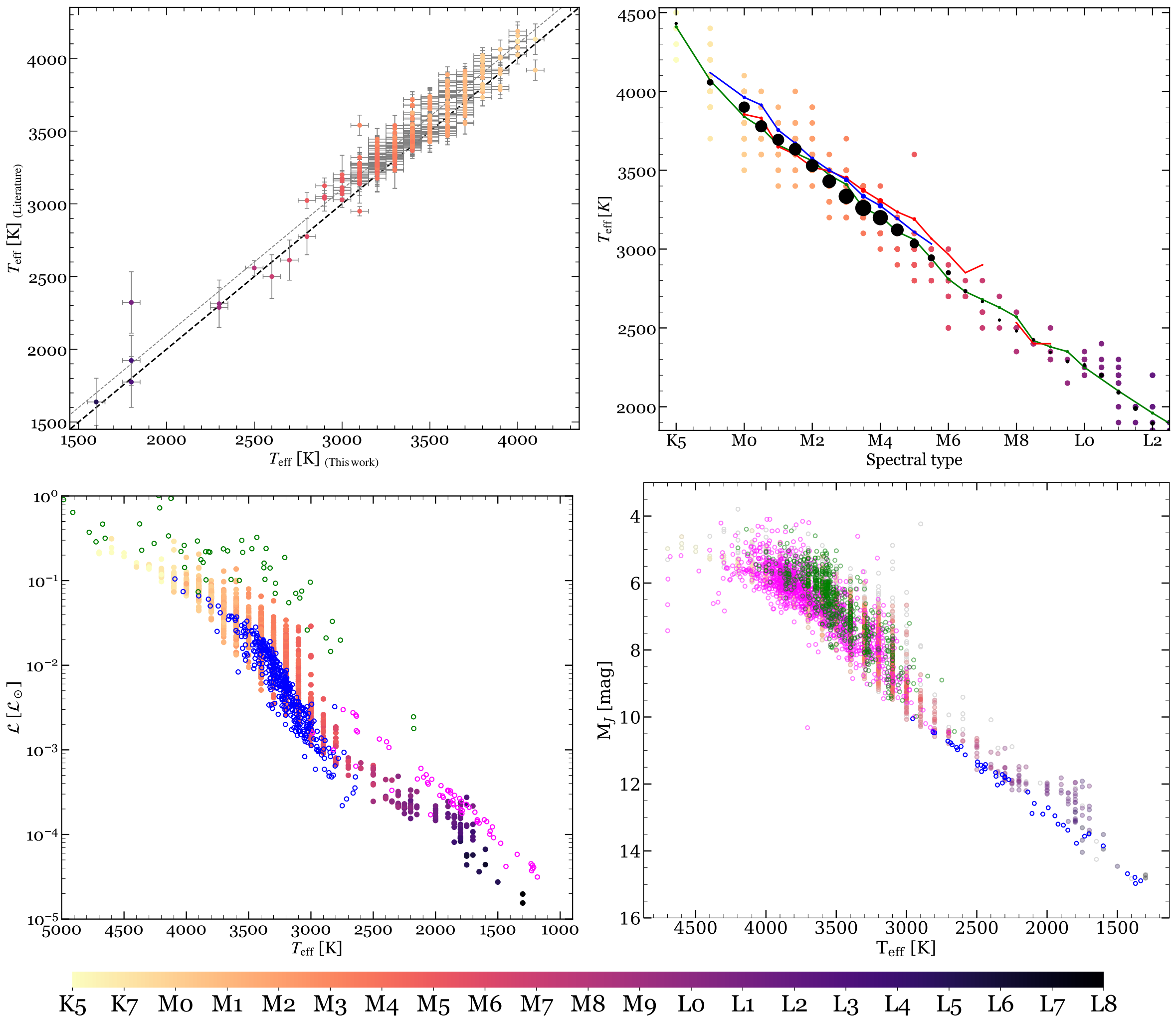

Figure 3.22: Four representative diagrams involving $T_{\rm eff}$. In the four panels our investigated stars are represented with filled circles colour-coded by spectral type. *Top left:* Comparison of $T_{\rm eff}$ from this work and from the literature. *Top right:* Individual (coloured points) and median (black circles) values of $T_{\rm eff}$ as a function of the spectral sequence shown in Table 3.6. The size of the black circles is proportional to the number of stars per spectral type. The green, red, and blue lines mark the mean values tabulated by Pecaut & Mamajek (2013), Rajpurohit et al. (2018a), and Passegger et al. (2019), respectively. *Bottom left:* $\mathcal{L}$ vs. $T_{\rm eff}$. As a comparison we also plot pre-main sequence stars with `BT-Settl` model fitting from Pecaut & Mamajek (2013, green empty circles), M dwarfs in the MEarth sample with stellar parameters from Newton et al. (2015, blue empty circles, inferred from the pseudo-equivalent width of Mg I near-infrared lines), and high-confidence moving group members from Farihi (2016, magenta empty circles) with parameters computed as in Filippazzo et al. (2015). *Bottom right:* $J$-band absolute magnitude vs. $T_{\rm eff}$. As a comparison we also plot the samples of Dahn et al. (2002, blue open circles), Lépine et al. (2013, green open circles), and Gaidos & Mann (2014, magenta empty circles).

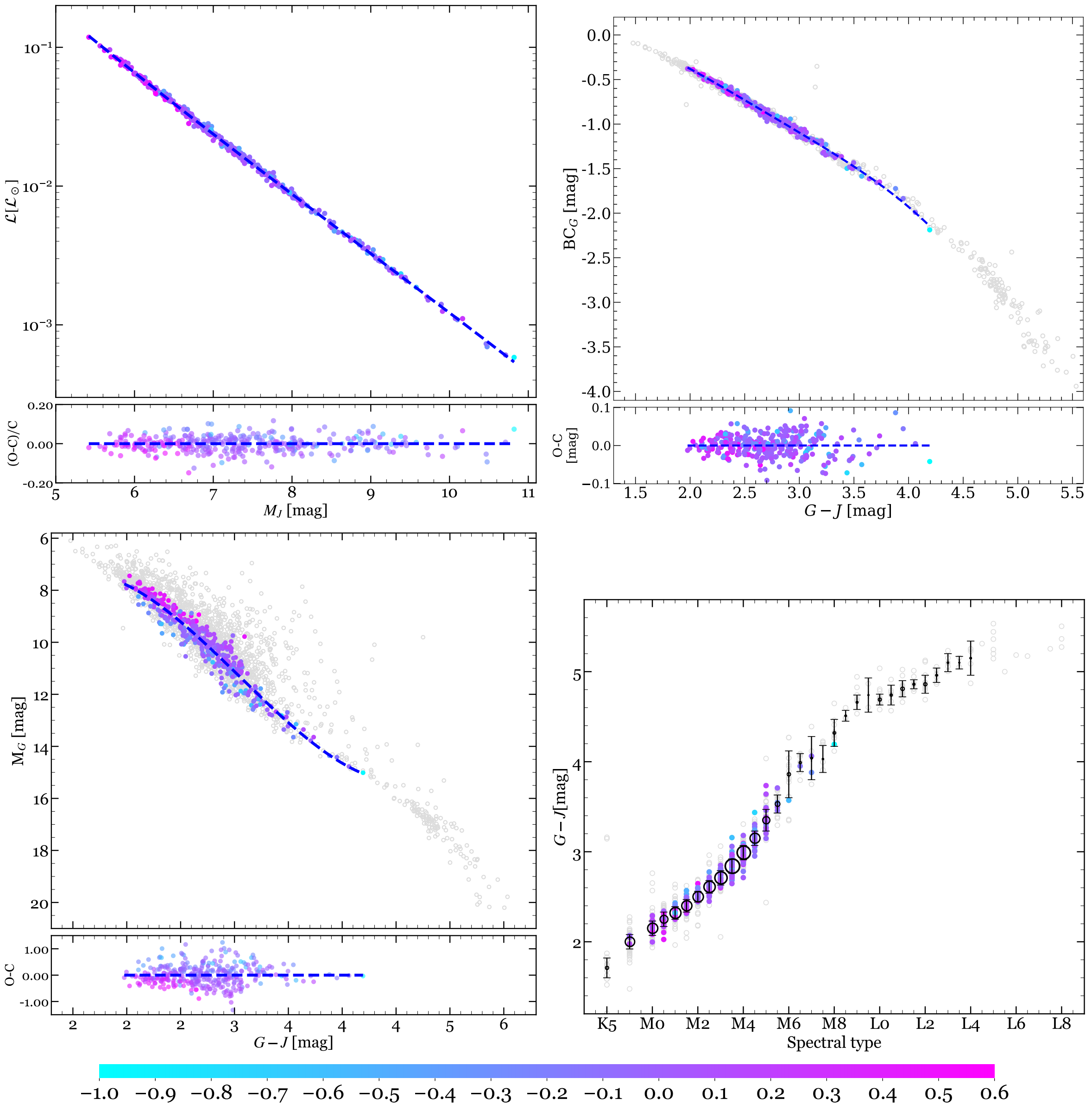

Figure 3.23: Revisiting empirical relations in Table 3.5 for stars with [Fe/H] values published in the literature, colour-coded by this parameter. *Top left:* $\mathcal{L}_{\mathrm{VOSA,BT-Settl\ CIFIST}}$ vs. $M_J$. *Top right:* $BC_G$ vs. $G-J$. *Bottom left:* $M_G$ vs. $G-J$. The blue dashed lines in the three panels represent new 3-, 4-, and 3-degree polynomial fits, respectively. *Bottom right:* $G-J$ vs. spectral type. The grey circles represent the mean values with a symbol size proportional to the sample size in each type.

In order to quantify the impact of metallicity within our empirical relations in Table 3.5, first we compiled values of spectroscopically derived iron abundances of 510 single stars in our sample from Mann et al. (2015, 2019), Majewski et al. (2017), and Passegger et al. (2019). The compiled [Fe/H] values ranged from −1.63 dex for the mid-M dwarf HD 285190 to +0.59 dex for the early-M dwarf LP 397–041, with a mean and dispersion of −0.04 ± 0.26 dex.

Figure 3.23 displays the relations parametrised in Table 3.5, as well as the colour-spectral type diagram discussed in Sect. 3.2.3, colour-coded by the metallicity values from the literature. In either of the top plots ($\mathcal{L}$ vs. $M_J$ and $BC_G$ vs. $G-J$), the distribution of residuals did not show any correlation with the metal content of the stars. Both representations benefit from the fact that deriving $\mathcal{L}$ does not rely on precise [Fe/H] measurements. In the bottom panels, the distribution of metallicity values in the $G-J$ versus spectral type diagram shows no significant dependence on metallicity. This lack of correlation is also apparent in the additional colour diagrams displayed in this chapter. However, the $M_G$ versus $G-J$ relation exhibits a notable correlation between metallicity and the residuals of the fit: more metallic stars appear brighter than less metallic stars of the same $G-J$ colour or, alternatively or simultaneously, more metallic stars appear redder than less metallic stars of the same $M_G$ absolute magnitude. This dependence is most likely the main source of uncertainty for photometric distances, as we noted in Sect. 3.2.3. By using standard broad passbands in the red optical or the near infrared, such as $r'$ or $J$, the effect of metallicity can be reduced compared to using wider, bluer passbands, such as $G$, which are more affected by the features that metallicity imprints on the spectra.

In the diagrams involving $T_{\rm eff}$, Mann et al. (2015) pointed out that the effect of metallicity can be severely masked due to the steeper dependence on the temperature. This is an important point to underline because the uncertainties in $T_{\rm eff}$ of models are a major source of uncertainties in the final products of the SED fitting. In other words, the approximation of near-solar metallicity implies an error that is always within the errors due to temperature uncertainties. We argue that, with the exception of absolute magnitude against colours and extreme cases (i.e. very metal-poor stars), the models described in this work can be treated as independent of the metal content of the star.

As an additional test, we used `VOSA` to perform a new SED fit of the CARMENES GTO stars in the sample using the `BT-Settl` grid of spectra ("no CIFIST"; Allard et al., 2012), which allowed us to explore iron abundances different from [Fe/H] = 0. In particular, we let [Fe/H] vary between −1.5 dex and +0.5 dex, with a step size of 0.5 dex, and constrained $T_{\rm eff}$ and $\log g$ as in Table 3.3. The [Fe/H] values derived from this new fit are compared to the spectroscopic values from the literature in Fig. 3.24. While the median of `VOSA` `BT-Settl` and published values are in fair agreement (−0.097 dex and +0.033 dex, respectively), the scatter of the `VOSA` [Fe/H] values is much greater than that of the literature ($\sigma_{\rm [Fe/H],VOSA}$ = 0.596 dex and $\sigma_{\rm [Fe/H],literature}$ = 0.216 dex). From the diagram, `VOSA` assigned artificially low [Fe/H] to stars with spectroscopically derived solar values, which reinforced our initial approach of setting [Fe/H] = 0. This is in line with the quality tests carried out by the `VOSA` team in 2017, in which they compared `VOSA` metallicities with those derived by Yee et al. (2017), Lindgren & Heiter (2017), and Rajpurohit et al. (2018b). In particular, they concluded that "metallicities [...] provided by `VOSA` are not reliable due to the minor contribution of [this parameter] to the SED shape"[12].

In Fig. 3.25 we compare the $\mathcal{L}$ and $T_{\rm eff}$ obtained for the GTO stars using `BT-Settl CIFIST` with [Fe/H] = 0 (used throughout this work) and `BT-Settl` with a free range in metallicity. While the derivation of bolometric luminosities in K dwarfs, with a normalised difference of $\Delta\mathcal{L}/\mathcal{L} = -0.0065 \pm 0.0046$, is marginally dependent on metallicity, the derivation in M dwarfs is independent: the normalised differences between $\mathcal{L}$ computed with the two methods is $\Delta\mathcal{L}/\mathcal{L} = 0.012 \pm 0.035$, consistent with zero. The $T_{\rm eff}$ difference is also consistent with zero, and its standard deviation is 101 K, identical to the $T_{\rm eff}$ step size in the M-dwarf domain.

[12] `http://svo2.cab.inta-csic.es/theory/vosa/`.

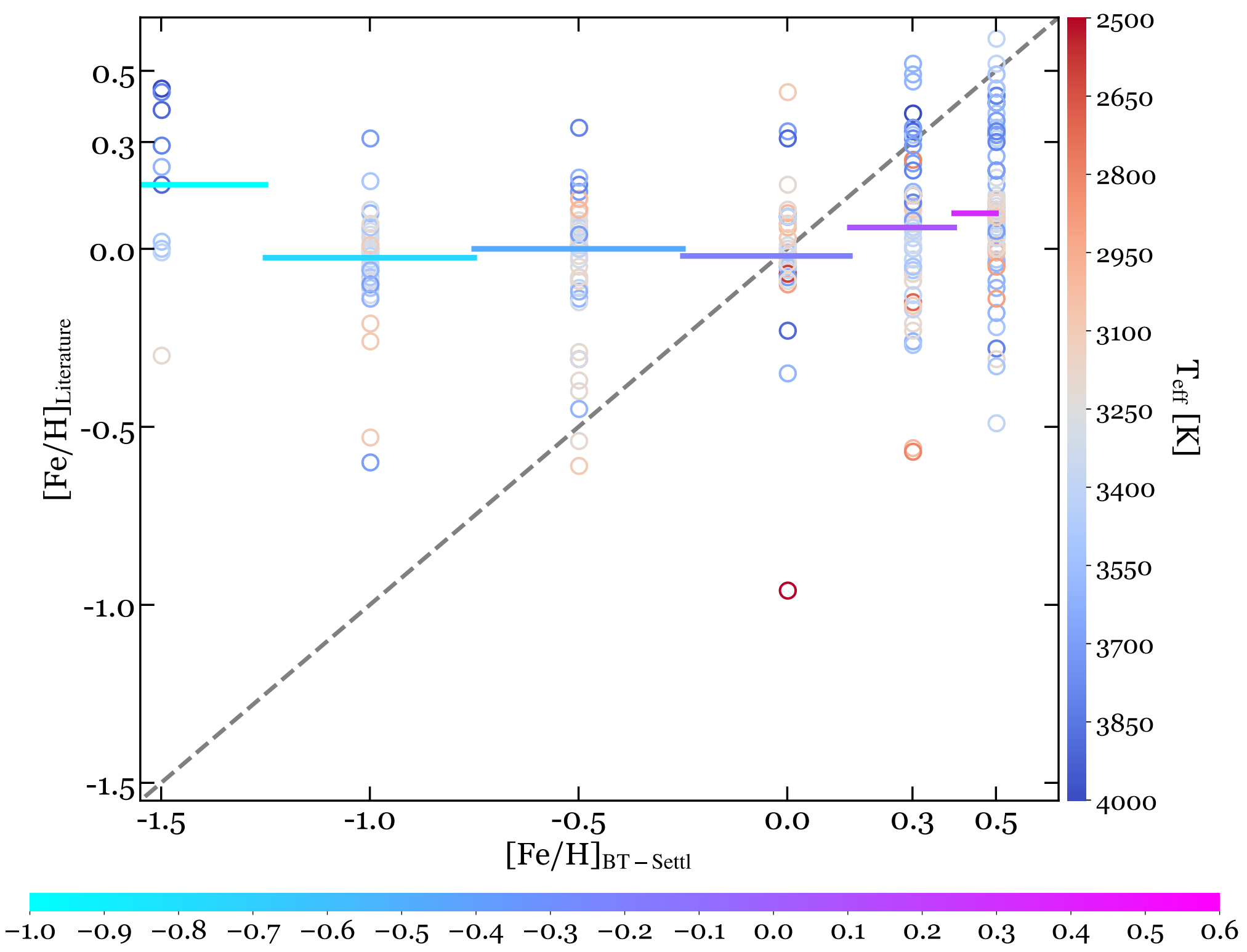

Figure 3.24: Comparison of metallicites from `VOSA` `BT-Settl` fit and from literature for CARMENES GTO M dwarfs, colour-coded by `BT-Settl CIFIST` $T_{\rm eff}$, and by [Fe/H], represented by the horizontal colourbar. Horizontal lines represent the median values in each `BT-Settl` [Fe/H] bin.

**Radii** (Fig. 3.26). We compared our $\mathcal{R}$, derived from `VOSA`'s $\mathcal{L}$ (`BT-Settl CIFIST`, [Fe/H] = 0) and $T_{\rm eff}$ using the Stefan-Boltzmann law, with the same works as in Fig. 3.21 (top left panel). Some of these works in turn compared their results with independent direct radius determinations (e.g. near-infrared interferometry – Boyajian et al., 2012; von Braun et al., 2014b). On average, our $\mathcal{R}$ are larger by $0.022 \pm 0.037\,\mathcal{R}_\odot$, meaning they are identical within the dispersion of the data. However, the standard deviation includes both random errors (in magnitudes, parallax, SED integration) and systematic errors (in passband $\lambda_{\rm eff}$ and $W_{\rm eff}$, `VOSA` minimisation procedure, CIFIST models), and the $\mathcal{R}$ difference appears systematically across the whole sample, so it is likely to be significant. Furthermore, because of the $T_{\rm eff}$ shift with respect to Passegger et al. (2018) and other spectral synthesis works, our $\mathcal{R}$ are also larger by about 5 % than those of Schweitzer et al. (2019), who used almost identical $\mathcal{L}$ to ours. For that reason, when $T_{\rm eff}$ from spectral synthesis on high-resolution spectra is available, we recommend using it together with our $\mathcal{L}$ for determining $\mathcal{R}$ (and $\mathcal{M}$), and use $T_{\rm eff}$ from `VOSA` when there is no spectral synthesis.

The offset in these diagrams, specially apparent in the bottom left panel, can include several causes. For one, non-resolved binarity is certainly a reason for a star to be overluminous (or oversized, in terms of the radius). As investigated in Chapter 4, this is the case for a sizeable amount of stars in the sample, but for the vast majority of cases there are only hints or clues of disguised multiplicity (specifically, via *Gaia* statistical indicators), but no actual confirmation. Because luminosities are empirical to a very large degree, we can be confident that they reflect the actual emitting power (i.e. luminosity) of the object. Also, their values are fundamentally unaffected by the uncertainty in synthetic models, because they almost do not rely on them. But probably a more decisive cause traces back to the models from which effective temperatures are inferred. Unfortunately, many stars lack detailed spectroscopic information, which would certainly help providing with more reliable effective temperatures (again, always preferred to those from fitted synthetic models), identifying potential compact binarity (i.e. spectroscopic binarity), and telling if a star does have signs to be young – all to better diagnose the causes for overluminosity.

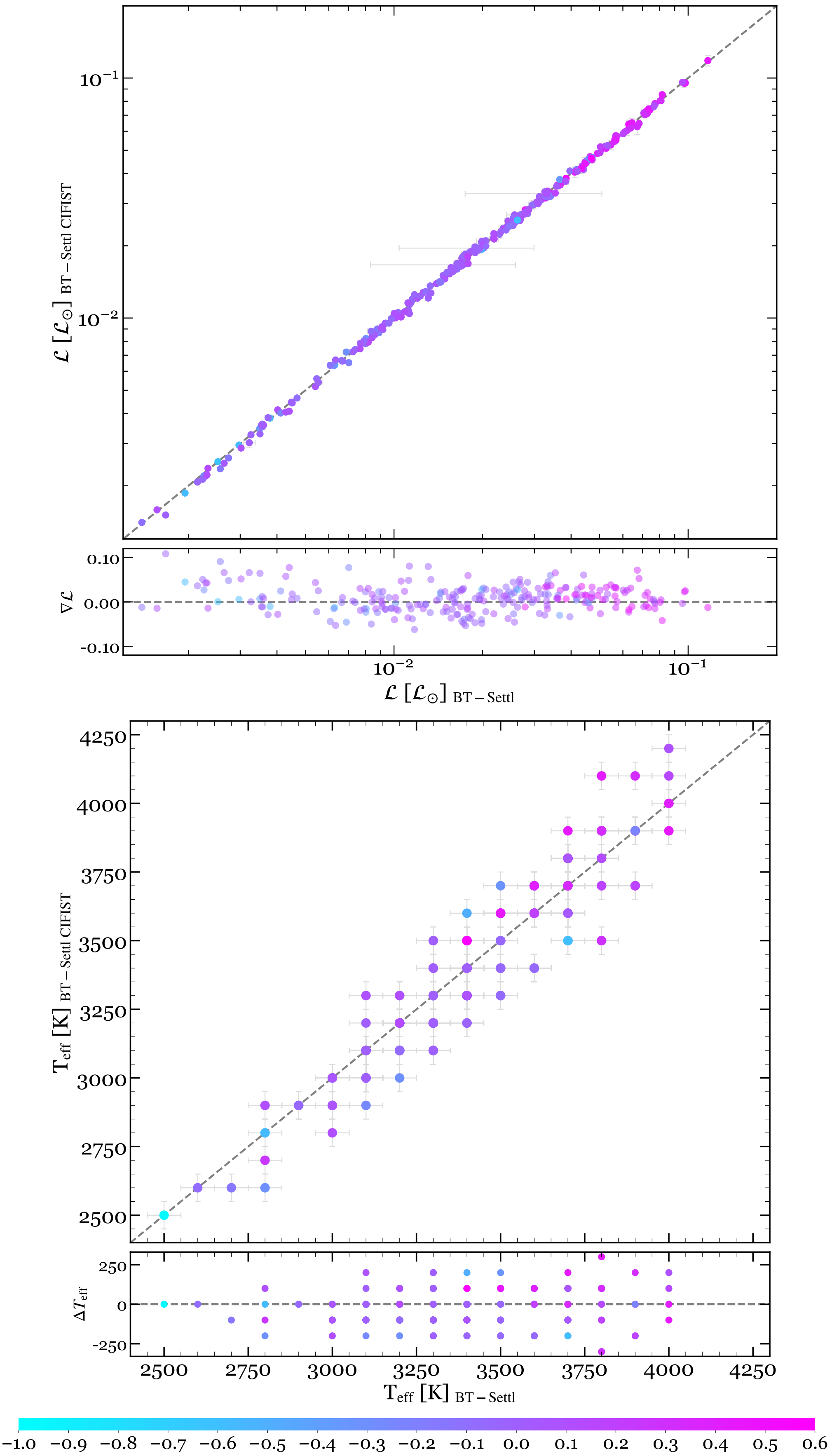

Figure 3.25: Comparison of previous and recomputed $\mathcal{L}$ (*top*) and $T_{\rm eff}$ (*bottom*) using `BT-Settl` with [Fe/H] = −1.5 to +0.5 for CARMENES GTO M dwarfs, colour-coded by metallicities published in the literature. The small bottom panels depict $\nabla L = \log \mathcal{L}_{\rm BT-Settl\,CIFIST} - \log \mathcal{L}_{\rm BT-Settl}$ and $\Delta T_{\rm eff} = T_{\rm eff,BT-Settl\,CIFIST} - T_{\rm eff,BT-Settl}$, respectively.

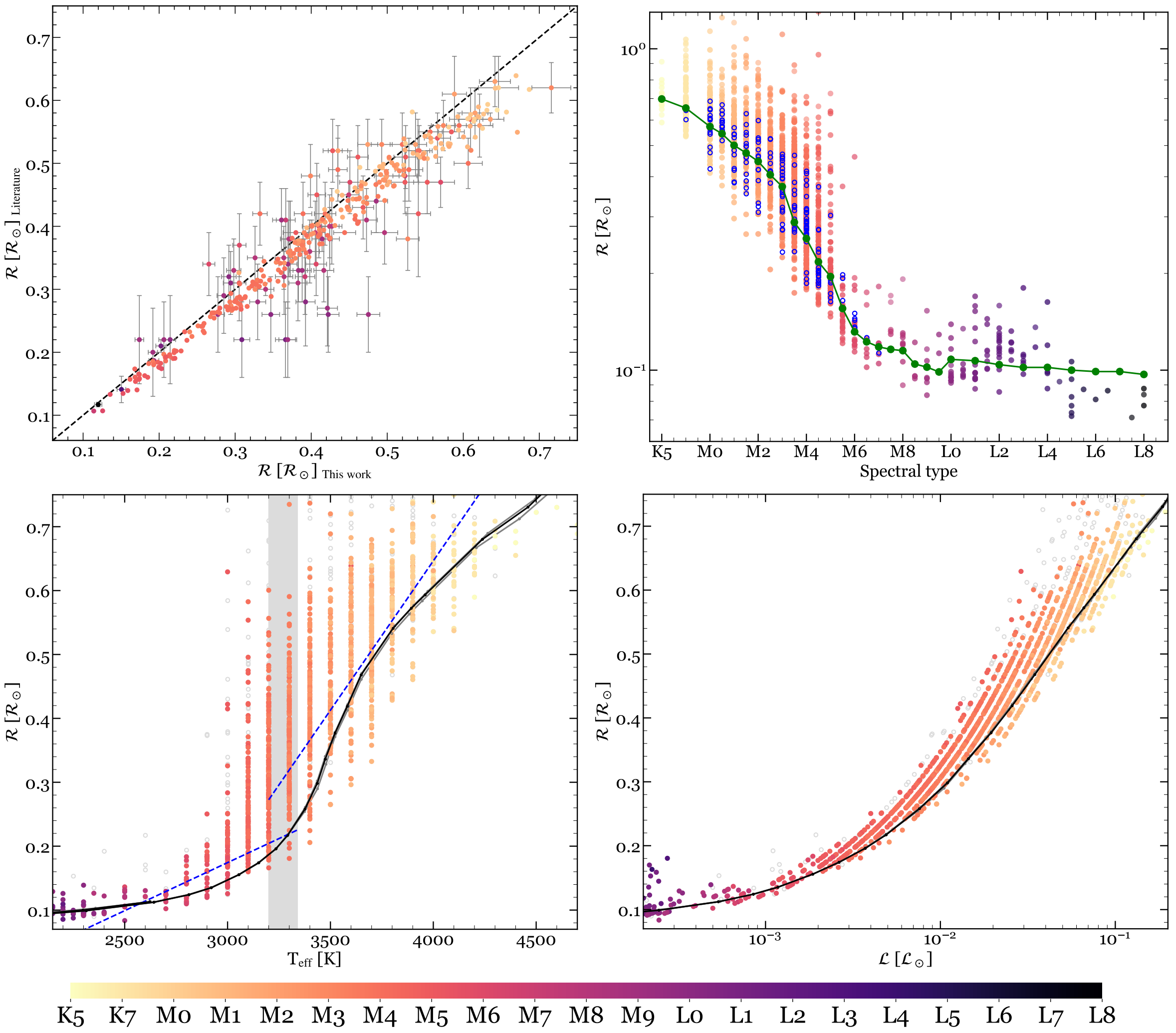

Figure 3.26: Four representative diagrams involving $\mathcal{R}$. In the four panels our investigated stars are represented with filled circles colour-coded by spectral type. *Top left:* Comparison of $\mathcal{R}$ from this work and from the literature, including Schweitzer et al. 2019 (with symbol-size error bars). *Top right:* Individual (coloured points) and median (black circles) values of $\mathcal{R}$ as a function of the spectral sequence shown in Table 3.6. The green line marks the median values from Pecaut & Mamajek (2013) and the blue circles are stars from Mann et al. (2015). *Bottom left:* $\mathcal{R}$ vs. $T_{\rm eff}$. The black and grey solid lines are the `NextGen` isochrones of the Lyon group for 1.0, 4.0, and 8.0 Ga (overlapping), the blue dashed lines are the linear fittings from Rabus et al. (2019), and the grey shaded area is the region where they reported a possible discontinuity. *Bottom right:* $\mathcal{R}$ vs. $\mathcal{L}$. The black solid lines are the same isochrones as in the bottom left panel.

In spite of the large spread at spectral types earlier than M4.5 V and some poorly sampled SEDs later than M7.0 V, the matches with the $\mathcal{R}$-spectral type relation of Pecaut & Mamajek (2013) and the values reported by Mann et al. (2015) are also good (top right panel). Our $\mathcal{R}$-$T_{\rm eff}$ diagram (bottom left panel) naturally reproduces the sigmoid shape predicted by the widely used theoretical models of Baraffe et al. (1998), but shifted by ~100 K towards cooler $T_{\rm eff}$ (see previous paragraph). More than two decades after that cornerstone work by the Lyon group, Rabus et al. (2019) fitted an $\mathcal{R}$-$T_{\rm eff}$ relation using two linear polynomials and identified a discontinuous behaviour that the authors attributed to the transition between partially and fully convective stars at 3200–3340 K or ~0.23 $\mathcal{M}_\odot$. Soon after, Cassisi & Salaris (2019) confirmed this discontinuity, but considered instead the contribution of the electron degeneracy to the gas equation of state as the physical phenomenon behind this behaviour (see also Chabrier & Baraffe, 1997). While the boundary between partially and fully convective stars is better exposed in for example the $NUV - G_{RP}$ versus spectral type diagram (see Fig. 3.15), in our data we did not find evidence for any discontinuity in the vicinity of 3250 K in the $\mathcal{R}$-$T_{\rm eff}$ diagram, but just a change of slope, as proven by Schweitzer et al. (2019, see their Fig. 11). The statistics in Rabus et al. (2019) were poorer than ours: they added around one hundred objects from Mann et al. (2015) to their sample of 22 low-mass dwarfs, while we have 1031 homogeneously investigated stars with $T_{\rm eff}$ in the 3000–3500 K interval. Furthermore, the continuity of $\mathcal{R}$ as a function of $\mathcal{L}$ is obvious in the bottom right panel of Fig. 3.26.

**Masses** (Fig. 3.27). We compared our $\mathcal{M}$, derived from our $\mathcal{R}$ and the $\mathcal{M}$-$\mathcal{R}$ relation of Schweitzer et al. (2019), with those from the literature (same works as in Fig. 3.21, top panel). This comparison is shown in the top panel of Fig. 3.27. Among our parameters, $\mathcal{M}$ is the one that shows more dissimilarities with respect to published values, although $\mathcal{M}_{\rm This\ work} - \mathcal{M}_{\rm lit} = 0.025 \pm 0.081\,\mathcal{M}_\odot$, consistent with a null difference (but probably significant as in $\mathcal{M}$ when random and systematic errors are taken into account). For example, the two stars for which our $\mathcal{M}$ deviate more than 80 % from published values are LP 229–17 (M3.5 V, $\mathcal{M}_{\rm This\ work} = 0.476 \pm 0.017\,\mathcal{M}_\odot$, $\mathcal{M}_{\rm lit} = 0.23 \pm 0.08\,\mathcal{M}_\odot$) and YZ CMi (M4.5 V, $\mathcal{M}_{\rm This\ work} = 0.368 \pm 0.008\,\mathcal{M}_\odot$, $\mathcal{M}_{\rm lit} = 0.19 \pm 0.08\,\mathcal{M}_\odot$), both from Gaidos & Mann (2014). The former star was tabulated as a spectroscopic binary by Houdebine et al. (2019), although we do not see any CARMENES radial-velocity variation attributable to binarity (Reiners et al., 2018b, see also Cortés-Contreras et al. 2017b for a lucky imaging analysis), while the latter star is a candidate member of the young $\beta$ Pictoris moving group (not in Table 3.4 – Montes et al., 2001; Alonso-Floriano et al., 2015a) with strong chromospheric activity (Kahler et al., 1982; Kowalski et al., 2010; Tal-Or et al., 2018), which may partly explain the differences. In planet-host stars, such changes can translate into significant differences in the published (minimum) masses of M-dwarf planets.

We also compared our values of $\mathcal{M}$ with those calculated from the $\mathcal{M}$-$M_K$ relations of Delfosse et al. (2000), valid for 4.5 mag ≤ $M_K$ ≤ 9.5 mag, and Benedict et al. (2016), valid for $M_K$ ≤ 10 mag, and the $\mathcal{M}$-$M_{K_s}$ relation of Mann et al. (2019), valid for 4 mag ≤ $M_{K_s}$ ≤ 11 mag, and “safe” for 4.5 mag ≤ $M_{K_s}$ ≤ 10.5 mag. For the relations of Delfosse et al. (2000), we converted our 2MASS $K_s$ magnitudes to CIT $K$ values (Elias et al., 1982) using the colour transformation provided by Carpenter (2001). The means of the mass differences were: $\mathcal{M}_{\rm This\ work} - \mathcal{M}_{\rm Del00} = -0.0080 \pm 0.0320\,\mathcal{M}_\odot$, $\mathcal{M}_{\rm This\ work} - \mathcal{M}_{\rm Ben06} = 0.0242 \pm 0.0474\,\mathcal{M}_\odot$, and $\mathcal{M}_{\rm This\ work} - \mathcal{M}_{\rm Man19} = 0.0042 \pm 0.0223\,\mathcal{M}_\odot$ Taking into account the standard deviations, Mann et al. (2019) provided the relation that best matched our $\mathcal{M}$. In the bottom panel of Fig. 3.27 we show this relation, valid in a wide mass range from 0.075 $\mathcal{M}_\odot$ to 0.70 $\mathcal{M}_\odot$. Since we fixed [Fe/H] = 0, we used the Mann et al. (2019) relation independent of metallicity ($f = 0$). Besides, the authors stated that the impact of [Fe/H] is sufficiently weak for the $f = 0$ relation to be safely used for most stars in the solar neighbourhood.

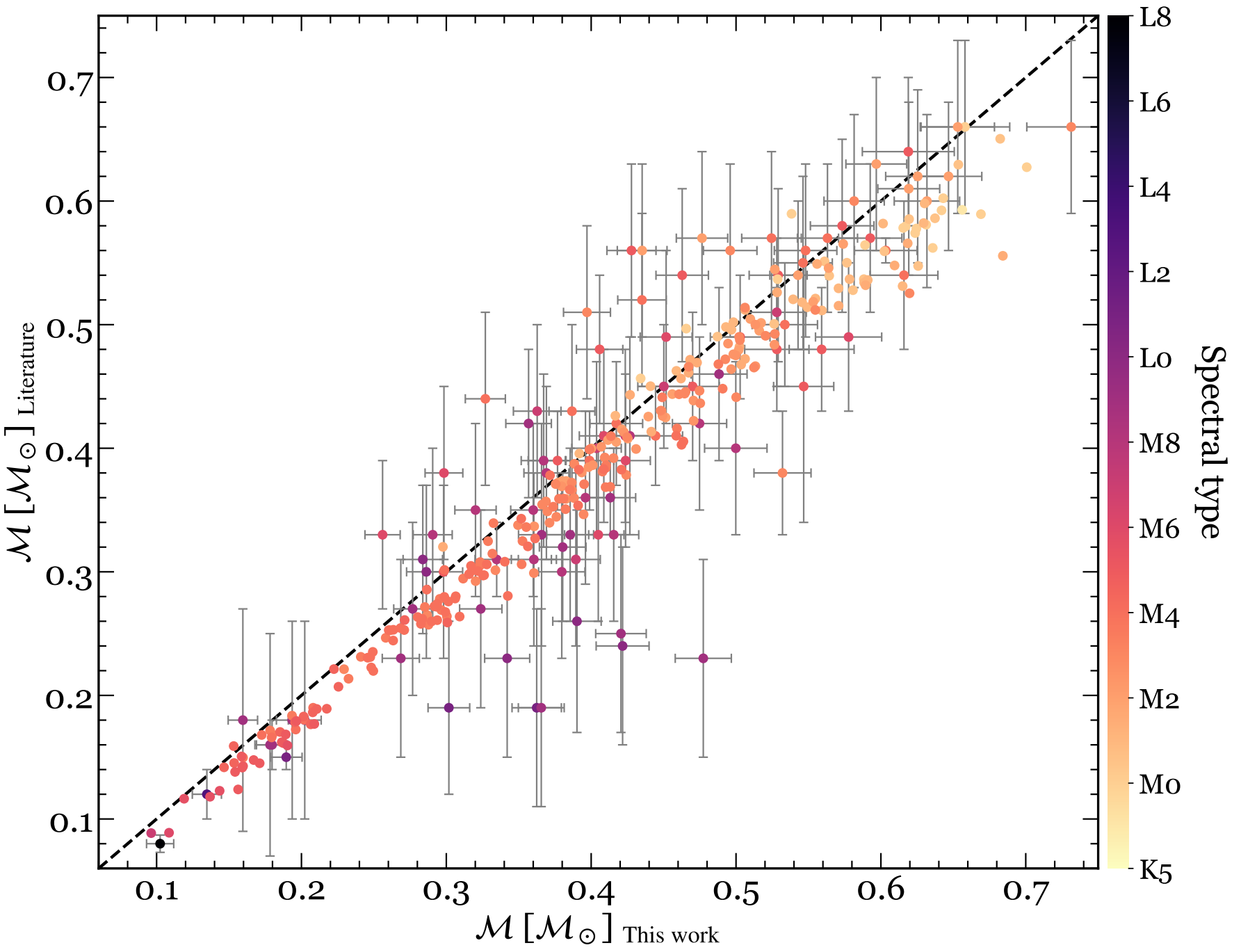

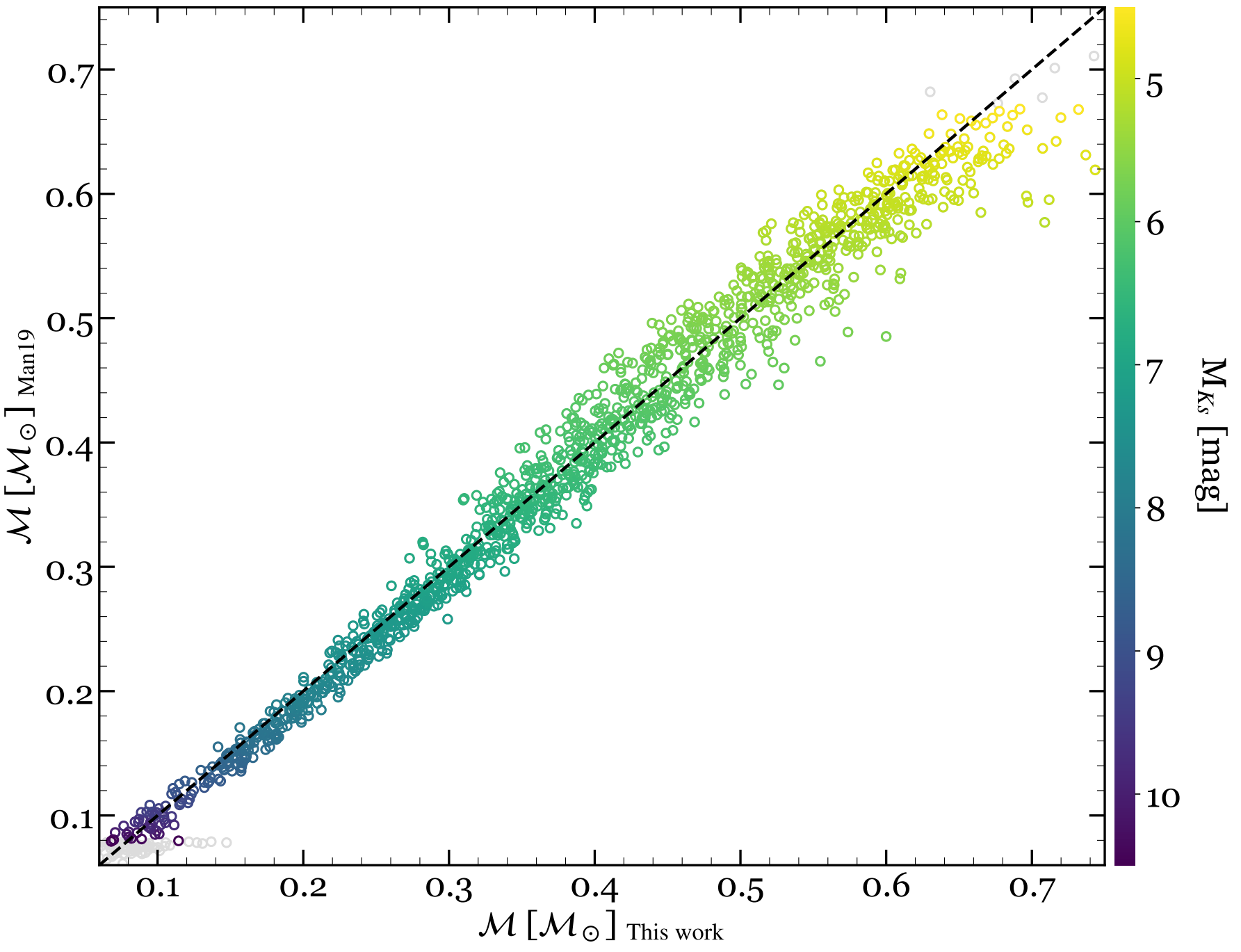

Figure 3.27: *Top:* Comparison of our masses with those from the literature. *Bottom:* Comparison with those derived from absolute magnitude $M_{K_s}$ using the metallicity-independent relation from Mann et al. (2019) only in its validity range (4.5 mag < $M_{K_s}$ < 10.5 mag).

**Colours** (Fig. 3.28). Although with the advent of *Gaia* the $V - J$ colour should be replaced by $G - J$, the former had been used extensively in the past. The match of our mean $V - J$ colours as a function of $T_{\rm eff}$ with those of Pecaut & Mamajek (2013) is once again excellent, but the relation significantly deviates from the values tabulated by Casagrande et al. (2008). However, as noted by them, the range of applicability of their colour-temperature-metallicity relations involving $V - J$ is narrow, between 0.61 mag and 2.44 mag. As a result, from the top left panel, extrapolating the $T_{\rm eff}$ versus $V - J$ relation of Casagrande et al. (2008) beyond 2.44 mag may result in $T_{\rm eff}$ systematically cooler by more than 300 K. In the top right panel, we revisit the $r' - i'$-spectral type diagram, which is an evolution of that with $R - I$ colour in the Johnson-Cousins passbands (Veeder, 1974; Bessell, 1979; Leggett, 1992; Boyajian et al., 2012; Mann et al., 2015; Houdebine et al., 2019). We reproduce the reversal at M7.0–8.0 V ($r' - i' \sim 2.8$ mag) observed by Hawley et al. (2002), Bochanski et al. (2007), and West et al. (2008), among many others. Therefore, we confirm that the $r' - i'$ colour alone cannot be used for spectral classification beyond M5.0 V. In the optical colour-colour diagram of the bottom left panel, our $g' - r'$ colours are slightly bluer than those of Davenport et al. (2014) for a fixed $r' - i'$, and significantly bluer, by about 0.5 mag, than those of Bochanski et al. (2007). Finally, in the bottom right panel, there is a good agreement with the location of the M-dwarf main sequence of Knapp et al. (2004) in the near-infrared $M_J$ versus $J - K_s$ diagram, but our data show instead the turnovers towards bluer and redder $J - K_s$ colours of late-K dwarfs and early-L dwarfs, respectively.

**Online table** We provide a summary table[13] with the compiled astro-photometric data and derived stellar parameters of all our targets. The description of the columns is given in Table D.1. The assembled catalogue in comma separated value (csv) format is available in its entirety in the electronic edition of this article. For each star or ultracool dwarf we tabulate its identifiers, equatorial coordinates, spectral type (and reference), parallax and distance (and reference), all magnitudes and their uncertainties, origin, quality flags (when available), $\mathcal{L}$, $T_{\rm eff}$, and $\log g$ from `VOSA`, $\mathcal{R}$ and $\mathcal{M}$ from the Stefan-Boltzmann law and the $\mathcal{M}$-$\mathcal{R}$ relation, *Gaia* DR2 identifier for primary and secondary sources (in the case of binary sources), four Boolean indices for close multiplicity ($\rho < 5$ arcsec), astrometric and photometric quality of the *Gaia* solution, and youth. Finally, most of the `Python` code developed by us for determining the parameters or preparing the plots shown in this work is available at GitHub[14].

## 3.4 Stellar characterisation

### 3.4.1 Different roads to radii and masses of the target stars

In Schweitzer et al. (2019) we homogeneously determined radii and masses for a sample of 293 nearby M dwarfs in Carmencita, presenting four routes to achieve it, represented by $\mathcal{M}_{\mathcal{M}-\mathcal{R}}$, $\mathcal{M}_{\log g}$, $\mathcal{M}_{K_s}$, $\mathcal{M}_{\rm P}$. Two groups of ingredients can provide these fundamental parameters, used both together and inde-pendently. These are broadband photometry from the ultraviolet to the mid-infrared (see Table 3.1), and photospheric parameters (effective temperature, $T_{\rm eff}$, surface gravity, $\log g$, and iron abundance, [Fe/H]), obtained by fitting `PHOENIX-ACES` synthetic spectra (Husser et al., 2013) to the CARMENES spectra of the visual channel as described in Passegger et al. (2016, 2018), and later extended to include the near-infrared channel, using `SteParSyn` (Passegger et al., 2019; Marfil et al., 2021). Figure 3.29 shows a schema of the possibilities.

[13] https://cdsarc.cds.unistra.fr/viz-bin/cat/J/A+A/642/A115.
[14] https://github.com/ccifuentesr/CARMENES-V.

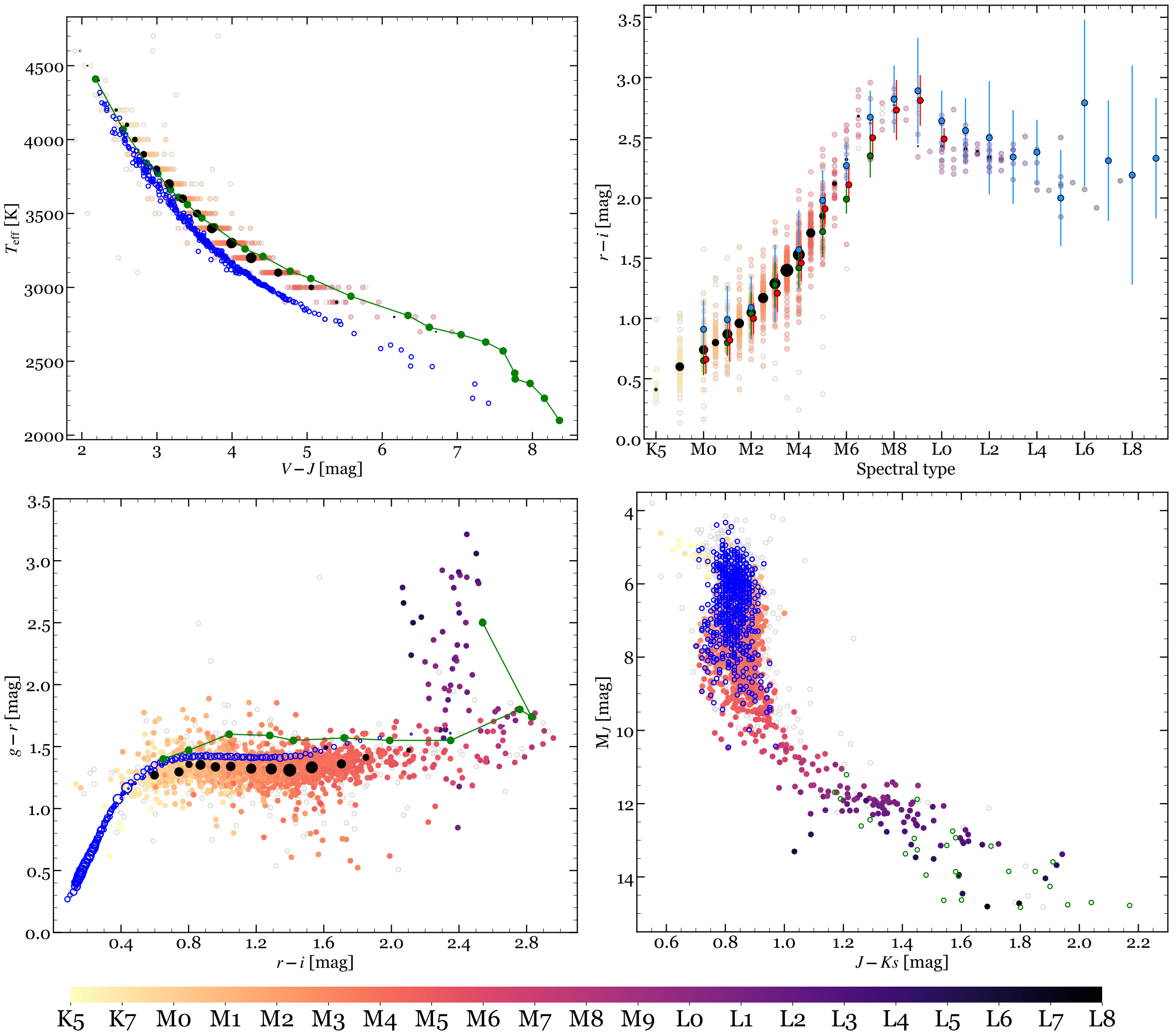

Figure 3.28: *Top left:* $T_{\rm eff}$ vs. $V-J$. The blue empty circles and green line are data from Casagrande et al. (2008) and Pecaut & Mamajek (2013), respectively. Black filled circles are the mean colours from 2700 K to 4600 K in our sample, with a size proportional to the number of stars. *Top right:* $r' - i'$ vs. spectral type. Open circles are the mean colours of Hawley et al. (2002, blue), Bochanski et al. (2007, "inactive" colours, green), and West et al. (2008, red). Black filled circles are the mean colours from K5 V to L2 in our sample, taken from Table C.2, with a size proportional to the number of stars. The error bars are the standard deviation (Hawley et al., 2002; Bochanski et al., 2007) and the intrinsic scatter of the stellar locus (West et al., 2008). *Bottom left:* $g'-r'$ vs. $r'-i'$. Black filled circles are the mean colours from K5 V to L2 in our sample, taken from Table C.2, with a size proportional to the number of stars. Blue empty circles are the mean colours by Davenport et al. (2014), with a size proportional to the numbers of stars, and the green line links the mean "inactive" colours by Bochanski et al. (2007) for spectral types M0 V to L0. *Bottom right:* $M_J$ vs. $J-K_s$. Green and blue empty circles are data from Knapp et al. (2004) and Lépine et al. (2013), respectively. In the four panels, the stars in our sample are colour-coded by spectral type, and the discarded stars are plotted with grey empty circles.

Table 3.8: Description of the online table.

| Parameter | Units | Column(s) | Description |
|---|---|---|---|
| Karmn | ... | 1 | Carmencita star identifier (JHHMMm+DDd)[a] |
| Name | ... | 2 | Discovery name or most common name[b] |
| RA, DE | hms | 3–4 | Right ascension and declination (equinox J2000, epoch 2015.5) |
| SpType, SpTnum | ... | 5–6 | Spectral type and its numerical format[c] |
| Ref_SpT | ... | 7 | Reference for the spectral type |
| Plx, ePlx | mas | 8–9 | Parallax and its uncertainty |
| Ref_Plx | ... | 10 | Reference for the parallax |
| d_pc, ed_pc | pc | 11–12 | Distance and its uncertainty |
| Ref_d | ... | 13 | Reference for the distance |
| Lbol, eLbol | $L_\odot$ | 14–15 | Luminosity and its uncertainty from VOSA |
| Teff | K | 16 | Effective temperature from VOSA[d] |
| logg | dex | 17 | Surface gravity from VOSA[d] |
| Radius, eRadius | $\mathcal{R}_\odot$ | 18–19 | Radius and its uncertainty |
| Mass, eMass | $\mathcal{M}_\odot$ | 20–21 | Mass and its uncertainty |
| NN_mag, eNN_mag | mag | 22–97 | Magnitude and its uncertainty for the NN passband[e] |
| Qf_NN, Ref_NN | mag | 22–97 | Quality flag (if available) and reference for the NN passband[e] |
| Gaia_id_1 | ... | 98 | *Gaia* DR2 identifier of single or primary star |
| Gaia_id_2 | ... | 99 | *Gaia* DR2 identifier of secondary star in close binary system |
| Multiple | ... | 100 | Boolean index for close multiple stars |
| Young | ... | 101 | Boolean index for overluminous young stars |
| ruwe | ... | 102 | Boolean index for stars with *Gaia* ruwe > 1.41 |
| Excess | ... | 103 | Boolean index for stars with photometric flux excess in *Gaia* $G_{B_P}$ and $G_{R_P}$ passbands |

[a] For the K dwarfs, we tabulate the SUPERBLINK catalogue identifier (Lépine & Shara, 2005; Lépine et al., 2013).
[b] For the ultracool dwarfs, we tabulate the *Gaia* UltraCool Dwarf Catalogue identifier (Smart et al., 2017, 2019).
[c] SpTnum = −2 for K5 V, −1 for K7 V, 0.0 for M0.0 V, 0.5 for M0.5 V... 10.0 for L0.0, etc.
[d] VOSA uncertainties are 50 K for $T_{\rm eff}$ (25 K for $T_{\rm eff} \lesssim 2400$ K) and 0.5 dex for log $g$.
[e] FUV, NUV: GALEX DR5 $FUV$ and $NUV$; BP, GG, RP: $G_{B_P}$, $G$, and $G_{R_P}$ from *Gaia* DR2; BT, VT: $B_T$ and $V_T$ from Tycho-2; B, V: $B$ and $V$ from UCAC4 or APASS9; u, g, r, i: $u'$, $g'$, $r'$, and $i'$ from SDSS9, UCAC4, APASS9, PanSTARRS-1 and/or CMC15; J, H, Ks: $J$, $H$, and $Ks$ from 2MASS; W1, W2, W3, W4: $W1$, $W2$, $W3$, and $W4$ from AllWISE or WISE.

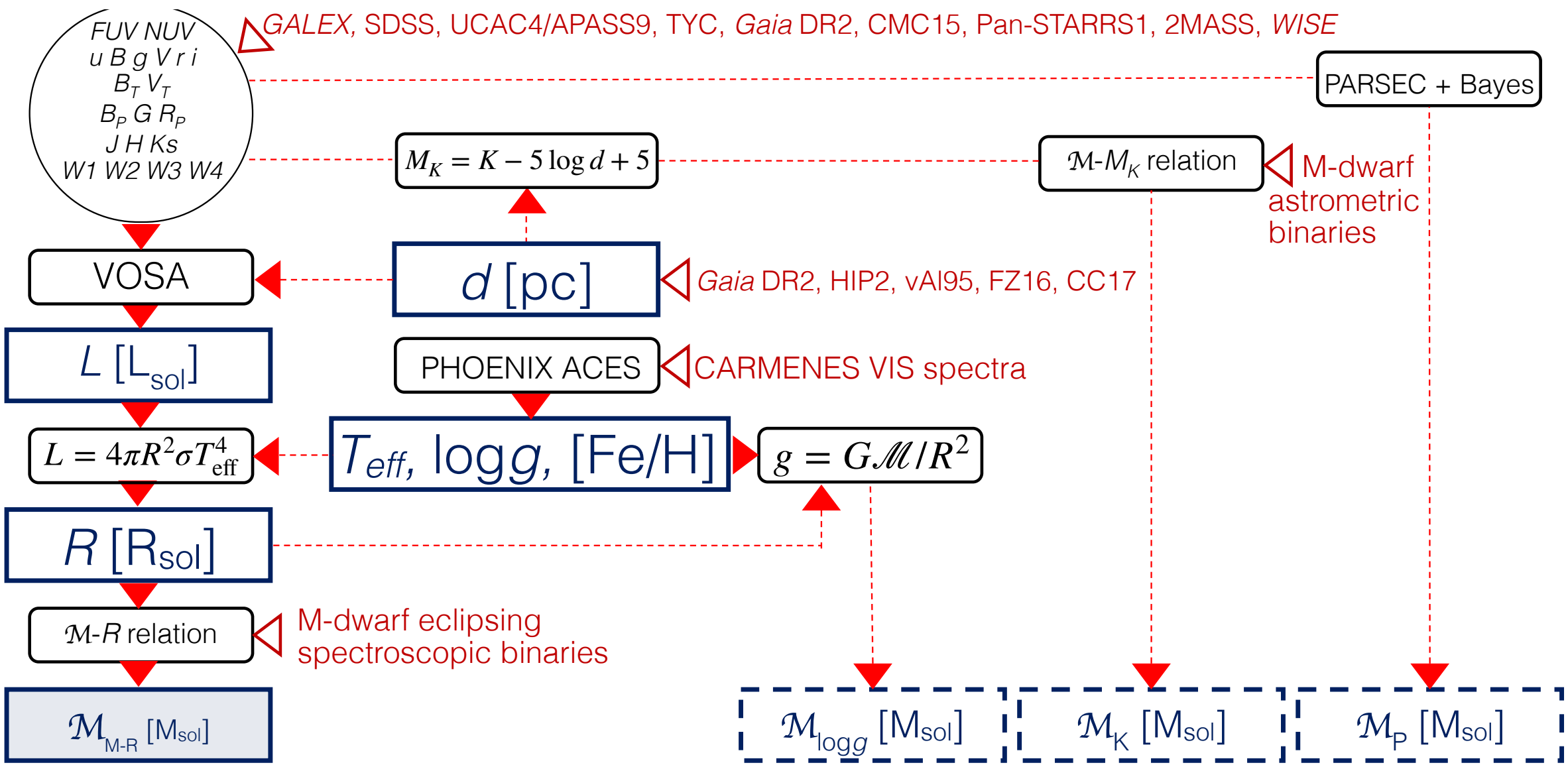

Figure 3.29: Flowchart of the different roads to masses (extracted from Schweitzer et al. 2019).

From integrated broadband photometry, together with the *Gaia* DR2 parallaxes, we obtained the bolometric luminosities $\mathcal{L}$ using `VOSA`, and Stefan-Boltzmann's law provides the radius, given that the effective temperature is known. The radius offers two routes for deriving masses. First, using an empirical relation between masses and radii from eclipsing binaries ($\mathcal{M}_{\mathcal{M}-\mathcal{R}}$). Second, from the definition of surface gravity, $\log g$ ($\mathcal{M}_{\log g}$), by fitting PHOENIX-ACES synthetic spectra (Husser et al., 2013) to the CARMENES spectra of the VIS channel, as in Passegger et al. (2016, 2018). From photometry alone ($r$ and $J$), another method consist of a Bayesian approach that uses `PARSEC` library of stellar evolution models. The absolute magnitudes of our stars, $M_J$, were compared to synthetic models, using the colour $r - J$ and assuming solar metallicity, and the mass ($\mathcal{M}_{\rm P}$) was derived from this comparison. Finally, among the mass-magnitude relations available in the literature using infrared filters (Delfosse et al., 2000; Benedict et al., 2016, e.g.), we used the $M_{K_s}$-$\mathcal{M}$ relation from Mann et al. (2019). This way we obtain the mass from absolute magnitude ($\mathcal{M}_{K_s}$). The comparison between the masses obtained using the four routes is shown in Fig. 3.30). Because the metallicity was also a product of the spectral analysis, the plots account for this parameter.

The four methods used in this work delivered masses for 293 M dwarfs from the CARMENES GTO sample that we found in good agreement for the majority of our targets. These are typical field stars that are not too young, but show discrepancies in the case of very young objects. particularly when using the masses $\mathcal{M}_{\mathcal{M}-\mathcal{R}}$ and $\mathcal{M}_{\mathcal{M}-{\rm K_s}}$. In the case of $\mathcal{M}_{\rm P}$, though, the relation is not restricted to a single age, and with $\mathcal{M}_{\log g}$ it is possible to use isochrones given that the age of the star is known. Between spectral types M0 V and M7 V our radii covered the range $0.1\,\mathcal{R}_\odot < \mathcal{R} < 0.6\,\mathcal{R}_\odot$ with an error of 2–3% and our masses cover $0.09\,\mathcal{M}_\odot < \mathcal{M} < 0.6\,\mathcal{M}_\odot$ with an error of 3–5%. The methods work best for a field star of at least a few hundred million years when it is possible to spectroscopically determine an effective temperature and when $\mathcal{L}$ is well determined with multiwavelength, broadband photometry, and an accurate parallax.

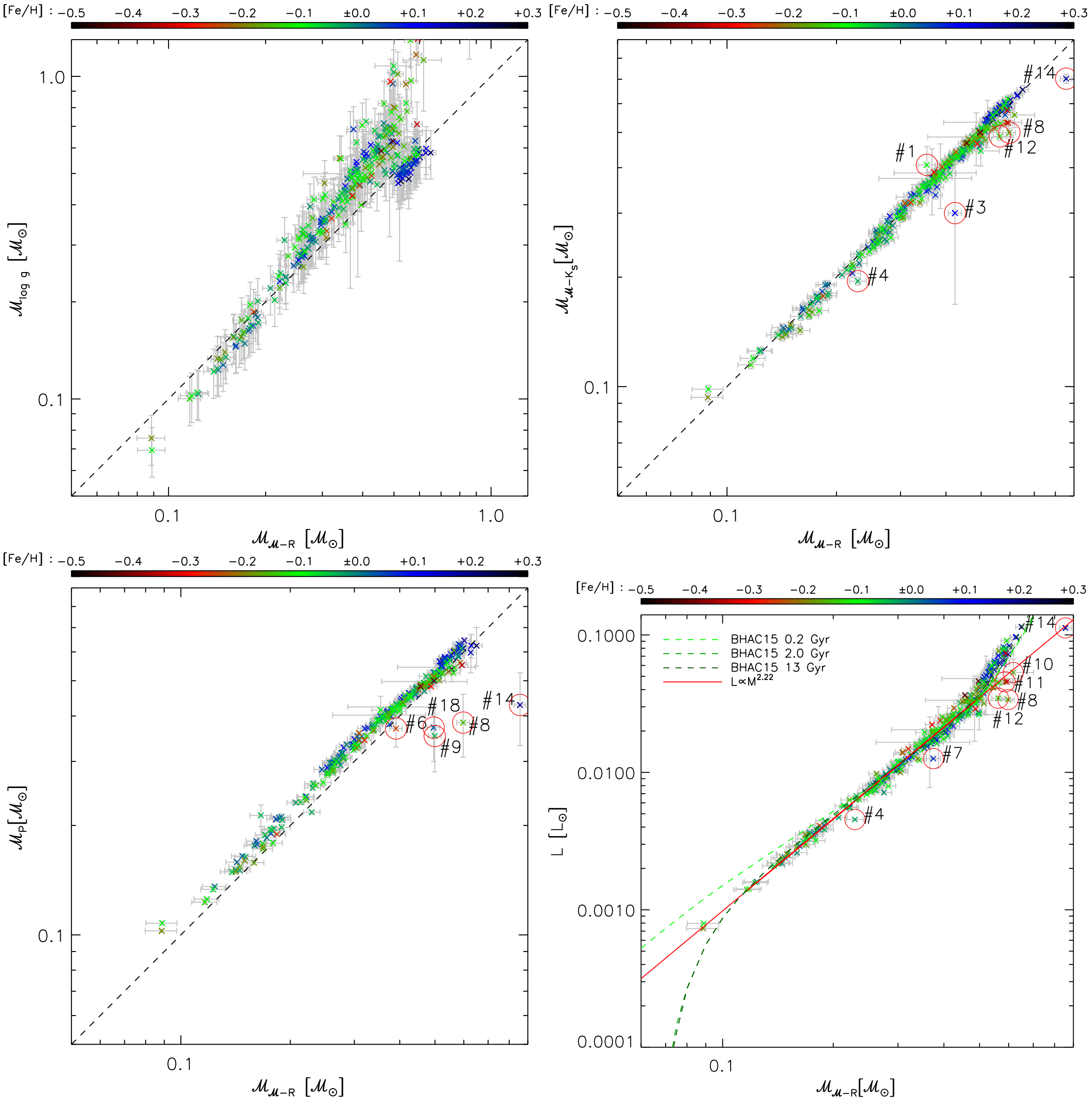

Figure 3.30: Comparison of masses for the stars studied in Schweitzer et al. (2019), obtained from detached, double-lined, double-eclipsing, main-sequence M dwarf binaries, $\mathcal{M}_{\mathcal{M}-\mathcal{R}}$, colour-coded by metallicity, against: Spectroscopic masses, using surface gravity, $\log g$, from fitting synthetic spectra $\mathcal{M}_{\log g}$ (*top left*), photometric masses from absolute magnitude $M_{K_s}$, as in Mann et al. (2019), $\mathcal{M}_{\mathcal{M}-\mathrm{K_s}}$ (*top right*), PARSEC-based masses with a Bayesian approach, $\mathcal{M}_{\mathrm{P}}$ (*bottom left*), and bolometric luminosities, with isochrones for 0.2, 2.0, and 13 Ga as dashed lines in increasingly darker green, from BHAC15 (*bottom right*). The black dashed lines marks the 1:1 relation, and the red solid line indicates the fit $\mathcal{L} \propto \mathcal{M}^{2.22\pm0.02}$. Obvious outliers are tagged in the fours plot, and discussed in detail in the original article.

### 3.4.2 Exomoons in the habitable zones of M dwarfs

Similarly to exoplanets, exomoons can also be targets for biosignature surveys. In Martínez-Rodríguez et al. (2019) we investigated the habitability, stability, and detectability of potential exomoons of exoplanets that orbit M dwarfs. While exoplanets in the CHZ of M dwarfs can be hostile environments to life, the existence of stable exomoons may alleviate this issue.

For 205 exoplanets orbiting 109 M dwarfs discovered with the radial velocity or transit methods, we compiled orbital, astrometric, photometric, and basic astrophysical parameters. For 192 of them, we estimated the most probable masses and radii using the models of Chen & Kipping (2017), assuming literature values for the remaining 13 planets. For the stars we derived luminosities, masses, and radii proceeding as in Cifuentes et al. (2020), and determined the inner and outer habitable zone boundaries for every star using a one-dimensional climate model. We found that 33 of the known planets orbit within the zone of habitability (Fig. 3.31). In Table C.3 we provide a complete characterisation of these systems, including whether they fall in the habitable zone[15].

For these we modelled non-eccentric, non-obliquous moons, and computed moon migration timescales for two different scenarios: strip-away from the planet, and fall-back onto the planet. Additionally, we took into consideration the planetary spin with two conditions: a maximum value equal to the orbital period (i.e. tidally locked to the star), and a minimum value of 3 h.

In the $P_0 = P_{\rm orb}$ scenario, all hypothetical moons fall back onto their planets after short time scales, except for four planets that survive because the migration time scales are longer than the Hubble time and, more interestingly, longer than protoplanetary disk dissipation times . These are Ross 1003 b, IL Aqr b and c, and CD-23 1056 b), and they may potentially harbour life and be targets for biosignature surveys. We further explored whether if their mantles would partially melt under the effects of tidal heating. Io is an example of this effect in the Solar System. Similarly to the Moon that is the main cause for the tides in the Earth oceans, Jupiter is responsible for bending the surface of Io back and forth, building up heat that makes the interior melt and boil. This may be the reason why Io is the most volcanically active object in the Solar System (see Tyler et al., 2015). Only a potential exomoon in IL Aqr c could suffer from melting. Even when most of these four planets are tidally locked to their host stars, it may still be possible for the hypothetical exomoons to retain the conditions for habitability. And even if their planets might not host liquid water because they are icy neptunians, some of their hypothetical exomoons could instead. Is in this sense that exomoons could alleviate the problem of habitability in many cases.

The two of observable effects that moons produce around exoplanets require that the planet transits. In one, the transit of the moon is superimposed of photometric transits, on the other the timing and length of the transits of the planet is perturbed (Kipping, 2009a,b). Hypothetical moons in suitable (stable) planets like the four exoplanets selected in this work would exhibit large orbital periods, which minimises the opportunities for transiting episodes. An opportunity may come first for the most massive moons, by the depth of its transit. In this eclipsing event, the massive moon might retain an atmosphere that could potentially be detected. Some of the missions that might be able to perform these detections are *CHEOPS* (Fortier et al., 2014), the upcoming *PLATO*[16] (Rauer et al., 2014), *Kepler*-class photometry (Kipping et al., 2009), or ESPRESSO (Pepe et al., 2010), via the Rossiter-McLaughlin effect as in Zhuang et al. (2012).

To end up with, it is worth mentioning the work by Tokadjian & Piro (2020), which is an extension of that by Martínez-Rodríguez et al. (2019). The work explores the potential for exoplanets to host exomoons using differential equations to model the orbital and rotational evolution of planet-star-moon systems. The authors identified 36 exoplanets that may be good candidates for hosting exomoons, including Kepler-22b, Kepler-1638 b, Kepler-991 b, Kepler-1635 b, and Kepler-1625 b. The authors suggest that the detection of exomoons may be an important constraint on exoplanet structure, and call for more detailed models of tidal dissipation in future research.

[15] The list of derived parameters, as well as the code produced in this work can be found in the GitHub public repository (`https://github.com/hector-mr/Exomoons_HZ_Mdwarfs`).

[16] PLAnetary Transits and Oscillation of stars (*PLATO*) is an European Space Agency (ESA) mission to be launched in the last quarter of 2026.

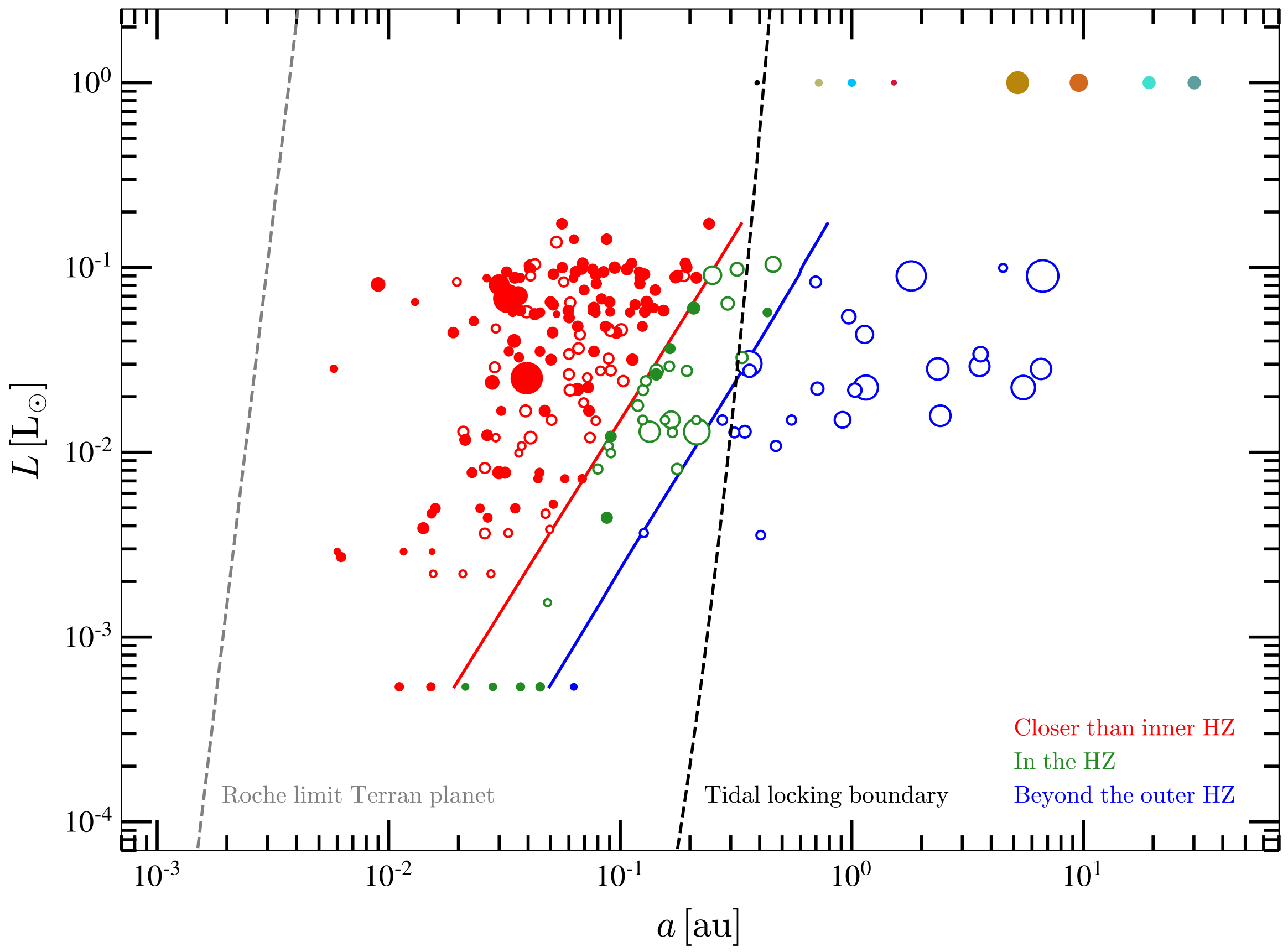

Figure 3.31: Conservative HZ for all the M dwarfs in the sample. Their hosted exoplanets are depicted with sizes proportional to their masses, with filled circles if they have been detected by transit and with open circles if they have been detected by RV measurements. Exoplanets closer than the inner HZ, in the HZ, and beyond the outer HZ are shown in red ("recent Venus"), green, and blue ("maximum greenhouse"), respectively. The dashed lines are the Roche limit for an Earth-like planet (Aggarwal & Oberbeck, 1974) in grey, and a "constant-time-lag" tidal locking model for a 10 $\mathcal{M}_\odot$-planet with rapid initial rotation after 1 Ga (Barnes, 2017), in black. The eight Solar System planets are depicted in the upper area (extracted from Martínez-Rodríguez et al. 2019).

### 3.4.3 GTO and Legacy

Many works in the last two years have benefited from astrophysical parameters derived in Schweitzer et al. (2019), Cifuentes et al. (2020), or have proceeded in the same manner as theirs with the latest astrometric, photometric and spectroscopic parameters available. In Table 3.9 we list a sample of a selection of published works, whose stellar description benefited from our characterisation, most of them corresponding to planet detections.

Table 3.9: Published planet-detection works with parameters derived by us.

| Work | System | $\mathcal{L}_\star$ [$10^{-4}\,\mathcal{L}_\odot$] | $\mathcal{M}_\star$ [$\mathcal{M}_\odot$] | $\mathcal{R}_\star$ [$\mathcal{R}_\odot$] |
|---|---|---|---|---|
| Suárez-Mascareño et al. [2023] | GJ 1002 | 14.06 ± 0.19 | 0.120 ± 0.010 | 0.137 ± 0.005 |
| Kossakowski et al. (2023) | GJ 806 | 29.44 ± 0.28 | 0.167 ± 0.011 | 0.1813 ± 0.0063 |
| Palle et al. (2023) | GJ 806 | 259.9 ± 0.9 | 0.413 ± 0.011 | 0.4144 ± 0.0038 |
| Blanco-Pozo et al. (2023) | GJ 1151 | 33.15 ± 0.18 | 0.1639 ± 0.0093 | 0.1781 ± 0.0042 |
| Chaturvedi et al. (2022) | TOI 1468 | 159.5 ± 0.9 | 0.339 ± 0.011 | 0.344 ± 0.005 |
| Kossakowski et al. (2022) | AD Leo | 235.9 ± 1.1 | 0.423 ± 0.012 | 0.423 ± 0.006 |
| Luque et al. (2022b) | G 9-40 | 109.6 ± 1.9 | 0.295 ± 0.014 | 0.303 ± 0.009 |
| Caballero et al. (2022) | Gl 486 | 121.20 ± 0.82 | 0.333 ± 0.019 | 0.339 ± 0.015 |
| Kemmer et al. (2022) | GJ 3929 | 115.5 ± 1.1 | 0.309 ± 0.014 | 0.315 ± 0.010 |
| Espinoza et al. (2022) | TOI 1759 | 876.7 ± 6.3 | 0.606 ± 0.020 | 0.597 ± 0.015 |
| Kossakowski et al. (2021) | TOI 1201 | 340.0 ± 5.7 | 0.512 ± 0.020 | 0.508 ± 0.016 |
| | TOI 393 | 268.3 ± 2.5 | 0.463 ± 0.018 | 0.462 ± 0.014 |
| Amado et al. (2021) | Gl 393 | 268.7 ± 5.4 | 0.426 ± 0.017 | 0.426 ± 0.013 |
| | G 264-012 | 106.6 ± 1.1 | 0.297 ± 0.024 | 0.305 ± 0.011 |
| Trifonov et al. (2021) | Gl 486 | 121.0 ± 2.3 | 0.333 ± 0.019 | 0.339 ± 0.015 |
| Bluhm et al. (2021) | TOI-1685 | 303.5 ± 5.4 | 0.495 ± 0.019 | 0.492 ± 0.015 |
| Soto et al. (2021) | LHS 1478 | 71.5 ± 1.2 | 0.236 ± 0.012 | 0.246 ± 0.008 |
| Luque et al. (2021) | TOI-776 | 490 ± 20 | 0.544 ± 0.028 | 0.538 ± 0.024 |
| Kemmer et al. (2020) | GJ 3473 | 150.0 ± 1.9 | 0.360 ± 0.016 | 0.364 ± 0.012 |
| Baroch et al. (2020) | YZ CMi | 113 ± 17 | ... | 0.369 ± 0.055 |

# Chapter 4

# Multiplicity

Among the many possibilities of observational astronomy, stellar multiplicity may well be regarded as one of the most appealing. The existence of double stars has been recognised historically, but their true nature remained elusive even long after the invention of the telescope. Michell (1784) proposed that the odds of so many systems to appear so close by pure chance was of 500 000 to 1, at the supposition that they had been scattered by mere chance[1]. It was later admitted that the apparently fixed stars had a proper motion, hence the idea that stars that lied close in the sky were mutually affected gained interest. Observing the proper motions of several pairs, Mayer (1778) published the first catalogue of double stars, which contained 72 systems. William Herschel catalogued 898 visual doubles, which the majority were later confirmed physical pairs (Herschel & Banks, 1782; Herschel, 1785, 1822). The terrain was thereafter meticulously traveled by many others (e.g. Struve, 1837; Herschel et al., 1874; Burnham, 1906; Aitken & Doolittle, 1932, and more recently Tokovinin 1997, 2018; Mason et al. 2009; Raghavan et al. 2010). Worley (1962) carried out one of the first systematic studies of M dwarfs, which became a progenitor of the main database of astrometric double and multiple systems of use today, the Washington Double Star catalogue (WDS; Worley & Douglass, 1997; Mason et al., 2001), whose lineage traces back over 100 years. WDS collects to date 155 438 systems, with precise astrometric history and orbital description for many of the pairs.

## 4.1 Introduction

Mizar and Alcor ($\xi$ Ursae Majoris) constitute an epitome of the advance of astronomical technique in this matter. Easily visible by the naked eye, at least as a single star, it is the fourth brightest star in the

[1]"The very great number of stars that have been discovered to be double, triple, &c. particularly by Mr. Herschel, if we apply the doctrine of chances, as I have heretofore done in my *Enquiry into the probable Parallax, &c. of the Fixed Stars* (Michell, 1767), cannot leave a doubt with any one [...] that by far the greatest part, if not all of them, are systems of stars so near to each other, as probably to be liable to be affected sensibly by their mutual gravitation; and it is therefore not unlikely, that the periods of the revolutions of some of these about their principals [...] may some time or other be discovered."

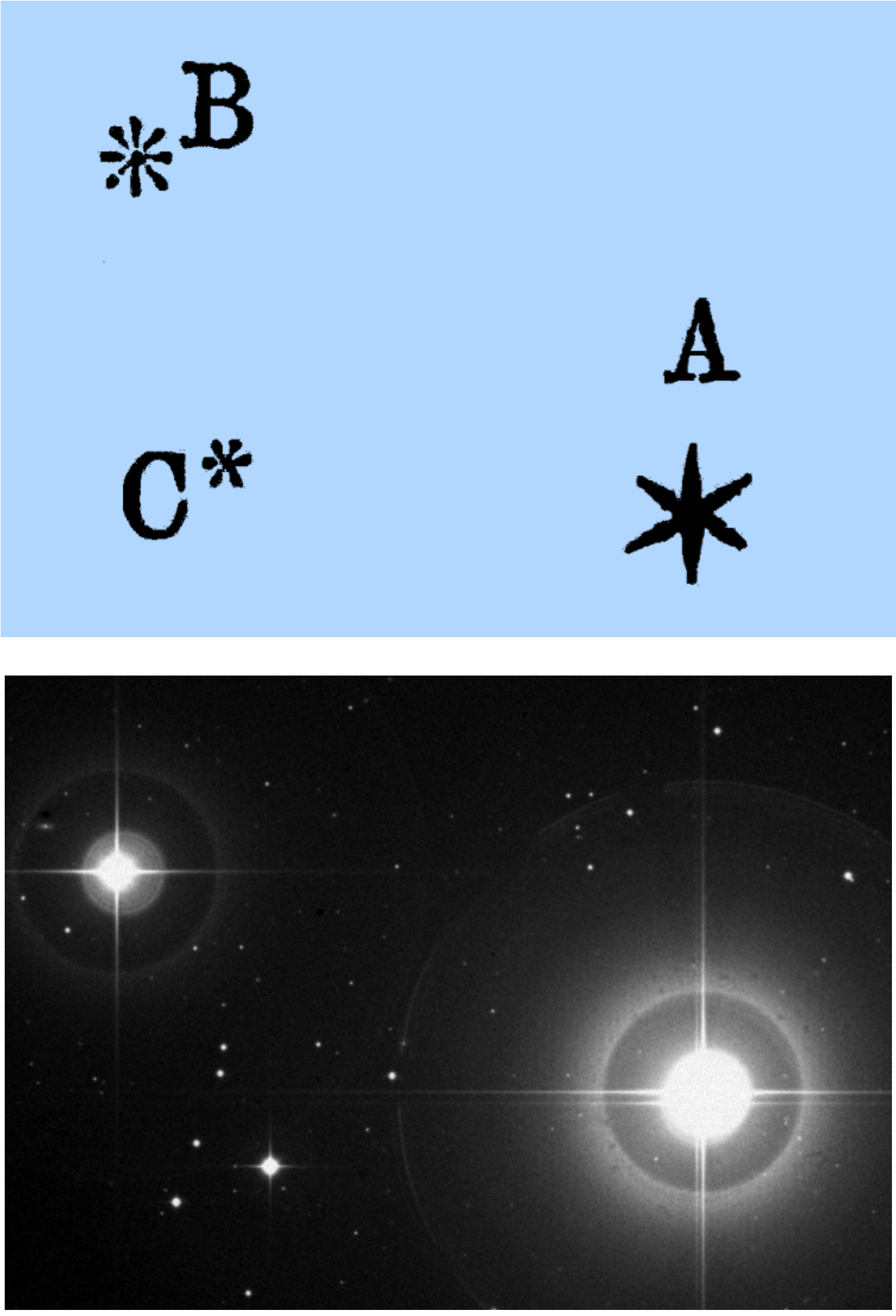

Figure 4.1: Castelli's asterism including Mizar (A), Alcor (B), and "Sidus Ludoviciana" (C). From Galileo's *Opere*, circa 1617 (*top*), and the same view from the Digitized Sky Survey (DSS) photographic plates, in red filter from the POSS-II F survey, 1991 (*bottom*).

Big Dipper. Although known to serve as a visual acuity test more than a thousand years ago (al-Sufi, 964 CE, and see Bohigian, 2008), it was the Benedictine monk Benedetto Castelli who first proposed to Galileo a systematic observation of the pair (Fig. 4.1)[2]. Through the recently invented telescope, Galileo separated the pair in 1617 (Siebert, 2005). Mizar and Alcor evolved from a visual binary, seemingly unconnected in ancient times, into a well-defined, physically bound sextuple system (Maury & Pickering, 1897; Ludendorff, 1908; Mamajek et al., 2010). Because of its relevance, stellar multiplicity has undergone intense scrutiny over the years to the point that some evidence suggest that all stars form as binaries. A successful theory of star formation should be able to explain why do we observe some stars that belong to multiple systems while many others do not, and how the multiple systems come to be as they are – for instance, how many pairs are found depending on the physical separation.

Stellar multiplicity is a common outcome of the stellar formation process (Heintz, 1969; Larson, 1972; Batten, 1973; Poveda et al., 1982; Abt, 1983; Pringle, 1989; Duquennoy & Mayor, 1991; Fischer & Marcy, 1992; Chapman et al., 1992; Mathieu et al., 2000; Tokovinin, 2008, and see Duchêne & Kraus 2013 for a review; cf. Lada 2006). The frequency of multiple systems increases with the primary stellar mass (Lada, 2006; Parker & Meyer, 2014). The observational evidence is that the percentage of OBA main sequence stars that belong to a multiple system is larger than 70–80 %, and perhaps 100 % for the O type (Shatsky & Tokovinin, 2002; Mason et al., 2009), for solar type stars is around 44–67 % (Duquennoy & Mayor, 1991; Raghavan et al., 2010; Duchêne & Kraus, 2013), and in the case of low-mass stars is 10–30 % (Burgasser et al., 2003, 2006; Joergens, 2008; Guszejnov et al., 2017). Nevertheless, in the latter the efficiency in terms of the fraction of the available mass that goes into stars can be considered high (Matzner & McKee, 2000). In the particular case of M dwarfs, early studies estimated in 33–42 % the fraction of them that are part of a multiple system (Fischer & Marcy, 1992; Reid et al., 1997). More specifically, the multiplicity rate of M dwarfs has been estimated to approximate to 30 and 20 for a stellar (M dwarf) and substellar (brown dwarf) companion, respectively (Chabrier, 2003b). Recent estimates suggest that the multiplicity of M dwarfs is 26–27 %, or even lower (Delgado-Donate et al., 2004; Ward-Duong et al., 2015; Cortés-Contreras et al., 2017a; Winters et al., 2019a).

[2]Sidus Ludoviciana or Ludwig's star (HD 116798) is a bright ($G$ = 7.5 mag), distant ($d \simeq 91.2$ pc), giant star (A8/F0 III), in Galileo's times mistaken for a planet by Johann Georg Liebknecht.

Larson (1972) first suggested that a rotating cloud would not collapse into a single star, but instead it would fragment into two or more condensations, resulting in the formation of a multiple system of stars. Now it is widely accepted that multiple systems result from the collapse and fragmentation of cloud cores (Boss & Bodenheimer, 1979; Larson, 1985; Boss, 1988; Pringle, 1989; Bate, 2000; White & Ghez, 2001; Padoan & Nordlund, 2004; Ward-Thompson et al., 2007; Reipurth et al., 2014; Tokovinin & Moe, 2020). Although it may not be the only mechanism, it is successful in explaining the mass-ratio distribution observed among the pre-main-sequence systems (Bonnell et al., 2001; Goodwin et al., 2007; Reggiani & Meyer, 2011). There is a possibility that all stars were born in multiple systems, but a majority could not remain together as they evolved (Kroupa, 2008; King et al., 2012b; Elliott et al., 2014; Sadavoy & Stahler, 2017, c.f. King et al. 2012a). This is supported by the fact that young stars are almost always found to be part of multiple systems. If the primordial population of stars was essentially of multiple systems, their diversity observed today, including the lack of multiplicity in many stars, implies that dynamical interaction must have a role in the early stages of their evolution (Goodwin, 2010; Reipurth et al., 2014). It is clear that multiple systems offer valuable insights into the ways in which stars interact and influence one another.

The dynamical interplay between the components turns into a competition for attaining stable orbits. The loss of angular momentum during the shrinkage of the closest pair is transferred to the widest, which can result in one or more of them being ejected from the original compact arrangement (Delgado-Donate et al., 2003, 2004; Tokovinin et al., 2006; Basri & Reiners, 2006; Caballero, 2007; Moeckel & Bate, 2010; Reipurth & Mikkola, 2012b, and see again Reipurth et al. 2014). This process normally unfolds during the very youth of the system with durations of less than $10^5$ years (Goodwin & Kroupa, 2005). These authors also noted that ejections would produce a significant population of close binaries that are not observed. This is an important point to remember, as the topic of unresolved binaries will be proven of great relevance in this work. Ejections are not the only mechanism to account for the large separations of some binary systems, for which the theory of collapse and fragmentation does not serve satisfactorily. Other channels proposed are turbulent fragmentation (Goodwin et al., 2004; Bate, 2009; Offner et al., 2010; Tobin et al., 2016), the dissolution of clusters (Kouwenhoven et al., 2010; Moeckel & Clarke, 2011), and random bindings from slow-moving near pre-stellar cores (Leigh & Geller, 2013; Tokovinin, 2017), this latter accounting for the large fraction of young wide pairs.

Systems that contain more than two components are, in principle, unstable (Harrington, 1972, and see again Goodwin & Kroupa 2005). However, the dynamical evolution might produce a hierarchical arrangement of binaries within binaries, or nested orbits, that conduces to stability (Evans, 1968; Bonnell et al., 2003; Tokovinin, 2014). In this sense, some works such as those of Basri & Reiners (2006), Caballero (2007), and Kouwenhoven et al. (2010) predicted or pointed out to a major prevalence of wide triples over wide binaries (and see Czavalinga et al., 2023). Indeed, in many instances typically one of the components of a wide binary is further resolved as a binary itself, therefore producing a triple system (see e.g. Cifuentes et al., 2021). Luckily, these can be treated individually as two-body problems (e.g. Evans, 1968). Our nearest stellar neighbour is a triple system of this kind: $\alpha$ Centauri consists of two Sun-like main stars (A and B) in mutual orbit. The third component (C) is Proxima Centauri, a red dwarf orbiting the pair at a separation that is 400 to 1100 times that of the pair AB[3]. Proxima orbits a double star system, but for all practical purposes, the star appears to be single. For young systems, without enough lifetime to have settled down into stable configurations, it is often unclear whether a given system can be treated as a young trapezia-like architecture, or as a mature cluster (see for instance the system of V1311 Ori in Tokovinin, 2022).

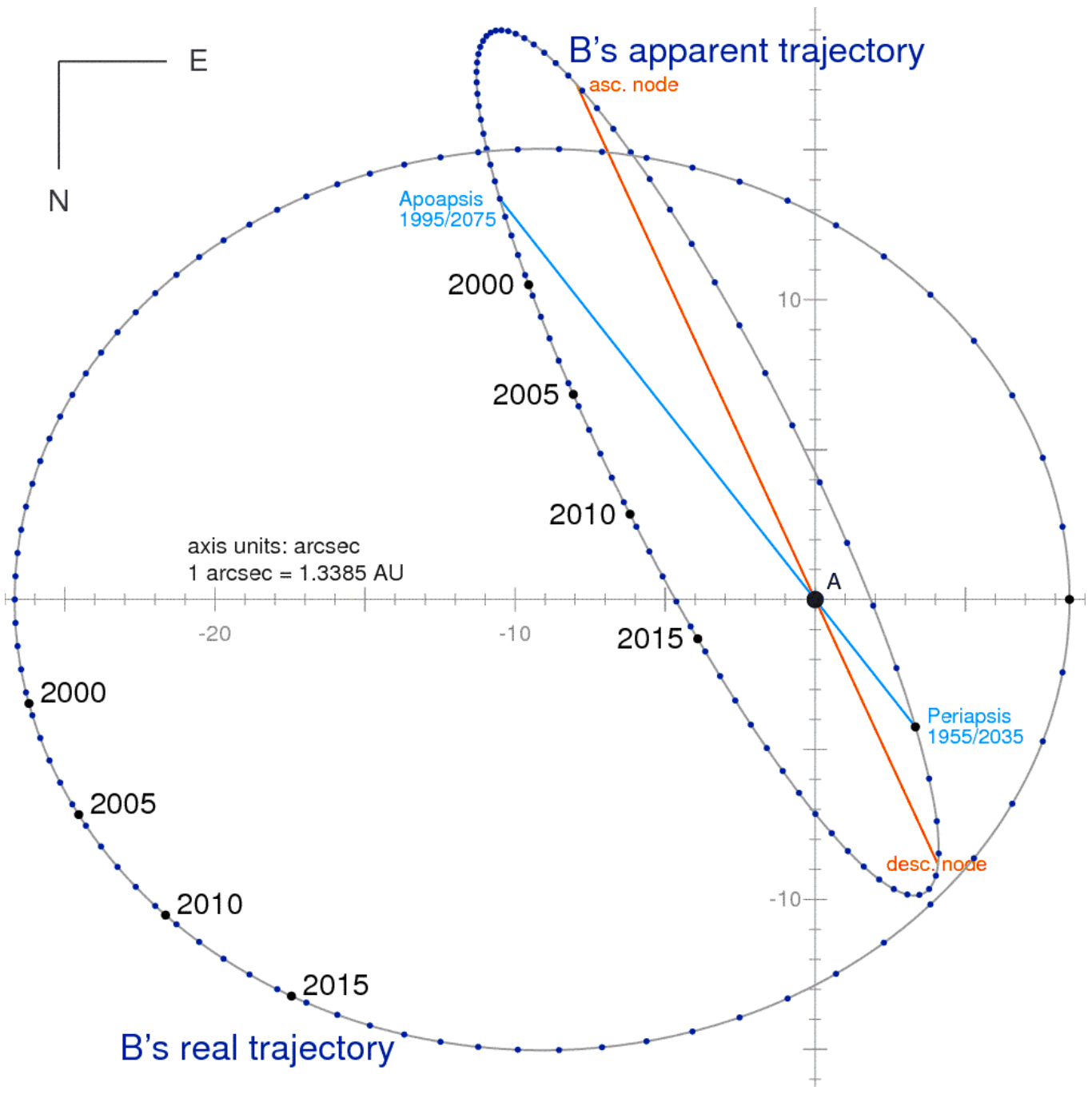

Figure 4.2: Apparent trajectory of $\alpha$ Centauri B orbiting $\alpha$ Centauri A, represented as the elongated ellipse, with the axes show the separation in arcseconds. The actual orbit of B is the less elongated ellipse with A at one of the two foci, as seen face-on (source: Wikimedia).

Physically bound binary systems can be found at a broad range of orbital separations. Only for the nearest stars it is feasible to physically discriminate the components in some cases, and produce detailed orbits in some others (Fig. 4.2). Still, some binaries are so closely packed that they are disguised as single objects by direct imaging, even with telescopes of the most powerful resolving capabilities. At one end, the so called contact binaries physically interact with one another (see the extreme case of the 5-minute-period white dwarf binary HM Cnc, Israel et al., 2002, and see Munday et al. 2022 for a two-decade worth of data review). The mutual pull is so strong that their gaseous envelopes fill their Roche lobes (their region of gravitational influence) and mass is transferred in the process (Lucy, 1968, and see Yakut & Eggleton 2005 for the case of low mass stars). The ultracool dwarf binary LP 413-53AB is another extreme and interesting example. Hsu et al. (2023) reported it as “the shortest-period ultracool binary discovered to date, and one of the smallest separation main sequence binaries known”, with an orbital period of around 20.5 days. The authors noted that, at the current separation (around one hundredth of an astronomical unit), their larger radii at young ages would have put them in contact. While very close binaries are undetectable by direct means such as imaging, they can be recognised in the Doppler shift of the spectral lines (spectroscopic binaries), in the periodic eclipsing of their light (eclipsing binaries), or in the measurable change in their motion (astrometric binaries). Stars in binary systems offer the opportunity to directly measure fundamental parameters, such as masses and/or radii (Mathieu et al., 2000; Zapatero Osorio et al., 2004; Torres et al., 2010; Schweitzer et al., 2019). They are also convenient laboratories to investigate other areas, such as the nature of clusters (e.g. Kouwenhoven & de Grijs, 2008), and even a measure of gravity (e.g. Banik & Zhao, 2018).

[3] Harrington (1972) concluded that for a triple system in a direct orbit like $\alpha$ Centauri to be stable, the ratio of the periastron distance in the outer orbit to the semi-major axis of the inner orbit must be at least 3.5.

The frontier that defines a close and a wide binary is blurry (e.g. Caballero, 2009, 2010; González-Payo et al., 2021, 2023). In the following sections we use the terms 'close' and 'wide' in a qualitative way, unless specified otherwise. As a general rule, we set a boundary between close and wide binaries at $\rho$ = 5 arcsec, which results from avoiding contamination in the spectra from nearby sources in the aperture of the optical fibre of CARMENES projected on the sky (Sect. 2.1). From the CARMENES perspective, there is a quantitative difference, motivated in Sect. 3.1.4.

Wide systems have orbital periods of centuria or millennia, and thus the prospect of following them during one single orbit is unrealistic in practice, and in the best cases, full-orbit description must rely on photographic materials (see Gatewood et al., 2003). Because of this, it is not possible to discriminate between actual bound multiple systems and disintegrating clusters. Nevertheless, the components of wide pairs can be exposed individually. At these large separations, most of the binaries are fated to dissolve in timescales of a few million years (see Sect. 4.4.4). The observed separations between components of a wide system are usually smaller than 0.1 parsec (but see for instance Shaya & Olling, 2011). The youth of the system is a very common characteristic of very wide pairs, as components still undergo a process of stabilisation (Poveda & Allen, 2004). Notable examples are the system of AU Mic and AT Mic (Kalas et al., 2004; Caballero, 2009), or 'the Family of V1311 Ori', as presented by Tokovinin (2022). However, these configurations are so fragile that they are susceptible to be unbound by the dynamical encounters with different sources of perturbation, such as stars, molecular clouds, clusters, and even Stellar-mass black holes (Hills, 1975; Retterer & King, 1982; Weinberg et al., 1987; Kroupa, 1995; Kroupa et al., 1999; Caballero, 2009; Jiang & Tremaine, 2010; Parker et al., 2011; Deacon & Kraus, 2020; Cournoyer-Cloutier et al., 2021; Ryu et al., 2022). Indeed, simulations find that no binary with a separation $\geq 10^4$ au (larger than the typical size of a pre-collapse core) must be produced by isolated star formation (see Goodwin, 2010) because it cannot survive in any cluster (Parker et al., 2009). The authors argue that in binaries with semi-major axis $50 \lesssim a \lesssim 10^4$ au the effect of dynamical influence can be severe, while for $a \lesssim 50$ au they evolve almost unaffected. At these very small separations, they also note that the field population reflects the sum of all star formation. It has been suggested that very close binary systems ($r < 10$ au) are not formed by fragmentation in situ, but are a product of wider multiple systems (Bate et al., 2003).

Stars that have a common origin in space must have the same chemical imprint. Wide binaries are demonstrated to be coeval (Makarov et al., 2008) and co-chemical (Gizis, 1997; Gratton et al., 2001; Desidera et al., 2004, 2006; Makarov et al., 2008; Kraus & Hillenbrand, 2009; Hawkins et al., 2020). Sufficiently resolved pairs can serve to prove this assumption and serve as pieces in the puzzle of the Galactic formation. Fitting these pieces and reassembling the original configuration is the goal of Galactic archaeology studies, which benefits from systems of wide pairs (Andrews et al., 2019; Hawkins et al., 2020). Important applications of these are the calibration of metallicities of M dwarfs (Bonfils et al., 2005a; Bean et al., 2006; Lépine et al., 2007; Rojas-Ayala et al., 2010; Montes et al., 2018), the age-metallicity (Rebassa-Mansergas et al., 2016), age-magnetic activity relation (Garcés et al., 2011; Chanamé & Ramírez, 2012), and also studies of dark matter in the Milky Way (Yoo et al., 2004; Chanamé & Gould, 2004).

Although intrinsically small and faint ($\mathcal{M} \lesssim 0.62\mathcal{M}_\odot$, $\mathcal{L} \lesssim 0.076\,\mathcal{L}_\odot$, Cifuentes et al., 2020), M dwarfs make up the majority of the stars in the Universe (Henry et al., 1994, 2006; Reid et al., 2004; Bochanski et al., 2010; Winn & Fabrycky, 2015; Reylé et al., 2021). The advances in instrumentation and image detection have opened the door to finer visions of M dwarfs. Among some of the top high-resolution ground-based spectrographs that look at the coolest stars is CARMENES (Quirrenbach et al., 2014) (see Table 1.1). The lack of detailed observations regarding individual stars carries an important observational selection effect, since very close multiple system remain undetected. In this sense, the nearest stars represent a valuable sample of study because it permits accurate photometric and astrometric measurements, which is specially true for the case of M dwarfs.

Even if the current population of M dwarfs had not evolved from a primordial one of mostly multiple systems, the frequency of binaries and higher order multiples makes their study a very important test for the theories of star formation. Also, the distributions of parameters of binary systems such as orbital periods and separations, binding energies, or masses, contain important insights into the formation process (Sterzik et al., 2003; Burgasser et al., 2007; Goodwin et al., 2007). The statistics on their frequency, their primary-companion mass ratio, and their physical separations can set meaningful constraints to the models of stellar formation and evolution, and whether the observed distribution is compatible with a primordial multiple-only population (Hartigan et al., 1994; White & Ghez, 2001; Parker et al., 2009; Reggiani & Meyer, 2011; Clark et al., 2012; Leigh & Geller, 2013; Reipurth et al., 2014; Parker & Meyer, 2014). The use of high contrast imaging techniques, like the Spectro-Polarimetric High-Contrast Exoplanet Research (SPHERE), provides a close and realistic observation of binary systems, even during the formation phase. This suggests that the interplay between different mechanisms would produce the variety of observed systems, with a possible continuum between the formation of substellar objects (massive planets and brown dwarfs) and of stellar companions (Gratton et al., 2022).

In this work we present a systematic study of the multiplicity of M dwarfs, in a volume-limited sample of more than two thousand M dwarfs with spectral subtypes form M0.0 V to M9.5 V, and separations from 0.3 au to about 206 000 au. Since the study fundamentally relies on the resolution of the systems by the *Gaia* mission, we distinguish between systems with resolved components (Sect. 4.3.1), and systems with unresolved components (Sect. 4.3.2) in *Gaia* DR3. The conclusions and prospects for a future work motivated by these investigations are summarised in Chapter 5.

## 4.2 Sample of study

The sample of our study is Carmencita, the CARMENES input catalogue. Carmencita contains 2216 late-K and M dwarfs, which were intentionally chosen to be independent of multiplicity, age, or metallicity. As explained in Sect. 2.2, these stars satisfy simple selection criteria based on their spectral determination, on their visibility from Calar Alto Observatory in Southern Spain ($\delta \gtrsim -23$ deg), and on their apparent brightness in the $J$-band magnitude, between 4.2 mag and 11.5 mag (cf. Alonso-Floriano et al., 2015b). The continuous update of the catalogue has included additional M dwarfs that are part of the Transiting Exoplanet Survey Satellite (TESS; Ricker et al., 2015) program.

We removed four late-K single stars from the original Carmencita catalogue for a more rigorous sample only populated by M dwarf objects. These are LP 415-17 (J04219+213), HD 168442 (J18198-019), TYC 4450-1440-1 (J20109+708), and Ross 176 (J20227+473), all of them classified as K7 V stars. Figure 4.3 shows the classification by spectral subtype of the 2212 M dwarfs in the sample.

A magnitude limited sample may be unintentionally overpopulated of brighter, unresolved systems (see Duquennoy & Mayor, 1991, and references therein). For this reason, we study the completeness in distance (or volume) in our sample. To address this calculation we assume a uniform distribution of the stars over the area of the celestial sphere. First, we define a 'completeness distance' that depends on the spectral type, $d_{\rm com}$, using the definition of absolute magnitude in the $J$ band, $M_J - J = -5 \log d_{\rm com} + 5$, where $J = J$(SpT) (Alonso-Floriano et al., 2015b), and $M_J = M_J$(SpT) using the results by Cifuentes et al. (2020). Therefore, the meaning of $d_{\rm com}$ is, for a given spectral type, the maximum distance for which we can be confident of containing all known M dwarfs identified in the literature[4]. The results are summarised in Table 4.1. Up to 30 parsec, our sample contains all M dwarfs with spectral types M4.5 V or earlier, and it is complete for all the domain of M dwarfs within 10 pc, This important fact has been checked with Reylé et al. (2021).

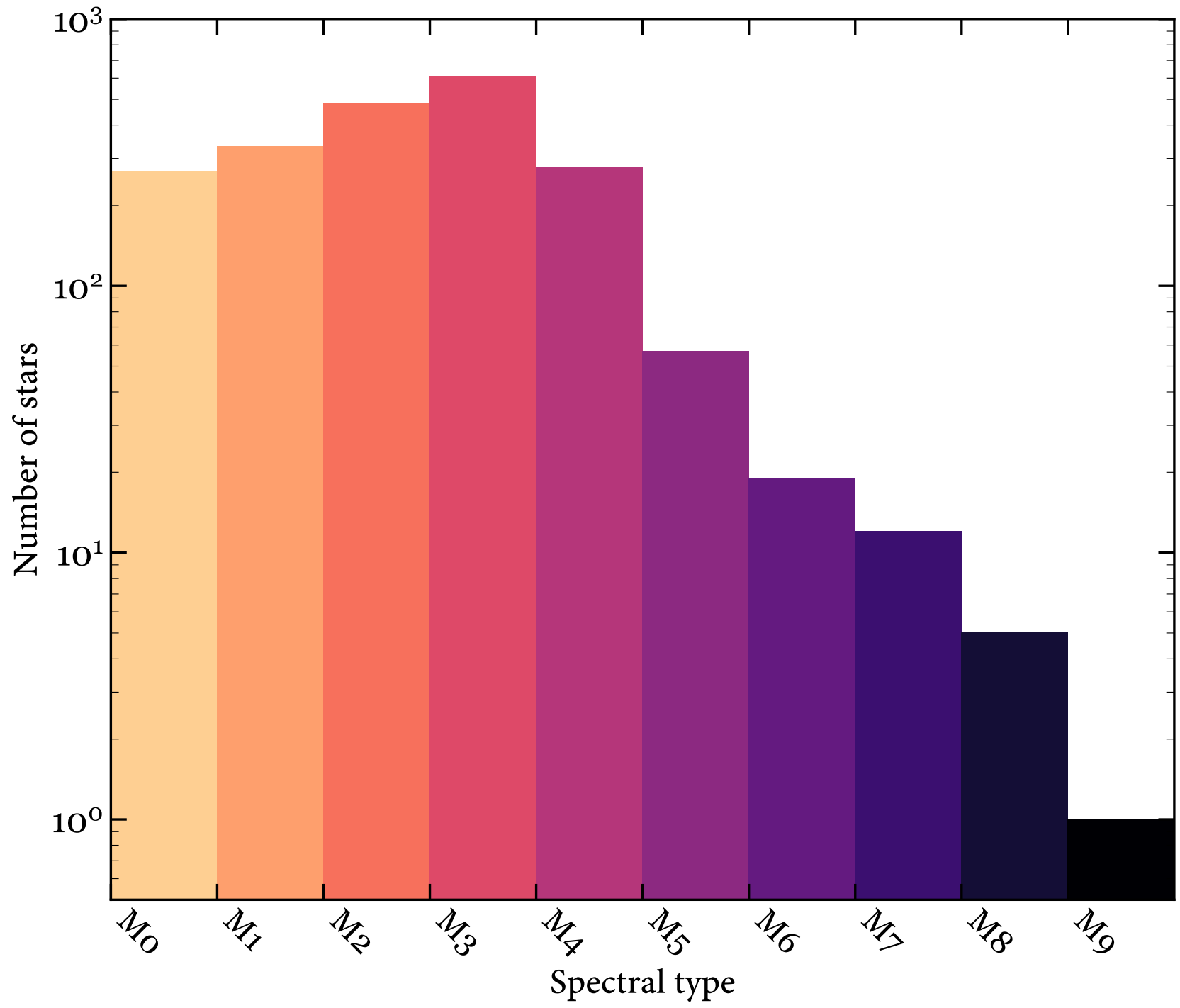

Figure 4.3: Distribution of spectral types of the stars in the sample. The M9.5 V object is Scholz's Star.

Table 4.1: Completeness distance of Carmencita as a function of spectral type.

| Spectral type | $d_{\rm com}$ (pc) | Number of stars |
|---|---|---|
| M0.0–0.5 V | 33.4 | 235 |
| M1.0–1.5 V | 35.3 | 331 |
| M2.0–2.5 V | 35.9 | 381 |
| M3.0–3.5 V | 33.9 | 560 |
| M4.0–4.5 V | 32.1 | 469 |
| M5.0–5.5 V | 24.2 | 159 |
| M6.0–6.5 V | 19.6 | 32 |
| M7.0–7.5 V | 15.3 | 16 |
| M8.0–8.5 V | 13.1 | 10 |
| M9.0–9.5 V | 9.9 | 2 |

[4]As a reminder, Carmencita does not include stars that are not visible from Calar Alto, where CARMENES is located. As a reminder, see again the distribution of Carmencita stars in Fig. 3.2.

## 4.3 Analysis

For the multiplicity analysis in our sample we made extensive use of the third data release of *Gaia* astrometry and photometry (DR3, Gaia Collaboration et al., 2022b), numerous public all-sky surveys from the ground and space, the Washington Double Star catalogue (WDS; Mason et al., 2001)[5], and Virtual Observatory tools such as the `Aladin` interactive sky atlas (Bonnarel et al., 2000), the Tool for OPerations on Catalogues And Tables (`TOPCAT`; Taylor, 2005), and the Virtual Observatory Spectral energy distribution Analyzer (`VOSA`; Bayo et al., 2008).

The WDS is the principal database for astrometric double and multiple star information, and is updated nightly. Using `TOPCAT` along with `Aladin`, we looked for entries in this catalogue for all the stars in our sample and found that WDS tabulates 558 individual pairs in it. For these, we obtained the data that includes the last measured epoch, and the corresponding positional angle and separation in the 'precise' format (i.e. non-rounded values), the number of observations, and the visual magnitudes for both components.

Every star in our sample has equatorial coordinates in the 2016.0 epoch and $G$ magnitude in the *Gaia* catalog. 95.1 % of them have the full, five-parameter astrometric solution: positions, parallaxes, and proper motions ($\alpha$, $\delta$, $\mu_\alpha \cos\delta$, $\mu_\delta$, $\varpi$). Among these, 61.2 % also have barycentric radial velocity, $V_r$, in the second or the third data releases (preferring the former in the case of availability in both). For the sources without astrometry or radial velocities provided by *Gaia*, we searched in the literature for published data, making sure that we maximise the completeness of the astrometric information in our sample.

For every star we looked for physical companions covering all ranges of separation: from compact configurations (typically binaries), only resolved employing dedicated techniques (lucky imaging, adaptive optics, speckle interferometry, etc.), to wider pairs that can be resolved using the *Gaia* astrometric solution, and in some cases other all-sky surveys, such as 2MASS or AllWISE. In this search we found a total of 800 systems in our sample. To ensure the correct interpretation of the data, sometimes within only a few arcseconds of apparent separation, we inspected individually each one of them, primarily using `Aladin` and Simbad, and we took great care to consider the existing literature and past characterisations.

Before delving into the analysis of the sample, it is crucial to note that the sample was subject to observational selection limitations. In the case of very close-in binary systems, only spectroscopic analysis can reveal the presence of pairs, as discussed in Sect. 4.3.1. Because observations of individual stars with sufficiently large telescopes and powerful instruments are not, in general, available, the very close pairs go unnoticed and do not contribute to the distributions of parameters. The consequence is that the period and mass ratio coverage lacks of uniform data at short periods. Of course, this is a fundamental concern of many investigations based on a large sample of stars.

### 4.3.1 Resolved systems in *Gaia*

*Gaia* DR3 contains data for 1 811 709 771 objects, with a complete astrometric description (positions, parallaxes, and proper motions) for 585.4 million of them. Figure 4.4 shows a relatively small sample of the stellar content identified in the catalogue within an area of $3 \times 3\,\mathrm{deg}^2$. The nominal mission of *Gaia* ended on July 2019, but the extension of their operations has been approved until 2025, when the propellant for the micropropulsion system is expected to be exhausted. In this condition, the astrometric precision for that relies on the attitude and spin rate cannot be guaranteed. By then, *Gaia* will have gained a decade worth of precise measurements, and at least two more data releases are expected to include the results from the extended mission: DR4 (not before the end of 2025) and DR5 (not before the end of 2030). For the moment, *Gaia* has already published (in DR1, DR2, EDR3, and DR3) the most precise astrometric catalogues ever known, with accuracies ranging from 10 to 1000 $\mu$arcsec for more than one billion stars.

[5] `http://www.astro.gsu.edu/wds/`.

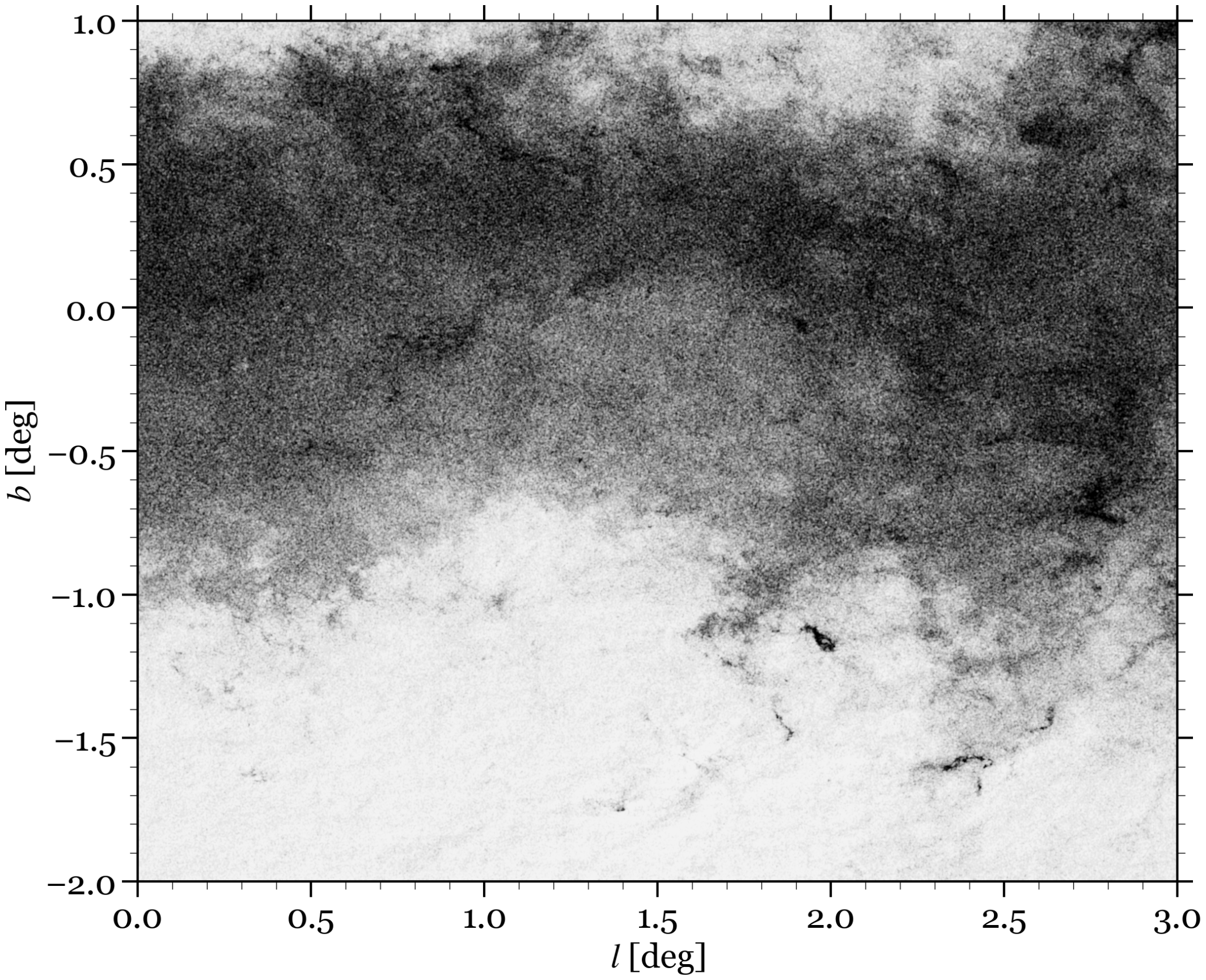

Figure 4.4: *Gaia* sources (approx. 3 million) in a $3 \times 3$ deg$^2$ area near the Galactic centre. Each white dot represents a source resolved by *Gaia*, in most cases with accurate astrometry, photometry, and low-resolution spectroscopy. Light-blocking dusty regions can be easily spotted in the crowded areas as darker patches.

For the search of systems that are resolved by *Gaia*, we divided the search in two steps. Firstly, we looked for any close source in the *Gaia* DR3 catalogue using the `TOPCAT` automatic positional cross-match tool, `CDS X-match`, with a search radius of 5 arcsec. By setting the 'find' option to "All", we made sure to keep every object found in the vicinity, regardless of their kind, parallactic distance, or proper motions, even when this information was not available.

Secondly, we performed a much wider, volume-limited, blind search in the same catalog, using the AQDL form available in the *Gaia* Archive[6]. In this case, we looked for resolved sources with the full 5-parameter astrometric solution that is compatible with physical binding, based on proper motions and parallactic distances (see Sect. 4.3.1). We retrieved all elements within $10^5$ au, which implies a projected separation of $10^4$ arcsec to $\sim$ $10^3$ arcsec (2.8 and ~0.3 degrees, respectively) for stars located in a range between 10 and 30 parsec, whose similarity in parallax was 10% or less, ignoring those stars without measured parallax (i.e. photometric distances). We did not impose any restriction on the proper motion because, given the small probability of finding a similar-distance source within the search radius, the benefit of finding a companion that may have perturbed motion caused by a very close companion (see Sect. 4.3.2) outweights the effort of checking individually all the potential pairs. This was done by accessing to images and catalogues via `Aladin`.

[6]https://gea.esac.esa.int/archive/.

The distance on the celestial sphere from a given source was measured along a great circle or geodesic, as opposed to straight lines in the Euclidean space. Given $\delta_1$ and $\delta_2$ the declination of two objects, and $\Delta\alpha$ the absolute difference between their right ascensions, the central angle is given by:

$$\Delta\sigma = \arccos\left(\sin\delta_1 \sin\delta_2 + \cos\delta_1 \cos\delta_2 \cos\Delta\alpha\right), \tag{4.1}$$

and the arc length is given by $s = d\Delta\sigma = d\rho$. For all these components in the systems resolved by *Gaia*, we calculate the projected separation, $\rho$, and the positional angle, $\theta$. With these variables we compute two parameters introduced by Montes et al. (2018), which are defined by pairs that are generically referred as components 1 and 2 in the formulae. These are the $\mu$ ratio, defined as:

$$(\mu\,\mathrm{ratio})^2 = \frac{(\mu_\alpha \cos\delta_1 - \mu_\alpha \cos\delta_2)^2 + (\mu_{\delta 1} - \mu_{\delta 2})^2}{(\mu_\alpha \cos\delta_1)^2 + (\mu_{\delta 1})^2}, \tag{4.2}$$

and the proper motion position angle difference:

$$\Delta PA = |PA_1 - PA_2|, \tag{4.3}$$

to which we add a third parameter as in Cifuentes et al. (2021) and González-Payo et al. (2023), the distance ratio defined as:

$$\Delta d = \left|\frac{d_1 - d_2}{d_1}\right|. \tag{4.4}$$

We apply a preliminary criteria for multiplicity, to distinguish between physical (bound) and optical (unbound) pairs.

$$\begin{cases} \mu\,\mathrm{ratio} < 0.15, \\ \Delta PA < 15\,\mathrm{deg}, \\ \Delta d < 0.10. \end{cases} \tag{4.5}$$

The first two criteria account for the similarity in their movement, in particular in the modulus and orientation of their proper motions. The third criterion ensures that the components of the pair are approximately equidistant, up to a conservative difference of 10 per cent. Jointly, they serve to test if a pair is a chance alignment.

While the criteria above are sufficiently valid for resolved, relatively wide pairs with well-defined astrometry, it is indeed troublesome in the case of pairs that are too close in the sky. For these, the astrometric solution may not be accounting for the relative orbital motion, but rather for the instantaneous, tangential movement of the components. The nominal operations of *Gaia* span 1028 days, from August 5, 2014 to May 28, 2017. DR3 benefits from a good time coverage, which is an advantage with respect to DR2 during the study of proper motion anomalies (see Kervella et al., 2019). Nevertheless, this astrometric coverage is insufficient in many cases and can translate into ill-defined proper motions that cannot account for the real motion of the components in compact systems. Because of this, very close-in, physically bound systems do not comply with the criteria above.

Figure 4.5 shows these extreme cases. We show the compliance of the pairs with the criteria of Eqn. 4.5 (top panel), and the comparison of distances (bottom panel). The error bars are comparatively very small for almost all cases due to the high precision of *Gaia*'s astrometry, and had been omitted without compromising the rigour.

The distances and total proper motions of the vast majority resolved pairs compare favorably against those of their resolved companions. The criteria for physical parity also hold true for them. Nevertheless, 31 components (shown as red circles in some or all of the panels in Fig. 4.5) present anomalies in their relative positional angles proper motions, and/or distances. We have investigated their particular characteristics in further detail, and the results are presented in Table D.3, including a probable main cause for this behaviour. We encourage paying special attention to some of these pairs, because their potential closeness could make their orbital motion to be defined easier, and further characteristics of the components, such as their dynamical masses, might be obtained sooner. The vast majority of relatively wide multiple systems that are resolved in *Gaia* are known to date and collected in the WDS catalog. In Fig. 4.6 we compare the values of separations tabulated in the WDS with those computed by us using *Gaia* DR3 astrometry. The 10 stars outside the limits represented by the dashed lines have a difference in their values larger than 25 %, a characteristic typically observed in very close pairs of stars that are more strongly influenced by their mutual gravitational attraction, therefore exhibiting more noticeable common movements.

In this work we discover several pairs with an astrometry that we find to be compatible with physical parity with our stars, but that are not found either in the WDS catalogue or in the literature. Some of them are isolated pairs (i.e. binaries), whereas some are components of known systems, thus producing systems of higher order. We tabulate these 117 new systems proposed in Table D.4, including 31 binaries, 15 triples, and 4 quadruples. Apart from *Gaia* DR3 astrometric solutions ($\alpha$, $\delta$, $\mu_{\rm total}$) and $G$ magnitudes, for each pair we tabulate the position angles ($\theta$) and the angular separations ($\rho$). In a few cases, their entries in Simbad includes the denomination 'Double or multiple star', but without published references. Additionally to describing the new pairs found, we also note the impact of these findings on the general scheme of its system, and in particular on the redefinition of the class (i.e. from single to binary, binary to triple, etc.).

### 4.3.2 Unresolved systems in *Gaia*

*Gaia* is the all-sky survey with the highest angular resolution available to date. The spatial resolution of *Gaia* was limited to 0.4–0.5 arcsec in the second data release (DR2, Gaia Collaboration et al., 2018b), and slightly improved in the early third data release (EDR3, Gaia Collaboration et al., 2021). While EDR3 represented a significant advance over DR2, the third data release (DR3, Gaia Collaboration et al., 2022b) exploited in this work maintains the same astrometric data as the previous release, EDR3. Nevertheless, it incorporates a wealth of new data products, that we exploit in this work. Fabricius et al. (2021) showed that EDR3 is complete for separations above 1.5–2.0 arcsec, with severe incompleteness below 0.7 arcsec. The authors addressed the improvement of the spatial resolution of the EDR3 as compared with DR2, testing the capacity of resolving visual double stars from the WDS catalogue in both releases. Below 0.7 arcsec objects can be discriminated, depending on the magnitude difference and the orientation along the dominating scan directions. Indeed, EDR3 improved the treatment of sources separated between 0.18 and 0.4 arcsec, which were erroneously considered duplicated sources in DR2.

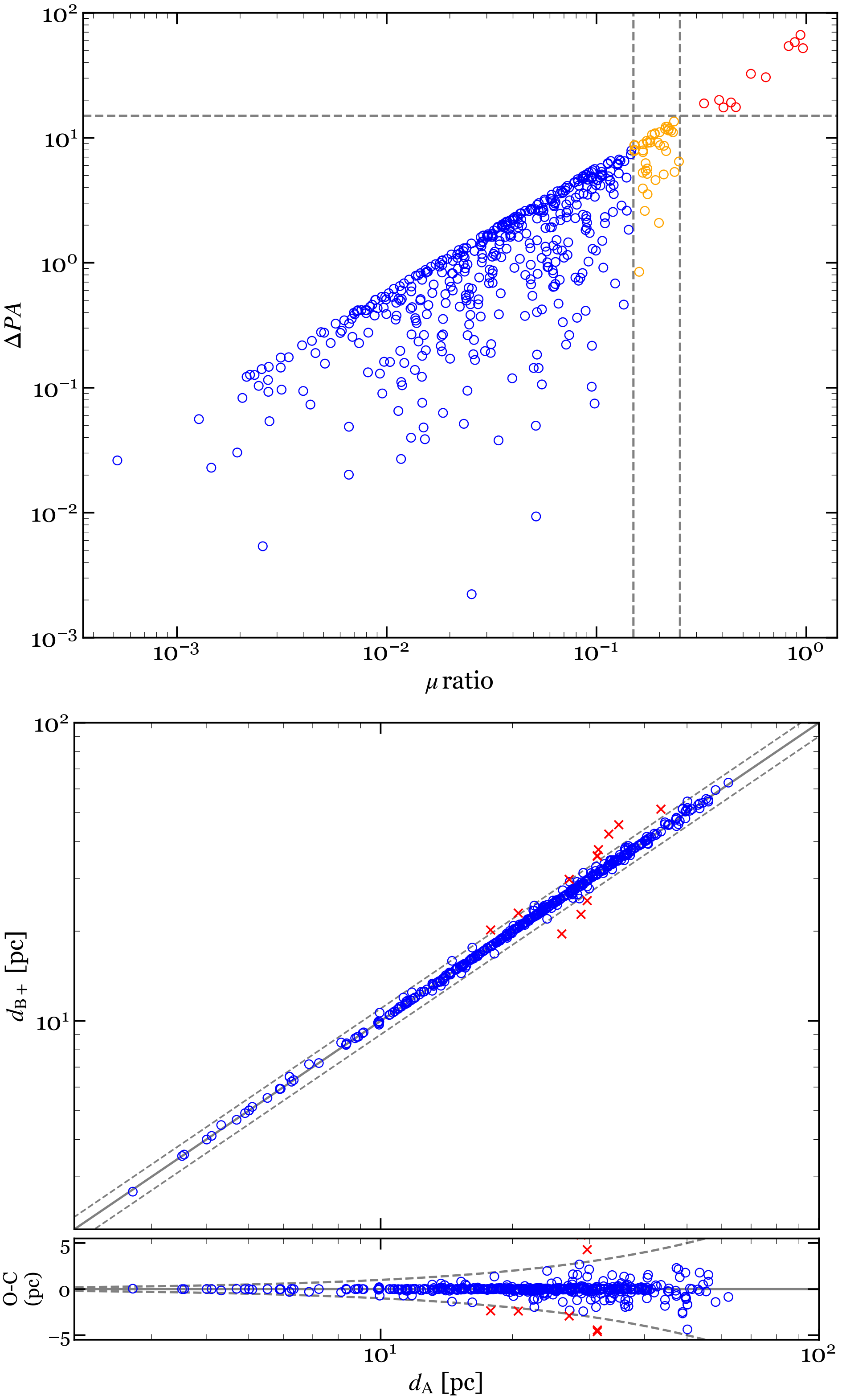

Figure 4.5: *Top*: $\Delta PA$ vs. $\mu$ ratio diagram, where the dashed grey lines set the upper limits of the criteria for physical parity (Eqn. 4.5). The red and orange empty circles are pairs that respectively do not comply, or do so partially, with that criteria. *Bottom*: Comparison of parallactic distances between the primaries and their resolved components. The dashed and dotted grey lines represent the 1:1 relation, and the differences in 10 %, respectively. The red crosses are stars out of this latter limit.

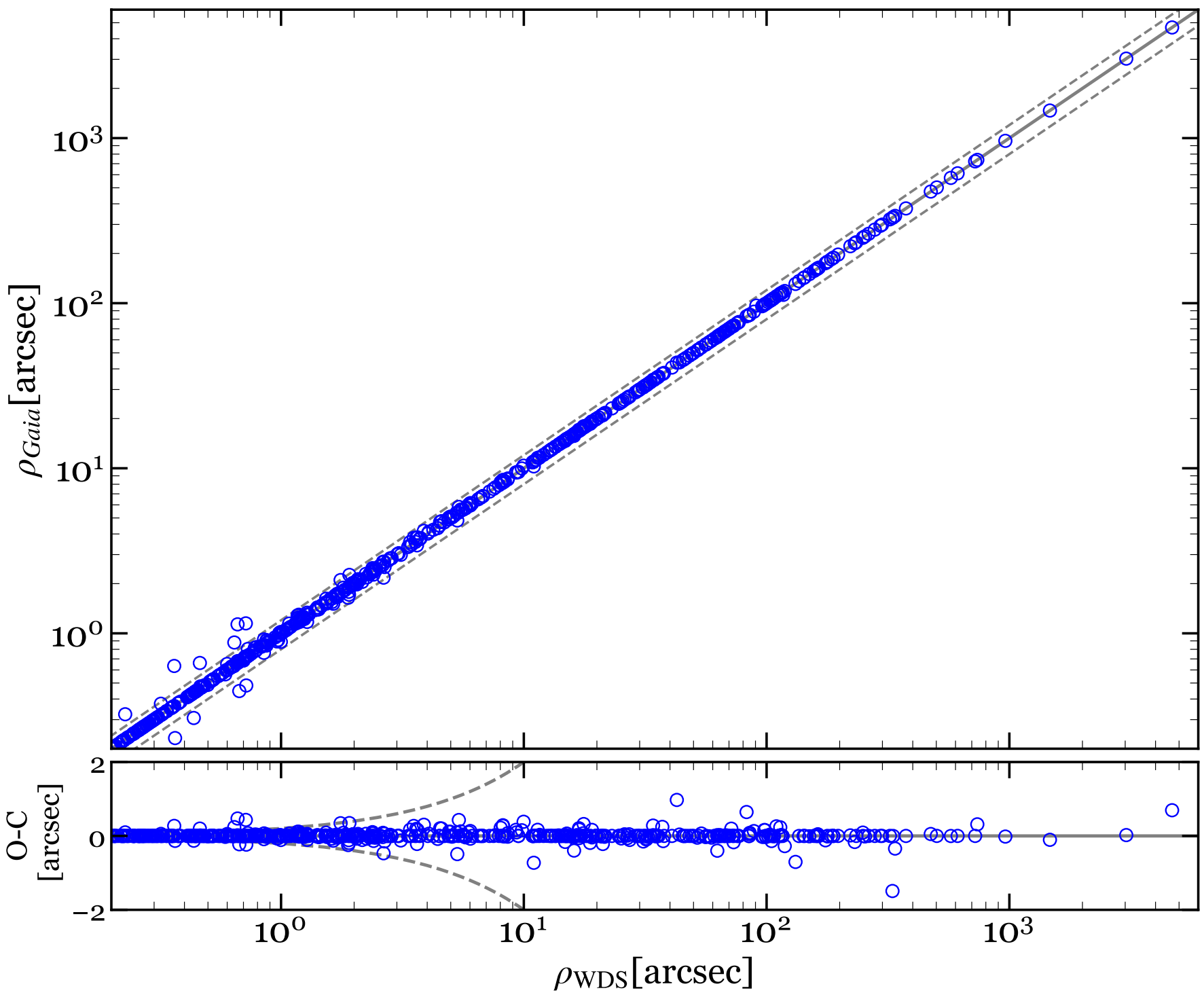

Figure 4.6: Comparison of projected separations tabulated by the WDS and measured by us using *Gaia* astrometry. The continuous and dashed grey lines represent the 1:1 relation, and the differences in 10 %, respectively.

This means that *Gaia* is unable to resolve the closest ($\rho \lesssim 0.2$ arcsec) pairs in our sample. For these, different techniques and technologies can successfully disentangle the components, such as adaptative optics, lucky imaging or, for instance, the ultra-violet vision of the Hubble Space Telescope. Therefore, we investigate in the literature for the identification of known spectroscopic binaries and triples, and the WDS for astrometric pairs. We note that this search is performed not only in the stars in our sample, but is also in their known companions.

In the case of spectroscopic systems, by definition, their components cannot be resolved by any earth- or space-based telescope, and the existence of two or more stars can only be deduced from meticulous observations in moderate or high resolution shots of their spectra (see Sect. 4.4.2). We find 132 spectroscopic multiples among the stars in our sample and their companions. We took special care with those cases for which spectroscopic binarity was reported in previous investigations (typically dated decades back), and there exist recent astrometric observations that resolve the components. Indeed, in 37 cases, the components of a binary detected by spectroscopic means are later on exposed individually, as a result of the advances in instrumentation. For these, we pay attention to the reported orbital periods and the magnitude differences reported, in order to consider the possibility that both measures, the oldest spectroscopic and the newest astrometric binaries, refer in fact to the same pair. These are not classified as 'SB', except in the case of doubt, in which case we prefer to stay conservative, and avoid losing a potential triple system (thought to be double) in a compact configuration. We list all the spectroscopic systems found in our sample in Table D.5, including their derived periods, the masses for the primary and secondary, the $q$ ratios, and the physical separations, $a$, when available. The mass ratio of a star in a multiple system indicates its mass as a fraction of the mass of the primary, $q = M_B/M_A$, where B represents a component other than the primary. For a system of equal-mass components, $q$ is 1.

Although *Gaia* is unable to resolve very close-in systems, they are not *invisible* to its eyes. Even when the precision is not enough to discriminate the components of a very close binary, from the vast amount of data that it collects it is possible to extract information that can be much revealing, and can boast from statistical robustness.

*Gaia* has three different channels: astrometric, photometric and spectroscopic. The astrometric measures the movement in the sky and heliocentric distance, the photometric measures fluxes in three passbands, and the spectroscopic derives information from the spectrum, such as radial velocities by cross-correlation with theoretical spectra. *Gaia* DR3 is the first release providing analysis of the RV time-series looking for orbital motion. All together provide a coherent view of the nature of the objects.

In *Gaia* DR3 every source has been observed an average of ~70 times, and varying from ~30 to ~240, depending on the sky coordinates. Among the different types of variable stars, the Gaia data reduction identified ~2.2 million eclipsing binaries and ellipsoidal systems, with various levels of characterisation (and see Rimoldini et al., 2022). The parameters were derived by fitting the light curves (LCs) to Gaussian or cosine functions, and also to synthetic LCs of binary systems, modelling stellar surfaces as equipotentials of the Roche potential. In some cases including radial velocity data were incorporated in the study. The parameters derived include the photometric period, the times of mid-eclipse, as well as eclipse durations and depths.

Astrometrically, the full description of a system can be achieved when the orbital period is smaller than the astrometric mission time interval. As mentioned earlier, the nominal operations of *Gaia* span 1028 days (2 years, 9 months and 24 days). If the orbital period surpasses this length of time, still it is feasible to determine the orbital parameters by fitting the so called orbital models and finding the optimal goodness of fit and significance values. A binary system can be described by seven parameters (the so called Campbell's elements): the orbital period ($P_{\rm orb}$), the epoch of periastron passage ($T_0$), the eccentricity ($e$), the semi-major axis ($a$), the inclination of the orbital plane with respect to the sky ($i$), the position angle of the ascending node ($\Omega$), and the periastron longitude measured from the ascending node ($\omega$). However, in the present data release of *Gaia* the handling of resolved binaries was not yet possible. Nevertheless, *Gaia* comes with numerous statistical parameters, built-in indicators, and data products that are useful in the identification of unresolved systems. In Table 4.2 we summarise the different approaches that we take to make the most of the available opportunities that the third data release of *Gaia* offers. These indicators are sensitive to deviations from the single star model that is assumed in the pre-processing stage. Among the stars without known close companions (less than a few hundred astronomical units or less) we find 344 stars that comply with any of these criteria for unresolved multiplicity. Among them, 272 are single and 72 are members of multiple systems, but sufficiently separated to discard the effect of the companion as the cause. These are preliminary candidates to binaries. There are also pairs in very close proximity, but resolved by *Gaia*, which lack in some instances of parallaxes and proper motions in the DR2/DR3 catalogues. Despite of this, we deem them as candidates, as the odds for a chance alignment are much lower than that of a physical connection (El-Badry et al., 2021; Chulkov & Malkov, 2022). In a similar fashion, Vrijmoet et al. (2020) provided an analysis of unresolved astrometric multiples using two decades of astrometric data from the RECONS program along with *Gaia* DR2 observations, and recognised perturbation in its astrometric residual in a few instances.

The Non-single star tables include unresolved astrometric, spectroscopic, and eclipsing binaries (see Gaia Collaboration et al., 2022a). These solutions are distributed in four tables: `nss_two_body_orbit` when the full orbital motion is known, `nss_acceleration_astro` and `nss_non_linear_spectro` when a trend is known only, and `nss_vim_fl` for variable fixed binaries. More details on the processing scheme, its validation, and the various types of solutions reached can be found in Chapter 7 of the *Gaia* documentation[7]. Additionally, we incorporate the table `vari_eclipsing_binary` (Mowlavi et al., 2022; Eyer et al., 2022) from the study of variability. The latter is the first *Gaia* catalogue of eclipsing binaries.

Table 4.2: Criteria for the detection of unresolved sources based on *Gaia* DR3 statistical indicators.

| Criterion | Selection | Description | Reference |
|---|---|---|---|
| 1 | `ruwe` > 2[a] | Measure of a poor behaviour of the centre of light. | *Gaia* Documentation |
| 2 | `ipd_gof_harmonic_amplitude` > 0.1<br>`ruwe` > 1.4 | Useful for identifying spurious solutions of resolved doubles, not correctly handled in the *Gaia* EDR3 astrometric processing. | Fabricius et al. (2021) |
| 3 | `rv_chisq_pvalue` < 0.01<br>`rv_renormalised_gof` > 4<br>`rv_nb_transits` ≥ 10 | Identification of variability in the radial velocity among all the *Gaia* measurement epochs. | Katz et al. (2022) |
| 4 | `radial_velocity_error`[b] ≥ 10 km s$^{-1}$ | Idem. | See Katz et al. (2022) |
| 5 | `ipd_frac_multi_peak` > 30[c] | Fraction (0–100) of windows for which the IPD[d] has identified a double peak. | *Gaia* Documentation |
| 6 | `duplicated_source` = 1 | Existence of a duplicated source during data processing[e] | *Gaia* Documentation |
| 7 | `non_single_star`[f] | Non-constant behaviour using binary orbit models. | Pourbaix et al. (2022) |

[a] This amplitude of the centroid perturbation correlates with the physical separation between companions and scales with the binary period and the mass ratio (Belokurov et al., 2020). Instead of the generally adopted value of 1.4 (e.g. Arenou et al., 2018; Lindegren et al., 2018; Cifuentes et al., 2020), we set a conservative minimum of 2.0 (see Fig. 4.9).

[b] For the bright stars ($G \lesssim 13$ mag) it is the uncertainty on the median of the epoch radial velocities, to which a constant shift of 0.11 km s$^{-1}$ was added to take into account a calibration floor contribution. For stars fainter than $G = 12$ mag, a method involving a cross-correlation function is used instead. This criterion applies exclusively to radial velocity values coming from *Gaia* DR2/DR3. A practical application of this metric for the discovery of unresolved pairs is developed in Sect. 4.4.2, where a selected subsample of stars with large values in this metric are spectroscopically inspected in detail, finding evidence of spectroscopic binarity in many of them using mid-resolution spectroscopy with FIES.

[c] The selected value of 30 (per cent) is based on the distribution of values and a conservative threshold. This fraction means that the detection might be a visually resolved double star (either visual or physical binary). In some cases for which this fraction is large, two or three individual detections are given by *Gaia* for the same star (i.e. separated measurements). In other, the data do not provide the full astrometric determination (i.e. lacks proper motions and parallax). Both cases are worth a further investigation for possible binarity.

[d] IPD stands for the Image Parameters Determination, which is the stage of *Gaia* data processing where the standard stellar model is fit to the image locations. The ultimate purpose of the pre-processing is to measure the key properties of the source within each window, specifically its location and flux (further details on pre-processing are given in the *Gaia* Documentation.

[e] This may indicate observational, cross-matching or processing problems, or stellar multiplicity, and probable astrometric or photometric problems in all cases. The duplicity criterion used for *Gaia* E/DR3 is an angular distance of 0.18 arcsec, while a limit of 0.4 arcsec was used for *Gaia* DR2.

[f] Flag indicating the availability of additional information in the various Non-Single Star tables. Three bits (*nnn*) indicate ($n$=1) whether it has been identified as astrometric, spectroscopic, or eclipsing binary, respectively.

We look for matches of our sample with these tables, finding 49, 6, 1, 0, and 2 coincidences, respectively. This translates into 55 individual stars proposed as unresolved pairs by *Gaia*. Of them, 24 are known pairs with very close-in orbits, and 31 are bona fide single stars, or in a few cases component in very wide pairs, in principle unaffected by their companion. Of the 2 eclipsing binary candidates, one turns out to be the known eclipsing binaries Castor C (Joy & Sanford, 1926; Gizis et al., 2002) and GJ 3547 (Shkolnik et al., 2010; Reiners et al., 2012). The `non_single_star` flag fails to identify a large portion of the confirmed very close binary pairs in our sample. Additionally to this automatic categorisation, we perform a further analysis employing a combination of statistical parameters offered by DR3, which we use to flag sources that do behave as non-single objects and might have been overlooked in the automatic categorisation.

[7] `https://gea.esac.esa.int/archive/documentation/GDR3`.

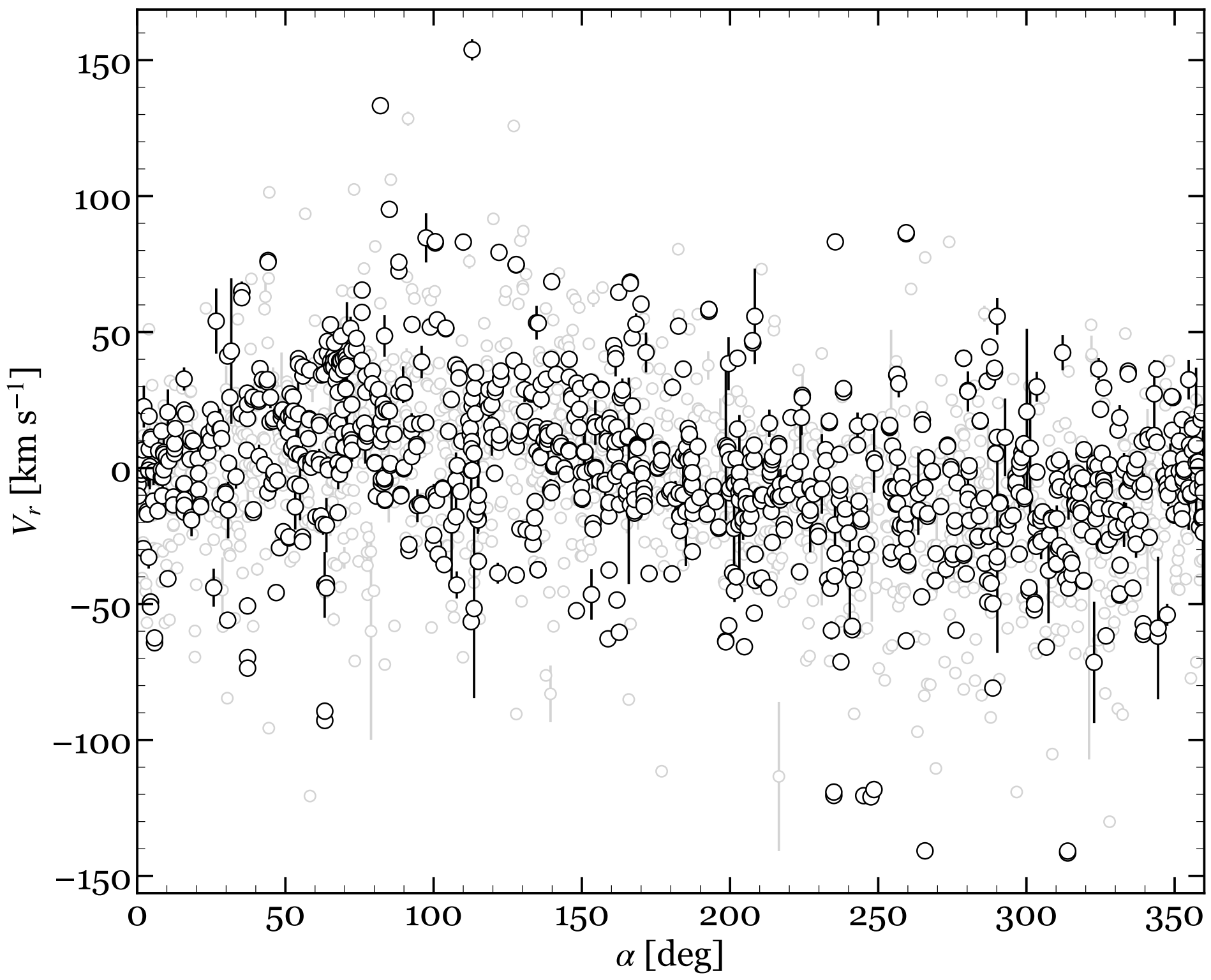

Figure 4.7: Barycentric radial velocity as a function of right ascension for the single (grey) and components of multiple systems (black), respectively.

With them we can identify potential candidates to multiple systems based on the variations of radial velocities among all the DR3 epochs. It is interesting to note the existence of stars in our sample with remarkably high values of radial velocity (Fig. 4.7), and also the existence of a wobble that is visible to a first sight in the same plot. Given that it is dependent on the right ascension of the sources, it is likely due to a positional calibration effect. The metrics that are used in this work are presented in Table 4.2, with a brief description and a general criterion recommended for each of them. More details on the specific origin of these metrics can be found in the *Gaia* DR3 documentation.

Spurious astrometric solutions can be also indicative of close binarity. The close binaries not resolved in DR2 were handled as single objects, with blended photometry and occasional spurious astrometric solutions. The approach in *Gaia* to address the consistency of these anomalous detections is to compare data from multiple transits (see Ziegler et al., 2018). In this sense, the third data release (DR3) greatly benefits from a larger number of observation epochs, which translates into the potential capability of detecting astrometric, spectroscopic or photometric (eclipsing) binaries. *Gaia* DR3 does not handle resolved binaries, but only the unresolved ones. Firstly, selects the candidates by adopting `ruwe` > 1.4. In these they added other criteria to filter out resolved system, to keep only the unresolved candidates. The spurious solutions were minimised by selecting those with a reasonable number of observations.

We apply the criteria 1–6 in Table 4.2 to all the stars in our sample without confirmed close companions. To those, we include the stars found to behave as non-single in criterion 7. By 'close' we mean here those with physical separations $s \leq 220$ au. This corresponds to a projected separation of 10.0 arcsec for a star at a distance of 22 pc, which is the median distance of the objects in our sample. Overall, we find that 344 stars in our sample comply with one or some of these criteria, but many binaries already known in the literature. But among these, 272 are single stars, without any known (close or wide) companion, and 72 are stars with known wide companions ($\rho > 5$ arcsec).

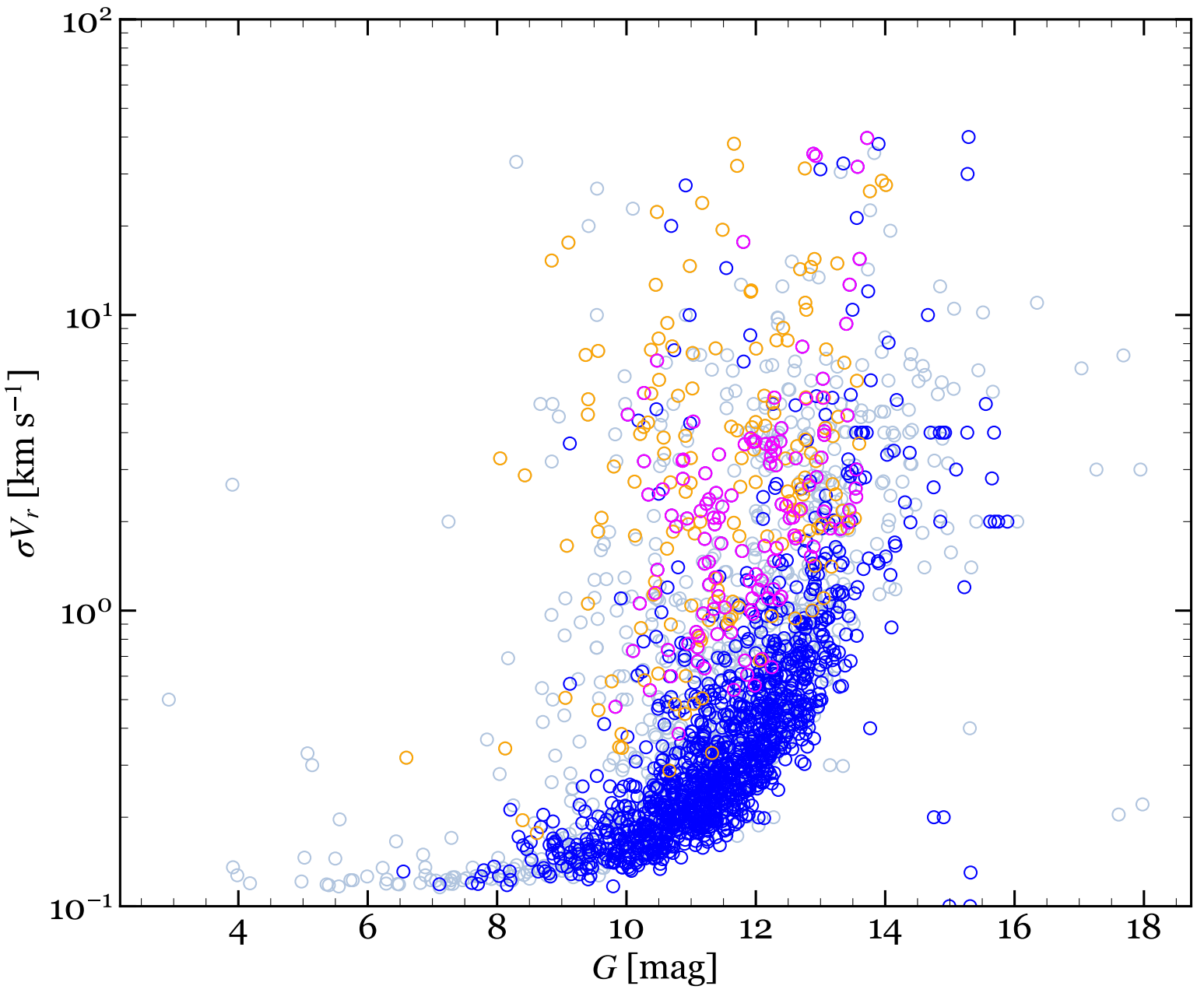

Figure 4.8: Uncertainty on the radial velocities as a function of $G$ magnitude, shown as light and medium blue for the single and multiple stars, respectively. Stars with *evidence* for variability in the radial velocities (as hinted by criterion 3 in Table 4.2) are shown as magenta and orange empty circles, if they are single and if they are already components of a multiple system, respectively.

*Gaia* DR3 contains radial velocities for 33.8 million stars, with a temperature interval expanded with respect to DR2 (see Katz et al., 2022). In Sect. 4.4.2 we show the spectral analysis of a selected subsample of bona fide single stars with notable deviations from the median value of the radial velocity in *Gaia* (as high as $\sim 8$ km s$^{-1}$), as measured by the `radial_velocity_error` parameter. We estimated their periods to be of the order of months, assuming a binary in a circular orbit with equal-mass components. We also considered the parameters `ruwe` and `astrometric_excess_noise` (and its significance) as a potential signature of orbital wobble of individual components in binary star systems (Lindegren et al., 2012, and see Gan et al. 2022 for a practical example). Measuring the periodic variations in radial velocities (RV) in several epochs is a powerful method of detecting and describing binary systems. Furthermore, it allows us to discriminate true orbital acceleration due to multiplicity from the acceleration due to an effect of perspective. Figure 4.8 shows the median of the epoch radial velocities as a function of $G$ magnitude for the stars in our sample with these available data in *Gaia*. The circles coloured in orange and magenta correspond to single and multiple stars, respectively, with evidence for variability in the radial velocities according to criterion 3 in Table 4.2. Therefore, the orange empty circles correspond to *candidates* to unresolved pairs.

## 4.4 Results and discussion

### 4.4.1 Multiplicity fraction

Approximately one-third of the M dwarfs in our sample belong to a multiple system, with binary and triple arrangements embody the majority of architectures. In this section we discuss their characteristics and provide an analysis of these statistics, including the common definitions of multiplicity and companion frequency rates.

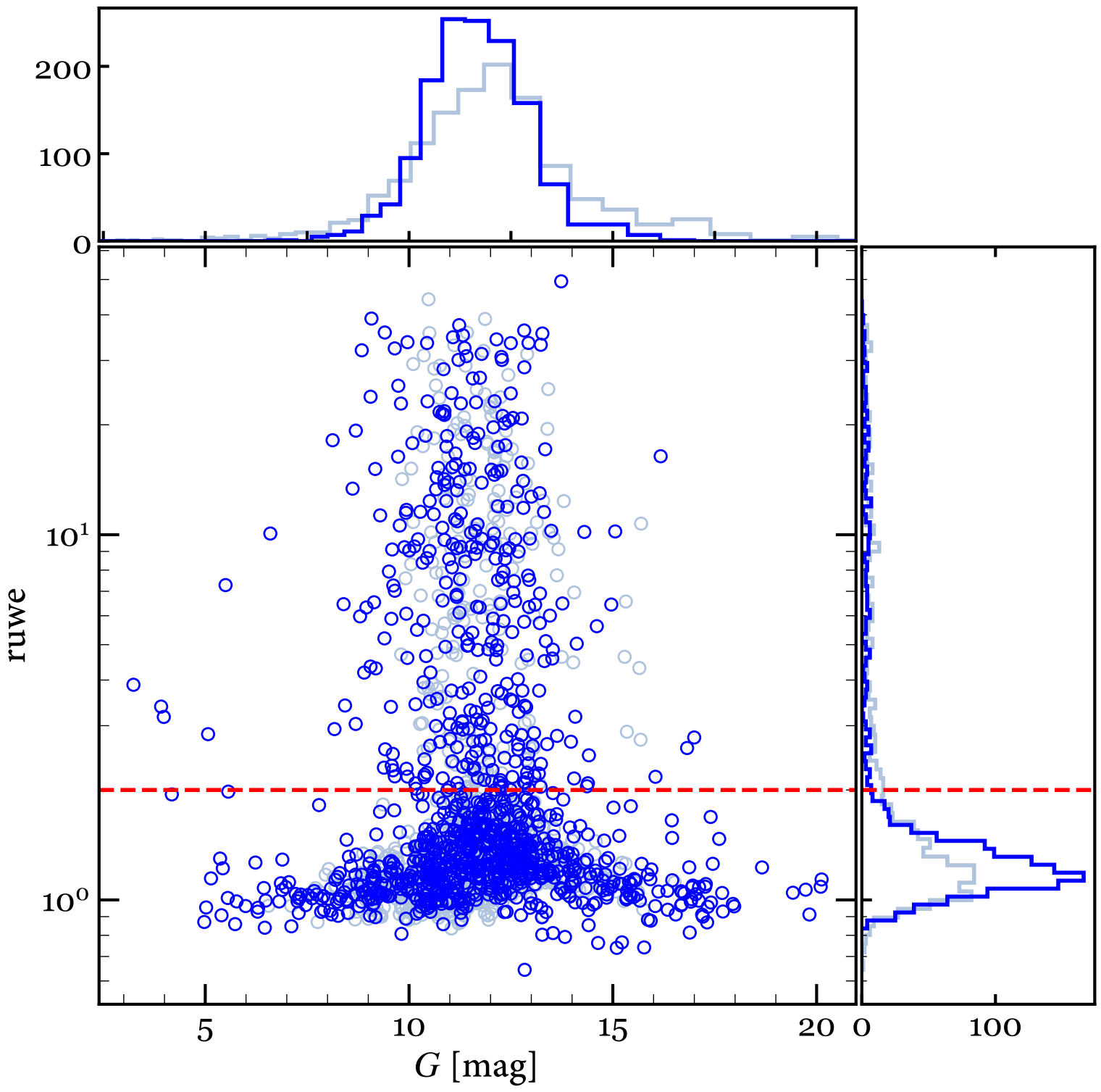

Figure 4.9: `ruwe` as a function of $G$ magnitude for the single stars (grey circles) and the stars belonging to multiple systems (blue circles). The histograms follows the same colour codes. The dashed red line marks a conservative value of `ruwe` = 2.0.

There are 834 M dwarfs in our sample that belong to a multiple system, from which binaries and triples embody the majority of architectures, representing 81.8 % and 15.8 %, respectively. The remaining 1378 are single stars, or 62.3 % of the sample. Figure 4.10 shows the distribution of the stars, showing in red the single stars that are candidates to unresolved binaries.

Among the 834 M dwarfs in multiple systems, 641 are primary components of their systems (i.e. the most massive). For all statistical purposes, we adopt this latter definition of stellar multiplicity. In this sense, the M dwarfs in our sample that belong to systems with AFGK-type primaries *are not* computed in these statistics, but they are analysed in Sect. 4.4.5. To quantitatively asses the multiplicity of M dwarfs, we follow the notation of Batten (1973), who denoted as $f_n$ the fraction of primaries that have $n$ companions. Therefore, the multiplicity frequency or multiplicity fraction (*MF*):

$$MF = \frac{\sum\limits_{n=1} f_n}{\sum\limits_{n=0} f_n} = \frac{B + T + Q + \ldots}{S + B + T + Q + \ldots} \tag{4.6}$$

is the number of non-single systems (i.e. binaries of higher order), where $S$, $B$, $T$, $Q$ denote the number of single, binary, triple, and quadruple systems, respectively. The companion star fraction (*CSF*), is a measure of the average number of companions per system:

$$CSF = \frac{\sum\limits_{i=1} (n-1) f_n}{\sum\limits_{i=0} f_n} = \frac{B + 2T + 3Q + \ldots}{S + B + T + Q + \ldots}, \tag{4.7}$$

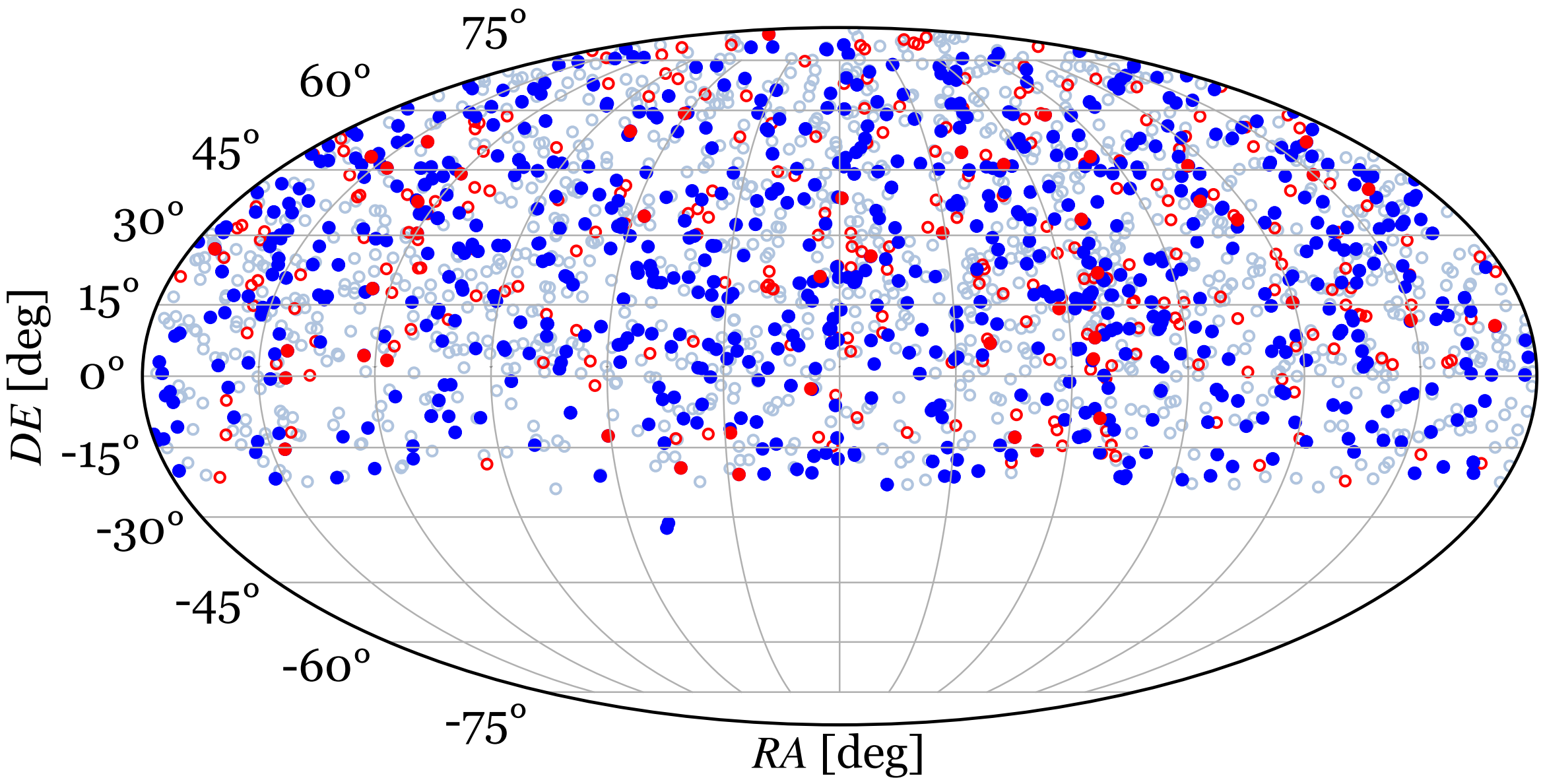

Figure 4.10: Location in the Mollweide-projection sky of all the stars in the sample of M dwarfs, including their companions, in equatorial coordinates. The coloured circles show single stars (empty grey), stars that belong to a multiple system (filled darker blue), and candidates to unresolved binarity for bona fide singles (empty red) and widely separated components of multiple systems (filled red).

which can be larger than 1. We compute these statistics globally, from M0.0–M9.5 V, and, in order to study the dependence with the mass, also for each individual spectral type, from M0.0 V to M9.5 V. We always impose that the M dwarf is the primary of each system, and we also account for observational biases by restricting the statistics to volume-limited subsamples, according to the constrains in Table 4.1. For the estimation of the uncertainties, we compute the 95 % confidence interval using the Wilson (Wilson, 1927) expression:

$$CI_{\rm Wilson} = \frac{k + \kappa^2/2}{n + \kappa^2} \pm \frac{\kappa n^{1/2}}{n + \kappa^2} \sqrt{[\hat{\epsilon}(1 - \hat{\epsilon}) + \kappa^2]/4n}, \tag{4.8}$$

where $n$ is the number of trials, $k$ is number of observed successes, $\hat{\epsilon} = k/n$, and $\kappa$ is the number of standard deviations corresponding to the desired confidence interval for a normal distribution. For a confidence interval of $100(1 - \alpha)$ %,

$$\kappa = \Phi^{-1}(1 - \alpha/s) = \sqrt{2}\mathrm{erf}^{-1}(1 - \alpha). \tag{4.9}$$

In our sample we find that the multiplicity fraction for all the range of M dwarfs, once observational biases have been taken into account, is $MF$ = 28.9%, and the companion star fraction is $CSF$ = 50.1%. The candidates to multiple systems found as described in Sect. 4.3.2 represent an additional 12.3% to the $MF$ value. With this, we find that the multiplicity of M dwarfs could be *as high as* 41%.

In Table 4.4 we tabulate these statistics together with the multiplicity fraction for M dwarfs reported in the literature during the last three decades, sorted by decreasing order of publication. To compare properly, we choose works for which similar spectral ranges in the M-dwarf domain were studied, although they are rarely identical. We indicate, when possible, the relevant constrains of the studies: spectral range, sample size, completeness volume, search separations, and methodology. We help the comparison by also providing the $MF$ in two consecutive spectral ranges: from M0.0 to M5.0 V, and from M5.5–M9.0 V.

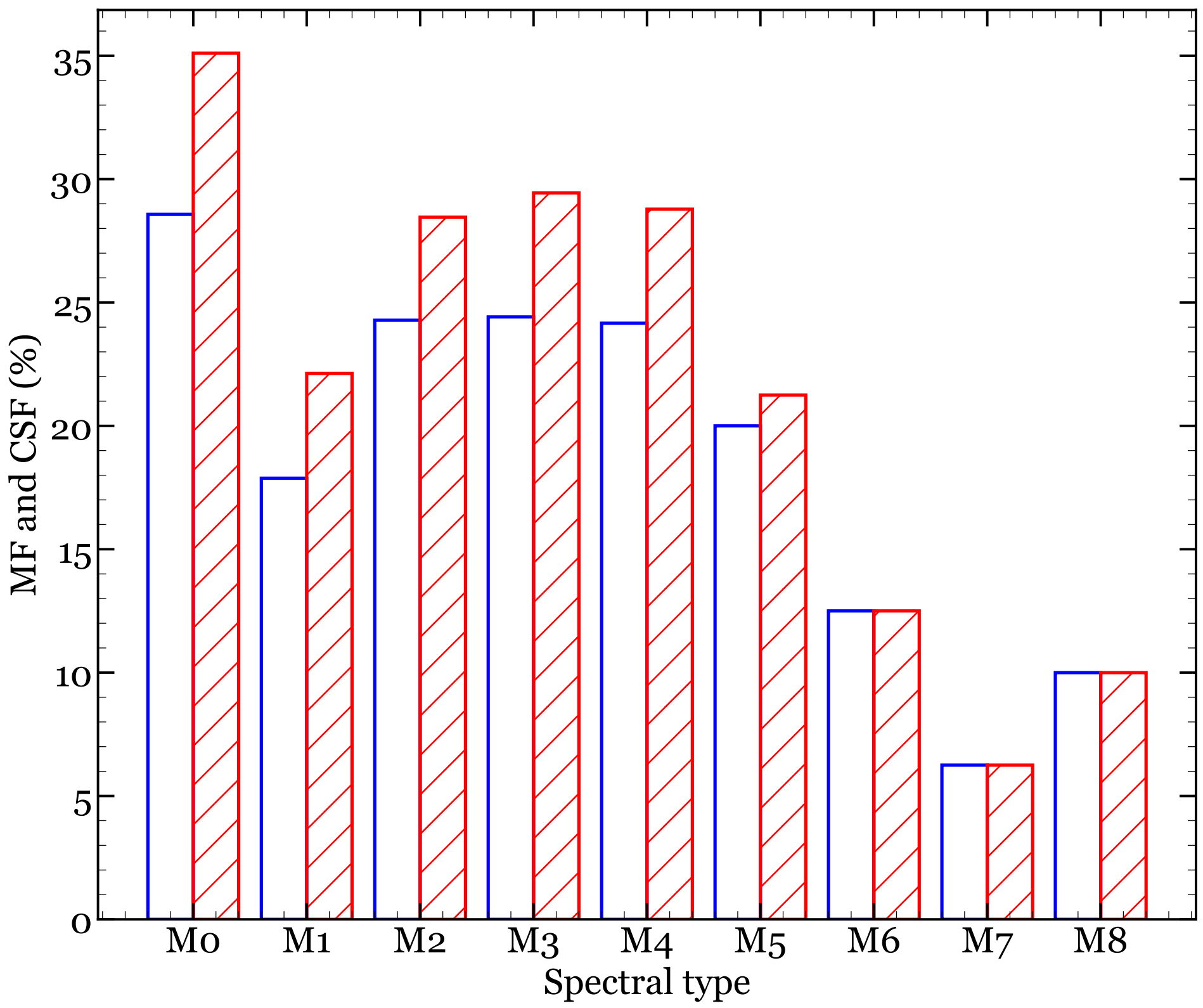

Figure 4.11: Multiplicity fraction (*MF*) and companion star fraction (*CSF*) as a function of spectral type, shown as empty blue and dashed red bar plots, respectively.

Our results are in agreement with the values reported in the literature, especially in those cases where a sizeable sample was studied. In this sense, our work doubles in size the largest sample among the compared studies.

Next, we study the dependence of multiplicity fraction on spectral type (Table 4.3). In Fig. 4.11 we show the *MF* and *CSF* percentages as a function of the spectral type, and in Table 4.4 we tabulate these fractions. The results for spectral type M7.0 V and later are omitted because of the absence of a substantial sample. The dependence of *MF* with primary mass shown is consistent with the expected values in the literature, as we introduced in Sect. 4.1. The *CSF* also reflects this dependence, meaning that the smaller the stars, the fewer the companions. With an asterisk ($MF^*$) we refer to the multiplicity fraction that takes into account the non-resolved candidates. We also include the mean values of the mass ratio, $q$, which behaviour is not monotonic (i.e. it does not indicate that the masses of the companions are comparatively smaller for smaller stars). This parameter $q$ has calculated carefully considering what *primary* means in each system. For instance, the primary component for Proxima Centauri is the double star $\alpha$ Centauri, as introduced in Sect. 4.1.

Table 4.3: *MF*, *CSF*, and mean mass ratio, $q$, as a function of spectral type for the 2215 stars in the sample.

| Spectral type | Sample size | *MF* [%] | *MF** [%] | *CSF* [%] | $q$ |
|---|---|---|---|---|---|
| M0.0–1.0 V | 245 | $28.6^{+6.0}_{-5.3}$ | $40.0^{+6.2}_{-5.9}$ | $35.5^{+6.2}_{-5.7}$ | 0.75 |
| M1.0–1.5 V | 330 | $17.9^{+4.5}_{-3.8}$ | $30.9^{+5.2}_{-4.7}$ | $22.1^{+4.8}_{-4.1}$ | 0.49 |
| M2.0–2.5 V | 383 | $24.5^{+4.5}_{-4.0}$ | $35.5^{+4.9}_{-4.6}$ | $27.9^{+4.7}_{-4.3}$ | 0.69 |
| M3.0–3.5 V | 557 | $24.4^{+3.7}_{-3.4}$ | $35.9^{+4.1}_{-3.9}$ | $29.6^{+3.9}_{-3.6}$ | 0.69 |
| M4.0–4.5 V | 476 | $23.9^{+4.0}_{-3.6}$ | $36.1^{+4.4}_{-4.2}$ | $27.9^{+4.2}_{-3.8}$ | 0.72 |
| M5.0–5.5 V | 160 | $20.0^{+6.9}_{-5.5}$ | $34.4^{+7.6}_{-6.9}$ | $21.2^{+7.0}_{-5.6}$ | 0.73 |
| M6.0–6.5 V | 32 | $12.5^{+15.6}_{-7.5}$ | $28.1^{+17.2}_{-12.6}$ | $12.5^{+15.6}_{-7.5}$ | … |
| M7.0–9.5 V | 28 | … | … | … | … |

Table 4.4: Multiplicity fraction (*MF*) for M dwarfs calculated in this work and published in the literature.

| Reference | Spectral range[a] | Sample size | $d_{\mathrm{lim}}$[b] [pc] | $s$ [au] | Multiplicity fraction[c] [%] | Methodology[d] |
|---|---|---|---|---|---|---|
| This work | M0.0–M5.0 | 2118 | ~24–33 | $\lesssim 10^5$ | $29.2^{+2.0}_{-1.9}$ ($41.2^{+2.1}_{-2.1}$) | See Sect. 4.3 |
| | M5.5–M9.0 | 94 | ~10–24 | $\lesssim 10^5$ | $23.4^{+9.5}_{-7.4}$ ($42.6^{+10.1}_{-9.5}$) | |
| | M0–M9 | 2212 | ~10–33 | $\lesssim 10^5$ | $28.9^{+1.9}_{-1.9}$ ($41.2^{+2.1}_{-2.0}$) | |
| Susemiehl & Meyer (2022)† | M | 1550 | 15 | $\lesssim 10^4$ | 22.9 ± 2.8 | † |
| Clark et al. (2022) | M | 1070 | 15 | $\lesssim 60$ | 29.2–31.3 | SI |
| Winters et al. (2019a) | M | 1120 | 25 | $\lesssim 10^4$ | 26.8 ± 1.4 | WI |
| Cortés-Contreras et al. (2017b) | M0–M5 | 425 | 14 (86 %)[c] | ~1.4–65.6 | 19.5 ± 2.3 | LI |
| Ward-Duong et al. (2015) | K7–M6 | 245 | 15 | ~3–10 000 | 23.5 ± 3.2 | AO, WI |
| Jódar et al. (2013) | K5–M4 | 451 | 25 | $\lesssim 80$ | $20.3^{+6.9}_{-5.2}$ | LI |
| Janson et al. (2012) | M0–M5 | 761 | 52 | ~3–227 | 27 ± 3 | LI |
| Bergfors et al. (2010) | M0–M6 | 108 | 52 | ~3–180 | 32 ± 6 | LI |
| Law et al. (2008) | M4.5–M6.0 | 108 | $\lesssim 20$ | $\lesssim 80$ | $13.6^{+6.5}_{-4.0}$ | LI |
| Reid et al. (1997) | K2–M6 | 106 | 8 | $\lesssim 1800$ | 32 | SI, WI, RV |
| Leinert et al. (1997) | M0–M6 | 34 | 5 | ~1–100 | 26 ± 9 | SI |
| Fischer & Marcy (1992) | M | 62 | 20 | $\lesssim 10^4$ | 42 ± 9 | SI, WI |
| Henry (1991) | M | 74 | 8 | … | 30–40 % | SI |

[a] "M" must be read as "all the M dwarfs within the volume limited by $d_{\mathrm{lim}}$", when no specific limitation on the spectral classification of the sample is given.
[b] The volume-complete samples are taken for each spectral type, based on the values of Table 4.1.
[c] The percentage in parentheses takes into account the candidates to unresolved binaries detected by *Gaia*.
[d] SI: Speckle interferometry; LI: Lucky imaging; WI: Wide-field imaging; AO: Adaptive optics; RV: Radial-velocity. These methods are referred to data acquisition.
[e] The authors indicate that the percentage may increase to at least 36 % by including the pairs at $\rho < 0.2$ arcsec and $\rho > 5$ arcsec.
† Data compiled from four different works. The sample size is the sum of the individual sample sizes, and the limiting distance is the smallest one. The methodology consisted on fitting a log-normal function to the orbital separation distribution.

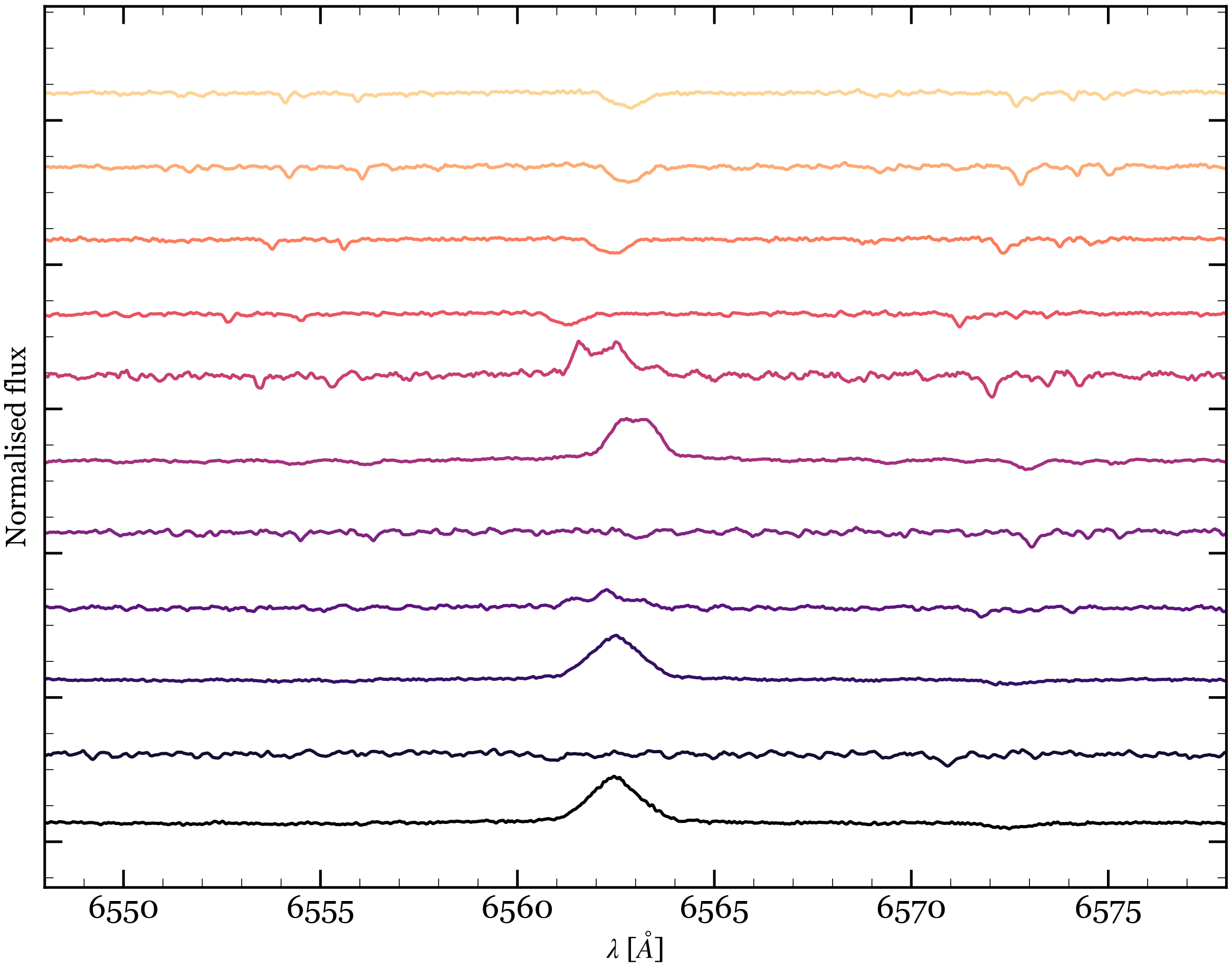

Figure 4.12: FIES spectra of early M dwarfs in our spectroscopic study, from hottest (*top*) to coolest (*bottom*). They are vertically offset for visualisation purposes.

### 4.4.2 Unresolved binaries resolved with FIES

Among the stars for which *Gaia* DR3 measures the largest deviations in radial velocity and the largest values of ruwe, we selected the brightest to study their close multiplicity using medium resolution spectra. We carried out the observations with the high-resolution FIbre-fed Echelle Spectrograph (FIES) mounted on the 2.56m Nordic Optical Telescope (NOT, Djupvik & Andersen, 2010), in the mid-resolution mode ($R$ = 46 000). The observations took place in service mode during four nights over seven months, from April to October, 2022, and are in process in a separate proposal during four nights granted in a campaign of semester 2022B. Additionally, several spectra were obtained in mid-February with High-Efficiency and high-Resolution Mercator Échelle Spectrograph (HERMES) at the 1.2-m Mercator telescope. Finally, a follow-up proposal for this programme has been approved, also with FIES, for the semester 2023A. We have obtained so far 30 spectra of 17 targets, which were processed with automatic reduction software FIEStool (Stempels & Telting, 2017). Figures 4.12 and 4.13 display some representative spectra taking with FIES.

From the study of cross-correlation functions (CCFs) and radial velocities, we discovered 7 SB1 (5 of them with more observations needed to confirm), 5 SB2, 2 ST3 (one of them also to be confirmed), and 2 high-rotation stars (whose multiplicity cannot be determined with the available data). Several CCF fitting plots can be found in Fig. 4.17. We summarise the findings in Table 4.5, including the observation epochs, the heliocentric radial and rotational velocities (for both components if possible), and the unresolved binarity criteria (1–7 in Table 4.2) that they meet. The errors associated to the heliocentric radial velocities were obtained as weighted means of the individual values deduced for each of the slices of the spectra.

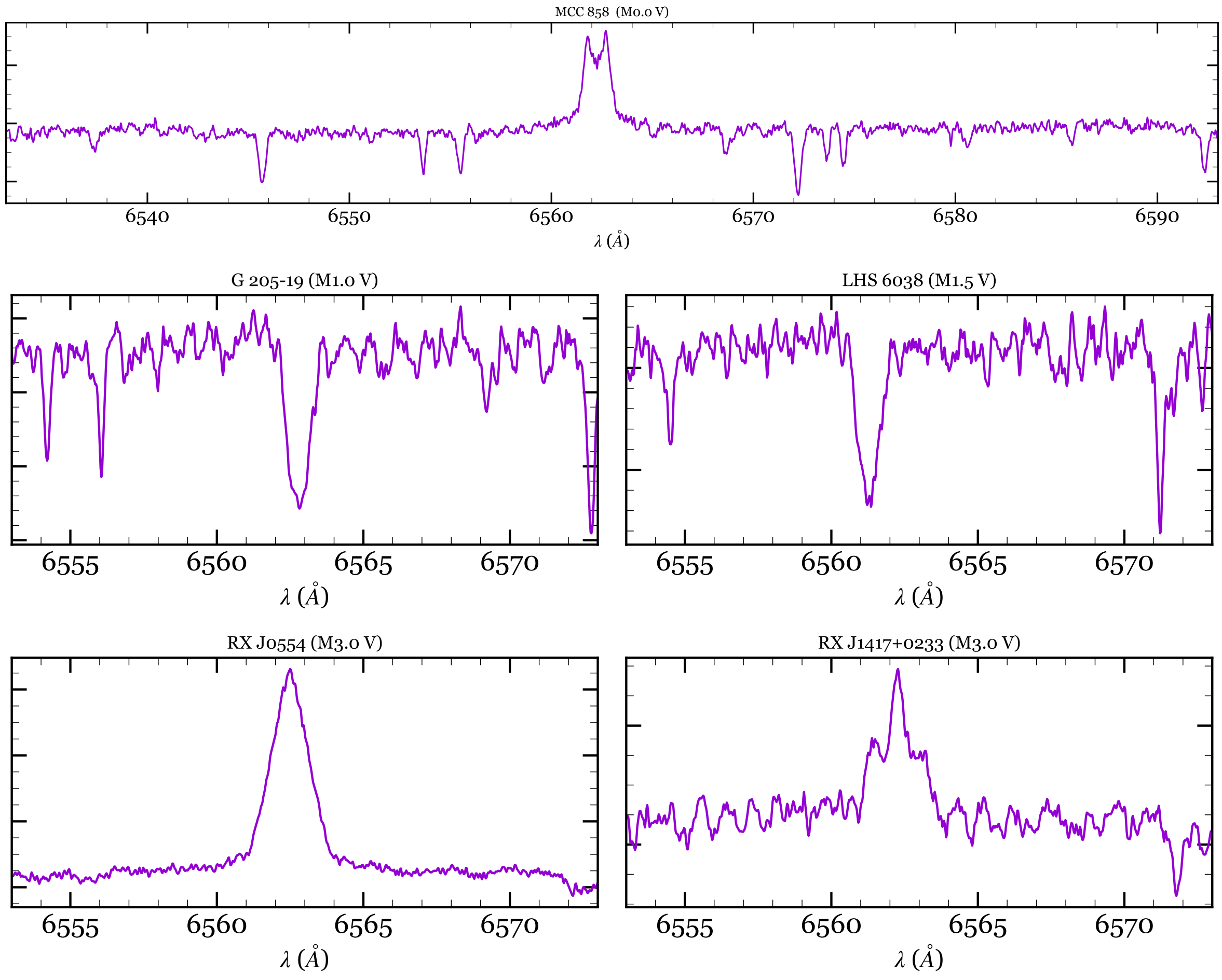

Figure 4.13: FIES mid-resolution spectra of different cases of early M dwarfs in our spectroscopic study, centred in H$\alpha$ ($\lambda$ = 6563 Å).

Heliocentric radial velocities were obtained using the CCF technique (see e.g. Gálvez et al., 2002). The spectra of the target were cross-correlated in 6 slices in the range 4500–7500 Å, using the routine `fxcor` in the software Image Reduction and Analysis Facility (IRAF), against spectrum of a radial velocity standard. We selected Barnard's Star, because is a well-studied single, non-active, low-rotator mid-M dwarf with known radial velocity (−110 km s$^{-1}$ Fouqué et al., 2018). We derived the radial velocity for each slice from the position of the peak of the CCF and calculated the uncertainties based on the fitted peak height and the antisymmetric noise as described by Tonry & Davis (1979). For SB2 cases, when we could discriminate the peaks in the CCF associated with both components, we fitted each peak separately.

In order to determine rotational velocities, $v \sin i$ we applied the cross-correlation technique, again using `fxcor`. The method is described in detail in Gálvez et al. (2002). In short, it is based on the fact that when a stellar spectrum with rotationally broadened lines is cross-correlated against a narrow-line spectrum, the width of the CCF will depend on the amount of rotational broadening of the former spectrum. We therefore used Barnard's Star to calibrate the relation between $v \sin i$ and the full-width half-maximum (FWHM) of the CCF. For this, we used the range 6500–7000 Å and a series of rotations between 1 and 50 km s$^{-1}$, fitting the relation with a 4-degree polynomial (Fig. 4.14). The rotational velocities measured

Table 4.5: Spectroscopic binaries and candidates detected from radial velocity variability and ruwe in *Gaia*.

| Karmn | Name | SB[a] | HJD (240000–JD) | $V_A$ [km s$^{-1}$] | $v \sin i_A$ [km s$^{-1}$] | $V_B$ [km s$^{-1}$] | $v \sin i_B$ [km s$^{-1}$] | Note[b] |
|---|---|---|---|---|---|---|---|---|
| J00374+515 | G 172-14 | Single or SB1 | 59864.57 | −55.68 ± 0.48 | < 2 | ... | ... | • |
| J01556+028 | LHS 6038 | SB2 | 59864.51 | −67.32 ± 0.32 | < 2 | −44.66 ± 0.73 | 3–4 | |
| J02069+451 | V374 And | SB2 | 59864.44 | 59.13 ± 0.80 | ∼4 | 64–65 | ... | • |
| | | | 59864.74 | 58.82 ± 1.00 | ... | ... | ... | |
| | | | 59945.46 | 55.27 ± 0.58 | < 3 | 68.75 ± 1.02 | < 3 | |
| J03026-181 | GJ 121.1 | Single or SB1 | 59864.54 | 19.84 ± 0.13 | < 2 | ... | ... | |
| J03147+485 | Ross 346 | SB2/ST3 | 59864.60 | 30.30 ± 0.45 | 2–3 | 8.72 ± 0.95 | 7–8 | • |
| | | | 59944.56 | 10.92 ± 0.45 | < 3 | 36.78 ± 0.96 | ∼3 | |
| | | | 59945.55 | 8.62 ± 0.36 | ... | 40.05 ± 0.72 | ... | |
| J05091+154 | Ross 388 | Single or SB1 | 59944.59 | −21.60 ± 0.24 | ∼3 | ... | ... | |
| J05547+109 | RX J0554.7+1055 | rot | 59864.62 | 10.51 ± 2.12 | 25–26 | ... | ... | • |
| | | | 59864.65 | 10.52 ± 1.52 | ... | ... | ... | |
| | | | 59864.65 | 11.16 ± 2.21 | ∼30 | ... | ... | |
| J06035+155 | TYC 1313-1482-1 | ST3? | 59864.75 | 49.58 ± 0.93 | 3–4 | ... | ... | • |
| | | | 59945.61 | 50.20 ± 1.00 | ∼3 | ... | ... | |
| J06105+024 | TYC 135-239-1 | Single or SB1 | 59864.72 | 35.76 ± 0.57 | 2–3 | ... | ... | • |
| J08158+346 | LP 311-008 | SB2 | 59864.68 | 23.92 ± 0.46 | 2–3 | 43.54 ± 0.66 | 2–3 | • |
| | | | 59945.68 | 23.03 ± 0.47 | < 3 | 42.22 ± 0.07 | < 3 | |
| J14175+025 | RX J1417.5+0233 | SB2 | 59697.48 | −47.56 ± 0.20 | 2–3 | −8.66 ± 0.51 | 4–5 | • |
| | | | 59723.47 | −22.76 ± 0.13 | ... | −45.61 ± 0.38 | ... | |
| | | | 59775.38 | −18.23 ± 0.16 | ... | −51.4 ± 0.45 | ... | |
| J14200+390 | IZ Boo | SB1?/rot | 59775.38 | −22.10 ± 2.87 | 34–36 | ... | ... | • |
| J18227+379 | G 205-19 | Single or SB1 | 59864.37 | −14.53 ± 0.40 | < 2 | ... | ... | • |
| J18394+690 | RX J1839.4+6903 | SB1 | 59723.52 | −33.54 ± 0.30 | 1–3 | ... | ... | • |
| | | | 59775.43 | −32.67 ± 0.28 | ... | ... | ... | |
| | | | 59864.40 | −2.19 ± 0.30 | ... | ... | ... | |
| | | | 59883.31 | 0.54 ± 0.28 | ... | ... | ... | |
| J18519+130 | 2MJ18515965+1300034 | Single or SB1 | 59864.30 | −17.19 ± 0.68 | 13–14 | ... | ... | |
| J22176+565 | 2MJ22173704+5633100 | SB2 | 59864.45 | −22.12 ± 0.53 | < 2 | −9.26 ± 0.9 | < 2 | |
| J23060+639 | MCC 858 | Single or SB1 | 59864.49 | −22.95 ± 0.72 | 3–4 | ... | ... | |

[a] SB1: single-lined binary; SB2: duble-lined binary; ST3: triple-lined triple; rot: high rotation star.
[b] J00374+515: Included in Non-single table nss_two_body_orbit as 'SB1'.
J02069+451: Clear asymmetry found, most probably from a secondary component, but difficult to determine. Classified as eclipsing binary candidate (Malkov et al., 2006) in the Combined General Catalogue of Variable Stars (GCVS; Samus et al., 2004).
J03147+485: Probable SB3, with $V_B \simeq 17$ km s$^{-1}$. Unclear with a single spectrum.
J05547+109: 12.40 ± 1.88 and 12.05 ± 1.20 are the values obtained with reference star (Barnard's) rotated to the object's $v \sin i$.
J06035+155: Probable triple. 10, 50, and 70 km s$^{-1}$ measured for the components.
J06105+024: Included in Non-single table nss_two_body_orbit as 'AstroSpectroSB1'.
J08158+346: Included in Non-single table nss_two_body_orbit as 'AstroSpectroSB1'.
J14175+025: Orbital solution attempted (Fig. 4.16).
J14200+390: −23.90 ± 4.38 is the value obtained with reference star (Barnard's) rotated to the object's $v \sin i$.
J18227+379: Probable higher order multiple.
J18394+690: Orbital solution attempted (Fig. 4.16).

this way are included in Table 4.5.

Most of our targets are not fast rotators (i.e. the rotation velocity is similar to that of Barnard's standard). Nevertheless, two objects are found to be moderate rotators: IZ Boo ($v \sin i \simeq 25$ km s$^{-1}$) and RX J0554.7+1055 ($v \sin i \simeq 35$ km s$^{-1}$). In these the lines widen and flatten, which impedes a correct fitting of the CCF to a gaussian profile (Fig. 4.15, *left*). In their spectra the lines appear as lumpy, which may look as the effect of two components, but that is not evident in the CCFs. For instance, measuring the FWHM of IZ Boo yields a rather broad range 35–45 km s$^{-1}$ for $v \sin i$. For these two stars, we performed the CCF with the spectrum of Barnard's Star, but rotated to the corresponding $v \sin i$ esti-mated for each high rotating star, as explained previously. This way we could directly measure their radial velocity and calculate the rotation velocity. With this procedure we also smoothed the profile, preventing from seeing possible close and blended companions, but a previous inspection suggest that, although very close companion or active regions may be the cause or the irregularities in the CCF, it is not possible to discriminate clearly any additional component.

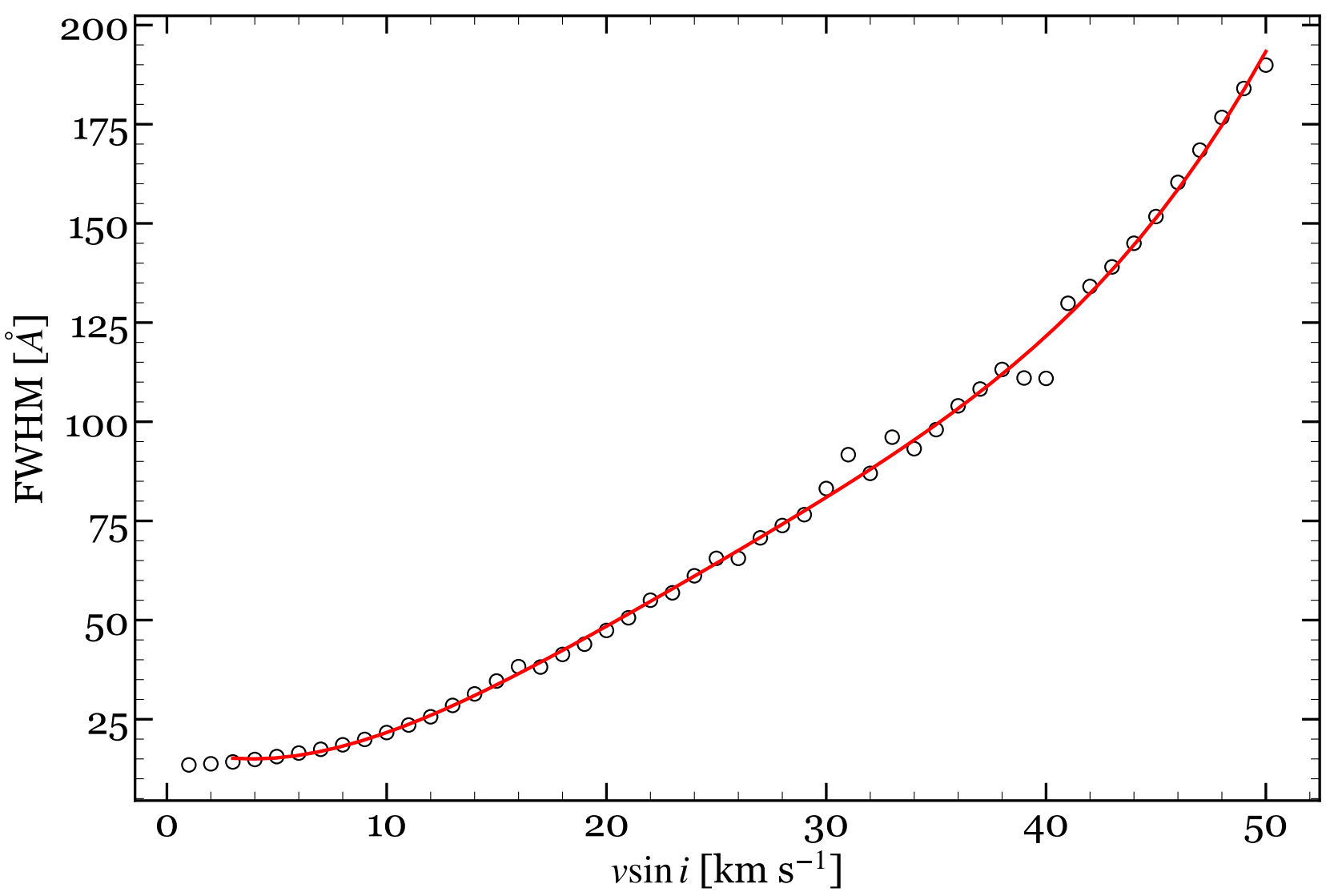

Figure 4.14: Rotational velocity, $V_{\rm rot}$, as a function of the full width half maximum (FWHM), calculated for Barnard's Star, serving as a yardstick for high-rotation objects.

Table 4.6: Prior distributions to perform the MCMC fit.

| Parameter | RX J1417.5+0233 | RX J1839.4+6903 | Units |
|---|---|---|---|
| $V_r$ | $\mathcal{U}(-40, -20)$ | $\mathcal{U}(-40, -20)$ | km s$^{-1}$ |
| $K$ | $\mathcal{G}\left(0, 10^4\right)$ | $\mathcal{G}\left(0, 10^4\right)$ | km s$^{-1}$ |
| $P_{\rm orb}$ | $\mathcal{G}(148, 20)$ | $\mathcal{G}(\mathrm{X}, \mathrm{Y})$ | d |
| $T_0$ | $\mathcal{U}(t_1, t_1 + P_{\rm orb})$ | $\mathcal{U}(t_1, t_1 + P_{\rm orb})$ | d |
| $\sqrt{e}\cos\omega$ | $\mathcal{G}_t(0, 0.2)$ | $\mathcal{G}_t(0, 0.2)$ | … |
| $\sqrt{e}\sin\omega$ | $\mathcal{G}_t(0, 0.2)$ | $\mathcal{G}_t(0, 0.2)$ | … |
| jitter | $\mathcal{U}(0, 10)$ | $\mathcal{U}(0, 10)$ | km s$^{-1}$ |

For two candidates, RX J1417.5+0233 and RXJ1839.4+6903, we have made an attempt of obtaining orbital solutions, using the three and four datapoints available to date for each resolved components, respectively (Fig. 4.16). We used the sum of $n$ Keplerian models, where $n$ is the number of detected components in the system. We explored the parameter space by sampling the posterior distribution with the Markov chain Monte Carlo (MCMC) code `emcee` (Foreman-Mackey et al., 2013), setting the number of walkers to be four times the number of parameters (i.e. 28). Next, we iterated $3{\times}10^3$ times per walker, which sums a total of 28 000 iterations. Given the scarcity of observations per target, we can only aim at providing a preliminary orbital solution. Thus, we restrict some of the orbital parameters by employing narrow gaussian distributions ($\mathcal{G}$, or $\mathcal{G}_t$ if it is truncated) as informative priors. This is the case for the radial velocity semi-amplitude ($K$), the orbital period ($P_{\rm orb}$), the eccentricity ($e$), and the argument of periastron ($\omega$). For the remaining parameters (epoch of periastron passage, $T_0$, longitude of periastron, $\omega$, eccentricity, $e$, and jitter) we selected uniform prior distributions ($\mathcal{U}$).

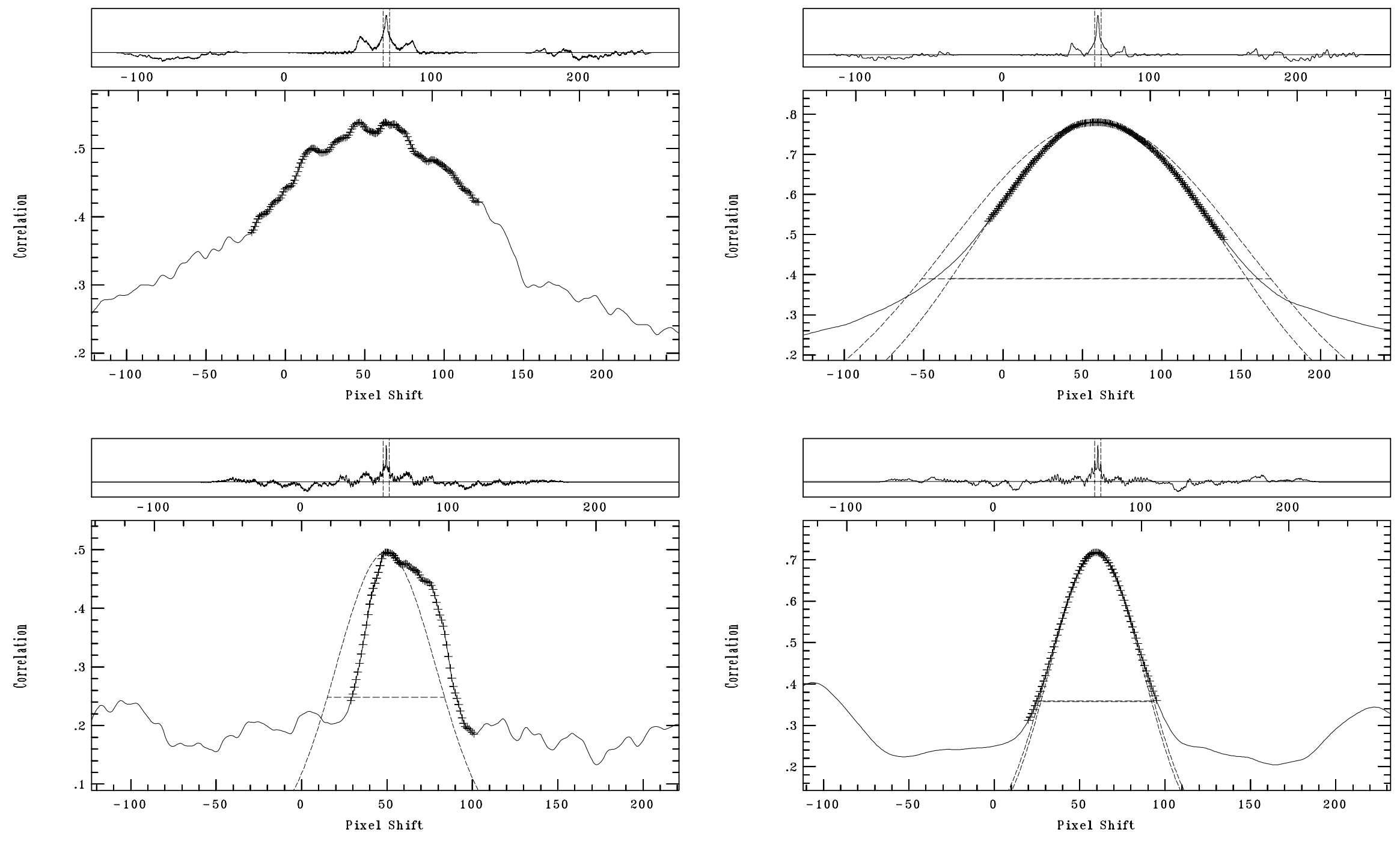

Figure 4.15: CCF of IZ Boo (*top*) and RX J0554.7+1055 (*bottom*). On the left are shown the CCFs with non-rotated Barnard's Star, while on the right are shown the CCFs with rotated Barnard's Star as standard, better suited for measuring radial velocities.

The values employed for these priors (see Table 4.6) are obtained using simple assumptions and basic physical laws. The orbital period is estimated using Kepler's Third Law (Eqn. 4.11). In it, the *maximum* separation, $a$, was estimated given the fact that *Gaia* DR3 is not resolving the pair, but is providing instead an altered (high) `ruwe` statistic (see Sect. 4.3.2), and therefore the physical separation must be $\rho \lesssim 0.15$ arcsec[8]. The definition $a = \rho d$ can be used inserting the parallactic distance, $d$, which is known. For the total mass, $\mathcal{M}_T$, we used the peak height ratio from the CCFs (Fig. 4.17) to approximate the mass of the secondary, given that the mass of the primary is known, or is approximated by the observed spectral type. Finally, the radial velocity deviation from the median measured by *Gaia*, $\Delta V_r$, can be related to the semi-amplitude, $K$ (Eqn. 1.1, with $\mathcal{M}_p = \mathcal{M}_\star$), as $K \simeq \frac{3}{2}\Delta V_r$. This expression comes from the fact that a full orbital revolution (equivalent to *two* semi-amplitudes, $2K$) can be related with the *Gaia* radial velocity variance since its 3-$\sigma$ (that is, $3\Delta V_r$), by definition corresponds with the 99.7% of the measured radial velocities. Finally, we computed the 1-$\sigma$ and 2-$\sigma$ uncertainty intervals of the model from the posterior marginal distribution of each parameter.

At this point it is worth mentioning that the current release of *Gaia* (DR3) contains mean BP/RP spectra for 219 million sources (most of them with $G < 17.6$ mag), and mean radial velocity spectrometer (RVS) spectra for 1 million well-behaved objects. *Gaia* DR4 will incorporate the epoch to these spectra, and will also provide all available variable-star and non-single-star solutions, include source classifications (probabilities) plus multiple astrophysical parameters (derived from BP/RP, RVS, and astrometry) for stars, unresolved binaries, galaxies, and quasars. Therefore, DR4 (projected to no earlier than late 2025) will be of additional help in the systematic investigations such as these.

[8]The minimum angular separation for a pair resolved by *Gaia* DR3 in our sample is $\simeq 0.4$ arcsec.

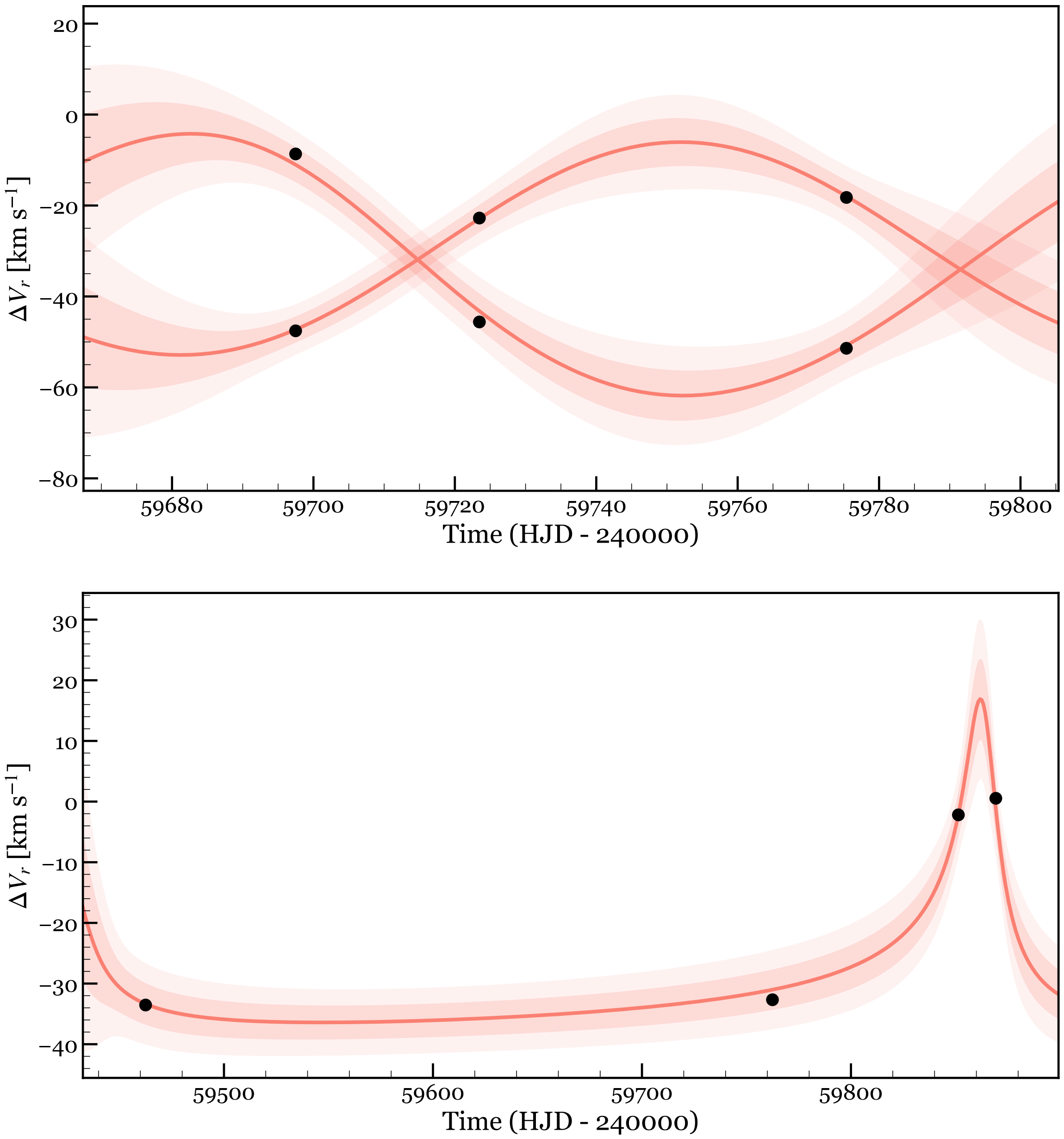

Figure 4.16: Preliminary orbital solutions for the SB2 RX J1417.5+0233 (*top*) and the SB1 RX J1839.4+6903 (*bottom*). The grey shaded regions are MCMC-derived 1- and 2-$\sigma$ uncertainty intervals.

### 4.4.3 Astrophysical parameters

When broadband multi-wavelength photometry is available for a star, it can be arranged to produce its spectral energy distribution (SED), which contains important information about the overall emitting power of the star, or bolometric luminosity (see Chapter 3). These empirical SEDs can be approximated by synthetic models and yield the best estimation for their luminosities and effective temperatures, among other parameters. Because these models normally reproduce the stellar photosphere of the stars, we do not include magnitudes from passbands with $\lambda_{\rm eff}$ ~420 nm (i.e. ultraviolet), because these are of chromospheric origin. When combined, photometric data from different passbands serves to narrow down fundamental properties, such as masses or radii. Colours of the stars are made from the difference of magnitudes in different passbands. An advantage of colours is that they do not depend on the distance to the star, which sometimes is unknown. Nevertheless, if the distance to the star is known with precision (by parallactic means), absolute magnitudes are powerful proxies for other fundamental properties, such as the bolometric luminosity (Cifuentes et al., 2020). For stars in the main sequence (MS), the close relation between the spectral classification and the absolute magnitudes, bolometric corrections, and colours, makes it possible to indirectly estimate parameters that are only accessible with direct inspection of the spectra.

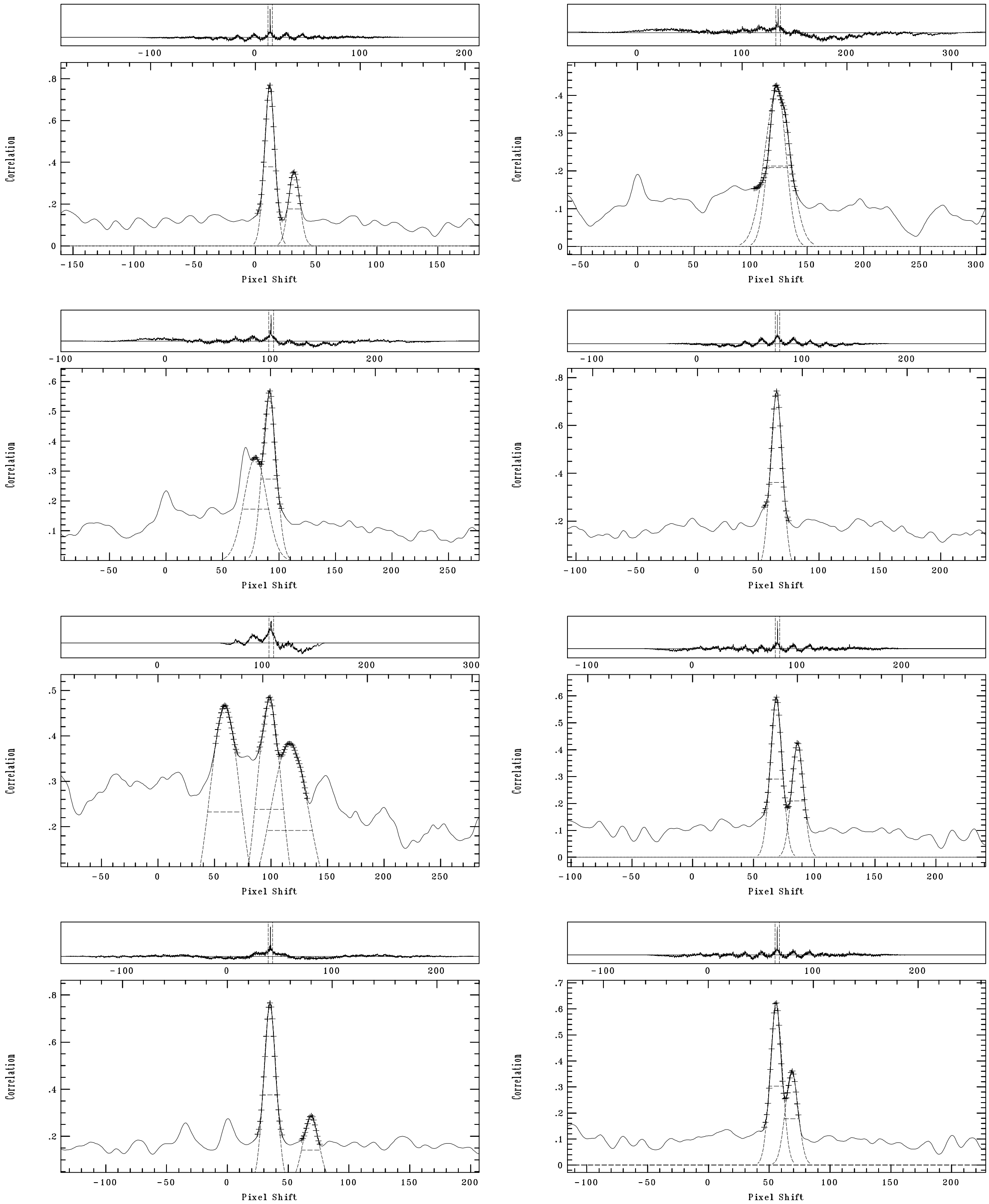

Figure 4.17: *From top to bottom, and left to right*: Cross-correlation functions of LHS 6038 (SB2), V374 And (SB2), Ross 346 (SB2/ST3), Ross 388 (single or SB1), TYC 1313-1482-1 (ST3?), LP 311-008 (SB2), RX J1417.5+0233 (SB2), and 2MASS J22173704+5633100 (SB2).

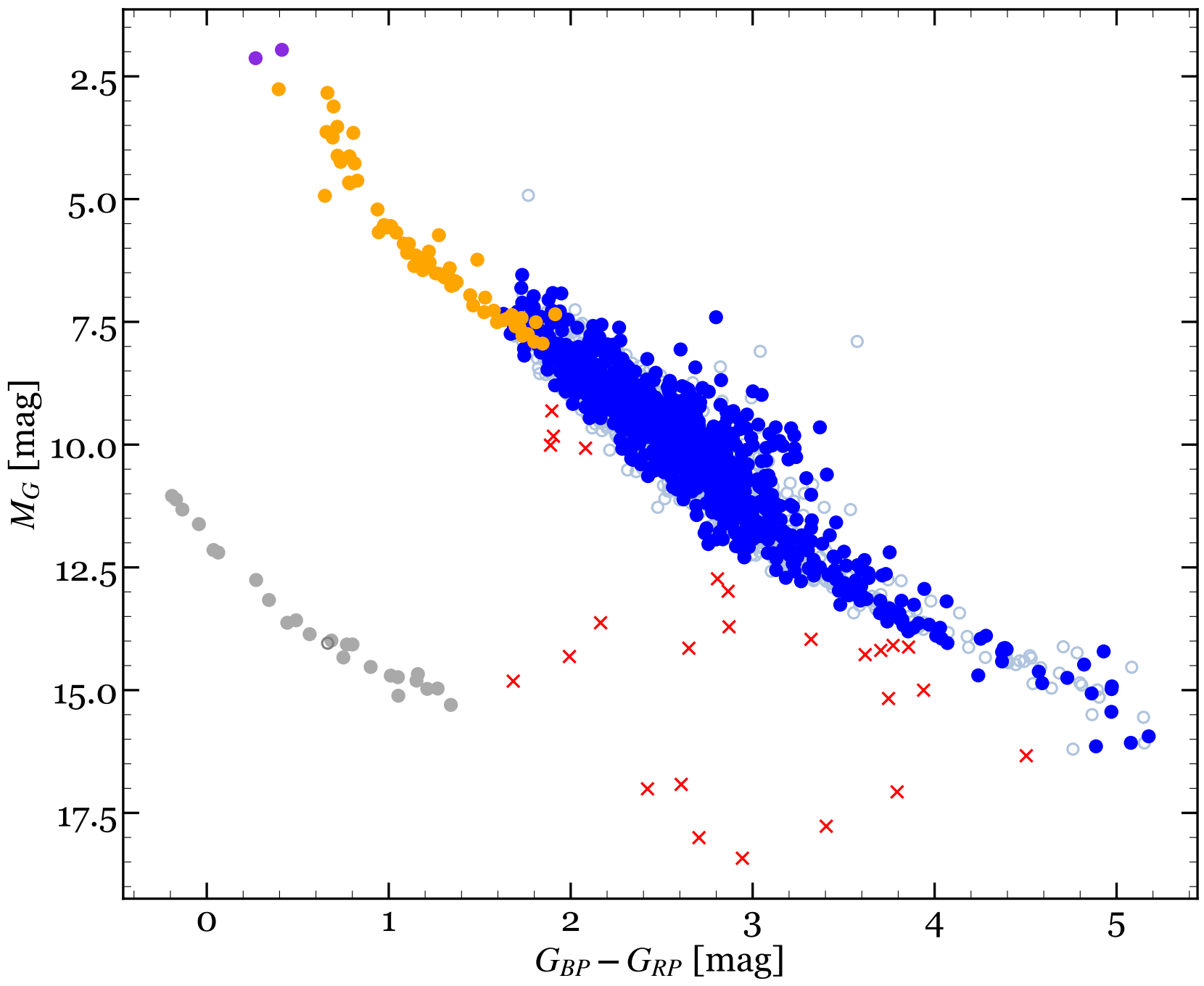

Figure 4.18: Absolute magnitude $M_G$ against $G_{BP} - G_{RP}$ colour for all the stars with full photometry available in *Gaia* DR3. It represents M dwarfs that are single (empty light blue) or part of a multiple system (filled blue), FGK primaries (filled orange), OBA primaries (filled violet), and known and candidates to white dwarfs (filled and empty grey, respectively). Red crosses correspond to components of multiple systems that are very close, very faint, or both, resulting in compromised photometry.

We compiled up to 10 magnitudes for each star, when available: three from *Gaia* ($G_{BP}$, $G$, $G_{RP}$), three from the Two Micron All Sky Survey (2MASS, Skrutskie et al., 2006) ($J$, $H$, $K_s$), and four form the Wide-field Infrared Survey Explorer All-Sky Data Release (AllWISE, Cutri & et al., 2014) ($W1$, $W2$, $W3$, $W4$). Our attempt to include these catalogues automatically by using the 'best_neighbour' automatic crossmatch from the *Gaia* Archive was proven unsuccessful. Indeed, given the singular characteristics of these systems (only partially resolved by 2MASS and AllWISE), the use of automatic crossmatch catalogues provided by *Gaia* (`gaiaedr3.allwise_best_neighbour` and `gaiaedr3.tmass_best_neighbour` in this case) was not found appropriate. Instead, we performed this search manually to ensure a correct discrimination of the components of the systems. This is of paramount importance at this point, because the very description of each system fundamentally relies on the fact of whether 2MASS (along with *Gaia*) is able to resolve the system or not.

**Unresolved binaries** By definition, an unresolved binary only appears as one source in the photometric surveys, with the flux corresponding to the aggregated fluxes of all the components. As seen in Sect. 4.3.2, the latest photometric surveys like *Gaia* are able to see with better detail and discriminate very close ob-jects previously considered as single. Still, a notable percentage of stars seen as single are most probably binaries and, to a lesser extent, triples or even multiples of higher order. Some works suggest that certain types of unresolved binarity can be discovered in apparently single stars using only photometric observations. Indeed, photometry has been employed for the identification of unresolved binaries, especially applied to clusters (e.g. Li et al., 2020; Malofeeva et al., 2023). In Fig. 4.18 we show the colour-magnitude diagram (CMD) for all the stars in our sample and their resolved companions, when the photometry in *Gaia* is available. This excludes the very close systems, and the very bright (Capella, Castor) and the very faint (e.g. GJ 570 D) stars and brown dwarfs. Several groups are highlighted: solar-like primaries (orange), OBA primaries (violet), and white dwarfs companions (light grey). All the candidates to white dwarfs based on their location in the CMD have been identified by Jiménez-Esteban et al. (2018). The authors investigated candidates to white dwarfs from the *Gaia* HR diagram using population synthesis simulator. They cut in the `phot_bp_rp_excess` factor in order to prevent against photometric errors in the $G_{BP}$ and $G_{RP}$ bands, especially relevant for the faintest sources. Very compact systems and resolved young stars are outliers above the main sequence. The outliers below the main sequence correspond to faint components of both very close and wide pairs. The former has its photometry affected by the brighter nearby source; for the latter, *Gaia* simply cannot provide a good measure of the flux using the blue filter, $G_{BP}$.

Using up to 10 magnitudes from the optical blue to the mid-infrared, and the latest parallactic distances available in *Gaia*, we derive effective temperatures and model-independent bolometric luminosities using `VOSA` to fit `BT-Settl CIFIST` models (Baraffe et al., 1998) to the observed spectral energy distribution, as in Cifuentes et al. (2020) (see Chapter 3). We only perform these calculations for the objects whose photometric data have not been compromised by the presence of a companion that is very close, or very bright, or both. These are the single stars (excluding the candidates to close binaries, as explained in Sect. 4.3.2) and, resolved components of multiple systems. For these, we assume the same condition for photometric contamination as Cifuentes et al. (2020) did. Briefly, for a star not to be affected by the presence of a nearby ($\rho \lesssim 5$ arcsec) object, the authors imposed that the flux of the latter, as measured in the $G$ passband, should not exceed more than 1% the flux of the former. This simply translates into a minimum difference in $G$ magnitudes, $\Delta G$, of 5 mag. Moreover, we also restrict this condition further, by choosing only those that are resolved in at least two of the three surveys (*Gaia*, 2MASS, AllWISE), so that the SED is as complete as possible, which minimises the uncertainty in the determination of the best fitting model. This way, we make sure that the presence of a close companion with a comparable or greater luminosity does not have a negative impact on the photometry and any property derived from it. We do not compute luminosities for known spectroscopic binaries in any case.

From the luminosities we homogeneously derived radii $\mathcal{R}$ and masses $\mathcal{M}$,. In particular, we derived the masses using the $\mathcal{M}$-$\mathcal{R}$ relation in (Schweitzer et al., 2019, Eq. 6), coming from the study of detached, double-lined, double-eclipsing, main-sequence, M-dwarf binaries from the literature. This relation is valid a wide range of metallicities for M dwarfs older than a few hundred million years (Dhital et al., 2010). For the companions to the objects in our sample that are outside the M dwarf range, we use the mean values by Pecaut et al. (2012) and Pecaut & Mamajek (2013). For the stars without spectral classification provided in the literature, we estimate the spectral types photometrically, using the average values of the absolute magnitudes, $M_G$ and $M_J$, which are included in Table 7 of Cifuentes et al. (2020).

For the stars for which it is not possible do derive masses directly from the bolometric luminosities, we use instead their absolute magnitudes in $G$. For the M dwarfs, first we fitted a 2-degree polynomial to the relation $M_G$ versus stellar mass, $\mathcal{M}$ (Fig. 4.19), in the range $M_G \in [7.5, 15.0]$ . We use the Bayesian information criterion (BIC) or Schwarz criterion to determine the preferred degree. We proceeded similarly for the $M_G$ versus stellar radius, $\mathcal{R}$, although it is not plotted due to its similarity. The coefficients for both fits are given in Table 4.7, along with the correlation parameters. Both relations are valid for main sequence M dwarfs with solar metallicities, and for $M_G \in [7.5, 15.0]$ (K7/M0.0 V to M7.5 V Cifuentes et al., 2020). This upper limit was chosen with the purpose of the robustness of the fitting in mind, because least-square fitting polynomials of a higher order are prone to degeneracy. Its usage up to $M_G$ = 16 mag still can be done in the case of necessity, but with extreme caution, considering that the masses of ultracool dwarfs are strongly dependent with age (Burgasser & Blake, 2009; Soderblom, 2010, and see Fig. 13 in Sahlmann et al. 2020). We purposefully did not include dwarfs cooler than $M_G$ = 15 mag, because precise masses for these ultracool objects are obtained via dynamical analysis in compact astrometric binary systems (e.g. Sahlmann et al., 2011; Dupuy & Liu, 2017; Brandt et al., 2019; Sahlmann et al., 2020), and therefore *Gaia* cannot provide individual measurements for their apparent magnitudes in $G$.

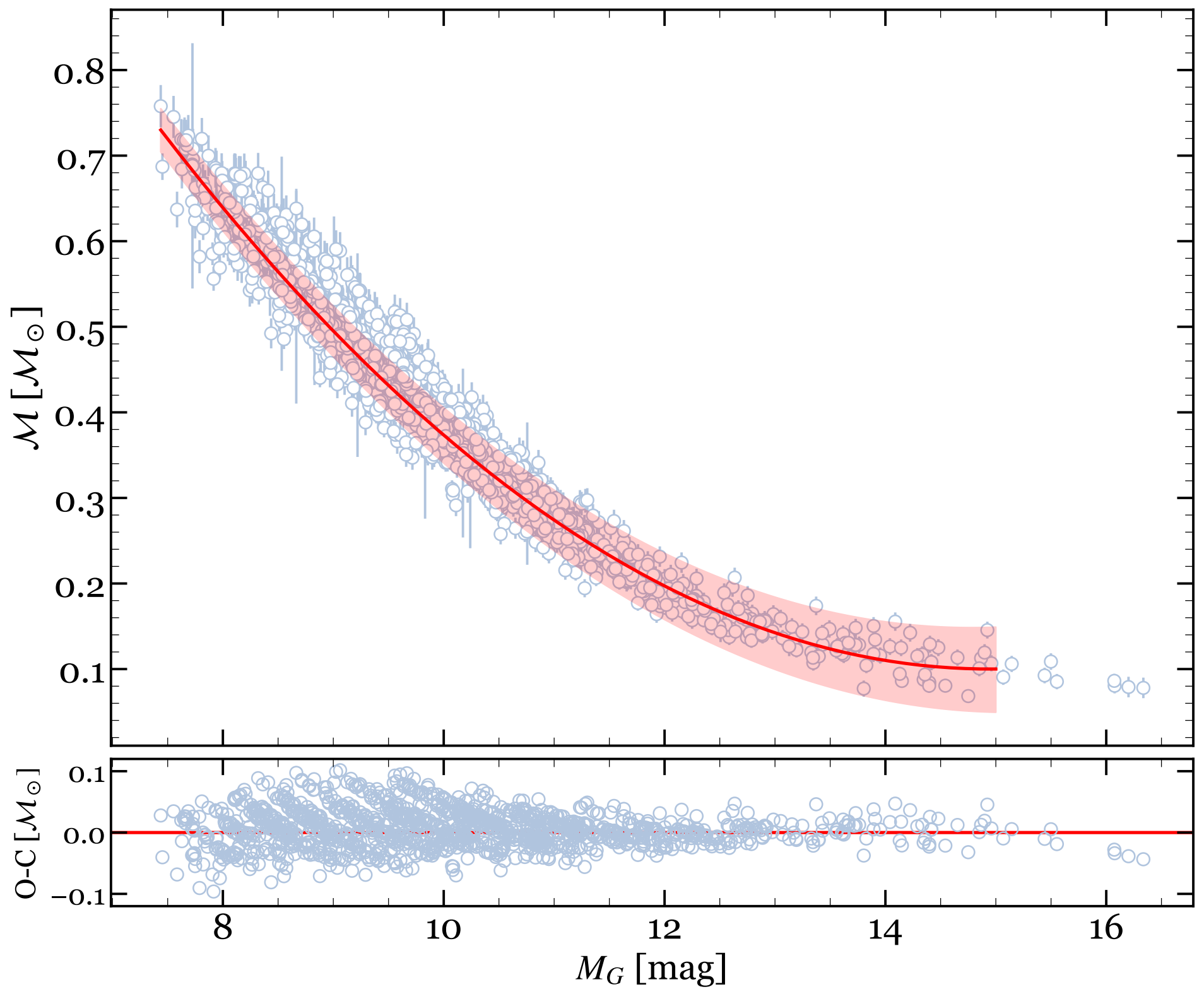

Figure 4.19: Stellar mass $\mathcal{M}$ as a function of the absolute magnitude $M_G$ for the M dwarfs domain. The red line represents the polynomial fit described in the main text, and the red shaded area accounts the 1-$\sigma$ level of uncertainty.

Table 4.7: Coefficients of the polynomial fits for $\mathcal{M}$ and $\mathcal{R}$ as a function of $M_G$.

| $Y$[a] [$\mathcal{M}_\odot$] | $X$ [$\mathcal{M}_\odot$] | $a$ [$\mathcal{M}_\odot^{-1}$] | $b$ [$\mathcal{M}_\odot^{-2}$] | $c$ | $r$ / $\rho$[b] | $\Delta X$ [mag] |
|---|---|---|---|---|---|---|
| $\mathcal{M}$ | $M_G$ | 2.595 ± 0.014 | −0.3338 ± 0.0026 | 0.01117 ± 0.00012 | 0.981 / 0.982 | [7.5, 15.0] |
| $\mathcal{R}$ | $M_G$ | 2.308 ± 0.007 | −0.2871 ± 0.0012 | 0.00933 ± 0.00005 | 0.981 / 0.982 | [7.5, 15.0] |

[a] The polynomial fits follow the form $Y = a + bX + cX^2$.
[b] Pearson's $r$ and Spearman's $\rho$ coefficients.

For this fit we only selected M dwarfs with parallactic distances from *Gaia*, and excluding spectroscopic and astrometric binaries known but not resolved by it, candidates to unresolved binaries (see Sect. 4.3.2), and overluminous young objects, for which the masses are underestimated. For the stars outside the range of validity of the relation above, we use the $M_G$-$\mathcal{M}$ relation by Pecaut & Mamajek (2013). We assume a minimum of 15% of uncertainty from these tabulated values. For objects cooler than L2, we do not provide masses. For white dwarfs, we assign half a solar mass.

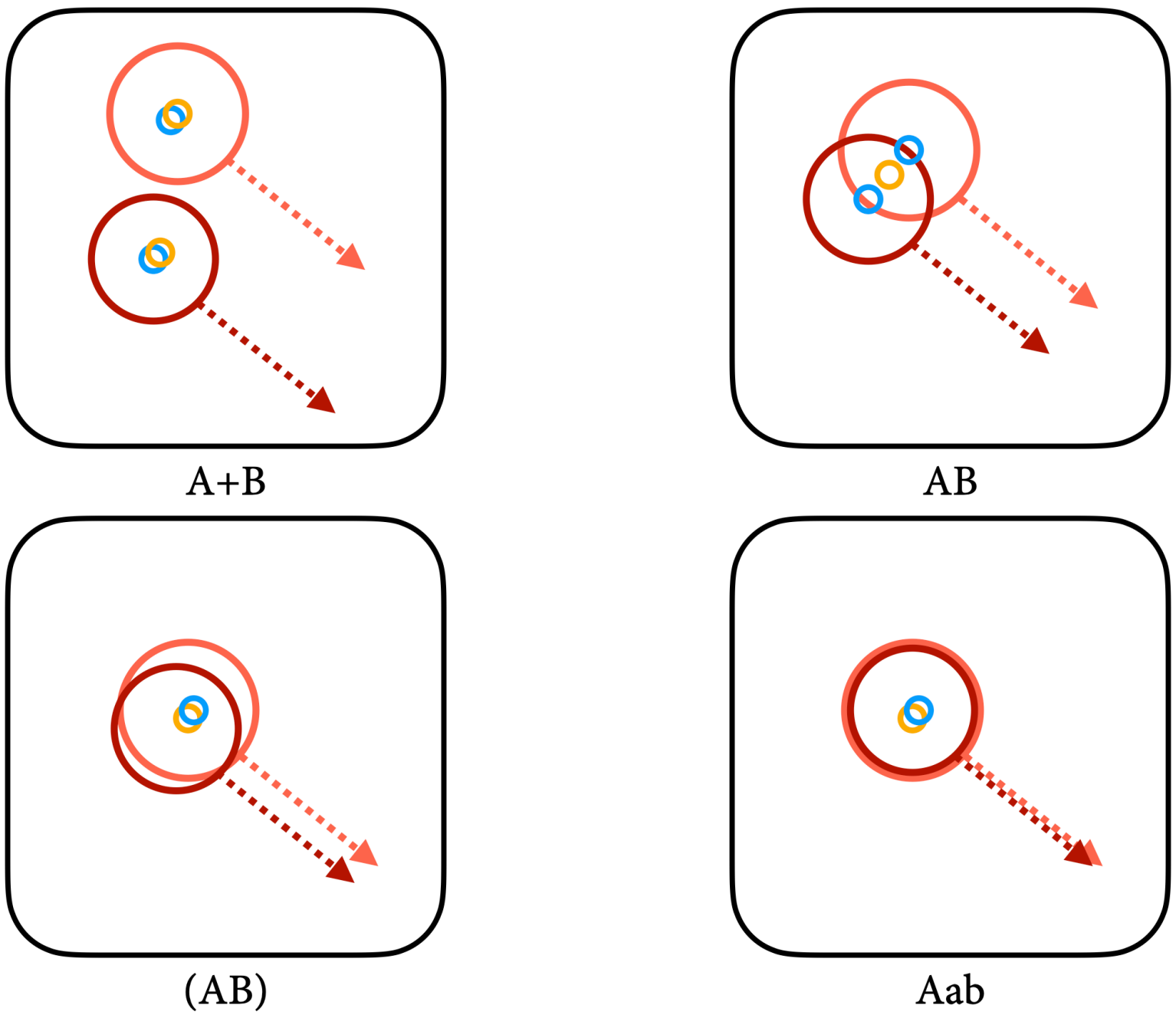

Figure 4.20: Nomenclature of multiple systems based on the resolution in *Gaia* DR3 and 2MASS, represented as blue and orange circles, respectively. *From left to right and top to bottom*: A+B (resolved in both surveys), AB (only resolved in *Gaia*), (AB) (not resolved either in *Gaia* or 2MASS, but resolved by adaptive optics, lucky imaging, or space imaging), and Aab (spectroscopic binaries, only resolved in spectral analysis). Systems may be a combination of the cases above.

### 4.4.4 Description of the systems

**About the main table of this Chapter** The structure of the table that contains all the sample plus the discovered components of multiple systems found (Table D.2, and see Table D.1 for the column-by-column description) aims to be both human- and machine-readable. For the former, the systems are displayed with each component of the system resolved by *Gaia* using a single row. The stars are sorted by right ascension, but making sure that stars that belong to the same system are consecutive, in order of decreasing brightness. The nomenclature of the system is described as in Figure 4.20. The most recent version of the complete table can be found in the GitHub repository: `https://github.com/ccifuentesr/`.

#### Projected physical separations

The cumulative number of binaries as a function of their orbital separation can be parametrised by power laws. Öpik (1924) observed that the distribution of the semimajor axis in binary systems, $f(a)$, follows the form $f(a) \propto a^{-1}$, which in terms of the cumulative distribution is $N(a) \propto \log a$. It can be generalised as $f(a) \propto a^{-\lambda}$, or $N(a) \propto a^{-\lambda+1}$, where $\lambda = 1$ recovers the original Öpik's law. This parametrisation states the intuitive idea that as we go to wider separations, the probability of finding a bound companion strongly diminishes. These wider components are more scarce because their fragile bindings are easily affected by the Galactic gravitational potential. Early investigations, such as that of van Albada (1968), suggested that binaries observed at separations greater than about 1000 au must be either "escaping" double stars (or run-away stars — see also Poveda et al. 1967), or one of the components is actually a double star, and therefore a triple system. We analyse several individual instances of this latter case using the precise astrometry of *Gaia* in Sect. 4.4.6.

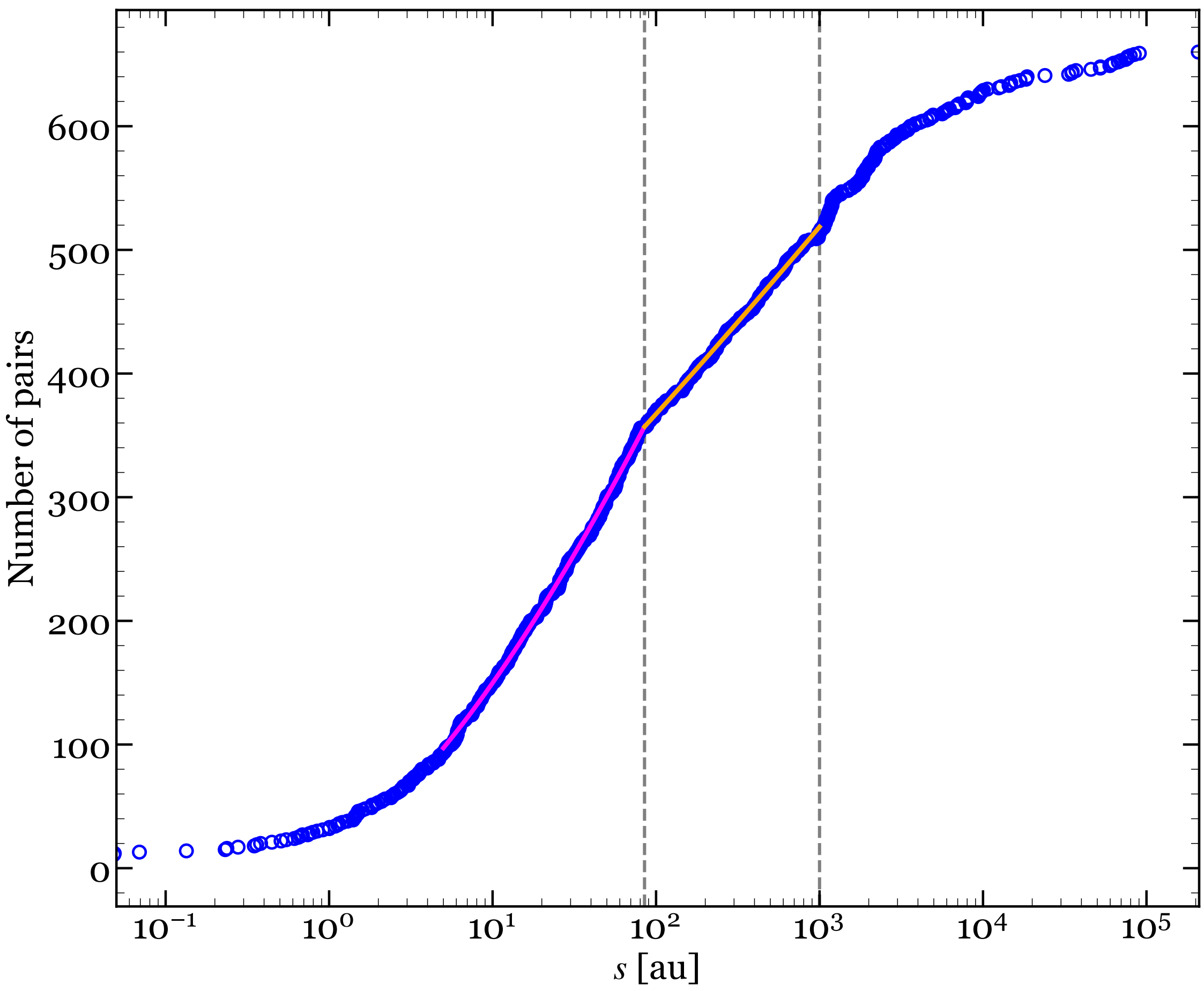

Figure 4.21: Cumulative number of pairs as a function of the physical separation, $s$. The grey dashed lines mark the values for $s = 85$ and $10^3$ au. The magenta and orange lines are power-law fits for for the ranges (5, 85) and (85, 1000) au, respectively.

Because of the small difference that exists between semi-major axis, $a$, and projected physical separation, $s$ ($E(\log a) - E(\log s) = 1.46$, Couteau, 1960) it is more practical to write this expression in terms of this observed separation, $F(\log s) \propto \log s$, which can be derived from the projected angular separation, $\rho$, by using the small angle approximation ($\tan\rho \sim \rho$), as $s = \rho d$, given in au. Öpik's law can be satisfactorily applied to limited ranges of separation, both in pre-main sequence and main sequence stars (Poveda et al., 1982; Poveda, 1988; Poveda et al., 1994; Poveda & Allen, 2004). However, it has been indeed argued (e.g. Allen et al., 1997) that it might be preferable to describe the distribution function depending on the separation, in accordance with the two main mechanisms of binary formation: disk fragmentation and first-collapse fragmentation (see again Sect. 4.1), for which the authors propose limits of $s < 25$ au and $s > 25$ au, respectively. There also exists an interesting age-dependence of this empirical law in wide binaries, in the sense that the upper limit decreases with increasing age (see again Poveda & Allen, 2004).

In Fig. 4.21 we present the cumulative number of pairs as a function of the physical separation, $s$. This distribution flattens at very close ($s \lesssim 1$ au) and very wide ($s \gtrsim 10^5$ au) orbital separations. The former is

Table 4.8: Coefficients of the power-law fit for the cumulative distribution of separations, $N(\log s)$.

| Range [au] | $a$ | $b$ | $\lambda$ | $R^2$ |
|---|---|---|---|---|
| [5, 85] | 23.11 ± 2.11 | 46.428 ± 1.214 | −0.3782 ± 0.0142 | 0.9992 |
| (85, $10^4$] | 126.9 ± 16.4 | 44.58 ± 6.54 | −0.1786 ± 0.0569 | 0.9982 |

**Note:** The polynomial fits follow the form $N(\log s) = a + b \log s^{\lambda-1}$.

caused by an observational bias regarding physical limitations of the data acquisition of the main high-resolution techniques (speckle, lucky imaging, adaptive optics, and also imaging). In Sect. 4.1 we already noted how investigations suggest that in binaries with semi-major axis $a \lesssim 50$ au the effect of dynamical influence is negligible towards their disruption, and they evolve essentially unaffected. In the era of *Gaia*, the flattens at large values of $s$ occurs mainly due to the difficulty of their formation and eventual survival, rather than observational limitations.

Next, we fit the cumulative distribution between these flat regions to two power laws. This is motivated by the necessity of discriminating between at least two scenarios corresponding to different formation mechanisms, and also motivated by the changes of slope at $s \sim 85$ au and $s \sim 10^3$ au that are visually noticeable. This observational evidence also alleviates the problems that a single theory faces when explaining the existence of binaries with very wide separations. Modelling the binary formation with simple prescriptions (Tokovinin & Moe, 2020) and performing a Bayesian approach on *Gaia* data (Hwang et al., 2022), the authors found that the distribution of eccentricities is approximately uniform at $\sim 10^2$ au, becomes thermal at values larger than $10^{2.5}$–$10^3$ au, and superthermal[9] at $>10^3$ au. Small mass ratios ($q << 1$) might indicate that stars are formed as binary systems either by the dynamical interaction in unstable molecular clouds or by the turbulent fragmentation of molecular cloud cores. We fit the number of pairs as a function of the separation, $N(\log s)$, in the ranges $s \in [10, 85]$ au, and $s \in [85, 1000]$ au, to a more general form of Öpik's law:

$$N(\log s) = a + b \log s^{-\lambda+1}, \tag{4.10}$$

where $a$, $b$, and $\lambda$ are free parameters. We tabulate the coefficients obtained for these two fits in Table 4.8, including the coefficient of determination, $R^2$.

In Fig. 4.22 we provide an additional view of the cumulative distributions, discriminating between three consecutive mass ranges: $\mathcal{M}/\mathcal{M}_\odot \leq 0.25$, $0.25 < \mathcal{M}/\mathcal{M}_\odot < 0.65$, and $\mathcal{M}/\mathcal{M}_\odot \geq 0.65$, which approximately correspond to M dwarfs later than M4.5, between M0.0 and M4.5, and earlier than M0.0, respectively. For visualisation purposes, we add grey dashed vertical lines as an early attempt to translate to the three ranges of mass the change in slope that is visible for the complete range of masses. These delimitations are qualitative in nature, but it can serve as a foundation for future investigations. The reason for this is the significance of the observed log-normal shape in the separation distribution and the continuing discussion about its possibility of characterising distinct populations. For instance, measuring the separation distributions of seven young star-forming regions, King et al. (2012a) noted that the separation distributions (and the multiplicity fractions) in them are very different to the field, specifically there is an excess of close binaries (the authors refer to less than 100 au) compared to the field population. They also suggested that the separation distribution is not statistically different between the different regions, except in the 19-100 au separation range. All in all, the multifactorial nature of the separation distribution makes it a relevant characteristic that any theory of star formation should be able to account for.

[9]Thermal and superthermal are terminologies that refer to the energy level of particles within a system, usually in the field of plasma physics. The energy of thermal particles follows the Maxwell-Boltzmann distribution and has an average energy that depends on the temperature of the system. Conversely, superthermal particles are characterised by energy levels that are significantly greater than the average energy of thermal particles.

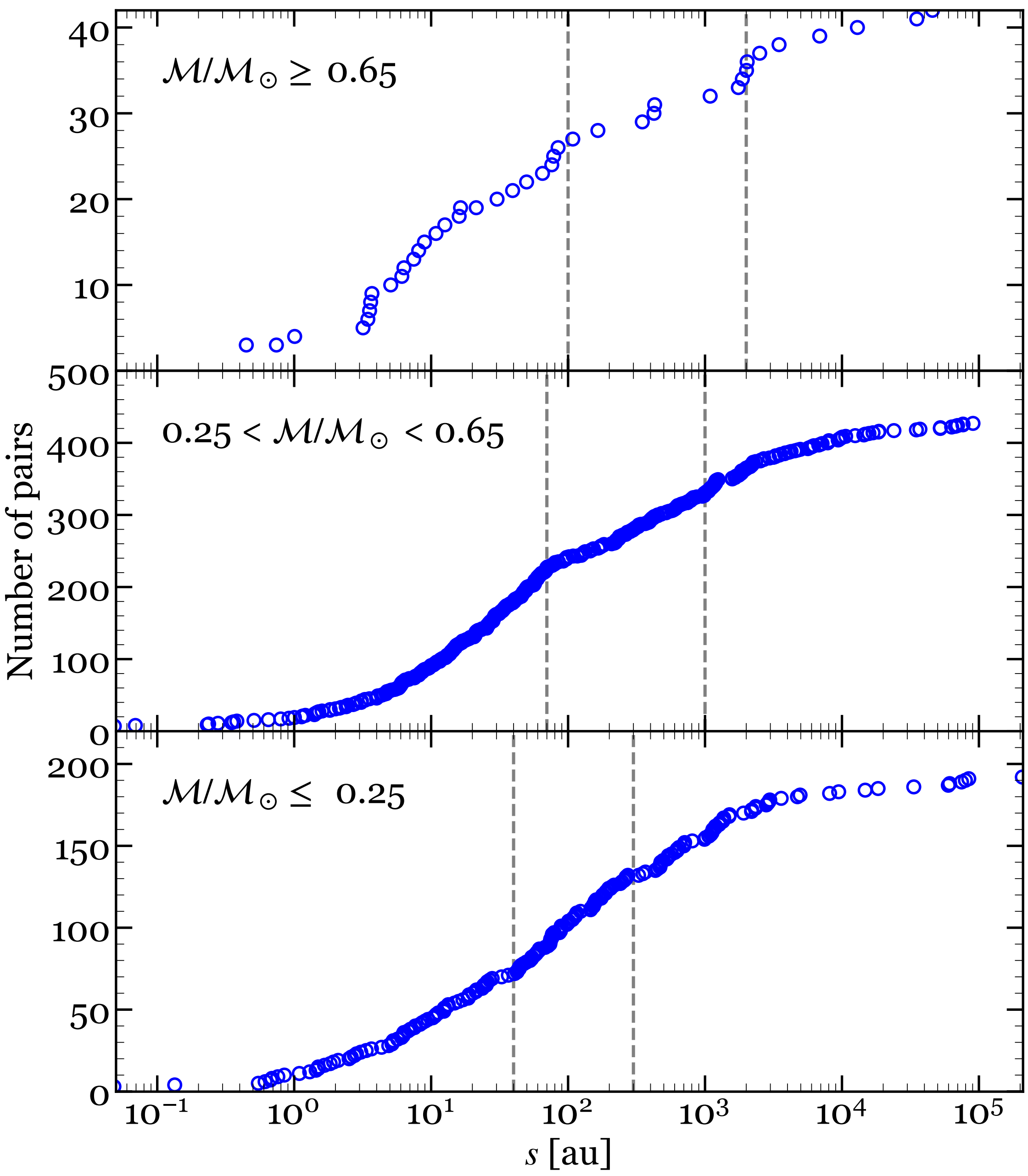

Figure 4.22: Cumulative number of pairs as a function of the physical separation, $s$. We show three equal-size non-overlapping ranges of primary mass: $\mathcal{M} \geq 0.65\,\mathcal{M}_\odot$ (*top*), $\mathcal{M} \in (0.25, 0.65)\,\mathcal{M}_\odot$ (*middle*), and $\mathcal{M} \leq 0.25\,\mathcal{M}_\odot$ (*bottom*). The grey dashed lines mark the values for $s \in [100, 2000]$ au (*top*), $s \in [70, 1000]$ au (*middle*), and $s \in [40, 300]$ au (*bottom*).

#### Orbital periods

The orbital period of a system constrains, to a certain extent, how much detail can be poured on its description. There are binary stars for which completing a single orbit around the common centre of mass takes millennia or even longer periods. In Fig. 4.23 (right panel) we show a distribution of the orbital periods calculated for all the pairs in our sample with available masses and orbital separations, for which we take special caution when defining the *total* mass of each system. In this sense, the definition of pair must be understood as introduced in Sect. 4.1: triple, quadruple, or even higher order systems can be treated as a binary or group of binaries in the majority of cases. This is, a close compact system can be adequately treated as a single star (with the aggregated masses) if considered from a distance much larger than the separation between its components. Masses of close binary systems need to be managed with caution, as dedicated estimation of the individual components is mandatory. The orbital periods, $P_{\rm orb}$, are calculated using the Kepler's Third Law:

$$P_{\rm orb} = 2\pi \sqrt{\frac{a^3}{G\mathcal{M}_T}}, \tag{4.11}$$

where $a \sim s = \rho d$ is the physical separation in au, and $\mathcal{M}_T$ is the total mass of the system. The majority of the orbital periods lie between 10 and $10^5$ years. Without the information about their orbital orientation, these periods could be over- as well as under-estimated. As mentioned earlier in this chapter, for these systems with orbital periods of hundreds or thousands of years, the prospect of following them during one single orbit is unfeasible, only wishing for *discrete* data available from decades to attempt a draft of their orbits (as an example, in Sect. 4.4.6 we showcase two attempts of orbital characterisations with a rather scarce amount of available data, in the most favourable case of spectroscopic binary, with periods of only a few months).

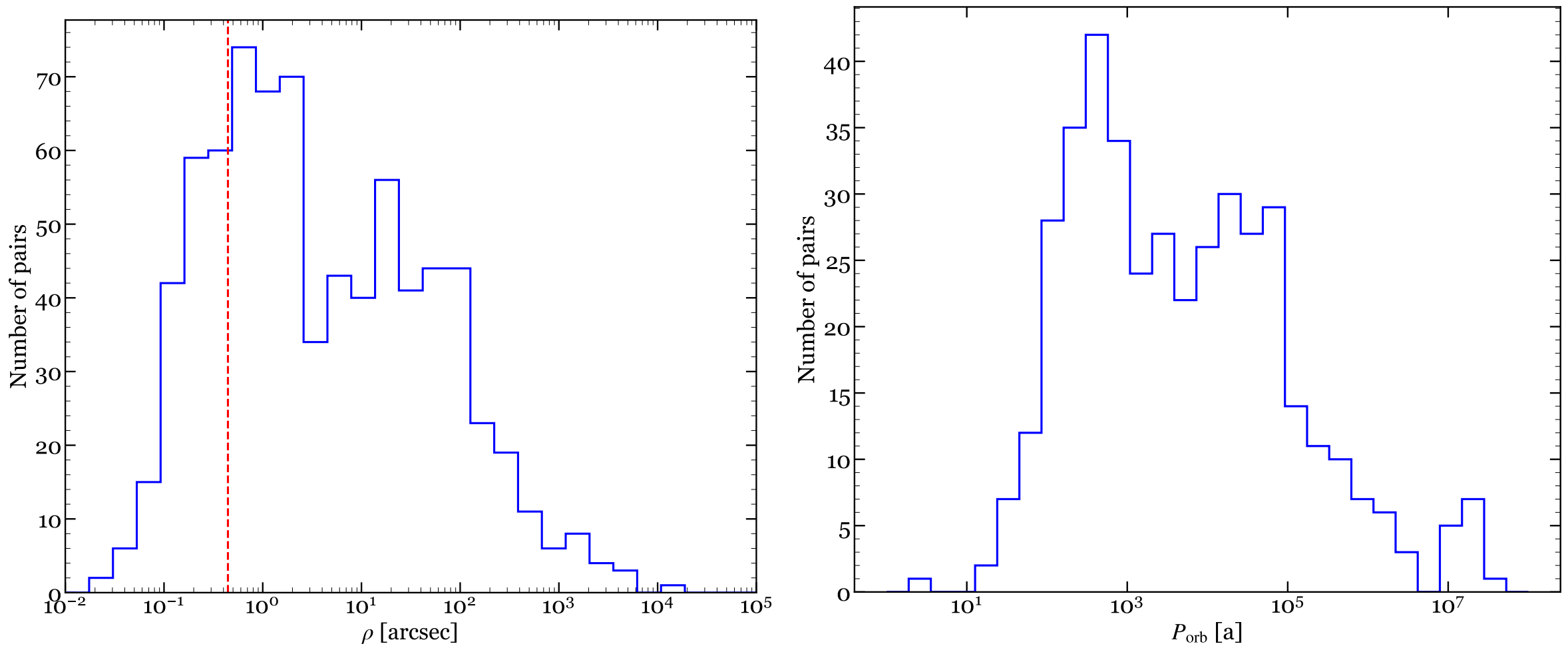

Figure 4.23: Distribution of projected separations in arcseconds (*left*) and orbital periods in years (*right*) for all the multiple systems in our sample. The red dashed vertical line represents the smallest separation that *Gaia* reports in a binary star in our sample.

### Binding energies

The gravitational potential binding energy of a binary system can be interpreted as the minimum energy required to separate the components to infinity, therefore it is a measure of the strength of their attachment. In a rather intuitive denomination, Heggie (1975) introduced the categories 'soft' and 'hard' binaries as a broad classification for the class of bound that the pairs experience. The former are fragile and easy to break, the latter are highly resilient to encounters. The general expression for the binding energy of a binary system of masses $\mathcal{M}_1$ and $\mathcal{M}_2$, physically separated $s$ (in au), can be put as:

$$-U_g^* = G\frac{\mathcal{M}_1\mathcal{M}_2}{s}. \tag{4.12}$$

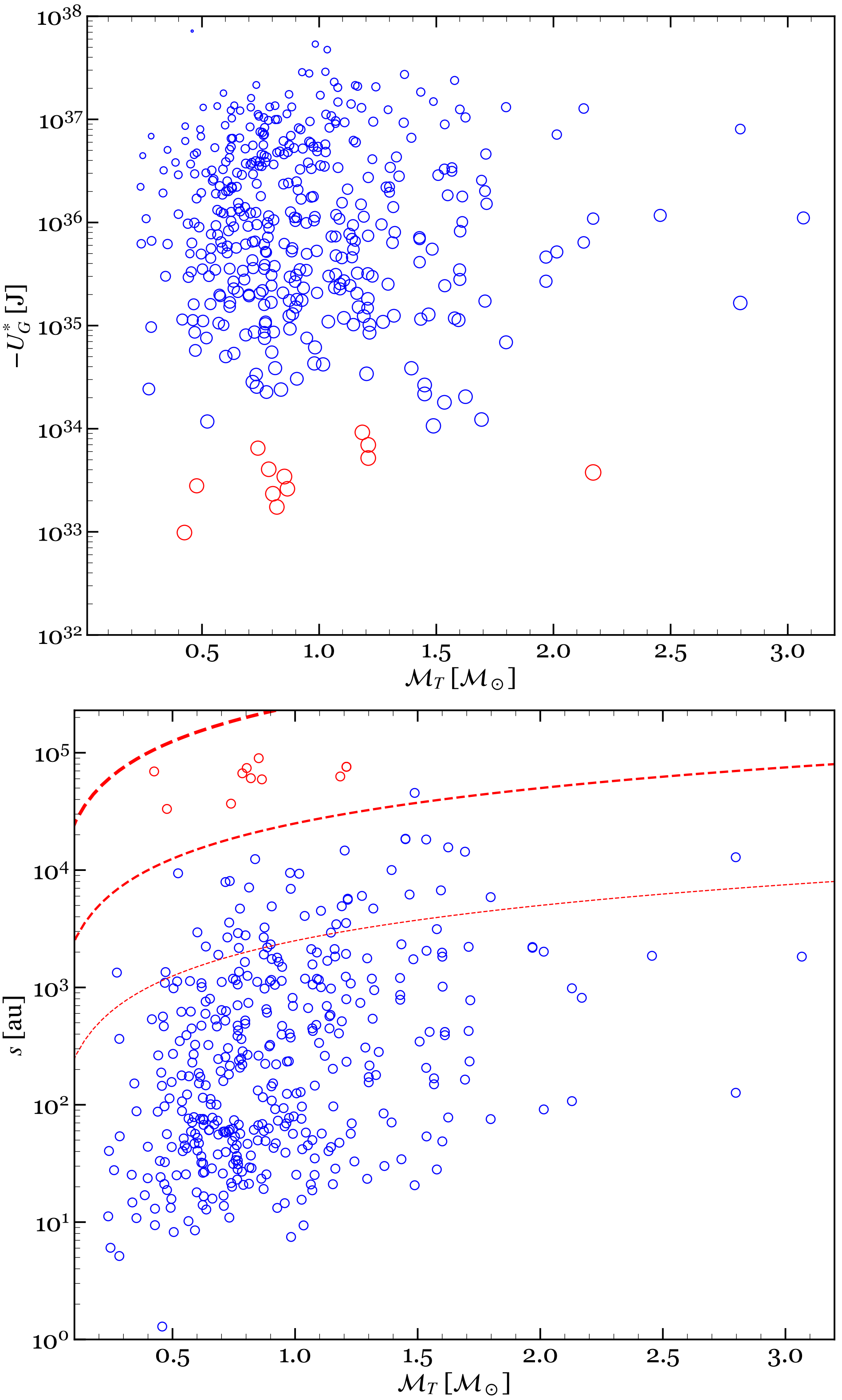

Figure 4.24: Binding energy (*top*) and physical separation (*bottom*) as a function of the total mass of the systems. In the top panel, sizes are proportional to the physical separation from the primary. Red empty circles highlight the most fragile pairs ($|U_g^*| \lesssim 10^{34}$ J). The red dashed lines represent the maximum separation, $s$, for expected survival in the case of 1, 10, and 100 Ga (in order of decreasing thickness).

Figure 4.24 (*top*) shows the binding energy with the approximation $r \sim s$ ($U_g^*$) as a function of the total mass $\mathcal{M}_{\rm total}$ of the system, with sizes proportional to the size of the orbit (i.e. the physical separation between the components). Although $U_g^*$ is defined by pairs, we compute the binding energy of triple or higher order of multiplicity systems as long as they dynamically behave like a binary, as in the case of the orbital periods. Figure 4.24 (*bottom*) shows the physical separation between components as a function of the total stellar mass. The most fragile systems (see Sect. 4.4.5) are highlighted in red in both panels.

The dynamical evolution of the stars over time sets an expiring date for the binding of multiple systems, especially in the case of wide pairs. While catastrophic encounters such as collisions are not common, there are a myriad of more subtle chances to disrupt their stability. The gravitational interaction with nearby clouds or stars can be of small intensity, but it is disruptive in the long term. Weinberg et al. (1987) were interested in the fate of wide binaries in the solar neighbourhood caused by these phenomena. Based on the Fokker-Planck coefficients[10], they estimate the average lifetime of a binary (Eqn. 28 of their work) as:

$$t_*(a_0) = 1.8\times10^4\,\mathrm{Ma}\left(\frac{n_*}{0.05\,\mathrm{pc}^{-3}}\right)^{-1}\left(\frac{\mathcal{M}_{\rm tot}}{\mathcal{M}_\odot}\right)\left(\frac{\mathcal{M}_*}{\mathcal{M}_\odot}\right)^{-2}\times\left(\frac{\langle 1/V_{\rm rel}\rangle^{-1}}{20\,\mathrm{km\,s^{-1}}}\right)\left(\frac{a_0}{0.1\,\mathrm{pc}}\right)^{-1}\ln^{-1}\Lambda, \tag{4.13}$$

where $n_*$ and $\mathcal{M}_*$ account for number density and mass of the perturbers, $V_{\rm rel}$ is the relative velocity between the binary and the perturber, $\mathcal{M}$ and $a$ refer to the total mass and semi-major axis of the binary, and $\Lambda$ is the Coulomb logarithm. The expression accounts for the stochastic gravitational perturbations and the encounters with passing stars, and can be greatly simplified (Close et al., 2007; Dhital et al., 2010) to the form:

$$s \simeq 1.212\frac{\mathcal{M}_{\rm tot}}{t_\star}. \tag{4.14}$$

Here, $s$ represents the maximum separation in astronomical units, for a total mass $\mathcal{M}_{\rm tot}$ in solar units, to survive for a given age $t_\star$ in Gigayears. These separations are represented for the ages of 1, 10, and 100 Ga as red dashed lines in Fig. 4.24. There are 10 pairs with expected survival periods less than 10 Gigayears, while the vast majority may well be stable for hundreds or thousands of Gigayears[11], assuming a negligible occurrence of catastrophic encounters.

### 4.4.5 Companions to M dwarfs

#### FGK primaries

Modeling the atmospheres of a star is a challenging task, and the coolest the star, the harder it becomes (see e.g. Valenti et al., 1998; Woolf & Wallerstein, 2006; Önehag et al., 2012). Early attempts such as ATLAS (Kurucz, 1970, 1979) or MARCS (Gustafsson et al., 1975, and Gustafsson et al. 2008), are examples of the laudable efforts on this matter despite of the technological limitations, and from which many more spectral libraries would derive. PHOENIX (Hauschildt et al., 1997, and the more recent Husser et al. 2013), NEXTGEN (Hauschildt et al., 1999a,b), or BT-Settl, are just a few notable examples. In order to reproduce the spectral features successfully, models need to take into account a wider range of atomic and molecular interactions and opacities, and introducing this complexity also comes with a greater computational cost. Improvements in high-resolution spectroscopy, high-angular-resolution interferometry, and processing power have made a major difference in this important topic of research. Gaining precision in the determination of stellar abundances translates into better constrain mantle composition and relative core size of the orbiting planets (Dorn et al., 2017).

[10] In a few words: The Fokker-Planck coefficients describe the rate of diffusion and drift of particles in a stochastic process.
[11] As a reminder, the age of the Universe is around 13.77 Ga (Planck Collaboration et al., 2020).

In this effort of calibrating the metallicity of the coolest stars, a Solar-type physical companion can pave a smoother passage, because their abundances are much easier to determine. Numerous works have estimated M dwarf metallicities using wide binary pairs (e.g. Rojas-Ayala et al., 2012; Montes et al., 2018; Birky et al., 2020; Ishikawa et al., 2022b). Among them, Montes et al. (2018) presented an relatively large sample of 192 wide visual binaries covering a reasonably large range in metallicity and spectral type, with atmospheric parameters of the primaries homogeneously derived using the STEPAR (Tabernero et al., 2019), and chemical abundances for 13 atomic species.

Among the M dwarfs in our sample that belong to multiple systems, we find that 54 have a Solar-type star as a primary. These are late-F to mid-K main sequence stars, including objects from F7 V to K5 V, and excluding the frontier K6/K7-M0 V. We tabulate the systems and their main astrometric properties in Table D.6. M dwarfs accompanied by a hotter, Sun-like star, are precious targets for chemical abundance studies, a promising field of investigation in the era of exoplanet discovery (see Sect. 1.3).

### White dwarfs

White dwarfs (WDs) are the final stage for almost all the stars in the Milky Way (Fontaine et al., 2001). The can be considered the fossils of the stellar evolution, and similarly to the preserved remains in paleontology, white dwarfs can be used as clocks that trace back into the past history of the Universe. Several works have made use of white dwarf companions as effective chronometers for M dwarfs (Monteiro et al., 2006; Fouesneau et al., 2019; Qiu et al., 2021; Kiman et al., 2021). When physically paired with M dwarfs, WDs are a valuable source of information about their low mass companions, because, contrary to these, their modelling turns out much easier (Bergeron et al., 1995; Renedo et al., 2010). For instance, these systems represent opportunities for age estimation (Fouesneau et al., 2019), because the age estimation for WDs is based on their cooling, which is well-understood physics (see Soderblom, 2010). Other important applications include the study of mass-loss rate and the characterisation of main sequence companions in multiple systems (Pyrzas et al., 2009, 2012; Parsons et al., 2012). Regarding the observed deficit of white dwarfs compared with predictions, Williams (2004) suggested that a large portion of the missing white dwarfs might be explained if these are part of unresolved systems (e.g. Morales-Rueda et al., 2005; Toonen et al., 2017).

We list in Table D.7 the 26 systems in our sample that have at least one white dwarf. Among them, 20 are binaries, and 6 are triples [12] One of the binaries contains a known eclipsing binary (CM Dra, M4.5 V). Based on their position on the CMD, we propose the object *Gaia* DR3 2005884249925303168 (physical companion of the LF 4 +54 152, M0.0 V) as a candidate to white dwarfs, in agreement with Jiménez-Esteban et al. (2018, 2023) and Gentile Fusillo et al. (2019).

[12] Clark et al. (2022) showcases an example of the estimation for the age for one of these systems, Wolf 672 A.

### Young systems

Stellar associations or moving groups are loose star clusters, containing from dozens to hundreds of stars with a common origin in space (comoving) and time (coeval), and unbound by gravity. Although sparsely located, they maintain the imprint of the mother cloud: similar motion in space, and also similar chemistry (e.g. Tabernero et al., 2017, but see Tabernero et al. (2012) about the chemical tagging of the Hyades supercluster members). Generally, 'core' stars are less dispersed members, more tied to their origin and indisputably part of a given moving group; 'stream' stars are more spatially spread members, in some cases ambiguously assigned to the group. For instance, the closest association is the Ursa Major Moving Group, which includes many of the visible stars in the constellation. Sirius was once thought to be a part of it because it moves like its members, but it turned out out be much younger (King et al., 2003). Ursa Major and the Hyades were the only moving groups known by the end of the 20th century. Dozens of co-moving associations of stars are catalogued in the solar neighbourhood. Among these associations, the YMG are of particular interest.

In regard to tracing the past, stellar associations are one of the best tale-tellers. Stars in associations convey, arguably more than any other arrangement, insights into the genesis of vast populations of stars, which ultimately reveal the tale of the Galactic population. When investigating their probable origin, younger associations are preferred because they retain a less distorted imprint of their origin, contrary to the older ones, which have necessarily endured many more perturbing effects of gravity with the environment and within the group itself. The result is that fraction of higher-order multiples decreases rapidly with age.

In Sect. 3.2.2 we identified several members of YMG motivated by the overluminosity observed in them. In this case we look for stars in our sample that belong to known stellar kinematic groups, with particular interest in the close binaries, because they provide model-independent information like no other members can. An important clarification to include here is that stars in young moving groups are *not* classified as multiple systems per se.

We compile and collate those with either solid evidence or with incontrovertible assignation, for which two or more independent authors declare the same membership, and no further studies have debated the results. We look in the literature but also use `SteParKin` code to assign a population based on the Galactocentric velocities, $U$, $V$, $W$, the radial velocities, $V_r$, and the parallaxes, $\varpi$. In Fig. 4.25 we show the location in the sky of the main stellar kinematic groups found in the sample. We list in Table 4.9 the associations with members among the Carmencita stars, the abbreviations used, and the reference for these assignations. Among the stars found in the sample, a few cases are pairs discovered for the first time, which means that are proposed as candidate members of the corresponding associations.

### Stars with planets

Circumstellar environments of close binary stars seem to difficult or suppress the genesis of planets, both orbiting either separately, or altogether as circumbinary planets (Wang et al., 2014b,a; Kraus et al., 2016; Zagaria et al., 2022, and see Standing et al. 2023 for the first transiting circumbinary planet also measured with radial velocities). Nevertheless, planet formation does take place in binary environments (Holman & Wiegert, 1999; Mathieu et al., 2000; Naoz et al., 2012, and see Winn & Fabrycky 2015), conveniently triggering planet migration dynamics in many instances (Fabrycky & Tremaine, 2007; Petrovich, 2015, and see particular cases in Cochran et al. 1997; Wu & Murray 2003; O'Connor et al. 2021), via the Lidov-Kozai mechanism (Lidov, 1962; Kozai, 1962) .

Table 4.9: Stellar kinematic groups and associations with members found in our sample

| Name | Abbreviation | In sample[a] | Reference(s)[b] |
|---|---|---|---|
| AB Doradus | AB Dor | 25 | 9:12,20,22,23 |
| Argus | Arg | 2 | 11,16,20 |
| $\beta$ Pictoris | $\beta$ Pic | 40 | 6,9:16,20,22,23 |
| Carina | Car | 9 | 10,13,22,23 |
| Castor[c] | Cas | 30 | 1,6,10,12,13,27 |
| Columba | Col | 5 | 10,23 |
| Hercules-Lyrae | Her-Lyr | 2 | 4,10 |
| Hyades/Hyades SC | Hya/HS | 58/77 | 8,18,22:27 |
| IC 2391 | IC 2391 | 16 | 1,10,22,23 |
| Local association | LA | 98 | 10,23 |
| Pleiades | Ple | 2 | 10 |
| Taurus | Tau | 7 | 3,7,17,19,21 |
| Tucana-Horologium | Tuc-Hor | 5 | 10 |
| TW Hydrae | TWA | 2 | 16 |
| Ursa Major | UMa | 69 | 10,27 |
| Upper Scorpius | USco | 1 | 2 |
| Young disk | YD | 4 | 1,27 |

[a] In cases of multiple assignations for a given object, we choose a compromise between homogeneity and recent assignation.

[b] 1: Montes et al. (2001); 2: Preibisch et al. (2001); 3: Bertout & Genova (2006); 4: López-Santiago et al. (2006); 5: Caballero (2009); 6: Caballero (2010); 7: Rebull et al. (2010); 8: Röser et al. (2011); 9: Schlieder et al. (2012a); 10: Shkolnik et al. (2012); 11: Malo et al. (2013); 12: Malo et al. (2014d); 13: Elliott et al. (2014); 14: Kraus et al. (2014); 15: Alonso-Floriano et al. (2015a); 16: Gagné et al. (2015); 17: Gómez de Castro et al. (2015); 18: Kopytova et al. (2016); 19: Duchêne et al. (2017); 20: Janson et al. (2017); 21: Kraus et al. (2017); 22: Gagné et al. (2018b); 23: Gagné & Faherty (2018); 24: Gagné et al. (2018a); 25: Lodieu et al. (2019); 26: Freund et al. (2020); 27: This work.

[c] Castor is often referred as a 'dynamical stream' of non coeval stars that share kinematics (e.g. Famaey et al., 2005). Nevertheless, given the young ages of their potential members, it is considered as the other groups.

All-sky surveys such as *Kepler* or *TESS* gather data from stars in bulk portions on the sky, assuming that all the stars are single, regardless of a binarity not resolved (see Lillo-Box et al., 2012). Statistical conclusions may ignore these observational selection effects (e.g. Mulders et al., 2021). Failing to account for the possibility of stellar binarity essentially biases the planet characterisation of these searches to smaller radii (Ciardi et al., 2015). The mass of a planet derived from the mass of a star that is assumed single will be subject to a large error (see Bouma et al., 2018).

In our sample, including the individual components of the multiple systems, we find that 81 stars have confirmed exoplanets, 26 of which are detected in a multiple system (Table D.8). Of these, 12 are multiplanetary systems (2 or more planets), including the 5-planet system of 55 Cnc[13], a K0 subgiant in a binary system with components separated by ~1000 au. Although these are usually detected orbiting the primary component of wide systems, they can be also detected orbiting spectroscopic binaries (e.g. EQ Peg A, and the candidate GJ 373).

[13] Butler et al. (1997), Marcy et al. (2002), McArthur et al. (2004), and Fischer et al. (2008).

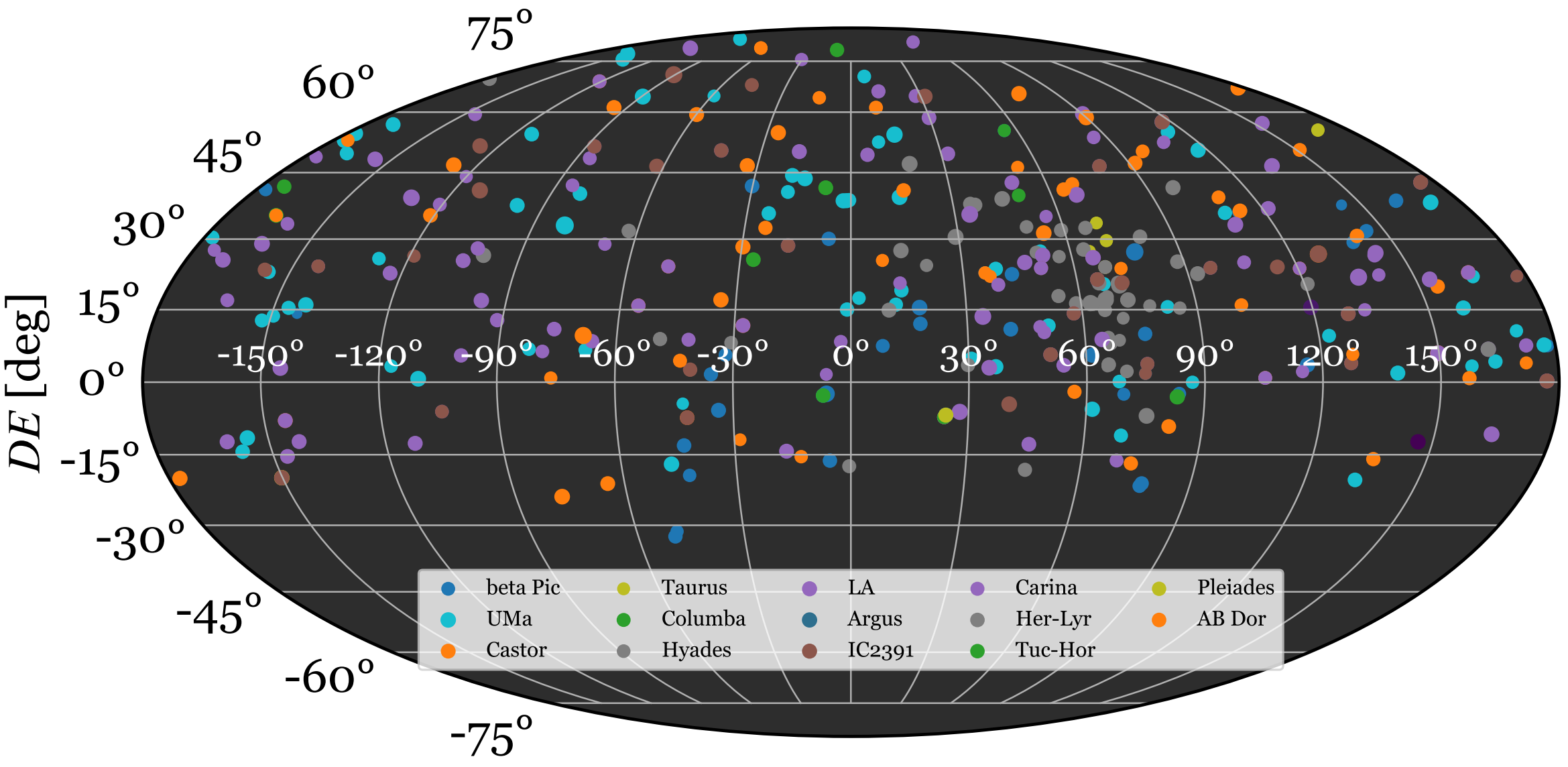

Figure 4.25: Location in the Mollweide-projection sky of the main stellar kinematic groups identified in the sample, in equatorial coordinates.

### Eclipsing binaries

The cases where the orbital plane of a binary system contains our line of sight are not common. These are referred as eclipsing binaries (EB) and are a particular case of variable stars, because the components periodically eclipse each other from our point of view. This causes a periodic decrease in the brightness that depends on the relative size of the components and their physical separation. Algol ($\beta$ Per) is the epitome of these systems, also being the first eclipsing binary discovered.

Eclipsing binaries are of particular interest because they are the source of empirical masses and radii[14] (Huang & Struve, 1956; Popper, 1980; Andersen, 1991). The mass of a star is perhaps its most fundamental characteristic, as the majority of its properties exhibit a strong dependence on its mass. Indeed, from binaries it is possible to accurately measure dynamical masses (in particular, in detached double-lined systems) and radii measurements (from eclipsing light curves), with accuracies of ~1–2% (Ribas, 2003a; Torres et al., 2010). Because these do not rely on models, these can be safely tested against each other, to evaluate the accuracy of model predictions. Many fields benefit from model-independent parameters, such as the non-trivial matter of determining ages (David et al., 2019). An important point to consider, specially with short-period, close-orbiting eclipsing binaries, is that the level of stellar activity can cause bias. It has been observed that inflated radii occur for young, magnetically active, or fast rotating stars (Caballero, 2010; Jackson et al., 2018; Kesseli et al., 2018; Parsons et al., 2018)[15].

When radial velocities can be measured for the components of an eclipsing binary, the size of the orbit can be calculated. The derivation of the mass and radius of Castor C serves as a good example (Torres et al., 2022). Interestingly, one of the components of the Castor system might be a flare star, which can be classified among peculiar variables. It is important to note that the properties derived for the components of these systems might not be of application to isolated stars, because of the effect that stars so close might have on one another. In our sample we identify a total of 8 eclipsing binaries, which are listed along with masses and radii in Table D.9.

[14] Stellar radii have been measured for a few hundreds of stars, via optical interferometry, lunar occultation, or eclipsing binaries, whereas masses are known empirically for an even smaller number.

[15] It is worth mentioning at this point that this issue has been adknowledged by Schweitzer et al. (2019) when deriving masses from the $\mathcal{M}$-$\mathcal{R}$ relation from eclipsing binaries (see Sect. 3.4.1), as the sample mostly contained old, inactive, and slowly rotating stars.

#### The widest systems

In our search for physical companions to our stars with separations up to $10^5$ au we find a number of components with $\rho > 10^4$ au. There are 32 stars of this kind in our sample. From these, 13 are known members of stellar kinematic groups and open clusters already discussed in Sect. 2.4. The remaining 19 are not recognised as members of kinematic groups or open clusters, and some of them are identified as pairs in this work for the first time. The binding energies for these systems range from $\sim$ 2 to 40 $10^{33}$ J. In Fig. 4.24, we highlighted in red these components, showing that the survival expectancy for these fragile configurations ranges from 1 to 10 Gigayears. They are collected in Table 4.10, along with the most relevant parameters.

### 4.4.6 One is the loneliest number

The content of this subsection has been adapted from the article *One Is the Loneliest Number: Multiplicity in Cool Dwarfs*, published in the *Research Notes of the AAS* (RNAAS) (Cifuentes et al. 2021, *RNAAS*, 5, 129).

As multiple systems settle down into stable configurations, they result in a variety of hierarchies and a wide range of separations between the components. We examine 11 known and 11 newly discovered multiple systems including at least one M dwarf with the latest astrometric data from *Gaia* third Data Release (DR3). We find several examples that the individual components of systems at very wide separations are often multiple systems themselves.

As introduced in Chapter 1, evidence suggests that most (if not all) stars form in multiple systems, and in systems of higher order, at least one component (usually the least massive) is ejected into a distant orbit during the pre-main sequence phase (Duchêne & Kraus, 2013). Low-mass objects are preferentially ejected in three-body dynamics, and Reipurth & Mikkola (2012a) noted that these expelled companions can also be binaries. Using N-body simulations, Reipurth & Mikkola (2012a) found that extreme hierarchical architectures can be reached on timescales of millions of years with the appropriate exchange of energy and momentum. In this regard, we also call the attention to early simulations such as those of Poveda et al. (1967). In their work, one component in a triple system is dynamically scattered into a very wide orbit at the expense of shrinking the orbit of the remaining binary. In many physical star systems, seemingly single distant components are resolved as multiple systems themselves, typically binaries, which implies that known double or triple systems in wide orbits would actually be hierarchical triples or quadruples in wide configurations (see Basri & Reiners, 2006; Caballero, 2007; González-Payo et al., 2021). These findings challenge the fundamentals of star formation because the separations of wide systems cannot be the direct product of a collapsing cloud core. In this section we examine the characteristics of known and new physically bound multiple systems containing at least one M dwarf by using the astrometry from *Gaia* DR3.

We study 53 stars in 22 multiple systems, 11 of which are known systems tabulated in the Washington Double Star catalogue (Mason et al., 2001, e.g. Caballero 2007; Caballero et al. 2012; Dhital et al. 2010; Janson et al. 2014b), 4 are known systems with new candidate members, and 11 are newly discovered systems. All of them contain at least one M dwarf in different hierarchical configurations, including 12 double, 5 triple and 4 quadruple systems, plus one quintuple system. They are located between 14.1 pc and 276.1 pc. *Gaia* information is complete for all stars except two, for which we retrieved proper motions

Table 4.10: Components of multiple systems at separations larger than $10^4$ au from the primary.

| Karmn | Name | Spectral type | Component[a] | WDS id. | WDS disc. | $s$ [au] | $\pi$ [mas] | $\mu_\alpha \cos\delta$ [mas a$^{-1}$] | $\mu_\delta$ [mas a$^{-1}$] | Note[b] |
|---|---|---|---|---|---|---|---|---|---|---|
| J00169+200 | GJ 3022 | M3.5 V | A | | | | 29.31 | 238.89 | 17.67 | |
| | G 131-47B | M3.5 V | B | 00169+2004 | CRC 43 | 37.0 | 29.14 | 231.63 | 28.37 | |
| | LP 404-54 | M5.0 V | C* | *New* | | 60680 | 28.87 | 232.74 | 22.23 | |
| J00341+253 | V493 And A | M0.0 V | A | | | | 20.10 | 82.97 | −97.36 | • |
| | V493 And B | K7 V | B | 00341+2524 | SKF 220 | 77.9 | 19.72 | 85.93 | −96.83 | |
| | UCAC4 578-001365 | M4.0 V | C* | *New* | | 15618 | 19.67 | 84.60 | −94.78 | • |
| J03454+729 | G 221-21 | M1.5 V | A | | | | 39.09 | 203.31 | −442.85 | • |
| | LP 31-210 | | B | 03454+7259 | LDS1581 | | ... | ... | ... | |
| | LP 31-200 | M3.5 V | C | 03454+7259 | WIS 99 | 14667 | 39.15 | 206.71 | −442.36 | |
| J04011+513 | Ross 25 | M3.8 V | A | | | | 39.82 | 365.81 | −804.17 | |
| | LSPM J0401+5131 | DC8 | B* | *New* | | 12402 | 39.84 | 367.43 | −803.19 | |
| | PM J05334+4809 | M0.0 V | A | | | | 30.27 | −54.57 | 38.76 | |
| J05341+475 | PM J05341+4732A | M2.5 V | B* | *New* | | 75945 | 30.05 | −58.35 | 36.85 | |
| | PM J05341+4732B | M3.0 V | C* | *New* | | 75899 | 30.04 | −47.45 | 39.68 | • |
| | UPM J0533+4809 | M3.5 V | D* | *New* | | 4219 | 30.23 | −52.55 | 38.28 | |
| | 1RXS J073138.4+455718 | M3.0 V | Aab | | | | 17.88 | −13.69 | −92.77 | • |
| J07310+460 | 1RXS J073101.9+460030 | M4.0 V | B* | *New* | | 23780 | 18.14 | −13.52 | −100.85 | |
| | G3 975312928903090560 | M4.5 V | C* | *New* | | 16739 | 18.39 | −12.17 | −100.00 | |
| J09579+118 | GJ 3576 | M4.0 V | A | | | | 39.57 | −422.75 | −158.52 | |
| | LP 489-1 | M5.0 V | B* | *New* | | 33201 | 40.84 | −456.75 | −13.67 | |
| | PM J13255+2738 | M1.0 V | A | | | | 22.00 | −0.56 | 71.21 | |
| J13260+275 | PM J13260+2735A | M3.0 V | B* | *New* | | 18494 | 21.87 | −2.47 | 72.85 | |
| | PM J13260+2735B | M2.5 V | C | 13260+2735 | KPP3896 | 18344 | 22.03 | 9.15 | 62.91 | |
| | HD 140232 | A8 V | A | | | | 18.72 | −61.65 | 54.32 | |
| | G3 1197801408884577408 | M3.5 V | B | 15419+1828 | DRS 17 | 127 | 18.64 | −61.76 | 54.37 | |
| J15416+184 | StKM 1-1264 | M1.5 V | C | 15419+1828 | TOK 302 | 12870 | 19.37 | −65.05 | 64.89 | |
| | $\sigma$ CrB A | F6 V | Aab | | | | 44.06 | −268.22 | −87.28 | • |
| | $\sigma$ CrB B | G1 V | B | 16147+3352 | STF2032 | 164 | 44.13 | −290.86 | −78.52 | |
| J16139+337 | $\sigma$ CrB C | M2.5 V | CD | 16147+3352 | STF2032 | 14346 | 44.27 | −286.97 | −89.82 | • |
| | HD 160269A | G0 IV/V | AB | 17350+6153 | BU 962 | 9.02 | 69.28 | 236.25 | −466.11 | • |
| J17355+616 | GJ 685 | M0.5 V | C | 17350+6153 | LDS2736 | 10561 | 69.89 | 261.92 | −514.50 | |
| | HD 230017A | M0.0 V | A | | | | 53.42 | 29.43 | 129.98 | |
| J18548+109 | HD 230017B | M3.5 V | B | 18550+1058 | VYS 8 | 70.7 | 53.64 | 29.42 | 84.30 | • |
| | PM J18542+1058 | M4.0 V | C* | *New* | | 10010 | 53.84 | 20.21 | 112.82 | • |
| J22018+164 | Ross 265 | M2.5 V | AB | 22018+1628 | YSC 165 | 5.75 | 61.79 | 391.82 | 156.08 | • |
| | Ross 268 | M3.5 V | C* | *New* | | 89858 | 56.76 | 361.91 | 84.11 | |
| J22058−119 | Wolf 1548 | M0.0 V | A | | | | 38.62 | −276.64 | −159.37 | • |
| | LP 759-25 | M6.0 V | B | 22059-1155 | WNO 57 | 59343 | 51.07 | −270.62 | −175.40 | |
| J23175+063 | GJ 4329 | M3.0 V | A | | | | 48.92 | 170.44 | −249.49 | |
| | GJ 4319 | M3.5 V | B* | *New* | | 36823 | 48.98 | 173.64 | −250.62 | |
| | V368 Cep | G9 V | A | | | | 52.78 | 203.45 | 72.31 | |
| J23194+790 | HD 220140B | M3.5 V | B | 23194+7900 | LDS2035 | 206 | 52.84 | 209.25 | 61.43 | • |
| J23228+787 | LP 12-90 | M5.0 V | C | 23194+7900 | MKR 1 | 18222 | 52.83 | 207.45 | 65.17 | |
| J23317−027 | AF Psc | M4.5 V | A | | | | 28.63 | 93.96 | −72.12 | • |
| J23301−026 | 2M J23301129-0237227 | M6.0 V | B | 23317-0245 | CAB24 | 66834 | 21.98 | 101.33 | −67.96 | |
| J23308+157 | LP 462-51 | M1.0 V | AB | 23309+1547 | LDS5096 | 45.5 | 44.00 | −164.96 | −109.58 | |
| | HD 221503 | K6 V | A | | | | 68.74 | 341.15 | −219.11 | |
| J23302−203 | GJ 1284 | M2.0 V | Bab | 22577-2937 | SHY110 | 206051 | 62.87 | 314.43 | −204.78 | • |
| J23327−167 | GJ 897 | M3.0 V | C | 23328-1651 | LDS 816 | 21895 | 64.8 | 352.0 | −216.1 | • |
| | G3 2395220664463236992 | | D | 23328-1645 | VOU 28 | | ... | ... | ... | |

[a] An asterisk (∗) indicates that the component is assigned for the first time in this work.

[b] J00341+253: Member of AB Doradus (Janson et al., 2017). C is a member candidate of AB Doradus (this work).

J03454+729: Component B (G 221-21) appears an 'Inexistent' in Simbad (Not an Object – Error, Artefact, ...).

J05341+475: Absent in WDS but see the detection at 2.33 arcsec by Ansdell et al. (2015).

J07310+460: A is SB2? (Fouqué et al., 2018). B is member of the Hyades supercluster (this work).

J16139+337: A is SB2 (Strassmeier & Rice, 2003; Raghavan et al., 2009). YSC 152 Ea Eb refers to the close pair CD (0.507 arcsec).

J17355+616: SB1 according to Duquennoy & Mayor (1991) but resolved in WDS (0.625 arcsec in 2013, BU 962 AB).

J18548+109: Member of Carina (Gagné & Faherty, 2018). C is a member candidate of Carina (this work).

J22018+164: Proper motions from van Leeuwen (2007).

J22058-119: Member of Castor (Caballero, 2010).

J23194+790: Member of Columba (Gagné & Faherty, 2018).

J23317-027: Member of $\beta$ Pic (Gagné et al., 2018b).

J23302-203: SB2 (Gizis et al., 2002; Jeffers et al., 2018; Cardona Guillén et al., 2021) and member of the Local Association (Cardona Guillén et al., 2021).

J23327-167: Parallax and proper motions from Tokovinin (2018).

and parallactic distances from the literature (Lépine & Shara, 2005; Dittmann et al., 2014). The spectral classifications of the primary components range from F7 V to M5.5 V, and of their physical companions from F9 V to L1. We computed the astrophysical properties as described in Chapter 3, and estimated spectral types for 22 stars from luminosities and absolute magnitudes.

We tested the physical binding of each system member candidate by applying the same criteria as in Sect. 4.3.1. Of the 22 investigated systems, 7 do not meet the conditions for actual parity. Of them, 3 are close pairs with *Gaia* proper motions perturbed by their relative orbital motion. The remaining 4 pairs are wide: one does not have parallax measured by *Gaia* for the secondary (WDS 06104+2234), one exhibits a high `ruwe` in *Gaia* (i.e. larger than 1.4, which is indicative of a problematic astrometric solution; KO 4), and only for the other two (KO 6 and SLW 1299) we disprove their physical connection based on the astrometric analysis.

We summarise the conclusions as follows:

- In 6 multiple systems the astrometric data from *Gaia* suggest additional multiplicity in at least one member.
- These are KO 1, KO 2, KO 4, 1RXS J073138.4+455718, HD 61606 A, 1RXS J074948.5–031712.
- With current data, the pairs LSPM J0651+1845/LSPM J0651+1843 and HD 77825/1RXSJ090406.8-155512 are the only wide binaries (112 and 220 arcsec, respectively) with no evidence of close binarity of the individual components in the *Gaia* solution.
- The young candidate member in $\beta$ Pictoris, PYC J07311+4556, is a wide member in a quadruple, perhaps quintuple, physical system. This current configuration is expected if the system is actually young ($18.5^{+2.0}_{+2.4}$ Ma, Miret-Roig et al., 2020b) and still undergoes a process of dynamical stabilisation.
- The least-bound system in this analysis is FMR 83 with $U_g^* \sim -7.1\ 10^{33}$ J (Rica & Caballero, 2012), while the pair with largest projected separation (54 800 au) and orbital period ($10.9\ 10^6$ a) is WDS 07400-0336 (Montes et al., 2018).

In Table 4.11 we list the names of the 53 stars, spectral types, angular separations ($\rho$), position angles ($\theta$), projected physical separations ($s$), and remarks[16]. Finally, dedicated charts with a detailed description of each system can be found in Appendix D.

[16]The full version of this table with all astro-photometric parameters and derived masses, binding energies, orbital periods, and reference codes can be downloaded in csv format from the GitHub repository (`https://github.com/ccifuentesr/cif21-multiplicity`). It has also been published in its entirety in the electronic edition of *Research Notes* (`https://iopscience.iop.org/article/10.3847/2515-5172/ac05ce`).

Table 4.11: Relative astrometry of the multiple systems investigated by Cifuentes et al. (2021).

| Star name | Comp. | Spectral type[a] | $\rho$ [arcsec] | $\theta$ [deg] | $s$ [au] | Remarks |
|---|---|---|---|---|---|---|
| LEHPM 494 | A | m5.5 V | | | | Confirmed known binary (KO 1) |
| 2M J00210589-4244433 | B | L0.6: V | 77.78 | 317.0 | 2083.4 | B might be double |
| NLTT 6496 | A | M4.5 V | | | | Confirmed known binary (KO 4) |
| NLTT 6491 | B | m4.5 V | 299.13 | 190.6 | 9391.9 | B might be double |
| LP 655-23 | A | M4.0 V | | | | Confirmed known binary (KO 2) |
| DENIS J043051.5-084900 | B | M8 V | 19.81 | 339.8 | 597.3 | A and B might be double |
| 2M J06101775+2234199 | A | M4.0 V+ | | | | *New* triple. JNN 269 is visual |
| LP 362-121 | BaBb | M6 V+m7 V | 65.16 | 89.2 | 1866.9 | |
| BD+37 1541 | A | f0: V | | | | *New* F+M binary |
| Karmn J06353865+3751139 B | B | m2.5 V | 3.88 | 201.5 | 848.7 | |
| LSPM J0651+1845 | A | m4.5 V | | | | Confirmed known binary (FMR 83) |
| LSPM J0651+1843 | B | m4.5 V | 111.72 | 150.6 | 7133.4 | |
| 1RXS J073138.4+455718 | AaAb | M3 V+m4.5 V | | | | *New* quadruple |
| 1RXS J073101.9+460030 | B | M4.0 V | 431.39 | 296.0 | 24127.1 | B might be double, probably young ($\beta$ Pictoris) |
| [SLS2012] PYC J07311+4556 | C | m4 V | 307.80 | 266.3 | 17214.6 | |
| HD 61606 A | A | K3 V | | | | Confirmed known triple |
| HD 61606 B | B | K7 V | 57.90 | 112.7 | 815.2 | No WDS entry but reported by Poveda et al. (2009) |
| BD-02 2198 | C | M1.0 V | 3894.18 | 296.7 | 54822.6 | C might be double |
| 1RXS J074948.5-031712 | A | M3.5 V | | | | *New* triple |
| 2M J07495087-0317194 | B | m3.5 V | 1.93 | 266.3 | 32.8 | |
| 2M J07494215-0320338 | C | M3.5 V | 234.86 | 214.0 | 4002.1 | C might be double |
| LP 209-28 | "A" | m3: V | | | | Disproved binary (KO 6) |
| LP 209-27 | "B" | m4 V | 666.68 | 208.5 | 69836.7 | |
| HD 77825 | A | K2 V | | | | *New* K+M binary |
| 1RXS J090406.8-155512 | B | M2.5 V | 220.02 | 262.9 | 6026.0 | |
| 2M J13181352+7322073 | A | m3 V | | | | *New* M+M binary |
| G2 1688578285187648128 | B | M3.5 V | 7.39 | 335.7 | 186.8 | |
| HD 130666 | A | G5 V | | | | *New* G+M binary |
| 2M J14474531+4934020 | B | m4.5 V | 29.54 | 336.6 | 3072.3 | |
| TYC 2565-684-1 | A | g1 V | | | | *New* G+M binary |
| 2M J15080798+3310222 | B | m3 V | 43.64 | 306.2 | 8644.3 | |
| HD 134494 | A | K0 IV | | | | Reclassified as sub-giant |
| BD+33 2548B | B | f9 V | 23.38 | 285.0 | 6455.7 | *New* m3-type companion to pair of evolved |
| G2 1288848427727490048 | C | m3 V | 5.85 | 180.3 | 1615.4 | solar-mass stars |
| HD 149162 | AaAbAc | K0 Ve+k6 V+m5 V | | | | Confirmed quintuple (LEP 79) |
| G 17-23 | B | M3.0 V | 252.03 | 138.4 | 11405.7 | |
| LSPM J1633+0311S | C | D: | 258.42 | 138.4 | 11694.8 | |
| G 125-15 | Aab | M4.5 V+M5 V | | | | Confirmed triple (GIC 158) |
| G 125-14 | B | M4.5 V | 45.78 | 347.4 | 1835.1 | |
| LP 395-8 A | Aab | M3.0 V+m0 V | | | | |
| LP 395-8 B | B | m3.5 V | 1.92 | 355.5 | 56.6 | *New* m9-type companion to trio of M dwarfs |
| G2 1829571684884360832 | C | m9: V | 11.02 | 307.4 | 325.1 | |
| HD 212168 | A | G0 V | | | | Confirmed known quadruple (DUN 38, KO 5) |
| CPD-75 1748B | BaBb | k3 V+ | 20.90 | 79.3 | 489.2 | |
| DENIS J222644.3-750342 | C | M8 V | 264.82 | 128.9 | 6198.7 | |
| SLW J2305+0613 A | "A" | M1.7 V | | | | Disproved triple (SLW 1300) |
| SLW J2305+0613 B | "B" | M3.2 V | 242.37 | 260.9 | | SLW 1300 is a visual pair |
| SLW J2305+0613 C | "C" | M3.7 V | 86.00 | 283.1 | 18656.7 | |
| HD 221356 | A | F7 V | | | | Confirmed known quadruple (KO 3, GZA 1) |
| 2MASSW J2331016-040618 | BC | M8.0 V+L3.0 V | 451.70 | 261.7 | 11668.3 | |
| 2M J23313095-0405234 | D | L1 V | 12.46 | 221.6 | 321.9 | |
| StKM 2-1787[b] | A | K4 V | | | | Confirmed known binary (VYS 11) |
| TYC 1174-955-2 | B | M2.5 V | 5.78 | 165.1 | 215.5 | |

[a] Lower case denotes photometrically derived spectral types.
[b] Not strictly a binary according to Montes et al. (2018), but we confirm physical parity.

# Chapter 5

## Conclusions and future work

We present in this thesis the most comprehensive photometric and astrometric homogeneous analysis to date of an M-dwarf sample in the close solar neighbourhood. Carmencita, the input catalogue of the CARMENES project, is a thoroughly revised and detailed catalogue with more than two thousand stars, with almost four hundred of them followed-up individually by radial-velocity and transit exoplanet surveys by the consortium. This work adds the value of a thorough, meticulous, individualised inspection of every object in Carmencita. It includes, for each star (and the physical companions, if any), a one-by-one visual examination, photometric and astrometric compilation, literature revision, catalogue data incorporation, and contextualisation in the M-dwarf big picture.

### 5.1 Summary

#### 5.1.1 Astrophysical parameters

In the past, the most reliable observable of a star was the temperature, which could be obtained either through spectral analysis or through colors. This was much easier than to get a reliable luminosity, which frequently also came from spectroscopic analyses of line widths. Multi-band photometry and precise parallaxes are generally available for the vast majority of nearby stars, making it possible to integrate the spectral energy distribution, with a reduced intervention of model assumptions. With this, luminosity can be now regarded as a very reliable *observable*. We should re-consider using bolometric luminosities as a more convenient proxy for other fundamental parameters such as masses and radii, instead of using effective temperatures o spectral types. $\mathcal{M}$ vs. $\mathcal{L}$ diagrams instead of $\mathcal{M}$ vs. $T_{\rm eff}$ or $\mathcal{M}$ vs. Spectral type could be proven much more useful. Saying, for example, that $\mathcal{L} = 0.1\,\mathcal{L}_\odot$ may be a better boundary for separating K7 V from M0.0 V, or talking about $\mathcal{L} = 0.01\,\mathcal{L}_\odot$ rather than M3.5 V stars. This is of special application in the case of ultracool dwarfs. The following is a summary of the main results regarding the determination of astrophysical parameters:

- We started with the latest version of Carmencita, the CARMENES input catalogue, to which we added **168** and **117** single, nearby, bright K and ultracool dwarfs, respectively. Although our main objective was investigating luminosities, colours, and spectral energy distributions of M dwarfs, our sample contained stars and ultracool dwarfs as early as K5 V and as late as L8, in order to avoid boundary issues.

- From public all-sky surveys, we collected **40 094** photometric magnitudes for the **2479** stars and ultracool dwarfs in **20** different passbands from the far ultraviolet, through the blue and red optical and near infrared, to the mid infrared. Except for the bluest passbands, the completeness of high-quality data is of the order of **97 %**. Thanks especially to *Gaia*, we could collect parallactic distances for **98 %** of the sample and identified close multiple systems unresolved by ground all-sky surveys and *WISE*.

- We estimated spectrophotometric distances for **31** single stars without parallactic distance, using our absolute magnitude-colour relations obtained from the GTO subsample.

- We computed bolometric luminosities, effective temperatures, and surface gravities for **1843** stars and ultracool dwarfs with parallactic distance and no physical companions at less than 5 arcsec or less than 5 mag fainter in *Gaia G* than our target. For that, we used `VOSA` and all high-quality photometric data redder than SDSS $u'$. Because of the limitations of the `BT-Settl CIFIST` models implemented in `VOSA`, we set the metallicity to solar. However, except for a few stars with poorly sampled spectral energy distributions, the luminosities are independent of models at least at the **99.5 %** level, which supersede any pre-*Gaia* determination.

- From their loci in the Hertzsprung-Russell diagram, we identified **36** overluminous stars that had been previously assigned to young stellar kinematic groups and associations. We estimated masses and radii using isochronal models.

- We examined colour-spectral type, colour-colour, colour-magnitude, luminosity-magnitude, and bolometric correction-colour diagrams. After discarding stars with young ages, close companions, and bad photometric or astrometric quality flags (i.e. *Gaia* `phot_bp_rp_excess_factor` and `ruwe`), we fitted empirical relations of absolute magnitude-colour, bolometric correction-colour, and luminosity-absolute magnitudes including widely available $G$, $r'$, and $J$ magnitudes and *Gaia* DR2 parallaxes. In addition, we also used the Stefan-Boltzman law and the $\mathcal{M}$-$\mathcal{R}$ relation of Schweitzer et al. (2019) to derive radii and masses of all well-behaved stars in our sample.

- We tabulated median $G$- and $J$-band bolometric corrections, $\mathcal{L}$, $T_{\rm eff}$, $\mathcal{R}$, and $\mathcal{M}$, as well as absolute magnitudes in **14** passbands, for stars and ultracool dwarfs with spectral types from K5 V to L2.0.

- From the HR diagram we paid special attention to the overluminous outliers, and to the low-mass tail of the main sequence ($\mathcal{L} < 0.1\,\mathcal{L}_\odot$, M6.0–M9.5 V). The overluminous were in most cases identified as bona fide or candidate members to young kinematic groups, and so we used `PARSEC` isochrones to derive probable ranges of masses and radii, based on the ages of their stellar associations. For the least massive stars in the sample, we proceeded analogously using `DUSTY00` isochrones, not assuming a given age but an array of them. With this, we updated the masses and radii of overluminous, young stars, and of the low-mass end of Carmencita, superseding the old values in Carmencita.

- In Schweitzer et al. (2019) we used these bolometric luminosities and photometric data, together with high resolution ($R$ >80 000) spectroscopic parameters, to determine radii and masses for **293** nearby, bright M dwarfs in Carmencita. It was the first time that such a large and homogeneous derivation of fundamental stellar parameters was carried out.

- In Martínez-Rodríguez et al. (2019) we took advantage of this work and the same precise determination of stellar luminosities to help investigating the potential habitability, stability, and detectability of exomoons around exoplanets orbiting M dwarfs.

- The averaged astrophysical parameters ($\mathcal{L}$, $T_{\rm eff}$, $\mathcal{R}$, $\mathcal{M}$), bolometric corrections ($BC_G$, $BC_J$), absolute magnitudes in **14** passbands (from *B* to *W*3), and **19** adjacent colours (from *FUV* to *W*4) derived for K5 V to L2 objects, have served as a quick guide and reference in many published works.

### 5.1.2 Multiplicity

- Some multiple systems, usually young, walk on the edge of stability. There is a myriad of occasions for a bound system to be perturbed and disappear, and even bound systems are bound to fail. Some of them, though, will endure the external factors by their own gravity.

- We carried out a blind search of equidistant and comoving companions to all the stars in Carmencita, up to physical separations of $10^5$ au, and for the first time, also considering the potential unresolved binaries at very close separations, by using several statistical indicators and data products in *Gaia* DR3.

- To the known physically bound systems reported in the literature, we add **48** newly discovered pairs, and propose **344** candidates to very compact binaries, to date unresolved. We incorporate these potential candidates to the discussion of multiplicity fractions and distribution of separations.

- When possible, we determined descriptive parameters of the multiplicity (angular and physical separation, positional angle, binding energies, orbital periods), fundamental parameters of the components (luminosities, masses, radii, effective temperatures, surface gravities), compile astrometry (positions, proper motions, parallaxes, radial velocities), photometry (in up to 10 passbands in *Gaia* DR2, 2MASS, AllWISE), and *Gaia* statistical indicators. Additionally, we give an individual description of the components, including their candidacy for unresolved binarity.

- We prove the adequacy of statistical data products in *Gaia* to pinpoint probable very compact binaries, or higher order systems, most of them only able to be resolved by spectroscopic scrutiny. We discovered compact multiplicity in **14** of these *Gaia* candidates, performing a systematic study with medium resolution spectra using FIES. These spectroscopic binaries included clear double- and triple-lined doubles and triples, several single-lined probable binaries, and two high rotators for which the multiplicity could not be properly confirmed. For one SB1 and one SB2 we were able to derive a potential orbital solution. Although in these studies the more amount of spectra collected, the more defined the results, it is possible to identify multiplicity in many cases from the very first spectrum.

- Approximately **38%** of the M dwarfs in Carmencita belong to an identified multiple system with a companion of *any* mass. In this broad definition of multiplicity, the number of single, binaries, triples, quadruples, and quintuples systems for every 100 M dwarfs in all the range of subtypes (from M0.0 V to M9.5 V) represented as S:B:T:Q:Q, is **62.3**:**40.1**:**13.3**:**2.8**:**0.8**.

- Defining the multiplicity of M dwarfs in this broad term (i.e. addressing the fact that a star *belongs* to a multiple system or not, regardless of the mass) is a sensible approach. For instance, all-sky surveys such as *TESS* or *Kepler* may draw conclusions, for example about planet occurrence, based on the systematic observation of brightness-limited samples of M dwarfs, without taking into consideration (at least in a first approximation) the effect of many unresolved binaries disguised as single objects.

- We computed the classically defined multiplicity fraction (*MF*) and the stellar companion fraction (*SCF*) in different un-biased samples:

  - For every individual spectral type (defined as, e.g., M0.0 V and M0.5 V).
  - For two consecutive spectral ranges: M0.0–M5.0 V, and M5.5–M9.5 V.
  - For all the range of M dwarfs (from M0.0 V to M9.5 V) in the volume-limited sample for which Carmencita is complete ($d < 10.0$ pc).

  In these definitions, the M dwarf must be the primary component in the system, that is, the most massive. This calculation has been performed with two general conditions: i) Confirmed multiplicity: including the *known* multiple systems (resolved in *Gaia* DR3 or reported in the literature with higher precision techniques) or newly discovered multiple systems (empirically compliant with astrometric criteria), at all ranges of separations, and ii) Confirmed and to be confirmed multiplicity: to the previous sample, also including the strong candidates to unresolved multiples. Regarding the *MF* and the *SCF*, once the completeness has been taking into account, we conclude that:

  - For M dwarfs in which multiplicity is confirmed, the *MF* and *SCF* are **28.9%** and **35.1%**, respectively. The MF becomes **41.2%** if the unresolved binary candidates are confirmed.
  - Binaries outnumber by far the higher order systems: For every 100 M dwarfs with a companion, the ratio S:B:T:Q:Q is **70.9:23.4:4.6:0.6:0.1**.
  - The MF and SCF *decreases* as a function of spectral type, as expected.
  - For M dwarfs later than M6 we do not provide conclusions for individual spectral types because the small sample size is not of statistical significance.

- These results are in agreement with similar studies published in the last three decades. This work, however, estimates that the actual multiplicity fraction of M dwarfs could be as high as ~**40%**.

- The mass fraction, $q$, does not behave monotonically, which means that the masses of the companions are not comparatively smaller for smaller stars.

- In the cumulative distribution of physical separations, changes of slope are apparent and measurable, following Öpik's law. Based on the numerous candidacy of many single stars to binary systems, we deem the scarcity of multiple systems at very close separations to an observational effect, rather than a real configuration. In a non-cumulative distribution of projected separations, the observed peaks can be traced back to the resolving precision of surveys like *Gaia*. On the opposite side, the flattening of the distribution at wider separations is a foreseeable effect of the easier gravitational disruption of wide systems, corresponding to smaller values in their measured binding energies.

- We provide homogeneous observational data for different formation mechanisms of close and wide binaries in stars of small mass. Particularly, we provide an extensive analysis of the published data in the literature for the close binaries.

- Many are statistical indicators in *Gaia* DR3 have been regarded as useful hints of non-resolved binarity. We have demonstrated the suitability of some of these as a criteria to gauge the existence of compact multiple systems, typically of two stars, in objects previously considered to be single. If all new candidates to unresolved multiples are confirmed, the multiplicity fraction of M dwarfs can increase by **12.3%**.

## 5.2 Future work

### 5.2.1 Carmencita

- The characterisation of the stars necessarily gains precision over time. We have been keen on incorporating every new value that supersedes previous (because of its methodology and its superior handling of the uncertainties), and propagate this improvement to the parameters derived by it. As in any catalogue, a tabulated value is not (or should not be) *the* value for the parameter, but the one that, hopefully, *is very close to* the real one, given the uncertainties. In this sense, all-sky photometric and astrometric surveys, specially *Gaia*, always represent a notable improvement with regards to astrometric and photometric data, and the fundamental parameters that rely on them. Future work in Carmencita means to keep providing an up-to-date, homogeneous, and accessible information resource for the consortium members and all the scientific community, always aiming for a quality-over-quantity approach.

- Carmencita can gain in statistical robustness by incorporating late type, ultracool dwarfs, specially of spectral types M6.0 V and cooler. The same can be said about the late K dwarfs, specially in the frontier of K7/M0.0 V. We enhanced Carmencita this way in Cifuentes et al. (2020) for statistical robustness purposes, but the added objects did not remain in Carmencita. For instance, multiplicity studies would benefit from populating the least massive tail of the sequence. In the same line, Carmencita could aim for a strictly volume-complete sample, rather by a brightness-limited sample, although the former can be inferred from the latter. In this sense, volume completed censuses, such as the 10 pc scrutiny performed by Reylé et al. (2021), can be a valuable reference. One step further in this direction would be to volume-limit the sample to both hemispheres (not only limited to the Calar Alto sky), including all M dwarfs known in the solar vicinity.

### 5.2.2 Astrophysical parameters

There are many ways of improving our $\mathcal{L}$, $\mathcal{R}$, and $\mathcal{M}$ determinations:

- CatWISE (Eisenhardt et al., 2020), a recent NeoWISE enhanced and contributed product (Mainzer et al., 2011), represents a step forward with respect to the AllWISE mid-infrared photometry used here.

- The Legacy Survey of Space and Time (LSST), to be carried out in the Vera C. Rubin Observatory, with its spectacularly large etendue and multi-band photometry in $u'g'r'i'z'y'$ passbands, will start full survey operations in October 2024, with data scheduled to become fully public after two years.

The first data release will supersede all previous UCAC, SDSS, and Pan-STARRS optical datasets (but see also J-PAS, Benítez et al., 2014).

- The ESA *Euclid* mission (launch scheduled to Q3 2023) will complement LSST in the near infrared at Galactic latitudes far from the ecliptic, especially for the latest M dwarfs.

- Thanks to the Transiting Exoplanet Survey Satellite (*TESS*) and the discovery of new detached M-dwarf eclipsing binaries, the $\mathcal{M}$-$\mathcal{R}$ relation will be more refined and probably determined for different intervals of age and metal abundances.

- The *Gaia* DR4, will improve $G$, $G_{BP}$, and $G_{RP}$ photometry and, especially, astrometry, with which we will have more accurate parallax and close multiplicity identifications.

- With new spectral-synthesis determinations of $T_{\rm eff}$, $\log g$, and [Fe/H] in late-type stars for calibration (e.g. with the equivalent width method or with deep learning – Marfil et al. 2018; Passegger et al. 2018, 2020; Marfil et al. 2021).

- From new studies that link kinematics, activity, and youth (and, therefore, radius and surface gravity).

- New grids of theoretical atmospheric models considering a much wider range of metallicities, which are a fundamental cornerstone of these determinations, will be available for `VOSA`.

- Among all the parameters in Carmencita, age is probably the one subject to a larger uncertainty, and with it, the values of masses and radii of the younger stars. It is accepted that age has an impact, but the actual process through which this happens remains unclear. We derived masses from isochrone models, fixing for a certain age boundaries given by the SKG assumed age. Masses could be compared by using this approach and others in which, for example, the age is a *free* parameter. Luckily, the spread of young stars in the HR diagram does represent an advantage from the point of view of discriminating the best-fitting model.

### 5.2.3 Multiplicity

- There are many unresolved binaries that are still overlooked. *Gaia* data have an enormous potential, and there is so much to look forward to with the following years of approved extension of the mission. Many data products of DR4 will address (as they already do in DR3) the topic of multiplicity, in particular the kind that keeps unresolved in *Gaia*. In successive data releases, *Gaia* provides more data products derived by the automatic pipelines from the original data, and the study of very close multiplicity can extensively gain in depth with *Gaia* DR4 new data, in particular with the help of radial-velocity measurements.

- The spectroscopic investigation of these suspects is the main way to prove that they are not single. These require time, constancy, and patience, because providing orbital descriptions of most of them is worth months or years of dutiful data collection.

- How multiple systems are distributed in the space of separations or mass ratios encodes fundamental clues about their very origin and interplay. Particularly, how very wide systems come to be, and how very wide pairs with *single* components may have reached that puzzling configuration.

- In more general terms, given the amount of evidence that suggests that stars are almost always born together, how *single* stars became single, is also a topic worth investigating. These discoveries have the potential of bringing forth a new picture of the formation and dynamical evolution of stellar systems.

- The formation of planetary systems does also benefit from understanding the prevalence of multiple systems. Planetary detection and statistics, mainly based on the transit technique, are biased towards single stars.

- Despite being neither a novelty, nor groundbreaking science, the abundance studies of M dwarfs in wide physical systems with FGK primaries continues to be a topic of profound impact for the theoretical study of M dwarfs atmospheres. Because binaries are assumed to be both coeval and to have the same chemical composition, the one of the higher mass star can be extrapolated to the M-dwarf companion. We are already involved in ongoing detailed spectroscopic study using the CARMENES spectrograph, for a selected sample of these systems, which were partially selected from the study developed in this work. The proposal includes observing a sample of wide visual binaries composed of an F-, G- or K- dwarf primary and an M-dwarf secondary. These data will allow us to test the metallicity (described as the iron abundance, [Fe/H]) and chemical abundances (C, O, Na, Mg, Al, Si, Ca, Sc, Ti, V, Cr, Mn, Co, Ni, Rb, Y, Ba, and Nd), derived for the primaries with the EW method with `STEPAR` (Tabernero et al., 2019) or, equivalently, via spectral synthesis with `STEPARSYN` (Tabernero et al., 2022a). The goal of this proposal is to derive a precise spectroscopic calibration of the M-dwarf stellar parameters and chemical abundances.

- The observed cumulative distributions of separations in multiple systems under different constrains can help test the hypothesis of whether all stars form in systems, which eventually are disrupted to produce the single stars that we observe in the field. In this regard, we can use N-body analyses to study the long-term stability of these systems, with the observed distributions of orbital periods, separations, binding energies, and masses serving as valuable observational constraints.

In this thesis we contribute to our understanding of the astrometry, photometry, and multiplicity of M dwarfs. The homogeneously derived fundamental parameters and observables provide an important benchmark for testing and improving theoretical predictions, and the findings underscore the importance of continued research into M dwarfs, specially in the field of exoplanetary exploration. The study takes an individualised approach, examining each star and their potential physical companions, resulting in the discovery of many previously unknown pairs. It is hoped that the data gathered in this study will serve as a valuable resource for astronomers and researchers in related fields, providing a foundation for future discoveries.

# Acknowledgements

I sincerely thank Prof. Adam J. Burgasser and Prof. Andrei A. Tokovinin for generously reading this work and for taking the time to provide an extremely valuable refereeing. I am grateful to Prof. E. E. Mamajek, Prof. E. Gaidos, and Dr. C. Reylé for their useful insights and suggestions at some point during the course of this thesis. F. J. Abellán de Paco, M. Cortés-Contreras, R. Dorda, G. Holgado, and I, among many others, have compiled data for Carmencita during MSc and PhD theses at the Universidad Complutense de Madrid.

This document was typeset using Leslie Lamport's LaTeX, based on Donald E. Knuth's TeX. Javier González Payo kindly provided me with the basis of this template. Robert Slimbach's Minion Pro serves as both the text and display typeface. Erik Spiekermann and Ralph du Carrois' `Fira Sans` is the monospaced font of choice.

This publication made use of `VOSA` and the Filter Profile Service, developed and maintained by the Spanish Virtual Observatory through grant AYA2017-84089, the SIMBAD database, the `Aladin` sky atlas, the VizieR catalogue access tool developed at Centre de Données astronomiques de Strasbourg (CDS) in Strasbourg Observatory, France, the Tool for OPerations on Catalogues And Tables (`TOPCAT`), and the NASA Exoplanet Archive, which is operated by the California Institute of Technology, under contract with the National Aeronautics and Space Administration under the Exoplanet Exploration Program.

Some of the material and figures included in this document have been already published in Astronomy and Astrophysics (A&A), Monthly Notices of the Royal Astronomical Society (MNRAS), and Research Notes of the American Astronomical Society (RNAAS). For scientific graphics not produced specifically for this work, due credit is given. Additional images are used under the Creative Commons[1] conditions in each particular case. No credit is mentioned for creations of my own.

All the code produced in this thesis has been written in Guido van Rossum's `Python` language. It is publicly available in my personal GitHub[2].

This thesis was mostly financed by the Spanish Ministry of Science and Innovation through grants AYA2016-79425-C3-1/2/3-P and BES-2017-080769.

CARMENES is an instrument for the Centro Astronómico Hispano-Alemán de Calar Alto (CAHA, Almería, Spain). CARMENES is funded by the German Max-Planck-Gesellschaft (MPG), the Spanish Consejo Superior de Investigaciones Científicas (CSIC), the European Union through FEDER/ERF

[1] `https://creativecommons.org`.
[2] `https://github.com/ccifuentesr`.

FICTS-2011-02 funds, and the members of the CARMENES Consortium (Max-Planck-Institut für Astronomie, Instituto de Astrofísica de Andalucía, Landessternwarte Königstuhl, Institut de Ciències de l'Espai, Insitut für Astrophysik Göttingen, Universidad Complutense de Madrid, Thüringer Landessternwarte Tautenburg, Instituto de Astrofísica de Canarias, Hamburger Sternwarte, Centro de Astrobiología and Centro Astronómico Hispano-Alemán), with additional contributions by the Spanish Ministry of Economy, the German Science Foundation through the Major Research Instrumentation Programme and DFG Research Unit FOR2544 "Blue Planets around Red Stars", the Klaus Tschira Stiftung, the states of Baden-Württemberg and Niedersachsen, and by the Junta de Andalucía.

# Conventions

- The terms 'M dwarf' and 'red dwarf' are used interchangeably.
- The calligraphic $\mathcal{M}$ is used for mass, in order to be easily differentiated from other instances, such as 'M dwarf', or '$M_G$'. For homogeneity, radius and luminosity are denoted as $\mathcal{R}$ and $\mathcal{L}$, respectively.
- The abbreviation 'a' is used for *year*, instead of 'yr'[3].
- Software names and parameters from electronic catalogues are shown in `monospaced` font.
- Sloan's passbands $u'g'r'i'z'$ may be simplified as *ugriz*[4].
- For simplicity purposes and to design table layouts with a better readability, some star catalog names are abbreviated (only in the tabular content): 2M for 2MASS J, G2- and G3- for *Gaia* DR2 and E/DR3, 1R for 1RXS J.
- In tabulated form, names of variable stars (e.g. 'EZ Psc', 'AD Leo') usually appear as single words (i.e. 'EZPsc', 'ADLeo').
- I predominantly use the plural pronoun 'we' instead of the singular 'I', except on a few occasions. I do this to acknowledge the collective effort of many who have worked before, and alongside me.

[3]The recognised symbol for a year is the letter 'a', rather than 'yr', which is often used in papers in English. The term 'year' should be referred to the Julian year of 365.25 days (31.5576 Ms) unless otherwise specified. See the IAU Style Manual (IAU Commission 5, IAU Transactions XXB, 1989, https://www.iau.org/publications/proceedings_rules/units/), or the NIST SP811 Special Publication 811 (https://www.nist.gov/pml/special-publication-811), and also see http://exoterrae.eu/annus.html.

[4]The *griz* photometric system (aka 'Gunn', 'Thuan-Gunn', or 'Oke-Gunn' photometric system) was developed in the decades of 1970 and 1980 to be used with electronic sensors. The *ugriz* photometric system used by SDSS is a follow-on but the passbands are not identical. The apostrophes are used to distinguish both, but are often skipped.

# Appendix A

# List of publications

## A.1 Peer-reviewed publications

Articles in this thesis are marked with a red star (⋆).

**2023**

35. "GJ 806 (TOI-4481): A bright nearby multi-planetary system with a transiting hot, low-density super-Earth"
E. Pallé E., J. Orell-Miquel, M. Brady, J. Bean, A. P. Hatzes, et al. (incl. **C. Cifuentes**)
*Astronomy & Astrophysics*, accepted. arXiv: 2301.06873.

34. "Searching for the nature of stars with debris disks and planets"
R. de la Reza, C. Chavero, S. Roca-Fábrega, F. Llorente de Andrés, P. Cruz, and **C. Cifuentes**
*Astronomy & Astrophysics*, 671, A136. March 2023.

33. "The CARMENES search for exoplanets around M dwarfs. A long-period planet around GJ 1151 measured with CARMENES and HARPS-N data"
J. Blanco-Pozo, M. Perger, M. Damasso, G. Anglada Escudé, I. Ribas, D. Baroch, J. A. Caballero, **C.Cifuentes**, et al.
*Astronomy & Astrophysics*, 671, A50. March 2023.

32. "The CARMENES search for exoplanets around M dwarfs. Guaranteed Time Observations Data Release 1 (2016-2020)"
I. Ribas, A. Reiners, M. Zechmeister, J. A. Caballero, et al. (incl. **C. Cifuentes**)
*Astronomy & Astrophysics*, 670, A139. February 2023.

31. "The CARMENES search for exoplanets around M dwarfs, Wolf 1069 b: Earth-mass planet in the habitable zone of a nearby, very low-mass star"
D. Kossakowski, M. Kürster, T. Trifonov, T. Henning, J. Kemmer, et al. (incl. **C. Cifuentes**)
*Astronomy & Astrophysics*, 670, A84. February 2023.

**2022**

**2021**

P. Bluhm, R. Luque, N. Espinoza, E. Pallé, J. A. Caballero, et al (incl. **C. Cifuentes**).
*Astronomy & Astrophysics* 639, A132. July 2020.

9. **"A He I upper atmosphere around the warm Neptune in GJ 3470 b"**
E. Pallé, S. Nortmann, N. Casasayas-Barris, M. Lampón, et al. (incl. **C. Cifuentes**)
*Astronomy & Astrophysics* 638, A61. June 2020.

**2019**

8. **"Exomoons in the habitable zones of M dwarfs"**
H. Martínez-Rodríguez, J. A. Caballero, **C. Cifuentes**, A. L. Piro, and R. Barnes.
*The Astrophysical Journal*, 887, 2, 261. December 2019.

7. **"A giant exoplanet around a very low mass star challenging formation models"**
J. C. Morales, A. J. Mustill, I. Ribas, M. B. Davies, A. Reiners, et al. (incl. **C. Cifuentes**)
*Science*, 365, 6460, 1441-1445. September 2019.

6. **"The CARMENES search for exoplanets around M dwarfs. Two temperate Earth-mass planet candidates around Teegarden's star"**
M. Zechmeister, S. Dreizler, I. Ribas, A. Reiners, J. A. Caballero, et al. (incl. **C. Cifuentes**)
*Astronomy & Astrophysics*. 627, A49. July 2019.

5. **"The CARMENES search for exoplanets around M dwarfs. Radii and masses of the target stars"**
A. Schweitzer, V. M. Passeger, **C. Cifuentes**, V. J. S. Béjar, M. Cortés-Contreras, et al.
*Astronomy & Astrophysics* 625, A68. May 2019.

**2018**

4. **"The CARMENES search for exoplanets around M dwarfs. A Neptune-mass planet traversing the habitable zone around HD 180617"**
A. Kaminski, T. Trifonov, J. A. Caballero, A. Quirrenbach, I. Ribas, et al. (incl. **C. Cifuentes**)
*Astronomy & Astrophysics*, 618, A115. October 2018.

3. **"The CARMENES search for exoplanets around M dwarfs. High-resolution optical and near-infrared spectroscopy of 324 survey stars"**
A. Reiners, M. Zechmeister, J. A. Caballero, I. Ribas, J. C. Morales, et al. (incl. **C. Cifuentes**)
*Astronomy & Astrophysics*, 612, A49. April 2018.

2. **"The CARMENES search for exoplanets around M dwarfs. First visual-channel radial-velocity measurements and orbital parameter updates of seven M-dwarf planetary systems"**
T. Trifonov, M. Kürster, M. Zechmeister, L. Tal-Or, J. A. Caballero, et al. (incl. **C. Cifuentes**)
*Astronomy & Astrophysics*, 609, A117. February 2018.

1. **"The CARMENES search for exoplanets around M dwarfs. HD 147379 b: A nearby Neptune in an early-M dwarf's temperate zone"**
A. Reiners, I. Ribas, M. Zechmeister, J. A. Caballero, et al. (incl. **C. Cifuentes**)
*Astronomy & Astrophysics*, 609, A49. January 2018.
*Selected as an A&A Highlighted paper of the week.*

## A.2 Non-refereed publications

### 2021

2. (★) "One Is the Loneliest Number: Multiplicity in Cool Dwarfs"
   **C. Cifuentes**, J. A. Caballero, and S. Agustí .
   *Research Notes of the AAS*, Vol. 5, Number 5. June 2021.

### 2018

1. "Cool dwarfs in wide multiple systems: A curious quintuple system of a compact sun-like triple and a close M dwarf-white dwarf pair at a wide separation"
   R. González-Peinado, J. A. Caballero, D. Montes, and **C. Cifuentes**.
   *The Observatory*, Vol. 138, p. 292-298. December 2018.

## A.3 Conference proceedings

### 2021

9. "Luminosities of cool stars"
   **C. Cifuentes**, J. A. Caballero, M. Cortés-Contreras, D. Montes, and the Carmenes Consortium.
   *The Star-Planet Connection, On-line Workshop*, October 25 – 28, 2021.

8. "CARMENES and the Frontiers of High-Resolution Spectroscopy for M dwarfs"
   Y. Shan, A. Reiners, P. J. Amado, V. J. S. Béjar, J. A. Caballero, **C. Cifuentes**, and the Carmenes Consortium. *Plato Online Mission Conference 2021*, October, 2021.

7. "Stellar atmospheric parameters of CARMENES GTO M dwarfs with spectral synthesis and SteParSyn"
   E. Marfil, H. M. Tabernero, D. Montes, J. A. Caballero, F. J. Lázaro-Barrasa, et al. (incl. **C. Cifuentes**)
   *The 20.5th Cambridge Workshop on Cool Stars, Stellar Systems, and the Sun (CS20.5)*. Virtually anywhere. March 2 – 4, 2021.

### 2020

6. "The CARMENES M-dwarf planet survey"
   A. Quirrenbach, and the CARMENES Consortium (incl. **C. Cifuentes**).
   *Proceedings of the SPIE*, Vol. 11447, id. 114473C. Ground-based and Airborne Instrumentation for Astronomy. SPIE Astronomical Telescopes + Instrumentation 2020 Digital Forum. Online. December 14 – 18, 2020.

5. "Colours and luminosities of M dwarfs in the CARMENES input catalogue"
   **C. Cifuentes**, J. A. Caballero, M. Cortés-Contreras, D. Montes, F. J. Abellán, et al.
   *Contributions to the XIV.0 Scientific Meeting (virtual) of the Spanish Astronomical Society*. Online. July 13 – 15 2020.

4. "Kinematics of M dwarfs in the CARMENES input catalogue"
   M. Cortés-Contreras, A. J. Domínguez-Fernández, J. A. Caballero, D. Montes, C. Cardona, V. J. S. Béjar, **C. Cifuentes**, et al.

*Contributions to the XIV.0 Scientific Meeting (virtual) of the Spanish Astronomical Society*. Online. July 13 – 15 2020.

3. **"The bimodal A(Li) distribution of Milky Way's thin disk stars and the Galactic scale events"**
S. Roca-Fábrega, F. Llorente de Andrés, **C. Cifuentes**, C. Chavero, R. de la Reza and B. Montesinos.
*Contributions to the XIV.0 Scientific Meeting (virtual) of the Spanish Astronomical Society*. Online. July 13 – 15 2020.

**2018**

2. **"Spectral energy distributions and luminosities of M dwarfs in the CARMENES search for exoplanets"**
**C. Cifuentes**, J. A. Caballero, M. Cortés-Contreras, D. Montes, A. Schweitzer, I. Ribas, P. J. Amado, and the CARMENES Consortium.
*Proceedings of the XIII Scientific Meeting of the Spanish Astronomical Society*. Highlights on Spanish Astrophysics X Salamanca, Spain. 16 – 20 July, 2018.

1. **"CARMENES: high-resolution spectra and precise radial velocities in the red and infrared"**
A. Quirrenbach, P. J. Amado, I. Ribas, A. Reiners, J. A. Caballero, et al. (incl. **C. Cifuentes**)
*Proceedings of the SPIE*, Vol. 10702, id. 107020W. SPIE Astronomical Telescopes + Instrumentation. Austin, Texas, USA. 10 – 15 June, 2018.

## A.4 Circulars

1. **"MPEC 2020-A99: 2020 AV2"**
P. Bacci, M. Maestripieri, M. Facchini, M. D. Grazia, L. Tesi, et al. (incl. **C. Cifuentes**).
*Minor Planet Electronic Circulars*, No. 2020-A99. January 2020.

## A.5 VizieR online data catalogues

24. **CARMENES search for exoplanets around M dwarfs** (Ribas+, 2023)
23. **GJ 1151 CARMENES and HARPS-N data** (Blanco-Pozo+, 2023)
22. **Wolf 1069 RV and stellar activity indices** (Kossakowski+, 2023)
21. **Gl514 RVs and Activity diagnostics** (Damasso+, 2022)
20. **TOI-1468 photometry and radial velocities** (Chaturvedi+, 2022)
19. **AD Leo RV and stellar activity indices** (Kossakowski+, 2022)
18. **GJ 3929 b RVs and activity indicators** (Kemmer+, 2022)
17. **VRI photometry and radial velocity of TOI-1759** (Espinoza+, 2022)
16. **CARMENES stellar atmospheric parameters** (Marfil+, 2021)
15. **TOI-1201 RV and activity index** (Kossakowski+, 2021)
14. **Evolution of Li in FGK dwarf stars** (Llorente de Andrés+, 2021)

13. CARMENES time-resolved CaII H&K catalog (Perdelwitz+, 2021)
12. G 264-012 and Gl 393 radial velocity curves (Amado+, 2021)
11. LP714-47 (TOI 442) radial velocity curve (Dreizler+, 2020)
10. GJ 3473 (TOI-488) radial velocity curve (Kemmer+, 2020)
9. CARMENES input catalogue of M dwarfs. V (**Cifuentes**+, 2020)
8. Compilation of planets around M dwarfs (Martínez-Rodríguez+, 2019)
7. GJ 3512 radial velocity and light curves (Morales+, 2019)
6. Teegarden's Star RV and H$\alpha$ curves (Zechmeister+, 2019)
5. Radii and masses of the CARMENES targets (Schweitzer+, 2019)
4. A Neptune-mass planet traversing the habitable zone around HD 180617 (Kaminski+, 2018)
3. 324 CARMENES M dwarfs velocities (Reiners+, 2018)
2. CARMENES radial velocity curves of 7 M-dwarf (Trifonov+, 2018)
1. HD147379 b velocity curve (Reiners+, 2018)

# Appendix B

## Long tables of Chapter 2

**B.1** Carmencita, the CARMENES input catalogue[1]

[1] This table displays eight columns of carmencita, which actually includes 177 columns, as described in Chapter 2.

Table B.1: Carmencita, the CARMENES input catalogue.

| Karmn | Name | α (2016.0) | δ (2016.0) | Spectral type | J [mag] | Multiplicity[a] | DR1[b] |
|---|---|---|---|---|---|---|---|
| J00012+139N | BD+13 5195A | 00:01:13.21 | +13:58:32.7 | M0.5 V | 7.798 | Triple | |
| J00012+139S | BD+13 5195B | 00:01:12.89 | +13:58:22.0 | M0.0 V | 8.359 | Triple | |
| J00026+383 | PM J00026+3821A | 00:02:40.00 | +38:21:44.1 | M4.0 V | 9.707 | Binary* | |
| J00033+046 | StKM 1-2199 | 00:03:18.97 | +04:41:11.6 | M1.5 V | 8.833 | | |
| J00056+458 | HD 38B | 00:05:42.30 | +45:48:35.0 | M0.0 V | 6.142 | Quadruple | • |
| J00051+457 | GJ 2 | 00:05:12.22 | +45:47:09.2 | M1.0 V | 6.704 | Quadruple | |
| J00067-075 | GJ 1002 | 00:06:42.32 | -07:32:47.3 | M5.5 V | 8.323 | | • |
| J00077+603 | G 217-32A | 00:07:43.28 | +60:22:54.0 | M4.0 V | 8.911 | Binary | |
| J00078+676 | PM J00078+6736 | 00:07:50.65 | +67:36:23.9 | M2.0 V | 8.355 | | |
| J00079+080 | GJ 3007 | 00:07:58.74 | +08:00:12.8 | M4.0 V | 9.392 | | |
| J00081+479 | 1R000806.3+475659 | 00:08:06.23 | +47:57:02.4 | M4.0 V | 8.523 | Binary (SB2) | |
| J00084+174 | GJ 3008 | 00:08:27.18 | +17:25:26.4 | M0.0 V | 7.807 | | |
| J00088+208 | GJ 3010 | 00:08:53.86 | +20:50:21.4 | M5.0 V | 8.870 | Binary | |
| J00110+052 | G 31-29 | 00:11:04.89 | +05:12:33.4 | M1.0 V | 8.530 | | |
| J00115+591 | LSPM J0011+5908 | 00:11:29.94 | +59:08:21.2 | M6.0 V | 9.945 | | • |
| J00118+229 | LP 348-40 | 00:11:53.17 | +22:59:01.2 | M3.5 V | 8.862 | | |
| J00119+330 | G 130-53 | 00:11:55.73 | +33:03:10.7 | M3.5 V | 9.066 | | |
| J00122+304 | 1R001213.6+302906 | 00:12:13.49 | +30:28:43.8 | M4.5 V | 10.242 | | |
| J00131+703 | TYC 4298-613-1 | 00:13:11.68 | +70:23:54.9 | M1.0 V | 8.259 | | |
| J00132+693 | GJ 11 A | 00:13:18.00 | +69:19:32.4 | M3.5 V | 8.556 | Binary | |
| J00133+275 | UPM J0013+2733 | 00:13:19.55 | +27:33:29.1 | M4.5 V | 10.431 | | |
| J00136+806 | GJ 3014 | 00:13:40.36 | +80:39:59.8 | M1.5 V | 7.756 | Binary | |
| J00137+806 | GJ 3015 | 00:13:44.59 | +80:39:52.3 | M5.0 V | 10.936 | Binary | |
| J00154-161 | GJ 1005 | 00:15:28.77 | -16:08:11.7 | M4.0 V | 7.215 | Binary | |
| J00156+722 | LP 49-338 | 00:15:37.59 | +72:17:03.5 | M2.0 V | 8.837 | | |
| J00158+135 | GJ 12 | 00:15:49.92 | +13:33:27.6 | M4.0 V | 8.619 | | |
| J00159-166 | 1R001557.5-163659 | 00:15:57.94 | -16:36:57.5 | M4.1 V | 8.736 | Binary | |
| J00162+198W | EZPsc | 00:16:15.44 | +19:51:25.3 | M4.0 V | 7.875 | Triple (SB2) | • |
| J00162+198E | GJ 1006 B | 00:16:16.96 | +19:51:38.5 | M4.0 V | 8.893 | Triple | • |
| J00169+051 | GJ 1007 | 00:16:56.20 | +05:07:16.4 | M4.0 V | 9.398 | | |
| J00169+200 | GJ 3022 | 00:16:57.03 | +20:03:55.7 | M3.5 V | 9.681 | Triple* | |
| J00173+291 | Ross 680 | 00:17:21.17 | +29:11:05.7 | M2.0 V | 8.152 | | |
| J00176-086 | GJ 3025 | 00:17:41.23 | -08:40:55.8 | M0.0 V | 8.095 | | |
| J00179+209 | LP 404-81 | 00:17:58.87 | +20:57:18.6 | M1.0 V | 8.250 | Binary | |
| J00182+102 | GJ 16 | 00:18:16.59 | +10:12:09.6 | M1.5 V | 7.564 | | |
| J00183+440 | HD 1326 | 00:18:27.17 | +44:01:29.2 | M1.0 V | 5.252 | Binary | • |
| J00184+440 | HD 1326B | 00:18:30.07 | +44:01:43.5 | M3.5 V | 6.789 | Binary | • |
| J00188+278 | GJ 3027 | 00:18:54.07 | +27:48:48.1 | M4.0 V | 9.535 | | |
| J00190-099 | GJ 1008 | 00:19:05.52 | -09:57:58.3 | M0.0 V | 7.376 | | |
| J00201-170 | GJ 2003 | 00:20:08.54 | -17:03:41.2 | M1+ Vk: | 8.545 | | |
| J00204+330 | GJ 3028 | 00:20:30.78 | +33:04:52.6 | M5.5 V | 10.284 | | • |
| J00207+596 | [I81] M 134 | 00:20:47.70 | +59:36:15.6 | M2.5 V | 8.936 | | |
| J00209+176 | StKM 1-25 | 00:20:57.24 | +17:38:14.7 | M0.0 V | 8.367 | | |
| J00210+557 | G 217-43 | 00:21:05.02 | +55:43:55.6 | M2.0 V | 9.194 | | |
| J00218+382 | G 171-51 | 00:21:54.79 | +38:16:24.2 | M3.0 V | 8.953 | | |
| J00219+492 | GJ 3030 | 00:21:58.20 | +49:12:37.3 | M2.5 V | 9.139 | Binary | |
| J00234+243 | GJ 1011 | 00:23:27.73 | +24:18:26.5 | M4.0 V | 9.753 | | |
| J00234+771 | GJ 1010 A | 00:23:24.80 | +77:11:22.2 | M2.5 V | 8.042 | Binary | |
| J00235+771 | GJ 1010 B | 00:23:27.78 | +77:11:27.5 | M4.5 V | 9.934 | Binary | |
| J00240+264 | LSPM J0024+2626 | 00:24:03.96 | +26:26:28.9 | M4.0 V | 10.222 | | |
| J00244+360 | G 130-67 | 00:24:26.30 | +36:03:54.0 | M1.0 V | 8.886 | | |
| J00245+300 | GJ 3033 | 00:24:35.60 | +30:02:29.7 | M5.0 V | 9.776 | | |
| J00253+228 | GJ 3034 | 00:25:20.33 | +22:53:03.7 | M4.71 | 9.716 | | |
| J00268+701 | GJ 21 | 00:26:52.27 | +70:08:30.4 | M1.0 V | 7.411 | | |
| J00271+496 | GJ 3035 | 00:27:07.39 | +49:41:49.3 | M4.0 V | 9.733 | | |
| J00279+223 | LP 349-25 | 00:27:56.46 | +22:19:29.7 | M8.0 V | 10.614 | Binary | |
| J00286-066 | GJ 1012 | 00:28:39.11 | -06:40:02.0 | dM4.0 | 8.038 | | • |
| J00288+503 | GJ 3036 | 00:28:54.66 | +50:22:35.3 | M3.7 V | 8.847 | Binary | |

Table B.1: Carmencita, the CARMENES input catalogue (continued).

| Karmn | Name | α (2016.0) | δ (2016.0) | Spectral type | J [mag] | Multiplicity[a] | DR1[b] |
|---|---|---|---|---|---|---|---|
| J00315-058 | GJ 1013 | 00:31:35.79 | -05:52:30.0 | M3.7 V | 8.762 | | |
| J00322+544 | G 217-56 | 00:32:15.34 | +54:28:55.3 | M4.5 V | 9.387 | | |
| J00324+672N | V547Cas | 00:32:34.33 | +67:14:03.6 | M2.0 V | 6.844 | Triple | |
| J00324+672S | GJ 22 B | 00:32:34.28 | +67:13:59.8 | M3.0 V | 7.172 | Triple | |
| J00325+074 | GJ 3039 | 00:32:34.91 | +07:29:25.7 | M4.0 V | 8.397 | Binary | |
| J00328-045 | GR* 50 | 00:32:53.21 | -04:34:09.4 | M4.5 V | 9.276 | Binary | |
| J00333+368 | G 132-4 | 00:33:21.17 | +36:50:28.6 | M3.0 V | 8.768 | | |
| J00341+253 | V493AndA | 00:34:08.48 | +25:23:48.5 | M0.0 V | 8.481 | Triple* | |
| J00346+711 | GJ 3040 | 00:34:39.40 | +71:11:36.6 | M3.5 V | 9.465 | | |
| J00357+025 | LP 585-55 | 00:35:43.30 | +02:33:10.9 | M5.0 V | 10.517 | Binary | |
| J00358+526 | G 172-11 | 00:35:54.70 | +52:41:09.3 | M2.5 V | 8.932 | Triple | |
| J00359+104 | GJ 1014 | 00:35:56.70 | +10:28:29.4 | M5.0 V | 10.222 | | • |
| J00361+455 | GJ 3042 | 00:36:08.05 | +45:30:55.3 | M2.0 V | 8.167 | Binary | |
| J00374+515 | G 172-14 | 00:37:25.07 | +51:33:06.8 | M0.5 V | 8.429 | | |
| J00380+169 | PM J00380+1656 | 00:38:03.75 | +16:56:01.3 | M3.0 V | 9.380 | | |
| J00382+523 | GJ 3044 | 00:38:15.16 | +52:19:53.3 | M0.0 V | 7.714 | | |
| J00385+514 | GJ 3045 | 00:38:33.48 | +51:27:58.4 | M3.0 V | 8.892 | | |
| J00389+306 | Wolf 1056 | 00:39:00.98 | +30:36:58.8 | M2.5 V | 7.453 | | • |
| J00395+149S | LP 465-061 | 00:39:33.91 | +14:54:19.6 | M4.0 V | 9.964 | Triple | |
| J00395+149N | LP 465-62 | 00:39:34.16 | +14:54:35.4 | M4.5 V | 9.826 | Triple | |
| J00395+605 | Wolf 10 | 00:39:33.51 | +60:33:10.9 | M2.42 V | 9.205 | | |
| J00403+612 | 2M00402129+6112490 | 00:40:21.40 | +61:12:48.2 | M2.0 V | 10.154 | | • |
| J00409+313 | GJ 3047 | 00:40:56.19 | +31:22:51.2 | M3.3 V | 9.491 | | |
| J00413+558 | GJ 1015 A | 00:41:21.44 | +55:50:03.2 | M4.2 V | 9.839 | Binary | |
| J00427+438 | PM J00427+4349 | 00:42:47.79 | +43:49:24.0 | M2.5 V | 8.487 | | |
| J00428+355 | FFAnd | 00:42:48.59 | +35:32:56.9 | M1.0 V | 7.164 | Binary (SB2) | |
| J00435+284 | GJ 1019 | 00:43:35.44 | +28:26:24.4 | M4.0 V | 10.392 | | |
| J00443+091 | GJ 3052 | 00:44:21.54 | +09:07:34.5 | M4.5 V | 9.501 | | |
| J00443+126 | GJ 3051 | 00:44:19.64 | +12:36:59.8 | M3.5 V | 8.868 | | |
| J00449-152 | GJ 3053 | 00:44:59.68 | -15:16:27.1 | M4.5 V | 9.612 | | |
| J00459+337 | G 132-25 | 00:45:57.01 | +33:47:11.3 | M4.5 V | 10.183 | Binary | |
| J00463+353 | PM J00463+3522 | 00:46:21.76 | +35:22:10.6 | M1.5 V | 8.933 | | |
| J00464+506 | G 172-22 | 00:46:30.66 | +50:38:35.3 | M4.0 V | 9.964 | | |
| J00468+160 | PM J00468+1603 | 00:46:53.16 | +16:03:01.8 | M2.0 V | 8.363 | | |
| J00484+753 | LSPM J0048+7518 | 00:48:30.66 | +75:18:47.2 | M3.0 V | 9.469 | | |
| J00487+270 | GJ 3057 | 00:48:45.34 | +27:01:04.4 | M2.5 V | 8.774 | | |
| J00489+445 | GJ 3058 | 00:48:58.46 | +44:35:06.9 | M3.0 V | 9.125 | Binary | |
| J00490+657 | PM J00490+6544 | 00:49:05.09 | +65:44:36.8 | M2.5 V | 9.304 | | |
| J00502+086 | RX J0050.2+0837 | 00:50:17.59 | +08:37:33.6 | M4.5 V | 9.745 | Binary (SB2) | |
| J00505+248 | FTPscA | 00:50:33.49 | +24:48:59.7 | M3.0 Vkee | 7.951 | Binary | |
| J00511+225 | BPM 84579 | 00:51:10.71 | +22:34:43.8 | M1.5 V | 8.262 | | |
| J00514+583 | Wolf 33 | 00:51:33.02 | +58:18:13.8 | M0.0 V | 7.831 | | |
| J00515-229 | HD 4967B | 00:51:35.91 | -22:54:35.4 | M5.5 V | 10.904 | Binary | • |
| J00520+205 | G 69-27 | 00:52:00.27 | +20:34:56.6 | M1.0 V | 8.490 | | |
| J00532+190 | LSPM J0053+1903 | 00:53:12.84 | +19:03:25.0 | M2.5 V | 8.792 | | |
| J00538+459 | G 172-28 | 00:53:53.72 | +45:56:41.6 | M0.0 V | 8.311 | | |
| J00540+691 | Ross 317 | 00:54:00.72 | +69:10:56.9 | M2.0 V | 9.462 | | |
| J00548+275 | G 69-32 | 00:54:48.49 | +27:31:03.9 | M4.5 V | 10.340 | | |
| J00566+174 | GJ 1024 | 00:56:39.14 | +17:27:30.3 | M3.8 V | 9.285 | | |
| J00570+450 | G 172-30 | 00:57:03.64 | +45:05:08.7 | M3.0 V | 8.101 | | • |
| J00577+058 | BD+05 127 | 00:57:44.49 | +05:51:20.6 | M0.0 V | 7.478 | | |
| J00580+393 | 1R005802.4+391912 | 00:58:01.01 | +39:19:11.5 | M4.5 V | 9.561 | | |
| J01008+669 | GJ 3068 | 01:00:48.83 | +66:56:53.8 | M3.5 V | 9.408 | | |
| J01009-044 | GJ 1025 | 01:00:57.71 | -04:26:49.5 | M4.0 V | 9.042 | | |
| J01013+613 | Wolf 44 | 01:01:20.86 | +61:21:43.7 | M2.0 V | 7.272 | | • |
| J01019+541 | GJ 3069 | 01:01:58.94 | +54:10:55.8 | M5.0 V | 9.778 | | • |
| J01023-104 | GJ 3072 | 01:02:21.16 | -10:25:28.7 | M0.0 V | 7.417 | | |
| J01025+716 | Ross 318 | 01:02:38.16 | +71:40:41.2 | M3.0 V | 6.301 | | • |
| J01026+623 | Wolf 46 | 01:02:40.55 | +62:20:43.6 | M1.5 V | 6.230 | Binary | • |

Table B.1: Carmencita, the CARMENES input catalogue (continued).

| Karmn | Name | α (2016.0) | δ (2016.0) | Spectral type | J [mag] | Multiplicity[a] | DR1[b] |
|---|---|---|---|---|---|---|---|
| J01033+623 | V388Cas | 01:03:21.51 | +62:21:57.2 | M5.0 V | 8.611 | Binary | |
| J01032+200 | GJ 1026 A | 01:03:14.92 | +20:05:53.1 | M2.0 V | 7.670 | Binary | |
| J01032+316 | GJ 3073 | 01:03:14.23 | +31:40:59.7 | M3.2 V | 9.561 | Binary | |
| J01032+712 | LP 29-70 | 01:03:16.13 | +71:13:11.8 | M4.0 V | 9.689 | Binary | • |
| J01036+408 | G 132-50 | 01:03:40.31 | +40:51:26.6 | M0.0 V | 8.134 | Quadruple | |
| J01037+408 | G 132-51A | 01:03:42.24 | +40:51:13.5 | M2.6 V | 9.372 | Quadruple | |
| J01041+108 | StKM 1-112 | 01:04:11.07 | +10:51:35.4 | M1.0 V | 8.918 | | |
| J01048-181 | GJ 1028 | 01:04:55.25 | -18:07:20.8 | dM5.0 | 9.387 | | • |
| J01056+284 | GJ 1029 | 01:05:39.97 | +28:29:30.6 | M5.5 V | 9.486 | Binary (SB2) | • |
| J01066+152 | GJ 1030 | 01:06:41.39 | +15:16:18.1 | M2.0 V | 8.005 | | |
| J01066+192 | LSPM J0106+1913 | 01:06:36.93 | +19:13:29.6 | M3.0 V | 9.343 | | • |
| J01069+804 | LP 12-502 | 01:06:56.00 | +80:27:34.0 | M4.5 V | 9.350 | | |
| J01078+128 | G 2-21 | 01:07:52.53 | +12:52:51.4 | M1.5 V | 8.787 | | |
| J01102-118 | LP 707-16 | 01:10:17.75 | -11:51:19.3 | M3.0 V | 8.625 | | |
| J01114+154 | GJ 3076 | 01:11:25.63 | +15:26:19.9 | M5.93 | 9.082 | Binary | |
| J01116+120 | LP 467-15 | 01:11:36.66 | +12:05:02.3 | M2.0 V | 8.914 | | |
| J01119+049N | GJ 3077 | 01:11:56.01 | +04:54:56.6 | M3.5 V | 8.804 | Triple | |
| J01119+049S | GJ 3078 | 01:11:58.39 | +04:54:04.0 | M4.0 V | 9.641 | Triple | |
| J01125-169 | YZCet | 01:12:31.98 | -16:59:46.2 | M4.0 V | 7.258 | | • |
| J01133+589 | Wolf 58 | 01:13:20.12 | +58:55:20.3 | M1.5 V | 8.409 | Binary* | |
| J01134-229 | GJ 1033 | 01:13:24.20 | -22:54:07.3 | M4.0 V | 9.896 | | |
| J01141+790 | PM J01141+7904 | 01:14:06.69 | +79:04:01.9 | M3.0 V | 9.667 | | |
| J01147+253 | LP 351-6 | 01:14:49.86 | +25:18:57.6 | M1.5 V | 8.828 | | |
| J01158+470 | 1R011549.5+470159 | 01:15:50.51 | +47:02:02.1 | M4.5 V | 10.210 | Quadruple | |
| J01161+601 | Wolf 59 | 01:16:10.94 | +60:09:09.6 | M0.5 V | 8.333 | | |
| J01178+054 | GJ 3084 | 01:17:53.34 | +05:28:16.0 | M0.5 V | 7.839 | | |
| J01178+286 | Ross 324 | 01:17:50.21 | +28:40:09.6 | M0.5 V | 8.323 | | |
| J01182-128 | GJ 56.1 | 01:18:16.20 | -12:54:10.2 | M2.0 V | 8.356 | | |
| J01198+841 | GJ 1035 | 01:19:41.88 | +84:09:40.4 | M5.0 V | 9.855 | | • |
| J01214+243 | Ross 788 | 01:21:29.78 | +24:19:50.2 | M0.0 V | 7.894 | | |
| J01221+221 | LP 351-34 | 01:22:10.58 | +22:09:00.6 | M4.0 V | 8.412 | Binary | |
| J01227+005 | GJ 3093 | 01:22:44.77 | +00:31:55.7 | M5.0 V | 9.201 | Binary | |
| J01256+097 | Wolf 66 A | 01:25:36.89 | +09:45:18.5 | M4.0 V | 8.952 | Binary | |
| J01317+209 | Wolf 1523 | 01:32:44.80 | +20:59:13.5 | M2.0 V | 8.825 | | |
| J01324-219 | GJ 3098 | 01:32:25.53 | -21:54:32.7 | M1.5 Vk: | 7.977 | | |
| J01339-176 | LP 768-113 | 01:33:58.05 | -17:38:26.8 | dM4.0 | 8.842 | | • |
| J01352-072 | 1R013514.2-071254 | 01:35:14.03 | -07:12:52.2 | M4.0 V | 8.964 | Single | • |
| J01369-067 | LP 648-20 | 01:36:55.36 | -06:47:39.6 | M3.5 V | 9.707 | Binary | |
| J01373+610 | TYC 4031-2527-1 | 01:37:21.44 | +61:05:27.6 | M1.5 V | 8.662 | | |
| J01383+572 | Ross 10 | 01:38:21.23 | +57:13:51.4 | M2.5 V | 8.233 | | |
| J01384+006 | GJ 3103 | 01:38:30.49 | +00:39:08.4 | M2.5 V | 8.189 | | |
| J01390-179 | BLCet | 01:39:05.17 | -17:56:53.9 | M5.0 V | 6.283 | Binary | |
| J01395+050 | GJ 3104 | 01:39:31.32 | +05:03:20.3 | M3.0 V | 9.141 | Binary | |
| J01402+317 | GJ 3105 | 01:40:17.16 | +31:47:30.4 | M4.21 | 9.437 | | |
| J01431+210 | RX J0143.1+2101 | 01:43:11.75 | +21:01:10.4 | M4.0 V | 9.249 | Binary | |
| J01432+278 | GJ 3108 | 01:43:16.64 | +27:50:31.1 | M1.0 V | 7.483 | | |
| J01433+043 | GJ 70 | 01:43:19.73 | +04:19:05.7 | M2.0 V | 7.370 | | • |
| J01437-060 | PM J01437-0602 | 01:43:45.20 | -06:02:40.6 | M3.5 V | 8.770 | Binary (SB2) | |
| J01449+163 | Wolf 1530 | 01:44:57.63 | +16:20:32.5 | M4.0 V | 9.584 | | |
| J01466-086 | GJ 3113 | 01:46:37.29 | -08:39:00.5 | M4.0 V | 8.832 | Binary | |
| J01453+465 | G 173-18 | 01:45:18.81 | +46:32:11.3 | M2.0 V | 8.058 | Binary (SB2) | |
| J01480+212 | Wolf 87 | 01:48:04.37 | +21:12:20.8 | M2.5 V | 8.514 | | |
| J01510-061 | GJ 3119 | 01:51:04.69 | -06:07:09.3 | M4.5 V | 9.413 | | |
| J01514+213 | Wolf 90 | 01:51:24.17 | +21:23:33.9 | M4.0 V | 9.489 | | |
| J01518+644 | GJ 3117 | 01:51:51.72 | +64:26:02.8 | M2.5 V | 7.838 | Binary | |
| J01518-108 | Ross 555 | 01:51:49.30 | -10:48:21.1 | M2.0 V | 8.375 | | • |
| J01531-210 | BD-21 332 | 01:53:11.67 | -21:05:42.2 | M2.0 V | 8.066 | Binary (SB2) | |
| J01538-149 | PM J01538-1459A | 01:53:50.97 | -14:59:51.5 | M3.0 V | 7.938 | Binary | |
| J01544+576 | 1R015426.6+574136 | 01:54:28.04 | +57:41:27.7 | M3.5 V | 8.525 | Binary | |

Table B.1: Carmencita, the CARMENES input catalogue (continued).

| Karmn | Name | $\alpha$ (2016.0) | $\delta$ (2016.0) | Spectral type | $J$ [mag] | Multiplicity[a] | DR1[b] |
|---|---|---|---|---|---|---|---|
| J01550+379 | LSPM J0155+3758 | 01:55:02.64 | +37:57:54.6 | M5.0 V | 10.469 | | • |
| J01556+028 | G 73-5 | 01:55:36.98 | +02:52:53.8 | M1.5 V | 8.965 | | |
| J01567+305 | LP 296-57 | 01:56:45.99 | +30:33:28.6 | M4.5 V | 10.323 | Binary | |
| J01592+035E | GJ 1041 A | 01:59:12.66 | +03:31:09.6 | M1.0 V | 7.906 | Triple | |
| J01592+035W | GJ 1041 B | 01:59:12.89 | +03:31:12.1 | M3.0 V | 7.998 | Triple (SB2) | |
| J01593+585 | V596Cas | 01:59:24.17 | +58:31:13.0 | M4.0 V | 7.790 | | |
| J02000+437 | GJ 3123 | 02:00:03.00 | +43:45:23.9 | M3.0 V | 9.258 | | |
| J02001+366 | GJ 3124 | 02:00:07.50 | +36:39:43.8 | M3.8 V | 9.805 | | |
| J02002+130 | TZAri | 02:00:14.16 | +13:02:38.7 | M3.5: V | 7.514 | | • |
| J02007-103 | GJ 3127 | 02:00:46.86 | -10:21:26.6 | M3.5 V | 9.890 | | |
| J02015+637 | GJ 3126 | 02:01:34.72 | +63:46:10.5 | M2.5 V | 7.265 | | • |
| J02019+735 | GJ 3125 | 02:01:55.13 | +73:32:30.2 | M4.5 V | 9.252 | Binary | |
| J02020+039 | Wolf 109 B | 02:02:03.11 | +03:56:37.4 | M2.0 V | 9.395 | Triple | |
| J02022+103 | GJ 3128 | 02:02:15.51 | +10:20:09.4 | M5.5 V | 9.842 | | • |
| J02026+105 | PM J02024+1034B | 02:02:28.15 | +10:34:51.9 | M4.5 V | 8.396 | Binary | |
| J02027+135 | GJ 3129 | 02:02:44.86 | +13:34:31.9 | M5.0 V | 9.652 | Binary (SB2) | |
| J02028+047 | RX J0202.8+0446 | 02:02:52.00 | +04:47:00.4 | M3.5 V | 8.975 | | |
| J02033-212 | GJ 3131 | 02:03:20.52 | -21:13:50.2 | M2.5 V | 7.609 | Triple* (SB2) | |
| J02044-018 | GJ 3132 | 02:04:26.87 | -01:53:06.0 | M4.5 V | 9.585 | | |
| J02050-176 | GJ 84 | 02:05:06.31 | -17:36:55.5 | M2.5 V | 6.542 | Binary | |
| J02055+056 | Wolf 116 | 02:05:30.35 | +05:41:43.0 | M1.0 V | 8.936 | | |
| J02069+451 | V374And | 02:06:57.61 | +45:10:56.9 | M0.0 V | 7.397 | Binary (SB2) | |
| J02070+496 | G 173-37 | 02:07:04.24 | +49:38:36.6 | M3.5 V | 8.366 | | • |
| J02071+642 | GJ 3134 | 02:07:10.89 | +64:17:08.7 | M4.5 V | 9.875 | | |
| J02082+802 | G 242-81 | 02:08:18.76 | +80:13:11.1 | M0.0 V | 8.453 | | |
| J02088+494 | GJ 3136 | 02:08:54.01 | +49:26:51.8 | M4.0 V | 8.423 | | • |
| J02096-143 | GJ 3139 | 02:09:36.70 | -14:21:38.2 | M2.5 V | 8.122 | | |
| J02116+185 | G 35-32 | 02:11:41.19 | +18:33:42.3 | M3.0 V | 8.672 | | |
| J02123+035 | Wolf 124 | 02:12:19.10 | +03:34:02.6 | M1.5 V | 6.830 | | • |
| J02133+368 | 1R021320.6+364837 | 02:13:20.68 | +36:48:51.6 | M4.5 V | 9.367 | Binary | |
| J02129+000 | GJ 3142 | 02:12:55.22 | +00:00:17.3 | M4.0 V | 9.055 | | |
| J02142-039 | LP 649-72 | 02:14:13.11 | -03:57:46.1 | M5.5 V | 10.481 | | |
| J02149+174 | GJ 1045 | 02:15:00.20 | +17:25:00.8 | M4.0 V | 9.966 | | |
| J02153+074 | Wolf 127 | 02:15:22.39 | +07:29:32.5 | M1.5 V | 8.632 | | |
| J02155+339 | GJ 3143 | 02:15:34.63 | +33:57:35.0 | M4.2 V | 9.320 | | |
| J02158-126 | GJ 3145 | 02:15:49.43 | -12:40:24.2 | M3.5 V | 9.051 | | |
| J02164+135 | GJ 3146 | 02:16:30.40 | +13:35:05.9 | M5.0 V | 9.871 | | • |
| J02171+354 | GJ 3147 | 02:17:10.74 | +35:26:28.4 | M7.0 V | 9.983 | | |
| J02185+207 | G 35-39 | 02:18:35.93 | +20:47:44.8 | M2.5 V | 8.844 | | |
| J02186+123 | RX J0218.6+1219 | 02:18:36.75 | +12:18:56.0 | M2.5 V | 8.797 | | |
| J02190+238 | GJ 3150 | 02:19:02.67 | +23:52:53.7 | M3.6 V | 9.777 | | |
| J02190+353 | Ross 19 | 02:19:03.89 | +35:21:11.8 | M3.5 V | 8.662 | | |
| J02204+377 | GJ 3151 | 02:20:25.74 | +37:47:29.6 | M2.5 V | 8.952 | Binary | |
| J02207+029 | GJ 3153 | 02:20:46.42 | +02:58:32.9 | M6.0 V | 10.064 | | |
| J02210+368 | GJ 1047 A | 02:21:05.05 | +36:52:55.2 | M3.0 V | 9.368 | Triple | |
| J02222+478 | GJ 96 | 02:22:14.99 | +47:52:48.8 | M0.5 V | 6.377 | | • |
| J02230+181 | StKM 1-261 | 02:23:06.15 | +18:10:31.6 | M0.5 V | 8.493 | | |
| J02234+227 | LP 353-51 | 02:23:26.76 | +22:44:05.0 | M0.5 V | 8.182 | | |
| J02247+259 | GJ 3156 | 02:24:46.00 | +25:58:31.5 | M0.5 V | 8.456 | | |
| J02254+246 | StKM 1-265 | 02:25:28.00 | +24:40:36.9 | M2.0 V | 8.876 | | |
| J02256+375 | GJ 3157 | 02:25:38.83 | +37:32:32.8 | M5e | 9.712 | | |
| J02272+545 | 1R022716.4+543258 | 02:27:17.31 | +54:32:46.1 | M4.5 V | 10.212 | Binary | |
| J02274+031 | PM J02274+0310 | 02:27:27.43 | +03:10:54.7 | M4.0 V | 9.978 | | |
| J02277+044 | HD 15285 | 02:27:46.05 | +04:25:58.8 | M1.0 V | 5.990 | Binary | |
| J02282+014 | GJ 3159 | 02:28:17.41 | +01:26:29.0 | M3.0 V | 9.281 | | |
| J02283+219 | TYC 1221-1171-1 | 02:28:22.17 | +21:59:45.3 | M0.5 V | 8.456 | | |
| J02285-200 | GJ 100 C | 02:28:32.62 | -20:02:22.5 | M2.5 V | 9.181 | Triple | |
| J02287+156 | LSPM J0228+1538 | 02:28:47.15 | +15:38:53.6 | M2.0 V | 8.792 | Binary | |
| J02289+120 | GJ 3160 | 02:28:54.68 | +12:05:22.3 | M2.5 V | 8.373 | Binary (SB2) | |

Table B.1: Carmencita, the CARMENES input catalogue (continued).

| Karmn | Name | $\alpha$ (2016.0) | $\delta$ (2016.0) | Spectral type | $J$ [mag] | Multiplicity[a] | DR1[b] |
|---|---|---|---|---|---|---|---|
| J02289+226 | StKM 2-213A | 02:28:58.41 | +22:36:24.5 | M2.0 V | 8.749 | Binary | |
| J02292+195 | LP 410-33 | 02:29:14.32 | +19:32:31.9 | M2.5 V | 8.959 | | |
| J02293+884 | GJ 3137 | 02:29:14.09 | +88:24:20.3 | M3.5 V | 9.065 | | |
| J02314+573 | Ross 21 | 02:31:29.87 | +57:22:43.3 | M3.5 V | 9.218 | | |
| J02330+078 | LP 530-26 | 02:33:04.78 | +07:49:41.0 | M2.0 V | 8.648 | | |
| J02336+249 | GJ 102 | 02:33:37.23 | +24:55:26.9 | M3.5 V | 8.472 | | • |
| J02337+150 | GJ 3165 | 02:33:47.96 | +15:00:17.8 | M3.0 V | 9.692 | | |
| J02340+417 | GJ 3164 | 02:34:00.46 | +41:46:50.4 | M3.0 V | 9.638 | | |
| J02345+566 | G 174-4 | 02:34:34.87 | +56:36:42.1 | M2.0 V | 8.893 | | |
| J02353+235 | GJ 3166 | 02:35:22.50 | +23:34:29.5 | M4.0 V | 9.536 | | |
| J02358+202 | GJ 104 | 02:35:53.59 | +20:13:09.3 | M2.0 V | 7.208 | | • |
| J02362+068 | BXCet | 02:36:17.20 | +06:52:41.1 | M4.0 V | 7.333 | Triple | |
| J02364+554 | GJ 3168 | 02:36:26.63 | +55:28:30.2 | M3.0 V | 9.339 | | |
| J02367+226 | G 36-26 | 02:36:44.08 | +22:40:20.2 | M5.0 V | 10.081 | | |
| J02367+320 | GJ 3169 | 02:36:47.30 | +32:04:18.9 | M3.5 V | 8.965 | Binary | |
| J02392+074 | GJ 3174 | 02:39:17.85 | +07:28:14.9 | M3.7 V | 9.881 | | |
| J02412-045 | G 75-35 | 02:41:15.51 | -04:32:18.8 | M4.5 V | 9.199 | | |
| J02419+435 | StKM 1-291 | 02:41:58.94 | +43:34:19.0 | M1.0 V | 8.244 | | |
| J02424+182 | LP 410-81 | 02:42:25.80 | +18:14:44.0 | M1.5 V | 8.873 | | |
| J02438-088 | Wolf 1132 | 02:43:53.88 | -08:49:57.8 | M1.5 V | 8.693 | | |
| J02441+492 | GJ 107 B | 02:44:10.79 | +49:13:53.0 | M1.5 V | 6.688 | Binary | |
| J02442+255 | VXAri | 02:44:16.53 | +25:31:18.3 | dM3 | 6.752 | | • |
| J02443+109W | MCC 401 | 02:44:21.45 | +10:57:40.2 | M1.0 V | 7.973 | Triple | |
| J02443+109E | 2M02442272+1057349 | 02:44:22.81 | +10:57:34.2 | M5.0 V | 10.301 | Triple | |
| J02456+449 | GJ 3178 | 02:45:40.30 | +44:56:53.5 | M0.5 V | 7.818 | Binary | |
| J02462-049 | GJ 3180 | 02:46:16.73 | -04:59:50.7 | M6.0 V | 10.970 | | |
| J02465+164 | GJ 3181 | 02:46:33.79 | +16:25:01.1 | M6.0 V | 10.971 | | • |
| J02486+621 | 2M02483695+6211228 | 02:48:37.25 | +62:11:21.3 | M5.5 V | 12.505 | | • |
| J02489-145W | PM J02489-1432W | 02:48:59.45 | -14:32:14.2 | M2.0 V | 9.528 | Binary | • |
| J02489-145E | PM J02489-1432E | 02:49:00.02 | -14:32:15.5 | M2.5 V | 9.733 | Binary | • |
| J02502+628 | G 246-12 | 02:50:16.95 | +62:51:16.4 | M2.5 V | 9.368 | | |
| J02518+062 | GJ 3184 | 02:51:51.15 | +06:13:39.6 | M3.0 V | 9.415 | | |
| J02518+294 | GJ 3183 | 02:51:49.63 | +29:29:10.4 | M4.0 V | 9.518 | Binary | |
| J02519+224 | RBS 365 | 02:51:54.22 | +22:27:28.2 | dM4.0 | 8.919 | | • |
| J02524+269 | GJ 3186 | 02:52:25.03 | +26:58:26.2 | M2.0 V | 7.937 | | |
| J02530+168 | Teegarden's Star | 02:53:04.71 | +16:51:51.7 | M7.0 V | 8.394 | | • |
| J02534+174 | LP 411-18 | 02:53:26.14 | +17:24:28.4 | M3.5 V | 8.716 | | |
| J02555+268 | HD 18143C | 02:55:36.11 | +26:52:17.6 | M4.0 V | 9.561 | Triple | |
| J02560-006 | LP 591-156 | 02:56:04.17 | -00:36:32.0 | M5.0 V | 10.421 | | • |
| J02562+239 | LSPM J0256+2359 | 02:56:14.06 | +23:59:07.4 | M5.0 V | 9.977 | Binary | |
| J02565+554W | Ross 364 | 02:56:35.79 | +55:26:06.8 | M1.0 V | 7.425 | Binary | |
| J02565+554E | Ross 365 | 02:56:36.49 | +55:26:22.3 | M2.5 V | 8.006 | Binary | • |
| J02573+765 | LP 14-53 | 02:57:21.43 | +76:33:04.9 | M4.0 V | 9.615 | | • |
| J02575+107 | Ross 791 | 02:57:32.96 | +10:47:17.9 | M4.0 V | 9.162 | | |
| J02581-128 | GJ 3189 | 02:58:10.53 | -12:52:57.4 | sdM3.0 | 8.952 | | |
| J02591+366 | Ross 331 | 02:59:11.44 | +36:36:35.7 | M3.7 V | 9.064 | Binary | |
| J02592+317 | GJ 3191 | 02:59:16.77 | +31:46:27.8 | M3.3 V | 9.523 | | |
| J02597+389 | G 134-63 | 02:59:46.67 | +38:55:34.6 | M4.5 V | 10.411 | Binary | |
| J03018-165S | GJ 3193 | 03:01:50.98 | -16:35:40.3 | M3.0 V | 7.294 | Binary | |
| J03018-165N | GJ 3192 | 03:01:50.63 | -16:35:35.2 | M2.5 V | 7.110 | Binary | |
| J03026-181 | GJ 9108 | 03:02:38.51 | -18:09:56.1 | M2.5 V | 8.208 | | |
| J03033-080 | StM 20 | 03:03:21.47 | -08:05:16.0 | M3.0 V | 9.122 | | |
| J03037-128 | GJ 3197 | 03:03:48.10 | -12:51:21.0 | M3.5 V | 9.283 | Binary | |
| J03036-128 | GJ 3196 | 03:03:40.99 | -12:50:33.7 | M3.5 V | 9.411 | Binary | |
| J03040-203 | GJ 3198 | 03:04:05.05 | -20:22:50.7 | M4.0 V | 8.634 | | |
| J03047+617 | GJ 3195 | 03:04:45.06 | +61:43:57.7 | M3.0 V | 8.877 | Binary | |
| J03075-039 | GJ 3202 | 03:07:33.51 | -03:58:23.2 | M0.0 V | 7.950 | Binary | |
| J03077+249 | LP 355-27 | 03:07:47.13 | +24:57:53.3 | M4.5 V | 10.132 | | |
| J03090+100 | GJ 1055 | 03:09:00.48 | +10:01:16.3 | M5.0 V | 9.926 | | • |

Table B.1: Carmencita, the CARMENES input catalogue (continued).

| Karmn | Name | α (2016.0) | δ (2016.0) | Spectral type | J [mag] | Multiplicity[a] | DR1[b] |
|---|---|---|---|---|---|---|---|
| J03095+457 | GJ 125 | 03:09:30.17 | +45:43:52.7 | M2.0 V | 6.730 | Binary | |
| J03102+059 | EKCet | 03:10:15.34 | +05:54:22.5 | M2.5 V | 8.363 | | |
| J03104+584 | GJ 3204 | 03:10:26.72 | +58:26:05.5 | M2.0 V | 8.332 | | |
| J03109+737 | GJ 1053 | 03:11:05.27 | +73:46:02.5 | M6.0 V | 9.850 | | • |
| J03110-046 | LP 652-62 | 03:11:04.90 | -04:36:41.1 | M3.0 V | 9.406 | | |
| J03112+011 | 1R031114.2+010655 | 03:11:15.60 | +01:06:30.6 | M5.5 V | 10.682 | | |
| J03118+196 | Wolf 132 | 03:11:48.26 | +19:40:11.3 | M0.5 V | 8.044 | | |
| J03119+615 | GJ 3206 | 03:11:56.89 | +61:31:14.6 | M0.0 V | 7.374 | Binary | |
| J03133+047 | CDCet | 03:13:24.78 | +04:46:30.7 | M4.5 V | 8.775 | | • |
| J03136+653 | LP 53-55 | 03:13:37.88 | +65:21:19.5 | M1.5 V | 8.728 | | |
| J03142+286 | GJ 3208 | 03:14:12.85 | +28:40:27.6 | M6.0 V | 10.993 | | • |
| J03145+594 | Ross 369A | 03:14:33.17 | +59:26:13.8 | M2.5 V | 8.448 | Binary* | |
| J03147+114 | RX J0314.7+1127 | 03:14:47.26 | +11:27:26.7 | M2.0 V | 9.352 | | |
| J03147+485 | Ross 346 | 03:14:45.11 | +48:31:05.5 | M1.5 V | 8.174 | | |
| J03162+581S | Ross 370A | 03:16:14.73 | +58:09:57.3 | M2.0 V | 7.344 | Binary | |
| J03162+581N | Ross 370B | 03:16:14.75 | +58:10:02.4 | M2.0 V | 7.501 | Binary | |
| J03167+389 | PM J03167+3855 | 03:16:46.02 | +38:55:27.7 | M3.5 V | 9.157 | | |
| J03172+453 | GJ 3213 | 03:17:11.82 | +45:22:20.9 | M3.0 V | 8.422 | Binary | |
| J03177+252 | GJ 3215 | 03:17:46.17 | +25:15:00.6 | M2.5 V | 8.492 | | |
| J03181+382 | HD 275122 | 03:18:08.09 | +38:14:58.3 | M1.5 V | 7.023 | | • |
| J03181+426 | Wolf 140 | 03:18:07.31 | +42:40:06.8 | M3.5 V | 9.254 | | |
| J03185+103 | StKM 1-354 | 03:18:35.36 | +10:18:43.2 | M1.5 V | 8.995 | | |
| J03186+326 | GJ 3216 | 03:18:38.59 | +32:39:55.3 | M1.0 V | 8.227 | | |
| J03187+606 | Ross 371 | 03:18:43.66 | +60:36:21.1 | M3.5 V | 9.450 | | |
| J03194+619 | G 246-33 | 03:19:29.28 | +61:56:01.5 | M4.0 V | 9.511 | Binary | |
| J03207+397 | LP 198-637 A | 03:20:45.42 | +39:42:59.3 | M1.5 V | 8.085 | Binary | |
| J03213+799 | GJ 133 | 03:21:24.29 | +79:58:06.8 | M2.0 V | 7.704 | | • |
| J03217-066 | GJ 3218 | 03:21:47.27 | -06:40:25.0 | M2.0 V | 7.857 | | • |
| J03220+029 | GJ 1058 | 03:22:04.46 | +02:56:22.7 | M4.5 V | 10.314 | | |
| J03224+271 | GJ 3219 | 03:22:28.40 | +27:09:20.8 | M0.0 V | 8.249 | | |
| J03230+420 | GJ 1059 | 03:23:02.16 | +42:00:15.8 | M5.0 V | 10.389 | | • |
| J03233+116 | GJ 3221 | 03:23:22.15 | +11:41:11.0 | M3.5 V | 8.386 | | |
| J03236+056 | 1R032338.7+054117 | 03:23:39.25 | +05:41:14.1 | M4.5 V | 9.867 | | |
| J03241+237 | GJ 140 A | 03:24:06.73 | +23:47:04.2 | M1.5 V | 7.128 | Triple | |
| J03242+237 | GJ 140 C | 03:24:13.10 | +23:46:17.3 | M2.5 V | 8.276 | Triple | |
| J03247+447 | PM J03247+4447A | 03:24:42.31 | +44:47:41.4 | M1.5 V | 8.566 | Binary | |
| J03257+058 | GJ 3224 | 03:25:42.04 | +05:51:48.2 | M4.5 V | 9.946 | Binary | |
| J03263+171 | PM J03263+1709 | 03:26:23.74 | +17:09:30.1 | M4.0 V | 9.774 | Binary | |
| J03267+192 | GJ 3225 | 03:26:44.97 | +19:14:37.7 | M4.5 V | 10.123 | | |
| J03272+273 | CKAri | 03:27:14.22 | +27:23:07.8 | M1.0 V | 8.637 | | |
| J03275+222 | ATO J051.8788+22.2102 | 03:27:30.94 | +22:12:36.9 | M4.5 V | 10.044 | | |
| J03284+352 | LSPM J0328+3515A | 03:28:29.35 | +35:15:18.6 | M2.0 V | 8.970 | Binary | |
| J03286-156 | GJ 3228 | 03:28:39.18 | -15:37:16.4 | M3.5 V | 9.855 | Binary | |
| J03303+346 | 1R033021.4+344044 | 03:30:23.37 | +34:40:31.7 | M4.0 V | 9.995 | Triple* | |
| J03288+264 | GJ 3227 | 03:28:49.84 | +26:29:10.2 | M4.0 V | 9.288 | | |
| J03308+542 | LSPM J0330+5413 | 03:30:48.61 | +54:13:55.1 | M5.0 V | 10.173 | | |
| J03309+706 | LP 31-368 | 03:30:56.01 | +70:41:06.4 | M3.5 V | 9.487 | Binary | |
| J03317+143 | GJ 143.3 | 03:31:47.18 | +14:19:07.0 | M2.5 V | 8.695 | | |
| J03325+287 | RX J0332.6+2843 | 03:32:35.85 | +28:43:54.1 | M4.5 V | 9.357 | Triple | |
| J03332+462 | HD 21845B | 03:33:14.16 | +46:15:16.2 | M0.0 V | 8.382 | Binary | |
| J03340+585 | Ross 563 | 03:34:01.11 | +58:35:52.3 | M0.5 V | 8.288 | | |
| J03346-048 | GJ 3235 | 03:34:40.07 | -04:50:38.5 | M3.8 V | 8.829 | Triple (ST3) | |
| J03361+313 | 1R033609.2+311853 | 03:36:08.85 | +31:18:37.4 | M4.5 V | 9.187 | | |
| J03366+034 | GJ 3237 | 03:36:40.97 | +03:29:17.6 | M5.0 V | 9.295 | | |
| J03372+691 | GJ 3236 | 03:37:14.55 | +69:10:47.9 | M3.8 V | 9.806 | Binary (EB/SB2) | |
| J03375+178N | GJ 3239 | 03:37:33.54 | +17:51:14.1 | M2.5 V | 9.100 | Quadruple (SB2) | |
| J03375+178S | GJ 3240 | 03:37:34.09 | +17:51:00.0 | M3.5 V | 9.186 | Quadruple (EB?/SB2) | |
| J03394+249 | KPTau | 03:39:29.85 | +24:58:05.7 | M3.5 V | 8.813 | | |
| J03396+254E | Wolf 204 | 03:39:36.47 | +25:28:11.6 | M3.0 V | 8.747 | Binary | |

Table B.1: Carmencita, the CARMENES input catalogue (continued).

| Karmn | Name | $\alpha$ (2016.0) | $\delta$ (2016.0) | Spectral type | $J$ [mag] | Multiplicity[a] | DR1[b] |
|---|---|---|---|---|---|---|---|
| J03396+254W | Wolf 205 | 03:39:40.77 | +25:28:39.1 | M3.5 V | 9.079 | Binary | |
| J03397+334 | HD 278874B | 03:39:47.79 | +33:28:30.7 | M3.0 V | 8.967 | Triple | |
| J03416+552 | TYC 3720-426-1 | 03:41:37.46 | +55:13:05.0 | M0.0 V | 8.347 | | |
| J03430+459 | LSPM J0343+4554A | 03:43:01.79 | +45:54:17.4 | M4.0 V | 9.668 | Binary | |
| J03433-095 | GJ 3247 | 03:43:22.53 | -09:33:46.1 | M4.5 V | 9.799 | | |
| J03438+166 | GJ 150.1 A | 03:43:52.74 | +16:40:14.2 | M0.0 V | 7.046 | Binary | |
| J03437+166 | GJ 150.1 B | 03:43:45.42 | +16:39:57.2 | M1.0 V | 7.533 | Binary | |
| J03445+349 | HD 278968 | 03:44:31.21 | +34:58:20.8 | M0.0 V | 7.905 | | |
| J03454+729 | G 221-21 | 03:45:28.68 | +72:59:25.2 | M1.5 V | 8.296 | Triple | |
| J03455+703 | PM J03455+7018 | 03:45:32.34 | +70:18:00.4 | M1.0 V | 8.932 | | |
| J03459+147 | G 6-33 | 03:45:54.96 | +14:42:47.2 | M1.5 V | 8.772 | | |
| J03463+262 | HD 23453 | 03:46:20.60 | +26:12:52.7 | M1.0 V | 6.689 | | • |
| J03467+821 | TYC 4521-1342-1 | 03:46:42.67 | +82:07:50.1 | M1.0 V | 8.987 | | |
| J03467-112 | GJ 3249 | 03:46:46.04 | -11:17:40.5 | M2.5 V | 9.002 | | |
| J03473+086 | GJ 3250 | 03:47:21.40 | +08:41:36.7 | M5.0 V | 9.849 | | • |
| J03473-019 | G 80-21 | 03:47:23.53 | -01:58:24.3 | M3.0 V | 7.804 | | • |
| J03479+027 | Ross 588 | 03:47:57.68 | +02:47:09.3 | M0.5 V | 7.964 | | |
| J03480+686 | GJ 153 C | 03:48:02.08 | +68:40:42.9 | M2 V | 7.379 | Triple | |
| J03486+735 | GJ 3248 | 03:48:39.73 | +73:32:30.9 | M1.0 V | 7.994 | | |
| J03505+634 | GJ 3251 | 03:50:33.69 | +63:27:14.9 | M1.5 V | 8.192 | | |
| J03507-060 | GJ 1065 | 03:50:43.81 | -06:06:03.6 | M3.0 V | 8.570 | | |
| J03510+142 | PM J03510+1413 | 03:51:00.87 | +14:13:38.7 | M4.5 V | 9.436 | Binary | |
| J03510-008 | GJ 3252 | 03:51:00.04 | -00:52:52.4 | M8.0 V | 11.302 | | |
| J03519+397 | TYC 2868-639-1 | 03:51:58.19 | +39:46:55.8 | M0.0 V | 8.277 | Binary | |
| J03526+170 | Wolf 227 | 03:52:42.24 | +17:00:53.8 | M5.0 V | 8.933 | Binary (SB2) | |
| J03531+625 | Ross 567 | 03:53:10.51 | +62:34:03.8 | M3.0 V | 7.782 | | • |
| J03543-146 | 2M03542008-1437388 | 03:54:20.02 | -14:37:37.2 | M6.5 V | 11.339 | | |
| J03544-091 | GJ 3256 | 03:54:25.52 | -09:09:29.2 | M1.0 V | 7.817 | Binary* | |
| J03548+163 | LP 413-108 | 03:54:53.37 | +16:18:55.9 | M4.0 V | 9.960 | | |
| J03565+319 | 1R035632.5+315746 | 03:56:33.26 | +31:57:23.8 | M3.5 V | 9.795 | | |
| J03567+039 | Ross 23 | 03:56:47.95 | +53:33:30.5 | M1.5 V | 7.809 | | |
| J03574-011 | HD 24916B | 03:57:28.68 | -01:09:25.7 | M2.5 V | 7.773 | Triple (SB) | |
| J03586+520 | Ross 24 | 03:58:36.92 | +52:01:21.7 | M1.0 V | 8.922 | | |
| J03588+125 | G 7-14 | 03:58:49.38 | +12:30:18.3 | M4.0 V | 9.757 | | |
| J03598+260 | Wolf 1322 | 03:59:54.53 | +26:05:19.5 | M3.0 V | 8.714 | | |
| J04011+513 | Ross 25 | 04:01:08.18 | +51:23:06.4 | M3.8 V | 9.665 | Binary* | |
| J04056+057 | GJ 3261 | 04:05:38.94 | +05:44:40.4 | M4 V | 8.813 | Triple* | |
| J04059+712E | LP 31-301 | 04:05:58.09 | +71:16:34.7 | M4.0 V | 9.527 | Triple | |
| J04059+712W | LP 31-302 A | 04:05:57.21 | +71:16:32.4 | M5.0 V | 10.099 | Triple | |
| J04061-055 | PM J04061-0534 | 04:06:06.90 | -05:34:46.9 | M3.5 V | 9.128 | | |
| J04077+142 | LP 474-123 | 04:07:44.14 | +14:13:22.1 | M0.0 V | 8.041 | Binary | |
| J04079+142 | LP 474-124 | 04:07:54.99 | +14:12:58.2 | M2.5 V | 9.215 | Binary | |
| J04081+743 | LP 32-16 | 04:08:13.87 | +74:22:51.6 | M3.5 V | 9.247 | | |
| J04083+691 | LP 31-433 | 04:08:24.60 | +69:10:57.9 | M4.5 V | 10.263 | | |
| J04086+336 | HD 281621 | 04:08:38.07 | +33:38:15.3 | M0.5 V | 7.016 | Binary | |
| J04093+057 | LP 534-29 | 04:09:22.50 | +05:46:25.2 | M4.5 V | 10.708 | Binary | |
| J04108-128 | LP 714-37 | 04:10:47.88 | -12:51:48.4 | M5.5 V | 11.008 | Binary | |
| J04112+495 | Ross 27 | 04:11:13.29 | +49:31:45.0 | M3.5 V | 9.486 | | |
| J04122+647 | GJ 3266 | 04:12:18.25 | +64:43:48.7 | M4.0 V | 9.156 | | |
| J04123+162 | LP 414-117 | 04:12:21.90 | +16:15:02.9 | M4.0 V | 9.736 | SKG (SB2) | |
| J04129+526 | Ross 28 | 04:12:58.22 | +52:36:28.9 | M4.0 V | 8.773 | Binary | |
| J04131+505 | Ross 29A | 04:13:09.43 | +50:31:38.0 | M4.0 V | 9.260 | Binary | |
| J04137+476 | LSPM J0413+4737E | 04:13:47.70 | +47:37:42.5 | M2.5 V | 8.951 | | |
| J04139+829 | GJ 3262 | 04:13:56.69 | +82:55:03.0 | M0.0 V | 7.955 | | |
| J04148+277 | G 39-3 | 04:14:53.80 | +27:45:26.0 | M3.5 V | 8.763 | | |
| J04153-076 | DYEri | 04:15:19.12 | -07:40:15.3 | M4.5 V | 6.747 | Triple | • |
| J04166-125 | GJ 2033 A | 04:16:41.59 | -12:33:19.3 | M1.0 V | 7.572 | Binary | |
| J04167-120 | LP 714-47 | 04:16:45.65 | -12:05:05.6 | M0.0 V | 9.493 | | • |
| J04173+088 | GJ 3270 | 04:17:18.68 | +08:49:16.0 | M5.0 V | 9.030 | | |

Table B.1: Carmencita, the CARMENES input catalogue (continued).

| Karmn | Name | α (2016.0) | δ (2016.0) | Spectral type | J [mag] | Multiplicity[a] | DR1[b] |
|---|---|---|---|---|---|---|---|
| J04177+410 | LSPM J0417+4103A | 04:17:44.44 | +41:03:10.1 | M3.5 V | 9.238 | Binary | |
| J04188+013 | HIP 20122 | 04:18:51.45 | +01:23:35.0 | M2.0 V | 8.877 | | |
| J04191+097 | UPM J0419+0944 | 04:19:08.15 | +09:44:50.2 | M3.0 V | 9.990 | | |
| J04191-074 | LP 654-39 | 04:19:06.41 | -07:27:43.6 | M3.5 V | 9.968 | | |
| J04198+425 | LSPM J0419+4233 | 04:19:52.90 | +42:33:07.4 | M8/9V | 11.094 | | • |
| J04199+364 | Ross 592 | 04:19:59.96 | +36:29:04.1 | M1.5 V | 8.225 | | |
| J04205+815 | PM J04205+8131 | 04:20:34.24 | +81:31:54.4 | M3.0 V | 9.482 | | |
| J04206+272 | XEST 16-045 | 04:20:39.20 | +27:17:31.4 | M4.5 V | 10.497 | | |
| J04207+152 | LP 415-363 | 04:20:48.16 | +15:14:08.2 | M4.0 V | 9.490 | Binary | |
| J04218+213 | GJ 3274 | 04:21:50.22 | +21:19:39.2 | M3.5 V | 9.080 | | |
| J04219+751 | GJ 3271 | 04:21:59.86 | +75:08:20.5 | M3.0 V | 8.545 | | • |
| J04221+192 | GJ 3275 | 04:22:08.13 | +19:15:21.4 | M3.0 V | 9.209 | | |
| J04224+036 | RX J0422.4+0337 | 04:22:25.19 | +03:37:08.5 | M3.5 V | 9.857 | Binary* | |
| J04224+740 | LP 31-339 | 04:22:28.52 | +74:01:21.5 | M1.5 V | 8.854 | | |
| J04225+105 | LSPM J0422+1031 | 04:22:32.25 | +10:31:19.3 | M3.5 V | 8.471 | | |
| J04225+390 | GJ 1070 | 04:22:34.31 | +39:00:34.0 | M5.0 V | 10.473 | | • |
| J04227+205 | LP 415-30 | 04:22:42.99 | +20:34:11.9 | M4.0 V | 10.458 | | |
| J04229+259 | G 8-31 | 04:22:59.30 | +25:59:10.4 | M4.5 V | 9.645 | | |
| J04234+495 | TYC 3337-1716-1 | 04:23:26.84 | +49:34:15.6 | M2.5 V | 8.329 | | |
| J04234+809 | 1R042323.2+805511 | 04:23:29.60 | +80:55:08.8 | M4.0 V | 9.412 | | |
| J04238+092 | LP 535-73 | 04:23:50.83 | +09:12:19.4 | M3.0 V | 9.117 | | |
| J04238+149 | INTau | 04:23:50.50 | +14:55:17.0 | M3.5 V | 9.293 | | |
| J04247-067 | 1R042441.9-064725 | 04:24:42.78 | -06:47:31.2 | M4.0 V | 9.566 | Triple (ST3) | |
| J04248+324 | GJ 3280 | 04:24:49.49 | +32:26:56.0 | M2.5 V | 8.818 | | |
| J04251+515 | PM J04251+5131 | 04:25:09.86 | +51:31:56.2 | M2.0 V | 8.803 | | |
| J04252+080S | GJ 3282 | 04:25:15.25 | +08:02:55.8 | M2.5 V | 8.908 | SKG (SB2) | |
| J04252+080N | GJ 3283 | 04:25:17.09 | +08:04:03.9 | M4.0 V | 10.422 | SKG | |
| J04252+172 | V805Tau | 04:25:13.67 | +17:16:05.1 | M3.5 V | 9.149 | SKG | |
| J04274+203 | TYC 1273-9-1 | 04:27:24.97 | +20:22:44.5 | M1.5 V | 8.769 | | |
| J04276+595 | GJ 3287 | 04:27:41.56 | +59:35:13.5 | M3.8 V | 9.975 | Binary | |
| J04278+117 | GJ 3291 | 04:27:53.89 | +11:46:46.8 | M4.2 V | 9.699 | | |
| J04284+176 | V1102Tau | 04:28:28.89 | +17:41:44.9 | M2.0 V | 8.592 | Binary | |
| J04290+219 | HD 28343 | 04:29:00.05 | +21:55:24.5 | M0.5 V | 5.674 | | |
| J04293+142 | GJ 3292 | 04:29:18.76 | +14:14:02.0 | M3.8 V | 9.350 | | • |
| J04294+262 | FW Tau | 04:29:29.71 | +26:16:52.8 | M5.5 V | 10.340 | Triple | |
| J04302+708 | PM J04302+7049 | 04:30:11.72 | +70:49:14.3 | M1.5 V | 8.753 | | |
| J04304+398 | V546Per | 04:30:25.67 | +39:50:50.4 | M5.0 V | 9.113 | | |
| J04308-088 | LP 655-23 | 04:30:52.04 | -08:49:22.0 | M4.0 V | 9.853 | Binary | |
| J04310+367 | PM J04310+3647A | 04:30:59.95 | +36:47:54.7 | M3.0 V | 9.445 | Binary | |
| J04311+589 | GJ 169.1 A | 04:31:14.21 | +58:58:04.7 | M4.0 V | 6.622 | Binary | |
| J04312+422 | PM J04312+4217 | 04:31:14.99 | +42:17:08.9 | M2.5 V | 8.349 | | • |
| J04313+241 | V927Tau | 04:31:23.83 | +24:10:52.6 | M4.5: V | 9.729 | Binary | |
| J04326+098 | LP 475-1095 | 04:32:37.96 | +09:51:06.5 | M1.5 V | 8.386 | | |
| J04329+001E | LP 595-23 | 04:32:56.07 | +00:06:14.7 | M0.5 V | 8.421 | Triple | |
| J04329+001S | G 82-28 | 04:32:55.38 | +00:06:28.3 | M4.0 V | 9.861 | Triple | |
| J04329+001N | LP 595-21 | 04:32:55.32 | +00:06:33.2 | M4.0 V | 10.300 | Triple | |
| J04333+239 | V697TauA | 04:33:23.91 | +23:59:26.0 | M3.0 V | 8.914 | Binary | |
| J04335+207 | GJ 3296 | 04:33:34.48 | +20:44:40.7 | M5.0 V | 9.769 | | |
| J04343+430 | PM J04343+4302 | 04:34:22.55 | +43:02:13.3 | M2.65 V | 9.616 | | |
| J04347-004 | LP 595-11 | 04:34:45.24 | -00:26:50.2 | M4.0 V | 9.307 | | • |
| J04350+086 | StKM 1-495 | 04:35:02.65 | +08:39:30.5 | M1.0 V | 8.828 | Binary | |
| J04352-161 | LP 775-31 | 04:35:16.33 | -16:06:52.2 | M8.0 V | 10.406 | Binary (SB2) | |
| J04366+112 | GJ 3302 | 04:36:39.94 | +11:13:22.8 | M4.0 V | 10.089 | | |
| J04369+593 | LP 84-34 | 04:36:58.75 | +59:21:57.7 | M2.0 V | 8.986 | | |
| J04369-162 | 1R043657.1-161258 | 04:36:57.49 | -16:13:07.0 | M3.5 V | 9.117 | | |
| J04373+193 | LP 415-1644 | 04:37:22.00 | +19:21:16.9 | M4.0 V | 10.182 | | |
| J04376+528 | HD 232979 | 04:37:41.47 | +52:53:29.4 | M0.0 V | 5.866 | | |
| J04376-024 | GJ 3305 | 04:37:37.51 | -02:29:29.7 | M1.1 V | 7.299 | Triple | |
| J04376-110 | GJ 173 | 04:37:41.62 | -11:02:23.1 | M1.5 V | 6.943 | | • |

Table B.1: Carmencita, the CARMENES input catalogue (continued).

| Karmn | Name | $\alpha$ (2016.0) | $\delta$ (2016.0) | Spectral type | $J$ [mag] | Multiplicity[a] | DR1[b] |
|---|---|---|---|---|---|---|---|
| J04382+282 | GJ 3304 A | 04:38:13.13 | +28:12:58.4 | M4.0 V | 8.173 | Binary | • |
| J04386-115 | LP 715-39 | 04:38:36.86 | -11:30:18.6 | M3.5 V | 8.672 | | |
| J04388+217 | G 8-48A | 04:38:53.72 | +21:47:51.7 | M3.5 V | 9.552 | Triple | |
| J04393+335 | PM J04393+3331 | 04:39:23.22 | +33:31:48.7 | M4.0 V | 9.919 | Triple | |
| J04395+162 | LP 415-302 | 04:39:31.54 | +16:15:30.2 | M5.5 V | 10.139 | | |
| J04398+251 | PM J04398+2509 | 04:39:48.86 | +25:09:25.4 | M3.5 V | 9.642 | | |
| J04403-055 | LP 655-48 | 04:40:23.63 | -05:30:06.1 | M6.0 V | 10.658 | | |
| J04404-091 | GJ 9163 A | 04:40:29.13 | -09:11:48.5 | M0.0 V | 7.133 | Binary | |
| J04406-128 | TOI-2457 | 04:40:40.16 | -12:53:26.6 | M0.0 V | 9.741 | | |
| J04407+022 | GJ 3307 | 04:40:42.67 | +02:13:52.7 | M2.0 V | 7.894 | | |
| J04413+327 | G 39-30A | 04:41:24.22 | +32:42:19.9 | M4.0 V | 9.463 | Binary | |
| J04414+132 | TYC 694-1183-1 | 04:41:29.78 | +13:13:16.0 | M0.5 V | 8.356 | Binary (SB) | |
| J04422+577 | LP 84-59 | 04:42:15.86 | +57:42:18.2 | M0.0 V | 8.456 | | |
| J04423+207 | LP 415-1896 | 04:42:23.76 | +20:46:34.9 | M2.0 V | 8.560 | | |
| J04425+204 | LP 415-345 | 04:42:30.40 | +20:27:10.8 | M3.0 V | 9.396 | SKG (SB2) | |
| J04429+095 | PM J04429+0935 | 04:42:55.14 | +09:35:53.7 | M1.0 V | 8.882 | Binary* | |
| J04429+189 | HD 285968 | 04:42:56.52 | +18:57:11.5 | M2.5 V | 6.462 | | |
| J04429+214 | PM J04429+2128 | 04:42:55.90 | +21:28:24.8 | M3.5 V | 7.958 | | • |
| J04433+296 | Haro 6-36 | 04:43:20.23 | +29:40:05.7 | M5.5 V | 10.402 | | • |
| J04444+278 | HD 283779 | 04:44:26.17 | +27:51:37.2 | M1.5 V | 7.908 | | |
| J04458-144 | PM J04458-1426 | 04:45:52.69 | -14:26:23.6 | M4.0 V | 9.088 | | |
| J04468-112 | PM J04468-1116A | 04:46:51.63 | -11:16:48.6 | M3.0 V | 8.144 | Binary | |
| J04471+021 | GJ 3313 | 04:47:11.55 | +02:09:39.6 | M0.0 V | 8.308 | | |
| J04472+206 | RX J0447.2+2038 | 04:47:12.35 | +20:38:09.2 | M5.0 V | 9.380 | | |
| J04480+170 | LP 416-43 | 04:48:00.98 | +17:03:21.1 | M0.5 V | 8.214 | SKG (SB) | • |
| J04488+100 | 1R044847.6+100302 | 04:48:47.32 | +10:03:01.4 | M3.0 V | 8.127 | Binary (SB2) | |
| J04494+484 | G 81-34 | 04:49:29.77 | +48:28:42.9 | M4.0 V | 9.059 | Binary | |
| J04499+236 | EM* LkCa 18A | 04:49:56.34 | +23:41:00.1 | M1.0 V | 8.242 | Binary | |
| J04499+711 | LP 32-204 | 04:49:56.29 | +71:09:46.5 | M3.5 V | 9.633 | | |
| J04502+459 | GJ 3315 | 04:50:15.59 | +45:58:46.2 | M1.0 V | 8.533 | | |
| J04504+199 | BPM 85800 | 04:50:25.49 | +19:59:09.1 | M1.5 V | 8.731 | | |
| J04508+221 | GJ 1072 | 04:50:51.65 | +22:07:14.7 | M5.0 V | 9.896 | | |
| J04508+261 | GJ 3316 | 04:50:51.21 | +26:07:22.3 | M2.5 V | 9.152 | | |
| J04520+064 | Wolf 1539 | 04:52:05.90 | +06:28:30.7 | M3.5 V | 7.814 | | |
| J04524-168 | LP 776-25 | 04:52:24.55 | -16:49:25.3 | M3.3 V | 7.740 | | • |
| J04525+407 | GJ 1073 | 04:52:36.26 | +40:42:06.6 | M5.0 V | 9.071 | | |
| J04536+623 | LP 84-48 | 04:53:40.79 | +62:18:59.9 | M3.5 V | 9.226 | | |
| J04538+158 | LSPM J0453+1549 | 04:53:50.10 | +15:49:12.6 | M2.5 V | 9.432 | | |
| J04538-177 | GJ 180 | 04:53:50.44 | -17:46:34.6 | M2.0 V | 7.413 | | • |
| J04544+650 | 1R045430.9+650451 | 04:54:29.98 | +65:04:39.5 | M4.0 V | 9.668 | | |
| J04559+046 | HD 31412B | 04:55:54.60 | +04:40:13.5 | M2.0 V | 8.501 | Triple | |
| J04560+432 | LP 202-2 | 04:56:04.12 | +43:13:53.0 | M4.0 V | 9.304 | | |
| J04587+509 | GJ 1074 | 04:58:46.84 | +50:56:32.4 | M1.0 V | 7.896 | | |
| J04588+498 | GJ 181 | 04:58:50.76 | +49:50:55.6 | M0.0 V | 6.925 | | |
| J04595+017 | GJ 182 | 04:59:34.88 | +01:46:59.2 | M0.0 V | 7.117 | | • |
| J05012+248 | Ross 794 | 05:01:15.79 | +24:52:18.2 | M2.0 V | 8.084 | | |
| J05013+226 | LSPM J0501+2237 | 05:01:17.95 | +22:36:55.3 | M4.5 V | 10.161 | | |
| J05018+037 | GJ 3321 | 05:01:50.71 | +03:45:53.1 | M1.5 V | 8.079 | | |
| J05019+011 | 1R050156.7+010845 | 05:01:56.70 | +01:08:41.4 | M4.0 V | 8.526 | | |
| J05019+099 | GJ 3322 A | 05:01:58.83 | +09:58:57.1 | M4.0 V | 7.212 | Binary (SB2) | • |
| J05019-069 | GJ 3323 | 05:01:56.83 | -06:56:54.9 | M4.0 V | 7.617 | | • |
| J05024-212 | HD 32450A | 05:02:28.28 | -21:15:28.4 | M2.0 V | 5.450 | Binary | |
| J05032+213 | HD 285190 | 05:03:16.21 | +21:23:54.0 | M1.5 V | 7.451 | Quadruple (SB2) | |
| J05033-173 | GJ 3325 | 05:03:19.83 | -17:22:31.8 | M3.0 V | 7.819 | | |
| J05034+531 | GJ 184 | 05:03:26.22 | +53:07:17.9 | M0.5 V | 7.001 | Binary | |
| J05042+110 | GJ 3326 | 05:04:14.69 | +11:03:27.0 | M5.0 V | 9.144 | | • |
| J05050+442 | UPM J0505+4414 | 05:05:06.06 | +44:14:03.3 | M5.0 V | 9.829 | | |
| J05051-120 | GJ 3327 | 05:05:11.55 | -12:00:30.9 | M3.0 V | 9.103 | | |
| J05060+043 | GJ 3328 | 05:06:04.44 | +04:20:16.1 | M1.0 V | 8.303 | | |

Table B.1: Carmencita, the CARMENES input catalogue (continued).

| Karmn | Name | $\alpha$ (2016.0) | $\delta$ (2016.0) | Spectral type | $J$ [mag] | Multiplicity[a] | DR1[b] |
|---|---|---|---|---|---|---|---|
| J05062+046 | RX J0506.2+0439 | 05:06:12.96 | +04:39:25.7 | M4.0 V | 8.909 | | |
| J05068-215E | GJ 3331 | 05:06:49.97 | -21:35:09.4 | M1.5 V | 7.046 | Triple | |
| J05068-215W | GJ 3332 | 05:06:49.48 | -21:35:04.3 | M3.5 V | 7.003 | Triple | • |
| J05072+375 | RX J0507.2+3731A | 05:07:14.33 | +37:30:42.1 | M5.0 V | 10.284 | Binary* | |
| J05076+275 | TYC 1853-1649-1 | 05:07:36.74 | +27:30:03.8 | M0.5 V | 8.321 | | |
| J05078+179 | Wolf 230 | 05:07:49.34 | +17:58:53.5 | M3.0 V | 8.023 | Triple (ST2) | |
| J05083+756 | LP 15-315 | 05:08:19.36 | +75:38:13.3 | M4.5 V | 9.391 | Binary | |
| J05084-210 | 2M05082729-2101444 | 05:08:27.34 | -21:01:44.6 | M5.0 V | 9.716 | Binary | |
| J05085-181 | GJ 190 | 05:08:35.61 | -18:10:41.8 | M3.5 V | 6.175 | Binary | |
| J05091+154 | Ross 388 | 05:09:10.11 | +15:27:22.8 | M3.5 V | 8.770 | | • |
| J05103+095 | G 97-23 | 05:10:18.12 | +09:30:01.8 | M2.0 V | 8.864 | | |
| J05103+272 | LSPM J0510+2714 | 05:10:19.83 | +27:13:51.8 | M7.0 V | 10.698 | | |
| J05103+488 | GJ 3336 A | 05:10:22.30 | +48:50:26.3 | M2.5 V | 7.827 | Binary | |
| J05106+297 | G 86-28 | 05:10:39.22 | +29:46:48.9 | M3.0 V | 8.600 | | |
| J05109+186 | GJ 3337 | 05:10:57.16 | +18:37:24.1 | M4.0 V | 9.935 | | |
| J05111+158 | StKM 1-549 | 05:11:09.78 | +15:48:57.0 | M1.0 V | 8.974 | | |
| J05114+101 | LP 477-36 | 05:11:29.68 | +10:07:12.2 | M1.0 V | 8.846 | | |
| J05127+196 | GJ 192 | 05:12:42.54 | +19:40:00.2 | M2.0 V | 7.299 | | |
| J05151-073 | GJ 3340 | 05:15:08.33 | -07:20:55.4 | M1.0 V | 8.355 | | |
| J05152+236 | UPM J0515+2336 | 05:15:17.58 | +23:36:25.2 | M5.0 V | 10.186 | | • |
| J05155+591 | LSPM J0515+5911 | 05:15:31.19 | +59:11:01.3 | M7.5 V | 11.320 | | |
| J05173+321 | G 86-37 | 05:17:20.09 | +32:07:29.7 | M3.5 V | 9.236 | Binary | |
| J05173+458 | Capella H | 05:17:24.00 | +45:50:16.1 | M1.0 V | 6.777 | Quadruple | |
| J05173+721 | TYC 4351-466-1 | 05:17:21.23 | +72:10:49.8 | M1.0 V | 8.689 | | |
| J05187+464 | PM J05187+4629 | 05:18:44.62 | +46:29:57.9 | M4.5 V | 9.957 | | |
| J05195+649 | 1R051929.3+645435 | 05:19:31.21 | +64:54:36.2 | M3.5 V | 8.950 | | |
| J05206+587N | GJ 3342 | 05:20:41.70 | +58:47:25.2 | M3.5 V | 9.295 | Binary | |
| J05206+587S | GJ 3343 | 05:20:41.05 | +58:47:12.0 | M3.5 V | 9.926 | Binary | |
| J05211+557 | GJ 3345 | 05:21:10.47 | +55:45:46.5 | M3.5 V | 9.229 | | |
| J05223+305 | PM J05223+3031 | 05:22:20.65 | +30:31:08.3 | M3.0 V | 9.406 | | |
| J05226+795 | TYC 4532-731-1 | 05:22:39.87 | +79:34:30.3 | M0.5 V | 8.200 | | |
| J05228+202 | PM J05228+2016 | 05:22:50.18 | +20:16:36.5 | M2.5 V | 8.635 | Binary | |
| J05243-160 | PM J05243-1601A | 05:24:19.14 | -16:01:15.8 | M4.5 V | 8.668 | Binary | |
| J05256-091 | LP 717-36 | 05:25:41.70 | -09:09:15.8 | M3.5 V | 8.454 | Binary | |
| J05280+096 | Ross 41 | 05:27:59.94 | +09:38:26.0 | M3.5 V | 8.311 | | |
| J05282+029 | GJ 1080 | 05:28:14.24 | +02:57:56.6 | M3.5 V | 8.979 | Triple (SB) | |
| J05289+125 | GJ 3348A | 05:28:56.61 | +12:31:50.4 | M4.0 V | 9.649 | Quintuple | • |
| J05294+155E | GJ 2043 | 05:29:26.92 | +15:34:36.2 | M0.0: V | 7.557 | Binary | |
| J05294+155W | GJ 2043 B | 05:29:26.02 | +15:34:43.4 | M4.0 V | 10.252 | Binary | |
| J05298+320 | Ross 406 | 05:29:52.40 | +32:04:40.4 | M2.5 V | 8.649 | | |
| J05298-034 | Wolf 1450 | 05:29:51.70 | -03:26:37.6 | M3.0 V | 8.276 | | |
| J05306+152 | LSPM J0530+1514 | 05:30:37.09 | +15:14:26.3 | M3.0 V | 8.985 | | |
| J05314-036 | HD 36395 | 05:31:28.21 | -03:41:11.5 | M1.5 V | 4.999 | | |
| J05320-030 | V1311Ori | 05:32:04.51 | -03:05:30.0 | M2.0 V | 7.879 | Quintuple | |
| J05322+098 | Ross 42 | 05:32:14.46 | +09:49:11.5 | M3.5 V | 7.423 | Binary (SB2) | • |
| J05328+338 | G 98-7 | 05:32:51.57 | +33:49:39.8 | M3.5 V | 9.391 | | |
| J05333+448 | GJ 1081 | 05:33:19.20 | +44:48:52.9 | M3.5 V | 8.197 | Binary | |
| J05337+019 | V371Ori | 05:33:44.55 | +01:56:41.0 | M3.0 V | 7.764 | Binary (SB1) | |
| J05339-023 | RX J0534.0-0221 | 05:33:59.83 | -02:21:33.3 | M3.0 V | 8.564 | | |
| J05341+475 | PM J05341+4732A | 05:34:10.56 | +47:32:02.8 | M2.5 V | 8.752 | Quadruple* | • |
| J05341+512 | GJ 3352 | 05:34:08.57 | +51:12:52.8 | M1.0 V | 8.001 | | |
| J05342+103N | Ross 45 | 05:34:15.05 | +10:19:08.0 | M3.0 V | 8.561 | Triple | |
| J05342+103S | Ross 45B | 05:34:15.00 | +10:19:03.0 | M4.5 V | 9.186 | Triple (SB) | |
| J05348+138 | Ross 46 | 05:34:51.99 | +13:52:40.4 | M3.0 V | 7.781 | | |
| J05360-076 | Wolf 1457 | 05:36:00.20 | -07:38:51.0 | dM4.0 | 8.464 | | |
| J05365+113 | V2689Ori | 05:36:30.99 | +11:19:39.4 | M0.0 V | 6.126 | Binary | • |
| J05366+112 | PM J05366+1117 | 05:36:38.46 | +11:17:47.8 | M4.0 V | 8.266 | Binary | • |
| J05394+406 | LSPM J0539+4038 | 05:39:25.71 | +40:38:29.5 | M8.0 V | 11.109 | | • |
| J05394+747 | LP 33-191 | 05:39:25.44 | +74:46:02.6 | M3.5 V | 9.330 | | • |

Table B.1: Carmencita, the CARMENES input catalogue (continued).

| Karmn | Name | $\alpha$ (2016.0) | $\delta$ (2016.0) | Spectral type | $J$ [mag] | Multiplicity[a] | DR1[b] |
|---|---|---|---|---|---|---|---|
| J05402+126 | V1402Ori | 05:40:16.07 | +12:38:56.4 | M1.5 V | 8.072 | Binary (SB2) | • |
| J05415+534 | HD 233153 | 05:41:30.74 | +53:29:15.0 | M1.0 V | 6.586 | Binary | |
| J05404+248 | V780Tau | 05:40:25.82 | +24:48:02.0 | M5.5 V | 8.978 | Binary | |
| J05419+153 | GJ 9188 | 05:41:58.95 | +15:20:13.3 | M0.0 V | 7.690 | | |
| J05421+124 | V1352Ori | 05:42:11.45 | +12:28:56.5 | M4.0 V | 7.124 | | • |
| J05422-054 | GJ 2045 | 05:42:12.53 | -05:27:40.3 | M5.0 V | 10.206 | | |
| J05425+154 | 1R054232.1+152459 | 05:42:31.70 | +15:25:00.2 | M3.5 V | 9.443 | | • |
| J05455-119 | PM J05455-1158 | 05:45:32.04 | -11:58:02.3 | M4.5 V | 9.590 | | |
| J05456+729 | PM J05456+7255 | 05:45:39.09 | +72:55:14.6 | M3.0 V | 9.395 | | |
| J05458+729 | PM J05458+7254 | 05:45:50.02 | +72:54:09.0 | M2.5 V | 9.338 | | |
| J05466+441 | Wolf 237 | 05:46:37.60 | +44:07:14.0 | M4.0 V | 8.459 | Triple (SB2) | |
| J05468+665 | TYC 4106-420-1 | 05:46:48.92 | +66:30:13.1 | M0.5 V | 8.421 | | |
| J05471-052 | GJ 3366 | 05:47:09.69 | -05:12:20.3 | M4.5 V | 10.039 | | |
| J05472-000 | GJ 3367 | 05:47:17.89 | -00:00:49.9 | M0 | 7.987 | | |
| J05484+077 | GJ 3368 | 05:48:24.15 | +07:45:34.3 | M4.0 V | 9.784 | | |
| J05511+122 | PM J05511+1216 | 05:51:10.51 | +12:16:09.4 | M4.0 V | 9.453 | | |
| J05530+047 | G 106-7 | 05:53:04.75 | +04:43:02.6 | M1.5 V | 8.950 | Binary* | |
| J05530+251 | LSPM J0553+2507 | 05:53:01.92 | +25:07:40.9 | M3.0 V | 8.552 | Binary | |
| J05532+242 | Ross 59 | 05:53:14.24 | +24:15:22.1 | M1.5 V | 7.485 | Binary (SB2) | |
| J05547+109 | RX J0554.7+1055 | 05:54:45.58 | +10:55:55.9 | M3.0 V | 8.832 | | |
| J05558+406 | PM J05558+4036 | 05:55:48.31 | +40:36:48.0 | M1.0 V | 8.776 | Binary* | • |
| J05566-103 | 1R055641.0-101837 | 05:56:40.63 | -10:18:35.8 | M3.5 V | 9.067 | | |
| J05587+259 | PM J05587+2557 | 05:58:47.68 | +25:57:40.1 | M1.0 V | 8.776 | | |
| J05588+213 | G 104-9 | 05:58:53.53 | +21:20:54.5 | M5.0 V | 9.968 | Binary | |
| J05596+585 | EGCam | 05:59:37.77 | +58:35:30.8 | M0.5 V | 7.068 | Binary | |
| J05599+585 | GJ 3372 | 05:59:55.68 | +58:34:11.2 | M4.2 V | 9.028 | Binary | |
| J06000+027 | GJ 3379 | 06:00:03.83 | +02:42:22.9 | M3.5 V | 6.905 | | |
| J06008+681 | GJ 3373 | 06:00:50.71 | +68:09:05.3 | M3.5 V | 8.922 | Binary | |
| J06007+681 | GJ 3374 | 06:00:47.81 | +68:08:11.6 | M4.0 V | 9.178 | Binary | • |
| J06011+595 | GJ 3378 | 06:01:10.82 | +59:35:35.0 | M4.0 V | 7.465 | | |
| J06017+130 | LSPM J0601+1305 | 06:01:45.54 | +13:05:00.7 | M2.5 V | 8.444 | | |
| J06023-203 | GJ 3382 | 06:02:22.56 | -20:19:35.3 | M3.5 V | 9.215 | | • |
| J06024+498 | GJ 3380 | 06:02:29.31 | +49:51:42.6 | M5.0 V | 9.350 | | |
| J06024+663 | LP 57-46 | 06:02:26.51 | +66:20:32.3 | M4.5 V | 9.855 | | |
| J06025+371 | PM J06025+3707 | 06:02:35.46 | +37:07:36.2 | M1.0 V | 8.433 | | • |
| J06034+478 | Wolf 261 | 06:03:29.48 | +47:48:06.0 | M4.2 V | 9.691 | | |
| J06035+155 | 1R060335.0+153132 | 06:03:34.74 | +15:31:30.4 | M0.0 V | 8.203 | | |
| J06035+168 | 1R060334.8+165128 | 06:03:34.49 | +16:51:45.4 | M4.0 V | 9.387 | | |
| J06039+261 | Ross 60 | 06:03:54.54 | +26:08:46.0 | M3.0 V | 9.839 | | |
| J06054+608 | LP 86-173 | 06:05:30.04 | +60:49:09.8 | M4.5 V | 9.096 | | |
| J06066+465 | PM J06066+4633A | 06:06:37.78 | +46:33:47.0 | M3.0 V | 9.231 | Binary | |
| J06071+335 | Ross 70 | 06:07:11.91 | +33:32:30.9 | M2.0 V | 8.992 | | |
| J06075+472 | 1R060732.5+471154 | 06:07:31.91 | +47:12:23.3 | M4.5 V | 9.723 | | |
| J06097+001 | HD 291290 | 06:09:46.30 | +00:09:30.8 | M0.0 V | 8.124 | | |
| J06102+225 | PM J06102+2234 | 06:10:17.81 | +22:34:17.2 | M4.0 V | 9.876 | Triple | |
| J06103+225 | LP 362-121 | 06:10:22.52 | +22:34:18.1 | M5.0 V | 10.644 | Triple | |
| J06103+722 | LSPM J0610+7212 | 06:10:18.09 | +72:11:58.0 | M2.5 V | 9.272 | | |
| J06103+821 | GJ 226 | 06:10:20.24 | +82:06:02.9 | M2.0 V | 6.869 | | |
| J06105+024 | TYC 135-239-1 | 06:10:31.41 | +02:25:30.3 | M0.0 V | 8.342 | | |
| J06105-218 | HD 42581 | 06:10:34.46 | -21:52:04.2 | M0.5 V | 5.104 | Binary | • |
| J06107+259 | Wolf 1058 | 06:10:46.51 | +25:55:53.3 | M1.5 V | 8.245 | | • |
| J06109+103 | Ross 79 | 06:10:54.88 | +10:18:50.6 | M2.5 V | 6.795 | Binary | |
| J06140+516 | GJ 3388 | 06:14:01.72 | +51:40:06.5 | M3.5 V | 8.860 | | |
| J06145+025 | G 106-35 | 06:14:34.76 | +02:30:19.7 | M3.0 V | 9.296 | | |
| J06171+051 | GJ 231.1 B | 06:17:10.42 | +05:07:05.3 | M3.5 V | 9.088 | Quadruple | |
| J06151-164 | LP 779-34 | 06:15:11.90 | -16:26:21.4 | M4.0 V | 9.283 | | |
| J06171+751 | TYC 4525-194-1 | 06:18:07.08 | +75:06:04.3 | M2.0 V | 8.041 | Triple (ST3?) | |
| J06171+838 | LSPM J0617+8353 | 06:17:04.81 | +83:53:32.5 | M3.5 V | 8.961 | | |
| J06185+250 | G 103-29 | 06:18:34.83 | +25:03:00.6 | M4.0 V | 9.954 | | |

Table B.1: Carmencita, the CARMENES input catalogue (continued).

| Karmn | Name | α (2016.0) | δ (2016.0) | Spectral type | J [mag] | Multiplicity[a] | DR1[b] |
|---|---|---|---|---|---|---|---|
| J06193-066 | Ross 417 | 06:19:20.74 | -06:39:32.1 | M3.0 V | 9.122 | | |
| J06194+139 | TYC 743-1836-1 | 06:19:29.61 | +13:57:02.3 | M0.5 V | 7.841 | | |
| J06212+442 | GJ 3391 | 06:21:13.26 | +44:14:26.7 | M2.0 V | 8.724 | Binary | |
| J06216+163 | LP 420-5 | 06:21:36.85 | +16:18:33.8 | M1.0 V | 8.482 | Binary | |
| J06217+163 | LP 420-6 | 06:21:44.17 | +16:19:19.9 | M2.5 V | 9.064 | Binary | |
| J06218-227 | GJ 2049 | 06:21:53.08 | -22:43:19.7 | M1 Vk | 7.832 | | |
| J06223+334 | TYC 2425-1286-1 | 06:22:20.65 | +33:26:54.6 | M1.0 V | 8.753 | | |
| J06236-096 | LP 720-10 | 06:23:38.41 | -09:38:51.5 | M3.5 V | 9.819 | Binary | |
| J06237+020 | TYC 141-24-1 | 06:23:46.49 | +05:02:40.1 | M1.5 V | 8.032 | | |
| J06238+456 | LP 160-22 | 06:23:51.20 | +45:40:00.0 | M5.0 V | 10.348 | Binary | |
| J06246+234 | Ross 64 | 06:24:41.93 | +23:25:50.8 | M4.0 V | 8.662 | | |
| J06258+561 | GJ 3393 | 06:25:53.21 | +56:10:16.9 | M4.0 V | 10.257 | | |
| J06262+238 | 1R062614.2+234942 | 06:26:14.52 | +23:49:36.4 | M1.5 V | 8.636 | | • |
| J06277+093 | Ross 603A | 06:27:43.80 | +09:23:51.3 | M2.0 V | 8.269 | Binary | |
| J06293-028 | V577Mon | 06:29:24.18 | -02:49:01.9 | M4.5 V | 6.376 | Binary | |
| J06298-027 | G 108-4 | 06:29:50.47 | -02:47:49.0 | M4.0 V | 9.468 | Binary (SB2) | |
| J06306+456 | PM J06306+4539 | 06:30:37.39 | +45:39:23.0 | M1.0 V | 8.893 | | |
| J06307+397 | PM J06307+3947 | 06:30:47.39 | +39:47:38.5 | M2.0 V | 9.405 | | |
| J06310+500 | GJ 3395 | 06:31:00.97 | +50:02:45.5 | M0.8 V | 7.873 | | |
| J06318+414 | GJ 3396 | 06:31:50.73 | +41:29:42.2 | M5.84 | 9.680 | | |
| J06322+378 | TYC 2928-1568-1 | 06:32:14.91 | +37:48:10.6 | M1.5 V | 8.220 | | |
| J06323-097 | PM J06323-0943 | 06:32:20.28 | -09:43:29.9 | M4.5 V | 9.848 | | • |
| J06325+641 | LP 57-192 | 06:32:31.27 | +64:06:12.2 | M4.0 V | 9.811 | | |
| J06345+315 | G 103-41 | 06:34:33.48 | +31:30:05.1 | M3.5 V | 8.705 | | |
| J06354-040 | 1R063531.2-040314 | 06:35:29.75 | -04:03:17.2 | M5.5 V | 9.272 | Binary | |
| J06361+116 | GJ 3398 | 06:36:06.16 | +11:36:49.5 | M5.0 V | 9.794 | | |
| J06361+201 | LP 420-4 | 06:36:12.05 | +20:08:10.3 | M2.5 V | 9.434 | | |
| J06371+175 | HD 260655 | 06:37:09.94 | +17:33:58.7 | M0.0 V | 6.674 | | |
| J06396-210 | LP 780-32 | 06:39:37.20 | -21:01:32.4 | dM4.0 | 8.507 | Binary* | |
| J06400+285 | GJ 3399 | 06:40:05.54 | +28:35:10.5 | M2.5 V | 8.270 | Binary | • |
| J06401-164 | LP 780-23 | 06:40:08.72 | -16:27:21.5 | M2.5 V | 9.121 | Binary* | • |
| J06414+157 | Wolf 289 | 06:41:28.23 | +15:45:42.6 | M4.0 V | 9.570 | | |
| J06421+035 | GJ 3404 | 06:42:11.24 | +03:34:48.5 | M3.0 V | 8.166 | Binary | |
| J06422+035 | GJ 3405 | 06:42:13.39 | +03:35:26.8 | M4.0 V | 9.112 | Binary | |
| J06435+166 | G 110-14 | 06:43:34.53 | +16:41:35.5 | M4.5 V | 9.776 | | • |
| J06438+511 | GJ 3406A | 06:43:49.95 | +51:08:06.0 | M2.5 V | 8.361 | Binary | |
| J06447+718 | GJ 2050 | 06:44:45.25 | +71:53:06.7 | M0.5 V | 7.846 | Binary | |
| J06461+325 | HD 263175B | 06:46:06.88 | +32:33:16.6 | M1.0 V | 8.992 | Binary | |
| J06467+159 | 1R064645.7+155739 | 06:46:45.62 | +15:57:41.8 | M1.0 V | 7.923 | | |
| J06474+054 | G 108-27 | 06:47:27.57 | +05:24:23.3 | M4.0 V | 9.453 | | |
| J06486+532 | LP 121-58 | 06:48:38.72 | +53:17:24.3 | M1.5 V | 8.952 | | |
| J06489+211 | 1R064855.9+210754 | 06:48:55.18 | +21:08:02.8 | M2.5 V | 9.367 | | |
| J06490+371 | GJ 1092 | 06:49:05.73 | +37:06:25.0 | M4.0 V | 9.561 | | |
| J06509-091 | LP 661-2 | 06:50:59.37 | -09:10:59.3 | M3.5 V | 9.400 | | |
| J06523-051 | HD 50281B | 06:52:17.42 | -05:11:24.3 | M2.0 V | 6.579 | Binary | |
| J06524+182 | GJ 3413 | 06:52:24.45 | +18:17:06.9 | M3.5 V | 9.052 | | |
| J06540+608 | GJ 3412 | 06:54:05.43 | +60:52:02.4 | M3.0 V | 7.128 | Binary | |
| J06548+332 | HD 265866 | 06:54:48.03 | +33:15:59.1 | M3.0 V | 6.104 | | |
| J06564+121 | TYC 756-1685-1 | 06:56:25.84 | +12:07:31.4 | M1.0 V | 7.995 | | |
| J06564+400 | GJ 3416 | 06:56:28.64 | +40:04:58.3 | M1.0 V | 8.009 | Binary | • |
| J06564+759 | LP 34-110 | 06:53:24.30 | +72:55:09.3 | M1.0 V | 8.918 | | |
| J06565+440 | G 107-36 | 06:56:31.24 | +44:01:45.0 | M4.5 V | 9.923 | | |
| J06574+740 | 1R065728.1+740529 | 06:57:25.77 | +74:05:26.1 | M4.0 V | 8.926 | | |
| J06579+623 | GJ 3417 A | 06:57:57.83 | +62:19:11.0 | M6.0 V | 8.585 | Binary | |
| J06582+511 | G 192-59 | 06:58:12.70 | +51:08:32.4 | M2.0 V | 8.976 | | • |
| J06594+193 | GJ 1093 | 06:59:29.84 | +19:20:41.5 | M5.0 V | 9.160 | | |
| J06594+195 | G 88-2 | 06:59:28.65 | +19:30:30.4 | M3.0 V | 8.939 | | |
| J06596+057 | PM J06596+0545 | 06:59:41.55 | +05:45:38.9 | M2.5 V | 8.947 | | • |
| J07001-190 | 1R070005.1-190115 | 07:00:07.00 | -19:01:25.1 | M5.0 V | 9.029 | Binary (SB2) | |

Table B.1: Carmencita, the CARMENES input catalogue (continued).

| Karmn | Name | $\alpha$ (2016.0) | $\delta$ (2016.0) | Spectral type | $J$ [mag] | Multiplicity[a] | DR1[b] |
|---|---|---|---|---|---|---|---|
| J07009-023 | PM J07009-0221 | 07:00:59.74 | -02:21:32.2 | M3.0 V | 9.301 | | |
| J07012+008 | PM J07012+0052 | 07:01:15.54 | +00:52:40.4 | M2.5 V | 8.794 | | • |
| J07033+346 | GJ 3423 | 07:03:23.09 | +34:41:53.6 | M4.0 V | 8.773 | | |
| J07034+767 | LP 16-379 | 07:03:29.98 | +76:46:21.9 | M3.5 V | 8.836 | | |
| J07039+527 | GJ 3421 | 07:03:56.92 | +52:41:51.8 | M5.0 V | 8.537 | Binary | • |
| J07042-105 | Ross 54 | 07:04:17.55 | -10:30:44.6 | M3.5 V | 7.313 | Binary | |
| J07044+682 | GJ 258 | 07:04:26.94 | +68:17:20.5 | M3.0 V | 8.170 | | |
| J07047+249 | Ross 874 | 07:04:49.44 | +24:59:50.6 | M1.5 V | 8.273 | | |
| J07051-101 | 1R070511.2-100801 | 07:05:12.10 | -10:07:51.6 | M5.0 V | 10.196 | | • |
| J07052+084 | G 108-52 | 07:05:12.39 | +08:25:45.7 | M2.0 V | 8.829 | | |
| J07076+486 | GJ 3426 | 07:07:37.70 | +48:41:08.6 | M4.3 V | 9.106 | | |
| J07078+672 | GJ 3425 | 07:07:49.68 | +67:12:03.7 | M1.5 V | 7.872 | | |
| J07081-228 | LP 840-16 | 07:08:06.53 | -22:48:51.0 | M2.0 V | 8.094 | | |
| J07086+307 | GJ 3429 | 07:08:39.72 | +30:42:51.6 | M0.5 V | 8.329 | | |
| J07095+698 | GJ 3427 | 07:09:31.85 | +69:50:53.1 | M3.0 V | 8.862 | | |
| J07100+385 | QYAur | 07:10:01.23 | +38:31:31.0 | M4.5 V | 6.731 | Binary (SB2) | |
| J07102+376 | GJ 3430 | 07:10:13.32 | +37:40:05.9 | M4.0 V | 10.297 | Binary | |
| J07105-087 | 1R071032.6-084232 | 07:10:31.37 | -08:42:46.7 | M3.5 V | 9.054 | | |
| J07111+434 | LP 206-11 | 07:11:11.97 | +43:29:49.0 | M5.5 V | 9.979 | Binary | |
| J07119+773 | TYC 4530-1414-1 | 07:11:57.13 | +77:21:57.4 | M1.5 V | 7.723 | Binary (SB1) | |
| J07121+522 | GJ 3432 | 07:12:11.14 | +52:16:20.4 | M1.0 V | 8.118 | | |
| J07129+357 | 1R071259.5+354655 | 07:12:59.62 | +35:47:03.1 | M2.5 V | 8.834 | | |
| J07140+507 | G 193-39 | 07:14:04.29 | +50:43:28.9 | M0.5 V | 8.443 | Binary | |
| J07163+271 | GJ 268.3 | 07:16:19.73 | +27:08:29.9 | M2.5 V | 7.013 | Binary | |
| J07163+331 | GJ 1096 | 07:16:17.89 | +33:09:03.4 | M5.0 V | 9.763 | | |
| J07172-050 | PM J07172-0501 | 07:17:17.54 | -05:01:09.8 | M3.5 V | 8.873 | | |
| J07174+195 | GJ 3437 | 07:17:29.57 | +19:34:12.4 | M3.2 V | 9.017 | | |
| J07181+392 | Ross 987 | 07:18:07.89 | +39:16:27.4 | M0.0 V | 7.209 | Binary | |
| J07182+137 | PM J07182+1342 | 07:18:12.86 | +13:42:16.2 | M3.5 V | 9.361 | | |
| J07195+328 | GJ 270 | 07:19:31.79 | +32:49:42.8 | M0.0 V | 7.184 | | |
| J07199+840 | TYC 4618-116-1 | 07:19:57.65 | +84:04:36.8 | M2.5 V | 8.305 | | |
| J07200-087 | Scholz's star | 07:20:03.21 | -08:46:51.9 | M9.5+T5 | 10.628 | Binary | |
| J07212+005 | TYC 178-2187-1 | 07:31:12.97 | +00:33:13.8 | M0.5 V | 8.305 | | |
| J07227+306 | GJ 3439 | 07:22:41.50 | +30:40:02.3 | M4.0 V | 9.506 | | |
| J07232+460 | GJ 272 | 07:23:14.71 | +46:05:10.9 | M0.5 V | 7.343 | | |
| J07274+052 | Luyten's Star | 07:27:25.11 | +05:12:33.8 | M3.5 V | 5.714 | Binary | |
| J07274+220 | Ross 878 | 07:27:28.31 | +22:02:35.6 | M1.5 V | 7.818 | | |
| J07282-187 | GJ 3442 | 07:28:13.09 | -18:47:25.2 | M4.5 V | 9.049 | | • |
| J07287-032 | GJ 1097 | 07:28:45.91 | -03:18:05.9 | M3.0 V | 7.544 | | |
| J07295+359 | 1R072931.4+355607 | 07:29:31.04 | +35:55:58.5 | M1.5 V | 8.644 | Triple | |
| J07307+481 | GJ 275.2 A | 07:30:42.46 | +48:11:38.2 | M4.0 V | 9.141 | Triple | • |
| J07310+460 | 1R073101.9+460030 | 07:31:01.27 | +46:00:24.8 | M4.0 V | 9.948 | Quadruple* | |
| J07319+362S | LynA | 07:31:57.38 | +36:13:06.2 | M2.5 V | 6.771 | Triple | |
| J07319+362N | BLLyn | 07:31:56.97 | +36:13:43.2 | M3.5 V | 7.571 | Triple | |
| J07319+392 | GJ 3445 | 07:31:56.74 | +39:13:34.0 | M2.48 V | 9.164 | | • |
| J07320+173E | GJ 3447 | 07:32:02.63 | +17:19:07.0 | M0.0 V | 8.169 | Triple | |
| J07320+173W | GJ 3448 | 07:32:01.87 | +17:19:09.4 | M3.2 V | 9.739 | Triple | |
| J07320+686 | GJ 9235 | 07:32:01.51 | +68:37:13.6 | M1.5 V | 7.745 | | |
| J07325+248 | G 88-37 | 07:32:30.71 | +24:53:42.4 | M3.0 V | 8.987 | | |
| J07342+009 | GJ 1099 | 07:34:17.57 | +00:58:59.7 | M2.5 V | 8.261 | | |
| J07344+629 | GJ 9236 | 07:34:26.27 | +62:56:27.6 | M0.5 V | 7.340 | | |
| J07346+223 | GJ 3453 | 07:34:39.31 | +22:20:13.9 | M1.0 V | 8.362 | | |
| J07346+318 | Castor C | 07:34:37.19 | +31:52:08.6 | M0.5 V | 6.073 | Sextuple (EB/DESB2) | |
| J07349+147 | TYC 777-141-1 | 07:34:56.25 | +14:45:52.5 | M3.0 V | 7.287 | Binary | |
| J07353+548 | GJ 3452 | 07:35:21.67 | +54:50:59.2 | M2.5 V | 7.772 | | |
| J07354+482 | LP 162-39 | 07:35:26.96 | +48:14:33.0 | M1.0 V | 8.620 | | |
| J07359+785 | LP 17-66 | 07:35:57.31 | +78:32:49.7 | M3.0 V | 9.214 | | • |
| J07361-031 | GJ 282 C | 07:36:07.15 | -03:06:43.4 | M1.0 V | 6.791 | Quadruple (SB1) | |
| J07364+070 | GJ 3454 | 07:36:25.37 | +07:04:38.2 | M4.5 V | 8.180 | Binary | |

Table B.1: Carmencita, the CARMENES input catalogue (continued).

| Karmn | Name | $\alpha$ (2016.0) | $\delta$ (2016.0) | Spectral type | $J$ [mag] | Multiplicity[a] | DR1[b] |
|---|---|---|---|---|---|---|---|
| J07365-006 | PM J07365-0039 | 07:36:30.27 | -00:39:37.3 | M3.5 V | 9.422 | | • |
| J07366+440 | G 111-20 | 07:36:39.13 | +44:04:43.5 | M3.5 V | 9.961 | Binary | |
| J07383+344 | TYC 2461-826-1 | 07:38:19.92 | +34:27:00.6 | M0.0 V | 8.485 | | |
| J07384+240 | 1R073829.3+240014 | 07:38:29.32 | +24:00:07.1 | M3.5 V | 8.928 | | |
| J07386-212 | GJ 3459 | 07:38:41.48 | -21:13:36.1 | dM3.0 | 7.848 | | |
| J07393+021 | Ross 880 | 07:39:22.88 | +02:10:57.3 | M0.0 V | 6.769 | | |
| J07395+334 | GJ 3457 | 07:39:35.62 | +33:27:42.6 | M2.0 V | 8.420 | Binary | • |
| J07403-174 | GJ 283 B | 07:40:20.66 | -17:24:54.5 | M6.5 V | 10.155 | Binary | • |
| J07418+050 | GJ 3461 | 07:41:52.56 | +05:02:23.1 | M3.0 V | 8.910 | Triple (SB2) | |
| J07421+500 | LP 162-55 | 07:42:10.07 | +50:04:28.5 | M2.5 V | 8.455 | | • |
| J07431+181 | GJ 3462 | 07:43:11.67 | +18:10:34.8 | M1.5 V | 8.117 | | |
| J07446+035 | YZCMi | 07:44:39.80 | +03:33:01.7 | M4.0 V | 6.581 | | |
| J07467+574 | G 193-65 | 07:46:41.97 | +57:26:49.5 | M4.5 V | 9.699 | | |
| J07470+760 | LP 17-75 | 07:47:06.48 | +76:03:13.1 | M4.0 V | 9.976 | | • |
| J07472+503 | 1R074714.1+502032 | 07:47:13.84 | +50:20:39.8 | M4.0 V | 8.855 | | |
| J07482+203 | Wolf 1421 | 07:48:18.04 | +20:21:49.4 | M1.5 V | 8.120 | | |
| J07493+849 | GJ 3456 | 07:49:17.72 | +84:58:32.5 | M3.0 V | 8.988 | | • |
| J07497-033 | PM J07497-0320 | 07:49:41.97 | -03:20:34.9 | M3.5 V | 8.891 | | |
| J07518+055 | GJ 3463 | 07:51:51.86 | +05:32:50.6 | M5.0 V | 9.966 | | |
| J07519-000 | GJ 1103 A | 07:51:54.95 | -00:00:24.4 | M4.5 V | 8.496 | | |
| J07523+162 | LP 423-31 | 07:52:24.13 | +16:12:09.4 | M6.0 V | 10.879 | | |
| J07545+085 | 1R075434.3+083213 | 07:54:33.90 | +08:32:25.5 | M2.5 V | 8.538 | Binary (SB1) | |
| J07525+063 | GJ 3465 | 07:52:33.63 | +06:18:22.0 | M3.0 V | 9.608 | | |
| J07545-096 | PM J07545-0941 | 07:54:32.60 | -09:41:47.9 | M3.5 V | 9.697 | Binary | |
| J07558+833 | GJ 1101 | 07:55:51.23 | +83:22:55.4 | M4.5: V | 8.744 | | |
| J07581+072 | GJ 3467 | 07:58:08.74 | +07:17:00.9 | M5.0 V | 9.272 | | |
| J07582+413 | GJ 1105 | 07:58:13.01 | +41:18:02.3 | M3.5 V | 7.734 | | • |
| J07583+496 | LP 163-47 | 07:58:23.26 | +49:39:41.3 | M4.0 V | 8.706 | | |
| J07585+155N | GJ 3468 | 07:58:30.88 | +15:30:12.6 | M4.5 V | 9.970 | Triple | • |
| J07585+155S | GJ 3469 A | 07:58:30.37 | +15:29:58.8 | M4.5 V | 10.429 | Triple | |
| J07590+153 | GJ 3470 | 07:59:05.63 | +15:23:28.3 | M2.0 V | 8.794 | | |
| J07591+173 | 1R075908.2+171957 | 07:59:07.07 | +17:19:46.8 | M4.0 V | 9.468 | | |
| J08005+258 | TYC 1930-667-1 | 08:00:34.87 | +25:53:32.6 | M2.0 V | 8.204 | | |
| J08017+237 | TYC 1926-794-1 | 08:01:43.44 | +23:42:25.3 | M1.5 V | 7.670 | | |
| J08023+033 | GJ 3473 | 08:02:22.45 | +03:20:13.6 | M4.0 V | 9.627 | Binary | |
| J08025-130 | LP 724-16 | 08:02:33.09 | -13:05:33.4 | M2.5 V | 9.421 | | |
| J08031+203 | PM J08031+2022 | 08:03:10.06 | +20:22:14.3 | M3.5 V | 9.242 | Binary | • |
| J08033+528 | G 194-7 | 08:03:20.18 | +52:50:27.3 | M1.5 V | 8.058 | Binary | |
| J08066+558 | GJ 3477 | 08:06:36.75 | +55:53:37.1 | M2.0 V | 8.050 | Binary | |
| J08068+367 | GJ 3479 | 08:06:48.21 | +36:45:32.5 | M3.0 V | 8.976 | | |
| J08069+422 | G 111-56 | 08:06:54.99 | +42:17:28.7 | M4.0 V | 9.724 | | |
| J08082+211 | GJ 3482 | 08:08:12.85 | +21:06:12.6 | M3.0 V | 6.860 | Triple | |
| J08083+585 | GJ 3480 | 08:08:17.82 | +58:31:08.4 | M3.0 V | 8.804 | | |
| J08089+328 | FPCncB | 08:08:55.38 | +32:49:01.4 | M3.0 V | 7.999 | Quadruple (SB2) | |
| J08095+219 | GJ 3484 | 08:09:30.59 | +21:54:16.2 | M2.0 V | 8.332 | | |
| J08103+095 | PM J08103+0935 | 08:10:20.65 | +09:35:15.4 | M2.5 V | 8.382 | | |
| J08105-138 | HD 68146B | 08:10:34.02 | -13:48:50.1 | M2.5 V | 8.276 | Binary | |
| J08108+039 | GJ 3485 | 08:10:53.75 | +03:58:28.2 | M4.0 V | 9.238 | | |
| J08117+531 | G 194-14 | 08:11:47.07 | +53:11:48.4 | M2.5 V | 9.293 | | |
| J08119+087 | Ross 619 | 08:11:58.72 | +08:45:01.4 | M4.5 V | 8.424 | | |
| J08126-215 | GJ 300 | 08:12:40.90 | -21:33:18.1 | M3.5 V | 7.601 | | |
| J08158+346 | LP 311-8 | 08:15:53.78 | +31:36:35.8 | M1.0 V | 8.935 | | • |
| J08161+013 | GJ 2066 | 08:16:07.58 | +01:18:10.2 | M2.0 V | 6.625 | | • |
| J08175+209 | LP 367-67 | 08:17:31.31 | +20:59:48.7 | M2.5 V | 8.989 | | |
| J08178+311 | GJ 3491 | 08:17:51.20 | +31:07:49.4 | M1.0 V | 7.943 | | • |
| J08202+055 | PM J08202+0532 | 08:20:13.29 | +05:32:08.2 | M2.0 V | 8.526 | | |
| J08258+690 | GJ 3497 | 08:25:50.76 | +69:01:40.7 | M7.0 V | 10.078 | | |
| J08282+201 | GJ 1110 | 08:28:12.37 | +20:08:11.4 | M4.0 V | 9.397 | | |
| J08283+350 | GJ 308 | 08:28:20.83 | +35:00:53.6 | M0.0 V | 7.633 | Binary | |

Table B.1: Carmencita, the CARMENES input catalogue (continued).

| Karmn | Name | $\alpha$ (2016.0) | $\delta$ (2016.0) | Spectral type | $J$ [mag] | Multiplicity[a] | DR1[b] |
|---|---|---|---|---|---|---|---|
| J08283+553 | PM J08283+5522 | 08:28:18.75 | +55:22:40.6 | M2.5 V | 9.235 | | |
| J08286+660 | 1R082839.4+660229 | 08:28:41.33 | +66:02:25.4 | M4.0 V | 9.197 | Binary | |
| J08293+039 | PM J08293+0355E | 08:29:21.81 | +03:55:08.2 | M2.5 V | 7.932 | | |
| J08298+267 | DXCnc | 08:29:48.02 | +26:46:23.8 | M6.5 V | 8.235 | | |
| J08313-060 | GJ 3501 | 08:31:21.14 | -06:02:02.8 | M2.0 V | 7.998 | Triple | • |
| J08314-060 | GJ 3502 | 08:31:26.75 | -06:02:13.6 | M3.0 V | 8.716 | Triple | • |
| J08313-104 | GJ 3503 | 08:31:22.81 | -10:29:58.9 | M4.0 V | 10.070 | | |
| J08315+730 | LP 35-219 | 08:31:32.36 | +73:03:50.1 | M4.0 V | 8.780 | | |
| J08316+193S | CUCnc | 08:31:37.32 | +19:23:37.5 | M3.5 V | 7.509 | Quintuple (EB) | |
| J08316+193N | CCncA | 08:31:37.17 | +19:23:47.6 | M4.0 V | 8.625 | Quintuple | • |
| J08317+057 | 1R083147.3+054504 | 08:31:47.89 | +05:45:17.0 | M1.0 V | 8.915 | | |
| J08321+844 | GJ 3496 | 08:32:13.99 | +84:24:34.8 | M3.5 V | 9.441 | | |
| J08325+451 | Wolf 312 | 08:32:35.78 | +45:10:16.3 | M2.5 V | 8.880 | | |
| J08334+185 | GJ 3505 | 08:33:25.06 | +18:31:34.9 | M4.5 V | 10.263 | | |
| J08344-011 | GJ 2070 | 08:34:26.13 | -01:08:46.3 | M3.0 V | 8.810 | | |
| J08353+141 | LSPM J0835+1408 | 08:35:19.75 | +14:08:31.9 | M4.5 V | 9.163 | Triple (ST3) | |
| J08358+680 | GJ 3506 | 08:35:46.65 | +68:04:00.1 | M3.0 V | 7.861 | | |
| J08364+264 | LP 311-37 | 08:36:26.45 | +26:28:18.9 | M2.0 V | 8.964 | Binary | |
| J08364+672 | GJ 310 | 08:36:22.51 | +67:17:42.9 | M0.5 V | 6.425 | Binary | • |
| J08371+151 | GJ 3508 | 08:37:07.82 | +15:07:31.2 | M3.0 V | 8.122 | | |
| J08375+035 | LSPM J0837+0333 | 08:37:30.28 | +03:33:43.1 | M4.0 V | 9.853 | | |
| J08387+516 | StKM 1-711 | 08:38:42.06 | +51:41:31.9 | M1.5 V | 8.695 | | |
| J08398+089 | GJ 3510 | 08:39:47.80 | +08:56:21.0 | M2.0 V | 9.470 | Binary | |
| J08402+314 | LSPM J0840+3127 | 08:40:16.24 | +31:27:08.7 | M3.5 V | 8.122 | | |
| J08404+184 | AZCnc | 08:40:28.77 | +18:24:01.5 | M6.0 V | 11.053 | | |
| J08409-234 | GJ 317 | 08:40:58.67 | -23:27:09.7 | M3.5 V | 7.934 | | • |
| J08410+676 | GJ 3509 | 08:41:01.83 | +67:39:33.1 | M4.0 V | 10.290 | | |
| J08413+594 | GJ 3512 | 08:41:19.58 | +59:29:30.0 | dM5.5 | 9.615 | | • |
| J08427+095 | GJ 319 A | 08:42:44.77 | +09:33:14.0 | M0.0 V | 6.687 | Quadruple (SB1) | |
| J08428+095 | GJ 319 C | 08:42:52.47 | +09:33:01.3 | M2.5 V | 8.122 | Quadruple | • |
| J08443-104 | GJ 3513 | 08:44:22.71 | -10:24:20.1 | M3.5 V | 9.799 | | |
| J08447+182 | G 9-19 | 08:44:45.08 | +18:12:59.2 | M3.5 V | 8.942 | | |
| J08449-066 | PM J08449-0637 | 08:44:55.59 | -06:37:28.2 | M3.5 V | 9.325 | Binary | |
| J08517+181 | Ross 622 | 08:51:42.78 | +18:07:29.1 | M1.5 V | 8.281 | | |
| J08526+283 | GJ 324 B | 08:52:40.28 | +28:18:54.9 | M4.5 V | 8.560 | Binary | |
| J08531-202 | PM J08531-2017 | 08:53:10.96 | -20:17:19.3 | M3.0 V | 9.323 | | |
| J08536-034 | GJ 3517 | 08:53:35.61 | -03:29:35.4 | M9.0 V | 11.212 | | • |
| J08537+149 | StKM 1-730 | 08:53:43.67 | +14:58:09.7 | M0.0 V | 8.165 | | |
| J08540-131 | GJ 326 A | 08:54:05.69 | -13:07:39.9 | M2.5 V | 8.048 | Binary | • |
| J08551+015 | Ross 623 | 08:55:07.67 | +01:32:30.7 | M0.0 V | 7.191 | | |
| J08555+664 | PM J08555+6628 | 08:55:31.46 | +66:28:06.6 | M3.0 V | 8.824 | | |
| J08563+126 | G 41-8 | 08:56:19.49 | +12:39:45.8 | M6.0 V | 9.585 | Binary | |
| J08570+116 | GJ 330 | 08:57:04.65 | +11:38:43.9 | M1.0 V | 7.311 | Binary | |
| J08572+194 | LP 426-35 | 08:57:15.55 | +19:24:15.2 | M3.5 V | 9.447 | | |
| J08582+197 | GJ 1116 A | 08:58:14.09 | +19:45:45.3 | M5.5 V | 7.791 | Binary | |
| J08588+210 | G 41-13 | 08:58:52.53 | +21:04:29.1 | M2.0 V | 10.058 | | |
| J08595+537 | G 194-47 | 08:59:35.41 | +53:43:47.5 | M3.5 V | 9.014 | Binary | |
| J08589+084 | GJ 3522 | 08:58:56.73 | +08:28:20.8 | M3.5 V | 6.507 | Binary | |
| J08599+729 | GJ 3520 | 08:59:59.70 | +72:57:35.8 | M5.0 V | 9.731 | | |
| J09003+218 | LP 368-128 | 09:00:22.95 | +21:49:55.4 | M6.5 V | 9.436 | | |
| J09005+465 | GJ 1119 | 09:00:31.74 | +46:35:02.7 | M4.5 V | 8.604 | | |
| J09008+052W | Ross 686 | 09:00:48.25 | +05:14:38.1 | M3.0 V | 8.605 | Binary | • |
| J09008+052E | Ross 687 | 09:00:50.05 | +05:14:26.3 | M3.0 V | 8.845 | Binary | • |
| J09011+019 | Ross 625 | 09:01:10.07 | +01:56:33.7 | M3.0 V | 7.932 | Triple | |
| J09023+084 | GJ 3528 | 09:02:20.55 | +08:28:03.3 | M3.0 V | 8.145 | | |
| J09023+177 | PM J09023+1746 | 09:02:22.91 | +17:46:31.8 | M4.0 V | 9.645 | | |
| J09028+680 | GJ 3526 | 09:02:53.41 | +68:03:52.1 | M4.0 V | 8.453 | | |
| J09029+716 | LSPM J0902+7138 | 09:02:55.82 | +71:38:11.0 | M1.5 V | 9.731 | | |
| J09033+056 | LP 546-37 | 09:03:20.91 | +05:40:08.5 | M7.0 V | 10.766 | | • |

Table B.1: Carmencita, the CARMENES input catalogue (continued).

| Karmn | Name | $\alpha$ (2016.0) | $\delta$ (2016.0) | Spectral type | $J$ [mag] | Multiplicity[a] | DR1[b] |
|---|---|---|---|---|---|---|---|
| J09037+520 | G 194-52 | 09:03:43.39 | +52:02:49.1 | M3.5 V | 8.964 | | • |
| J09038+129 | LP 486-43 | 09:03:53.41 | +12:59:24.8 | M2.0 V | 8.919 | | • |
| J09040-159 | 1R090406.8-155512 | 09:04:05.44 | -15:55:19.0 | M2.5 V | 9.156 | Binary* | |
| J09050+028 | GJ 3530 | 09:05:04.11 | +02:50:03.9 | M1.5 V | 8.177 | Binary | |
| J09057+186 | LP 426-56 | 09:05:43.02 | +18:36:27.6 | M2.5 V | 8.957 | | |
| J09062+128 | GJ 3531 | 09:06:13.79 | +12:51:30.1 | M3.5 V | 9.243 | | |
| J09070-221 | GJ 3533 | 09:07:02.40 | -22:08:56.6 | M4.5 V | 9.533 | | |
| J09087+665 | GJ 3532 | 09:08:46.58 | +66:35:36.4 | M2.5 V | 9.168 | | |
| J09091+227 | 2M09090798+2247413 | 09:09:07.88 | +22:47:40.1 | M4.5 V | 10.474 | | |
| J09093+401 | GJ 1121 | 09:09:23.06 | +40:05:55.5 | M4.0 V | 10.135 | | |
| J09095+328 | GJ 336 | 09:09:30.18 | +32:48:59.1 | M0.5 V | 7.035 | Binary | |
| J09096+067 | GJ 3537 | 09:09:39.07 | +06:42:11.8 | M3.0 V | 9.294 | | |
| J09099+004 | G 46-24 | 09:09:59.43 | +00:23:39.5 | M1.0 V | 8.892 | | |
| J09115+126 | LP 487-10 | 09:11:32.14 | +12:37:18.3 | M2.5 V | 9.408 | | |
| J09115+466 | GJ 336.1 | 09:11:30.27 | +46:37:00.7 | M0.5 V | 7.900 | | |
| J09120+279 | GJ 3540 | 09:12:02.44 | +27:54:16.2 | M3.0 V | 8.430 | Binary (SB2) | |
| J09133+688 | G 234-57A | 09:13:23.43 | +68:52:27.1 | M2.5 V | 7.775 | Binary | |
| J09140+196 | LP 427-16 | 09:14:03.02 | +19:40:03.2 | M3.0 V | 8.424 | Binary (SB1) | |
| J09143+526 | HD 79210 | 09:14:20.05 | +52:41:02.7 | M0.0 V | 4.889 | Triple (SB1) | • |
| J09144+526 | HD 79211 | 09:14:21.91 | +52:41:00.3 | M0.0 V | 4.779 | Triple | • |
| J09156-105 | G 161-7 | 09:15:35.96 | -10:35:50.2 | M5.0 V | 8.605 | Binary | • |
| J09160+293 | G 47-31 | 09:16:05.04 | +29:19:36.3 | M2.0 V | 8.915 | | • |
| J09161+018 | RX J0916.1+0153 | 09:16:10.24 | +01:53:07.2 | M4.0 V | 8.770 | | |
| J09163-186 | GJ 3543 | 09:16:20.29 | -18:37:30.6 | M1.5 V | 7.351 | | |
| J09165+841 | GJ 3536 | 09:16:24.75 | +84:11:06.4 | M1.5 V | 8.618 | | • |
| J09168+248 | 2M09165078+2448559 | 09:16:50.70 | +24:48:54.0 | M4.5 V | 10.466 | | • |
| J09177+462 | RX J0917.7+4612 | 09:17:44.52 | +46:12:24.4 | M2.5 V | 8.126 | Binary | |
| J09177+584 | GJ 3542 | 09:17:46.04 | +58:25:02.7 | M5.0 V | 10.261 | | |
| J09187+267 | GJ 3548 | 09:18:45.99 | +26:45:05.4 | M1.5 V | 8.296 | Binary | |
| J09193+385S | GJ 1122 | 09:19:18.62 | +38:31:15.9 | M5.0 V | 9.924 | Binary | |
| J09193+385N | G 115-69 | 09:19:18.71 | +38:31:23.3 | M5.0 V | 10.048 | Binary | |
| J09193+620 | GJ 3547 | 09:19:22.21 | +62:03:10.7 | M1.0 V | 8.168 | Binary (EB/SB2) | |
| J09200+308 | TYC 2493-1386-1 | 09:20:00.38 | +30:52:39.1 | M1.5 V | 8.309 | Triple | |
| J09201+037 | 1R092010.8+034731 | 09:20:10.74 | +03:47:27.0 | M3.5 V | 9.310 | | |
| J09209+033 | GJ 3553 | 09:20:58.28 | +03:21:48.3 | M4.0 V | 9.363 | | |
| J09213+731 | GJ 3550 | 09:21:16.03 | +73:06:33.1 | M5.0 V | 10.365 | | |
| J09218+435 | GJ 3554 | 09:21:48.61 | +43:30:26.5 | M4.5 V | 9.431 | Binary | |
| J09218-023 | RAVE J092148.1-021943 | 09:21:48.32 | -02:19:43.2 | M2.5 V | 8.441 | | |
| J09228+467 | G 115-72 | 09:22:51.31 | +46:46:58.8 | M1.0 V | 8.533 | Binary | |
| J09231+223 | BD+22 2086B | 09:23:06.01 | +22:18:25.6 | M0.0 V | 8.497 | Binary | |
| J09238+001 | GJ 3555 | 09:23:52.36 | +00:08:13.8 | M1.0 V | 8.506 | | |
| J09248+306 | LSPM J0924+3041 | 09:24:50.68 | +30:41:34.2 | M3.5 V | 9.490 | | |
| J09256+634 | G 235-25 | 09:25:39.51 | +63:29:14.9 | M4.5 V | 9.818 | Binary | |
| J09275+506 | GJ 3556 | 09:27:30.19 | +50:39:10.1 | M2.5 V | 8.481 | | |
| J09286-121 | LP 727-31 | 09:28:41.63 | -12:10:02.0 | M2.5 V | 8.837 | | |
| J09288-073 | Ross 439 | 09:28:53.16 | -07:22:27.2 | M2.5 V | 8.446 | Binary | |
| J09289-073 | GJ 347 B | 09:28:55.53 | -07:22:23.3 | M4.5 V | 10.370 | Binary | |
| J09291+259 | LP 370-26 | 09:29:09.83 | +25:58:05.0 | M5.0 V | 10.906 | | |
| J09300+396 | GJ 3558 | 09:30:01.84 | +39:37:21.1 | M2.5 V | 8.467 | | |
| J09302+265 | LSPM J0930+2630 | 09:30:14.24 | +26:30:22.7 | M3.0 V | 8.866 | | |
| J09307+003 | GJ 1125 | 09:30:43.98 | +00:19:12.8 | M3.5 V | 7.697 | | |
| J09308+024 | 1R093051.2+022741 | 09:30:50.81 | +02:27:21.5 | M4.0 V | 9.415 | | |
| J09313-134 | Ross 440 | 09:31:20.26 | -13:29:18.9 | M3.0 V | 6.361 | Binary | • |
| J09315+202 | Ross 84 | 09:31:33.05 | +20:16:43.6 | M2.0 V | 8.783 | | |
| J09319+363 | GJ 353 | 09:31:56.06 | +36:19:04.4 | M0.0 V | 7.121 | | |
| J09328+269 | HD 82443B | 09:32:48.07 | +26:59:39.9 | M5.5 V | 10.356 | Binary | |
| J09352+612 | GJ 3560 | 09:35:13.45 | +61:14:37.6 | M2.5 V | 8.463 | | |
| J09360-061 | GJ 3561 | 09:36:04.10 | -06:07:01.2 | M3.5 V | 9.854 | | |
| J09360-216 | GJ 357 | 09:36:01.80 | -21:39:54.7 | M2.5 V | 7.337 | | |

Table B.1: Carmencita, the CARMENES input catalogue (continued).

| Karmn | Name | α (2016.0) | δ (2016.0) | Spectral type | J [mag] | Multiplicity[a] | DR1[b] |
|---|---|---|---|---|---|---|---|
| J09362+375 | GJ 9303 | 09:36:04.14 | +37:33:08.9 | M0.0 V | 8.085 | Triple (SB2) | |
| J09370+405 | GJ 3562 | 09:37:03.29 | +40:34:37.7 | M3.8 V | 9.766 | | • |
| J09394+146 | LP 428-20 | 09:39:29.77 | +14:38:48.5 | M3.5 V | 9.393 | | |
| J09394+317 | G 117-34 | 09:39:24.04 | +31:45:13.7 | M1.5 V | 8.486 | Binary | |
| J09410+220 | Ross 92 | 09:41:02.58 | +22:01:20.5 | M4.5 V | 9.627 | | |
| J09411+132 | Ross 85 | 09:41:09.64 | +13:12:32.1 | M1.5 V | 6.971 | | |
| J09423+559 | GJ 363 | 09:42:21.84 | +55:58:53.1 | M3.0 V | 8.374 | | |
| J09425+700 | GJ 360 | 09:42:32.74 | +70:01:57.6 | M2.5 V | 6.917 | Binary | • |
| J09428+700 | GJ 362 | 09:42:49.63 | +70:02:17.6 | M3.5 V | 7.326 | Binary | • |
| J09425-192 | GJ 3563 | 09:42:35.14 | -19:14:08.6 | M2.5 V | 8.298 | | |
| J09430+237 | LP 370-35 | 09:43:01.12 | +23:49:18.3 | M1.0 V | 8.994 | Triple | • |
| J09439+269 | Ross 93 | 09:43:54.92 | +26:58:06.8 | M3.5 V | 8.035 | | • |
| J09447-182 | GJ 1129 | 09:44:45.55 | -18:12:51.7 | M3.5 V | 8.122 | | |
| J09449-123 | G 161-71 | 09:44:53.83 | -12:20:53.7 | M5.0 V | 8.496 | | • |
| J09461-044 | GJ 3566 | 09:46:08.66 | -04:25:40.6 | M4.0 V | 9.688 | Binary | • |
| J09468+760 | Ross 434 | 09:46:48.87 | +76:02:22.1 | M1.5 V | 7.437 | | • |
| J09473+263 | Ross 94 | 09:47:22.15 | +26:18:06.9 | M0.0 V | 8.141 | | |
| J09475+129 | GJ 3568 | 09:47:34.71 | +12:56:42.8 | M4.0 V | 9.267 | | • |
| J09506-138 | LP 728-70 | 09:50:40.70 | -13:48:40.2 | M4.0 V | 8.579 | Binary (SB2) | |
| J09488+156 | G 43-2 | 09:48:50.18 | +15:38:48.5 | M3.0 V | 9.303 | | |
| J09511-123 | GJ 369 | 09:51:10.88 | -12:20:10.8 | dM0.5 | 6.988 | | |
| J09526-156 | LP 728-71 | 09:52:41.65 | -15:36:15.9 | M3.5 V | 9.320 | | |
| J09527+554 | G 195-43 | 09:52:45.26 | +55:28:16.2 | M1.5 V | 8.989 | Binary | • |
| J09531-036 | GJ 372 | 09:53:11.67 | -03:41:31.8 | M2.0 V | 6.998 | Binary (SB2) | |
| J09535+507 | LP 126-73 | 09:53:32.64 | +50:45:04.2 | M1.5 V | 8.733 | | |
| J09539+209 | GJ 3571 | 09:53:54.78 | +20:56:53.1 | M4.0 V | 9.208 | | |
| J09557+353 | Wolf 330 | 09:55:43.55 | +35:21:36.8 | M3.5 V | 8.850 | | |
| J09561+627 | GJ 373 | 09:56:07.96 | +62:47:09.1 | M0.5 V | 6.030 | Binary (SB?) | |
| J09564+226 | GJ 3573 | 09:56:26.42 | +22:38:56.9 | M4.0 V | 9.621 | | |
| J09579+118 | GJ 3576 | 09:57:57.54 | +11:48:26.3 | M4.0 V | 10.092 | Binary* | • |
| J09587+555 | G 196-1 | 09:58:46.62 | +55:32:59.0 | M1.0 V | 8.899 | | |
| J09589+059 | LP 549-6 | 09:58:56.31 | +05:57:58.8 | M4.5 V | 9.937 | | |
| J09593+438W | GJ 3577 | 09:59:18.65 | +43:50:21.9 | M3.5 V | 9.682 | Binary | |
| J09593+438E | GJ 3578 | 09:59:20.78 | +43:50:22.1 | M5.0 V | 9.917 | Binary | |
| J09597+472 | GJ 3579 | 09:59:46.11 | +47:12:06.9 | M4.0 V | 9.756 | | |
| J09597+721 | PM J09597+7211 | 09:59:45.30 | +72:12:01.3 | M3.5 V | 9.059 | | |
| J10004+272 | GJ 375.2 | 10:00:26.69 | +27:16:03.5 | M0.5 V | 8.382 | Triple* | |
| J10007+323 | Wolf 335 | 10:00:43.01 | +32:18:23.1 | M1.0 V | 8.780 | | |
| J10020+697 | LP 37-57 | 10:02:05.67 | +69:45:25.6 | M4.0 V | 9.768 | | |
| J10023+480 | GJ 378 | 10:02:20.74 | +48:04:56.1 | M1.0 V | 6.949 | | |
| J10027+149 | GJ 3582 | 10:02:42.62 | +14:59:09.2 | M4.61 | 9.649 | | |
| J10028+484 | G 195-55 | 10:02:48.79 | +48:27:28.7 | M5.5 V | 9.963 | Binary | • |
| J10035+059 | GJ 3583 | 10:03:32.76 | +05:57:45.4 | M3.5 V | 9.290 | | |
| J10040+187 | GJ 9312 | 10:04:05.80 | +18:47:41.5 | M0.5 V | 8.365 | Binary | |
| J10043+503 | G 196-3 | 10:04:21.23 | +50:23:10.1 | M2.5 V | 8.081 | Binary | |
| J10067+417 | GJ 3585 | 10:06:43.45 | +41:42:46.1 | M1.0 V | 8.209 | | |
| J10068-127 | PM J10068-1246 | 10:06:51.98 | -12:46:54.3 | M4.5 V | 9.749 | | |
| J10069+126 | LP 489-35 | 10:06:57.49 | +12:40:51.6 | M1.5 V | 8.776 | | |
| J10079+692 | GJ 1131 | 10:07:56.60 | +69:14:46.1 | M4.0 V | 10.096 | | |
| J10087+027 | LP 549-23 | 10:08:44.52 | +02:43:49.6 | M3.0 V | 8.590 | | |
| J10087+355 | Wolf 346 | 10:08:42.37 | +35:32:51.3 | M1.5 V | 9.167 | | |
| J10088+692 | TYC 4384-1735-1 | 10:08:52.40 | +69:16:35.8 | M0.5 V | 8.711 | | |
| J10094+512 | GJ 3586 | 10:09:29.22 | +51:17:06.4 | M4.65 | 9.299 | | |
| J10094+544 | PM J10094+5424 | 10:09:26.88 | +54:24:22.5 | M2.0 V | 8.644 | | • |
| J10120-026 | GJ 381 | 10:12:05.23 | -02:41:14.8 | M2.5 V | 7.021 | Binary | |
| J10117+353 | Wolf 351 | 10:11:44.11 | +35:18:40.1 | M4.0 V | 10.112 | | |
| J10122-037 | ANSex | 10:12:17.50 | -03:44:48.3 | M1.5 V | 5.888 | | |
| J10125+570 | LP 92-48 | 10:12:34.10 | +57:03:40.5 | M3.5 V | 7.759 | | |
| J10130+233 | G 54-18 | 10:13:00.45 | +23:20:45.9 | M3.5 V | 9.194 | Binary | • |

Table B.1: Carmencita, the CARMENES input catalogue (continued).

| Karmn | Name | $\alpha$ (2016.0) | $\delta$ (2016.0) | Spectral type | $J$ [mag] | Multiplicity[a] | DR1[b] |
|---|---|---|---|---|---|---|---|
| J10143+210 | DKLeo | 10:14:19.03 | +21:04:26.8 | M0.5 V | 7.074 | Binary | • |
| J10133+467 | LP 167-17 | 10:13:20.56 | +46:47:24.5 | M5.5 V | 10.873 | | |
| J10148+213 | G 54-19 | 10:14:52.91 | +21:23:42.5 | M4.5 V | 9.725 | | |
| J10151+314 | GJ 3590 | 10:15:06.87 | +31:25:07.0 | M4.0 V | 9.326 | Binary | |
| J10155-164 | PM J10155-1628E | 10:15:34.86 | -16:28:20.4 | M4.0 V | 9.363 | | |
| J10158+174 | LSPM J1015+1729 | 10:15:54.26 | +17:29:27.2 | M3.5 V | 8.696 | | |
| J10167-119 | GJ 386 | 10:16:45.49 | -11:57:52.1 | dM3.0 | 7.323 | | |
| J10182-204 | LP 790-2 | 10:18:13.39 | -20:28:39.3 | M4.5 V | 8.999 | Triple (SB2) | |
| J10185-117 | LP 729-54 | 10:18:34.77 | -11:43:04.2 | M4.0 V | 9.007 | Binary | • |
| J10196+198 | ADLeo | 10:19:35.72 | +19:52:11.3 | M3.0 V | 5.449 | | • |
| J10200+289 | G 118-51 | 10:20:00.23 | +28:57:09.5 | M3.0 V | 9.159 | | • |
| J10206+492 | GJ 3595 | 10:20:37.15 | +49:17:43.2 | M3.0 V | 9.396 | | • |
| J10238+438 | LP 212-62 | 10:23:52.13 | +43:53:33.2 | M5.0 V | 10.039 | | |
| J10240+366 | PM J10240+3639 | 10:24:05.04 | +36:39:30.2 | M3.5 V | 9.429 | | |
| J10243+119 | GJ 3598 | 10:24:20.16 | +11:57:23.5 | M2.5 V | 8.847 | Binary | |
| J10251-102 | GJ 390 | 10:25:10.09 | -10:13:41.3 | M1.0 V | 6.895 | | |
| J10255+263 | GJ 3599 | 10:25:29.93 | +26:23:09.8 | M3.0 V | 9.030 | | |
| J10260+504W | GJ 3600 | 10:26:01.99 | +50:27:00.0 | M4.0 V | 9.268 | Binary | • |
| J10260+504E | GJ 3601 | 10:26:02.64 | +50:27:13.0 | M4.0 V | 9.404 | Binary | |
| J10273+799 | PM J10273+7959 | 10:27:22.67 | +79:59:50.0 | M2.0 V | 8.669 | | |
| J10278+028 | G 44-19 | 10:27:49.12 | +02:51:35.5 | M3.5 V | 9.438 | | |
| J10284+482 | GJ 3602 | 10:28:28.86 | +48:14:17.7 | M3.5 V | 9.055 | Binary* | |
| J10286+322 | GJ 3604 | 10:28:40.73 | +32:14:21.7 | M2.5 V | 9.043 | Binary | |
| J10289+008 | Ross 446 | 10:28:54.91 | +00:50:15.9 | dM2 | 6.176 | | |
| J10303+328 | GJ 3607 | 10:30:23.19 | +32:50:07.3 | M3.0 V | 8.867 | | |
| J10315+570 | GJ 397.1 B | 10:31:30.64 | +57:05:20.4 | M5.0 V | 9.738 | Triple | • |
| J10320+033 | PM J10320+0318 | 10:32:02.31 | +03:18:54.9 | M2.0 V | 8.365 | | |
| J10345+463 | GJ 3610 | 10:34:29.53 | +46:18:07.2 | M3.0 V | 9.195 | Binary | |
| J10350-094 | LP 670-17 | 10:35:01.36 | -09:24:41.5 | dM3.0 | 8.276 | | |
| J10354+694 | GJ 3612 | 10:35:21.96 | +69:26:48.5 | M4.0 V | 7.898 | Binary (SB2) | |
| J10359+288 | RX J1035.9+2853 | 10:35:57.12 | +28:53:30.3 | M3.0 V | 9.245 | | • |
| J10360+051 | RYSex | 10:36:00.52 | +05:07:14.8 | M4.0 V | 8.463 | | • |
| J10364+415 | G 146-48 | 10:36:27.19 | +41:30:02.8 | M2.5 V | 8.647 | Binary | |
| J10367+153 | PM J10367+1521A | 10:36:44.96 | +15:21:38.6 | M3.5 V | 8.748 | Triple | • |
| J10368+509 | LP 127-502 | 10:36:48.58 | +50:55:00.7 | M4.5 V | 9.866 | Binary (SB2?) | |
| J10379+127 | LP 490-42 A | 10:37:55.03 | +12:46:37.6 | M3.0 V | 8.730 | Binary | |
| J10384+485 | GJ 3613 | 10:38:29.47 | +48:31:43.4 | M3.0 V | 9.495 | | |
| J10385+354 | LP 262-400 | 10:38:32.66 | +35:29:53.7 | M2.5 V | 8.676 | | |
| J10389+250 | StKM 1-873 | 10:38:56.64 | +25:05:39.0 | M2.0 V | 8.982 | | |
| J10396-069 | GJ 399 | 10:39:39.79 | -06:55:27.2 | dM2.5 | 7.664 | | |
| J10403+015 | TYC 254-88-1 | 10:40:21.42 | +01:34:36.6 | M1.0 V | 8.787 | | |
| J10416+376 | GJ 1134 | 10:41:35.91 | +37:36:33.4 | M4.0 V | 8.493 | | • |
| J10430-092 | PM J10430-0912 | 10:43:00.72 | -09:12:35.0 | M5.5 V | 9.667 | Binary | |
| J10443+124 | LP 490-63 | 10:44:18.52 | +12:25:11.5 | M3.5 V | 9.422 | | • |
| J10448+324 | GJ 3616 A | 10:44:52.40 | +32:24:41.3 | M3.0 V | 9.494 | Triple | |
| J10453+385 | GJ 400 A | 10:45:21.43 | +38:30:44.8 | M0.5 V | 6.354 | Binary | |
| J10456-191 | GJ 401 | 10:45:36.98 | -19:07:01.3 | M0.5 V | 8.069 | Binary | |
| J10460+096 | GJ 3619 | 10:46:03.81 | +09:41:47.1 | M3.5 V | 9.441 | | |
| J10472+404 | LP 213-67 | 10:47:12.19 | +40:26:43.2 | M6.5 V | 11.384 | Triple | |
| J10474+025 | Ross 895 | 10:47:24.47 | +02:35:32.7 | M2.0 V | 8.878 | Binary | |
| J10482-113 | GJ 3622 | 10:48:13.24 | -11:20:34.1 | M6.5 V | 8.857 | | |
| J10485+191 | GJ 3623 | 10:48:32.84 | +19:09:00.3 | M3.0 V | 9.491 | | |
| J10497+355 | GJ 1138 | 10:49:44.71 | +35:32:34.3 | M5.0 V | 8.537 | Binary | • |
| J10504+331 | GJ 3626 | 10:50:26.08 | +33:05:54.0 | M3.5 V | 8.899 | Binary (SB1) | |
| J10506+517 | GJ 3628 | 10:50:37.91 | +51:45:01.6 | M4.1 V | 9.828 | Binary | |
| J10508+068 | EELeo | 10:50:51.11 | +06:48:16.2 | M4.0 V | 7.319 | Binary | • |
| J10513+361 | GJ 3629 | 10:51:20.33 | +36:07:24.5 | M3.0 V | 9.422 | Binary | |
| J10520+005 | GJ 3630 | 10:52:02.83 | +00:32:38.5 | M4.0 V | 9.426 | Triple (ST3/SQ4) | • |
| J10520+139 | GJ 403 | 10:52:03.01 | +13:59:54.5 | M4.0 V | 8.607 | | |

Table B.1: Carmencita, the CARMENES input catalogue (continued).

| Karmn | Name | $\alpha$ (2016.0) | $\delta$ (2016.0) | Spectral type | $J$ [mag] | Multiplicity[a] | DR1[b] |
|---|---|---|---|---|---|---|---|
| J10522+059 | GJ 3631 | 10:52:13.50 | +05:55:08.9 | M5.5 V | 9.834 | | |
| J10546-073 | LP 671-8 | 10:54:41.77 | -07:18:39.4 | M4.0 V | 8.877 | Binary | |
| J10555-093 | GJ 3632 | 10:55:34.19 | -09:21:18.7 | M3.5 V | 9.419 | | |
| J10563+042 | PM J10563+0415 | 10:56:22.32 | +04:15:44.6 | M2.5 V | 9.179 | | |
| J10564+070 | CNLeo | 10:56:24.77 | +07:00:09.8 | M6.0 V | 7.085 | | |
| J10576+695 | Ross 447 | 10:57:36.16 | +69:35:48.8 | M0.0 V | 7.514 | | |
| J10584-107 | LP 731-76 | 10:58:27.78 | -10:46:31.8 | M5.0 V | 9.512 | | • |
| J11000+228 | Ross 104 | 11:00:03.76 | +22:49:54.1 | M2.5 V | 6.314 | | |
| J11003+728 | G 254-11A | 11:00:23.29 | +72:52:17.6 | M2.0 V | 8.934 | Binary | • |
| J11008+120 | GJ 3636 | 11:00:50.58 | +12:04:08.7 | M5.5 V | 10.676 | | • |
| J11013+030 | GJ 3637 | 11:01:20.81 | +03:00:10.8 | M5.0 V | 9.708 | | |
| J11014+568 | StKM 1-902 | 11:01:26.72 | +56:52:04.0 | M1.0 V | 8.994 | | |
| J11023+165E | GJ 1141 A | 11:02:19.29 | +16:30:27.1 | M1.0 V | 8.270 | Binary | |
| J11023+165W | GJ 1141 B | 11:02:18.02 | +16:30:30.8 | M1.0 V | 8.353 | Binary | |
| J11026+219 | DSLeo | 11:02:38.51 | +21:58:00.9 | M1.0 V | 6.522 | | |
| J11030+037 | Wolf 360 | 11:03:04.39 | +03:44:19.2 | M2.5 V | 9.307 | | |
| J11031+152 | LP 431-50 | 11:03:08.01 | +15:17:50.3 | M3.5 V | 8.890 | | • |
| J11031+366 | GJ 3639 | 11:03:09.74 | +36:39:09.1 | M3.5 V | 9.464 | | |
| J11033+359 | HD 95735 | 11:03:19.43 | +35:56:55.2 | M1.5 V | 4.203 | | |
| J11036+136 | LP 491-51 | 11:03:21.05 | +13:37:58.2 | M4.0 V | 8.759 | Binary (SB1) | |
| J11042+400 | GJ 3640 | 11:04:15.64 | +40:00:15.1 | M0.0 V | 8.025 | | • |
| J11044+304 | LSPM J1104+3027 | 11:04:28.35 | +30:27:30.9 | M3.0 V | 10.929 | | |
| J11054+435 | GJ 412 A | 11:05:22.09 | +43:31:51.4 | M1.0 V | 5.538 | Binary | |
| J11055+435 | WXUMa | 11:05:24.50 | +43:31:33.3 | M5.5 V | 8.742 | Binary | • |
| J11055+450 | GJ 3641 | 11:05:33.91 | +45:00:27.9 | M0.0 V | 8.107 | Binary | • |
| J11057+102 | GJ 3643 | 11:05:43.80 | +10:13:58.1 | M3.0 V | 8.643 | | • |
| J11075+437 | PM J11075+4345 | 11:07:31.88 | +43:45:56.3 | M3.0 V | 9.941 | | |
| J11081-052 | GJ 1142 A | 11:08:06.48 | -05:13:54.2 | M3.0 V | 8.797 | Binary | |
| J11108+479 | GJ 3646 | 11:10:51.05 | +47:56:53.7 | M4.0 V | 10.082 | | |
| J11110+304E | HD 97101 | 11:11:05.90 | +30:26:42.5 | K7 V | 5.764 | Binary | |
| J11110+304W | HD 97101B | 11:11:03.29 | +30:26:38.0 | M2.0 V | 6.592 | Binary | |
| J11113+434 | GJ 9351 A | 11:11:18.88 | +43:24:55.8 | M2.5 V | 7.333 | Binary | |
| J11118+335 | GJ 3647 | 11:11:51.52 | +33:32:13.1 | M3.5 V | 8.297 | Binary | • |
| J11126+189 | GJ 3649 | 11:12:38.95 | +18:56:05.5 | dM1.5 | 7.445 | | |
| J11131+002 | Wolf 370 | 11:13:09.63 | +00:14:16.7 | M0.0 V | 7.471 | Binary | |
| J11151+734 | HD 97584B | 11:15:09.56 | +73:28:38.0 | M2.5 V | 7.880 | Binary | • |
| J11152+194 | GJ 3652 | 11:15:12.62 | +19:27:04.3 | M3.5 V | 8.919 | | |
| J11152-181 | GJ 421 C | 11:15:15.67 | -18:07:47.8 | M3.0 V | 9.643 | Triple | |
| J11154+410 | G 122-8 | 11:15:26.72 | +41:05:12.5 | M3.5 V | 8.974 | | |
| J11159+553 | GJ 3653 | 11:15:53.69 | +55:19:49.2 | M0.5 V | 8.090 | | |
| J11195+466 | LP 169-22 | 11:19:31.09 | +46:41:33.4 | M5.5 V | 10.087 | | |
| J11200+658 | GJ 424 | 11:19:57.14 | +65:50:50.3 | M0.0 V | 6.306 | Binary | |
| J11201-104 | LP 733-99 | 11:20:05.89 | -10:29:46.4 | M2.0 V | 7.814 | | |
| J11214-204 | HD 98712B | 11:21:26.84 | -20:27:11.5 | M2.5 V | 6.638 | Binary | |
| J11216+061 | GJ 1146 | 11:21:37.67 | +06:08:00.7 | M3.5 V | 9.766 | | • |
| J11231+258 | GJ 3657 | 11:23:06.77 | +25:53:31.7 | M5.0 V | 10.294 | | |
| J11233+448 | GJ 3658 | 11:23:20.14 | +44:48:36.5 | M2.0 V | 9.146 | | |
| J11237+085 | Wolf 386 | 11:23:43.49 | +08:33:51.6 | M0.5 V | 7.994 | | |
| J11238+106 | LSPM J1123+1037 | 11:23:50.12 | +10:37:07.0 | M0.5 V | 7.787 | Quadruple | |
| J11239-183 | Ross 1002 | 11:23:56.61 | -18:21:49.5 | M3.0 V | 9.172 | | |
| J11247+675 | Ross 448 | 11:24:46.53 | +67:33:08.5 | M1.0 V | 8.659 | | |
| J11249+024 | StKM 1-941 | 11:24:58.52 | +02:28:26.8 | M1.0 V | 8.660 | | |
| J11240+381 | 1R112405.0+380809 | 11:24:04.53 | +38:08:10.7 | M4.5 V | 9.928 | Binary | |
| J11254+782 | GJ 3660 | 11:25:26.06 | +78:15:52.9 | d/sdM4 | 8.733 | Binary | |
| J11266+379 | PM J11266+3756 | 11:26:37.40 | +37:56:22.8 | M2.0 V | 8.572 | Binary | |
| J11276+039 | GJ 3664 | 11:27:38.47 | +03:58:36.1 | M0.0 V | 7.848 | | |
| J11289+101 | Wolf 398 | 11:28:55.45 | +10:10:48.2 | M4.0 V | 8.478 | | |
| J11302+076 | K2-18 | 11:30:14.43 | +07:35:16.1 | dM2.5 | 9.763 | | |
| J11306-080 | LP 672-42 | 11:30:41.44 | -08:05:38.9 | M3.5 V | 8.033 | | • |

Table B.1: Carmencita, the CARMENES input catalogue (continued).

| Karmn | Name | α (2016.0) | δ (2016.0) | Spectral type | J [mag] | Multiplicity[a] | DR1[b] |
|---|---|---|---|---|---|---|---|
| J11307+549 | StKM 1-950 | 11:30:43.80 | +54:57:29.1 | M1.0 V | 8.846 | Binary* | • |
| J11311-149 | GJ 3668 | 11:31:08.84 | -14:57:43.2 | M5.0 V | 9.359 | | • |
| J11315+022 | LP 552-68 | 11:31:32.21 | +02:13:34.8 | M2.5 V | 8.999 | | |
| J11317+226 | Ross 903 | 11:31:42.72 | +22:40:02.2 | M0.5 V | 7.171 | | |
| J11351-056 | GJ 3672 | 11:35:07.66 | -05:39:38.2 | M4.5 V | 10.269 | | |
| J11355+389 | GJ 3673 | 11:35:30.94 | +38:55:33.4 | M3.5 V | 9.034 | Binary | |
| J11376+587 | Ross 112 | 11:37:38.30 | +58:42:38.2 | M2.5 V | 8.978 | | |
| J11404+770 | LP 19-403 | 11:40:27.09 | +77:04:19.0 | M2.0 V | 8.379 | | |
| J11417+427 | Ross 1003 | 11:41:43.80 | +42:45:05.7 | M4.0 V | 7.608 | | |
| J11420+147 | Ross 115 | 11:42:01.43 | +14:46:39.9 | M3.0 V | 8.859 | | |
| J11421+267 | Ross 905 | 11:42:12.16 | +26:42:10.6 | M2.5 V | 6.900 | | • |
| J11423+230 | LP 375-23 | 11:42:18.14 | +23:01:37.3 | M0.5 V | 8.649 | Binary | |
| J11433+253 | GJ 3682 | 11:43:23.27 | +25:18:13.2 | M4.0 V | 9.507 | | • |
| J11451+183 | LP 433-47 | 11:45:11.57 | +18:20:53.8 | M4.0 V | 9.162 | | • |
| J11467-140 | GJ 443 | 11:46:43.69 | -14:01:04.4 | M3.0 V | 7.965 | | |
| J11470+700 | GJ 3684 A | 11:47:04.37 | +70:01:57.7 | M4.0 V | 9.309 | Binary | |
| J11474+667 | 1R114728.8+664405 | 11:47:28.27 | +66:44:02.6 | M5.0 V | 9.684 | | • |
| J11476+002 | GJ 3685 | 11:47:40.41 | +00:15:18.5 | M4.0 V | 8.991 | Binary | |
| J11476+786 | GJ 445 | 11:47:45.46 | +78:41:35.9 | M4.0 V | 6.724 | | • |
| J11477+008 | FIVir | 11:47:45.05 | +00:47:56.8 | dM4 | 6.505 | | • |
| J11483-112 | GJ 3688 | 11:48:18.61 | -11:17:15.0 | M3.0 V | 9.028 | | • |
| J11485+076 | G 10-52 | 11:48:35.63 | +07:41:37.8 | M3.5 V | 9.476 | | • |
| J11496+220 | BPM 87650 | 11:49:40.33 | +22:03:52.5 | M0.0 V | 8.403 | | |
| J11509+483 | GJ 1151 | 11:50:55.24 | +48:22:23.2 | M4.5 V | 8.488 | | |
| J11511+352 | GJ 450 | 11:51:06.98 | +35:16:23.3 | M1.5 V | 6.419 | | |
| J11519+075 | RX J1151.9+0731 | 11:51:56.69 | +07:31:25.2 | M2.5 V | 8.812 | Binary | • |
| J11529+244 | GJ 3691 | 11:52:57.55 | +24:28:46.9 | M4.1 V | 9.937 | | • |
| J11532-073 | GJ 452 | 11:53:15.91 | -07:22:35.8 | M2.5 V | 8.303 | | |
| J11533+430 | TYC 3016-577-1 | 11:53:23.24 | +43:02:56.3 | M1.0 V | 8.417 | | |
| J11538+069 | GJ 3693 | 11:53:53.01 | +06:59:41.9 | M8.0 V | 11.256 | | |
| J11541+098 | Ross 119 | 11:54:07.98 | +09:48:10.0 | M4.0 V | 8.699 | | |
| J11549-021 | PM J11549-0206 | 11:54:56.83 | -02:06:08.4 | M3.0 V | 9.547 | | |
| J11551+009 | Ross 129 | 11:55:06.42 | +00:58:26.1 | M1.5 V | 8.170 | | |
| J11557-189 | GJ 3694 | 11:55:44.87 | -18:54:36.6 | M3.5 V | 9.967 | | |
| J11557-227 | LP 851-346 | 11:55:42.42 | -22:25:01.8 | M7.5 V | 10.930 | | |
| J11575+118 | Ross 122 | 11:57:32.06 | +11:49:43.7 | M2.0 V | 8.429 | | |
| J11582+425 | GJ 3696 | 11:58:17.81 | +42:34:23.0 | M4.0 V | 9.594 | | |
| J11521+039 | StM 162 | 11:52:09.87 | +03:57:21.4 | M4.0 V | 8.382 | Binary | |
| J11589+426 | GJ 3697 | 11:58:58.99 | +42:39:40.8 | M2.0 V | 8.638 | Binary | |
| J11585+595 | G 197-38 | 11:58:33.53 | +59:33:22.2 | M0.0 V | 8.296 | Binary | |
| J12006-138 | GJ 3698 | 12:00:36.36 | -13:49:36.6 | M3.5 V | 8.852 | Binary | |
| J12016-122 | GJ 3700 | 12:01:40.72 | -12:13:57.6 | M3.0 V | 8.685 | Binary | |
| J12054+695 | Ross 689 | 12:05:28.29 | +69:32:21.7 | M4.0 V | 8.740 | | |
| J12057+784 | LSPM J1205+7825 | 12:05:45.99 | +78:25:51.6 | M2.5 V | 8.575 | | |
| J12023+285 | GJ 455 | 12:02:17.12 | +28:35:13.4 | sdM3.5 | 9.132 | Binary (SB2) | |
| J12088+303 | 1R120847.7+302120 | 12:08:49.63 | +30:21:00.5 | M2.5 V | 8.988 | | • |
| J12093+210 | StM 165 | 12:09:21.70 | +21:03:05.8 | M2.5 V | 9.472 | | |
| J12100-150 | GJ 3707 | 12:10:05.54 | -15:04:28.4 | dM3.5 | 7.768 | | |
| J12063-132 | 1R120622.6-131453 | 12:06:22.22 | -13:14:57.2 | M3.5 V | 8.702 | Binary | |
| J12109+410 | GJ 9393 | 12:10:56.90 | +41:03:31.7 | M0.0 V | 7.851 | | |
| J12104-131 | LP 734-34 | 12:10:28.65 | -13:10:29.5 | M4.5 V | 9.292 | Binary (SB2) | • |
| J12111-199 | GJ 3708 | 12:11:11.52 | -19:57:41.0 | dM3.0 | 7.895 | Binary | |
| J12112-199 | GJ 3709 | 12:11:16.71 | -19:58:24.7 | M2.5 V | 8.596 | Binary | |
| J12121+488 | GJ 3713 | 12:12:11.68 | +48:48:58.3 | M2.5 V | 9.258 | Binary | • |
| J12122+714 | LP 39-66 | 12:12:13.89 | +71:25:26.2 | M3.0 V | 8.815 | | |
| J12123+544S | HD 238090 | 12:12:21.29 | +54:29:10.2 | M0.0 V | 6.875 | Binary | |
| J12123+544N | GJ 458 B | 12:12:21.58 | +54:29:24.6 | M3.0 V | 9.171 | Binary | |
| J12124+121 | PM J12124+1211 | 12:12:25.99 | +12:11:38.2 | M2.0 V | 9.391 | | |
| J12124+396 | GJ 3714 | 12:12:29.65 | +39:40:25.2 | M1.0 V | 8.121 | Binary | • |

Table B.1: Carmencita, the CARMENES input catalogue (continued).

| Karmn | Name | $\alpha$ (2016.0) | $\delta$ (2016.0) | Spectral type | $J$ [mag] | Multiplicity[a] | DR1[b] |
|---|---|---|---|---|---|---|---|
| J12133+166 | IVCom | 12:13:19.91 | +16:41:32.3 | M1.5 V | 8.867 | | |
| J12144+245 | GJ 3717 | 12:14:26.04 | +24:35:20.5 | M2.0 V | 8.754 | | |
| J12151+487 | GJ 458.2 | 12:15:08.46 | +48:43:56.4 | M0.5 V | 7.610 | | |
| J12154+391 | GJ 3718 | 12:15:27.93 | +39:11:15.4 | M1.5 V | 8.607 | | |
| J12156+526 | StKM 2-809 | 12:15:39.55 | +52:39:08.7 | M4.0 V | 8.588 | | |
| J12142+006 | GJ 1154 | 12:14:15.53 | +00:37:21.8 | M4.5 V | 8.456 | | |
| J12162+508 | RX J1216.2+5053 | 12:16:14.90 | +50:53:36.7 | M4.0 V | | Binary | |
| J12168+029 | GJ 1155 A | 12:16:51.16 | +02:58:09.0 | M3.0 V | 9.234 | Binary | • |
| J12168+248 | PM J12168+2451E | 12:16:52.62 | +24:51:06.0 | M1.5 V | 8.993 | | |
| J12189+111 | GLVir | 12:18:58.02 | +11:07:37.0 | M4.5 V | 8.525 | | |
| J12169+311 | GJ 3719 | 12:16:58.27 | +31:09:22.6 | M3.0 V | 9.909 | Binary (SB2) | |
| J12191+318 | LP 320-626 | 12:19:05.55 | +31:50:43.5 | M3.5 V | 8.289 | Triple (SB2) | |
| J12194+283 | Wolf 408 | 12:19:23.31 | +28:22:57.8 | M0.5 V | 7.666 | | • |
| J12198+527 | StKM 1-1007 | 12:19:47.76 | +52:46:43.1 | M0.0 V | 8.281 | | |
| J12199+364 | G 123-36 | 12:29:55.22 | +36:26:40.0 | M1.0 V | 8.784 | | |
| J12204+005 | GJ 461 | 12:20:25.59 | +00:35:00.4 | M0.0 V | 6.861 | Binary | |
| J12214+306W | G 148-48 | 12:21:26.79 | +30:38:31.4 | M5.0 V | 9.987 | Binary | |
| J12214+306E | LP 320-416 | 12:21:26.49 | +30:38:33.6 | M4.5 V | 10.057 | Binary | |
| J12217+682 | LP 39-245 | 12:21:46.41 | +68:16:07.3 | M3.0 V | 8.828 | | |
| J12223+251 | Wolf 409 | 12:22:20.46 | +25:10:08.6 | M0.5 V | 8.409 | | |
| J12225+123 | BD+28 2110 | 12:22:33.90 | +27:36:16.5 | M0.0 V | 8.055 | | |
| J12230+640 | Ross 690 | 12:22:58.52 | +64:01:56.9 | M3.0 V | 7.937 | | |
| J12235+279 | Wolf 411 | 12:23:34.55 | +27:54:49.6 | M0.0 V | 8.480 | | |
| J12228-040 | G 13-33 | 12:22:50.33 | -04:04:47.5 | M4.5 V | 9.662 | Binary | |
| J12238+125 | HD 107888 | 12:23:53.60 | +12:34:46.2 | M0.0 V | 7.483 | | • |
| J12248-182 | Ross 695 | 12:24:53.73 | -18:15:09.2 | dM2.0 | 7.734 | | |
| J12235+671 | GJ 3722 | 12:23:33.85 | +67:11:16.4 | M2.5 V | 7.598 | Binary | |
| J12251+604 | LP 95-135 | 12:25:05.64 | +60:25:06.0 | M1.0 V | 9.872 | Binary | |
| J12269+270 | CXCom | 12:26:57.47 | +27:00:49.7 | M4.5 V | 10.197 | Triple | • |
| J12274+374 | G 148-61 | 12:27:29.12 | +37:26:35.1 | M1.5 V | 8.873 | | |
| J12277-032 | G 13-39A | 12:27:44.38 | -03:15:01.3 | M3.5 V | 8.763 | Binary | |
| J12288-106N | Ross 948A | 12:28:52.85 | -10:39:48.6 | M2.0 V | 7.682 | Binary | |
| J12288-106S | Ross 948B | 12:28:52.65 | -10:39:50.8 | M2.0 V | 7.652 | Binary | |
| J12289+084 | Wolf 414 | 12:28:56.90 | +08:25:27.3 | M3.5 V | 7.844 | Binary | |
| J12290+417 | GJ 3729 | 12:29:02.65 | +41:43:45.9 | M3.5 V | 8.786 | Binary (SB2(3?)) | |
| J12292+535 | GJ 1159 A | 12:29:12.22 | +53:32:47.0 | M4.0 V | 9.983 | Binary | |
| J12294+229 | GJ 3730 | 12:29:26.92 | +22:59:46.4 | M4.0 V | 9.823 | | |
| J12299-054W | GJ 3731 | 12:29:53.57 | -05:27:29.2 | M3.5 V | 8.818 | Triple (SB2) | |
| J12299-054E | GJ 3732 | 12:29:54.05 | -05:27:25.2 | M4.0 V | 9.792 | Triple | |
| J12312+086 | Wolf 417 | 12:31:15.12 | +08:48:29.8 | M0.5 V | 6.782 | | |
| J12323+315 | GJ 3733 | 12:32:19.79 | +31:36:03.2 | M3.0 V | 9.374 | | |
| J12324+203 | GJ 3734 | 12:32:26.37 | +20:23:28.0 | M2.5 V | 9.092 | | • |
| J12327+682 | LP 39-249 | 12:32:44.66 | +68:15:42.1 | M0.0 V | 8.353 | | |
| J12332+090 | Wolf 424 A | 12:33:15.48 | +09:01:19.5 | M5.0 V | 6.995 | Binary | |
| J12349+322 | PM J12349+3214 | 12:34:54.03 | +32:14:29.2 | M3.5 V | 9.459 | | |
| J12350+098 | Wolf 427 | 12:35:00.22 | +09:49:37.5 | M2.5 V | 7.995 | | |
| J12363-043 | GJ 3736 | 12:36:22.35 | -04:22:41.7 | M3.0 V | 9.579 | | |
| J12368-019 | PM J12368-0159 | 12:36:51.96 | -01:59:02.0 | M3.5 V | 9.440 | | • |
| J12373-208 | LP 795-38 | 12:37:21.52 | -20:52:42.4 | dM4.0 | 8.972 | | |
| J12387-043 | GJ 1162 | 12:38:46.47 | -04:19:20.2 | M4.3 V | 9.329 | | |
| J12388+116 | Wolf 433 | 12:38:51.18 | +11:41:42.1 | M3.5 V | 7.581 | | |
| J12390+470 | G 123-049A | 12:39:05.24 | +47:02:21.4 | M2.5 V | | Binary | • |
| J12397+255 | GJ 3739 | 12:39:43.32 | +25:30:43.3 | M4.0 V | 10.242 | | |
| J12416+482 | GJ 3741 | 12:41:38.24 | +48:14:22.3 | M1.0 V | 8.376 | | • |
| J12417+567 | RX J1241.7+5645 | 12:41:47.59 | +56:45:13.7 | M3.5 V | 9.483 | | |
| J12428+418 | G 123-55 | 12:42:49.10 | +41:53:47.8 | M4.0 V | 8.118 | | |
| J12436+251 | GJ 1163 | 12:43:35.62 | +25:06:21.1 | M3.5 V | 8.972 | | |
| J12440-111 | LP 735-29 | 12:44:00.22 | -11:10:32.8 | M4.5 V | 9.516 | | |
| J12470+466 | Ross 991 | 12:46:59.81 | +46:37:28.6 | M2.5 V | 8.104 | | • |

Table B.1: Carmencita, the CARMENES input catalogue (continued).

| Karmn | Name | α (2016.0) | δ (2016.0) | Spectral type | J [mag] | Multiplicity[a] | DR1[b] |
|---|---|---|---|---|---|---|---|
| J12471-035 | GJ 3747 | 12:47:09.24 | -03:34:18.0 | M3.0 V | 8.767 | | |
| J12479+097 | Wolf 437 | 12:47:55.53 | +09:44:57.7 | M3.5 V | 7.195 | | |
| J12364+352 | G 123-45 | 12:36:28.17 | +35:11:59.0 | M4.5 V | 9.113 | Binary (SB1) | |
| J12485+495 | RX J1248.5+4933 | 12:48:34.67 | +49:33:53.8 | M3.5 V | 8.684 | | |
| J12481+472 | GJ 3749 | 12:48:10.07 | +47:13:23.2 | M3.5 V | 9.607 | Binary | • |
| J12495+094 | Wolf 439 | 12:49:33.74 | +09:28:31.6 | M3.5 V | 9.091 | | |
| J12505+269 | GJ 3755 | 12:50:34.35 | +26:55:20.3 | M3.8 V | 9.947 | | |
| J12508-213 | APMPM J1251-2121 | 12:50:53.16 | -21:21:18.9 | M7.5 V | 11.160 | | |
| J12490+661 | DPDra | 12:49:01.61 | +66:06:35.2 | M3.0 V | 6.880 | Triple (ST3) | |
| J12513+221 | GJ 1166A | 12:51:23.71 | +22:06:15.7 | M3.0 V | 9.132 | Triple | |
| J12576+352E | BFCVn | 12:57:39.89 | +35:13:27.9 | M0.0 V | 7.401 | Quadruple | |
| J12576+352W | GJ 490 B | 12:57:38.94 | +35:13:16.9 | M4.5 V | 8.872 | Quadruple | |
| J12594+077 | GJ 3757 | 12:59:23.31 | +07:43:54.8 | M5.0 V | 10.740 | | |
| J13000-056 | Ross 972 | 13:00:03.58 | -05:37:47.1 | M3.0 V | 8.655 | | |
| J13005+056 | FNVir | 13:00:32.51 | +05:41:11.6 | M4.5 V | 8.553 | | |
| J12583+405 | LP 41-165 | 12:58:21.97 | +40:33:20.7 | M1.5 V | 8.699 | Binary | |
| J13007+123 | Wolf 462 | 13:00:45.87 | +12:22:32.1 | M0.0 V | 6.437 | Triple | |
| J13019+335 | G 164-38 | 13:01:55.96 | +33:35:23.0 | M1.0 V | 8.969 | | • |
| J13027+415 | G 123-84 | 13:02:46.65 | +41:31:06.6 | M3.5 V | 9.033 | | |
| J13047+559 | GJ 497 A | 13:04:46.29 | +55:54:10.6 | M0.5 V | 7.853 | Binary | |
| J13054+371 | GJ 3760 A | 13:05:29.44 | +37:08:07.6 | M2.5 V | 8.216 | Binary | |
| J13068+308 | GJ 3762 | 13:06:50.53 | +30:50:46.4 | M6.0 V | 10.226 | | |
| J13084+169 | GJ 9428 | 13:08:24.64 | +16:58:18.5 | M1.0 V | 8.492 | | |
| J13088+163 | GJ 3763 | 13:08:49.96 | +16:22:00.9 | M2.5 V | 9.264 | | |
| J13089+490 | GJ 3765 | 13:08:55.46 | +49:04:50.3 | M0.5 V | 7.897 | | |
| J13102+477 | G 177-25 | 13:10:11.62 | +47:45:08.8 | M5.0 V | 9.584 | | |
| J13113+285 | GJ 3766 | 13:11:21.13 | +28:32:34.4 | M5.0 V | 10.841 | | |
| J13118+253 | GJ 3767 | 13:11:51.24 | +25:20:48.1 | M5.0 V | 10.736 | | |
| J13119+658 | PM J13119+6550 | 13:11:59.17 | +65:50:01.3 | M3.0 V | 9.706 | | • |
| J13130+201 | GJ 1168 | 13:13:04.08 | +20:11:29.0 | M3.5 V | 8.867 | | |
| J13140+038 | GJ 3772 | 13:14:05.05 | +03:53:59.0 | M3.8 V | 9.490 | | |
| J13095+289 | GJ 1167 A | 13:09:34.56 | +28:59:03.1 | M4.8V | 9.476 | | • |
| J13142+792 | G 255-29A | 13:14:14.06 | +79:14:45.9 | M1.0 V | 8.603 | Binary | |
| J13165+278 | GJ 1169 | 13:16:31.95 | +27:52:33.5 | M4.0 V | 9.267 | | |
| J13167-123 | LP 737-14 | 13:16:45.10 | -12:20:21.2 | M3.5 V | 9.489 | | |
| J13143+133 | LP 497-33 | 13:14:20.08 | +13:19:57.9 | M6.0 V | 9.754 | Binary | |
| J13168+170 | HD 115404B | 13:16:52.28 | +17:00:55.7 | M0.5 V | 6.532 | Binary | |
| J13168+231 | GJ 3774 | 13:16:53.38 | +23:10:05.5 | M1.5 V | 8.424 | | |
| J13179+362 | GJ 1170 | 13:17:58.30 | +36:17:51.5 | M1.0 V | 8.112 | | |
| J13180+022 | GJ 3775 | 13:18:01.42 | +02:14:00.4 | M3.5 V | 8.787 | Binary | |
| J13182+733 | PM J13182+7322 | 13:18:13.82 | +73:22:05.6 | M3.5 V | 9.541 | Binary* | |
| J13195+351W | GJ 507 A | 13:19:34.10 | +35:06:24.2 | M0.5 V | 6.383 | Quadruple | |
| J13195+351E | GJ 507 B | 13:19:35.18 | +35:06:12.4 | M3.0 V | 8.287 | Quadruple (SB1) | |
| J13197+477 | HD 115953 | 13:19:45.90 | +47:46:40.5 | M2.0 V | 5.338 | Triple | |
| J13196+333 | Ross 1007 | 13:19:39.74 | +33:20:45.2 | M1.5 V | 7.266 | | |
| J13209+342 | Ross 1008 | 13:20:58.67 | +34:16:39.4 | M1.0 V | 7.398 | | • |
| J13215+035 | LSPM J1321+0332 | 13:21:30.15 | +03:33:02.1 | M1.0 V | 8.877 | | |
| J13215+037 | GJ 3778 | 13:21:34.69 | +03:45:54.5 | M2.0 V | 8.531 | | • |
| J13229+244 | Ross 1020 | 13:22:56.03 | +24:27:49.8 | M4.0 V | 8.728 | | |
| J13235+292 | HD 116495A | 13:23:32.21 | +29:14:18.9 | M0.0 V | 6.259 | Binary | |
| J13239+694 | LP 40-109 | 13:23:56.16 | +69:27:03.5 | M0.0 V | 8.426 | | • |
| J13247-050 | G 14-52 | 13:24:46.56 | -05:04:24.8 | M4.0 V | 9.465 | | |
| J13251-114 | PM J13251-1126 | 13:25:11.61 | -11:26:33.7 | M3.0 V | 9.156 | | |
| J13254+377 | BD+38 2445 | 13:25:28.07 | +37:43:10.9 | M0.0 V | 8.274 | Binary | |
| J13255+688 | 2M13253177+6850106 | 13:25:31.76 | +68:50:09.8 | M0.0 V | 10.039 | | |
| J13260+275 | PM J13260+2735A | 13:26:02.70 | +27:35:03.7 | M3.0 V | 9.249 | Triple* | |
| J13282+300 | BD+30 2400 | 13:28:17.53 | +30:02:43.0 | M0.0 V | 8.378 | Triple | • |
| J13283-023W | Ross 486A | 13:28:21.24 | -02:21:45.0 | dM3.0 | 7.515 | Binary | |
| J13283-023E | Ross 486B | 13:28:21.68 | -02:21:39.6 | M4.0 V | 9.240 | Binary | |

Table B.1: Carmencita, the CARMENES input catalogue (continued).

| Karmn | Name | $\alpha$ (2016.0) | $\delta$ (2016.0) | Spectral type | $J$ [mag] | Multiplicity[a] | DR1[b] |
|---|---|---|---|---|---|---|---|
| J13293+114 | GJ 513 | 13:29:21.67 | +11:26:07.6 | M3.0 V | 8.368 | | |
| J13294-143 | 1R132923.9-142206 | 13:29:24.21 | -14:22:13.0 | M3.5 V | 9.061 | | • |
| J13299+102 | Ross 490 | 13:30:01.01 | +10:22:20.6 | M1.0 V | 5.902 | | • |
| J13300-087 | Ross 476 | 13:30:01.59 | -08:42:33.0 | M4.0 V | 9.599 | Binary | |
| J13305+191 | GJ 1171 | 13:30:30.49 | +19:09:13.5 | M5.0 V | 10.065 | | • |
| J13318+233 | GJ 3790 | 13:31:50.26 | +23:23:21.1 | M2.5 V | 8.608 | | |
| J13317+292 | DGCVn | 13:31:46.33 | +29:16:34.3 | M4.0 V | 7.561 | Binary | |
| J13319+311 | GJ 9448 B | 13:31:58.08 | +31:08:05.3 | M0.0 V | 7.564 | Binary | |
| J13326+309 | LP 323-169 | 13:32:38.85 | +30:59:05.3 | M4.5 V | 9.620 | Binary | |
| J13327+168 | VWCom | 13:32:44.93 | +16:48:35.8 | M2.5 V | 7.643 | Binary | |
| J13335+704 | PM J13335+7029 | 13:33:33.27 | +70:29:41.6 | M3.5 V | 9.227 | | |
| J13343+046 | Wolf 1487 | 13:34:21.67 | +04:40:00.7 | M0.0 V | 7.213 | | |
| J13348+201 | GJ 3793 | 13:34:49.40 | +20:11:35.7 | M3.5 V | 9.671 | | |
| J13348+745 | GJ 9453 B | 13:34:49.83 | +74:30:12.6 | M3.5 V | 9.574 | Binary | |
| J13358+146 | G 150-17 | 13:35:50.73 | +14:41:07.4 | M2.5 V | 8.638 | | |
| J13369+229 | Ross 1021 | 13:36:55.35 | +22:57:58.2 | M2.5 V | 8.979 | | |
| J13378+481 | GJ 520 A | 13:37:50.84 | +48:08:14.8 | M0.0 V | 6.943 | Triple | |
| J13376+481 | GJ 520 C | 13:37:40.09 | +48:07:52.0 | M4.0 V | 10.122 | Triple | |
| J13386+258 | Ross 1022 | 13:38:36.37 | +25:49:50.7 | M3.0 V | 8.751 | | |
| J13386-115 | 1R133841.3-113137 | 13:38:41.05 | -11:32:09.1 | M4.5 V | 9.714 | | |
| J13388-022 | Ross 488 | 13:38:53.11 | -02:15:48.5 | M2.0 V | 8.595 | | |
| J13401+437 | Ross 1026 | 13:40:07.19 | +43:46:42.8 | M3.5 V | 8.544 | | |
| J13413-091 | PM J13413-0907 | 13:41:21.33 | -09:07:16.4 | M2.5 V | 9.444 | | |
| J13414+489 | StM 186 | 13:41:27.84 | +48:54:43.4 | M3.5 V | 9.002 | Binary | |
| J13394+461 | GJ 521 | 13:39:24.04 | +46:11:17.6 | M1.5 V | 7.054 | Binary | |
| J13415+148 | GJ 3799 | 13:41:31.58 | +14:49:27.6 | M1.5 V | 8.855 | | |
| J13417+582 | StM 187 | 13:41:46.48 | +58:15:18.9 | M3.5 V | 8.733 | Binary | |
| J13427+332 | Ross 1015 | 13:42:43.13 | +33:17:12.9 | M3.5 V | 7.787 | | |
| J13430+090 | GJ 3802 | 13:43:00.91 | +09:04:21.8 | M3.0 V | 9.093 | | |
| J13434+111 | TYC 896-760-1 | 13:43:25.03 | +11:06:42.1 | M0.5 V | 8.428 | | |
| J13421-160 | GJ 3800 | 13:42:09.28 | -16:00:24.1 | M4e | 8.971 | Binary | • |
| J13444+516 | Ross 492 | 13:44:27.00 | +51:41:08.6 | M2.5 V | 8.989 | Binary | |
| J13445+249 | LP 379-98 A | 13:44:33.38 | +24:57:03.6 | M1.0 V | 8.699 | Binary | |
| J13450+176 | Wolf 497 | 13:45:05.59 | +17:46:38.2 | M0.0 V | 6.997 | | |
| J13455+609 | MCC 699 | 13:45:31.28 | +60:58:58.3 | M0.5 V | 8.150 | | |
| J13457+148 | HD 119850 | 13:45:45.74 | +14:53:06.2 | M1.5 V | 5.181 | | • |
| J13458-179 | GJ 3804 | 13:45:50.37 | -17:58:14.5 | dM3.5 | 7.745 | | |
| J13477+214 | BD+22 2632A | 13:47:42.49 | +21:27:36.3 | M0.0 V | 8.215 | Binary | • |
| J13481-137 | LP 738-14 | 13:48:06.52 | -13:44:39.8 | M4.5 V | 10.413 | Binary | • |
| J13482+236 | GJ 1179 A | 13:48:11.69 | +23:36:50.7 | M5.0 V | 10.082 | Binary | |
| J13485+563 | Ross 493 | 13:48:34.32 | +56:20:09.6 | M1.5 V | 8.612 | | |
| J13488+041 | Wolf 1494 | 13:48:48.61 | +04:05:59.4 | M4.5 V | 9.755 | | |
| J13490+026 | Wolf 1495 A | 13:49:01.16 | +02:47:23.6 | M1.5 V | 7.839 | Binary | |
| J13507-216 | GJ 3810 | 13:50:43.96 | -21:41:33.0 | M3.0 V | 8.871 | Binary | |
| J13503-216 | LP 798-41 | 13:50:23.73 | -21:37:25.9 | M3.5 V | 9.458 | Binary | |
| J13508+367 | Ross 1019 | 13:50:51.18 | +36:44:18.5 | M4.0 V | 9.299 | | |
| J13518+127 | RX J1351.8+1247 | 13:51:53.02 | +12:47:07.0 | M2.0 V | 8.788 | | |
| J13526+144 | Wolf 515 | 13:52:36.26 | +14:25:15.7 | M2.0 V | 8.013 | Binary | |
| J13528+656 | GJ 533.1 | 13:52:48.61 | +65:37:17.7 | M1.5 V | 8.523 | Binary | |
| J13528+668 | GJ 3815 | 13:52:49.25 | +66:48:57.2 | M5.0 V | 10.539 | | |
| J13536+776 | LP 21-224 | 13:53:39.91 | +77:37:07.9 | M4.0 V | 8.635 | | |
| J13529+536 | LP 97-259 | 13:52:55.76 | +56:36:17.0 | M1.0 V | 8.648 | | |
| J13534+129 | Ross 835 | 13:53:27.37 | +12:56:22.9 | M0.0 V | 6.945 | Binary | |
| J13537+521 | PM J13537+5210A | 13:53:45.89 | +52:10:27.3 | M3.5 V | 9.128 | Binary | |
| J13537+788 | GJ 534.2 | 13:53:45.76 | +78:51:08.7 | M0.0 V | 7.639 | | • |
| J13582+125 | Ross 837 | 13:58:13.56 | +12:34:55.4 | M3.5 V | 8.269 | | |
| J13582-120 | LP 739-2 | 13:58:15.80 | -12:02:58.4 | M4.5 V | 9.728 | | |
| J13583-132 | LP 739-3 | 13:58:19.96 | -13:16:26.0 | M4.0 V | 9.488 | | |
| J13587-000 | GJ 3818 | 13:58:43.20 | -00:04:54.3 | M4.0 V | 9.963 | | • |

Table B.1: Carmencita, the CARMENES input catalogue (continued).

| Karmn | Name | α (2016.0) | δ (2016.0) | Spectral type | J [mag] | Multiplicity[a] | DR1[b] |
|---|---|---|---|---|---|---|---|
| J13591-198 | GJ 3820 | 13:59:09.78 | -19:50:06.6 | M4.5 V | 8.334 | | |
| J14010-026 | HD 122303 | 14:01:02.31 | -02:39:07.9 | M1.0 V | 6.516 | | |
| J14019+154 | GJ 536.1 A | 14:01:58.86 | +15:29:40.7 | M0.0 V | 7.728 | Binary | • |
| J14019+432 | PM J14019+4316A | 14:01:58.67 | +43:16:41.1 | M2.5 V | 9.281 | Binary | • |
| J14023+136 | GJ 3822 | 14:02:19.73 | +13:41:20.4 | M0.5 V | 7.563 | | |
| J14024-210 | GJ 3821 | 14:02:29.43 | -21:00:42.9 | M3.5 V | 9.163 | | |
| J14025+463S | GJ 537 A | 14:02:34.03 | +46:20:23.0 | M0.5 V | 6.269 | Binary | |
| J14025+463N | GJ 537 B | 14:02:34.19 | +46:20:26.4 | M0.5 V | 6.264 | Binary | |
| J14039+242 | LSPM J1403+2440 | 14:03:54.74 | +24:40:44.2 | M2.5 V | 8.935 | | |
| J14041+207 | StKM 1-1119 | 14:04:09.06 | +20:44:30.9 | M1.0 V | 8.590 | Triple | |
| J14062+693 | NLTT 36313 | 14:06:14.66 | +69:18:38.8 | M3.0 V | 8.731 | | |
| J14082+805 | GJ 540 | 14:08:14.37 | +80:35:41.5 | M1.0 V | 7.179 | | |
| J14083+758 | GJ 3826 | 14:08:20.91 | +75:51:13.2 | M1.5 V | 8.349 | | |
| J14121-005 | GJ 3828 A | 14:12:10.22 | -00:35:00.3 | M2.5 V | 9.352 | Triple | • |
| J14130-120 | GQVir | 14:13:04.19 | -12:01:32.9 | M4.5 V | 9.040 | Binary (SB2) | |
| J14142-153 | GJ 3832 | 14:14:16.87 | -15:21:15.9 | M3.5 V | 9.687 | Quadruple | |
| J14144+234 | GJ 3834 | 14:14:26.33 | +23:27:26.4 | M3.5 V | 9.516 | | |
| J14152+450 | Ross 992 | 14:15:15.96 | +45:00:49.7 | M3.0 V | 8.014 | | |
| J14153+153 | LP 439-350 | 14:15:20.63 | +15:23:01.5 | M2.0 V | 8.654 | | |
| J14155+046 | GJ 1182 | 14:15:31.75 | +04:39:19.1 | M5.69 | 9.433 | Binary (SB2) | • |
| J14157+594 | LP 97-674 A | 14:15:42.13 | +59:27:29.1 | M2.2 V | 8.850 | Binary | |
| J14159+362 | G 165-58 | 14:15:56.39 | +36:16:42.2 | M3.5 V | 8.940 | | • |
| J14161+233 | LP 81-30 | 14:16:11.25 | +23:23:26.5 | M1.0 V | 8.713 | | |
| J14170+105 | GJ 3838 | 14:17:04.56 | +10:35:34.3 | M1.5 V | 8.185 | | |
| J14170+317 | GJ 3839 | 14:17:02.14 | +31:42:44.7 | M4.5 V | 8.443 | Binary (SB2/ST3) | |
| J14171+088 | PM J14171+0851 | 14:17:07.18 | +08:51:37.1 | M4.5 V | 9.109 | Binary (SB2) | |
| J14174+454 | GJ 541.2 | 14:17:24.45 | +45:26:39.8 | M0.0 V | 7.389 | Binary | |
| J14173+454 | RX J1417.3+4525 | 14:17:22.17 | +45:25:45.7 | dM5.0 | 9.467 | Binary | |
| J14175+025 | RX J1417.5+0233 | 14:17:30.18 | +02:33:42.5 | M3.0 V | 9.274 | | • |
| J14177+214 | LP 381-94 | 14:17:47.76 | +21:25:58.8 | M1.0 V | 8.534 | | |
| J14179-005 | GJ 3840 | 14:17:58.75 | -00:31:33.8 | M2.5 V | 9.040 | | |
| J14189+386 | LP 220-78 | 14:18:58.09 | +38:38:22.5 | M1.0 V | 8.500 | | |
| J14191-073 | Wolf 534 | 14:19:09.81 | -07:18:24.0 | M3.0 V | 9.665 | | |
| J14194+029 | LP 560-1 | 14:19:29.37 | +02:54:34.1 | M5.0 V | 9.954 | | |
| J14200+390 | IZBoo | 14:20:04.66 | +39:03:02.5 | M3.0 V | 8.572 | | |
| J14201-096 | Ross 848 | 14:20:06.71 | -09:37:26.8 | M4.0 V | 8.740 | | |
| J14212-011 | GJ 3843 | 14:21:15.31 | -01:07:29.8 | M4.0 V | 8.948 | | |
| J14215-079 | PM J14215-0755 | 14:21:33.96 | -07:55:18.0 | M4.0 V | 9.456 | | |
| J14219+376 | LP 270-68 | 14:21:55.61 | +37:39:45.5 | M1.5 V | 8.982 | | |
| J14227+164 | LP 440-13 | 14:22:43.19 | +16:24:47.2 | M5.0 V | 10.303 | | |
| J14231-222 | GJ 3845 | 14:23:07.49 | -22:17:16.7 | M4.5 V | 10.402 | | |
| J14249+088 | GJ 3846 | 14:24:56.58 | +08:53:18.0 | M3.0 V | 8.420 | | |
| J14210+275 | GJ 3844 | 14:21:03.13 | +27:35:34.5 | M2.5 V | 8.928 | Binary | |
| J14251+518 | HD 126660B | 14:25:11.17 | +51:49:46.6 | M2.5 V | 7.883 | Binary | |
| J14255-118 | LP 740-10 | 14:25:33.79 | -11:48:51.5 | M4.0 V | 9.353 | | |
| J14257+236W | GJ 548 A | 14:25:44.40 | +23:36:43.6 | M0.0 V | 6.769 | Binary | • |
| J14257+236E | GJ 548 B | 14:25:47.58 | +23:36:55.8 | M0.5 V | 6.889 | Binary | |
| J14259+142 | V358Boo | 14:25:55.87 | +14:12:09.6 | M0.0 V | 8.147 | | • |
| J14269+241 | LSPM J1426+2408 | 14:26:58.63 | +24:08:56.9 | M1.0 V | 8.908 | | • |
| J14279-003S | GJ 1183 A | 14:27:55.69 | -00:22:30.5 | M4.65 | 9.305 | Binary | |
| J14279-003N | GJ 1183 B | 14:27:56.01 | -00:22:18.3 | M4.70 | 9.345 | Binary | |
| J14280+139 | LP 500-35 | 14:28:03.75 | +13:56:05.4 | M7.0 V | 11.014 | | |
| J14283+053 | LP 560-27 | 14:28:21.12 | +05:19:00.3 | M3.0 V | 8.724 | Binary | |
| J14282+053 | LP 560-26 | 14:28:17.18 | +05:18:44.8 | M3.5 V | 8.932 | Binary | |
| J14294+155 | Ross 130 | 14:29:28.53 | +15:32:18.3 | M2.0 V | 7.229 | | |
| J14299+295 | GJ 3853 | 14:29:59.26 | +29:33:54.9 | M4.0 V | 10.072 | | |
| J14306+597 | GJ 3855 | 14:30:36.08 | +59:43:27.4 | M6.5 V | 10.790 | | • |
| J14307-086 | HD 127339 | 14:30:46.35 | -08:38:50.6 | M0.5 V | 6.616 | | |
| J14310-122 | Wolf 1478 | 14:31:00.72 | -12:17:52.3 | dM3.5 | 7.803 | | |

Table B.1: Carmencita, the CARMENES input catalogue (continued).

| Karmn | Name | $\alpha$ (2016.0) | $\delta$ (2016.0) | Spectral type | $J$ [mag] | Multiplicity[a] | DR1[b] |
|---|---|---|---|---|---|---|---|
| J14312+754 | LSPM J1431+7526 | 14:31:12.59 | +75:26:41.7 | M4.0 V | 9.792 | | • |
| J14320+738 | G 255-55 | 14:32:01.99 | +73:49:23.3 | M2.0 V | 8.139 | | • |
| J14321+081 | LP 560-35 | 14:32:07.99 | +08:11:31.4 | M6.0 V | 10.108 | | |
| J14321+160 | GJ 3856 | 14:32:11.00 | +16:00:48.2 | M5.0 V | 9.288 | | |
| J14322+496 | GJ 3858 | 14:32:13.58 | +49:39:04.2 | M3.5 V | 9.277 | | • |
| J14331+610 | G 224-13A | 14:33:06.93 | +61:00:43.9 | M2.5 V | 8.171 | Binary | |
| J14342-125 | HNLib | 14:34:16.42 | -12:31:00.9 | M3.5 V | 6.838 | | |
| J14366+143 | StKM 1-1170 | 14:36:38.98 | +14:21:52.6 | M1.0 V | 8.676 | | |
| J14368+583 | GJ 3861 | 14:36:54.24 | +58:20:43.6 | M3.0 V | 8.079 | Binary (SB2) | • |
| J14371+756 | LSPM J1437+7536N | 14:37:09.94 | +75:36:54.7 | M2.0 V | 8.650 | Binary | |
| J14376+677 | G 239-22 | 14:37:39.32 | +67:45:34.9 | M1.5 V | 8.917 | Binary | |
| J14388+422 | GPM 219.718548+42.229288 | 14:38:51.66 | +42:13:43.9 | M1.5 V | 8.793 | | |
| J14415+064 | LP 560-66 | 14:41:32.72 | +06:27:41.2 | M1.5 V | 8.843 | | |
| J14423+660 | GJ 9492 | 14:42:20.79 | +66:03:20.2 | M2.0 V | 7.306 | Binary | |
| J14438+667 | NLTT 38291 | 14:43:50.67 | +66:44:34.6 | M1.0 V | 8.935 | | |
| J14472+570 | RX J1447.2+5701 | 14:47:13.67 | +57:01:54.4 | M4.0 V | 9.914 | | |
| J14485+101 | LP 501-17 | 14:48:32.79 | +10:06:55.7 | M3.5 V | 9.475 | | |
| J14501+323 | LP 326-34 | 14:50:11.14 | +32:18:13.8 | M3.5 V | 9.144 | | |
| J14511+311 | LP 326-38 | 14:51:09.97 | +31:06:37.4 | M4.0 V | 8.408 | | |
| J14524+123 | GJ 3871 | 14:52:28.47 | +12:23:29.2 | M2.5 V | 7.967 | | |
| J14525+001 | Wolf 555 | 14:52:32.27 | +00:10:02.8 | M2.5 V | 8.999 | | |
| J14469+170 | Ross 994 | 14:46:59.22 | +17:05:07.5 | M1.5 V | 8.600 | Binary | |
| J14538+235 | Ross 52A | 14:53:50.59 | +23:33:22.5 | M3.5 V | 7.438 | Binary | • |
| J14544+161 | CEBoo | 14:54:29.55 | +16:06:01.9 | M1.0: V | 6.633 | Triple | |
| J14544+355 | Ross 1041 | 14:54:28.10 | +35:32:43.5 | M3.5 V | 8.243 | | |
| J14548+099 | Ross 1028b | 14:54:53.15 | +09:56:30.1 | M2.0 V | 8.217 | | |
| J14549+411 | GJ 3875 | 14:54:54.61 | +41:08:50.5 | M4.5 V | 10.249 | Binary (SB) | • |
| J14557+072 | G 66-42 | 14:55:47.82 | +07:17:47.7 | M0.5 V | 8.412 | | |
| J14564+168 | G 136-35 | 14:56:28.16 | +16:48:29.1 | M1.5 V | 8.960 | Binary (SB) | |
| J14574-214 | HD 131976 | 14:57:27.70 | -21:25:07.9 | M1.0 V | 4.550 | Quadruple (SB2) | |
| J14575+313 | Ross 53A | 14:57:31.43 | +31:23:25.9 | M2.0 V | 8.444 | Binary | |
| J14578+566 | GJ 1187 | 14:57:54.28 | +56:39:14.1 | M5.5 V | 10.207 | | |
| J15009+454 | GJ 572 | 15:00:55.91 | +45:25:39.8 | M0.5 V | 6.198 | Binary | |
| J15011+354 | Ross 1042 | 15:01:11.98 | +35:27:10.5 | M2.0 V | 8.670 | | |
| J15013+055 | GJ 3885 | 15:01:20.20 | +05:32:48.5 | M3.0 V | 8.326 | | |
| J15018+550 | GJ 3891 | 15:05:48.60 | +55:04:45.8 | M3.5 V | 9.239 | | |
| J15030+704 | LP 41-431 | 15:03:01.68 | +70:26:14.0 | M3.0 V | 8.871 | | |
| J15043+294 | GJ 575.1 | 15:04:22.56 | +29:28:39.9 | M2.5 V | 9.220 | | • |
| J15043+603 | Ross 1051 | 15:04:17.09 | +60:23:07.4 | M1.0 V | 7.701 | | |
| J15049-211 | GJ 3888 | 15:04:57.53 | -21:07:03.9 | M4.5 V | 10.174 | | |
| J15060+453 | PM J15060+4521 | 15:06:03.02 | +45:21:52.8 | M1.5 V | 8.999 | | |
| J15073+249 | GJ 579 | 15:07:22.59 | +24:56:15.8 | M0.0 V | 7.296 | | |
| J15011+071 | Ross 1030a | 15:01:10.20 | +07:09:46.5 | M3.5 V | 8.682 | Binary | |
| J15079+762 | LSPM J1507+7613 | 15:07:56.64 | +76:14:01.8 | M4.5 V | 9.235 | Triple | |
| J15081+623 | LSPM J1508+6221 | 15:08:11.53 | +62:21:56.4 | M4.0 V | 9.296 | Binary | |
| J15095+031 | Ross 1047 | 15:09:34.95 | +03:10:08.3 | M3.0 V | 7.720 | | |
| J15100+193 | GJ 3893 | 15:10:04.82 | +19:21:20.2 | M4.3 V | 9.056 | | |
| J15118-102 | GJ 3894 | 15:11:49.55 | -10:14:22.1 | M4.5 V | 9.651 | | • |
| J15119+179 | GJ 3895 | 15:11:55.49 | +17:57:07.4 | M4.0 V | 9.556 | | |
| J15147+645 | LP 67-339 | 15:14:45.76 | +64:33:50.3 | M3.5 V | 9.785 | Binary | |
| J15151+333 | LP 272-63 | 15:15:06.93 | +33:17:57.0 | M2.0 V | 9.207 | | |
| J15156+638 | PM J15156+6349 | 15:15:37.69 | +63:49:50.8 | M1.5 V | 8.992 | | |
| J15166+391 | LP 222-65 | 15:16:40.44 | +39:10:47.4 | M7.0 V | 10.796 | | |
| J15126+457 | GJ 3898 | 15:12:37.60 | +45:43:52.3 | M4.0 V | 8.977 | Binary | |
| J15188+292 | StKM 1-1229 | 15:18:49.75 | +29:15:06.4 | M1.0 V | 8.624 | Triple* | |
| J15193+678 | GJ 3902 | 15:19:17.35 | +67:51:24.0 | M3.0 V | 9.568 | | |
| J15194-077 | HOLib | 15:19:25.51 | -07:43:21.7 | M3.0 V | 6.706 | | |
| J15197+046 | PM J15197+0439 | 15:19:45.88 | +04:39:36.0 | M4.0 V | 9.546 | | |
| J15210+309 | PM J15210+3057 | 15:21:00.56 | +30:57:00.9 | M2.5 V | 8.963 | | |

Table B.1: Carmencita, the CARMENES input catalogue (continued).

| Karmn | Name | α (2016.0) | δ (2016.0) | Spectral type | J [mag] | Multiplicity[a] | DR1[b] |
|---|---|---|---|---|---|---|---|
| J15214+042 | TYC 344-504-1 | 15:21:25.39 | +04:14:49.9 | M1.5 V | 8.553 | | • |
| J15218+209 | OTSer | 15:21:53.03 | +20:58:42.0 | M1.0 V | 6.610 | | |
| J15219+185 | LP 442-37 | 15:21:56.83 | +18:35:47.9 | M1.5 V | 8.714 | | |
| J15238+174 | Ross 508 | 15:23:50.70 | +17:27:37.3 | M4.92 | 9.105 | | |
| J15191-127 | GJ 3900 | 15:19:10.93 | -12:45:09.3 | M4.0 V | 8.507 | Binary (SB2) | • |
| J15238+561 | StKM 1-1240 | 15:23:54.01 | +56:09:31.5 | M1.0 V | 8.794 | Triple (SB2) | |
| J15238+584 | G 224-65 | 15:23:51.06 | +58:28:11.2 | M4.0 V | 9.905 | | |
| J15273+415 | TYC 3055-1525-1 | 15:27:19.05 | +41:30:09.9 | M1.5 V | 8.363 | Binary | |
| J15276+408 | G 179-29 | 15:27:38.85 | +40:52:02.2 | M1.0 V | 8.480 | | |
| J15280+257 | GJ 587.1 | 15:28:01.43 | +25:47:22.9 | M0.0 V | 8.143 | | |
| J15290+467 | RX J1529.0+4646 | 15:29:02.77 | +46:46:23.5 | M4.5 V | 9.942 | Binary | |
| J15305+094 | LP 502-56 | 15:30:30.13 | +09:26:04.4 | M5.5 V | 9.569 | | |
| J15297+428 | GJ 3907 | 15:29:44.63 | +42:52:38.9 | M4.68 | 9.587 | Binary | |
| J15336+462 | GJ 3911 | 15:33:39.33 | +46:15:06.1 | M3.6 V | 9.417 | | |
| J15319+288 | GJ 3910 | 15:31:53.50 | +28:51:10.2 | M4.0 V | 9.673 | Binary (SB1) | • |
| J15339+379 | GJ 588.1 | 15:33:54.82 | +37:54:48.4 | M0.5 V | 8.226 | Binary | |
| J15340+513 | LP 135-414 | 15:34:03.52 | +51:22:06.8 | M4.5 V | 9.370 | | |
| J15345+142 | Ross 512 | 15:34:29.76 | +14:16:15.5 | M4.0 V | 9.642 | | |
| J15349-143 | 2MUCD 11346 | 15:34:55.92 | -14:18:54.5 | M8.6V | 11.380 | | |
| J15353+177S | Ross 513 | 15:35:19.17 | +17:42:44.3 | M2.5 V | 8.694 | Binary | |
| J15353+177N | Ross 513B | 15:35:18.98 | +17:43:01.7 | M4.5 V | 10.282 | Binary | |
| J15357+221 | GJ 3913 | 15:35:45.31 | +22:09:01.6 | M3.5 V | 8.681 | | |
| J15368+375 | BKCrB | 15:36:49.97 | +37:34:48.0 | M0.0 V | 8.397 | Binary | |
| J15369-141 | Ross 802 | 15:36:58.13 | -14:08:11.8 | dM4.0 | 8.432 | | |
| J15386+371 | G 179-42 | 15:38:36.68 | +37:07:28.0 | M3.5 V | 9.979 | | |
| J15400+434N | GJ 1194 A | 15:40:05.37 | +43:29:34.1 | M3.0 V | 8.312 | Binary | • |
| J15400+434S | GJ 1194 B | 15:40:05.54 | +43:29:30.0 | M4.0 V | 8.893 | Binary | |
| J15412+759 | UU UMi | 15:41:20.12 | +75:59:22.5 | M3.5 V | 8.260 | Binary (SB2) | |
| J15416+184 | StKM 1-1264 | 15:41:37.17 | +18:28:09.2 | M1.5 V | 8.955 | Triple | |
| J15421-194 | GJ 595 | 15:42:04.27 | -19:28:34.9 | M3.0 V | 7.920 | Binary (SB1) | • |
| J15474-108 | GJ 3916 | 15:47:24.21 | -10:53:53.2 | M2.5 V | 7.582 | Triple (ST3) | |
| J15476+226 | 1R154741.3+224108 | 15:47:40.51 | +22:41:16.0 | M4.5 V | 9.543 | | |
| J15474+451 | LP 177-102 | 15:47:27.03 | +45:07:54.5 | M4.0 V | 9.082 | Binary (EB/SB2) | • |
| J15480+043 | RX J1548.0+0421 | 15:48:02.78 | +04:21:38.4 | M2.5 V | 9.058 | Binary* | |
| J15488+305 | PM J15488+3030 | 15:48:48.60 | +30:30:38.7 | M3.0 V | 8.958 | | |
| J15493+250 | G 168-13 | 15:49:20.38 | +25:03:48.5 | M2.0 V | 8.884 | | |
| J15496+510 | GJ 3920 | 15:49:35.62 | +51:03:02.0 | M2.0 V | 8.573 | | |
| J15499+796 | LP 23-35 | 15:49:53.83 | +79:39:53.6 | M5.0 V | 9.721 | | |
| J15501+009 | Wolf 587 | 15:50:11.41 | +00:57:32.0 | M3.0 V | 8.677 | | |
| J15512+306 | TYC 2572-633-1 | 15:51:14.63 | +30:40:42.2 | M1.5 V | 8.974 | | |
| J15513+295 | GJ 3923 | 15:51:21.51 | +29:30:59.2 | M3.5 V | 8.961 | | • |
| J15496+348 | GJ 3919 | 15:49:37.37 | +34:49:07.8 | M4.0 V | 8.728 | Binary | |
| J15531+347N | Ross 806 | 15:53:06.69 | +34:45:05.9 | M3.0 V | 8.001 | Triple | |
| J15531+347S | GJ 3926 | 15:53:06.97 | +34:44:39.5 | M3.0 V | 8.994 | Triple | |
| J15538+641 | NLTT 41533 | 15:53:48.31 | +64:09:36.2 | M0.5 V | 8.441 | | |
| J15555+352 | GJ 3928 | 15:55:31.51 | +35:12:05.2 | M4.0 V | 8.928 | Binary | |
| J15557+686 | RX J1555.7+6840 | 15:55:47.10 | +68:40:16.0 | M2.5 V | 8.593 | | |
| J15569+376 | RX J1556.9+3738 | 15:56:58.12 | +37:38:14.3 | M2.5 V | 9.417 | | |
| J15578+090 | LSPM J1557+0901 | 15:57:48.42 | +09:01:07.6 | M4.0 V | 9.282 | | |
| J15581+494 | V1022Her | 15:58:10.41 | +49:27:05.5 | M1.0 V | 8.732 | | |
| J15583+354 | GJ 3929 | 15:58:18.61 | +35:24:29.4 | M3.5 V | 8.694 | | |
| J15587+346 | StM 258 | 15:58:45.75 | +34:48:53.8 | M3.5 V | 8.809 | | |
| J15597+440 | RX J1559.7+4403 | 15:59:47.20 | +44:03:59.6 | M2.0 V | 8.509 | Binary | • |
| J15598-082 | GJ 606 | 15:59:53.60 | -08:15:12.0 | M1.0 V | 7.185 | | |
| J16008+403 | GJ 3933 | 16:00:50.40 | +40:19:38.7 | M3.0 V | 9.216 | | |
| J16017+301 | GJ 607 | 16:01:43.15 | +30:10:52.9 | M3.0 V | 8.666 | | • |
| J16018+304 | GJ 3935 | 16:01:52.43 | +30:27:37.0 | M2.5 V | 9.530 | Binary | |
| J16017+304 | GJ 3936 | 16:01:44.35 | +30:27:42.9 | M4.5 V | 10.427 | Binary | |
| J16028+205 | GJ 609 | 16:02:49.85 | +20:35:01.2 | M4.0 V | 8.132 | | |

Table B.1: Carmencita, the CARMENES input catalogue (continued).

| Karmn | Name | $\alpha$ (2016.0) | $\delta$ (2016.0) | Spectral type | $J$ [mag] | Multiplicity[a] | DR1[b] |
|---|---|---|---|---|---|---|---|
| J16033+175 | PM J16033+1735 | 16:03:20.60 | +17:35:55.0 | M2.0 V | 8.821 | | |
| J16043-062 | GJ 3937 | 16:04:19.91 | -06:16:59.9 | M4.5 V | 10.452 | | • |
| J16046+263 | BPM 91242 | 16:04:36.85 | +26:20:44.6 | M0.5 V | 8.251 | | |
| J16048+391 | HD 144579B | 16:04:50.08 | +39:09:36.6 | M4.0 V | 9.903 | Binary | |
| J16054+769 | GJ 3941 | 16:05:26.51 | +76:54:58.5 | M3.0 V | 8.632 | | |
| J16062+290 | LP 329-30 | 16:06:13.41 | +29:02:01.6 | M2.0 V | 8.473 | | |
| J16066+083 | GJ 611.3 | 16:06:40.66 | +08:23:19.6 | M1.0 V | 8.422 | Triple | |
| J16074+059 | GJ 3939 | 16:07:27.90 | +05:57:56.0 | M3.8 V | 9.525 | | |
| J16082-104 | GJ 1198 | 16:08:14.56 | -10:26:35.1 | M4.5 V | 10.257 | | |
| J16090+529 | GJ 3942 | 16:09:03.50 | +52:56:39.0 | M0.5 V | 7.185 | | |
| J16092+093 | LP 504-59 | 16:09:15.90 | +09:21:11.0 | M3.0 V | 7.969 | | |
| J16102-193 | K2-33 | 16:10:14.73 | -19:19:09.8 | dM3.0 | 11.095 | | |
| J16120+033 | TYC 371-1053-1 | 16:12:04.68 | +03:18:19.8 | M2.0 V | 8.127 | Binary | • |
| J16126-188 | LP 804-27 | 16:12:41.82 | -18:52:35.2 | M3.0 V | 7.555 | | • |
| J16139+337 | sig CrB C | 16:13:55.90 | +33:46:22.7 | M2.5 V | 8.598 | Quintuple | |
| J16144-028 | LP 624-54 | 16:14:25.19 | -02:50:54.9 | M5.0 V | 11.303 | | |
| J16145+191 | GJ 1200 | 16:14:30.36 | +19:06:16.7 | M3.5 V | 8.990 | | |
| J16147+048 | GJ 3946 | 16:14:43.36 | +04:52:09.7 | M3.5 V | 9.476 | | |
| J16155+244 | GJ 3947 | 16:15:32.10 | +24:27:50.9 | M1.5 V | 8.645 | | |
| J16167+672S | HD 147379 | 16:16:41.37 | +67:14:21.2 | M0.0 V | 5.779 | Binary | |
| J16167+672N | EWDra | 16:16:43.98 | +67:15:23.9 | M3.0 V | 6.908 | Binary | |
| J16180+062 | 1R161804.9+061702 | 16:18:05.01 | +06:17:12.0 | M3.0 V | 8.950 | | • |
| J16170+552 | CRDra | 16:17:05.52 | +55:16:01.9 | M1.0 V | 6.600 | Binary | • |
| J16204-042 | GJ 618.1 A | 16:20:24.34 | -04:16:02.6 | M0.0 V | 7.950 | Binary | |
| J16220+228 | V1169Her | 16:22:01.12 | +22:50:22.8 | M1.5 V | 8.904 | | |
| J16247+229 | LSPM J1624+2254 | 16:24:43.71 | +22:54:19.3 | M1.0 V | 8.016 | | |
| J16254+543 | GJ 625 | 16:25:25.41 | +54:18:12.0 | M1.5 V | 6.608 | | |
| J16241+483 | GJ 623 | 16:24:11.16 | +48:21:03.1 | M3.0 V | 6.638 | Binary | |
| J16255+323 | LP 330-13 | 16:25:32.91 | +32:18:34.1 | M2.0 V | 8.868 | | |
| J16259+834 | TYC 4647-2406-1 | 16:25:59.21 | +83:24:23.2 | M1.5 V | 8.642 | | • |
| J16255+260 | GJ 3953 | 16:25:32.12 | +26:01:38.0 | M3.0 V | 14.776 | Binary | |
| J16268-173 | GJ 3954 | 16:26:47.75 | -17:23:40.5 | M5e | 9.548 | Binary | |
| J16280+155 | GJ 3955 | 16:28:02.04 | +15:33:52.1 | M3.6 V | 9.375 | Binary | |
| J16303-126 | V2306Oph | 16:30:17.96 | -12:40:04.3 | M3.0 V | 5.950 | | |
| J16313+408 | GJ 3959 | 16:31:18.59 | +40:51:56.6 | M6.0 V | 9.461 | | |
| J16315+175 | GJ 1202 | 16:31:34.69 | +17:33:35.9 | M4.0 V | 8.926 | | |
| J16327+126 | GJ 1203 | 16:32:44.36 | +12:36:43.8 | M3.0 V | 8.429 | | • |
| J16328+098 | GJ 3960 | 16:32:53.09 | +09:50:28.5 | M3.5 V | 8.988 | | • |
| J16342+543 | LP 137-37 | 16:34:13.28 | +54:23:48.5 | M1.0 V | 8.720 | | |
| J16302-146 | GJ 2121 | 16:30:12.52 | -14:39:53.0 | M2.5 V | 8.416 | Binary | • |
| J16343+571 | CMDra | 16:34:18.14 | +57:10:03.3 | M4.5 V | 8.501 | Triple (EB/SB2) | |
| J16360+088 | GJ 1204 | 16:36:05.09 | +08:48:46.7 | M4.0 V | 9.419 | | |
| J16395+505 | G 202-68 | 16:39:30.81 | +50:33:57.0 | M1.0 V | 8.628 | | • |
| J16401+007 | GJ 3967 | 16:40:06.17 | +00:42:16.3 | M5.0 V | 9.116 | | |
| J16403+676 | GJ 3971 | 16:40:19.87 | +67:36:10.6 | M7.0 V | 9.854 | | |
| J16408+363 | Ross 812 | 16:40:48.72 | +36:19:02.9 | M2.0 V | 8.069 | | |
| J16420+192 | PM J16420+1916 | 16:42:00.72 | +19:16:11.6 | M2.5 V | 8.692 | | |
| J16462+164 | GJ 3972 | 16:46:13.35 | +16:28:33.2 | M3.0 V | 7.951 | | |
| J16465+345 | LP 276-22 | 16:46:31.03 | +34:34:49.0 | M6.0 V | 10.533 | | |
| J16354+350 | V1200Her | 16:35:27.61 | +35:00:55.3 | M4.0 V | 8.615 | Binary | |
| J16487-157 | GJ 3973 | 16:48:45.95 | -15:44:23.7 | M1.5 Vk: | 7.656 | | • |
| J16508-048 | GJ 3975 | 16:50:52.99 | -04:50:38.8 | M3.5 V | 9.465 | | |
| J16509+224 | GJ 3976 | 16:50:57.98 | +22:27:12.1 | M5.0 V | 9.136 | | |
| J16528+630 | GSC 04194-01561 | 16:52:49.82 | +63:04:41.2 | M4.5 V | 9.592 | | |
| J16529+400 | GJ 3979 | 16:52:54.92 | +40:05:04.6 | M3.0 V | 9.389 | | |
| J16542+119 | Ross 644 | 16:54:11.44 | +11:54:57.9 | M0.0 V | 7.886 | | |
| J16487+106 | LSPM J1648+1038 | 16:48:46.40 | +10:38:51.1 | M2.5 V | 7.820 | Binary | |
| J16554-083S | V1054Oph | 16:55:27.89 | -08:20:25.2 | M3.0 V | 5.270 | Quintuple (ST3) | |
| J16554-083N | Wolf 629 | 16:55:24.34 | -08:19:35.7 | M3.5 V | 7.555 | Quintuple | |

Table B.1: Carmencita, the CARMENES input catalogue (continued).

| Karmn | Name | α (2016.0) | δ (2016.0) | Spectral type | J [mag] | Multiplicity[a] | DR1[b] |
|---|---|---|---|---|---|---|---|
| J16555-083 | GJ 644 C | 16:55:34.38 | -08:23:54.7 | M7.0 V | 9.776 | Quintuple | • |
| J16570-043 | GJ 1207 | 16:57:06.25 | -04:21:02.3 | dM4 | 7.971 | | |
| J16573+124 | GJ 3980 | 16:57:23.12 | +13:28:06.6 | M4.0 V | 10.036 | | • |
| J16573+271 | PM J16573+2708 | 16:57:22.27 | +27:08:31.2 | M2.0 V | 9.356 | Binary | • |
| J16574+777 | GJ 3986 | 16:57:29.79 | +77:43:06.3 | M3.0 V | 8.982 | | |
| J16577+132 | GJ 647 | 16:57:46.15 | +13:17:31.1 | M0.0 V | 7.872 | Binary | |
| J16578+473 | HD 153557B | 16:57:53.40 | +47:22:06.7 | M1.5 V | 6.874 | Triple | |
| J16581+257 | Ross 860 | 16:58:08.71 | +25:44:30.8 | M1.0 V | 6.448 | | |
| J16584+139 | GJ 3981 A | 16:58:24.81 | +13:58:10.9 | M4.0 V | 8.801 | Binary | |
| J16587+688 | LP 43-338 | 16:58:42.17 | +68:53:55.8 | M1.5 V | 8.749 | | • |
| J16591+209 | V1234Her | 16:59:09.58 | +20:58:18.3 | M3.5 V | 8.338 | Binary | |
| J17003+253 | GJ 3983 | 17:00:20.18 | +25:21:05.1 | M3.2 V | 9.404 | | |
| J17006+063 | G 139-4 | 17:00:38.65 | +06:18:42.0 | M1.0 V | 8.586 | | |
| J17010+082 | GJ 3984 | 17:01:01.85 | +08:12:23.5 | M3.8 V | 9.437 | | |
| J17027-060 | GJ 3987 | 17:02:49.45 | -06:04:07.6 | M0 | 7.802 | | |
| J17033+514 | GJ 3988 | 17:03:24.10 | +51:24:32.6 | M5.0 V | 8.768 | | |
| J17038+321 | LP 331-57 A | 17:03:53.09 | +32:11:47.6 | M2.0 V | 7.886 | Triple (SB2?) | |
| J17043+169 | GJ 1209 | 17:04:22.49 | +16:55:37.5 | M3.0 V | 8.571 | | • |
| J17052-050 | HD 154363B | 17:05:12.80 | -05:05:57.4 | M1.5 V | 6.780 | Binary | |
| J17058+260 | LP 387-37 | 17:05:52.55 | +26:05:27.4 | M1.5 V | 8.887 | Binary | |
| J17071+215 | Ross 863 | 17:07:06.96 | +21:33:14.1 | M3.0 V | 7.875 | | • |
| J17082+516 | G 203-44 | 17:08:12.71 | +51:38:10.6 | M1.0 V | 8.794 | | |
| J17076+073 | GJ 1210 | 17:07:40.31 | +07:22:00.6 | M5.0 V | 9.284 | Binary | • |
| J17098+119 | GJ 3990 | 17:09:52.36 | +11:55:32.8 | M4.0 V | 9.611 | | |
| J17095+436 | GJ 3991 | 17:09:32.03 | +43:40:48.4 | M3.5 V | 7.380 | Binary (SB1) | |
| J17104+279 | StM 336 | 17:10:25.49 | +27:58:38.6 | M2.5 V | 8.987 | Binary | |
| J17121+456 | HD 155876 | 17:12:08.21 | +45:39:32.6 | M3.0 V | 5.552 | Binary | |
| J17115+384 | Wolf 654 | 17:11:35.04 | +38:26:33.2 | M4.0 V | 7.630 | | |
| J17118-018 | GJ 660 | 17:11:51.82 | -01:51:10.5 | M3.0 V | 7.462 | Binary | • |
| J17146+269 | GJ 3994 | 17:14:39.87 | +26:55:44.4 | M2.0 V | 8.861 | | |
| J17153+049 | GJ 1214 | 17:15:19.56 | +04:57:38.1 | M4.5 V | 9.750 | | |
| J17136-084 | V2367Oph | 17:13:40.00 | -08:25:21.4 | M3.5 V | 8.120 | Binary (SB2) | |
| J17158+190 | GJ 3997 | 17:15:49.95 | +19:00:00.3 | M0.5 V | 7.212 | Binary | |
| J17160+110 | GJ 3998 | 17:16:00.49 | +11:03:22.1 | M1.0 V | 7.634 | | |
| J17166+080 | GJ 2128 | 17:16:40.68 | +08:03:29.1 | M2.5 V | 7.933 | | |
| J17177+116 | GJ 1215 | 17:17:43.71 | +11:40:05.3 | M5.0 V | 9.817 | Binary | |
| J17177-118 | GJ 3999 | 17:17:45.28 | -11:48:59.0 | M3.0 V | 8.818 | Binary | • |
| J17183+181 | GJ 4002 | 17:18:22.44 | +18:08:52.4 | M3.0 V | 9.050 | Binary | |
| J17183-017 | GJ 4001 | 17:18:21.82 | -01:46:55.4 | M0.0 V | 7.554 | Binary | |
| J17199+265 | V647Her | 17:19:53.95 | +26:30:08.7 | M3.5 V | 7.273 | Binary | |
| J17198+265 | V639Her | 17:19:52.70 | +26:30:08.3 | M4.5 V | 8.229 | Binary | |
| J17198+417 | GJ 671 | 17:19:53.12 | +41:42:36.4 | M2.5 V | 7.712 | | |
| J17207+492 | GJ 1216 | 17:20:47.19 | +49:15:01.4 | M4.0 V | 10.124 | | • |
| J17219+214 | GJ 4003 | 17:21:54.44 | +21:25:51.3 | M4.0 V | 9.344 | | |
| J17225+055 | PM J17225+0531 | 17:22:33.80 | +05:31:15.1 | M2.5 V | 8.966 | | |
| J17242-043 | GJ 4005 | 17:24:16.69 | -04:21:54.1 | M2.5 V | 8.515 | | |
| J17276+144 | GJ 1219 | 17:27:38.65 | +14:28:56.4 | M4.0 V | 9.693 | | |
| J17285+374 | Wolf 750 | 17:28:30.41 | +37:27:04.1 | M3.5 V | 9.387 | | |
| J17303+055 | Wolf 751 | 17:30:22.76 | +05:32:50.7 | M1.0 V | 6.240 | | |
| J17312+820 | GJ 1220 | 17:31:14.95 | +82:05:27.8 | M4.0 V | 9.572 | | |
| J17316+047 | Wolf 755 | 17:31:37.97 | +01:47:48.0 | M1.5 V | 8.586 | | • |
| J17321+504 | GJ 4011 | 17:32:07.82 | +50:24:41.7 | M2.5 V | 9.008 | | |
| J17328+543 | G 226-64 | 17:32:53.06 | +54:20:19.6 | M2.0 V | 8.935 | | |
| J17338+169 | V1274Her | 17:33:53.03 | +16:55:11.0 | M6.0 V | 8.895 | | |
| J17340+446 | RX J1734.0+4447B | 17:34:05.47 | +44:47:08.4 | M3.5 V | | Binary | |
| J17355+616 | GJ 685 | 17:35:35.07 | +61:40:45.4 | M0.5 V | 6.884 | Triple | • |
| J17364+683 | GJ 687 | 17:36:24.97 | +68:20:00.6 | M3.0 V | 5.335 | | |
| J17376+220 | GJ 4015 | 17:37:36.52 | +22:05:45.2 | M4.1 V | 9.764 | | • |
| J17378+185 | GJ 686 | 17:37:54.39 | +18:35:45.9 | M1.5 V | 6.360 | | • |

Table B.1: Carmencita, the CARMENES input catalogue (continued).

| Karmn | Name | $\alpha$ (2016.0) | $\delta$ (2016.0) | Spectral type | $J$ [mag] | Multiplicity[a] | DR1[b] |
|---|---|---|---|---|---|---|---|
| J17386+612 | GJ 4021 | 17:38:40.87 | +61:14:00.0 | M4.0 V | 10.245 | | |
| J17388+080 | GJ 4016 | 17:38:51.16 | +08:01:31.7 | M2.5 V | 8.702 | | • |
| J17395+277S | GJ 4018 | 17:39:30.73 | +27:45:40.8 | M0.5 V | 8.097 | Triple | |
| J17395+277N | GJ 4019 | 17:39:32.26 | +27:46:33.5 | M3.0 V | 8.828 | Triple | |
| J17419+407 | G 204-25 | 17:41:57.04 | +40:44:44.3 | M1.5 V | 8.917 | | |
| J17421-088 | Wolf 1471 | 17:42:09.85 | -08:49:07.8 | M3.0 V | 9.812 | | |
| J17423+574 | 2M17422264+5726521 | 17:42:22.63 | +57:26:52.0 | M0.0 V | 10.484 | | |
| J17425-166 | GJ 690.1 | 17:42:32.15 | -16:38:35.9 | M2.5 V | 9.441 | | |
| J17430+057 | GJ 4024 | 17:43:01.06 | +05:47:20.5 | M1.0 V | 7.441 | Binary | |
| J17431+854 | G 259-20 | 17:43:07.85 | +85:26:25.5 | M2.0 V | 8.736 | Binary | |
| J17432-185 | Ross 133 | 17:43:17.37 | -18:31:27.7 | M1.5 V | 8.872 | | |
| J17439+433 | GJ 694 | 17:43:55.98 | +43:22:33.3 | M2.5 V | 6.812 | | |
| J17460+246 | GJ 4026 | 17:46:04.22 | +24:39:13.1 | M4.0 V | 8.814 | | |
| J17455+468 | GJ 694.2 | 17:45:33.50 | +46:51:19.8 | M0.0 V | 7.636 | Binary | |
| J17464+277 | GJ 695 C | 17:46:24.65 | +27:42:49.8 | M3.5 V | 5.772 | Quadruple | |
| J17469+228 | StKM 1-1528 | 17:46:55.96 | +22:48:01.6 | M1.0 V | 8.674 | | |
| J17481+159 | PM J17481+1558 | 17:48:11.19 | +15:58:48.4 | M3.0 V | 9.610 | | |
| J17502+237 | GJ 4030 | 17:50:14.74 | +23:45:58.1 | M3.3 V | 9.320 | | |
| J17515+147 | GJ 4031 | 17:51:30.58 | +14:45:32.4 | M3.0 V | 9.653 | | |
| J17521+647 | G 227-20 | 17:52:11.83 | +64:46:02.9 | M0.5 V | 8.249 | | |
| J17464-087 | Wolf 1473 | 17:46:29.28 | -08:42:43.4 | M3.5 V | 8.198 | Binary (SB2) | |
| J17530+169 | GJ 4032 A | 17:53:00.32 | +16:54:59.5 | M3.0 V | 8.699 | Binary | |
| J17542+073 | GJ 1222 | 17:54:16.48 | +07:22:40.2 | M4.0 V | 8.772 | | |
| J17547+128 | LSPM J1754+1251 | 17:54:43.24 | +12:51:18.6 | M2.0 V | 8.738 | | |
| J17570+157 | GJ 4038 | 17:57:03.33 | +15:46:39.3 | M3.0 V | 8.329 | Binary | • |
| J17572+707 | LP 44-162 | 17:57:15.43 | +70:42:06.4 | M7.5 V | 11.452 | | |
| J17578+046 | Barnard's Star | 17:57:47.64 | +04:44:21.9 | M3.5 V | 5.244 | | |
| J17578+465 | GJ 4040 | 17:57:50.94 | +46:35:28.4 | M2.5 V | 7.847 | Binary | |
| J17589+807 | LP 24-152 | 17:58:55.00 | +80:42:53.9 | M3.5 V | 8.994 | | • |
| J18010+508 | Wolf 1403 | 18:01:05.86 | +50:49:38.9 | M1.5 V | 8.927 | | • |
| J18012+355 | G 182-34 | 18:01:16.14 | +35:35:41.4 | M3.5 V | 9.695 | | |
| J18022+642 | LP 71-82 | 18:02:17.09 | +64:15:38.1 | M5.0 V | 8.541 | | |
| J18027+375 | GJ 1223 | 18:02:46.49 | +37:30:44.7 | dM5.0 | 9.720 | | |
| J18031+179 | Wolf 792 | 18:03:06.02 | +17:54:20.0 | M1.0 V | 8.664 | | • |
| J18037+247 | Ross 820 | 18:03:47.68 | +25:45:16.7 | M0.0 V | 8.021 | | • |
| J18042+359 | GJ 4041 | 18:04:17.70 | +35:57:21.5 | M0.5 V | 7.728 | Binary | |
| J18051-030 | HD 165222 | 18:05:08.19 | -03:01:58.1 | M0.0 V | 6.161 | | |
| J18063+728 | NLTT 46021 | 18:06:18.11 | +72:49:20.1 | M0.0 V | 8.488 | | |
| J18074+184 | Wolf 806 | 18:07:27.99 | +18:27:54.1 | M1.0 V | 8.646 | | • |
| J18075-159 | GJ 1224 | 18:07:32.15 | -15:57:52.6 | M4.0 V | 8.639 | | |
| J18096+318 | LP 334-11 | 18:09:40.77 | +31:52:16.0 | M1.0 V | 8.232 | | |
| J18103+512 | Wolf 1412 | 18:10:23.37 | +51:15:55.0 | M2.0 V | 8.996 | | • |
| J18109+220 | StKM 1-1582 | 18:10:56.18 | +22:01:31.2 | M0.0 V | 8.437 | | |
| J18116+061 | LP 569-163 | 18:11:36.49 | +06:06:27.9 | M3.0 V | 9.266 | Binary | |
| J18131+260 | V1334Her | 18:13:06.83 | +26:01:51.3 | M4.0 V | 8.899 | | |
| J18134+054 | LP 569-15 | 18:13:28.12 | +05:26:54.8 | M1.5 V | 8.432 | Binary | |
| J18135+055 | LP 569-16 | 18:13:33.09 | +05:32:08.4 | M4.0 V | 9.702 | Binary | • |
| J18157+189 | HD 348274 | 18:15:43.54 | +18:56:12.7 | M0.0 V | 7.757 | | |
| J18160+139 | GJ 708.2 | 18:16:02.36 | +13:54:40.2 | M0.0 V | 7.343 | | |
| J18163+015 | GJ 708.3 | 18:16:17.74 | +01:31:17.1 | M3.0 V | 8.738 | | |
| J18165+048 | G 140-51 | 18:16:31.37 | +04:52:52.3 | M5.0 V | 9.798 | | |
| J18165+455 | GJ 709 | 18:16:31.07 | +45:33:33.7 | M0.5 V | 7.264 | | |
| J18174+483 | V401Dra | 18:17:25.05 | +48:22:03.1 | dM2.0 | 7.770 | | • |
| J18180+387E | GJ 4048 | 18:18:03.73 | +38:46:16.2 | M3.0 V | 8.040 | Binary | |
| J18180+387W | GJ 4049 | 18:18:02.88 | +38:46:17.4 | M4.0 V | | Binary | • |
| J18189+661 | GJ 4053 | 18:18:58.38 | +66:11:26.2 | M4.5 V | 8.740 | | • |
| J18193-057 | GJ 4051 | 18:19:21.68 | -05:46:33.5 | M2.0 V | 9.233 | | |
| J18195+420 | PM J18195+4201 | 18:19:34.48 | +42:01:37.4 | M1.5 V | 8.788 | | • |
| J18209-010 | GJ 1226 A | 18:20:56.62 | -01:03:14.2 | M3.5 V | 8.754 | Binary | |

Table B.1: Carmencita, the CARMENES input catalogue (continued).

| Karmn | Name | α (2016.0) | δ (2016.0) | Spectral type | J [mag] | Multiplicity[a] | DR1[b] |
|---|---|---|---|---|---|---|---|
| J18221+063 | Ross 136 | 18:22:05.40 | +06:20:39.4 | M4.0 V | 8.671 | | |
| J18224+620 | GJ 1227 | 18:22:24.92 | +62:02:42.0 | M4.0 V | 8.640 | | |
| J18227+379 | G 205-19 | 18:22:43.47 | +37:57:41.1 | M1.0 V | 8.542 | | |
| J18234+281 | Ross 708A | 18:23:28.24 | +28:10:01.0 | M3.5 V | 8.306 | Binary | • |
| J18240+016 | GJ 4056 | 18:24:05.35 | +01:41:12.3 | M2.5 V | 8.297 | | • |
| J18248+282 | Ross 710 | 18:24:52.32 | +28:17:22.3 | M1.5 V | 8.821 | Binary | |
| J18250+246 | HD 336196 | 18:25:04.75 | +24:37:57.3 | M0.0 V | 7.899 | | |
| J18255+383 | GJ 4058 | 18:25:31.82 | +38:21:00.8 | M0.0 V | 8.285 | | |
| J18264+113 | GJ 4059 | 18:26:24.58 | +11:20:52.9 | M3.5 V | 8.918 | Binary | |
| J18292+638 | TYC 4222-2195-1 | 18:29:13.36 | +63:51:10.9 | M1.5 V | 8.721 | | |
| J18296+338 | 2M18294012+3350130 | 18:29:40.02 | +33:50:13.3 | M3.0 V | 9.841 | | |
| J18312+068 | LP 570-92 | 18:31:16.21 | +06:50:08.2 | M1.0 V | 7.579 | | |
| J18319+406 | GJ 4062 | 18:31:58.24 | +40:41:17.6 | M3.5 V | 8.065 | | |
| J18346+401 | GJ 4063 | 18:34:36.73 | +40:07:23.1 | M4.0 V | 7.184 | | |
| J18352+243 | G 184-13A | 18:35:13.42 | +24:18:39.2 | M2.5 V | 8.590 | Binary | |
| J18352+414 | GJ 4064 | 18:35:18.38 | +41:29:14.5 | M2.0 V | 8.446 | | • |
| J18353+457 | GJ 720 A | 18:35:19.08 | +45:44:44.4 | M0.5 V | 6.881 | Binary | • |
| J18354+457 | GJ 720 B | 18:35:27.99 | +45:45:46.7 | M2.5 V | 8.886 | Binary | |
| J18356+329 | LSPM J1835+3259 | 18:35:37.79 | +32:59:41.2 | M8.5 V | 10.270 | | |
| J18358+800 | GJ 4068 | 18:35:52.58 | +80:05:42.9 | M4.0 V | 8.989 | | • |
| J18362+567 | G 227-39 | 18:36:12.72 | +56:44:37.8 | M0.0 V | 8.354 | | |
| J18363+136 | Ross 149 | 18:36:19.43 | +13:36:30.8 | M4.0 V | 8.186 | | • |
| J18387+047 | LP 570-22 | 18:38:47.63 | +04:46:01.7 | M0.5 V | 8.320 | Binary | |
| J18394+690 | RX J1839.4+6903 | 18:39:25.71 | +69:03:06.8 | M2.0 V | 8.533 | | |
| J18395+298 | LP 335-12 | 18:39:33.18 | +29:52:12.9 | M6.5 V | 11.011 | | • |
| J18395+301 | LP 335-13 | 18:39:32.00 | +30:09:52.1 | M0.0 V | 8.065 | | |
| J18399+334 | GJ 4067 | 18:39:59.94 | +33:24:58.9 | M3.5 V | 9.131 | | |
| J18400+726 | LP 44-334 | 18:40:02.20 | +72:40:57.2 | M6.5 V | 10.974 | Binary | |
| J18387-144 | GJ 2138 | 18:38:44.87 | -14:29:35.1 | M2.5 V | 7.661 | Binary | |
| J18402-104 | Wolf 1466 | 18:40:17.68 | -10:28:04.1 | M1.0 V | 8.262 | | |
| J18405+595 | G 227-43 | 18:40:35.77 | +59:30:53.6 | M2.0 V | 8.777 | | |
| J18409+315 | BD+31 3330B | 18:40:55.33 | +31:31:37.3 | M1.0 V | 8.210 | Triple | |
| J18409-133 | Ross 720 | 18:40:57.20 | -13:22:57.3 | dM1.0 | 7.397 | | |
| J18411+247S | GJ 1230 A | 18:41:10.35 | +24:47:15.8 | M4.5 V | 7.528 | Triple (SB2) | |
| J18411+247N | GJ 1230 B | 18:41:10.39 | +24:47:20.5 | M4.5 V | 8.860 | Triple | • |
| J18416+397 | GJ 4069 | 18:41:37.04 | +39:42:09.3 | M4.0 V | 9.216 | | |
| J18419+318 | Ross 145 | 18:41:58.66 | +31:49:49.9 | M3.0 V | 7.523 | | |
| J18427+139 | V816Her | 18:42:44.94 | +13:54:22.4 | M5.0 V | 8.361 | | |
| J18427+596N | HD 173739 | 18:42:43.94 | +59:38:18.1 | M3.0 V | 5.189 | Binary | |
| J18427+596S | HD 173740 | 18:42:43.94 | +59:38:06.5 | M3.5 V | 5.721 | Binary | • |
| J18432+253 | TYC 2112-920-1 | 18:43:14.51 | +25:22:43.1 | M1.0 V | 8.771 | | |
| J18433+406 | V492Lyr | 18:43:21.96 | +40:40:30.8 | M7.5 V | 11.313 | Binary | • |
| J18451+063 | 1R184510.6+062016 | 18:45:10.23 | +06:20:14.7 | M1.0 V | 7.656 | | • |
| J18453+188 | G 184-24 | 18:45:22.79 | +18:51:54.3 | M4.0 V | 9.273 | | |
| J18480-145 | GJ 4077 | 18:48:01.01 | -14:34:55.0 | dM2.5 | 8.375 | | |
| J18482+076 | G 141-36 | 18:48:17.94 | +07:41:25.2 | M5.0 V | 8.853 | | |
| J18487+615 | LSPM J1848+6135 | 18:48:47.07 | +61:35:06.3 | M2.0 V | 8.967 | | |
| J18498-238 | V1216Sgr | 18:49:50.11 | -23:50:13.6 | M3.5 V | 6.222 | | • |
| J18499+186 | G 184-31 | 18:49:54.34 | +18:40:24.2 | M4.5 V | 9.380 | | • |
| J18500+030 | Ross 142 | 18:50:00.63 | +03:05:10.7 | M0.5 V | 7.724 | | |
| J18507+479 | GJ 4083 | 18:50:45.66 | +47:58:17.3 | M3.5 V | 8.686 | | • |
| J18515+027 | BD+02 3698 | 18:51:35.59 | +02:46:19.2 | M0.5 V | 8.432 | | |
| J18516+244 | GJ 4084 | 18:51:40.44 | +24:27:31.7 | M3.0 V | 8.925 | | |
| J18518+165 | HD 229793 | 18:51:50.93 | +16:34:52.1 | M0.0 V | 7.162 | | |
| J18519+130 | 1R185200.0+130005 | 18:51:59.66 | +13:00:01.1 | M2.0 V | 8.345 | | |
| J18534+028 | G 141-46 | 18:53:25.68 | +02:50:49.4 | M2.5 V | 8.834 | | |
| J18548+008 | BD+00 4050 | 18:54:53.14 | +00:51:44.8 | M0.0 V | 7.789 | | |
| J18548+109 | HD 230017B | 18:54:53.86 | +10:58:45.1 | M3.5 V | 7.139 | Triple* | |
| J18554+084 | GJ 735 | 18:55:27.51 | +08:24:07.9 | M3.0 V | 6.311 | Binary (SB2) | |

Table B.1: Carmencita, the CARMENES input catalogue (continued).

| Karmn | Name | $\alpha$ (2016.0) | $\delta$ (2016.0) | Spectral type | $J$ [mag] | Multiplicity[a] | DR1[b] |
|---|---|---|---|---|---|---|---|
| J18563+544 | GJ 4091 B | 18:56:18.29 | +54:29:45.5 | M2.0 V | | Quadruple (SB2) | |
| J18564+463 | GJ 4089 | 18:56:26.40 | +46:22:58.5 | M4.1 V | 9.598 | | |
| J18571+075 | GJ 4088 | 18:57:10.40 | +07:34:14.7 | M2.0 V | 8.331 | | |
| J18576+535 | G 229-20A | 18:57:38.42 | +53:31:14.4 | M3.5 V | 8.906 | Triple | |
| J18580+059 | HD 176029 | 18:57:59.93 | +05:54:09.7 | M1.0 V | 6.239 | | |
| J18596+079 | GJ 4092 | 18:59:39.00 | +07:59:11.2 | M0.5 V | 7.971 | | |
| J19025+704 | GJ 4093 | 19:02:30.83 | +70:25:53.6 | M2.5 V | 8.958 | | |
| J19025+754 | LSPM J1902+7525 | 19:02:32.79 | +75:25:06.6 | M1.5 V | 9.797 | | • |
| J19032+034 | G 141-57 | 19:03:13.40 | +03:24:01.4 | M3.0 V | 8.665 | | |
| J19032+639 | GJ 4094 | 19:03:17.46 | +63:59:35.9 | M3.5 V | 7.785 | Triple | |
| J19032-135 | GJ 741 | 19:03:16.10 | -13:34:12.9 | M4.0 V | 10.375 | | |
| J19041+211 | 1R190405.9+211030 | 19:04:06.24 | +21:10:32.7 | M2.0 V | 8.750 | Binary* | |
| J19044+590 | LSPM J1907+5905 | 19:07:24.98 | +59:05:11.9 | M3.0 V | 8.457 | Binary | |
| J19045+240 | TYC 2122-1204-1 | 19:04:31.24 | +24:01:54.8 | M2.0 V | 8.789 | | |
| J19072+208 | HD 349726 | 19:07:12.66 | +20:52:31.9 | M2.0 V | 7.278 | Binary | |
| J19070+208 | Ross 730 | 19:07:05.02 | +20:53:11.4 | M2.0 V | 7.295 | Binary | |
| J19082+265 | GJ 1231 | 19:08:15.55 | +26:34:57.6 | M5.0 V | 10.361 | | |
| J19084+322 | GJ 4098 | 19:08:29.67 | +32:16:48.0 | M3.0 V | 7.905 | | • |
| J19077+325 | GJ 747 | 19:07:44.54 | +32:32:59.3 | M3.0 V | 7.242 | Binary | • |
| J19093+382 | GJ 4099 | 19:09:19.01 | +39:12:00.7 | M2.0 V | 7.967 | Triple | |
| J19095+391 | LSPM J1909+3910 | 19:09:31.54 | +39:10:48.4 | M2.0 V | 8.841 | Triple | |
| J19093-147 | Ross 727 | 19:09:20.08 | -14:45:02.4 | M2.5 V | 8.425 | | • |
| J19098+176 | GJ 1232 | 19:09:50.16 | +17:39:59.6 | M4.0 V | 8.819 | | |
| J19106+015 | GJ 4100 | 19:10:38.38 | +01:32:07.3 | M1.5 V | 8.929 | Binary | |
| J19116+050 | TYC 471-1564-1 | 19:11:47.89 | +05:00:37.4 | M1.0 V | 8.413 | | |
| J19124+355 | GJ 4105 | 19:12:29.70 | +35:33:50.2 | M2.0 V | 8.399 | | • |
| J19122+028 | Wolf 1062 | 19:12:16.50 | +02:53:02.6 | M3.5 V | 7.087 | Binary | |
| J19146+193N | Ross 733 | 19:14:38.46 | +19:19:10.7 | M4.5 V | 7.579 | Binary | |
| J19146+193S | Ross 734 | 19:14:38.51 | +19:18:29.9 | M4.0 V | 9.101 | Binary | |
| J19169+051N | V1428Aql | 19:16:54.64 | +05:09:46.7 | M2.5 V | 5.583 | Binary | |
| J19169+051S | V1298Aql | 19:16:56.97 | +05:08:39.7 | M8.0 V | 9.908 | Binary | |
| J19185+580 | LSPM J1918+5803 | 19:18:30.48 | +58:03:16.6 | M1.0 V | 8.979 | Binary* | |
| J19205-076 | GJ 754.1 B | 19:20:33.38 | -07:39:46.6 | M2.5 V | 8.221 | Binary | • |
| J19206+731S | 2M19204172+7311434 | 19:20:41.75 | +73:11:42.4 | M4.0 V | 10.604 | Binary* | • |
| J19206+731N | 2M19204172+7311467 | 19:20:41.76 | +73:11:45.5 | M4.5 V | 10.795 | Binary* | |
| J19216+208 | GJ 1235 | 19:21:37.62 | +20:51:40.0 | M4.0 V | 8.796 | | |
| J19218+286 | V2078Cyg | 19:21:52.48 | +28:40:02.3 | M1.5 V | 8.352 | | |
| J19220+070 | GJ 1236 | 19:22:01.28 | +07:02:23.2 | M4.0 V | 8.524 | | • |
| J19228+307 | GSC 02654-01527 | 19:22:48.64 | +30:45:13.5 | M0.0 V | 7.859 | | |
| J19234+666 | NLTT 47788 | 19:23:24.51 | +66:39:54.2 | M1.0 V | 8.864 | | • |
| J19215+425 | 1R192132.1+423041 | 19:21:32.18 | +42:30:54.8 | M2.0 V | 8.624 | Binary | |
| J19242+797 | TYC 4592-101-1 | 19:24:15.66 | +79:43:37.3 | M1.0 V | 8.665 | Binary | |
| J19237+797 | PM J19237+7944 | 19:23:46.46 | +79:44:37.6 | M1.5 V | 8.806 | Binary | |
| J19242+755 | GJ 1238 | 19:24:17.89 | +75:33:21.3 | M6.0 V | 9.908 | | |
| J19251+283 | Ross 164 | 19:25:08.74 | +28:21:19.8 | M3.5 V | 8.442 | | |
| J19255+096 | LSPM J1925+0938 | 19:25:31.00 | +09:38:19.4 | M8/9V | 11.214 | | |
| J19260+244 | G 185-23 | 19:26:01.84 | +24:26:18.7 | M4.5 V | 9.625 | | |
| J19268+167 | GJ 4110 | 19:26:49.42 | +16:42:58.3 | M3.5 V | 9.000 | | • |
| J19284+289 | TYC 2137-1575-1 | 19:28:25.51 | +28:54:09.6 | M0.0 V | 8.016 | | • |
| J19289+066 | PM J19289+0638 | 19:28:55.70 | +06:38:24.5 | M1.5 V | 8.977 | | |
| J19312+361 | G 125-15 | 19:31:12.38 | +36:07:28.2 | M4.5 V | 9.609 | Triple (SB2) | |
| J19326+005 | GJ 761.2 | 19:32:38.15 | +00:34:39.5 | M0.0 V | 7.635 | | |
| J19336+395 | Ross 1063 | 19:33:39.75 | +39:31:29.9 | M1.5 V | 8.120 | | |
| J19346+045 | HD 184489 | 19:34:40.40 | +04:35:02.0 | M0.0 V | 6.714 | | |
| J19349+532 | Wolf 1108 | 19:34:55.56 | +53:15:31.3 | M2.5 V | 8.554 | | |
| J19351+084S | GJ 4114 | 19:35:06.24 | +08:27:37.9 | M0.5 V | 7.329 | Triple | |
| J19351+084N | GJ 4115 | 19:35:06.33 | +08:27:43.4 | M2.5 V | 8.864 | Triple | • |
| J19358+413 | Ross 1064 | 19:35:51.00 | +41:19:06.7 | M0.0 V | 7.563 | | |
| J19395+718 | GJ 4120 | 19:39:32.22 | +71:52:12.1 | M0.5 V | 8.023 | | |

Table B.1: Carmencita, the CARMENES input catalogue (continued).

| Karmn | Name | $\alpha$ (2016.0) | $\delta$ (2016.0) | Spectral type | $J$ [mag] | Multiplicity[a] | DR1[b] |
|---|---|---|---|---|---|---|---|
| J19419+031 | GJ 1242 | 19:41:53.93 | +03:09:08.3 | M2.0 V | 9.304 | | |
| J19354+377 | RX J1935.4+3746 | 19:35:29.02 | +37:46:06.7 | M3.5 V | 7.562 | Binary (SB1) | |
| J19422-207 | 2M19421282-2045477 | 19:42:12.81 | -20:45:50.4 | M5.1 V | 9.598 | | |
| J19420-210 | LP 869-19 | 19:42:00.74 | -21:04:09.4 | M3.5 V | 8.692 | Binary (SB2) | |
| J19457+271 | Ross 165A | 19:45:45.41 | +27:07:11.7 | M4.0 V | 7.998 | Binary | |
| J19457+323 | GJ 4122 | 19:45:50.25 | +32:23:16.8 | M1.5 V | 7.572 | | |
| J19463+320 | HD 331161A | 19:46:24.51 | +32:00:55.2 | M0.5 V | 6.883 | Binary | • |
| J19464+320 | HD 331161B | 19:46:24.83 | +32:00:51.1 | M2.5 V | 7.323 | Binary | |
| J19468-019 | PM J19468-0157 | 19:46:50.63 | -01:57:40.1 | M3.0 V | 8.512 | | |
| J19470+352 | LSPM J1947+3516 | 19:47:03.30 | +35:16:55.5 | M2.0 V | 8.839 | | |
| J19486+359 | G 125-34 | 19:48:40.88 | +35:55:12.6 | M3.5 V | 8.909 | | |
| J19500+325 | GJ 4124 | 19:50:03.10 | +32:35:05.5 | M3.0 V | 8.647 | Binary | |
| J19502+317 | GJ 4125 | 19:50:16.11 | +31:47:05.7 | M2.1 V | 9.178 | Binary | |
| J19510+104 | GJ 9671 B | 19:51:00.97 | +10:24:38.0 | M4.0 V | 8.888 | Binary | |
| J19511+464 | GJ 1243 | 19:51:09.60 | +46:29:04.5 | M4.0 V | 8.586 | | |
| J19512+622 | G 260-35 | 19:51:11.69 | +62:17:15.7 | M2.0 V | 8.348 | | |
| J19535+341 | GJ 4127 | 19:53:32.73 | +34:08:32.2 | M1.5 V | 8.329 | | |
| J19539+444W | V1581Cyg | 19:53:55.14 | +44:24:44.4 | M5.5 V | 7.791 | Triple | • |
| J19539+444E | GJ 1245 B | 19:53:55.66 | +44:24:46.5 | M5.5 V | 8.275 | Triple | |
| J19540+325 | GJ 4128 | 19:54:02.88 | +32:33:55.8 | M2.5 V | 8.946 | | |
| J19546+202 | TYC 1624-397-1 | 19:54:37.52 | +20:13:05.5 | M0.0 V | 8.068 | | |
| J19558+512 | Wolf 1122 | 19:55:53.62 | +51:16:27.7 | M1.5 V | 8.658 | | |
| J19565+591 | GJ 9677 A | 19:56:33.06 | +59:09:40.3 | M0.0 V | 7.423 | Binary | |
| J19564+591 | GJ 9677 B | 19:56:23.96 | +59:09:19.7 | M3.5 V | 9.653 | Binary | |
| J19573-125 | GJ 773 B | 19:57:23.71 | -12:33:58.5 | M5.0 V | 10.212 | Binary | |
| J19582+020 | GJ 4129 | 19:58:15.38 | +02:02:02.5 | M2.5 V | 8.371 | | |
| J19582+650 | G 260-38 | 19:58:16.31 | +65:02:25.0 | M3.5 V | 8.740 | | |
| J20011+002 | LP 634-16 | 20:01:06.21 | +00:16:12.0 | M2.0 V | 8.106 | | |
| J20005+593 | 1R200031.8+592127 | 20:00:31.98 | +59:21:29.9 | M4.1 V | 9.636 | Binary | |
| J20034+298 | GJ 777 B | 20:03:27.42 | +29:51:51.1 | M4.5 V | 9.554 | Binary | |
| J20037+644 | 1R200348.4+642542 | 20:03:47.80 | +64:25:45.3 | M0.0 V | 8.267 | | |
| J20038+059 | GJ 1248 | 20:03:50.47 | +05:59:31.4 | M2.5 V | 8.632 | | |
| J20039-081 | GJ 4132 | 20:03:58.37 | -08:07:51.4 | M4.0 V | 9.184 | | |
| J20050+544 | V1513Cyg | 20:05:00.07 | +54:25:48.8 | sdM1 | 8.830 | Triple (SB1) | |
| J20057+529 | Wolf 1131 | 20:05:44.59 | +52:58:21.1 | M4.0 V | 9.095 | | |
| J20079-015 | GJ 4136 | 20:07:57.91 | -01:32:31.5 | M3.0 V | 9.586 | | |
| J20082+333 | GJ 1250 | 20:08:18.35 | +33:18:19.0 | M4.77 | 9.961 | | |
| J20093-012 | PM J20093-0113 | 20:09:18.20 | -01:13:44.3 | M5.0 V | 9.403 | | |
| J20105+065 | LP 574-21 | 20:10:34.49 | +06:32:10.7 | M4.0 V | 8.021 | Triple (SB2) | |
| J20112+161 | GJ 783.2 B | 20:11:12.80 | +16:11:14.4 | M4.0 V | 9.627 | Binary | |
| J20112+379 | LSPM J2011+3757 | 20:11:12.85 | +37:57:48.1 | M1.5 V | 8.593 | | • |
| J20129+342 | LP 283-5 | 20:12:54.72 | +34:16:54.5 | M1.0 V | 8.213 | Binary | |
| J20132+029 | [R78b] 440 | 20:13:12.94 | +02:56:02.3 | M1.0 V | 8.739 | Binary | |
| J20138+133 | Ross 754 | 20:13:52.27 | +13:23:20.0 | M1.0 V | 8.309 | | |
| J20139+066 | GJ 784.2 A | 20:13:58.71 | +06:41:06.8 | M3.3 V | 9.090 | Binary | |
| J20151+635 | LP 106-240 | 20:15:10.64 | +63:31:16.4 | M0.0 V | 8.384 | | |
| J20165+351 | G 210-11 | 20:16:32.04 | +35:10:37.6 | M2.0 V | 8.905 | | |
| J20187+158 | GJ 4143 | 20:18:44.75 | +15:50:45.3 | M2.5 V | 8.174 | | |
| J20195+080 | GJ 4144 | 20:19:34.60 | +08:00:27.1 | M3.0 V | 9.190 | Binary* | |
| J20198+229 | LP 395-8 A | 20:19:49.36 | +22:56:38.1 | M3.0 V | 8.166 | Quadruple* (SB2) | |
| J20220+216 | TYC 1643-120-1 | 20:22:01.62 | +21:47:19.7 | M2.0 V | 8.741 | Binary* | |
| J20223+322 | PM J20223+3217 | 20:22:18.82 | +32:17:15.1 | M3.5 V | 8.810 | | |
| J20229+106 | G 143-48 | 20:22:55.79 | +10:40:44.5 | M3.0 V | 9.953 | | |
| J20232+671 | LP 73-196 | 20:23:18.57 | +67:10:12.9 | M5.0 V | 10.075 | Binary | • |
| J20260+585 | Wolf 1069 | 20:26:05.84 | +58:34:31.4 | dM5.0 | 9.029 | | |
| J20287-114 | LP 755-19 | 20:28:43.80 | -11:28:32.4 | M1.5 V | 8.394 | | |
| J20269+275 | GJ 4146 | 20:26:56.29 | +27:31:03.1 | M2.0 V | 8.762 | Binary | |
| J20298+096 | HUDel | 20:29:49.04 | +09:41:22.2 | M4.5 V | 8.228 | Binary | |
| J20305+654 | GJ 793 | 20:30:33.18 | +65:27:02.9 | M2.5 V | 6.735 | | |

Table B.1: Carmencita, the CARMENES input catalogue (continued).

| Karmn | Name | $\alpha$ (2016.0) | $\delta$ (2016.0) | Spectral type | $J$ [mag] | Multiplicity[a] | DR1[b] |
|---|---|---|---|---|---|---|---|
| J20301+798 | 1R203011.0+795040 | 20:30:07.75 | +79:50:47.4 | M3.0 V | 8.480 | Binary (SB2) | • |
| J20336+617 | GJ 1254 | 20:33:41.53 | +61:45:28.2 | M4.0 V | 8.287 | | |
| J20314+385 | Ross 188 | 20:31:25.91 | +38:33:55.8 | M5.0 V | 9.193 | Binary | |
| J20337+233 | GJ 4148 | 20:33:43.10 | +23:22:15.4 | M3.0 V | 9.110 | Binary | |
| J20339+643 | GJ 4150 | 20:33:58.81 | +64:19:08.1 | M3.5 V | 9.228 | | |
| J20347+033 | GJ 4149 | 20:34:43.33 | +03:20:43.6 | M2.5 V | 8.446 | | • |
| J20349+592 | Wolf 1074 | 20:34:54.81 | +59:17:26.0 | M4.0 V | 9.318 | | |
| J20367+388 | GJ 4152 | 20:36:46.27 | +38:50:30.3 | M3.5 V | 9.270 | | • |
| J20373+219 | Wolf 1351 | 20:37:20.77 | +21:56:47.8 | M0.5 V | 8.157 | Binary | |
| J20403+616 | TYC 4246-488-1 | 20:40:18.59 | +61:41:28.4 | M1.0 V | 8.154 | | |
| J20405+154 | GJ 1256 | 20:40:35.33 | +15:30:09.3 | M4.5 V | 8.641 | | |
| J20407+199 | GJ 797 B | 20:40:44.65 | +19:54:08.2 | M2.5 V | 8.160 | Triple | |
| J20409-101 | GJ 4155 | 20:40:56.32 | -10:06:43.1 | M1.5 V | 8.535 | Binary (SB) | |
| J20418-324 | ATMicA | 20:41:51.44 | -32:26:13.3 | M4.5 V | 5.807 | SKG | |
| J20451-313 | AUMic | 20:45:09.88 | -31:20:33.0 | M0.5 V | 5.436 | SKG | |
| J20429-189 | Ross 751 | 20:42:57.83 | -18:55:19.8 | M1.5 V | 7.557 | | • |
| J20433+047 | LP 575-35 | 20:43:24.35 | +04:45:52.9 | M5.0 V | 10.083 | | |
| J20435+240 | Wolf 1360 | 20:43:34.69 | +24:07:40.2 | M2.5 V | 8.241 | | |
| J20436+642 | G 262-26 | 20:43:42.13 | +64:16:52.5 | M0.0 V | 8.399 | | |
| J20436-001 | GJ 4159 | 20:43:41.71 | -00:10:37.5 | M1.0 V | 8.484 | | |
| J20433+553 | Wolf 1084 | 20:43:20.89 | +55:21:20.5 | M5.0 V | 9.563 | Binary | |
| J20443+197 | HD 352860 | 20:44:21.98 | +19:44:49.8 | M0.0 V | 7.359 | Binary | |
| J20445+089S | GJ 4160 | 20:44:30.94 | +08:54:12.6 | M1.5 V | 8.136 | Triple | |
| J20445+089N | GJ 4161 | 20:44:30.67 | +08:54:27.1 | M0.5 V | 8.607 | Triple (SB2) | |
| J20450+444 | GJ 806 | 20:45:04.75 | +44:30:01.0 | M1.5 V | 7.329 | | |
| J20496-003 | Wolf 882 | 20:49:39.86 | -00:21:06.7 | M3.5 V | 9.099 | | |
| J20488+197 | GJ 4163 | 20:48:52.27 | +19:43:01.6 | M3.3 V | 9.236 | Binary | |
| J20525-169 | LP 816-60 | 20:52:32.67 | -16:58:28.4 | M4.0 V | 7.090 | | |
| J20519+691 | GJ 4170 | 20:52:00.55 | +69:10:07.6 | M1.0 V | 8.452 | Binary | • |
| J20533+621 | HD 199305 | 20:53:19.79 | +62:09:03.4 | M1.0 V | 5.429 | | • |
| J20535+106 | GJ 4169 | 20:53:32.53 | +10:36:55.0 | M5.0 V | 9.348 | | |
| J20549+675 | LP 74-35 | 20:54:54.90 | +67:35:09.8 | M2.0 V | 8.749 | | |
| J20532-023 | LP 636-19 | 20:53:14.87 | -02:21:21.7 | M3.0 V | 9.329 | Binary | |
| J20556-140N | GJ 810 A | 20:55:39.31 | -14:02:15.6 | M4.0 V | 8.117 | Triple (SB2) | • |
| J20556-140S | GJ 810 B | 20:55:38.68 | -14:04:02.4 | M5.0 V | 9.717 | Triple | |
| J20567-104 | Wolf 896 | 20:56:46.56 | -10:27:12.7 | dM2.5 | 7.766 | | • |
| J20568-048 | FRAqr | 20:56:49.39 | -04:50:52.6 | M4.0 V | 7.816 | Triple (SB2) | |
| J20574+223 | Wolf 1373 | 20:57:26.24 | +22:21:42.4 | M3.0 V | 8.410 | | |
| J20586+342 | GJ 4173 | 20:58:42.31 | +34:16:24.7 | M0.5 V | 8.073 | | • |
| J21001+495 | G 212-14 | 21:00:09.24 | +49:35:20.8 | M2.0 V | 8.546 | | • |
| J21000+400 | V1396Cyg | 21:00:06.23 | +40:04:08.6 | M2.0 VM0.5V | 6.668 | Binary | • |
| J21012+332 | GJ 4176 | 21:01:16.51 | +33:14:30.7 | M4.0 V | 8.439 | Quadruple | |
| J21013+332 | GJ 4177 | 21:01:21.03 | +33:14:25.9 | M4.0 V | 8.936 | Quadruple | |
| J21019-063 | Wolf 906 | 21:01:58.39 | -06:19:14.5 | M2.5 V | 7.563 | | |
| J21027+349 | G 211-9 | 21:02:46.40 | +34:54:31.2 | M4.5 V | 9.846 | | |
| J21044+455 | TYC 3588-5589-1 | 21:04:28.94 | +45:35:42.3 | M2.0 V | 8.993 | | |
| J21048-169 | Ross 769 | 21:04:52.36 | -16:58:04.6 | M1.5 V | 8.285 | | |
| J21014+207 | GJ 4175 | 21:01:24.40 | +20:43:31.6 | M3.5 V | 9.941 | Binary | |
| J21055+061 | PM J21055+0609N | 21:05:32.09 | +06:09:16.2 | M3.0 V | 8.694 | Binary* | |
| J21057+502 | PM J21057+5015E | 21:05:45.54 | +50:15:44.1 | M3.5 V | 9.543 | Binary* | • |
| J21059+044 | GJ 4180 | 21:05:56.52 | +04:25:38.0 | M2.5 V | 8.574 | | |
| J21074+468 | PM J21074+4651 | 21:07:28.14 | +46:51:55.4 | M2.0 V | 9.488 | | |
| J21076-130 | 1R210736.5-130500 | 21:07:36.86 | -13:04:59.6 | M3.0 V | 8.734 | | |
| J21087-044S | GJ 9721 | 21:08:45.41 | -04:25:36.7 | M1.0 V | 7.146 | Triple | |
| J21087-044N | GJ 9721B | 21:08:44.75 | -04:25:18.3 | M3.0 V | 9.481 | Triple | |
| J21092-133 | Wolf 918 | 21:09:18.21 | -13:18:40.9 | M1.0 V | 7.688 | | |
| J21100-193 | 1R211004.9-192005 | 21:10:05.46 | -19:19:59.1 | M2.0 V | 8.112 | Binary | |
| J21123+359 | LP 285-9 | 21:12:22.66 | +35:55:24.7 | M1.5 V | 8.490 | | |
| J21127-073 | PM J21127-0719 | 21:12:45.71 | -07:19:56.5 | M3.5 V | 9.902 | | |

Table B.1: Carmencita, the CARMENES input catalogue (continued).

| Karmn | Name | $\alpha$ (2016.0) | $\delta$ (2016.0) | Spectral type | $J$ [mag] | Multiplicity[a] | DR1[b] |
|---|---|---|---|---|---|---|---|
| J21109+294 | Ross 824 | 21:10:54.47 | +29:25:18.6 | M1.5 V | 7.799 | Binary | |
| J21137+087 | LSPM J2113+0846N | 21:13:44.77 | +08:46:09.1 | M2.0 V | 8.384 | Triple | |
| J21138+180 | Ross 772 | 21:13:53.01 | +18:05:58.8 | M3.0 V | 8.970 | | |
| J21145+508 | LSPM J2114+5052 | 21:14:32.61 | +50:52:31.6 | M2.5 V | 8.456 | | |
| J21147+380 | GJ 822.1 C | 21:14:47.09 | +38:01:21.0 | M2.5 V | 8.337 | Quadruple | |
| J21152+257 | GJ 4184 | 21:15:12.76 | +25:47:41.2 | M3.0 V | 8.403 | | |
| J21160+298E | Ross 776 | 21:16:06.06 | +29:51:51.5 | M3.3 V | 8.448 | Triple | |
| J21160+298W | Ross 826 | 21:16:04.10 | +29:51:46.7 | M3.3 V | 9.295 | Triple | |
| J21164+025 | LSPM J2116+0234 | 21:16:27.55 | +02:34:50.8 | M3.0 V | 8.219 | | |
| J21173+208N | Ross 773A | 21:17:23.09 | +20:53:59.2 | M3.0 V | 8.683 | Binary | |
| J21173+208S | Ross 773B | 21:17:23.00 | +20:54:03.4 | M4.0 V | 8.911 | Binary | • |
| J21173+640 | G 262-38 | 21:17:22.75 | +64:02:39.1 | M5.0 V | 10.043 | | |
| J21176-089N | GJ 4187 | 21:17:36.07 | -08:54:11.7 | M2.5 V | 8.467 | Triple | |
| J21176-089S | GJ 4188 | 21:17:39.55 | -08:54:49.6 | M3.0 V | 9.517 | Triple | • |
| J21185+302 | 1R211833.8+301434 | 21:18:33.83 | +30:14:34.3 | M1.5 V | 8.643 | | |
| J21221+229 | TYC 2187-512-1 | 21:22:06.41 | +22:55:55.0 | M1.0 V | 7.400 | | |
| J21243+085 | GJ 4192 | 21:24:19.08 | +08:30:03.6 | M3.5 V | 9.667 | | |
| J21245+400 | LSPM J2124+4003 | 21:24:33.07 | +40:04:06.7 | M5.5 V | 10.339 | | |
| J21267+037 | GJ 828.1 | 21:26:42.41 | +03:44:12.9 | M0.0 V | 7.794 | | |
| J21272-068 | Wolf 920 | 21:27:16.89 | -06:50:45.8 | M0.5 V | 8.046 | | |
| J21275+340 | V2160Cyg | 21:27:32.60 | +34:01:25.7 | M1.5 V | 8.031 | | • |
| J21277+072 | Ross 778 | 21:27:46.19 | +07:17:45.9 | M1.0 V | 8.373 | | |
| J21280+179 | LP 457-38 | 21:28:05.57 | +17:54:02.7 | M1.5 V | 8.907 | | |
| J21283-223 | GJ 4197 | 21:28:18.04 | -22:18:36.5 | M2.5 V | 8.501 | | |
| J21296+176 | Ross 775 | 21:29:37.94 | +17:38:41.9 | M3.0 V | 6.249 | Binary (EB?/SB2) | |
| J21313-097 | BBCap | 21:31:19.92 | -09:47:27.5 | M4.5 V | 7.316 | Binary | |
| J21323+245 | GJ 4201 A | 21:32:22.33 | +24:33:41.9 | M4.0 V | 8.476 | Binary | |
| J21338+017S | GJ 4203 | 21:33:49.11 | +01:46:44.9 | M4.0 V | 9.646 | Binary | |
| J21338+017N | GJ 4204 | 21:33:49.13 | +01:46:50.0 | M4.0 V | 9.977 | Binary | |
| J21338-068 | Wolf 923 | 21:33:49.04 | -06:51:18.4 | M4.0 V | 9.562 | | |
| J21348+515 | Wolf 926 | 21:34:51.13 | +51:32:18.6 | M2.5 V | 8.038 | | |
| J21366+394 | V2168Cyg | 21:36:38.29 | +39:27:18.0 | M0.0 V | 7.111 | Binary | |
| J21369+561 | Ross 215 | 21:36:58.82 | +56:07:07.2 | M1.5 V | 8.695 | | |
| J21374-059 | PM J21374-0555 | 21:37:29.02 | -05:55:05.7 | M3.0 V | 8.779 | Binary | |
| J21378+530 | Ross 199 | 21:37:50.77 | +53:04:49.9 | M0.0 V | 7.753 | | |
| J21376+016 | 1R213740.3+013711 | 21:37:40.28 | +01:37:12.7 | M4.5 V | 8.802 | Binary | • |
| J21380+277 | GJ 835 | 21:38:00.96 | +27:43:24.4 | M0.0 V | 6.809 | Triple | |
| J21402+370 | LP 286-1 | 21:40:12.27 | +37:03:24.1 | M0.5 V | 8.460 | | |
| J21399+276 | GJ 4210 | 21:39:54.70 | +27:36:39.8 | M2.0 V | 8.211 | Binary | |
| J21421-121 | Ross 206 | 21:42:07.58 | -12:09:59.3 | M3.0 V | 8.922 | | |
| J21419+276 | GJ 4212 | 21:41:58.03 | +27:41:14.2 | M4.0 V | 9.825 | Binary | |
| J21441+170S | GJ 4214 | 21:44:09.31 | +17:03:35.0 | M4.0 V | 9.313 | Binary | |
| J21441+170N | GJ 4215 | 21:44:08.25 | +17:04:37.3 | M5.2 V | 10.078 | Binary | |
| J21442+066 | GJ 4213 | 21:44:12.64 | +06:38:26.6 | M3.0 V | 8.271 | Binary (SB1) | |
| J21449+442 | G 215-12 | 21:44:53.78 | +44:16:58.4 | M1.5 V | 8.100 | Binary | |
| J21450+198 | G 126-32B | 21:45:04.90 | +19:53:31.7 | M1.5 V | | Triple* | |
| J21450-057 | Wolf 937 | 21:45:00.59 | -05:47:20.3 | M3.0 V | 9.034 | | |
| J21454-059 | Wolf 939 | 21:45:24.74 | -05:54:11.5 | M3.5 V | 9.604 | | |
| J21463+382 | LSPM J2146+3813 | 21:46:22.29 | +38:13:03.1 | M5.0 V | 7.949 | | |
| J21466+668 | G 264-12 | 21:46:41.30 | +66:48:14.0 | M4.0 V | 8.837 | | |
| J21466-001 | Wolf 940 | 21:46:41.24 | -00:10:31.9 | dM4.0 | 8.364 | Binary | |
| J21469+466 | Wolf 944 | 21:46:56.68 | +46:38:06.1 | M4.0 V | 9.089 | | |
| J21472-047 | PM J21472-0444 | 21:47:17.73 | -04:44:40.6 | M4.5 V | 9.416 | | |
| J21474+627 | TYC 4266-736-1 | 21:47:24.39 | +62:45:13.7 | M0.0 V | 8.771 | Binary | • |
| J21478+502 | Wolf 945 | 21:47:53.64 | +50:14:54.6 | M4.0 V | 9.311 | | • |
| J21479+058 | Ross 779 | 21:47:57.57 | +05:49:16.4 | M2.0 V | 8.626 | | • |
| J21481+014 | GJ 4221 | 21:48:10.46 | +01:26:42.0 | M3.0 V | 9.824 | | |
| J21482+279 | GJ 4225 | 21:48:15.02 | +27:55:31.0 | M2.0 V | 8.508 | | |
| J21512+128 | GJ 4227 | 21:51:18.24 | +12:50:33.2 | M4.0 V | 9.349 | | • |

Table B.1: Carmencita, the CARMENES input catalogue (continued).

| Karmn | Name | $\alpha$ (2016.0) | $\delta$ (2016.0) | Spectral type | $J$ [mag] | Multiplicity[a] | DR1[b] |
|---|---|---|---|---|---|---|---|
| J21518+136 | GJ 4228 | 21:51:48.53 | +13:36:13.8 | M4.0 V | 9.311 | Binary | |
| J21516+592 | TYC 3980-1081-1 | 21:51:38.15 | +59:17:40.0 | M4.0 V | 6.529 | Binary | |
| J21521+274 | GJ 4232 | 21:52:12.12 | +27:24:48.2 | M5.0 V | 9.984 | | |
| J21539+417 | GJ 839 | 21:53:59.55 | +41:46:38.7 | M0.0 V | 7.572 | | |
| J21566+197 | Ross 263 | 21:56:37.83 | +19:46:06.8 | M3.0 V | 8.928 | | |
| J21569-019 | GJ 4239 | 21:56:56.61 | -01:53:59.5 | M5.0 V | 9.880 | | |
| J21574+081 | Wolf 953 | 21:57:26.64 | +08:08:15.5 | M1.5 V | 7.723 | | |
| J21584+755 | GJ 842.2 | 21:58:25.51 | +75:35:21.0 | M0.5 V | 7.601 | | |
| J21585+612 | LSPM J2158+6117 | 21:58:36.41 | +61:17:07.8 | M6.0 V | 11.292 | | |
| J21593+418 | GJ 4246 | 21:59:22.08 | +41:51:25.4 | M3.0 V | 8.982 | | |
| J22012+283 | V374Peg | 22:01:13.58 | +28:18:25.6 | M3.5 V | 7.635 | | |
| J21521+056 | GJ 4231 | 21:52:10.52 | +05:37:33.5 | M2.4 V | 8.248 | Binary | |
| J22012+323 | Wolf 1154 A | 22:01:14.13 | +32:23:13.9 | M1.5 V | 8.829 | Triple | |
| J22018+164 | Ross 265 | 22:01:49.49 | +16:28:05.2 | M2.5 V | 7.009 | Triple* | |
| J22020-194 | GJ 843 | 22:02:01.85 | -19:28:58.0 | dM3.5 | 8.046 | | |
| J22021+014 | HD 209290 | 22:02:09.79 | +01:23:56.4 | M0.5 V | 6.196 | | • |
| J22033+674 | GJ 4251 | 22:03:22.74 | +67:29:55.1 | M4.5 V | 9.410 | | |
| J22051+051 | Wolf 983 | 22:05:07.30 | +05:08:14.2 | M4.0 V | 9.518 | | |
| J22035+036 | 1R220330.8+034001 | 22:03:33.39 | +03:40:21.4 | M4.0 V | 9.742 | Binary | • |
| J22057+656 | GJ 4258 | 22:05:44.55 | +65:38:58.9 | M1.5 V | 8.422 | Binary | • |
| J22058-119 | Wolf 1548 | 22:05:51.00 | -11:54:53.7 | M0.0 V | 7.221 | Binary | |
| J22060+393 | GJ 4256 | 22:06:00.76 | +39:17:55.6 | M3.0 V | 8.912 | | |
| J22067+034 | Wolf 990 | 22:06:46.87 | +03:24:58.7 | M4.0 V | 9.409 | | |
| J22088+117 | PM J22088+1144 | 22:08:50.44 | +11:44:12.4 | M4.5 V | 9.901 | | • |
| J22095+118 | LP 519-38 | 22:09:31.86 | +11:52:51.6 | M3.0 V | 9.901 | | |
| J22096-046 | Wolf 1329 | 22:09:41.56 | -04:38:27.0 | M3.5 V | 6.510 | | |
| J22097+410 | GJ 4260 | 22:09:43.64 | +41:02:09.5 | M3.0 V | 8.755 | | |
| J22102+587 | UCAC4 744-073158 | 22:10:15.18 | +58:42:21.9 | M2.0 V | 9.860 | | |
| J22107+079 | Wolf 1003 A | 22:10:44.98 | +07:54:32.9 | M0.5 V | 7.936 | Binary | |
| J22112+410 | G 214-14 | 22:11:16.65 | +41:00:58.6 | M0.0 V | 8.228 | | |
| J22112-025 | GJ 4262 | 22:11:13.95 | -02:32:38.2 | M2.0 V | 8.688 | | • |
| J22114+409 | 1R221124.3+410000 | 22:11:24.04 | +40:59:59.8 | M5.5 V | 9.725 | | |
| J22115+184 | Ross 271 | 22:11:30.46 | +18:25:37.2 | M2.0 V | 6.725 | | |
| J22117-207 | WT 2221 | 22:11:42.27 | -20:44:19.3 | M3.5 V | 9.676 | Binary | |
| J22125+085 | Wolf 1014 | 22:12:36.06 | +08:33:00.9 | M3.0 V | 8.277 | | |
| J22129+550 | LF 4 +54 152 | 22:12:56.63 | +55:04:50.8 | M0.0 V | 8.125 | Binary* | |
| J22134-147 | Wolf 1556 | 22:13:28.46 | -14:44:58.8 | M3.5 V | 9.449 | | • |
| J22135+259 | GJ 4264 | 22:13:35.90 | +25:58:08.1 | M4e | 9.511 | | • |
| J22137-176 | GJ 1265 | 22:13:43.82 | -17:41:13.6 | dM4.5 | 8.955 | | |
| J22138+052 | Wolf 1019 | 22:13:53.53 | +05:16:34.9 | M1.5 V | 8.541 | Binary | • |
| J22154+662 | GJ 4267 | 22:15:26.16 | +66:13:31.1 | M3.0 V | 8.749 | | |
| J22142+255 | 1R221419.3+253411 | 22:14:17.88 | +25:34:05.6 | M4.3 V | 10.177 | Binary | |
| J22160+546 | GJ 4269 | 22:16:02.99 | +54:40:00.5 | M4.0 V | 9.718 | Binary | |
| J22163+709 | GJ 1266 | 22:16:23.03 | +70:56:39.2 | M2.0 V | 8.746 | | • |
| J22173-088N | FGAqr | 22:17:18.47 | -08:48:16.9 | M4.0 V | 9.024 | Triple | |
| J22173-088S | Wolf 1561 B | 22:17:18.16 | -08:48:23.4 | M5.0 V | 9.459 | Triple | |
| J22176+565 | PM J22176+5633 | 22:17:37.21 | +56:33:11.1 | M1.5 V | 8.992 | | |
| J22202+067 | Wolf 1034 | 22:20:13.57 | +06:43:36.5 | M2.5 V | 9.500 | | |
| J22228+280 | GJ 4275 | 22:22:51.34 | +28:01:46.4 | M3.8 V | 9.758 | | |
| J22231-176 | GJ 4274 | 22:23:07.34 | -17:36:37.8 | M4.5 V | 8.242 | | |
| J22234+324 | Wolf 1225 A | 22:23:29.44 | +32:27:29.9 | M3.0 V | 6.898 | Binary | |
| J22249+520 | GJ 1268 | 22:24:56.33 | +52:00:25.6 | M5.1 V | 10.155 | | |
| J22250+356 | Wolf 1231 | 22:25:01.66 | +35:40:05.3 | M2.0 V | 8.537 | | |
| J22252+594 | GJ 4276 | 22:25:17.32 | +59:24:44.9 | M3.5 V | 8.745 | | |
| J22262+030 | Wolf 1201 | 22:26:15.24 | +03:00:11.3 | M4.0 V | 9.623 | Binary | |
| J22264+583 | PM J22264+5823 | 22:26:24.73 | +58:23:03.9 | M3.0 V | 9.461 | | • |
| J22270+068 | GJ 4279 | 22:27:03.00 | +06:49:32.1 | M4.0 V | 9.020 | | |
| J22279+576 | HD 239960A | 22:27:58.11 | +57:41:38.5 | M3.0 V | 5.575 | Binary | |
| J22287+189 | GJ 9784 | 22:28:46.13 | +18:55:52.1 | M1.0 V | 7.819 | | |

Table B.1: Carmencita, the CARMENES input catalogue (continued).

| Karmn | Name | $\alpha$ (2016.0) | $\delta$ (2016.0) | Spectral type | $J$ [mag] | Multiplicity[a] | DR1[b] |
|---|---|---|---|---|---|---|---|
| J22289-134 | GJ 4281 | 22:28:53.99 | -13:25:36.2 | M6.5 V | 10.768 | | • |
| J22290+016 | LP 640-74 | 22:29:05.91 | +01:39:45.0 | M0.5 V | 7.616 | | |
| J22298+414 | GJ 1270 | 22:29:50.68 | +41:28:55.8 | M4.0 V | 8.849 | | |
| J22300+488 | PM J22300+4851A | 22:30:04.10 | +48:51:33.8 | M4.5 V | 9.524 | Binary | |
| J22330+093 | GJ 863 | 22:33:02.81 | +09:22:43.0 | M1.0 V | 7.208 | | |
| J22333-096 | GJ 4282 A | 22:33:22.77 | -09:36:53.6 | M2.6 V | 8.534 | Binary | |
| J22347+040 | GJ 4283 | 22:34:46.11 | +04:02:39.7 | M3.0 V | 8.964 | | |
| J22348-010 | GJ 4284 | 22:34:54.85 | -01:04:54.5 | M4.5 V | 10.388 | | |
| J22353+746 | G 242-3 | 22:35:20.75 | +74:41:20.1 | M0.0 V | 8.425 | | • |
| J22373+299 | LP 344-27 | 22:37:23.33 | +29:59:05.6 | M1.5 V | 8.959 | | |
| J22361-008 | HD 214100 | 22:36:09.75 | -00:50:39.8 | M1.0 V | 7.038 | Binary | • |
| J22374+395 | GJ 4287 | 22:37:29.83 | +39:22:45.9 | M0.0 V | 6.639 | Triple | |
| J22387+252 | G 127-42 | 22:38:44.65 | +25:13:30.5 | M3.5 V | 9.769 | | |
| J22385-152 | EZAqr | 22:38:36.17 | -15:17:22.7 | M5.5 V | 6.553 | Triple (ST3) | |
| J22387-206S | FKAqr | 22:38:46.09 | -20:37:17.4 | M1.5 V | 5.669 | Quadruple (SB2) | |
| J22387-206N | FLAqr | 22:38:45.77 | -20:36:52.9 | M3.5 V | 7.344 | Quadruple (SB1) | |
| J22406+445 | GJ 4290 | 22:40:42.53 | +44:35:47.0 | M3.5 V | 9.216 | | |
| J22415+188 | GJ 9793 | 22:41:35.30 | +18:49:28.9 | M0.0 V | 7.883 | | |
| J22415+260 | 1R224134.7+260210 | 22:41:35.76 | +26:02:13.8 | M3.5 V | 9.044 | | |
| J22426+176 | GJ 1271 | 22:42:39.99 | +17:40:17.6 | M2.5 V | 8.062 | | |
| J22433+221 | GJ 4292 | 22:43:23.65 | +22:08:18.1 | M5.0 V | 10.143 | | |
| J22437+192 | RX J2243.7+1916 | 22:43:43.73 | +19:16:52.4 | M3.0 V | 9.242 | | |
| J22441+405 | TYC 3218-905-1 | 22:44:06.16 | +40:29:58.1 | M1.0 V | 8.199 | Binary | |
| J22440+405 | TYC 3218-907-1 | 22:44:04.51 | +40:29:58.5 | M1.0 V | 8.167 | Binary | |
| J22457+016 | LP 641-4 | 22:45:46.48 | +01:41:22.0 | M1.0 V | 8.915 | Binary | |
| J22464-066 | GJ 4294 | 22:46:27.09 | -06:39:35.0 | M5.0 V | 10.790 | | |
| J22468+443 | ELac | 22:46:48.68 | +44:19:55.0 | M4.0 V | 6.106 | | |
| J22476+184 | LP 461-11 | 22:47:39.35 | +18:26:40.4 | M2.5 V | 9.102 | | |
| J22479+318 | GJ 4297 | 22:47:54.67 | +31:52:18.4 | M3.0 V | 9.087 | | |
| J22489+183 | PM J22489+1819 | 22:48:54.56 | +18:19:56.9 | M4.5 V | 9.957 | | |
| J22503-070 | HD 216133 | 22:50:19.31 | -07:05:22.7 | M0.5 V | 6.932 | | |
| J22506+348 | GJ 1274 | 22:50:38.79 | +34:51:26.8 | M2+ V | 8.280 | | |
| J22507+286 | GJ 4300 | 22:50:45.77 | +28:36:07.7 | M2.5 V | 8.810 | | • |
| J22509+499 | 1R225056.4+495906 | 22:50:55.28 | +49:59:13.2 | M4.0 V | 9.804 | | |
| J22518+317 | GTPeg | 22:51:54.19 | +31:45:14.4 | M3.5 V | 7.697 | | |
| J22524+099 | GJ 9801 B | 22:52:30.31 | +09:54:04.9 | M3.0 V | 9.657 | Triple | |
| J22526+750 | LP 48-305 | 22:52:40.21 | +75:04:16.7 | M4.5 V | 9.089 | | • |
| J22543+609 | Ross 226 | 22:54:19.95 | +60:59:41.5 | M3.0 V | 8.836 | | |
| J22547-054 | GJ 4302 | 22:54:47.14 | -05:28:21.1 | M4.0 V | 9.650 | | |
| J22532-142 | ILAqr | 22:53:17.79 | -14:16:00.1 | M4.0 V | 5.934 | | |
| J22559+057 | GJ 4304 | 22:55:57.20 | +05:45:14.0 | M1.0 V | 8.130 | Binary | • |
| J22559+178 | GJ 4306 | 22:55:59.87 | +17:48:38.0 | M1.0 V | 7.319 | | |
| J22565+165 | HD 216899 | 22:56:33.65 | +16:33:07.8 | M1.5 V | 5.360 | | |
| J22576+373 | G 189-53A | 22:57:40.20 | +37:19:17.8 | M3.0 V | 8.975 | Binary | • |
| J23028+436 | 1R230251.9+433814 | 23:02:52.29 | +43:38:15.5 | M4.0 V | 9.316 | | |
| J23036+097 | PM J23036+0942 | 23:03:37.56 | +09:42:59.0 | M3.5 V | 9.992 | | |
| J23045+667 | GJ 1278 | 23:04:31.03 | +66:45:50.5 | M0.5 V | 7.094 | | |
| J23051+519 | PM J23051+5159 | 23:05:06.47 | +51:59:11.6 | M3.5 V | 9.680 | | • |
| J23060+639 | GJ 9809 | 23:06:05.27 | +63:55:33.4 | M0.3 V | 7.815 | | • |
| J23051+452 | LSPM J2305+4517 | 23:05:09.00 | +45:17:32.9 | M3.5 V | 9.297 | Binary* | |
| J23063+126 | LP 521-79 | 23:06:24.19 | +12:36:25.6 | M0.5 V | 8.375 | Quadruple (SB2/3?) | |
| J23064-050 | 2MUCD 12171 | 23:06:30.37 | -05:02:36.7 | M7.5 V | 11.354 | | |
| J23065+717 | GJ 4311 | 23:06:39.95 | +71:43:32.5 | M2.0 V | 8.336 | | |
| J23075+686 | GJ 4312 | 23:07:33.27 | +68:40:06.1 | M3.0 V | 8.624 | | |
| J23081+033 | GJ 889.1 | 23:08:07.49 | +03:19:48.5 | M0.0 V | 7.865 | | |
| J23083-154 | HKAqr | 23:08:19.67 | -15:24:36.1 | M0.0 V | 7.979 | | |
| J23088+065 | StKM 1-2100 | 23:08:52.60 | +06:33:39.9 | M0.0 V | 8.047 | | |
| J23089+551 | G 233-42 | 23:09:58.62 | +55:06:48.1 | M5.0 V | 10.541 | Binary | • |
| J23096-019 | GJ 4314 | 23:09:39.64 | -01:58:29.9 | M3.5 V | 8.671 | Triple (SB2) | |

Table B.1: Carmencita, the CARMENES input catalogue (continued).

| Karmn | Name | α (2016.0) | δ (2016.0) | Spectral type | J [mag] | Multiplicity[a] | DR1[b] |
|---|---|---|---|---|---|---|---|
| J23107-192 | GJ 1281 | 23:10:42.22 | -19:13:57.9 | M2.5 V | 8.976 | | |
| J23113+085 | G 28-46 | 23:11:23.45 | +08:30:56.4 | M3.0 V | 8.466 | Binary* | |
| J23121-141 | GJ 4316 | 23:12:11.11 | -14:06:23.2 | M3.0 V | 9.064 | | |
| J23166+196 | GJ 893.4 | 23:16:39.52 | +19:37:14.2 | M2.0 V | 8.134 | Binary | |
| J23142-196N | GJ 2154 A | 23:14:17.13 | -19:38:38.5 | M0.5 Vk | 7.471 | Binary | |
| J23142-196S | GJ 2154 B | 23:14:16.97 | -19:38:45.4 | M4.0 V | 9.411 | Binary | |
| J23174+196 | GJ 4326 | 23:17:28.54 | +19:36:45.1 | M2.0 V | 8.020 | Binary | |
| J23161+067 | GJ 4319 | 22:34:26.25 | -00:28:12.9 | M3.0 V | 10.369 | | • |
| J23174+382 | GJ 4327 | 23:17:24.09 | +38:12:34.8 | M3.0 V | 7.761 | Binary (SB2) | |
| J23175+063 | GJ 4329 | 23:17:34.73 | +06:23:24.5 | M3.0 V | 8.776 | Binary* | |
| J23182+462 | Ross 244 | 23:18:18.44 | +46:17:23.6 | M0.5 V | 7.890 | | |
| J23182+795 | LP 12-69 | 23:18:19.85 | +79:34:45.8 | M3.0 V | 9.712 | | |
| J23193+154 | StKM 1-2115 | 23:19:21.03 | +15:24:13.5 | M1.0 V | 8.887 | | |
| J23194+790 | HD 220140B | 23:19:25.64 | +79:00:04.8 | M3.5 V | 8.036 | SKG | |
| J23228+787 | LP 12-90 | 23:22:54.97 | +78:47:39.6 | M5.0 V | 10.418 | SKG | |
| J23215+568 | LSPM J2321+5651 | 23:21:32.23 | +56:51:19.7 | M1.0 V | 8.903 | | |
| J23216+172 | GJ 4333 | 23:21:36.85 | +17:17:03.3 | M3.5 V | 7.391 | | |
| J23220+569 | G 217-6 | 23:22:01.52 | +56:59:21.4 | M3.0 V | 9.473 | | |
| J23229+372 | PM J23229+3717 | 23:22:58.46 | +37:17:13.3 | M2.0 V | 8.797 | | |
| J23234+155 | LP 522-65 | 23:23:24.72 | +15:34:09.2 | M2.0 V | 8.255 | | |
| J23245+578 | Ross 302 | 23:24:30.39 | +57:51:11.0 | M1.5 V | 6.795 | | |
| J23249+506 | PM J23549+5036 | 23:54:56.57 | +50:36:15.0 | M3.0 V | 8.798 | | • |
| J23252+009 | Wolf 1038 | 23:25:16.81 | +00:57:44.1 | M1.0 V | 8.964 | | |
| J23256+531 | GJ 4334 | 23:25:42.10 | +53:08:11.1 | M5.0 V | 9.878 | | |
| J23262+088 | GJ 2155 | 23:26:12.93 | +08:53:41.2 | M0.0 V | 7.763 | | |
| J23261+170 | PM J23261+1700 | 23:26:12.00 | +17:00:06.9 | M4.0 V | 9.356 | Binary | |
| J23265+121 | GJ 4336 | 23:26:33.26 | +12:09:37.9 | M2.5 V | 8.962 | | • |
| J23262+278 | V595Peg | 23:26:17.02 | +27:52:02.8 | M3.0 V | 8.455 | Binary | |
| J23293+414N | GJ 4337 | 23:29:24.06 | +41:28:06.0 | M3.0 V | 7.925 | Triple | |
| J23293+414S | GJ 4338 | 23:29:23.18 | +41:27:51.4 | M4.2 V | 8.017 | Triple | |
| J23317-027 | AFPsc | 23:31:45.03 | -02:44:40.7 | M4.5 V | 9.507 | Binary | |
| J23301-026 | 2M23301129-0237227 | 23:30:11.41 | -02:37:23.9 | M6.0 V | 10.648 | Binary | |
| J23308+157 | LP 462-51 | 23:30:53.52 | +15:47:38.8 | M1.0 V | 8.413 | Binary | |
| J23318+199E | EQPegA | 23:31:52.83 | +19:56:13.2 | M3.5 V | 6.162 | Quadruple (SB2) | |
| J23318+199W | EQPegB | 23:31:53.20 | +19:56:14.3 | M4.0 V | 7.101 | Quadruple (SB1) | |
| J23323+540 | G 217-12 | 23:32:20.61 | +54:01:48.5 | M2.0 V | 8.920 | | |
| J23302-203 | GJ 1284 | 23:30:13.80 | -20:23:30.7 | M2.0 V | 7.200 | Quintuple (SB2) | |
| J23327-167 | GJ 897 | 23:32:46.98 | -16:45:12.3 | M3.0 V | 6.712 | Quintuple | |
| J23340+001 | Wolf 1039 | 23:34:02.28 | +00:10:30.9 | M2.5 V | 7.664 | | |
| J23350+252 | Ross 298 | 23:35:03.91 | +25:15:00.7 | M3.0 V | 9.138 | | |
| J23351-023 | GJ 1286 | 23:35:11.30 | -02:23:34.1 | M5.0 V | 9.148 | | |
| J23354+300 | GJ 4342 | 23:35:24.05 | +30:03:40.6 | M3.5 V | 9.399 | | |
| J23357+419 | GJ 4346 | 23:35:45.48 | +41:58:06.3 | M1.0 V | 8.108 | | |
| J23364+554 | Ross 303 | 23:36:26.53 | +55:29:42.1 | M1.5 V | 8.448 | | |
| J23376-128 | LP 763-3 | 23:37:38.56 | -12:50:33.3 | M5.5 V | 11.462 | | • |
| J23381-162 | GJ 4352 | 23:38:07.84 | -16:14:11.4 | M2.0 V | 7.813 | | |
| J23386+391 | GJ 4354 | 23:38:41.41 | +39:09:17.9 | M3.3 V | 9.580 | | |
| J23350+016 | GJ 900 | 23:35:00.64 | +01:36:19.9 | M0.0 V | 6.881 | Binary | • |
| J23389+210 | GJ 4356 | 23:38:56.00 | +21:01:24.5 | M4.3 V | 9.941 | Binary | |
| J23401+606 | GJ 4358 A | 23:40:07.91 | +60:41:14.4 | M1.0 V | 8.136 | Binary | |
| J23414+200 | TYC 1727-1708-1 | 23:41:28.99 | +20:02:31.0 | M0.5 V | 8.487 | Binary | |
| J23419+441 | HHAnd | 23:41:55.20 | +44:10:13.4 | M5.0 V | 6.884 | | |
| J23423+349 | PM J23423+3458 | 23:42:22.21 | +34:58:25.5 | M4.0 V | 9.315 | | • |
| J23428+308 | GJ 1288 | 23:42:52.33 | +30:49:16.9 | M5.0 V | 9.637 | | |
| J23431+365 | GJ 1289 | 23:43:07.56 | +36:32:10.7 | M4.5 V | 8.110 | | |
| J23438+325 | GJ 905.2 A | 23:43:52.84 | +32:35:37.8 | M1.5 V | 7.863 | Triple | |
| J23438+610 | G 217-18 | 23:43:51.95 | +61:02:07.5 | M3.0 V | 9.392 | | |
| J23443+216 | GJ 1290 | 23:44:21.41 | +21:36:06.4 | M3.4 V | 9.070 | | • |
| J23439+647 | Ross 676 | 23:44:00.86 | +64:44:30.4 | M0.3 V | 8.149 | Binary (SB2) | |

Table B.1: Carmencita, the CARMENES input catalogue (continued).

| Karmn | Name | α (2016.0) | δ (2016.0) | Spectral type | J [mag] | Multiplicity[a] | DR1[b] |
|---|---|---|---|---|---|---|---|
| J23455-161 | GJ 4360 A | 23:45:30.82 | -16:10:28.8 | M5.0 V | 9.206 | Binary | |
| J23462+284 | PM J23462+2826 | 23:46:14.17 | +28:26:05.0 | M3.5 V | 8.957 | | • |
| J23480+490 | Ross 249 | 23:48:04.16 | +49:00:58.4 | M3.0 V | 8.765 | | |
| J23489+098 | [R78b] 377 | 23:48:58.98 | +09:51:53.3 | M1.0 V | 8.891 | Binary | |
| J23492+024 | BRPsc | 23:49:13.58 | +02:23:48.9 | dM1 | 5.827 | | |
| J23492+100 | GJ 4363 | 23:49:15.05 | +10:05:32.7 | M3.0 V | 9.459 | | |
| J23496+083 | GJ 4364 | 23:49:37.78 | +08:21:29.1 | M1.0 V | 8.276 | | |
| J23505-095 | GJ 4367 | 23:50:32.33 | -09:33:39.5 | dM4.0 | 8.943 | | |
| J23506+099 | GJ 4368 | 23:50:36.86 | +09:56:57.5 | M3.0 V | 7.672 | Binary | |
| J23509+384 | GJ 4369 | 23:50:53.91 | +38:29:30.1 | M3.8 V | 9.800 | | |
| J23517+069 | GJ 4370 | 23:51:44.72 | +06:58:11.8 | M3.0 V | 8.841 | Binary | • |
| J23523-146 | GJ 4371 | 23:52:23.92 | -14:41:28.2 | M4.5 V | 10.436 | | |
| J23535+121 | PM J23535+1206S | 23:53:35.69 | +12:06:14.8 | M2.5 V | 8.670 | Binary | |
| J23541+516 | GJ 4373 | 23:54:10.95 | +51:41:11.0 | M3.5 V | 9.574 | | • |
| J23544+081 | GJ 4374 | 23:54:26.50 | +08:09:42.6 | M3.0 V | 9.243 | | |
| J23548+385 | RX J2354.8+3831 | 23:54:51.28 | +38:31:34.8 | M4.0 V | 8.937 | | |
| J23554-039 | GJ 4376 | 23:55:26.51 | -03:58:59.8 | M3.5 V | 9.866 | | |
| J23560+150 | LP 523-78 | 23:56:00.14 | +15:01:37.9 | M2.5 V | 9.382 | | |
| J23556-061 | GJ 912 | 23:55:39.27 | -06:08:39.4 | M2.5Vk | 7.600 | Binary (SB1) | |
| J23569+230 | G 129-45 | 23:56:54.75 | +23:05:04.2 | M1.5 V | 8.483 | Binary | |
| J23573-129E | GJ 4378 | 23:57:20.81 | -12:58:48.6 | M4.0 V | 8.636 | Quadruple | |
| J23573-129W | GJ 4379 | 23:57:19.59 | -12:58:40.3 | M3.0 V | 9.128 | Quadruple (SB2) | • |
| J23577+197 | GJ 4380 | 23:57:45.32 | +19:46:03.4 | M3.7 V | 9.035 | | |
| J23577+233 | GJ 1292 | 23:57:45.32 | +23:18:00.1 | M3.5 V | 7.800 | | • |
| J23578+386 | GJ 4381 | 23:57:49.68 | +38:37:44.4 | M3.0 V | 8.691 | Binary | |
| J23582-174 | LP 764-40 A | 23:58:13.95 | -17:24:34.6 | M2.0 V | 8.311 | Binary | |
| J23585+076 | Wolf 1051 | 23:58:32.74 | +07:39:25.0 | M3.0 V | 7.907 | Triple (ST3) | |
| J23587+467 | GJ 913 | 23:58:44.50 | +46:43:44.9 | M0.0 V | 6.659 | Binary | |
| J23590+208 | G 129-51 | 23:59:00.74 | +20:51:37.2 | M2.5 V | 9.072 | Binary* | |
| J23598+477 | GJ 4385 | 23:59:50.89 | +47:45:41.3 | M5.0 V | 10.866 | | |

[a] An asterisk (∗) means that further multiplicity is possible, generally hinted by *Gaia*.

[b] The star is part of the first data release (DR1) of CARMENES, which includes all the spectra collected in four years of GTO operations (Ribas et al., 2023).

# Appendix C

## Long tables of Chapter 3

Table C.1: Star candidates belonging to multiple systems not tabulated by the Washington Double Star Catalog (WDS).

| Identifier | Name[a] | Spectral type | $\alpha$ (J2015.5) | $\delta$ (J2015.5) | $\pi$ [mas] | $\mu_\alpha \cos\delta$ [mas a$^{-1}$] | $\mu_\delta$ [mas a$^{-1}$] | $\mu_{total}$ [mas a$^{-1}$] | $G$ [mag] | $\theta$ [deg] | $\rho$ [arcsec] |
|---|---|---|---|---|---|---|---|---|---|---|---|
| J00026+383 | 2M J00024011+3821453 | M4.0 V | 00:02:40.00 | +38:21:44.1 | 24.54 ± 0.24 | -70.31 ± 0.27 | -22.34 ± 0.19 | 73.77 ± 0.27 | 13.1900 ± 0.0012 | 34.0 | 1.419 |
| | | | 00:02:40.06 | +38:21:45.4 | 24.16 ± 0.38 | -57.16 ± 0.54 | -35.59 ± 0.20 | 67.33 ± 0.47 | 13.3648 ± 0.0014 | | |
| I01007+2356 | PM J01007+2356 | K7 V | 01:00:46.85 | +23:56:54.4 | 24.75 ± 0.04 | 129.88 ± 0.06 | 8.41 ± 0.06 | 130.15 ± 0.06 | 10.7186 ± 0.0007 | 4.7 | 1.480 |
| | | | 01:00:46.85 | +23:56:55.9 | … | … | … | … | 14.6491 ± 0.0089 | | |
| J01074-025 | RAVE J010727.5-023326 | K5 V | 01:07:27.46 | −02:33:27.4 | 6.77 ± 0.05 | -54.38 ± 0.09 | -62.33 ± 0.05 | 82.72 ± 0.07 | 12.1930 ± 0.0003 | 165.2 | 1.363 |
| | | | 01:07:27.48 | −02:33:28.8 | 6.42 ± 0.08 | -54.13 ± 0.17 | -61.36 ± 0.07 | 81.82 ± 0.12 | 14.5322 ± 0.0020 | | |
| J02026+105 | RX J0202.4+1034 | M4.5 V | 02:02:28.15 | +10:34:51.9 | 70.43 ± 0.53 | -54.60 ± 1.07 | -96.95 ± 0.77 | 111.27 ± 0.85 | 11.8652 ± 0.0012 | 25.3 | 0.904 |
| | | | 02:02:28.18 | +10:34:52.7 | 68.79 ± 1.20 | -101.45 ± 2.02 | -58.95 ± 1.43 | 117.33 ± 1.89 | 12.3296 ± 0.0064 | | |
| J02287+156 | BPM 85139 | M2.0 V | 02:28:47.14 | +15:38:53.6 | 28.53 ± 0.11 | 170.91 ± 0.18 | -9.17 ± 0.17 | 171.15 ± 0.18 | 11.5139 ± 0.0037 | 147.7 | 0.814 |
| | | | 02:28:47.17 | +15:38:52.9 | … | … | … | … | 13.0258 ± 0.0055 | | |
| J02289+226 | BPM 85140 | M2.0 V | 02:28:58.41 | +22:36:24.5 | 17.16 ± 0.04 | 148.78 ± 0.08 | -48.74 ± 0.07 | 156.56 ± 0.08 | 11.3170 ± 0.0006 | 147.4 | 3.022 |
| | | | 02:28:58.52 | +22:36:21.9 | 17.19 ± 0.53 | 134.47 ± 1.13 | -36.20 ± 0.99 | 139.26 ± 1.12 | 12.0611 ± 0.0003 | | |
| J03207+397 | LP 198-637 | M1.5 V | 03:20:45.41 | +39:42:59.4 | 31.60 ± 0.51 | 126.71 ± 1.33 | -129.25 ± 0.86 | 181.00 ± 1.12 | 10.9868 ± 0.0020 | 278.1 | 0.783 |
| | | | 03:20:45.35 | +39:42:59.7 | … | … | … | … | 11.2553 ± 0.0065 | | |
| I03276+0956 | GJ 3226 | K7 V | 03:27:38.21 | +09:56:05.3 | 22.80 ± 0.09 | 77.33 ± 0.15 | -24.81 ± 0.14 | 81.22 ± 0.15 | 10.5483 ± 0.0014 | 296.8 | 1.541 |
| | | | 03:27:38.12 | +09:56:06.0 | 23.82 ± 0.12 | 57.31 ± 0.19 | -13.65 ± 0.21 | 58.91 ± 0.19 | 10.5825 ± 0.0008 | | |
| J03284+352 | LSPM J0328+3515 | M2.0 V | 03:28:29.35 | +35:15:18.7 | 20.90 ± 0.09 | 99.21 ± 0.15 | -121.12 ± 0.08 | 156.56 ± 0.11 | 12.1366 ± 0.0007 | 204.2 | 1.230 |
| | | | 03:28:29.31 | +35:15:17.5 | 21.14 ± 0.09 | 95.48 ± 0.15 | -108.47 ± 0.08 | 144.51 ± 0.12 | 12.1711 ± 0.0010 | | |
| J03544-091 | StKM 1-430 | M1.0 V | 03:54:25.52 | −09:09:29.2 | 47.39 ± 0.04 | -95.44 ± 0.06 | 110.84 ± 0.05 | 146.27 ± 0.06 | 10.5351 ± 0.0006 | 153.2 | 3.177 |
| | | | 03:54:25.62 | −09:09:32.1 | 47.40 ± 0.06 | -96.46 ± 0.09 | 98.93 ± 0.08 | 138.17 ± 0.09 | 11.8800 ± 0.0012 | | |
| J05530+047 | G 106-007 | M1.5 V | 05:53:04.74 | +04:43:02.7 | 24.70 ± 0.10 | 258.57 ± 0.28 | -295.56 ± 0.20 | 392.71 ± 0.24 | 11.3303 ± 0.0011 | 278.1 | 1.517 |
| | | | 05:53:04.64 | +04:43:02.9 | … | … | … | … | 15.8285 ± 0.0108 | | |
| I07245+1836 | PM J07245+1836 | K7 V | 07:24:32.30 | +18:36:31.3 | 19.99 ± 0.05 | 53.55 ± 0.09 | -36.35 ± 0.08 | 64.72 ± 0.09 | 10.8267 ± 0.0005 | 326.9 | 1.836 |
| | | | 07:24:32.23 | +18:36:32.9 | … | … | … | … | 12.6023 ± 0.0024 | | |
| J07418+050[b] | G 050-001 | M2.5 V+ | 07:41:52.56 | +05:02:23.1 | 36.09 ± 0.07 | -248.35 ± 0.13 | -87.34 ± 0.10 | 263.26 ± 0.12 | 11.6216 ± 0.0009 | 136.1 | 1.006 |
| | | | 07:41:52.61 | +05:02:22.4 | … | … | … | … | 16.1257 ± 0.0263 | | |
| J07545-096 | 2M J07543272-0941478 | M3.5 V | 07:54:32.61 | −09:41:47.9 | 27.81 ± 0.10 | -91.49 ± 0.16 | -13.16 ± 0.11 | 92.43 ± 0.16 | 12.6737 ± 0.0014 | 129.9 | 1.233 |
| | | | 07:54:32.67 | −09:41:48.7 | … | … | … | … | 13.9321 ± 0.0029 | | |
| I08192+5752 | PM J08192+5752 | K7 V | 08:19:14.01 | +57:52:26.8 | 19.67 ± 0.04 | 39.66 ± 0.06 | -79.76 ± 0.06 | 89.08 ± 0.06 | 10.7381 ± 0.0005 | 93.3 | 1.475 |
| | | | 08:19:14.19 | +57:52:26.6 | 19.71 ± 0.14 | 40.66 ± 0.23 | -72.23 ± 0.46 | 82.89 ± 0.42 | 14.1599 ± 0.0067 | | |
| J09050+028 | LP 546-48 | M1.5 V | 09:05:04.12 | +02:50:03.8 | 42.58 ± 0.25 | -312.21 ± 0.39 | 29.17 ± 0.42 | 313.57 ± 0.39 | 10.9288 ± 0.0021 | 253.0 | 1.214 |
| | | | 09:05:04.04 | +02:50:03.5 | … | … | … | … | 12.1972 ± 0.0021 | | |
| J09527+554 | G 195-043 | M1.5 V | 09:52:45.24 | +55:28:16.3 | 28.51 ± 0.36 | 298.92 ± 0.59 | -201.23 ± 0.64 | 360.34 ± 0.61 | 11.3624 ± 0.0008 | 331.1 | 2.716 |
| | | | 09:52:45.14 | +55:28:18.9 | 27.15 ± 0.12 | 285.07 ± 0.23 | -190.75 ± 0.15 | 343.00 ± 0.21 | 16.4864 ± 0.0034 | | |
| I10526+0029 | PM J10526+0029 | K7 V | 10:52:39.52 | +00:29:01.5 | 25.18 ± 0.08 | -91.32 ± 0.10 | -31.13 ± 0.08 | 96.48 ± 0.09 | 10.1687 ± 0.0016 | 251.0 | 1.594 |

Table C.1: Star candidates belonging to multiple systems not tabulated by the Washington Double Star Catalog (continued).

| Identifier | Name[a] | Spectral type | $\alpha$ (J2015.5) | $\delta$ (J2015.5) | $\pi$ [mas] | $\mu_\alpha \cos\delta$ [mas a$^{-1}$] | $\mu_\delta$ [mas a$^{-1}$] | $\mu_{\rm total}$ [mas a$^{-1}$] | $G$ [mag] | $\theta$ [deg] | $\rho$ [arcsec] |
|---|---|---|---|---|---|---|---|---|---|---|---|
| | | | 10:52:39.42 | +00:29:01.0 | … | … | … | … | 12.7067 ± 0.0042 | | |
| I11585+4626[c] | PM J11585+4626 | K7 V | 11:58:33.82 | +46:26:28.9 | 16.20 ± 0.06 | -129.71 ± 0.07 | 1.77 ± 0.06 | 129.72 ± 0.07 | 10.9651 ± 0.0011 | 333.9 | 1.540 |
| | | | 11:58:33.77 | +46:26:30.4 | 14.50 ± 0.14 | -141.05 ± 0.31 | 1.26 ± 0.17 | 141.05 ± 0.31 | 11.5845 ± 0.0015 | | |
| J12191+318[b] | LP 320-626 | M4.0 V+ | 12:19:05.57 | +31:50:43.6 | … | … | … | … | 11.1940 ± 0.0006 | 225.2 | 1.764 |
| | | | 12:19:05.48 | +31:50:42.2 | 35.13 ± 0.09 | -295.73 ± 0.10 | 5.02 ± 0.11 | 295.77 ± 0.10 | 13.9526 ± 0.0023 | | |
| J12390+470 | G 123-049 | M2.0 V | 12:39:05.24 | +47:02:21.4 | … | … | … | … | 11.1336 ± 0.0044 | 110.4 | 0.463 |
| | | | 12:39:05.28 | +47:02:21.2 | 43.50 ± 0.05 | 384.45 ± 0.07 | -118.41 ± 0.08 | 402.27 ± 0.07 | 11.2091 ± 0.0009 | | |
| J12513+221 | GJ 1166A | M3.0 V | 12:51:23.72 | +22:06:15.7 | 30.32 ± 0.51 | -177.34 ± 0.98 | 50.54 ± 0.79 | 184.40 ± 0.96 | 12.1313 ± 0.0019 | 91.8 | 1.263 |
| | | | 12:51:23.81 | +22:06:15.6 | … | … | … | … | 13.3117 ± 0.0038 | | |
| J13282+300 | BD+30 2400 | M0.0 V | 13:28:17.54 | +30:02:43.1 | 25.33 ± 0.08 | -186.41 ± 0.25 | -183.87 ± 0.13 | 261.84 ± 0.20 | 10.5043 ± 0.0006 | 320.1 | 1.243 |
| | | | 13:28:17.48 | +30:02:44.1 | … | … | … | … | 14.1569 ± 0.0103 | | |
| J13445+249 | LP 379-098 | M1.0 V | 13:44:33.39 | +24:57:03.7 | 22.31 ± 0.33 | -245.17 ± 0.54 | -96.38 ± 0.39 | 263.43 ± 0.52 | 11.5057 ± 0.0021 | 355.8 | 0.879 |
| | | | 13:44:33.39 | +24:57:04.6 | … | … | … | … | 11.8891 ± 0.0098 | | |
| J13490+026 | Wolf 1495 | M1.5 V | 13:49:01.18 | +02:47:23.3 | … | … | … | … | 10.7706 ± 0.0210 | 315.7 | 0.680 |
| | | | 13:49:01.15 | +02:47:23.8 | 55.78 ± 0.75 | 149.68 ± 1.64 | -333.14 ± 1.56 | 365.22 ± 1.57 | 10.8252 ± 0.0075 | | |
| I15380+3224[c] | PM J15380+3224 | K7 V | 15:38:04.49 | +32:24:31.9 | 16.48 ± 0.23 | -63.14 ± 0.31 | -78.46 ± 0.38 | 100.71 ± 0.36 | 11.2639 ± 0.0021 | 143.8 | 1.061 |
| | | | 15:38:04.53 | +32:24:31.0 | 15.10 ± 0.25 | -74.75 ± 0.39 | -83.43 ± 0.48 | 112.02 ± 0.44 | 11.3307 ± 0.0019 | | |
| J16573+271 | 2M J16572235+2708304 | M2.0 V | 16:57:22.27 | +27:08:31.1 | 27.12 ± 0.12 | -34.09 ± 0.21 | 44.34 ± 0.26 | 55.93 ± 0.24 | 12.3752 ± 0.0011 | 117.6 | 1.048 |
| | | | 16:57:22.34 | +27:08:30.6 | … | … | … | … | 13.2959 ± 0.0025 | | |
| I17068+3212 | PM J17068+3212 | K7 V | 17:06:48.88 | +32:11:59.3 | 31.93 ± 0.02 | 53.18 ± 0.03 | -74.71 ± 0.04 | 91.70 ± 0.04 | 10.7788 ± 0.0003 | 31.4 | 3.279 |
| | | | 17:06:49.00 | +32:12:02.2 | 31.93 ± 0.03 | 46.05 ± 0.09 | -82.76 ± 0.06 | 94.71 ± 0.07 | 12.6244 ± 0.0004 | | |
| J18116+061 | NLTT 46076 | M3.0 V | 18:11:36.49 | +06:06:27.8 | … | … | … | … | 11.9662 ± 0.0074 | 139.6 | 0.625 |
| | | | 18:11:36.51 | +06:06:27.3 | … | … | … | … | 13.5146 ± 0.0077 | | |
| J18400+726 | LP 044-334 | M6.5 V | 18:40:02.20 | +72:40:57.1 | 51.04 ± 0.52 | -43.74 ± 0.83 | 184.49 ± 1.09 | 189.60 ± 1.08 | 15.3854 ± 0.0114 | 110.3 | 0.821 |
| | | | 18:40:02.32 | +72:40:56.5 | … | … | … | … | 15.7040 ± 0.0035 | | |
| I18447+6241 | PM J18447+6241 | K7 V | 18:44:47.49 | +62:41:08.3 | 22.52 ± 0.03 | -33.82 ± 0.06 | 56.65 ± 0.05 | 65.98 ± 0.05 | 10.7927 ± 0.0006 | 277.5 | 1.342 |
| | | | 18:44:47.30 | +62:41:08.7 | … | … | … | … | 14.7260 ± 0.0167 | | |
| I21088+1247 | BD+12 4554 | K7 V | 21:08:51.84 | +12:47:36.9 | 23.81 ± 0.04 | 85.71 ± 0.06 | -67.99 ± 0.05 | 109.40 ± 0.06 | 10.4390 ± 0.0005 | 2.5 | 1.833 |
| | | | 21:08:51.85 | +12:47:38.7 | … | … | … | … | 14.4780 ± 0.0077 | | |
| I21415+4925 | PM J21415+4925 | K7 V | 21:41:31.36 | +49:25:38.1 | 29.93 ± 0.02 | 33.99 ± 0.04 | -85.44 ± 0.04 | 91.95 ± 0.04 | 9.9125 ± 0.0003 | 308.8 | 1.609 |
| | | | 21:41:31.26 | +49:25:39.3 | 30.10 ± 0.10 | 50.32 ± 0.60 | -88.08 ± 0.83 | 101.44 ± 0.78 | 13.2790 ± 0.0044 | | |
| J22012+323 | TYC 2723-908-1 | M1.5 V | 22:01:14.12 | +32:23:13.9 | 32.55 ± 0.08 | 118.50 ± 0.09 | 62.63 ± 0.15 | 134.03 ± 0.11 | 11.4391 ± 0.0012 | 239.8 | 1.295 |
| | | | 22:01:14.04 | +32:23:13.1 | 32.10 ± 0.20 | 107.88 ± 0.27 | 53.40 ± 0.31 | 120.37 ± 0.28 | 12.8902 ± 0.0022 | | |
| I22142+1712 | PM J22142+1712 | K7 V | 22:14:12.84 | +17:12:24.4 | 17.18 ± 0.05 | -10.52 ± 0.08 | -84.77 ± 0.07 | 85.42 ± 0.07 | 10.7551 ± 0.0006 | 268.2 | 1.510 |
| | | | 22:14:12.73 | +17:12:24.4 | … | … | … | … | 15.4819 ± 0.0194 | | |
| I22569+0031 | PM J22569+0031 | K7 V | 22:56:54.65 | +00:31:23.6 | … | … | … | … | 10.9107 ± 0.0145 | 22.4 | 0.907 |

Table C.1: Star candidates belonging to multiple systems not tabulated by the Washington Double Star Catalog (continued).

| Identifier | Name[a] | Spectral type | $\alpha$ (J2015.5) | $\delta$ (J2015.5) | $\pi$ [mas] | $\mu_\alpha \cos\delta$ [mas a$^{-1}$] | $\mu_\delta$ [mas a$^{-1}$] | $\mu_{\rm total}$ [mas a$^{-1}$] | $G$ [mag] | $\theta$ [deg] | $\rho$ [arcsec] |
|---|---|---|---|---|---|---|---|---|---|---|---|
| | | | 22:56:54.67 | +00:31:24.4 | 17.57 ± 0.68 | 13.04 ± 1.16 | -82.42 ± 0.87 | 83.44 ± 0.88 | 11.0021 ± 0.0058 | | |
| I22596+2154 | PM J22596+2154 | K7 V | 22:59:41.42 | +21:54:05.8 | 26.30 ± 0.05 | 127.97 ± 0.09 | -59.09 ± 0.06 | 140.96 ± 0.09 | 10.1878 ± 0.0006 | 37.3 | 2.093 |
| | | | 22:59:41.51 | +21:54:07.6 | … | … | … | … | 13.3417 ± 0.0037 | | |
| J23051+452 | LSPM J2305+4517 | M3.5 V | 23:05:08.99 | +45:17:32.9 | 21.94 ± 0.17 | 184.58 ± 0.26 | 67.37 ± 0.27 | 196.49 ± 0.26 | 12.3370 ± 0.0021 | 80.7 | 0.785 |
| | | | 23:05:09.06 | +45:17:33.1 | … | … | … | … | 14.3079 ± 0.0029 | | |
| J23489+098 | [R78b] 377 | M1.0 V | 23:48:58.97 | +09:51:53.4 | 21.08 ± 0.05 | 147.68 ± 0.09 | -52.74 ± 0.05 | 156.82 ± 0.08 | 11.3848 ± 0.0009 | 20.1 | 1.945 |
| | | | 23:48:59.02 | +09:51:55.2 | 20.76 ± 0.08 | 141.19 ± 0.18 | -43.69 ± 0.06 | 147.80 ± 0.17 | 13.9181 ± 0.0013 | | |
| J23590+208 | G 129-051 | M2.5 V | 23:59:00.73 | +20:51:37.3 | 14.96 ± 0.79 | 228.92 ± 1.25 | -104.85 ± 0.57 | 251.79 ± 1.16 | 12.0177 ± 0.0072 | 170.7 | 0.521 |
| | | | 23:59:00.73 | +20:51:36.7 | … | … | … | … | 12.3273 ± 0.0186 | | |

[a] Primaries "A" are always brighter than secondaries "B" in the $G$ band.

[b] Previously identified as spectroscopic binaries in Reipurth & Mikkola (2012a) and Jeffers et al. (2018).

[c] Common proper motion pairs with $\Delta\pi > 5\,\%$ labelled in Fig. 3.8

Table C.2: Average colours for K5 V to L6 sources. The number in parentheses indicates the number of useful data points.

| Spectral type | $FUV-NUV$ [mag] | $NUV-u$ [mag] | $u-B_T$ [mag] | $B_T-B$ [mag] | $B-g$ [mag] | $g-G_{BP}$ [mag] | $G_{BP}-V_T$ [mag] | $V_T-V$ [mag] | $V-G$ [mag] | $G-r$ [mag] |
|---|---|---|---|---|---|---|---|---|---|---|
| K5 V | 3.59 ± 0.95 (2) | 5.37 (1) | … | 0.24 ± 0.40 (6) | 0.63 ± 0.23 (11) | 0.32 ± 0.11 (11) | 0.04 ± 0.08 (8) | 0.26 ± 0.07 (4) | 0.45 ± 0.04 (9) | −0.03 ± 0.21 (13) |
| K7 V | 3.26 ± 0.65 (7) | 4.23 ± 0.75 (36) | 2.66 ± 0.83 (59) | 0.49 ± 0.43 (98) | 0.62 ± 0.16 (107) | 0.48 ± 0.13 (101) | 0.02 ± 0.12 (94) | 0.21 ± 0.16 (96) | 0.61 ± 0.10 (100) | −0.06 ± 0.12 (103) |
| M0.0 V | 2.83 ± 0.61 (22) | 4.05 ± 0.83 (37) | 2.87 ± 0.73 (54) | 0.46 ± 0.36 (108) | 0.65 ± 0.14 (113) | 0.51 ± 0.12 (117) | 0.08 ± 0.09 (113) | 0.15 ± 0.11 (109) | 0.71 ± 0.09 (115) | −0.17 ± 0.14 (120) |
| M0.5 V | 2.40 ± 0.42 (14) | 4.01 ± 0.96 (22) | 2.85 ± 0.82 (36) | 0.43 ± 0.36 (70) | 0.66 ± 0.13 (72) | 0.53 ± 0.11 (72) | 0.07 ± 0.14 (69) | 0.17 ± 0.16 (68) | 0.77 ± 0.10 (72) | −0.18 ± 0.11 (71) |
| M1.0 V | 2.29 ± 0.44 (11) | 4.72 ± 0.82 (43) | 2.48 ± 0.70 (42) | 0.35 ± 0.40 (86) | 0.72 ± 0.14 (135) | 0.51 ± 0.10 (140) | 0.09 ± 0.16 (87) | 0.15 ± 0.16 (87) | 0.82 ± 0.09 (135) | −0.23 ± 0.10 (140) |
| M1.5 V | 2.24 ± 0.49 (13) | 4.95 ± 0.98 (43) | 2.53 ± 0.66 (32) | 0.28 ± 0.47 (63) | 0.72 ± 0.13 (115) | 0.51 ± 0.15 (121) | 0.12 ± 0.17 (65) | 0.10 ± 0.20 (65) | 0.89 ± 0.11 (115) | −0.30 ± 0.13 (121) |
| M2.0 V | 1.90 ± 0.66 (12) | 5.30 ± 0.78 (36) | 2.08 ± 0.69 (29) | 0.36 ± 0.59 (54) | 0.76 ± 0.05 (118) | 0.49 ± 0.11 (124) | 0.15 ± 0.24 (55) | 0.11 ± 0.26 (53) | 0.95 ± 0.10 (117) | −0.37 ± 0.10 (124) |
| M2.5 V | 1.66 ± 0.65 (9) | 5.37 ± 1.07 (37) | 2.35 ± 0.85 (21) | 0.35 ± 0.55 (30) | 0.79 ± 0.07 (138) | 0.48 ± 0.09 (145) | 0.12 ± 0.18 (31) | 0.16 ± 0.18 (29) | 1.04 ± 0.09 (137) | −0.47 ± 0.11 (146) |
| M3.0 V | 1.33 ± 0.45 (12) | 5.17 ± 0.93 (38) | 1.82 ± 0.37 (8) | 0.42 ± 0.37 (20) | 0.80 ± 0.11 (157) | 0.47 ± 0.09 (173) | 0.09 ± 0.12 (20) | 0.17 ± 0.11 (20) | 1.13 ± 0.09 (156) | −0.55 ± 0.11 (173) |
| M3.5 V | 1.31 ± 0.36 (23) | 4.72 ± 1.13 (46) | 2.18 ± 0.80 (7) | 0.20 ± 0.59 (14) | 0.85 ± 0.07 (217) | 0.46 ± 0.09 (237) | 0.13 ± 0.19 (12) | 0.14 ± 0.19 (13) | 1.24 ± 0.09 (215) | −0.66 ± 0.10 (237) |
| M4.0 V | 1.49 ± 0.56 (29) | 3.91 ± 1.36 (36) | 2.43 ± 0.63 (2) | 0.05 ± 0.50 (5) | 0.90 ± 0.07 (167) | 0.46 ± 0.08 (203) | 0.23 ± 0.24 (6) | 0.04 ± 0.24 (6) | 1.38 ± 0.09 (165) | −0.77 ± 0.09 (203) |
| M4.5 V | 1.04 ± 0.65 (15) | 2.96 ± 0.90 (18) | … | … | 0.97 ± 0.08 (95) | 0.47 ± 0.10 (116) | … | … | 1.54 ± 0.10 (96) | −0.92 ± 0.10 (116) |
| M5.0 V | 1.12 ± 0.42 (8) | 2.75 ± 0.77 (12) | … | … | 0.98 ± 0.19 (35) | 0.46 ± 0.16 (60) | … | … | 1.71 ± 0.16 (38) | −1.10 ± 0.16 (60) |
| M5.5 V | 1.52 ± 0.34 (2) | 2.80 ± 0.58 (7) | … | … | 1.13 ± 0.21 (14) | 0.49 ± 0.13 (26) | … | … | 2.04 ± 0.12 (14) | −1.37 ± 0.10 (26) |
| M6.0 V | 0.84 ± 0.15 (2) | 0.89 ± 1.78 (5) | … | … | 1.24 ± 0.04 (2) | 0.48 ± 0.20 (14) | … | … | 2.01 (1) | −1.73 ± 0.24 (14) |
| M6.5 V | … | … | … | … | 1.07 ± 0.20 (2) | 0.49 ± 0.15 (7) | … | … | 2.72 ± 0.21 (2) | −2.00 ± 0.17 (7) |
| M7.0 V | … | 2.90 (1) | … | … | 1.30 (1) | 0.46 ± 0.19 (5) | … | … | 2.80 (1) | −2.08 ± 0.28 (5) |
| M7.5 V | … | … | … | … | … | 0.42 ± 0.16 (3) | … | … | … | −2.06 ± 0.12 (3) |
| M8.0 V | … | … | … | … | … | 0.56 ± 0.15 (4) | … | … | … | −2.23 ± 0.08 (6) |
| M8.5 V | … | … | … | … | … | 0.52 ± 0.01 (2) | … | … | … | −2.16 ± 0.13 (3) |
| M9.0 V | … | … | … | … | … | 0.72 (1) | … | … | … | −2.01 ± 0.07 (2) |
| M9.5 V | … | … | … | … | … | … | … | … | … | −1.96 (1) |
| L0.0 | … | … | … | … | … | … | … | … | … | −1.98 ± 0.16 (10) |
| L0.5 | … | … | … | … | … | … | … | … | … | −1.92 ± 0.18 (5) |
| L1.0 | … | … | … | … | … | … | … | … | … | −1.98 ± 0.12 (14) |
| L1.5 | … | … | … | … | … | … | … | … | … | −1.99 ± 0.08 (4) |
| L2.0 | … | … | … | … | … | … | … | … | … | −1.88 ± 0.08 (11) |
| L2.5 | … | … | … | … | … | … | … | … | … | −1.90 ± 0.04 (4) |
| L3.0 | … | … | … | … | … | … | … | … | … | −1.88 ± 0.12 (3) |
| L3.5 | … | … | … | … | … | … | … | … | … | −2.02 ± 0.15 (2) |
| L4.0 | … | … | … | … | … | … | … | … | … | −1.86 ± 0.10 (3) |
| L4.5 | … | … | … | … | … | … | … | … | … | … |
| L5.0 | … | … | … | … | … | … | … | … | … | −2.01 (1) |
| L5.5 | … | … | … | … | … | … | … | … | … | −2.04 (1) |
| L6.0 | … | … | … | … | … | … | … | … | … | −2.09 (1) |

Table C.2: Average colours for K5 V to L6 sources. The number in parentheses indicates the number of useful data points (continued).

| Spectral type | $r-i$ [mag] | $i-G_{RP}$ [mag] | $G_{RP}-J$ [mag] | $J-H$ [mag] | $H-K_S$ [mag] | $K_S-W1$ [mag] | $W1-W2$ [mag] | $W2-W3$ [mag] | $W3-W4$ [mag] |
|---|---|---|---|---|---|---|---|---|---|
| K5 V | 0.41 ± 0.07 (13) | 0.40 ± 0.20 (12) | 0.95 ± 0.06 (13) | 0.57 ± 0.03 (14) | 0.14 ± 0.03 (14) | 0.08 ± 0.06 (14) | 0.04 ± 0.19 (15) | −0.02 ± 0.20 (15) | 0.01 ± 0.13 (10) |
| K7 V | 0.60 ± 0.12 (106) | 0.34 ± 0.12 (99) | 1.12 ± 0.05 (104) | 0.63 ± 0.03 (112) | 0.17 ± 0.03 (112) | 0.11 ± 0.04 (109) | −0.03 ± 0.06 (109) | 0.05 ± 0.06 (112) | 0.06 ± 0.11 (112) |
| M0.0 V | 0.74 ± 0.14 (119) | 0.36 ± 0.11 (117) | 1.22 ± 0.05 (118) | 0.63 ± 0.04 (120) | 0.19 ± 0.03 (119) | 0.12 ± 0.05 (116) | 0.02 ± 0.09 (117) | 0.02 ± 0.08 (119) | 0.10 ± 0.09 (118) |
| M0.5 V | 0.80 ± 0.13 (70) | 0.36 ± 0.08 (71) | 1.28 ± 0.05 (72) | 0.63 ± 0.03 (72) | 0.21 ± 0.03 (71) | 0.12 ± 0.05 (70) | 0.05 ± 0.09 (70) | 0.03 ± 0.07 (71) | 0.11 ± 0.09 (73) |
| M1.0 V | 0.87 ± 0.11 (140) | 0.36 ± 0.09 (139) | 1.32 ± 0.04 (139) | 0.62 ± 0.04 (140) | 0.21 ± 0.03 (138) | 0.14 ± 0.04 (138) | 0.05 ± 0.09 (138) | 0.06 ± 0.06 (138) | 0.09 ± 0.11 (141) |
| M1.5 V | 0.96 ± 0.14 (124) | 0.37 ± 0.11 (121) | 1.37 ± 0.05 (119) | 0.63 ± 0.04 (121) | 0.21 ± 0.03 (121) | 0.14 ± 0.04 (122) | 0.06 ± 0.08 (121) | 0.06 ± 0.06 (123) | 0.09 ± 0.11 (125) |
| M2.0 V | 1.05 ± 0.12 (125) | 0.38 ± 0.11 (126) | 1.44 ± 0.05 (127) | 0.61 ± 0.05 (128) | 0.23 ± 0.03 (128) | 0.15 ± 0.04 (126) | 0.09 ± 0.07 (126) | 0.07 ± 0.05 (126) | 0.10 ± 0.12 (122) |
| M2.5 V | 1.17 ± 0.11 (147) | 0.40 ± 0.10 (144) | 1.51 ± 0.05 (147) | 0.59 ± 0.04 (150) | 0.24 ± 0.03 (147) | 0.16 ± 0.05 (147) | 0.11 ± 0.05 (149) | 0.08 ± 0.04 (149) | 0.12 ± 0.13 (144) |
| M3.0 V | 1.29 ± 0.11 (173) | 0.41 ± 0.10 (172) | 1.58 ± 0.05 (172) | 0.59 ± 0.04 (172) | 0.25 ± 0.03 (171) | 0.16 ± 0.04 (173) | 0.13 ± 0.05 (172) | 0.09 ± 0.05 (172) | 0.13 ± 0.14 (162) |
| M3.5 V | 1.40 ± 0.13 (240) | 0.44 ± 0.12 (235) | 1.66 ± 0.05 (237) | 0.58 ± 0.05 (242) | 0.26 ± 0.03 (241) | 0.17 ± 0.04 (241) | 0.15 ± 0.05 (239) | 0.10 ± 0.05 (239) | 0.15 ± 0.15 (226) |
| M4.0 V | 1.53 ± 0.15 (205) | 0.47 ± 0.13 (201) | 1.76 ± 0.06 (202) | 0.58 ± 0.04 (206) | 0.27 ± 0.03 (206) | 0.18 ± 0.04 (201) | 0.17 ± 0.04 (200) | 0.12 ± 0.04 (199) | 0.19 ± 0.17 (169) |
| M4.5 V | 1.71 ± 0.13 (116) | 0.49 ± 0.09 (116) | 1.87 ± 0.06 (116) | 0.56 ± 0.04 (116) | 0.29 ± 0.03 (116) | 0.20 ± 0.04 (114) | 0.19 ± 0.03 (114) | 0.15 ± 0.04 (115) | 0.23 ± 0.21 (88) |
| M5.0 V | 1.85 ± 0.21 (59) | 0.59 ± 0.14 (59) | 2.02 ± 0.12 (61) | 0.58 ± 0.03 (62) | 0.31 ± 0.03 (62) | 0.21 ± 0.03 (62) | 0.20 ± 0.03 (61) | 0.16 ± 0.04 (60) | 0.28 ± 0.28 (43) |
| M5.5 V | 2.12 ± 0.15 (26) | 0.64 ± 0.13 (26) | 2.14 ± 0.08 (27) | 0.57 ± 0.04 (27) | 0.33 ± 0.02 (26) | 0.23 ± 0.03 (26) | 0.21 ± 0.05 (27) | 0.17 ± 0.05 (27) | 0.23 ± 0.16 (21) |
| M6.0 V | 2.32 ± 0.25 (13) | 0.83 ± 0.12 (12) | 2.33 ± 0.16 (12) | 0.60 ± 0.08 (15) | 0.37 ± 0.03 (15) | 0.23 ± 0.02 (15) | 0.21 ± 0.04 (15) | 0.21 ± 0.07 (15) | 0.33 ± 0.22 (5) |
| M6.5 V | 2.68 ± 0.14 (7) | 0.82 ± 0.10 (7) | 2.49 ± 0.07 (7) | 0.61 ± 0.03 (7) | 0.37 ± 0.02 (7) | 0.24 ± 0.04 (7) | 0.20 ± 0.03 (7) | 0.23 ± 0.03 (7) | 0.10 ± 0.15 (4) |
| M7.0 V | 2.62 ± 0.26 (5) | 0.99 ± 0.15 (5) | 2.52 ± 0.17 (5) | 0.58 ± 0.06 (5) | 0.38 ± 0.04 (5) | 0.26 ± 0.09 (5) | 0.21 ± 0.03 (5) | 0.23 ± 0.08 (5) | 0.21 ± 0.07 (3) |
| M7.5 V | 2.66 ± 0.09 (3) | 0.91 ± 0.05 (3) | 2.52 ± 0.09 (3) | 0.64 ± 0.01 (3) | 0.37 ± 0.03 (3) | 0.25 ± 0.05 (3) | 0.21 ± 0.01 (3) | 0.27 ± 0.06 (3) | … |
| M8.0 V | 2.77 ± 0.10 (6) | 1.03 ± 0.10 (6) | 2.75 ± 0.12 (9) | 0.66 ± 0.03 (9) | 0.41 ± 0.03 (9) | 0.25 ± 0.04 (9) | 0.24 ± 0.05 (9) | 0.28 ± 0.04 (8) | … |
| M8.5 V | 2.70 ± 0.05 (2) | 1.12 ± 0.02 (2) | 2.93 ± 0.05 (4) | 0.64 ± 0.05 (4) | 0.46 ± 0.01 (4) | 0.33 ± 0.04 (4) | 0.28 ± 0.05 (4) | 0.35 ± 0.08 (4) | 0.32 ± 0.05 (3) |
| M9.0 V | 2.43 ± 0.09 (2) | 1.17 ± 0.02 (3) | 3.06 ± 0.08 (5) | 0.67 ± 0.06 (5) | 0.46 ± 0.04 (5) | 0.35 ± 0.03 (5) | 0.29 ± 0.03 (5) | 0.49 ± 0.08 (5) | 0.30 (1) |
| M9.5 V | 2.39 (1) | 1.18 (1) | 3.12 ± 0.11 (2) | 0.84 ± 0.07 (2) | 0.57 ± 0.03 (2) | 0.34 ± 0.02 (2) | 0.28 ± 0.02 (2) | 0.45 ± 0.08 (2) | … |
| L0.0 | 2.43 ± 0.16 (10) | 1.16 ± 0.03 (9) | 3.07 ± 0.05 (11) | 0.75 ± 0.05 (12) | 0.50 ± 0.05 (12) | 0.33 ± 0.04 (12) | 0.27 ± 0.04 (12) | 0.49 ± 0.15 (12) | … |
| L0.5 | 2.38 ± 0.19 (5) | 1.17 ± 0.02 (4) | 3.15 ± 0.06 (6) | 0.73 ± 0.06 (7) | 0.53 ± 0.06 (7) | 0.33 ± 0.02 (7) | 0.26 ± 0.02 (7) | 0.45 ± 0.07 (7) | … |
| L1.0 | 2.41 ± 0.10 (14) | 1.20 ± 0.04 (14) | 3.18 ± 0.10 (15) | 0.82 ± 0.05 (15) | 0.51 ± 0.06 (15) | 0.36 ± 0.05 (15) | 0.24 ± 0.04 (15) | 0.46 ± 0.10 (14) | … |
| L1.5 | 2.39 ± 0.03 (5) | 1.23 ± 0.06 (4) | 3.23 ± 0.06 (7) | 0.80 ± 0.08 (8) | 0.52 ± 0.04 (8) | 0.40 ± 0.09 (8) | 0.27 ± 0.03 (8) | 0.55 ± 0.11 (6) | … |
| L2.0 | 2.34 ± 0.08 (12) | 1.16 ± 0.03 (11) | 3.24 ± 0.10 (12) | 0.85 ± 0.09 (14) | 0.53 ± 0.07 (14) | 0.40 ± 0.06 (14) | 0.27 ± 0.02 (14) | 0.54 ± 0.15 (14) | … |
| L2.5 | 2.31 ± 0.05 (5) | 1.20 ± 0.04 (4) | 3.31 ± 0.07 (5) | 0.91 ± 0.04 (6) | 0.54 ± 0.08 (6) | 0.45 ± 0.06 (6) | 0.26 ± 0.03 (6) | 0.56 ± 0.32 (6) | … |
| L3.0 | 2.34 ± 0.11 (5) | 1.25 ± 0.01 (2) | 3.48 ± 0.09 (3) | 1.00 ± 0.08 (6) | 0.64 ± 0.07 (6) | 0.58 ± 0.09 (6) | 0.33 ± 0.07 (6) | 0.43 ± 0.17 (6) | … |
| L3.5 | 2.40 ± 0.11 (2) | 1.35 ± 0.04 (2) | 3.37 ± 0.03 (2) | 0.93 ± 0.05 (2) | 0.58 ± 0.05 (2) | 0.58 ± 0.06 (2) | 0.30 ± 0.01 (2) | 0.53 ± 0.13 (2) | … |
| L4.0 | 2.22 ± 0.13 (4) | 1.31 ± 0.07 (3) | 3.50 ± 0.15 (4) | 0.90 ± 0.09 (5) | 0.57 ± 0.07 (5) | 0.53 ± 0.09 (5) | 0.27 ± 0.02 (5) | 0.32 ± 0.14 (5) | … |
| L4.5 | 2.06 (1) | … | … | 1.13 (1) | 0.69 (1) | 0.82 (1) | 0.37 (1) | 0.66 (1) | … |
| L5.0 | 2.15 ± 0.21 (5) | 1.53 (1) | 3.75 (1) | 1.01 ± 0.16 (5) | 0.59 ± 0.14 (5) | 0.70 ± 0.09 (5) | 0.38 ± 0.11 (6) | 0.52 ± 0.33 (6) | … |
| L5.5 | 2.13 (1) | … | … | 0.88 (1) | 0.56 (1) | 0.67 (1) | 0.28 (1) | 0.62 (1) | … |
| L6.0 | 2.07 (1) | 1.71 (1) | 3.49 (1) | 0.69 (1) | 0.34 (1) | 0.75 (1) | 0.34 (1) | 0.76 (1) | … |

Table C.3: Basic properties of M-dwarf hosted exoplanets.

| Name | Planet | $a$ [au] | $P_{orb}$ [days] | $\mathcal{M}_p$ [a] [$\mathcal{M}_\oplus$] | $\mathcal{R}_p$ [a] [$\mathcal{R}_\oplus$] | References[b] | $S_{eff}$ [$S_\odot$] | $T_{eq,p}$ [K] | Potentially HZ |
|---|---|---|---|---|---|---|---|---|---|
| GX And | b | $0.072^{+0.003}_{-0.004}$ | $11.4407^{+0.0017}_{-0.0016}$ | $> 3.03^{+0.46}_{-0.44}$ | (> 1.53) | Pin18 | 4.89 ± 0.10 | 413.9 ± 8.7 | |
| K2-149 | b | $0.083^{+0.027}_{-0.027}$ | $11.332^{+0.0013}_{-0.0013}$ | (3.85) | $1.64^{+0.20}_{-0.18}$ | Hir18 | 9.82 ± 0.25 | 493 ± 13 | |
| CD-44 170 | b | $0.101^{+0.009}_{-0.013}$ | $15.8190^{+0.0049}_{-0.0026}$ | $> 13^{+4.1}_{-6.6}$ | (> 3.63) | Tuo14 | 4.51 ± 0.35 | 406 ± 32 | |
| LHS 1140 | b | $0.0875^{+0.0041}_{-0.0041}$ | $24.73712^{+0.00025}_{-0.00025}$ | $6.65^{+1.82}_{-1.82}$ | $1.43^{+0.10}_{-0.10}$ | Dit17 | 0.5780 ± 0.0072 | 242.7 ± 3.0 | Yes |
| | c | $0.02675^{+0.00070}_{-0.00070}$ | $3.777931^{+0.000006}_{-0.000006}$ | $1.81^{+0.39}_{-0.39}$ | $1.282^{+0.024}_{-0.024}$ | Men19 | 6.184 ± 0.077 | 438.9 ± 5.4 | |
| 2MASS J01021226-6145216 | b | $0.03946^{+0.00095}_{-0.00168}$ | $3.7955213^{+0.0000011}_{-0.0000011}$ | (1750) | $12.106^{+0.179}_{-0.179}$ | Bak18 | 16.18 ± 1.14 | 558 ± 39 | |
| BD+61 195 | b | $0.0905^{+0.0011}_{-0.0011}$ | $13.8508^{+0.0053}_{-0.0051}$ | $> 5.63^{+0.67}_{-0.68}$ | (> 2.25) | Per19 | 5.52 ± 0.12 | 426.7 ± 9.0 | |
| YZ Cet | b | $0.01557^{+0.00052}_{-0.00052}$ | $1.96876^{+0.00021}_{-0.00021}$ | $> 0.75^{+0.13}_{-0.13}$ | (> 0.93) | Ast17b | 9.08 ± 0.14 | 483.1 ± 7.6 | |
| | c | $0.0209^{+0.0007}_{-0.0007}$ | $3.06008^{+0.00022}_{-0.00022}$ | $> 0.98^{+0.14}_{-0.14}$ | (> 1.00) | Ast17b | 5.037 ± 0.079 | 417.0 ± 6.5 | |
| | d | $0.02764^{+0.00093}_{-0.00093}$ | $4.65627^{+0.00042}_{-0.00042}$ | $> 1.14^{+0.17}_{-0.17}$ | (> 1.05) | Ast17b | 2.880 ± 0.045 | 362.6 ± 5.7 | |
| K2-150 | b | $0.0727^{+0.0027}_{-0.0027}$ | $10.59357^{+0.00084}_{-0.00084}$ | (5.26) | $2.00^{+0.27}_{-0.21}$ | Hir18 | 4.24 ± 0.10 | 399.5 ± 9.8 | |
| K2-151 | b | $0.0365^{+0.0014}_{-0.0014}$ | $3.835592^{+0.000023}_{-0.000023}$ | (2.44) | $1.35^{+0.16}_{-0.14}$ | Hir18 | 24.48 ± 0.34 | 619.1 ± 8.7 | |
| BD-17 400 | b | $0.0197^{+0.0005}_{-0.0005}$ | $1.22003^{+0.00006}_{-0.00004}$ | $> 1.78^{+0.34}_{-0.33}$ | (> 1.20) | Ast17a | 215.2 ± 3.4 | 1066 ± 17 | |
| | c | $0.057^{+0.001}_{-0.001}$ | $5.974^{+0.001}_{-0.001}$ | $> 4.18^{+0.61}_{-0.59}$ | (> 1.88) | Ast17a | 25.70 ± 0.41 | 626.7 ± 9.9 | |
| | d | $0.698^{+0.018}_{-0.019}$ | $257.8^{+3.6}_{-3.5}$ | $> 10.5^{+2.3}_{-2.1}$ | (> 3.21) | Ast17a | 0.1714 ± 0.0027 | 179.1 ± 2.8 | |
| GJ 96 | b | $0.291^{+0.005}_{-0.005}$ | $73.94^{+0.33}_{-0.38}$ | $> 19.66^{+2.20}_{-2.30}$ | (> 4.74) | Hob18 | 0.754 ± 0.011 | 259.4 ± 3.8 | Yes |
| CD-23 1056 | b | $0.25^{+0.01}_{-0.01}$ | $53.435^{+0.042}_{-0.042}$ | $> 114^{+22}_{-22}$ | (> 13.1) | Sta17 | 1.45 ± 0.12 | 306 ± 25 | Yes |
| LP 413-32 B | b | $0.164^{+0.03}_{-0.03}$ | $31.393463^{+0.000067}_{-0.000069}$ | (4.01) | $1.70^{+0.36}_{-0.36}$ | Fei19 | 1.354 ± 0.066 | 300 ± 15 | Yes |
| BD-21 784 | b | $0.053^{+0.004}_{-0.007}$ | $5.2354^{+0.0027}_{-0.0065}$ | $> 10.2^{+7.2}_{-4.1}$ | (> 3.10) | Tuo14 | 48.8 ± 2.9 | 736 ± 43 | |
| LPM 178 | b | $0.06070^{+0.00001}_{-0.00001}$ | $8.63300^{+0.00155}_{-0.00155}$ | $> 10.60^{+0.06}_{-0.06}$ | (> 3.29) | Bon13b | 5.89 ± 0.10 | 433.6 ± 7.5 | |
| | c | $0.1254^{+0.0001}_{-0.0001}$ | $25.6450^{+0.0235}_{-0.0235}$ | $> 6.8^{+0.9}_{-0.9}$ | (> 2.52) | Bon13b | 1.380 ± 0.024 | 301.6 ± 5.2 | Yes |
| | d | $1.0304^{+0.0086}_{-0.0086}$ | $660.89^{+7.56}_{-7.56}$ | $> 29.4^{+2.9}_{-2.9}$ | (> 5.91) | Bon13b | 0.02044 ± 0.00035 | 105.2 ± 1.8 | |
| Melotte 25 VA 50 | b | $0.0299^{+0.0005}_{-0.0005}$ | $3.484552^{+0.000031}_{-0.000037}$ | (11.6) | $3.43^{+0.95}_{-0.31}$ | Man16a | 8.68 ± 0.15 | 477.7 ± 8.1 | |
| HG 8-15 | b | $0.0562^{+0.0013}_{-0.0014}$ | $6.342^{+0.002}_{-0.002}$ | $4.7^{+0.5}_{-0.3}$ | $1.8^{+0.2}_{-0.1}$ | Die18 | 31.61 ± 0.49 | 660 ± 10 | |
| | c | $0.0946^{+0.0031}_{-0.0030}$ | $13.850^{+0.006}_{-0.006}$ | $6.5^{+1.5}_{-0.5}$ | $2.6^{+0.7}_{-0.2}$ | Die18 | 11.16 ± 0.17 | 509 ± 7.8 | |
| | d | $0.1937^{+0.0064}_{-0.0059}$ | $40.718^{+0.005}_{-0.005}$ | $4.9^{+1.7}_{-0.6}$ | $1.9^{+0.7}_{-0.2}$ | Die18 | 22.661 ± 0.041 | 355.5 ± 5.5 | |
| LP 834-042 | b | $0.14339^{+0.00003}_{-0.00003}$ | $30.5987^{+0.0083}_{-0.0084}$ | $> 23.54^{+0.88}_{-0.89}$ | (> 5.14) | Ast17a | 1.341 ± 0.019 | 299.5 ± 4.2 | Yes |
| | c | $0.36175^{+0.00048}_{-0.00047}$ | $122.6196^{+0.2429}_{-0.2371}$ | $> 21.09^{+1.24}_{-1.26}$ | (> 4.88) | Ast17a | 0.2108 ± 0.0030 | 188.6 ± 2.7 | |

Table C.3: Basic properties of M-dwarf hosted exoplanets (continued).

| Name | Planet | $a$ [au] | $P_{orb}$ [days] | $\mathcal{M}_p$ [a] [$\mathcal{M}_\oplus$] | $\mathcal{R}_p$ [a] [$\mathcal{R}_\oplus$] | References[b] | $S_{eff}$ [$S_\odot$] | $T_{eq,p}$ [K] | Potentially HZ |
|---|---|---|---|---|---|---|---|---|---|
| | d | $0.19394^{+0.00017}_{-0.00018}$ | $48.1345^{+0.0628}_{-0.0661}$ | $> 7.60^{+1.05}_{-1.05}$ | (> 2.63) | Ast17a | 0.733 ± 0.010 | 257.5 ± 3.7 | Yes |
| | e | $0.08208^{+0.00003}_{-0.00004}$ | $13.2543^{+0.0078}_{-0.0104}$ | $> 3.28^{+0.64}_{-0.64}$ | (> 1.62) | Ast17a | 4.094 ± 0.058 | 395.9 ± 5.6 | |
| LEHPM 3808 | b | $0.0306^{+0.0033}_{-0.0057}$ | $3.360080^{+0.000065}_{-0.000070}$ | (1.98) | $1.247^{+0.089}_{-0.083}$ | Gun19 | 17.89 ± 1.44 | 572 ± 46 | |
| | c | $0.0472^{+0.0030}_{-0.0033}$ | $5.660172^{+0.000035}_{-0.000035}$ | (6.88) | $2.42^{+0.13}_{-0.13}$ | Gun19 | 7.52 ± 0.14 | 460.9 ± 8.3 | |
| | d | $0.0733^{+0.0042}_{-0.0042}$ | $11.38014^{+0.00011}_{-0.00010}$ | (5.48) | $2.13^{+0.12}_{-0.12}$ | Gun19 | 3.119 ± 0.052 | 369.9 ± 6.2 | |
| LP 358-499 | b | $0.033^{+0.002}_{-0.002}$ | $3.0712^{+0.0001}_{-0.0001}$ | (2.29) | $1.31^{+0.08}_{-0.08}$ | Wel17 | 32.26 ± 0.39 | 663.3 ± 8.0 | |
| | c | $0.045^{+0.003}_{-0.003}$ | $4.8682^{+0.0001}_{-0.0003}$ | (3.07) | $1.48^{+0.09}_{-0.09}$ | Wel17 | 17.35 ± 0.21 | 568.0 ± 6.8 | |
| | d | $0.077^{+0.005}_{-0.005}$ | $11.0234^{+0.0008}_{-0.0003}$ | (5.15) | $2.02^{+0.13}_{-0.13}$ | Wel17 | 5.925 ± 0.071 | 434.2 ± 5.2 | |
| HD 285968 | b | $0.066^{+0.001}_{-0.001}$ | $8.776^{+0.001}_{-0.002}$ | $> 9.06^{+1.54}_{-0.70}$ | (> 2.87) | Tri18 | 8.37 ± 0.17 | 473.5 ± 9.5 | |
| Wolf 1539 | b | $2.41^{+0.04}_{-0.04}$ | $2288^{+59}_{-59}$ | $> 260.61^{+22.25}_{-22.25}$ | (> 13.9) | How10 | 0.002714 ± 0.000052 | 63.5 ± 1.2 | |
| LPM 198 | b | $0.103^{+0.006}_{-0.014}$ | $17.380^{+0.018}_{-0.020}$ | $> 8.3^{+3.5}_{-5.3}$ | (> 2.81) | Tuo14 | 2.29 ± 0.18 | 342.3 ± 27.1 | Yes |
| | c | $0.1290^{+0.0070}_{-0.0017}$ | $24.329^{+0.052}_{-0.066}$ | $> 6.4^{+3.7}_{-4.1}$ | (> 2.43) | Tuo14 | 1.459 ± 0.064 | 306 ± 13 | |
| LP 656-38 | b | $0.03282^{+0.00054}_{-0.00056}$ | $5.3636^{+0.0007}_{-0.0007}$ | $> 2.02^{+0.26}_{-0.25}$ | (> 1.23) | Ast17a | 3.392 ± 0.049 | 377.7 ± 5.5 | |
| | c | $0.1264^{+0.0021}_{-0.0022}$ | $40.54^{+0.21}_{-0.19}$ | $> 2.31^{+0.50}_{-0.49}$ | (> 1.32) | Ast17a | 0.2287 ± 0.0033 | 192.5 ± 2.8 | |
| Kapteyn's | b | $0.168^{+0.006}_{-0.008}$ | $48.616^{+0.036}_{-0.036}$ | $> 4.8^{+0.9}_{-1.0}$ | (> 1.98) | Ang14 | 0.454 ± 0.014 | 228.4 ± 6.9 | Yes |
| | c | $0.311^{+0.038}_{-0.014}$ | $121.54^{+0.25}_{-0.25}$ | $> 7.0^{+1.2}_{-1.0}$ | (> 2.54) | Ang14 | 0.132 ± 0.012 | 167.9 ± 13.8 | |
| LP 892-26 | b | 0.089 | $14.207^{+0.007}_{-0.007}$ | $> 6.60^{+0.01}_{-0.01}$ | (> 2.43) | Ast15 | 4.1 | 395 | |
| NGTS-1 | b | $0.0326^{+0.0047}_{-0.0045}$ | $2.647298^{+0.000020}_{-0.000020}$ | $258.1^{+20.9}_{-23.8}$ | $14.91^{+6.84}_{-3.70}$ | Bay18 | 63.9 ± 1.4 | 786.8 ± 16.9 | |
| HATS-6 | b | $0.03623^{+0.00042}_{-0.00057}$ | $3.3252725^{+0.0000021}_{-0.0000021}$ | $101.4^{+22.2}_{-22.2}$ | $11.19^{+0.21}_{-0.21}$ | Har15 | 53.5 ± 1.2 | 752.7 ± 16.8 | |
| BD-06 1339 | b | $0.0428^{+0.0007}_{-0.0007}$ | $3.8728^{+0.0004}_{-0.0004}$ | $> 8.5^{+1.3}_{-1.3}$ | (> 2.81) | LoC13 | 56.7 ± 1.3 | 763.9 ± 17.8 | |
| | c | $0.457^{+0.007}_{-0.007}$ | $125.95^{+0.44}_{-0.44}$ | $> 53^{+8}_{-8}$ | (> 8.42) | LoC13 | 0.498 ± 0.012 | 233.8 ± 5.4 | Yes |
| HD 42581 | b | $0.97^{+0.12}_{-0.09}$ | $471^{+22}_{-12}$ | $> 32^{+17}_{-16}$ | (> 6.21) | Tuo14 | 0.0576 ± 0.0035 | 136.3 ± 8.3 | |
| PM J06168+2435 | b | $0.0962^{+0.0054}_{-0.0061}$ | $14.5665^{+0.0016}_{-0.0020}$ | (7.64) | $2.67^{+0.46}_{-0.42}$ | Sch16 | 4.79 ± 0.10 | 411.8 ± 8.9 | |
| Luyten's star | b | $0.091101^{+0.000019}_{-0.000017}$ | $18.6498^{+0.0059}_{-0.0052}$ | $> 2.89^{+0.27}_{-0.26}$ | (> 1.45) | Ast17a | 1.193 ± 0.042 | 291 ± 10 | Yes |
| | c | $0.036467^{+0.000002}_{-0.000002}$ | $4.7234^{+0.0004}_{-0.0004}$ | $> 1.18^{+0.16}_{-0.16}$ | (> 1.06) | Ast17a | 7.44 ± 0.26 | 460 ± 16 | |
| LP 424-4 | b | $0.0348^{+0.0014}_{-0.0014}$ | $3.33714^{+0.00017}_{-0.00017}$ | $14.0^{+1.7}_{-1.7}$ | $4.2^{+0.6}_{-0.6}$ | Bon12 | 33.07 ± 0.41 | 667.4 ± 8.2 | |
| NGC 2632 JS 183 | b | $0.0653^{+0.0039}_{-0.0045}$ | $10.13389^{+0.00068}_{-0.00077}$ | $0.361^{+0.069}_{-0.069}$ | $3.47^{+0.78}_{-0.53}$ | Obe16 | 5.13 ± 0.27 | 419 ± 22 | |
| K2-146 | b | $0.0266^{+0.0010}_{-0.0010}$ | $2.644646^{+0.000043}_{-0.000043}$ | (5.48) | $2.2^{+0.23}_{-0.23}$ | Hir18 | 17.47 ± 0.38 | 569 ± 13 | |

Table C.3: Basic properties of M-dwarf hosted exoplanets (continued).

| Name | Planet | $a$ [au] | $P_{orb}$ [days] | $\mathcal{M}_p$[a] [$\mathcal{M}_\oplus$] | $\mathcal{R}_p$[a] [$\mathcal{R}_\oplus$] | References[b] | $S_{eff}$ [$S_\odot$] | $T_{eq,p}$ [K] | Potentially HZ |
|---|---|---|---|---|---|---|---|---|---|
| LP 844-8 | b | $1.15^{+0.05}_{-0.05}$ | $692^{+2}_{-2}$ | $> 572.07^{+15.89}_{-15.89}$ | (> 13.3) | Ang12 | 0.01695 ± 0.00029 | 100.4 ± 8.9 | |
| | c | 5.5 | $7100^{+8000}_{-1500}$ | 509 | (> 13.5) | Ang12 | 0.00074 | 45 | |
| NGC 2632 JS 597 | b | $0.05023^{+0.00042}_{-0.00043}$ | $5.840002^{+0.000676}_{-0.000602}$ | (5.83) | $2.231^{+0.151}_{-0.145}$ | Liv19 | 12.57 ± 0.38 | 524 ± 16 | |
| | c | $0.11283^{+0.00095}_{-0.00097}$ | $19.660302^{+0.003496}_{-0.003337}$ | (7.64) | $2.668^{+0.201}_{-0.194}$ | Liv19 | 2.490 ± 0.075 | 350 ± 11 | |
| BD+02 2098 | b | $4.5^{+0.2}_{-0.2}$ | $4100^{+300}_{-300}$ | $> 2.30^{+0.13}_{-0.13}$ | (> 1.32) | Rob13 | 0.004911 ± 0.000060 | 73.68 ± 0.90 | |
| K2-117 | b | $0.019^{+0.001}_{-0.001}$ | $1.291505^{+0.000040}_{-0.000040}$ | (4.84) | $1.96^{+0.12}_{-0.12}$ | Dre17 | 123.1 ± 2.6 | 927 ± 20 | |
| | c | $0.051^{+0.002}_{-0.002}$ | $5.444820^{+0.000397}_{-0.000417}$ | (5.26) | $2.03^{+0.13}_{-0.13}$ | Dre17 | 17.09 ± 0.36 | 565 ± 12 | |
| BD+48 1829 | b | $0.039435^{+0.00023}_{-0.00023}$ | $3.822^{+0.001}_{-0.001}$ | $> 13.02^{+2.03}_{-2.01}$ | (> 3.62) | Hob19 | 37.08 ± 0.67 | 687 ± 13 | |
| LTT 3758 | b | $0.0153^{+0.0005}_{-0.0005}$ | $1.628931^{+0.000027}_{-0.000027}$ | $1.66^{+0.23}_{-0.23}$ | $1.43^{+0.16}_{-0.16}$ | Sou17, Bon18b | 19.88 ± 0.25 | 587.7 ± 7.3 | |
| | c | $0.0476^{+0.0017}_{-0.0017}$ | $8.929^{+0.010}_{-0.010}$ | $> 2.64^{+0.44}_{-0.44}$ | (> 1.40) | Bon18b | 2.054 ± 0.026 | 333.2 ± 4.2 | |
| K2-239 | b | $0.0441^{+0.0008}_{-0.0008}$ | $5.240^{+0.001}_{-0.001}$ | $1.4^{+0.4}_{-0.4}$ | $1.1^{+0.1}_{-0.1}$ | Die18 | 3.697 ± 0.048 | 385.9 ± 5.0 | |
| | c | $0.0576^{+0.0009}_{-0.0009}$ | $7.775^{+0.001}_{-0.001}$ | $0.9^{+0.3}_{-0.3}$ | $1.1^{+0.1}_{-0.1}$ | Die18 | 2.167 ± 0.026 | 337.7 ± 4.4 | |
| | d | $0.0685^{+0.0012}_{-0.0012}$ | $10.115^{+0.001}_{-0.001}$ | $1.3^{+0.4}_{-0.4}$ | $1.1^{+0.1}_{-0.1}$ | Die18 | 1.532 ± 0.048 | 309.7 ± 4.0 | |
| LP 905-36 | b | $0.0287^{+0.0010}_{-0.0011}$ | $2.64561^{+0.00066}_{-0.00066}$ | $7.0^{+0.9}_{-0.8}$[c] | (> 2.54) | Bon11 | 35.02 ± 0.56 | 677.0 ± 10.7 | |
| Lalande 21185 | b | 0.0695 | $9.8693^{+0.0016}_{-0.0016}$ | 3.8 | (> 1.75) | But17 | 3.84 ± 0.27 | 390 ± 27 | |
| Innes' star | b | $0.119^{+0.014}_{-0.009}$ | $26.161^{+0.082}_{-0.098}$ | $> 9.9^{+5.6}_{-4.0}$ | (> 3.13) | Tuo14 | 1.264 ± 0.056 | 295 ± 13 | Yes |
| K2-22 | b | 0.009 | $0.381078^{+0.000003}_{-0.000003}$ | (20.2) | $4.75^{+0.35}_{-0.36}$ | Dre17 | 1000 | 1565 | |
| PM J11293-0127 | b | $0.0769^{+0.0039}_{-0.0039}$ | $10.05449^{+0.00026}_{-0.00026}$ | $8.4^{+2.1}_{-2.1}$ | $2.18^{+0.3}_{-0.3}$ | Alm15, Sin16 | 10.20 ± 0.20 | 497.4 ± 9.9 | |
| | c | $0.1399^{+0.0070}_{-0.0070}$ | $24.64354^{+0.00117}_{-0.00117}$ | $2.1^{+2.1}_{-1.3}$ | $1.85^{+0.27}_{-0.27}$ | Alm15, Sin16 | 3.083 ± 0.062 | 368.8 ± 7.4 | |
| | d | $0.2076^{+0.0104}_{-0.0104}$ | $44.55983^{+0.00590}_{-0.00590}$ | $11.1^{+3.5}_{-3.5}$ | $1.51^{+0.23}_{-0.23}$ | Alm15, Sin16 | 1.400 ± 0.028 | 302.8 ± 6.0 | Yes |
| PM J11302+0735 | b | $0.1429^{+0.0060}_{-0.0065}$ | $32.939614^{+0.000101}_{-0.000084}$ | $8.43^{+1.44}_{-1.35}$ | $2.38^{+0.22}_{-0.22}$ | Clo17, Sar18 | 1.293 ± 0.021 | 296.8 ± 4.8 | Yes |
| | c | $0.060^{+0.003}_{-0.003}$ | $8.962^{+0.008}_{-0.008}$ | $> 7.51^{+1.33}_{-1.33}$ | (> 2.66) | Clo17 | 7.33 ± 0.12 | 458.0 ± 7.3 | |
| CD-31 9113 | b | $0.060^{+0.004}_{-0.008}$ | $7.3697^{+0.0034}_{-0.0036}$ | $> 5.3^{+2.0}_{-1.9}$ | (> 2.19) | Tuo14 | 9.43 ± 0.65 | 487.7 ± 33.4 | |
| | c | 3.6 | $3693^{+253}_{-253}$ | 44.6 | (> 7.73) | Del13 | 0.0026 | 63 | |
| Ross 1003 | b | $0.166^{+0.001}_{-0.001}$ | $41.380^{+0.002}_{-0.001}$ | $> 96.70^{+1.41}_{-1.02}$ | (> 11.9) | Tri18 | 0.5440 ± 0.0061 | 239.0 ± 2.7 | Yes |
| | c | $0.912^{+0.005}_{-0.002}$ | $532.58^{+4.14}_{-2.52}$ | $> 68.06^{+4.91}_{-2.19}$ | (> 9.40) | Tri18 | 0.01802 ± 0.00021 | 102.0 ± 1.2 | |
| Ross 905 | b | $0.028^{+0.001}_{-0.001}$ | $2.644^{+0.001}_{-0.001}$ | $> 21.36^{+0.20}_{-0.21}$ | $4.170^{+0.168}_{-0.168}$ | Mac14, Tri18 | 30.48 ± 0.45 | 653.9 ± 9.7 | |
| LP 613-39 | b | $0.0910^{+0.0130}_{-0.0160}$ | $18.4498^{+0.0015}_{-0.0015}$ | (5.83) | $2.25^{+0.53}_{-0.96}$ | Sch16 | 1.470 ± 0.054 | 306.4 ± 11.3 | Yes |

Table C.3: Basic properties of M-dwarf hosted exoplanets (continued).

| Name | Planet | $a$ [au] | $P_{orb}$ [days] | $\mathcal{M}_p$[a] [$\mathcal{M}_\oplus$] | $\mathcal{R}_p$[a] [$\mathcal{R}_\oplus$] | References[b] | $S_{eff}$ [$S_\odot$] | $T_{eq,p}$ [K] | Potentially HZ |
|---|---|---|---|---|---|---|---|---|---|
| FI Vir | b | $0.0496^{+0.0017}_{-0.0017}$ | $9.8658^{+0.0070}_{-0.0070}$ | $> 1.4^{+0.21}_{-0.21}$ | (> 1.11) | Bon18b | 1.556 ± 0.023 | 310.8 ± 4.5 | |
| K2-152 | b | $0.1735^{+0.0054}_{-0.0054}$ | $32.6527^{+0.0035}_{-0.0035}$ | (8.30) | $2.81^{+0.34}_{-0.30}$ | Hir18 | 2.944 ± 0.054 | 364.6 ± 6.6 | |
| K2-153 | b | $0.0601^{+0.0021}_{-0.0021}$ | $7.51554^{+0.00098}_{-0.00098}$ | (4.89) | $2.00^{+0.26}_{-0.22}$ | Hir18 | 14.82 ± 0.33 | 546 ± 12 | |
| K2-137 | b | $0.0058^{+0.0006}_{-0.0006}$ | $0.179715^{+0.000006}_{-0.000006}$ | (0.65) | $0.89^{+0.09}_{-0.09}$ | Smi18 | 841 ± 18 | 1499 ± 33 | |
| K2-154 | b | $0.0408^{+0.0012}_{-0.0012}$ | $3.67635^{+0.00017}_{-0.00017}$ | (6.08) | $2.23^{+0.37}_{-0.24}$ | Hir18 | 59.04 ± 0.96 | 772 ± 13 | |
| | c | $0.0683^{+0.0021}_{-0.0021}$ | $7.95478^{+0.00063}_{-0.00063}$ | (5.48) | $2.10^{+0.25}_{-0.23}$ | Hir18 | 21.07 ± 0.34 | 596.3 ± 9.7 | |
| Ross 1020 | b | $0.026^{+0.001}_{-0.001}$ | $3.023^{+0.001}_{-0.001}$ | $> 7.0^{+0.5}_{-0.5}$ | (> 2.54) | Luq18 | 12.16 ± 0.16 | 519.7 ± 6.6 | |
| HD 122303 | b | $0.067^{+0.001}_{-0.001}$ | $8.708^{+0.002}_{-0.001}$ | $> 6.51^{+0.69}_{-0.40}$ | (> 2.42) | Tri18 | 9.66 ± 0.16 | 490.6 ± 8.2 | |
| Proxima Cen | b | $0.0485^{+0.0041}_{-0.0041}$ | $11.186^{+0.001}_{-0.001}$ | $> 1.27^{+0.19}_{-0.17}$ | (> 1.07) | Ang16 | 0.653 ± 0.017 | 250.2 ± 6.4 | Yes |
| K2-240 | b | $0.0513^{+0.0009}_{-0.0009}$ | $6.034^{+0.001}_{-0.001}$ | $5.0^{+0.5}_{-0.2}$ | $2.0^{+0.2}_{-0.1}$ | Die18 | 23.88 ± 0.58 | 615 ± 15 | |
| | c | $0.1159^{+0.0020}_{-0.0020}$ | $20.523^{+0.001}_{-0.001}$ | $4.6^{+0.7}_{-0.3}$ | $1.8^{+0.3}_{-0.1}$ | Die18 | 4.68 ± 0.11 | 409 ± 10 | |
| HO Lib | b | $0.041^{+0.001}_{-0.001}$ | $5.368^{+0.001}_{-0.001}$ | $> 15.20^{+0.22}_{-0.27}$ | (> 4.03) | Tri18 | 7.14 ± 0.10 | 455.0 ± 6.4 | |
| | c | $0.074^{+0.001}_{-0.001}$ | $12.919^{+0.003}_{-0.002}$ | $> 5.652^{+0.386}_{-0.239}$ | (> 2.20) | Tri18 | 2.192 ± 0.031 | 338.6 ± 4.7 | |
| | d | $0.029^{+0.001}_{-0.001}$ | $3.153^{+0.001}_{-0.006}$ | $> 1.657^{+0.240}_{-0.161}$ | (> 1.18) | Tri18 | 14.27 ± 0.20 | 541.0 ± 7.6 | |
| K2-286 | b | $0.1768^{+0.0175}_{-0.0205}$ | $27.359^{+0.005}_{-0.005}$ | (5.26) | $2.1^{+0.2}_{-0.2}$ | Die19 | 2.891 ± 0.063 | 362.9 ± 7.9 | |
| MCC 759 | b | $0.0608^{+0.0068}_{-0.0068}$ | $6.905^{+0.040}_{-0.040}$ | $> 7.14^{+0.59}_{-0.59}$ | (> 2.67) | Per17 | 17.5 ± 0.2 | 569.1 ± 6.4 | |
| USco J161014.7-191909 | b | $0.0409^{+0.0021}_{-0.0023}$ | $5.424865^{+0.000035}_{-0.000031}$ | < 11.76 | $5.04^{+0.34}_{-0.37}$ | Dav16, Man16b | 60.9 ± 1.9 | 778 ± 25 | |
| LP 804-27 | b | 0.36 | $111.7^{+0.7}_{-0.7}$ | > 667.415 | (> 13.2) | App10 | 0.23 | 193 | |
| HD 147379 | b | $0.3193^{+0.0002}_{-0.0002}$ | $86.54^{+0.07}_{-0.06}$ | $> 24.7^{+1.8}_{-2.4}$ | (> 5.28) | Rei18 | 0.959 ± 0.023 | 275.4 ± 6.5 | Yes |
| MCC 767 | b | $0.078361^{+0.000044}_{-0.000046}$ | $14.628^{+0.012}_{-0.013}$ | $> 2.82^{+0.51}_{-0.51}$ | (> 1.45) | Sua17 | 2.412 ± 0.033 | 346.8 ± 4.8 | |
| V2306 Oph | b | $0.0375^{+0.0012}_{-0.0013}$ | $4.8869^{+0.0005}_{-0.0005}$ | $> 1.91^{+0.26}_{-0.25}$ | (> 1.23) | Ast17a | 7.70 ± 0.12 | 463.6 ± 7.1 | |
| | c | $0.0890^{+0.0029}_{-0.0031}$ | $17.8719^{+0.0059}_{-0.0059}$ | $> 3.41^{+0.43}_{-0.41}$ | (> 1.66) | Ast17a | 1.366 ± 0.021 | 300.9 ± 4.6 | Yes |
| | d | $0.470^{+0.015}_{-0.017}$ | $217.21^{+0.55}_{-0.52}$ | $> 7.70^{+1.12}_{-1.06}$ | (> 2.66) | Ast17a | 0.04899 ± 0.00077 | 130.9 ± 2.1 | |
| BD+25 3173 | b | $1.135^{+0.035}_{-0.035}$ | $598.3^{+4.2}_{-4.2}$ | $> 104.244^{+10.170}_{-10.170}$ | (> 12.5) | Jon10 | 0.03372 ± 0.00036 | 119.3 ± 1.3 | |
| LHS 3275 | b | $0.01411^{+0.00032}_{-0.00032}$ | $1.58040456^{+0.00000016}_{-0.00000016}$ | $6.55^{+0.98}_{-0.98}$ | $2.85^{+0.20}_{-0.20}$ | Cha09, Har13 | 19.51 ± 0.25 | 584.9 ± 7.6 | |
| BD+11 3149 | b | $0.029^{+0.001}_{-0.001}$ | $2.64977^{+0.00081}_{-0.00077}$ | $> 2.47^{+0.27}_{-0.27}$ | (> 1.36) | Aff16 | 55.6 ± 1.5 | 760 ± 20 | |
| | c | $0.089^{+0.003}_{-0.003}$ | $13.740^{+0.016}_{-0.016}$ | $> 6.26^{+0.79}_{-0.76}$ | (> 2.36) | Aff16 | 5.90 ± 0.16 | 434 ± 11 | |
| HD 156384C | b | $0.0505^{+0.0053}_{-0.0044}$ | $7.2004^{+0.0017}_{-0.0017}$ | $> 5.6^{+1.4}_{-1.3}$ | (> 2.23) | Ang13 | 5.86 ± 0.14 | 433 ± 10 | |

Table C.3: Basic properties of M-dwarf hosted exoplanets (continued).

| Name | Planet | $a$ [au] | $P_{orb}$ [days] | $\mathcal{M}_p$[a] [$\mathcal{M}_\oplus$] | $\mathcal{R}_p$[a] [$\mathcal{R}_\oplus$] | References[b] | $S_{eff}$ [$S_\odot$] | $T_{eq,p}$ [K] | Potentially HZ |
|---|---|---|---|---|---|---|---|---|---|
| | c | $0.125^{+0.012}_{-0.013}$ | $28.14^{+0.03}_{-0.03}$ | $> 3.8^{+1.5}_{-1.2}$ | (> 1.78) | Ang13 | 0.956 ± 0.016 | 275.2 ± 4.7 | Yes |
| | d | $0.276^{+0.024}_{-0.030}$ | $91.61^{+0.81}_{-0.89}$ | $> 5.1^{+1.8}_{-1.7}$ | (> 2.08) | Ang13 | 0.1962 ± 0.0052 | 185.2 ± 4.9 | |
| | e | $0.213^{+0.020}_{-0.020}$ | $62.24^{+0.55}_{-0.55}$ | $> 2.7^{+1.6}_{-1.4}$ | (> 1.47) | Ang13 | 0.3294 ± 0.0050 | 210.9 ± 3.2 | Yes |
| | f | $0.156^{+0.014}_{-0.017}$ | $39.026^{+0.194}_{-0.211}$ | $> 2.7^{+1.4}_{-1.2}$ | (> 1.42) | Ang13 | 0.614 ± 0.015 | 246.4 ± 6.0 | Yes |
| | g | $0.549^{+0.052}_{-0.058}$ | $256.2^{+13.8}_{-7.9}$ | $> 4.6^{+2.6}_{-2.3}$ | (> 1.98) | Ang13 | 0.04958 ± 0.00092 | 131.3 ± 2.4 | |
| CD-46 11540 | b | 0.039 | $4.6938^{+0.0070}_{-0.0070}$ | > 11.09 | (> 3.32) | Bon07 | 11 | 507 | |
| CD-51 10924 | b | $1.8049^{+0.0302}_{-0.0302}$ | $1051.1^{+0.5}_{-0.5}$ | $> 1497.9^{+2.9}_{-2.9}$ | (> 12.8) | Sah16 | 0.02763 ± 0.00050 | 113.5 ± 2.0 | |
| | c | $6.6675^{+0.4684}_{-0.4684}$ | $7462.9^{+105.4}_{-101.4}$ | $> 2190^{+32}_{-32}$ | (> 12.6) | Sah16 | 0.002024 ± 0.000037 | 59.0 ± 1.1 | |
| | d | $0.0410^{+0.0007}_{-0.0007}$ | $3.6005^{+0.0002}_{-0.0002}$ | $> 4.4^{+0.3}_{-0.3}$ | (> 1.94) | Sah16 | 53.50 ± 0.96 | 753 ± 14 | |
| | e | $0.1882^{+0.0032}_{-0.0032}$ | $35.39^{+0.03}_{-0.04}$ | $> 8.1^{+0.7}_{-0.7}$ | (> 2.71) | Sah16 | 2.541 ± 0.046 | 351.4 ± 6.3 | |
| BD+68 946 | b | $0.71^{+0.01}_{-0.01}$ | $38.140^{+0.015}_{-0.015}$ | $> 19^{+3}_{-3}$ | (> 4.57) | Sta17 | 0.0439 ± 0.0010 | 127.4 ± 2.8 | |
| CD-44 11909 | b | $0.08^{+0.014}_{-0.004}$ | $17.478^{+0.062}_{-0.040}$ | $> 4.4^{+3.7}_{-2.4}$ | (> 1.92) | Tuo14 | 1.268 ± 0.159 | 295 ± 37 | Yes |
| | c | $0.176^{+0.030}_{-0.009}$ | $57.32^{+0.45}_{-0.48}$ | $> 8.7^{+5.8}_{-4.6}$ | (> 2.81) | Tuo14 | 0.262 ± 0.031 | 199 ± 24 | Yes |
| BD+18 3421 | b | $0.091^{+0.004}_{-0.004}$ | $15.53209^{+0.00166}_{-0.00167}$ | $> 7.1^{+0.9}_{-0.9}$ | (> 2.58) | Aff19 | 3.34 ± 0.22 | 376 ± 25 | |
| Barnard's star | b | $0.404^{+0.018}_{-0.018}$ | $232.80^{+0.38}_{-0.41}$ | $> 3.23^{+0.44}_{-0.44}$ | (> 1.58) | Rib18 | 0.02180 ± 0.00044 | 106.9 ± 2.2 | |
| Kepler-83 | b | 0.0785 | $9.770469^{+0.000022}_{-0.000022}$ | (7.03) | $2.53^{+0.16}_{-0.16}$ | Row14 | 15 | 547 | |
| | c | 0.0514 | $5.169796^{+0.000016}_{-0.000016}$ | (4.74) | $1.94^{+0.13}_{-0.13}$ | Row14 | 35 | 676 | |
| | d | 0.1270 | $20.090227^{+0.000102}_{-0.000102}$ | (6.08) | $2.30^{+0.35}_{-0.35}$ | Row14 | 5.7 | 430 | |
| Kepler-446 | b | $0.0159^{+0.0007}_{-0.0007}$ | $1.5654090^{+0.0000033}_{-0.0000033}$ | $4.5^{+0.5}_{-0.5}$ | $1.50^{+0.25}_{-0.25}$ | Mui15 | 19.61 ± 0.39 | 586 ± 12 | |
| | c | $0.0248^{+0.0011}_{-0.0011}$ | $3.0361790^{+0.0000055}_{-0.0000055}$ | $3^{+1}_{-1}$ | $1.11^{+0.18}_{-0.18}$ | Mui15 | 8.11 ± 0.16 | 469.6 ± 9.3 | |
| | d | $0.0352^{+0.0016}_{-0.0016}$ | $5.148921^{+0.000022}_{-0.000022}$ | $4^{+1}_{-1}$ | $1.35^{+0.22}_{-0.22}$ | Mui15 | 4.008 ± 0.079 | 393.8 ± 7.8 | |
| Kepler-303 | b | 0.0265 | $1.937055^{+0.000004}_{-0.000004}$ | (0.66) | $0.89^{+0.05}_{-0.05}$ | Row14 | 125 | 930 | |
| | c | 0.0628 | $7.061149^{+0.000019}_{-0.000019}$ | (1.55) | $1.14^{+0.09}_{-0.09}$ | Row14 | 22 | 604 | |
| Kepler-617 | b | 0.0233 | $1.682696148^{+0.000001112}_{-0.000001112}$ | (2.32) | $1.32^{+0.07}_{-0.07}$ | Mor16 | 94 | 867 | |
| Kepler-236 | b | 0.0698 | $8.295611^{+0.000032}_{-0.000032}$ | (3.40) | $1.57^{+0.12}_{-0.12}$ | Row14 | 15 | 552 | |
| | c | 0.1416 | $23.968127^{+0.000174}_{-0.000174}$ | (4.94) | $2.00^{+0.17}_{-0.17}$ | Row14 | 3.8 | 388 | |
| Kepler-235 | b | 0.035 | $3.340222^{+0.000005}_{-0.000005}$ | (5.83) | $2.23^{+0.10}_{-0.10}$ | Row14 | 71 | 810 | |
| | c | 0.037 | $7.824904^{+0.000055}_{-0.000055}$ | (2.02) | $1.28^{+0.08}_{-0.08}$ | Row14 | 64 | 788 | |

Table C.3: Basic properties of M-dwarf hosted exoplanets (continued).

| Name | Planet | $a$ [au] | $P_{orb}$ [days] | $\mathcal{M}_p$[a] $[\mathcal{M}_\oplus]$ | $\mathcal{R}_p$[a] $[\mathcal{R}_\oplus]$ | References[b] | $S_{eff}$ $[S_\odot]$ | $T_{eq,p}$ [K] | Potentially HZ |
|---|---|---|---|---|---|---|---|---|---|
| | d | 0.122 | $20.060548^{+0.000099}_{-0.000099}$ | (5.26) | $2.05^{+0.10}_{-0.10}$ | Row14 | 5.9 | 434 | |
| | e | 0.213 | $46.183669^{+0.000425}_{-0.000425}$ | (5.95) | $2.22^{+0.29}_{-0.29}$ | Row14 | 1.9 | 328 | |
| Kepler-52 | b | … | $7.877407^{+0.000020}_{-0.000020}$ | (6.08) | $2.34^{+0.22}_{-0.22}$ | Row14 | 22 | 604 | |
| | c | … | $16.384888^{+0.000080}_{-0.000080}$ | (4.01) | $1.71^{+0.08}_{-0.08}$ | Row14 | 8.4 | 474 | |
| | d | … | $36.445171^{+0.000253}_{-0.000253}$ | (4.74) | $1.95^{+0.21}_{-0.21}$ | Row14 | 2.9 | 363 | |
| Kepler-155 | b | 0.056 | $5.931194^{+0.000008}_{-0.000008}$ | (5.26) | $2.09^{+0.21}_{-0.21}$ | Row14 | 55 | 758 | |
| | c | 0.242 | $52.661793^{+0.000236}_{-0.000236}$ | (5.83) | $2.24^{+0.15}_{-0.15}$ | Row14 | 2.9 | 365 | |
| V1428 Aql | b | $0.3357^{+0.0099}_{-0.0100}$ | $105.90^{+0.09}_{-0.10}$ | $> 12.2^{+1.0}_{-1.4}$ | (> 3.40) | Kam18 | 0.2891 ± 0.0035 | 204.1 ± 2.5 | Yes |
| Kepler-138 | b | $0.0746^{+0.0021}_{-0.0021}$ | $10.3126^{+0.0004}_{-0.0006}$ | $0.066^{+0.059}_{-0.037}$ | $0.522^{+0.032}_{-0.032}$ | Jhu15 | 10.34 ± 0.14 | 499.1 ± 6.5 | |
| | c | $0.0905^{+0.0026}_{-0.0026}$ | $13.7813^{+0.0001}_{-0.0001}$ | $1.970^{+1.912}_{-1.120}$ | $1.197^{+0.070}_{-0.070}$ | Jhu15 | 7.026 ± 0.092 | 453.1 ± 5.9 | |
| | d | $0.1277^{+0.0036}_{-0.0036}$ | $23.0881^{+0.0009}_{-0.0008}$ | $0.640^{+0.674}_{-0.387}$ | $2.1^{+2.2}_{-1.2}$ | Jhu15 | 3.531 ± 0.046 | 381.5 ± 5.0 | |
| LSPM J1928+4437 | b | 0.0116 | $1.2137672^{+0.0000046}_{-0.0000046}$ | (0.39) | $0.78^{+0.22}_{-0.22}$ | Mui12b | 22 | 600 | |
| | c | 0.006 | $0.45328509^{+0.00000097}_{-0.00000097}$ | (0.33) | $0.73^{+0.20}_{-0.20}$ | Mui12b | 81 | 834 | |
| | d | 0.0154 | $1.865169^{+0.000014}_{-0.000014}$ | (0.13) | $0.57^{+0.18}_{-0.18}$ | Mui12b | 12 | 521 | |
| Kepler-49 | b | 0.0642 | $7.203871^{+0.000008}_{-0.000008}$ | (6.21) | $2.35^{+0.09}_{-0.09}$ | Row14 | 23 | 609 | |
| | c | 0.0847 | $10.912732^{+0.000021}_{-0.000021}$ | (5.10) | $2.06^{+0.09}_{-0.09}$ | Row14 | 13 | 530 | |
| | d | 0.0323 | $2.576549^{+0.000003}_{-0.000003}$ | (3.55) | $1.60^{+0.07}_{-0.07}$ | Row14 | 90 | 858 | |
| | e | 0.1208 | $18.596108^{+0.000079}_{-0.000079}$ | (3.33) | $1.56^{+0.08}_{-0.08}$ | Row14 | 6.5 | 444 | |
| LSPM J1930+4149 | b | $0.0514^{+0.0028}_{-0.0028}$ | $8.689090^{+0.000024}_{-0.000024}$ | (1.28) | $1.08^{+0.15}_{-0.15}$ | Age17 | 1.982 ± 0.039 | 330.2 ± 6.6 | |
| Kepler-45 | b | 0.03 | $2.455239^{+0.000004}_{-0.000004}$ | $160.497^{+28.604}_{-28.604}$ | $10.76^{+1.23}_{-1.23}$ | Jon12 | 90 | 856 | |
| Kepler-231 | b | 0.0788 | $10.065275^{+0.000027}_{-0.000027}$ | (4.10) | $1.73^{+0.12}_{-0.12}$ | Row14 | 13 | 530 | |
| | c | 0.1215 | $19.271566^{+0.000099}_{-0.000099}$ | (4.74) | $1.93^{+0.19}_{-0.19}$ | Row14 | 5.5 | 427 | |
| Kepler-54 | b | 0.0655 | $8.010260^{+0.000030}_{-0.000030}$ | (5.15) | $2.07^{+0.12}_{-0.12}$ | Row14 | 11 | 509 | |
| | c | 0.0862 | $12.072389^{+0.000100}_{-0.000100}$ | (4.23) | $1.75^{+0.12}_{-0.12}$ | Row14 | 6.5 | 444 | |
| | d | 0.1246 | $20.995694^{+0.000143}_{-0.000143}$ | (3.26) | $1.53^{+0.08}_{-0.08}$ | Row14 | 3.1 | 369 | |
| Kepler-252 | b | 0.0631 | $6.668391^{+0.000031}_{-0.000031}$ | (1.90) | $1.23^{+0.12}_{-0.12}$ | Row14 | 36 | 680 | |
| | c | 0.0873 | $10.848463^{+0.000015}_{-0.000015}$ | (5.37) | $2.15^{+0.13}_{-0.13}$ | Row14 | 19 | 578 | |
| Kepler-32 | b | 0.05 | $5.90124^{+0.00010}_{-0.00010}$ | (5.89) | $2.2^{+0.2}_{-0.2}$ | Fab12 | 26 | 629 | |

Table C.3: Basic properties of M-dwarf hosted exoplanets (continued).

| Name | Planet | $a$ [au] | $P_{orb}$ [days] | $\mathcal{M}_p$[a] $[\mathcal{M}_\oplus]$ | $\mathcal{R}_p$[a] $[\mathcal{R}_\oplus]$ | References[b] | $S_{eff}$ $[S_\odot]$ | $T_{eq,p}$ [K] | Potentially HZ |
|---|---|---|---|---|---|---|---|---|---|
| | c | 0.09 | $8.7522^{+0.0003}_{-0.0003}$ | (4.94) | $2.0^{+0.2}_{-0.2}$ | Fab12 | 8.0 | 469 | |
| | d | 0.13 | $22.7802^{+0.0005}_{-0.0005}$ | (7.96) | $2.7^{+0.2}_{-0.2}$ | Fab12 | 3.9 | 390 | |
| | e | 0.033 | $2.8960^{+0.0003}_{-0.0003}$ | (4.36) | $1.85^{+0.01}_{-0.01}$ | Fab12 | 60 | 7734 | |
| | f | 0.013 | $0.74296^{+0.00007}_{-0.00007}$ | (0.47) | $0.82^{+0.07}_{-0.07}$ | Fab12 | 385 | 1233 | |
| Kepler-125 | b | 0.0427 | $4.164389^{+0.000003}_{-0.000003}$ | (6.34) | $2.37^{+0.10}_{-0.10}$ | Row14 | 31 | 655 | |
| | c | 0.0531 | $5.774464^{+0.000047}_{-0.000047}$ | (0.32) | $0.74^{+0.05}_{-0.05}$ | Row14 | 20 | 587 | |
| Kepler-186 | b | $0.0343^{+0.0046}_{-0.0046}$ | $3.8867907^{+0.0000062}_{-0.0000063}$ | (1.21) | $1.07^{+0.12}_{-0.12}$ | Qui14 | 48.6 ± 1.2 | 735 ± 19 | |
| | c | $0.0451^{+0.0070}_{-0.0070}$ | $7.267302^{+0.000012}_{-0.000011}$ | (2.02) | $1.25^{+0.14}_{-0.14}$ | Qui14 | 28.09 ± 0.71 | 641 ± 16 | |
| | d | $0.0781^{+0.0100}_{-0.0100}$ | $13.342996^{+0.000025}_{-0.000024}$ | (2.65) | $1.40^{+0.16}_{-0.16}$ | Qui14 | 9.37 ± 0.24 | 487 ± 12 | |
| | e | $0.110^{+0.015}_{-0.015}$ | $22.407704^{+0.000074}_{-0.000072}$ | (2.05) | $1.27^{+0.15}_{-0.14}$ | Qui14 | 4.72 ± 0.12 | 410 ± 10 | |
| | f | $0.432^{+0.171}_{-0.053}$ | $129.9441^{+0.0013}_{-0.0012}$ | (1.65) | $1.17^{+0.08}_{-0.08}$ | Tor15 | 0.306 ± 0.084 | 207 ± 57 | Yes |
| Kepler-445 | b | $0.0229^{+0.0010}_{-0.0010}$ | $2.984151^{+0.000011}_{-0.000011}$ | (3.55) | $1.58^{+0.23}_{-0.23}$ | Mui15 | 14.79 ± 0.40 | 546 ± 15 | |
| | c | $0.0318^{+0.0013}_{-0.0013}$ | $4.871229^{+0.000011}_{-0.000011}$ | (6.74) | $2.51^{+0.36}_{-0.36}$ | Mui15 | 7.70 ± 0.21 | 464 ± 13 | |
| | d | $0.0448^{+0.0019}_{-0.0019}$ | $8.15275^{+0.00040}_{-0.00040}$ | (1.86) | $1.25^{+0.19}_{-0.19}$ | Mui15 | 3.87 ± 0.11 | 390 ± 11 | |
| Kepler-267 | b | 0.0370 | $3.353728^{+0.000007}_{-0.000007}$ | (5.04) | $1.98^{+0.09}_{-0.09}$ | Row14 | 43 | 712 | |
| | c | 0.0597 | $6.877450^{+0.000016}_{-0.000016}$ | (5.59) | $2.13^{+0.11}_{-0.11}$ | Row14 | 16 | 560 | |
| | d | 0.1539 | $28.464515^{+0.000180}_{-0.000180}$ | (5.71) | $2.27^{+0.12}_{-0.12}$ | Row14 | 2.5 | 349 | |
| HD 204961 | b | $3.56^{+0.28}_{-0.28}$ | $3657^{+104}_{-104}$ | $> 216^{+29}_{-28}$ | (> 13.8) | Wit14 | 0.002304 ± 0.000088 | 61.0 ± 2.3 | |
| | c | $0.163^{+0.006}_{-0.006}$ | $35.68^{+0.03}_{-0.03}$ | $> 5.40^{+0.95}_{-0.95}$ | (> 2.16) | Wit14 | 1.099 ± 0.042 | 285 ± 11 | Yes |
| BD-05 5715 | b | $2.35^{+0.22}_{-0.22}$ | $1882^{+250}_{-250}$ | $> 318^{+99}_{-99}$ | (> 13.5) | Sta17 | 0.005115 ± 0.000077 | 74.4 ± 1.1 | |
| | c | 6.5640 | 8775 | > 245.7 | (> 13.9) | Mon14 | 0.00066 | 45 | |
| LP 819-052 | b | $0.026^{+0.001}_{-0.001}$ | $3.651^{+0.001}_{-0.001}$ | $> 6.4^{+0.5}_{-0.5}$ | (> 2.36) | Luq18 | 5.378 ± 0.068 | 423.8 ± 5.4 | |
| LP 700-6 | b | $0.0214^{+0.0013}_{-0.0013}$ | $2.260455^{+0.000041}_{-0.000041}$ | (6.08) | $2.32^{+0.24}_{-0.24}$ | Hir16 | 25.50 ± 0.50 | 625 ± 12 | |
| K2-21 | b | $0.076^{+0.002}_{-0.003}$ | $9.325038^{+0.000379}_{-0.000403}$ | (4.45) | $1.84^{+0.10}_{-0.10}$ | Dre17 | 16.93 ± 0.32 | 565 ± 11 | |
| | c | $0.107^{+0.003}_{-0.004}$ | $15.50192^{+0.00092}_{-0.00093}$ | (7.18) | $2.49^{+0.17}_{-0.19}$ | Dre17 | 8.54 ± 0.14 | 475.8 ± 8.0 | |
| LHS 3844 | b | $0.00623^{+0.00015}_{-0.00015}$ | $0.46292792^{+0.0000016}_{-0.0000016}$ | (2.25) | $1.32^{+0.02}_{-0.02}$ | Van18 | 69.82 ± 0.74 | 804.5 ± 8.6 | |
| IL Aqr | b | $0.214^{+0.001}_{-0.001}$ | $61.082^{+0.006}_{-0.010}$ | $> 760.9^{+1.0}_{-1.0}$ | (> 13.1) | Tri18 | 0.2823 ± 0.0066 | 202.9 ± 4.8 | Yes |
| | c | $0.134^{+0.001}_{-0.001}$ | $30.126^{+0.011}_{-0.003}$ | $> 241.5^{+0.7}_{-0.6}$ | (> 13.7) | Tri18 | 0.720 ± 0.017 | 256.4 ± 6.0 | Yes |

Table C.3: Basic properties of M-dwarf hosted exoplanets (continued).

| Name | Planet | $a$ [au] | $P_{\rm orb}$ [days] | $\mathcal{M}_{\rm p}$[a] [$\mathcal{M}_\oplus$] | $\mathcal{R}_{\rm p}$[a] [$\mathcal{R}_\oplus$] | References[b] | $S_{\rm eff}$ [$S_\odot$] | $T_{\rm eq,p}$ [K] | Potentially HZ |
|---|---|---|---|---|---|---|---|---|---|
| | d | $0.021^{+0.001}_{-0.001}$ | $1.938^{+0.001}_{-0.001}$ | $> 6.910^{+0.220}_{-0.270}$ | (> 2.54) | Tri18 | 29.31 ± 0.68 | 648 ± 15 | |
| | e | $0.345^{+0.001}_{-0.001}$ | $124.4^{+0.3}_{-0.7}$ | $> 15.43^{+1.29}_{-1.27}$ | (> 3.92) | Tri18 | 0.1086 ± 0.0025 | 159.8 ± 3.7 | |
| 2MUCD 12171 | b | $0.01111^{+0.00034}_{-0.00034}$ | $1.51087081^{+0.00000060}_{-0.00000060}$ | $0.85^{+0.72}_{-0.72}$ | $1.086^{+0.035}_{-0.035}$ | Gil17 | 4.357 ± 0.046 | 402.1 ± 4.3 | |
| | c | $0.01521^{+0.00047}_{-0.00047}$ | $2.4218233^{+0.0000017}_{-0.0000017}$ | $1.38^{+0.61}_{-0.61}$ | $1.056^{+0.035}_{-0.035}$ | Gil17 | 2.325 ± 0.025 | 343.7 ± 3.7 | |
| | d | $0.02144^{+0.00066}_{-0.00063}$ | $4.049610^{+0.000063}_{-0.000063}$ | $0.41^{+0.27}_{-0.27}$ | $0.772^{+0.030}_{-0.030}$ | Gil17 | 1.170 ± 0.013 | 289.5 ± 3.1 | Yes |
| | e | $0.02817^{+0.00083}_{-0.00087}$ | $6.099615^{+0.000011}_{-0.000011}$ | $0.62^{+0.58}_{-0.58}$ | $0.918^{+0.039}_{-0.039}$ | Gil17 | 0.6777 ± 0.0073 | 252.5 ± 2.7 | Yes |
| | f | $0.0371^{+0.0011}_{-0.0011}$ | $9.206690^{+0.000015}_{-0.000015}$ | $0.68^{+0.18}_{-0.18}$ | $1.045^{+0.038}_{-0.038}$ | Gil17 | 0.3907 ± 0.0042 | 220.0 ± 2.3 | Yes |
| | g | $0.0451^{+0.0014}_{-0.0014}$ | $12.35294^{+0.00012}_{-0.00012}$ | $1.34^{+0.88}_{-0.88}$ | $1.127^{+0.041}_{-0.041}$ | Gil17 | 0.2644 ± 0.0028 | 199.6 ± 2.1 | Yes |
| | h | $0.063^{+0.027}_{-0.013}$ | $18.767^{+0.004}_{-0.003}$ | (0.39) | $0.752^{+0.032}_{-0.031}$ | Gil17, Lug17 | 0.136 ± 0.030 | 169 ± 38 | |

[a] The planetary masses and radii shown in parentheses correspond to our adopted mass-radius relation.

[b] Aff16: Affer et al. (2016); Aff19: Affer et al. (2019); Age17: Angelo et al. (2017); Alm15: Almenara et al. (2015); Ang12: Anglada-Escudé et al. (2012); Ang13: Anglada-Escudé et al. (2013); Ang14: Anglada-Escudé et al. (2014); Ang16: Anglada-Escudé et al. (2016); App10: Apps et al. (2010); Ast15: Astudillo-Defru et al. (2015); Ast17a: Astudillo-Defru et al. (2017a); Ast17b: Astudillo-Defru et al. (2017b); Bak18: Bakos et al. (2018); Bay18: Bayliss et al. (2018); Bid14: Biddle et al. (2014); Bon07: Bonfils et al. (2007); Bon11: Bonfils et al. (2011); Bon13b: Bonfils et al. (2013b); Bon17: Bonfils et al. (2018c); Bon18a: Bonfils et al. (2018b); Bon18b: Bonfils et al. (2018a); But17: Butler et al. (2017); Cha09: Charbonneau et al. (2009); Clo17: Cloutier et al. (2017); Dav16: David et al. (2016); Del13: Delfosse et al. (2013); Die18a: (Díez Alonso et al., 2018b); Die18b: (Díez Alonso et al., 2018a); Die19: (Díez Alonso et al., 2019); Dit17: Dittmann et al. (2017); Dre17: Dressing et al. (2017); Fab12: Fabrycky et al. (2012); Fei19: Feinstein et al. (2019); Gil17: Gillon et al. (2017); Gil17: Gillon et al. (2017); Gun19: Günther et al. (2019); Har13: Harpsøe et al. (2013); Har15: Hartman et al. (2015); Hir16: Hirano et al. (2016); Hir18: Hirano et al. (2018); Hob18: Hobson et al. (2018); Hob19: Hobson et al. (2019); How10: Howard et al. (2010); Jhu15: Jontof-Hutter et al. (2015); Jon10: Johnson et al. (2010b); Jon12: Johnson et al. (2012); Kam18: Kaminski et al. (2018); Liv19: Livingston et al. (2019); LoC13: Lo Curto et al. (2013); Lug17: Luger et al. (2017); Luq18: Luque et al. (2018); Mac14: Maciejewski et al. (2014); Man16a: Mann et al. (2016a); Man16b: Mann et al. (2016b); Men19: Ment et al. (2019); Mon14: Montet et al. (2014); Mor16: Morton et al. (2016); Mui12: Muirhead et al. (2012); Mui15: Muirhead et al. (2015); Obe16: Obermeier et al. (2016); Per17: Perger et al. (2017); Per19: Perger et al. (2019); Pin18: Pinamonti et al. (2018); Qui14: Quirrenbach et al. (2014); Rei18: Reiners et al. (2018a); Rob13: Astropy Collaboration et al. (2013); Row14: Rowe et al. (2014); Sah16: Sahlmann et al. (2016); San15: Sanchís-Ojeda et al. (2015); Sar18: Sarkis et al. (2018); Sch16: Schlieder et al. (2016); Sin16: Sinukoff et al. (2016); Sou17: Southworth et al. (2017); Smi18: Smith (2018); Sta17: Stassun et al. (2017); Sua17: Suárez Mascareño et al. (2017c); Tor15: Torres et al. (2015); Tri18: Trifonov et al. (2018); Tuo14: Tuomi et al. (2014); Van18: Vanderspek et al. (2019); Wel17: Wells et al. (2018); Wit14: Wittenmyer et al. (2014).

[c] Bonfils et al. (2011) gave true mass for LP 905-36 b calculated with a Markov chain analysis, $\mathcal{M}_p = 8.4^{+4.0}_{-1.5}\,\mathcal{M}_\oplus$, but we adopt the minimum mass, $\mathcal{M}_2 \sin i$, for consistency.

# Appendix D

## Long tables of Chapter 4

Table D.1: Description of the online table used in Chapter 4.

| Parameter | Units | Column(s) | Description |
|---|---|---|---|
| | | | *Identification* |
| ID_star, ID_system | ... | 1, 2 | Star and system identifiers[a] |
| Name | ... | 3 | Discovery name or most common name[b] |
| Karmn | ... | 4 | Carmencita star identifier (JHHMMm+DDd)[d] |
| GJ | ... | 5 | Gliese-Jahreiss catalogue number[d] |
| RA_J2016, DE_J2016 | hms, dms | 6, 7 | Right ascension and declination in the epoch J2016.0 |
| SpT, SpTnum, SpT_ref | ... | 8–10 | Spectral type, its numerical format, and the reference[e] |
| N_planet | ... | 11 | Number of confirmed planets[f] |
| | | | *Multiplicity* |
| Type, Class | ... | 12, 13 | Type of system and multiplicity class[g] |
| Component, System | ... | 14, 15 | Component designation and resolution of the components in the multiple system[h] |
| SB, SB_ref | ... | 16, 17 | Type of spectroscopic system and reference[h] |
| Remarks | ... | 18 | Annotations and remarks |
| WDS_ ... | ... | 19–28 | Washington double star catalogue data[i] |
| theta_deg | deg | 29 | Positional angle[j] |
| rho_arcsec | arcsec | 30 | Projected separation |
| s_au | au | 31 | Physical separation |
| muratio, deltaPA, deltad | ... | 32–34 | $\mu$ratio, $\Delta PA$, $\Delta d$ criteria for physical parity |
| crit_ ... | Boolean | 35–41 | Criteria for unresolved companions[k] |
| Candidate | Boolean | 42 | Candidate to unresolved companion[l] |
| | | | *Stellar parameters* |
| Teff_K, eTeff_K, Teff_K_ref | K | 43–45 | Effective temperature |
| logg, elogg | dex | 46, 47 | Surface gravity |
| L_Lsol, eL_Lsol, L_ref | $\mathcal{L}_\odot$ | 48–50 | Luminosity and uncertainty |
| R1_Rsol, eR1_Rsol | $\mathcal{R}_\odot$ | 51, 52 | Radius and uncertainty |
| M1_Msol, eM1_Msol, RM1_ref | $\mathcal{M}_\odot$ | 53–55 | Mass and uncertainty |
| M2_Msol, eM2_Msol, RM2_ref | $\mathcal{M}_\odot$ | 56–58 | Mass and uncertainty for the unresolved companion |
| Mt_Msol, eMt_Msol | $\mathcal{M}_\odot$ | 59, 60 | Total mass of the system |
| q | ... | 61 | Mass ratio |
| Ug_J, eUg_J | J | 62, 63 | Binding energy |
| Porb_d, ePorb_d, Porb_d_ref | d | 64–66 | Orbital period |
| | | | *Astrometry and photometry* |
| _id | ... | 67–70 | Catalog identifiers |
| ra, ra_error | deg | 71, 72 | Barycentric right ascension and uncertainty in the epoch J2016.0 |
| dec, dec_error | deg | 73, 74 | Barycentric declination and uncertainty in the epoch J2016.0 |
| parallax, parallax_error | mas | 75–77 | Parallax and uncertainty |
| d_pc, ed_pc, d_ref | pc | 78, 79 | Distance, uncertainty, and reference |
| pm, pm_error, pm_ref | mas a$^{-1}$ | 80–82 | Total proper motion and reference |
| pmra, pmra_error | mas a$^{-1}$ | 83, 84 | Proper motion in right ascension, uncertainty, and reference |
| pmdec, pmdec_error | mas a$^{-1}$ | 85, 86 | Proper motion in declination, uncertainty, and reference |
| rv, rv_error, rv_ref | km s$^{-1}$ | 87–89 | Radial velocity, uncertainty, and reference |
| ruwe | ... | 90 | Renormalised unite weight error from *Gaia* DR3 |
| l, b | deg | 91, 92 | Galactic longitude and latitude in the epoch J2016.0 |
| RA_J2000, DE_J2000 | deg | 93, 94 | Right ascension and declination in the epoch J2016.0 |
| NN_mag, eNN_mag | mag | 95–116 | Photometric magnitudes and quality flags in up to 10 passbands[m] |
| | | | *Statistical indicators (Gaia DR3)* |
| [Statistics DR3] | ... | 117–130 | Statistical indicators in *Gaia* DR3[n] |

a) `ID_star` is a unique identifier that sorts the table by right ascension but priorising that components of the same system (equal `ID_system`) are together and sorted by decreasing brightness.

b) Name of the star, obeying the following priority ($n$ designates a natural number): Proper name, variable in constellation (V* V$n$ Con; but not suspected variables, SV*), Henry Draper (HD $n$, with $n \leq 225300$), Gliese-Jahreiss (GJ $n$, only if $n < 3000$), Bonner Durchmusterung (BD±$n$ $n$), Luyten (LP $n$-$n$), Giclas (G $n$-$n$, only if unique Giclas designation), Luyten (LHS $n$), other designations in chronological order (Haro, StKM/StM, 1RXS/RXS, HIP, LSPM, PM, NLTT, GSC, TYC, MCC, R78b, I81, 2MUCD), catalog identifier (*Gaia* DR3, 2MASS, UCAC4).

c) "HHMMm+DDd" are the truncated equatorial coordinates. For stars in a close binary system, a position of the star in the system is added as "N", "S", "E" or "W".

d) Gliese-Jahreiss designation (GJ $n$) is given regardless of the denomination in the `Name` column, and regardless of the $n$ number.

e) `SpTnum` = 10.0 for O0.0 V, 20.0 for B0.0 V, 30.0 for A0.0 V, 40.0 for F0.0 V, 50.0 for G0.0 V, 60.0 for K0.0 V, 70.0 for M0.0 V, 70.5 for M0.5 V, 80.0 for L0.0. Reference 'This work' refers to a photometric estimation of the spectral type based on the absolute magnitudes of Cifuentes et al. (2020).

f) Information extracted from the NASA Exoplanet Archive.

g) For the cases were it exists evidence of an unresolved component, we include an asterisk (∗) at the end of the type (i.e. 'Single*' and 'Multiple*') and the class (e.g. 'Single*' or 'Binary*').

h) SB1: Single-lined double; SB2: Double-lined double; ST2: Double-lined triple; ST3: Triple-lined triple; SQ: Triple- or quadruple-lined quadruple.

i) `id`: WDS identifier; `disc`: Discoverer code plus number, if assigned; `comp`: Component designations; `obs1`, `obs2`: First and last observation years; `pa2`: Positional angle in the more recent measurement; `sep2`: Separation in the more recent measurement; `mag1`, `mag2`: Magnitudes of the two components. More details can be found in the WDS website: http://www.astro.gsu.edu/wds/.

j) Measured eastward from the north in the epoch 2016.0.

k) Criteria for physical parity as described in Eqn. 4.5 (`crit_parity`) and in Table 4.2 (1–6: `crit_ruwe`, `crit_ipd`, `crit_rv`, `crit_rv_error`, `crit_non_single`, `crit_DR3_non_single`). Orbital: Orbital model for an astrometric binary; OrbitalTargetedSearch: Orbital model for a priori known systems, with a subset containing suffix 'Validated'; SB1: Single-lined spectroscopic binary; SB2: Double-lined spectroscopic binary; SB2C: Double-lined spectroscopic binary with circular orbit; AstroSpectroSB1: Combined astrometric + single lined spectroscopic orbital model.

l) It refers to unknown unresolved binaries. That is, if an unresolved companion already exists (e.g. an spectroscopic binary), the value is 'false'.

m) `BP`, `G`, `RP`: $G_{B_P}$, $G$, and $G_{R_P}$ from *Gaia* DR3; `J`, `H`, `Ks`: $J$, $H$, and $K_s$ from 2MASS; `W1`, `W2`, `W3`, `W4`: $W1$, $W2$, $W3$, and $W4$ from AllWISE. The photometric uncertainties in the *Gaia* passbands have been calculated by us $\Delta\lambda = |-2.5/\ln 10 \times \Delta F_\lambda / F_\lambda|$, where $F_\lambda$ and $\Delta F_\lambda$ re the flux and its error in the $\lambda$ passband, using the errors in the corresponding fluxes, and the zero points as provided by VizieR.

n) Statistics from DR3 related to the criteria for unresolved binarity: `astrometric_excess_noise`, `astrometric_excess_noise_sig`, `phot_bp_rp_excess_factor`, `phot_bp_n_blended_transits`, `phot_rp_n_blended_transits`, `phot_variable_flag`, `rv_chisq_pvalue`, `rv_amplitude_robust`, `rv_nb_transits`, `renormalised_gof`, `astrometric_n_obs_al`, `astrometric_n_good_obs_al`, `ipd_gof_harmonic_amplitude`, `duplicated_source` (see Sect. 4.3.2).

Table D.2: Complete sample with the description of multiple systems.

| WDS id | WDS disc | Name | Karmn | Spectral type | $\alpha$ (2016.0) | $\delta$ (2016.0) | System | Component | $\rho$ [arcsec] | $\theta$ [deg] | $\varpi$ [mas] | $\mu_{\rm total}$ [mas a$^{-1}$] | $\mathcal{L}$ [$10^{-4}$ $\mathcal{L}_\odot$] | $\mathcal{M}$ [$\mathcal{M}_\odot$] |
|---|---|---|---|---|---|---|---|---|---|---|---|---|---|---|
| 00012+1357 | LSC 2 | BD+13 5195A | J00012+139N | M0.5 V | 00:01:13.21 | +13:58:32.7 | (AB)+C | AB | 0.247 | 170.0 | 27.34 | 150.86 | … | … |
| 00012+1357 | WNO 12 | BD+13 5195B | J00012+139S | M0.0 V | 00:01:12.89 | +13:58:22.0 | … | C | 11.657 | 203.8 | 27.71 | 152.02 | 1019.35 ± 5.62 | 0.692 ± 0.026 |
| … | … | PM J00026+3821A | J00026+383 | M4.0 V | 00:02:40.00 | +38:21:44.1 | AB | A | … | … | 24.62 | 74.86 | … | 0.357 ± 0.033 |
| … | … | PM J00026+3821B | … | M3.5 V | 00:02:40.05 | +38:21:45.3 | … | B* | 1.415 | 28.1 | 25.39 | 71.14 | … | 0.330 ± 0.034 |
| … | … | StKM 1-2199 | J00033+046 | M1.5 V | 00:03:18.97 | +04:41:11.6 | … | … | … | … | 34.88 | 88.03 | 394.76 ± 2.03 | 0.511 ± 0.029 |
| … | … | HD 38A | … | K6 V | 00:05:42.38 | +45:48:40.9 | Aab+B+C | Aab | … | … | 26.59 | 903.28 | … | … |
| 00057+4549 | STT 547 | HD 38B | J00056+458 | M0.0 V | 00:05:42.30 | +45:48:35.0 | … | B | 6.034 | 188.4 | 86.80 | 859.02 | … | 0.640 ± 0.027 |
| 00057+4549 | STT 547 | GJ 2 | J00051+457 | M1.0 V | 00:05:12.22 | +45:47:09.2 | … | C | 328.479 | 253.8 | 86.82 | 883.80 | 435.76 ± 2.26 | 0.492 ± 0.012 |
| … | … | GJ 1002 | J00067-075 | M5.5 V | 00:06:42.32 | -07:32:47.3 | … | … | … | … | 86.93 | 2059.86 | 14.07 ± 0.07 | 0.107 ± 0.009 |
| … | … | G 217-32A | J00077+603 | M4.0 V | 00:07:43.28 | +60:22:54.0 | AB | A | … | … | 206.35 | 321.59 | … | 0.235 ± 0.037 |
| 00077+6022 | JNN 247 | G 217-32B | … | M5.0 V | 00:07:43.40 | +60:22:53.8 | … | B | 0.848 | 100.4 | 65.00 | 368.61 | … | 0.193 ± 0.039 |
| … | … | PM J00078+6736 | J00078+676 | M2.0 V | 00:07:50.65 | +67:36:23.9 | … | … | … | … | 65.21 | 106.84 | 436.13 ± 4.29 | 0.492 ± 0.030 |
| … | … | GJ 3007 | J00079+080 | M4.0 V | 00:07:58.74 | +08:00:12.8 | … | … | … | … | 39.65 | 544.03 | 193.29 ± 1.70 | 0.399 ± 0.016 |
| … | … | 1R000806.3+475659 | J00081+479 | M4.0 V | 00:08:06.23 | +47:57:02.4 | Aab | Aab(2) | … | … | 32.26 | 124.13 | … | … |
| … | … | GJ 3008 | J00084+174 | M0.0 V | 00:08:27.18 | +17:25:26.4 | … | … | … | … | 36.80 | 113.87 | 598.00 ± 2.32 | 0.587 ± 0.028 |
| 00089+2050 | BEU 1 | GJ 3010 | J00088+208 | M5.0 V | 00:08:53.86 | +20:50:21.4 | (AB) | AB | 0.152 | 74.2 | 53.62 | 264.70 | … | 0.300 ± 0.035 |
| … | … | G 31-29 | J00110+052 | M1.0 V | 00:11:04.89 | +05:12:33.4 | … | … | … | … | 45.99 | 265.51 | 874.37 ± 12.45 | 0.645 ± 0.027 |
| … | … | LSPM J0011+5908 | J00115+591 | M6.0 V | 00:11:29.94 | +59:08:21.2 | … | … | … | … | 55.26 | 1477.10 | 11.29 ± 0.05 | 0.116 ± 0.009 |
| … | … | LP 348-40 | J00118+229 | M3.5 V | 00:11:53.17 | +22:59:01.2 | … | … | … | … | 26.97 | 243.70 | 164.18 ± 1.10 | 0.390 ± 0.016 |
| … | … | G 130-53 | J00119+330 | M3.5 V | 00:11:55.73 | +33:03:10.7 | … | … | … | … | 107.39 | 683.41 | 153.33 ± 0.89 | 0.376 ± 0.016 |
| … | … | 1R001213.6+302906 | J00122+304 | M4.5 V | 00:12:13.49 | +30:28:43.8 | … | … | … | … | 48.83 | 61.99 | 584.88 ± 34.92 | 0.489 ± 0.031 |
| … | … | TYC 4298-613-1 | J00131+703 | M1.0 V | 00:13:11.68 | +70:23:54.9 | … | … | … | … | 47.75 | 145.71 | 849.73 ± 3.70 | 0.650 ± 0.027 |
| 00133+6919 | KUI 1 | GJ 11 A | J00132+693 | M3.5 V | 00:13:18.00 | +69:19:32.4 | AB | A | 0.849 | 276.7 | 13.37 | 784.08 | … | 0.322 ± 0.034 |
| … | … | GJ 11 B | … | M3.0 V | 00:13:18.21 | +69:19:32.3 | … | B | 1.100 | 98.7 | 31.53 | 774.33 | … | … |
| … | … | UPM J0013+2733 | J00133+275 | M4.5 V | 00:13:19.55 | +27:33:29.1 | … | … | … | … | 48.73 | 116.63 | 152.59 ± 0.82 | 0.428 ± 0.018 |
| … | … | GJ 3014 | J00136+806 | M1.5 V | 00:13:40.36 | +80:39:59.8 | AB | A | … | … | … | 311.44 | 461.06 ± 2.94 | 0.522 ± 0.029 |
| 00137+8038 | LDS1503 | GJ 3015 | J00137+806 | M5.0 V | 00:13:44.59 | +80:39:52.3 | … | B | 12.757 | 126.2 | 24.19 | 316.44 | 20.27 ± 0.15 | 0.164 ± 0.011 |
| 00155-1608 | HEI299 | GJ 1005 | J00154-161 | M4.0 V | 00:15:28.77 | -16:08:11.7 | (AB) | AB | 0.417 | 236.6 | 51.98 | 951.27 | … | 0.212 ± 0.039 |
| … | … | LP 49-338 | J00156+722 | M2.0 V | 00:15:37.59 | +72:17:03.5 | … | … | … | … | 51.98 | 358.92 | 293.55 ± 1.59 | 0.468 ± 0.018 |
| … | … | GJ 12 | J00158+135 | M4.0 V | 00:15:49.92 | +13:33:27.6 | … | … | … | … | 200.53 | 700.38 | 74.68 ± 0.32 | 0.255 ± 0.012 |
| 00160-1637 | BWL 2 | 1R001557.5-163659 | J00159-166 | M4.1 V | 00:15:57.94 | -16:36:57.5 | (AB) | AB | 0.051 | 140.9 | 39.03 | 117.74 | … | 0.322 ± 0.034 |
| … | … | EZPsc | J00162+198W | M4.0 V | 00:16:15.44 | +19:51:25.3 | Aab+B | Aab(2) | … | … | 30.35 | 1044.65 | … | 0.491 ± 0.003 |
| 00164+1950 | LDS 863 | GJ 1006 B | J00162+198E | M4.0 V | 00:16:16.96 | +19:51:38.5 | … | B | 25.141 | 58.5 | 82.19 | 1030.67 | 88.32 ± 0.59 | 0.274 ± 0.011 |
| … | … | GJ 1007 | J00169+051 | M4.0 V | 00:16:56.20 | +05:07:16.4 | … | … | … | … | 56.10 | 630.38 | 78.38 ± 0.53 | 0.280 ± 0.013 |
| … | … | GJ 3022 | J00169+200 | M3.5 V | 00:16:57.03 | +20:03:55.7 | AB+C | A | … | … | 65.11 | 239.54 | … | 0.319 ± 0.034 |
| 00169+2004 | CRC 43 | G 131-47B | … | M3.5 V | 00:16:57.10 | +20:03:55.4 | … | B | 1.077 | 106.5 | 65.05 | 233.36 | … | 0.313 ± 0.034 |
| … | … | LP 404-54 | … | M5.0 V | 00:15:13.81 | +19:47:40.5 | … | C* | 1751.981 | 236.2 | 56.16 | 233.80 | 33.49 ± 0.20 | 0.188 ± 0.011 |
| … | … | Ross 680 | J00173+291 | M2.0 V | 00:17:21.17 | +29:11:05.7 | … | … | … | … | 29.31 | 812.80 | 458.04 ± 2.43 | 0.522 ± 0.029 |
| … | … | GJ 3025 | J00176-086 | M0.0 V | 00:17:41.23 | -08:40:55.8 | … | … | … | … | 29.14 | 304.40 | 1011.74 ± 9.86 | 0.680 ± 0.026 |
| … | … | LP 404-81 | J00179+209 | M1.0 V | 00:17:58.87 | +20:57:18.6 | A+B | A | … | … | 28.87 | 445.15 | 681.37 ± 2.55 | 0.612 ± 0.027 |

Table D.2: Complete sample with the description of multiple systems (continued).

| WDS id | WDS disc | Name | Karmn | Spectral type | $\alpha$ (2016.0) | $\delta$ (2016.0) | System | Component | $\rho$ [arcsec] | $\theta$ [deg] | $\varpi$ [mas] | $\mu_{\rm total}$ [mas a$^{-1}$] | $\mathcal{L}$ [$10^{-4}\ \mathcal{L}_\odot$] | $\mathcal{M}$ [$\mathcal{M}_\odot$] |
|---|---|---|---|---|---|---|---|---|---|---|---|---|---|---|
| 00180+2057 | LDS 864 | LP 404-80 | … | M3.0 V | 00:17:58.33 | +20:57:13.0 | … | B | 9.426 | 233.7 | 43.77 | 443.59 | 432.11 ± 2.19 | 0.510 ± 0.019 |
| … | … | GJ 16 | J00182+102 | M1.5 V | 00:18:16.59 | +10:12:09.6 | … | … | … | … | 31.20 | 29.63 | 411.22 ± 3.15 | 0.511 ± 0.029 |
| … | … | HD 1326 | J00183+440 | M1.0 V | 00:18:27.17 | +44:01:29.2 | A+B | A | … | … | 36.10 | 2920.70 | 239.65 ± 2.18 | 0.395 ± 0.012 |
| 00184+4401 | GRB 34 | HD 1326B | J00184+440 | M3.5 V | 00:18:30.07 | +44:01:43.5 | … | B | 34.349 | 65.5 | 36.06 | 2882.50 | 34.75 ± 0.14 | 0.164 ± 0.010 |
| … | … | GJ 3027 | J00188+278 | M4.0 V | 00:18:54.07 | +27:48:48.1 | … | … | … | … | 59.63 | 410.00 | 75.64 ± 0.35 | 0.275 ± 0.013 |
| … | … | GJ 1008 | J00190-099 | M0.0 V | 00:19:05.52 | -09:57:58.3 | … | … | … | … | 280.71 | 303.67 | 839.44 ± 3.66 | 0.672 ± 0.026 |
| … | … | GJ 2003 | J00201-170 | M1+ Vk: | 00:20:08.54 | -17:03:41.2 | … | … | … | … | 280.69 | 142.53 | 334.46 ± 1.68 | 0.446 ± 0.017 |
| … | … | GJ 3028 | J00204+330 | M5.5 V | 00:20:30.78 | +33:04:52.6 | … | … | … | … | 53.06 | 1379.44 | 14.53 ± 0.07 | 0.147 ± 0.010 |
| … | … | [I81] M 134 | J00207+596 | M2.5 V | 00:20:47.70 | +59:36:15.6 | … | … | … | … | 49.48 | 117.69 | 329.12 ± 1.80 | 0.497 ± 0.019 |
| … | … | StKM 1-25 | J00209+176 | M0.0 V | 00:20:57.24 | +17:38:14.7 | … | … | … | … | 43.79 | 91.16 | 687.19 ± 3.27 | 0.618 ± 0.027 |
| … | … | G 217-43 | J00210+557 | M2.0 V | 00:21:05.02 | +55:43:55.6 | … | … | … | … | 81.57 | 443.46 | 691.48 ± 3.33 | 0.615 ± 0.027 |
| … | … | G 171-51 | J00218+382 | M3.0 V | 00:21:54.79 | +38:16:24.2 | … | … | … | … | 35.12 | 700.25 | 232.41 ± 1.94 | 0.468 ± 0.019 |
| … | … | GJ 3030 | J00219+492 | M2.5 V | 00:21:58.20 | +49:12:37.3 | AB | A | … | … | 33.45 | 211.68 | … | 0.430 ± 0.031 |
| 00220+4913 | SKF1600 | LP 149-56 B | … | M5.0 V | 00:21:58.00 | +49:12:38.4 | … | B | 2.276 | 301.0 | 22.71 | 211.19 | … | 0.197 ± 0.039 |
| … | … | GJ 1011 | J00234+243 | M4.0 V | 00:23:27.73 | +24:18:26.5 | … | … | … | … | 40.34 | 254.16 | 47.37 ± 0.19 | 0.212 ± 0.011 |
| … | … | GJ 1010 A | J00234+771 | M2.5 V | 00:23:24.80 | +77:11:22.2 | A+B | A | … | … | 33.87 | 838.64 | 376.84 ± 1.92 | 0.503 ± 0.019 |
| 00243+7711 | LDS1506 | GJ 1010 B | J00235+771 | M4.5 V | 00:23:27.78 | +77:11:27.5 | … | B | 11.253 | 61.7 | 33.75 | 838.76 | 55.38 ± 0.28 | 0.232 ± 0.012 |
| … | … | LSPM J0024+2626 | J00240+264 | M4.0 V | 00:24:03.96 | +26:26:28.9 | … | … | … | … | 60.20 | 151.89 | 165.78 ± 3.31 | 0.418 ± 0.018 |
| … | … | G 130-67 | J00244+360 | M1.0 V | 00:24:26.30 | +36:03:54.0 | … | … | … | … | 52.27 | 292.46 | 593.32 ± 3.03 | 0.580 ± 0.028 |
| … | … | GJ 3033 | J00245+300 | M5.0 V | 00:24:35.60 | +30:02:29.7 | … | … | … | … | 52.29 | 585.98 | 62.23 ± 0.32 | 0.265 ± 0.013 |
| … | … | GJ 3034 | J00253+228 | M4.71 | 00:25:20.33 | +22:53:03.7 | … | … | … | … | 26.04 | 519.80 | 46.24 ± 0.20 | 0.209 ± 0.011 |
| … | … | GJ 21 | J00268+701 | M1.0 V | 00:26:52.27 | +70:08:30.4 | … | … | … | … | 27.74 | 200.45 | 479.69 ± 2.44 | 0.542 ± 0.029 |
| … | … | GJ 3035 | J00271+496 | M4.0 V | 00:27:07.39 | +49:41:49.3 | … | … | … | … | 51.32 | 430.86 | 88.43 ± 0.39 | 0.299 ± 0.014 |
| 00279+2220 | FRV 1 | LP 349-25 | J00279+223 | M8.0 Ve | 00:27:56.46 | +22:19:29.7 | (AB) | AB | 0.117 | 17.3 | 61.90 | 434.79 | … | 0.106 ± 0.047 |
| … | … | GJ 1012 | J00286-066 | dM4.0 | 00:28:39.11 | -06:40:02.0 | … | … | … | … | 61.87 | 865.14 | 152.11 ± 1.09 | 0.347 ± 0.012 |
| 00289+5023 | DAE 1 | GJ 3036 | J00288+503 | M3.7 V | 00:28:54.66 | +50:22:35.3 | (AB) | AB | 0.320 | 85.0 | 44.81 | 467.17 | … | 0.238 ± 0.037 |
| … | … | GJ 1013 | J00315-058 | M3.7 V | 00:31:35.79 | -05:52:30.0 | … | … | … | … | 70.78 | 1112.30 | 88.57 ± 0.45 | 0.280 ± 0.013 |
| … | … | G 217-56 | J00322+544 | M4.5 V | 00:32:15.34 | +54:28:55.3 | … | … | … | … | 74.71 | 487.74 | 96.07 ± 0.39 | 0.313 ± 0.014 |
| 00321+6715 | MCY 1 | V547Cas | J00324+672N | M2.0 V | 00:32:34.33 | +67:14:03.6 | (AB)+C | AB | 0.310 | 276.0 | 74.21 | 1782.54 | … | 0.421 ± 0.031 |
| 00321+6715 | VYS 2 | GJ 22 B | J00324+672S | M3.0 V | 00:32:34.28 | +67:13:59.8 | … | C | 3.790 | 184.0 | 71.02 | 1721.48 | … | 0.264 ± 0.036 |
| … | … | GJ 3039 | J00325+074 | M4.0 V | 00:32:34.91 | +07:29:25.7 | AB | A | … | … | 49.95 | 121.86 | … | 0.506 ± 0.029 |
| 00326+0729 | MCT 1 | GJ 3039 | … | M3.0 V | 00:32:34.89 | +07:29:26.4 | … | B | 0.720 | 333.6 | 101.09 | 123.57 | … | 0.378 ± 0.032 |
| 00329-0434 | JNN 12 | GR* 50 | J00328-045 | M4.5 V | 00:32:53.21 | -04:34:09.4 | (AB) | AB | 0.978 | 194.4 | 100.40 | 154.70 | … | 0.260 ± 0.036 |
| … | … | G 132-4 | J00333+368 | M3.0 V | 00:33:21.17 | +36:50:28.6 | … | … | … | … | 28.43 | 513.00 | 249.90 ± 1.09 | 0.486 ± 0.019 |
| … | … | V493AndA | J00341+253 | M0.0 V | 00:34:08.48 | +25:23:48.5 | AB+C | A | … | … | 28.23 | 127.91 | … | 0.680 ± 0.026 |
| 00341+2524 | SKF 220 | V493AndB | … | K7 V | 00:34:08.59 | +25:23:48.2 | … | B | 1.536 | 102.3 | 52.85 | 129.46 | … | 0.679 ± 0.026 |
| … | … | UCAC4 578-001365 | … | M4.0 V | 00:34:20.04 | +25:28:12.9 | … | C* | 307.226 | 30.6 | 43.03 | 127.04 | 71.47 ± 1.95 | 0.266 ± 0.013 |
| … | … | GJ 3040 | J00346+711 | M3.5 V | 00:34:39.40 | +71:11:36.6 | … | … | … | … | 20.10 | 626.48 | 86.52 ± 0.41 | 0.276 ± 0.013 |
| 00357+0233 | LAW 7 | LP 585-55 | J00357+025 | M5.0 V | 00:35:43.30 | +02:33:10.9 | (AB) | AB | 0.446 | 104.3 | 19.72 | 239.17 | … | 0.185 ± 0.040 |
| 00358+5241 | RAO 682 | G 172-11 | J00358+526 | M2.5 V | 00:35:54.70 | +52:41:09.3 | A+(BC) | A | … | … | 19.67 | 798.14 | … | 0.419 ± 0.031 |

Table D.2: Complete sample with the description of multiple systems (continued).

| WDS id | WDS disc | Name | Karmn | Spectral type | $\alpha$ (2016.0) | $\delta$ (2016.0) | System | Component | $\rho$ [arcsec] | $\theta$ [deg] | $\varpi$ [mas] | $\mu_{\rm total}$ [mas a$^{-1}$] | $\mathcal{L}$ [$10^{-4}$ $\mathcal{L}_\odot$] | $\mathcal{M}$ [$\mathcal{M}_\odot$] |
|---|---|---|---|---|---|---|---|---|---|---|---|---|---|---|
| 00358+5241 | GIC 11 | G 217-59 | … | M3.0 V | 00:35:55.04 | +52:41:33.7 | … | BC | 24.573 | 55.7 | 51.80 | 795.48 | … | 0.215 ± 0.038 |
| … | … | GJ 1014 | J00359+104 | M5.0 V | 00:35:56.70 | +10:28:29.4 | … | … | … | … | 36.15 | 1151.32 | 21.52 ± 0.11 | 0.157 ± 0.010 |
| … | … | GJ 3042 | J00361+455 | M2.0 V | 00:36:08.05 | +45:30:55.3 | … | … | … | … | 41.17 | 285.12 | 390.74 ± 175.97 | 0.439 ± 0.067 |
| … | … | G 172-14 | J00374+515 | M0.5 V | 00:37:25.07 | +51:33:06.8 | … | … | … | … | 41.18 | 506.35 | 732.28 ± 3.39 | 0.618 ± 0.027 |
| … | … | PM J00380+1656 | J00380+169 | M3.0 V | 00:38:03.75 | +16:56:01.3 | … | … | … | … | 69.47 | 142.34 | 117.75 ± 0.52 | 0.326 ± 0.014 |
| … | … | GJ 3044 | J00382+523 | M0.0 V | 00:38:15.16 | +52:19:53.3 | … | … | … | … | 43.90 | 156.40 | 752.94 ± 2.82 | 0.639 ± 0.027 |
| … | … | GJ 3045 | J00385+514 | M3.0 V | 00:38:33.48 | +51:27:58.4 | … | … | … | … | 30.73 | 231.31 | 134.33 ± 0.68 | 0.328 ± 0.014 |
| … | … | Wolf 1056 | J00389+306 | M2.5 V | 00:39:00.98 | +30:36:58.8 | … | … | … | … | 47.59 | 1558.15 | 251.40 ± 1.11 | 0.418 ± 0.011 |
| 00395+1454 | GIC 12 | LP 465-061 | J00395+149S | M4.0 V | 00:39:33.91 | +14:54:19.6 | A+(BC) | A | … | … | 43.44 | 333.74 | … | 0.310 ± 0.034 |
| 00395+1454 | JNN 249 | LP 465-62 | J00395+149N | M4.5 V | 00:39:34.16 | +14:54:35.4 | … | BC | 16.188 | 12.5 | 55.83 | 328.89 | … | 0.308 ± 0.034 |
| … | … | Wolf 10 | J00395+605 | M2.42 V | 00:39:33.51 | +60:33:10.9 | … | … | … | … | 78.90 | 332.44 | 421.38 ± 2.44 | 0.495 ± 0.029 |
| … | … | 2M00402129+6112490 | J00403+612 | M2.0 V | 00:40:21.40 | +61:12:48.2 | … | … | … | … | 34.82 | 73.93 | 374.04 ± 1.95 | 0.470 ± 0.012 |
| … | … | GJ 3047 | J00409+313 | M3.3 V | 00:40:56.19 | +31:22:51.2 | … | … | … | … | 34.26 | 334.19 | 112.15 ± 0.64 | 0.340 ± 0.015 |
| … | … | GJ 1015 A | J00413+558 | M4.2 V | 00:41:21.44 | +55:50:03.2 | A+B | A | … | … | 27.12 | 332.50 | … | 0.280 ± 0.035 |
| 00415+5550 | GIC 13 | GJ 1015 B | … | DBQ5 | 00:41:22.64 | +55:50:07.2 | … | B | 10.854 | 68.4 | 19.33 | 324.16 | … | 0.500 ± 0.100 |
| … | … | PM J00427+4349 | J00427+438 | M2.5 V | 00:42:47.79 | +43:49:24.0 | … | … | … | … | 44.39 | 55.19 | 259.09 ± 1.16 | 0.466 ± 0.018 |
| … | … | FFAnd | J00428+355 | M1.0 V | 00:42:48.59 | +35:32:56.9 | Aab | Aab(2) | … | … | 46.82 | 273.37 | … | … |
| … | … | GJ 1019 | J00435+284 | M4.0 V | 00:43:35.44 | +28:26:24.4 | … | … | … | … | 43.73 | 1071.52 | 43.22 ± 0.18 | 0.202 ± 0.011 |
| … | … | GJ 3052 | J00443+091 | M4.5 V | 00:44:21.54 | +09:07:34.5 | … | … | … | … | 43.66 | 813.09 | 132.31 ± 1.10 | 0.371 ± 0.016 |
| … | … | GJ 3051 | J00443+126 | M3.5 V | 00:44:19.64 | +12:36:59.8 | … | … | … | … | 47.61 | 337.85 | 158.32 ± 0.82 | 0.382 ± 0.016 |
| … | … | GJ 3053 | J00449-152 | M4.5 V | 00:44:59.68 | -15:16:27.1 | … | … | … | … | 46.13 | 676.15 | 44.18 ± 0.23 | 0.219 ± 0.012 |
| 00459+3347 | LAW 8 | G 132-25 | J00459+337 | M4.5 V | 00:45:57.01 | +33:47:11.3 | (AB) | AB | 0.262 | 127.6 | 47.93 | 263.15 | … | 0.196 ± 0.039 |
| … | … | PM J00463+3522 | J00463+353 | M1.5 Ve | 00:46:21.76 | +35:22:10.6 | … | … | … | … | 40.01 | 121.34 | 346.23 ± 1.67 | 0.463 ± 0.030 |
| … | … | G 172-22 | J00464+506 | M4.0 V | 00:46:30.66 | +50:38:35.3 | … | … | … | … | 50.65 | 470.57 | 149.72 ± 0.70 | 0.396 ± 0.017 |
| … | … | PM J00468+1603 | J00468+160 | M2.0 V | 00:46:53.16 | +16:03:01.8 | … | … | … | … | 66.83 | 128.89 | 277.92 ± 1.34 | 0.454 ± 0.018 |
| … | … | LSPM J0048+7518 | J00484+753 | M3.0 V | 00:48:30.66 | +75:18:47.2 | … | … | … | … | 49.93 | 226.34 | 263.78 ± 1.21 | 0.470 ± 0.018 |
| … | … | GJ 3057 | J00487+270 | M2.5 V | 00:48:45.34 | +27:01:04.4 | … | … | … | … | 34.50 | 328.24 | 228.43 ± 1.31 | 0.410 ± 0.016 |
| … | … | GJ 3058 | J00489+445 | M3.0 V | 00:48:58.46 | +44:35:06.9 | AB | A | … | … | 30.95 | 182.00 | … | 0.391 ± 0.032 |
| 00490+4435 | MCT 2 | GJ 3058 | … | M3.0 V | 00:48:58.37 | +44:35:06.6 | … | B | 1.006 | 256.7 | 49.71 | 181.78 | … | 0.367 ± 0.033 |
| … | … | PM J00490+6544 | J00490+657 | M2.5 V | 00:49:05.09 | +65:44:36.8 | … | … | … | … | 30.72 | 132.46 | 312.69 ± 2.16 | 0.483 ± 0.018 |
| … | … | RX J0050.2+0837 | J00502+086 | M4.5 V | 00:50:17.59 | +08:37:33.6 | Aab | Aab(2) | … | … | 49.80 | 75.62 | … | … |
| 00505+2450 | LDS3203 | FTPscA | J00505+248 | M3.0 V | 00:50:33.49 | +24:48:59.7 | AB | A | 1.020 | 322.0 | 45.13 | 205.83 | … | 0.335 ± 0.034 |
| … | … | FTPscB | … | … | 00:50:33.45 | +24:49:00.4 | … | B | 0.887 | 320.8 | 29.56 | 202.83 | … | … |
| … | … | BPM 84579 | J00511+225 | M1.5 V | 00:51:10.71 | +22:34:43.8 | … | … | … | … | 29.76 | 120.57 | 548.80 ± 5.39 | 0.550 ± 0.028 |
| … | … | Wolf 33 | J00514+583 | M0.0 V | 00:51:33.02 | +58:18:13.8 | … | … | … | … | 30.44 | 1618.53 | 455.66 ± 1.43 | 0.496 ± 0.018 |
| … | … | HD 4967 | … | K5 V | 00:51:34.73 | -22:54:40.7 | A+B | A | … | … | 16.31 | 672.69 | … | 0.722 ± 0.025 |
| 00516-2255 | LDS1082 | HD 4967B | J00515-229 | M5.5 V | 00:51:35.91 | -22:54:35.4 | … | B | 17.073 | 72.1 | 66.73 | 680.57 | 13.34 ± 0.08 | 0.129 ± 0.010 |
| … | … | G 69-27 | J00520+205 | M1.0 V | 00:52:00.27 | +20:34:56.6 | … | … | … | … | … | 211.39 | 659.59 ± 2.59 | 0.601 ± 0.027 |
| … | … | LSPM J0053+1903 | J00532+190 | M2.5 V | 00:53:12.84 | +19:03:25.0 | … | … | … | … | 37.57 | 161.71 | 234.53 ± 1.34 | 0.442 ± 0.017 |
| … | … | G 172-28 | J00538+459 | M0.0 V | 00:53:53.72 | +45:56:41.6 | … | … | … | … | 52.67 | 313.34 | 699.65 ± 3.60 | 0.622 ± 0.027 |

Table D.2: Complete sample with the description of multiple systems (continued).

| WDS id | WDS disc | Name | Karmn | Spectral type | $\alpha$ (2016.0) | $\delta$ (2016.0) | System | Component | $\rho$ [arcsec] | $\theta$ [deg] | $\varpi$ [mas] | $\mu_{\rm total}$ [mas a$^{-1}$] | $\mathcal{L}$ [$10^{-4}$ $\mathcal{L}_\odot$] | $\mathcal{M}$ [$\mathcal{M}_\odot$] |
|---|---|---|---|---|---|---|---|---|---|---|---|---|---|---|
| … | … | Ross 317 | J00540+691 | M2.0 V | 00:54:00.72 | +69:10:56.9 | … | … | … | … | 65.09 | 273.99 | 214.65 ± 4.47 | 0.396 ± 0.016 |
| … | … | G 69-32 | J00548+275 | M4.5 V | 00:54:48.49 | +27:31:03.9 | … | … | … | … | 65.07 | 341.78 | 74.02 ± 0.34 | 0.291 ± 0.014 |
| … | … | GJ 1024 | J00566+174 | M3.8 V | 00:56:39.14 | +17:27:30.3 | … | … | … | … | 31.82 | 743.02 | 91.08 ± 0.54 | 0.284 ± 0.013 |
| … | … | G 172-30 | J00570+450 | M3.0 V | 00:57:03.64 | +45:05:08.7 | … | … | … | … | 44.66 | 629.52 | 149.81 ± 0.84 | 0.330 ± 0.011 |
| … | … | BD+05 127 | J00577+058 | M0.0 V | 00:57:44.49 | +05:51:20.6 | … | … | … | … | 34.56 | 55.54 | 576.84 ± 3.35 | 0.592 ± 0.028 |
| … | … | 1R005802.4+391912 | J00580+393 | M4.5 V | 00:58:01.01 | +39:19:11.5 | … | … | … | … | 34.27 | 109.93 | 43.49 ± 0.24 | 0.217 ± 0.012 |
| … | … | GJ 3068 | J01008+669 | M3.5 V | 01:00:48.83 | +66:56:53.8 | … | … | … | … | 36.46 | 242.48 | 135.21 ± 0.71 | 0.351 ± 0.015 |
| … | … | GJ 1025 | J01009-044 | M4.0 V | 01:00:57.71 | -04:26:49.5 | … | … | … | … | 54.85 | 1323.95 | 50.94 ± 0.35 | 0.221 ± 0.012 |
| … | … | Wolf 44 | J01013+613 | M2.0 V | 01:01:20.86 | +61:21:43.7 | … | … | … | … | 76.11 | 888.89 | 202.20 ± 0.84 | 0.370 ± 0.011 |
| … | … | GJ 3069 | J01019+541 | M5.0 V | 01:01:58.94 | +54:10:55.8 | … | … | … | … | 56.15 | 325.64 | 19.52 ± 0.08 | 0.141 ± 0.008 |
| … | … | GJ 3072 | J01023-104 | M0.0 V | 01:02:21.16 | -10:25:28.7 | … | … | … | … | 68.39 | 178.83 | … | 0.677 ± 0.026 |
| … | … | Ross 318 | J01025+716 | M3.0 V | 01:02:38.16 | +71:40:41.2 | … | … | … | … | 44.32 | 1787.93 | 307.28 ± 1.99 | 0.472 ± 0.013 |
| … | … | Wolf 46 | J01026+623 | M1.5 V | 01:02:40.55 | +62:20:43.6 | A+B | A | … | … | 80.79 | 736.67 | 487.34 ± 3.21 | 0.536 ± 0.029 |
| 01026+6221 | WNO 51 | V388Cas | J01033+623 | M5.0 V | 01:03:21.51 | +62:21:57.2 | … | B | 294.459 | 75.5 | 94.97 | 735.44 | 46.11 ± 0.24 | 0.231 ± 0.012 |
| … | … | GJ 1026 A | J01032+200 | M2.0 V | 01:03:14.92 | +20:05:53.1 | AB | A | … | … | 90.01 | 674.09 | … | 0.413 ± 0.031 |
| 01032+2006 | LDS 873 | GJ 1026 B | … | M3.5 V | 01:03:15.07 | +20:05:54.5 | … | B | 2.482 | 56.6 | 47.38 | 676.83 | … | 0.304 ± 0.034 |
| 01032+3141 | RAO 183 | GJ 3073 | J01032+316 | M3.2 V | 01:03:14.23 | +31:40:59.7 | (AB) | AB | 0.202 | 147.7 | 121.46 | 209.40 | … | 0.405 ± 0.069 |
| 01032+7113 | JNN 250 | LP 29-70 | J01032+712 | M4.0 V | 01:03:16.13 | +71:13:11.8 | (AB) | AB | 0.147 | 34.2 | 101.42 | 510.53 | … | 0.333 ± 0.034 |
| 01037+4051 | BWL 5 | G 132-50 | J01036+408 | M0.0 Ve | 01:03:40.31 | +40:51:26.6 | (AB)+C+D | AB | 0.264 | 308.9 | 101.37 | 195.87 | … | 0.685 ± 0.026 |
| 01037+4051 | LDS3225 | G 132-51A | J01037+408 | M2.6 V | 01:03:42.24 | +40:51:13.5 | … | C | 25.562 | 120.9 | 62.77 | 204.43 | 215.28 ± 1.82 | 0.450 ± 0.018 |
| 01037+4051 | LDS3225 | LP 194-20 | … | M3.8 V | 01:03:42.46 | +40:51:13.1 | … | D | 27.859 | 118.9 | 62.77 | 208.49 | … | 0.294 ± 0.035 |
| … | … | StKM 1-112 | J01041+108 | M1.0 V | 01:04:11.07 | +10:51:35.4 | … | … | … | … | 24.90 | 71.97 | 538.49 ± 3.95 | 0.565 ± 0.028 |
| … | … | GJ 1028 | J01048-181 | dM5.0 | 01:04:55.25 | -18:07:20.8 | … | … | … | … | 36.32 | 1381.30 | 22.21 ± 0.16 | 0.137 ± 0.009 |
| … | … | GJ 1029 | J01056+284 | M5.5 V | 01:05:39.97 | +28:29:30.6 | Aab | Aab(2) | … | … | … | 1926.77 | … | 0.034 ± 0.002 |
| … | … | GJ 1030 | J01066+152 | M2.0 V | 01:06:41.39 | +15:16:18.1 | … | … | … | … | 32.81 | 273.61 | 475.81 ± 3.06 | 0.525 ± 0.029 |
| … | … | LSPM J0106+1913 | J01066+192 | M3.0 V | 01:06:36.93 | +19:13:29.6 | … | … | … | … | 32.53 | 226.73 | 160.77 ± 0.79 | 0.341 ± 0.011 |
| … | … | LP 12-502 | J01069+804 | M4.5 V | 01:06:56.00 | +80:27:34.0 | … | … | … | … | 32.31 | 208.20 | 110.16 ± 0.61 | 0.360 ± 0.016 |
| … | … | G 2-21 | J01078+128 | M1.5 V | 01:07:52.53 | +12:52:51.4 | … | … | … | … | 28.68 | 276.82 | 394.17 ± 1.71 | 0.502 ± 0.029 |
| … | … | LP 707-16 | J01102-118 | M3.0 V | 01:10:17.75 | -11:51:19.3 | … | … | … | … | 102.28 | 248.87 | 180.86 ± 1.42 | 0.410 ± 0.017 |
| 01114+1526 | BEU 2 | GJ 3076 | J01114+154 | M5.93 | 01:11:25.63 | +15:26:19.9 | (AB) | AB | 0.441 | 284.6 | 79.93 | 231.87 | … | 0.226 ± 0.043 |
| … | … | LP 467-15 | J01116+120 | M2.0 V | 01:11:36.66 | +12:05:02.3 | … | … | … | … | 44.98 | 292.92 | 180.76 ± 0.78 | 0.362 ± 0.015 |
| 01119+0455 | RAO 683 | GJ 3077 | J01119+049N | M3.5 V | 01:11:56.01 | +04:54:56.6 | (AB)+C | AB | 0.440 | 357.7 | 40.45 | 665.19 | … | 0.293 ± 0.035 |
| 01119+0455 | GIC 20 | GJ 3078 | J01119+049S | M4.0 V | 01:11:58.39 | +04:54:04.0 | … | C | 63.506 | 146.0 | 46.93 | 648.10 | 46.79 ± 0.22 | 0.211 ± 0.011 |
| … | … | YZCet | J01125-169 | M4.0 Ve | 01:12:31.98 | -16:59:46.2 | … | … | … | … | 35.23 | 1363.33 | 22.88 ± 0.12 | 0.137 ± 0.009 |
| … | … | Wolf 58 | J01133+589 | M1.5 Ve | 01:13:20.12 | +58:55:20.3 | AB | A | … | … | 54.35 | 210.93 | … | 0.539 ± 0.029 |
| … | … | G3- 414108140954108672 | … | M5.0 V | 01:13:19.97 | +58:55:18.6 | … | B* | 2.082 | 213.9 | 58.00 | 200.51 | … | 0.179 ± 0.040 |
| … | … | GJ 1033 | J01134-229 | M4.0 V | 01:13:24.20 | -22:54:07.3 | … | … | … | … | 48.79 | 153.27 | 76.88 ± 0.41 | 0.277 ± 0.013 |
| … | … | PM J01141+7904 | J01141+790 | M3.0 V | 01:14:06.69 | +79:04:01.9 | … | … | … | … | 64.67 | 89.71 | 333.30 ± 4.19 | 0.440 ± 0.031 |
| … | … | LP 351-6 | J01147+253 | M1.5 V | 01:14:49.86 | +25:18:57.6 | … | … | … | … | 64.57 | 422.33 | 502.15 ± 2.64 | 0.535 ± 0.029 |
| 01158+4702 | LAW 9 | 1R011549.5+470159 | J01158+470 | M4.5 V | 01:15:50.51 | +47:02:02.1 | (AB)+(CD) | AB | 0.267 | 267.4 | 269.06 | 186.53 | … | … |

Table D.2: Complete sample with the description of multiple systems (continued).

| WDS id | WDS disc | Name | Karmn | Spectral type | $\alpha$ (2016.0) | $\delta$ (2016.0) | System | Component | $\rho$ [arcsec] | $\theta$ [deg] | $\varpi$ [mas] | $\mu_{total}$ [mas a$^{-1}$] | $\mathcal{L}$ [$10^{-4}$ $\mathcal{L}_\odot$] | $\mathcal{M}$ [$\mathcal{M}_\odot$] |
|---|---|---|---|---|---|---|---|---|---|---|---|---|---|---|
| 01158+4702 | FMR 42 | LP 151-21 | ... | M5.0 V | 01:15:49.20 | +47:02:25.7 | ... | CD | 27.195 | 330.4 | 36.03 | 198.56 | ... | 0.190 ± 0.039 |
| ... | ... | Wolf 59 | J01161+601 | M0.5 V | 01:16:10.94 | +60:09:09.6 | ... | ... | ... | ... | 35.97 | 428.48 | 515.71 ± 2.23 | 0.563 ± 0.028 |
| ... | ... | GJ 3084 | J01178+054 | M0.5 V | 01:17:53.34 | +05:28:16.0 | ... | ... | ... | ... | 44.17 | 640.24 | 384.29 ± 2.26 | 0.479 ± 0.018 |
| ... | ... | Ross 324 | J01178+286 | M0.5 V | 01:17:50.21 | +28:40:09.6 | ... | ... | ... | ... | 23.93 | 431.54 | 207.05 ± 0.91 | 0.366 ± 0.015 |
| ... | ... | GJ 56.1 | J01182-128 | M2.0 V | 01:18:16.20 | -12:54:10.2 | ... | ... | ... | ... | 30.47 | 697.94 | 397.62 ± 2.32 | 0.500 ± 0.029 |
| ... | ... | GJ 1035 | J01198+841 | M5.0 V | 01:19:41.88 | +84:09:40.4 | ... | ... | ... | ... | ... | 1090.97 | 32.12 ± 0.16 | 0.183 ± 0.011 |
| ... | ... | Ross 788 | J01214+243 | M0.0 V | 01:21:29.78 | +24:19:50.2 | ... | ... | ... | ... | 21.43 | 340.62 | 859.10 ± 3.73 | 0.661 ± 0.026 |
| 01222+2209 | CRC 44 | LP 351-34 | J01221+221 | M4.0 V | 01:22:10.58 | +22:09:00.6 | (AB) | AB | 0.248 | 359.9 | 39.08 | 287.16 | ... | 0.207 ± 0.041 |
| ... | ... | GJ 3093 | J01227+005 | M5.0 V | 01:22:44.77 | +00:31:55.7 | AB | A | ... | ... | 55.25 | 559.58 | ... | 0.220 ± 0.038 |
| 01227+0032 | LDS3270 | GJ 3094 | ... | M5.0 V | 01:22:44.83 | +00:31:56.5 | ... | B | 1.173 | 43.2 | 59.85 | 580.70 | ... | 0.167 ± 0.041 |
| ... | ... | Wolf 66 A | J01256+097 | M4.0 V | 01:25:36.89 | +09:45:18.5 | AB | A | ... | ... | 42.10 | 455.85 | ... | 0.270 ± 0.036 |
| 01256+0945 | JOD 1 | Wolf 66 B | ... | ... | 01:25:36.86 | +09:45:18.0 | ... | B | 0.598 | 213.6 | 69.21 | ... | ... | ... |
| ... | ... | Wolf 1523 | J01317+209 | M2.0 V | 01:32:44.80 | +20:59:13.5 | ... | ... | ... | ... | 36.89 | 441.64 | 303.38 ± 1.72 | 0.448 ± 0.017 |
| ... | ... | GJ 3098 | J01324-219 | M1.5 Vk: | 01:32:25.53 | -21:54:32.7 | ... | ... | ... | ... | 87.60 | 1065.06 | 369.99 ± 1.62 | 0.498 ± 0.018 |
| ... | ... | LP 768-113 | J01339-176 | dM4.0 | 01:33:58.05 | -17:38:26.8 | ... | ... | ... | ... | 65.52 | 190.97 | 89.10 ± 0.46 | 0.267 ± 0.011 |
| ... | ... | 1R013514.2-071254 | J01352-072 | M4.0 V | 01:35:14.03 | -07:12:52.2 | ... | ... | ... | ... | 55.78 | 108.44 | ... | ... |
| ... | ... | EXCet | ... | K0/1 V | 01:37:35.65 | -06:45:39.1 | A+B | A | ... | ... | 69.20 | 197.86 | 4565.07 ± 17.60 | 0.880 ± 0.132 |
| 01376-0645 | CAB 3 | LP 648-20 | J01369-067 | M3.5 V | 01:36:55.36 | -06:47:39.6 | ... | B | 612.075 | 258.6 | 54.56 | 200.55 | 101.79 ± 0.61 | 0.322 ± 0.014 |
| ... | ... | TYC 4031-2527-1 | J01373+610 | M1.5 V | 01:37:21.44 | +61:05:27.6 | ... | ... | ... | ... | ... | 66.75 | 1011.45 ± 12.79 | 0.660 ± 0.026 |
| ... | ... | Ross 10 | J01383+572 | M2.5 V | 01:38:21.23 | +57:13:51.4 | ... | ... | ... | ... | 38.68 | 398.10 | 130.70 ± 0.59 | 0.324 ± 0.014 |
| ... | ... | GJ 3103 | J01384+006 | M2.5 V | 01:38:30.49 | +00:39:08.4 | ... | ... | ... | ... | 53.67 | 543.94 | 359.01 ± 1.63 | 0.486 ± 0.030 |
| ... | ... | BLCet | J01390-179 | M5.0 V | 01:39:05.17 | -17:56:53.9 | AB | A | ... | ... | 68.02 | 3428.81 | ... | 0.129 ± 0.044 |
| 01388-1758 | LDS 838 | UCet | ... | M6.0 V | 01:39:05.20 | -17:56:51.7 | ... | B | 2.259 | 10.5 | 26.82 | 3231.91 | ... | 0.118 ± 0.045 |
| 01395+0503 | JOD 2 | GJ 3104 | J01395+050 | M3.0 V | 01:39:31.32 | +05:03:20.3 | (AB) | AB | 0.271 | 209.0 | 41.56 | 187.30 | ... | 0.491 ± 0.030 |
| ... | ... | GJ 3105 | J01402+317 | M4.21 | 01:40:17.16 | +31:47:30.4 | ... | ... | ... | ... | 41.70 | 465.64 | 135.34 ± 2.27 | 0.375 ± 0.016 |
| 01431+2101 | JNN 251 | RX J0143.1+2101 | J01431+210 | M4.0 V | 01:43:11.75 | +21:01:10.4 | (AB) | AB | 0.355 | 325.8 | 23.02 | 96.35 | ... | 0.409 ± 0.031 |
| ... | ... | GJ 3108 | J01432+278 | M1.0 V | 01:43:16.64 | +27:50:31.1 | ... | ... | ... | ... | 75.89 | 551.81 | 842.50 ± 4.11 | 0.648 ± 0.027 |
| ... | ... | GJ 70 | J01433+043 | M2.0 V | 01:43:19.73 | +04:19:05.7 | ... | ... | ... | ... | 48.22 | 873.90 | 225.82 ± 1.22 | 0.396 ± 0.011 |
| ... | ... | PM J01437-0602 | J01437-060 | M3.5 V | 01:43:45.20 | -06:02:40.6 | (Aab) | Aab(2) | ... | ... | 32.54 | 58.38 | ... | ... |
| ... | ... | Wolf 1530 | J01449+163 | M4.0 V | 01:44:57.63 | +16:20:32.5 | ... | ... | ... | ... | 367.71 | 849.89 | 52.94 ± 0.24 | 0.226 ± 0.012 |
| 01466-0839 | JOD 3 | GJ 3113 | J01466-086 | M4.0 V | 01:46:37.29 | -08:39:00.5 | (AB) | AB | 0.182 | 5.3 | 373.84 | 451.35 | ... | 0.387 ± 0.033 |
| ... | ... | G 173-18 | J01453+465 | M2.0 V | 01:45:18.81 | +46:32:11.3 | Aab | Aab(2) | ... | ... | 47.42 | 444.34 | ... | ... |
| ... | ... | Wolf 87 | J01480+212 | M2.5 V | 01:48:04.37 | +21:12:20.8 | ... | ... | ... | ... | 23.40 | 409.52 | 206.73 ± 1.43 | 0.413 ± 0.017 |
| ... | ... | GJ 3119 | J01510-061 | M4.5 Ve | 01:51:04.69 | -06:07:09.3 | ... | ... | ... | ... | 41.55 | 607.06 | 25.29 ± 0.12 | 0.172 ± 0.011 |
| ... | ... | Wolf 90 | J01514+213 | M4.0 V | 01:51:24.17 | +21:23:33.9 | ... | ... | ... | ... | 26.04 | 347.55 | 61.73 ± 0.33 | 0.246 ± 0.012 |
| ... | ... | GJ 3117 | J01518+644 | M2.5 V | 01:51:51.72 | +64:26:02.8 | AB | A | ... | ... | 45.42 | 306.37 | 336.69 ± 162.31 | 0.467 ± 0.119 |
| 01519+6426 | GIC 27 | GJ 3118 | ... | DA5.6 | 01:51:51.69 | +64:25:49.3 | ... | B | 13.503 | 180.7 | 88.31 | 301.94 | ... | 0.500 ± 0.100 |
| ... | ... | Ross 555 | J01518-108 | M2.0 V | 01:51:49.30 | -10:48:21.1 | ... | ... | ... | ... | 46.70 | 783.75 | 200.70 ± 0.86 | 0.383 ± 0.015 |
| ... | ... | BD-21 332 | J01531-210 | M2.0 Ve | 01:53:11.67 | -21:05:42.2 | Aab | Aab(2) | ... | ... | 47.49 | 281.99 | ... | ... |
| ... | ... | PM J01538-1459A | J01538-149 | M3.0 V | 01:53:50.97 | -14:59:51.5 | AB | A | ... | ... | 61.67 | 114.18 | ... | 0.516 ± 0.029 |

Table D.2: Complete sample with the description of multiple systems (continued).

| WDS id | WDS disc | Name | Karmn | Spectral type | $\alpha$ (2016.0) | $\delta$ (2016.0) | System | Component | $\rho$ [arcsec] | $\theta$ [deg] | $\varpi$ [mas] | $\mu_{total}$ [mas a$^{-1}$] | $\mathcal{L}$ [$10^{-4}$ $\mathcal{L}_\odot$] | $\mathcal{M}$ [$\mathcal{M}_\odot$] |
|---|---|---|---|---|---|---|---|---|---|---|---|---|---|---|
| 01538-1500 | BRG 7 | PM J01538-1459B | … | M1.5 V | 01:53:50.79 | -14:59:50.5 | … | B | 2.854 | 291.4 | 39.10 | 115.33 | … | 0.512 ± 0.029 |
| … | … | 1R015426.6+574136 | J01544+576 | M3.5 Ve | 01:54:28.04 | +57:41:27.7 | A+B | A | … | … | 39.42 | 199.20 | 255.20 ± 1.76 | 0.492 ± 0.019 |
| 01545+5741 | NSN 1 | LSPM J0154+5741N | … | M4.5 V | 01:54:28.05 | +57:41:36.4 | … | B | 8.669 | 0.4 | 53.73 | 207.78 | 44.47 ± 0.31 | 0.205 ± 0.011 |
| … | … | LSPM J0155+3758 | J01550+379 | M5.0 V | 01:55:02.64 | +37:57:54.6 | … | … | … | … | 94.71 | 532.97 | 35.81 ± 0.17 | 0.195 ± 0.011 |
| … | … | G 73-5 | J01556+028 | M1.5 V | 01:55:36.98 | +02:52:53.8 | … | … | … | … | 59.11 | 413.23 | 637.34 ± 10.28 | 0.587 ± 0.028 |
| 01568+3033 | KO 4 | LP 296-57 | J01567+305 | M4.5 V | 01:56:45.99 | +30:33:28.6 | A+B | A | 299.135 | 190.5 | 57.82 | 211.06 | 97.48 ± 0.90 | 0.337 ± 0.015 |
| … | … | LP 296-56 | … | M5.0 V | 01:56:41.74 | +30:28:34.6 | … | B | 299.134 | 190.6 | 57.80 | 208.62 | 32.82 ± 0.25 | 0.186 ± 0.011 |
| … | … | GJ 1041 A | J01592+035E | M1.0 V | 01:59:12.66 | +03:31:09.6 | A+Bab | A | … | … | 58.24 | 264.35 | … | 0.555 ± 0.028 |
| 01592+0330 | LDS3331 | GJ 1041 B | J01592+035W | M3.0 V | 01:59:12.89 | +03:31:12.1 | … | Bab(2) | 4.176 | 54.2 | 60.77 | 269.39 | … | … |
| … | … | V596Cas | J01593+585 | M4.0 V | 01:59:24.17 | +58:31:13.0 | … | … | … | … | 31.21 | 374.04 | 171.26 ± 1.35 | 0.425 ± 0.018 |
| … | … | GJ 3123 | J02000+437 | M3.0 V | 02:00:03.00 | +43:45:23.9 | … | … | … | … | 29.64 | 299.03 | 371.03 ± 2.34 | 0.474 ± 0.030 |
| … | … | GJ 3124 | J02001+366 | M3.8 V | 02:00:07.50 | +36:39:43.8 | … | … | … | … | 29.56 | 266.52 | 112.27 ± 0.62 | 0.318 ± 0.014 |
| … | … | TZAri | J02002+130 | M3.5: V | 02:00:14.16 | +13:02:38.7 | … | … | … | … | 44.86 | 2083.39 | 26.40 ± 0.13 | 0.148 ± 0.009 |
| … | … | GJ 3127 | J02007-103 | M3.5 V | 02:00:46.86 | -10:21:26.6 | … | … | … | … | 44.87 | 521.54 | 131.85 ± 0.76 | 0.370 ± 0.016 |
| … | … | GJ 3126 | J02015+637 | M2.5 V | 02:01:34.72 | +63:46:10.5 | … | … | … | … | 49.33 | 275.12 | 260.90 ± 1.79 | 0.419 ± 0.012 |
| … | … | GJ 3125 | J02019+735 | M4.5 Ve | 02:01:55.13 | +73:32:30.2 | AB | A | … | … | 26.74 | … | … | … |
| 02019+7332 | JNN 252 | G3- 558611490992292480 | … | M5.5 V | 02:01:55.18 | +73:32:30.4 | … | B | 0.307 | 45.3 | 31.85 | 281.78 | … | 0.152 ± 0.042 |
| … | … | Wolf 109 | … | K5 V | 02:02:02.81 | +03:56:20.0 | A+BC | A | … | … | 26.64 | 466.14 | 1184.99 ± 6.10 | 0.700 ± 0.105 |
| 02021+0355 | OSV 1 | Wolf 109 B | J02020+039 | M2.0 V | 02:02:03.11 | +03:56:37.4 | … | B | 17.961 | 14.4 | 41.87 | 465.77 | 299.82 ± 8.81 | 0.406 ± 0.032 |
| … | … | G3- 2517912315149114240 | … | … | 02:02:03.16 | +03:56:36.9 | … | C | 17.739 | 16.8 | 41.90 | … | … | … |
| … | … | GJ 3128 | J02022+103 | M5.5 V | 02:02:15.51 | +10:20:09.4 | … | … | … | … | 76.29 | 729.74 | 11.76 ± 0.06 | 0.119 ± 0.009 |
| … | … | PM J02024+1034B | J02026+105 | M4.5 V | 02:02:28.15 | +10:34:51.9 | A+B | A | … | … | 28.13 | 109.34 | … | 0.266 ± 0.036 |
| 02025+1035 | WSS 1 | PM J02024+1034A | … | M5.5 V | 02:02:28.18 | +10:34:52.7 | … | B | 0.911 | 23.3 | 38.11 | 117.33 | … | 0.226 ± 0.038 |
| … | … | GJ 3129 | J02027+135 | M5.0 V | 02:02:44.86 | +13:34:31.9 | Aab | Aab(2) | … | … | 24.61 | 473.30 | … | … |
| … | … | RX J0202.8+0446 | J02028+047 | M3.5 Ve | 02:02:52.00 | +04:47:00.4 | … | … | … | … | 223.73 | 150.90 | 158.65 ± 1.39 | 0.383 ± 0.016 |
| … | … | GJ 3131 | J02033-212 | M2.5 V | 02:03:20.52 | -21:13:50.2 | AabB | Aab(2) | … | … | 58.02 | 471.89 | … | … |
| … | … | GJ 3131 B | … | … | 02:03:20.25 | -21:13:51.5 | … | B* | 4.008 | 251.9 | 34.35 | 450.04 | … | 0.1475 ± 0.058 |
| … | … | GJ 3132 | J02044-018 | M4.5 V | 02:04:26.87 | -01:53:06.0 | … | … | … | … | 84.06 | 822.41 | 68.42 ± 0.33 | 0.260 ± 0.013 |
| 02051-1737 | BEU 3 | GJ 84 | J02050-176 | M2.5 V | 02:05:06.31 | -17:36:55.5 | AabB | AabB(1) | 0.244 | 100.7 | 66.76 | 1286.46 | … | … |
| … | … | Wolf 116 | J02055+056 | M1.0 V | 02:05:30.35 | +05:41:43.0 | … | … | … | … | … | 301.59 | 661.35 ± 4.74 | 0.597 ± 0.027 |
| … | … | V374And | J02069+451 | M0.0 V | 02:06:57.61 | +45:10:56.9 | Aab | Aab(2) | … | … | 48.92 | 537.61 | … | 0.680 |
| … | … | G 173-37 | J02070+496 | M3.5 Ve | 02:07:04.24 | +49:38:36.6 | … | … | … | … | 84.39 | 487.87 | 135.17 ± 0.59 | 0.318 ± 0.010 |
| … | … | GJ 3134 | J02071+642 | M4.5 V | 02:07:10.89 | +64:17:08.7 | … | … | … | … | 27.83 | 278.20 | 57.47 ± 0.28 | 0.236 ± 0.012 |
| … | … | G 242-81 | J02082+802 | M0.0 V | 02:08:18.76 | +80:13:11.1 | … | … | … | … | 28.92 | 556.27 | 711.03 ± 4.97 | 0.630 ± 0.027 |
| … | … | GJ 3136 | J02088+494 | M4.0 Ve | 02:08:54.01 | +49:26:51.8 | … | … | … | … | … | 375.08 | 176.80 ± 0.74 | 0.415 ± 0.032 |
| … | … | GJ 3139 | J02096-143 | M2.5 V | 02:09:36.70 | -14:21:38.2 | … | … | … | … | 108.26 | 623.31 | 330.39 ± 2.15 | 0.461 ± 0.030 |
| … | … | G 35-32 | J02116+185 | M3.0 V | 02:11:41.19 | +18:33:42.3 | … | … | … | … | 69.64 | 334.94 | 204.62 ± 1.09 | 0.411 ± 0.017 |
| … | … | Wolf 124 | J02123+035 | M1.5 V | 02:12:19.10 | +03:34:02.6 | … | … | … | … | 68.79 | 2557.19 | 322.05 ± 1.30 | 0.470 ± 0.012 |
| 02133+3649 | JNN 18 | 1R021320.6+364837 | J02133+368 | M4.5 V | 02:13:20.68 | +36:48:51.6 | (AB) | AB | 0.213 | 82.4 | 26.04 | 92.11 | … | 0.213 ± 0.038 |
| … | … | GJ 3142 | J02129+000 | M4.0 V | 02:12:55.22 | +00:00:17.3 | … | … | … | … | 48.08 | 556.17 | 76.74 ± 0.40 | 0.277 ± 0.013 |

Table D.2: Complete sample with the description of multiple systems (continued).

| WDS id | WDS disc | Name | Karmn | Spectral type | $\alpha$ (2016.0) | $\delta$ (2016.0) | System | Component | $\rho$ [arcsec] | $\theta$ [deg] | $\varpi$ [mas] | $\mu_{total}$ [mas a$^{-1}$] | $\mathcal{L}$ [$10^{-4}$ $\mathcal{L}_\odot$] | $\mathcal{M}$ [$\mathcal{M}_\odot$] |
|---|---|---|---|---|---|---|---|---|---|---|---|---|---|---|
| ... | ... | LP 649-72 | J02142-039 | M5.5 V | 02:14:13.11 | -03:57:46.1 | ... | ... | ... | ... | 46.67 | 532.92 | 12.80 ± 0.07 | 0.136 ± 0.010 |
| ... | ... | GJ 1045 | J02149+174 | M4.0 V | 02:15:00.20 | +17:25:00.8 | ... | ... | ... | ... | 46.75 | 584.12 | 68.69 ± 0.31 | 0.261 ± 0.013 |
| ... | ... | Wolf 127 | J02153+074 | M1.5 V | 02:15:22.39 | +07:29:32.5 | ... | ... | ... | ... | 54.45 | 577.56 | 360.68 ± 2.55 | 0.491 ± 0.018 |
| ... | ... | GJ 3143 | J02155+339 | M4.2 V | 02:15:34.63 | +33:57:35.0 | ... | ... | ... | ... | 107.30 | 414.45 | 249.16 ± 1.31 | 0.396 ± 0.032 |
| ... | ... | GJ 3145 | J02158-126 | M3.5 V | 02:15:49.43 | -12:40:24.2 | ... | ... | ... | ... | 25.80 | 543.01 | 257.57 ± 2.08 | 0.494 ± 0.020 |
| ... | ... | GJ 3146 | J02164+135 | M5.0 Ve | 02:16:30.40 | +13:35:05.9 | ... | ... | ... | ... | 51.47 | 664.72 | 11.55 ± 0.06 | 0.128 ± 0.010 |
| ... | ... | GJ 3147 | J02171+354 | M7.0 V | 02:17:10.74 | +35:26:28.4 | ... | ... | ... | ... | 69.83 | 607.26 | 13.54 ± 0.06 | 0.141 ± 0.010 |
| ... | ... | G 35-39 | J02185+207 | M2.5 V | 02:18:35.93 | +20:47:44.8 | ... | ... | ... | ... | 52.27 | 292.61 | 179.77 ± 1.28 | 0.361 ± 0.015 |
| ... | ... | RX J0218.6+1219 | J02186+123 | M2.5 V | 02:18:36.75 | +12:18:56.0 | ... | ... | ... | ... | 32.62 | 185.61 | 411.18 ± 2.31 | 0.498 ± 0.029 |
| ... | ... | GJ 3150 | J02190+238 | M3.6 V | 02:19:02.67 | +23:52:53.7 | ... | ... | ... | ... | 58.51 | 307.13 | 86.91 ± 0.40 | 0.296 ± 0.014 |
| ... | ... | Ross 19 | J02190+353 | M3.5 V | 02:19:03.89 | +35:21:11.8 | ... | ... | ... | ... | 50.51 | 795.17 | 151.70 ± 0.86 | 0.374 ± 0.016 |
| 02206+3748 | LDS3370 | GJ 3151 | J02204+377 | M2.5 V | 02:20:25.74 | +37:47:29.6 | (AB) | AB | 2.000 | 89.0 | 49.07 | 349.30 | ... | 0.414 ± 0.031 |
| ... | ... | GJ 3153 | J02207+029 | M6.0 V | 02:20:46.42 | +02:58:32.9 | ... | ... | ... | ... | 95.16 | 330.57 | 42.92 ± 0.21 | 0.216 ± 0.012 |
| ... | ... | GJ 1047 A | J02210+368 | M3.0 V | 02:21:05.05 | +36:52:55.2 | AB+C | A | ... | ... | 70.02 | 927.46 | ... | 0.278 ± 0.035 |
| 02212+3653 | WNO 7 | GJ 1047 B | ... | M4.5 V | 02:21:05.01 | +36:52:56.4 | ... | B | 1.237 | 340.4 | 65.49 | 925.86 | ... | 0.259 ± 0.036 |
| 02212+3653 | GIC 31 | GJ 1047 C | ... | M3.5 V | 02:21:02.88 | +36:52:38.6 | ... | C | 30.910 | 237.5 | 80.05 | 921.31 | 60.94 ± 0.26 | 0.228 ± 0.012 |
| ... | ... | GJ 96 | J02222+478 | M0.5 V | 02:22:14.99 | +47:52:48.8 | ... | ... | ... | ... | 45.42 | 220.07 | 657.93 ± 3.86 | 0.601 ± 0.027 |
| ... | ... | StKM 1-261 | J02230+181 | M0.5 V | 02:23:06.15 | +18:10:31.6 | ... | ... | ... | ... | 38.80 | 218.67 | ... | ... |
| ... | ... | LP 353-51 | J02234+227 | M0.5 V | 02:23:26.76 | +22:44:05.0 | ... | ... | ... | ... | 32.11 | 149.70 | 650.86 ± 2.39 | 0.599 ± 0.027 |
| ... | ... | GJ 3156 | J02247+259 | M0.5 V | 02:24:46.00 | +25:58:31.5 | ... | ... | ... | ... | 36.57 | 197.30 | 533.43 ± 2.34 | 0.558 ± 0.028 |
| ... | ... | StKM 1-265 | J02254+246 | M2.0 V | 02:25:28.00 | +24:40:36.9 | ... | ... | ... | ... | 109.95 | 89.94 | 576.99 ± 3.71 | 0.557 ± 0.028 |
| ... | ... | GJ 3157 | J02256+375 | M5e | 02:25:38.83 | +37:32:32.8 | ... | ... | ... | ... | 96.74 | 293.78 | 151.70 ± 0.79 | 0.399 ± 0.017 |
| 02273+5433 | LAW10 | 1R022716.4+543258 | J02272+545 | M4.5 V | 02:27:17.31 | +54:32:46.1 | (AB) | AB | 0.677 | 185.8 | 49.95 | 151.50 | ... | 0.210 ± 0.038 |
| ... | ... | PM J02274+0310 | J02274+031 | M4.0 V | 02:27:27.43 | +03:10:54.7 | ... | ... | ... | ... | 33.43 | 122.39 | 80.21 ± 0.76 | 0.283 ± 0.013 |
| 02278+0426 | A 2329 | HD 15285 | J02277+044 | M1.0 V | 02:27:46.05 | +04:25:58.8 | AB | AB | 0.630 | 109.3 | 44.22 | 255.09 | ... | ... |
| ... | ... | GJ 3159 | J02282+014 | M3.0 V | 02:28:17.41 | +01:26:29.0 | ... | ... | ... | ... | 57.33 | 310.15 | 147.38 ± 1.06 | 0.345 ± 0.015 |
| ... | ... | TYC 1221-1171-1 | J02283+219 | M0.5 V | 02:28:22.17 | +21:59:45.3 | ... | ... | ... | ... | 40.87 | 56.52 | 1211.57 ± 45.27 | 0.699 ± 0.026 |
| 02290-1959 | RST2280 | HD 15468 | ... | K4.5 Vk | 02:29:02.40 | -19:58:40.8 | (AB)+C | AB | 0.182 | 145.3 | 53.35 | 646.40 | ... | 0.699 ± 0.105 |
| 02290-1959 | UC 744 | GJ 100 C | J02285-200 | M2.5 V | 02:28:32.62 | -20:02:22.5 | ... | C | 474.569 | 242.1 | 43.84 | 649.19 | 126.10 ± 0.65 | 0.318 ± 0.014 |
| ... | ... | LSPM J0228+1538 | J02287+156 | M2.0 V | 02:28:47.15 | +15:38:53.6 | AB | A | ... | ... | 43.82 | 171.10 | ... | 0.520 ± 0.029 |
| 02288+1539 | RAO 536 | StKM 2-212B | ... | ... | 02:28:47.18 | +15:38:52.9 | ... | B | 0.804 | 146.7 | 43.86 | ... | ... | ... |
| ... | ... | GJ 3160 | J02289+120 | M2.5 V | 02:28:54.68 | +12:05:22.3 | Aab | Aab(2) | ... | ... | 48.98 | 81.71 | ... | ... |
| ... | ... | StKM 2-213A | J02289+226 | M2.0 V | 02:28:58.41 | +22:36:24.5 | A+B | A | ... | ... | 83.66 | 156.39 | ... | 0.721 ± 0.025 |
| 02290+2236 | CFN 2 | StKM 2-213B | ... | M0.5 V | 02:28:58.52 | +22:36:21.9 | ... | B | 3.012 | 149.6 | ... | 140.86 | ... | 0.609 ± 0.028 |
| ... | ... | LP 410-33 | J02292+195 | M2.5 Ve | 02:29:14.32 | +19:32:31.9 | ... | ... | ... | ... | 36.90 | 270.77 | 612.28 ± 3.80 | 0.558 ± 0.028 |
| ... | ... | GJ 3137 | J02293+884 | M3.5 V | 02:29:14.09 | +88:24:20.3 | ... | ... | ... | ... | 35.77 | 437.01 | 159.16 ± 0.65 | 0.360 ± 0.015 |
| ... | ... | Ross 21 | J02314+573 | M3.5 V | 02:31:29.87 | +57:22:43.3 | ... | ... | ... | ... | 28.20 | 1115.77 | 119.49 ± 0.92 | 0.329 ± 0.014 |
| ... | ... | LP 530-26 | J02330+078 | M2.0 V | 02:33:04.78 | +07:49:41.0 | ... | ... | ... | ... | 34.63 | 231.24 | 522.43 ± 5.62 | 0.541 ± 0.029 |
| ... | ... | GJ 102 | J02336+249 | M3.5 Ve | 02:33:37.23 | +24:55:26.9 | ... | ... | ... | ... | 36.56 | 677.52 | 55.55 ± 0.23 | 0.215 ± 0.009 |
| ... | ... | GJ 3165 | J02337+150 | M3.0 V | 02:33:47.96 | +15:00:17.8 | ... | ... | ... | ... | 42.04 | 438.16 | 91.68 ± 0.43 | 0.285 ± 0.013 |

Table D.2: Complete sample with the description of multiple systems (continued).

| WDS id | WDS disc | Name | Karmn | Spectral type | $\alpha$ (2016.0) | $\delta$ (2016.0) | System | Component | $\rho$ [arcsec] | $\theta$ [deg] | $\varpi$ [mas] | $\mu_{total}$ [mas a$^{-1}$] | $\mathcal{L}$ [$10^{-4}$ $\mathcal{L}_\odot$] | $\mathcal{M}$ [$\mathcal{M}_\odot$] |
|---|---|---|---|---|---|---|---|---|---|---|---|---|---|---|
| … | … | GJ 3164 | J02340+417 | M3.0 V | 02:34:00.46 | +41:46:50.4 | … | … | … | … | 58.33 | 297.00 | 98.83 ± 0.42 | 0.297 ± 0.013 |
| … | … | G 174-4 | J02345+566 | M2.0 V | 02:34:34.87 | +56:36:42.1 | … | … | … | … | 44.81 | 240.69 | 583.07 ± 5.10 | 0.554 ± 0.028 |
| … | … | GJ 3166 | J02353+235 | M4.0 V | 02:35:22.50 | +23:34:29.5 | … | … | … | … | 23.62 | 111.79 | 202.38 ± 5.05 | 0.435 ± 0.018 |
| … | … | GJ 104 | J02358+202 | M2.0 V | 02:35:53.59 | +20:13:09.3 | … | … | … | … | 51.61 | 286.55 | 397.25 ± 1.73 | 0.487 ± 0.012 |
| 02361+0653 | GKI 1 | HD 16160 | … | K3 V | 02:36:06.81 | +06:53:36.0 | (AB)+C | AB | 1.730 | 289.5 | 50.97 | 2312.10 | 2771.37 ± 81.28 | 0.780 ± 0.117 |
| 02361+0653 | PLQ 32 | BXCet | J02362+068 | M4.0 V | 02:36:17.20 | +06:52:41.1 | … | C | 164.247 | 318.4 | 28.44 | 2312.99 | 81.17 ± 0.42 | 0.261 ± 0.012 |
| … | … | GJ 3168 | J02364+554 | M3.0 V | 02:36:26.63 | +55:28:30.2 | … | … | … | … | … | 358.33 | 228.52 ± 2.77 | 0.464 ± 0.019 |
| … | … | G 36-26 | J02367+226 | M5.0 V | 02:36:44.08 | +22:40:20.2 | … | … | … | … | 38.66 | 376.42 | 25.09 ± 0.10 | 0.172 ± 0.011 |
| … | … | GJ 3169 | J02367+320 | M3.5 V | 02:36:47.30 | +32:04:18.9 | AB | A | … | … | 17.21 | 336.16 | … | 0.285 ± 0.035 |
| 02369+3204 | LDS3404 | GJ 3170 | … | M3.0 V | 02:36:47.44 | +32:04:20.0 | … | B | 2.064 | 58.2 | 16.81 | 338.06 | … | 0.257 ± 0.036 |
| … | … | GJ 3174 | J02392+074 | M3.7 V | 02:39:17.85 | +07:28:14.9 | … | … | … | … | 25.11 | 499.67 | 60.61 ± 0.30 | 0.243 ± 0.012 |
| … | … | G 75-35 | J02412-045 | M4.5 V | 02:41:15.51 | -04:32:18.8 | … | … | … | … | 46.01 | 357.49 | 44.25 ± 0.36 | 0.219 ± 0.012 |
| … | … | StKM 1-291 | J02419+435 | M1.0 V | 02:41:58.94 | +43:34:19.0 | … | … | … | … | 50.01 | 49.84 | 737.12 ± 30.43 | 0.620 ± 0.028 |
| … | … | LP 410-81 | J02424+182 | M1.5 V | 02:42:25.80 | +18:14:44.0 | … | … | … | … | 32.61 | 241.21 | 705.27 ± 5.64 | 0.607 ± 0.027 |
| … | … | Wolf 1132 | J02438-088 | M1.5 V | 02:43:53.88 | -08:49:57.8 | … | … | … | … | 100.15 | 973.37 | 241.51 ± 1.45 | 0.398 ± 0.016 |
| … | … | HD 16895 | … | F8 V | 02:44:12.53 | +49:13:41.0 | A+B | A | … | … | 45.42 | 346.40 | … | 1.203 ± 0.180 |
| 02442+4914 | STF 296 | GJ 107 B | J02441+492 | M1.5 V | 02:44:10.79 | +49:13:53.0 | … | B | 20.932 | 305.1 | 45.24 | 322.95 | … | 0.509 ± 0.029 |
| … | … | VXAri | J02442+255 | dM3 | 02:44:16.53 | +25:31:18.3 | … | … | … | … | 27.86 | 937.72 | 170.73 ± 0.88 | 0.349 ± 0.011 |
| 02444+1057 | JNN 300 | MCC 401 | J02443+109W | M1.0 V | 02:44:21.45 | +10:57:40.2 | (AB)+C | AB | 0.249 | 322.6 | 32.85 | 86.18 | … | … |
| 02444+1057 | SKF 20 | 2M02442272+1057349 | J02443+109E | M5.0 Ve | 02:44:22.81 | +10:57:34.2 | … | C | 20.917 | 106.7 | 71.22 | 93.99 | 217.08 ± 1.41 | 0.346 ± 0.033 |
| 02457+4456 | LDS5393 | GJ 3178 | J02456+449 | M0.5 V | 02:45:40.30 | +44:56:53.5 | A+B | A | 17.980 | 66.2 | 138.34 | 428.98 | 707.84 ± 2.71 | 0.613 ± 0.027 |
| … | … | GJ 3179 | … | M5.0 V | 02:45:41.85 | +44:57:00.8 | … | B | 17.984 | 66.2 | 138.44 | 440.90 | 20.99 ± 0.47 | 0.155 ± 0.010 |
| … | … | GJ 3180 | J02462-049 | M6.0 V | 02:46:16.73 | -04:59:50.7 | … | … | … | … | 34.18 | 2529.50 | 14.45 ± 0.08 | 0.125 ± 0.009 |
| … | … | GJ 3181 | J02465+164 | M6.0 V | 02:46:33.79 | +16:25:01.1 | … | … | … | … | 68.34 | 1007.59 | 9.43 ± 0.05 | 0.104 ± 0.009 |
| … | … | 2M02483695+6211228 | J02486+621 | M5.5 V | 02:48:37.25 | +62:11:21.3 | … | … | … | … | 51.38 | 153.80 | 18.05 ± 0.19 | 0.142 ± 0.010 |
| … | … | PM J02489-1432W | J02489-145W | M2.0 V | 02:48:59.45 | -14:32:14.2 | A+B | A | … | … | 51.27 | 170.54 | 342.53 ± 2.92 | 0.463 ± 0.013 |
| 02490-1432 | KPP2871 | PM J02489-1432E | J02489-145E | M2.5 V | 02:49:00.02 | -14:32:15.5 | … | B | 8.354 | 99.0 | 50.05 | 180.26 | 273.58 ± 1.47 | 0.432 ± 0.013 |
| … | … | G 246-12 | J02502+628 | M2.5 V | 02:50:16.95 | +62:51:16.4 | … | … | … | … | 81.20 | 303.05 | 207.62 ± 0.98 | 0.414 ± 0.017 |
| … | … | GJ 3184 | J02518+062 | M3.0 V | 02:51:51.15 | +06:13:39.6 | … | … | … | … | 33.49 | 484.90 | 155.46 ± 1.01 | 0.355 ± 0.015 |
| 02518+2929 | CRC 45 | GJ 3183 | J02518+294 | M4.0 V | 02:51:49.63 | +29:29:10.4 | (AB) | AB | 0.890 | 243.6 | 25.40 | 168.63 | … | 0.251 ± 0.036 |
| … | … | RBS 365 | J02519+224 | dM4.0 | 02:51:54.22 | +22:27:28.2 | … | … | … | … | 46.89 | 155.52 | 268.93 ± 1.71 | 0.404 ± 0.032 |
| … | … | GJ 3186 | J02524+269 | M2.0 V | 02:52:25.03 | +26:58:26.2 | … | … | … | … | 89.68 | 231.37 | 600.88 ± 2.55 | 0.576 ± 0.028 |
| … | … | Teegarden's Star | J02530+168 | M7.0 V | 02:53:04.71 | +16:51:51.7 | … | … | … | … | 89.37 | 5122.57 | 8.55 ± 0.05 | 0.104 ± 0.047 |
| … | … | LP 411-18 | J02534+174 | M3.5 V | 02:53:26.14 | +17:24:28.4 | … | … | … | … | 130.20 | 252.85 | 180.66 ± 3.73 | 0.410 ± 0.017 |
| … | … | HD 18143A | … | … | 02:55:39.38 | +26:52:20.5 | A+B+C | A | … | … | 20.78 | 327.70 | … | 1.079 ± 0.021 |
| 02556+2652 | STF 326 | HD 18143B | … | K7.5 V | 02:55:39.14 | +26:52:16.9 | … | B | 4.781 | 221.3 | 21.35 | 327.98 | … | 0.719 ± 0.025 |
| 02556+2652 | LDS 883 | HD 18143C | J02555+268 | M4.0 V | 02:55:36.11 | +26:52:17.6 | … | C | 43.787 | 266.3 | 41.71 | 328.06 | 106.92 ± 0.47 | 0.331 ± 0.015 |
| … | … | LP 591-156 | J02560-006 | M5.0 V | 02:56:04.17 | -00:36:32.0 | … | … | … | … | 45.45 | 266.17 | 22.52 ± 0.11 | 0.161 ± 0.010 |
| 02562+2359 | JNN 253 | LSPM J0256+2359 | J02562+239 | M5.0 V | 02:56:14.06 | +23:59:07.4 | (AB) | AB | 0.107 | 98.4 | 59.84 | 183.53 | … | 0.255 ± 0.036 |
| … | … | Ross 364 | J02565+554W | M1.0 V | 02:56:35.79 | +55:26:06.8 | A+B | A | … | … | 71.60 | 858.30 | 720.47 ± 3.03 | 0.615 ± 0.027 |

Table D.2: Complete sample with the description of multiple systems (continued).

| WDS id | WDS disc | Name | Karmn | Spectral type | $\alpha$ (2016.0) | $\delta$ (2016.0) | System | Component | $\rho$ [arcsec] | $\theta$ [deg] | $\varpi$ [mas] | $\mu_{\rm total}$ [mas a$^{-1}$] | $\mathcal{L}$ [$10^{-4}\ \mathcal{L}_{\odot}$] | $\mathcal{M}$ [$\mathcal{M}_{\odot}$] |
|---|---|---|---|---|---|---|---|---|---|---|---|---|---|---|
| 02565+5526 | LDS5401 | Ross 365 | J02565+554E | M2.5 V | 02:56:36.49 | +55:26:22.3 | ... | B | 16.600 | 21.0 | 24.10 | 844.15 | ... | 0.483 ± 0.030 |
| ... | ... | LP 14-53 | J02573+765 | M4.0 V | 02:57:21.43 | +76:33:04.9 | ... | ... | ... | ... | 26.57 | 798.30 | 72.79 ± 0.33 | 0.238 ± 0.012 |
| ... | ... | Ross 791 | J02575+107 | M4.0 V | 02:57:32.96 | +10:47:17.9 | ... | ... | ... | ... | 26.54 | 1803.80 | 84.20 ± 0.48 | 0.272 ± 0.013 |
| ... | ... | GJ 3189 | J02581-128 | sdM3.0 | 02:58:10.53 | -12:52:57.4 | ... | ... | ... | ... | 36.96 | 605.17 | 51.63 ± 0.28 | 0.195 ± 0.011 |
| ... | ... | Ross 331 | J02591+366 | M3.7 V | 02:59:11.44 | +36:36:35.7 | AB | A | ... | ... | 40.82 | 644.87 | ... | 0.366 ± 0.033 |
| 02592+3637 | CRC 46 | Ross 331B | ... | M6.0 V | 02:59:11.45 | +36:36:37.6 | ... | B | 1.889 | 3.9 | 43.62 | 648.91 | ... | 0.131 ± 0.043 |
| ... | ... | GJ 3191 | J02592+317 | M3.3 V | 02:59:16.77 | +31:46:27.8 | ... | ... | ... | ... | 37.01 | 190.13 | 192.89 ± 1.02 | 0.424 ± 0.017 |
| 02598+3856 | LAW 11 | G 134-63 | J02597+389 | M4.5 V | 02:59:46.67 | +38:55:34.6 | (AB) | AB | 0.876 | 17.3 | 42.79 | 250.28 | ... | 0.218 ± 0.038 |
| ... | ... | GJ 3193 | J03018-165S | M3.0 V | 03:01:50.98 | -16:35:40.3 | A+B | A | ... | ... | 260.99 | 456.80 | ... | 0.285 ± 0.035 |
| 03019-1633 | RST2292 | GJ 3192 | J03018-165N | M2.5 V | 03:01:50.63 | -16:35:35.2 | ... | B | 7.211 | 315.0 | 50.49 | 450.46 | ... | 0.261 ± 0.037 |
| ... | ... | GJ 9108 | J03026-181 | M2.5 V | 03:02:38.51 | -18:09:56.1 | ... | ... | ... | ... | 44.43 | 435.93 | 292.37 ± 1.87 | 0.467 ± 0.018 |
| ... | ... | StM 20 | J03033-080 | M3.0 V | 03:03:21.47 | -08:05:16.0 | ... | ... | ... | ... | 44.51 | 125.55 | 576.61 ± 44.39 | 0.533 ± 0.031 |
| ... | ... | GJ 3197 | J03037-128 | M3.5 V | 03:03:48.10 | -12:51:21.0 | A+B | A | ... | ... | 44.52 | 259.64 | 152.70 ± 1.05 | 0.375 ± 0.016 |
| 03037-1251 | LDS3450 | GJ 3196 | J03036-128 | M3.5 V | 03:03:40.99 | -12:50:33.7 | ... | B | 114.293 | 294.5 | 61.87 | 255.81 | 133.60 ± 0.94 | 0.349 ± 0.015 |
| ... | ... | GJ 3198 | J03040-203 | M4.0 V | 03:04:05.05 | -20:22:50.7 | ... | ... | ... | ... | 40.08 | 688.31 | 147.75 ± 0.97 | 0.368 ± 0.016 |
| ... | ... | HD 18757 | ... | G1.5 V | 03:04:11.26 | +61:42:09.9 | A+B | A | ... | ... | 49.32 | 1000.79 | 10444.81 ± 35.05 | 1.030 ± 0.155 |
| 03042+6142 | LDS9142 | GJ 3195 | J03047+617 | M3.0 V | 03:04:45.06 | +61:43:57.7 | ... | B | 263.299 | 65.8 | 49.31 | 1000.09 | 229.77 ± 1.05 | 0.437 ± 0.017 |
| 03076-0358 | HDS396 | GJ 3202 | J03075-039 | M0.0 V | 03:07:33.51 | -03:58:23.2 | (AB) | AB | 0.082 | 162.4 | 54.90 | 470.78 | ... | ... |
| ... | ... | LP 355-27 | J03077+249 | M4.5 V | 03:07:47.13 | +24:57:53.3 | ... | ... | ... | ... | 61.06 | 261.83 | 76.21 ± 0.36 | 0.295 ± 0.014 |
| ... | ... | GJ 1055 | J03090+100 | M5.0 V | 03:09:00.48 | +10:01:16.3 | ... | ... | ... | ... | 86.96 | 652.48 | 22.79 ± 0.12 | 0.151 ± 0.010 |
| 03095+4544 | HDS 404 | GJ 125 | J03095+457 | M2.0 V | 03:09:30.17 | +45:43:52.7 | (AB) | AB | 0.510 | 20.2 | 43.16 | 581.18 | ... | 0.622 ± 0.027 |
| ... | ... | EKCet | J03102+059 | M2.5 V | 03:10:15.34 | +05:54:22.5 | ... | ... | ... | ... | 43.25 | 576.45 | 228.27 ± 1.29 | 0.410 ± 0.016 |
| ... | ... | GJ 3204 | J03104+584 | M2.0 V | 03:10:26.72 | +58:26:05.5 | ... | ... | ... | ... | 33.98 | 222.41 | 365.27 ± 1.81 | 0.494 ± 0.018 |
| ... | ... | GJ 1053 | J03109+737 | M6.0 V | 03:11:05.27 | +73:46:02.5 | ... | ... | ... | ... | 32.94 | 2118.60 | 21.91 ± 0.13 | 0.147 ± 0.010 |
| ... | ... | LP 652-62 | J03110-046 | M3.0 V | 03:11:04.90 | -04:36:41.1 | ... | ... | ... | ... | 145.69 | 304.49 | 230.58 ± 1.35 | 0.438 ± 0.017 |
| ... | ... | 1R031114.2+010655 | J03112+011 | M5.5 V | 03:11:15.60 | +01:06:30.6 | ... | ... | ... | ... | 139.70 | 115.06 | 19.17 ± 0.10 | 0.159 ± 0.010 |
| ... | ... | Wolf 132 | J03118+196 | M0.5 V | 03:11:48.26 | +19:40:11.3 | ... | ... | ... | ... | 52.64 | 314.74 | 672.18 ± 5.42 | 0.585 ± 0.028 |
| 03119+6131 | HDS 407 | GJ 3206 | J03119+615 | M0.0 V | 03:11:56.89 | +61:31:14.6 | (AB) | AB | 0.522 | 185.7 | 23.95 | 86.96 | ... | ... |
| ... | ... | CDCet | J03133+047 | M4.5 V | 03:13:24.78 | +04:46:30.7 | ... | ... | ... | ... | 42.72 | 1744.02 | 29.75 ± 0.17 | 0.162 ± 0.009 |
| ... | ... | LP 53-55 | J03136+653 | M1.5 V | 03:13:37.88 | +65:21:19.5 | ... | ... | ... | ... | 42.85 | 193.07 | 256.11 ± 1.40 | 0.410 ± 0.016 |
| ... | ... | GJ 3208 | J03142+286 | M6.0 V | 03:14:12.85 | +28:40:27.6 | ... | ... | ... | ... | 58.62 | 812.01 | 11.51 ± 0.06 | 0.128 ± 0.010 |
| ... | ... | Ross 369A | J03145+594 | M2.5 V | 03:14:33.17 | +59:26:13.8 | AB | A | ... | ... | 42.46 | 250.95 | ... | 0.444 ± 0.031 |
| ... | ... | Ross 369B | ... | M3.0 V | 03:14:33.10 | +59:26:13.2 | ... | B* | 0.775 | 221.0 | 42.53 | 237.34 | ... | 0.368 ± 0.033 |
| ... | ... | RX J0314.7+1127 | J03147+114 | M2.0 V | 03:14:47.26 | +11:27:26.7 | ... | ... | ... | ... | 18.29 | 75.42 | ... | 0.587 ± 0.028 |
| ... | ... | Ross 346 | J03147+485 | M1.5 V | 03:14:45.11 | +48:31:05.5 | ... | ... | ... | ... | 38.82 | 367.80 | 500.54 ± 3.43 | 0.536 ± 0.029 |
| ... | ... | Ross 370A | J03162+581S | M2.0 V | 03:16:14.73 | +58:09:57.3 | A+B | A | ... | ... | 79.45 | 550.45 | 304.01 ± 2.52 | 0.449 ± 0.017 |
| 03162+5810 | MLB 115 | Ross 370B | J03162+581N | M2.0 V | 03:16:14.75 | +58:10:02.4 | ... | B | 5.041 | 2.0 | 58.12 | 523.34 | 264.69 ± 2.33 | 0.417 ± 0.016 |
| ... | ... | PM J03167+3855 | J03167+389 | M3.5 V | 03:16:46.02 | +38:55:27.7 | ... | ... | ... | ... | 55.24 | 63.83 | 209.88 ± 6.93 | 0.444 ± 0.019 |
| 03172+4522 | PRV 2 | GJ 3213 | J03172+453 | M3.0 V | 03:17:11.82 | +45:22:20.9 | (AB) | AB | 0.042 | 206.6 | 44.43 | 270.09 | ... | 0.391 ± 0.032 |
| ... | ... | GJ 3215 | J03177+252 | M2.5 V | 03:17:46.17 | +25:15:00.6 | ... | ... | ... | ... | 82.66 | 865.99 | 297.53 ± 1.85 | 0.444 ± 0.017 |

Table D.2: Complete sample with the description of multiple systems (continued).

| WDS id | WDS disc | Name | Karmn | Spectral type | $\alpha$ (2016.0) | $\delta$ (2016.0) | System | Component | $\rho$ [arcsec] | $\theta$ [deg] | $\varpi$ [mas] | $\mu_{\rm total}$ [mas a$^{-1}$] | $\mathcal{L}$ [$10^{-4}$ $\mathcal{L}_\odot$] | $\mathcal{M}$ [$\mathcal{M}_\odot$] |
|---|---|---|---|---|---|---|---|---|---|---|---|---|---|---|
| … | … | HD 275122 | J03181+382 | M1.5 V | 03:18:08.09 | +38:14:58.3 | … | … | … | … | 33.81 | 725.86 | 623.67 ± 3.01 | 0.579 ± 0.028 |
| … | … | Wolf 140 | J03181+426 | M3.5 V | 03:18:07.31 | +42:40:06.8 | … | … | … | … | 59.55 | 252.44 | 156.30 ± 0.82 | 0.380 ± 0.016 |
| … | … | StKM 1-354 | J03185+103 | M1.5 V | 03:18:35.36 | +10:18:43.2 | … | … | … | … | 37.51 | 167.76 | 484.76 ± 2.26 | 0.540 ± 0.029 |
| … | … | GJ 3216 | J03186+326 | M1.0 V | 03:18:38.59 | +32:39:55.3 | … | … | … | … | 27.48 | 231.71 | 549.17 ± 2.70 | 0.567 ± 0.028 |
| … | … | Ross 371 | J03187+606 | M3.5 V | 03:18:43.66 | +60:36:21.1 | … | … | … | … | 116.27 | 597.29 | 348.03 ± 6.16 | 0.459 ± 0.030 |
| 03194+6156 | JNN 254 | G 246-33 | J03194+619 | M4.0 V | 03:19:29.28 | +61:56:01.5 | (AB) | AB | 0.386 | 239.8 | 44.35 | 292.21 | … | 0.278 ± 0.039 |
| … | … | LP 198-637 A | J03207+397 | M1.5 V | 03:20:45.42 | +39:42:59.3 | AB | A | … | … | 67.07 | 177.60 | … | 0.558 ± 0.028 |
| 03208+3943 | RAO 187 | LP 198-637 B | … | M1.5 V | 03:20:45.35 | +39:42:59.6 | … | B | 0.795 | 293.2 | 36.82 | 180.92 | … | 0.522 ± 0.029 |
| … | … | GJ 133 | J03213+799 | M2.0 V | 03:21:24.29 | +79:58:06.8 | … | … | … | … | 36.50 | 501.04 | 244.96 ± 1.26 | 0.401 ± 0.011 |
| … | … | GJ 3218 | J03217-066 | M2.0 V | 03:21:47.27 | -06:40:25.0 | … | … | … | … | 17.63 | 326.95 | 333.84 ± 2.06 | 0.455 ± 0.012 |
| … | … | GJ 1058 | J03220+029 | M4.5 V | 03:22:04.46 | +02:56:22.7 | … | … | … | … | 39.91 | 823.48 | 27.82 ± 0.13 | 0.157 ± 0.010 |
| … | … | GJ 3219 | J03224+271 | M0.0 V | 03:22:28.40 | +27:09:20.8 | … | … | … | … | 73.62 | 230.09 | 876.64 ± 4.42 | 0.665 ± 0.026 |
| … | … | GJ 1059 | J03230+420 | M5.0 V | 03:23:02.16 | +42:00:15.8 | … | … | … | … | 73.58 | 741.20 | 20.55 ± 0.10 | 0.142 ± 0.010 |
| … | … | GJ 3221 | J03233+116 | M3.5 Ve | 03:23:22.15 | +11:41:11.0 | … | … | … | … | 37.35 | 291.26 | 188.84 ± 0.91 | 0.394 ± 0.016 |
| … | … | 1R032338.7+054117 | J03236+056 | M4.5 V | 03:23:39.25 | +05:41:14.1 | … | … | … | … | 53.11 | 108.50 | 210.30 ± 1.42 | 0.507 ± 0.021 |
| … | … | GJ 140 A | J03241+237 | M1.5 V | 03:24:06.73 | +23:47:04.2 | AB+C | A | … | … | 46.06 | 246.23 | … | 0.583 ± 0.028 |
| 03242+2347 | WOR 4 | GJ 140 B | … | M3.0 V | 03:24:06.67 | +23:47:06.6 | … | B | 2.593 | 339.7 | 62.64 | 231.12 | … | 0.467 ± 0.030 |
| 03242+2347 | LDS 884 | GJ 140 C | J03242+237 | M2.5 V | 03:24:13.10 | +23:46:17.3 | … | C | 99.222 | 118.2 | 41.57 | 233.16 | 316.38 ± 1.52 | 0.486 ± 0.018 |
| … | … | PM J03247+4447A | J03247+447 | M1.5 V | 03:24:42.31 | +44:47:41.4 | AB | A | … | … | 28.36 | 105.26 | … | 0.552 ± 0.028 |
| 03247+4448 | KPP3523 | PM J03247+4447B | … | M3.5 V | 03:24:42.23 | +44:47:39.7 | … | B | 1.912 | 206.5 | 38.98 | 89.93 | … | 0.340 ± 0.033 |
| 03257+0551 | JNN 255 | GJ 3224 | J03257+058 | M4.5 V | 03:25:42.04 | +05:51:48.2 | (AB) | AB | 0.225 | 162.3 | 26.07 | 246.27 | … | 0.204 ± 0.038 |
| 03263+1709 | JNN 256 | PM J03263+1709 | J03263+171 | M4.0 V | 03:26:23.74 | +17:09:30.1 | (AB) | AB | 0.945 | 222.9 | 35.80 | 108.79 | … | 0.319 ± 0.034 |
| … | … | GJ 3225 | J03267+192 | M4.5 V | 03:26:44.97 | +19:14:37.7 | … | … | … | … | 31.70 | 160.56 | 38.96 ± 0.19 | 0.204 ± 0.011 |
| … | … | CKAri | J03272+273 | M1.0 Ve | 03:27:14.22 | +27:23:07.8 | … | … | … | … | 31.54 | 116.46 | 543.83 ± 3.68 | 0.567 ± 0.028 |
| … | … | ATO J051.8788+22.2102 | J03275+222 | M4.5 V | 03:27:30.94 | +22:12:36.9 | … | … | … | … | 71.66 | 104.42 | 191.98 ± 1.10 | 0.452 ± 0.018 |
| … | … | LSPM J0328+3515A | J03284+352 | M2.0 V | 03:28:29.35 | +35:15:18.6 | AB | A | … | … | 57.99 | 156.47 | … | 0.524 ± 0.029 |
| 03284+3515 | KPP4352 | LSPM J0328+3515B | … | M1.5 V | 03:28:29.32 | +35:15:17.4 | … | B | 1.225 | 200.3 | 61.36 | 144.49 | … | 0.520 ± 0.029 |
| … | … | GJ 3228 | J03286-156 | M3.5 Ve | 03:28:39.18 | -15:37:16.4 | A+B | A | … | … | 31.47 | 189.19 | 85.08 ± 3.44 | 0.293 ± 0.015 |
| 03286-1537 | LDS3498 | GJ 3229 | … | M3.5 V | 03:28:39.36 | -15:37:32.6 | … | B | 16.424 | 170.6 | 67.88 | 204.04 | 70.48 ± 0.31 | 0.264 ± 0.013 |
| … | … | 1R033021.4+344044 | J03303+346 | M4.0 V | 03:30:23.37 | +34:40:31.7 | A+BC | A | … | … | 58.47 | 56.66 | … | … |
| … | … | G3- 221088045766546816 | … | M1/2 V | 03:30:16.90 | +34:39:50.2 | … | B* | 89.956 | 242.5 | 26.64 | 53.59 | … | 0.411 ± 0.031 |
| … | … | G3- 221088050062935680 | … | M3.0 V | 03:30:16.80 | +34:39:49.2 | … | C* | 91.483 | 242.3 | 48.27 | 52.34 | … | 0.359 ± 0.033 |
| … | … | GJ 3227 | J03288+264 | M4.0 V | 03:28:49.84 | +26:29:10.2 | … | … | … | … | 48.26 | 262.15 | 65.65 ± 0.37 | 0.254 ± 0.012 |
| … | … | LSPM J0330+5413 | J03308+542 | M5.0 V | 03:30:48.61 | +54:13:55.1 | … | … | … | … | 48.34 | 152.03 | 11.58 ± 0.05 | 0.118 ± 0.009 |
| 03309+7041 | JNN 257 | LP 31-368 | J03309+706 | M3.5 V | 03:30:56.01 | +70:41:06.4 | (AB) | AB | 0.354 | 315.2 | 30.28 | 607.40 | … | 0.296 ± 0.035 |
| … | … | GJ 143.3 | J03317+143 | M2.5 V | 03:31:47.18 | +14:19:07.0 | … | … | … | … | 30.42 | 675.27 | 204.13 ± 1.16 | 0.386 ± 0.016 |
| 03326+2844 | JNN 24 | RX J0332.6+2843 | J03325+287 | M4.5 V | 03:32:35.85 | +28:43:54.1 | (ABC) | ABC | 0.098 | 282.4 | 39.77 | 100.09 | … | … |
| … | … | V577Per | … | K2 V | 03:33:13.60 | +46:15:23.7 | A+B | A | … | … | 27.45 | 188.42 | 1014.02 ± 28.06 | 0.820 ± 0.012 |
| 03332+4615 | ES 560 | HD 21845B | J03332+462 | M0.0 V | 03:33:14.16 | +46:15:16.2 | … | B | 9.502 | 142.3 | 55.31 | 185.83 | … | 0.688 ± 0.026 |
| … | … | Ross 563 | J03340+585 | M0.5 V | 03:34:01.11 | +58:35:52.3 | … | … | … | … | 32.96 | 348.57 | 1173.61 ± 7.05 | 0.686 ± 0.026 |

Table D.2: Complete sample with the description of multiple systems (continued).

| WDS id | WDS disc | Name | Karmn | Spectral type | $\alpha$ (2016.0) | $\delta$ (2016.0) | System | Component | $\rho$ [arcsec] | $\theta$ [deg] | $\varpi$ [mas] | $\mu_{\rm total}$ [mas a$^{-1}$] | $\mathcal{L}$ [$10^{-4}\,\mathcal{L}_\odot$] | $\mathcal{M}$ [$\mathcal{M}_\odot$] |
|---|---|---|---|---|---|---|---|---|---|---|---|---|---|---|
| ... | ... | GJ 3235 | J03346-048 | M3.8 V | 03:34:40.07 | -04:50:38.5 | Aabc | Aabc(3) | ... | ... | 59.55 | 520.01 | ... | ... |
| ... | ... | 1R033609.2+311853 | J03361+313 | M4.5 V | 03:36:08.85 | +31:18:37.4 | ... | ... | ... | ... | 25.90 | 175.56 | 52.98 ± 0.26 | 0.242 ± 0.012 |
| ... | ... | GJ 3237 | J03366+034 | M5.0 V | 03:36:40.97 | +03:29:17.6 | ... | ... | ... | ... | 21.22 | 171.39 | 186.43 ± 1.30 | 0.476 ± 0.020 |
| ... | ... | GJ 3236 | J03372+691 | M3.8 V | 03:37:14.55 | +69:10:47.9 | Aab | Aab(EB) | ... | ... | 27.51 | 196.46 | ... | 0.657 ± 0.023 |
| ... | ... | GJ 3239 | J03375+178N | M2.5 V | 03:37:33.54 | +17:51:14.1 | Aab+Bab | Aab(2) | ... | ... | 42.04 | 172.97 | ... | ... |
| 03376+1751 | LDS3512 | GJ 3240 | J03375+178S | M3.5 V | 03:37:34.09 | +17:51:00.0 | ... | Bab(EB?) | 16.150 | 150.6 | 112.39 | 172.61 | ... | ... |
| ... | ... | KPTau | J03394+249 | M3.5 Ve | 03:39:29.85 | +24:58:05.7 | ... | ... | ... | ... | 21.15 | 228.67 | 110.09 ± 0.73 | 0.315 ± 0.014 |
| ... | ... | Wolf 204 | J03396+254E | M3.0 V | 03:39:36.47 | +25:28:11.6 | A+B | A | ... | ... | 43.80 | 626.22 | 290.22 ± 2.81 | 0.494 ± 0.019 |
| 03396+2530 | LDS9158 | Wolf 205 | J03396+254W | M3.5 V | 03:39:40.77 | +25:28:39.1 | ... | B | 64.329 | 64.7 | 41.87 | 622.15 | 231.86 ± 1.81 | 0.439 ± 0.017 |
| ... | ... | HD 278874 | ... | K2 V | 03:39:48.92 | +33:28:24.3 | Aab+B | Aab(2) | ... | ... | 29.00 | 36.21 | ... | ... |
| 03398+3328 | ES 327 | HD 278874B | J03397+334 | M3.0 V | 03:39:47.79 | +33:28:30.7 | ... | B | 15.472 | 294.3 | ... | 38.83 | 568.68 ± 3.97 | 0.521 ± 0.029 |
| ... | ... | TYC 3720-426-1 | J03416+552 | M0.0 Ve | 03:41:37.46 | +55:13:05.0 | ... | ... | ... | ... | 11.38 | 151.56 | 976.90 ± 4.67 | 0.673 ± 0.026 |
| ... | ... | LSPM J0343+4554A | J03430+459 | M4.0 V | 03:43:01.79 | +45:54:17.4 | AB | A | ... | ... | 11.73 | 199.94 | ... | 0.235 ± 0.037 |
| 03430+4554 | JNN 259 | LSPM J0343+4554B | ... | M4.5 V | 03:43:01.72 | +45:54:17.9 | ... | B | 0.886 | 305.6 | 65.13 | 213.60 | ... | 0.231 ± 0.037 |
| ... | ... | GJ 3247 | J03433-095 | M4.5 V | 03:43:22.53 | -09:33:46.1 | ... | ... | ... | ... | 95.22 | 519.04 | 63.40 ± 0.91 | 0.267 ± 0.013 |
| ... | ... | GJ 150.1 A | J03438+166 | M0.0 V | 03:43:52.74 | +16:40:14.2 | A+B | A | ... | ... | 43.83 | 353.41 | 754.91 ± 2.48 | 0.633 ± 0.027 |
| 03439+1640 | GIC 44 | GJ 150.1 B | J03437+166 | M1.0 V | 03:43:45.42 | +16:39:57.2 | ... | B | 106.541 | 260.8 | 49.53 | 347.26 | 453.19 ± 1.98 | 0.530 ± 0.029 |
| ... | ... | HD 278968 | J03445+349 | M0.0 V | 03:44:31.21 | +34:58:20.8 | ... | ... | ... | ... | ... | 254.27 | 784.51 ± 2.74 | 0.649 ± 0.027 |
| ... | ... | G 221-21 | J03454+729 | M1.5 V | 03:45:28.68 | +72:59:25.2 | (AB)+C | A | ... | ... | 27.48 | 487.29 | ... | 0.549 ± 0.028 |
| 03454+7259 | LDS1581 | LP 31-210 | ... | ... | ... | ... | ... | B | 8.100 | 179.8 | 27.48 | ... | ... | ... |
| 03454+7259 | WIS 99 | LP 31-200 | ... | M3.5 Ve | 03:43:44.04 | +72:53:42.2 | ... | C | 574.149 | 233.5 | 25.61 | 488.28 | 51.72 ± 0.25 | 0.223 ± 0.012 |
| ... | ... | PM J03455+7018 | J03455+703 | M1.0 V | 03:45:32.34 | +70:18:00.4 | ... | ... | ... | ... | 37.82 | 137.73 | 395.25 ± 2.11 | 0.512 ± 0.029 |
| ... | ... | G 6-33 | J03459+147 | M1.5 V | 03:45:54.96 | +14:42:47.2 | ... | ... | ... | ... | 73.21 | 283.12 | 662.74 ± 3.05 | 0.597 ± 0.027 |
| ... | ... | HD 23453 | J03463+262 | M1.0 V | 03:46:20.60 | +26:12:52.7 | ... | ... | ... | ... | 37.35 | 433.72 | 739.35 ± 3.48 | 0.626 ± 0.027 |
| ... | ... | TYC 4521-1342-1 | J03467+821 | M1.0 V | 03:46:42.67 | +82:07:50.1 | ... | ... | ... | ... | 27.05 | 109.54 | 742.46 ± 4.37 | 0.623 ± 0.027 |
| ... | ... | GJ 3249 | J03467-112 | M2.5 V | 03:46:46.04 | -11:17:40.5 | ... | ... | ... | ... | 25.83 | 582.16 | 185.89 ± 3.10 | 0.391 ± 0.016 |
| ... | ... | GJ 3250 | J03473+086 | M5.0 V | 03:47:21.40 | +08:41:36.7 | ... | ... | ... | ... | 25.78 | 803.41 | 24.90 ± 0.19 | 0.159 ± 0.010 |
| ... | ... | G 80-21 | J03473-019 | M3.0 V | 03:47:23.53 | -01:58:24.3 | ... | ... | ... | ... | 62.65 | 328.20 | 317.72 ± 2.22 | 0.502 ± 0.039 |
| ... | ... | Ross 588 | J03479+027 | M0.5 V | 03:47:57.68 | +02:47:09.3 | ... | ... | ... | ... | 38.83 | 579.87 | 342.39 ± 2.13 | 0.451 ± 0.017 |
| 03480+6840 | KUI 13 | HD 23189 | ... | K2 V | 03:48:01.40 | +68:40:26.4 | A+(BC) | A | ... | ... | 38.91 | 279.83 | 1046.22 ± 5.04 | 0.820 ± 0.123 |
| 03480+6840 | KUI 13 | GJ 153 C | J03480+686 | M2 V | 03:48:02.08 | +68:40:42.9 | ... | BC | 17.206 | 13.2 | 25.43 | 288.83 | ... | 0.467 ± 0.030 |
| ... | ... | GJ 3248 | J03486+735 | M1.0 V | 03:48:39.73 | +73:32:30.9 | ... | ... | ... | ... | 25.57 | 482.03 | 256.76 ± 1.16 | 0.411 ± 0.016 |
| ... | ... | GJ 3251 | J03505+634 | M1.5 V | 03:50:33.69 | +63:27:14.9 | ... | ... | ... | ... | 27.92 | 230.99 | 1332.76 ± 6.09 | 0.708 ± 0.026 |
| ... | ... | GJ 1065 | J03507-060 | M3.0 V | 03:50:43.81 | -06:06:03.6 | ... | ... | ... | ... | 41.91 | 1441.79 | 51.45 ± 0.39 | 0.222 ± 0.012 |
| ... | ... | PM J03510+1413 | J03510+142 | M4.5 V | 03:51:00.87 | +14:13:38.7 | A+B | A | ... | ... | 41.95 | 102.59 | 374.12 ± 2.75 | 0.434 ± 0.031 |
| 03510+1414 | JLM 1 | UPM J0350+1414 | ... | M3.5 V | 03:50:59.58 | +14:14:00.5 | ... | B | 28.818 | 319.2 | 49.65 | 94.89 | 159.59 ± 1.63 | 0.469 ± 0.020 |
| ... | ... | GJ 3252 | J03510-008 | M8.0 V | 03:51:00.04 | -00:52:52.4 | ... | ... | ... | ... | 58.04 | 470.09 | 8.22 ± 0.05 | 0.125 ± 0.010 |
| ... | ... | HD 275867 | ... | K2 V | 03:52:00.35 | +39:47:43.7 | A+B | A | ... | ... | 58.03 | 62.07 | 2050.28 ± 9.78 | 0.820 ± 0.123 |
| 03520+3947 | GRV 197 | TYC 2868-639-1 | J03519+397 | M0.0 V | 03:51:58.19 | +39:46:55.8 | ... | B | 53.919 | 207.5 | 38.92 | 64.40 | 859.01 ± 5.45 | 0.662 ± 0.026 |
| ... | ... | Wolf 227 | J03526+170 | M5.0 V | 03:52:42.24 | +17:00:53.8 | (Aab) | Aab(2) | ... | ... | 26.76 | 777.27 | ... | ... |

Table D.2: Complete sample with the description of multiple systems (continued).

| WDS id | WDS disc | Name | Karmn | Spectral type | $\alpha$ (2016.0) | $\delta$ (2016.0) | System | Component | $\rho$ [arcsec] | $\theta$ [deg] | $\varpi$ [mas] | $\mu_{\rm total}$ [mas a$^{-1}$] | $\mathcal{L}$ [$10^{-4}\ \mathcal{L}_\odot$] | $\mathcal{M}$ [$\mathcal{M}_\odot$] |
|---|---|---|---|---|---|---|---|---|---|---|---|---|---|---|
| … | … | Ross 567 | J03531+625 | M3.0 V | 03:53:10.51 | +62:34:03.8 | … | … | … | … | 39.09 | 270.93 | 167.54 ± 0.76 | 0.347 ± 0.011 |
| … | … | 2M03542008-1437388 | J03543-146 | M6.5 V | 03:54:20.02 | -14:37:37.2 | … | … | … | … | … | 111.80 | 10.98 ± 0.31 | 0.136 ± 0.010 |
| … | … | GJ 3256 | J03544-091 | M1.0 V | 03:54:25.52 | -09:09:29.2 | AB | A | … | … | 39.15 | 146.38 | … | 0.506 ± 0.029 |
| … | … | GJ 3256 B | … | M3.0 V | 03:54:25.61 | -09:09:32.0 | … | B* | 3.182 | 153.6 | 33.28 | 138.31 | … | 0.344 ± 0.033 |
| … | … | LP 413-108 | J03548+163 | M4.0 V | 03:54:53.37 | +16:18:55.9 | … | … | … | … | 27.71 | 133.30 | 276.38 ± 2.68 | 0.405 ± 0.032 |
| … | … | 1R035632.5+315746 | J03565+319 | M3.5 V | 03:56:33.26 | +31:57:23.8 | … | … | … | … | 69.51 | 139.28 | 376.01 ± 2.18 | 0.455 ± 0.030 |
| … | … | Ross 23 | J03567+039 | M1.5 V | 03:56:47.95 | +53:33:30.5 | … | … | … | … | 24.18 | 501.26 | 642.36 ± 2.93 | 0.585 ± 0.028 |
| … | … | HD 24916 | … | K4 V | 03:57:28.50 | -01:09:36.4 | A+Bab | A | … | … | 44.67 | 234.33 | … | 0.721 ± 0.108 |
| 03575-0110 | BU 543 | HD 24916B | J03574-011 | M2.5 V | 03:57:28.68 | -01:09:25.7 | … | Bab | 10.968 | 14.1 | 46.64 | 251.75 | … | … |
| … | … | Ross 24 | J03586+520 | M1.0 V | 03:58:36.92 | +52:01:21.7 | … | … | … | … | 80.06 | 429.52 | 352.25 ± 2.53 | 0.485 ± 0.018 |
| … | … | G 7-14 | J03588+125 | M4.0 V | 03:58:49.38 | +12:30:18.3 | … | … | … | … | 59.43 | 403.12 | 119.85 ± 0.67 | 0.352 ± 0.015 |
| … | … | Wolf 1322 | J03598+260 | M3.0 V | 03:59:54.53 | +26:05:19.5 | … | … | … | … | 56.99 | 774.40 | 251.17 ± 4.91 | 0.488 ± 0.020 |
| … | … | Ross 25 | J04011+513 | M3.8 V | 04:01:08.18 | +51:23:06.4 | A+B | A | … | … | 55.83 | 883.46 | 125.09 ± 0.63 | 0.337 ± 0.015 |
| … | … | LSPM J0401+5131 | … | DC8 | 04:01:02.14 | +51:31:17.2 | … | B* | 494.032 | 353.4 | 55.98 | 883.24 | … | 0.500 ± 0.100 |
| … | … | GJ 3261 | J04056+057 | M4e | 04:05:38.94 | +05:44:40.4 | AB+C | A | … | … | 62.67 | 47.98 | … | 0.630 ± 0.027 |
| 04056+0545 | MCT 3 | G3- 3296932486866670720 | … | … | 04:05:38.89 | +05:44:40.1 | … | B | 0.817 | 251.9 | 25.34 | … | … | … |
| … | … | ATO J061.4727+05.5235 | … | M3.0 V | 04:05:53.46 | +05:31:24.6 | … | C* | 824.788 | 164.8 | 100.73 | 49.85 | 227.47 ± 1.68 | 0.494 ± 0.020 |
| … | … | LP 31-301 | J04059+712E | M4.0 V | 04:05:58.09 | +71:16:34.7 | A+BC | A | … | … | 24.98 | 415.87 | 69.28 ± 0.38 | 0.262 ± 0.013 |
| 04058+7117 | LDS1589 | LP 31-302 A | J04059+712W | M5.0 V | 04:05:57.21 | +71:16:32.4 | … | B | 4.828 | 240.7 | 24.87 | 408.76 | … | 0.148 ± 0.042 |
| 04058+7117 | LEP 123 | LP 31-302 B | … | M6.0 V | 04:05:57.02 | +71:16:31.9 | … | C | 5.832 | 241.4 | 68.12 | 413.74 | … | 0.130 ± 0.044 |
| … | … | PM J04061-0534 | J04061-055 | M3.5 V | 04:06:06.90 | -05:34:46.9 | … | … | … | … | 30.98 | 147.05 | 63.82 ± 0.27 | 0.250 ± 0.012 |
| … | … | LP 474-123 | J04077+142 | M0.0 V | 04:07:44.14 | +14:13:22.1 | A+B | A | … | … | 31.02 | 233.82 | 1038.15 ± 5.12 | 0.694 ± 0.026 |
| 04077+1413 | LDS5187 | LP 474-124 | J04079+142 | M2.5 V | 04:07:54.99 | +14:12:58.2 | … | B | 159.493 | 98.6 | 103.50 | 234.83 | 292.69 ± 1.49 | 0.496 ± 0.019 |
| … | … | LP 32-16 | J04081+743 | M3.5 V | 04:08:13.87 | +74:22:51.6 | … | … | … | … | 81.20 | 899.05 | 73.58 ± 0.32 | 0.253 ± 0.012 |
| … | … | LP 31-433 | J04083+691 | M4.5 V | 04:08:24.60 | +69:10:57.9 | … | … | … | … | 56.21 | 279.79 | 58.43 ± 0.85 | 0.256 ± 0.013 |
| … | … | HD 281621 | J04086+336 | M0.5 V | 04:08:38.07 | +33:38:15.3 | (AB) | AB | … | … | 47.41 | 541.74 | … | 0.543 ± 0.029 |
| 04094+0546 | LAW 12 | LP 534-29 | J04093+057 | M4.5 V | 04:09:22.50 | +05:46:25.2 | (AB) | AB | 0.247 | 130.0 | 47.54 | 254.91 | … | 0.153 ± 0.042 |
| … | … | LP 714-37 | J04108-128 | M5.5 V | 04:10:47.88 | -12:51:48.4 | AB | A | … | … | 22.74 | 421.99 | … | 0.136 ± 0.043 |
| 04108-1252 | PHB 1 | LP 714-37 B | … | M6.5 V | 04:10:47.98 | -12:51:49.4 | … | B | 1.722 | 123.7 | 20.83 | 418.94 | … | 0.104 ± 0.047 |
| … | … | Ross 27 | J04112+495 | M3.5 V | 04:11:13.29 | +49:31:45.0 | … | … | … | … | 44.40 | 480.41 | 108.19 ± 0.51 | 0.312 ± 0.014 |
| … | … | GJ 3266 | J04122+647 | M4.0 V | 04:12:18.25 | +64:43:48.7 | … | … | … | … | 65.43 | 657.85 | 43.32 ± 0.19 | 0.202 ± 0.011 |
| … | … | LP 414-117 | J04123+162 | M4.0 V | 04:12:21.90 | +16:15:02.9 | Aab+B | Aab(2) | … | … | 52.78 | 156.84 | … | 0.652 ± 0.052 |
| … | … | LSPM J0409+1622 | … | M5.5 V | 04:09:57.30 | +16:22:41.3 | … | B* | 2131.567 | 282.5 | 65.49 | 161.15 | … | 0.151 ± 0.042 |
| 04130+5237 | PRV1 | Ross 28 | J04129+526 | M4.0 V | 04:12:58.22 | +52:36:28.9 | (AB) | AB | 0.057 | 294.1 | 35.49 | 893.94 | … | 0.247 ± 0.037 |
| … | … | Ross 29A | J04131+505 | M4.0 V | 04:13:09.43 | +50:31:38.0 | AB | A | … | … | 37.57 | 440.74 | … | 0.256 ± 0.036 |
| 04132+5032 | CHR 15 | Ross 29B | … | M4.5 V | 04:13:09.56 | +50:31:39.1 | … | B | 1.643 | 51.1 | 43.65 | 432.33 | … | 0.213 ± 0.038 |
| … | … | LSPM J0413+4737E | J04137+476 | M2.5 V | 04:13:47.70 | +47:37:42.5 | … | … | … | … | 39.82 | 161.35 | 604.36 ± 6.30 | 0.553 ± 0.028 |
| … | … | GJ 3262 | J04139+829 | M0.0 V | 04:13:56.69 | +82:55:03.0 | … | … | … | … | 39.84 | 264.59 | 1547.11 ± 19.88 | 0.759 ± 0.025 |
| … | … | G 39-3 | J04148+277 | M3.5 Ve | 04:14:53.80 | +27:45:26.0 | … | … | … | … | 16.09 | 260.87 | 225.24 ± 1.16 | 0.461 ± 0.018 |
| … | … | HD 26965 | … | K0 V | 04:15:13.91 | -07:40:05.1 | A+B+C | A | … | … | … | 4089.84 | … | 0.845 ± 0.127 |

Table D.2: Complete sample with the description of multiple systems (continued).

| WDS id | WDS disc | Name | Karmn | Spectral type | $\alpha$ (2016.0) | $\delta$ (2016.0) | System | Component | $\rho$ [arcsec] | $\theta$ [deg] | $\varpi$ [mas] | $\mu_{\rm total}$ [mas a$^{-1}$] | $\mathcal{L}$ [$10^{-4}\,\mathcal{L}_\odot$] | $\mathcal{M}$ [$\mathcal{M}_\odot$] |
|---|---|---|---|---|---|---|---|---|---|---|---|---|---|---|
| 04153-0739 | STF 518 | HD 26976 | … | DA2.9 | 04:15:19.39 | -07:40:22.6 | … | B | 83.337 | 102.2 | 15.88 | 4018.59 | … | 0.500 ± 0.100 |
| 04153-0739 | STF 518 | DYEri | J04153-076 | M4.5 V | 04:15:19.12 | -07:40:15.3 | … | C | 78.097 | 97.5 | 54.57 | 4083.72 | 69.46 ± 0.39 | 0.266 ± 0.014 |
| … | … | GJ 2033 A | J04166-125 | M1.0 V | 04:16:41.59 | -12:33:19.3 | AB | A | … | … | 54.63 | 251.29 | … | 0.534 ± 0.029 |
| 04167-1233 | HDS 544 | GJ 2033 B | … | M2.5 V | 04:16:41.80 | -12:33:19.8 | … | B | 2.990 | 99.3 | 54.59 | 200.50 | … | 0.424 ± 0.031 |
| … | … | LP 714-47 | J04167-120 | M0.0 V | 04:16:45.65 | -12:05:05.6 | … | … | … | … | 70.87 | 203.40 | 735.47 ± 3.97 | 0.619 ± 0.027 |
| … | … | GJ 3270 | J04173+088 | M5.0 V | 04:17:18.68 | +08:49:16.0 | … | … | … | … | 32.13 | 400.23 | 68.11 ± 0.37 | 0.278 ± 0.013 |
| … | … | LSPM J0417+4103A | J04177+410 | M3.5 V | 04:17:44.44 | +41:03:10.1 | AB | A | … | … | 32.40 | 219.80 | … | 0.404 ± 0.032 |
| 04177+4103 | NSN 546 | LSPM J0417+4103B | … | M4.5 V | 04:17:44.24 | +41:03:09.2 | … | B | 2.427 | 249.6 | 62.82 | 231.80 | … | 0.224 ± 0.038 |
| … | … | HIP 20122 | J04188+013 | M2.0 V | 04:18:51.45 | +01:23:35.0 | … | … | … | … | 43.00 | 59.65 | 725.30 ± 11.14 | 0.593 ± 0.028 |
| … | … | UPM J0419+0944 | J04191+097 | M3.0 V | 04:19:08.15 | +09:44:50.2 | … | … | … | … | 71.82 | 138.27 | 90.91 ± 0.47 | 0.284 ± 0.013 |
| … | … | LP 654-39 | J04191-074 | M3.5 V | 04:19:06.41 | -07:27:43.6 | … | … | … | … | 47.80 | 182.82 | 155.23 ± 0.90 | 0.404 ± 0.017 |
| … | … | LSPM J0419+4233 | J04198+425 | M8/9V | 04:19:52.90 | +42:33:07.4 | … | … | … | … | 44.84 | 1535.49 | 4.74 ± 0.03 | 0.109 ± 0.010 |
| … | … | Ross 592 | J04199+364 | M1.5 V | 04:19:59.96 | +36:29:04.1 | … | … | … | … | 42.63 | 497.03 | 449.81 ± 2.23 | 0.525 ± 0.029 |
| … | … | PM J04205+8131 | J04205+815 | M3.0 V | 04:20:34.24 | +81:31:54.4 | … | … | … | … | 46.34 | 140.83 | 396.54 ± 1.96 | 0.490 ± 0.030 |
| … | … | XEST 16-045 | J04206+272 | M4.5 V | 04:20:39.20 | +27:17:31.4 | … | … | … | … | 82.90 | 28.01 | 1483.86 ± 14.82 | … |
| 04207+1514 | JNN261 | LP 415-363 | J04207+152 | M4.0 V | 04:20:48.16 | +15:14:08.2 | (AB) | AB | 0.220 | 91.2 | 28.24 | 181.83 | … | 0.338 ± 0.034 |
| … | … | GJ 3274 | J04218+213 | M3.5 V | 04:21:50.22 | +21:19:39.2 | … | … | … | … | 28.77 | 257.43 | 148.13 ± 0.87 | 0.369 ± 0.016 |
| … | … | GJ 3271 | J04219+751 | M3.0 V | 04:21:59.86 | +75:08:20.5 | … | … | … | … | 75.61 | 728.35 | 220.08 ± 0.93 | 0.402 ± 0.016 |
| … | … | GJ 3275 | J04221+192 | M3.0 V | 04:22:08.13 | +19:15:21.4 | … | … | … | … | 50.45 | 106.52 | 405.81 ± 2.66 | 0.487 ± 0.030 |
| … | … | TYC 78-257-1 | … | … | 04:21:04.26 | +03:16:07.9 | A+B | A | … | … | 50.64 | 139.56 | 2519.01 ± 12.78 | 0.7418 ± 0.022 |
| … | … | RX J0422.4+0337 | J04224+036 | M3.5 V | 04:22:25.19 | +03:37:08.5 | … | B* | 1748.607 | 43.9 | 25.97 | 143.01 | 208.87 ± 1.22 | 0.443 ± 0.018 |
| … | … | LP 31-339 | J04224+740 | M1.5 V | 04:22:28.52 | +74:01:21.5 | … | … | … | … | 27.07 | 327.30 | 432.58 ± 1.90 | 0.521 ± 0.029 |
| … | … | LSPM J0422+1031 | J04225+105 | M3.5 V | 04:22:32.25 | +10:31:19.3 | … | … | … | … | 44.12 | 247.15 | 255.46 ± 1.48 | 0.423 ± 0.012 |
| … | … | GJ 1070 | J04225+390 | M5.0 V | 04:22:34.31 | +39:00:34.0 | … | … | … | … | 199.61 | 844.50 | 34.15 ± 0.18 | 0.190 ± 0.011 |
| … | … | LP 415-30 | J04227+205 | M4.0 V | 04:22:42.99 | +20:34:11.9 | … | … | … | … | 199.69 | 115.40 | 183.41 ± 2.78 | 0.441 ± 0.018 |
| … | … | G 8-31 | J04229+259 | M4.5 V | 04:22:59.30 | +25:59:10.4 | … | … | … | … | 199.45 | 246.98 | 58.33 ± 0.26 | 0.238 ± 0.012 |
| … | … | TYC 3337-1716-1 | J04234+495 | M2.5 V | 04:23:26.84 | +49:34:15.6 | … | … | … | … | 45.93 | 214.40 | 330.65 ± 2.20 | 0.498 ± 0.019 |
| … | … | 1R042323.2+805511 | J04234+809 | M4.0 V | 04:23:29.60 | +80:55:08.8 | … | … | … | … | 45.62 | 121.25 | 422.78 ± 17.95 | 0.471 ± 0.030 |
| … | … | LP 535-73 | J04238+092 | M3.0 V | 04:23:50.83 | +09:12:19.4 | … | … | … | … | 19.11 | 114.41 | 673.23 ± 4.05 | 0.556 ± 0.028 |
| … | … | INTau | J04238+149 | M3.5 V | 04:23:50.50 | +14:55:17.0 | … | … | … | … | 68.56 | 117.51 | 411.84 ± 8.32 | 0.471 ± 0.030 |
| … | … | 1R042441.9-064725 | J04247-067 | M4.0 V | 04:24:42.78 | -06:47:31.2 | Aabc | Aabc(3) | … | … | 39.46 | 147.33 | … | … |
| … | … | GJ 3280 | J04248+324 | M2.5 V | 04:24:49.49 | +32:26:56.0 | … | … | … | … | 32.80 | 259.61 | 257.74 ± 1.12 | 0.437 ± 0.017 |
| … | … | PM J04251+5131 | J04251+515 | M2.0 V | 04:25:09.86 | +51:31:56.2 | … | … | … | … | 32.78 | 122.30 | 541.62 ± 27.70 | 0.536 ± 0.030 |
| … | … | GJ 3282 | J04252+080S | M2.5 V | 04:25:15.25 | +08:02:55.8 | Aab+B | Aab(2) | … | … | 52.84 | 140.04 | … | … |
| 04252+0803 | LDS3584 | GJ 3283 | J04252+080N | M4.0 V | 04:25:17.09 | +08:04:03.9 | … | B | 73.393 | 21.9 | 24.78 | 146.66 | … | 0.300 ± 0.035 |
| 04244+1705 | OCC 615 | HD 27848 | … | … | 04:24:22.38 | +17:04:43.8 | AB+C+DE+F | AB | … | … | 40.25 | 103.39 | … | … |
| … | … | V991Tau | … | K4 V | 04:25:00.35 | +16:59:05.2 | … | C | 641.168 | 121.9 | 29.97 | 94.41 | … | 0.771 ± 0.116 |
| 04252+1716 | AST 4 | V805Tau | J04252+172 | M3.5 V | 04:25:13.67 | +17:16:05.1 | … | DE | 1002.145 | 47.1 | 97.45 | 110.65 | … | 0.575 ± 0.028 |
| … | … | LP 415-881 | … | M7.0 V | 04:26:19.16 | +17:03:01.7 | … | F | 1677.610 | 93.4 | 42.21 | 105.70 | … | 0.144 ± 0.042 |
| … | … | TYC 1273-9-1 | J04274+203 | M1.5 V | 04:27:24.97 | +20:22:44.5 | … | … | … | … | 25.00 | 84.54 | 292.41 ± 1.29 | 0.440 ± 0.017 |

Table D.2: Complete sample with the description of multiple systems (continued).

| WDS id | WDS disc | Name | Karmn | Spectral type | $\alpha$ (2016.0) | $\delta$ (2016.0) | System | Component | $\rho$ [arcsec] | $\theta$ [deg] | $\varpi$ [mas] | $\mu_{\rm total}$ [mas a$^{-1}$] | $\mathcal{L}$ [$10^{-4}\ \mathcal{L}_\odot$] | $\mathcal{M}$ [$\mathcal{M}_\odot$] |
|---|---|---|---|---|---|---|---|---|---|---|---|---|---|---|
| 04277+5935 | BWL16 | GJ 3287 | J04276+595 | M3.8 V | 04:27:41.56 | +59:35:13.5 | (AB) | AB | 4.239 | 184.1 | 7.52 | 225.85 | … | 0.255 ± 0.036 |
| … | … | GJ 3291 | J04278+117 | M4.2 V | 04:27:53.89 | +11:46:46.8 | … | … | … | … | 36.85 | 593.07 | 141.19 ± 0.72 | 0.360 ± 0.015 |
| … | … | V1102Tau | J04284+176 | M2.0 V | 04:28:28.89 | +17:41:44.9 | AB | A | … | … | 48.41 | 118.08 | … | 0.645 ± 0.027 |
| 04285+1742 | GUE 5 | HG 7-232B | … | M3.0 V | 04:28:29.01 | +17:41:45.3 | … | B | 1.663 | 74.7 | 51.41 | 108.17 | … | 0.381 ± 0.032 |
| … | … | HD 28343 | J04290+219 | M0.5 V | 04:29:00.05 | +21:55:24.5 | … | … | … | … | 27.92 | 187.05 | 1182.29 ± 6.36 | 0.727 ± 0.025 |
| … | … | GJ 3292 | J04293+142 | M3.8 V | 04:29:18.76 | +14:14:02.0 | … | … | … | … | 26.97 | 300.06 | 146.35 ± 1.28 | 0.391 ± 0.017 |
| 04295+2617 | SMN 11 | FW Tau | J04294+262 | M5.5 V | 04:29:29.71 | +26:16:52.8 | (ABC) | ABC | 0.105 | 317.0 | 27.85 | 17.58 | … | … |
| … | … | PM J04302+7049 | J04302+708 | M1.5 V | 04:30:11.72 | +70:49:14.3 | … | … | … | … | 32.99 | 118.86 | 327.29 ± 1.78 | 0.467 ± 0.018 |
| … | … | V546Per | J04304+398 | M5.0 V | 04:30:25.67 | +39:50:50.4 | … | … | … | … | 48.00 | 628.65 | 35.57 ± 0.14 | 0.194 ± 0.011 |
| … | … | LP 655-23 | J04308-088 | M4.0 V | 04:30:52.04 | -08:49:22.0 | A+B | A | … | … | 49.97 | 161.83 | 145.83 ± 1.06 | 0.391 ± 0.017 |
| 04309-0849 | KO 2 | Koenigstuhl 2B | … | M8.0 V | 04:30:51.58 | -08:49:03.5 | … | B | 19.809 | 339.8 | 22.20 | 157.34 | 7.72 ± 0.15 | 0.146 ± 0.010 |
| … | … | PM J04310+3647A | J04310+367 | M3.0 V | 04:30:59.95 | +36:47:54.7 | AB | A | … | … | 57.14 | 63.11 | … | 0.401 ± 0.032 |
| … | … | PM J04310+3647B | … | M3.0 V | 04:30:59.91 | +36:47:54.0 | … | B | 0.793 | 215.4 | 46.52 | 60.11 | … | 0.366 ± 0.033 |
| … | … | GJ 169.1 A | J04311+589 | M4.0 V | 04:31:14.21 | +58:58:04.7 | A+B | A | … | … | 23.99 | 2424.35 | 95.39 ± 0.62 | 0.276 ± 0.011 |
| 04312+5858 | STI2051 | GJ 169.1 B | … | DC5 | 04:31:15.33 | +58:58:10.1 | … | B | 10.263 | 58.2 | 22.36 | 2361.13 | … | 0.500 ± 0.100 |
| … | … | PM J04312+4217 | J04312+422 | M2.5 Ve | 04:31:14.99 | +42:17:08.9 | … | … | … | … | 25.66 | 122.32 | 344.48 ± 2.07 | 0.509 ± 0.019 |
| 04314+2411 | LEI 14 | V927Tau | J04313+241 | M4.5: V+ | 04:31:23.83 | +24:10:52.6 | (AB) | AB | 0.260 | 292.0 | 28.24 | 19.77 | … | … |
| … | … | LP 475-1095 | J04326+098 | M1.5 Ve | 04:32:37.96 | +09:51:06.5 | … | … | … | … | 41.13 | 124.68 | 1291.73 ± 34.82 | 0.686 ± 0.026 |
| … | … | LP 595-23 | J04329+001E | M0.5 V | 04:32:56.07 | +00:06:14.7 | A+B+C | A | … | … | 29.04 | 184.91 | 499.69 ± 2.41 | 0.548 ± 0.028 |
| 04329+0007 | LDS 121 | G 82-28 | J04329+001S | M4.0 Ve | 04:32:55.38 | +00:06:28.3 | … | B | 17.079 | 322.7 | 29.68 | 190.36 | 114.99 ± 0.64 | 0.322 ± 0.014 |
| 04329+0007 | LDS 121 | LP 595-21 | J04329+001N | M4.0 V | 04:32:55.32 | +00:06:33.2 | … | C | 21.662 | 328.7 | 28.88 | 202.02 | 73.62 ± 0.75 | 0.271 ± 0.013 |
| … | … | V697TauA | J04333+239 | M3.0 V | 04:33:23.91 | +23:59:26.0 | AB | A | … | … | 19.96 | 111.52 | … | 0.484 ± 0.030 |
| … | … | V697TauB | … | M1.5 V | 04:33:23.87 | +23:59:26.5 | … | B | 0.767 | 309.5 | 18.36 | 107.60 | … | … |
| … | … | GJ 3296 | J04335+207 | M5.0 V | 04:33:34.48 | +20:44:40.7 | … | … | … | … | 19.31 | 557.27 | 29.93 ± 0.12 | 0.176 ± 0.011 |
| … | … | PM J04343+4302 | J04343+430 | M2.65 V | 04:34:22.55 | +43:02:13.3 | … | … | … | … | 21.20 | 94.90 | 297.59 ± 1.43 | 0.464 ± 0.013 |
| … | … | LP 595-11 | J04347-004 | M4.0 V | 04:34:45.24 | -00:26:50.2 | … | … | … | … | 40.74 | 244.65 | 93.57 ± 0.78 | 0.308 ± 0.014 |
| 04350+0840 | RAO 549 | StKM 1-495 | J04350+086 | M1.0 Ve | 04:35:02.65 | +08:39:30.5 | (AB) | AB | 0.320 | 97.0 | 43.84 | 90.95 | … | … |
| … | … | LP 775-31 | J04352-161 | M8.0 Ve | 04:35:16.33 | -16:06:52.2 | Aab | Aab(2) | … | … | 52.83 | 353.47 | … | … |
| … | … | GJ 3302 | J04366+112 | M4.0 V | 04:36:39.94 | +11:13:22.8 | … | … | … | … | 35.79 | 635.73 | 84.09 ± 0.40 | 0.291 ± 0.013 |
| … | … | LP 84-34 | J04369+593 | M2.0 V | 04:36:58.75 | +59:21:57.7 | … | … | … | … | 20.92 | 190.55 | 310.86 ± 1.24 | 0.482 ± 0.018 |
| … | … | 1R043657.1-161258 | J04369-162 | M3.5 V | 04:36:57.49 | -16:13:07.0 | … | … | … | … | 21.90 | 87.52 | 500.96 ± 4.03 | 0.498 ± 0.029 |
| … | … | LP 415-1644 | J04373+193 | M4.0 V | 04:37:22.00 | +19:21:16.9 | … | … | … | … | 89.05 | 97.37 | … | … |
| … | … | HD 232979 | J04376+528 | M0.0 V | 04:37:41.47 | +52:53:29.4 | … | … | … | … | 41.54 | 564.14 | 757.04 ± 4.92 | 0.643 ± 0.027 |
| 04376-0228 | WAL 32 | HD 29391 | … | F0 V | 04:37:36.18 | -02:28:25.8 | A+(BC) | A | 66.700 | 162.5 | … | 77.72 | … | 1.750 ± 0.050 |
| 04376-0228 | KAS 1 | GJ 3305 | J04376-024 | M1.1 V | 04:37:37.51 | -02:29:29.7 | … | BC | 66.962 | 162.6 | 38.87 | 72.38 | … | 0.706 ± 0.026 |
| … | … | GJ 173 | J04376-110 | M1.5 V | 04:37:41.62 | -11:02:23.1 | … | … | … | … | 91.84 | 299.84 | 329.57 ± 1.45 | 0.451 ± 0.012 |
| … | … | GJ 3304 A | J04382+282 | M4.0 V | 04:38:13.13 | +28:12:58.4 | AB | A | … | … | 33.16 | 404.45 | … | 0.267 ± 0.036 |
| 04382+2813 | BEU 6 | GJ 3304 B | … | M4.5 V | 04:38:13.05 | +28:12:59.0 | … | B | 1.202 | 304.5 | 31.95 | 394.70 | … | 0.228 ± 0.037 |
| … | … | LP 715-39 | J04386-115 | M3.5 V | 04:38:36.86 | -11:30:18.6 | … | … | … | … | 24.53 | 361.03 | 207.07 ± 1.39 | 0.414 ± 0.017 |
| … | … | G 8-48A | J04388+217 | M3.5 V | 04:38:53.72 | +21:47:51.7 | AB+C | A | … | … | 25.50 | 285.83 | … | 0.414 ± 0.031 |

Table D.2: Complete sample with the description of multiple systems (continued).

| WDS id | WDS disc | Name | Karmn | Spectral type | $\alpha$ (2016.0) | $\delta$ (2016.0) | System | Component | $\rho$ [arcsec] | $\theta$ [deg] | $\varpi$ [mas] | $\mu_{\rm total}$ [mas a$^{-1}$] | $\mathcal{L}$ [$10^{-4}$ $\mathcal{L}_\odot$] | $\mathcal{M}$ [$\mathcal{M}_\odot$] |
|---|---|---|---|---|---|---|---|---|---|---|---|---|---|---|
| 04390+2149 | JNN 262 | LP 358-478 | … | M5.5 Ve | 04:38:53.79 | +21:47:51.0 | … | B | 1.238 | 124.6 | 181.24 | 284.57 | … | 0.380 ± 0.032 |
| 04390+2149 | LDS1183 | G 8-48B | … | M5.0 V | 04:38:54.69 | +21:47:45.0 | … | C | 15.002 | 116.3 | 181.27 | 274.20 | 28.16 ± 0.40 | 0.198 ± 0.012 |
| … | … | V583Aur | … | K5 V | 04:39:25.47 | +33:32:43.9 | A+(BC) | A | … | … | 44.83 | 52.35 | 3249.50 ± 20.07 | … |
| 04393+3331 | JNN 263 | PM J04393+3331 | J04393+335 | M4.0 V | 04:39:23.22 | +33:31:48.7 | … | BC | 61.936 | 207.1 | … | 50.28 | … | … |
| … | … | LP 415-302 | J04395+162 | M5.5 Ve | 04:39:31.54 | +16:15:30.2 | … | … | … | … | 23.25 | 799.07 | 16.60 ± 0.07 | 0.135 ± 0.010 |
| … | … | PM J04398+2509 | J04398+251 | M3.5 V | 04:39:48.86 | +25:09:25.4 | … | … | … | … | 38.08 | 105.73 | 88.18 ± 0.43 | 0.279 ± 0.013 |
| … | … | LP 655-48 | J04403-055 | M6.0 V | 04:40:23.63 | -05:30:06.1 | … | … | … | … | 38.17 | 358.15 | 6.32 ± 0.03 | 0.107 ± 0.009 |
| … | … | GJ 9163 A | J04404-091 | M0.0 V | 04:40:29.13 | -09:11:48.5 | AB | A | … | … | 38.12 | 179.28 | … | 0.545 ± 0.028 |
| 04406-0912 | WOR 17 | GJ 9163 B | … | M1.0 V | 04:40:29.15 | -09:11:46.8 | … | B | 1.715 | 9.6 | 22.94 | 124.61 | … | 0.535 ± 0.029 |
| … | … | TOI-2457 | J04406-128 | M0.0 V | 04:40:40.16 | -12:53:26.6 | … | … | … | … | 19.49 | 24.78 | 870.01 ± 5.24 | 0.679 ± 0.026 |
| … | … | GJ 3307 | J04407+022 | M2.0 V | 04:40:42.67 | +02:13:52.7 | … | … | … | … | 73.51 | 177.22 | 590.81 ± 5.52 | 0.564 ± 0.028 |
| … | … | G 39-30A | J04413+327 | M4.0 V | 04:41:24.22 | +32:42:19.9 | AB | A | … | … | 26.59 | 310.42 | … | 0.409 ± 0.031 |
| 04413+3242 | JNN 264 | G 39-30B | … | M4.0 V | 04:41:24.22 | +32:42:21.4 | … | B | 1.473 | 359.3 | 52.82 | 301.62 | … | 0.298 ± 0.035 |
| … | … | TYC 694-1183-1 | J04414+132 | M0.5 V | 04:41:29.78 | +13:13:16.0 | Aab | Aab | … | … | 46.64 | 100.97 | … | … |
| … | … | LP 84-59 | J04422+577 | M0.0 V | 04:42:15.86 | +57:42:18.2 | … | … | … | … | 16.85 | 576.53 | 790.91 ± 8.46 | 0.641 ± 0.027 |
| … | … | LP 415-1896 | J04423+207 | M2.0 V | 04:42:23.76 | +20:46:34.9 | … | … | … | … | 94.31 | 299.49 | 477.00 ± 2.52 | 0.534 ± 0.029 |
| … | … | LP 415-345 | J04425+204 | M3.0 V | 04:42:30.40 | +20:27:10.8 | Aab+B+C | Aab(2) | … | … | 27.10 | 97.45 | … | … |
| … | … | LP 415-3051 | … | M3.0 V | 04:42:58.58 | +20:36:16.8 | … | B* | 674.349 | 35.9 | 39.11 | 91.26 | 160.48 ± 1.04 | 0.385 ± 0.016 |
| … | … | G2-3411054848866601472 | … | M6.0 V | 04:43:55.36 | +20:08:40.5 | … | C* | 1631.351 | 132.8 | 35.68 | 93.82 | 12.77 ± 0.21 | 0.148 ± 0.010 |
| … | … | PM J04429+0935 | J04429+095 | M1.0 V | 04:42:55.14 | +09:35:53.7 | A+B | A | … | … | 25.37 | 82.04 | 568.80 ± 3.59 | 0.574 ± 0.028 |
| … | … | G3- 3293060625388613248 | … | M6.5 V | 04:42:53.91 | +09:35:50.3 | … | B* | 18.386 | 259.1 | … | 77.51 | 9.72 ± 0.11 | 0.126 ± 0.010 |
| … | … | HD 285968 | J04429+189 | M2.5 V | 04:42:56.52 | +18:57:11.5 | … | … | … | … | 100.92 | 1295.36 | 358.08 ± 2.37 | 0.468 ± 0.012 |
| … | … | PM J04429+2128 | J04429+214 | M3.5 V | 04:42:55.90 | +21:28:24.8 | … | … | … | … | 33.44 | 123.58 | 181.62 ± 1.59 | 0.363 ± 0.011 |
| … | … | Haro 6-36 | J04433+296 | M5.5 V | 04:43:20.23 | +29:40:05.7 | … | … | … | … | 36.01 | 22.58 | 2282.44 ± 48.94 | … |
| … | … | HD 283779 | J04444+278 | M1.5 V | 04:44:26.17 | +27:51:37.2 | … | … | … | … | 89.19 | 437.44 | 293.26 ± 1.16 | 0.441 ± 0.017 |
| … | … | PM J04458-1426 | J04458-144 | M4.0 V | 04:45:52.69 | -14:26:23.6 | … | … | … | … | 75.67 | 142.89 | 218.23 ± 1.50 | 0.453 ± 0.018 |
| … | … | PM J04468-1116A | J04468-112 | M3.0 V | 04:46:51.63 | -11:16:48.6 | AB | A | … | … | 76.31 | 153.74 | … | 0.372 ± 0.032 |
| 04469-1117 | JNN 28 | PM J04468-1116B | … | M6.0 V | 04:46:51.53 | -11:16:48.2 | … | B | 1.518 | 285.0 | 48.75 | 147.07 | … | 0.289 ± 0.035 |
| … | … | GJ 3313 | J04471+021 | M0.0 V | 04:47:11.55 | +02:09:39.6 | … | … | … | … | 22.45 | 193.33 | 538.44 ± 3.03 | 0.569 ± 0.028 |
| … | … | RX J0447.2+2038 | J04472+206 | M5.0 V | 04:47:12.35 | +20:38:09.2 | … | … | … | … | 21.89 | 132.32 | 123.97 ± 0.75 | 0.428 ± 0.032 |
| … | … | LP 416-43 | J04480+170 | M0.5 V | 04:48:00.98 | +17:03:21.1 | Aab+B | Aab | … | … | 33.65 | 82.10 | … | 0.13180 …0.004 |
| … | … | UCAC4 536-010184 | … | M4.5 V | 04:49:36.67 | +17:01:58.6 | … | B* | 1374.841 | 93.4 | 21.57 | 87.10 | 75.26 ± 0.47 | 0.293 ± 0.014 |
| … | … | 1R044847.6+100302 | J04488+100 | M3.0 Ve | 04:48:47.32 | +10:03:01.4 | Aab | Aab(2) | … | … | 94.37 | 101.41 | … | … |
| 04494+4828 | JNN265 | G 81-34 | J04494+484 | M4.0 V | 04:49:29.77 | +48:28:42.9 | (AB) | AB | 0.740 | 250.0 | 11.13 | 264.56 | … | 0.304 ± 0.035 |
| … | … | EM* LkCa 18A | J04499+236 | M1.0 V | 04:49:56.34 | +23:41:00.1 | AB | A | … | … | 7.43 | 173.62 | … | 0.647 ± 0.027 |
| 04499+2341 | KPP3177 | EM* LkCa 18B | … | M1.5 V | 04:49:56.51 | +23:41:00.1 | … | B | 2.380 | 90.0 | 83.99 | 162.99 | … | 0.509 ± 0.029 |
| … | … | LP 32-204 | J04499+711 | M3.5 V | 04:49:56.29 | +71:09:46.5 | … | … | … | … | 47.26 | 183.39 | 132.65 ± 0.70 | 0.348 ± 0.015 |
| … | … | GJ 3315 | J04502+459 | M1.0 V | 04:50:15.59 | +45:58:46.2 | … | … | … | … | 102.59 | 305.58 | 397.19 ± 1.23 | 0.512 ± 0.029 |
| … | … | BPM 85800 | J04504+199 | M1.5 V | 04:50:25.49 | +19:59:09.1 | … | … | … | … | 49.43 | 145.65 | 471.15 ± 2.40 | 0.533 ± 0.029 |
| … | … | GJ 1072 | J04508+221 | M5.0 V | 04:50:51.65 | +22:07:14.7 | … | … | … | … | 49.11 | 759.23 | 21.98 ± 0.10 | 0.159 ± 0.010 |

Table D.2: Complete sample with the description of multiple systems (continued).

| WDS id | WDS disc | Name | Karmn | Spectral type | $\alpha$ (2016.0) | $\delta$ (2016.0) | System | Component | $\rho$ [arcsec] | $\theta$ [deg] | $\varpi$ [mas] | $\mu_{total}$ [mas a$^{-1}$] | $\mathcal{L}$ [$10^{-4}$ $\mathcal{L}_\odot$] | $\mathcal{M}$ [$\mathcal{M}_\odot$] |
|---|---|---|---|---|---|---|---|---|---|---|---|---|---|---|
| ... | ... | GJ 3316 | J04508+261 | M2.5 V | 04:50:51.21 | +26:07:22.3 | ... | ... | ... | ... | 16.29 | 623.45 | 172.89 ± 0.97 | 0.376 ± 0.016 |
| ... | ... | Wolf 1539 | J04520+064 | M3.5 V | 04:52:05.90 | +06:28:30.7 | ... | ... | ... | ... | 42.69 | 342.44 | 158.33 ± 1.02 | 0.350 ± 0.012 |
| ... | ... | LP 776-25 | J04524-168 | M3.3 V | 04:52:24.55 | -16:49:25.3 | ... | ... | ... | ... | 24.93 | 243.44 | 295.75 ± 1.51 | 0.499 ± 0.019 |
| ... | ... | GJ 1073 | J04525+407 | M5.0 V | 04:52:36.26 | +40:42:06.6 | ... | ... | ... | ... | 24.96 | 1606.93 | 58.14 ± 0.25 | 0.238 ± 0.012 |
| ... | ... | LP 84-48 | J04536+623 | M3.5 V | 04:53:40.79 | +62:18:59.9 | ... | ... | ... | ... | 22.66 | 361.91 | 96.58 ± 0.35 | 0.293 ± 0.013 |
| ... | ... | LSPM J0453+1549 | J04538+158 | M2.5 V | 04:53:50.10 | +15:49:12.6 | ... | ... | ... | ... | 29.47 | 176.11 | 166.00 ± 0.73 | 0.368 ± 0.015 |
| ... | ... | GJ 180 | J04538-177 | M2.0 V | 04:53:50.44 | -17:46:34.6 | ... | ... | ... | ... | 35.55 | 763.06 | 236.80 ± 1.12 | 0.398 ± 0.011 |
| ... | ... | 1R045430.9+650451 | J04544+650 | M4.0 V | 04:54:29.98 | +65:04:39.5 | ... | ... | ... | ... | 20.48 | 118.11 | 73.45 ± 0.30 | 0.270 ± 0.013 |
| 04559+0440 | EGN 4 | HD 31412 | ... | F9.5 V | 04:55:56.03 | +04:40:10.5 | (AB)+C | AB | 0.443 | 17.7 | 19.43 | 233.46 | ... | 1.114 ± 0.167 |
| 04559+0440 | LDS9181 | HD 31412B | J04559+046 | M2.0 V | 04:55:54.60 | +04:40:13.5 | ... | C | 21.657 | 277.9 | 19.48 | 234.00 | 791.43 ± 5.69 | 0.600 ± 0.027 |
| ... | ... | LP 202-2 | J04560+432 | M4.0 V | 04:56:04.12 | +43:13:53.0 | ... | ... | ... | ... | 28.74 | 420.36 | 66.13 ± 0.38 | 0.255 ± 0.012 |
| ... | ... | GJ 1074 | J04587+509 | M1.0 V | 04:58:46.84 | +50:56:32.4 | ... | ... | ... | ... | 28.68 | 604.82 | 513.22 ± 2.02 | 0.558 ± 0.028 |
| ... | ... | GJ 181 | J04588+498 | M0.0 V | 04:58:50.76 | +49:50:55.6 | ... | ... | ... | ... | 105.43 | 159.39 | 796.11 ± 3.90 | 0.642 ± 0.027 |
| ... | ... | GJ 182 | J04595+017 | M0.0 Ve | 04:59:34.88 | +01:46:59.2 | ... | ... | ... | ... | 71.90 | 102.65 | 1422.03 ± 6.60 | 0.739 ± 0.025 |
| ... | ... | Ross 794 | J05012+248 | M2.0 V | 05:01:15.79 | +24:52:18.2 | ... | ... | ... | ... | 6.02 | 456.24 | 456.28 ± 2.43 | 0.519 ± 0.029 |
| ... | ... | LSPM J0501+2237 | J05013+226 | M4.5 V | 05:01:17.95 | +22:36:55.3 | ... | ... | ... | ... | 60.82 | 356.57 | 23.39 ± 0.10 | 0.165 ± 0.010 |
| ... | ... | GJ 3321 | J05018+037 | M1.5 V | 05:01:50.71 | +03:45:53.1 | ... | ... | ... | ... | 39.10 | 186.47 | 337.11 ± 1.66 | 0.474 ± 0.018 |
| ... | ... | 1R050156.7+010845 | J05019+011 | M4.0 V | 05:01:56.70 | +01:08:41.4 | ... | ... | ... | ... | 52.77 | 97.03 | 340.51 ± 2.04 | 0.432 ± 0.031 |
| ... | ... | GJ 3322 A | J05019+099 | M4.0 V | 05:01:58.83 | +09:58:57.1 | AabB | Aab(2) | ... | ... | 31.90 | 125.62 | ... | 0.022 |
| 05020+0959 | HDS 654 | GJ 3322 B | ... | M2.5 V | 05:01:58.89 | +09:58:55.9 | ... | B | 1.398 | 146.7 | 52.85 | 128.14 | ... | 0.426 ± 0.031 |
| ... | ... | GJ 3323 | J05019-069 | M4.0 Ve | 05:01:56.83 | -06:56:54.9 | ... | ... | ... | ... | 38.24 | 767.60 | 36.40 ± 0.24 | 0.173 ± 0.009 |
| ... | ... | HD 32450A | J05024-212 | M2.0 V | 05:02:28.28 | -21:15:28.4 | AB | A | ... | ... | 42.99 | 316.71 | ... | 0.611 ± 0.027 |
| 05025-2115 | DON 91 | HD 32450B | ... | M3.0 V | 05:02:28.26 | -21:15:27.5 | ... | B | 0.888 | 345.8 | 20.16 | 260.93 | ... | 0.372 ± 0.032 |
| ... | ... | HD 285190 | J05032+213 | M1.5 V+ | 05:03:16.21 | +21:23:54.0 | Aab+(BC) | Aab(2) | ... | ... | 43.63 | 169.67 | ... | ... |
| 05033+2125 | LAW 13 | LP 359-186 | ... | M5.0 V | 05:03:05.77 | +21:22:33.8 | ... | BC | 166.404 | 241.2 | 19.84 | 177.43 | ... | 0.251 ± 0.045 |
| ... | ... | GJ 3325 | J05033-173 | M3.0 V | 05:03:19.83 | -17:22:31.8 | ... | ... | ... | ... | 57.90 | 499.86 | 92.97 ± 0.49 | 0.259 ± 0.010 |
| ... | ... | GJ 184 | J05034+531 | M0.5 V | 05:03:26.22 | +53:07:17.9 | A+B | A | ... | ... | 44.27 | 2014.91 | 515.22 ± 2.40 | 0.563 ± 0.028 |
| 05034+5308 | WDK 1 | BD+52 911B | ... | ... | 05:03:25.60 | +53:07:18.7 | ... | B | 5.600 | 278.5 | 24.62 | 2030.03 | ... | 0.100 ± 0.050 |
| ... | ... | GJ 3326 | J05042+110 | M5.0 V | 05:04:14.69 | +11:03:27.0 | ... | ... | ... | ... | 24.57 | 204.64 | 30.94 ± 0.14 | 0.180 ± 0.011 |
| ... | ... | UPM J0505+4414 | J05050+442 | M5.0 V | 05:05:06.06 | +44:14:03.3 | ... | ... | ... | ... | 38.34 | 99.41 | 27.86 ± 0.15 | 0.169 ± 0.010 |
| ... | ... | GJ 3327 | J05051-120 | M3.0 V | 05:05:11.55 | -12:00:30.9 | ... | ... | ... | ... | 39.39 | 260.57 | 154.14 ± 0.96 | 0.354 ± 0.015 |
| ... | ... | GJ 3328 | J05060+043 | M1.0 V | 05:06:04.44 | +04:20:16.1 | ... | ... | ... | ... | 33.18 | 413.12 | 393.82 ± 2.32 | 0.511 ± 0.029 |
| ... | ... | RX J0506.2+0439 | J05062+046 | M4.0 V | 05:06:12.96 | +04:39:25.7 | ... | ... | ... | ... | 79.04 | 94.79 | 280.33 ± 1.85 | 0.398 ± 0.032 |
| ... | ... | GJ 3331 | J05068-215E | M1.5 V | 05:06:49.97 | -21:35:09.4 | A+BC | A | ... | ... | 43.28 | 49.64 | 931.51 ± 8.79 | 0.627 ± 0.027 |
| 05069-2135 | DON 93 | GJ 3332 | J05068-215W | M3.5 V | 05:06:49.48 | -21:35:04.3 | ... | B | 8.489 | 306.7 | 80.56 | 65.51 | ... | 0.516 ± 0.029 |
| 05069-2135 | DON 93 | BD-21 1074C | ... | M2.5 V | 05:06:49.55 | -21:35:04.7 | ... | C | 7.560 | 308.5 | 63.13 | 69.22 | ... | 0.424 ± 0.031 |
| ... | ... | RX J0507.2+3731A | J05072+375 | M5.0 V | 05:07:14.33 | +37:30:42.1 | AB | A | ... | ... | 74.07 | 102.11 | ... | 0.187 ± 0.041 |
| ... | ... | RX J0507.2+3731B | ... | M5.0 V | 05:07:14.37 | +37:30:42.1 | ... | B* | 0.476 | 93.2 | 54.28 | 96.09 | ... | 0.166 ± 0.041 |
| ... | ... | TYC 1853-1649-1 | J05076+275 | M0.5 V | 05:07:36.74 | +27:30:03.8 | ... | ... | ... | ... | 38.54 | 107.00 | ... | ... |
| 05078+1759 | CRC 48 | Wolf 230 | J05078+179 | M3.0 V | 05:07:49.34 | +17:58:53.5 | Aabc | Aabc(2) | 0.063 | 84.4 | 48.89 | 263.54 | ... | ... |

Table D.2: Complete sample with the description of multiple systems (continued).

| WDS id | WDS disc | Name | Karmn | Spectral type | $\alpha$ (2016.0) | $\delta$ (2016.0) | System | Component | $\rho$ [arcsec] | $\theta$ [deg] | $\varpi$ [mas] | $\mu_{\rm total}$ [mas a$^{-1}$] | $\mathcal{L}$ [$10^{-4}$ $\mathcal{L}_\odot$] | $\mathcal{M}$ [$\mathcal{M}_\odot$] |
|---|---|---|---|---|---|---|---|---|---|---|---|---|---|---|
| 05083+7538 | JNN 266 | LP 15-315 | J05083+756 | M4.5 V | 05:08:19.36 | +75:38:13.3 | (AB) | AB | 0.095 | 352.3 | 83.69 | 237.23 | ... | 0.201 ± 0.039 |
| 05085-2102 | YMG 9 | 2M05082729-2101444 | J05084-210 | M5.0 V | 05:08:27.34 | -21:01:44.6 | (AB) | AB | 0.029 | 272.6 | 51.00 | 36.58 | ... | 0.414 ± 0.031 |
| 05086-1810 | WSI 72 | GJ 190 | J05085-181 | M3.5 V | 05:08:35.61 | -18:10:41.8 | (AB) | AB | 0.041 | 224.5 | 27.86 | 1487.47 | ... | 0.465 ± 0.030 |
| ... | ... | Ross 388 | J05091+154 | M3.5 V | 05:09:10.11 | +15:27:22.8 | ... | ... | ... | ... | 27.85 | 649.24 | 395.01 ± 2.83 | 0.488 ± 0.030 |
| ... | ... | G 97-23 | J05103+095 | M2.0 V | 05:10:18.12 | +09:30:01.8 | ... | ... | ... | ... | 62.46 | 353.27 | 578.29 ± 4.23 | 0.556 ± 0.028 |
| ... | ... | LSPM J0510+2714 | J05103+272 | M7.0 V | 05:10:19.83 | +27:13:51.8 | ... | ... | ... | ... | 47.50 | 665.89 | 6.92 ± 0.04 | 0.113 ± 0.010 |
| ... | ... | GJ 3336 A | J05103+488 | M2.5 V | 05:10:22.30 | +48:50:26.3 | AB | A | ... | ... | 60.19 | 428.14 | ... | 0.464 ± 0.030 |
| 05104+4850 | HEI 321 | GJ 3336 B | ... | M2.0 V | 05:10:22.48 | +48:50:25.5 | ... | B | 2.009 | 112.5 | 40.99 | 426.80 | ... | 0.457 ± 0.030 |
| ... | ... | G 86-28 | J05106+297 | M3.0 V | 05:10:39.22 | +29:46:48.9 | ... | ... | ... | ... | 44.25 | 253.57 | 83.14 ± 0.42 | 0.270 ± 0.013 |
| ... | ... | GJ 3337 | J05109+186 | M4.0 V | 05:10:57.16 | +18:37:24.1 | ... | ... | ... | ... | 69.89 | 690.53 | 68.29 ± 0.32 | 0.260 ± 0.013 |
| ... | ... | StKM 1-549 | J05111+158 | M1.0 Ve | 05:11:09.78 | +15:48:57.0 | ... | ... | ... | ... | 53.08 | 63.81 | 1171.33 ± 52.24 | 0.690 ± 0.027 |
| ... | ... | LP 477-36 | J05114+101 | M1.0 V | 05:11:29.68 | +10:07:12.2 | ... | ... | ... | ... | 39.54 | 221.80 | 936.70 ± 10.56 | 0.650 ± 0.027 |
| ... | ... | GJ 192 | J05127+196 | M2.0 V | 05:12:42.54 | +19:40:00.2 | ... | ... | ... | ... | 42.04 | 368.87 | 269.16 ± 1.49 | 0.421 ± 0.011 |
| ... | ... | GJ 3340 | J05151-073 | M1.0 V | 05:15:08.33 | -07:20:55.4 | ... | ... | ... | ... | 42.10 | 508.34 | 343.64 ± 1.78 | 0.479 ± 0.018 |
| ... | ... | UPM J0515+2336 | J05152+236 | M5.0 V | 05:15:17.58 | +23:36:25.2 | ... | ... | ... | ... | 186.05 | 65.40 | 150.58 ± 4.90 | 0.349 ± 0.016 |
| ... | ... | LSPM J0515+5911 | J05155+591 | M7.5 V | 05:15:31.19 | +59:11:01.3 | ... | ... | ... | ... | 119.57 | 1012.71 | 9.43 ± 0.11 | 0.136 ± 0.010 |
| 05173+3208 | RAO 201 | G 86-37 | J05173+321 | M3.5 V | 05:17:20.09 | +32:07:29.7 | (AB) | AB | 0.170 | 214.5 | 118.82 | 308.44 | ... | 0.441 ± 0.031 |
| ... | ... | Capella | ... | G3 III: | ... | ... | Aab+BC | Aab(2) | ... | ... | 46.83 | 433.47 | ... | 4.909 ± 0.022 |
| 05167+4600 | FRH 1 | Capella H | J05173+458 | M1.0 V | 05:17:24.00 | +45:50:16.1 | ... | B | 722.840 | 141.8 | 48.52 | 437.93 | ... | 0.537 ± 0.029 |
| 05167+4600 | ST 3 | Capella L | ... | M4.0 V | 05:17:24.04 | +45:50:12.6 | ... | C | 3.457 | 173.5 | 43.50 | 420.38 | ... | 0.272 ± 0.036 |
| ... | ... | TYC 4351-466-1 | J05173+721 | M1.0 V | 05:17:21.23 | +72:10:49.8 | ... | ... | ... | ... | 108.27 | 117.79 | 416.93 ± 1.59 | 0.500 ± 0.018 |
| ... | ... | PM J05187+4629 | J05187+464 | M4.5 V | 05:18:44.62 | +46:29:57.9 | ... | ... | ... | ... | 72.17 | 115.11 | 321.34 ± 4.10 | 0.412 ± 0.031 |
| ... | ... | 1R051929.3+645435 | J05195+649 | M3.5 V | 05:19:31.21 | +64:54:36.2 | ... | ... | ... | ... | 72.24 | 151.99 | 251.64 ± 2.88 | 0.399 ± 0.032 |
| ... | ... | GJ 3342 | J05206+587N | M3.5 V | 05:20:41.70 | +58:47:25.2 | A+B | A | ... | ... | 98.23 | 524.57 | 224.15 ± 2.80 | 0.459 ± 0.019 |
| 05207+5848 | GIC 46 | GJ 3343 | J05206+587S | M3.5 V | 05:20:41.05 | +58:47:12.0 | ... | B | 14.107 | 200.9 | 74.14 | 519.48 | 101.44 ± 0.59 | 0.322 ± 0.014 |
| ... | ... | GJ 3345 | J05211+557 | M3.5 V | 05:21:10.47 | +55:45:46.5 | ... | ... | ... | ... | 48.19 | 309.78 | 222.36 ± 1.46 | 0.429 ± 0.017 |
| ... | ... | PM J05223+3031 | J05223+305 | M3.0 V | 05:22:20.65 | +30:31:08.3 | ... | ... | ... | ... | 44.04 | 113.73 | 210.68 ± 0.94 | 0.445 ± 0.018 |
| ... | ... | TYC 4532-731-1 | J05226+795 | M0.5 V | 05:22:39.87 | +79:34:30.3 | ... | ... | ... | ... | 36.19 | 100.16 | 555.01 ± 2.35 | 0.575 ± 0.028 |
| ... | ... | PM J05228+2016 | J05228+202 | M2.5 Ve | 05:22:50.18 | +20:16:36.5 | AB | A | ... | ... | 50.43 | 46.07 | ... | 0.487 ± 0.030 |
| ... | ... | G3- 3402193923314210944 | ... | ... | 05:22:50.43 | +20:16:36.6 | ... | B | 3.548 | 87.2 | 50.52 | ... | ... | ... |
| ... | ... | PM J05243-1601A | J05243-160 | M4.5 V | 05:24:19.14 | -16:01:15.8 | AB | A | ... | ... | 50.58 | 39.99 | ... | 0.371 ± 0.033 |
| 05243-1601 | BRG 21 | PM J05243-1601B | ... | M3.5 V | 05:24:19.17 | -16:01:15.6 | ... | B | 0.444 | 66.9 | 43.80 | 34.25 | ... | 0.339 ± 0.033 |
| ... | ... | LP 717-36 | J05256-091 | M3.5 V | 05:25:41.70 | -09:09:15.8 | AB | A | ... | ... | 44.02 | 205.70 | ... | 0.353 ± 0.033 |
| 05257-0909 | DAE 2 | G3- 3014628959425010688 | ... | M3.5 V | 05:25:41.73 | -09:09:15.3 | ... | B | 0.723 | 41.9 | ... | 183.57 | ... | 0.310 ± 0.034 |
| ... | ... | Ross 41 | J05280+096 | M3.5 V | 05:27:59.94 | +09:38:26.0 | ... | ... | ... | ... | 86.10 | 780.65 | 67.93 ± 0.31 | 0.221 ± 0.009 |
| ... | ... | GJ 1080 | J05282+029 | M3.5 V | 05:28:14.24 | +02:57:56.6 | Aab+B | Aab | ... | ... | 33.42 | 1157.54 | ... | ... |
| 05282+0258 | SKF 270 | GJ 1080 B | ... | M6.0 V | 05:28:14.27 | +02:57:51.9 | ... | B | 4.698 | 175.2 | 66.20 | 1168.06 | ... | 0.122 ± 0.044 |
| ... | ... | HD 35956 | ... | G0 V | 05:28:51.74 | +12:32:59.6 | Aab+B+CD | Aab(1) | ... | ... | 56.12 | 231.78 | ... | ... |
| 05289+1233 | TOK 94 | HD 35956B | ... | M1.0 V | 05:28:52.11 | +12:33:01.4 | ... | B | 5.830 | 71.4 | 20.70 | 237.81 | ... | 0.564 ± 0.028 |
| 05289+1233 | LDS6186 | GJ 3348A | J05289+125 | M4.0 V | 05:28:56.61 | +12:31:50.4 | ... | C | 99.391 | 134.1 | 108.33 | 230.59 | ... | 0.263 ± 0.060 |

Table D.2: Complete sample with the description of multiple systems (continued).

| WDS id | WDS disc | Name | Karmn | Spectral type | $\alpha$ (2016.0) | $\delta$ (2016.0) | System | Component | $\rho$ [arcsec] | $\theta$ [deg] | $\varpi$ [mas] | $\mu_{total}$ [mas a$^{-1}$] | $\mathcal{L}$ [$10^{-4}$ $\mathcal{L}_\odot$] | $\mathcal{M}$ [$\mathcal{M}_\odot$] |
|---|---|---|---|---|---|---|---|---|---|---|---|---|---|---|
| 05289+1233 | RAO 552 | GJ 3348B | … | … | 05:28:56.61 | +12:31:50.2 | … | D | 99.461 | 134.2 | 33.57 | … | … | … |
| … | … | GJ 2043 | J05294+155E | M0.0: V | 05:29:26.92 | +15:34:36.2 | A+B | A | … | … | 27.45 | 152.85 | 456.65 ± 3.51 | 0.543 ± 0.029 |
| 05296+1534 | LDS6187 | GJ 2043 B | J05294+155W | M4.0 V | 05:29:26.02 | +15:34:43.4 | … | B | 14.862 | 298.9 | 97.32 | 138.81 | 32.76 ± 0.13 | 0.173 ± 0.010 |
| … | … | Ross 406 | J05298+320 | M2.5 V | 05:29:52.40 | +32:04:40.4 | … | … | … | … | 42.84 | 709.93 | 245.33 ± 0.93 | 0.426 ± 0.017 |
| … | … | Wolf 1450 | J05298-034 | M3.0 V | 05:29:51.70 | -03:26:37.6 | … | … | … | … | 42.82 | 559.50 | 215.44 ± 1.01 | 0.422 ± 0.017 |
| … | … | LSPM J0530+1514 | J05306+152 | M3.0 V | 05:30:37.09 | +15:14:26.3 | … | … | … | … | 80.34 | 182.76 | 147.29 ± 0.80 | 0.345 ± 0.015 |
| … | … | HD 36395 | J05314-036 | M1.5 V | 05:31:28.21 | -03:41:11.5 | … | … | … | … | 47.04 | 2226.53 | 657.08 ± 16.58 | 0.590 ± 0.028 |
| 05321-0305 | JNN 39 | V1311Ori | J05320-030 | M2.0 V+ | 05:32:04.51 | -03:05:30.0 | (AB)+C+D+E | AB | 0.171 | 293.0 | 19.12 | 41.38 | … | 0.700 ± 0.027 |
| 05321-0305 | JNN 39 | PM J05319-0303W | … | M5.0 V | 05:31:57.88 | -03:03:37.6 | … | C | 150.006 | 318.5 | 22.63 | 54.71 | 337.69 ± 5.08 | 0.394 ± 0.032 |
| 05321-0305 | JNN 39 | 2M05315816-0303397 | … | M3.5 V | 05:31:58.17 | -03:03:40.7 | … | D | 144.834 | 319.0 | 82.31 | 54.64 | 175.72 ± 1.03 | 0.461 ± 0.019 |
| 05321-0305 | JNN 39 | ESO-HA 737 | … | M5.0 V | 05:32:05.97 | -03:01:16.8 | … | E | 254.187 | 4.9 | 46.25 | 51.30 | 110.16 ± 0.67 | 0.360 ± 0.016 |
| … | … | Ross 42 | J05322+098 | M3.5 V | 05:32:14.46 | +09:49:11.5 | Aab | Aab(2) | … | … | 75.57 | 281.96 | … | 0.400 |
| … | … | G 98-7 | J05328+338 | M3.5 V | 05:32:51.57 | +33:49:39.8 | … | … | … | … | 24.47 | 493.99 | 53.78 ± 0.23 | 0.213 ± 0.011 |
| 05333+4449 | AST 3 | GJ 1081 | J05333+448 | M3.5 V | 05:33:19.20 | +44:48:52.9 | (AB) | AB | 0.323 | 46.7 | 63.36 | 381.20 | … | 0.363 ± 0.033 |
| … | … | V371Ori | J05337+019 | M3.0 Ve | 05:33:44.55 | +01:56:41.0 | Aab | Aab(1) | … | … | 39.91 | 278.07 | … | … |
| … | … | RX J0534.0-0221 | J05339-023 | M3.0 V | 05:33:59.83 | -02:21:33.3 | … | … | … | … | 30.56 | 59.18 | 660.15 ± 4.91 | 0.552 ± 0.028 |
| … | … | PM J05334+4809 | … | M0.0 V | 05:33:28.97 | +48:09:26.2 | A+BC+D | A | … | … | 76.20 | 66.93 | 653.27 ± 2.87 | 0.610 ± 0.027 |
| … | … | PM J05341+4732A | J05341+475 | M2.5 V | 05:34:10.56 | +47:32:02.8 | … | B* | 2282.136 | 169.4 | 74.95 | 69.01 | … | 0.492 ± 0.030 |
| … | … | PM J05341+4732B | … | M3.0 V | 05:34:10.62 | +47:32:05.2 | … | C* | 2279.958 | 169.3 | 75.18 | 61.85 | … | 0.367 ± 0.033 |
| … | … | UPM J0533+4809 | … | M3.5 V | 05:33:16.22 | +48:09:23.3 | … | D* | 127.551 | 268.7 | 36.27 | 65.01 | 119.19 ± 2.21 | 0.351 ± 0.016 |
| … | … | GJ 3352 | J05341+512 | M1.0 V | 05:34:08.57 | +51:12:52.8 | … | … | … | … | 20.90 | 227.66 | 623.11 ± 2.49 | 0.592 ± 0.028 |
| … | … | Ross 45 | J05342+103N | M3.0 V | 05:34:15.05 | +10:19:08.0 | A+Bab | A | … | … | 37.69 | 387.58 | 258.09 ± 1.69 | 0.465 ± 0.018 |
| 05342+1019 | LDS6189 | Ross 45B | J05342+103S | M4.5 V | 05:34:15.00 | +10:19:03.0 | … | Bab | 5.028 | 188.0 | 50.35 | 379.22 | … | … |
| … | … | Ross 46 | J05348+138 | M3.0 V | 05:34:51.99 | +13:52:40.4 | … | … | … | … | 35.20 | 413.93 | 156.71 ± 0.61 | 0.343 ± 0.011 |
| … | … | Wolf 1457 | J05360-076 | dM4.0 | 05:36:00.20 | -07:38:51.0 | … | … | … | … | 38.86 | 479.46 | 126.63 ± 0.76 | 0.314 ± 0.011 |
| … | … | V2689Ori | J05365+113 | M0.0 V | 05:36:30.99 | +11:19:39.4 | A+B | A | … | … | 37.12 | 56.42 | 821.94 ± 4.56 | 0.654 ± 0.027 |
| 05365+1120 | TOK255 | PM J05366+1117 | J05366+112 | M4.0 Ve | 05:36:38.46 | +11:17:47.8 | … | B | 156.630 | 135.4 | 34.81 | 60.93 | 90.40 ± 0.73 | 0.273 ± 0.010 |
| … | … | LSPM J0539+4038 | J05394+406 | M8.0 Ve | 05:39:25.71 | +40:38:29.5 | … | … | … | … | 39.76 | 1055.41 | 5.89 ± 0.03 | 0.113 ± 0.010 |
| … | … | LP 33-191 | J05394+747 | M3.5 V | 05:39:25.44 | +74:46:02.6 | … | … | … | … | 34.70 | 147.59 | 120.02 ± 0.57 | 0.330 ± 0.014 |
| … | … | V1402Ori | J05402+126 | M1.5 Ve | 05:40:16.07 | +12:38:56.4 | Aab | Aab(2) | … | … | 57.08 | 261.39 | … | … |
| … | … | V538Aur | … | K1 V | 05:41:20.34 | +53:28:43.4 | A+B | A | … | … | … | 523.48 | 4863.31 ± 137.03 | 0.860 ± 0.129 |
| 05413+5329 | ENG 22 | HD 233153 | J05415+534 | M1.0 V | 05:41:30.74 | +53:29:15.0 | … | B | 98.035 | 71.2 | 32.06 | 515.97 | 573.08 ± 2.66 | 0.568 ± 0.028 |
| 05404+2448 | WNO45 | V780Tau | J05404+248 | M5.5 V | 05:40:25.82 | +24:48:02.0 | (AB) | AB | 0.658 | 92.2 | 32.27 | 390.93 | … | 0.132 ± 0.043 |
| … | … | GJ 9188 | J05419+153 | M0.0 V | 05:41:58.95 | +15:20:13.3 | … | … | … | … | 46.20 | 90.07 | 702.52 ± 2.67 | 0.621 ± 0.027 |
| … | … | V1352Ori | J05421+124 | M4.0 V | 05:42:11.45 | +12:28:56.5 | … | … | … | … | 45.57 | 2540.13 | 63.27 ± 0.27 | 0.229 ± 0.010 |
| … | … | GJ 2045 | J05422-054 | M5.0 V | 05:42:12.53 | -05:27:40.3 | … | … | … | … | 97.89 | 969.56 | 18.33 ± 0.08 | 0.143 ± 0.010 |
| … | … | 1R054232.1+152459 | J05425+154 | M3.5 V | 05:42:31.70 | +15:25:00.2 | … | … | … | … | 47.51 | 106.30 | … | … |
| … | … | PM J05455-1158 | J05455-119 | M4.5 V | 05:45:32.04 | -11:58:02.3 | … | … | … | … | 47.22 | 86.59 | 92.76 ± 0.46 | 0.307 ± 0.014 |
| … | … | PM J05456+7255 | J05456+729 | M3.0 V | 05:45:39.09 | +72:55:14.6 | … | … | … | … | 33.79 | 139.63 | 218.16 ± 1.03 | 0.425 ± 0.017 |
| … | … | PM J05458+7254 | J05458+729 | M2.5 V | 05:45:50.02 | +72:54:09.0 | … | … | … | … | 35.40 | 139.41 | 224.62 ± 1.09 | 0.432 ± 0.017 |

Table D.2: Complete sample with the description of multiple systems (continued).

| WDS id | WDS disc | Name | Karmn | Spectral type | $\alpha$ (2016.0) | $\delta$ (2016.0) | System | Component | $\rho$ [arcsec] | $\theta$ [deg] | $\varpi$ [mas] | $\mu_{\rm total}$ [mas a$^{-1}$] | $\mathcal{L}$ [$10^{-4}$ $\mathcal{L}_\odot$] | $\mathcal{M}$ [$\mathcal{M}_\odot$] |
|---|---|---|---|---|---|---|---|---|---|---|---|---|---|---|
| 05466+4407 | CRC 49 | Wolf 237 | J05466+441 | M4.0 V | 05:46:37.60 | +44:07:14.0 | AabB | AabB(2) | 3.712 | 222.7 | 27.67 | 670.09 | … | … |
| … | … | TYC 4106-420-1 | J05468+665 | M0.5 V | 05:46:48.92 | +66:30:13.1 | … | … | … | … | 39.50 | 121.53 | 635.41 ± 5.89 | 0.574 ± 0.028 |
| … | … | GJ 3366 | J05471-052 | M4.5 V | 05:47:09.69 | -05:12:20.3 | … | … | … | … | … | 768.26 | 40.79 ± 0.17 | 0.195 ± 0.011 |
| … | … | GJ 3367 | J05472-000 | M0 | 05:47:17.89 | -00:00:49.9 | … | … | … | … | 58.20 | 102.95 | 707.09 ± 3.86 | 0.618 ± 0.027 |
| … | … | GJ 3368 | J05484+077 | M4.0 V | 05:48:24.15 | +07:45:34.3 | … | … | … | … | 58.23 | 282.11 | 86.78 ± 0.40 | 0.296 ± 0.014 |
| … | … | PM J05511+1216 | J05511+122 | M4.0 V | 05:51:10.51 | +12:16:09.4 | … | … | … | … | 46.64 | 89.83 | 190.38 ± 3.85 | 0.372 ± 0.016 |
| … | … | G 106-7 | J05530+047 | M1.5 V | 05:53:04.75 | +04:43:02.6 | AB | A | … | … | 57.76 | 391.42 | … | 0.595 ± 0.028 |
| … | … | G 106-7B | … | M5.5 V | 05:53:04.65 | +04:43:02.8 | … | B* | 1.520 | 278.2 | 50.83 | 387.78 | … | 0.139 ± 0.043 |
| 05530+2508 | RAO 206 | LSPM J0553+2507 | J05530+251 | M3.0 V | 05:53:01.92 | +25:07:40.9 | (AB) | AB | 0.259 | 2.7 | 175.31 | 191.68 | … | 0.367 ± 0.033 |
| … | … | Ross 59 | J05532+242 | M1.5 V | 05:53:14.24 | +24:15:22.1 | Aab | Aab(2) | … | … | 39.31 | 625.08 | … | 0.076 ± 0.002 |
| … | … | RX J0554.7+1055 | J05547+109 | M3.0 V | 05:54:45.58 | +10:55:55.9 | … | … | … | … | 27.22 | 156.47 | 269.87 ± 1.84 | 0.476 ± 0.019 |
| … | … | PM J05558+4036 | J05558+406 | M1.0 V | 05:55:48.31 | +40:36:48.0 | AB | A | … | … | 26.30 | 122.73 | … | 0.544 ± 0.028 |
| … | … | G3- 3458612029599345664 | … | M3.0 V | 05:55:48.41 | +40:36:47.7 | … | B* | 1.200 | 103.0 | 25.91 | 138.62 | … | 0.368 ± 0.033 |
| … | … | 1R055641.0-101837 | J05566-103 | M3.5 V | 05:56:40.63 | -10:18:35.8 | … | … | … | … | 25.96 | 130.74 | 85.20 ± 0.41 | 0.293 ± 0.014 |
| … | … | PM J05587+2557 | J05587+259 | M1.0 V | 05:58:47.68 | +25:57:40.1 | … | … | … | … | 77.14 | 98.96 | 485.71 ± 2.23 | 0.550 ± 0.028 |
| 05588+2121 | JNN268 | G 104-9 | J05588+213 | M5.0 V | 05:58:53.53 | +21:20:54.5 | (AB) | AB | 0.455 | 99.8 | 68.41 | 487.62 | … | 0.164 ± 0.041 |
| … | … | EGCam | J05596+585 | M0.5 V | 05:59:37.77 | +58:35:30.8 | A+B | A | … | … | 65.44 | 254.17 | 440.26 ± 1.82 | 0.527 ± 0.029 |
| 05599+5834 | GIC 61 | GJ 3372 | J05599+585 | M4.2 V | 05:59:55.68 | +58:34:11.2 | … | B | 161.111 | 119.6 | 63.36 | 252.69 | 60.62 ± 0.23 | 0.243 ± 0.012 |
| … | … | GJ 3379 | J06000+027 | M3.5 Ve | 06:00:03.83 | +02:42:22.9 | … | … | … | … | 29.12 | 311.77 | 63.55 ± 0.29 | 0.234 ± 0.009 |
| … | … | GJ 3373 | J06008+681 | M3.5 V | 06:00:50.71 | +68:09:05.3 | A+B | A | … | … | 30.27 | 1175.81 | 152.66 ± 0.71 | 0.375 ± 0.016 |
| 06007+6809 | LDS1201 | GJ 3374 | J06007+681 | M4.0 V | 06:00:47.81 | +68:08:11.6 | … | B | 56.075 | 196.8 | 30.05 | 1174.60 | 117.60 ± 0.59 | 0.326 ± 0.014 |
| … | … | GJ 3378 | J06011+595 | M4.0 V | 06:01:10.82 | +59:35:35.0 | … | … | … | … | 30.04 | 934.54 | 83.30 ± 0.41 | 0.249 ± 0.010 |
| … | … | LSPM J0601+1305 | J06017+130 | M2.5 Ve | 06:01:45.54 | +13:05:00.7 | … | … | … | … | 30.23 | 145.98 | 224.93 ± 1.01 | 0.432 ± 0.017 |
| … | … | GJ 3382 | J06023-203 | M3.5 V | 06:02:22.56 | -20:19:35.3 | … | … | … | … | 40.50 | 556.62 | 67.38 ± 0.39 | 0.241 ± 0.012 |
| … | … | GJ 3380 | J06024+498 | M5.0 V | 06:02:29.31 | +49:51:42.6 | … | … | … | … | 45.19 | 852.65 | 21.41 ± 0.11 | 0.134 ± 0.009 |
| … | … | LP 57-46 | J06024+663 | M4.5 V | 06:02:26.51 | +66:20:32.3 | … | … | … | … | 45.13 | 582.95 | 55.95 ± 0.23 | 0.250 ± 0.013 |
| … | … | PM J06025+3707 | J06025+371 | M1.0 V | 06:02:35.46 | +37:07:36.2 | … | … | … | … | 83.97 | 138.30 | 578.82 ± 3.73 | 0.547 ± 0.028 |
| … | … | Wolf 261 | J06034+478 | M4.2 V | 06:03:29.48 | +47:48:06.0 | … | … | … | … | 66.44 | 562.68 | 125.47 ± 0.59 | 0.361 ± 0.016 |
| … | … | 1R060335.0+153132 | J06035+155 | M0.0 V | 06:03:34.74 | +15:31:30.4 | … | … | … | … | 87.53 | 68.32 | 1903.04 ± 77.82 | 0.815 ± 0.025 |
| … | … | 1R060334.8+165128 | J06035+168 | M4.0 V | 06:03:34.49 | +16:51:45.4 | … | … | … | … | 87.35 | 93.78 | 124.65 ± 28.81 | 0.359 ± 0.047 |
| … | … | Ross 60 | J06039+261 | M3.0 V | 06:03:54.54 | +26:08:46.0 | … | … | … | … | 87.97 | 640.80 | 85.01 ± 0.42 | 0.274 ± 0.013 |
| … | … | LP 86-173 | J06054+608 | M4.5 V | 06:05:30.04 | +60:49:09.8 | … | … | … | … | 47.26 | 840.47 | 80.39 ± 0.41 | 0.249 ± 0.012 |
| … | … | PM J06066+4633A | J06066+465 | M3.0 V | 06:06:37.78 | +46:33:47.0 | AB | A | … | … | 29.07 | 67.55 | … | 0.398 ± 0.032 |
| 06066+4634 | KPP3658 | PM J06066+4633B | … | M3.5 V | 06:06:37.77 | +46:33:45.2 | … | B | 1.790 | 182.3 | 81.50 | 84.00 | … | 0.331 ± 0.034 |
| … | … | Ross 70 | J06071+335 | M2.0 V | 06:07:11.91 | +33:32:30.9 | … | … | … | … | 81.46 | 421.94 | 371.91 ± 2.20 | 0.490 ± 0.030 |
| … | … | 1R060732.5+471154 | J06075+472 | M4.5 V | 06:07:31.91 | +47:12:23.3 | … | … | … | … | 97.60 | 192.04 | 136.35 ± 0.74 | 0.377 ± 0.016 |
| … | … | HD 291290 | J06097+001 | M0.0 V | 06:09:46.30 | +00:09:30.8 | … | … | … | … | 44.56 | 192.85 | 1002.68 ± 6.20 | 0.693 ± 0.026 |
| … | … | PM J06102+2234 | J06102+225 | M4.0 V | 06:10:17.81 | +22:34:17.2 | A+BC | A | … | … | 172.68 | 158.78 | 121.97 ± 0.61 | 0.355 ± 0.015 |
| … | … | LP 362-121 | J06103+225 | M5.0 V | 06:10:22.52 | +22:34:18.1 | … | B | 65.155 | 89.2 | 77.08 | 166.63 | … | 0.175 ± 0.041 |
| 06104+2234 | LAW 14 | G3- 3425067888342287616 | … | … | 06:10:22.50 | +22:34:17.9 | … | C | 64.941 | 89.4 | … | … | … | … |

Table D.2: Complete sample with the description of multiple systems (continued).

| WDS id | WDS disc | Name | Karmn | Spectral type | $\alpha$ (2016.0) | $\delta$ (2016.0) | System | Component | $\rho$ [arcsec] | $\theta$ [deg] | $\varpi$ [mas] | $\mu_{total}$ [mas a$^{-1}$] | $\mathcal{L}$ [$10^{-4}$ $\mathcal{L}_{\odot}$] | $\mathcal{M}$ [$\mathcal{M}_{\odot}$] |
|---|---|---|---|---|---|---|---|---|---|---|---|---|---|---|
| ... | ... | LSPM J0610+7212 | J06103+722 | M2.5 V | 06:10:18.09 | +72:11:58.0 | ... | ... | ... | ... | 46.40 | 160.32 | 290.05 ± 1.13 | 0.494 ± 0.019 |
| ... | ... | GJ 226 | J06103+821 | M2.0 V | 06:10:20.24 | +82:06:02.9 | ... | ... | ... | ... | 35.10 | 1337.19 | 239.14 ± 1.04 | 0.406 ± 0.011 |
| ... | ... | TYC 135-239-1 | J06105+024 | M0.0 V | 06:10:31.41 | +02:25:30.3 | ... | ... | ... | ... | 35.10 | 62.81 | 1178.64 ± 64.36 | 0.712 ± 0.027 |
| 06106-2152 | NAJ 1 | HD 42581 | J06105-218 | M0.5 V | 06:10:34.46 | -21:52:04.2 | (AB) | AB | 4.891 | 179.7 | 43.25 | 731.87 | 545.28 ± 13.88 | 0.563 ± 0.028 |
| ... | ... | Wolf 1058 | J06107+259 | M1.5 V | 06:10:46.51 | +25:55:53.3 | ... | ... | ... | ... | 32.56 | 594.86 | 569.45 ± 7.01 | 0.561 ± 0.028 |
| 06109+1020 | KAM1 | Ross 79 | J06109+103 | M2.5 V | 06:10:54.88 | +10:18:50.6 | (AB) | AB | 1.157 | 355.9 | 56.93 | 934.18 | ... | 0.454 ± 0.032 |
| ... | ... | GJ 3388 | J06140+516 | M3.5 V | 06:14:01.72 | +51:40:06.5 | ... | ... | ... | ... | 38.56 | 386.61 | 86.95 ± 0.37 | 0.277 ± 0.013 |
| ... | ... | G 106-35 | J06145+025 | M3.0 V | 06:14:34.76 | +02:30:19.7 | ... | ... | ... | ... | 44.68 | 493.68 | 193.70 ± 1.51 | 0.425 ± 0.017 |
| 06173+0506 | CAT 1 | HD 43587 | ... | G0 V | 06:17:15.90 | +05:06:02.6 | (AB)+(CD) | AB(1) | 0.840 | 70.9 | 214.04 | 269.29 | ... | 1.300 ± 0.000 |
| 06173+0506 | PRV 3 | GJ 231.1 B | J06171+051 | M3.5 V | 06:17:10.42 | +05:07:05.3 | ... | CD | 103.189 | 307.4 | 33.04 | 261.65 | ... | 0.300 ± 0.035 |
| ... | ... | LP 779-34 | J06151-164 | M4.0 V | 06:15:11.90 | -16:26:21.4 | ... | ... | ... | ... | 24.71 | 379.97 | 46.90 ± 0.23 | 0.211 ± 0.011 |
| ... | ... | TYC 4525-194-1 | J06171+751 | M2.0 Ve | 06:18:07.08 | +75:06:04.3 | Aabc | Aabc(3?) | ... | ... | 52.11 | 77.87 | ... | ... |
| ... | ... | LSPM J0617+8353 | J06171+838 | M3.5 V | 06:17:04.81 | +83:53:32.5 | ... | ... | ... | ... | 24.82 | 188.78 | 157.08 ± 1.90 | 0.336 ± 0.014 |
| ... | ... | G 103-29 | J06185+250 | M4.0 V | 06:18:34.83 | +25:03:00.6 | ... | ... | ... | ... | 58.06 | 324.78 | 172.94 ± 0.82 | 0.401 ± 0.017 |
| ... | ... | Ross 417 | J06193-066 | M3.0 V | 06:19:20.74 | -06:39:32.1 | ... | ... | ... | ... | 47.73 | 625.97 | 92.86 ± 0.54 | 0.287 ± 0.013 |
| ... | ... | TYC 743-1836-1 | J06194+139 | M0.5 V | 06:19:29.61 | +13:57:02.3 | ... | ... | ... | ... | 40.54 | 115.03 | 768.13 ± 2.94 | 0.636 ± 0.027 |
| ... | ... | GJ 3391 | J06212+442 | M2.0 Ve | 06:21:13.26 | +44:14:26.7 | AB | A | ... | ... | 27.78 | 294.38 | ... | 0.571 ± 0.028 |
| 06212+4415 | CRC 50 | G3- 962141157559432064 | ... | M5.0 V | 06:21:13.21 | +44:14:25.5 | ... | B | 1.334 | 203.2 | 27.97 | 288.78 | ... | 0.173 ± 0.040 |
| ... | ... | LP 420-5 | J06216+163 | M1.0 V | 06:21:36.85 | +16:18:33.8 | A+B | A | ... | ... | 62.84 | 281.57 | 355.02 ± 2.85 | 0.487 ± 0.018 |
| 06215+1618 | LDS894 | LP 420-6 | J06217+163 | M2.5 V | 06:21:44.17 | +16:19:19.9 | ... | B | 115.018 | 66.4 | 32.63 | 283.64 | 203.11 ± 1.08 | 0.385 ± 0.016 |
| ... | ... | GJ 2049 | J06218-227 | M1 Vk | 06:21:53.08 | -22:43:19.7 | ... | ... | ... | ... | 60.71 | 687.88 | 893.70 ± 4.81 | 0.651 ± 0.027 |
| ... | ... | TYC 2425-1286-1 | J06223+334 | M1.0 V | 06:22:20.65 | +33:26:54.6 | ... | ... | ... | ... | 74.14 | 113.71 | 1721.47 ± 26.29 | 0.761 ± 0.025 |
| ... | ... | LP 720-10 | J06236-096 | M3.5 V | 06:23:38.41 | -09:38:51.5 | AB | A | ... | ... | 74.21 | 62.21 | ... | 0.299 ± 0.035 |
| 06236-0938 | JNN 270 | LP 720-10 B | ... | M8.0 V | 06:23:38.29 | -09:38:51.4 | ... | B | 1.836 | 273.0 | 192.01 | 63.64 | ... | 0.100 ± 0.050 |
| ... | ... | TYC 141-24-1 | J06237+020 | M1.5 V | 06:23:46.49 | +05:02:40.1 | ... | ... | ... | ... | 50.41 | 78.91 | 624.56 ± 3.30 | 0.581 ± 0.028 |
| 06239+4540 | RAO 210 | LP 160-22 | J06238+456 | M5.0 V | 06:23:51.20 | +45:40:00.0 | (AB) | AB | 0.602 | 273.4 | 50.39 | 282.93 | ... | 0.167 ± 0.041 |
| ... | ... | Ross 64 | J06246+234 | M4.0 V | 06:24:41.93 | +23:25:50.8 | ... | ... | ... | ... | 129.30 | 749.61 | 34.45 ± 0.23 | 0.171 ± 0.009 |
| ... | ... | GJ 3393 | J06258+561 | M4.0 V | 06:25:53.21 | +56:10:16.9 | ... | ... | ... | ... | 53.27 | 516.23 | 45.39 ± 0.20 | 0.207 ± 0.011 |
| ... | ... | 1R062614.2+234942 | J06262+238 | M1.5 Ve | 06:26:14.52 | +23:49:36.4 | ... | ... | ... | ... | 66.52 | 136.66 | 431.13 ± 3.02 | 0.522 ± 0.029 |
| ... | ... | Ross 603A | J06277+093 | M2.0 V | 06:27:43.80 | +09:23:51.3 | AB | A | ... | ... | 104.96 | 245.06 | ... | 0.456 ± 0.030 |
| 06277+0924 | CRC 51 | Ross 603B | ... | M3.5 V | 06:27:43.74 | +09:23:50.5 | ... | B | 1.170 | 221.8 | 52.89 | 264.74 | ... | 0.327 ± 0.034 |
| ... | ... | V577Mon | J06293-028 | M4.5 V | 06:29:24.18 | -02:49:01.9 | AB | A | ... | ... | 32.62 | 1098.83 | ... | 0.230 ± 0.037 |
| 06293-0248 | B2601 | Ross 614 B | ... | M4.5 V | 06:29:24.19 | -02:49:01.7 | ... | B | 0.313 | 48.6 | 38.05 | 1098.83 | ... | 0.228 ± 0.037 |
| ... | ... | G 108-4 | J06298-027 | M4.0 V | 06:29:50.47 | -02:47:49.0 | Aab | Aab(2) | ... | ... | 44.36 | 264.66 | ... | ... |
| ... | ... | PM J06306+4539 | J06306+456 | M1.0 V | 06:30:37.39 | +45:39:23.0 | ... | ... | ... | ... | 21.63 | 131.61 | 363.81 ± 1.42 | 0.466 ± 0.017 |
| ... | ... | PM J06307+3947 | J06307+397 | M2.0 V | 06:30:47.39 | +39:47:38.5 | ... | ... | ... | ... | 43.10 | 129.54 | 170.51 ± 0.61 | 0.351 ± 0.015 |
| ... | ... | GJ 3395 | J06310+500 | M0.8 V | 06:31:00.97 | +50:02:45.5 | ... | ... | ... | ... | 44.93 | 201.46 | 471.63 ± 1.90 | 0.537 ± 0.029 |
| ... | ... | GJ 3396 | J06318+414 | M5.84 | 06:31:50.73 | +41:29:42.2 | ... | ... | ... | ... | 61.43 | 208.66 | 131.98 ± 0.58 | 0.357 ± 0.070 |
| ... | ... | TYC 2928-1568-1 | J06322+378 | M1.5 V | 06:32:14.91 | +37:48:10.6 | ... | ... | ... | ... | 30.33 | 160.15 | 888.61 ± 25.23 | 0.636 ± 0.027 |
| ... | ... | PM J06323-0943 | J06323-097 | M4.5 V | 06:32:20.28 | -09:43:29.9 | ... | ... | ... | ... | 30.32 | 56.94 | 103.08 ± 2.55 | 0.325 ± 0.015 |

Table D.2: Complete sample with the description of multiple systems (continued).

| WDS id | WDS disc | Name | Karmn | Spectral type | $\alpha$ (2016.0) | $\delta$ (2016.0) | System | Component | $\rho$ [arcsec] | $\theta$ [deg] | $\varpi$ [mas] | $\mu_{\rm total}$ [mas a$^{-1}$] | $\mathcal{L}$ [$10^{-4}\,\mathcal{L}_\odot$] | $\mathcal{M}$ [$\mathcal{M}_\odot$] |
|---|---|---|---|---|---|---|---|---|---|---|---|---|---|---|
| … | … | LP 57-192 | J06325+641 | M4.0 V | 06:32:31.27 | +64:06:12.2 | … | … | … | … | 32.58 | 560.21 | 165.63 ± 0.81 | 0.418 ± 0.017 |
| … | … | G 103-41 | J06345+315 | M3.5 V | 06:34:33.48 | +31:30:05.1 | … | … | … | … | 35.44 | 192.39 | 175.03 ± 0.69 | 0.403 ± 0.017 |
| 06354-0403 | JNN271 | 1R063531.2-040314 | J06354-040 | M5.5 V | 06:35:29.75 | -04:03:17.2 | (AB) | AB | 0.043 | 165.0 | 31.96 | 121.86 | … | 0.186 ± 0.039 |
| … | … | GJ 3398 | J06361+116 | M5.0 V | 06:36:06.16 | +11:36:49.5 | … | … | … | … | 34.90 | 876.52 | 52.60 ± 0.23 | 0.225 ± 0.012 |
| … | … | LP 420-4 | J06361+201 | M2.5 V | 06:36:12.05 | +20:08:10.3 | … | … | … | … | 43.90 | 236.63 | 164.69 ± 0.67 | 0.366 ± 0.015 |
| … | … | HD 260655 | J06371+175 | M0.0 V | 06:37:09.94 | +17:33:58.7 | … | … | … | … | … | 835.76 | 364.95 ± 1.58 | 0.440 ± 0.011 |
| … | … | LP 780-32 | J06396-210 | dM4.0 | 06:39:37.20 | -21:01:32.4 | AB | A | … | … | 31.67 | 225.54 | 136.81 ± 4.35 | 0.326 ± 0.013 |
| … | … | G3- 2926756741750933120 | … | M4.0 V | 06:39:37.22 | -21:01:31.9 | … | B* | 0.556 | 35.1 | 105.69 | 146.66 | … | 0.266 ± 0.036 |
| 06401+2835 | CRC 52 | GJ 3399 | J06400+285 | M2.5 V | 06:40:05.54 | +28:35:10.5 | (AB) | AB | 0.269 | 91.3 | 25.93 | 245.94 | … | 0.476 ± 0.030 |
| … | … | LP 780-23 | J06401-164 | M2.5 V | 06:40:08.72 | -16:27:21.5 | AB | A | … | … | 173.57 | 319.03 | … | … |
| … | … | LP 780-23 B | … | … | 06:40:08.72 | -16:27:21.7 | … | B* | 0.199 | 187.9 | 37.23 | … | … | … |
| … | … | Wolf 289 | J06414+157 | M4.0 V | 06:41:28.23 | +15:45:42.6 | … | … | … | … | 91.65 | 318.30 | 186.02 ± 0.83 | 0.416 ± 0.017 |
| … | … | GJ 3404 | J06421+035 | M3.0 V | 06:42:11.24 | +03:34:48.5 | A+B | A | … | … | 67.60 | 261.94 | 182.11 ± 0.93 | 0.369 ± 0.012 |
| 06423+0334 | GIC 65 | GJ 3405 | J06422+035 | M4.0 V | 06:42:13.39 | +03:35:26.8 | … | B | 50.055 | 40.0 | 37.87 | 257.50 | 71.60 ± 0.37 | 0.267 ± 0.013 |
| … | … | G 110-14 | J06435+166 | M4.5 V | 06:43:34.53 | +16:41:35.5 | … | … | … | … | 51.62 | 208.68 | 93.06 ± 2.89 | 0.329 ± 0.016 |
| … | … | GJ 3406A | J06438+511 | M2.5 V | 06:43:49.95 | +51:08:06.0 | AB | A | … | … | 52.14 | 874.58 | … | 0.359 ± 0.033 |
| 06438+5108 | LDS6200 | GJ 3406B | … | M4.5 V | 06:43:49.77 | +51:08:05.8 | … | B | 1.700 | 265.8 | 77.54 | 916.52 | … | 0.259 ± 0.036 |
| 06448+7153 | BAG 22 | GJ 2050 | J06447+718 | M0.5 V | 06:44:45.25 | +71:53:06.7 | (AB) | AB | 0.630 | 64.1 | 31.69 | 558.83 | … | 0.560 ± 0.028 |
| … | … | HD 263175 | … | K3 V | 06:46:04.47 | +32:33:22.0 | A+B | A | … | … | 47.29 | 467.06 | 2218.02 ± 7.94 | 0.780 ± 0.117 |
| 06461+3233 | LDS6201 | HD 263175B | J06461+325 | M1.0 V | 06:46:06.88 | +32:33:16.6 | … | B | 30.855 | 100.1 | 28.96 | 472.87 | 278.81 ± 1.20 | 0.429 ± 0.016 |
| … | … | 1R064645.7+155739 | J06467+159 | M1.0 Ve | 06:46:45.62 | +15:57:41.8 | … | … | … | … | 59.48 | 102.50 | 541.77 ± 2.45 | 0.565 ± 0.028 |
| … | … | G 108-27 | J06474+054 | M4.0 V | 06:47:27.57 | +05:24:23.3 | … | … | … | … | 39.95 | 308.45 | 175.68 ± 0.87 | 0.404 ± 0.017 |
| … | … | LP 121-58 | J06486+532 | M1.5 V | 06:48:38.72 | +53:17:24.3 | … | … | … | … | 26.93 | 357.19 | 641.38 ± 3.40 | 0.584 ± 0.028 |
| … | … | 1R064855.9+210754 | J06489+211 | M2.5 V | 06:48:55.18 | +21:08:02.8 | … | … | … | … | 27.11 | 72.41 | 330.27 ± 1.71 | 0.461 ± 0.030 |
| … | … | GJ 1092 | J06490+371 | M4.0 V | 06:49:05.73 | +37:06:25.0 | … | … | … | … | 43.05 | 1603.32 | 43.79 ± 0.16 | 0.203 ± 0.011 |
| … | … | LP 661-2 | J06509-091 | M3.5 V | 06:50:59.37 | -09:10:59.3 | … | … | … | … | 43.06 | 584.19 | 151.38 ± 1.27 | 0.373 ± 0.016 |
| 06523-0510 | WSI 125 | HD 50281 | … | K3.5 V | 06:52:17.47 | -05:10:25.4 | (AB)+C | AB | 0.175 | 149.6 | 37.07 | 543.70 | … | 0.745 ± 0.112 |
| 06523-0510 | WNO 17 | HD 50281B | J06523-051 | M2.0 V+ | 06:52:17.42 | -05:11:24.3 | … | C | 58.833 | 180.6 | 17.17 | 576.48 | … | 0.445 ± 0.031 |
| … | … | GJ 3413 | J06524+182 | M3.5 V | 06:52:24.45 | +18:17:06.9 | … | … | … | … | 40.77 | 175.88 | 169.69 ± 1.04 | 0.397 ± 0.016 |
| 06541+6052 | HEI 334 | GJ 3412 | J06540+608 | M3.0 V | 06:54:05.43 | +60:52:02.4 | (AB) | AB(1) | 0.351 | 253.1 | 40.35 | 1093.13 | … | … |
| … | … | HD 265866 | J06548+332 | M3.0 V | 06:54:48.03 | +33:15:59.1 | … | … | … | … | 41.83 | 828.58 | 163.26 ± 0.94 | 0.342 ± 0.012 |
| … | … | TYC 756-1685-1 | J06564+121 | M1.0 V | 06:56:25.84 | +12:07:31.4 | … | … | … | … | 40.34 | 98.10 | 1366.51 ± 39.35 | 0.715 ± 0.026 |
| 06565+4004 | KUI27 | GJ 3415 | … | K4.5 V | 06:56:28.28 | +40:04:20.6 | A+B | A | 37.808 | 6.8 | 48.95 | 452.03 | 1944.33 ± 7.69 | 0.730 ± 0.110 |
| … | … | GJ 3416 | J06564+400 | M1.0 V | 06:56:28.64 | +40:04:58.3 | … | B | 37.895 | 6.2 | 117.73 | 446.75 | 642.93 ± 2.58 | 0.593 ± 0.028 |
| … | … | LP 34-110 | J06564+759 | M1.0 V | 06:53:24.30 | +72:55:09.3 | … | … | … | … | 50.45 | 346.36 | 1170.26 ± 17.73 | 0.693 ± 0.026 |
| … | … | G 107-36 | J06565+440 | M4.5 V | 06:56:31.24 | +44:01:45.0 | … | … | … | … | 36.34 | 708.19 | 101.14 ± 0.51 | 0.321 ± 0.014 |
| … | … | 1R065728.1+740529 | J06574+740 | M4.0 Ve | 06:57:25.77 | +74:05:26.1 | … | … | … | … | 43.22 | 96.24 | … | … |
| … | … | GJ 3417 A | J06579+623 | M6.0 V | 06:57:57.83 | +62:19:11.0 | AB | A | … | … | 43.45 | 614.04 | … | 0.207 ± 0.038 |
| 06579+6220 | HEN 2 | GJ 3417 B | … | M5.0 V | 06:57:57.66 | +62:19:10.5 | … | B | 1.284 | 246.9 | 242.97 | 573.10 | … | 0.129 ± 0.044 |
| … | … | G 192-59 | J06582+511 | M2.0 V | 06:58:12.70 | +51:08:32.4 | … | … | … | … | 242.97 | 364.19 | 441.91 ± 1.94 | 0.517 ± 0.029 |

Table D.2: Complete sample with the description of multiple systems (continued).

| WDS id | WDS disc | Name | Karmn | Spectral type | $\alpha$ (2016.0) | $\delta$ (2016.0) | System | Component | $\rho$ [arcsec] | $\theta$ [deg] | $\varpi$ [mas] | $\mu_{\rm total}$ [mas a$^{-1}$] | $\mathcal{L}$ [$10^{-4}$ $\mathcal{L}_\odot$] | $\mathcal{M}$ [$\mathcal{M}_\odot$] |
|---|---|---|---|---|---|---|---|---|---|---|---|---|---|---|
| … | … | GJ 1093 | J06594+193 | M5.0 Ve | 06:59:29.84 | +19:20:41.5 | … | … | … | … | 24.17 | 1275.43 | 16.35 ± 0.09 | 0.120 ± 0.009 |
| … | … | G 88-2 | J06594+195 | M3.0 V | 06:59:28.65 | +19:30:30.4 | … | … | … | … | 35.23 | 269.52 | 158.26 ± 0.73 | 0.382 ± 0.016 |
| … | … | PM J06596+0545 | J06596+057 | M2.5 Ve | 06:59:41.55 | +05:45:38.9 | … | … | … | … | 39.37 | 70.32 | 344.17 ± 2.14 | 0.508 ± 0.019 |
| … | … | 1R070005.1-190115 | J07001-190 | M5.0 V | 07:00:07.00 | -19:01:25.1 | Aab | Aab(2) | … | … | 46.17 | 170.88 | … | … |
| … | … | PM J07009-0221 | J07009-023 | M3.0 V | 07:00:59.74 | -02:21:32.2 | … | … | … | … | 48.58 | 68.30 | 226.16 ± 1.15 | 0.433 ± 0.017 |
| … | … | PM J07012+0052 | J07012+008 | M2.5 V | 07:01:15.54 | +00:52:40.4 | … | … | … | … | 35.91 | 104.42 | 476.53 ± 10.10 | 0.522 ± 0.029 |
| … | … | GJ 3423 | J07033+346 | M4.0 Ve | 07:03:23.09 | +34:41:53.6 | … | … | … | … | 30.18 | 151.83 | 74.58 ± 0.28 | 0.253 ± 0.010 |
| … | … | LP 16-379 | J07034+767 | M3.5 V | 07:03:29.98 | +76:46:21.9 | … | … | … | … | 38.99 | 233.75 | 122.79 ± 0.57 | 0.334 ± 0.015 |
| 07039+5242 | BEU 8 | GJ 3421 | J07039+527 | M5.0 V | 07:03:56.92 | +52:41:51.8 | (AB) | AB | 0.165 | 335.5 | 31.64 | 1143.04 | … | 0.202 ± 0.039 |
| 07043-1031 | BEU 9 | Ross 54 | J07042-105 | M3.5 V | 07:04:17.55 | -10:30:44.6 | (AB) | AB(2) | 0.108 | 289.7 | 40.67 | 826.29 | … | … |
| … | … | GJ 258 | J07044+682 | M3.0 V | 07:04:26.94 | +68:17:20.5 | … | … | … | … | 52.35 | 349.83 | 217.05 ± 1.10 | 0.392 ± 0.011 |
| … | … | Ross 874 | J07047+249 | M1.5 V | 07:04:49.44 | +24:59:50.6 | … | … | … | … | 79.16 | 326.17 | 430.16 ± 1.58 | 0.518 ± 0.029 |
| … | … | 1R070511.2-100801 | J07051-101 | M5.0 V | 07:05:12.10 | -10:07:51.6 | … | … | … | … | 56.64 | 146.49 | 31.70 ± 0.14 | 0.182 ± 0.011 |
| … | … | G 108-52 | J07052+084 | M2.0 V | 07:05:12.39 | +08:25:45.7 | … | … | … | … | 39.56 | 425.52 | 172.20 ± 0.98 | 0.353 ± 0.015 |
| … | … | GJ 3426 | J07076+486 | M4.3 V | 07:07:37.70 | +48:41:08.6 | … | … | … | … | 100.02 | 305.76 | 36.29 ± 0.12 | 0.183 ± 0.011 |
| … | … | GJ 3425 | J07078+672 | M1.5 V | 07:07:49.68 | +67:12:03.7 | … | … | … | … | 63.74 | 280.71 | 403.42 ± 3.41 | 0.509 ± 0.029 |
| … | … | LP 840-16 | J07081-228 | M2.0 V | 07:08:06.53 | -22:48:51.0 | … | … | … | … | 65.31 | 471.01 | 259.09 ± 1.15 | 0.413 ± 0.016 |
| … | … | GJ 3429 | J07086+307 | M0.5 V | 07:08:39.72 | +30:42:51.6 | … | … | … | … | 43.89 | 197.24 | 549.10 ± 3.55 | 0.569 ± 0.028 |
| … | … | GJ 3427 | J07095+698 | M3.0 V | 07:09:31.85 | +69:50:53.1 | … | … | … | … | … | 324.11 | 238.73 ± 1.05 | 0.446 ± 0.018 |
| … | … | QYAur | J07100+385 | M4.5 V | 07:10:01.23 | +38:31:31.0 | Aab | Aab(2) | … | … | 38.76 | 1041.98 | … | 0.399 ± 0.001 |
| … | … | GJ 3430 | J07102+376 | M4.0 V | 07:10:13.32 | +37:40:05.9 | A+B | A | … | … | … | 350.22 | 53.23 ± 1.28 | 0.226 ± 0.012 |
| 07103+3739 | GIC 69 | GJ 3431 | … | DQ8 | 07:10:14.04 | +37:40:15.0 | … | B | 12.472 | 43.1 | 33.46 | 357.87 | … | 0.500 ± 0.100 |
| … | … | 1R071032.6-084232 | J07105-087 | M3.5 V | 07:10:31.37 | -08:42:46.7 | … | … | … | … | 65.98 | 129.38 | 131.45 ± 0.72 | 0.346 ± 0.015 |
| … | … | LP 206-11 | J07111+434 | M5.5 V | 07:11:11.97 | +43:29:49.0 | AB | A | … | … | 66.06 | 675.37 | … | 0.120 ± 0.045 |
| 07112+4330 | MTG1 | LP 206-11 B | … | … | 07:11:12.02 | +43:29:49.3 | … | B | 0.634 | 65.5 | 41.99 | … | … | … |
| … | … | TYC 4530-1414-1 | J07119+773 | M1.5 V | 07:11:57.13 | +77:21:57.4 | Aab | Aab(1) | … | … | 40.33 | 88.25 | … | … |
| … | … | GJ 3432 | J07121+522 | M1.0 V | 07:12:11.14 | +52:16:20.4 | … | … | … | … | 53.03 | 295.08 | 594.22 ± 2.57 | 0.574 ± 0.028 |
| … | … | 1R071259.5+354655 | J07129+357 | M2.5 Ve | 07:12:59.62 | +35:47:03.1 | … | … | … | … | 52.90 | 66.50 | 243.95 ± 1.05 | 0.451 ± 0.018 |
| … | … | G 193-39 | J07140+507 | M0.5 V | 07:14:04.29 | +50:43:28.9 | AB | A | … | … | 46.49 | 299.13 | … | 0.566 ± 0.028 |
| 07141+5044 | KPP3199 | G 193-39B | … | M5.0 V | 07:14:04.09 | +50:43:28.4 | … | B | 1.892 | 256.2 | 38.83 | 301.87 | … | 0.189 ± 0.039 |
| 07163+2709 | BEU 10 | GJ 268.3 | J07163+271 | M2.5 V | 07:16:19.73 | +27:08:29.9 | (AB) | AB(2) | 0.060 | 154.5 | 66.91 | 201.19 | … | 0.727 ± 0.012 |
| … | … | GJ 1096 | J07163+331 | M5.0 V | 07:16:17.89 | +33:09:03.4 | … | … | … | … | 38.87 | 448.75 | 39.39 ± 0.20 | 0.206 ± 0.011 |
| … | … | PM J07172-0501 | J07172-050 | M3.5 V | 07:17:17.54 | -05:01:09.8 | … | … | … | … | 44.69 | 574.06 | 49.73 ± 0.35 | 0.218 ± 0.011 |
| … | … | GJ 3437 | J07174+195 | M3.2 V | 07:17:29.57 | +19:34:12.4 | … | … | … | … | 36.59 | 417.61 | 162.91 ± 0.70 | 0.364 ± 0.015 |
| 07181+3916 | RAO 216 | Ross 987 | J07181+392 | M0.0 V | 07:18:07.89 | +39:16:27.4 | (AB) | AB | 0.107 | 67.3 | 25.87 | 244.37 | … | 0.524 ± 0.029 |
| … | … | PM J07182+1342 | J07182+137 | M3.5 V | 07:18:12.86 | +13:42:16.2 | … | … | … | … | 28.59 | 44.09 | 165.89 ± 0.75 | 0.392 ± 0.016 |
| … | … | GJ 270 | J07195+328 | M0.0 V | 07:19:31.79 | +32:49:42.8 | … | … | … | … | 69.20 | 536.19 | 745.86 ± 3.85 | 0.630 ± 0.027 |
| … | … | TYC 4618-116-1 | J07199+840 | M2.5 V | 07:19:57.65 | +84:04:36.8 | … | … | … | … | 41.00 | 89.65 | 112.02 ± 0.62 | 0.298 ± 0.013 |
| 07200-0847 | BUG 17 | Scholz's star | J07200-087 | M9.5+T5 | 07:20:03.21 | -08:46:51.9 | (AB) | AB | 0.381 | 63.5 | 114.35 | 125.25 | … | 0.116 ± 0.055 |
| … | … | TYC 178-2187-1 | J07212+005 | M0.5 V | 07:31:12.97 | +00:33:13.8 | … | … | … | … | 114.29 | 100.59 | 658.56 ± 4.71 | 0.607 ± 0.027 |

Table D.2: Complete sample with the description of multiple systems (continued).

| WDS id | WDS disc | Name | Karmn | Spectral type | $\alpha$ (2016.0) | $\delta$ (2016.0) | System | Component | $\rho$ [arcsec] | $\theta$ [deg] | $\varpi$ [mas] | $\mu_{\rm total}$ [mas a$^{-1}$] | $\mathcal{L}$ [$10^{-4}$ $\mathcal{L}_\odot$] | $\mathcal{M}$ [$\mathcal{M}_\odot$] |
|---|---|---|---|---|---|---|---|---|---|---|---|---|---|---|
| … | … | GJ 3439 | J07227+306 | M4.0 V | 07:22:41.50 | +30:40:02.3 | … | … | … | … | 44.35 | 719.15 | 156.09 ± 0.87 | 0.379 ± 0.016 |
| … | … | GJ 272 | J07232+460 | M0.5 V | 07:23:14.71 | +46:05:10.9 | … | … | … | … | 91.61 | 266.76 | 542.28 ± 2.73 | 0.558 ± 0.028 |
| 07274+0514 | WDK 2 | Luyten's Star | J07274+052 | M3.5 V | 07:27:25.11 | +05:12:33.8 | (AB) | AB | 0.170 | 327.0 | 179.06 | 3735.42 | 108.89 ± 1.00 | 0.297 ± 0.012 |
| … | … | Ross 878 | J07274+220 | M1.5 V | 07:27:28.31 | +22:02:35.6 | … | … | … | … | 27.43 | 298.23 | 477.33 ± 2.02 | 0.527 ± 0.029 |
| … | … | GJ 3442 | J07282-187 | M4.5 V | 07:28:13.09 | -18:47:25.2 | … | … | … | … | 39.87 | 616.89 | 57.38 ± 1.07 | 0.236 ± 0.012 |
| … | … | GJ 1097 | J07287-032 | M3.0 V | 07:28:45.91 | -03:18:05.9 | … | … | … | … | 39.93 | 902.58 | 187.98 ± 1.04 | 0.367 ± 0.011 |
| 07295+3556 | JNN 57 | 1R072931.4+355607 | J07295+359 | M1.5 V | 07:29:31.04 | +35:55:58.5 | (AB)+C | AB | 0.074 | 25.9 | 19.55 | 122.54 | … | 0.654 ± 0.027 |
| 07295+3556 | UC 1586 | 2M07293670+3554531 | … | M3.0 V | 07:29:36.67 | +35:54:51.3 | … | C | 95.884 | 134.5 | 37.92 | 108.79 | … | 0.385 ± 0.032 |
| … | … | GJ 275.2 A | J07307+481 | M4.0 V | 07:30:42.46 | +48:11:38.2 | A+BC | A | 0.054 | | … | 1287.99 | … | 0.207 ± 0.038 |
| 07307+4813 | GIC 75 | GJ 275.2 B | … | DA | 07:30:46.99 | +48:10:05.7 | … | B | 103.053 | 153.9 | 87.03 | 1281.73 | … | 0.500 ± 0.100 |
| 07307+4813 | WNO 49 | G 107-70B | … | DA | 07:30:46.96 | +48:10:06.3 | … | C | 102.385 | 153.9 | 87.24 | … | … | 0.500 ± 0.100 |
| … | … | 1R073138.4+455718 | … | M3.0 V | 07:31:38.47 | +45:57:15.8 | Aab+B+C | Aab | … | … | 197.96 | 93.78 | … | … |
| … | … | 1R073101.9+460030 | J07310+460 | M4.0 V | 07:31:01.27 | +46:00:24.8 | … | B* | 431.391 | 296.0 | 30.20 | 101.75 | 438.36 ± 4.02 | 0.470 ± 0.030 |
| … | … | G3- 975312928903090560 | … | M4.5 V | 07:31:09.03 | +45:56:55.6 | … | C* | 307.796 | 266.3 | 128.99 | 100.74 | 65.93 ± 0.49 | 0.273 ± 0.013 |
| … | … | LynA | J07319+362S | M2.5 V | 07:31:57.38 | +36:13:06.2 | A+BC | A | … | … | 49.45 | 351.29 | 388.95 ± 2.56 | 0.468 ± 0.030 |
| 07319+3613 | LDS6206 | BLLyn | J07319+362N | M3.5 V | 07:31:56.97 | +36:13:43.2 | … | B | 37.364 | 352.4 | 34.38 | 370.92 | 187.91 ± 0.75 | 0.397 ± 0.010 |
| 07319+3613 | BEU 11 | LynB | … | M5.0 V | 07:31:57.36 | +36:13:04.6 | … | C | 1.560 | 190.0 | 33.53 | 347.61 | … | 0.205 ± 0.038 |
| … | … | GJ 3445 | J07319+392 | M2.48 V | 07:31:56.74 | +39:13:34.0 | … | … | … | … | 35.44 | 293.58 | 471.99 ± 13.25 | 0.496 ± 0.030 |
| … | … | GJ 3447 | J07320+173E | M0.0 V | 07:32:02.63 | +17:19:07.0 | (AB)+C | A | … | … | 31.25 | 285.23 | … | 0.620 ± 0.027 |
| 07320+1720 | HDS1065 | G 88-36B | … | … | … | … | … | B | 5.090 | 115.8 | 75.35 | 285.75 | … | … |
| 07320+1720 | WOR 27 | GJ 3448 | J07320+173W | M3.2 V | 07:32:01.87 | +17:19:09.4 | … | C | 11.127 | 282.5 | 59.31 | 281.89 | 159.27 ± 0.87 | 0.360 ± 0.015 |
| … | … | GJ 9235 | J07320+686 | M1.5 V | 07:32:01.51 | +68:37:13.6 | … | … | … | … | 110.83 | 210.11 | 732.07 ± 3.32 | 0.608 ± 0.027 |
| … | … | G 88-37 | J07325+248 | M3.0 V | 07:32:30.71 | +24:53:42.4 | … | … | … | … | 63.55 | 231.15 | 280.93 ± 1.15 | 0.486 ± 0.019 |
| … | … | GJ 1099 | J07342+009 | M2.5 V | 07:34:17.57 | +00:58:59.7 | … | … | … | … | 61.14 | 599.58 | 180.91 ± 0.86 | 0.385 ± 0.016 |
| … | … | GJ 9236 | J07344+629 | M0.5 V | 07:34:26.27 | +62:56:27.6 | … | … | … | … | 42.20 | 507.61 | 257.85 ± 0.95 | 0.388 ± 0.015 |
| … | … | GJ 3453 | J07346+223 | M1.0 V | 07:34:39.31 | +22:20:13.9 | … | … | … | … | 58.49 | 199.08 | 496.87 ± 2.77 | 0.546 ± 0.028 |
| 07346+3153 | STF1110 | Castor | … | A1 V | 07:34:35.95 | +31:53:18.6 | (AabBab)+Cab | AabBab(2+2) | 5.380 | 52.6 | 33.48 | 240.28 | … | 4.933 ± 0.023 |
| 07346+3153 | STF1110 | Castor C | J07346+318 | M0.5 V | 07:34:37.19 | +31:52:08.6 | … | Cab(DEB) | 71.756 | 167.2 | 293.60 | 223.55 | … | 1.19840 ± 0.007 |
| … | … | TYC 777-141-1 | J07349+147 | M3.0 Ve | 07:34:56.25 | +14:45:52.5 | AB | A | … | … | 51.60 | 108.94 | … | 0.446 ± 0.031 |
| 07349+1446 | CRC 53 | G3- 3165346543026736512 | … | … | 07:34:56.18 | +14:45:52.9 | … | B | 1.018 | 291.7 | 94.22 | 109.05 | … | … |
| … | … | GJ 3452 | J07353+548 | M2.5 V | 07:35:21.67 | +54:50:59.2 | … | … | … | … | 52.72 | 114.33 | 196.34 ± 0.93 | 0.364 ± 0.011 |
| … | … | LP 162-39 | J07354+482 | M1.0 V | 07:35:26.96 | +48:14:33.0 | … | … | … | … | 59.87 | 222.98 | 532.91 ± 2.85 | 0.558 ± 0.028 |
| … | … | LP 17-66 | J07359+785 | M3.0 V | 07:35:57.31 | +78:32:49.7 | … | … | … | … | 38.14 | 238.75 | 194.03 ± 0.74 | 0.426 ± 0.017 |
| … | … | V869Mon | … | K3 V | 07:39:59.40 | -03:35:55.5 | A+B+Cab | A | … | … | 42.33 | 286.81 | … | 0.787 ± 0.118 |
| 07400-0336 | BGH 3 | HD 61606B | … | K7 V | 07:40:02.97 | -03:36:17.8 | … | B | 57.903 | 112.7 | 165.21 | 294.21 | 1009.21 ± 4.63 | 0.640 ± 0.096 |
| … | … | GJ 282 C | J07361-031 | M1.0 V | 07:36:07.15 | -03:06:43.4 | … | Cab(1) | 3894.176 | 296.7 | 59.06 | 302.35 | … | 0.742 ± 0.058 |
| 07364+0705 | HEN3 | GJ 3454 | J07364+070 | M4.5 V | 07:36:25.37 | +07:04:38.2 | (AB) | AB | 0.722 | 279.0 | 44.60 | 383.93 | … | 0.176 ± 0.040 |
| … | … | PM J07365-0039 | J07365-006 | M3.5 V | 07:36:30.27 | -00:39:37.3 | … | … | … | … | 41.03 | 107.56 | 190.81 ± 2.05 | 0.422 ± 0.017 |
| … | … | G 111-20 | J07366+440 | M3.5 V | 07:36:39.13 | +44:04:43.5 | A+B | A | … | … | 51.72 | 348.25 | … | 0.432 ± 0.031 |
| 07367+4405 | NSN 578 | G 111-20B | … | M6.5 V | 07:36:38.87 | +44:04:46.5 | … | B | 4.122 | 317.5 | 76.40 | 349.53 | … | 0.105 ± 0.047 |

Table D.2: Complete sample with the description of multiple systems (continued).

| WDS id | WDS disc | Name | Karmn | Spectral type | $\alpha$ (2016.0) | $\delta$ (2016.0) | System | Component | $\rho$ [arcsec] | $\theta$ [deg] | $\varpi$ [mas] | $\mu_{\rm total}$ [mas a$^{-1}$] | $\mathcal{L}$ [$10^{-4}\ \mathcal{L}_\odot$] | $\mathcal{M}$ [$\mathcal{M}_\odot$] |
|---|---|---|---|---|---|---|---|---|---|---|---|---|---|---|
| … | … | TYC 2461-826-1 | J07383+344 | M0.0 V | 07:38:19.92 | +34:27:00.6 | … | … | … | … | … | 121.21 | 991.07 ± 17.24 | 0.689 ± 0.026 |
| … | … | 1R073829.3+240014 | J07384+240 | M3.5 V | 07:38:29.32 | +24:00:07.1 | … | … | … | … | 44.68 | 180.35 | 142.06 ± 0.66 | 0.361 ± 0.015 |
| … | … | GJ 3459 | J07386-212 | dM3.0 | 07:38:41.48 | -21:13:36.1 | … | … | … | … | 39.62 | 659.14 | 113.53 ± 0.55 | 0.284 ± 0.010 |
| … | … | Ross 880 | J07393+021 | M0.0 V | 07:39:22.88 | +02:10:57.3 | … | … | … | … | 42.32 | 286.45 | 775.50 ± 4.16 | 0.638 ± 0.027 |
| … | … | GJ 3457 | J07395+334 | M2.0 V | 07:39:35.62 | +33:27:42.6 | A+B | A | … | … | 35.15 | 236.68 | 818.81 ± 17.83 | 0.610 ± 0.027 |
| 07397+3328 | LDS3755 | GJ 3458 | … | M6.0 V | 07:39:36.51 | +33:27:49.5 | … | B | 13.099 | 58.2 | 35.26 | 237.05 | 11.91 ± 0.17 | 0.155 ± 0.011 |
| … | … | GJ 283 A | … | DZQA6 | 07:40:22.06 | -17:24:57.8 | A+B | A | … | … | 83.88 | 1261.34 | … | 0.500 ± 0.100 |
| 07402-1724 | LUY5693 | GJ 283 B | J07403-174 | M6.5 Ve | 07:40:20.66 | -17:24:54.5 | … | B | 20.338 | 99.5 | 65.48 | 1270.98 | 8.29 ± 0.04 | 0.107 ± 0.047 |
| … | … | GJ 3461 | J07418+050 | M3.0 V | 07:41:52.56 | +05:02:23.1 | AabB | Aab(EB) | … | … | 32.35 | 263.05 | … | … |
| 07419+0502 | CFN 5 | G 50-1B | … | … | 07:41:52.61 | +05:02:22.4 | … | B | 1.014 | 130.9 | 89.82 | … | … | … |
| … | … | LP 162-55 | J07421+500 | M2.5 V | 07:42:10.07 | +50:04:28.5 | … | … | … | … | 47.73 | 195.83 | 257.35 ± 5.89 | 0.464 ± 0.019 |
| … | … | GJ 3462 | J07431+181 | M1.5 V | 07:43:11.67 | +18:10:34.8 | … | … | … | … | 71.32 | 472.06 | 938.46 ± 6.50 | 0.649 ± 0.027 |
| … | … | YZCMi | J07446+035 | M4.0 Ve | 07:44:39.80 | +03:33:01.7 | … | … | … | … | 40.18 | 565.33 | 111.34 ± 0.82 | 0.415 ± 0.031 |
| … | … | G 193-65 | J07467+574 | M4.5 V | 07:46:41.97 | +57:26:49.5 | … | … | … | … | 54.87 | 234.80 | 96.31 ± 0.54 | 0.335 ± 0.015 |
| … | … | LP 17-75 | J07470+760 | M4.0 V | 07:47:06.48 | +76:03:13.1 | … | … | … | … | 80.02 | 414.88 | 96.09 ± 0.41 | 0.313 ± 0.014 |
| … | … | 1R074714.1+502032 | J07472+503 | M4.0 Ve | 07:47:13.84 | +50:20:39.8 | … | … | … | … | 147.00 | 81.37 | 78.57 ± 0.36 | 0.271 ± 0.014 |
| … | … | Wolf 1421 | J07482+203 | M1.5 V | 07:48:18.04 | +20:21:49.4 | … | … | … | … | 35.23 | 1756.92 | 187.13 ± 0.76 | 0.347 ± 0.014 |
| … | … | GJ 3456 | J07493+849 | M3.0 V | 07:49:17.72 | +84:58:32.5 | … | … | … | … | 38.99 | 383.89 | 212.85 ± 1.06 | 0.420 ± 0.017 |
| … | … | PM J07497-0320 | J07497-033 | M3.5 V | 07:49:41.97 | -03:20:34.9 | … | … | … | … | 58.90 | 171.22 | 117.19 ± 1.15 | 0.348 ± 0.015 |
| … | … | GJ 3463 | J07518+055 | M5.0 V | 07:51:51.86 | +05:32:50.6 | … | … | … | … | 264.13 | 603.72 | 30.09 ± 0.13 | 0.177 ± 0.011 |
| … | … | GJ 1103 A | J07519-000 | M4.5 V | 07:51:54.95 | -00:00:24.4 | … | … | … | … | 49.34 | 801.43 | 46.94 ± 0.32 | 0.211 ± 0.011 |
| … | … | LP 423-31 | J07523+162 | M6.0 V | 07:52:24.13 | +16:12:09.4 | … | … | … | … | 73.97 | 395.30 | 19.48 ± 0.12 | 0.174 ± 0.011 |
| … | … | 1R075434.3+083213 | J07545+085 | M2.5 V | 07:54:33.90 | +08:32:25.5 | Aab | Aab(1) | … | … | 145.62 | 213.07 | … | … |
| … | … | GJ 3465 | J07525+063 | M3.0 V | 07:52:33.63 | +06:18:22.0 | … | … | … | … | 85.22 | 214.56 | 215.86 ± 1.17 | 0.450 ± 0.018 |
| … | … | PM J07545-0941 | J07545-096 | M3.5 V | 07:54:32.60 | -09:41:47.9 | AB | A | … | … | 23.42 | 90.90 | … | 0.392 ± 0.032 |
| 07545-0942 | CFN 6 | PM J07545-0941B | … | … | 07:54:32.66 | -09:41:48.7 | … | B | 1.227 | 130.6 | 23.57 | … | … | … |
| … | … | GJ 1101 | J07558+833 | M4.5: V | 07:55:51.23 | +83:22:55.4 | … | … | … | … | 88.72 | 666.30 | 74.17 ± 0.31 | 0.262 ± 0.013 |
| … | … | GJ 3467 | J07581+072 | M5.0 V | 07:58:08.74 | +07:17:00.9 | … | … | … | … | 83.48 | 327.27 | 47.63 ± 0.31 | 0.213 ± 0.011 |
| … | … | GJ 1105 | J07582+413 | M3.5 V | 07:58:13.01 | +41:18:02.3 | … | … | … | … | … | 721.00 | 87.32 ± 0.48 | 0.255 ± 0.010 |
| … | … | LP 163-47 | J07583+496 | M4.0 V | 07:58:23.26 | +49:39:41.3 | … | … | … | … | 17.88 | 811.01 | 110.22 ± 0.59 | 0.336 ± 0.015 |
| … | … | GJ 3468 | J07585+155N | M4.5 V | 07:58:30.88 | +15:30:12.6 | A+(BC) | A | … | … | 18.14 | 139.95 | 139.97 ± 0.81 | 0.382 ± 0.016 |
| 07586+1530 | LDS3768 | GJ 3469 A | J07585+155S | M4.5 V | 07:58:30.37 | +15:29:58.8 | … | B | 15.708 | 208.1 | 18.39 | 143.68 | … | 0.225 ± 0.038 |
| 07586+1530 | SKF2835 | GJ 3469 B | … | M5.0 V | 07:58:30.34 | +15:29:57.8 | … | C | 16.815 | 207.9 | 83.38 | 136.71 | … | 0.190 ± 0.039 |
| … | … | GJ 3470 | J07590+153 | M2.0 V | 07:59:05.63 | +15:23:28.3 | … | … | … | … | 83.40 | 194.26 | 391.79 ± 2.01 | 0.493 ± 0.030 |
| … | … | 1R075908.2+171957 | J07591+173 | M4.0 V | 07:59:07.07 | +17:19:46.8 | … | … | … | … | 83.52 | 97.01 | … | … |
| … | … | TYC 1930-667-1 | J08005+258 | M2.0 V | 08:00:34.87 | +25:53:32.6 | … | … | … | … | 25.87 | 122.18 | 987.25 ± 8.40 | 0.650 ± 0.027 |
| … | … | TYC 1926-794-1 | J08017+237 | M1.5 V | 08:01:43.44 | +23:42:25.3 | … | … | … | … | 37.05 | 186.08 | 640.27 ± 23.97 | 0.576 ± 0.028 |
| … | … | GJ 3473 | J08023+033 | M4.0 V | 08:02:22.45 | +03:20:13.6 | A+B | A | … | … | … | 554.68 | 149.78 ± 0.72 | 0.343 ± 0.011 |
| 08024+0320 | LDS5160 | GJ 3474 | … | M6.0 V | 08:02:20.22 | +03:19:37.4 | … | B | 49.260 | 222.8 | 34.13 | 551.81 | 11.52 ± 0.09 | 0.128 ± 0.010 |
| … | … | LP 724-16 | J08025-130 | M2.5 V | 08:02:33.09 | -13:05:33.4 | … | … | … | … | 42.76 | 291.96 | 291.77 ± 11.59 | 0.466 ± 0.020 |

Table D.2: Complete sample with the description of multiple systems (continued).

| WDS id | WDS disc | Name | Karmn | Spectral type | $\alpha$ (2016.0) | $\delta$ (2016.0) | System | Component | $\rho$ [arcsec] | $\theta$ [deg] | $\varpi$ [mas] | $\mu_{\rm total}$ [mas a$^{-1}$] | $\mathcal{L}$ [$10^{-4}$ $\mathcal{L}_\odot$] | $\mathcal{M}$ [$\mathcal{M}_\odot$] |
|---|---|---|---|---|---|---|---|---|---|---|---|---|---|---|
| 08032+2022 | JNN 60 | PM J08031+2022 | J08031+203 | M3.5 V | 08:03:10.06 | +20:22:14.3 | (AB) | AB | 0.206 | 17.8 | 36.89 | 110.90 | … | 0.455 ± 0.032 |
| 08033+5251 | HDS1149 | G 194-7 | J08033+528 | M1.5 V | 08:03:20.18 | +52:50:27.3 | (AB) | AB | 0.287 | 218.2 | 64.00 | 767.31 | … | 0.629 ± 0.045 |
| 08066+5554 | CRC 54 | GJ 3477 | J08066+558 | M2.0 V | 08:06:36.75 | +55:53:37.1 | (AB) | AB | 0.174 | 236.5 | 84.70 | 154.63 | … | 0.550 ± 0.029 |
| … | … | GJ 3479 | J08068+367 | M3.0 V | 08:06:48.21 | +36:45:32.5 | … | … | … | … | 38.70 | 417.51 | 114.76 ± 0.55 | 0.322 ± 0.014 |
| … | … | G 111-56 | J08069+422 | M4.0 V | 08:06:54.99 | +42:17:28.7 | … | … | … | … | 64.12 | 353.75 | 70.46 ± 0.32 | 0.264 ± 0.013 |
| 08082+2106 | COU 91 | GJ 3482 | J08082+211 | M3.0 V | 08:08:12.85 | +21:06:12.6 | A+BC | A | … | … | 66.31 | 463.21 | 999.07 ± 6.46 | 0.705 ± 0.026 |
| 08082+2106 | CRC 55 | GJ 3481 | … | K7 Ve | 08:08:13.29 | +21:06:03.9 | … | BC(2) | 10.655 | 144.6 | 66.43 | 453.31 | … | … |
| … | … | GJ 3480 | J08083+585 | M3.0 V | 08:08:17.82 | +58:31:08.4 | … | … | … | … | 61.37 | 185.60 | 105.26 ± 0.39 | 0.307 ± 0.014 |
| 08090+3249 | BDT 2 | FPCnc | … | K6 V | 08:08:56.33 | +32:49:08.3 | (AB)+Cab | AB | 0.207 | 69.6 | 61.80 | 203.10 | … | … |
| 08090+3249 | DYR 1 | FPCncB | J08089+328 | M3.0 V | 08:08:55.38 | +32:49:01.4 | … | Cab(2) | 13.927 | 240.0 | 28.63 | 214.09 | … | … |
| … | … | GJ 3484 | J08095+219 | M2.0 V | 08:09:30.59 | +21:54:16.2 | … | … | … | … | 78.16 | 326.02 | 306.03 ± 1.21 | 0.478 ± 0.018 |
| … | … | PM J08103+0935 | J08103+095 | M2.5 V | 08:10:20.65 | +09:35:15.4 | … | … | … | … | 33.75 | 86.55 | 426.80 ± 2.74 | 0.505 ± 0.029 |
| … | … | HD 68146 | … | F6.5 V | 08:10:39.55 | -13:47:56.2 | A+BC | A | … | … | 38.67 | 257.61 | … | 1.232 ± 0.185 |
| 08107-1348 | LDS204 | HD 68146B | J08105-138 | M2.5 V | 08:10:34.02 | -13:48:50.1 | … | B | 96.934 | 236.2 | 71.03 | 259.91 | … | 0.463 ± 0.030 |
| 08107-1348 | JOD 4 | G3- 5725122965265271680 | … | M4.0 V | 08:10:33.96 | -13:48:49.9 | … | C | 97.537 | 236.6 | 70.99 | 245.49 | … | 0.274 ± 0.036 |
| … | … | GJ 3485 | J08108+039 | M4.0 V | 08:10:53.75 | +03:58:28.2 | … | … | … | … | 70.27 | 358.58 | 142.46 ± 0.72 | 0.361 ± 0.015 |
| … | … | G 194-14 | J08117+531 | M2.5 V | 08:11:47.07 | +53:11:48.4 | … | … | … | … | 116.60 | 334.28 | 359.61 ± 2.32 | 0.471 ± 0.030 |
| … | … | Ross 619 | J08119+087 | M4.5 V | 08:11:58.72 | +08:45:01.4 | … | … | … | … | 36.37 | 5205.34 | 27.95 ± 0.17 | 0.153 ± 0.009 |
| … | … | GJ 300 | J08126-215 | M3.5 V | 08:12:40.90 | -21:33:18.1 | … | … | … | … | 22.93 | 694.12 | 82.76 ± 0.41 | 0.260 ± 0.012 |
| … | … | LP 311-8 | J08158+346 | M1.0 V | 08:15:53.78 | +31:36:35.8 | … | … | … | … | 23.17 | 214.62 | 1138.48 ± 17.87 | 0.693 ± 0.026 |
| … | … | GJ 2066 | J08161+013 | M2.0 V | 08:16:07.58 | +01:18:10.2 | … | … | … | … | 26.56 | 381.04 | 278.67 ± 1.55 | 0.436 ± 0.012 |
| … | … | LP 367-67 | J08175+209 | M2.5 V | 08:17:31.31 | +20:59:48.7 | … | … | … | … | 51.56 | 427.91 | 371.26 ± 2.93 | 0.484 ± 0.030 |
| … | … | GJ 3491 | J08178+311 | M1.0 V | 08:17:51.20 | +31:07:49.4 | … | … | … | … | 95.66 | 229.91 | 493.50 ± 1.95 | 0.543 ± 0.029 |
| … | … | PM J08202+0532 | J08202+055 | M2.0 V | 08:20:13.29 | +05:32:08.2 | … | … | … | … | 66.34 | 75.55 | 312.10 ± 1.84 | 0.455 ± 0.017 |
| … | … | GJ 3497 | J08258+690 | M7.0 V | 08:25:50.76 | +69:01:40.7 | … | … | … | … | 28.47 | 1451.45 | 16.85 ± 0.22 | 0.148 ± 0.010 |
| … | … | GJ 1110 | J08282+201 | M4.0 V | 08:28:12.37 | +20:08:11.4 | … | … | … | … | 27.73 | 685.79 | 139.95 ± 0.58 | 0.336 ± 0.014 |
| … | … | GJ 308 | J08283+350 | M0.0 V | 08:28:20.83 | +35:00:53.6 | AB | A | … | … | 109.34 | 1086.44 | … | 0.473 ± 0.046 |
| 08286+3502 | WOR19 | G3- 903444927988618752 | … | … | 08:28:20.82 | +35:00:54.0 | … | B | 0.487 | 340.4 | 109.25 | … | … | … |
| … | … | PM J08283+5522 | J08283+553 | M2.5 V | 08:28:18.75 | +55:22:40.6 | … | … | … | … | 36.11 | 106.35 | 233.73 ± 0.96 | 0.441 ± 0.017 |
| 08286+6602 | JNN 273 | 1R082839.4+660229 | J08286+660 | M4.0 V | 08:28:41.33 | +66:02:25.4 | (AB) | AB | 0.294 | 146.4 | … | 101.92 | … | 0.274 ± 0.038 |
| … | … | PM J08293+0355E | J08293+039 | M2.5 V | 08:29:21.81 | +03:55:08.2 | … | … | … | … | 48.58 | 116.89 | 357.26 ± 2.01 | 0.475 ± 0.012 |
| … | … | DXCnc | J08298+267 | M6.5 V | 08:29:48.02 | +26:46:23.8 | … | … | … | … | 31.28 | 1270.86 | 7.67 ± 0.03 | 0.103 ± 0.048 |
| 08313-0601 | BAG 49 | GJ 3501 | J08313-060 | M2.0 V | 08:31:21.14 | -06:02:02.8 | (AB)+C | AB | 0.121 | 76.8 | 166.98 | 436.18 | … | 0.587 ± 0.028 |
| 08313-0601 | LDS 221 | GJ 3502 | J08314-060 | M3.0 V | 08:31:26.75 | -06:02:13.6 | … | C | 84.405 | 97.3 | 43.26 | 441.13 | 307.15 ± 1.57 | 0.479 ± 0.018 |
| … | … | GJ 3503 | J08313-104 | M4.0 V | 08:31:22.81 | -10:29:58.9 | … | … | … | … | 39.36 | 676.43 | 36.19 ± 0.16 | 0.196 ± 0.011 |
| … | … | LP 35-219 | J08315+730 | M4.0 V | 08:31:32.36 | +73:03:50.1 | … | … | … | … | 70.88 | 669.51 | 93.09 ± 0.46 | 0.276 ± 0.012 |
| 08317+1924 | BEU 12 | CUCnc | J08316+193S | M3.5 V | 08:31:37.32 | +19:23:37.5 | (AabB)+CD | AabB(EB) | 0.345 | 222.9 | 32.94 | 258.23 | … | 0.834 ± 0.001 |
| 08317+1924 | LDS 905 | CCncA | J08316+193N | M4.0 V | 08:31:37.17 | +19:23:47.6 | … | C | 10.363 | 348.5 | 68.03 | 266.18 | … | 0.259 ± 0.036 |
| 08317+1924 | DEL 1 | CCncB | … | M5.0 V | 08:31:37.16 | +19:23:46.7 | … | D | 9.435 | 346.9 | 41.80 | 242.46 | … | 0.208 ± 0.038 |
| … | … | 1R083147.3+054504 | J08317+057 | M1.0 Ve | 08:31:47.89 | +05:45:17.0 | … | … | … | … | 57.83 | 93.30 | 1049.84 ± 16.60 | 0.674 ± 0.026 |

Table D.2: Complete sample with the description of multiple systems (continued).

| WDS id | WDS disc | Name | Karmn | Spectral type | $\alpha$ (2016.0) | $\delta$ (2016.0) | System | Component | $\rho$ [arcsec] | $\theta$ [deg] | $\varpi$ [mas] | $\mu_{\rm total}$ [mas a$^{-1}$] | $\mathcal{L}$ [$10^{-4}\ \mathcal{L}_\odot$] | $\mathcal{M}$ [$\mathcal{M}_\odot$] |
|---|---|---|---|---|---|---|---|---|---|---|---|---|---|---|
| … | … | GJ 3496 | J08321+844 | M3.5 V | 08:32:13.99 | +84:24:34.8 | … | … | … | … | 67.62 | 432.48 | 89.12 ± 0.41 | 0.281 ± 0.013 |
| … | … | Wolf 312 | J08325+451 | M2.5 V | 08:32:35.78 | +45:10:16.3 | … | … | … | … | 107.85 | 155.67 | 252.21 ± 1.47 | 0.432 ± 0.017 |
| … | … | GJ 3505 | J08334+185 | M4.5 V | 08:33:25.06 | +18:31:34.9 | … | … | … | … | 52.96 | 648.79 | 46.99 ± 0.30 | 0.211 ± 0.011 |
| … | … | GJ 2070 | J08344-011 | M3.0 V | 08:34:26.13 | -01:08:46.3 | … | … | … | … | … | 479.24 | 79.46 ± 0.40 | 0.264 ± 0.013 |
| … | … | LSPM J0835+1408 | J08353+141 | M4.5 V | 08:35:19.75 | +14:08:31.9 | Aabc | Aabc(3) | … | … | 30.09 | 171.72 | … | … |
| … | … | GJ 3506 | J08358+680 | M3.0 V | 08:35:46.65 | +68:04:00.1 | … | … | … | … | 31.43 | 1008.72 | 179.89 ± 0.71 | 0.362 ± 0.011 |
| … | … | LP 311-37 | J08364+264 | M2.0 V | 08:36:26.45 | +26:28:18.9 | AB | A | … | … | 27.09 | 199.63 | … | 0.473 ± 0.030 |
| 08364+2628 | KPP3209 | LP 311-37 B | … | M6.5 V | 08:36:26.29 | +26:28:17.9 | … | B | 2.445 | 246.7 | … | 202.70 | … | 0.107 ± 0.047 |
| 08364+6718 | YR 13 | GJ 310 | J08364+672 | M0.5 V | 08:36:22.51 | +67:17:42.9 | (AB) | AB | 0.259 | 239.8 | 76.39 | 1072.67 | … | 0.661 ± 0.026 |
| … | … | GJ 3508 | J08371+151 | M3.0 V | 08:37:07.82 | +15:07:31.2 | … | … | … | … | 76.28 | 903.19 | 265.52 ± 1.65 | 0.472 ± 0.018 |
| … | … | LSPM J0837+0333 | J08375+035 | M4.0 V | 08:37:30.28 | +03:33:43.1 | … | … | … | … | 112.99 | 186.28 | 56.64 ± 0.27 | 0.234 ± 0.012 |
| … | … | StKM 1-711 | J08387+516 | M1.5 V | 08:38:42.06 | +51:41:31.9 | … | … | … | … | 62.47 | 141.61 | 573.78 ± 2.66 | 0.567 ± 0.028 |
| … | … | GJ 3510 | J08398+089 | M2.0 V | 08:39:47.80 | +08:56:21.0 | A+B | A | … | … | 32.08 | 284.72 | 173.99 ± 1.05 | 0.355 ± 0.015 |
| 08398+0856 | LDS3806 | GJ 3511 | … | M4.5 V | 08:39:48.25 | +08:56:21.1 | … | B | 6.645 | 88.7 | 32.03 | 282.64 | 47.42 ± 0.25 | 0.212 ± 0.011 |
| … | … | LSPM J0840+3127 | J08402+314 | M3.5 V | 08:40:16.24 | +31:27:08.7 | … | … | … | … | 32.16 | 238.18 | 98.82 ± 0.54 | 0.274 ± 0.010 |
| … | … | AZCnc | J08404+184 | M6.0 V | 08:40:28.77 | +18:24:01.5 | … | … | … | … | 34.02 | 925.95 | 8.55 ± 0.05 | 0.117 ± 0.010 |
| … | … | GJ 317 | J08409-234 | M3.5 V | 08:40:58.67 | -23:27:09.7 | … | … | … | … | … | 928.02 | 223.23 ± 2.04 | 0.406 ± 0.014 |
| … | … | GJ 3509 | J08410+676 | M4.0 V | 08:41:01.83 | +67:39:33.1 | … | … | … | … | 28.12 | 761.15 | 68.04 ± 0.29 | 0.259 ± 0.013 |
| … | … | GJ 3512 | J08413+594 | dM5.5 | 08:41:19.58 | +59:29:30.0 | … | … | … | … | 46.09 | 1305.77 | 16.88 ± 0.08 | 0.122 ± 0.009 |
| … | … | GJ 319 A | J08427+095 | M0.0 V | 08:42:44.77 | +09:33:14.0 | AB+C | A | … | … | 36.61 | 670.38 | … | 0.610 |
| 08427+0935 | ST 8 | GJ 319 B | … | M4.0 V | 08:42:44.84 | +09:33:15.1 | … | B | 1.512 | 40.5 | 36.45 | 610.58 | … | 0.270 ± 0.036 |
| 08427+0935 | LUY6218 | GJ 319 C | J08428+095 | M2.5 V | 08:42:52.47 | +09:33:01.3 | … | C | 114.573 | 96.3 | 29.45 | 653.39 | 208.03 ± 1.09 | 0.415 ± 0.017 |
| … | … | GJ 3513 | J08443-104 | M3.5 V | 08:44:22.71 | -10:24:20.1 | … | … | … | … | 27.25 | 603.95 | 150.76 ± 0.80 | 0.372 ± 0.016 |
| … | … | G 9-19 | J08447+182 | M3.5 V | 08:44:45.08 | +18:12:59.2 | … | … | … | … | 31.87 | 510.61 | 372.62 ± 2.28 | 0.456 ± 0.030 |
| 08449-0637 | JNN 63 | PM J08449-0637 | J08449-066 | M3.5 V | 08:44:55.59 | -06:37:28.2 | (AB) | AB | 0.274 | 23.2 | 40.34 | 137.75 | … | … |
| … | … | Ross 622 | J08517+181 | M1.5 V | 08:51:42.78 | +18:07:29.1 | … | … | … | … | 56.59 | 901.53 | 216.97 ± 0.98 | 0.376 ± 0.015 |
| … | … | 55 Cnc | … | K0 IV-V | 08:52:35.22 | +28:19:47.2 | A+B | A | … | … | 50.35 | 538.90 | 6005.81 ± 100.88 | … |
| 08526+2820 | LDS6219 | GJ 324 B | J08526+283 | M4.5 V | 08:52:40.28 | +28:18:54.9 | … | B | 84.826 | 128.1 | 56.01 | 539.75 | 79.01 ± 0.53 | 0.259 ± 0.011 |
| … | … | PM J08531-2017 | J08531-202 | M3.0 V | 08:53:10.96 | -20:17:19.3 | … | … | … | … | 56.23 | 123.21 | 152.91 ± 0.91 | 0.352 ± 0.015 |
| … | … | GJ 3517 | J08536-034 | M9.0 Ve | 08:53:35.61 | -03:29:35.4 | … | … | … | … | 64.43 | 553.84 | 2.87 ± 0.02 | 0.079 ± 0.012 |
| … | … | StKM 1-730 | J08537+149 | M0.0 V | 08:53:43.67 | +14:58:09.7 | … | … | … | … | 41.73 | 83.64 | 804.48 ± 10.57 | 0.618 ± 0.027 |
| … | … | GJ 326 A | J08540-131 | M2.5 V | 08:54:05.69 | -13:07:39.9 | AB | A | … | … | 40.82 | 668.71 | … | 0.318 ± 0.034 |
| 08539-1308 | ST 9 | GJ 326 B | … | M3.5 V | 08:54:05.63 | -13:07:39.8 | … | B | 0.919 | 275.8 | 47.74 | 631.53 | … | 0.305 ± 0.034 |
| … | … | Ross 623 | J08551+015 | M0.0 V | 08:55:07.67 | +01:32:30.7 | … | … | … | … | 39.61 | 1046.84 | 970.55 ± 4.09 | 0.682 ± 0.026 |
| … | … | PM J08555+6628 | J08555+664 | M3.0 V | 08:55:31.46 | +66:28:06.6 | … | … | … | … | 44.34 | 94.03 | 264.23 ± 1.69 | 0.470 ± 0.018 |
| 08563+1239 | JNN 274 | G 41-8 | J08563+126 | M6.0 V | 08:56:19.49 | +12:39:45.8 | (AB) | AB | 1.824 | 210.3 | 44.52 | 251.17 | … | 0.171 ± 0.040 |
| 08571+1139 | HDS1296 | GJ 330 | J08570+116 | M1.0 V | 08:57:04.65 | +11:38:43.9 | (AB) | AB | 0.703 | 237.3 | 44.15 | 311.03 | … | 0.509 ± 0.029 |
| … | … | LP 426-35 | J08572+194 | M3.5 V | 08:57:15.55 | +19:24:15.2 | … | … | … | … | 45.38 | 196.36 | 276.39 ± 67.04 | 0.376 ± 0.043 |
| … | … | GJ 1116 A | J08582+197 | M5.5 V | 08:58:14.09 | +19:45:45.3 | AB | A | … | … | 28.06 | 937.77 | … | 0.111 ± 0.046 |
| 08582+1945 | LDS3836 | GJ 1116 B | … | M7.0 V | 08:58:14.21 | +19:45:46.7 | … | B | 2.175 | 50.9 | 147.72 | 773.57 | … | 0.127 ± 0.044 |

Table D.2: Complete sample with the description of multiple systems (continued).

| WDS id | WDS disc | Name | Karmn | Spectral type | $\alpha$ (2016.0) | $\delta$ (2016.0) | System | Component | $\rho$ [arcsec] | $\theta$ [deg] | $\varpi$ [mas] | $\mu_{total}$ [mas a$^{-1}$] | $\mathcal{L}$ [$10^{-4}$ $\mathcal{L}_\odot$] | $\mathcal{M}$ [$\mathcal{M}_\odot$] |
|---|---|---|---|---|---|---|---|---|---|---|---|---|---|---|
| … | … | G 41-13 | J08588+210 | M2.0 V | 08:58:52.53 | +21:04:29.1 | … | … | … | … | 123.20 | 363.62 | … | 0.322 ± 0.034 |
| 08596+5344 | CRC 56 | G 194-47 | J08595+537 | M3.5 V | 08:59:35.41 | +53:43:47.5 | (AB) | AB | 0.485 | 220.7 | 19.69 | 324.42 | … | 0.290 ± 0.035 |
| … | … | GJ 3522 | J08589+084 | M3.5 V | 08:58:56.73 | +08:28:20.8 | AB | A(2) | … | … | 21.98 | 492.32 | … | 0.235 ± 0.001 |
| 08589+0829 | DEL 2 | G3- 585250248258396416 | … | … | 08:58:56.76 | +08:28:20.9 | … | B | 0.447 | 81.7 | 111.87 | … | … | … |
| … | … | GJ 3520 | J08599+729 | M5.0 V | 08:59:59.70 | +72:57:35.8 | … | … | … | … | 32.01 | 962.14 | 35.71 ± 0.13 | 0.195 ± 0.011 |
| … | … | LP 368-128 | J09003+218 | M6.5 V | 09:00:22.95 | +21:49:55.4 | … | … | … | … | 46.06 | 784.81 | 7.76 ± 0.04 | 0.103 ± 0.048 |
| … | … | GJ 1119 | J09005+465 | M4.5 Ve | 09:00:31.74 | +46:35:02.7 | … | … | … | … | 44.51 | 705.49 | 49.39 ± 0.21 | 0.204 ± 0.009 |
| … | … | Ross 686 | J09008+052W | M3.0 V | 09:00:48.25 | +05:14:38.1 | A+B | A | … | … | 81.37 | 333.91 | 262.86 ± 1.62 | 0.469 ± 0.018 |
| 09008+0516 | OSV 2 | Ross 687 | J09008+052E | M3.0 V | 09:00:50.05 | +05:14:26.3 | … | B | 29.459 | 113.7 | 42.56 | 328.91 | 207.33 ± 1.75 | 0.414 ± 0.017 |
| 09012+0157 | CRC 57 | Ross 625 | J09011+019 | M3.0 V | 09:01:10.07 | +01:56:33.7 | (AB)+C | AB(2) | 0.182 | 148.8 | 44.00 | 387.95 | … | … |
| 09012+0157 | CRC 57 | Ross 625B | … | M6.0 V | 09:01:10.16 | +01:56:31.0 | … | C | 3.051 | 155.6 | 50.82 | 404.21 | … | 0.111 ± 0.046 |
| … | … | GJ 3528 | J09023+084 | M3.0 V | 09:02:20.55 | +08:28:03.3 | … | … | … | … | … | 651.46 | 352.55 ± 2.17 | 0.473 ± 0.030 |
| … | … | PM J09023+1746 | J09023+177 | M4.0 V | 09:02:22.91 | +17:46:31.8 | … | … | … | … | 35.56 | 141.53 | 84.46 ± 0.35 | 0.292 ± 0.014 |
| … | … | GJ 3526 | J09028+680 | M4.0 V | 09:02:53.41 | +68:03:52.1 | … | … | … | … | 50.50 | 391.40 | 86.12 ± 0.39 | 0.271 ± 0.012 |
| … | … | LSPM J0902+7138 | J09029+716 | M1.5 V | 09:02:55.82 | +71:38:11.0 | … | … | … | … | 54.28 | 174.76 | 357.82 ± 1.44 | 0.473 ± 0.030 |
| … | … | LP 546-37 | J09033+056 | M7.0 V | 09:03:20.91 | +05:40:08.5 | … | … | … | … | 279.25 | 377.92 | … | … |
| … | … | G 194-52 | J09037+520 | M3.5 V | 09:03:43.39 | +52:02:49.1 | … | … | … | … | 39.62 | 335.51 | 129.83 ± 0.66 | 0.344 ± 0.015 |
| … | … | LP 486-43 | J09038+129 | M2.0 V | 09:03:53.41 | +12:59:24.8 | … | … | … | … | 39.79 | 214.36 | 761.64 ± 7.64 | 0.599 ± 0.027 |
| … | … | V405Hya | … | K2 V | 09:04:20.57 | -15:54:51.8 | A+B | A | … | … | 58.91 | 112.03 | 2581.76 ± 12.41 | 0.820 ± 0.123 |
| … | … | 1R090406.8-155512 | J09040-159 | M2.5 V | 09:04:05.44 | -15:55:19.0 | … | B* | 220.022 | 82.9 | 68.21 | 113.77 | 245.78 ± 1.55 | 0.453 ± 0.018 |
| … | … | GJ 3530 | J09050+028 | M1.5 V | 09:05:04.11 | +02:50:03.9 | AB | A | … | … | 60.06 | 314.11 | … | 0.487 ± 0.030 |
| 09051+0250 | RAO 232 | GJ 3530 | … | M3.5 V | 09:05:04.03 | +02:50:03.5 | … | B | 1.220 | 253.9 | 60.25 | 336.22 | … | 0.335 ± 0.034 |
| … | … | LP 426-56 | J09057+186 | M2.5 V | 09:05:43.02 | +18:36:27.6 | … | … | … | … | 60.69 | 397.28 | 603.56 ± 11.58 | 0.554 ± 0.028 |
| … | … | GJ 3531 | J09062+128 | M3.5 V | 09:06:13.79 | +12:51:30.1 | … | … | … | … | 20.63 | 407.05 | 244.99 ± 6.18 | 0.481 ± 0.020 |
| … | … | GJ 3533 | J09070-221 | M4.5 V | 09:07:02.40 | -22:08:56.6 | … | … | … | … | 51.92 | 508.98 | 51.19 ± 0.27 | 0.238 ± 0.012 |
| … | … | GJ 3532 | J09087+665 | M2.5 V | 09:08:46.58 | +66:35:36.4 | … | … | … | … | 40.66 | 264.00 | 129.40 ± 3.29 | 0.322 ± 0.015 |
| … | … | 2M09090798+2247413 | J09091+227 | M4.5 V | 09:09:07.88 | +22:47:40.1 | … | … | … | … | 47.87 | 109.41 | 53.93 ± 0.78 | 0.245 ± 0.013 |
| … | … | GJ 1121 | J09093+401 | M4.0 V | 09:09:23.06 | +40:05:55.5 | … | … | … | … | 74.41 | 807.61 | 54.83 ± 0.61 | 0.230 ± 0.012 |
| … | … | GJ 336 | J09095+328 | M0.5 V | 09:09:30.18 | +32:48:59.1 | AB | A | … | … | 38.85 | 715.71 | … | 0.656 ± 0.027 |
| 09095+3250 | COU 1561 | BD+33 1814B | … | M0.0 V | 09:09:30.17 | +32:48:59.5 | … | B | 0.483 | 353.0 | 76.94 | … | … | … |
| … | … | GJ 3537 | J09096+067 | M3.0 V | 09:09:39.07 | +06:42:11.8 | … | … | … | … | 34.35 | 91.36 | 146.99 ± 0.88 | 0.367 ± 0.016 |
| … | … | G 46-24 | J09099+004 | M1.0 V | 09:09:59.43 | +00:23:39.5 | … | … | … | … | 34.48 | 506.47 | 280.68 ± 1.35 | 0.430 ± 0.017 |
| … | … | LP 487-10 | J09115+126 | M2.5 V | 09:11:32.14 | +12:37:18.3 | … | … | … | … | 70.62 | 378.85 | 721.32 ± 12.48 | 0.578 ± 0.028 |
| … | … | GJ 336.1 | J09115+466 | M0.5 V | 09:11:30.27 | +46:37:00.7 | … | … | … | … | 56.11 | 352.70 | 643.41 ± 2.45 | 0.597 ± 0.027 |
| … | … | GJ 3540 | J09120+279 | M3.0 V | 09:12:02.44 | +27:54:16.2 | Aab | Aab(2) | … | … | 159.66 | 489.68 | … | … |
| … | … | G 234-57A | J09133+688 | M2.5 V | 09:13:23.43 | +68:52:27.1 | AB | A | … | … | 52.93 | 288.04 | 672.86 ± 12.64 | 0.494 ± 0.030 |
| 09134+6853 | RAO 568 | G 234-57B | … | M2.0 V | 09:13:23.34 | +68:52:26.7 | … | B | 0.596 | 231.8 | 30.52 | 275.37 | … | 0.463 ± 0.030 |
| … | … | LP 427-16 | J09140+196 | M3.0 V | 09:14:03.02 | +19:40:03.2 | Aab | Aab(1) | … | … | 42.97 | 217.28 | … | … |
| … | … | HD 79210 | J09143+526 | M0.0 V | 09:14:20.05 | +52:41:02.7 | Aab+B | Aab(1) | … | … | 42.71 | 1647.20 | … | … |
| 09144+5241 | STF1321 | HD 79211 | J09144+526 | M0.0 V | 09:14:21.91 | +52:41:00.3 | … | B | 17.075 | 97.9 | 38.06 | 1705.85 | 772.68 ± 20.80 | 0.632 ± 0.027 |

Table D.2: Complete sample with the description of multiple systems (continued).

| WDS id | WDS disc | Name | Karmn | Spectral type | $\alpha$ (2016.0) | $\delta$ (2016.0) | System | Component | $\rho$ [arcsec] | $\theta$ [deg] | $\varpi$ [mas] | $\mu_{total}$ [mas a$^{-1}$] | $\mathcal{L}$ [$10^{-4}$ $\mathcal{L}_\odot$] | $\mathcal{M}$ [$\mathcal{M}_\odot$] |
|---|---|---|---|---|---|---|---|---|---|---|---|---|---|---|
| 09156-1036 | MTG 2 | G 161-7 | J09156-105 | M5.0 V | 09:15:35.96 | -10:35:50.2 | (AB) | AB | 0.146 | 62.1 | 38.00 | 440.48 | … | 0.159 ± 0.042 |
| … | … | G 47-31 | J09160+293 | M2.0 V | 09:16:05.04 | +29:19:36.3 | … | … | … | … | 89.07 | 544.57 | 452.92 ± 2.55 | 0.522 ± 0.029 |
| … | … | RX J0916.1+0153 | J09161+018 | M4.0 Ve | 09:16:10.24 | +01:53:07.2 | … | … | … | … | 73.86 | 115.19 | 106.69 ± 0.72 | 0.323 ± 0.016 |
| … | … | GJ 3543 | J09163-186 | M1.5 V | 09:16:20.29 | -18:37:30.6 | … | … | … | … | 65.88 | 347.01 | 327.45 ± 1.87 | 0.441 ± 0.011 |
| … | … | GJ 3536 | J09165+841 | M1.5 V | 09:16:24.75 | +84:11:06.4 | … | … | … | … | 38.94 | 658.85 | 338.59 ± 163.73 | 0.449 ± 0.116 |
| … | … | 2M09165078+2448559 | J09168+248 | M4.5 V | 09:16:50.70 | +24:48:54.0 | … | … | … | … | 105.29 | 141.95 | 102.87 ± 0.56 | 0.347 ± 0.015 |
| 09177+4612 | JNN 68 | RX J0917.7+4612 | J09177+462 | M2.5 V | 09:17:44.52 | +46:12:24.4 | (AB) | AB | 0.190 | 348.0 | 64.68 | 131.04 | … | 0.591 ± 0.028 |
| … | … | GJ 3542 | J09177+584 | M5.0 V | 09:17:46.04 | +58:25:02.7 | … | … | … | … | 64.64 | 1171.23 | 25.97 ± 0.11 | 0.162 ± 0.010 |
| 09188+2647 | LDS6226 | GJ 3548 | J09187+267 | M1.5 V | 09:18:45.99 | +26:45:05.4 | A+B | A | 76.454 | 302.6 | 64.71 | 404.42 | 455.75 ± 2.34 | 0.518 ± 0.029 |
| … | … | GJ 3549 | … | M5.0 Ve | 09:18:41.19 | +26:45:46.6 | … | B | 76.454 | 302.7 | 33.65 | 402.96 | 24.30 ± 0.15 | 0.169 ± 0.010 |
| … | … | GJ 1122 | J09193+385S | M5.0 V | 09:19:18.62 | +38:31:15.9 | A+B | A | … | … | 31.81 | 234.55 | 52.79 ± 0.23 | 0.242 ± 0.012 |
| 09193+3831 | GIC 84 | G 115-69 | J09193+385N | M5.0 V | 09:19:18.71 | +38:31:23.3 | … | B | 7.500 | 7.5 | … | 238.98 | … | 0.215 ± 0.038 |
| 09194+6203 | RAO 569 | GJ 3547 | J09193+620 | M1.0 Ve | 09:19:22.21 | +62:03:10.7 | (AB) | AB | 0.790 | 189.0 | 33.00 | 481.74 | … | … |
| 09200+3053 | RAO 570 | TYC 2493-1386-1 | J09200+308 | M1.5 Ve | 09:20:00.38 | +30:52:39.1 | (AB)+C | AB | 0.430 | 71.3 | 59.84 | 81.54 | … | … |
| 09200+3053 | CRB 80 | 2M09195883+3052156 | … | M5.0 V | 09:19:58.74 | +30:52:15.0 | … | C | 32.038 | 221.2 | 79.45 | 82.62 | 27.70 ± 0.52 | 0.196 ± 0.012 |
| … | … | 1R092010.8+034731 | J09201+037 | M3.5 V | 09:20:10.74 | +03:47:27.0 | … | … | … | … | 79.66 | 139.64 | 167.55 ± 5.65 | 0.394 ± 0.018 |
| … | … | GJ 3553 | J09209+033 | M4.0 V | 09:20:58.28 | +03:21:48.3 | … | … | … | … | 42.19 | 1179.28 | 82.45 ± 0.49 | 0.269 ± 0.013 |
| … | … | GJ 3550 | J09213+731 | M5.0 V | 09:21:16.03 | +73:06:33.1 | … | … | … | … | 115.49 | 974.40 | 21.48 ± 0.08 | 0.146 ± 0.010 |
| 09218+4330 | LAW 17 | GJ 3554 | J09218+435 | M4.5 V | 09:21:48.61 | +43:30:26.5 | (AB) | AB | 0.710 | 124.0 | 33.06 | 323.07 | … | 0.297 ± 0.035 |
| … | … | RAVE J092148.1-021943 | J09218-023 | M2.5 V | 09:21:48.32 | -02:19:43.2 | … | … | … | … | 64.94 | 177.10 | 90.07 ± 0.54 | 0.265 ± 0.012 |
| … | … | G 115-72 | J09228+467 | M1.0 V | 09:22:51.31 | +46:46:58.8 | A+B | A | … | … | 65.67 | 238.54 | 730.83 ± 3.53 | 0.607 ± 0.027 |
| 09229+4647 | KPP3219 | G 115-72B | … | … | 09:22:51.44 | +46:47:02.0 | … | B | 3.412 | 22.7 | 48.74 | 226.53 | … | 0.262 ± 0.036 |
| … | … | BD+22 2086A | … | K5 V | 09:23:06.19 | +22:18:17.6 | A+B | A | … | … | 40.70 | 218.39 | 2015.37 ± 8.48 | 0.700 ± 0.105 |
| 09231+2218 | HDS1348 | BD+22 2086B | J09231+223 | M0.0 V | 09:23:06.01 | +22:18:25.6 | … | B | 8.301 | 343.4 | 85.80 | 223.16 | … | 0.641 ± 0.027 |
| … | … | GJ 3555 | J09238+001 | M1.0 V | 09:23:52.36 | +00:08:13.8 | … | … | … | … | 61.32 | 305.88 | 493.22 ± 2.98 | 0.546 ± 0.028 |
| … | … | LSPM J0924+3041 | J09248+306 | M3.5 V | 09:24:50.68 | +30:41:34.2 | … | … | … | … | 28.50 | 198.06 | 225.22 ± 1.24 | 0.460 ± 0.018 |
| 09256+6329 | JNN 275 | G 235-25 | J09256+634 | M4.5 V | 09:25:39.51 | +63:29:14.9 | (AB) | AB | 0.126 | 92.7 | 196.26 | 420.18 | … | 0.288 ± 0.035 |
| … | … | GJ 3556 | J09275+506 | M2.5 V | 09:27:30.19 | +50:39:10.1 | … | … | … | … | 194.14 | 265.31 | 621.40 ± 3.79 | 0.563 ± 0.028 |
| … | … | LP 727-31 | J09286-121 | M2.5 V | 09:28:41.63 | -12:10:02.0 | … | … | … | … | 35.93 | 391.43 | 499.92 ± 3.92 | 0.534 ± 0.029 |
| … | … | Ross 439 | J09288-073 | M2.5 V | 09:28:53.16 | -07:22:27.2 | A+B | A | … | … | 55.73 | 732.01 | 192.22 ± 0.83 | 0.398 ± 0.016 |
| 09288-0722 | GIC 87 | GJ 347 B | J09289-073 | M4.5 V | 09:28:55.53 | -07:22:23.3 | … | B | 35.610 | 83.6 | 147.66 | 731.82 | 30.18 ± 0.13 | 0.177 ± 0.011 |
| … | … | LP 370-26 | J09291+259 | M5.0 V | 09:29:09.83 | +25:58:05.0 | … | … | … | … | … | 1082.47 | 17.21 ± 0.09 | 0.149 ± 0.010 |
| … | … | GJ 3558 | J09300+396 | M2.5 V | 09:30:01.84 | +39:37:21.1 | … | … | … | … | 69.46 | 205.27 | 255.60 ± 1.03 | 0.462 ± 0.018 |
| … | … | LSPM J0930+2630 | J09302+265 | M3.0 V | 09:30:14.24 | +26:30:22.7 | … | … | … | … | 157.27 | 206.22 | 237.43 ± 1.19 | 0.473 ± 0.019 |
| … | … | GJ 1125 | J09307+003 | M3.5 V | 09:30:43.98 | +00:19:12.8 | … | … | … | … | 99.16 | 790.95 | 115.62 ± 0.55 | 0.292 ± 0.010 |
| … | … | 1R093051.2+022741 | J09308+024 | M4.0 V | 09:30:50.81 | +02:27:21.5 | … | … | … | … | 45.14 | 88.96 | 114.20 ± 1.96 | 0.343 ± 0.015 |
| 09313-1329 | KUI 41 | Ross 440 | J09313-134 | M3.0 V | 09:31:20.26 | -13:29:18.9 | AB | A | 0.379 | 102.5 | 45.15 | 700.39 | … | 0.391 ± 0.033 |
| … | … | G3- 5690031948585514368 | … | … | 09:31:20.22 | -13:29:18.9 | … | B | 0.502 | 265.9 | 42.63 | … | … | … |
| … | … | Ross 84 | J09315+202 | M2.0 V | 09:31:33.05 | +20:16:43.6 | … | … | … | … | 41.75 | 807.29 | 228.23 ± 2.14 | 0.410 ± 0.016 |
| … | … | GJ 353 | J09319+363 | M0.0 V | 09:31:56.06 | +36:19:04.4 | … | … | … | … | 48.96 | 564.65 | 423.82 ± 2.63 | 0.505 ± 0.018 |

Table D.2: Complete sample with the description of multiple systems (continued).

| WDS id | WDS disc | Name | Karmn | Spectral type | $\alpha$ (2016.0) | $\delta$ (2016.0) | System | Component | $\rho$ [arcsec] | $\theta$ [deg] | $\varpi$ [mas] | $\mu_{\rm total}$ [mas a$^{-1}$] | $\mathcal{L}$ [$10^{-4}\,\mathcal{L}_\odot$] | $\mathcal{M}$ [$\mathcal{M}_\odot$] |
|---|---|---|---|---|---|---|---|---|---|---|---|---|---|---|
| 09327+2659 | LDS3903 | DXLeo | … | G9 V(k) | 09:32:43.58 | +26:59:14.8 | A+B | A | 65.558 | 64.4 | 47.60 | 287.22 | 4708.24 ± 16.38 | 0.900 ± 0.135 |
| … | … | HD 82443B | J09328+269 | M5.5 V | 09:32:48.07 | +26:59:39.9 | … | B | 65.013 | 67.3 | 81.36 | 284.17 | 29.79 ± 0.19 | 0.189 ± 0.011 |
| … | … | GJ 3560 | J09352+612 | M2.5 V | 09:35:13.45 | +61:14:37.6 | … | … | … | … | 23.19 | 123.91 | 254.60 ± 1.09 | 0.434 ± 0.017 |
| … | … | GJ 3561 | J09360-061 | M3.5 V | 09:36:04.10 | -06:07:01.2 | … | … | … | … | … | 747.42 | 92.25 ± 0.44 | 0.306 ± 0.014 |
| … | … | GJ 357 | J09360-216 | M2.5 V | 09:36:01.80 | -21:39:54.7 | … | … | … | … | 53.66 | 1000.01 | 161.20 ± 0.84 | 0.341 ± 0.011 |
| … | … | HD 82939 | … | G5 V | 09:36:15.78 | +37:31:44.1 | A+Bab | A | … | … | 23.47 | 134.64 | 2187.00 ± 15.44 | 0.980 ± 0.147 |
| 09361+3733 | SKF 254 | GJ 9303 | J09362+375 | M0.0 V | 09:36:04.14 | +37:33:08.9 | … | Bab(2) | 162.314 | 301.5 | 159.91 | 134.34 | … | … |
| … | … | GJ 3562 | J09370+405 | M3.8 V | 09:37:03.29 | +40:34:37.7 | … | … | … | … | 36.51 | 189.48 | 78.38 ± 0.37 | 0.280 ± 0.013 |
| … | … | LP 428-20 | J09394+146 | M3.5 V | 09:39:29.77 | +14:38:48.5 | … | … | … | … | 36.63 | 180.10 | 96.32 ± 0.50 | 0.293 ± 0.013 |
| … | … | G 117-34 | J09394+317 | M1.5 V | 09:39:24.04 | +31:45:13.7 | AB | A | … | … | 42.09 | 284.64 | … | 0.459 ± 0.030 |
| 09394+3145 | NSN 596 | G 117-34B | … | M7.5 V | 09:39:23.89 | +31:45:14.1 | … | B | 2.004 | 282.9 | 42.22 | 312.79 | … | 0.100 ± 0.049 |
| … | … | Ross 92 | J09410+220 | M4.5 V | 09:41:02.58 | +22:01:20.5 | … | … | … | … | 25.74 | 671.41 | 28.92 ± 0.12 | 0.173 ± 0.010 |
| … | … | Ross 85 | J09411+132 | M1.5 V | 09:41:09.64 | +13:12:32.1 | … | … | … | … | 34.17 | 675.39 | 339.21 ± 2.01 | 0.449 ± 0.012 |
| … | … | GJ 363 | J09423+559 | M3.0 V | 09:42:21.84 | +55:58:53.1 | … | … | … | … | 65.22 | 873.10 | 168.44 ± 0.72 | 0.362 ± 0.012 |
| … | … | GJ 360 | J09425+700 | M2.5 Ve | 09:42:32.74 | +70:01:57.6 | A+B | A | … | … | 49.31 | 724.44 | 348.80 ± 1.48 | 0.498 ± 0.013 |
| 09427+7004 | OSV 3 | GJ 362 | J09428+700 | M3.5 Ve | 09:42:49.63 | +70:02:17.6 | … | B | 88.780 | 76.9 | 39.69 | 721.57 | 240.39 ± 0.84 | 0.420 ± 0.013 |
| … | … | GJ 3563 | J09425-192 | M2.5 V | 09:42:35.14 | -19:14:08.6 | … | … | … | … | 47.85 | 522.62 | 208.21 ± 0.89 | 0.415 ± 0.017 |
| … | … | LP 370-35 | J09430+237 | M1.0 V | 09:43:01.12 | +23:49:18.3 | A+B+C | A | … | … | 46.81 | 269.52 | 541.75 ± 223.84 | 0.560 ± 0.028 |
| 09430+2349 | GOM 35 | LSPM J0943+2349S | … | M6.0 V | 09:43:00.94 | +23:49:13.2 | … | B | 5.731 | 206.0 | … | 267.91 | … | 0.108 ± 0.047 |
| 09430+2349 | LDS3917 | LP 370-34 | … | M7.0 V | 09:42:56.96 | +23:51:16.1 | … | C | 130.856 | 334.1 | 43.48 | 267.86 | 7.49 ± 0.10 | 0.108 ± 0.009 |
| … | … | Ross 93 | J09439+269 | M3.5 V | 09:43:54.92 | +26:58:06.8 | … | … | … | … | 39.41 | 586.64 | 179.35 ± 0.87 | 0.362 ± 0.011 |
| … | … | GJ 1129 | J09447-182 | M3.5 Ve | 09:44:45.55 | -18:12:51.7 | … | … | … | … | 19.09 | 1609.66 | 93.86 ± 0.54 | 0.268 ± 0.010 |
| … | … | G 161-71 | J09449-123 | M5.0 V | 09:44:53.83 | -12:20:53.7 | … | … | … | … | 42.23 | 333.75 | 89.06 ± 0.45 | 0.387 ± 0.030 |
| … | … | GJ 3566 | J09461-044 | M4.0 V | 09:46:08.66 | -04:25:40.6 | AB | A | … | … | 34.31 | 578.71 | … | 0.229 ± 0.037 |
| 09461-0425 | JNN 276 | G 161-74B | … | M5.0 V | 09:46:08.66 | -04:25:39.4 | … | B | 1.191 | 358.9 | 42.36 | 576.97 | … | 0.170 ± 0.040 |
| … | … | Ross 434 | J09468+760 | M1.5 V | 09:46:48.87 | +76:02:22.1 | … | … | … | … | 41.11 | 1004.25 | 423.76 ± 1.80 | 0.522 ± 0.029 |
| … | … | Ross 94 | J09473+263 | M0.0 V | 09:47:22.15 | +26:18:06.9 | … | … | … | … | 43.51 | 411.82 | 1107.65 ± 4.51 | 0.711 ± 0.026 |
| … | … | GJ 3568 | J09475+129 | M4.0 V | 09:47:34.71 | +12:56:42.8 | … | … | … | … | 157.89 | 243.06 | 123.28 ± 0.61 | 0.334 ± 0.015 |
| … | … | LP 728-70 | J09506-138 | M4.0 V | 09:50:40.70 | -13:48:40.2 | Aab | Aab(2) | … | … | 31.90 | 187.98 | … | … |
| … | … | G 43-2 | J09488+156 | M3.0 V | 09:48:50.18 | +15:38:48.5 | … | … | … | … | 157.88 | 229.71 | 82.28 ± 0.39 | 0.269 ± 0.013 |
| … | … | GJ 369 | J09511-123 | dM0.5 | 09:51:10.88 | -12:20:10.8 | … | … | … | … | 134.90 | 1847.30 | 461.67 ± 2.25 | 0.508 ± 0.012 |
| … | … | LP 728-71 | J09526-156 | M3.5 V | 09:52:41.65 | -15:36:15.9 | … | … | … | … | 30.57 | 179.24 | 81.39 ± 0.41 | 0.267 ± 0.013 |
| … | … | G 195-43 | J09527+554 | M1.5 V | 09:52:45.26 | +55:28:16.2 | AB | A | … | … | 63.89 | 354.45 | … | 0.562 ± 0.029 |
| 09527+5528 | RAO 238 | G 195-43B | … | M6.0 V | 09:52:45.16 | +55:28:18.8 | … | B | 2.716 | 342.5 | 74.39 | 342.98 | … | 0.117 ± 0.045 |
| … | … | GJ 372 | J09531-036 | M2.0 V | 09:53:11.67 | -03:41:31.8 | Aab | Aab(2) | … | … | 68.74 | 473.34 | … | 1.043 ± 0.057 |
| … | … | LP 126-73 | J09535+507 | M1.5 V | 09:53:32.64 | +50:45:04.2 | … | … | … | … | 38.50 | 524.29 | 322.93 ± 1.82 | 0.463 ± 0.018 |
| … | … | GJ 3571 | J09539+209 | M4.0 Ve | 09:53:54.78 | +20:56:53.1 | … | … | … | … | 28.49 | 523.78 | 25.69 ± 0.11 | 0.161 ± 0.010 |
| … | … | Wolf 330 | J09557+353 | M3.5 Ve | 09:55:43.55 | +35:21:36.8 | … | … | … | … | 29.90 | 314.77 | 151.82 ± 0.81 | 0.374 ± 0.016 |
| … | … | GJ 373 | J09561+627 | M0.5 V | 09:56:07.96 | +62:47:09.1 | Aab | Aab(?) | … | … | 37.25 | 657.83 | … | … |
| … | … | GJ 3573 | J09564+226 | M4.0 V | 09:56:26.42 | +22:38:56.9 | … | … | … | … | 64.24 | 532.49 | 64.35 ± 0.31 | 0.251 ± 0.012 |

Table D.2: Complete sample with the description of multiple systems (continued).

| WDS id | WDS disc | Name | Karmn | Spectral type | $\alpha$ (2016.0) | $\delta$ (2016.0) | System | Component | $\rho$ [arcsec] | $\theta$ [deg] | $\varpi$ [mas] | $\mu_{total}$ [mas a$^{-1}$] | $\mathcal{L}$ [$10^{-4}\ \mathcal{L}_\odot$] | $\mathcal{M}$ [$\mathcal{M}_\odot$] |
|---|---|---|---|---|---|---|---|---|---|---|---|---|---|---|
| … | … | GJ 3576 | J09579+118 | M4.0 V | 09:57:57.54 | +11:48:26.3 | A+B | A | … | … | 40.29 | 451.50 | 80.70 ± 0.53 | 0.305 ± 0.014 |
| … | … | LP 489-1 | … | M5.0 V | 09:56:45.21 | +11:34:23.6 | … | B* | 1356.049 | 231.6 | 40.17 | 456.96 | 28.92 ± 0.15 | 0.173 ± 0.010 |
| … | … | G 196-1 | J09587+555 | M1.0 V | 09:58:46.62 | +55:32:59.0 | … | … | … | … | 51.74 | 242.72 | 610.72 ± 8.15 | 0.575 ± 0.028 |
| … | … | LP 549-6 | J09589+059 | M4.5 V | 09:58:56.31 | +05:57:58.8 | … | … | … | … | 51.73 | 193.34 | 32.71 ± 0.15 | 0.185 ± 0.011 |
| … | … | GJ 3577 | J09593+438W | M3.5 V | 09:59:18.65 | +43:50:21.9 | A+B | A | … | … | 26.36 | 242.55 | 138.56 ± 0.66 | 0.380 ± 0.016 |
| 09593+4350 | GIC 91 | GJ 3578 | J09593+438E | M5.0 V | 09:59:20.78 | +43:50:22.1 | … | B | 23.063 | 89.4 | … | 242.07 | 111.28 ± 0.56 | 0.338 ± 0.015 |
| … | … | GJ 3579 | J09597+472 | M4.0 V | 09:59:46.11 | +47:12:06.9 | … | … | … | … | 27.21 | 284.91 | 86.60 ± 0.41 | 0.296 ± 0.014 |
| … | … | PM J09597+7211 | J09597+721 | M3.5 V | 09:59:45.30 | +72:12:01.3 | … | … | … | … | 40.36 | 97.75 | 114.77 ± 0.56 | 0.322 ± 0.014 |
| … | … | GJ 375.2 | J10004+272 | M0.5 V | 10:00:26.69 | +27:16:03.5 | AC+B | A | … | … | 56.73 | 118.42 | … | 0.679 ± 0.026 |
| 10004+2716 | TOK 273 | 2M10003572+2717054 | … | M6.5 V | 10:00:35.70 | +27:17:07.5 | … | B | 136.040 | 61.9 | 66.44 | 118.34 | 9.54 ± 0.11 | 0.125 ± 0.010 |
| … | … | G3- 740041664172917376 | … | M7.5 V | 10:00:26.56 | +27:16:05.7 | … | C* | 2.882 | 320.9 | 34.03 | 117.20 | … | 0.101 ± 0.049 |
| … | … | Wolf 335 | J10007+323 | M1.0 V | 10:00:43.01 | +32:18:23.1 | … | … | … | … | 83.68 | 1208.02 | 377.44 ± 1.78 | 0.475 ± 0.018 |
| … | … | LP 37-57 | J10020+697 | M4.0 V | 10:02:05.67 | +69:45:25.6 | … | … | … | … | 29.43 | 232.04 | 44.28 ± 0.16 | 0.219 ± 0.012 |
| … | … | GJ 378 | J10023+480 | M1.0 V | 10:02:20.74 | +48:04:56.1 | … | … | … | … | 29.42 | 1604.96 | 592.56 ± 2.53 | 0.579 ± 0.028 |
| … | … | GJ 3582 | J10027+149 | M4.61 | 10:02:42.62 | +14:59:09.2 | … | … | … | … | 29.41 | 283.17 | 59.04 ± 0.50 | 0.240 ± 0.012 |
| 10028+4828 | CRC 58 | G 195-55 | J10028+484 | M5.5 V | 10:02:48.79 | +48:27:28.7 | (AB) | AB | 0.200 | 306.0 | 29.44 | 425.96 | … | 0.169 ± 0.041 |
| … | … | GJ 3583 | J10035+059 | M3.5 V | 10:03:32.76 | +05:57:45.4 | … | … | … | … | 37.04 | 587.97 | 155.79 ± 0.95 | 0.379 ± 0.016 |
| 10041+1848 | TOK 889 | GJ 9312 | J10040+187 | M0.5 V | 10:04:05.80 | +18:47:41.5 | (AB) | AB | 0.033 | 113.1 | 31.49 | 275.15 | … | … |
| … | … | G 196-3 | J10043+503 | M2.5 V | 10:04:21.23 | +50:23:10.1 | A+B | A | … | … | 38.63 | 246.66 | 441.92 ± 1.80 | 0.506 ± 0.029 |
| 10044+5023 | REB 1 | G 196-3B | … | L3beta | 10:04:20.41 | +50:22:56.1 | … | B | 16.070 | 209.1 | 32.09 | 246.29 | 1.11 ± 0.03 | … |
| … | … | GJ 3585 | J10067+417 | M1.0 V | 10:06:43.45 | +41:42:46.1 | … | … | … | … | 30.29 | 479.34 | 415.23 ± 2.01 | 0.499 ± 0.018 |
| … | … | PM J10068-1246 | J10068-127 | M4.5 V | 10:06:51.98 | -12:46:54.3 | … | … | … | … | 57.18 | 81.30 | 76.34 ± 16.29 | 0.296 ± 0.037 |
| … | … | LP 489-35 | J10069+126 | M1.5 V | 10:06:57.49 | +12:40:51.6 | … | … | … | … | 57.18 | 324.45 | 311.51 ± 1.90 | 0.455 ± 0.017 |
| … | … | GJ 1131 | J10079+692 | M4.0 V | 10:07:56.60 | +69:14:46.1 | … | … | … | … | 56.59 | 909.36 | 38.08 ± 0.15 | 0.188 ± 0.011 |
| … | … | LP 549-23 | J10087+027 | M3.0 V | 10:08:44.52 | +02:43:49.6 | … | … | … | … | 48.74 | 456.50 | 317.65 ± 1.64 | 0.452 ± 0.030 |
| … | … | Wolf 346 | J10087+355 | M1.5 V | 10:08:42.37 | +35:32:51.3 | … | … | … | … | 41.31 | 234.01 | 114.71 ± 0.49 | 0.302 ± 0.013 |
| … | … | TYC 4384-1735-1 | J10088+692 | M0.5 V | 10:08:52.40 | +69:16:35.8 | … | … | … | … | 100.90 | 197.42 | 864.16 ± 3.58 | 0.661 ± 0.026 |
| … | … | GJ 3586 | J10094+512 | M4.65 | 10:09:29.22 | +51:17:06.4 | … | … | … | … | 45.28 | 937.45 | 47.12 ± 0.21 | 0.212 ± 0.011 |
| … | … | PM J10094+5424 | J10094+544 | M2.0 V | 10:09:26.88 | +54:24:22.5 | … | … | … | … | 99.88 | 149.51 | 401.27 ± 1.94 | 0.501 ± 0.029 |
| 10121-0241 | DEL 3 | GJ 381 | J10120-026 | M2.5 V | 10:12:05.23 | -02:41:14.8 | (AB) | AB(2) | 0.072 | 270.1 | 62.87 | 798.92 | … | … |
| … | … | Wolf 351 | J10117+353 | M4.0 V | 10:11:44.11 | +35:18:40.1 | … | … | … | … | … | 281.48 | 56.73 ± 0.23 | 0.235 ± 0.012 |
| … | … | ANSex | J10122-037 | M1.5 V | 10:12:17.50 | -03:44:48.3 | … | … | … | … | 45.05 | 287.61 | 418.61 ± 2.83 | 0.496 ± 0.013 |
| … | … | LP 92-48 | J10125+570 | M3.5 V | 10:12:34.10 | +57:03:40.5 | … | … | … | … | 73.70 | 634.64 | 116.41 ± 0.56 | 0.290 ± 0.010 |
| 10130+2321 | RAO 241 | G 54-18 | J10130+233 | M3.5 V | 10:13:00.45 | +23:20:45.9 | (AB) | AB | 0.614 | 16.0 | 55.33 | 301.16 | … | 0.366 ± 0.033 |
| 10143+2104 | BWL 25 | DKLeo | J10143+210 | M0.5 V | 10:14:19.03 | +21:04:26.8 | (AB) | AB(1) | 0.095 | 320.2 | 77.48 | 225.49 | … | … |
| … | … | LP 167-17 | J10133+467 | M5.5 Ve | 10:13:20.56 | +46:47:24.5 | … | … | … | … | 55.29 | 266.25 | 25.25 ± 0.14 | 0.186 ± 0.011 |
| … | … | G 54-19 | J10148+213 | M4.5 V | 10:14:52.91 | +21:23:42.5 | … | … | … | … | 50.39 | 288.59 | 57.68 ± 0.93 | 0.254 ± 0.013 |
| … | … | GJ 3590 | J10151+314 | M4.0 V | 10:15:06.87 | +31:25:07.0 | AB | A | … | … | 42.50 | 219.97 | … | 0.344 ± 0.033 |
| 10151+3125 | CRC 59 | G 118-43B | … | M3.5 V | 10:15:06.76 | +31:25:08.1 | … | B | 1.829 | 306.4 | 105.98 | 217.31 | … | 0.306 ± 0.034 |
| … | … | PM J10155-1628E | J10155-164 | M4.0 V | 10:15:34.86 | -16:28:20.4 | … | … | … | … | 25.74 | 479.16 | 228.94 ± 1.36 | 0.464 ± 0.019 |

Table D.2: Complete sample with the description of multiple systems (continued).

| WDS id | WDS disc | Name | Karmn | Spectral type | $\alpha$ (2016.0) | $\delta$ (2016.0) | System | Component | $\rho$ [arcsec] | $\theta$ [deg] | $\varpi$ [mas] | $\mu_{\rm total}$ [mas a$^{-1}$] | $\mathcal{L}$ [$10^{-4}$ $\mathcal{L}_\odot$] | $\mathcal{M}$ [$\mathcal{M}_\odot$] |
|---|---|---|---|---|---|---|---|---|---|---|---|---|---|---|
| … | … | LSPM J1015+1729 | J10158+174 | M3.5 V | 10:15:54.26 | +17:29:27.2 | … | … | … | … | 25.84 | 290.75 | 93.03 ± 0.54 | 0.287 ± 0.013 |
| … | … | GJ 386 | J10167-119 | dM3.0 | 10:16:45.49 | -11:57:52.1 | … | … | … | … | 47.00 | 734.48 | 320.63 ± 1.83 | 0.475 ± 0.013 |
| … | … | LP 790-2 | J10182-204 | M4.5 V | 10:18:13.39 | -20:28:39.3 | Aab+B | Aab(2) | … | … | … | 412.59 | … | 0.427 ± 0.001 |
| 10183-2029 | LDS3975 | LP 790-1 | … | M4.5 V | 10:18:11.66 | -20:28:19.3 | … | B | 31.519 | 309.3 | 51.49 | 407.26 | 46.68 ± 0.25 | 0.226 ± 0.012 |
| … | … | LP 729-54 | J10185-117 | M4.0 V | 10:18:34.77 | -11:43:04.2 | A+B | A | … | … | 46.97 | 421.93 | 172.24 ± 1.46 | 0.367 ± 0.012 |
| 10185-1143 | LDS3977 | LP 729-55 | … | M3.5 Ve | 10:18:35.83 | -11:43:06.2 | … | B | 15.788 | 97.1 | 47.03 | 422.28 | 22.77 ± 0.13 | 0.162 ± 0.010 |
| … | … | ADLeo | J10196+198 | M3.0 V | 10:19:35.72 | +19:52:11.3 | … | … | … | … | 81.02 | 500.51 | 238.49 ± 1.73 | 0.431 ± 0.011 |
| … | … | G 118-51 | J10200+289 | M3.0 V | 10:20:00.23 | +28:57:09.5 | … | … | … | … | 86.62 | 511.22 | 107.58 ± 0.39 | 0.291 ± 0.013 |
| … | … | GJ 3595 | J10206+492 | M3.0 V | 10:20:37.15 | +49:17:43.2 | … | … | … | … | 60.51 | 392.53 | 151.44 ± 0.63 | 0.350 ± 0.015 |
| … | … | LP 212-62 | J10238+438 | M5.0 V | 10:23:52.13 | +43:53:33.2 | … | … | … | … | 84.33 | 178.72 | 40.45 ± 0.21 | 0.224 ± 0.012 |
| … | … | PM J10240+3639 | J10240+366 | M3.5 V | 10:24:05.04 | +36:39:30.2 | … | … | … | … | 84.30 | 148.05 | 193.59 ± 2.42 | 0.425 ± 0.017 |
| 10243+1157 | RAO 245 | GJ 3598 | J10243+119 | M2.5 V | 10:24:20.16 | +11:57:23.5 | (AB) | AB | 0.110 | 325.0 | 58.63 | 177.93 | … | 0.376 ± 0.032 |
| … | … | GJ 390 | J10251-102 | M1.0 V | 10:25:10.09 | -10:13:41.3 | … | … | … | … | 27.55 | 703.74 | 461.94 ± 2.35 | 0.530 ± 0.029 |
| … | … | GJ 3599 | J10255+263 | M3.0 V | 10:25:29.93 | +26:23:09.8 | … | … | … | … | 27.65 | 591.17 | 161.58 ± 0.82 | 0.386 ± 0.016 |
| … | … | GJ 3600 | J10260+504W | M4.0 V | 10:26:01.99 | +50:27:00.0 | A+B | A | … | … | 27.86 | 654.77 | 97.45 ± 0.45 | 0.295 ± 0.013 |
| 10261+5029 | LDS1241 | GJ 3601 | J10260+504E | M4.0 V | 10:26:02.64 | +50:27:13.0 | … | B | 14.379 | 25.7 | 69.59 | 659.62 | 84.52 ± 0.35 | 0.292 ± 0.014 |
| … | … | PM J10273+7959 | J10273+799 | M2.0 V | 10:27:22.67 | +79:59:50.0 | … | … | … | … | 90.99 | 128.32 | 424.24 ± 7.40 | 0.512 ± 0.029 |
| … | … | G 44-19 | J10278+028 | M3.5 V | 10:27:49.12 | +02:51:35.5 | … | … | … | … | 76.05 | 502.39 | 193.30 ± 3.42 | 0.425 ± 0.018 |
| … | … | GJ 3602 | J10284+482 | M3.5 V | 10:28:28.86 | +48:14:17.7 | AB | A | … | … | 50.45 | 615.83 | … | 0.270 ± 0.038 |
| … | … | G 146-35B | … | … | 10:28:28.87 | +48:14:17.8 | … | B* | 0.196 | 48.9 | 50.33 | … | … | … |
| … | … | GJ 3604 | J10286+322 | M2.5 V | 10:28:40.73 | +32:14:21.7 | AB | A | … | … | 63.33 | 499.85 | … | 0.307 ± 0.034 |
| 10287+3214 | KPP3229 | G 118-61B | … | M4.0 V | 10:28:40.84 | +32:14:23.1 | … | B | 1.989 | 45.4 | 29.44 | 514.32 | … | 0.282 ± 0.035 |
| … | … | Ross 446 | J10289+008 | dM2 | 10:28:54.91 | +00:50:15.9 | … | … | … | … | 47.11 | 948.29 | 257.06 ± 1.71 | 0.414 ± 0.012 |
| … | … | GJ 3607 | J10303+328 | M3.0 V | 10:30:23.19 | +32:50:07.3 | … | … | … | … | 51.14 | 553.01 | 303.87 ± 2.05 | 0.506 ± 0.019 |
| 10318+5706 | LDS2314 | GJ 397.1 | … | K5 V | 10:31:43.09 | +57:06:59.9 | A+(BC) | A | 141.959 | 225.6 | 58.18 | 185.37 | 872.48 ± 3.36 | 0.700 ± 0.105 |
| 10318+5706 | RAO 252 | GJ 397.1 B | J10315+570 | M5.0 V | 10:31:30.64 | +57:05:20.4 | … | BC | 142.037 | 225.6 | 76.16 | 161.31 | … | … |
| … | … | PM J10320+0318 | J10320+033 | M2.0 V | 10:32:02.31 | +03:18:54.9 | … | … | … | … | 56.82 | 40.58 | 860.94 ± 9.67 | 0.620 ± 0.027 |
| … | … | GJ 3610 | J10345+463 | M3.0 V | 10:34:29.53 | +46:18:07.2 | A+B | A | … | … | 26.96 | 424.70 | 174.98 ± 1.20 | 0.403 ± 0.017 |
| 10345+4618 | LDS3999 | GJ 3611 | … | M4.5 V | 10:34:25.22 | +46:18:20.8 | … | B | 46.701 | 286.9 | 27.12 | 429.93 | 24.23 ± 0.13 | 0.168 ± 0.010 |
| … | … | LP 670-17 | J10350-094 | dM3.0 | 10:35:01.36 | -09:24:41.5 | … | … | … | … | 63.61 | 293.87 | 192.01 ± 1.23 | 0.371 ± 0.011 |
| … | … | GJ 3612 | J10354+694 | M4.0 V | 10:35:21.96 | +69:26:48.5 | Aab | Aab(2) | … | … | … | 1773.72 | … | 0.389 ± 0.003 |
| … | … | RX J1035.9+2853 | J10359+288 | M3.0 V | 10:35:57.12 | +28:53:30.3 | … | … | … | … | 39.60 | 141.60 | 243.89 ± 7.08 | 0.480 ± 0.020 |
| … | … | RYSex | J10360+051 | M4.0 V | 10:36:00.52 | +05:07:14.8 | … | … | … | … | 104.76 | 666.43 | 135.43 ± 0.75 | 0.359 ± 0.011 |
| … | … | G 146-48 | J10364+415 | M2.5 V | 10:36:27.19 | +41:30:02.8 | AB | A | … | … | 52.12 | 336.63 | … | 0.494 ± 0.030 |
| 10364+4130 | RAO 253 | G 146-48B | … | M5.5 V | 10:36:27.02 | +41:30:01.8 | … | B | 2.122 | 241.3 | 95.05 | 345.11 | … | 0.155 ± 0.041 |
| 10367+1522 | DAE 3 | PM J10367+1521A | J10367+153 | M3.5 V | 10:36:44.96 | +15:21:38.6 | A+BC | A | 0.730 | 228.6 | 54.62 | 135.56 | … | 0.295 ± 0.035 |
| 10367+1522 | DAE 3 | PM J10367+1521B | … | M4.5 V | 10:36:44.92 | +15:21:37.9 | … | BC | 0.858 | 218.7 | 39.57 | 118.34 | … | … |
| … | … | LP 127-502 | J10368+509 | M4.5 V | 10:36:48.58 | +50:55:00.7 | Aab | Aab(2?) | … | … | 55.92 | 319.05 | … | … |
| … | … | LP 490-42 A | J10379+127 | M3.0 V | 10:37:55.03 | +12:46:37.6 | AB | A | … | … | 40.84 | 223.11 | … | 0.363 ± 0.033 |
| 10379+1247 | CRC 60 | LP 490-42 B | … | M3.0 V | 10:37:55.05 | +12:46:36.8 | … | B | 0.826 | 155.9 | 26.86 | 249.10 | … | 0.376 ± 0.032 |

Table D.2: Complete sample with the description of multiple systems (continued).

| WDS id | WDS disc | Name | Karmn | Spectral type | $\alpha$ (2016.0) | $\delta$ (2016.0) | System | Component | $\rho$ [arcsec] | $\theta$ [deg] | $\varpi$ [mas] | $\mu_{total}$ [mas a$^{-1}$] | $\mathcal{L}$ [$10^{-4}$ $\mathcal{L}_\odot$] | $\mathcal{M}$ [$\mathcal{M}_\odot$] |
|---|---|---|---|---|---|---|---|---|---|---|---|---|---|---|
| … | … | GJ 3613 | J10384+485 | M3.0 V | 10:38:29.47 | +48:31:43.4 | … | … | … | … | 66.49 | 215.04 | 225.94 ± 1.13 | 0.461 ± 0.018 |
| … | … | LP 262-400 | J10385+354 | M2.5 V | 10:38:32.66 | +35:29:53.7 | … | … | … | … | 36.75 | 515.87 | 219.75 ± 1.05 | 0.427 ± 0.017 |
| … | … | StKM 1-873 | J10389+250 | M2.0 V | 10:38:56.64 | +25:05:39.0 | … | … | … | … | 36.72 | 191.10 | 296.60 ± 1.12 | 0.443 ± 0.017 |
| … | … | GJ 399 | J10396-069 | dM2.5 | 10:39:39.79 | -06:55:27.2 | … | … | … | … | 44.64 | 725.55 | 313.94 ± 2.17 | 0.459 ± 0.012 |
| … | … | TYC 254-88-1 | J10403+015 | M1.0 V | 10:40:21.42 | +01:34:36.6 | … | … | … | … | 54.49 | 70.12 | 622.21 ± 4.13 | 0.590 ± 0.028 |
| … | … | GJ 1134 | J10416+376 | M4.0 V | 10:41:35.91 | +37:36:33.4 | … | … | … | … | 27.77 | 1500.56 | 54.72 ± 0.31 | 0.215 ± 0.009 |
| 10430-0913 | WSI 112 | PM J10430-0912 | J10430-092 | M5.5 V | 10:43:00.72 | -09:12:35.0 | (AB) | AB | 0.575 | 136.2 | 27.72 | 2009.04 | … | 0.153 ± 0.042 |
| … | … | LP 490-63 | J10443+124 | M3.5 V | 10:44:18.52 | +12:25:11.5 | … | … | … | … | 26.56 | 274.62 | 207.59 ± 1.00 | 0.441 ± 0.018 |
| … | … | GJ 3616 A | J10448+324 | M3.0 V | 10:44:52.40 | +32:24:41.3 | AB+C | A | … | … | 36.36 | 214.33 | … | 0.336 ± 0.034 |
| 10449+3224 | CRC 61 | GJ 3616 B | … | M4.0 V | 10:44:52.43 | +32:24:40.1 | … | B | 1.296 | 160.3 | 62.28 | 226.22 | … | 0.287 ± 0.035 |
| 10449+3224 | LDS1258 | GJ 3617 | … | M5.0 V | 10:44:54.79 | +32:24:23.4 | … | C | 35.225 | 120.7 | 66.84 | 211.36 | … | 0.184 ± 0.040 |
| … | … | GJ 400 A | J10453+385 | M0.5 V | 10:45:21.43 | +38:30:44.8 | AB | A | … | … | 56.01 | 155.18 | … | 0.642 ± 0.027 |
| 10454+3831 | HO 532 | G3- 776067093937332992 | … | M2.5 V | 10:45:21.39 | +38:30:44.4 | … | B | 0.684 | 232.7 | 60.10 | 159.02 | … | 0.393 ± 0.032 |
| … | … | GJ 401 | J10456-191 | M0.5 V | 10:45:36.98 | -19:07:01.3 | AB | A | … | … | 42.55 | 1963.77 | … | 0.506 ± 0.029 |
| 10457-1907 | LDS4013 | GJ 401 B | … | DQ | 10:45:36.93 | -19:06:54.7 | … | B | 6.655 | 354.5 | 22.90 | 1962.56 | … | 0.500 ± 0.100 |
| … | … | GJ 3619 | J10460+096 | M3.5 V | 10:46:03.81 | +09:41:47.1 | … | … | … | … | 45.85 | 287.36 | 148.75 ± 0.86 | 0.370 ± 0.016 |
| 10472+4027 | CLO 5 | LP 213-67 | J10472+404 | M6.5 Ve | 10:47:12.19 | +40:26:43.2 | (AB)+C | AB | 0.101 | 345.6 | 46.20 | 300.14 | … | 0.135 ± 0.043 |
| 10472+4027 | LDS4016 | LP 213-68 | … | M8.0 V | 10:47:13.36 | +40:26:48.7 | … | C | 14.382 | 67.5 | 44.88 | 303.57 | 8.00 ± 0.17 | 0.149 ± 0.010 |
| 10474+0236 | HDS1542 | Ross 895 | J10474+025 | M2.0 V | 10:47:24.47 | +02:35:32.7 | (AB) | AB | 0.528 | 258.3 | 46.80 | 181.27 | … | 0.423 ± 0.032 |
| … | … | GJ 3622 | J10482-113 | M6.5 V | 10:48:13.24 | -11:20:34.1 | … | … | … | … | 40.15 | 1635.97 | 7.46 ± 0.04 | 0.102 ± 0.048 |
| … | … | GJ 3623 | J10485+191 | M3.0 V | 10:48:32.84 | +19:09:00.3 | … | … | … | … | 58.25 | 256.84 | 122.49 ± 0.53 | 0.333 ± 0.014 |
| 10498+3533 | BEU 14 | GJ 1138 | J10497+355 | M5.0 V | 10:49:44.71 | +35:32:34.3 | (AB) | AB | 0.281 | 54.2 | 41.44 | 1261.32 | … | 0.219 ± 0.038 |
| … | … | GJ 3626 | J10504+331 | M3.5 V | 10:50:26.08 | +33:05:54.0 | Aab | Aab(1) | … | … | 23.15 | 638.33 | … | 0.436 ± 0.017 |
| … | … | LZUMa | … | G5 V | 10:50:39.98 | +51:47:58.8 | A+B | A | … | … | 53.90 | 196.65 | 3514.87 ± 816.83 | 0.980 ± 0.147 |
| 10507+5148 | LDS3019 | GJ 3628 | J10506+517 | M4.1 V | 10:50:37.91 | +51:45:01.6 | … | B | 178.251 | 186.2 | 25.24 | 191.40 | 111.80 ± 0.55 | 0.339 ± 0.015 |
| 10509+0648 | RAO256 | EELeo | J10508+068 | M4.0 V | 10:50:51.11 | +06:48:16.2 | (AB) | AB | 0.098 | 214.2 | 76.01 | 1184.61 | 79.95 ± 0.56 | 0.246 ± 0.011 |
| 10513+3607 | BWL26 | GJ 3629 | J10513+361 | M3.0 V | 10:51:20.33 | +36:07:24.5 | (AB) | AB | 0.206 | 119.6 | 37.15 | 198.40 | … | 0.414 ± 0.031 |
| … | … | GJ 3630 | J10520+005 | M4.0 V | 10:52:02.83 | +00:32:38.5 | Aabc | Aabc(3/4) | … | … | 19.48 | 407.59 | … | … |
| … | … | GJ 403 | J10520+139 | M4.0 V | 10:52:03.01 | +13:59:54.5 | … | … | … | … | 86.65 | 1142.17 | 88.32 ± 0.39 | 0.279 ± 0.013 |
| … | … | GJ 3631 | J10522+059 | M5.5 V | 10:52:13.50 | +05:55:08.9 | … | … | … | … | 47.40 | 698.13 | 30.73 ± 0.15 | 0.179 ± 0.011 |
| … | … | LP 671-8 | J10546-073 | M4.0 V | 10:54:41.77 | -07:18:39.4 | AB | A | … | … | 129.75 | 410.62 | … | 0.303 ± 0.035 |
| 10547-0719 | CRC 62 | G3- 3763681220170173952 | … | … | 10:54:41.83 | -07:18:39.1 | … | B | 0.844 | 66.0 | 99.56 | … | … | … |
| … | … | GJ 3632 | J10555-093 | M3.5 V | 10:55:34.19 | -09:21:18.7 | … | … | … | … | 35.08 | 522.15 | 87.27 ± 0.41 | 0.278 ± 0.013 |
| … | … | PM J10563+0415 | J10563+042 | M2.5 V | 10:56:22.32 | +04:15:44.6 | … | … | … | … | 42.74 | 113.88 | 245.51 ± 1.66 | 0.453 ± 0.018 |
| … | … | CNLeo | J10564+070 | M6.0 V | 10:56:24.77 | +07:00:09.8 | … | … | … | … | 47.29 | 4715.33 | 10.10 ± 0.09 | 0.108 ± 0.047 |
| … | … | Ross 447 | J10576+695 | M0.0 V | 10:57:36.16 | +69:35:48.8 | … | … | … | … | 55.19 | 632.69 | 934.57 ± 3.20 | 0.680 ± 0.026 |
| … | … | LP 731-76 | J10584-107 | M5.0 V | 10:58:27.78 | -10:46:31.8 | … | … | … | … | 30.03 | 210.47 | 42.28 ± 0.21 | 0.197 ± 0.009 |
| … | … | Ross 104 | J11000+228 | M2.5 V | 11:00:03.76 | +22:49:54.1 | … | … | … | … | 29.95 | 511.84 | 195.79 ± 93.22 | 0.370 ± 0.095 |
| … | … | G 254-11A | J11003+728 | M2.0 V | 11:00:23.29 | +72:52:17.6 | AB | A | … | … | 33.38 | 423.20 | … | 0.486 ± 0.030 |
| 11004+7252 | NSN 608 | G 254-11B | … | M4.0 V | 11:00:23.11 | +72:52:18.3 | … | B | 1.061 | 309.1 | 71.54 | 414.68 | … | 0.280 ± 0.035 |

Table D.2: Complete sample with the description of multiple systems (continued).

| WDS id | WDS disc | Name | Karmn | Spectral type | $\alpha$ (2016.0) | $\delta$ (2016.0) | System | Component | $\rho$ [arcsec] | $\theta$ [deg] | $\varpi$ [mas] | $\mu_{\rm total}$ [mas a$^{-1}$] | $\mathcal{L}$ [$10^{-4}$ $\mathcal{L}_\odot$] | $\mathcal{M}$ [$\mathcal{M}_\odot$] |
|---|---|---|---|---|---|---|---|---|---|---|---|---|---|---|
| … | … | GJ 3636 | J11008+120 | M5.5 V | 11:00:50.58 | +12:04:08.7 | … | … | … | … | 74.24 | 187.54 | 31.99 ± 0.27 | 0.197 ± 0.011 |
| … | … | GJ 3637 | J11013+030 | M5.0 V | 11:01:20.81 | +03:00:10.8 | … | … | … | … | 39.49 | 1146.90 | 39.25 ± 0.22 | 0.191 ± 0.011 |
| … | … | StKM 1-902 | J11014+568 | M1.0 V | 11:01:26.72 | +56:52:04.0 | … | … | … | … | 39.36 | 194.90 | 755.37 ± 3.68 | 0.629 ± 0.027 |
| … | … | GJ 1141 A | J11023+165E | M1.0 V | 11:02:19.29 | +16:30:27.1 | A+B | A | … | … | 45.40 | 164.09 | 513.84 ± 3.05 | 0.550 ± 0.028 |
| 11024+1631 | LDS 917 | GJ 1141 B | J11023+165W | M1.0 V | 11:02:18.02 | +16:30:30.8 | … | B | 18.685 | 281.4 | 45.23 | 168.37 | 481.57 ± 2.83 | 0.536 ± 0.029 |
| … | … | DSLeo | J11026+219 | M1.0 Ve | 11:02:38.51 | +21:58:00.9 | … | … | … | … | 201.41 | 151.83 | 563.47 ± 2.47 | 0.575 ± 0.028 |
| … | … | Wolf 360 | J11030+037 | M2.5 V | 11:03:04.39 | +03:44:19.2 | … | … | … | … | 53.58 | 244.83 | … | … |
| … | … | LP 431-50 | J11031+152 | M3.5 Ve | 11:03:08.01 | +15:17:50.3 | … | … | … | … | 41.63 | 427.34 | … | … |
| … | … | GJ 3639 | J11031+366 | M3.5 V | 11:03:09.74 | +36:39:09.1 | … | … | … | … | 54.97 | 190.86 | 119.20 ± 0.51 | 0.351 ± 0.015 |
| … | … | HD 95735 | J11033+359 | M1.5 V | 11:03:19.43 | +35:56:55.2 | … | … | … | … | 34.55 | 4811.68 | 230.42 ± 6.91 | 0.398 ± 0.014 |
| … | … | LP 491-51 | J11036+136 | M4.0 V | 11:03:21.05 | +13:37:58.2 | Aab | Aab(1) | … | … | 49.20 | 207.70 | … | … |
| … | … | GJ 3640 | J11042+400 | M0.0 V | 11:04:15.64 | +40:00:15.1 | … | … | … | … | 51.79 | 225.85 | 976.99 ± 4.46 | 0.678 ± 0.026 |
| … | … | LSPM J1104+3027 | J11044+304 | M3.0 V | 11:04:28.35 | +30:27:30.9 | … | … | … | … | 78.17 | 168.16 | 182.11 ± 0.94 | 0.356 ± 0.023 |
| … | … | GJ 412 A | J11054+435 | M1.0 V | 11:05:22.09 | +43:31:51.4 | A+B | A | … | … | 46.74 | 4505.31 | 230.89 ± 1.87 | 0.382 ± 0.011 |
| 11055+4332 | VBS 18 | WXUMa | J11055+435 | M5.5 V | 11:05:24.50 | +43:31:33.3 | … | B | 31.854 | 124.7 | 53.45 | 4444.91 | 9.57 ± 0.05 | 0.115 ± 0.045 |
| … | … | GJ 3641 | J11055+450 | M0.0 V | 11:05:33.91 | +45:00:27.9 | AB | A | … | … | 53.43 | 254.23 | … | 0.632 ± 0.027 |
| 11056+4501 | NSN 609 | G 176-8B | … | M6.0 V | 11:05:33.92 | +45:00:30.5 | … | B | 2.560 | 2.5 | 35.90 | 269.33 | … | 0.120 ± 0.045 |
| … | … | GJ 3643 | J11057+102 | M3.0 V | 11:05:43.80 | +10:13:58.1 | … | … | … | … | 35.70 | 930.73 | 168.50 ± 1.07 | 0.371 ± 0.015 |
| … | … | PM J11075+4345 | J11075+437 | M3.0 V | 11:07:31.88 | +43:45:56.3 | … | … | … | … | 48.60 | 117.73 | 101.43 ± 0.44 | 0.301 ± 0.014 |
| … | … | GJ 1142 A | J11081-052 | M3.0 V | 11:08:06.48 | -05:13:54.2 | A+B | A | … | … | … | 443.17 | 277.25 ± 1.62 | 0.482 ± 0.019 |
| 11080-0509 | LDS 852 | GJ 1142 B | … | DA3 | 11:07:59.89 | -05:09:33.1 | … | B | 278.996 | 339.3 | 44.61 | 446.11 | … | 0.500 ± 0.100 |
| … | … | GJ 3646 | J11108+479 | M4.0 V | 11:10:51.05 | +47:56:53.7 | … | … | … | … | 42.63 | 598.83 | 76.50 ± 0.39 | 0.276 ± 0.013 |
| … | … | HD 97101 | J11110+304E | K7 V | 11:11:05.90 | +30:26:42.5 | A+B | A | … | … | 142.10 | 623.58 | 1236.01 ± 8.02 | 0.640 ± 0.096 |
| 11111+3027 | STT 231 | HD 97101B | J11110+304W | M2.0 V | 11:11:03.29 | +30:26:38.0 | … | B | 34.153 | 262.4 | 36.49 | 639.04 | 511.07 ± 1.72 | 0.539 ± 0.029 |
| … | … | GJ 9351 A | J11113+434 | M2.5 V | 11:11:18.88 | +43:24:55.8 | AB | A | … | … | 57.01 | 782.79 | … | 0.432 ± 0.031 |
| 11114+4325 | KUI 55 | GJ 9351 B | … | M2.5 V | 11:11:18.56 | +43:24:55.3 | … | B | 3.568 | 262.5 | … | 764.96 | … | 0.440 ± 0.031 |
| … | … | GJ 3647 | J11118+335 | M3.5 V | 11:11:51.52 | +33:32:13.1 | AB | A | … | … | 28.96 | 211.07 | … | 0.320 ± 0.034 |
| … | … | LSPM J1111+3332E | … | … | … | … | … | B | … | … | 41.27 | 197.47 | … | … |
| … | … | GJ 3649 | J11126+189 | dM1.5 | 11:12:38.95 | +18:56:05.5 | … | … | … | … | 41.20 | 21.84 | 456.48 ± 2.23 | 0.510 ± 0.012 |
| 11132+0014 | YSC 208 | Wolf 370 | J11131+002 | M0.0 V | 11:13:09.63 | +00:14:16.7 | (AB) | AB | 0.292 | 339.0 | 61.22 | 456.20 | … | 0.670 ± 0.032 |
| 11152+7329 | STF1516 | HD 97584 | … | K4 V | 11:15:10.39 | +73:28:32.5 | A+B | A | 67.600 | 102.9 | 78.57 | 418.80 | 2152.23 ± 5.97 | 0.730 ± 0.110 |
| … | … | HD 97584B | J11151+734 | M2.5 V | 11:15:09.56 | +73:28:38.0 | … | B | 6.481 | 327.1 | 34.41 | 396.51 | 209.63 ± 0.93 | 0.416 ± 0.017 |
| … | … | GJ 3652 | J11152+194 | M3.5 V | 11:15:12.62 | +19:27:04.3 | … | … | … | … | 65.47 | 512.69 | 134.04 ± 0.78 | 0.350 ± 0.015 |
| … | … | GJ 421 A | … | K7 V | 11:15:20.90 | -18:08:49.2 | A+B+C | A | … | … | 35.30 | 748.09 | … | 0.695 ± 0.104 |
| 11154-1807 | LDS 342 | GJ 421 B | … | K5 V | 11:15:19.61 | -18:08:52.0 | … | B | 18.567 | 261.4 | 34.94 | 752.44 | 1010.88 ± 4.59 | 0.700 ± 0.105 |
| 11154-1807 | LDS 342 | GJ 421 C | J11152-181 | M3.0 V | 11:15:15.67 | -18:07:47.8 | … | C | 96.651 | 309.5 | 49.98 | 753.07 | 108.79 ± 0.51 | 0.313 ± 0.014 |
| … | … | G 122-8 | J11154+410 | M3.5 V | 11:15:26.72 | +41:05:12.5 | … | … | … | … | … | 247.04 | 204.74 ± 1.36 | 0.411 ± 0.017 |
| … | … | GJ 3653 | J11159+553 | M0.5 V | 11:15:53.69 | +55:19:49.2 | … | … | … | … | 40.73 | 200.30 | 733.95 ± 3.11 | 0.605 ± 0.027 |
| … | … | LP 169-22 | J11195+466 | M5.5 V | 11:19:31.09 | +46:41:33.4 | … | … | … | … | 38.25 | 682.13 | 12.86 ± 0.07 | 0.126 ± 0.010 |
| 11200+6551 | TAM 1 | GJ 424 | J11200+658 | M0.0 V | 11:19:57.14 | +65:50:50.3 | (AB) | AB | 0.132 | 334.0 | 38.34 | 2952.33 | … | 0.531 ± 0.029 |

Table D.2: Complete sample with the description of multiple systems (continued).

| WDS id | WDS disc | Name | Karmn | Spectral type | $\alpha$ (2016.0) | $\delta$ (2016.0) | System | Component | $\rho$ [arcsec] | $\theta$ [deg] | $\varpi$ [mas] | $\mu_{total}$ [mas a$^{-1}$] | $\mathcal{L}$ [$10^{-4}$ $\mathcal{L}_{\odot}$] | $\mathcal{M}$ [$\mathcal{M}_{\odot}$] |
|---|---|---|---|---|---|---|---|---|---|---|---|---|---|---|
| … | … | LP 733-99 | J11201-104 | M2.0 V | 11:20:05.89 | -10:29:46.4 | … | … | … | … | 32.20 | 198.88 | 414.88 ± 2.99 | 0.506 ± 0.029 |
| … | … | HD 98712A | … | K7 V | 11:21:26.87 | -20:27:15.3 | AB | A | … | … | 47.74 | 191.29 | … | 0.650 ± 0.097 |
| 11214-2027 | STN 22 | HD 98712B | J11214-204 | M2.5 V | 11:21:26.84 | -20:27:11.5 | … | B | 3.811 | 354.4 | 36.46 | 255.49 | … | 0.421 ± 0.031 |
| … | … | GJ 1146 | J11216+061 | M3.5 V | 11:21:37.67 | +06:08:00.7 | … | … | … | … | 64.20 | 1758.39 | 72.59 ± 0.34 | 0.235 ± 0.012 |
| … | … | GJ 3657 | J11231+258 | M5.0 V | 11:23:06.77 | +25:53:31.7 | … | … | … | … | 28.93 | 1061.30 | 25.81 ± 0.14 | 0.162 ± 0.010 |
| … | … | GJ 3658 | J11233+448 | M2.0 V | 11:23:20.14 | +44:48:36.5 | … | … | … | … | 99.46 | 351.11 | 366.15 ± 2.60 | 0.478 ± 0.030 |
| … | … | Wolf 386 | J11237+085 | M0.5 V | 11:23:43.49 | +08:33:51.6 | … | … | … | … | 82.00 | 1021.78 | 420.33 ± 1.98 | 0.521 ± 0.029 |
| … | … | GJ 426.1 A | … | F1 IV | 11:23:55.62 | +10:31:44.9 | (AabB)+C | Aab | … | … | 36.94 | 169.81 | … | … |
| 11239+1032 | STF1536 | GJ 426.1 B | … | F5 V | 11:23:55.75 | +10:31:44.7 | … | B | 2.050 | 95.1 | 34.04 | 185.02 | … | 1.330 ± 0.200 |
| 11239+1032 | STF1536 | LSPM J1123+1037 | J11238+106 | M0.5 V | 11:23:50.12 | +10:37:07.0 | … | C | 332.098 | 345.9 | 32.04 | 173.50 | 767.99 ± 13.39 | 0.634 ± 0.027 |
| … | … | Ross 1002 | J11239-183 | M3.0 V | 11:23:56.61 | -18:21:49.5 | … | … | … | … | 27.91 | 612.81 | 102.04 ± 0.60 | 0.302 ± 0.014 |
| … | … | Ross 448 | J11247+675 | M1.0 V | 11:24:46.53 | +67:33:08.5 | … | … | … | … | 28.05 | 393.02 | 362.59 ± 1.63 | 0.493 ± 0.018 |
| … | … | StKM 1-941 | J11249+024 | M1.0 V | 11:24:58.52 | +02:28:26.8 | … | … | … | … | 72.12 | 275.17 | 322.14 ± 1.49 | 0.437 ± 0.016 |
| … | … | 1R112405.0+380809 | J11240+381 | M4.5 V | 11:24:04.53 | +38:08:10.7 | A+B | A | … | … | 72.99 | 121.95 | 49.52 ± 0.23 | 0.233 ± 0.012 |
| 11241+3808 | RED23 | 2MUCD 10984 | … | M8.5 V | 11:24:05.06 | +38:08:05.3 | … | B | 8.250 | 130.6 | 53.17 | 126.02 | … | 0.111 ± 0.054 |
| … | … | GJ 3660 | J11254+782 | d/sdM4 | 11:25:26.06 | +78:15:52.9 | A+B | A | … | … | 53.19 | 672.13 | 253.30 ± 1.11 | 0.433 ± 0.017 |
| 11254+7817 | LDS 920 | GJ 3661 | … | M3.0 V | 11:25:13.53 | +78:16:55.0 | … | B | 72.909 | 328.4 | 39.82 | 671.49 | 181.53 ± 0.76 | 0.363 ± 0.015 |
| … | … | PM J11266+3756 | J11266+379 | M2.0 V | 11:26:37.40 | +37:56:22.8 | AB | A | … | … | 40.43 | 150.84 | … | 0.554 ± 0.028 |
| 11266+3756 | KPP3238 | PM J11266+3756B | … | M1.5 V | 11:26:37.31 | +37:56:22.5 | … | B | 1.145 | 254.9 | 39.47 | 150.58 | … | 0.498 ± 0.029 |
| … | … | GJ 3664 | J11276+039 | M0.0 V | 11:27:38.47 | +03:58:36.1 | … | … | … | … | 32.57 | 98.48 | 1045.91 ± 5.46 | 0.689 ± 0.026 |
| … | … | Wolf 398 | J11289+101 | M4.0 V | 11:28:55.45 | +10:10:48.2 | … | … | … | … | 219.33 | 935.04 | 91.65 ± 0.57 | 0.266 ± 0.011 |
| … | … | K2-18 | J11302+076 | dM2.5 | 11:30:14.43 | +07:35:16.1 | … | … | … | … | 43.22 | 155.46 | 272.31 ± 1.61 | 0.433 ± 0.013 |
| … | … | LP 672-42 | J11306-080 | M3.5 V | 11:30:41.44 | -08:05:38.9 | … | … | … | … | 102.75 | 437.49 | 160.57 ± 0.70 | 0.343 ± 0.011 |
| … | … | StKM 1-950 | J11307+549 | M1.0 V | 11:30:43.80 | +54:57:29.1 | A+B | A | … | … | 44.27 | 111.40 | 835.36 ± 5.44 | 0.641 ± 0.027 |
| … | … | G3- 844502037681090688 | … | M6.0 V | 11:30:42.66 | +54:56:57.0 | … | B* | 33.615 | 197.0 | 37.39 | 112.07 | 13.69 ± 0.13 | 0.131 ± 0.010 |
| … | … | GJ 3668 | J11311-149 | M5.0 V | 11:31:08.84 | -14:57:43.2 | … | … | … | … | 37.89 | 1433.08 | 31.60 ± 0.22 | 0.182 ± 0.011 |
| … | … | LP 552-68 | J11315+022 | M2.5 V | 11:31:32.21 | +02:13:34.8 | … | … | … | … | 143.54 | 792.86 | 313.43 ± 1.51 | 0.484 ± 0.018 |
| … | … | Ross 903 | J11317+226 | M0.5 V | 11:31:42.72 | +22:40:02.2 | … | … | … | … | 29.96 | 585.58 | 567.56 ± 3.30 | 0.568 ± 0.028 |
| … | … | GJ 3672 | J11351-056 | M4.5 V | 11:35:07.66 | -05:39:38.2 | … | … | … | … | 25.46 | 994.67 | 35.58 ± 0.19 | 0.194 ± 0.011 |
| 11355-3856 | CRC 63 | GJ 3673 | J11355+389 | M3.5 V | 11:35:30.94 | +38:55:33.4 | AB | A | 0.339 | 65.6 | 75.49 | 754.70 | … | 0.386 ± 0.033 |
| … | … | G 122-34B | … | M3.0 V | 11:35:30.95 | +38:55:33.4 | … | B | 0.196 | 87.8 | 71.80 | 754.70 | … | 0.346 ± 0.034 |
| … | … | Ross 112 | J11376+587 | M2.5 V | 11:37:38.30 | +58:42:38.2 | … | … | … | … | 48.81 | 414.06 | 611.04 ± 5.80 | 0.553 ± 0.028 |
| … | … | LP 19-403 | J11404+770 | M2.0 V | 11:40:27.09 | +77:04:19.0 | … | … | … | … | … | 542.43 | 558.91 ± 2.64 | 0.552 ± 0.028 |
| … | … | Ross 1003 | J11417+427 | M4.0 V | 11:41:43.80 | +42:45:05.7 | … | … | … | … | 52.12 | 582.49 | 149.18 ± 0.65 | 0.350 ± 0.014 |
| … | … | Ross 115 | J11420+147 | M3.0 V | 11:42:01.43 | +14:46:39.9 | … | … | … | … | 36.63 | 353.99 | 258.47 ± 1.31 | 0.465 ± 0.018 |
| … | … | Ross 905 | J11421+267 | M2.5 V | 11:42:12.16 | +26:42:10.6 | … | … | … | … | 415.18 | 1209.56 | 241.90 ± 1.20 | 0.414 ± 0.012 |
| … | … | LP 375-23 | J11423+230 | M0.5 V | 11:42:18.14 | +23:01:37.3 | A+B | A | … | … | 42.67 | 209.12 | 562.02 ± 2.63 | 0.582 ± 0.028 |
| 11423+2302 | LDS 926 | LP 375-24 | … | M4.0 V | 11:42:20.98 | +23:03:14.8 | … | B | 105.050 | 22.0 | 68.96 | 209.27 | 96.86 ± 0.44 | 0.294 ± 0.013 |
| … | … | GJ 3682 | J11433+253 | M4.0 V | 11:43:23.27 | +25:18:13.2 | … | … | … | … | 148.20 | 229.78 | 265.46 ± 4.18 | 0.444 ± 0.018 |
| … | … | LP 433-47 | J11451+183 | M4.0 V | 11:45:11.57 | +18:20:53.8 | … | … | … | … | 31.51 | 411.30 | 227.61 ± 1.26 | 0.463 ± 0.018 |

Table D.2: Complete sample with the description of multiple systems (continued).

| WDS id | WDS disc | Name | Karmn | Spectral type | $\alpha$ (2016.0) | $\delta$ (2016.0) | System | Component | $\rho$ [arcsec] | $\theta$ [deg] | $\varpi$ [mas] | $\mu_{\rm total}$ [mas a$^{-1}$] | $\mathcal{L}$ [$10^{-4}\,\mathcal{L}_\odot$] | $\mathcal{M}$ [$\mathcal{M}_\odot$] |
|---|---|---|---|---|---|---|---|---|---|---|---|---|---|---|
| … | … | GJ 443 | J11467-140 | M3.0 V | 11:46:43.69 | -14:01:04.4 | … | … | … | … | 31.61 | 1059.90 | 379.08 ± 2.64 | 0.472 ± 0.030 |
| … | … | GJ 3684 A | J11470+700 | M4.0 V | 11:47:04.37 | +70:01:57.7 | AB | A | … | … | 46.68 | 347.22 | … | 0.324 ± 0.034 |
| 11471+7002 | NSN 621 | GJ 3684 B | … | M3.5 V | 11:47:04.25 | +70:01:58.4 | … | B | 0.873 | 318.6 | 67.26 | 334.67 | … | 0.307 ± 0.034 |
| … | … | 1R114728.8+664405 | J11474+667 | M5.0 V | 11:47:28.27 | +66:44:02.6 | … | … | … | … | 24.22 | 110.02 | 80.25 ± 0.79 | 0.289 ± 0.010 |
| … | … | GJ 3685 | J11476+002 | M4.0 V | 11:47:40.41 | +00:15:18.5 | A+B | A | … | … | 38.99 | 330.04 | 123.65 ± 0.94 | 0.339 ± 0.011 |
| 11477+0016 | LDS5207 | GJ 3686 | … | M5e | 11:47:41.77 | +00:15:04.4 | … | B | 24.744 | 124.5 | 38.99 | 322.04 | … | 0.126 ± 0.044 |
| … | … | GJ 445 | J11476+786 | M4.0 V | 11:47:45.46 | +78:41:35.9 | … | … | … | … | 83.76 | 889.55 | 79.81 ± 0.51 | 0.254 ± 0.012 |
| … | … | FIVir | J11477+008 | dM4 | 11:47:45.05 | +00:47:56.8 | … | … | … | … | … | 1365.51 | 38.56 ± 0.26 | 0.173 ± 0.010 |
| … | … | GJ 3688 | J11483-112 | M3.0 V | 11:48:18.61 | -11:17:15.0 | … | … | … | … | … | 733.60 | 281.29 ± 2.31 | 0.425 ± 0.031 |
| … | … | G 10-52 | J11485+076 | M3.5 V | 11:48:35.63 | +07:41:37.8 | … | … | … | … | 43.80 | 221.18 | 98.96 ± 0.47 | 0.318 ± 0.014 |
| … | … | BPM 87650 | J11496+220 | M0.0 V | 11:49:40.33 | +22:03:52.5 | … | … | … | … | 392.75 | 132.52 | 1871.25 ± 15.15 | 0.808 ± 0.024 |
| … | … | GJ 1151 | J11509+483 | M4.5 V | 11:50:55.24 | +48:22:23.2 | … | … | … | … | 65.84 | 1820.46 | 34.51 ± 0.18 | 0.168 ± 0.009 |
| … | … | GJ 450 | J11511+352 | M1.5 V | 11:51:06.98 | +35:16:23.3 | … | … | … | … | 32.82 | 372.60 | 323.22 ± 1.66 | 0.440 ± 0.012 |
| 11519+0731 | BWL 66 | RX J1151.9+0731 | J11519+075 | M2.5 Ve | 11:51:56.69 | +07:31:25.2 | (AB) | AB(2) | 0.514 | 107.1 | 71.23 | 150.54 | … | … |
| … | … | GJ 3691 | J11529+244 | M4.1 V | 11:52:57.55 | +24:28:46.9 | … | … | … | … | 18.05 | 325.98 | 40.47 ± 0.16 | 0.209 ± 0.011 |
| … | … | GJ 452 | J11532-073 | M2.5 V | 11:53:15.91 | -07:22:35.8 | … | … | … | … | 203.89 | 534.44 | 284.29 ± 1.82 | 0.460 ± 0.018 |
| … | … | TYC 3016-577-1 | J11533+430 | M1.0 V | 11:53:23.24 | +43:02:56.3 | … | … | … | … | 203.83 | 177.20 | 548.56 ± 1.98 | 0.567 ± 0.028 |
| … | … | GJ 3693 | J11538+069 | M8.0 V | 11:53:53.01 | +06:59:41.9 | … | … | … | … | 34.61 | 947.89 | 8.66 ± 0.06 | 0.129 ± 0.010 |
| … | … | Ross 119 | J11541+098 | M4.0 Ve | 11:54:07.98 | +09:48:10.0 | … | … | … | … | 34.62 | 806.20 | 62.41 ± 0.39 | 0.231 ± 0.012 |
| … | … | PM J11549-0206 | J11549-021 | M3.0 V | 11:54:56.83 | -02:06:08.4 | … | … | … | … | 55.99 | 107.77 | 140.39 ± 0.69 | 0.359 ± 0.015 |
| … | … | Ross 129 | J11551+009 | M1.5 V | 11:55:06.42 | +00:58:26.1 | … | … | … | … | 38.84 | 741.06 | 547.58 ± 3.89 | 0.554 ± 0.028 |
| … | … | GJ 3694 | J11557-189 | M3.5 V | 11:55:44.87 | -18:54:36.6 | … | … | … | … | 40.18 | 609.07 | 81.34 ± 0.39 | 0.286 ± 0.013 |
| … | … | LP 851-346 | J11557-227 | M7.5 V | 11:55:42.42 | -22:25:01.8 | … | … | … | … | 40.29 | 418.69 | 6.49 ± 0.04 | 0.119 ± 0.010 |
| … | … | Ross 122 | J11575+118 | M2.0 V | 11:57:32.06 | +11:49:43.7 | … | … | … | … | 41.46 | 723.41 | 426.28 ± 2.00 | 0.511 ± 0.029 |
| … | … | GJ 3696 | J11582+425 | M4.0 V | 11:58:17.81 | +42:34:23.0 | … | … | … | … | 84.18 | 398.77 | 94.75 ± 0.45 | 0.310 ± 0.014 |
| 11522+0357 | CRC 64 | StM 162 | J11521+039 | M4.0 V | 11:52:09.87 | +03:57:21.4 | (AB) | AB | 0.337 | 26.0 | 84.16 | 158.65 | … | 0.296 ± 0.035 |
| 11590+4240 | PKO 32 | GJ 3697 | J11589+426 | M2.0 V | 11:58:58.99 | +42:39:40.8 | (AB) | AB | 0.303 | 232.8 | 59.60 | 335.59 | … | … |
| 11588+5934 | WOR 21 | G 197-38 | J11585+595 | M0.0 V | 11:58:33.53 | +59:33:22.2 | (AB) | AB | 0.903 | 286.5 | 59.68 | 712.92 | … | … |
| … | … | GJ 3698 | J12006-138 | M3.5 V | 12:00:36.36 | -13:49:36.6 | A+B | A | … | … | 74.88 | 477.06 | 277.14 ± 2.30 | 0.419 ± 0.031 |
| 12007-1348 | LDS4166 | GJ 3699 | … | M4.5 V | 12:00:35.98 | -13:49:40.6 | … | B | 6.791 | 233.9 | … | 468.87 | … | 0.194 ± 0.039 |
| … | … | GJ 3700 | J12016-122 | M3.0 V | 12:01:40.72 | -12:13:57.6 | A+B | A | … | … | 59.92 | 237.56 | 305.10 ± 2.25 | 0.507 ± 0.020 |
| 12017-1214 | CRC 65 | L 829-10 B | … | M7.0 V | 12:01:40.90 | -12:13:52.0 | … | B | 6.122 | 24.4 | 39.42 | 245.96 | … | 0.101 ± 0.049 |
| … | … | Ross 689 | J12054+695 | M4.0 V | 12:05:28.29 | +69:32:21.7 | … | … | … | … | 69.14 | 460.17 | 103.39 ± 0.50 | 0.297 ± 0.012 |
| … | … | LSPM J1205+7825 | J12057+784 | M2.5 V | 12:05:45.99 | +78:25:51.6 | … | … | … | … | 69.13 | 204.94 | 255.48 ± 1.24 | 0.435 ± 0.017 |
| … | … | GJ 455 | J12023+285 | sdM3.5 | 12:02:17.12 | +28:35:13.4 | Aab | Aab(2) | … | … | 60.95 | 788.79 | … | … |
| … | … | 1R120847.7+302120 | J12088+303 | M2.5 V | 12:08:49.63 | +30:21:00.5 | … | … | … | … | 53.89 | 93.75 | 313.28 ± 1.38 | 0.484 ± 0.018 |
| … | … | StM 165 | J12093+210 | M2.5 V | 12:09:21.70 | +21:03:05.8 | … | … | … | … | 43.66 | 161.21 | 329.22 ± 6.30 | 0.452 ± 0.031 |
| … | … | GJ 3707 | J12100-150 | dM3.5 | 12:10:05.54 | -15:04:28.4 | … | … | … | … | 43.62 | 715.19 | 163.68 ± 1.08 | 0.362 ± 0.013 |
| 12064-1315 | JNN77 | 1R120622.6-131453 | J12063-132 | M3.5 V | 12:06:22.22 | -13:14:57.2 | (AB) | AB | 0.425 | 54.1 | 43.58 | 102.51 | … | 0.466 ± 0.031 |
| … | … | GJ 9393 | J12109+410 | M0.0 V | 12:10:56.90 | +41:03:31.7 | … | … | … | … | 43.23 | 229.44 | 600.49 ± 2.24 | 0.602 ± 0.027 |

Table D.2: Complete sample with the description of multiple systems (continued).

| WDS id | WDS disc | Name | Karmn | Spectral type | $\alpha$ (2016.0) | $\delta$ (2016.0) | System | Component | $\rho$ [arcsec] | $\theta$ [deg] | $\varpi$ [mas] | $\mu_{total}$ [mas a$^{-1}$] | $\mathcal{L}$ [$10^{-4}$ $\mathcal{L}_\odot$] | $\mathcal{M}$ [$\mathcal{M}_\odot$] |
|---|---|---|---|---|---|---|---|---|---|---|---|---|---|---|
| … | … | LP 734-34 | J12104-131 | M4.5 V | 12:10:28.65 | -13:10:29.5 | Aab | Aab(2) | … | … | 139.34 | 422.76 | … | 0.013 |
| … | … | GJ 3708 | J12111-199 | dM3.0 | 12:11:11.52 | -19:57:41.0 | A+B | A | … | … | 35.49 | 281.54 | 166.78 ± 0.86 | 0.343 ± 0.010 |
| 12113-1958 | LDS 390 | GJ 3709 | J12112-199 | M2.5 V | 12:11:16.71 | -19:58:24.7 | … | B | 85.283 | 120.8 | 95.52 | 278.50 | 85.89 ± 0.43 | 0.275 ± 0.013 |
| … | … | GJ 3713 | J12121+488 | M2.5 V | 12:12:11.68 | +48:48:58.3 | AB | A | … | … | 110.23 | 371.62 | … | 0.392 ± 0.032 |
| 12122+4849 | SKF1634 | G 122-74B | … | M4.5 V | 12:12:11.74 | +48:48:55.5 | … | B | 2.848 | 167.9 | 52.82 | 382.48 | … | 0.221 ± 0.038 |
| … | … | LP 39-66 | J12122+714 | M3.0 V | 12:12:13.89 | +71:25:26.2 | … | … | … | … | 72.86 | 409.18 | 240.42 ± 1.06 | 0.477 ± 0.019 |
| … | … | HD 238090 | J12123+544S | M0.0 V | 12:12:21.29 | +54:29:10.2 | A+B | A | … | … | 76.19 | 250.12 | 695.72 ± 2.31 | 0.620 ± 0.027 |
| 12123+5429 | VYS 5 | GJ 458 B | J12123+544N | M3.0 V | 12:12:21.58 | +54:29:24.6 | … | B | 14.655 | 10.1 | 50.13 | 255.62 | 67.73 ± 0.28 | 0.259 ± 0.013 |
| … | … | PM J12124+1211 | J12124+121 | M2.0 V | 12:12:25.99 | +12:11:38.2 | … | … | … | … | 61.65 | 73.33 | 350.00 ± 2.08 | 0.481 ± 0.030 |
| 12125+3940 | HDS1727 | GJ 3714 | J12124+396 | M1.0 V | 12:12:29.65 | +39:40:25.2 | (AB) | AB | 0.483 | 333.9 | 30.46 | 224.65 | … | 0.560 ± 0.028 |
| … | … | IVCom | J12133+166 | M1.5 V | 12:13:19.91 | +16:41:32.3 | … | … | … | … | 49.58 | 679.56 | 427.50 ± 2.41 | 0.507 ± 0.018 |
| … | … | GJ 3717 | J12144+245 | M2.0 V | 12:14:26.04 | +24:35:20.5 | … | … | … | … | 42.35 | 371.53 | 359.35 ± 1.92 | 0.486 ± 0.030 |
| … | … | GJ 458.2 | J12151+487 | M0.5 V | 12:15:08.46 | +48:43:56.4 | … | … | … | … | 41.50 | 231.72 | 792.43 ± 6.12 | 0.637 ± 0.027 |
| … | … | GJ 3718 | J12154+391 | M1.5 V | 12:15:27.93 | +39:11:15.4 | … | … | … | … | 41.67 | 337.93 | 418.22 ± 2.52 | 0.514 ± 0.029 |
| … | … | StKM 2-809 | J12156+526 | M4.0 Ve | 12:15:39.55 | +52:39:08.7 | … | … | … | … | 55.58 | 105.25 | 328.98 ± 4.21 | 0.477 ± 0.012 |
| … | … | GJ 1154 | J12142+006 | M4.5 Ve | 12:14:15.53 | +00:37:21.8 | … | … | … | … | 39.16 | 992.84 | 34.61 ± 0.25 | 0.206 ± 0.011 |
| … | … | G3- 1547529669400723968 | … | M2/3 V | 12:16:14.94 | +50:53:38.4 | AB | A | … | … | 42.67 | 86.59 | … | 0.362 ± 0.033 |
| 12163+5054 | CRC 66 | RX J1216.2+5053 | J12162+508 | M4.0 V | 12:16:14.90 | +50:53:36.7 | … | B | 1.785 | 192.0 | 54.19 | 86.61 | … | 0.308 ± 0.034 |
| … | … | GJ 1155 A | J12168+029 | M3.0 V | 12:16:51.16 | +02:58:09.0 | AB | A | … | … | 54.31 | 700.55 | … | 0.348 ± 0.033 |
| 12169+0258 | LDS 935 | GJ 1155 B | … | DA | 12:16:51.17 | +02:58:06.9 | … | B | 2.126 | 177.9 | 44.32 | 711.07 | … | 0.500 ± 0.100 |
| … | … | PM J12168+2451E | J12168+248 | M1.5 V | 12:16:52.62 | +24:51:06.0 | … | … | … | … | 44.27 | 51.85 | 298.84 ± 1.41 | 0.445 ± 0.017 |
| … | … | GLVir | J12189+111 | M4.5 Ve | 12:18:58.02 | +11:07:37.0 | … | … | … | … | 25.30 | 1285.97 | 22.07 ± 0.12 | 0.152 ± 0.009 |
| … | … | GJ 3719 | J12169+311 | M3.0 V | 12:16:58.27 | +31:09:22.6 | Aab | Aab(2) | … | … | 32.83 | 133.17 | … | … |
| … | … | LP 320-626 | J12191+318 | M3.5 V | 12:19:05.55 | +31:50:43.5 | AabB | Aab(2) | … | … | 35.26 | 321.46 | … | … |
| 12191+3151 | RAO 274 | LP 320-626 | … | M3.5 V | 12:19:05.47 | +31:50:42.2 | … | B | 1.739 | 220.6 | 25.34 | 295.65 | … | … |
| … | … | Wolf 408 | J12194+283 | M0.5 V | 12:19:23.31 | +28:22:57.8 | … | … | … | … | 34.81 | 652.53 | 814.53 ± 4.58 | 0.640 ± 0.027 |
| … | … | StKM 1-1007 | J12198+527 | M0.0 V | 12:19:47.76 | +52:46:43.1 | … | … | … | … | 78.29 | 209.58 | 684.47 ± 3.22 | 0.617 ± 0.027 |
| … | … | G 123-36 | J12199+364 | M1.0 V | 12:29:55.22 | +36:26:40.0 | … | … | … | … | 26.25 | 204.34 | 438.43 ± 4.56 | 0.527 ± 0.029 |
| … | … | GJ 461 | J12204+005 | M0.0 V | 12:20:25.59 | +00:35:00.4 | AB | A | … | … | 74.20 | 60.60 | … | 0.609 ± 0.027 |
| 12204+0035 | RST5366 | G3- 3699797155055781760 | … | M2.5 V | 12:20:25.60 | +00:35:01.3 | … | B | 0.882 | 14.6 | 24.78 | 129.05 | … | 0.418 ± 0.031 |
| … | … | G 148-48 | J12214+306W | M5.0 V | 12:21:26.79 | +30:38:31.4 | A+B | A | … | … | 24.73 | 339.77 | 18.53 ± 0.21 | 0.144 ± 0.010 |
| 12215+3038 | GIC 106 | LP 320-416 | J12214+306E | M4.5 V | 12:21:26.49 | +30:38:33.6 | … | B | 4.497 | 298.9 | 86.90 | 317.63 | … | 0.141 ± 0.042 |
| … | … | LP 39-245 | J12217+682 | M3.0 V | 12:21:46.41 | +68:16:07.3 | … | … | … | … | 35.62 | 429.42 | 168.82 ± 0.84 | 0.371 ± 0.015 |
| … | … | Wolf 409 | J12223+251 | M0.5 V | 12:22:20.46 | +25:10:08.6 | … | … | … | … | 63.21 | 723.19 | 559.65 ± 2.91 | 0.584 ± 0.028 |
| … | … | BD+28 2110 | J12225+123 | M0.0 V | 12:22:33.90 | +27:36:16.5 | … | … | … | … | 54.83 | 148.54 | 894.83 ± 4.86 | 0.667 ± 0.026 |
| … | … | Ross 690 | J12230+640 | M3.0 V | 12:22:58.52 | +64:01:56.9 | … | … | … | … | 31.04 | 769.71 | 335.88 ± 1.52 | 0.484 ± 0.014 |
| … | … | Wolf 411 | J12235+279 | M0.0 V | 12:23:34.55 | +27:54:49.6 | … | … | … | … | 31.04 | 183.00 | 608.70 ± 2.78 | 0.599 ± 0.027 |
| 12228-0405 | BWL 29 | G 13-33 | J12228-040 | M4.5 V | 12:22:50.33 | -04:04:47.5 | (AB) | AB | 0.107 | 144.8 | 25.19 | 263.33 | … | 0.184 ± 0.044 |
| … | … | HD 107888 | J12238+125 | M0.0 V | 12:23:53.60 | +12:34:46.2 | … | … | … | … | 35.78 | 173.35 | 775.49 ± 3.08 | 0.637 ± 0.027 |
| … | … | Ross 695 | J12248-182 | dM2.0 | 12:24:53.73 | -18:15:09.2 | … | … | … | … | 90.69 | 2555.76 | 96.99 ± 0.37 | 0.270 ± 0.010 |

Table D.2: Complete sample with the description of multiple systems (continued).

| WDS id | WDS disc | Name | Karmn | Spectral type | $\alpha$ (2016.0) | $\delta$ (2016.0) | System | Component | $\rho$ [arcsec] | $\theta$ [deg] | $\varpi$ [mas] | $\mu_{\rm total}$ [mas a$^{-1}$] | $\mathcal{L}$ [$10^{-4}\ \mathcal{L}_\odot$] | $\mathcal{M}$ [$\mathcal{M}_\odot$] |
|---|---|---|---|---|---|---|---|---|---|---|---|---|---|---|
| 12236+6711 | HDS1745 | GJ 3722 | J12235+671 | M2.5 V | 12:23:33.85 | +67:11:16.4 | (AB) | AB | 0.621 | 33.3 | 40.70 | 268.66 | ... | 0.350 ± 0.033 |
| 12251+6025 | GIC 108 | LP 95-135 | J12251+604 | M1.0 V | 12:25:05.64 | +60:25:06.0 | A+B | A | 19.987 | 65.4 | 102.30 | 198.04 | 516.78 ± 2.59 | 0.552 ± 0.028 |
| ... | ... | LP 95-136 | ... | M1.5 V | 12:25:08.10 | +60:25:14.3 | ... | B | 19.987 | 65.5 | 32.37 | 200.57 | 433.44 ± 2.23 | 0.520 ± 0.029 |
| ... | ... | HD 108421A | ... | K2 V | 12:27:13.83 | +27:01:24.9 | AB+C | A | ... | ... | 32.40 | 265.14 | ... | 0.770 ± 0.116 |
| 12272+2701 | STF1643 | HD 108421B | ... | K4 V | 12:27:13.84 | +27:01:27.6 | ... | B | 2.756 | 4.1 | 27.75 | 256.19 | ... | 0.729 ± 0.109 |
| 12272+2701 | LEP 54 | CXCom | J12269+270 | M4.5 V | 12:26:57.47 | +27:00:49.7 | ... | C | 221.398 | 260.9 | 36.36 | 265.95 | 77.61 ± 1.50 | 0.298 ± 0.014 |
| ... | ... | G 148-61 | J12274+374 | M1.5 V | 12:27:29.12 | +37:26:35.1 | ... | ... | ... | ... | 51.58 | 249.10 | 307.45 ± 1.17 | 0.452 ± 0.017 |
| ... | ... | G 13-39A | J12277-032 | M3.5 V | 12:27:44.38 | -03:15:01.3 | AB | A | ... | ... | 32.86 | 295.11 | ... | 0.396 ± 0.032 |
| 12277-0315 | CRC 67 | G 13-39B | ... | M4.5 V | 12:27:44.41 | -03:14:59.9 | ... | B | 1.469 | 14.9 | 32.65 | 319.40 | ... | 0.221 ± 0.038 |
| ... | ... | Ross 948A | J12288-106N | M2.0 V | 12:28:52.85 | -10:39:48.6 | A+B | A | ... | ... | 47.50 | 279.65 | ... | 0.469 ± 0.030 |
| 12288-1040 | RST3792 | Ross 948B | J12288-106S | M2.0 V | 12:28:52.65 | -10:39:50.8 | ... | B | 3.771 | 233.3 | 53.14 | 280.43 | ... | 0.461 ± 0.030 |
| 12290+0826 | WSI 113 | Wolf 414 | J12289+084 | M3.5 V | 12:28:56.90 | +08:25:27.3 | (AB) | AB | 0.072 | 69.8 | 53.02 | 683.10 | ... | 0.356 ± 0.033 |
| 12290+4144 | BWL 31 | GJ 3729 | J12290+417 | M3.5 V | 12:29:02.65 | +41:43:45.9 | (AB) | AabB(2/3?) | 0.050 | 255.5 | 50.42 | 286.43 | ... | ... |
| ... | ... | GJ 1159 A | J12292+535 | M4.0 V | 12:29:12.22 | +53:32:47.0 | A+B | A | ... | ... | 190.33 | 1228.81 | 89.20 ± 0.40 | 0.300 ± 0.014 |
| 12294+5333 | LDS3046 | GJ 1159 B | ... | M6.0 V | 12:29:11.96 | +53:33:08.3 | ... | B | 21.394 | 353.7 | 296.31 | 1230.60 | 9.73 ± 0.06 | 0.116 ± 0.009 |
| ... | ... | GJ 3730 | J12294+229 | M4.0 V | 12:29:26.92 | +22:59:46.4 | ... | ... | ... | ... | 34.94 | 173.83 | 82.96 ± 0.48 | 0.289 ± 0.013 |
| ... | ... | GJ 3731 | J12299-054W | M3.5 V | 12:29:53.57 | -05:27:29.2 | Aab+B | Aab(2) | ... | ... | 34.64 | 646.28 | ... | ... |
| 12298-0527 | B2737 | GJ 3732 | J12299-054E | M4.0 V | 12:29:54.05 | -05:27:25.2 | ... | B | 8.153 | 60.5 | 48.29 | 633.07 | ... | 0.257 ± 0.036 |
| ... | ... | Wolf 417 | J12312+086 | M0.5 V | 12:31:15.12 | +08:48:29.8 | ... | ... | ... | ... | 20.21 | 823.04 | 627.57 ± 3.32 | 0.599 ± 0.027 |
| ... | ... | GJ 3733 | J12323+315 | M3.0 V | 12:32:19.79 | +31:36:03.2 | ... | ... | ... | ... | 124.34 | 473.16 | 276.47 ± 5.11 | 0.482 ± 0.019 |
| ... | ... | GJ 3734 | J12324+203 | M2.5 V | 12:32:26.37 | +20:23:28.0 | ... | ... | ... | ... | 114.09 | 49.52 | 179.03 ± 1.13 | 0.408 ± 0.017 |
| ... | ... | LP 39-249 | J12327+682 | M0.0 V | 12:32:44.66 | +68:15:42.1 | ... | ... | ... | ... | 19.71 | 215.82 | 878.12 ± 3.28 | 0.664 ± 0.026 |
| ... | ... | Wolf 424 A | J12332+090 | M5.0 V | 12:33:15.48 | +09:01:19.5 | AB | A | ... | ... | 59.63 | 1808.82 | ... | 0.140 ± 0.043 |
| 12335+0901 | REU1 | Wolf 424 B | ... | M5.5 V | 12:33:15.52 | +09:01:18.6 | ... | B | 1.149 | 145.3 | 51.25 | 1722.48 | ... | 0.143 ± 0.042 |
| ... | ... | PM J12349+3214 | J12349+322 | M3.5 V | 12:34:54.03 | +32:14:29.2 | ... | ... | ... | ... | 35.47 | 79.89 | 254.04 ± 1.51 | 0.491 ± 0.019 |
| ... | ... | Wolf 427 | J12350+098 | M2.5 V | 12:35:00.22 | +09:49:37.5 | ... | ... | ... | ... | 67.14 | 551.02 | 331.44 ± 1.51 | 0.482 ± 0.012 |
| ... | ... | GJ 3736 | J12363-043 | M3.0 V | 12:36:22.35 | -04:22:41.7 | ... | ... | ... | ... | 86.78 | 502.65 | 174.92 ± 0.99 | 0.378 ± 0.016 |
| ... | ... | PM J12368-0159 | J12368-019 | M3.5 V | 12:36:51.96 | -01:59:02.0 | ... | ... | ... | ... | 40.60 | 176.91 | 194.26 ± 1.17 | 0.426 ± 0.017 |
| ... | ... | LP 795-38 | J12373-208 | dM4.0 | 12:37:21.52 | -20:52:42.4 | ... | ... | ... | ... | 39.19 | 429.67 | 202.80 ± 1.02 | 0.401 ± 0.014 |
| ... | ... | GJ 1162 | J12387-043 | M4.3 V | 12:38:46.47 | -04:19:20.2 | ... | ... | ... | ... | 42.12 | 773.98 | 91.94 ± 0.47 | 0.305 ± 0.014 |
| ... | ... | Wolf 433 | J12388+116 | M3.5 V | 12:38:51.18 | +11:41:42.1 | ... | ... | ... | ... | 91.27 | 1182.67 | 276.45 ± 1.60 | 0.446 ± 0.013 |
| ... | ... | G 123-049A | J12390+470 | M2.5 Ve | 12:39:05.24 | +47:02:21.4 | (AB) | A | ... | ... | 39.53 | ... | ... | ... |
| 12391+4702 | CFN 9 | G 123-049B | ... | M2.0 V | 12:39:05.30 | +47:02:21.2 | ... | B | 0.661 | 115.1 | 45.90 | 402.46 | ... | 0.444 ± 0.031 |
| ... | ... | GJ 3739 | J12397+255 | M4.0 V | 12:39:43.32 | +25:30:43.3 | ... | ... | ... | ... | 62.55 | 240.19 | 59.91 ± 0.29 | 0.242 ± 0.012 |
| ... | ... | GJ 3741 | J12416+482 | M1.0 V | 12:41:38.24 | +48:14:22.3 | ... | ... | ... | ... | ... | 267.98 | 396.01 ± 2.39 | 0.507 ± 0.029 |
| ... | ... | RX J1241.7+5645 | J12417+567 | M3.5 V | 12:41:47.59 | +56:45:13.7 | ... | ... | ... | ... | 23.21 | 116.23 | 136.82 ± 0.67 | 0.354 ± 0.015 |
| ... | ... | G 123-55 | J12428+418 | M4.0 Ve | 12:42:49.10 | +41:53:47.8 | ... | ... | ... | ... | 39.11 | 550.33 | 163.84 ± 0.79 | 0.382 ± 0.012 |
| ... | ... | GJ 1163 | J12436+251 | M3.5 Ve | 12:43:35.62 | +25:06:21.1 | ... | ... | ... | ... | 39.29 | 370.26 | 184.47 ± 2.43 | 0.414 ± 0.017 |
| ... | ... | LP 735-29 | J12440-111 | M4.5 V | 12:44:00.22 | -11:10:32.8 | ... | ... | ... | ... | 40.36 | 502.79 | 41.38 ± 0.19 | 0.211 ± 0.011 |
| ... | ... | Ross 991 | J12470+466 | M2.5 V | 12:46:59.81 | +46:37:28.6 | ... | ... | ... | ... | 40.19 | 817.84 | 351.80 ± 1.70 | 0.469 ± 0.030 |

Table D.2: Complete sample with the description of multiple systems (continued).

| WDS id | WDS disc | Name | Karmn | Spectral type | $\alpha$ (2016.0) | $\delta$ (2016.0) | System | Component | $\rho$ [arcsec] | $\theta$ [deg] | $\varpi$ [mas] | $\mu_{\rm total}$ [mas a$^{-1}$] | $\mathcal{L}$ [$10^{-4}\ \mathcal{L}_\odot$] | $\mathcal{M}$ [$\mathcal{M}_\odot$] |
|---|---|---|---|---|---|---|---|---|---|---|---|---|---|---|
| … | … | GJ 3747 | J12471-035 | M3.0 V | 12:47:09.24 | -03:34:18.0 | … | … | … | … | 64.80 | 508.13 | 200.56 ± 1.28 | 0.407 ± 0.016 |
| … | … | Wolf 437 | J12479+097 | M3.5 V | 12:47:55.53 | +09:44:57.7 | … | … | … | … | 47.22 | 1108.26 | 122.75 ± 0.79 | 0.311 ± 0.013 |
| … | … | G 123-45 | J12364+352 | M4.5 V | 12:36:28.17 | +35:11:59.0 | Aab | Aab(1) | … | … | 24.83 | 375.94 | … | … |
| … | … | RX J1248.5+4933 | J12485+495 | M3.5 Ve | 12:48:34.67 | +49:33:53.8 | … | … | … | … | 57.12 | 109.89 | 377.59 ± 11.59 | 0.469 ± 0.030 |
| 12482+4714 | LHS2637 | GJ 3749 | J12481+472 | M3.5 V | 12:48:10.07 | +47:13:23.2 | (AB) | AB | 0.830 | 169.0 | 35.35 | 593.19 | … | 0.336 ± 0.033 |
| … | … | Wolf 439 | J12495+094 | M3.5 V | 12:49:33.74 | +09:28:31.6 | … | … | … | … | 26.86 | 426.63 | 159.06 ± 1.01 | 0.383 ± 0.016 |
| … | … | GJ 3755 | J12505+269 | M3.8 V | 12:50:34.35 | +26:55:20.3 | … | … | … | … | 81.57 | 261.34 | … | … |
| … | … | APMPM J1251-2121 | J12508-213 | M7.5 Ve | 12:50:53.16 | -21:21:18.9 | … | … | … | … | 32.85 | 558.72 | 13.35 ± 0.14 | 0.182 ± 0.012 |
| 12490+6607 | DEL 4 | DPDra | J12490+661 | M3.0 V | 12:49:01.61 | +66:06:35.2 | (AabB) | AabB(3) | 0.269 | 221.4 | 29.98 | 447.63 | … | 0.463 ± 0.001 |
| … | … | GJ 1166A | J12513+221 | M3.0 V | 12:51:23.71 | +22:06:15.7 | A+BC | A | … | … | 45.28 | 182.91 | 342.08 ± 8.14 | 0.507 ± 0.020 |
| 12515+2207 | LDS 940 | LP 377-78 B | … | M3.5 V | 12:51:28.60 | +22:07:06.3 | … | B | 84.764 | 53.3 | 45.47 | 180.19 | … | 0.336 ± 0.033 |
| 12515+2207 | RAO 284 | GJ 1166 B | … | M3.5 V | 12:51:23.80 | +22:06:15.7 | … | C | 1.268 | 91.6 | 78.48 | 180.35 | … | 0.304 ± 0.034 |
| 12576+3514 | BWL 35 | BFCVn | J12576+352E | M0.0 V | 12:57:39.89 | +35:13:27.9 | (AB)+(CD) | AB | 0.082 | 156.6 | 78.46 | 297.85 | … | 0.562 ± 0.028 |
| 12576+3514 | BWL 35 | GJ 490 B | J12576+352W | M4.5 Ve | 12:57:38.94 | +35:13:16.9 | … | CD | 16.046 | 226.7 | 37.64 | 317.47 | … | 0.336 ± 0.033 |
| … | … | GJ 3757 | J12594+077 | M5.0 V | 12:59:23.31 | +07:43:54.8 | … | … | … | … | 37.59 | 677.91 | 21.86 ± 0.11 | 0.159 ± 0.010 |
| … | … | Ross 972 | J13000-056 | M3.0 V | 13:00:03.58 | -05:37:47.1 | … | … | … | … | 42.01 | 346.78 | 185.01 ± 1.56 | 0.415 ± 0.017 |
| … | … | FNVir | J13005+056 | M4.5 Ve | 13:00:32.51 | +05:41:11.6 | … | … | … | … | 65.58 | 967.81 | 36.32 ± 0.28 | 0.188 ± 0.010 |
| 12584+4033 | HDS1819 | LP 41-165 | J12583+405 | M1.5 V | 12:58:21.97 | +40:33:20.7 | (AB) | AB | 0.187 | 195.2 | 65.60 | 268.38 | … | 0.613 ± 0.029 |
| 13008+1223 | BEU 16 | Wolf 462 | J13007+123 | M0.0 V | 13:00:45.87 | +12:22:32.1 | (ABC) | AB | 0.525 | 60.0 | 28.08 | 629.61 | … | 0.551 ± 0.028 |
| 13008+1223 | GDM 1 | Ross 458C | … | T8.5p | … | … | … | C | 12.060 | 220.2 | 34.85 | 639.45 | … | … |
| … | … | G 164-38 | J13019+335 | M1.0 V | 13:01:55.96 | +33:35:23.0 | … | … | … | … | 33.35 | 449.63 | 384.64 ± 1.94 | 0.480 ± 0.018 |
| … | … | G 123-84 | J13027+415 | M3.5 V | 13:02:46.65 | +41:31:06.6 | … | … | … | … | 36.68 | 577.05 | 245.25 ± 1.15 | 0.482 ± 0.019 |
| … | … | GJ 497 A | J13047+559 | M0.5 V | 13:04:46.29 | +55:54:10.6 | AB | A | … | … | 43.80 | 167.46 | … | 0.687 ± 0.026 |
| 13048+5555 | WOR 23 | GJ 497 B | … | M1.0 V | 13:04:46.36 | +55:54:08.7 | … | B | 2.019 | 161.7 | 37.11 | 169.95 | … | 0.544 ± 0.028 |
| … | … | GJ 3760 A | J13054+371 | M2.5 V | 13:05:29.44 | +37:08:07.6 | AB | A | … | … | 39.85 | 391.69 | … | 0.464 ± 0.030 |
| 13055+3708 | HDS1830 | GJ 3760 B | … | M2.0 V | 13:05:29.46 | +37:08:07.2 | … | B | 0.477 | 141.9 | 123.64 | 345.22 | … | 0.463 ± 0.030 |
| … | … | GJ 3762 | J13068+308 | M6.0 V | 13:06:50.53 | +30:50:46.4 | … | … | … | … | 26.29 | 508.51 | 20.81 ± 0.08 | 0.154 ± 0.010 |
| … | … | GJ 9428 | J13084+169 | M1.0 V | 13:08:24.64 | +16:58:18.5 | … | … | … | … | 26.23 | 124.83 | 439.07 ± 18.86 | 0.520 ± 0.029 |
| … | … | GJ 3763 | J13088+163 | M2.5 V | 13:08:49.96 | +16:22:00.9 | … | … | … | … | 42.82 | 558.87 | 227.02 ± 9.60 | 0.434 ± 0.020 |
| … | … | GJ 3765 | J13089+490 | M0.5 V | 13:08:55.46 | +49:04:50.3 | … | … | … | … | 42.77 | 114.23 | 361.58 ± 1.29 | 0.464 ± 0.017 |
| … | … | G 177-25 | J13102+477 | M5.0 V | 13:10:11.62 | +47:45:08.8 | … | … | … | … | 36.74 | 885.73 | 30.24 ± 0.11 | 0.162 ± 0.009 |
| … | … | GJ 3766 | J13113+285 | M5.0 V | 13:11:21.13 | +28:32:34.4 | … | … | … | … | 154.70 | 603.28 | 21.63 ± 0.09 | 0.158 ± 0.010 |
| … | … | GJ 3767 | J13118+253 | M5.0 V | 13:11:51.24 | +25:20:48.1 | … | … | … | … | 25.98 | 493.64 | 41.10 ± 0.22 | 0.211 ± 0.011 |
| … | … | PM J13119+6550 | J13119+658 | M3.0 V | 13:11:59.17 | +65:50:01.3 | … | … | … | … | … | 152.67 | 279.05 ± 1.26 | 0.408 ± 0.011 |
| … | … | GJ 1168 | J13130+201 | M3.5 V | 13:13:04.08 | +20:11:29.0 | … | … | … | … | 35.20 | 627.69 | 198.19 ± 1.02 | 0.430 ± 0.017 |
| … | … | GJ 3772 | J13140+038 | M3.8 V | 13:14:05.05 | +03:53:59.0 | … | … | … | … | 41.90 | 736.12 | 88.67 ± 0.46 | 0.280 ± 0.013 |
| … | … | GJ 1167 A | J13095+289 | M4.8V | 13:09:34.56 | +28:59:03.1 | … | … | … | … | 34.78 | 395.26 | 35.72 ± 0.19 | 0.195 ± 0.011 |
| … | … | G 255-29A | J13142+792 | M1.0 V | 13:14:14.06 | +79:14:45.9 | AB | A | … | … | 33.89 | 297.31 | … | 0.631 ± 0.027 |
| 13143+7915 | KPP3265 | G 255-29B | … | M2.0 V | 13:14:13.78 | +79:14:47.1 | … | B | 1.383 | 325.5 | 57.13 | 295.36 | … | 0.478 ± 0.030 |
| … | … | GJ 1169 | J13165+278 | M4.0 V | 13:16:31.95 | +27:52:33.5 | … | … | … | … | 56.81 | 777.15 | 67.09 ± 0.30 | 0.240 ± 0.012 |

Table D.2: Complete sample with the description of multiple systems (continued).

| WDS id | WDS disc | Name | Karmn | Spectral type | $\alpha$ (2016.0) | $\delta$ (2016.0) | System | Component | $\rho$ [arcsec] | $\theta$ [deg] | $\varpi$ [mas] | $\mu_{total}$ [mas a$^{-1}$] | $\mathcal{L}$ [$10^{-4}$ $\mathcal{L}_\odot$] | $\mathcal{M}$ [$\mathcal{M}_\odot$] |
|---|---|---|---|---|---|---|---|---|---|---|---|---|---|---|
| … | … | LP 737-14 | J13167-123 | M3.5 V | 13:16:45.10 | -12:20:21.2 | … | … | … | … | 83.50 | 289.65 | 111.96 ± 3.17 | 0.339 ± 0.016 |
| 13143+1320 | LAW 2 | LP 497-33 | J13143+133 | M6.0 V | 13:14:20.08 | +13:19:57.9 | (AB) | AB(2) | 0.172 | 96.4 | 37.42 | 306.81 | … | … |
| … | … | HD 115404 | … | K2 V | 13:16:51.76 | +17:00:57.6 | A+B | A | … | … | 83.46 | 689.14 | 2979.43 ± 25.18 | 0.820 ± 0.123 |
| 13169+1701 | BU 800 | HD 115404B | J13168+170 | M0.5 V | 13:16:52.28 | +17:00:55.7 | … | B | 7.671 | 104.6 | 51.13 | 701.23 | … | 0.541 ± 0.029 |
| … | … | GJ 3774 | J13168+231 | M1.5 V | 13:16:53.38 | +23:10:05.5 | … | … | … | … | 35.92 | 283.60 | 536.26 ± 2.91 | 0.552 ± 0.028 |
| … | … | GJ 1170 | J13179+362 | M1.0 V | 13:17:58.30 | +36:17:51.5 | … | … | … | … | 33.68 | 345.03 | 501.91 ± 1.74 | 0.553 ± 0.028 |
| 13180+0214 | CRC 68 | GJ 3775 | J13180+022 | M3.5 V | 13:18:01.42 | +02:14:00.4 | AB | A | 0.701 | 243.0 | 54.45 | 290.74 | … | 0.218 ± 0.038 |
| … | … | G3- 3688439268658769408 | … | M4.5 V | 13:18:01.41 | +02:14:00.2 | … | B | 0.342 | 225.1 | 33.87 | 290.74 | … | 0.211 ± 0.038 |
| … | … | PM J13182+7322 | J13182+733 | M3.5 V | 13:18:13.82 | +73:22:05.6 | A+B | A | … | … | 58.90 | 129.46 | 137.10 ± 0.64 | 0.354 ± 0.015 |
| … | … | PM J13182+7322B | … | M7.0 V | 13:18:13.11 | +73:22:12.4 | … | B* | 7.387 | 335.7 | 46.55 | 122.66 | … | 0.100 ± 0.049 |
| 13196+3507 | BAG 11 | GJ 507 A | J13195+351W | M0.5 V | 13:19:34.10 | +35:06:24.2 | (AB)+Cab | AB(2) | 0.073 | 182.7 | 76.10 | 870.64 | … | 0.684 ± 0.022 |
| 13196+3507 | HJ 529 | GJ 507 B | J13195+351E | M3.0 V | 13:19:35.18 | +35:06:12.4 | … | Cab(1) | 17.793 | 131.6 | 41.62 | 883.65 | … | … |
| 13198+4747 | CHR 193 | HD 115953 | J13197+477 | M2.0 V | 13:19:45.90 | +47:46:40.5 | AB+C | AB(2) | 0.088 | 97.2 | 47.32 | 227.71 | … | … |
| 13198+4747 | HU 644 | HD 115953B | … | M2.0 V | 13:19:45.97 | +47:46:40.6 | … | C | 0.666 | 85.5 | 112.67 | 227.71 | … | 0.473 ± 0.030 |
| … | … | Ross 1007 | J13196+333 | M1.5 V | 13:19:39.74 | +33:20:45.2 | … | … | … | … | 75.62 | 330.74 | 545.68 ± 2.66 | 0.553 ± 0.028 |
| … | … | Ross 1008 | J13209+342 | M1.0 V | 13:20:58.67 | +34:16:39.4 | … | … | … | … | 18.87 | 567.80 | 483.53 ± 214.18 | 0.541 ± 0.029 |
| … | … | LSPM J1321+0332 | J13215+035 | M1.0 V | 13:21:30.15 | +03:33:02.1 | … | … | … | … | 18.86 | 192.61 | 365.15 ± 2.88 | 0.467 ± 0.017 |
| … | … | GJ 3778 | J13215+037 | M2.0 V | 13:21:34.69 | +03:45:54.5 | … | … | … | … | 36.50 | 519.37 | 320.53 ± 1.96 | 0.490 ± 0.019 |
| … | … | Ross 1020 | J13229+244 | M4.0 V | 13:22:56.03 | +24:27:49.8 | … | … | … | … | 36.51 | 1061.67 | 86.85 ± 0.58 | 0.272 ± 0.012 |
| … | … | HD 116495A | J13235+292 | M0.0 V | 13:23:32.21 | +29:14:18.9 | AB | A | … | … | 37.64 | 531.64 | … | 0.710 ± 0.026 |
| 13235+2914 | HO 260 | HD 116495B | … | K6 V | 13:23:32.33 | +29:14:19.0 | … | B | 1.659 | 88.4 | 38.51 | 515.54 | … | 0.654 ± 0.027 |
| … | … | LP 40-109 | J13239+694 | M0.0 V | 13:23:56.16 | +69:27:03.5 | … | … | … | … | 40.46 | 180.01 | 744.36 ± 2.65 | 0.638 ± 0.027 |
| … | … | G 14-52 | J13247-050 | M4.0 V | 13:24:46.56 | -05:04:24.8 | … | … | … | … | 40.38 | 309.86 | 128.77 ± 2.63 | 0.302 ± 0.014 |
| … | … | PM J13251-1126 | J13251-114 | M3.0 V | 13:25:11.61 | -11:26:33.7 | … | … | … | … | 51.44 | 238.84 | 498.98 ± 3.50 | 0.492 ± 0.018 |
| … | … | BD+38 2445 | J13254+377 | M0.0 V | 13:25:28.07 | +37:43:10.9 | AB | A | … | … | 51.51 | 205.97 | … | 0.661 ± 0.026 |
| 13260+3743 | SKF 942 | BD+38 2445B | … | M4.5 V | 13:25:28.31 | +37:43:10.4 | … | B | 2.847 | 99.7 | 72.27 | 213.36 | … | 0.258 ± 0.037 |
| … | … | 2M13253177+6850106 | J13255+688 | M0.0 V | 13:25:31.76 | +68:50:09.8 | … | … | … | … | 42.54 | 46.15 | 823.44 ± 3.59 | 0.644 ± 0.027 |
| … | … | PM J13255+2738 | … | M1.0 V | 13:25:35.66 | +27:38:08.9 | A+BC | A | … | … | 39.91 | 71.21 | 570.45 ± 2.86 | 0.579 ± 0.028 |
| … | … | PM J13260+2735A | J13260+275 | M3.0 V | 13:26:02.70 | +27:35:03.7 | … | B* | 404.411 | 117.2 | 39.95 | 72.89 | … | 0.481 ± 0.030 |
| 13260+2735 | KPP3896 | PM J13260+2735B | … | M2.5 V | 13:26:02.63 | +27:35:02.5 | … | C | 404.111 | 117.4 | 44.08 | 63.57 | … | 0.391 ± 0.032 |
| … | … | BD+30 2400 | J13282+300 | M0.0 V | 13:28:17.53 | +30:02:43.0 | AB+C | A | … | … | 46.99 | 260.99 | … | 0.724 ± 0.025 |
| 13283+3003 | CFN 11 | BD+30 2400B | … | … | 13:28:17.48 | +30:02:44.1 | … | B | 1.247 | 324.0 | 47.09 | … | … | … |
| 13284+3005 | LDS1390 | LP 323-115 | … | M7.0 V | 13:28:20.69 | +30:03:15.8 | … | C | 52.481 | 51.3 | 72.89 | 259.70 | 10.25 ± 0.14 | 0.142 ± 0.010 |
| … | … | Ross 486A | J13283-023W | dM3.0 | 13:28:21.24 | -02:21:45.0 | A+B | A | … | … | 30.71 | 517.21 | 259.47 ± 1.78 | 0.424 ± 0.013 |
| 13283-0222 | B 2753 | Ross 486B | J13283-023E | M4.0 V | 13:28:21.68 | -02:21:39.6 | … | B | 8.472 | 50.2 | 43.82 | 503.74 | 48.50 ± 0.84 | 0.231 ± 0.012 |
| … | … | GJ 513 | J13293+114 | M3.0 V | 13:29:21.67 | +11:26:07.6 | … | … | … | … | 30.10 | 1235.48 | 248.46 ± 1.65 | 0.440 ± 0.013 |
| … | … | 1R132923.9-142206 | J13294-143 | M3.5 V | 13:29:24.21 | -14:22:13.0 | … | … | … | … | 231.12 | 116.41 | 198.11 ± 1.45 | 0.430 ± 0.017 |
| … | … | Ross 490 | J13299+102 | M1.0 V | 13:30:01.01 | +10:22:20.6 | … | … | … | … | 223.48 | 1556.96 | 424.22 ± 2.99 | 0.487 ± 0.012 |
| … | … | Wolf 485 | … | DA3.5 | 13:30:12.44 | -08:34:37.0 | A+B | A | … | … | 30.23 | 1207.51 | … | 0.500 ± 0.100 |
| 13303-0834 | LDS 448 | Ross 476 | J13300-087 | M4.0 V | 13:30:01.59 | -08:42:33.0 | … | B | 502.447 | 198.7 | 54.97 | 1210.61 | 49.55 ± 0.26 | 0.234 ± 0.012 |

Table D.2: Complete sample with the description of multiple systems (continued).

| WDS id | WDS disc | Name | Karmn | Spectral type | $\alpha$ (2016.0) | $\delta$ (2016.0) | System | Component | $\rho$ [arcsec] | $\theta$ [deg] | $\varpi$ [mas] | $\mu_{\rm total}$ [mas a$^{-1}$] | $\mathcal{L}$ [$10^{-4}$ $\mathcal{L}_\odot$] | $\mathcal{M}$ [$\mathcal{M}_\odot$] |
|---|---|---|---|---|---|---|---|---|---|---|---|---|---|---|
| … | … | GJ 1171 | J13305+191 | M5.0 V | 13:30:30.49 | +19:09:13.5 | … | … | … | … | 35.54 | 1378.97 | 28.83 ± 0.15 | 0.172 ± 0.010 |
| … | … | GJ 3790 | J13318+233 | M2.5 V | 13:31:50.26 | +23:23:21.1 | … | … | … | … | 35.35 | 284.20 | 201.30 ± 1.26 | 0.407 ± 0.016 |
| 13318+2917 | BEU 17 | DGCVn | J13317+292 | M4.0 V | 13:31:46.33 | +29:16:34.3 | (AB) | AB(2) | 0.190 | 85.4 | 47.96 | 285.30 | … | … |
| 13320+3108 | WOR 24 | GJ 9448 B | J13319+311 | M0.0 V | 13:31:58.08 | +31:08:05.3 | AB | A | 0.134 | 125.5 | 41.82 | 132.92 | … | 0.668 ± 0.039 |
| … | … | G3- 1468339715235126784 | … | … | 13:31:58.10 | +31:08:05.6 | … | B | 0.384 | 42.1 | 53.52 | … | … | … |
| 13327+3100 | BWL36 | LP 323-169 | J13326+309 | M4.5 V | 13:32:38.85 | +30:59:05.3 | (AB) | AB | 0.137 | 213.8 | 70.11 | 208.43 | … | 0.274 ± 0.036 |
| … | … | VWCom | J13327+168 | M2.5 V | 13:32:44.93 | +16:48:35.8 | AB | A | … | … | … | 354.79 | … | 0.401 ± 0.032 |
| 13328+1649 | VYS 6 | GJ 516 B | … | M4.0 Ve | 13:32:45.08 | +16:48:37.4 | … | B | 2.713 | 54.0 | 43.52 | 376.03 | … | 0.361 ± 0.033 |
| … | … | PM J13335+7029 | J13335+704 | M3.5 V | 13:33:33.27 | +70:29:41.6 | … | … | … | … | 43.32 | 132.93 | 123.10 ± 0.45 | 0.334 ± 0.015 |
| … | … | Wolf 1487 | J13343+046 | M0.0 V | 13:34:21.67 | +04:40:00.7 | … | … | … | … | 42.28 | 197.24 | 953.40 ± 7.14 | 0.683 ± 0.026 |
| … | … | GJ 3793 | J13348+201 | M3.5 V | 13:34:49.40 | +20:11:35.7 | … | … | … | … | 41.85 | 191.37 | 625.12 ± 25.37 | 0.531 ± 0.029 |
| 13349+7430 | LDS1775 | GJ 9453 | … | K5 V | 13:34:51.91 | +74:30:01.1 | A+B | A | 14.800 | 323.3 | 69.84 | 439.09 | 2438.88 ± 15.15 | 0.700 ± 0.105 |
| … | … | GJ 9453 B | J13348+745 | M3.5 V | 13:34:49.83 | +74:30:12.6 | … | B | 14.192 | 324.0 | 44.55 | 437.06 | 331.23 ± 1.72 | 0.444 ± 0.031 |
| … | … | G 150-17 | J13358+146 | M2.5 V | 13:35:50.73 | +14:41:07.4 | … | … | … | … | 70.57 | 275.10 | 324.75 ± 1.75 | 0.453 ± 0.030 |
| … | … | Ross 1021 | J13369+229 | M2.5 V | 13:36:55.35 | +22:57:58.2 | … | … | … | … | 49.51 | 198.31 | 206.46 ± 1.04 | 0.413 ± 0.017 |
| … | … | GJ 520 A | J13378+481 | M0.0 V | 13:37:50.84 | +48:08:14.8 | AB+C | A | … | … | 48.03 | 277.90 | … | 0.663 ± 0.026 |
| 13379+4808 | ES 608 | GJ 520 B | … | M1.5 V | 13:37:50.77 | +48:08:16.3 | … | B | 1.621 | 335.4 | 123.78 | 244.96 | … | 0.541 ± 0.029 |
| 13379+4808 | LDS5784 | GJ 520 C | J13376+481 | M4.0 V | 13:37:40.09 | +48:07:52.0 | … | C | 109.969 | 258.0 | 84.22 | 257.67 | 54.87 ± 0.21 | 0.230 ± 0.012 |
| … | … | Ross 1022 | J13386+258 | M3.0 V | 13:38:36.37 | +25:49:50.7 | … | … | … | … | 37.22 | 583.81 | 242.76 ± 1.86 | 0.450 ± 0.018 |
| … | … | 1R133841.3-113137 | J13386-115 | M4.5 V | 13:38:41.05 | -11:32:09.1 | … | … | … | … | 29.57 | 172.73 | 69.46 ± 0.39 | 0.262 ± 0.013 |
| … | … | Ross 488 | J13388-022 | M2.0 V | 13:38:53.11 | -02:15:48.5 | … | … | … | … | 45.81 | 311.57 | 187.75 ± 1.05 | 0.369 ± 0.015 |
| … | … | Ross 1026 | J13401+437 | M3.5 V | 13:40:07.19 | +43:46:42.8 | … | … | … | … | … | 1125.83 | 159.71 ± 1.13 | 0.384 ± 0.016 |
| … | … | PM J13413-0907 | J13413-091 | M2.5 V | 13:41:21.33 | -09:07:16.4 | … | … | … | … | 56.79 | 103.29 | 149.97 ± 0.83 | 0.348 ± 0.015 |
| 13415+4855 | RAO 298 | StM 186 | J13414+489 | M3.5 V | 13:41:27.84 | +48:54:43.4 | (AB) | AB | 0.355 | 351.9 | 97.91 | 193.13 | … | … |
| 13394+4611 | JOD 6 | GJ 521 | J13394+461 | M1.5 V+ | 13:39:24.04 | +46:11:17.6 | (AB) | AB | 0.447 | 15.8 | 29.94 | 391.46 | … | 0.525 ± 0.029 |
| … | … | GJ 3799 | J13415+148 | M1.5 V | 13:41:31.58 | +14:49:27.6 | … | … | … | … | 28.89 | 258.55 | 512.12 ± 10.23 | 0.549 ± 0.028 |
| … | … | StM 187 | J13417+582 | M3.5 Ve | 13:41:46.48 | +58:15:18.9 | AB | A | … | … | 29.06 | 101.05 | … | 0.358 ± 0.033 |
| 13418+5815 | JNN 94 | G3- 1658968054798747008 | … | M4.0 V | 13:41:46.39 | +58:15:18.6 | … | B | 0.762 | 252.9 | 56.11 | 81.41 | … | 0.269 ± 0.036 |
| … | … | Ross 1015 | J13427+332 | M3.5 V | 13:42:43.13 | +33:17:12.9 | … | … | … | … | 49.55 | 721.13 | 91.12 ± 0.42 | 0.261 ± 0.010 |
| … | … | GJ 3802 | J13430+090 | M3.0 V | 13:43:00.91 | +09:04:21.8 | … | … | … | … | 55.08 | 352.29 | 159.05 ± 3.65 | 0.383 ± 0.017 |
| … | … | TYC 896-760-1 | J13434+111 | M0.5 V | 13:43:25.03 | +11:06:42.1 | … | … | … | … | 52.62 | 147.97 | 795.57 ± 5.07 | 0.633 ± 0.027 |
| 13422-1600 | WSI 114 | GJ 3800 | J13421-160 | M4e | 13:42:09.28 | -16:00:24.1 | (AB) | AB | 0.737 | 17.7 | 117.03 | 497.80 | … | 0.289 ± 0.035 |
| 13446+5142 | KUI64 | Ross 492 | J13444+516 | M2.5 V | 13:44:27.00 | +51:41:08.6 | (AB) | AB | 1.050 | 51.1 | 24.59 | 760.00 | … | 0.343 ± 0.048 |
| … | … | LP 379-98 A | J13445+249 | M1.0 V | 13:44:33.38 | +24:57:03.6 | AB | A | … | … | 86.90 | 262.46 | … | 0.599 ± 0.027 |
| 13446+2457 | RAO 301 | LP 379-98 B | … | M1.0 V | 13:44:33.38 | +24:57:04.5 | … | B | 0.887 | 356.4 | 85.54 | 256.64 | … | 0.537 ± 0.029 |
| … | … | Wolf 497 | J13450+176 | M0.0 V | 13:45:05.59 | +17:46:38.2 | … | … | … | … | 32.66 | 1887.91 | 451.52 ± 1.60 | 0.486 ± 0.012 |
| … | … | MCC 699 | J13455+609 | M0.5 V | 13:45:31.28 | +60:58:58.3 | … | … | … | … | 37.86 | 60.61 | 1002.74 ± 4.52 | 0.682 ± 0.026 |
| … | … | HD 119850 | J13457+148 | M1.5 V | 13:45:45.74 | +14:53:06.2 | … | … | … | … | 28.88 | 2296.01 | 378.79 ± 7.77 | 0.498 ± 0.014 |
| … | … | GJ 3804 | J13458-179 | dM3.5 | 13:45:50.37 | -17:58:14.5 | … | … | … | … | 29.10 | 632.04 | 140.59 ± 0.88 | 0.328 ± 0.011 |
| … | … | BD+22 2632A | J13477+214 | M0.0 V | 13:47:42.49 | +21:27:36.3 | AB | A | … | … | 36.49 | 119.73 | … | 0.619 ± 0.027 |

Table D.2: Complete sample with the description of multiple systems (continued).

| WDS id | WDS disc | Name | Karmn | Spectral type | $\alpha$ (2016.0) | $\delta$ (2016.0) | System | Component | $\rho$ [arcsec] | $\theta$ [deg] | $\varpi$ [mas] | $\mu_{\rm total}$ [mas a$^{-1}$] | $\mathcal{L}$ [$10^{-4}$ $\mathcal{L}_\odot$] | $\mathcal{M}$ [$\mathcal{M}_\odot$] |
|---|---|---|---|---|---|---|---|---|---|---|---|---|---|---|
| 13477+2128 | HDS1939 | BD+22 2632B | … | M3.5 V | 13:47:42.54 | +21:27:35.2 | … | B | 1.303 | 150.8 | 36.07 | 119.20 | … | 0.342 ± 0.033 |
| … | … | LP 738-14 | J13481-137 | M4.5 V | 13:48:06.52 | -13:44:39.8 | AB | A | … | … | 70.90 | 858.15 | … | 0.183 ± 0.040 |
| 13481-1345 | DEA1 | LP 738-14 B | … | T5.5 | … | … | … | B | 67.340 | 291.5 | 38.24 | 857.92 | … | … |
| … | … | GJ 1179 A | J13482+236 | M5.0 V | 13:48:11.69 | +23:36:50.7 | A+B | A | … | … | 35.92 | 1487.63 | 16.78 ± 0.09 | 0.136 ± 0.010 |
| 13484+2337 | LDS4410 | GJ 1179 B | … | DC9 | 13:48:01.28 | +23:34:48.4 | … | B | 188.131 | 229.5 | 56.13 | 1490.91 | … | 0.500 ± 0.100 |
| … | … | Ross 493 | J13485+563 | M1.5 V | 13:48:34.32 | +56:20:09.6 | … | … | … | … | 82.04 | 366.09 | 586.60 ± 2.63 | 0.570 ± 0.028 |
| … | … | Wolf 1494 | J13488+041 | M4.5 V | 13:48:48.61 | +04:05:59.4 | … | … | … | … | 53.55 | 180.33 | 44.88 ± 0.22 | 0.221 ± 0.012 |
| … | … | Wolf 1495 A | J13490+026 | M1.5 V | 13:49:01.16 | +02:47:23.6 | AB | A | … | … | 40.69 | 365.48 | … | 0.420 ± 0.031 |
| 13490+0247 | RAO 302 | Wolf 1495 B | … | … | 13:49:01.19 | +02:47:23.2 | … | B | 0.679 | 133.1 | 27.74 | 365.22 | … | … |
| … | … | GJ 3810 | J13507-216 | M3.0 V | 13:50:43.96 | -21:41:33.0 | A+B | A | … | … | 45.42 | 379.05 | 181.12 ± 1.23 | 0.385 ± 0.016 |
| 13507-2140 | LDS 461 | LP 798-41 | J13503-216 | M3.5 V | 13:50:23.73 | -21:37:25.9 | … | B | 375.007 | 311.2 | 51.29 | 378.00 | 106.82 ± 0.64 | 0.331 ± 0.015 |
| … | … | Ross 1019 | J13508+367 | M4.0 V | 13:50:51.18 | +36:44:18.5 | … | … | … | … | 77.90 | 447.25 | 76.46 ± 0.33 | 0.276 ± 0.013 |
| … | … | RX J1351.8+1247 | J13518+127 | M2.0 Ve | 13:51:53.02 | +12:47:07.0 | … | … | … | … | 24.41 | 91.68 | 337.77 ± 1.60 | 0.475 ± 0.018 |
| … | … | Wolf 515 | J13526+144 | M2.0 V | 13:52:36.26 | +14:25:15.7 | A+B | A | … | … | 24.34 | 284.44 | … | 0.517 ± 0.029 |
| 13526+1425 | JOD 7 | Wolf 515 B | … | M3.5 V | 13:52:36.26 | +14:25:16.9 | … | B | 1.204 | 359.8 | 64.80 | 294.25 | … | 0.304 ± 0.034 |
| … | … | GJ 533.1 | J13528+656 | M1.5 V | 13:52:48.61 | +65:37:17.7 | (AB) | AB | … | … | 44.80 | 562.38 | … | 0.514 ± 0.029 |
| … | … | GJ 3815 | J13528+668 | M5.0 V | 13:52:49.25 | +66:48:57.2 | … | … | … | … | 61.00 | 723.77 | 21.07 ± 0.08 | 0.155 ± 0.010 |
| … | … | LP 21-224 | J13536+776 | M4.0 Ve | 13:53:39.91 | +77:37:07.9 | … | … | … | … | 91.02 | 224.76 | 86.22 ± 0.31 | 0.295 ± 0.019 |
| … | … | LP 97-259 | J13529+536 | M1.0 V | 13:52:55.76 | +56:36:17.0 | … | … | … | … | 90.95 | 211.10 | 472.19 ± 1.89 | 0.544 ± 0.028 |
| 13535+1257 | BEU18 | Ross 835 | J13534+129 | M0.0 V | 13:53:27.37 | +12:56:22.9 | (AB) | AB | 0.278 | 9.0 | 35.96 | 628.21 | … | … |
| … | … | PM J13537+5210A | J13537+521 | M3.5 V | 13:53:45.89 | +52:10:27.3 | AB | A | … | … | 42.23 | 134.34 | … | 0.441 ± 0.031 |
| 13538+5210 | JNN 96 | PM J13537+5210B | … | M2.5 V | 13:53:45.88 | +52:10:28.3 | … | B | 1.027 | 351.7 | 74.45 | 130.52 | … | 0.439 ± 0.031 |
| … | … | GJ 534.2 | J13537+788 | M0.0 V | 13:53:45.76 | +78:51:08.7 | … | … | … | … | 74.45 | 277.86 | 750.31 ± 2.89 | 0.628 ± 0.027 |
| … | … | Ross 837 | J13582+125 | M3.5 Ve | 13:58:13.56 | +12:34:55.4 | … | … | … | … | 39.53 | 792.17 | 63.69 ± 0.37 | 0.213 ± 0.009 |
| … | … | LP 739-2 | J13582-120 | M4.5 V | 13:58:15.80 | -12:02:58.4 | … | … | … | … | 39.28 | 340.49 | 49.75 ± 0.24 | 0.234 ± 0.012 |
| … | … | LP 739-3 | J13583-132 | M4.0 V | 13:58:19.96 | -13:16:26.0 | … | … | … | … | 72.97 | 348.28 | 53.73 ± 0.26 | 0.228 ± 0.012 |
| … | … | GJ 3818 | J13587-000 | M4.0 V | 13:58:43.20 | -00:04:54.3 | … | … | … | … | 73.92 | 543.64 | 103.66 ± 0.53 | 0.305 ± 0.014 |
| … | … | GJ 3820 | J13591-198 | M4.5 V | 13:59:09.78 | -19:50:06.6 | … | … | … | … | 109.98 | 590.77 | 75.29 ± 0.62 | 0.263 ± 0.010 |
| … | … | HD 122303 | J14010-026 | M1.0 V | 14:01:02.31 | -02:39:07.9 | … | … | … | … | 109.98 | 1019.56 | 433.61 ± 2.26 | 0.505 ± 0.012 |
| … | … | GJ 536.1 A | J14019+154 | M0.0 V | 14:01:58.86 | +15:29:40.7 | AB | A | … | … | 59.31 | 105.24 | … | 0.593 ± 0.028 |
| 14019+1530 | ALD 112 | GJ 536.1 B | … | M0.5 V | 14:01:58.89 | +15:29:39.1 | … | B | 1.621 | 162.3 | 60.03 | 118.04 | … | 0.588 ± 0.028 |
| … | … | PM J14019+4316A | J14019+432 | M2.5 V | 14:01:58.67 | +43:16:41.1 | AB | A | … | … | 36.00 | 82.70 | … | 0.389 ± 0.032 |
| 14020+4317 | KPP3910 | PM J14019+4316B | … | M3.0 V | 14:01:58.69 | +43:16:43.1 | … | B | 1.960 | 7.1 | 43.49 | 83.90 | … | 0.382 ± 0.032 |
| … | … | GJ 3822 | J14023+136 | M0.5 V | 14:02:19.73 | +13:41:20.4 | … | … | … | … | 72.73 | 171.99 | 648.18 ± 2.71 | 0.594 ± 0.028 |
| … | … | GJ 3821 | J14024-210 | M3.5 V | 14:02:29.43 | -21:00:42.9 | … | … | … | … | 55.21 | 618.88 | 126.58 ± 0.86 | 0.339 ± 0.015 |
| … | … | GJ 537 A | J14025+463S | M0.5 V | 14:02:34.03 | +46:20:23.0 | A+B | A | … | … | 55.19 | 576.57 | … | 0.524 ± 0.029 |
| 14024+4620 | SWI 1 | GJ 537 B | J14025+463N | M0.5 V | 14:02:34.19 | +46:20:26.4 | … | B | 3.742 | 25.3 | 31.83 | 583.81 | … | 0.506 ± 0.029 |
| … | … | LSPM J1403+2440 | J14039+242 | M2.5 V | 14:03:54.74 | +24:40:44.2 | … | … | … | … | 40.56 | 161.57 | 286.52 ± 1.25 | 0.462 ± 0.018 |
| … | … | BD+21 2602 | … | K4 V | 14:04:09.85 | +20:45:32.5 | A+BC | A | … | … | 27.87 | 127.65 | 1762.73 ± 6.48 | 0.730 ± 0.110 |
| 14042+2046 | J 1128 | StKM 1-1119 | J14041+207 | M1.0 Ve | 14:04:09.06 | +20:44:30.9 | … | B | 62.660 | 190.2 | 29.63 | 132.08 | … | … |

Table D.2: Complete sample with the description of multiple systems (continued).

| WDS id | WDS disc | Name | Karmn | Spectral type | $\alpha$ (2016.0) | $\delta$ (2016.0) | System | Component | $\rho$ [arcsec] | $\theta$ [deg] | $\varpi$ [mas] | $\mu_{total}$ [mas a$^{-1}$] | $\mathcal{L}$ [$10^{-4}$ $\mathcal{L}_\odot$] | $\mathcal{M}$ [$\mathcal{M}_\odot$] |
|---|---|---|---|---|---|---|---|---|---|---|---|---|---|---|
| 14042+2046 | J 1128 | G3- 1247168140942467200 | … | … | 14:04:09.06 | +20:44:31.2 | … | C | 62.307 | 190.3 | 31.03 | … | … | … |
| … | … | NLTT 36313 | J14062+693 | M3.0 V | 14:06:14.66 | +69:18:38.8 | … | … | … | … | 14.16 | 199.13 | 310.28 ± 1.42 | 0.452 ± 0.030 |
| … | … | GJ 540 | J14082+805 | M1.0 V | 14:08:14.37 | +80:35:41.5 | … | … | … | … | 22.00 | 583.70 | 646.17 ± 2.90 | 0.586 ± 0.028 |
| … | … | GJ 3826 | J14083+758 | M1.5 V | 14:08:20.91 | +75:51:13.2 | … | … | … | … | 21.87 | 546.27 | 385.66 ± 5.43 | 0.509 ± 0.019 |
| 14121-0035 | WSI 129 | GJ 3828 A | J14121-005 | M2.5 V | 14:12:10.22 | -00:35:00.3 | (AB)+C | AB | 0.529 | 170.5 | 22.03 | 750.90 | … | 0.361 ± 0.033 |
| 14121-0035 | LDS4442 | GJ 3828 B | … | M6.5 V | 14:12:11.36 | -00:35:12.6 | … | C | 21.040 | 125.9 | 24.70 | 752.91 | 4.85 ± 0.06 | 0.100 ± 0.050 |
| … | … | GQVir | J14130-120 | M4.5 V | 14:13:04.19 | -12:01:32.9 | Aab | Aab(2) | … | … | 25.40 | 725.08 | … | … |
| 14144-1521 | CVN 25 | HD 124498A | … | K7 V | 14:14:21.24 | -15:21:24.8 | AB+C | A | 0.306 | 208.4 | … | 209.20 | … | 0.733 ± 0.026 |
| 14144-1521 | RST3869 | HD 124498B | … | M2.0 V | 14:14:21.34 | -15:21:24.4 | … | B | 1.553 | 78.4 | 24.10 | 229.25 | … | 0.471 ± 0.030 |
| 14144-1521 | LDS 483 | GJ 3832 | J14142-153 | M3.5 V | 14:14:16.87 | -15:21:15.9 | … | C | 63.743 | 278.0 | 75.88 | 231.00 | 150.21 ± 0.89 | 0.397 ± 0.017 |
| … | … | GJ 3834 | J14144+234 | M3.5 V | 14:14:26.33 | +23:27:26.4 | … | … | … | … | 75.60 | 492.02 | 162.69 ± 0.78 | 0.388 ± 0.016 |
| … | … | Ross 992 | J14152+450 | M3.0 V | 14:15:15.96 | +45:00:49.7 | … | … | … | … | 52.32 | 710.71 | 253.74 ± 0.91 | 0.429 ± 0.011 |
| … | … | LP 439-350 | J14153+153 | M2.0 V | 14:15:20.63 | +15:23:01.5 | … | … | … | … | 41.45 | 186.82 | 291.05 ± 1.87 | 0.466 ± 0.018 |
| … | … | GJ 1182 | J14155+046 | M5.69 | 14:15:31.75 | +04:39:19.1 | Aab | Aab(2) | … | … | 25.37 | 1070.70 | … | 0.258 ± 0.002 |
| … | … | LP 97-674 A | J14157+594 | M2.2 V | 14:15:42.13 | +59:27:29.1 | A+B | A | … | … | 131.10 | 184.40 | 360.20 ± 1.68 | 0.491 ± 0.018 |
| 14157+5928 | LDS2707 | LP 97-674 B | … | M3.0 V | 14:15:41.61 | +59:27:25.9 | … | B | 5.061 | 231.1 | 62.15 | 189.23 | … | 0.413 ± 0.031 |
| … | … | G 165-58 | J14159+362 | M3.5 V | 14:15:56.39 | +36:16:42.2 | … | … | … | … | 62.28 | 303.35 | 71.03 ± 0.26 | 0.248 ± 0.012 |
| … | … | LP 81-30 | J14161+233 | M1.0 V | 14:16:11.25 | +23:23:26.5 | … | … | … | … | 66.83 | 203.00 | 387.93 ± 1.77 | 0.509 ± 0.029 |
| … | … | GJ 3838 | J14170+105 | M1.5 V | 14:17:04.56 | +10:35:34.3 | … | … | … | … | 51.65 | 301.67 | 359.45 ± 2.81 | 0.487 ± 0.030 |
| 14170+3143 | DEL 5 | GJ 3839 | J14170+317 | M4.5 Ve | 14:17:02.14 | +31:42:44.7 | (AB) | AB(2/3) | 0.247 | 157.5 | 25.35 | 606.11 | … | … |
| … | … | PM J14171+0851 | J14171+088 | M4.5 V | 14:17:07.18 | +08:51:37.1 | Aab | Aab(2) | … | … | 39.77 | 139.55 | … | … |
| … | … | GJ 541.2 | J14174+454 | M0.0 V | 14:17:24.45 | +45:26:39.8 | A+B | A | … | … | 54.69 | 47.49 | 711.76 ± 2.58 | 0.625 ± 0.027 |
| 14174+4527 | VVO 12 | RX J1417.3+4525 | J14173+454 | dM5.0 | 14:17:22.17 | +45:25:45.7 | … | B | 59.202 | 203.9 | 31.24 | 47.89 | 76.76 ± 0.58 | 0.273 ± 0.010 |
| … | … | RX J1417.5+0233 | J14175+025 | M3.0 V | 14:17:30.18 | +02:33:42.5 | … | … | … | … | … | 70.45 | 271.08 ± 2.02 | 0.477 ± 0.019 |
| … | … | LP 381-94 | J14177+214 | M1.0 V | 14:17:47.76 | +21:25:58.8 | … | … | … | … | 44.58 | 185.25 | 1112.43 ± 14.00 | 0.686 ± 0.026 |
| … | … | GJ 3840 | J14179-005 | M2.5 V | 14:17:58.75 | -00:31:33.8 | … | … | … | … | 60.10 | 391.98 | 161.51 ± 0.97 | 0.363 ± 0.015 |
| … | … | LP 220-78 | J14189+386 | M1.0 V | 14:18:58.09 | +38:38:22.5 | … | … | … | … | 59.99 | 737.76 | 651.44 ± 3.70 | 0.598 ± 0.027 |
| … | … | Wolf 534 | J14191-073 | M3.0 V | 14:19:09.81 | -07:18:24.0 | … | … | … | … | 48.98 | 1351.08 | 62.01 ± 0.27 | 0.215 ± 0.011 |
| … | … | LP 560-1 | J14194+029 | M5.0 V | 14:19:29.37 | +02:54:34.1 | … | … | … | … | 48.97 | 246.22 | 60.87 ± 0.31 | 0.261 ± 0.013 |
| … | … | IZBoo | J14200+390 | M3.0 Ve | 14:20:04.66 | +39:03:02.5 | … | … | … | … | 17.49 | 71.53 | 685.27 ± 3.52 | 0.565 ± 0.028 |
| … | … | Ross 848 | J14201-096 | M4.0 V | 14:20:06.71 | -09:37:26.8 | … | … | … | … | 25.39 | 1036.46 | 85.27 ± 0.63 | 0.274 ± 0.013 |
| … | … | GJ 3843 | J14212-011 | M4.0 V | 14:21:15.31 | -01:07:29.8 | … | … | … | … | 25.49 | 648.81 | 67.32 ± 0.37 | 0.258 ± 0.013 |
| … | … | PM J14215-0755 | J14215-079 | M4.0 V | 14:21:33.96 | -07:55:18.0 | … | … | … | … | 39.74 | 130.49 | 128.13 ± 1.02 | 0.341 ± 0.015 |
| … | … | LP 270-68 | J14219+376 | M1.5 V | 14:21:55.61 | +37:39:45.5 | … | … | … | … | 42.55 | 469.43 | 374.34 ± 1.55 | 0.501 ± 0.019 |
| … | … | LP 440-13 | J14227+164 | M5.0 V | 14:22:43.19 | +16:24:47.2 | … | … | … | … | 47.21 | 195.81 | 113.14 ± 0.65 | 0.341 ± 0.015 |
| … | … | GJ 3845 | J14231-222 | M4.5 V | 14:23:07.49 | -22:17:16.7 | … | … | … | … | 47.21 | 545.98 | 55.75 ± 0.29 | 0.232 ± 0.012 |
| … | … | GJ 3846 | J14249+088 | M3.0 V | 14:24:56.58 | +08:53:18.0 | … | … | … | … | 47.19 | 570.88 | 108.73 ± 0.69 | 0.293 ± 0.013 |
| 14211+2736 | CRC 69 | GJ 3844 | J14210+275 | M2.5 V | 14:21:03.13 | +27:35:34.5 | (AB) | AB | 0.677 | 85.1 | 43.83 | 301.67 | … | 0.468 ± 0.030 |
| 14252+5151 | STT 580 | HD 126660 | … | F7 V | 14:25:11.39 | +51:50:56.3 | A+B | A | 70.160 | 182.5 | 51.10 | 464.15 | … | 1.210 ± 0.182 |
| … | … | HD 126660B | J14251+518 | M2.5 V | 14:25:11.17 | +51:49:46.6 | … | B | 69.674 | 181.7 | 55.39 | 469.65 | 222.34 ± 0.94 | 0.392 ± 0.011 |

Table D.2: Complete sample with the description of multiple systems (continued).

| WDS id | WDS disc | Name | Karmn | Spectral type | $\alpha$ (2016.0) | $\delta$ (2016.0) | System | Component | $\rho$ [arcsec] | $\theta$ [deg] | $\varpi$ [mas] | $\mu_{\rm total}$ [mas a$^{-1}$] | $\mathcal{L}$ [$10^{-4}\ \mathcal{L}_\odot$] | $\mathcal{M}$ [$\mathcal{M}_\odot$] |
|---|---|---|---|---|---|---|---|---|---|---|---|---|---|---|
| … | … | LP 740-10 | J14255-118 | M4.0 V | 14:25:33.79 | -11:48:51.5 | … | … | … | … | 57.92 | 296.82 | 272.54 ± 2.93 | 0.509 ± 0.020 |
| 14257+2338 | BU 1442 | GJ 548 A | J14257+236W | M0.0 V | 14:25:44.40 | +23:36:43.6 | A+B | A | 45.200 | 74.4 | 40.07 | 1369.12 | 847.75 ± 4.67 | 0.648 ± 0.027 |
| … | … | GJ 548 B | J14257+236E | M0.5 V | 14:25:47.58 | +23:36:55.8 | … | B | 45.350 | 74.4 | … | 1371.68 | 733.70 ± 4.01 | 0.618 ± 0.027 |
| … | … | V358Boo | J14259+142 | M0.0 Ve | 14:25:55.87 | +14:12:09.6 | … | … | … | … | 74.80 | 66.90 | … | 0.622 ± 0.093 |
| … | … | LSPM J1426+2408 | J14269+241 | M1.0 V | 14:26:58.63 | +24:08:56.9 | … | … | … | … | 30.17 | 160.22 | 526.05 ± 3.24 | 0.555 ± 0.028 |
| … | … | GJ 1183 A | J14279-003S | M4.65 | 14:27:55.69 | -00:22:30.5 | A+B | A | … | … | 46.25 | 363.55 | 80.53 ± 0.49 | 0.284 ± 0.013 |
| 14279-0032 | GIC 120 | GJ 1183 B | J14279-003N | M4.70 | 14:27:56.01 | -00:22:18.3 | … | B | 13.073 | 21.8 | 45.89 | 366.95 | 76.86 ± 0.49 | 0.297 ± 0.014 |
| … | … | LP 500-35 | J14280+139 | M7.0 V | 14:28:03.75 | +13:56:05.4 | … | … | … | … | 108.79 | 615.68 | 8.32 ± 0.04 | 0.115 ± 0.010 |
| … | … | LP 560-27 | J14283+053 | M3.0 V | 14:28:21.12 | +05:19:00.3 | A+B | A | … | … | 46.01 | 377.91 | 177.07 ± 1.08 | 0.381 ± 0.016 |
| 14284+0520 | LDS 961 | LP 560-26 | J14282+053 | M3.5 V | 14:28:17.18 | +05:18:44.8 | … | B | 60.904 | 255.2 | 30.66 | 378.25 | 136.89 ± 1.44 | 0.378 ± 0.016 |
| … | … | Ross 130 | J14294+155 | M2.0 V | 14:29:28.53 | +15:32:18.3 | … | … | … | … | 49.46 | 1673.55 | 397.97 ± 1.96 | 0.500 ± 0.012 |
| … | … | GJ 3853 | J14299+295 | M4.0 V | 14:29:59.26 | +29:33:54.9 | … | … | … | … | 41.10 | 522.71 | 85.23 ± 0.34 | 0.293 ± 0.014 |
| … | … | GJ 3855 | J14306+597 | M6.5 V | 14:30:36.08 | +59:43:27.4 | … | … | … | … | 22.11 | 823.39 | 5.77 ± 0.03 | 0.101 ± 0.009 |
| … | … | HD 127339 | J14307-086 | M0.5 V | 14:30:46.35 | -08:38:50.6 | … | … | … | … | 22.01 | 1293.18 | 965.25 ± 4.78 | 0.682 ± 0.026 |
| … | … | Wolf 1478 | J14310-122 | dM3.5 | 14:31:00.72 | -12:17:52.3 | … | … | … | … | 77.44 | 570.53 | 124.83 ± 0.83 | 0.306 ± 0.011 |
| … | … | LSPM J1431+7526 | J14312+754 | M4.0 V | 14:31:12.59 | +75:26:41.7 | … | … | … | … | 30.93 | 201.28 | 34.49 ± 0.12 | 0.178 ± 0.010 |
| … | … | G 255-55 | J14320+738 | M2.0 V | 14:32:01.99 | +73:49:23.3 | … | … | … | … | 184.00 | 219.96 | 472.73 ± 1.98 | 0.525 ± 0.029 |
| … | … | LP 560-35 | J14321+081 | M6.0 V | 14:32:07.99 | +08:11:31.4 | … | … | … | … | 89.33 | 477.20 | 18.32 ± 0.09 | 0.127 ± 0.011 |
| … | … | GJ 3856 | J14321+160 | M5.0 V | 14:32:11.00 | +16:00:48.2 | … | … | … | … | 33.42 | 191.86 | 66.54 ± 0.28 | 0.256 ± 0.013 |
| … | … | GJ 3858 | J14322+496 | M3.5 V | 14:32:13.58 | +49:39:04.2 | … | … | … | … | 33.34 | 592.90 | 91.46 ± 0.58 | 0.285 ± 0.013 |
| … | … | G 224-13A | J14331+610 | M2.5 V | 14:33:06.93 | +61:00:43.9 | AB | A | … | … | 55.03 | 207.18 | … | 0.413 ± 0.031 |
| 14331+6101 | CRC 70 | G 224-13B | … | M3.0 V | 14:33:06.80 | +61:00:43.6 | … | B | 0.971 | 251.5 | 59.50 | 206.02 | … | 0.376 ± 0.032 |
| … | … | HNLib | J14342-125 | M3.5 V | 14:34:16.42 | -12:31:00.9 | … | … | … | … | 84.22 | 691.25 | 100.75 ± 0.69 | 0.291 ± 0.013 |
| … | … | StKM 1-1170 | J14366+143 | M1.0 V | 14:36:38.98 | +14:21:52.6 | … | … | … | … | 84.31 | 60.77 | 1283.35 ± 7.68 | 0.709 ± 0.026 |
| … | … | GJ 3861 | J14368+583 | M3.0 Ve | 14:36:54.24 | +58:20:43.6 | Aab | Aab(2) | … | … | 39.36 | 937.75 | … | … |
| … | … | LSPM J1437+7536N | J14371+756 | M2.0 V | 14:37:09.94 | +75:36:54.7 | A+B | A | … | … | 31.32 | 219.27 | 167.49 ± 0.55 | 0.347 ± 0.014 |
| 14372+7537 | LDS1803 | LSPM J1437+7536S | … | M4.0 V | 14:37:13.27 | +75:36:41.5 | … | B | 18.130 | 136.7 | 61.48 | 214.13 | 112.68 ± 0.42 | 0.299 ± 0.013 |
| … | … | G 239-22 | J14376+677 | M1.5 V | 14:37:39.32 | +67:45:34.9 | (AB) | AB | … | … | 54.93 | 281.05 | … | 0.567 ± 0.028 |
| … | … | GPM 219.718548+42.229288 | J14388+422 | M1.5 V | 14:38:51.66 | +42:13:43.9 | … | … | … | … | … | 134.84 | 547.06 ± 5.10 | 0.506 ± 0.029 |
| … | … | LP 560-66 | J14415+064 | M1.5 V | 14:41:32.72 | +06:27:41.2 | … | … | … | … | 47.43 | 439.95 | 498.76 ± 2.83 | 0.546 ± 0.028 |
| … | … | GJ 9492 | J14423+660 | M2.0 V | 14:42:20.79 | +66:03:20.2 | A+B | A | … | … | 47.41 | 301.51 | … | 0.412 ± 0.031 |
| 14424+6603 | GKI 4 | GJ 9492 B | … | L0 | 14:42:21.17 | +66:03:20.1 | … | B | 2.290 | 93.6 | 59.18 | 337.31 | … | 0.104 ± 0.052 |
| … | … | NLTT 38291 | J14438+667 | M1.0 V | 14:43:50.67 | +66:44:34.6 | … | … | … | … | 37.51 | 251.26 | 652.89 ± 2.54 | 0.595 ± 0.028 |
| … | … | RX J1447.2+5701 | J14472+570 | M4.0 V | 14:47:13.67 | +57:01:54.4 | … | … | … | … | 40.66 | 87.07 | … | … |
| … | … | LP 501-17 | J14485+101 | M3.5 V | 14:48:32.79 | +10:06:55.7 | … | … | … | … | 40.49 | 361.92 | 286.77 ± 1.97 | 0.491 ± 0.019 |
| … | … | LP 326-34 | J14501+323 | M3.5 V | 14:50:11.14 | +32:18:13.8 | … | … | … | … | 36.18 | 193.39 | 314.21 ± 1.46 | 0.444 ± 0.031 |
| … | … | LP 326-38 | J14511+311 | M4.0 Ve | 14:51:09.97 | +31:06:37.4 | … | … | … | … | 61.47 | 389.31 | 103.92 ± 51.81 | 0.305 ± 0.083 |
| … | … | GJ 3871 | J14524+123 | M2.5 V | 14:52:28.47 | +12:23:29.2 | … | … | … | … | 75.48 | 235.83 | 420.46 ± 1.77 | 0.502 ± 0.013 |
| … | … | Wolf 555 | J14525+001 | M2.5 V | 14:52:32.27 | +00:10:02.8 | … | … | … | … | 34.76 | 352.08 | 446.84 ± 7.13 | 0.506 ± 0.029 |
| 14470+1705 | RAO 320 | Ross 994 | J14469+170 | M1.5 V | 14:46:59.22 | +17:05:07.5 | (AB) | AB | 1.096 | 185.4 | 45.62 | 548.58 | … | 0.563 ± 0.028 |

Table D.2: Complete sample with the description of multiple systems (continued).

| WDS id | WDS disc | Name | Karmn | Spectral type | $\alpha$ (2016.0) | $\delta$ (2016.0) | System | Component | $\rho$ [arcsec] | $\theta$ [deg] | $\varpi$ [mas] | $\mu_{total}$ [mas a$^{-1}$] | $\mathcal{L}$ [$10^{-4}$ $\mathcal{L}_\odot$] | $\mathcal{M}$ [$\mathcal{M}_\odot$] |
|---|---|---|---|---|---|---|---|---|---|---|---|---|---|---|
| … | … | Ross 52A | J14538+235 | M3.5 V | 14:53:50.59 | +23:33:22.5 | AB | A | … | … | 21.17 | 697.94 | … | 0.327 ± 0.034 |
| 14540+2335 | REU 2 | Ross 52B | … | M4.5 V | 14:53:50.66 | +23:33:22.5 | … | B | 0.880 | 89.2 | 20.92 | 761.38 | … | 0.238 ± 0.037 |
| 14545+1606 | FRT 1 | CEBoo | J14544+161 | M1.0: V | 14:54:29.55 | +16:06:01.9 | ABab | A | 4.917 | 40.5 | 44.08 | 303.00 | … | 0.477 ± 0.030 |
| 14545+1606 | MEL 2 | GJ 569 B | … | M8.5 V | 14:54:29.77 | +16:06:05.6 | … | Bab | 4.908 | 40.8 | 53.91 | 367.41 | … | 0.125 ± 0.016 |
| … | … | Ross 1041 | J14544+355 | M3.5 V | 14:54:28.10 | +35:32:43.5 | … | … | … | … | 106.84 | 854.39 | 155.28 ± 0.80 | 0.348 ± 0.012 |
| … | … | Ross 1028b | J14548+099 | M2.0 V | 14:54:53.15 | +09:56:30.1 | … | … | … | … | 57.79 | 508.62 | 519.13 ± 3.76 | 0.563 ± 0.028 |
| … | … | GJ 3875 | J14549+411 | M4.5 V | 14:54:54.61 | +41:08:50.5 | Aab | Aab | … | … | 53.23 | 272.51 | … | … |
| … | … | G 66-42 | J14557+072 | M0.5 V | 14:55:47.82 | +07:17:47.7 | … | … | … | … | 64.01 | 300.42 | 1152.71 ± 10.90 | 0.706 ± 0.026 |
| … | … | G 136-35 | J14564+168 | M1.5 V | 14:56:28.16 | +16:48:29.1 | Aab | Aab | … | … | 68.93 | 324.13 | … | … |
| … | … | KXLib | … | K4 V | 14:57:29.18 | -21:25:23.3 | A+BabC | A | … | … | 37.35 | 2008.68 | … | 0.746 ± 0.112 |
| 14575-2125 | HN 28 | HD 131976 | J14574-214 | M1.0 V | 14:57:27.70 | -21:25:07.9 | … | Bab(2) | 25.762 | 306.6 | 31.39 | 1933.94 | … | … |
| 14575-2125 | BUG 4 | GJ 570 D | … | T8 | 14:57:14.96 | -21:21:47.8 | … | C | 234.000 | 317.0 | 92.47 | 1972.88 | … | … |
| … | … | Ross 53A | J14575+313 | M2.0 V | 14:57:31.43 | +31:23:25.9 | AB | A | … | … | 95.96 | 1354.84 | … | 0.632 ± 0.027 |
| 14575+3124 | HDS2112 | Ross 53B | … | M0.5 V | 14:57:31.41 | +31:23:26.6 | … | B | 0.783 | 338.6 | 34.07 | 1354.46 | … | 0.610 ± 0.027 |
| … | … | GJ 1187 | J14578+566 | M5.5 V | 14:57:54.28 | +56:39:14.1 | … | … | … | … | 34.14 | 705.09 | 14.76 ± 0.08 | 0.126 ± 0.009 |
| … | … | GJ 572 | J15009+454 | M0.5 V | 15:00:55.91 | +45:25:39.8 | AB | A | … | … | 28.86 | 398.22 | … | 0.620 ± 0.027 |
| 15009+4526 | HDS2118 | BD+45 2247B | … | M4.5 V | 15:00:55.96 | +45:25:37.9 | … | B | 2.004 | 165.3 | 28.88 | 409.02 | … | 0.243 ± 0.037 |
| … | … | Ross 1042 | J15011+354 | M2.0 V | 15:01:11.98 | +35:27:10.5 | … | … | … | … | 49.12 | 312.63 | 304.92 ± 1.38 | 0.477 ± 0.018 |
| … | … | GJ 3885 | J15013+055 | M3.0 V | 15:01:20.20 | +05:32:48.5 | … | … | … | … | 49.84 | 450.43 | 154.16 ± 0.75 | 0.332 ± 0.011 |
| … | … | GJ 3891 | J15018+550 | M3.5 V | 15:05:48.60 | +55:04:45.8 | … | … | … | … | 89.60 | 481.59 | 128.37 ± 1.32 | 0.365 ± 0.016 |
| … | … | LP 41-431 | J15030+704 | M3.0 V | 15:03:01.68 | +70:26:14.0 | … | … | … | … | 89.53 | 444.88 | 259.62 ± 1.09 | 0.466 ± 0.018 |
| … | … | GJ 575.1 | J15043+294 | M2.5 V | 15:04:22.56 | +29:28:39.9 | … | … | … | … | 37.92 | 294.68 | 275.62 ± 1.35 | 0.452 ± 0.017 |
| … | … | Ross 1051 | J15043+603 | M1.0 V | 15:04:17.09 | +60:23:07.4 | … | … | … | … | 24.81 | 681.03 | 380.79 ± 1.60 | 0.505 ± 0.019 |
| … | … | GJ 3888 | J15049-211 | M4.5 V | 15:04:57.53 | -21:07:03.9 | … | … | … | … | … | 686.05 | 71.29 ± 0.34 | 0.266 ± 0.013 |
| … | … | PM J15060+4521 | J15060+453 | M1.5 V | 15:06:03.02 | +45:21:52.8 | … | … | … | … | … | 117.83 | 749.45 ± 3.01 | 0.609 ± 0.027 |
| … | … | GJ 579 | J15073+249 | M0.0 V | 15:07:22.59 | +24:56:15.8 | … | … | … | … | 38.93 | 979.55 | 762.02 ± 3.02 | 0.641 ± 0.027 |
| 15012+0710 | RAO 326 | Ross 1030a | J15011+071 | M3.5 V | 15:01:10.20 | +07:09:46.5 | (AB) | AB | 0.104 | 201.7 | 58.15 | 503.18 | … | 0.488 ± 0.030 |
| 15079+7612 | MET 10 | HD 135363 | … | G5 V | 15:07:55.68 | +76:12:05.4 | (AB)+C | AB | 0.363 | 133.1 | 43.02 | 208.70 | … | 0.797 ± 0.119 |
| 15079+7612 | LEP 72 | LSPM J1507+7613 | J15079+762 | M4.5 V | 15:07:56.64 | +76:14:01.8 | … | C | 116.445 | 1.7 | 37.03 | 207.10 | 244.03 ± 1.19 | 0.371 ± 0.032 |
| … | … | LSPM J1508+6221 | J15081+623 | M4.0 V | 15:08:11.53 | +62:21:56.4 | AB | A | … | … | 37.28 | 233.61 | … | 0.313 ± 0.034 |
| 15082+6222 | CRC 71 | G3- 1619631553142673664 | … | M3.5 V | 15:08:11.66 | +62:21:56.7 | … | B | 0.900 | 69.6 | 31.91 | 233.02 | … | 0.312 ± 0.034 |
| … | … | Ross 1047 | J15095+031 | M3.0 V | 15:09:34.95 | +03:10:08.3 | … | … | … | … | 34.50 | 770.81 | 257.95 ± 1.25 | 0.425 ± 0.011 |
| … | … | GJ 3893 | J15100+193 | M4.3 V | 15:10:04.82 | +19:21:20.2 | … | … | … | … | 31.86 | 452.25 | 97.41 ± 0.47 | 0.315 ± 0.014 |
| … | … | GJ 3894 | J15118-102 | M4.5 V | 15:11:49.55 | -10:14:22.1 | … | … | … | … | 34.85 | 999.70 | 42.58 ± 0.20 | 0.215 ± 0.012 |
| … | … | GJ 3895 | J15119+179 | M4.0 V | 15:11:55.49 | +17:57:07.4 | … | … | … | … | 37.24 | 699.72 | 126.47 ± 0.60 | 0.339 ± 0.015 |
| 15148+6434 | RAO 332 | LP 67-339 | J15147+645 | M3.5 V | 15:14:45.76 | +64:33:50.3 | (AB) | AB | 0.359 | 298.2 | 58.95 | 568.23 | … | 0.202 ± 0.039 |
| … | … | LP 272-63 | J15151+333 | M2.0 V | 15:15:06.93 | +33:17:57.0 | … | … | … | … | 42.69 | 362.69 | 177.81 ± 0.72 | 0.359 ± 0.015 |
| … | … | PM J15156+6349 | J15156+638 | M1.5 V | 15:15:37.69 | +63:49:50.8 | … | … | … | … | 72.00 | 126.66 | 399.71 ± 2.40 | 0.514 ± 0.029 |
| … | … | LP 222-65 | J15166+391 | M7.0 V | 15:16:40.44 | +39:10:47.4 | … | … | … | … | 35.19 | 212.01 | 18.42 ± 0.09 | 0.183 ± 0.011 |
| 15126+4544 | MCT 8 | GJ 3898 | J15126+457 | M4.0 Ve | 15:12:37.60 | +45:43:52.3 | (AB) | AB | 0.509 | 220.0 | 35.23 | 513.64 | … | 0.284 ± 0.035 |

Table D.2: Complete sample with the description of multiple systems (continued).

| WDS id | WDS disc | Name | Karmn | Spectral type | $\alpha$ (2016.0) | $\delta$ (2016.0) | System | Component | $\rho$ [arcsec] | $\theta$ [deg] | $\varpi$ [mas] | $\mu_{\rm total}$ [mas a$^{-1}$] | $\mathcal{L}$ [$10^{-4}$ $\mathcal{L}_\odot$] | $\mathcal{M}$ [$\mathcal{M}_\odot$] |
|---|---|---|---|---|---|---|---|---|---|---|---|---|---|---|
| … | … | StKM 1-1229 | J15188+292 | M1.0 V | 15:18:49.75 | +29:15:06.4 | AB+C | A | … | … | 73.46 | 98.17 | … | 0.608 ± 0.029 |
| … | … | G3- 1275127175448008448 | … | M0.0 V | 15:18:49.78 | +29:15:06.7 | … | B* | 0.588 | 50.9 | 36.57 | 78.58 | … | 0.625 ± 0.028 |
| 15189+2915 | LDS5168 | UCAC4 597-051773 | … | M3.5 V | 15:18:48.67 | +29:14:05.7 | … | C | 62.256 | 193.1 | 49.53 | 86.82 | 116.01 ± 0.67 | 0.346 ± 0.015 |
| … | … | GJ 3902 | J15193+678 | M3.0 V | 15:19:17.35 | +67:51:24.0 | … | … | … | … | 60.00 | 614.29 | 140.60 ± 0.55 | 0.359 ± 0.015 |
| … | … | HOLib | J15194-077 | M3.0 V | 15:19:25.51 | -07:43:21.7 | … | … | … | … | 54.30 | 1225.14 | 123.26 ± 0.65 | 0.295 ± 0.010 |
| … | … | PM J15197+0439 | J15197+046 | M4.0 V | 15:19:45.88 | +04:39:36.0 | … | … | … | … | 51.46 | 102.52 | 59.26 ± 0.27 | 0.240 ± 0.012 |
| … | … | PM J15210+3057 | J15210+309 | M2.5 V | 15:21:00.56 | +30:57:00.9 | … | … | … | … | 52.66 | 99.25 | 235.14 ± 1.50 | 0.416 ± 0.016 |
| … | … | TYC 344-504-1 | J15214+042 | M1.5 V | 15:21:25.39 | +04:14:49.9 | … | … | … | … | 33.21 | 95.70 | 430.88 ± 2.33 | 0.512 ± 0.029 |
| … | … | OTSer | J15218+209 | M1.0 V | 15:21:53.03 | +20:58:42.0 | … | … | … | … | 23.90 | 151.81 | 454.40 ± 2.11 | 0.518 ± 0.029 |
| … | … | LP 442-37 | J15219+185 | M1.5 V | 15:21:56.83 | +18:35:47.9 | … | … | … | … | 46.78 | 213.40 | 561.41 ± 2.42 | 0.564 ± 0.028 |
| … | … | Ross 508 | J15238+174 | M4.92 | 15:23:50.70 | +17:27:37.3 | … | … | … | … | 31.56 | 1318.87 | 38.05 ± 0.19 | 0.202 ± 0.011 |
| 15192-1245 | CRC 72 | GJ 3900 | J15191-127 | M4.0 V | 15:19:10.93 | -12:45:09.3 | (AB) | AabB(2) | 0.135 | 8.2 | 31.35 | 761.62 | … | … |
| … | … | StKM 1-1240 | J15238+561 | M1.0 Ve | 15:23:54.01 | +56:09:31.5 | Aab+B | Aab(2) | … | … | 316.48 | 88.93 | … | … |
| 15239+5610 | VVO 14 | PM J15237+5609 | … | M0.0 V | 15:23:46.47 | +56:09:06.0 | … | B | 67.990 | 248.1 | 57.04 | 87.64 | 851.63 ± 3.28 | 0.665 ± 0.026 |
| … | … | G 224-65 | J15238+584 | M4.0 V | 15:23:51.06 | +58:28:11.2 | … | … | … | … | 47.80 | 348.95 | 78.80 ± 0.31 | 0.281 ± 0.013 |
| 15270+4128 | SKF1924 | TYC 3055-1525-1 | J15273+415 | M1.5 V | 15:27:19.05 | +41:30:09.9 | A+B | A | 34.317 | 1.0 | 27.78 | 120.44 | … | … |
| … | … | PM J15273+4130N | … | M6.5 V | 15:27:19.11 | +41:30:44.2 | … | B | 34.282 | 1.1 | 72.32 | 122.65 | 13.08 ± 0.12 | 0.150 ± 0.010 |
| … | … | G 179-29 | J15276+408 | M1.0 V | 15:27:38.85 | +40:52:02.2 | … | … | … | … | 74.78 | 304.94 | 630.36 ± 2.43 | 0.594 ± 0.028 |
| … | … | GJ 587.1 | J15280+257 | M0.0 V | 15:28:01.43 | +25:47:22.9 | … | … | … | … | 42.11 | 112.49 | 564.66 ± 3.12 | 0.576 ± 0.028 |
| 15290+4646 | JNN 103 | RX J1529.0+4646 | J15290+467 | M4.5 V | 15:29:02.77 | +46:46:23.5 | (AB) | AB | 0.222 | 202.0 | 32.99 | 123.10 | … | 0.230 ± 0.037 |
| … | … | RX J1532.6+4653 | … | M1.0 V | 15:32:37.17 | +46:53:04.6 | … | … | … | … | 30.21 | 123.80 | 507.59 ± 1.94 | 0.548 ± 0.028 |
| … | … | LP 502-56 | J15305+094 | M5.5 V | 15:30:30.13 | +09:26:04.4 | … | … | … | … | 41.00 | 256.53 | 13.99 ± 0.07 | 0.115 ± 0.008 |
| 15297+4252 | JNN 278 | GJ 3907 | J15297+428 | M4.68 | 15:29:44.63 | +42:52:38.9 | (AB) | AB | 0.560 | 3.7 | 76.78 | 760.24 | … | 0.228 ± 0.037 |
| … | … | GJ 3911 | J15336+462 | M3.6 V | 15:33:39.33 | +46:15:06.1 | … | … | … | … | 30.78 | 199.81 | 132.74 ± 0.56 | 0.348 ± 0.015 |
| … | … | GJ 3910 | J15319+288 | M4.0 V | 15:31:53.50 | +28:51:10.2 | Aab | Aab(1) | … | … | 22.10 | 544.00 | … | … |
| … | … | GJ 588.1 | J15339+379 | M0.5 V | 15:33:54.82 | +37:54:48.4 | AB | A | … | … | 69.07 | 81.90 | … | 0.583 ± 0.028 |
| 15339+3755 | BEU 20 | G3- 1375767330164975616 | … | M2.5 V | 15:33:54.89 | +37:54:48.3 | … | B | 0.834 | 99.3 | 68.81 | 82.68 | … | 0.422 ± 0.031 |
| … | … | LP 135-414 | J15340+513 | M4.5 V | 15:34:03.52 | +51:22:06.8 | … | … | … | … | 31.29 | 335.59 | 65.77 ± 0.25 | 0.273 ± 0.013 |
| … | … | Ross 512 | J15345+142 | M4.0 V | 15:34:29.76 | +14:16:15.5 | … | … | … | … | 61.24 | 685.85 | 99.46 ± 0.49 | 0.318 ± 0.014 |
| … | … | 2MUCD 11346 | J15349-143 | M8.6V | 15:34:55.92 | -14:18:54.5 | … | … | … | … | 61.20 | 976.35 | 3.78 ± 0.03 | 0.104 ± 0.052 |
| … | … | Ross 513 | J15353+177S | M2.5 V | 15:35:19.17 | +17:42:44.3 | A+B | A | … | … | 6.33 | 1222.21 | 106.84 ± 0.45 | 0.290 ± 0.013 |
| 15354+1743 | LDS 977 | Ross 513B | J15353+177N | M4.5 V | 15:35:18.98 | +17:43:01.7 | … | B | 17.637 | 351.2 | 29.31 | 1221.56 | 23.02 ± 0.09 | 0.152 ± 0.010 |
| … | … | GJ 3913 | J15357+221 | M3.5 V | 15:35:45.31 | +22:09:01.6 | … | … | … | … | 57.01 | 719.72 | 158.07 ± 1.07 | 0.382 ± 0.016 |
| … | … | BKCrB | J15368+375 | M0.0 V | 15:36:49.97 | +37:34:48.0 | AB | A | … | … | 57.03 | 291.14 | … | 0.6220 ± 0.093 |
| 15368+3735 | LDS4572 | LP 273-44 | … | … | … | … | … | B | 15.278 | 89.3 | 75.59 | 287.84 | … | … |
| … | … | Ross 802 | J15369-141 | dM4.0 | 15:36:58.13 | -14:08:11.8 | … | … | … | … | 52.73 | 773.31 | 113.90 ± 0.63 | 0.295 ± 0.011 |
| … | … | G 179-42 | J15386+371 | M3.5 V | 15:38:36.68 | +37:07:28.0 | … | … | … | … | 52.66 | 367.18 | 140.78 ± 0.67 | 0.359 ± 0.015 |
| … | … | GJ 1194 A | J15400+434N | M3.0 V | 15:40:05.37 | +43:29:34.1 | A+B | A | … | … | 70.32 | 1229.91 | 105.50 ± 0.51 | 0.288 ± 0.013 |
| 15400+4330 | VBS 25 | GJ 1194 B | J15400+434S | M4.0 V | 15:40:05.54 | +43:29:30.0 | … | B | 4.487 | 155.3 | 38.25 | 1208.54 | … | 0.189 ± 0.039 |
| … | … | UU UMi | J15412+759 | M3.5 V | 15:41:20.12 | +75:59:22.5 | Aab | Aab(2) | … | … | 35.83 | 1077.66 | … | 0.502 ± 0.071 |

Table D.2: Complete sample with the description of multiple systems (continued).

| WDS id | WDS disc | Name | Karmn | Spectral type | $\alpha$ (2016.0) | $\delta$ (2016.0) | System | Component | $\rho$ [arcsec] | $\theta$ [deg] | $\varpi$ [mas] | $\mu_{\rm total}$ [mas a$^{-1}$] | $\mathcal{L}$ [$10^{-4}\,\mathcal{L}_\odot$] | $\mathcal{M}$ [$\mathcal{M}_\odot$] |
|---|---|---|---|---|---|---|---|---|---|---|---|---|---|---|
| … | … | HD 140232 | … | A8 Vam | 15:41:54.65 | +18:27:51.4 | AB+C | A | … | … | 100.31 | 82.17 | … | 1.810 ± 0.272 |
| 15419+1828 | DRS 17 | G3- 1197801408884577408 | … | M3.5 V | 15:41:54.81 | +18:27:51.5 | … | B | 2.359 | 87.6 | 63.36 | 82.28 | … | 0.319 ± 0.034 |
| 15419+1828 | TOK 302 | StKM 1-1264 | J15416+184 | M1.5 Ve | 15:41:37.17 | +18:28:09.2 | … | C | 249.250 | 274.1 | 91.90 | 91.89 | 1106.36 ± 19.33 | 0.668 ± 0.026 |
| … | … | GJ 595 | J15421-194 | M3.0 V | 15:42:04.27 | -19:28:34.9 | Aab | Aab(1) | … | … | … | 2261.94 | … | … |
| … | … | GJ 3916 | J15474-108 | M2.5 V | 15:47:24.21 | -10:53:53.2 | Aabc | Aabc(3) | … | … | 59.09 | 482.66 | … | 0.672 ± 0.048 |
| … | … | 1R154741.3+224108 | J15476+226 | M4.5 V | 15:47:40.51 | +22:41:16.0 | … | … | … | … | 69.04 | 186.83 | 164.85 ± 2.27 | 0.446 ± 0.019 |
| … | … | LP 177-102 | J15474+451 | M4.0 V | 15:47:27.03 | +45:07:54.5 | Aab | Aab(EB) | … | … | 50.47 | 318.39 | … | 0.516 ± 0.012 |
| … | … | RX J1548.0+0421 | J15480+043 | M2.5 V | 15:48:02.78 | +04:21:38.4 | A+B | A | … | … | 42.51 | 58.85 | 299.27 ± 1.99 | 0.502 ± 0.019 |
| … | … | UCAC4 472-052890 | … | M4.0 V | 15:47:54.89 | +04:18:02.9 | … | B* | 245.615 | 208.7 | 78.56 | 58.38 | 94.76 ± 0.46 | 0.310 ± 0.014 |
| … | … | PM J15488+3030 | J15488+305 | M3.0 V | 15:48:48.60 | +30:30:38.7 | … | … | … | … | 63.38 | 103.80 | 277.11 ± 1.48 | 0.482 ± 0.019 |
| … | … | G 168-13 | J15493+250 | M2.0 V | 15:49:20.38 | +25:03:48.5 | … | … | … | … | 54.79 | 390.77 | 256.11 ± 1.34 | 0.435 ± 0.017 |
| … | … | GJ 3920 | J15496+510 | M2.0 V | 15:49:35.62 | +51:03:02.0 | … | … | … | … | 46.87 | 480.73 | 254.48 ± 1.48 | 0.434 ± 0.017 |
| … | … | LP 23-35 | J15499+796 | M5.0 V | 15:49:53.83 | +79:39:53.6 | … | … | … | … | 46.78 | 252.39 | 73.49 ± 0.69 | 0.291 ± 0.010 |
| … | … | Wolf 587 | J15501+009 | M3.0 V | 15:50:11.41 | +00:57:32.0 | … | … | … | … | 159.92 | 189.62 | 144.99 ± 0.83 | 0.342 ± 0.015 |
| … | … | TYC 2572-633-1 | J15512+306 | M1.5 Ve | 15:51:14.63 | +30:40:42.2 | … | … | … | … | 20.54 | 181.48 | 905.27 ± 30.30 | 0.640 ± 0.027 |
| … | … | GJ 3923 | J15513+295 | M3.5 V | 15:51:21.51 | +29:30:59.2 | … | … | … | … | 54.18 | 498.11 | 127.33 ± 0.73 | 0.340 ± 0.015 |
| 15496+3449 | BWL 41 | GJ 3919 | J15496+348 | M4.0 Ve | 15:49:37.37 | +34:49:07.8 | (AB) | AB | 0.208 | 98.6 | 56.08 | 966.80 | … | 0.293 ± 0.035 |
| 15531+3445 | RAO 340 | Ross 806 | J15531+347N | M3.0 V | 15:53:06.69 | +34:45:05.9 | (AB)+C | AB | 0.433 | 304.9 | 56.11 | 559.77 | … | 0.481 ± 0.030 |
| 15531+3445 | LDS6309 | GJ 3926 | J15531+347S | M3.0 V | 15:53:06.97 | +34:44:39.5 | … | C | 26.607 | 172.6 | 25.37 | 570.00 | 176.19 ± 0.83 | 0.405 ± 0.017 |
| … | … | NLTT 41533 | J15538+641 | M0.5 V | 15:53:48.31 | +64:09:36.2 | … | … | … | … | 26.80 | 228.67 | 764.47 ± 3.93 | 0.636 ± 0.027 |
| … | … | GJ 3928 | J15555+352 | M4.0 V | 15:55:31.51 | +35:12:05.2 | AB | A | … | … | 30.98 | 279.97 | … | 0.366 ± 0.033 |
| 15555+3512 | MCT 9 | G 180-11B | … | M5.0 V | 15:55:31.39 | +35:12:04.7 | … | B | 1.630 | 252.4 | 91.48 | 266.99 | … | 0.184 ± 0.040 |
| … | … | RX J1555.7+6840 | J15557+686 | M2.5 Ve | 15:55:47.10 | +68:40:16.0 | … | … | … | … | 91.36 | 140.05 | 367.74 ± 1.63 | 0.484 ± 0.030 |
| … | … | RX J1556.9+3738 | J15569+376 | M2.5 V | 15:56:58.12 | +37:38:14.3 | … | … | … | … | 26.17 | 94.77 | 476.95 ± 10.61 | 0.506 ± 0.029 |
| … | … | LSPM J1557+0901 | J15578+090 | M4.0 V | 15:57:48.42 | +09:01:07.6 | … | … | … | … | … | 212.20 | 98.54 ± 0.47 | 0.317 ± 0.014 |
| … | … | V1022Her | J15581+494 | M1.0 Ve | 15:58:10.41 | +49:27:05.5 | … | … | … | … | 29.19 | 189.79 | 694.71 ± 4.37 | 0.590 ± 0.028 |
| … | … | GJ 3929 | J15583+354 | M3.5 V | 15:58:18.61 | +35:24:29.4 | … | … | … | … | 31.81 | 348.81 | 118.00 ± 0.45 | 0.292 ± 0.011 |
| … | … | StM 258 | J15587+346 | M3.5 V | 15:58:45.75 | +34:48:53.8 | … | … | … | … | 76.64 | 90.55 | 315.58 ± 8.53 | 0.436 ± 0.031 |
| … | … | RX J1559.7+4403 | J15597+440 | M2.0 Ve | 15:59:47.20 | +44:03:59.6 | A+B | A | … | … | 48.38 | 69.95 | 1092.28 ± 19.82 | 0.665 ± 0.026 |
| 15598+4404 | JNN 106 | PM J15597+4403B | … | M8.0 V | 15:59:46.70 | +44:04:01.0 | … | B | 5.622 | 284.3 | 29.05 | 68.31 | … | 0.106 ± 0.047 |
| … | … | GJ 606 | J15598-082 | M1.0 V | 15:59:53.60 | -08:15:12.0 | … | … | … | … | 31.07 | 204.06 | 381.38 ± 2.40 | 0.476 ± 0.012 |
| … | … | GJ 3933 | J16008+403 | M3.0 V | 16:00:50.40 | +40:19:38.7 | … | … | … | … | 87.50 | 410.82 | 134.10 ± 0.54 | 0.350 ± 0.015 |
| … | … | GJ 607 | J16017+301 | M3.0 V | 16:01:43.15 | +30:10:52.9 | … | … | … | … | 86.28 | 358.19 | 186.29 ± 0.78 | 0.391 ± 0.016 |
| … | … | GJ 3935 | J16018+304 | M2.5 V | 16:01:52.43 | +30:27:37.0 | A+B | A | … | … | 100.52 | 190.95 | 178.80 ± 0.75 | 0.383 ± 0.016 |
| 16019+3027 | LDS4626 | GJ 3936 | J16017+304 | M4.5 V | 16:01:44.35 | +30:27:42.9 | … | B | 104.711 | 273.3 | 93.81 | 188.24 | 63.82 ± 0.73 | 0.219 ± 0.011 |
| … | … | GJ 609 | J16028+205 | M4.0 V | 16:02:49.85 | +20:35:01.2 | … | … | … | … | 66.99 | 1571.06 | 72.64 ± 0.30 | 0.241 ± 0.010 |
| … | … | PM J16033+1735 | J16033+175 | M2.0 V | 16:03:20.60 | +17:35:55.0 | … | … | … | … | 40.33 | 55.81 | 878.93 ± 12.93 | 0.622 ± 0.027 |
| … | … | GJ 3937 | J16043-062 | M4.5 V | 16:04:19.91 | -06:16:59.9 | … | … | … | … | 20.89 | 869.82 | 26.06 ± 0.25 | 0.175 ± 0.011 |
| … | … | BPM 91242 | J16046+263 | M0.5 V | 16:04:36.85 | +26:20:44.6 | … | … | … | … | 25.39 | 129.04 | 637.04 ± 2.38 | 0.602 ± 0.027 |
| … | … | HD 144579 | … | G8 V | 16:04:56.01 | +39:09:24.3 | A+B | A | … | … | 22.54 | 573.29 | 4140.87 ± 12.13 | 0.940 ± 0.141 |

Table D.2: Complete sample with the description of multiple systems (continued).

| WDS id | WDS disc | Name | Karmn | Spectral type | $\alpha$ (2016.0) | $\delta$ (2016.0) | System | Component | $\rho$ [arcsec] | $\theta$ [deg] | $\varpi$ [mas] | $\mu_{\rm total}$ [mas a$^{-1}$] | $\mathcal{L}$ [$10^{-4}$ $\mathcal{L}_\odot$] | $\mathcal{M}$ [$\mathcal{M}_\odot$] |
|---|---|---|---|---|---|---|---|---|---|---|---|---|---|---|
| 16048+3910 | WNO 47 | HD 144579B | J16048+391 | M4.0 V | 16:04:50.08 | +39:09:36.6 | … | B | 69.998 | 280.1 | 169.88 | 564.44 | 30.52 ± 0.11 | 0.166 ± 0.010 |
| … | … | GJ 3941 | J16054+769 | M3.0 V | 16:05:26.51 | +76:54:58.5 | … | … | … | … | 168.77 | 353.09 | 276.28 ± 1.48 | 0.482 ± 0.019 |
| … | … | LP 329-30 | J16062+290 | M2.0 V | 16:06:13.41 | +29:02:01.6 | … | … | … | … | 169.30 | 405.78 | 869.00 ± 5.18 | 0.619 ± 0.027 |
| 16067+0823 | JOD 9 | GJ 611.3 | J16066+083 | M1.0 V | 16:06:40.66 | +08:23:19.6 | (AB)+C | AB | 0.599 | 291.0 | 23.58 | 513.49 | … | 0.610 ± 0.027 |
| 16067+0823 | JOY 6 | G3- 4451575895400385536 | … | … | 16:06:40.51 | +08:23:20.4 | … | C | 2.407 | 290.4 | 23.81 | … | … | … |
| … | … | GJ 3939 | J16074+059 | M3.8 V | 16:07:27.90 | +05:57:56.0 | … | … | … | … | 87.20 | 366.88 | 158.64 ± 1.06 | 0.383 ± 0.016 |
| … | … | GJ 1198 | J16082-104 | M4.5 V | 16:08:14.56 | -10:26:35.1 | … | … | … | … | 85.41 | 1353.47 | 50.15 ± 0.23 | 0.219 ± 0.012 |
| … | … | GJ 3942 | J16090+529 | M0.5 V | 16:09:03.50 | +52:56:39.0 | … | … | … | … | 85.43 | 213.40 | 649.64 ± 2.37 | 0.599 ± 0.027 |
| … | … | LP 504-59 | J16092+093 | M3.0 Ve | 16:09:15.90 | +09:21:11.0 | … | … | … | … | 42.22 | 405.12 | 177.74 ± 0.98 | 0.355 ± 0.011 |
| … | … | K2-33 | J16102-193 | dM3.0 | 16:10:14.73 | -19:19:09.8 | … | … | … | … | 66.54 | 25.81 | 1054.64 ± 10.66 | … |
| … | … | TYC 371-1053-1 | J16120+033 | M2.0 V | 16:12:04.68 | +03:18:19.8 | A+B | A | … | … | 48.75 | 66.33 | 774.53 ± 11.50 | 0.602 ± 0.027 |
| 16121+0318 | SKF2840 | 1R161204.8+031850 | … | M3.5 V | 16:12:05.06 | +03:18:52.4 | … | B | 33.100 | 10.0 | 40.30 | 73.41 | 107.25 ± 2.13 | 0.273 ± 0.013 |
| … | … | LP 804-27 | J16126-188 | M3.0 V | 16:12:41.82 | -18:52:35.2 | … | … | … | … | 33.38 | 217.11 | … | 0.440 ± 0.031 |
| … | … | sig CrB A | … | F6 V | 16:14:40.51 | +33:51:29.6 | Aab+B+(CD) | Aab(2) | … | … | 77.48 | 282.06 | … | 0.231 ± 0.001 |
| 16147+3352 | STF2032 | sig CrB B | … | G1 V | 16:14:40.01 | +33:51:25.8 | … | B | 7.231 | 238.5 | 58.54 | 301.27 | 9935.25 ± 17.97 | 1.030 ± 0.155 |
| 16147+3352 | YSC 152 | sig CrB C | J16139+337 | M2.5 V | 16:13:55.90 | +33:46:22.7 | … | CD | 635.058 | 241.1 | 40.65 | 300.69 | … | 0.432 ± 0.031 |
| … | … | LP 624-54 | J16144-028 | M5.0 V | 16:14:25.19 | -02:50:54.9 | … | … | … | … | 23.44 | 367.49 | 7.51 ± 0.05 | 0.108 ± 0.009 |
| … | … | GJ 1200 | J16145+191 | M3.5 V | 16:14:30.36 | +19:06:16.7 | … | … | … | … | 52.54 | 2035.67 | 103.97 ± 0.48 | 0.305 ± 0.014 |
| … | … | GJ 3946 | J16147+048 | M3.5 V | 16:14:43.36 | +04:52:09.7 | … | … | … | … | 31.93 | 416.37 | 166.18 ± 0.86 | 0.392 ± 0.016 |
| … | … | GJ 3947 | J16155+244 | M1.5 V | 16:15:32.10 | +24:27:50.9 | … | … | … | … | 33.71 | 256.12 | 361.06 ± 1.56 | 0.491 ± 0.018 |
| … | … | HD 147379 | J16167+672S | M0.0 V | 16:16:41.37 | +67:14:21.2 | A+B | A | … | … | 33.71 | 504.96 | 962.91 ± 5.28 | 0.680 ± 0.026 |
| 16167+6714 | ENG 57 | EWDra | J16167+672N | M3.0 V | 16:16:43.98 | +67:15:23.9 | … | B | 64.529 | 13.6 | 35.64 | 491.15 | 294.12 ± 1.14 | 0.449 ± 0.014 |
| … | … | 1R161804.9+061702 | J16180+062 | M3.0 Ve | 16:18:05.01 | +06:17:12.0 | … | … | … | … | 34.17 | 92.76 | 320.46 ± 4.55 | 0.450 ± 0.031 |
| 16171+5516 | BLA 3 | CRDra | J16170+552 | M1.0 V | 16:17:05.52 | +55:16:01.9 | (AB) | AB(2) | 0.097 | 8.5 | 119.58 | 439.85 | … | … |
| … | … | GJ 618.1 A | J16204-042 | M0.0 V | 16:20:24.34 | -04:16:02.6 | A+B | A | … | … | 68.45 | 416.69 | 1014.40 ± 5.86 | 0.691 ± 0.026 |
| 16204-0416 | WIL 3 | GJ 618.1 B | … | L2.4V | 16:20:25.72 | -04:16:32.0 | … | B | 35.907 | 144.8 | 57.57 | 415.28 | 1.08 ± 0.06 | 0.075 ± 0.011 |
| … | … | V1169Her | J16220+228 | M1.5 Ve | 16:22:01.12 | +22:50:22.8 | … | … | … | … | 65.50 | 85.58 | 1157.76 ± 16.87 | 0.677 ± 0.026 |
| … | … | LSPM J1624+2254 | J16247+229 | M1.0 V | 16:24:43.71 | +22:54:19.3 | … | … | … | … | 41.09 | 180.71 | 587.98 ± 13.14 | 0.572 ± 0.028 |
| … | … | GJ 625 | J16254+543 | M1.5 V | 16:25:25.41 | +54:18:12.0 | … | … | … | … | 57.20 | 465.07 | 144.88 ± 0.69 | 0.303 ± 0.010 |
| 16240+4822 | HEN 1 | GJ 623 | J16241+483 | M3.0 V | 16:24:11.16 | +48:21:03.1 | (AB) | AB(1) | 0.327 | 55.9 | 18.09 | 1254.76 | … | … |
| … | … | LP 330-13 | J16255+323 | M2.0 V | 16:25:32.91 | +32:18:34.1 | … | … | … | … | 42.42 | 184.66 | 254.78 ± 1.44 | 0.434 ± 0.017 |
| … | … | TYC 4647-2406-1 | J16259+834 | M1.5 V | 16:25:59.21 | +83:24:23.2 | … | … | … | … | 32.04 | 106.46 | 463.34 ± 1.81 | 0.533 ± 0.029 |
| 16255+2602 | RAO 347 | GJ 3953 | J16255+260 | M3.0 V | 16:25:32.12 | +26:01:38.0 | (AB) | AB | 0.285 | 54.8 | 56.68 | 190.85 | … | … |
| 16268-1724 | WSI 131 | GJ 3954 | J16268-173 | M5e | 16:26:47.75 | -17:23:40.5 | (AB) | AB | 0.691 | 172.7 | 52.59 | 524.83 | … | 0.185 ± 0.039 |
| 16280+1533 | JNN279 | GJ 3955 | J16280+155 | M3.6 V | 16:28:02.04 | +15:33:52.1 | (AB) | AB(2) | 0.558 | 35.1 | … | 307.67 | … | … |
| … | … | V2306Oph | J16303-126 | M3.0 V | 16:30:17.96 | -12:40:04.3 | … | … | … | … | 20.17 | 1187.66 | 109.31 ± 0.79 | 0.289 ± 0.011 |
| … | … | GJ 3959 | J16313+408 | M6.0 V | 16:31:18.59 | +40:51:56.6 | … | … | … | … | 20.93 | 338.19 | 26.95 ± 0.12 | 0.158 ± 0.010 |
| … | … | GJ 1202 | J16315+175 | M4.0 V | 16:31:34.69 | +17:33:35.9 | … | … | … | … | 19.84 | 873.35 | 107.96 ± 0.43 | 0.292 ± 0.013 |
| … | … | GJ 1203 | J16327+126 | M3.0 V | 16:32:44.36 | +12:36:43.8 | … | … | … | … | 39.33 | 776.89 | 210.00 ± 1.02 | 0.395 ± 0.011 |
| … | … | GJ 3960 | J16328+098 | M3.5 V | 16:32:53.09 | +09:50:28.5 | … | … | … | … | 158.72 | 276.61 | 86.50 ± 0.48 | 0.276 ± 0.013 |

Table D.2: Complete sample with the description of multiple systems (continued).

| WDS id | WDS disc | Name | Karmn | Spectral type | $\alpha$ (2016.0) | $\delta$ (2016.0) | System | Component | $\rho$ [arcsec] | $\theta$ [deg] | $\varpi$ [mas] | $\mu_{\rm total}$ [mas a$^{-1}$] | $\mathcal{L}$ [$10^{-4}$ $\mathcal{L}_\odot$] | $\mathcal{M}$ [$\mathcal{M}_\odot$] |
|---|---|---|---|---|---|---|---|---|---|---|---|---|---|---|
| … | … | LP 137-37 | J16342+543 | M1.0 V | 16:34:13.28 | +54:23:48.5 | … | … | … | … | 59.73 | 591.36 | 480.63 ± 1.91 | 0.541 ± 0.029 |
| 16302-1440 | WSI 132 | GJ 2121 | J16302-146 | M2.5 V | 16:30:12.52 | -14:39:53.0 | (AB) | AB | 0.105 | 100.5 | 41.52 | 571.32 | … | 0.435 ± 0.031 |
| … | … | CMDra | J16343+571 | M4.5 V | 16:34:18.14 | +57:10:03.3 | Aab+B | Aab(EB) | … | … | 50.62 | 1623.35 | … | 0.445 ± 0.001 |
| 16345+5709 | LDS1436 | GJ 630.1 B | … | DQ8 | 16:34:19.37 | +57:10:28.1 | … | B | 26.726 | 21.9 | 36.80 | 1634.85 | … | 0.500 ± 0.100 |
| … | … | GJ 1204 | J16360+088 | M4.0 V | 16:36:05.09 | +08:48:46.7 | … | … | … | … | 87.33 | 542.06 | 57.04 ± 0.34 | 0.235 ± 0.012 |
| … | … | G 202-68 | J16395+505 | M1.0 V | 16:39:30.81 | +50:33:57.0 | … | … | … | … | 30.80 | 514.38 | 239.84 ± 0.81 | 0.396 ± 0.016 |
| … | … | GJ 3967 | J16401+007 | M5.0 V | 16:40:06.17 | +00:42:16.3 | … | … | … | … | 89.13 | 236.37 | 39.62 ± 0.24 | 0.206 ± 0.011 |
| … | … | GJ 3971 | J16403+676 | M7.0 V | 16:40:19.87 | +67:36:10.6 | … | … | … | … | 46.44 | 456.21 | 27.66 ± 0.29 | 0.196 ± 0.011 |
| … | … | Ross 812 | J16408+363 | M2.0 V | 16:40:48.72 | +36:19:02.9 | … | … | … | … | 19.70 | 212.79 | 421.80 ± 2.12 | 0.508 ± 0.029 |
| … | … | PM J16420+1916 | J16420+192 | M2.5 V | 16:42:00.72 | +19:16:11.6 | … | … | … | … | 19.73 | 77.66 | 481.17 ± 3.60 | 0.519 ± 0.029 |
| … | … | GJ 3972 | J16462+164 | M3.0 V | 16:46:13.35 | +16:28:33.2 | … | … | … | … | 43.35 | 590.54 | 248.23 ± 1.07 | 0.413 ± 0.012 |
| … | … | LP 276-22 | J16465+345 | M6.0 V | 16:46:31.03 | +34:34:49.0 | … | … | … | … | … | 542.48 | 10.78 ± 0.04 | 0.134 ± 0.010 |
| 16355+3501 | BWL 44 | V1200Her | J16354+350 | M4.0 Ve | 16:35:27.61 | +35:00:55.3 | (AB) | AB | 0.092 | 25.6 | 30.84 | 200.15 | … | 0.330 ± 0.034 |
| … | … | GJ 3973 | J16487-157 | M1.5 Vk: | 16:48:45.95 | -15:44:23.7 | … | … | … | … | 32.67 | 220.40 | 510.00 ± 2.33 | 0.545 ± 0.028 |
| … | … | GJ 3975 | J16508-048 | M3.5 V | 16:50:52.99 | -04:50:38.8 | … | … | … | … | 40.69 | 786.19 | 80.11 ± 0.58 | 0.265 ± 0.013 |
| … | … | GJ 3976 | J16509+224 | M5.0 V | 16:50:57.98 | +22:27:12.1 | … | … | … | … | 45.62 | 401.79 | 30.59 ± 0.12 | 0.178 ± 0.011 |
| … | … | GSC 04194-01561 | J16528+630 | M4.5 V | 16:52:49.82 | +63:04:41.2 | … | … | … | … | 38.11 | 212.34 | 45.22 ± 0.22 | 0.222 ± 0.012 |
| … | … | GJ 3979 | J16529+400 | M3.0 V | 16:52:54.92 | +40:05:04.6 | … | … | … | … | 116.04 | 233.33 | 278.28 ± 7.40 | 0.412 ± 0.032 |
| … | … | Ross 644 | J16542+119 | M0.0 V | 16:54:11.44 | +11:54:57.9 | … | … | … | … | 53.07 | 628.96 | 474.85 ± 2.52 | 0.507 ± 0.018 |
| 16488+1039 | CRC 73 | LSPM J1648+1038 | J16487+106 | M2.5 V | 16:48:46.40 | +10:38:51.1 | (AB) | AB(2) | 0.102 | 176.3 | 53.76 | 177.27 | … | … |
| 16555-0820 | KUI 75 | V1054Oph | J16554-083S | M3.0 V | 16:55:27.89 | -08:20:25.2 | (AB)+C+D | AB(2) | 0.225 | 106.9 | 55.42 | 1202.11 | … | 0.640 ± 0.021 |
| 16555-0820 | LDS 573 | Wolf 629 | J16554-083N | M3.5 V | 16:55:24.34 | -08:19:35.7 | … | C | 72.292 | 313.2 | 43.18 | 1214.87 | 57.29 ± 0.40 | 0.207 ± 0.010 |
| 16555-0820 | WNO 55 | GJ 644 C | J16555-083 | M7.0 V | 16:55:34.38 | -08:23:54.7 | … | D | 230.576 | 155.3 | 37.64 | 1191.21 | 5.64 ± 0.03 | 0.101 ± 0.049 |
| … | … | GJ 1207 | J16570-043 | dM4 | 16:57:06.25 | -04:21:02.3 | … | … | … | … | 32.88 | 611.15 | 67.66 ± 0.60 | 0.257 ± 0.009 |
| … | … | GJ 3980 | J16573+124 | M4.0 V | 16:57:23.12 | +13:28:06.6 | … | … | … | … | 32.95 | 225.63 | 94.15 ± 1.39 | 0.309 ± 0.014 |
| … | … | PM J16573+2708 | J16573+271 | M2.0 V | 16:57:22.27 | +27:08:31.2 | AB | A | … | … | 60.84 | 55.64 | … | 0.425 ± 0.031 |
| 16574+2709 | CFN 13 | PM J16573+2708B | … | … | 16:57:22.34 | +27:08:30.6 | … | B | 1.044 | 121.2 | 44.12 | … | … | … |
| … | … | GJ 3986 | J16574+777 | M3.0 V | 16:57:29.79 | +77:43:06.3 | … | … | … | … | 91.42 | 254.66 | 208.29 ± 0.87 | 0.415 ± 0.017 |
| 16578+1317 | HDS2399 | GJ 647 | J16577+132 | M0.0 V | 16:57:46.15 | +13:17:31.1 | (AB) | AB | 0.102 | 193.4 | 66.88 | 104.81 | … | 0.728 ± 0.026 |
| … | … | V1090Her | … | K3 V | 16:57:52.95 | +47:22:04.4 | A+B+C | A | … | … | 66.93 | 308.70 | … | 0.775 ± 0.116 |
| 16579+4722 | A 1874 | HD 153557B | J16578+473 | M1.5 V | 16:57:53.40 | +47:22:06.7 | … | B | 5.091 | 63.0 | 54.34 | 301.44 | … | 0.475 ± 0.030 |
| 16579+4722 | STFA 32 | V1089Her | … | K3 V | 16:57:42.01 | +47:21:47.9 | … | C | 112.378 | 261.6 | 18.99 | 296.94 | … | 0.764 ± 0.115 |
| … | … | Ross 860 | J16581+257 | M1.0 V | 16:58:08.71 | +25:44:30.8 | … | … | … | … | … | 521.01 | 450.03 ± 2.13 | 0.497 ± 0.012 |
| 16584+1358 | YSC61 | GJ 3981 A | J16584+139 | M4.0 V | 16:58:24.81 | +13:58:10.9 | AB | A | 0.549 | 236.4 | 71.47 | 399.08 | … | 0.263 ± 0.037 |
| … | … | GJ 3981 B | … | … | 16:58:24.79 | +13:58:10.8 | … | B | 0.298 | 263.8 | 32.26 | … | … | … |
| … | … | LP 43-338 | J16587+688 | M1.5 V | 16:58:42.17 | +68:53:55.8 | … | … | … | … | 79.85 | 565.45 | 372.78 ± 1.86 | 0.500 ± 0.019 |
| … | … | V1234Her | J16591+209 | M3.5 Ve | 16:59:09.58 | +20:58:18.3 | AB | A | … | … | 79.84 | 120.04 | … | 0.370 ± 0.033 |
| 16592+2058 | JNN 111 | PM J16591+2058B | … | M4.0 V | 16:59:09.62 | +20:58:17.9 | … | B | 0.653 | 129.7 | 73.26 | 139.45 | … | 0.268 ± 0.036 |
| … | … | GJ 3983 | J17003+253 | M3.2 V | 17:00:20.18 | +25:21:05.1 | … | … | … | … | 18.72 | 182.98 | 124.65 ± 0.46 | 0.316 ± 0.014 |
| … | … | G 139-4 | J17006+063 | M1.0 V | 17:00:38.65 | +06:18:42.0 | … | … | … | … | 18.64 | 315.70 | 549.03 ± 2.84 | 0.568 ± 0.028 |

Table D.2: Complete sample with the description of multiple systems (continued).

| WDS id | WDS disc | Name | Karmn | Spectral type | $\alpha$ (2016.0) | $\delta$ (2016.0) | System | Component | $\rho$ [arcsec] | $\theta$ [deg] | $\varpi$ [mas] | $\mu_{total}$ [mas a$^{-1}$] | $\mathcal{L}$ [$10^{-4}$ $\mathcal{L}_\odot$] | $\mathcal{M}$ [$\mathcal{M}_\odot$] |
|---|---|---|---|---|---|---|---|---|---|---|---|---|---|---|
| ... | ... | GJ 3984 | J17010+082 | M3.8 V | 17:01:01.85 | +08:12:23.5 | ... | ... | ... | ... | 19.37 | 306.52 | 95.24 ± 0.55 | 0.291 ± 0.013 |
| ... | ... | GJ 3987 | J17027-060 | M0 | 17:02:49.45 | -06:04:07.6 | ... | ... | ... | ... | 103.18 | 151.44 | 488.47 ± 2.60 | 0.554 ± 0.028 |
| ... | ... | GJ 3988 | J17033+514 | M5.0 V | 17:03:24.10 | +51:24:32.6 | ... | ... | ... | ... | 62.52 | 622.70 | 39.68 ± 0.19 | 0.183 ± 0.009 |
| ... | ... | LP 331-57 A | J17038+321 | M2.0 V | 17:03:53.09 | +32:11:47.6 | AabB | Aab(2?) | ... | ... | 27.85 | 216.04 | ... | ... |
| 17039+3212 | DAE 6 | LP 331-57 B | ... | M4.0 V | 17:03:53.14 | +32:11:46.4 | ... | B | 1.393 | 151.3 | 35.23 | 185.92 | ... | 0.232 ± 0.037 |
| ... | ... | GJ 1209 | J17043+169 | M3.0 V | 17:04:22.49 | +16:55:37.5 | ... | ... | ... | ... | 45.07 | 1140.03 | 147.01 ± 0.75 | 0.345 ± 0.015 |
| ... | ... | HD 154363 | ... | K4/5 V | 17:05:02.41 | -05:04:17.7 | A+B | A | ... | ... | 34.44 | 1461.66 | 1307.20 ± 6.44 | 0.730 ± 0.110 |
| 17050-0504 | LDS 585 | HD 154363B | J17052-050 | M1.5 V | 17:05:12.80 | -05:05:57.4 | ... | B | 184.423 | 122.7 | 34.58 | 1457.06 | 337.16 ± 1.58 | 0.478 ± 0.012 |
| ... | ... | LP 387-37 | J17058+260 | M1.5 V | 17:05:52.55 | +26:05:27.4 | AB | A | ... | ... | 37.61 | 291.03 | ... | 0.553 ± 0.028 |
| 17059+2606 | LDS4721 | LP 387-36 | ... | DC7 | 17:05:52.54 | +26:05:46.7 | ... | B | 19.244 | 359.6 | 41.32 | 285.38 | ... | 0.500 ± 0.100 |
| ... | ... | Ross 863 | J17071+215 | M3.0 V | 17:07:06.96 | +21:33:14.1 | ... | ... | ... | ... | 47.42 | 464.00 | 204.83 ± 0.82 | 0.379 ± 0.012 |
| ... | ... | G 203-44 | J17082+516 | M1.0 V | 17:08:12.71 | +51:38:10.6 | ... | ... | ... | ... | 48.16 | 376.70 | 743.43 ± 2.72 | 0.617 ± 0.027 |
| 17077+0722 | YSC62 | GJ 1210 | J17076+073 | M5.0 V | 17:07:40.31 | +07:22:00.6 | (AB) | AB | 0.122 | 283.5 | 59.12 | 618.93 | ... | 0.163 ± 0.041 |
| ... | ... | GJ 3990 | J17098+119 | M4.0 V | 17:09:52.36 | +11:55:32.8 | ... | ... | ... | ... | 21.14 | 371.52 | 68.88 ± 1.09 | 0.261 ± 0.013 |
| ... | ... | GJ 3991 | J17095+436 | M3.5 V | 17:09:32.03 | +43:40:48.4 | Aab | Aab(1) | ... | ... | 18.00 | 430.81 | ... | ... |
| ... | ... | StM 336 | J17104+279 | M2.5 V | 17:10:25.49 | +27:58:38.6 | A+B | A | ... | ... | 54.11 | 91.91 | 215.47 ± 0.94 | 0.397 ± 0.016 |
| 17104+2759 | SKF 366 | CTI 170958.5+275905 | ... | M5.5 V | 17:10:28.32 | +27:58:08.4 | ... | B | 48.112 | 128.8 | 59.14 | 91.23 | 12.16 ± 0.07 | 0.122 ± 0.009 |
| ... | ... | HD 155876 | J17121+456 | M3.0 V | 17:12:08.21 | +45:39:32.6 | AB | A | ... | ... | 43.71 | 1661.81 | ... | 0.376 ± 0.033 |
| 17121+4540 | KUI 79 | G3- 1364668825433435776 | ... | ... | 17:12:08.21 | +45:39:33.7 | ... | B | 1.133 | 357.1 | 43.63 | ... | ... | ... |
| ... | ... | Wolf 654 | J17115+384 | M4.0 V | 17:11:35.04 | +38:26:33.2 | ... | ... | ... | ... | 30.81 | 217.81 | 184.71 ± 0.77 | 0.372 ± 0.013 |
| ... | ... | GJ 660 | J17118-018 | M3.0 V | 17:11:51.82 | -01:51:10.5 | AB | A | ... | ... | 36.01 | 639.91 | ... | 0.282 ± 0.043 |
| 17119-0151 | LPM629 | G3- 4367834134894197888 | ... | ... | 17:11:51.82 | -01:51:10.3 | ... | B | 0.216 | 357.7 | 36.22 | ... | ... | ... |
| ... | ... | GJ 3994 | J17146+269 | M2.0 V | 17:14:39.87 | +26:55:44.4 | ... | ... | ... | ... | 39.04 | 336.83 | 384.19 ± 6.98 | 0.490 ± 0.030 |
| ... | ... | GJ 1214 | J17153+049 | M4.5 V | 17:15:19.56 | +04:57:38.1 | ... | ... | ... | ... | 22.55 | 948.00 | 37.50 ± 0.22 | 0.200 ± 0.011 |
| 17121+4540 | KUI 79 | V2367Oph | J17136-084 | M3.5 V | 17:13:40.00 | -08:25:21.4 | Aab | Aab(2) | 0.662 | 289.4 | 49.49 | 600.61 | ... | ... |
| ... | ... | GJ 3997 | J17158+190 | M0.5 V | 17:15:49.95 | +19:00:00.3 | AB | A | ... | ... | 52.28 | 151.20 | ... | 0.499 ± 0.029 |
| 17158+1900 | JOD 11 | BD+19 3268B | ... | M4.5 V | 17:15:49.84 | +19:00:00.3 | ... | B | 1.471 | 271.0 | 26.36 | 89.24 | ... | 0.244 ± 0.037 |
| ... | ... | GJ 3998 | J17160+110 | M1.0 V | 17:16:00.49 | +11:03:22.1 | ... | ... | ... | ... | 63.17 | 373.65 | 469.14 ± 2.22 | 0.537 ± 0.029 |
| ... | ... | GJ 2128 | J17166+080 | M2.5 V | 17:16:40.68 | +08:03:29.1 | ... | ... | ... | ... | 36.36 | 286.72 | 237.65 ± 1.04 | 0.402 ± 0.011 |
| ... | ... | GJ 1215 | J17177+116 | M5.0 V | 17:17:43.71 | +11:40:05.3 | (AB) | AB | ... | ... | 23.76 | 566.80 | ... | 0.154 ± 0.041 |
| ... | ... | GJ 3999 | J17177-118 | M3.0 V | 17:17:45.28 | -11:48:59.0 | A+B | A | ... | ... | 23.55 | 292.10 | 312.91 ± 1.80 | 0.484 ± 0.018 |
| 17178-1149 | LDS 593 | GJ 4000 | ... | M3.0 V | 17:17:44.51 | -11:48:31.1 | ... | B | 30.072 | 337.8 | 75.29 | 292.36 | 269.55 ± 1.51 | 0.506 ± 0.020 |
| 17183+1809 | BWL45 | GJ 4002 | J17183+181 | M3.0 V | 17:18:22.44 | +18:08:52.4 | (AB) | AB | 0.452 | 324.5 | 47.33 | 244.52 | ... | 0.329 ± 0.034 |
| 17184-0147 | BAG51 | GJ 4001 | J17183-017 | M0.0 V | 17:18:21.82 | -01:46:55.4 | (AB) | AB | 0.521 | 291.3 | 51.77 | 160.45 | ... | 0.615 ± 0.027 |
| ... | ... | V647Her | J17199+265 | M3.5 V | 17:19:53.95 | +26:30:08.7 | A+B | A | ... | ... | 35.28 | 411.40 | 199.89 ± 1.03 | 0.432 ± 0.018 |
| 17199+2629 | OSV 5 | V639Her | J17198+265 | M4.5 V | 17:19:52.70 | +26:30:08.3 | ... | B | 16.860 | 268.9 | 35.52 | 420.84 | 77.24 ± 0.40 | 0.298 ± 0.014 |
| ... | ... | GJ 671 | J17198+417 | M2.5 V | 17:19:53.12 | +41:42:36.4 | ... | ... | ... | ... | 99.79 | 877.18 | 180.28 ± 0.77 | 0.353 ± 0.011 |
| ... | ... | GJ 1216 | J17207+492 | M4.0 V | 17:20:47.19 | +49:15:01.4 | ... | ... | ... | ... | 22.74 | 1296.78 | 39.02 ± 0.17 | 0.190 ± 0.011 |
| ... | ... | GJ 4003 | J17219+214 | M4.0 V | 17:21:54.44 | +21:25:51.3 | ... | ... | ... | ... | 57.00 | 293.77 | 61.49 ± 0.30 | 0.245 ± 0.012 |
| ... | ... | PM J17225+0531 | J17225+055 | M2.5 V | 17:22:33.80 | +05:31:15.1 | ... | ... | ... | ... | 36.19 | 99.21 | 234.22 ± 1.56 | 0.441 ± 0.017 |

Table D.2: Complete sample with the description of multiple systems (continued).

| WDS id | WDS disc | Name | Karmn | Spectral type | $\alpha$ (2016.0) | $\delta$ (2016.0) | System | Component | $\rho$ [arcsec] | $\theta$ [deg] | $\varpi$ [mas] | $\mu_{total}$ [mas a$^{-1}$] | $\mathcal{L}$ [$10^{-4}$ $\mathcal{L}_\odot$] | $\mathcal{M}$ [$\mathcal{M}_\odot$] |
|---|---|---|---|---|---|---|---|---|---|---|---|---|---|---|
| … | … | GJ 4005 | J17242-043 | M2.5 V | 17:24:16.69 | -04:21:54.1 | … | … | … | … | 69.64 | 264.75 | 253.01 ± 1.52 | 0.460 ± 0.018 |
| … | … | GJ 1219 | J17276+144 | M4.0 V | 17:27:38.65 | +14:28:56.4 | … | … | … | … | 69.64 | 1169.53 | 64.43 ± 0.25 | 0.235 ± 0.012 |
| … | … | Wolf 750 | J17285+374 | M3.5 V | 17:28:30.41 | +37:27:04.1 | … | … | … | … | 43.60 | 188.78 | 107.24 ± 0.44 | 0.310 ± 0.014 |
| … | … | Wolf 751 | J17303+055 | M1.0 V | 17:30:22.76 | +05:32:50.7 | … | … | … | … | 26.86 | 251.09 | 521.06 ± 2.99 | 0.563 ± 0.028 |
| … | … | GJ 1220 | J17312+820 | M4.0 V | 17:31:14.95 | +82:05:27.8 | … | … | … | … | 28.44 | 564.93 | 33.06 ± 0.12 | 0.186 ± 0.011 |
| … | … | Wolf 755 | J17316+047 | M1.5 V | 17:31:37.97 | +01:47:48.0 | … | … | … | … | … | 208.41 | 342.86 ± 1.78 | 0.478 ± 0.018 |
| … | … | GJ 4011 | J17321+504 | M2.5 V | 17:32:07.82 | +50:24:41.7 | … | … | … | … | 37.95 | 526.57 | 273.67 ± 2.17 | 0.479 ± 0.019 |
| … | … | G 226-64 | J17328+543 | M2.0 V | 17:32:53.06 | +54:20:19.6 | … | … | … | … | 46.30 | 506.37 | 215.09 ± 0.82 | 0.374 ± 0.015 |
| … | … | V1274Her | J17338+169 | M6.0 V | 17:33:53.03 | +16:55:11.0 | … | … | … | … | 59.00 | 187.78 | 94.16 ± 1.97 | 0.317 ± 0.066 |
| … | … | RX J1734.0+4447A | … | M2.0 V | 17:34:05.43 | +44:47:08.9 | AB | A | … | … | 73.72 | 115.42 | … | 0.381 ± 0.032 |
| 17341+4447 | CRC 74 | RX J1734.0+4447B | J17340+446 | M3.5 V | 17:34:05.47 | +44:47:08.4 | … | B | 0.633 | 146.4 | 7.19 | 94.46 | … | 0.328 ± 0.034 |
| 17350+6153 | BU 962 | HD 160269A | … | G0 IV/V | 17:35:00.16 | +61:52:20.8 | (AB)+C | AB(1) | 0.625 | 300.5 | 80.21 | 522.56 | … | … |
| 17350+6153 | LDS2736 | GJ 685 | J17355+616 | M0.5 V | 17:35:35.07 | +61:40:45.4 | … | C | 738.164 | 160.3 | 33.65 | 577.33 | 589.33 ± 2.28 | 0.579 ± 0.028 |
| … | … | GJ 687 | J17364+683 | M3.0 V | 17:36:24.97 | +68:20:00.6 | … | … | … | … | 33.85 | 1309.76 | 224.17 ± 1.78 | 0.434 ± 0.024 |
| … | … | GJ 4015 | J17376+220 | M4.1 V | 17:37:36.52 | +22:05:45.2 | … | … | … | … | 68.61 | 310.38 | 71.87 ± 0.29 | 0.267 ± 0.013 |
| … | … | GJ 686 | J17378+185 | M1.5 V | 17:37:54.39 | +18:35:45.9 | … | … | … | … | 44.06 | 1351.97 | 298.36 ± 1.47 | 0.429 ± 0.012 |
| 17387+6114 | HOM 1 | HD 160934B | J17386+612 | M4.0 V | 17:38:39.61 | +61:14:16.8 | (AB)+C | AB | 0.020 | 18.4 | 44.13 | 42.26 | … | … |
| 17387+6114 | VVO 17 | GJ 4021 | … | M4.0 V | 17:38:40.87 | +61:14:00.0 | … | C | 19.052 | 151.4 | 44.27 | 49.61 | 109.72 ± 0.47 | 0.336 ± 0.015 |
| … | … | GJ 4016 | J17388+080 | M2.5 V | 17:38:51.16 | +08:01:31.7 | … | … | … | … | 68.95 | 236.37 | 210.10 ± 1.22 | 0.392 ± 0.016 |
| 17397+2746 | LDS 999 | GJ 4018 | J17395+277S | M0.5 V | 17:39:30.73 | +27:45:40.8 | A+(BC) | A | 56.439 | 21.0 | 58.88 | 201.22 | 739.04 ± 3.54 | 0.621 ± 0.027 |
| 17397+2746 | JOD 12 | GJ 4019 | J17395+277N | M3.0 V | 17:39:32.26 | +27:46:33.5 | … | BC | 56.440 | 21.0 | 37.54 | 202.31 | … | 0.450 ± 0.031 |
| … | … | G 204-25 | J17419+407 | M1.5 V | 17:41:57.04 | +40:44:44.3 | … | … | … | … | 39.23 | 271.77 | 633.73 ± 2.22 | 0.583 ± 0.028 |
| … | … | Wolf 1471 | J17421-088 | M3.0 V | 17:42:09.85 | -08:49:07.8 | … | … | … | … | 92.88 | 964.49 | 80.31 ± 0.33 | 0.249 ± 0.012 |
| … | … | 2M17422264+5726521 | J17423+574 | M0.0 V | 17:42:22.63 | +57:26:52.0 | … | … | … | … | 92.90 | 14.84 | 710.85 ± 4.08 | 0.619 ± 0.027 |
| … | … | GJ 690.1 | J17425-166 | M2.5 V | 17:42:32.15 | -16:38:35.9 | … | … | … | … | 34.87 | 699.66 | 103.69 ± 0.44 | 0.286 ± 0.013 |
| 17430+0547 | HDS2506 | GJ 4024 | J17430+057 | M1.0 V | 17:43:01.06 | +05:47:20.5 | (AB) | AB | 0.255 | 99.2 | 51.77 | 257.98 | … | … |
| … | … | G 259-20 | J17431+854 | M2.0 V | 17:43:07.85 | +85:26:25.5 | A+B | A | … | … | 49.63 | 291.24 | 260.89 ± 5.02 | 0.414 ± 0.017 |
| 17432+8526 | LUH 12 | 2M17430860+8526594 | … | L5 | 17:43:07.13 | +85:26:55.2 | … | B | 29.739 | 358.3 | 34.08 | 291.00 | 0.85 ± 0.02 | … |
| … | … | Ross 133 | J17432-185 | M1.5 V | 17:43:17.37 | -18:31:27.7 | … | … | … | … | 33.87 | 574.92 | 386.59 ± 9.76 | 0.509 ± 0.020 |
| … | … | GJ 694 | J17439+433 | M2.5 V | 17:43:55.98 | +43:22:33.3 | … | … | … | … | 19.17 | 603.29 | 251.60 ± 1.26 | 0.458 ± 0.018 |
| … | … | GJ 4026 | J17460+246 | M4.0 V | 17:46:04.22 | +24:39:13.1 | … | … | … | … | 41.45 | 629.21 | 97.04 ± 0.46 | 0.294 ± 0.013 |
| 17456+4651 | JOD 13 | GJ 694.2 | J17455+468 | M0.0 V | 17:45:33.50 | +46:51:19.8 | (AB) | AB | 0.670 | 77.0 | 154.35 | 36.97 | … | 0.593 ± 0.028 |
| 17465+2743 | TRN 2 | HD 161797 | … | G5 IV | 17:46:27.17 | +27:43:02.2 | (AB)+C+D | AB | 1.780 | 254.8 | 127.48 | 833.79 | … | … |
| 17465+2743 | STF2220 | GJ 695 B | … | M2.5 V | 17:46:24.69 | +27:42:49.5 | … | C | 35.364 | 248.9 | 41.56 | 899.45 | … | 0.422 ± 0.031 |
| 17465+2743 | AC 7 | GJ 695 C | J17464+277 | M3.5 V | 17:46:24.65 | +27:42:49.8 | … | D | 35.759 | 249.7 | 34.60 | 719.40 | … | 0.375 ± 0.032 |
| … | … | StKM 1-1528 | J17469+228 | M1.0 V | 17:46:55.96 | +22:48:01.6 | … | … | … | … | 27.90 | 93.31 | 570.74 ± 2.15 | 0.575 ± 0.028 |
| … | … | PM J17481+1558 | J17481+159 | M3.0 V | 17:48:11.19 | +15:58:48.4 | … | … | … | … | 55.66 | 119.73 | 163.47 ± 0.69 | 0.365 ± 0.015 |
| … | … | GJ 4030 | J17502+237 | M3.3 V | 17:50:14.74 | +23:45:58.1 | … | … | … | … | 27.90 | 563.73 | 124.77 ± 0.67 | 0.337 ± 0.015 |
| … | … | GJ 4031 | J17515+147 | M3.0 V | 17:51:30.58 | +14:45:32.4 | … | … | … | … | 232.14 | 215.46 | 195.83 ± 0.99 | 0.428 ± 0.017 |
| … | … | G 227-20 | J17521+647 | M0.5 V | 17:52:11.83 | +64:46:02.9 | … | … | … | … | 88.82 | 347.39 | 730.57 ± 2.63 | 0.623 ± 0.027 |

Table D.2: Complete sample with the description of multiple systems (continued).

| WDS id | WDS disc | Name | Karmn | Spectral type | $\alpha$ (2016.0) | $\delta$ (2016.0) | System | Component | $\rho$ [arcsec] | $\theta$ [deg] | $\varpi$ [mas] | $\mu_{total}$ [mas a$^{-1}$] | $\mathcal{L}$ [$10^{-4}$ $\mathcal{L}_\odot$] | $\mathcal{M}$ [$\mathcal{M}_\odot$] |
|---|---|---|---|---|---|---|---|---|---|---|---|---|---|---|
| … | … | Wolf 1473 | J17464-087 | M3.5 V | 17:46:29.28 | -08:42:43.4 | Aab | Aab(2) | … | … | 34.47 | 430.41 | … | 0.396 ± 0.002 |
| … | … | GJ 4032 A | J17530+169 | M3.0 V | 17:53:00.32 | +16:54:59.5 | AB | A | … | … | 59.46 | 331.97 | … | 0.319 ± 0.034 |
| 17530+1655 | CRC 27 | GJ 4032 B | … | M4.0 V | 17:53:00.38 | +16:54:59.2 | … | B | 0.852 | 112.4 | 54.02 | 313.37 | … | 0.279 ± 0.035 |
| … | … | GJ 1222 | J17542+073 | M4.0 V | 17:54:16.48 | +07:22:40.2 | … | … | … | … | 64.90 | 664.62 | 103.80 ± 0.75 | 0.303 ± 0.011 |
| … | … | LSPM J1754+1251 | J17547+128 | M2.0 V | 17:54:43.24 | +12:51:18.6 | … | … | … | … | 33.21 | 153.20 | 214.77 ± 1.25 | 0.374 ± 0.015 |
| … | … | GJ 4038 | J17570+157 | M3.0 V | 17:57:03.33 | +15:46:39.3 | AB | A | … | … | 44.16 | 302.88 | … | 0.349 ± 0.033 |
| 17571+1547 | MCT 10 | G 183-13B | … | M3.0 V | 17:57:03.41 | +15:46:39.1 | … | B | 1.221 | 101.5 | 67.29 | 291.04 | … | 0.356 ± 0.033 |
| … | … | LP 44-162 | J17572+707 | M7.5 V | 17:57:15.43 | +70:42:06.4 | … | … | … | … | 67.35 | 326.76 | 11.99 ± 0.14 | 0.171 ± 0.011 |
| … | … | Barnard's Star | J17578+046 | M3.5 V | 17:57:47.64 | +04:44:21.9 | … | … | … | … | 66.16 | 10393.35 | 35.91 ± 0.28 | 0.175 ± 0.009 |
| … | … | GJ 4040 | J17578+465 | M2.5 V | 17:57:50.94 | +46:35:28.4 | AB | A | … | … | 48.37 | 578.34 | 206.46 ± 0.86 | 0.387 ± 0.012 |
| 17578+4635 | BDK9 | G 204-39B | … | T7 | … | … | … | B | 197.004 | 130.6 | 88.69 | 594.57 | … | … |
| … | … | LP 24-152 | J17589+807 | M3.5 V | 17:58:55.00 | +80:42:53.9 | … | … | … | … | 69.82 | 668.97 | 372.51 ± 24.02 | 0.459 ± 0.031 |
| … | … | Wolf 1403 | J18010+508 | M1.5 V | 18:01:05.86 | +50:49:38.9 | … | … | … | … | 46.57 | 211.58 | 608.52 ± 2.55 | 0.584 ± 0.028 |
| … | … | G 182-34 | J18012+355 | M3.5 Ve | 18:01:16.14 | +35:35:41.4 | … | … | … | … | 32.24 | 572.26 | 168.87 ± 0.65 | 0.396 ± 0.016 |
| … | … | LP 71-82 | J18022+642 | M5.0 V | 18:02:17.09 | +64:15:38.1 | … | … | … | … | 63.37 | 431.12 | 32.26 ± 0.14 | 0.169 ± 0.010 |
| … | … | GJ 1223 | J18027+375 | dM5.0 | 18:02:46.49 | +37:30:44.7 | … | … | … | … | 84.01 | 1154.51 | 24.22 ± 0.10 | 0.144 ± 0.009 |
| … | … | Wolf 792 | J18031+179 | M1.0 V | 18:03:06.02 | +17:54:20.0 | … | … | … | … | 57.99 | 352.27 | 408.48 ± 8.02 | 0.495 ± 0.019 |
| … | … | Ross 820 | J18037+247 | M0.0 V | 18:03:47.68 | +25:45:16.7 | … | … | … | … | 52.65 | 234.74 | 822.30 ± 4.61 | 0.648 ± 0.027 |
| 18043+3557 | JOD 14 | GJ 4041 | J18042+359 | M0.5 V | 18:04:17.70 | +35:57:21.5 | (AB) | AB | 0.297 | 339.0 | 55.10 | 276.68 | … | 0.587 ± 0.028 |
| … | … | HD 165222 | J18051-030 | M0.0 V | 18:05:08.19 | -03:01:58.1 | … | … | … | … | 95.52 | 659.78 | 330.03 ± 1.99 | 0.441 ± 0.011 |
| … | … | NLTT 46021 | J18063+728 | M0.0 V | 18:06:18.11 | +72:49:20.1 | … | … | … | … | 65.75 | 236.68 | 733.44 ± 2.58 | 0.630 ± 0.027 |
| … | … | Wolf 806 | J18074+184 | M1.0 V | 18:07:27.99 | +18:27:54.1 | … | … | … | … | 29.68 | 238.82 | 592.35 ± 2.53 | 0.573 ± 0.028 |
| … | … | GJ 1224 | J18075-159 | M4.0 Ve | 18:07:32.15 | -15:57:52.6 | … | … | … | … | 49.80 | 709.11 | 28.55 ± 0.17 | 0.157 ± 0.009 |
| … | … | LP 334-11 | J18096+318 | M1.0 V | 18:09:40.77 | +31:52:16.0 | … | … | … | … | … | 201.53 | 391.05 ± 1.30 | 0.484 ± 0.018 |
| … | … | Wolf 1412 | J18103+512 | M2.0 V | 18:10:23.37 | +51:15:55.0 | … | … | … | … | 161.41 | 162.19 | 205.63 ± 1.05 | 0.365 ± 0.015 |
| … | … | StKM 1-1582 | J18109+220 | M0.0 V | 18:10:56.18 | +22:01:31.2 | … | … | … | … | 153.88 | 41.29 | 1491.76 ± 7.71 | 0.754 ± 0.025 |
| … | … | LP 569-163 | J18116+061 | M3.0 V | 18:11:36.49 | +06:06:27.9 | (AB) | A | … | … | 153.97 | 118.07 | … | … |
| 18116+0606 | CFN 14 | LP 569-163 B | … | … | 18:11:36.51 | +06:06:27.3 | … | B | 0.629 | 144.2 | 114.92 | … | … | … |
| … | … | V1334Her | J18131+260 | M4.0 V | 18:13:06.83 | +26:01:51.3 | … | … | … | … | 37.84 | 224.18 | 117.10 ± 0.61 | 0.332 ± 0.017 |
| … | … | LP 569-15 | J18134+054 | M1.5 V | 18:13:28.12 | +05:26:54.8 | A+B | A | … | … | 27.27 | 238.09 | 632.76 ± 3.58 | 0.586 ± 0.028 |
| 18136+0527 | LDS1007 | LP 569-16 | J18135+055 | M4.0 V | 18:13:33.09 | +05:32:08.4 | … | B | 322.188 | 13.3 | … | 238.14 | 167.43 ± 0.89 | 0.394 ± 0.016 |
| … | … | HD 348274 | J18157+189 | M0.0 V | 18:15:43.54 | +18:56:12.7 | … | … | … | … | 43.23 | 429.87 | 690.14 ± 7.95 | 0.605 ± 0.027 |
| … | … | GJ 708.2 | J18160+139 | M0.0 V | 18:16:02.36 | +13:54:40.2 | … | … | … | … | 29.78 | 510.75 | 662.36 ± 3.16 | 0.613 ± 0.027 |
| … | … | GJ 708.3 | J18163+015 | M3.0 V | 18:16:17.74 | +01:31:17.1 | … | … | … | … | 55.75 | 757.81 | 117.98 ± 0.81 | 0.306 ± 0.014 |
| … | … | G 140-51 | J18165+048 | M5.0 V | 18:16:31.37 | +04:52:52.3 | … | … | … | … | 55.77 | 435.77 | 36.49 ± 0.18 | 0.178 ± 0.009 |
| … | … | GJ 709 | J18165+455 | M0.5 V | 18:16:31.07 | +45:33:33.7 | … | … | … | … | 55.72 | 337.61 | 620.48 ± 2.60 | 0.592 ± 0.028 |
| … | … | V401Dra | J18174+483 | dM2.0 | 18:17:25.05 | +48:22:03.1 | … | … | … | … | 96.23 | 66.55 | 447.04 ± 2.36 | 0.510 ± 0.029 |
| … | … | GJ 4048 | J18180+387E | M3.0 V | 18:18:03.73 | +38:46:16.2 | AB | A | … | … | 56.20 | 1089.90 | 114.55 ± 0.41 | 0.284 ± 0.010 |
| 18180+3846 | GIC 151 | GJ 4049 | J18180+387W | M4.0 V | 18:18:02.88 | +38:46:17.4 | … | B | 9.971 | 277.4 | … | 1116.56 | … | 0.204 ± 0.038 |
| … | … | GJ 4053 | J18189+661 | M4.5 Ve | 18:18:58.38 | +66:11:26.2 | … | … | … | … | 37.16 | 620.58 | 24.68 ± 0.14 | 0.150 ± 0.010 |

Table D.2: Complete sample with the description of multiple systems (continued).

| WDS id | WDS disc | Name | Karmn | Spectral type | $\alpha$ (2016.0) | $\delta$ (2016.0) | System | Component | $\rho$ [arcsec] | $\theta$ [deg] | $\varpi$ [mas] | $\mu_{\rm total}$ [mas a$^{-1}$] | $\mathcal{L}$ [$10^{-4}$ $\mathcal{L}_\odot$] | $\mathcal{M}$ [$\mathcal{M}_\odot$] |
|---|---|---|---|---|---|---|---|---|---|---|---|---|---|---|
| … | … | GJ 4051 | J18193-057 | M2.0 V | 18:19:21.68 | -05:46:33.5 | … | … | … | … | 50.77 | 544.91 | 252.78 ± 1.19 | 0.432 ± 0.017 |
| … | … | PM J18195+4201 | J18195+420 | M1.5 V | 18:19:34.48 | +42:01:37.4 | … | … | … | … | 50.95 | 59.88 | 554.43 ± 2.06 | 0.557 ± 0.028 |
| … | … | GJ 1226 A | J18209-010 | M3.5 V | 18:20:56.62 | -01:03:14.2 | AB | A | … | … | 44.84 | 1094.63 | … | 0.285 ± 0.035 |
| 18210-0101 | WOR 35 | GJ 1226 B | … | M4.0 V | 18:20:56.63 | -01:03:12.8 | … | B | 1.435 | 6.1 | 33.31 | 1107.99 | … | 0.270 ± 0.036 |
| … | … | Ross 136 | J18221+063 | M4.0 V | 18:22:05.40 | +06:20:39.4 | … | … | … | … | 51.06 | 1195.56 | 109.60 ± 0.63 | 0.295 ± 0.011 |
| … | … | GJ 1227 | J18224+620 | M4.0 V | 18:22:24.92 | +62:02:42.0 | … | … | … | … | 50.55 | 1561.71 | 32.78 ± 0.17 | 0.161 ± 0.010 |
| … | … | G 205-19 | J18227+379 | M1.0 V | 18:22:43.47 | +37:57:41.1 | … | … | … | … | 100.93 | 401.06 | 793.86 ± 13.72 | 0.619 ± 0.027 |
| … | … | Ross 708A | J18234+281 | M3.5 V | 18:23:28.24 | +28:10:01.0 | AB | A | … | … | 54.27 | 199.56 | … | 0.320 ± 0.034 |
| 18235+2810 | JOD 15 | Ross 708B | … | M5.5 V | 18:23:28.25 | +28:09:59.7 | … | B | 1.264 | 175.3 | 54.57 | 216.99 | … | 0.158 ± 0.041 |
| … | … | GJ 4056 | J18240+016 | M2.5 V | 18:24:05.35 | +01:41:12.3 | … | … | … | … | 60.73 | 281.95 | 221.36 ± 1.04 | 0.403 ± 0.016 |
| 18249+2817 | RAO 391 | Ross 710 | J18248+282 | M1.5 V | 18:24:52.32 | +28:17:22.3 | (AB) | AB | 0.143 | 313.3 | 95.57 | 189.54 | … | 0.573 ± 0.028 |
| … | … | HD 336196 | J18250+246 | M0.0 V | 18:25:04.75 | +24:37:57.3 | … | … | … | … | 95.56 | 450.44 | 660.88 ± 4.68 | 0.603 ± 0.027 |
| … | … | GJ 4058 | J18255+383 | M0.0 V | 18:25:31.82 | +38:21:00.8 | … | … | … | … | 28.75 | 749.11 | 450.70 ± 1.48 | 0.543 ± 0.029 |
| … | … | GJ 4059 | J18264+113 | M3.5 V | 18:26:24.58 | +11:20:52.9 | A+B | A | … | … | 28.84 | 276.07 | … | 0.423 ± 0.031 |
| 18264+1121 | NI 38 | LSPM J1826+1120S | … | DA | 18:26:24.42 | +11:20:45.1 | … | B | 8.196 | 196.8 | 70.97 | 281.07 | … | 0.500 ± 0.100 |
| … | … | TYC 4222-2195-1 | J18292+638 | M1.5 V | 18:29:13.36 | +63:51:10.9 | … | … | … | … | 26.18 | 217.06 | 467.23 ± 14.51 | 0.536 ± 0.029 |
| … | … | 2M18294012+3350130 | J18296+338 | M3.0 V | 18:29:40.02 | +33:50:13.3 | … | … | … | … | 82.00 | 75.41 | 254.45 ± 0.99 | 0.491 ± 0.019 |
| … | … | LP 570-92 | J18312+068 | M1.0 V | 18:31:16.21 | +06:50:08.2 | … | … | … | … | 54.33 | 182.49 | 430.36 ± 2.13 | 0.509 ± 0.019 |
| … | … | GJ 4062 | J18319+406 | M3.5 V | 18:31:58.24 | +40:41:17.6 | … | … | … | … | 131.60 | 423.15 | 160.86 ± 0.61 | 0.342 ± 0.011 |
| … | … | GJ 4063 | J18346+401 | M4.0 V | 18:34:36.73 | +40:07:23.1 | … | … | … | … | 42.77 | 214.33 | 218.01 ± 0.93 | 0.413 ± 0.016 |
| … | … | G 184-13A | J18352+243 | M2.5 V | 18:35:13.42 | +24:18:39.2 | AB | A | … | … | 42.77 | 294.62 | … | 0.415 ± 0.031 |
| 18352+2419 | KPP3331 | G 184-13B | … | M3.5 V | 18:35:13.44 | +24:18:41.7 | … | B | 2.466 | 4.5 | 167.29 | 302.61 | … | 0.331 ± 0.034 |
| … | … | GJ 4064 | J18352+414 | M2.0 V | 18:35:18.38 | +41:29:14.5 | … | … | … | … | … | 376.89 | 219.65 ± 0.91 | 0.378 ± 0.015 |
| … | … | GJ 720 A | J18353+457 | M0.5 V | 18:35:19.08 | +45:44:44.4 | A+B | A | … | … | 83.07 | 580.29 | 700.35 ± 2.75 | 0.617 ± 0.027 |
| 18355+4546 | LDS6329 | GJ 720 B | J18354+457 | M2.5 V | 18:35:27.99 | +45:45:46.7 | … | B | 112.139 | 56.2 | 98.19 | 582.11 | 93.70 ± 0.52 | 0.288 ± 0.013 |
| … | … | LSPM J1835+3259 | J18356+329 | M8.5 V | 18:35:37.79 | +32:59:41.2 | … | … | … | … | … | 758.63 | 2.93 ± 0.01 | 0.114 ± 0.054 |
| … | … | GJ 4068 | J18358+800 | M4.0 Ve | 18:35:52.58 | +80:05:42.9 | … | … | … | … | 33.70 | 244.24 | 97.33 ± 0.37 | 0.315 ± 0.014 |
| … | … | G 227-39 | J18362+567 | M0.0 V | 18:36:12.72 | +56:44:37.8 | … | … | … | … | 68.30 | 403.55 | 1038.72 ± 5.01 | 0.698 ± 0.026 |
| … | … | Ross 149 | J18363+136 | M4.0 V | 18:36:19.43 | +13:36:30.8 | … | … | … | … | 48.48 | 331.47 | 104.77 ± 0.49 | 0.300 ± 0.011 |
| 18388+0446 | RAO 395 | LP 570-22 | J18387+047 | M0.5 V | 18:38:47.63 | +04:46:01.7 | (AB) | AB | 0.427 | 153.3 | 73.06 | 187.09 | … | 0.523 ± 0.031 |
| … | … | RX J1839.4+6903 | J18394+690 | M2.0 Ve | 18:39:25.71 | +69:03:06.8 | … | … | … | … | 73.38 | 200.86 | 684.85 ± 4.04 | 0.581 ± 0.028 |
| … | … | LP 335-12 | J18395+298 | M6.5 Ve | 18:39:33.18 | +29:52:12.9 | … | … | … | … | 55.02 | 238.81 | 6.98 ± 0.04 | 0.113 ± 0.010 |
| … | … | LP 335-13 | J18395+301 | M0.0 V | 18:39:32.00 | +30:09:52.1 | … | … | … | … | 66.13 | 192.79 | 667.98 ± 2.19 | 0.619 ± 0.027 |
| … | … | GJ 4067 | J18399+334 | M3.5 V | 18:39:59.94 | +33:24:58.9 | … | … | … | … | 80.84 | 289.65 | 157.05 ± 0.66 | 0.381 ± 0.016 |
| … | … | LP 44-334 | J18400+726 | M6.5 V | 18:40:02.20 | +72:40:57.2 | AB | A | … | … | 36.77 | 187.35 | … | 0.107 ± 0.047 |
| 18400+7241 | CFN 15 | LP 44-334 B | … | … | 18:40:02.32 | +72:40:56.6 | … | B | 0.832 | 140.8 | 36.93 | … | … | … |
| 18387-1429 | HDS2641 | GJ 2138 | J18387-144 | M2.5 V | 18:38:44.87 | -14:29:35.1 | (AB) | AB | 0.107 | 358.0 | 47.05 | 579.86 | … | 0.418 ± 0.031 |
| … | … | Wolf 1466 | J18402-104 | M1.0 V | 18:40:17.68 | -10:28:04.1 | … | … | … | … | 45.03 | 574.97 | 247.66 ± 0.97 | 0.403 ± 0.016 |
| … | … | G 227-43 | J18405+595 | M2.0 V | 18:40:35.77 | +59:30:53.6 | … | … | … | … | 92.99 | 309.24 | 375.34 ± 1.68 | 0.492 ± 0.030 |
| … | … | BD+31 3330 | … | K2.5 V | 18:40:54.99 | +31:31:45.7 | A+BC | A | … | … | 92.96 | 841.17 | 2351.22 ± 7.15 | 0.820 ± 0.123 |

Table D.2: Complete sample with the description of multiple systems (continued).

| WDS id | WDS disc | Name | Karmn | Spectral type | $\alpha$ (2016.0) | $\delta$ (2016.0) | System | Component | $\rho$ [arcsec] | $\theta$ [deg] | $\varpi$ [mas] | $\mu_{total}$ [mas a$^{-1}$] | $\mathcal{L}$ [$10^{-4}$ $\mathcal{L}_\odot$] | $\mathcal{M}$ [$\mathcal{M}_\odot$] |
|---|---|---|---|---|---|---|---|---|---|---|---|---|---|---|
| 18409+3132 | HJ1337 | BD+31 3330B | J18409+315 | M1.0 V | 18:40:55.33 | +31:31:37.3 | … | B | 9.446 | 152.2 | 81.74 | … | … | … |
| … | … | BD+31 3330C | … | … | 18:40:55.31 | +31:31:37.5 | … | C | 9.210 | 153.1 | 55.69 | … | … | … |
| … | … | Ross 720 | J18409-133 | dM1.0 | 18:40:57.20 | -13:22:57.3 | … | … | … | … | 62.91 | 677.86 | 506.17 ± 3.41 | 0.547 ± 0.028 |
| … | … | GJ 1230 A | J18411+247S | M4.5 V | 18:41:10.35 | +24:47:15.8 | Aab+B | Aab(2) | … | … | … | 512.79 | … | 0.463 ± 0.001 |
| 18411+2447 | LDS6330 | GJ 1230 B | J18411+247N | M4.5 Ve | 18:41:10.39 | +24:47:20.5 | … | B | 4.725 | 5.6 | 40.89 | 501.94 | … | 0.145 ± 0.042 |
| … | … | GJ 4069 | J18416+397 | M4.0 V | 18:41:37.04 | +39:42:09.3 | … | … | … | … | 48.52 | 315.35 | 44.23 ± 0.20 | 0.204 ± 0.011 |
| … | … | Ross 145 | J18419+318 | M3.0 V | 18:41:58.66 | +31:49:49.9 | … | … | … | … | 54.10 | 300.44 | 165.64 ± 0.63 | 0.341 ± 0.011 |
| … | … | V816Her | J18427+139 | M5.0 V | 18:42:44.94 | +13:54:22.4 | … | … | … | … | 49.25 | 355.60 | 74.44 ± 0.38 | 0.272 ± 0.013 |
| 18428+5938 | STF2398 | HD 173739 | J18427+596N | M3.0 V | 18:42:43.94 | +59:38:18.1 | AB | A | 11.400 | 180.8 | 98.84 | 2221.02 | 152.45 ± 2.35 | 0.336 ± 0.012 |
| … | … | HD 173740 | J18427+596S | M3.5 V | 18:42:43.94 | +59:38:06.5 | … | B | 11.565 | 179.9 | 77.74 | 2330.18 | 90.22 ± 0.31 | 0.266 ± 0.012 |
| … | … | TYC 2112-920-1 | J18432+253 | M1.0 V | 18:43:14.51 | +25:22:43.1 | … | … | … | … | 40.44 | 215.40 | 528.52 ± 2.47 | 0.562 ± 0.028 |
| … | … | V492Lyr | J18433+406 | M7.5 Ve | 18:43:21.96 | +40:40:30.8 | (AB) | AB | … | … | 36.26 | 603.49 | … | 0.101 ± 0.048 |
| … | … | 1R184510.6+062016 | J18451+063 | M1.0 Ve | 18:45:10.23 | +06:20:14.7 | … | … | … | … | 43.87 | 86.44 | 493.76 ± 2.21 | 0.545 ± 0.028 |
| … | … | G 184-24 | J18453+188 | M4.0 V | 18:45:22.79 | +18:51:54.3 | … | … | … | … | 60.91 | 297.39 | 55.06 ± 0.34 | 0.231 ± 0.012 |
| … | … | GJ 4077 | J18480-145 | dM2.5 | 18:48:01.01 | -14:34:55.0 | … | … | … | … | 35.56 | 334.96 | 188.75 ± 0.87 | 0.361 ± 0.012 |
| … | … | G 141-36 | J18482+076 | M5.0 V | 18:48:17.94 | +07:41:25.2 | … | … | … | … | 37.64 | 450.85 | 21.83 ± 0.11 | 0.139 ± 0.009 |
| … | … | LSPM J1848+6135 | J18487+615 | M2.0 V | 18:48:47.07 | +61:35:06.3 | … | … | … | … | 69.28 | 174.31 | 223.68 ± 0.94 | 0.405 ± 0.016 |
| … | … | V1216Sgr | J18498-238 | M3.5 Ve | 18:49:50.11 | -23:50:13.6 | … | … | … | … | 69.89 | 668.14 | 40.79 ± 0.27 | 0.177 ± 0.009 |
| … | … | G 184-31 | J18499+186 | M4.5 V | 18:49:54.34 | +18:40:24.2 | … | … | … | … | 219.79 | 299.35 | 34.36 ± 0.16 | 0.190 ± 0.011 |
| … | … | Ross 142 | J18500+030 | M0.5 V | 18:50:00.63 | +03:05:10.7 | … | … | … | … | 48.70 | 448.36 | 614.37 ± 3.10 | 0.590 ± 0.028 |
| … | … | GJ 4083 | J18507+479 | M3.5 V | 18:50:45.66 | +47:58:17.3 | … | … | … | … | 122.55 | 281.90 | 187.07 ± 0.74 | 0.392 ± 0.016 |
| … | … | BD+02 3698 | J18515+027 | M0.5 V | 18:51:35.59 | +02:46:19.2 | … | … | … | … | 31.44 | 326.23 | 933.18 ± 6.23 | 0.669 ± 0.026 |
| … | … | GJ 4084 | J18516+244 | M3.0 V | 18:51:40.44 | +24:27:31.7 | … | … | … | … | 30.95 | 357.42 | 128.17 ± 0.76 | 0.320 ± 0.014 |
| … | … | HD 229793 | J18518+165 | M0.0 V | 18:51:50.93 | +16:34:52.1 | … | … | … | … | 49.47 | 532.16 | 536.53 ± 2.37 | 0.568 ± 0.028 |
| … | … | 1R185200.0+130005 | J18519+130 | M2.0 Ve | 18:51:59.66 | +13:00:01.1 | … | … | … | … | 35.98 | 157.31 | 622.29 ± 10.42 | 0.566 ± 0.028 |
| … | … | G 141-46 | J18534+028 | M2.5 V | 18:53:25.68 | +02:50:49.4 | … | … | … | … | 35.95 | 299.46 | 168.64 ± 2.71 | 0.349 ± 0.015 |
| … | … | BD+00 4050 | J18548+008 | M0.0 V | 18:54:53.14 | +00:51:44.8 | … | … | … | … | 26.55 | 145.01 | 822.44 ± 3.73 | 0.649 ± 0.027 |
| … | … | HD 230017A | … | M0.0 V | 18:54:53.67 | +10:58:42.4 | A+B+C | A | … | … | 46.82 | 133.27 | 1266.03 ± 10.71 | 0.730 ± 0.025 |
| 18550+1058 | VYS 8 | HD 230017B | J18548+109 | M3.5 V | 18:54:53.86 | +10:58:45.1 | … | B | 3.792 | 45.1 | 12.31 | 89.29 | … | 0.364 ± 0.033 |
| … | … | PM J18542+1058 | … | M4.0 V | 18:54:17.14 | +10:58:11.0 | … | C* | 538.890 | 266.7 | 49.23 | 114.61 | 89.09 ± 0.57 | 0.300 ± 0.014 |
| … | … | GJ 735 | J18554+084 | M3.0 V | 18:55:27.51 | +08:24:07.9 | Aab | Aab(2) | … | … | … | 114.60 | … | … |
| … | … | MCC 806A | … | … | 18:56:16.00 | +54:31:42.3 | AB+C | A | … | … | 47.83 | … | … | … |
| 18563+5432 | HDS2682 | MCC 806B | … | … | 18:56:15.98 | +54:31:42.4 | … | B | 0.232 | 296.8 | 45.26 | … | … | … |
| 18563+5432 | GIC154 | GJ 4091 B | J18563+544 | M2.0 V | 18:56:18.29 | +54:29:45.5 | … | C(2) | 118.495 | 170.3 | 64.73 | 349.10 | … | … |
| … | … | GJ 4089 | J18564+463 | M4.1 V | 18:56:26.40 | +46:22:58.5 | … | … | … | … | 45.14 | 309.79 | 75.77 ± 0.30 | 0.275 ± 0.013 |
| … | … | GJ 4088 | J18571+075 | M2.0 V | 18:57:10.40 | +07:34:14.7 | … | … | … | … | 34.13 | 191.40 | 485.85 ± 2.64 | 0.548 ± 0.028 |
| … | … | G 229-20A | J18576+535 | M3.5 V | 18:57:38.42 | +53:31:14.4 | AB+C | A | … | … | 104.91 | 261.44 | … | 0.368 ± 0.033 |
| 18576+5331 | LDS4802 | LP 141-13 | … | M3.0 V | 18:57:38.34 | +53:31:12.2 | … | B | 2.290 | 197.0 | 66.60 | 245.56 | … | 0.359 ± 0.033 |
| 18576+5331 | LDS4802 | LP 141-14 | … | DC | 18:57:39.78 | +53:30:32.5 | … | C | 43.653 | 163.9 | 46.53 | 246.40 | … | 0.500 ± 0.100 |
| … | … | HD 176029 | J18580+059 | M1.0 V | 18:57:59.93 | +05:54:09.7 | … | … | … | … | 119.92 | 1236.30 | 666.70 ± 2.97 | 0.604 ± 0.027 |

Table D.2: Complete sample with the description of multiple systems (continued).

| WDS id | WDS disc | Name | Karmn | Spectral type | $\alpha$ (2016.0) | $\delta$ (2016.0) | System | Component | $\rho$ [arcsec] | $\theta$ [deg] | $\varpi$ [mas] | $\mu_{\rm total}$ [mas a$^{-1}$] | $\mathcal{L}$ [$10^{-4}$ $\mathcal{L}_\odot$] | $\mathcal{M}$ [$\mathcal{M}_\odot$] |
|---|---|---|---|---|---|---|---|---|---|---|---|---|---|---|
| … | … | GJ 4092 | J18596+079 | M0.5 V | 18:59:39.00 | +07:59:11.2 | … | … | … | … | 119.89 | 408.58 | 884.16 ± 6.31 | 0.660 ± 0.026 |
| … | … | GJ 4093 | J19025+704 | M2.5 V | 19:02:30.83 | +70:25:53.6 | … | … | … | … | 120.40 | 215.31 | 225.54 ± 1.33 | 0.433 ± 0.017 |
| … | … | LSPM J1902+7525 | J19025+754 | M1.5 V | 19:02:32.79 | +75:25:06.6 | … | … | … | … | 31.33 | 204.59 | 345.93 ± 1.24 | 0.454 ± 0.017 |
| … | … | G 141-57 | J19032+034 | M3.0 V | 19:03:13.40 | +03:24:01.4 | … | … | … | … | 36.01 | 205.36 | 241.87 ± 1.17 | 0.449 ± 0.018 |
| 19033+6400 | JOD 16 | GJ 4094 | J19032+639 | M3.5 V | 19:03:17.46 | +63:59:35.9 | (AB)+C | AB | 0.236 | 141.0 | 46.83 | 63.09 | … | … |
| 19033+6400 | JOD 16 | StKM 1-1676B | … | M2.0 V | 19:03:17.96 | +63:59:37.4 | … | C | 3.578 | 65.1 | 31.63 | 144.33 | … | 0.455 ± 0.030 |
| … | … | GJ 741 | J19032-135 | M4.0 V | 19:03:16.10 | -13:34:12.9 | … | … | … | … | 33.64 | 695.97 | 33.87 ± 0.15 | 0.189 ± 0.011 |
| … | … | 1R190405.9+211030 | J19041+211 | M2.0 Ve | 19:04:06.24 | +21:10:32.7 | (AB) | A | … | … | 76.54 | 131.32 | … | 0.544 ± 0.029 |
| … | … | PM J19041+2110 B | … | … | 19:04:06.24 | +21:10:33.2 | … | B* | 0.526 | 353.4 | 48.23 | … | … | … |
| … | … | LSPM J1907+5905 | J19044+590 | M3.0 V | 19:07:24.98 | +59:05:11.9 | AB | A | … | … | 47.49 | 175.37 | … | 0.434 ± 0.031 |
| 19074+5905 | KPP3335 | LSPM J1907+5905B | … | M6.0 V | 19:07:24.72 | +59:05:14.7 | … | B | 3.394 | 323.7 | 62.16 | 166.16 | … | 0.132 ± 0.043 |
| … | … | TYC 2122-1204-1 | J19045+240 | M2.0 V | 19:04:31.24 | +24:01:54.8 | … | … | … | … | 48.35 | 147.85 | 365.16 ± 1.70 | 0.484 ± 0.030 |
| … | … | HD 349726 | J19072+208 | M2.0 V | 19:07:12.66 | +20:52:31.9 | A+B | A | … | … | 47.15 | 584.53 | 145.15 ± 0.83 | 0.311 ± 0.010 |
| 19072+2053 | LDS1017 | Ross 730 | J19070+208 | M2.0 V | 19:07:05.02 | +20:53:11.4 | … | B | 114.053 | 290.2 | 47.03 | 592.12 | 141.07 ± 0.94 | 0.309 ± 0.010 |
| … | … | GJ 1231 | J19082+265 | M5.0 V | 19:08:15.55 | +26:34:57.6 | … | … | … | … | 52.25 | 537.21 | 34.33 ± 0.18 | 0.190 ± 0.011 |
| … | … | GJ 4098 | J19084+322 | M3.0 V | 19:08:29.67 | +32:16:48.0 | … | … | … | … | 546.98 | 300.16 | 153.99 ± 0.62 | 0.332 ± 0.011 |
| 19074+3230 | KUI 90 | GJ 747 | J19077+325 | M3.0 V | 19:07:44.54 | +32:32:59.3 | (AB) | AB | 0.135 | 248.4 | 71.50 | 1678.52 | … | 0.288 ± 0.035 |
| 19093+3912 | JOD 17 | GJ 4099 | J19093+382 | M2.0 V | 19:09:19.01 | +39:12:00.7 | (AB)+C | AB | 0.299 | 162.5 | 73.75 | 210.22 | … | 0.605 ± 0.027 |
| 19093+3912 | FMR 144 | LSPM J1909+3910 | J19095+391 | M2.0 V | 19:09:31.54 | +39:10:48.4 | … | C | 162.605 | 116.4 | 31.23 | 213.14 | 334.29 ± 1.43 | 0.501 ± 0.019 |
| … | … | Ross 727 | J19093-147 | M2.5 V | 19:09:20.08 | -14:45:02.4 | … | … | … | … | 27.05 | 501.02 | 282.20 ± 1.78 | 0.458 ± 0.018 |
| … | … | GJ 1232 | J19098+176 | M4.0 V | 19:09:50.16 | +17:39:59.6 | … | … | … | … | 33.26 | 768.23 | 44.00 ± 0.22 | 0.193 ± 0.010 |
| … | … | GJ 4100 | J19106+015 | M1.5 V | 19:10:38.38 | +01:32:07.3 | A+B | A | … | … | 128.31 | 255.58 | 513.69 ± 2.64 | 0.546 ± 0.028 |
| 19106+0132 | GWP 2820 | GJ 4100 B | … | M4.5 Ve | 19:10:37.90 | +01:32:18.7 | … | B | 13.516 | 327.8 | 81.74 | 255.11 | … | 0.135 ± 0.043 |
| … | … | TYC 471-1564-1 | J19116+050 | M1.0 V | 19:11:47.89 | +05:00:37.4 | … | … | … | … | 37.63 | 126.80 | 548.80 ± 2.56 | 0.572 ± 0.028 |
| … | … | GJ 4105 | J19124+355 | M2.0 V | 19:12:29.70 | +35:33:50.2 | … | … | … | … | 35.97 | 265.65 | 209.22 ± 0.96 | 0.391 ± 0.016 |
| 19121+0254 | AST1 | Wolf 1062 | J19122+028 | M3.5 V | 19:12:16.50 | +02:53:02.6 | (AB) | AB | 0.192 | 333.0 | 46.25 | 1846.10 | … | 0.371 ± 0.033 |
| … | … | Ross 733 | J19146+193N | M4.5 V | 19:14:38.46 | +19:19:10.7 | A+B | A | … | … | 129.22 | 752.72 | 424.44 ± 2.28 | 0.482 ± 0.030 |
| 19147+1918 | LDS1020 | Ross 734 | J19146+193S | M4.0 V | 19:14:38.51 | +19:18:29.9 | … | B | 40.772 | 179.1 | 30.64 | 755.49 | 104.67 ± 0.42 | 0.327 ± 0.015 |
| … | … | V1428Aql | J19169+051N | M2.5 V | 19:16:54.64 | +05:09:46.7 | A+B | A | … | … | 30.89 | 1453.23 | 329.58 ± 2.93 | 0.475 ± 0.013 |
| 19169+0510 | LDS6334 | V1298Aql | J19169+051S | M8.0 V | 19:16:56.97 | +05:08:39.7 | … | B | 75.500 | 152.5 | 125.45 | 1491.52 | 4.29 ± 0.03 | 0.103 ± 0.052 |
| … | … | LSPM J1918+5803 | J19185+580 | M1.0 V | 19:18:30.48 | +58:03:16.6 | AB | A | … | … | 45.88 | 178.63 | … | 0.583 ± 0.028 |
| … | … | G3- 2143230844400445696 | … | M8.0 V | 19:18:30.33 | +58:03:18.1 | … | B* | 1.925 | 323.8 | 43.49 | 190.05 | … | 0.100 ± 0.050 |
| … | … | GJ 754.1 B | J19205-076 | M2.5 V | 19:20:33.38 | -07:39:46.6 | A+B | A | … | … | 22.07 | 191.46 | … | 0.286 ± 0.035 |
| 19206-0740 | LDS 678 | GJ 754.1 A | … | DBQA5 | 19:20:34.86 | -07:40:02.7 | … | B | 27.168 | 126.2 | … | 173.04 | … | 0.500 ± 0.100 |
| … | … | 2M19204172+7311434 | J19206+731S | M4.0 V | 19:20:41.75 | +73:11:42.4 | A+B | A | … | … | … | 74.49 | 72.37 ± 0.71 | 0.228 ± 0.014 |
| … | … | 2M19204172+7311467 | J19206+731N | M4.5 V | 19:20:41.76 | +73:11:45.5 | … | B* | 3.182 | 0.0 | 56.04 | 82.50 | … | 0.240 ± 0.037 |
| … | … | GJ 1235 | J19216+208 | M4.0 Ve | 19:21:37.62 | +20:51:40.0 | … | … | … | … | 34.09 | 1738.79 | 44.99 ± 0.25 | 0.195 ± 0.010 |
| … | … | V2078Cyg | J19218+286 | M1.5 V | 19:21:52.48 | +28:40:02.3 | … | … | … | … | 34.04 | 901.75 | 403.94 ± 1.51 | 0.514 ± 0.029 |
| … | … | GJ 1236 | J19220+070 | M4.0 V | 19:22:01.28 | +07:02:23.2 | … | … | … | … | 43.12 | 863.35 | 64.33 ± 0.27 | 0.235 ± 0.012 |
| … | … | GSC 02654-01527 | J19228+307 | M0.0 V | 19:22:48.64 | +30:45:13.5 | … | … | … | … | 54.73 | 115.48 | 1107.82 ± 12.63 | 0.704 ± 0.026 |

Table D.2: Complete sample with the description of multiple systems (continued).

| WDS id | WDS disc | Name | Karmn | Spectral type | $\alpha$ (2016.0) | $\delta$ (2016.0) | System | Component | $\rho$ [arcsec] | $\theta$ [deg] | $\varpi$ [mas] | $\mu_{\rm total}$ [mas a$^{-1}$] | $\mathcal{L}$ [$10^{-4}\,\mathcal{L}_{\odot}$] | $\mathcal{M}$ [$\mathcal{M}_{\odot}$] |
|---|---|---|---|---|---|---|---|---|---|---|---|---|---|---|
| ... | ... | NLTT 47788 | J19234+666 | M1.0 V | 19:23:24.51 | +66:39:54.2 | ... | ... | ... | ... | 62.91 | 209.84 | 506.72 ± 2.18 | 0.549 ± 0.028 |
| 19215+4231 | JNN 122 | 1R192132.1+423041 | J19215+425 | M2.0 Ve | 19:21:32.18 | +42:30:54.8 | (AB) | AB | 0.126 | 154.4 | 68.25 | 151.93 | ... | 0.541 ± 0.029 |
| ... | ... | TYC 4592-101-1 | J19242+797 | M1.0 V | 19:24:15.66 | +79:43:37.3 | A+B | A | ... | ... | 57.34 | 88.20 | 741.53 ± 3.46 | 0.620 ± 0.027 |
| 19239+7944 | LDS1906 | PM J19237+7944 | J19237+797 | M1.5 V | 19:23:46.46 | +79:44:37.6 | ... | B | 98.614 | 307.7 | 50.40 | 87.77 | 624.07 ± 2.96 | 0.588 ± 0.028 |
| ... | ... | GJ 1238 | J19242+755 | M6.0 V | 19:24:17.89 | +75:33:21.3 | ... | ... | ... | ... | 88.01 | 698.05 | 16.06 ± 0.07 | 0.119 ± 0.008 |
| ... | ... | Ross 164 | J19251+283 | M3.5 V | 19:25:08.74 | +28:21:19.8 | ... | ... | ... | ... | 87.98 | 431.01 | 166.62 ± 86.70 | 0.352 ± 0.099 |
| ... | ... | LSPM J1925+0938 | J19255+096 | M8/9V | 19:25:31.00 | +09:38:19.4 | ... | ... | ... | ... | 130.85 | 255.76 | 11.63 ± 0.13 | 0.102 ± 0.048 |
| ... | ... | G 185-23 | J19260+244 | M4.5 V | 19:26:01.84 | +24:26:18.7 | ... | ... | ... | ... | 35.57 | 206.57 | 65.28 ± 0.27 | 0.272 ± 0.013 |
| ... | ... | GJ 4110 | J19268+167 | M3.5 Ve | 19:26:49.42 | +16:42:58.3 | ... | ... | ... | ... | 29.76 | 210.36 | 162.10 ± 0.99 | 0.387 ± 0.016 |
| ... | ... | TYC 2137-1575-1 | J19284+289 | M0.0 V | 19:28:25.51 | +28:54:09.6 | ... | ... | ... | ... | 56.26 | 41.87 | 691.21 ± 2.53 | 0.624 ± 0.027 |
| ... | ... | PM J19289+0638 | J19289+066 | M1.5 V | 19:28:55.70 | +06:38:24.5 | ... | ... | ... | ... | 56.22 | 101.34 | 448.06 ± 2.52 | 0.526 ± 0.029 |
| ... | ... | G 125-15 | J19312+361 | M4.5 V | 19:31:12.38 | +36:07:28.2 | Aab+B | Aab(2) | ... | ... | 61.26 | 167.85 | ... | ... |
| 19312+3607 | GIC 158 | G 125-14 | ... | M4.5 V | 19:31:11.56 | +36:08:12.8 | ... | B | 45.782 | 347.4 | 67.54 | 167.38 | 91.67 ± 0.42 | 0.305 ± 0.014 |
| ... | ... | GJ 761.2 | J19326+005 | M0.0 V | 19:32:38.15 | +00:34:39.5 | ... | ... | ... | ... | 119.75 | 219.55 | 740.37 ± 3.82 | 0.636 ± 0.027 |
| ... | ... | Ross 1063 | J19336+395 | M1.5 V | 19:33:39.75 | +39:31:29.9 | ... | ... | ... | ... | 28.21 | 477.69 | 259.33 ± 1.40 | 0.438 ± 0.017 |
| ... | ... | HD 184489 | J19346+045 | M0.0 V | 19:34:40.40 | +04:35:02.0 | ... | ... | ... | ... | 67.55 | 609.74 | 763.97 ± 3.40 | 0.652 ± 0.027 |
| ... | ... | Wolf 1108 | J19349+532 | M2.5 V | 19:34:55.56 | +53:15:31.3 | ... | ... | ... | ... | 67.35 | 567.87 | 132.32 ± 63.09 | 0.326 ± 0.085 |
| 19352+0825 | BAG 27 | GJ 4114 | J19351+084S | M0.5 V | 19:35:06.24 | +08:27:37.9 | (AB)+C | AB | 0.311 | 26.6 | 57.60 | 84.67 | ... | 0.644 ± 0.050 |
| 19352+0825 | VYS9 | GJ 4115 | J19351+084N | M2.5 V | 19:35:06.33 | +08:27:43.4 | ... | C | 5.635 | 14.5 | 28.10 | 65.00 | 257.20 ± 3.79 | 0.436 ± 0.017 |
| ... | ... | Ross 1064 | J19358+413 | M0.0 V | 19:35:51.00 | +41:19:06.7 | ... | ... | ... | ... | 43.73 | 316.28 | 1284.29 ± 16.38 | 0.712 ± 0.026 |
| ... | ... | GJ 4120 | J19395+718 | M0.5 V | 19:39:32.22 | +71:52:12.1 | ... | ... | ... | ... | 42.38 | 482.57 | 480.45 ± 3.07 | 0.510 ± 0.018 |
| ... | ... | GJ 1242 | J19419+031 | M2.0 V | 19:41:53.93 | +03:09:08.3 | ... | ... | ... | ... | 37.06 | 549.63 | 145.37 ± 1.11 | 0.322 ± 0.014 |
| ... | ... | RX J1935.4+3746 | J19354+377 | M3.5 V | 19:35:29.02 | +37:46:06.7 | Aab | Aab(1) | ... | ... | 55.40 | 162.48 | ... | ... |
| ... | ... | 2M19421282-2045477 | J19422-207 | M5.1 V | 19:42:12.81 | -20:45:50.4 | ... | ... | ... | ... | 37.06 | 154.60 | 48.52 ± 0.23 | 0.209 ± 0.012 |
| ... | ... | LP 869-19 | J19420-210 | M3.5 V | 19:42:00.74 | -21:04:09.4 | Aab | Aab(2) | ... | ... | 21.14 | 262.23 | ... | ... |
| ... | ... | Ross 165A | J19457+271 | M4.0 V | 19:45:45.41 | +27:07:11.7 | AB | A | ... | ... | 32.93 | 1225.43 | ... | 0.245 ± 0.037 |
| 19458+2710 | KUI 95 | Ross 165B | ... | M4.0 V | 19:45:45.55 | +27:07:12.8 | ... | B | 2.270 | 60.4 | 25.34 | 1220.47 | ... | 0.206 ± 0.038 |
| ... | ... | GJ 4122 | J19457+323 | M1.5 V | 19:45:50.25 | +32:23:16.8 | ... | ... | ... | ... | 59.74 | 441.14 | 277.71 ± 1.07 | 0.428 ± 0.016 |
| ... | ... | HD 331161A | J19463+320 | M0.5 V | 19:46:24.51 | +32:00:55.2 | A+B | A | ... | ... | 73.71 | 607.80 | 531.00 ± 2.21 | 0.552 ± 0.028 |
| 19464+3201 | KAM 3 | HD 331161B | J19464+320 | M2.5 V | 19:46:24.83 | +32:00:51.1 | ... | B | 5.707 | 135.1 | 91.80 | 638.05 | ... | 0.443 ± 0.031 |
| ... | ... | PM J19468-0157 | J19468-019 | M3.0 V | 19:46:50.63 | -01:57:40.1 | ... | ... | ... | ... | 39.68 | 119.23 | 325.04 ± 7.43 | 0.493 ± 0.020 |
| ... | ... | LSPM J1947+3516 | J19470+352 | M2.0 V | 19:47:03.30 | +35:16:55.5 | ... | ... | ... | ... | 39.59 | 164.93 | 503.14 ± 2.32 | 0.534 ± 0.029 |
| ... | ... | G 125-34 | J19486+359 | M3.5 V | 19:48:40.88 | +35:55:12.6 | ... | ... | ... | ... | 53.78 | 293.20 | 69.53 ± 0.30 | 0.262 ± 0.013 |
| 19500+3235 | JOD 18 | GJ 4124 | J19500+325 | M3.0 V | 19:50:03.10 | +32:35:05.5 | (AB) | AB | 0.222 | 345.1 | 64.24 | 242.56 | ... | ... |
| 19503+3147 | RAO 418 | GJ 4125 | J19502+317 | M2.1 V | 19:50:16.11 | +31:47:05.7 | (AB) | AB | 0.640 | 39.6 | 64.21 | 364.83 | ... | 0.400 ± 0.032 |
| ... | ... | HD 187691 | ... | F8 V | 19:51:01.91 | +10:24:54.4 | A+B | A | ... | ... | 175.79 | 277.69 | 27700.41 ± 534.64 | 1.180 ± 0.177 |
| 19510+1025 | J 124 | GJ 9671 B | J19510+104 | M4.0 V | 19:51:00.97 | +10:24:38.0 | ... | B | 21.477 | 220.2 | 61.00 | 288.38 | 147.95 ± 1.04 | 0.369 ± 0.016 |
| ... | ... | GJ 1243 | J19511+464 | M4.0 Ve | 19:51:09.60 | +46:29:04.5 | ... | ... | ... | ... | 28.03 | 322.76 | 71.83 ± 0.28 | 0.271 ± 0.012 |
| ... | ... | G 260-35 | J19512+622 | M2.0 V | 19:51:11.69 | +62:17:15.7 | ... | ... | ... | ... | 83.14 | 290.26 | 634.34 ± 2.32 | 0.577 ± 0.028 |
| ... | ... | GJ 4127 | J19535+341 | M1.5 V | 19:53:32.73 | +34:08:32.2 | ... | ... | ... | ... | 40.23 | 228.28 | 538.09 ± 1.97 | 0.552 ± 0.028 |

Table D.2: Complete sample with the description of multiple systems (continued).

| WDS id | WDS disc | Name | Karmn | Spectral type | $\alpha$ (2016.0) | $\delta$ (2016.0) | System | Component | $\rho$ [arcsec] | $\theta$ [deg] | $\varpi$ [mas] | $\mu_{total}$ [mas a$^{-1}$] | $\mathcal{L}$ [$10^{-4}$ $\mathcal{L}_\odot$] | $\mathcal{M}$ [$\mathcal{M}_\odot$] |
|---|---|---|---|---|---|---|---|---|---|---|---|---|---|---|
| 19539+4425 | MCY3 | V1581Cyg | J19539+444W | M5.5 V | 19:53:55.14 | +44:24:44.4 | (AB)+C | AB | 0.410 | 183.3 | 29.53 | 619.31 | ... | 0.135 ± 0.043 |
| 19539+4425 | GIC159 | GJ 1245 B | J19539+444E | M5.5 V | 19:53:55.66 | +44:24:46.5 | ... | C | 5.945 | 69.6 | 82.36 | 593.94 | 12.99 ± 0.10 | 0.127 ± 0.010 |
| ... | ... | GJ 4128 | J19540+325 | M2.5 V | 19:54:02.88 | +32:33:55.8 | ... | ... | ... | ... | 39.24 | 236.14 | 858.84 ± 5.52 | 0.613 ± 0.027 |
| ... | ... | TYC 1624-397-1 | J19546+202 | M0.0 V | 19:54:37.52 | +20:13:05.5 | ... | ... | ... | ... | 45.42 | 73.50 | 780.28 ± 3.60 | 0.646 ± 0.027 |
| ... | ... | Wolf 1122 | J19558+512 | M1.5 V | 19:55:53.62 | +51:16:27.7 | ... | ... | ... | ... | 52.78 | 585.88 | 358.27 ± 1.67 | 0.489 ± 0.018 |
| ... | ... | GJ 9677 A | J19565+591 | M0.0 V | 19:56:33.06 | +59:09:40.3 | A+B | A | ... | ... | ... | 431.05 | 1393.72 ± 13.18 | 0.747 ± 0.025 |
| 19566+5910 | GIC 161 | GJ 9677 B | J19564+591 | M3.5 V | 19:56:23.96 | +59:09:19.7 | ... | B | 72.939 | 253.7 | 74.79 | 444.25 | 144.84 ± 0.53 | 0.342 ± 0.015 |
| ... | ... | HD 188807 | ... | K7 V | 19:57:19.54 | -12:34:13.0 | A+B | A | ... | ... | 56.26 | 523.41 | 1086.17 ± 4.34 | 0.640 ± 0.096 |
| 19573-1234 | LDS4824 | GJ 773 B | J19573-125 | M5.0 V | 19:57:23.71 | -12:33:58.5 | ... | B | 62.695 | 76.6 | 35.81 | 525.78 | 34.56 ± 0.19 | 0.206 ± 0.011 |
| ... | ... | GJ 4129 | J19582+020 | M2.5 V | 19:58:15.38 | +02:02:02.5 | ... | ... | ... | ... | 41.75 | 887.32 | 164.81 ± 0.70 | 0.366 ± 0.015 |
| ... | ... | G 260-38 | J19582+650 | M3.5 V | 19:58:16.31 | +65:02:25.0 | ... | ... | ... | ... | ... | 394.41 | 157.85 ± 0.69 | 0.382 ± 0.016 |
| ... | ... | LP 634-16 | J20011+002 | M2.0 V | 20:01:06.21 | +00:16:12.0 | ... | ... | ... | ... | ... | 224.69 | 358.05 ± 2.11 | 0.485 ± 0.030 |
| ... | ... | 1R200031.8+592127 | J20005+593 | M4.1 V | 20:00:31.98 | +59:21:29.9 | AB | A | ... | ... | 58.84 | 132.56 | ... | 0.309 ± 0.034 |
| 20006+5922 | JNN124 | G3-2237179062111268096 | ... | ... | 20:00:32.02 | +59:21:29.7 | ... | B | 0.374 | 121.4 | 100.69 | ... | ... | ... |
| ... | ... | HD 190360 | ... | G7 IV-V | 20:03:38.25 | +29:53:40.1 | A+B | A | ... | ... | 100.56 | 861.92 | 11241.75 ± 182.61 | ... |
| 20036+2954 | LDS6339 | GJ 777 B | J20034+298 | M4.5 V | 20:03:27.42 | +29:51:51.1 | ... | B | 178.054 | 232.3 | 80.10 | 860.49 | 49.73 ± 0.20 | 0.234 ± 0.012 |
| ... | ... | 1R200348.4+642542 | J20037+644 | M0.0 V | 20:03:47.80 | +64:25:45.3 | ... | ... | ... | ... | 92.61 | 98.74 | 1190.73 ± 12.94 | 0.722 ± 0.025 |
| ... | ... | GJ 1248 | J20038+059 | M2.5 V | 20:03:50.47 | +05:59:31.4 | ... | ... | ... | ... | 91.46 | 940.69 | 96.37 ± 0.46 | 0.258 ± 0.012 |
| ... | ... | GJ 4132 | J20039-081 | M4.0 V | 20:03:58.37 | -08:07:51.4 | ... | ... | ... | ... | 283.84 | 558.85 | 109.86 ± 0.84 | 0.336 ± 0.015 |
| ... | ... | V1513Cyg | J20050+544 | sdM1 | 20:05:00.07 | +54:25:48.8 | (AabB) | Aab(1) | ... | ... | 64.93 | 1470.28 | ... | ... |
| 20050+5426 | MGN1 | Wolf 1130 B | ... | T8p | ... | ... | ... | B | 188.540 | 115.0 | 283.84 | ... | ... | ... |
| ... | ... | Wolf 1131 | J20057+529 | M4.0 V | 20:05:44.59 | +52:58:21.1 | ... | ... | ... | ... | 31.24 | 277.53 | 145.54 ± 0.76 | 0.365 ± 0.015 |
| ... | ... | GJ 4136 | J20079-015 | M3.0 V | 20:07:57.91 | -01:32:31.5 | ... | ... | ... | ... | 69.63 | 405.55 | 95.66 ± 0.43 | 0.292 ± 0.013 |
| ... | ... | GJ 1250 | J20082+333 | M4.77 | 20:08:18.35 | +33:18:19.0 | ... | ... | ... | ... | 52.99 | 506.19 | 55.31 ± 13.13 | 0.216 ± 0.031 |
| ... | ... | PM J20093-0113 | J20093-012 | M5.0 V | 20:09:18.20 | -01:13:44.3 | ... | ... | ... | ... | 71.05 | 376.60 | 27.03 ± 0.14 | 0.144 ± 0.012 |
| ... | ... | LP 574-21 | J20105+065 | M4.0 V | 20:10:34.49 | +06:32:10.7 | Aab+B | Aab(2) | ... | ... | 20.81 | 208.13 | ... | ... |
| 20106+0632 | LUH 15 | 2M20103539+0634367 | ... | M8.5 V | 20:10:35.44 | +06:34:33.5 | ... | B | 143.531 | 5.6 | 58.60 | 213.72 | 3.00 ± 0.03 | 0.114 ± 0.054 |
| ... | ... | HD 191785 | ... | K0 V | 20:11:05.61 | +16:11:23.2 | A+B | A | ... | ... | 131.28 | 575.39 | 4626.61 ± 18.13 | 0.880 ± 0.132 |
| 20111+1611 | GIC 163 | GJ 783.2 B | J20112+161 | M4.0 V | 20:11:12.80 | +16:11:14.4 | ... | B | 103.843 | 94.8 | 41.89 | 577.77 | 78.99 ± 0.32 | 0.281 ± 0.013 |
| ... | ... | LSPM J2011+3757 | J20112+379 | M1.5 V | 20:11:12.85 | +37:57:48.1 | ... | ... | ... | ... | 336.03 | 187.54 | 672.18 ± 3.43 | 0.591 ± 0.028 |
| ... | ... | LP 283-5 | J20129+342 | M1.0 V | 20:12:54.72 | +34:16:54.5 | A+B | A | ... | ... | 83.79 | 405.26 | 930.29 ± 3.50 | 0.683 ± 0.026 |
| 20130+3416 | LDS1035 | LP 283-4 | ... | M1.0 V | 20:12:54.41 | +34:16:37.6 | ... | B | 17.283 | 192.8 | 47.08 | 412.00 | 793.89 ± 3.97 | 0.633 ± 0.027 |
| ... | ... | StKM 1-1767a | ... | K5 V | 20:13:11.14 | +02:56:20.5 | A+B | A | ... | ... | 50.91 | 108.28 | 1551.77 ± 7.12 | 0.700 ± 0.105 |
| 20132+0256 | CRI 26 | [R78b] 440 | J20132+029 | M1.0 V | 20:13:12.94 | +02:56:02.3 | ... | B | 32.497 | 123.9 | 27.68 | 110.73 | 707.14 ± 4.16 | 0.613 ± 0.027 |
| ... | ... | Ross 754 | J20138+133 | M1.0 V | 20:13:52.27 | +13:23:20.0 | ... | ... | ... | ... | 55.15 | 422.35 | 750.02 ± 3.23 | 0.623 ± 0.027 |
| ... | ... | GJ 784.2 A | J20139+066 | M3.3 V | 20:13:58.71 | +06:41:06.8 | A+B | A | ... | ... | 65.76 | 634.75 | 172.59 ± 1.07 | 0.400 ± 0.017 |
| 20139+0641 | GIC 164 | V1412Aql | ... | DC7 | 20:13:55.41 | +06:42:35.5 | ... | B | 101.455 | 330.9 | 33.49 | 635.70 | ... | 0.500 ± 0.100 |
| ... | ... | LP 106-240 | J20151+635 | M0.0 V | 20:15:10.64 | +63:31:16.4 | ... | ... | ... | ... | 50.24 | 377.93 | 949.14 ± 377.17 | 0.683 ± 0.026 |
| ... | ... | G 210-11 | J20165+351 | M2.0 V | 20:16:32.04 | +35:10:37.6 | ... | ... | ... | ... | 40.12 | 333.21 | 281.11 ± 1.13 | 0.457 ± 0.018 |
| ... | ... | GJ 4143 | J20187+158 | M2.5 V | 20:18:44.75 | +15:50:45.3 | ... | ... | ... | ... | 53.42 | 184.10 | 247.21 ± 0.94 | 0.427 ± 0.017 |

Table D.2: Complete sample with the description of multiple systems (continued).

| WDS id | WDS disc | Name | Karmn | Spectral type | $\alpha$ (2016.0) | $\delta$ (2016.0) | System | Component | $\rho$ [arcsec] | $\theta$ [deg] | $\varpi$ [mas] | $\mu_{\rm total}$ [mas a$^{-1}$] | $\mathcal{L}$ [$10^{-4}\,\mathcal{L}_\odot$] | $\mathcal{M}$ [$\mathcal{M}_\odot$] |
|---|---|---|---|---|---|---|---|---|---|---|---|---|---|---|
| … | … | GJ 4144 | J20195+080 | M3.0 V | 20:19:34.60 | +08:00:27.1 | AB | A | … | … | 53.64 | 210.99 | … | 0.427 ± 0.031 |
| … | … | G3- 4250232535851803136 | … | M3.0 V | 20:19:34.55 | +08:00:26.9 | … | B* | 0.777 | 258.3 | 53.84 | 200.82 | … | 0.387 ± 0.032 |
| … | … | LP 395-8 A | J20198+229 | M3.0 V | 20:19:49.36 | +22:56:38.1 | AabB+C | Aab(2) | … | … | 22.22 | 135.40 | … | 0.124 ± 0.001 |
| 20198+2257 | KPP4191 | LP 395-8 B | … | M3.5 V | 20:19:49.35 | +22:56:40.0 | … | B | 1.918 | 355.5 | 89.91 | 137.73 | … | 0.318 ± 0.034 |
| … | … | G3- 1829571684884360832 | … | … | 20:19:48.73 | +22:56:44.8 | … | C* | 11.019 | 307.4 | … | 142.76 | … | 0.1504 ± 0.059 |
| … | … | TYC 1643-120-1 | J20220+216 | M2.0 V | 20:22:01.62 | +21:47:19.7 | A+B | A | … | … | … | 131.82 | 355.01 ± 2.22 | 0.477 ± 0.030 |
| … | … | PM J20220+2147B | … | M4/5 V | 20:22:01.98 | +21:47:21.8 | … | B* | 5.445 | 67.5 | 23.49 | 133.78 | … | 0.158 ± 0.041 |
| … | … | PM J20223+3217 | J20223+322 | M3.5 V | 20:22:18.82 | +32:17:15.1 | … | … | … | … | 51.18 | 148.03 | 160.50 ± 0.89 | 0.385 ± 0.016 |
| … | … | G 143-48 | J20229+106 | M3.0 V | 20:22:55.79 | +10:40:44.5 | … | … | … | … | 39.00 | 548.17 | 151.33 ± 0.77 | 0.373 ± 0.016 |
| … | … | LP 73-196 | J20232+671 | M5.0 V | 20:23:18.57 | +67:10:12.9 | AB | A | … | … | 40.35 | 278.28 | … | 0.178 ± 0.040 |
| 20233+6710 | LAW 19 | LP 73-196 B | … | M5.5 V | 20:23:18.43 | +67:10:12.1 | … | B | 1.145 | 225.6 | 40.36 | 291.47 | … | 0.156 ± 0.041 |
| … | … | Wolf 1069 | J20260+585 | dM5.0 | 20:26:05.84 | +58:34:31.4 | … | … | … | … | 40.39 | 602.40 | 29.24 ± 0.27 | 0.159 ± 0.010 |
| … | … | LP 755-19 | J20287-114 | M1.5 Ve | 20:28:43.80 | -11:28:32.4 | … | … | … | … | 90.02 | 186.32 | 213.38 ± 1.20 | 0.448 ± 0.018 |
| 20269+2731 | RAO 432 | GJ 4146 | J20269+275 | M2.0 V | 20:26:56.29 | +27:31:03.1 | (AB) | AB | 0.203 | 335.5 | 35.33 | 281.66 | … | 0.511 ± 0.030 |
| 20298+0941 | AST 2 | HUDel | J20298+096 | M4.5 V | 20:29:49.04 | +09:41:22.2 | (AB) | AB(1) | 0.161 | 93.8 | 60.67 | 676.97 | … | 0.412 ± 0.007 |
| … | … | GJ 793 | J20305+654 | M2.5 V | 20:30:33.18 | +65:27:02.9 | … | … | … | … | 41.51 | 526.09 | 195.56 ± 1.25 | 0.374 ± 0.011 |
| 20287-1129 | BWL55 | 1R203011.0+795040 | J20301+798 | M3.0 Ve | 20:30:07.75 | +79:50:47.4 | Aab | Aab(2) | 5.611 | 34.0 | 169.22 | 108.40 | … | … |
| … | … | GJ 1254 | J20336+617 | M4.0 V | 20:33:41.53 | +61:45:28.2 | … | … | … | … | 23.86 | 1058.07 | 175.89 ± 1.38 | 0.378 ± 0.013 |
| 20314+3833 | JNN 284 | Ross 188 | J20314+385 | M5.0 V | 20:31:25.91 | +38:33:55.8 | (AB) | AB | 0.118 | 252.4 | 46.05 | 748.29 | … | 0.256 ± 0.036 |
| … | … | GJ 4148 | J20337+233 | M3.0 V | 20:33:43.10 | +23:22:15.4 | AB | A | … | … | … | 304.69 | … | 0.447 ± 0.031 |
| 20337+2322 | JNN 285 | G 186-29B | … | M4.0 V | 20:33:43.11 | +23:22:14.5 | … | B | 0.909 | 172.5 | 27.50 | 322.24 | … | 0.301 ± 0.035 |
| … | … | GJ 4150 | J20339+643 | M3.5 V | 20:33:58.81 | +64:19:08.1 | … | … | … | … | 53.84 | 441.98 | 237.64 ± 1.65 | 0.445 ± 0.018 |
| … | … | GJ 4149 | J20347+033 | M2.5 V | 20:34:43.33 | +03:20:43.6 | … | … | … | … | 27.74 | 558.88 | 287.35 ± 1.47 | 0.462 ± 0.018 |
| … | … | Wolf 1074 | J20349+592 | M4.0 V | 20:34:54.81 | +59:17:26.0 | … | … | … | … | … | 244.58 | 125.22 ± 0.51 | 0.337 ± 0.015 |
| … | … | GJ 4152 | J20367+388 | M3.5 V | 20:36:46.27 | +38:50:30.3 | … | … | … | … | 44.56 | 227.57 | 73.84 ± 0.28 | 0.253 ± 0.012 |
| … | … | Wolf 1351 | J20373+219 | M0.5 V | 20:37:20.77 | +21:56:47.8 | A+B | A | … | … | 44.52 | 279.80 | 922.02 ± 64.01 | 0.630 ± 0.029 |
| 20373+2157 | MCT 11 | GJ 4153 B | … | M4.5 V | 20:37:23.97 | +21:56:21.4 | … | B | 51.773 | 120.7 | 36.01 | 301.36 | 66.67 ± 0.33 | 0.256 ± 0.013 |
| … | … | TYC 4246-488-1 | J20403+616 | M1.0 V | 20:40:18.59 | +61:41:28.4 | … | … | … | … | 113.22 | 117.58 | 537.60 ± 3.27 | 0.558 ± 0.028 |
| … | … | GJ 1256 | J20405+154 | M4.5 V | 20:40:35.33 | +15:30:09.3 | … | … | … | … | 113.25 | 1477.37 | 42.26 ± 0.19 | 0.191 ± 0.010 |
| 20408+1956 | LDS1045 | HD 197076 | … | G5 V | 20:40:45.28 | +19:56:12.9 | A+(BC) | A | … | … | 52.47 | 334.03 | … | 1.001 ± 0.150 |
| 20408+1956 | RAO 23 | GJ 797 B | J20407+199 | M2.5 V | 20:40:44.65 | +19:54:08.2 | … | BC | 125.058 | 184.1 | 80.18 | 339.09 | … | … |
| … | … | GJ 4155 | J20409-101 | M1.5 V | 20:40:56.32 | -10:06:43.1 | Aab | Aab | … | … | 43.47 | 155.85 | … | … |
| … | … | ATMicA | J20418-324 | M4.5 V | 20:41:51.44 | -32:26:13.3 | AB+C | A | … | … | 120.20 | 484.79 | … | 0.420 ± 0.031 |
| 20452-3120 | LDS 720 | ATMicB | … | M4.5 Ve | 20:41:51.54 | -32:26:15.1 | … | B | 2.102 | 145.0 | 36.05 | 423.10 | … | 0.414 ± 0.031 |
| 20452-3120 | LDS 720 | AUMic | J20451-313 | M0.5 V | 20:45:09.88 | -31:20:33.0 | … | C | 4681.175 | 32.9 | 36.15 | 457.00 | 990.15 ± 7.63 | 0.654 ± 0.027 |
| … | … | Ross 751 | J20429-189 | M1.5 V | 20:42:57.83 | -18:55:19.8 | … | … | … | … | 48.39 | 1048.64 | 556.52 ± 2.76 | 0.564 ± 0.028 |
| … | … | LP 575-35 | J20433+047 | M5.0 V | 20:43:24.35 | +04:45:52.9 | … | … | … | … | 97.24 | 469.44 | … | 0.159 ± 0.041 |
| … | … | Wolf 1360 | J20435+240 | M2.5 Ve | 20:43:34.69 | +24:07:40.2 | … | … | … | … | 29.02 | 150.57 | 348.02 ± 1.56 | 0.475 ± 0.030 |
| … | … | G 262-26 | J20436+642 | M0.0 V | 20:43:42.13 | +64:16:52.5 | … | … | … | … | 29.06 | 326.18 | 334.83 ± 1.32 | 0.446 ± 0.017 |
| … | … | GJ 4159 | J20436-001 | M1.0 V | 20:43:41.71 | -00:10:37.5 | … | … | … | … | 36.20 | 440.43 | 464.18 ± 2.22 | 0.548 ± 0.028 |

Table D.2: Complete sample with the description of multiple systems (continued).

| WDS id | WDS disc | Name | Karmn | Spectral type | $\alpha$ (2016.0) | $\delta$ (2016.0) | System | Component | $\rho$ [arcsec] | $\theta$ [deg] | $\varpi$ [mas] | $\mu_{\rm total}$ [mas a$^{-1}$] | $\mathcal{L}$ [$10^{-4}$ $\mathcal{L}_\odot$] | $\mathcal{M}$ [$\mathcal{M}_\odot$] |
|---|---|---|---|---|---|---|---|---|---|---|---|---|---|---|
| 20433+5521 | LL01 | Wolf 1084 | J20433+553 | M5.0 V | 20:43:20.89 | +55:21:20.5 | (AB) | AB(2) | 0.086 | 20.2 | 31.29 | 1919.02 | … | 0.268 |
| 20444+1945 | CAR 2 | HD 352860 | J20443+197 | M0.0 V | 20:44:21.98 | +19:44:49.8 | (AB) | AB | 0.226 | 81.8 | 34.23 | 555.61 | … | … |
| … | … | GJ 4160 | J20445+089S | M1.5 V | 20:44:30.94 | +08:54:12.6 | A+Bab | A | … | … | 55.78 | 223.80 | … | 0.565 ± 0.028 |
| 20446+0854 | LDS1046 | GJ 4161 | J20445+089N | M0.5 V | 20:44:30.67 | +08:54:27.1 | … | Bab(2) | 15.094 | 344.5 | 38.88 | 230.26 | … | … |
| … | … | GJ 806 | J20450+444 | M1.5 V | 20:45:04.75 | +44:30:01.0 | … | … | … | … | 106.28 | 511.70 | 259.29 ± 0.98 | 0.413 ± 0.011 |
| … | … | Wolf 882 | J20496-003 | M3.5 V | 20:49:39.86 | -00:21:06.7 | … | … | … | … | 55.19 | 366.38 | 206.12 ± 0.95 | 0.439 ± 0.018 |
| 20488+1943 | JNN 286 | GJ 4163 | J20488+197 | M3.3 V | 20:48:52.27 | +19:43:01.6 | (AB) | AB | 0.233 | 178.6 | 55.22 | 257.20 | … | 0.377 ± 0.035 |
| … | … | LP 816-60 | J20525-169 | M4.0 V | 20:52:32.67 | -16:58:28.4 | … | … | … | … | 169.06 | 311.33 | 62.69 ± 0.40 | 0.219 ± 0.010 |
| 20520+6910 | JOD 19 | GJ 4170 | J20519+691 | M1.0 V | 20:52:00.55 | +69:10:07.6 | (AB) | AB | 0.470 | 171.6 | 168.95 | 231.52 | … | 0.474 ± 0.030 |
| … | … | HD 199305 | J20533+621 | M1.0 V | 20:53:19.79 | +62:09:03.4 | … | … | … | … | 26.27 | 773.10 | 525.60 ± 4.57 | 0.562 ± 0.028 |
| … | … | GJ 4169 | J20535+106 | M5.0 V | 20:53:32.53 | +10:36:55.0 | … | … | … | … | 26.31 | 665.25 | 40.84 ± 0.19 | 0.210 ± 0.011 |
| … | … | LP 74-35 | J20549+675 | M2.0 V | 20:54:54.90 | +67:35:09.8 | … | … | … | … | 95.18 | 198.43 | 414.46 ± 1.80 | 0.509 ± 0.029 |
| 20532-0221 | JNN126 | LP 636-19 | J20532-023 | M3.0 V | 20:53:14.87 | -02:21:21.7 | (AB) | AB | 0.138 | 327.3 | 95.18 | 195.08 | … | 0.445 ± 0.031 |
| … | … | GJ 810 A | J20556-140N | M4.0 V | 20:55:39.31 | -14:02:15.6 | Aab+B | Aab(2) | … | … | 58.80 | 1493.30 | … | 0.121 ± 0.009 |
| 20555-1400 | LDS6418 | GJ 810 B | J20556-140S | M5.0 V | 20:55:38.68 | -14:04:02.4 | … | B | 107.268 | 184.9 | 32.78 | 1497.19 | 26.26 ± 0.13 | 0.153 ± 0.009 |
| … | … | Wolf 896 | J20567-104 | dM2.5 | 20:56:46.56 | -10:27:12.7 | … | … | … | … | 32.79 | 1125.07 | 312.25 ± 1.75 | 0.466 ± 0.013 |
| … | … | FRAqr | J20568-048 | M4.0 V | 20:56:49.39 | -04:50:52.6 | Aab+B | Aab(2) | … | … | 50.15 | 826.32 | … | … |
| 20568-0449 | LDS6420 | Ross 193B | … | DC10 | 20:56:48.62 | -04:50:43.1 | … | B | 15.006 | 309.4 | 94.40 | 807.72 | … | 0.500 ± 0.100 |
| … | … | Wolf 1373 | J20574+223 | M3.0 V | 20:57:26.24 | +22:21:42.4 | … | … | … | … | 42.42 | 798.71 | 133.63 ± 0.87 | 0.308 ± 0.013 |
| … | … | GJ 4173 | J20586+342 | M0.5 V | 20:58:42.31 | +34:16:24.7 | … | … | … | … | 94.19 | 324.95 | 639.77 ± 2.64 | 0.597 ± 0.027 |
| … | … | G 212-14 | J21001+495 | M2.0 V | 21:00:09.24 | +49:35:20.8 | … | … | … | … | 32.70 | 296.48 | 532.43 ± 2.22 | 0.541 ± 0.029 |
| 21000+4004 | KUI 103 | V1396Cyg | J21000+400 | M2.0 V | 21:00:06.23 | +40:04:08.6 | (AB) | AB(2) | 0.885 | 58.2 | 50.81 | 652.26 | … | … |
| 21013+3315 | CRC 75 | GJ 4176 | J21012+332 | M4.0 V | 21:01:16.51 | +33:14:30.7 | (AB)+(CD) | AB | 0.232 | 6.3 | 30.67 | 364.63 | … | 0.406 ± 0.032 |
| 21011+3315 | JNN 288 | GJ 4177 | J21013+332 | M4.0 V | 21:01:21.03 | +33:14:25.9 | … | CD | 56.912 | 94.8 | 31.47 | 333.46 | … | 0.341 ± 0.034 |
| … | … | Wolf 906 | J21019-063 | M2.5 V | 21:01:58.39 | -06:19:14.5 | … | … | … | … | 27.88 | 498.67 | 296.13 ± 1.30 | 0.451 ± 0.012 |
| … | … | G 211-9 | J21027+349 | M4.5 V | 21:02:46.40 | +34:54:31.2 | … | … | … | … | 27.87 | 362.28 | 174.68 ± 0.82 | 0.460 ± 0.019 |
| … | … | TYC 3588-5589-1 | J21044+455 | M2.0 V | 21:04:28.94 | +45:35:42.3 | … | … | … | … | 94.07 | 173.29 | 587.57 ± 2.34 | 0.557 ± 0.028 |
| … | … | Ross 769 | J21048-169 | M1.5 V | 21:04:52.36 | -16:58:04.6 | … | … | … | … | 59.62 | 2236.05 | 261.50 ± 1.50 | 0.415 ± 0.016 |
| 21014+2043 | JNN 289 | GJ 4175 | J21014+207 | M3.5 V | 21:01:24.40 | +20:43:31.6 | (AB) | AB | 0.377 | 25.4 | 58.91 | 557.25 | … | 0.248 ± 0.037 |
| … | … | PM J21055+0609N | J21055+061 | M3.0 V | 21:05:32.09 | +06:09:16.2 | A+B | A | … | … | 53.83 | 52.64 | 248.97 ± 1.36 | 0.485 ± 0.019 |
| … | … | PM J21055+0609S | … | M5.5 V | 21:05:32.17 | +06:09:11.2 | … | B* | 5.094 | 166.0 | 47.71 | 61.85 | … | 0.139 ± 0.043 |
| … | … | PM J21057+5015E | J21057+502 | M3.5 V | 21:05:45.54 | +50:15:44.1 | A+B | A | … | … | 38.59 | 100.51 | 271.86 ± 1.50 | 0.508 ± 0.020 |
| … | … | PM J21057+5015W | … | M3.5 V | 21:05:42.60 | +50:15:58.1 | … | B* | 31.465 | 296.4 | 30.52 | 97.27 | 180.42 ± 0.92 | 0.437 ± 0.018 |
| … | … | GJ 4180 | J21059+044 | M2.5 V | 21:05:56.52 | +04:25:38.0 | … | … | … | … | 24.95 | 215.22 | 408.63 ± 2.12 | 0.492 ± 0.030 |
| … | … | PM J21074+4651 | J21074+468 | M2.0 V | 21:07:28.14 | +46:51:55.4 | … | … | … | … | 24.97 | 100.93 | 438.11 ± 1.84 | 0.503 ± 0.029 |
| … | … | 1R210736.5-130500 | J21076-130 | M3.0 Ve | 21:07:36.86 | -13:04:59.6 | … | … | … | … | 45.14 | 106.44 | 203.86 ± 1.17 | 0.437 ± 0.018 |
| 21088-0426 | HDS3013 | GJ 9721 | J21087-044S | M1.0 V | 21:08:45.41 | -04:25:36.7 | (AB)+C | AB | 0.192 | 340.0 | 57.57 | 85.47 | … | … |
| 21088-0426 | VVO 18 | GJ 9721B | J21087-044N | M3.0 V | 21:08:44.75 | -04:25:18.3 | … | C | 20.894 | 331.9 | 69.26 | 53.58 | 186.56 ± 0.95 | 0.417 ± 0.017 |
| … | … | Wolf 918 | J21092-133 | M1.0 V | 21:09:18.21 | -13:18:40.9 | … | … | … | … | 65.74 | 2119.70 | 190.69 ± 0.97 | 0.351 ± 0.014 |
| … | … | 1R211004.9-192005 | J21100-193 | M2.0 V | 21:10:05.46 | -19:19:59.1 | A+B | A | … | … | 36.42 | 128.41 | 896.94 ± 5.59 | 0.600 ± 0.027 |

Table D.2: Complete sample with the description of multiple systems (continued).

| WDS id | WDS disc | Name | Karmn | Spectral type | $\alpha$ (2016.0) | $\delta$ (2016.0) | System | Component | $\rho$ [arcsec] | $\theta$ [deg] | $\varpi$ [mas] | $\mu_{\rm total}$ [mas a$^{-1}$] | $\mathcal{L}$ [$10^{-4}$ $\mathcal{L}_\odot$] | $\mathcal{M}$ [$\mathcal{M}_\odot$] |
|---|---|---|---|---|---|---|---|---|---|---|---|---|---|---|
| 21100-1920 | LDS 734 | UCAC4 354-189365 | ... | M5.0 V | 21:10:04.71 | -19:20:32.0 | ... | B | 34.542 | 197.8 | 38.69 | 129.86 | 620.35 ± 29.47 | 0.497 ± 0.030 |
| ... | ... | LP 285-9 | J21123+359 | M1.5 V | 21:12:22.66 | +35:55:24.7 | ... | ... | ... | ... | 34.89 | 201.98 | 534.67 ± 3.26 | 0.554 ± 0.028 |
| ... | ... | PM J21127-0719 | J21127-073 | M3.5 V | 21:12:45.71 | -07:19:56.5 | ... | ... | ... | ... | 47.53 | 111.19 | 185.68 ± 0.99 | 0.416 ± 0.017 |
| 21109+2925 | BAG29 | Ross 824 | J21109+294 | M1.5 V | 21:10:54.47 | +29:25:18.6 | (AB) | AB | 0.188 | 174.4 | 43.34 | 371.86 | ... | 0.658 ± 0.039 |
| 21137+0846 | RAO 449 | LSPM J2113+0846N | J21137+087 | M2.0 V | 21:13:44.77 | +08:46:09.1 | (AB)+C | AB | 0.377 | 322.0 | 69.12 | 178.26 | ... | 0.482 ± 0.030 |
| 21137+0846 | UC 4412 | LSPM J2113+0846S | ... | M1.5 V | 21:13:44.56 | +08:46:01.1 | ... | C | 8.563 | 200.9 | 63.10 | 166.83 | 375.14 ± 2.70 | 0.493 ± 0.030 |
| ... | ... | Ross 772 | J21138+180 | M3.0 V | 21:13:53.01 | +18:05:58.8 | ... | ... | ... | ... | 53.58 | 449.37 | 347.31 ± 1.71 | 0.464 ± 0.030 |
| ... | ... | LSPM J2114+5052 | J21145+508 | M2.5 V | 21:14:32.61 | +50:52:31.6 | ... | ... | ... | ... | 93.88 | 211.51 | 467.76 ± 1.99 | 0.512 ± 0.029 |
| 21148+3803 | AGC 13 | GJ 9728 B | ... | ... | 21:14:47.68 | +38:02:50.6 | (AB)+(CD) | AB | 1.050 | 189.7 | 93.87 | 472.40 | ... | 2.6300 ± 0.120 |
| 21148+3803 | JOD 20 | GJ 822.1 C | J21147+380 | M2.5 V | 21:14:47.09 | +38:01:21.0 | ... | CD | 89.862 | 184.4 | 72.70 | 453.21 | ... | 0.437 ± 0.031 |
| ... | ... | GJ 4184 | J21152+257 | M3.0 V | 21:15:12.76 | +25:47:41.2 | ... | ... | ... | ... | 71.72 | 300.40 | 471.35 ± 2.78 | 0.503 ± 0.029 |
| 21161+2951 | BWL 56 | Ross 776 | J21160+298E | M3.3 V | 21:16:06.06 | +29:51:51.5 | (AB)+C | AB | 0.058 | 163.1 | 71.68 | 203.76 | ... | 0.380 ± 0.032 |
| 21161+2951 | LDS1053 | Ross 826 | J21160+298W | M3.3 V | 21:16:04.10 | +29:51:46.7 | ... | C | 26.031 | 259.3 | 42.59 | 214.61 | 109.38 ± 0.46 | 0.335 ± 0.015 |
| ... | ... | LSPM J2116+0234 | J21164+025 | M3.0 V | 21:16:27.55 | +02:34:50.8 | ... | ... | ... | ... | 30.23 | 250.53 | 241.76 ± 1.46 | 0.410 ± 0.012 |
| ... | ... | Ross 773A | J21173+208N | M3.0 V | 21:17:23.09 | +20:53:59.2 | A+B | A | ... | ... | 74.39 | 420.04 | 414.91 ± 2.68 | 0.479 ± 0.030 |
| 21174+2053 | KUI 106 | Ross 773B | J21173+208S | M4.0 V | 21:17:23.00 | +20:54:03.4 | ... | B | 4.337 | 341.2 | ... | 422.14 | ... | 0.369 ± 0.033 |
| ... | ... | G 262-38 | J21173+640 | M5.0 V | 21:17:22.75 | +64:02:39.1 | ... | ... | ... | ... | 33.73 | 342.80 | 114.87 ± 2.12 | 0.322 ± 0.015 |
| 21176-0854 | JOD 21 | GJ 4187 | J21176-089N | M2.5 V | 21:17:36.07 | -08:54:11.7 | (AB)+C | AB | 0.479 | 188.8 | 51.31 | 84.74 | ... | 0.461 ± 0.031 |
| 21176-0854 | VVO 19 | GJ 4188 | J21176-089S | M3.0 V | 21:17:39.55 | -08:54:49.6 | ... | C | 63.991 | 126.4 | 51.37 | 78.19 | 204.92 ± 1.71 | 0.411 ± 0.017 |
| ... | ... | 1R211833.8+301434 | J21185+302 | M1.5 Ve | 21:18:33.83 | +30:14:34.3 | ... | ... | ... | ... | 83.49 | 63.45 | 790.66 ± 3.57 | 0.628 ± 0.027 |
| ... | ... | TYC 2187-512-1 | J21221+229 | M1.0 V | 21:22:06.41 | +22:55:55.0 | ... | ... | ... | ... | 34.10 | 160.33 | 416.86 ± 2.32 | 0.483 ± 0.012 |
| ... | ... | GJ 4192 | J21243+085 | M3.5 V | 21:24:19.08 | +08:30:03.6 | ... | ... | ... | ... | 37.15 | 240.54 | 279.24 ± 42.07 | 0.374 ± 0.037 |
| ... | ... | LSPM J2124+4003 | J21245+400 | M5.5 V | 21:24:33.07 | +40:04:06.7 | ... | ... | ... | ... | 213.13 | 697.68 | 19.12 ± 0.10 | 0.159 ± 0.010 |
| ... | ... | GJ 828.1 | J21267+037 | M0.0 V | 21:26:42.41 | +03:44:12.9 | ... | ... | ... | ... | 214.57 | 66.87 | 1033.35 ± 4.90 | 0.699 ± 0.026 |
| ... | ... | Wolf 920 | J21272-068 | M0.5 V | 21:27:16.89 | -06:50:45.8 | ... | ... | ... | ... | 21.74 | 400.29 | 286.41 ± 1.74 | 0.411 ± 0.016 |
| ... | ... | V2160Cyg | J21275+340 | M1.5 V | 21:27:32.60 | +34:01:25.7 | ... | ... | ... | ... | 36.38 | 333.36 | 680.43 ± 4.48 | 0.603 ± 0.027 |
| ... | ... | Ross 778 | J21277+072 | M1.0 V | 21:27:46.19 | +07:17:45.9 | ... | ... | ... | ... | 39.37 | 684.31 | 480.70 ± 3.03 | 0.541 ± 0.029 |
| ... | ... | LP 457-38 | J21280+179 | M1.5 V | 21:28:05.57 | +17:54:02.7 | ... | ... | ... | ... | 36.81 | 174.63 | 644.14 ± 3.08 | 0.586 ± 0.028 |
| ... | ... | GJ 4197 | J21283-223 | M2.5 V | 21:28:18.04 | -22:18:36.5 | ... | ... | ... | ... | 36.74 | 333.87 | 262.88 ± 1.49 | 0.469 ± 0.018 |
| 21313-0947 | BLA 9 | Ross 775 | J21313-097 | M4.5 V | 21:31:19.92 | -09:47:27.5 | (AB) | AB | 0.156 | 122.8 | 66.21 | 1248.01 | ... | 0.228 ± 0.001 |
| ... | ... | BBCap | J21296+176 | M3.0 V | 21:29:37.94 | +17:38:41.9 | Aab | Aab(EB?) | ... | ... | 112.99 | 1075.79 | ... | ... |
| ... | ... | GJ 4201 A | J21323+245 | M4.0 V | 21:32:22.33 | +24:33:41.9 | AB | A | ... | ... | 55.82 | 229.20 | ... | 0.324 ± 0.034 |
| 21324+2434 | MCT 12 | GJ 4201 B | ... | M4.0 V | 21:32:22.23 | +24:33:41.1 | ... | B | 1.548 | 240.8 | 56.17 | 216.79 | ... | 0.300 ± 0.035 |
| ... | ... | GJ 4203 | J21338+017S | M4.0 V | 21:33:49.11 | +01:46:44.9 | A+B | A | ... | ... | 64.15 | 727.05 | 55.98 ± 0.44 | 0.233 ± 0.012 |
| 21338+0147 | BLZ 1 | GJ 4204 | J21338+017N | M4.0 V | 21:33:49.13 | +01:46:50.0 | ... | B | 5.173 | 3.2 | 53.81 | 723.26 | ... | 0.207 ± 0.038 |
| ... | ... | Wolf 923 | J21338-068 | M4.0 V | 21:33:49.04 | -06:51:18.4 | ... | ... | ... | ... | 50.26 | 518.43 | 58.05 ± 0.41 | 0.238 ± 0.012 |
| ... | ... | Wolf 926 | J21348+515 | M2.5 V | 21:34:51.13 | +51:32:18.6 | ... | ... | ... | ... | 29.05 | 555.58 | 274.19 ± 1.45 | 0.445 ± 0.012 |
| ... | ... | V2168Cyg | J21366+394 | M0.0 V | 21:36:38.29 | +39:27:18.0 | AB | A | ... | ... | ... | 263.42 | ... | 0.630 ± 0.027 |
| 21366+3928 | VYS 10 | GJ 834 B | ... | M2.5 V | 21:36:38.20 | +39:27:17.8 | ... | B | 1.003 | 260.8 | 62.49 | 245.73 | ... | 0.435 ± 0.031 |
| ... | ... | Ross 215 | J21369+561 | M1.5 V | 21:36:58.82 | +56:07:07.2 | ... | ... | ... | ... | 62.53 | 153.57 | 679.51 ± 3.30 | 0.600 ± 0.027 |

Table D.2: Complete sample with the description of multiple systems (continued).

| WDS id | WDS disc | Name | Karmn | Spectral type | $\alpha$ (2016.0) | $\delta$ (2016.0) | System | Component | $\rho$ [arcsec] | $\theta$ [deg] | $\varpi$ [mas] | $\mu_{total}$ [mas a$^{-1}$] | $\mathcal{L}$ [$10^{-4}$ $\mathcal{L}_\odot$] | $\mathcal{M}$ [$\mathcal{M}_\odot$] |
|---|---|---|---|---|---|---|---|---|---|---|---|---|---|---|
| 21375-0555 | JNN 133 | PM J21374-0555 | J21374-059 | M3.0 V | 21:37:29.02 | -05:55:05.7 | AB | A | 0.219 | 172.0 | 26.58 | 169.19 | … | … |
| … | … | G3- 2670835278558316160 | … | … | 21:37:29.01 | -05:55:05.5 | … | B | 0.261 | 316.3 | 75.25 | … | … | … |
| … | … | Ross 199 | J21378+530 | M0.0 V | 21:37:50.77 | +53:04:49.9 | … | … | … | … | 51.82 | 317.05 | 1297.77 ± 13.35 | 0.728 ± 0.025 |
| 21376+0137 | JNN291 | 1R213740.3+013711 | J21376+016 | M4.5 V | 21:37:40.28 | +01:37:12.7 | (AB) | AB | 0.420 | 345.0 | 60.30 | 99.36 | … | 0.445 ± 0.031 |
| 21379+2743 | HDS3080 | GJ 835 | J21380+277 | M0.0 V | 21:38:00.96 | +27:43:24.4 | (AB)C | AB | 0.167 | 109.0 | … | 485.01 | … | 0.558 ± 0.028 |
| 21379+2743 | SKF 245 | BD+27 4120B | … | M5.0 V | 21:38:01.04 | +27:43:27.6 | … | C | 3.336 | 16.3 | 46.97 | 487.70 | … | 0.195 ± 0.039 |
| … | … | LP 286-1 | J21402+370 | M0.5 V | 21:40:12.27 | +37:03:24.1 | … | … | … | … | 47.24 | 490.75 | 536.90 ± 1.81 | 0.569 ± 0.028 |
| 21399+2737 | HDS3083 | GJ 4210 | J21399+276 | M2.0 V | 21:39:54.70 | +27:36:39.8 | (AB) | AB | 0.085 | 298.4 | 46.20 | 353.56 | … | 0.508 ± 0.029 |
| … | … | Ross 206 | J21421-121 | M3.0 V | 21:42:07.58 | -12:09:59.3 | … | … | … | … | 92.65 | 692.97 | 412.28 ± 3.03 | 0.483 ± 0.030 |
| 21420+2741 | RAO 465 | GJ 4212 | J21419+276 | M4.0 V | 21:41:58.03 | +27:41:14.2 | (AB) | AB | 2.562 | 302.9 | 62.14 | 296.96 | … | 0.296 ± 0.035 |
| … | … | GJ 4214 | J21441+170S | M4.0 V | 21:44:09.31 | +17:03:35.0 | A+B | A | … | … | 61.85 | 252.08 | 75.44 ± 0.31 | 0.274 ± 0.013 |
| 21440+1705 | LDS6358 | GJ 4215 | J21441+170N | M5.2 V | 21:44:08.25 | +17:04:37.3 | … | B | 64.149 | 346.4 | 48.93 | 245.38 | 35.79 ± 0.15 | 0.195 ± 0.011 |
| … | … | GJ 4213 | J21442+066 | M3.0 Ve | 21:44:12.64 | +06:38:26.6 | Aab | Aab(1) | … | … | 40.10 | 364.35 | … | … |
| 21448+4425 | COU 2234 | G 215-12 | J21449+442 | M1.5 V | 21:44:53.78 | +44:16:58.4 | (AB) | AB | 0.107 | 84.3 | 48.93 | 670.31 | … | 0.534 ± 0.029 |
| … | … | G 126-32A | … | M1.0 V | 21:45:04.94 | +19:53:31.8 | ABC | A | … | … | 29.31 | 219.44 | … | 0.527 ± 0.029 |
| 21451+1954 | RAO 466 | G 126-32B | J21450+198 | M1.5 V | 21:45:04.90 | +19:53:31.7 | … | B | 0.564 | 253.2 | 31.05 | 240.74 | … | 0.484 ± 0.030 |
| … | … | G 126-32C | … | … | 21:45:05.04 | +19:53:36.9 | … | C* | 5.317 | 16.4 | 31.93 | 232.18 | … | 0.143 ± 0.058 |
| … | … | Wolf 937 | J21450-057 | M3.0 V | 21:45:00.59 | -05:47:20.3 | … | … | … | … | 27.36 | 467.42 | 101.66 ± 0.51 | 0.302 ± 0.014 |
| … | … | Wolf 939 | J21454-059 | M3.5 V | 21:45:24.74 | -05:54:11.5 | … | … | … | … | 27.36 | 371.17 | 128.06 ± 0.66 | 0.341 ± 0.015 |
| … | … | LSPM J2146+3813 | J21463+382 | M5.0 V | 21:46:22.29 | +38:13:03.1 | … | … | … | … | 32.59 | 201.01 | 45.87 ± 0.20 | 0.202 ± 0.009 |
| … | … | G 264-12 | J21466+668 | M4.0 V | 21:46:41.30 | +66:48:14.0 | … | … | … | … | 43.57 | 443.63 | 103.78 ± 0.45 | 0.284 ± 0.011 |
| … | … | Wolf 940 | J21466-001 | dM4.0 | 21:46:41.24 | -00:10:31.9 | (AB) | A | … | … | 43.57 | 920.67 | 95.47 ± 0.54 | 0.272 ± 0.010 |
| 21467-0010 | BNG 2 | Wolf 940 B | … | T8+/Y? | … | … | … | B | 31.640 | 250.5 | 28.71 | 919.00 | … | … |
| … | … | Wolf 944 | J21469+466 | M4.0 V | 21:46:56.68 | +46:38:06.1 | … | … | … | … | 38.92 | 278.80 | 93.00 ± 0.64 | 0.307 ± 0.014 |
| … | … | PM J21472-0444 | J21472-047 | M4.5 V | 21:47:17.73 | -04:44:40.6 | … | … | … | … | 57.18 | 257.05 | 30.29 ± 0.15 | 0.177 ± 0.011 |
| … | … | TYC 4266-736-1 | J21474+627 | M0.0 V | 21:47:24.39 | +62:45:13.7 | A+B | A | … | … | 26.83 | 173.75 | 870.84 ± 6.24 | 0.667 ± 0.026 |
| 21474+6245 | WIS 354 | LSPM J2147+6246 | … | M6.0 V | 21:47:25.40 | +62:46:22.4 | … | B | 69.038 | 5.8 | 26.58 | 175.12 | 11.90 ± 0.13 | 0.131 ± 0.010 |
| … | … | Wolf 945 | J21478+502 | M4.0 V | 21:47:53.64 | +50:14:54.6 | … | … | … | … | 33.90 | 813.93 | 113.31 ± 0.95 | 0.320 ± 0.014 |
| … | … | Ross 779 | J21479+058 | M2.0 V | 21:47:57.57 | +05:49:16.4 | … | … | … | … | 33.90 | 458.26 | 567.25 ± 2.45 | 0.566 ± 0.028 |
| … | … | GJ 4221 | J21481+014 | M3.0 V | 21:48:10.46 | +01:26:42.0 | … | … | … | … | 33.94 | 226.43 | 184.76 ± 0.99 | 0.415 ± 0.017 |
| … | … | GJ 4225 | J21482+279 | M2.0 V | 21:48:15.02 | +27:55:31.0 | … | … | … | … | 37.09 | 721.44 | 170.31 ± 0.68 | 0.351 ± 0.015 |
| … | … | GJ 4227 | J21512+128 | M4.0 V | 21:51:18.24 | +12:50:33.2 | … | … | … | … | 37.15 | 686.04 | 142.00 ± 0.63 | 0.361 ± 0.015 |
| 21518+1336 | CRC 40 | GJ 4228 | J21518+136 | M4.0 V | 21:51:48.53 | +13:36:13.8 | (AB) | AB | 10.909 | 121.6 | 51.13 | 211.25 | … | 0.240 ± 0.037 |
| … | … | TYC 3980-1081-1 | J21516+592 | M4.0 V | 21:51:38.15 | +59:17:40.0 | A+B | A | … | … | 31.99 | 79.89 | 231.04 ± 4.09 | 0.467 ± 0.019 |
| 21516+5918 | KPP4552 | UCAC4 747-070768 | … | DAH | 21:51:39.93 | +59:17:34.5 | … | B | 14.642 | 111.9 | 45.53 | 88.85 | … | 0.500 ± 0.100 |
| … | … | GJ 4232 | J21521+274 | M5.0 V | 21:52:12.12 | +27:24:48.2 | … | … | … | … | 45.24 | 806.23 | 89.39 ± 0.42 | 0.281 ± 0.013 |
| … | … | GJ 839 | J21539+417 | M0.0 V | 21:53:59.55 | +41:46:38.7 | … | … | … | … | 104.44 | 527.45 | 1031.11 ± 4.48 | 0.695 ± 0.026 |
| … | … | Ross 263 | J21566+197 | M3.0 V | 21:56:37.83 | +19:46:06.8 | … | … | … | … | 54.55 | 488.18 | 188.98 ± 1.49 | 0.394 ± 0.016 |
| … | … | GJ 4239 | J21569-019 | M5.0 V | 21:56:56.61 | -01:53:59.5 | … | … | … | … | 32.49 | 1419.70 | 27.60 ± 0.13 | 0.168 ± 0.010 |
| … | … | Wolf 953 | J21574+081 | M1.5 V | 21:57:26.64 | +08:08:15.5 | … | … | … | … | 133.81 | 386.95 | 565.14 ± 3.17 | 0.560 ± 0.028 |

Table D.2: Complete sample with the description of multiple systems (continued).

| WDS id | WDS disc | Name | Karmn | Spectral type | $\alpha$ (2016.0) | $\delta$ (2016.0) | System | Component | $\rho$ [arcsec] | $\theta$ [deg] | $\varpi$ [mas] | $\mu_{\rm total}$ [mas a$^{-1}$] | $\mathcal{L}$ [$10^{-4}$ $\mathcal{L}_\odot$] | $\mathcal{M}$ [$\mathcal{M}_\odot$] |
|---|---|---|---|---|---|---|---|---|---|---|---|---|---|---|
| … | … | GJ 842.2 | J21584+755 | M0.5 V | 21:58:25.51 | +75:35:21.0 | … | … | … | … | 123.65 | 232.45 | 747.29 ± 3.89 | 0.623 ± 0.027 |
| … | … | LSPM J2158+6117 | J21585+612 | M6.0 V | 21:58:36.41 | +61:17:07.8 | … | … | … | … | 29.46 | 815.50 | 12.65 ± 0.07 | 0.115 ± 0.009 |
| … | … | GJ 4246 | J21593+418 | M3.0 V | 21:59:22.08 | +41:51:25.4 | … | … | … | … | 62.59 | 423.27 | 159.25 ± 0.59 | 0.360 ± 0.015 |
| … | … | V374Peg | J22012+283 | M3.5 Ve | 22:01:13.58 | +28:18:25.6 | … | … | … | … | 64.94 | 375.25 | 99.29 ± 0.53 | 0.319 ± 0.011 |
| 21522+0538 | JOD 23 | GJ 4231 | J21521+056 | M2.4 V | 21:52:10.52 | +05:37:33.5 | (AB) | AB | 0.132 | 263.6 | 28.73 | 191.80 | … | 0.508 ± 0.058 |
| … | … | Wolf 1154 A | J22012+323 | M1.5 V | 22:01:14.13 | +32:23:13.9 | A(BC) | A | … | … | 29.02 | 133.62 | … | 0.493 ± 0.030 |
| 22012+3223 | GRV1283 | Wolf 1154 B | … | M3.5 V | 22:01:14.04 | +32:23:13.2 | … | B | 1.302 | 235.5 | 35.60 | 119.08 | … | 0.323 ± 0.034 |
| … | … | 2M22011701+3222062 | … | T2.5 | … | … | … | C | … | … | 47.42 | 125.77 | … | … |
| 22018+1628 | YSC 165 | Ross 265 | J22018+164 | M2.5 V | 22:01:49.49 | +16:28:05.2 | (AB)+C | AB | 0.355 | 36.3 | 45.79 | 421.76 | … | 0.505 ± 0.031 |
| … | … | Ross 268 | … | M3.5 V | 22:06:23.02 | +17:22:22.1 | … | C* | 5100.550 | 50.2 | 61.13 | 371.56 | 116.63 ± 0.52 | 0.347 ± 0.015 |
| … | … | GJ 843 | J22020-194 | dM3.5 | 22:02:01.85 | -19:28:58.0 | … | … | … | … | 30.17 | 933.09 | 145.39 ± 1.16 | 0.327 ± 0.011 |
| … | … | HD 209290 | J22021+014 | M0.5 V | 22:02:09.79 | +01:23:56.4 | … | … | … | … | 23.62 | 531.35 | 638.60 ± 5.17 | 0.596 ± 0.028 |
| … | … | GJ 4251 | J22033+674 | M4.5 V | 22:03:22.74 | +67:29:55.1 | … | … | … | … | 41.26 | 608.37 | 102.62 ± 0.49 | 0.303 ± 0.014 |
| … | … | Wolf 983 | J22051+051 | M4.0 V | 22:05:07.30 | +05:08:14.2 | … | … | … | … | 104.92 | 479.50 | 161.43 ± 0.75 | 0.386 ± 0.016 |
| 22035+0340 | JNN293 | 1R220330.8+034001 | J22035+036 | M4.0 V | 22:03:33.39 | +03:40:21.4 | (AB) | AB | 0.412 | 351.9 | 47.75 | 126.77 | … | 0.248 ± 0.037 |
| … | … | GJ 4258 | J22057+656 | M1.5 V | 22:05:44.55 | +65:38:58.9 | A+B | A | … | … | … | 380.08 | 373.21 ± 2.57 | 0.470 ± 0.012 |
| 22058+6539 | NI 44 | G 264-18B | … | M4.5 V | 22:05:45.29 | +65:38:53.9 | … | B | 6.782 | 137.7 | … | 377.33 | 57.83 ± 0.37 | 0.237 ± 0.012 |
| … | … | Wolf 1548 | J22058-119 | M0.0 V | 22:05:51.00 | -11:54:53.7 | A+B | A | … | … | 100.79 | 319.26 | 1476.70 ± 47.95 | 0.747 ± 0.026 |
| 22059-1155 | WNO 57 | LP 759-25 | … | M6.0 Ve | 22:05:35.44 | -11:04:31.8 | … | B | 3030.622 | 355.7 | 101.97 | 322.49 | 10.00 ± 0.06 | 0.118 ± 0.009 |
| … | … | GJ 4256 | J22060+393 | M3.0 V | 22:06:00.76 | +39:17:55.6 | … | … | … | … | 102.94 | 464.23 | 164.44 ± 1.22 | 0.366 ± 0.015 |
| … | … | Wolf 990 | J22067+034 | M4.0 V | 22:06:46.87 | +03:24:58.7 | … | … | … | … | 52.43 | 569.30 | 78.81 ± 0.53 | 0.281 ± 0.013 |
| … | … | PM J22088+1144 | J22088+117 | M4.5 V | 22:08:50.44 | +11:44:12.4 | … | … | … | … | 66.45 | 103.82 | 203.54 ± 1.19 | 0.498 ± 0.020 |
| … | … | LP 519-38 | J22095+118 | M3.0 V | 22:09:31.86 | +11:52:51.6 | … | … | … | … | 46.85 | 208.15 | 258.81 ± 1.54 | 0.465 ± 0.018 |
| … | … | Wolf 1329 | J22096-046 | M3.5 V | 22:09:41.56 | -04:38:27.0 | … | … | … | … | 46.58 | 1132.80 | 283.03 ± 1.83 | 0.448 ± 0.014 |
| … | … | GJ 4260 | J22097+410 | M3.0 V | 22:09:43.64 | +41:02:09.5 | … | … | … | … | 38.21 | 485.43 | 249.93 ± 1.37 | 0.457 ± 0.018 |
| … | … | UCAC4 744-073158 | J22102+587 | M2.0 V | 22:10:15.18 | +58:42:21.9 | … | … | … | … | 60.96 | 29.80 | 307.54 ± 1.61 | 0.479 ± 0.018 |
| … | … | Wolf 1003 A | J22107+079 | M0.5 V | 22:10:44.98 | +07:54:32.9 | AB | A | … | … | 50.80 | 255.11 | … | 0.596 ± 0.028 |
| 22107+0755 | WOR 10 | Wolf 1003 B | … | M1.0 V | 22:10:44.98 | +07:54:33.8 | … | B | 0.894 | 5.4 | 40.46 | 241.13 | … | 0.568 ± 0.028 |
| … | … | G 214-14 | J22112+410 | M0.0 V | 22:11:16.65 | +41:00:58.6 | … | … | … | … | 40.35 | 313.80 | 486.65 ± 2.11 | 0.563 ± 0.028 |
| … | … | GJ 4262 | J22112-025 | M2.0 V | 22:11:13.95 | -02:32:38.2 | … | … | … | … | 82.89 | 433.31 | 262.26 ± 2.07 | 0.441 ± 0.017 |
| … | … | 1R221124.3+410000 | J22114+409 | M5.5 V | 22:11:24.04 | +40:59:59.8 | … | … | … | … | 40.38 | 113.38 | 26.43 ± 0.10 | 0.160 ± 0.009 |
| … | … | Ross 271 | J22115+184 | M2.0 V | 22:11:30.46 | +18:25:37.2 | … | … | … | … | 29.80 | 375.38 | 435.24 ± 3.02 | 0.506 ± 0.029 |
| … | … | WT 2221 | J22117-207 | M3.5 V | 22:11:42.27 | -20:44:19.3 | A+B | A | … | … | 177.93 | 160.10 | 340.88 ± 2.19 | 0.443 ± 0.031 |
| 22117-2044 | FMR 18 | WT 2220 | … | M3.0 V | 22:11:41.51 | -20:44:12.0 | … | B | 12.876 | 304.6 | 45.59 | 159.43 | 55.47 ± 0.38 | 0.232 ± 0.012 |
| … | … | Wolf 1014 | J22125+085 | M3.0 V | 22:12:36.06 | +08:33:00.9 | … | … | … | … | 142.05 | 679.53 | 186.37 ± 0.78 | 0.362 ± 0.011 |
| … | … | LF 4 +54 152 | J22129+550 | M0.0 V | 22:12:56.63 | +55:04:50.8 | AB | A | … | … | 77.94 | 109.13 | … | 0.6220 ± 0.093 |
| … | … | G3- 2005884249925303168 | … | M6.5 V | 22:13:00.46 | +55:05:48.3 | … | B* | 66.265 | 29.8 | 35.01 | 108.97 | … | 0.109 ± 0.046 |
| … | … | Wolf 1556 | J22134-147 | M3.5 V | 22:13:28.46 | -14:44:58.8 | … | … | … | … | 29.20 | 408.23 | 244.01 ± 1.94 | 0.480 ± 0.019 |
| … | … | GJ 4264 | J22135+259 | M4e | 22:13:35.90 | +25:58:08.1 | … | … | … | … | 84.86 | 227.21 | 96.90 ± 0.41 | 0.314 ± 0.014 |
| … | … | GJ 1265 | J22137-176 | dM4.5 | 22:13:43.82 | -17:41:13.6 | … | … | … | … | 80.20 | 909.89 | 36.06 ± 0.24 | 0.175 ± 0.010 |

Table D.2: Complete sample with the description of multiple systems (continued).

| WDS id | WDS disc | Name | Karmn | Spectral type | $\alpha$ (2016.0) | $\delta$ (2016.0) | System | Component | $\rho$ [arcsec] | $\theta$ [deg] | $\varpi$ [mas] | $\mu_{total}$ [mas a$^{-1}$] | $\mathcal{L}$ [$10^{-4}\ \mathcal{L}_\odot$] | $\mathcal{M}$ [$\mathcal{M}_\odot$] |
|---|---|---|---|---|---|---|---|---|---|---|---|---|---|---|
| 22139+0517 | JOD 24 | Wolf 1019 | J22138+052 | M1.5 V | 22:13:53.53 | +05:16:34.9 | (AB) | AB | 3.053 | 317.9 | 61.98 | 189.46 | … | 0.507 ± 0.029 |
| … | … | GJ 4267 | J22154+662 | M3.0 V | 22:15:26.16 | +66:13:31.1 | … | … | … | … | 61.82 | 208.93 | 142.86 ± 0.78 | 0.362 ± 0.015 |
| 22143+2534 | BWL57 | 1R221419.3+253411 | J22142+255 | M4.3 V | 22:14:17.88 | +25:34:05.6 | (AB) | AB | 0.141 | 304.0 | 61.76 | 166.68 | … | 0.352 ± 0.033 |
| … | … | V447Lac | … | K1 V | 22:15:54.53 | +54:40:23.5 | A+B | A | … | … | 69.30 | 223.87 | 4489.55 ± 18.72 | 0.860 ± 0.129 |
| 22159+5440 | GIC 177 | GJ 4269 | J22160+546 | M4.0 V | 22:16:02.99 | +54:40:00.5 | … | B | 76.900 | 107.4 | 39.68 | 221.35 | … | 0.272 ± 0.036 |
| … | … | GJ 1266 | J22163+709 | M2.0 V | 22:16:23.03 | +70:56:39.2 | … | … | … | … | 33.02 | 863.79 | 253.12 ± 1.30 | 0.408 ± 0.016 |
| 22173-0847 | LDS 782 | FGAqr | J22173-088N | M4.0 V | 22:17:18.47 | -08:48:16.9 | A+(BC) | A | 7.900 | 215.6 | 66.55 | 544.62 | 40.71 ± 0.47 | 0.195 ± 0.011 |
| 22173-0847 | BEU 22 | Wolf 1561 B | J22173-088S | M5.0 V | 22:17:18.16 | -08:48:23.4 | … | BC | 7.900 | 215.6 | 49.78 | 555.64 | … | 0.158 ± 0.041 |
| … | … | PM J22176+5633 | J22176+565 | M1.5 V | 22:17:37.21 | +56:33:11.1 | … | … | … | … | 47.97 | 122.47 | 1199.15 ± 82.71 | 0.681 ± 0.028 |
| … | … | Wolf 1034 | J22202+067 | M2.5 V | 22:20:13.57 | +06:43:36.5 | … | … | … | … | 69.02 | 399.86 | 76.51 ± 0.43 | 0.242 ± 0.012 |
| 22212+3745 | NSN 753 | LSPM J2221+3744A | … | M1.5 V | 22:21:13.22 | +37:44:50.8 | AB | A | 2.053 | 248.8 | 28.86 | 159.43 | … | 0.509 ± 0.029 |
| 22212+3745 | NSN 753 | LSPM J2221+3744B | … | M2.0 V | 22:21:13.06 | +37:44:50.2 | … | B | 1.980 | 250.3 | 25.95 | 165.62 | … | 0.466 ± 0.030 |
| … | … | GJ 4275 | J22228+280 | M3.8 V | 22:22:51.34 | +28:01:46.4 | … | … | … | … | 54.62 | 402.72 | 135.75 ± 0.64 | 0.376 ± 0.016 |
| … | … | GJ 4274 | J22231-176 | M4.5 Ve | 22:23:07.34 | -17:36:37.8 | … | … | … | … | 44.10 | 781.91 | 35.04 ± 0.24 | 0.176 ± 0.009 |
| … | … | Wolf 1225 A | J22234+324 | M3.0 V | 22:23:29.44 | +32:27:29.9 | AB | A | … | … | 44.37 | 350.39 | … | 0.437 ± 0.031 |
| 22234+3228 | WOR 11 | Wolf 1225 B | … | M3.4 V | 22:23:29.34 | +32:27:29.5 | … | B | 1.252 | 254.0 | 44.37 | 306.69 | … | 0.434 ± 0.031 |
| … | … | GJ 1268 | J22249+520 | M5.1 V | 22:24:56.33 | +52:00:25.6 | … | … | … | … | 27.71 | 490.89 | 28.40 ± 0.13 | 0.171 ± 0.010 |
| … | … | Wolf 1231 | J22250+356 | M2.0 V | 22:25:01.66 | +35:40:05.3 | … | … | … | … | 27.65 | 163.04 | 211.93 ± 1.09 | 0.394 ± 0.016 |
| … | … | GJ 4276 | J22252+594 | M3.5 V | 22:25:17.32 | +59:24:44.9 | … | … | … | … | 37.08 | 333.29 | 205.64 ± 1.27 | 0.400 ± 0.014 |
| … | … | Wolf 1201 | J22262+030 | M4.0 V | 22:26:15.24 | +03:00:11.3 | A+B | A | … | … | 23.81 | 668.79 | 115.11 ± 0.82 | 0.344 ± 0.015 |
| 22263+0301 | LDS4967 | GJ 4277 B | … | M6.0 V | 22:26:14.97 | +03:00:00.4 | … | B | 11.565 | 200.5 | 48.14 | 675.07 | 21.66 ± 0.13 | 0.158 ± 0.010 |
| … | … | PM J22264+5823 | J22264+583 | M3.0 V | 22:26:24.73 | +58:23:03.9 | … | … | … | … | 34.85 | 147.29 | 161.33 ± 0.78 | 0.386 ± 0.016 |
| … | … | GJ 4279 | J22270+068 | M4.0 V | 22:27:03.00 | +06:49:32.1 | … | … | … | … | 35.26 | 186.47 | 65.35 ± 0.35 | 0.254 ± 0.012 |
| 22280+5742 | KR60 | HD 239960A | J22279+576 | M3.0 V | 22:27:58.11 | +57:41:38.5 | AB | A | 2.052 | 206.5 | 84.44 | 758.87 | … | 0.306 ± 0.034 |
| … | … | DOCep | … | M4.0 V | 22:27:57.93 | +57:41:38.8 | … | B | 1.420 | 280.0 | 30.90 | 1159.08 | … | 0.199 ± 0.039 |
| … | … | GJ 9784 | J22287+189 | M1.0 V | 22:28:46.13 | +18:55:52.1 | … | … | … | … | 29.77 | 210.47 | 619.64 ± 2.66 | 0.597 ± 0.027 |
| … | … | GJ 4281 | J22289-134 | M6.5 V | 22:28:53.99 | -13:25:36.2 | … | … | … | … | 34.07 | 1095.16 | 6.89 ± 0.04 | 0.104 ± 0.047 |
| … | … | LP 640-74 | J22290+016 | M0.5 V | 22:29:05.91 | +01:39:45.0 | … | … | … | … | 29.27 | 200.07 | 754.74 ± 4.65 | 0.630 ± 0.027 |
| … | … | GJ 1270 | J22298+414 | M4.0 V | 22:29:50.68 | +41:28:55.8 | … | … | … | … | 31.00 | 1274.43 | 75.93 ± 0.32 | 0.252 ± 0.010 |
| … | … | PM J22300+4851A | J22300+488 | M4.5 V | 22:30:04.10 | +48:51:33.8 | AB | A | … | … | 38.41 | 96.16 | … | 0.350 ± 0.033 |
| 22300+4851 | JNN 294 | PM J22300+4851B | … | M4.5 V | 22:30:03.87 | +48:51:33.1 | … | B | 2.344 | 252.4 | 36.46 | 106.07 | … | 0.238 ± 0.037 |
| … | … | GJ 863 | J22330+093 | M1.0 V | 22:33:02.81 | +09:22:43.0 | … | … | … | … | 33.19 | 556.38 | 360.49 ± 1.97 | 0.458 ± 0.012 |
| … | … | GJ 4282 A | J22333-096 | M2.6 V | 22:33:22.77 | -09:36:53.6 | AB | A | … | … | 36.41 | 158.89 | … | 0.380 ± 0.032 |
| 22334-0937 | MCT 13 | GJ 4282 B | … | M2.5 V | 22:33:22.87 | -09:36:53.7 | … | B | 1.399 | 95.5 | 49.58 | 139.59 | … | 0.382 ± 0.032 |
| … | … | GJ 4283 | J22347+040 | M3.0 V | 22:34:46.11 | +04:02:39.7 | … | … | … | … | 49.10 | 160.66 | 109.81 ± 0.55 | 0.295 ± 0.013 |
| … | … | GJ 4284 | J22348-010 | M4.5 V | 22:34:54.85 | -01:04:54.5 | … | … | … | … | 37.39 | 1118.60 | 48.94 ± 0.23 | 0.216 ± 0.011 |
| … | … | G 242-3 | J22353+746 | M0.0 V | 22:35:20.75 | +74:41:20.1 | … | … | … | … | 50.76 | 262.63 | 1095.50 ± 22.56 | 0.696 ± 0.026 |
| … | … | LP 344-27 | J22373+299 | M1.5 V | 22:37:23.33 | +29:59:05.6 | … | … | … | … | 49.91 | 299.20 | 513.19 ± 2.74 | 0.544 ± 0.028 |
| 22360-0050 | JOD 25 | HD 214100 | J22361-008 | M1.0 V | 22:36:09.75 | -00:50:39.8 | (AB) | AB | 0.631 | 291.7 | 56.69 | 632.79 | … | 0.610 ± 0.027 |
| 22375+3923 | HDS3211 | GJ 4287 | J22374+395 | M0.0 V | 22:37:29.83 | +39:22:45.9 | (AB)+C | AB | 0.261 | 11.0 | 36.56 | 365.03 | … | … |

Table D.2: Complete sample with the description of multiple systems (continued).

| WDS id | WDS disc | Name | Karmn | Spectral type | $\alpha$ (2016.0) | $\delta$ (2016.0) | System | Component | $\rho$ [arcsec] | $\theta$ [deg] | $\varpi$ [mas] | $\mu_{\rm total}$ [mas a$^{-1}$] | $\mathcal{L}$ [$10^{-4}$ $\mathcal{L}_\odot$] | $\mathcal{M}$ [$\mathcal{M}_\odot$] |
|---|---|---|---|---|---|---|---|---|---|---|---|---|---|---|
| 22375+3923 | KIR 5 | G 216-7B | … | M9.5 | 22:37:32.51 | +39:22:33.7 | … | C | 33.332 | 111.5 | 32.79 | 351.15 | 2.42 ± 0.02 | 0.078 ± 0.012 |
| … | … | G 127-42 | J22387+252 | M3.5 V | 22:38:44.65 | +25:13:30.5 | … | … | … | … | 32.75 | 284.85 | 161.19 ± 0.76 | 0.386 ± 0.016 |
| 22385-1519 | BLA 10 | EZAqr | J22385-152 | M5.5 V | 22:38:36.17 | -15:17:22.7 | (AabB) | AabB(3) | 0.095 | 245.8 | 58.41 | 3259.86 | … | 0.333 ± 0.005 |
| 22388-2037 | HJ 3126 | FKAqr | J22387-206S | M1.5 V | 22:38:46.09 | -20:37:17.4 | Aab+Bab | Aab(2) | 24.887 | 349.6 | 42.66 | 456.11 | … | 0.489 ± 0.008 |
| … | … | FLAqr | J22387-206N | M3.5 V | 22:38:45.77 | -20:36:52.9 | … | Bab(1) | 24.883 | 349.6 | 30.22 | 429.84 | … | 0.296 ± 0.007 |
| … | … | GJ 4290 | J22406+445 | M3.5 V | 22:40:42.53 | +44:35:47.0 | … | … | … | … | 31.97 | 478.30 | 407.24 ± 12.61 | 0.471 ± 0.030 |
| … | … | GJ 9793 | J22415+188 | M0.0 V | 22:41:35.30 | +18:49:28.9 | … | … | … | … | 37.15 | 268.53 | 1143.69 ± 20.68 | 0.702 ± 0.026 |
| … | … | 1R224134.7+260210 | J22415+260 | M3.5 V | 22:41:35.76 | +26:02:13.8 | … | … | … | … | 33.51 | 63.05 | 324.20 ± 2.00 | 0.447 ± 0.031 |
| … | … | GJ 1271 | J22426+176 | M2.5 V | 22:42:39.99 | +17:40:17.6 | … | … | … | … | 26.76 | 1218.10 | 345.97 ± 2.37 | 0.465 ± 0.030 |
| … | … | GJ 4292 | J22433+221 | M5.0 V | 22:43:23.65 | +22:08:18.1 | … | … | … | … | 64.58 | 390.11 | 45.02 ± 0.18 | 0.221 ± 0.012 |
| … | … | RX J2243.7+1916 | J22437+192 | M3.0 V | 22:43:43.73 | +19:16:52.4 | … | … | … | … | 25.90 | 106.47 | 409.73 ± 13.14 | 0.481 ± 0.030 |
| … | … | TYC 3218-905-1 | J22441+405 | M1.0 V | 22:44:06.16 | +40:29:58.1 | A+B | A | … | … | 68.92 | 147.00 | 398.71 ± 1.45 | 0.489 ± 0.018 |
| 22441+4029 | LDS1064 | TYC 3218-907-1 | J22440+405 | M1.0 V | 22:44:04.51 | +40:29:58.5 | … | B | 18.823 | 271.0 | 36.07 | 133.27 | 399.40 ± 1.66 | 0.489 ± 0.018 |
| … | … | LP 641-4 | J22457+016 | M1.0 V | 22:45:46.48 | +01:41:22.0 | AB | A | … | … | 59.63 | 210.48 | … | 0.520 ± 0.029 |
| 22458+0141 | NSN 760 | LP 641-4 B | … | M6.5 V | 22:45:46.51 | +01:41:19.5 | … | B | 2.533 | 170.7 | 38.29 | 194.61 | … | 0.105 ± 0.047 |
| … | … | GJ 4294 | J22464-066 | M5.0 V | 22:46:27.09 | -06:39:35.0 | … | … | … | … | 38.48 | 879.92 | 24.67 ± 0.13 | 0.170 ± 0.011 |
| … | … | ELac | J22468+443 | M4.0 Ve | 22:46:48.68 | +44:19:55.0 | … | … | … | … | 26.50 | 842.23 | 125.77 ± 0.85 | 0.337 ± 0.012 |
| … | … | LP 461-11 | J22476+184 | M2.5 V | 22:47:39.35 | +18:26:40.4 | … | … | … | … | 47.32 | 433.03 | 241.62 ± 9.93 | 0.478 ± 0.022 |
| … | … | GJ 4297 | J22479+318 | M3.0 V | 22:47:54.67 | +31:52:18.4 | … | … | … | … | 134.08 | 495.59 | 118.15 ± 0.48 | 0.307 ± 0.014 |
| … | … | PM J22489+1819 | J22489+183 | M4.5 V | 22:48:54.56 | +18:19:56.9 | … | … | … | … | 147.50 | 130.21 | 49.91 ± 0.22 | 0.219 ± 0.011 |
| … | … | HD 216133 | J22503-070 | M0.5 V | 22:50:19.31 | -07:05:22.7 | … | … | … | … | 48.59 | 148.74 | 591.79 ± 2.75 | 0.590 ± 0.028 |
| … | … | GJ 1274 | J22506+348 | M2+ V | 22:50:38.79 | +34:51:26.8 | … | … | … | … | 48.73 | 896.32 | 274.95 ± 1.20 | 0.452 ± 0.017 |
| … | … | GJ 4300 | J22507+286 | M2.5 V | 22:50:45.77 | +28:36:07.7 | … | … | … | … | 59.20 | 199.23 | 239.43 ± 1.20 | 0.447 ± 0.018 |
| … | … | 1R225056.4+495906 | J22509+499 | M4.0 V | 22:50:55.28 | +49:59:13.2 | … | … | … | … | 59.24 | 123.60 | 135.56 ± 0.59 | 0.402 ± 0.017 |
| … | … | GTPeg | J22518+317 | M3.5 Ve | 22:51:54.19 | +31:45:14.4 | … | … | … | … | 59.75 | 526.02 | 262.27 ± 1.26 | 0.419 ± 0.031 |
| 22524+0950 | LDS6388 | HD 216385 | … | F6 V | 22:52:24.64 | +09:50:09.1 | A+(BC) | A | 249.637 | 19.6 | 56.72 | 523.35 | 52617.38 ± 1081.20 | 1.250 ± 0.188 |
| 22524+0950 | RAO30 | GJ 9801 B | J22524+099 | M3.0 V | 22:52:30.31 | +09:54:04.9 | … | BC | 250.246 | 19.6 | 48.11 | 530.18 | … | 0.345 ± 0.034 |
| … | … | LP 48-305 | J22526+750 | M4.5 V | 22:52:40.21 | +75:04:16.7 | … | … | … | … | 47.91 | 180.65 | 68.82 ± 0.39 | 0.279 ± 0.013 |
| … | … | Ross 226 | J22543+609 | M3.0 V | 22:54:19.95 | +60:59:41.5 | … | … | … | … | 28.24 | 717.82 | 134.41 ± 0.61 | 0.350 ± 0.015 |
| … | … | GJ 4302 | J22547-054 | M4.0 V | 22:54:47.14 | -05:28:21.1 | … | … | … | … | … | 696.98 | 117.96 ± 5.89 | 0.349 ± 0.018 |
| … | … | ILAqr | J22532-142 | M4.0 V | 22:53:17.79 | -14:16:00.1 | Aab | Aab(2) | … | … | 36.72 | 1170.88 | … | 0.318 ± 0.034 |
| … | … | GJ 4304 | J22559+057 | M1.0 V | 22:55:57.20 | +05:45:14.0 | A+B | A | … | … | … | 446.04 | 586.61 ± 3.24 | 0.572 ± 0.028 |
| 22560+0546 | LDS5021 | GJ 4305 | … | DA8.1 | 22:55:56.09 | +05:45:18.4 | … | B | 17.168 | 284.6 | 32.13 | 439.47 | … | 0.500 ± 0.100 |
| … | … | GJ 4306 | J22559+178 | M1.0 V | 22:55:59.87 | +17:48:38.0 | … | … | … | … | 27.85 | 112.80 | 554.72 ± 2.39 | 0.563 ± 0.028 |
| … | … | HD 216899 | J22565+165 | M1.5 V | 22:56:33.65 | +16:33:07.8 | … | … | … | … | 76.26 | 1073.03 | 525.73 ± 5.17 | 0.548 ± 0.028 |
| … | … | G 189-53A | J22576+373 | M3.0 V | 22:57:40.20 | +37:19:17.8 | AB | A | … | … | 79.23 | 658.43 | … | 0.373 ± 0.032 |
| 22577+3719 | KPP3388 | G 189-53B | … | M5.0 V | 22:57:40.10 | +37:19:15.6 | … | B | 2.522 | 210.7 | 36.03 | 668.95 | … | 0.203 ± 0.039 |
| … | … | 1R230251.9+433814 | J23028+436 | M4.0 V | 23:02:52.29 | +43:38:15.5 | … | … | … | … | 42.80 | 137.69 | 42.02 ± 0.15 | 0.199 ± 0.011 |
| … | … | PM J23036+0942 | J23036+097 | M3.5 V | 23:03:37.56 | +09:42:59.0 | … | … | … | … | 31.22 | 107.10 | 277.95 ± 1.46 | 0.483 ± 0.019 |
| … | … | GJ 1278 | J23045+667 | M0.5 V | 23:04:31.03 | +66:45:50.5 | … | … | … | … | 40.41 | 315.35 | 1030.77 ± 5.18 | 0.692 ± 0.026 |

Table D.2: Complete sample with the description of multiple systems (continued).

| WDS id | WDS disc | Name | Karmn | Spectral type | $\alpha$ (2016.0) | $\delta$ (2016.0) | System | Component | $\rho$ [arcsec] | $\theta$ [deg] | $\varpi$ [mas] | $\mu_{\rm total}$ [mas a$^{-1}$] | $\mathcal{L}$ [$10^{-4}$ $\mathcal{L}_\odot$] | $\mathcal{M}$ [$\mathcal{M}_\odot$] |
|---|---|---|---|---|---|---|---|---|---|---|---|---|---|---|
| ... | ... | PM J23051+5159 | J23051+519 | M3.5 V | 23:05:06.47 | +51:59:11.6 | ... | ... | ... | ... | 58.58 | 121.96 | 318.43 ± 1.50 | 0.442 ± 0.031 |
| ... | ... | GJ 9809 | J23060+639 | M0.3 V | 23:06:05.27 | +63:55:33.4 | ... | ... | ... | ... | 58.61 | 183.38 | 681.87 ± 3.61 | 0.603 ± 0.027 |
| ... | ... | LSPM J2305+4517 | J23051+452 | M3.5 V | 23:05:09.00 | +45:17:32.9 | AB | A | ... | ... | 48.62 | 198.80 | ... | 0.485 ± 0.030 |
| ... | ... | G2-1935209944573613568 | ... | ... | 23:05:09.06 | +45:17:33.1 | ... | B* | 0.688 | 78.0 | 43.64 | ... | ... | ... |
| 23064+1235 | HDS3291 | LP 521-79 | J23063+126 | M0.5 V | 23:06:24.19 | +12:36:25.6 | (AB)+C | AB(2/3?) | 0.467 | 336.3 | 50.41 | 320.69 | ... | ... |
| 23064+1235 | GIC 188 | G 67-47 | ... | M3 IV(e) | 23:06:25.69 | +12:36:55.9 | ... | C | 37.439 | 36.0 | 27.66 | 322.41 | 130.34 ± 1.14 | ... |
| ... | ... | 2MUCD 12171 | J23064-050 | M7.5 Ve | 23:06:30.37 | -05:02:36.7 | ... | ... | ... | ... | 26.82 | 1046.82 | 5.35 ± 0.03 | 0.106 ± 0.009 |
| ... | ... | GJ 4311 | J23065+717 | M2.0 V | 23:06:39.95 | +71:43:32.5 | ... | ... | ... | ... | 26.25 | 1329.33 | 228.60 ± 1.36 | 0.410 ± 0.016 |
| ... | ... | GJ 4312 | J23075+686 | M3.0 V | 23:07:33.27 | +68:40:06.1 | ... | ... | ... | ... | 59.84 | 1143.30 | 120.11 ± 0.62 | 0.309 ± 0.014 |
| ... | ... | GJ 889.1 | J23081+033 | M0.0 V | 23:08:07.49 | +03:19:48.5 | ... | ... | ... | ... | 40.13 | 557.18 | 316.45 ± 1.35 | 0.433 ± 0.016 |
| ... | ... | HKAqr | J23083-154 | M0.0 Ve | 23:08:19.67 | -15:24:36.1 | ... | ... | ... | ... | 141.89 | 108.68 | 673.80 ± 3.36 | 0.608 ± 0.027 |
| ... | ... | StKM 1-2100 | J23088+065 | M0.0 V | 23:08:52.60 | +06:33:39.9 | ... | ... | ... | ... | 62.55 | 103.94 | 1130.74 ± 31.91 | 0.702 ± 0.026 |
| ... | ... | G 233-42 | J23089+551 | M5.0 V | 23:09:58.62 | +55:06:48.1 | A+B | A | ... | ... | 80.74 | 411.27 | 21.25 ± 0.08 | 0.156 ± 0.010 |
| 23100+5507 | NSN 11 | LSPM J2309+5506E | ... | DA | 23:09:59.29 | +55:06:50.2 | ... | B | 6.150 | 69.6 | 79.80 | 410.18 | ... | 0.500 ± 0.100 |
| ... | ... | GJ 4314 | J23096-019 | M3.5 V | 23:09:39.64 | -01:58:29.9 | AabB | Aab(2) | ... | ... | 36.83 | 458.40 | ... | ... |
| 23097-0158 | CRC 76 | G 28-44B | ... | M4.0 V | 23:09:39.68 | -01:58:28.3 | ... | B | 1.697 | 22.5 | 58.35 | 483.31 | ... | 0.286 ± 0.035 |
| ... | ... | GJ 1281 | J23107-192 | M2.5 V | 23:10:42.22 | -19:13:57.9 | ... | ... | ... | ... | 86.87 | 1435.04 | 174.71 ± 1.18 | 0.355 ± 0.015 |
| ... | ... | G 28-46 | J23113+085 | M3.0 V | 23:11:23.45 | +08:30:56.4 | AB | A | ... | ... | 24.92 | 458.92 | 239.41 ± 33.33 | 0.426 ± 0.035 |
| ... | ... | G 28-46B | ... | ... | 23:11:23.47 | +08:30:56.4 | ... | B* | 0.248 | 77.7 | 24.83 | ... | ... | ... |
| ... | ... | GJ 4316 | J23121-141 | M3.0 V | 23:12:11.11 | -14:06:23.2 | ... | ... | ... | ... | 48.52 | 720.48 | 113.47 ± 0.53 | 0.320 ± 0.014 |
| 23167+1937 | HDS3316 | GJ 893.4 | J23166+196 | M2.0 V | 23:16:39.52 | +19:37:14.2 | (AB) | AB | 0.129 | 7.8 | 32.01 | 252.63 | ... | 0.622 ± 0.038 |
| ... | ... | GJ 2154 A | J23142-196N | M0.5 Vk | 23:14:17.13 | -19:38:38.5 | A+B | A | ... | ... | 30.06 | 444.84 | 748.87 ± 4.05 | 0.622 ± 0.027 |
| 23142-1938 | LDS5060 | GJ 2154 B | J23142-196S | M4.0 V | 23:14:16.97 | -19:38:45.4 | ... | B | 7.223 | 198.2 | 59.41 | 451.31 | 105.55 ± 0.84 | 0.288 ± 0.013 |
| 23175+1937 | BEU23 | GJ 4326 | J23174+196 | M2.0 V | 23:17:28.54 | +19:36:45.1 | (AB) | AB | 0.145 | 220.2 | 42.18 | 375.65 | ... | 0.400 ± 0.032 |
| ... | ... | GJ 4319 | J23161+067 | M3.0 V | 22:34:26.25 | -00:28:12.9 | ... | ... | ... | ... | 55.50 | 297.50 | 171.64 ± 0.91 | 0.399 ± 0.016 |
| ... | ... | GJ 4327 | J23174+382 | M3.0 V | 23:17:24.09 | +38:12:34.8 | Aab | Aab(2) | ... | ... | 39.11 | 475.88 | ... | ... |
| ... | ... | GJ 4329 | J23175+063 | M3.0 V | 23:17:34.73 | +06:23:24.5 | A+B | A | ... | ... | 123.06 | 302.15 | 197.08 ± 0.91 | 0.403 ± 0.016 |
| ... | ... | GJ 4319 | ... | M3.5 V | 23:16:08.65 | +06:44:32.2 | ... | B* | 1803.432 | 314.7 | 118.16 | 304.89 | 124.16 ± 0.78 | 0.336 ± 0.015 |
| ... | ... | Ross 244 | J23182+462 | M0.5 V | 23:18:18.44 | +46:17:23.6 | ... | ... | ... | ... | 40.31 | 355.18 | 660.25 ± 2.52 | 0.602 ± 0.027 |
| ... | ... | LP 12-69 | J23182+795 | M3.0 V | 23:18:19.85 | +79:34:45.8 | ... | ... | ... | ... | 39.31 | 511.17 | 226.14 ± 1.14 | 0.461 ± 0.018 |
| ... | ... | StKM 1-2115 | J23193+154 | M1.0 V | 23:19:21.03 | +15:24:13.5 | ... | ... | ... | ... | 45.12 | 149.64 | 638.12 ± 3.17 | 0.591 ± 0.028 |
| ... | ... | V368Cep | ... | G9 V | 23:19:27.78 | +79:00:13.8 | A+B+C | A | ... | ... | 74.22 | 215.92 | 3480.02 ± 11.81 | 0.900 ± 0.135 |
| 23194+7900 | LDS2035 | HD 220140B | J23194+790 | M3.5 V | 23:19:25.64 | +79:00:04.8 | ... | B | 10.905 | 214.1 | 47.72 | 218.08 | 305.75 ± 1.31 | 0.428 ± 0.031 |
| 23194+7900 | MKR 1 | LP 12-90 | J23228+787 | M5.0 V | 23:22:54.97 | +78:47:39.6 | ... | C | 962.738 | 141.1 | 45.66 | 217.44 | 30.51 ± 0.17 | 0.207 ± 0.012 |
| ... | ... | LSPM J2321+5651 | J23215+568 | M1.0 V | 23:21:32.23 | +56:51:19.7 | ... | ... | ... | ... | 53.66 | 190.84 | 597.83 ± 2.98 | 0.579 ± 0.028 |
| ... | ... | GJ 4333 | J23216+172 | M3.5 V | 23:21:36.85 | +17:17:03.3 | ... | ... | ... | ... | 48.61 | 1483.04 | 172.03 ± 0.95 | 0.375 ± 0.015 |
| ... | ... | G 217-6 | J23220+569 | M3.0 V | 23:22:01.52 | +56:59:21.4 | ... | ... | ... | ... | 109.85 | 398.39 | 237.37 ± 0.99 | 0.445 ± 0.018 |
| ... | ... | PM J23229+3717 | J23229+372 | M2.0 V | 23:22:58.46 | +37:17:13.3 | ... | ... | ... | ... | 32.79 | 89.99 | 244.44 ± 1.39 | 0.425 ± 0.017 |
| ... | ... | LP 522-65 | J23234+155 | M2.0 V | 23:23:24.72 | +15:34:09.2 | ... | ... | ... | ... | 32.34 | 467.44 | 358.36 ± 1.41 | 0.490 ± 0.018 |
| ... | ... | Ross 302 | J23245+578 | M1.5 V | 23:24:30.39 | +57:51:11.0 | ... | ... | ... | ... | 32.42 | 288.73 | 547.80 ± 3.12 | 0.559 ± 0.028 |

Table D.2: Complete sample with the description of multiple systems (continued).

| WDS id | WDS disc | Name | Karmn | Spectral type | $\alpha$ (2016.0) | $\delta$ (2016.0) | System | Component | $\rho$ [arcsec] | $\theta$ [deg] | $\varpi$ [mas] | $\mu_{\rm total}$ [mas a$^{-1}$] | $\mathcal{L}$ [$10^{-4}\ \mathcal{L}_\odot$] | $\mathcal{M}$ [$\mathcal{M}_\odot$] |
|---|---|---|---|---|---|---|---|---|---|---|---|---|---|---|
| … | … | PM J23549+5036 | J23249+506 | M3.0 V | 23:54:56.57 | +50:36:15.0 | … | … | … | … | 31.40 | 54.62 | 135.28 ± 0.69 | 0.351 ± 0.015 |
| … | … | Wolf 1038 | J23252+009 | M1.0 V | 23:25:16.81 | +00:57:44.1 | … | … | … | … | 61.79 | 657.67 | 556.23 ± 3.39 | 0.570 ± 0.028 |
| … | … | GJ 4334 | J23256+531 | M5.0 V | 23:25:42.10 | +53:08:11.1 | … | … | … | … | 56.76 | 1065.05 | 93.76 ± 0.53 | 0.330 ± 0.015 |
| … | … | GJ 2155 | J23262+088 | M0.0 V | 23:26:12.93 | +08:53:41.2 | … | … | … | … | 77.44 | 561.15 | 708.98 ± 3.60 | 0.624 ± 0.027 |
| 23262+1700 | JNN138 | PM J23261+1700 | J23261+170 | M4.0 V | 23:26:12.00 | +17:00:06.9 | (AB) | AB | 0.273 | 1.7 | 94.56 | 133.35 | … | 0.280 ± 0.035 |
| … | … | GJ 4336 | J23265+121 | M2.5 V | 23:26:33.26 | +12:09:37.9 | … | … | … | … | 48.64 | 744.79 | 234.80 ± 0.97 | 0.442 ± 0.017 |
| 23263+2752 | JNN139 | V595Peg | J23262+278 | M3.0 V | 23:26:17.02 | +27:52:02.8 | (AB) | AB | 0.109 | 328.7 | 37.24 | 60.29 | … | 0.484 ± 0.030 |
| 23294+4128 | GIC 193 | GJ 4337 | J23293+414N | M3.0 V | 23:29:24.06 | +41:28:06.0 | A+(BC) | A | 17.701 | 214.1 | 45.91 | 411.68 | 218.46 ± 0.86 | 0.453 ± 0.018 |
| 23294+4128 | BWL 59 | GJ 4338 | J23293+414S | M4.2 V | 23:29:23.18 | +41:27:51.4 | … | BC | 17.707 | 214.0 | 42.61 | 405.83 | … | 0.353 ± 0.033 |
| … | … | AFPsc | J23317-027 | M4.5 V | 23:31:45.03 | -02:44:40.7 | A+B | A | … | … | 42.60 | 118.45 | 253.93 ± 1.55 | 0.365 ± 0.033 |
| 23317-0245 | CAB24 | 2M23301129-0237227 | J23301-026 | M6.0 V | 23:30:11.41 | -02:37:23.9 | … | B | 1469.115 | 287.3 | 38.62 | 122.01 | 147.11 ± 1.53 | 0.420 ± 0.018 |
| 23309+1547 | LDS5096 | LP 462-51 | J23308+157 | M1.0 V | 23:30:53.52 | +15:47:38.8 | (AB) | AB | 2.000 | 320.0 | 51.07 | 198.04 | … | 0.500 ± 0.029 |
| … | … | EQPegA | J23318+199E | M3.5 Ve | 23:31:52.83 | +19:56:13.2 | Aab+Bab | Aab(2) | … | … | 26.84 | 581.09 | … | … |
| 23317+1956 | WIR 1 | EQPegB | J23318+199W | M4.0 Ve | 23:31:53.20 | +19:56:14.3 | … | Bab(1) | 5.376 | 77.6 | 60.92 | 552.72 | … | … |
| … | … | G 217-12 | J23323+540 | M2.0 V | 23:32:20.61 | +54:01:48.5 | … | … | … | … | 49.05 | 217.31 | 430.72 ± 2.03 | 0.513 ± 0.029 |
| … | … | HD 221503 | … | K6 V | 23:32:49.78 | -16:50:47.8 | A+Bab+CD | A | … | … | 56.59 | 405.45 | 1267.48 ± 6.60 | 0.690 ± 0.104 |
| 22577-2937 | SHY10 | GJ 1284 | J23302-203 | M2.0 Ve | 23:30:13.80 | -20:23:30.7 | … | Bab(2) | 12953.999 | 189.8 | 26.79 | 375.23 | … | 0.133 ± 0.004 |
| 23328-1651 | LDS 816 | GJ 897 | J23327-167 | M3.0 Ve | 23:32:46.98 | -16:45:12.3 | … | C | 337.883 | 353.2 | 26.75 | 681.85 | … | … |
| 23328-1645 | VOU 28 | G3- 2395220664463236992 | … | … | 23:32:46.98 | -16:45:11.7 | … | D | 0.724 | 356.4 | 25.07 | … | … | … |
| … | … | Wolf 1039 | J23340+001 | M2.5 V | 23:34:02.28 | +00:10:30.9 | … | … | … | … | 113.44 | 1365.84 | 262.17 ± 1.30 | 0.433 ± 0.012 |
| … | … | Ross 298 | J23350+252 | M3.0 V | 23:35:03.91 | +25:15:00.7 | … | … | … | … | 43.33 | 208.10 | 87.68 ± 0.34 | 0.261 ± 0.012 |
| … | … | GJ 1286 | J23351-023 | M5.0 Ve | 23:35:11.30 | -02:23:34.1 | … | … | … | … | 23.54 | 1148.37 | 14.75 ± 0.08 | 0.113 ± 0.009 |
| … | … | GJ 4342 | J23354+300 | M3.5 V | 23:35:24.05 | +30:03:40.6 | … | … | … | … | 31.00 | 328.99 | 244.14 ± 5.49 | 0.480 ± 0.020 |
| … | … | GJ 4346 | J23357+419 | M1.0 V | 23:35:45.48 | +41:58:06.3 | … | … | … | … | 31.30 | 718.17 | 742.92 ± 4.99 | 0.615 ± 0.027 |
| … | … | Ross 303 | J23364+554 | M1.5 V | 23:36:26.53 | +55:29:42.1 | … | … | … | … | 42.20 | 499.31 | 546.70 ± 2.27 | 0.558 ± 0.028 |
| … | … | LP 763-3 | J23376-128 | M5.5 Ve | 23:37:38.56 | -12:50:33.3 | … | … | … | … | 44.58 | 361.40 | 23.37 ± 0.53 | 0.193 ± 0.012 |
| … | … | GJ 4352 | J23381-162 | M2.0 V | 23:38:07.84 | -16:14:11.4 | … | … | … | … | 78.88 | 291.26 | 198.05 ± 1.11 | 0.364 ± 0.011 |
| … | … | GJ 4354 | J23386+391 | M3.3 V | 23:38:41.41 | +39:09:17.9 | … | … | … | … | 85.41 | 567.55 | 93.24 ± 0.43 | 0.288 ± 0.013 |
| 23350+0136 | MEL9 | GJ 900 | J23350+016 | M0.0 V | 23:35:00.64 | +01:36:19.9 | (AB) | AB | 1.294 | 357.0 | 24.15 | 341.22 | … | … |
| … | … | GJ 4356 | J23389+210 | M4.3 V | 23:38:56.00 | +21:01:24.5 | A+B | A | … | … | 24.17 | 328.86 | 217.21 ± 1.41 | 0.482 ± 0.019 |
| 23389+2101 | LDS5108 | GJ 4357 | … | DA | 23:38:56.63 | +21:01:21.1 | … | B | 9.495 | 110.7 | 62.88 | 325.25 | … | 0.500 ± 0.100 |
| … | … | GJ 4358 A | J23401+606 | M1.0 V | 23:40:07.91 | +60:41:14.4 | AB | A | … | … | 18.39 | 299.68 | … | 0.476 ± 0.030 |
| 23401+6041 | SKF 283 | GJ 4358 B | … | M3.0 V | 23:40:08.17 | +60:41:12.9 | … | B | 2.408 | 127.7 | 18.55 | 302.70 | … | 0.356 ± 0.033 |
| … | … | TYC 1727-1708-1 | J23414+200 | M0.5 V | 23:41:28.99 | +20:02:31.0 | A+B | A | … | … | 31.94 | 225.26 | … | 0.592 ± 0.028 |
| 23415+2002 | UC 5005 | LSPM J2341+2002S | … | M2.0 V | 23:41:28.97 | +20:02:27.0 | … | B | 4.024 | 184.1 | 48.27 | 222.83 | … | 0.430 ± 0.031 |
| … | … | HHAnd | J23419+441 | M5.0 V | 23:41:55.20 | +44:10:13.4 | … | … | … | … | 97.68 | 1595.62 | 22.62 ± 0.17 | 0.141 ± 0.009 |
| … | … | PM J23423+3458 | J23423+349 | M4.0 V | 23:42:22.21 | +34:58:25.5 | … | … | … | … | 38.61 | 150.76 | 204.09 ± 2.57 | 0.466 ± 0.019 |
| … | … | GJ 1288 | J23428+308 | M5.0 V | 23:42:52.33 | +30:49:16.9 | … | … | … | … | 55.83 | 450.80 | 31.32 ± 0.16 | 0.181 ± 0.011 |
| … | … | GJ 1289 | J23431+365 | M4.5 Ve | 23:43:07.56 | +36:32:10.7 | … | … | … | … | 24.25 | 953.91 | 55.99 ± 0.35 | 0.217 ± 0.010 |
| 23439+3232 | JOD 26 | GJ 905.2 A | J23438+325 | M1.5 V | 23:43:52.84 | +32:35:37.8 | (AB)+C | AB(2) | 0.109 | 308.0 | 39.28 | 223.77 | … | … |

Table D.2: Complete sample with the description of multiple systems (continued).

| WDS id | WDS disc | Name | Karmn | Spectral type | $\alpha$ (2016.0) | $\delta$ (2016.0) | System | Component | $\rho$ [arcsec] | $\theta$ [deg] | $\varpi$ [mas] | $\mu_{\rm total}$ [mas a$^{-1}$] | $\mathcal{L}$ [$10^{-4}\,\mathcal{L}_\odot$] | $\mathcal{M}$ [$\mathcal{M}_\odot$] |
|---|---|---|---|---|---|---|---|---|---|---|---|---|---|---|
| 23439+3232 | LDS1070 | GJ 905.2 B | … | DA3.8 | 23:43:50.45 | +32:32:45.8 | … | C | 174.652 | 190.0 | 45.59 | 224.05 | … | 0.500 ± 0.100 |
| … | … | G 217-18 | J23438+610 | M3.0 V | 23:43:51.95 | +61:02:07.5 | … | … | … | … | 45.63 | 787.75 | 79.75 ± 0.31 | 0.264 ± 0.013 |
| … | … | GJ 1290 | J23443+216 | M3.4 V | 23:44:21.41 | +21:36:06.4 | … | … | … | … | 43.79 | 462.48 | 133.34 ± 0.63 | 0.349 ± 0.015 |
| … | … | Ross 676 | J23439+647 | M0.3 V | 23:44:00.86 | +64:44:30.4 | Aab | Aab(2) | … | … | 24.96 | 553.71 | … | … |
| … | … | GJ 4360 A | J23455-161 | M5.0 V | 23:45:30.82 | -16:10:28.8 | AB | A | … | … | 90.49 | 370.59 | … | … |
| 23455-1610 | MTG5 | GJ 4360 B | … | … | 23:45:30.81 | -16:10:29.1 | … | B | 0.323 | 194.5 | 89.73 | … | … | … |
| … | … | PM J23462+2826 | J23462+284 | M3.5 V | 23:46:14.17 | +28:26:05.0 | … | … | … | … | 18.32 | 103.73 | 88.61 ± 0.35 | 0.280 ± 0.013 |
| … | … | Ross 249 | J23480+490 | M3.0 V | 23:48:04.16 | +49:00:58.4 | … | … | … | … | 55.39 | 630.08 | 135.21 ± 0.71 | 0.291 ± 0.013 |
| … | … | [R78b] 377 | J23489+098 | M1.0 V | 23:48:58.98 | +09:51:53.3 | ABC | A | … | … | 25.68 | 156.75 | … | 0.636 ± 0.027 |
| 23490+0952 | KPP4328 | [R78b] 377B | … | M3.5 V | 23:48:59.02 | +09:51:55.2 | … | B | 1.949 | 19.7 | 25.65 | 148.46 | … | 0.317 ± 0.034 |
| … | … | BRPsc | J23492+024 | dM1 | 23:49:13.58 | +02:23:48.9 | … | … | … | … | 138.23 | 1387.11 | 266.00 ± 1.61 | 0.425 ± 0.012 |
| … | … | GJ 4363 | J23492+100 | M3.0 V | 23:49:15.05 | +10:05:32.7 | … | … | … | … | 65.65 | 381.33 | 234.93 ± 1.69 | 0.471 ± 0.019 |
| … | … | GJ 4364 | J23496+083 | M1.0 V | 23:49:37.78 | +08:21:29.1 | … | … | … | … | 65.38 | 144.91 | 518.39 ± 3.41 | 0.559 ± 0.028 |
| … | … | GJ 4367 | J23505-095 | dM4.0 | 23:50:32.33 | -09:33:39.5 | … | … | … | … | 63.77 | 764.31 | 107.06 ± 0.95 | 0.295 ± 0.011 |
| … | … | GJ 4368 | J23506+099 | M3.0 V | 23:50:36.86 | +09:56:57.5 | AB | A | … | … | 52.05 | 700.59 | … | 0.482 ± 0.030 |
| 23506+0957 | BEU 24 | GJ 4368 B | … | M4.0 V | 23:50:36.94 | +09:56:57.3 | … | B | 1.170 | 101.0 | 46.94 | 649.02 | … | 0.282 ± 0.035 |
| … | … | GJ 4369 | J23509+384 | M3.8 V | 23:50:53.91 | +38:29:30.1 | … | … | … | … | 42.64 | 219.87 | 88.24 ± 0.52 | 0.299 ± 0.014 |
| … | … | GJ 4370 | J23517+069 | M3.0 V | 23:51:44.72 | +06:58:11.8 | AB | A | … | … | 42.61 | 292.30 | … | 0.358 ± 0.033 |
| 23517+0658 | CRC 77 | G 30-26B | … | M3.0 V | 23:51:44.86 | +06:58:11.3 | … | B | 2.181 | 103.0 | 38.72 | 299.55 | … | 0.355 ± 0.033 |
| … | … | GJ 4371 | J23523-146 | M4.5 V | 23:52:23.92 | -14:41:28.2 | … | … | … | … | 72.68 | 476.88 | 54.20 ± 1.55 | 0.245 ± 0.013 |
| … | … | StKM 2-1787 | … | K4 V | 23:53:35.58 | +12:06:20.4 | A+B | A | … | … | 249.39 | 120.69 | 1065.90 ± 4.89 | 0.730 ± 0.110 |
| 23536+1207 | VYS 11 | PM J23535+1206S | J23535+121 | M2.5 Ve | 23:53:35.69 | +12:06:14.8 | … | B | 5.785 | 165.1 | 249.97 | 119.44 | … | 0.573 ± 0.028 |
| … | … | GJ 4373 | J23541+516 | M3.5 V | 23:54:10.95 | +51:41:11.0 | … | … | … | … | 44.91 | 270.79 | 138.78 ± 0.60 | 0.356 ± 0.015 |
| … | … | GJ 4374 | J23544+081 | M3.0 V | 23:54:26.50 | +08:09:42.6 | … | … | … | … | 91.84 | 291.13 | 461.83 ± 23.78 | 0.504 ± 0.030 |
| … | … | RX J2354.8+3831 | J23548+385 | M4.0 V | 23:54:51.28 | +38:31:34.8 | … | … | … | … | 45.03 | 157.68 | 103.47 ± 0.55 | 0.312 ± 0.010 |
| … | … | GJ 4376 | J23554-039 | M3.5 V | 23:55:26.51 | -03:58:59.8 | … | … | … | … | 71.11 | 522.42 | … | 0.410 ± 0.031 |
| … | … | LP 523-78 | J23560+150 | M2.5 V | 23:56:00.14 | +15:01:37.9 | … | … | … | … | 29.85 | 206.28 | 234.43 ± 1.10 | 0.442 ± 0.017 |
| … | … | GJ 912 | J23556-061 | M2.5Vk | 23:55:39.27 | -06:08:39.4 | Aab | Aab(1) | … | … | 60.89 | 611.70 | … | 0.580 ± 0.020 |
| … | … | G 129-45 | J23569+230 | M1.5 V | 23:56:54.75 | +23:05:04.2 | A+B | A | … | … | 29.87 | 273.13 | 901.25 ± 4.07 | 0.679 ± 0.026 |
| 23570+2305 | GIC 196 | G 129-46 | … | M1.73 V | 23:56:55.00 | +23:04:58.7 | … | B | 6.488 | 148.8 | 77.03 | 271.30 | 451.20 ± 2.52 | 0.531 ± 0.029 |
| 23573-1259 | BWL 67 | GJ 4378 | J23573-129E | M4.0 V | 23:57:20.81 | -12:58:48.6 | (AB)+Cab | AB | 0.755 | 342.6 | 38.34 | 211.92 | … | 0.352 ± 0.033 |
| 23573-1259 | LDS 830 | GJ 4379 | J23573-129W | M3.0 V | 23:57:19.59 | -12:58:40.3 | … | Cab(2) | 19.719 | 294.8 | 59.39 | 209.62 | … | … |
| … | … | GJ 4380 | J23577+197 | M3.7 V | 23:57:45.32 | +19:46:03.4 | … | … | … | … | 38.26 | 503.23 | 147.64 ± 0.78 | 0.368 ± 0.016 |
| … | … | GJ 1292 | J23577+233 | M3.5 V | 23:57:45.32 | +23:18:00.1 | … | … | … | … | 59.50 | 1444.53 | 286.19 ± 10.55 | 0.490 ± 0.021 |
| 23578+3838 | MCT 14 | GJ 4381 | J23578+386 | M3.0 V | 23:57:49.68 | +38:37:44.4 | (AB) | AB | 0.476 | 210.3 | 45.09 | 218.86 | … | 0.419 ± 0.031 |
| … | … | LP 764-40 A | J23582-174 | M2.0 Ve | 23:58:13.95 | -17:24:34.6 | AB | A | … | … | 25.46 | 220.68 | … | 0.482 ± 0.030 |
| 23582-1725 | DAE 8 | LP 764-40 B | … | M2.0 Ve | 23:58:13.94 | -17:24:32.5 | … | B | 2.080 | 357.0 | 28.11 | 226.62 | … | 0.483 ± 0.030 |
| … | … | Wolf 1051 | J23585+076 | M3.0 V | 23:58:32.74 | +07:39:25.0 | Aabc | Aabc(3) | … | … | 19.28 | 325.95 | … | 0.789 ± 0.018 |
| 23587+4644 | BAG 34 | GJ 913 | J23587+467 | M0.0 V | 23:58:44.50 | +46:43:44.9 | Aab | AB(2?) | 0.068 | 82.6 | 27.88 | 652.95 | … | … |
| … | … | G 129-51 | J23590+208 | M2.5 V | 23:59:00.74 | +20:51:37.2 | AB | A | … | … | 60.38 | 300.02 | … | 0.537 ± 0.029 |

Table D.2: Complete sample with the description of multiple systems (continued).

| WDS id | WDS disc | Name | Karmn | Spectral type | $\alpha$ (2016.0) | $\delta$ (2016.0) | System | Component | $\rho$ [arcsec] | $\theta$ [deg] | $\varpi$ [mas] | $\mu_{\rm total}$ [mas a$^{-1}$] | $\mathcal{L}$ [$10^{-4}$ $\mathcal{L}_{\odot}$] | $\mathcal{M}$ [$\mathcal{M}_{\odot}$] |
|---|---|---|---|---|---|---|---|---|---|---|---|---|---|---|
| … | … | G 129-51B | … | M2.0 V | 23:59:00.75 | +20:51:36.7 | … | B* | 0.563 | 165.6 | 46.89 | 301.61 | … | 0.455 ± 0.030 |
| … | … | GJ 4385 | J23598+477 | M5.0 V | 23:59:50.89 | +47:45:41.3 | … | … | … | … | 47.64 | 894.90 | 21.06 ± 0.09 | 0.155 ± 0.010 |

Table D.3: Components of multiple system that do not comply with one or more of the criteria for physical parity.

| Name[a] | Component[b] | $\alpha$ (J2016.0) | $\delta$ (J2016.0) | $\mu$ratio | $\Delta PA$ | $\Delta d$ | $\rho$[c] [arcsec] | Cause[d] |
|---|---|---|---|---|---|---|---|---|
| PM J00026+3821 B | B* | 00:02:40.05 | +38:21:45.3 | 0.214 | 12.254 | 0.031 | 1.415 | 1 |
| G 217-32 B | B | 00:07:43.40 | +60:22:53.8 | 0.196 | 9.161 | 0.0032 | 0.848 | 1 |
| LP 296-56 | B | 01:56:41.74 | +30:28:34.6 | 0.012 | 0.027 | 0.163 | 299.1 | 3 |
| PM J02024+1034 A | B | 02:02:28.18 | +10:34:52.7 | 0.545 | 32.513 | 0.012 | 0.911 | 1 |
| LP 198-637 B | B | 03:20:45.35 | +39:42:59.6 | 0.152 | 8.738 | 0.0051 | 0.795 | 1 |
| PM J03247+4447 B | B | 03:24:42.23 | +44:47:39.7 | 0.191 | 4.590 | 0.0048 | 1.912 | 1 |
| Gaia EDR3 3296932486866670720 | B | 04:05:38.89 | +05:44:40.1 | 0.331 | 1.614 | … | 0.817 | 1, 2 |
| GJ 2033 B | B | 04:16:41.80 | -12:33:19.8 | 0.334 | 11.137 | 0.0069 | 2.990 | 1 |
| HG 7-232 B | B | 04:28:29.01 | +17:41:45.3 | 0.227 | 11.403 | 0.047 | 1.663 | 1 |
| V697 Tau B | B | 04:33:23.87 | +23:59:26.5 | 0.200 | 11.095 | 0.150 | 0.767 | 1 |
| GJ 3305 (J04376-024) | BC | 04:37:37.51 | -02:29:29.7 | 0.274 | 14.659 | 0.077 | 0.098 | 1 |
| PM J04393+3331 (J04393+335) | BC | 04:39:23.22 | +33:31:48.7 | 0.147 | 7.913 | 0.333 | 0.126 | 1 |
| GJ 9163 B | B | 04:40:29.15 | -09:11:46.8 | 0.441 | 1.954 | 0.0064 | 1.715 | 1 |
| GJ 3322 B | B | 05:01:58.89 | +09:58:55.9 | 0.152 | 8.744 | 0.0015 | 1.398 | 1 |
| HD 32450 B | B | 05:02:28.26 | -21:15:27.5 | 0.247 | 6.461 | 0.0063 | 0.888 | 1 |
| LP 359-186 | BC | 05:03:05.77 | +21:22:33.8 | 0.046 | 0.799 | 0.104 | 0.302 | 1, 2 |
| GJ 3332 (J05068-215W) | B | 05:06:49.48 | -21:35:04.3 | 0.881 | 58.265 | 0.0018 | 8.489 | 1 |
| BD-21 1074 C | C | 05:06:49.55 | -21:35:04.7 | 0.309 | 8.384 | 0.0029 | 1.1 | 1 |
| GJ 3343 (J05206+587S) | B | 05:20:41.05 | +58:47:12.0 | 0.021 | 1.037 | 0.104 | 14 | 3 |
| PM J05243-1601 B | B | 05:24:19.17 | -16:01:15.6 | 0.964 | 52.141 | 0.0066 | 0.4 | 1 |
| GJ 3348A (J05289+125) | C | 05:28:56.61 | +12:31:50.4 | 0.023 | 1.307 | 0.169 | 99.39 | 2 |
| PM J05319-0303W | C | 05:31:57.88 | -03:03:37.6 | 0.254 | 4.659 | 0.034 | 150.0 | 5 |
| 2MASS J05315816-0303397 | D | 05:31:58.17 | -03:03:40.7 | 0.257 | 5.562 | 0.048 | 144.8 | 5 |
| ESO-HA 737 | E | 05:32:05.97 | -03:01:16.8 | 0.209 | 5.090 | 0.046 | 254.2 | 5 |
| PM J06066+4633 B | B | 06:06:37.77 | +46:33:45.2 | 0.198 | 2.085 | 0.00038 | 1.790 | 1 |

Table D.3: Components of multiple system that do not comply with one or more of the criteria for physical parity (continued).

| Name[a] | Component[b] | $\alpha$ (J2016.0) | $\delta$ (J2016.0) | $\mu$ratio | $\Delta PA$ | $\Delta d$ | $\rho$[c] [arcsec] | Cause[d] |
|---|---|---|---|---|---|---|---|---|
| LP 362-121 (J06103+225) | B | 06:10:22.52 | +22:34:18.1 | 0.057 | 1.898 | 0.258 | 65.16 | 2 |
| Ross 603 B | B | 06:27:43.74 | +09:23:50.5 | 0.167 | 8.901 | 0.0052 | 1.170 | 1 |
| Gaia EDR3 2926756741750933120 | B* | 06:39:37.22 | -21:01:31.9 | 0.590 | 11.207 | 0.025 | 0.556 | 1 |
| 2MASS J07293670+3554531 | C | 07:29:36.67 | +35:54:51.3 | 0.171 | 6.262 | 0.0063 | 95.88 | 3 |
| Castor C (J07346+318) | Cab | 07:34:37.19 | +31:52:08.6 | 0.220 | 11.459 | 0.034 | TBD | 1 |
| FP Cnc B (J08089+328) | Cab | 08:08:55.38 | +32:49:01.4 | 0.189 | 10.740 | 0.022 | < 0.92 au | 1 |
| GJ 1116 B | B | 08:58:14.21 | +19:45:46.7 | 0.236 | 5.329 | 0.011 | 2.175 | 1 |
| LP 489-1 | B* | 09:56:45.21 | +11:34:23.6 | 0.326 | 18.840 | 0.032 | 1356.0 | TBD |
| GJ 397.1 B (J10315+570) | BC | 10:31:30.64 | +57:05:20.4 | 0.166 | 3.931 | … | 0.284 | 1, 2 |
| PM J10367+1521 B | BC | 10:36:44.92 | +15:21:37.9 | 0.174 | 5.147 | … | 0.132 | 1, 2 |
| GJ 3616 B | B | 10:44:52.43 | +32:24:40.1 | 0.067 | 2.466 | 0.129 | 1.296 | 1 |
| GJ 3617 | C | 10:44:54.79 | +32:24:23.4 | 0.079 | 4.417 | 0.125 | 35.23 | TBD |
| GJ 3628 (J10506+517) | B | 10:50:37.91 | +51:45:01.6 | 0.219 | 12.290 | 0.013 | 178.3 | 3 |
| HD 98712 B (J11214-204) | B | 11:21:26.84 | -20:27:11.5 | 0.266 | 5.851 | 0.046 | 3.811 | 1 |
| GJ 426.1 B | B | 11:23:55.75 | +10:31:44.7 | 0.213 | 11.741 | 0.020 | 2.050 | 1 |
| Gaia EDR3 3699797155055781760 | B | 12:20:25.60 | +00:35:01.3 | 0.641 | 30.528 | 0.0055 | 0.882 | 1 |
| GJ 490 B (J12576+352W) | CD | 12:57:38.94 | +35:13:16.9 | 0.106 | 5.090 | 0.117 | 0.171 | 1 |
| PM J13260+2735 B | C | 13:26:02.63 | +27:35:02.5 | 0.201 | 8.729 | 0.0015 | 1.485 | 1 |
| GJ 520 B | B | 13:37:50.77 | +48:08:16.3 | 0.166 | 5.256 | 0.000046 | 1.621 | 1 |
| Gaia EDR3 1658968054798747008 | B | 13:41:46.39 | +58:15:18.6 | 0.360 | 13.784 | 0.0078 | 0.762 | 1 |
| GJ 536.1 B | B | 14:01:58.89 | +15:29:39.1 | 0.167 | 7.680 | 0.0020 | 1.621 | 1 |
| PM J14019+4316 B | B | 14:01:58.69 | +43:16:43.1 | 0.151 | 8.700 | 0.00081 | 1.960 | 1 |
| HD 124498 B | B | 14:14:21.34 | -15:21:24.4 | 0.175 | 9.130 | 0.077 | 1.553 | 1 |
| G 224-13 B | B | 14:33:06.80 | +61:00:43.6 | 0.190 | 10.854 | 0.0018 | 0.971 | 1 |
| GJ 9492 B | B | 14:42:21.17 | +66:03:20.1 | 0.221 | 11.795 | 0.0013 | 2.290 | 1 |

Table D.3: Components of multiple system that do not comply with one or more of the criteria for physical parity (continued).

| Name[a] | Component[b] | $\alpha$ (J2016.0) | $\delta$ (J2016.0) | $\mu$ratio | $\Delta PA$ | $\Delta d$ | $\rho$[c] [arcsec] | Cause[d] |
|---|---|---|---|---|---|---|---|---|
| GJ 569 B | Bab | 14:54:29.77 | +16:06:05.6 | 0.215 | 7.832 | 0.067 | 0.097 | 1 |
| Gaia EDR3 1275127175448008448 | B* | 15:18:49.78 | +29:15:06.7 | 0.302 | 8.736 | 0.037 | 0.588 | 1 |
| Gaia EDR3 1375767330164975616 | B | 15:33:54.89 | +37:54:48.3 | 0.234 | 13.506 | 0.0022 | 0.834 | 1 |
| LP 331-57 B | B | 17:03:53.14 | +32:11:46.4 | 0.175 | 3.536 | 0.0055 | 1.393 | 1 |
| BD+19 3268 B | B | 17:15:49.84 | +19:00:00.3 | 0.699 | 3.435 | 0.0044 | 1.471 | 1 |
| RX J1734.0+4447 B (J17340+446) | B | 17:34:05.47 | +44:47:08.4 | 0.403 | 17.500 | 0.058 | 0.633 | 1 |
| GJ 4021 | C | 17:38:40.87 | +61:14:00.0 | 0.232 | 11.069 | 0.016 | 19.05 | 1 |
| GJ 695 C (J17464+277) | E | 17:46:24.65 | +27:42:49.8 | 0.160 | 0.847 | 0.0039 | 0.575 | 1 |
| HD 230017 B (J18548+109) | B | 18:54:53.86 | +10:58:45.1 | 0.512 | 6.480 | 0.0041 | 3.792 | 1 |
| PM J18542+1058 | C* | 18:54:17.14 | +10:58:11.0 | 0.170 | 2.601 | 0.0078 | 538.9 | 1 |
| StKM 1-1676 B | C | 19:03:17.96 | +63:59:37.4 | 0.824 | 54.189 | … | 3.578 | 1 |
| GJ 4115 (J19351+084N) | C | 19:35:06.33 | +08:27:43.4 | 0.462 | 17.584 | 0.062 | 5.635 | 1 |
| GJ 1245 B (J19539+444E) | C | 19:53:55.66 | +44:24:46.5 | 0.174 | 9.449 | 0.0068 | 5.945 | 1 |
| LP 395-8 B | B | 20:19:49.35 | +22:56:40.0 | 0.184 | 10.601 | 0.000044 | 1.918 | 1 |
| GJ 4153 B | B | 20:37:23.97 | +21:56:21.4 | 0.072 | 0.221 | 0.217 | 51.77 | 2 |
| AT Mic B | B | 20:41:51.54 | -32:26:15.1 | 0.298 | 13.947 | 0.012 | 2.102 | 1 |
| PM J21055+0609S | B* | 21:05:32.17 | +06:09:11.2 | 0.175 | 5.680 | 0.000069 | 5.094 | 1 |
| GJ 9721 B (J21087-044N) | C | 21:08:44.75 | -04:25:18.3 | 0.652 | 12.094 | 0.012 | 20.89 | 1 |
| GJ 4188 (J21176-089S) | C | 21:17:39.55 | -08:54:49.6 | 0.166 | 7.927 | 0.098 | 63.99 | 1 |
| GJ 4201 B | B | 21:32:22.23 | +24:33:41.1 | 0.181 | 9.559 | 0.0029 | 1.548 | 1 |
| GJ 834 B | B | 21:36:38.20 | +39:27:17.8 | 0.181 | 9.172 | 0.0041 | 1.003 | 1 |
| UCAC4 747-070768 | B | 21:51:39.93 | +59:17:34.5 | 0.384 | 20.069 | 0.041 | 14.64 | 1 |
| Ross 268 | C* | 22:06:23.02 | +17:22:22.1 | 0.210 | 8.636 | 0.081 | 5100.6 | 1 |
| LP 759-25 | B | 22:05:35.44 | -11:04:31.8 | 0.053 | 3.002 | 0.323 | 3030.6 | 3 |
| Wolf 1225 B | B | 22:23:29.34 | +32:27:29.5 | 0.271 | 12.352 | 0.0040 | 1.252 | 1 |

Table D.3: Components of multiple system that do not comply with one or more of the criteria for physical parity (continued).

| Name[a] | Component[b] | $\alpha$ (J2016.0) | $\delta$ (J2016.0) | $\mu$ratio | $\Delta PA$ | $\Delta d$ | $\rho$[c] [arcsec] | Cause[d] |
|---|---|---|---|---|---|---|---|---|
| DO Cep | B | 22:27:57.93 | +57:41:38.8 | 0.438 | 19.178 | 0.0023 | 1.420 | 1 |
| GJ 4282 B | B | 22:33:22.87 | -09:36:53.7 | 0.172 | 5.466 | 0.0022 | 1.399 | 1 |
| 2M J23301129-0237227 (J23301-026) | B | 23:30:11.41 | -02:37:23.9 | 0.069 | 3.658 | 0.232 | 1469.1 | 4 |
| EQ Peg B (J23318+199W) | Bab | 23:31:53.20 | +19:56:14.3 | 0.152 | 8.006 | 0.0015 | 5.376 | 1 |
| GJ 897 (J23327-167) | C | 23:32:46.98 | -16:45:12.3 | 0.938 | 66.572 | ...[e] | 337.9 | 2 |
| LSPM J2350+1013 | C* | 23:50:24.02 | +10:13:47.9 | 0.152 | 7.831 | 0.051 | 1818.3 | TBD |

[a] Karmn identification in parentheses, when available.
[b] Component within the system it belongs to. An asterisk (*) denotes new components found.
[c] Projected separation of the closest component.
[d] Causes: 1: Presence of a very close companion ($\lesssim$ 220 au), also applicable to the primary; 2: Parallactic distances and proper motions from a source different than *Gaia* DR3, with larger uncertainties, or missing values; 3: Component A is candidate to binary; 4: Candidate to binary system; 5: Young stellar object candidate. These systems undergo a process of dynamical stabilisation; TBD: To be determined.
[e] Parallax from Hipparcos (van Leeuwen, 2007) removed because of very large relative error.

Table D.4: Star candidates belonging to multiple systems not tabulated by WDS.

| Name | Karmn | Spectral type | Component[a] | Class[b] | $\alpha$ (J2016.0) | $\delta$ (J2016.0) | $\pi$ [mas] | $\mu_{\rm total}$ [mas a$^{-1}$] | $G$ [mag] | $\theta$ [deg] | $\rho$ [arcsec] | Notes[c] |
|---|---|---|---|---|---|---|---|---|---|---|---|---|
| PM J00026+3821A | J00026+383 | M4.0 V | A | Binary* | 00:02:40.00 | +38:21:44.1 | 24.622 ± 0.210 | 74.86 ± 0.21 | 13.193 | | | |
| PM J00026+3821B | | … | B* | | 00:02:40.05 | +38:21:45.3 | 25.392 ± 0.368 | 71.14 ± 0.46 | 13.388 | 28.1 | 1.415 | • |
| GJ 3022 | J00169+200 | M3.5 V | A | Triple* | 00:16:57.03 | +20:03:55.7 | 29.310 ± 0.097 | 239.54 ± 0.21 | 13.183 | | | |
| G 131-47B | | M3.5 V | B | | 00:16:57.10 | +20:03:55.4 | 29.139 ± 0.071 | 233.36 ± 0.32 | 13.261 | 106.5 | 1.077 | |
| LP 404-54 | | M5.0 V | C* | | 00:15:13.81 | +19:47:40.5 | 28.873 ± 0.034 | 233.80 ± 0.05 | 14.874 | 236.2 | 1752.0 | |
| V493 And A | J00341+253 | M0.0 V | A | Triple* | 00:34:08.48 | +25:23:48.5 | 20.101 ± 0.039 | 127.91 ± 0.05 | 11.226 | | | • |
| V493 And B | | K7 V | B | | 00:34:08.59 | +25:23:48.2 | 19.720 ± 0.037 | 129.46 ± 0.06 | 11.276 | 102.3 | 1.536 | |
| UCAC4 578-001365 | | M4.0 V | C* | | 00:34:20.04 | +25:28:12.9 | 19.672 ± 0.234 | 127.04 ± 0.35 | 15.058 | 30.6 | 307.2 | |
| Wolf 58 | J01133+589 | M1.5 Ve | A | Binary* | 01:13:20.12 | +58:55:20.3 | 36.025 ± 0.017 | 210.93 ± 0.02 | 10.894 | | | • |
| Gaia EDR3 414108140954108672 | | M5.0 V | B* | | 01:13:19.97 | +58:55:18.6 | 35.974 ± 0.088 | 200.51 ± 0.11 | 14.516 | 213.9 | 2.082 | |
| GJ 3131 | J02033-212 | M2.5 V | Aab | Triple* | 02:03:20.52 | -21:13:50.2 | 46.672 ± 0.041 | 471.89 ± 0.05 | 10.169 | | | |
| GJ 3131 B | | … | B* | | 02:03:20.25 | -21:13:51.5 | 46.747 ± 0.307 | 450.04 ± 0.47 | 18.661 | 251.9 | 4.008 | |
| Ross 369A | J03145+594 | M2.5 V | A | Binary* | 03:14:33.17 | +59:26:13.8 | 36.823 ± 0.126 | 250.95 ± 0.18 | 11.571 | | | |
| Ross 369B | | M3.0 V | B* | | 03:14:33.10 | +59:26:13.2 | 36.496 ± 0.167 | 237.34 ± 0.24 | 12.236 | 221.0 | 0.775 | |
| 1RXS J033021.4+344044 | J03303+346 | M4.0 V | A | Triple* | 03:30:23.37 | +34:40:31.7 | … | 56.66 ± 2.14 | 13.371 | | | • |
| Gaia EDR3 221088045766546816 | | M1.5 V | B* | | 03:30:16.90 | +34:39:50.2 | 11.375 ± 0.057 | 53.59 ± 0.08 | 14.389 | 242.5 | 89.96 | |
| Gaia EDR3 221088050062935680 | | M3.0 V | C* | | 03:30:16.80 | +34:39:49.2 | 11.725 ± 0.234 | 52.34 ± 0.36 | 14.788 | 242.3 | 91.48 | |
| GJ 3256 | J03544-091 | M1.0 V | A | Binary* | 03:54:25.52 | -09:09:29.2 | 47.406 ± 0.023 | 146.38 ± 0.03 | 10.540 | | | |
| GJ 3256 B | | M3.0 V | B* | | 03:54:25.61 | -09:09:32.0 | 47.545 ± 0.035 | 138.31 ± 0.04 | 11.889 | 153.6 | 3.182 | • |
| Ross 25 | J04011+513 | M4.0 V | A | Binary* | 04:01:08.18 | +51:23:06.4 | 39.816 ± 0.021 | 883.46 ± 0.03 | 12.435 | | | |
| LSPM J0401+5131 | | DC8 | B* | | 04:01:02.14 | +51:31:17.2 | 39.836 ± 0.077 | 883.24 ± 0.12 | 17.113 | 353.4 | 494.0 | |
| GJ 3261 | J04056+057 | M4.0 Ve | A | Triple* | 04:05:38.94 | +05:44:40.4 | 16.094 ± 0.103 | 47.98 ± 0.15 | 12.028 | | | |
| Gaia EDR3 3296932486866670720 | | … | B | | 04:05:38.89 | +05:44:40.1 | … | … | 12.814 | 251.9 | 0.817 | |
| ATO J061.4727+05.5235 | | M3.0 V | C* | | 04:05:53.46 | +05:31:24.6 | 15.876 ± 0.026 | 49.85 ± 0.03 | 14.012 | 164.8 | 824.8 | |
| LP 414-117 | J04123+162 | M4.0 V | Aab | Triple* | 04:12:21.90 | +16:15:02.9 | 28.240 ± 0.091 | 156.84 ± 0.15 | 12.669 | | | • |
| LSPM J0409+1622 | | M5.5 V | B* | | 04:09:57.30 | +16:22:41.3 | 28.769 ± 0.051 | 161.15 ± 0.08 | 15.525 | 282.5 | 2131.6 | |
| TYC 78-257-1 | | … | A | Binary* | 04:21:04.26 | +03:16:07.9 | 26.974 ± 0.022 | 139.56 ± 0.03 | 9.154 | | | |
| RX J0422.4+0337 | J04224+036 | M3.5 V | B* | | 04:22:25.19 | +03:37:08.5 | 27.846 ± 0.026 | 143.01 ± 0.04 | 12.774 | 43.9 | 1748.6 | |
| LP 415-345 | J04425+204 | M3.0 V | Aab | Quadruple* | 04:42:30.40 | +20:27:10.8 | 20.476 ± 0.022 | 97.45 ± 0.03 | 12.146 | | | • |

Table D.4: Star candidates belonging to multiple systems not tabulated by WDS (continued).

| Name | Karmn | Spectral type | Component[a] | Class[b] | $\alpha$ (J2016.0) | $\delta$ (J2016.0) | $\pi$ [mas] | $\mu_{total}$ [mas a$^{-1}$] | $G$ [mag] | $\theta$ [deg] | $\rho$ [arcsec] | Notes[c] |
|---|---|---|---|---|---|---|---|---|---|---|---|---|
| LP 415-3051 | | M3.0 V | B | | 04:42:58.58 | +20:36:16.8 | 19.426 ± 0.020 | 91.26 ± 0.03 | 13.767 | 35.9 | 674.3 | |
| Gaia DR2 3411054848866601472 | | M6.0 V | C* | | 04:43:55.36 | +20:08:40.5 | 19.484 ± 0.118 | 93.82 ± 0.18 | 17.282 | 132.8 | 1631.4 | |
| PM J04429+0935 | J04429+095 | M1.0 V | A | Binary* | 04:42:55.14 | +09:35:53.7 | 28.741 ± 0.024 | 82.04 ± 0.03 | 11.142 | | | |
| Gaia EDR3 3293060625388613248 | | M6.5 V | B | | 04:42:53.91 | +09:35:50.3 | 28.684 ± 0.086 | 77.51 ± 0.12 | 16.753 | 259.1 | 18.39 | |
| LP 416-43 | J04480+170 | M0.5 V | Aab | Triple* | 04:48:00.98 | +17:03:21.1 | 20.164 ± 0.088 | 82.10 ± 0.14 | 10.526 | | | |
| UCAC4 536-010184 | | M4.5 V | B* | | 04:49:36.67 | +17:01:58.6 | 19.838 ± 0.027 | 87.10 ± 0.04 | 14.761 | 93.4 | 1374.8 | |
| RX J0507.2+3731A | J05072+375 | M5.0 V | A | Binary* | 05:07:14.33 | +37:30:42.1 | 43.800 ± 4.000 | 102.11 ± 4.90 | 13.950 | | | • |
| RX J0507.2+3731B | | M5.0 V | B* | | 05:07:14.37 | +37:30:42.1 | 44.022 ± 0.461 | 96.09 ± 0.82 | 14.300 | 93.2 | 0.476 | |
| PM J05334+4809 | | M0.0 V | A | Quadruple* | 05:33:28.97 | +48:09:26.2 | 30.270 ± 0.018 | 66.93 ± 0.02 | 10.788 | | | |
| PM J05341+4732A | J05341+475 | M2.5 V | B | | 05:34:10.56 | +47:32:02.8 | 30.050 ± 0.027 | 69.01 ± 0.03 | 11.635 | 169.4 | 2282.1 | |
| PM J05341+4732B | | M3.0 V | C | | 05:34:10.62 | +47:32:05.2 | 30.039 ± 0.023 | 61.85 ± 0.03 | 12.671 | 169.3 | 2280.0 | |
| UPM J0533+4809 | | M3.5 V | D | | 05:33:16.22 | +48:09:23.3 | 30.229 ± 0.235 | 65.01 ± 0.31 | 13.236 | 268.7 | 127.6 | |
| G 106-7 | J05530+047 | M1.5 V | A | Binary* | 05:53:04.75 | +04:43:02.6 | 24.715 ± 0.056 | 391.42 ± 0.10 | 11.325 | | | |
| G 106-7B | | M5.5 V | B* | | 05:53:04.65 | +04:43:02.8 | 24.818 ± 0.208 | 387.78 ± 0.40 | 16.120 | 278.2 | 1.520 | • |
| PM J05558+4036 | J05558+406 | M1.0 V | A | Binary* | 05:55:48.31 | +40:36:48.0 | 27.777 ± 0.043 | 122.73 ± 0.05 | 11.423 | | | |
| Gaia EDR3 3458612029599345664 | | M3.0 V | B* | | 05:55:48.41 | +40:36:47.7 | 27.966 ± 0.110 | 138.62 ± 0.12 | 12.814 | 103.0 | 1.200 | |
| LP 780-32 | J06396-210 | dM4.0 V | A | Binary* | 06:39:37.20 | -21:01:32.4 | 63.738 ± 0.886 | 225.54 ± 0.94 | 12.267 | | | • |
| Gaia EDR3 2926756741750933120 | | M4.0 V | B* | | 06:39:37.22 | -21:01:31.9 | 65.308 ± 0.325 | 146.66 ± 0.34 | 12.022 | 35.1 | 0.556 | |
| LP 780-23 | J06401-164 | M2.5 V | A | Binary* | 06:40:08.72 | -16:27:21.5 | ... | 319.03 ± 11.31 | 12.315 | | | • |
| LP 780-23 B | | ... | B* | | 06:40:08.72 | -16:27:21.7 | ... | ... | 12.438 | 187.9 | 0.199 | |
| 1RXS J073138.4+455718 | | M3.0 V | Aab | Quadruple* | 07:31:38.47 | +45:57:15.8 | 17.880 ± 0.416 | 93.78 ± 0.48 | 12.766 | | | |
| 1RXS J073101.9+460030 | J07310+460 | M4.0 V | B | | 07:31:01.27 | +46:00:24.8 | 18.141 ± 0.052 | 101.75 ± 0.06 | 12.896 | 296.0 | 431.4 | |
| Gaia EDR3 975312928903090560 | | M4.5 V | C* | | 07:31:09.03 | +45:56:55.6 | 18.388 ± 0.034 | 100.74 ± 0.04 | 15.219 | 266.3 | 307.8 | |
| GJ 3461 | J07418+050 | M3.0 V | Aab | Binary* | 07:41:52.56 | +05:02:23.1 | 36.109 ± 0.030 | 263.05 ± 0.04 | 11.624 | | | |
| G 50-1B | | ... | B* | | 07:41:52.61 | +05:02:22.4 | ... | ... | 16.860 | 130.9 | 1.014 | |
| 1RXS J090406.8-155512 | J09040-159 | M2.5 V | A | Binary* | 09:04:05.44 | -15:55:19.0 | 36.628 ± 0.021 | 113.77 ± 0.03 | 11.754 | | | |
| V405 Hya | | K2.0 V | B* | | 09:04:20.57 | -15:54:51.8 | 36.512 ± 0.022 | 112.03 ± 0.03 | 8.464 | 82.9 | 220.0 | |
| GJ 3576 | J09579+118 | M4.0 V | A | Binary* | 09:57:57.54 | +11:48:26.3 | 39.565 ± 0.048 | 451.50 ± 0.06 | 13.184 | | | |
| LP 489-1 | | M5.0 V | B* | | 09:56:45.21 | +11:34:23.6 | 40.843 ± 0.032 | 456.96 ± 0.04 | 14.267 | 231.6 | 1356.0 | |

Table D.4: Star candidates belonging to multiple systems not tabulated by WDS (continued).

| Name | Karmn | Spectral type | Component[a] | Class[b] | $\alpha$ (J2016.0) | $\delta$ (J2016.0) | $\pi$ [mas] | $\mu_{\rm total}$ [mas a$^{-1}$] | $G$ [mag] | $\theta$ [deg] | $\rho$ [arcsec] | Notes[c] |
|---|---|---|---|---|---|---|---|---|---|---|---|---|
| GJ 375.2 | J10004+272 | M0.5 V | A | Triple* | 10:00:26.69 | +27:16:03.5 | 27.767 ± 0.016 | 118.42 ± 0.02 | 10.528 | | | |
| 2MASS J10003572+2717054 | | M6.5 V | B | | 10:00:35.70 | +27:17:07.5 | 27.723 ± 0.103 | 118.34 ± 0.13 | 16.928 | 61.9 | 136.0 | |
| Gaia EDR3 740041664172917376 | | M7.5 V | C* | | 10:00:26.56 | +27:16:05.7 | 26.561 ± 0.349 | 117.20 ± 0.34 | 17.609 | 320.9 | 2.882 | |
| GJ 3602 | J10284+482 | M3.5 V | A | Binary* | 10:28:28.86 | +48:14:17.7 | 48.600 ± 3.300 | 615.83 ± 11.31 | 12.616 | | | • |
| G 146-35B | | … | B* | | 10:28:28.87 | +48:14:17.8 | … | … | 12.633 | 48.9 | 0.196 | |
| StKM 1-950 | J11307+549 | M1.0 V | A | Binary* | 11:30:43.80 | +54:57:29.1 | 24.778 ± 0.018 | 111.40 ± 0.02 | 11.019 | | | |
| Gaia EDR3 844502037681090688 | | M6.0 V | B* | | 11:30:42.66 | +54:56:57.0 | 24.731 ± 0.056 | 112.07 ± 0.07 | 16.672 | 197.0 | 33.61 | |
| PM J13182+7322 | J13182+733 | M3.5 V | A | Binary* | 13:18:13.82 | +73:22:05.6 | 39.534 ± 0.021 | 129.46 ± 0.04 | 12.396 | | | |
| PM J13182+7322B | | M7.0 V | B* | | 13:18:13.11 | +73:22:12.4 | 39.276 ± 0.098 | 122.66 ± 0.16 | 16.888 | 335.7 | 7.387 | |
| PM J13255+2738 | | M1.0 V | A | Triple* | 13:25:35.66 | +27:38:08.9 | 21.996 ± 0.019 | 71.21 ± 0.03 | 11.689 | | | |
| PM J13260+2735A | J13260+275 | M3.0 V | B* | | 13:26:02.70 | +27:35:03.7 | 21.868 ± 0.070 | 72.89 ± 0.11 | 12.410 | 117.2 | 404.4 | |
| PM J13260+2735B | | M2.5 V | C | | 13:26:02.63 | +27:35:02.5 | 22.030 ± 0.057 | 63.57 ± 0.09 | 13.131 | 117.4 | 404.1 | |
| BD+30 2400 | J13282+300 | M0.0 V | A | Triple* | 13:28:17.53 | +30:02:43.0 | 24.698 ± 0.058 | 260.99 ± 0.07 | 10.509 | | | |
| BD+30 2400B | | … | B | | 13:28:17.48 | +30:02:44.1 | … | … | 14.577 | 324.0 | 1.247 | |
| LP 323-115 | | M7.0 V | C* | | 13:28:20.69 | +30:03:15.8 | 24.101 ± 0.117 | 259.70 ± 0.14 | 17.314 | 51.3 | 52.48 | |
| StKM 1-1229 | J15188+292 | M1.0 V | A | Triple* | 15:18:49.75 | +29:15:06.4 | 20.173 ± 0.523 | 98.17 ± 0.53 | 11.682 | | | |
| Gaia EDR3 1275127175448008448 | | M0.0 V | B* | | 15:18:49.78 | +29:15:06.7 | 20.926 ± 0.337 | 78.58 ± 0.30 | 11.489 | 50.9 | 0.588 | |
| UCAC4 597-051773 | | M3.5 V | C | | 15:18:48.67 | +29:14:05.7 | 19.839 ± 0.018 | 86.82 ± 0.02 | 14.157 | 193.1 | 62.26 | |
| RX J1548.0+0421 | J15480+043 | M2.5 V | A | Binary* | 15:48:02.78 | +04:21:38.4 | 34.436 ± 0.027 | 58.85 ± 0.04 | 11.718 | | | |
| UCAC4 472-052890 | | M4.0 V | B* | | 15:47:54.89 | +04:18:02.9 | 34.579 ± 0.023 | 58.38 ± 0.03 | 13.186 | 208.7 | 245.6 | |
| HD 230017A | | M0.0 V | A | Triple* | 18:54:53.67 | +10:58:42.4 | 53.423 ± 0.108 | 133.27 ± 0.16 | 8.799 | | | |
| HD 230017B | J18548+109 | M3.5 V | B | | 18:54:53.86 | +10:58:45.1 | 53.644 ± 0.043 | 89.29 ± 0.06 | 11.441 | 45.1 | 3.792 | |
| PM J18542+1058 | | M4.0 V | C* | | 18:54:17.14 | +10:58:11.0 | 53.838 ± 0.024 | 114.61 ± 0.03 | 12.362 | 266.7 | 538.9 | |
| 1RXS J190405.9+211030 | J19041+211 | M2.0 Ve | A | Binary* | 19:04:06.24 | +21:10:32.7 | 27.742 ± 0.079 | 131.32 ± 0.10 | 11.429 | | | |
| PM J19041+2110 B | | … | B* | | 19:04:06.24 | +21:10:33.2 | … | … | 21.153 | 353.4 | 0.526 | |
| LSPM J1918+5803 | J19185+580 | M1.0 V | A | Binary* | 19:18:30.48 | +58:03:16.6 | 26.266 ± 0.012 | 178.63 ± 0.02 | 11.271 | | | |
| Gaia EDR3 2143230844400445696 | | M8.0 V | B* | | 19:18:30.33 | +58:03:18.1 | 26.310 ± 0.231 | 190.05 ± 0.40 | 17.978 | 323.8 | 1.925 | |
| 2MASS J19204172+7311434 | J19206+731S | M4.0 V | A | Binary* | 19:20:41.75 | +73:11:42.4 | 32.782 ± 0.014 | 74.49 ± 0.02 | 13.598 | | | |
| 2MASS J19204172+7311467 | J19206+731N | M4.5 V | B* | | 19:20:41.76 | +73:11:45.5 | 32.791 ± 0.014 | 82.50 ± 0.02 | 13.830 | 0.0 | 3.182 | • |

Table D.4: Star candidates belonging to multiple systems not tabulated by WDS (continued).

| Name | Karmn | Spectral type | Component[a] | Class[b] | $\alpha$ (J2016.0) | $\delta$ (J2016.0) | $\pi$ [mas] | $\mu_{\rm total}$ [mas a$^{-1}$] | $G$ [mag] | $\theta$ [deg] | $\rho$ [arcsec] | Notes[c] |
|---|---|---|---|---|---|---|---|---|---|---|---|---|
| GJ 4144 | J20195+080 | M3.0 V | A | Binary* | 20:19:34.60 | +08:00:27.1 | 26.833 ± 0.129 | 210.99 ± 0.19 | 12.393 | | | |
| Gaia EDR3 4250232535851803136 | | M3.0 V | B* | | 20:19:34.55 | +08:00:26.9 | 26.582 ± 0.128 | 200.82 ± 0.18 | 12.760 | 258.3 | 0.777 | |
| LP 395-8 A | J20198+229 | M3.0 V | Aab | Quadruple* | 20:19:49.36 | +22:56:38.1 | 33.897 ± 0.026 | 135.40 ± 0.03 | 11.023 | | | |
| LP 395-8 B | | M3.5 V | B | | 20:19:49.35 | +22:56:40.0 | 33.896 ± 0.053 | 137.73 ± 0.06 | 12.875 | 355.5 | 1.918 | |
| Gaia EDR3 1829571684884360832 | | … | C* | | 20:19:48.73 | +22:56:44.8 | 33.938 ± 0.342 | 142.76 ± 0.36 | 19.419 | 307.4 | 11.02 | • |
| TYC 1643-120-1 | J20220+216 | M2.0 V | A | Binary* | 20:22:01.62 | +21:47:19.7 | 37.091 ± 0.028 | 131.82 ± 0.03 | 11.290 | | | |
| PM J20220+2147B | | M4/5 V | B* | | 20:22:01.98 | +21:47:21.8 | 37.147 ± 0.042 | 133.78 ± 0.04 | 14.815 | 67.5 | 5.445 | |
| PM J21055+0609N | J21055+061 | M3.0 V | A | Binary* | 21:05:32.09 | +06:09:16.2 | 44.368 ± 0.033 | 52.64 ± 0.04 | 11.459 | | | |
| PM J21055+0609S | | M5.5 V | B* | | 21:05:32.17 | +06:09:11.2 | 44.365 ± 0.033 | 61.85 ± 0.04 | 14.855 | 166.0 | 5.094 | |
| PM J21057+5015E | J21057+502 | M3.5 V | A | Binary* | 21:05:45.54 | +50:15:44.1 | 27.714 ± 0.015 | 100.51 ± 0.02 | 12.453 | | | |
| PM J21057+5015W | | M3.5 V | B* | | 21:05:42.60 | +50:15:58.1 | 27.653 ± 0.016 | 97.27 ± 0.03 | 12.983 | 296.4 | 31.47 | |
| G 126-32A | | M1.0 V | A | Triple* | 21:45:04.94 | +19:53:31.8 | 27.661 ± 0.127 | 219.44 ± 0.14 | 11.554 | | | |
| G 126-32B | J21450+198 | M1.5 V | B | | 21:45:04.90 | +19:53:31.7 | 26.818 ± 0.296 | 240.74 ± 0.41 | 11.946 | 253.2 | 0.564 | |
| G 126-32C | | … | C* | | 21:45:05.04 | +19:53:36.9 | 26.252 ± 0.803 | 232.18 ± 1.02 | 19.824 | 16.4 | 5.317 | |
| a Ross 265 | J22018+164 | M2.5 V | AB | Triple* | 22:01:49.49 | +16:28:05.2 | 61.790 ± 2.230 | 421.76 ± 3.03 | 9.971 | | | • |
| Ross 268 | | M3.5 V | C* | | 22:06:23.02 | +17:22:22.1 | 56.763 ± 0.032 | 371.56 ± 0.04 | 11.830 | 50.2 | 5100.6 | |
| LF 4 +54 152 | J22129+550 | M0.0 V | A | Binary* | 22:12:56.63 | +55:04:50.8 | 18.385 ± 0.153 | 109.13 ± 0.25 | 10.222 | | | |
| Gaia EDR3 2005884249925303168 | | M6.5 V | B* | | 22:13:00.46 | +55:05:48.3 | 18.550 ± 0.079 | 108.97 ± 0.12 | 17.702 | 29.8 | 66.26 | • |
| LSPM J2305+4517 | J23051+452 | M3.5 V | A | Binary* | 23:05:09.00 | +45:17:32.9 | 22.102 ± 0.157 | 198.80 ± 0.18 | 12.356 | | | |
| Gaia DR2 1935209944573613568 | | … | B* | | 23:05:09.06 | +45:17:33.1 | … | … | 14.308 | 78.0 | 0.688 | • |
| G 28-46 | J23113+085 | M3.0 V | A | Binary* | 23:11:23.45 | +08:30:56.4 | 49.800 ± 3.300 | 458.92 ± 11.31 | 11.857 | | | • |
| G 28-46B | | … | B* | | 23:11:23.47 | +08:30:56.4 | … | … | 11.863 | 77.7 | 0.248 | • |
| GJ 4329 | J23175+063 | M3.0 V | A | Binary* | 23:17:34.73 | +06:23:24.5 | 48.918 ± 0.027 | 302.15 ± 0.04 | 11.429 | | | |
| GJ 4319 | | M3.5 V | B* | | 23:16:08.65 | +06:44:32.2 | 48.976 ± 0.026 | 304.89 ± 0.04 | 11.994 | 314.7 | 1803.4 | |
| G 129-51 | J23590+208 | M2.5 V | A | Binary* | 23:59:00.74 | +20:51:37.2 | 19.284 ± 0.448 | 300.02 ± 0.54 | 12.265 | | | |
| G 129-51B | | M2.0 V | B* | | 23:59:00.75 | +20:51:36.7 | 19.478 ± 0.146 | 301.61 ± 0.18 | 12.862 | 165.6 | 0.563 | • |

[a] An asterisk (*) denotes the new component found in each system.
[b] An asterisk (*) denotes the upgraded multiplicity order as the new found component is included.

[c] PM J00026+3821B: Component included in Table A.1 of Cifuentes et al. (2020), but not in WDS. V493 And A: Candidate to binary (criteria 1 and 2 in Table 4.2), thus a quadruple system. Wolf 58: The star has confirmed planet(s). 1RXS J033021.4+344044: Proper motions does not come from *Gaia*. GJ 3256 B: Component included in Table A.1 of Cifuentes et al. (2020), but not in WDS. LP 414-117: Hyades? LP 415-345: Hyades? RX J0507.2+3731A: Parallax and proper motions do not come from *Gaia*. G 106-7B: Component included in Table A.1 of Cifuentes et al. (2020), but not in WDS. LP 780-32: The star has confirmed planet(s). LP 780-23: Proper motions does not come from *Gaia*. GJ 3602: Parallax and proper motions do not come from *Gaia*. 2MASS J19204172+7311467: The star has confirmed planet(s). Gaia EDR3 1829571684884360832: Blended background object in the epoch of *Gaia*. Ross 265: Parallax and proper motions do not come from *Gaia*. *Gaia* EDR3 2005884249925303168: White dwarf candidate (Jiménez-Esteban et al., 2018). *Gaia* DR2 1935209944573613568: Component included in Table A.1 of Cifuentes et al. (2020), but not in WDS. G 28-46: Parallax and proper motions do not come from *Gaia*. G 28-46B: Sabotta et al. (2021) finds a 2225-day periodicity. G 129-51B: Component included in Table A.1 of Cifuentes et al. (2020), but not in WDS.

Table D.5: Spectroscopic binaries in Carmencita.

| Name | Karmn | Type[a] | $P_{\rm orb}$ [d] | $a$ [au] | Reference[b] |
|---|---|---|---|---|---|
| HD 38A | | SB | … | … | Str93 |
| 1R000806.3+475659 | J00081+479 | SB2 | 4.4 | 0.04 | Shk10 / Fou18 |
| EZPsc | J00162+198W | SB2 | 3.95652 ± 0.00008 | 0.038623 | Bar18 |
| FFAnd | J00428+355 | SB2 | … | … | Bop77 / Giz02 |
| RX J0050.2+0837 | J00502+086 | SB2 | … | … | Jeff18 |
| GJ 1029 | J01056+284 | SB2 | 95.76 ± 0.18 | 0.1331 | Bar18 / Win20a |
| PM J01437-0602 | J01437-060 | SB2 | … | … | Fou18 |
| G 173-18 | J01453+465 | SB2 | 157.9 | 0.56 | Giz02 / Shk10 |
| BD-21 332 | J01531-210 | SB2 | 2.9 | 0.04 | Shk10 |
| GJ 1041 B | J01592+035W | SB2 | 356.4 | 0.93 | Shk10 |
| GJ 3129 | J02027+135 | SB2 | 27.8 | 0.13 | Jen09 / Shk10 |
| GJ 3131 | J02033-212 | SB2 | … | … | Jeff18 |
| V374And | J02069+451 | SB2 | 897.0 ± 2.3 | … | Spe19 |
| GJ 3160 | J02289+120 | SB2 | … | … | Jeff18 |
| GJ 3235 | J03346-048 | ST3 | … | … | Jeff18 |
| GJ 3236 | J03372+691 | EB/SB2 | 0.7712600 ± 0.0000023 | 0.0143 | Irw09 / Shk10 |
| GJ 3239 | J03375+178N | SB2 | 33.2 | 0.18 | Shk10 / Fou18 |
| GJ 3240 | J03375+178S | EB?/SB2 | 0.4 | 0.01 | Shk10 |
| HD 278874 | | SB2 | … | … | Mon18 |
| Wolf 227 | J03526+170 | SB2 | … | … | Bon13 / Jeff18 |
| HD 24916B | J03574-011 | SB | … | … | Zak79 |
| LP 414-117 | J04123+162 | SB2 | 128.114 | 64.56 | Ben08 |
| 1R042441.9-064725 | J04247-067 | ST3 | 1.9 | 0.02 | Shk10 |
| GJ 3282 | J04252+080S | SB2 | … | … | Jeff18 |
| LP 775-31 | J04352-161 | SB2 | … | … | Rei09 |
| TYC 694-1183-1 | J04414+132 | SB | … | … | Mer09 |
| LP 415-345 | J04425+204 | SB2 | … | … | Sta97 |
| LP 416-43 | J04480+170 | SB | 8.49474 ± 0.00007 | 6.2 | Gri85 |
| 1R044847.6+100302 | J04488+100 | SB2 | … | … | Jeff18 |
| GJ 3322 A | J05019+099 | SB2 | … | … | Del98 / Del99b |
| HD 285190 | J05032+213 | SB2 | … | … | Jeff18 / Fou18 |
| Wolf 230 | J05078+179 | ST2 | 15.04547 ± 0.00041 | 0.082527 | Jeff18 / Win20 |
| Capella | | SB2 | 104.02173 ± 0.00022 | 0.7357 | She85 / Tor09 |
| GJ 1080 | J05282+029 | SB | … | … | Jen09 |
| HD 35956 | | SB1 | … | … | Kat13 |
| Ross 42 | J05322+098 | SB2 | 61.1 | 0.23 | Mar87 / Shk10 / Rei12 |
| V371Ori | J05337+019 | SB1 | … | … | Rei12 / Bar21 |
| Ross 45B | J05342+103S | SB | … | … | Rei12 |
| V1402Ori | J05402+126 | SB2 | 138.9 | 0.52 | Shk10 |
| Wolf 237 | J05466+441 | SB2 | … | … | Jeff18 |
| Ross 59 | J05532+242 | SB2 | 721 ± 2 | 0.6656 | Bar18 |
| TYC 4525-194-1 | J06171+751 | ST3? | … | … | Fou18 |
| G 108-4 | J06298-027 | SB2 | 46.9 | 0.16 | Shk10 |
| 1R070005.1-190115 | J07001-190 | SB2 | 6.56025 ± 0.00030 | 0.002797 | Bar21 |
| QYAur | J07100+385 | SB2 | 10.42673 ± 0.00010 | 0.0687 | TP86 / Win20a |
| TYC 4530-1414-1 | J07119+773 | SB1 | … | … | Jeff18 |
| 1R073138.4+455718 | | SB2? | … | … | Fou18 |
| Castor | | SB2+SB2 | 167570 ± 840 | 104.9 | Vin40 / Tor22 |
| Castor C | J07346+318 | EB/DESB2 | 0.8142822 | 0.01809 | Joy26 / Giz02 |
| GJ 282 C | J07361-031 | SB1 | 6591 ± 177 | 6.224 | Bar21 |
| GJ 3461 | J07418+050 | SB2 | … | … | Jeff18 |
| 1R075434.3+083213 | J07545+085 | SB1 | … | … | Jeff18 |
| FPCncB | J08089+328 | SB2 | 395.5 | 0.92 | Shk10 |

Table D.5: Spectroscopic binaries in Carmencita (continued).

| Name | Karmn | Type[a] | $P_{orb}$ [d] | $a$ [au] | Reference[b] |
|---|---|---|---|---|---|
| CUCnc | J08316+193S | EB | 2.771468 ± 0.000004 | 0.036240 | Del99a / Del99b / Tor10 |
| LSPM J0835+1408 | J08353+141 | ST3 | … | … | Ski18 |
| GJ 319 A | J08427+095 | SB1 | 20.9491 ± 0.0019 | 0.021 | Duq88 |
| GJ 3540 | J09120+279 | SB2 | … | … | Jeff18 |
| LP 427-16 | J09140+196 | SB1 | … | … | Bar21 |
| HD 79210 | J09143+526 | SB1 | … | … | Jeff18 |
| GJ 3547 | J09193+620 | EB/SB2 | 20.2 | 0.16 | Shk10 / Skr21 |
| GJ 9303 | J09362+375 | SB2 | … | … | Mal14a |
| LP 728-70 | J09506-138 | SB2 | … | … | Jeff18 |
| GJ 372 | J09531-036 | SB2 | 47.709 ± 0.053 | 39.06 | Har96 / Jeff18 / Bar18 |
| GJ 373 | J09561+627 | SB? | … | … | Rei12 |
| LP 790-2 | J10182-204 | SB2 | 5.922845 ± 0.000061 | 0.048242 | Bar18 |
| GJ 3612 | J10354+694 | SB2 | 119.41 ± 0.04 | 0.3464 | Bar18 |
| LP 127-502 | J10368+509 | SB2? | … | … | Fou18 |
| GJ 3626 | J10504+331 | SB1 | 2996 ± 31 | 0.2355 | Bar21 |
| GJ 3630 | J10520+005 | ST3/SQ4 | 28.1 | 0.11 | Shk10 |
| LP 491-51 | J11036+136 | SB1 | … | … | Jeff18 |
| GJ 426.1 A | | SB | … | … | Abt76 |
| GJ 455 | J12023+285 | SB2 | … | … | Giz02 |
| LP 734-34 | J12104-131 | SB2 | 33.6551 ± 0.0046 | 0.0479 | Win20a |
| GJ 3719 | J12169+311 | SB2 | … | … | Fou18 |
| LP 320-626 | J12191+318 | SB2 | … | … | Jeff18 |
| GJ 3729 | J12290+417 | SB2/3? | … | … | Shk12 / Bow15 / Fou18 |
| GJ 3731 | J12299-054W | SB2 | … | … | Jeff18 |
| G 123-45 | J12364+352 | SB1 | 34.7557 ± 0.0041 | 0.0219 | Win20a |
| DPDra | J12490+661 | ST3 | 54.075 ± 0.006 | … | Del99b |
| GJ 507 B | J13195+351E | SB1 | … | … | Jeff18 |
| GQVir | J14130-120 | SB2 | … | … | Jeff18 |
| GJ 1182 | J14155+046 | SB2 | 154.20 ± 0.10 | 0.35835 | Bar18 |
| GJ 3839 | J14170+317 | SB2/ST3 | … | … | Del99b / Rei12 / Fou18 |
| PM J14171+0851 | J14171+088 | SB2 | … | … | Jeff18 |
| GJ 3861 | J14368+583 | SB2 | … | … | Giz02 / Jeff18 |
| GJ 3875 | J14549+411 | SB | … | … | Ski18 |
| G 136-35 | J14564+168 | SB | … | … | Ski18 |
| HD 131976 | J14574-214 | SB2 | … | … | Mar87 |
| GJ 3900 | J15191-127 | SB2 | … | … | Bon13 |
| StKM 1-1240 | J15238+561 | SB2 | … | … | Fou18 |
| GJ 3910 | J15319+288 | SB1 | … | … | Jen09 |
| UU UMi | J15412+759 | SB2 | 5240 ± 410 | 4.7 | Bar18 / Bar21 |
| GJ 595 | J15421-194 | SB1 | … | … | Nid02 |
| GJ 3916 | J15474-108 | ST3 | 3028 ± 23 | 3.59 | Bar21 |
| LP 177-102 | J15474+451 | EB/SB2 | 3.5500184 | 0.036535 | Moc02 / Har11 / Bir12 |
| sig CrB A | | SB2 | 1.13979142 ± 0.00000008 | 0.013106 | Str03 / Rag09 |
| CMDra | J16343+571 | EB/SB2 | 1.268 | 0.017502 | Mor09 / Tor10 / Scha19 |
| V1054Oph | J16554-083S | ST3 | 2.965522 ± 0.000014 | 0.05 | Joy47 / Pet84 / Del98 |
| LP 331-57 A | J17038+321 | SB2? | … | … | Shk09/Fou18 |
| GJ 3991 | J17095+436 | SB1 | … | … | Rei97a / Del99b |
| V2367Oph | J17136-084 | SB2 | … | … | Rei12 / Jeff18 |
| Wolf 1473 | J17464-087 | SB2 | 83.926 ± 0.032 | 0.27559 | Mal14a / Win20a |
| GJ 1230 A | J18411+247S | SB2 | 5.06880 ± 0.00005 | … | GR96 / Del99b |
| GJ 735 | J18554+084 | SB2 | … | … | Mar87 / Giz02 / Kar04 |
| GJ 4091 B | J18563+544 | SB2 | … | … | Fou18 |
| G 125-15 | J19312+361 | SB2 | 0.7 | 0.01 | Shk10 |

Table D.5: Spectroscopic binaries in Carmencita (continued).

| Name | Karmn | Type[a] | $P_{orb}$ [d] | $a$ [au] | Reference[b] |
|---|---|---|---|---|---|
| RX J1935.4+3746 | J19354+377 | SB1 | ... | ... | Shk09/Jeff18 |
| LP 869-19 | J19420-210 | SB2 | ... | ... | Mal14a |
| V1513Cyg | J20050+544 | SB1 | ... | ... | Joy47 / Mac08 |
| LP 574-21 | J20105+065 | SB2 | 40.1 | 0.19 | Shk10 |
| LP 395-8 A | J20198+229 | SB2 | 1.1293392 ± 0.0000067 | 0.01057 | Bar18 |
| 1R203011.0+795040 | J20301+798 | SB2 | ... | ... | Jeff18 |
| GJ 4155 | J20409-101 | SB | ... | ... | Rei12 |
| GJ 4161 | J20445+089N | SB2 | ... | ... | Jeff18 |
| GJ 810 A | J20556-140N | SB2 | 812 ± 51 | 0.841 | Bar18 |
| FRAqr | J20568-048 | SB2 | ... | ... | Jeff18 |
| Ross 775 | J21296+176 | EB?/SB2 | 53.221 ± 0.004 | ... | Mar87 / Del99b |
| GJ 4213 | J21442+066 | SB1 | ... | ... | Jeff18 |
| EZAqr | J22385-152 | ST3 | 3.78652 ± 0.00001 | 1189 | Del99c |
| FKAqr | J22387-206S | SB2 | 4.08322 ± 0.00004 | 0.039372 | Her65 / Del99b |
| FLAqr | J22387-206N | SB1 | 1.795 ± 0.017 | 0.00353 | Davi14 |
| LP 521-79 | J23063+126 | SB2/3? | 58.9 | 0.31 | Shk10 |
| GJ 4314 | J23096-019 | SB2 | ... | ... | Jeff18 |
| GJ 4327 | J23174+382 | SB2 | ... | ... | Jeff18 |
| EQPegA | J23318+199E | SB2 | ... | ... | Del99b |
| EQPegB | J23318+199W | SB1 | ... | ... | Del99b |
| GJ 1284 | J23302-203 | SB2 | 11.838033 ± 0.000076 | 6.48275 | Giz02 / Jeff18 / Car21 |
| Ross 676 | J23439+647 | SB2 | 3.0 | 0.04 | Shk10 |
| GJ 912 | J23556-061 | SB1 | 5188 ± 58 | 0.505 | Bar21 |
| GJ 4379 | J23573-129W | SB2 | ... | ... | Giz02 / Jeff18 |
| Wolf 1051 | J23585+076 | ST3 | 4634 ± 17 | 0.4436 | Bar21 |

[a] A spectroscopic binary (SB) that displays one or two lines in the spectrum is a single- or double-lined spectroscopic binary, respectively, and abbreviated as SB1 or SB2. A spectroscopic triple (ST) showing two or three lines is a double- or triple-lined spectroscopic triple, and abbreviated as ST2 or ST3. Spectroscopic quadruples (SQ) are much more rare, and typically are three- or four-lined systems, denominated SQ3 and SQ4, respectively.

[b] And07: Andrade (2007); Bar18: Baroch et al. (2018); Bar21: Baroch et al. (2021); Ben00: Benedict et al. (2000); Bon13: Bonfils et al. (2013a); Bop74: Bopp (1974); Bow15: Bowler et al. (2015); BS08: Bender & Simon (2008); Cat06: Catala et al. (2006); Che16: Chelli et al. (2016); Cor14: Cortes-Contreras et al. (2014); Cor17: Cortés-Contreras et al. (2017a); Cur06: CURTIS (1906); Dav14: Davenport et al. (2014); Del98: Delfosse et al. (1998b); Del99a: Delfosse et al. (1999b); Del99b: Delfosse et al. (1999a); Ell15: Elliott et al. (2015); Fou18: Fouqué et al. (2018); Giz00a: Gizis et al. (2000a); Giz00b: Gizis et al. (2000c); Giz02: Gizis et al. (2002); Gli91: Gliese & Jahreiß (1991); GR96: Gizis & Reid (1996); Har11: Hartman et al. (2011); HC81: Harrington et al. (1981a); HM65: Herbig & Moorhead (1965); Ire08: Ireland et al. (2008); Irw09: Irwin et al. (2009); Jan12 : Janson et al. (2012) ; Jan14a: Janson et al. (2014c); Jan14b: Janson et al. (2014a); Jef18: Jeffers et al. (2018); Jen09: Jenkins et al. (2009); Jod13: Jódar et al. (2013); Joy26: Joy & Sanford (1926); Joy47: Joy (1947); Kar04: Karataş et al. (2004); Kat13: Katoh et al. (2013); Kon10 : Konopacky et al. (2010) ; Lan01: Lane et al. (2001); Law06: Law et al. (2006); Lin12: Lindegren et al. (2012); Malo14a: Malo et al. (2014b); Malo14b: Malo et al. (2014c); Mar00: Martín et al. (2000); Mar87: Marcy et al. (1987); Maz01: Mazeh et al. (2001); Mer09: Mermilliod et al. (2009); Mon18: Montes et al. (2018); Mor09: Morales et al. (2009); Nid02: Nidever et al. (2002); Pet84: Pettersen et al. (1984); Pou04: Pourbaix et al. (2004); Rag09: Raghavan et al. (2009); Rei09: Reiners & Basri (2009); Rei12: Reipurth & Mikkola (2012a); Rei97a: Reid & Gizis (1997); Sam07: Samus' et al. (2017); Scha19: Schanche et al. (2019); Schl12b: Schlieder et al. (2012b); Schw19: Schweitzer et al. (2019); Sha17: Shan et al. (2017); Sha21: Shara et al. (2021); Shk10: Shkolnik et al. (2010); Shk12: Shkolnik et al. (2012); Ski18: Skinner et al. (2018); Spe19: Sperauskas et al. (2019); Sta97: Stauffer et al. (1997); Ste14: Stebbins (1914); Str77: Strand (1977); Str93: Strassmeier et al. (1993); Tok10: Tokovinin et al. (2010); Tok97: Tokovinin (1997); Tor15: Torres et al. (2015); TP86: Tomkin & Pettersen (1986); Vin40: Vinter Hansen et al. (1940); Zak79: Zakhozhaj (1979).

Table D.6: Multiple systems containing M dwarfs and FGK primaries.

| Name | Karmn | Spectral type | Component | Class | $\alpha$ | $\delta$ | $\pi$ [mas] | $\mu_{total}$ [mas $a^{-1}$] | $G$ [mag] | $\theta$ [deg] | $\rho$ [arcsec] | Notes[a] |
|---|---|---|---|---|---|---|---|---|---|---|---|---|
| HD 4967 | | K5 V | A | | 00:51:34.73 | −22:54:40.7 | 65.090 ± 0.025 | 672.69 ± 0.05 | 8.416 | | | |
| HD 4967B | J00515-229 | M5.5 V | B | | 00:51:35.91 | −22:54:35.4 | 65.073 ± 0.050 | 680.57 ± 0.09 | 14.366 | 72.1 | 17.1 | |
| EXCet | | K0/1 V | A | | 01:37:35.65 | −06:45:39.1 | 41.564 ± 0.024 | 197.86 ± 0.03 | 7.466 | | | |
| LP 648-20 | J01369-067 | M3.5 V | B | | 01:36:55.36 | −06:47:39.6 | 41.699 ± 0.045 | 200.55 ± 0.05 | 12.709 | 258.6 | 612.075 | |
| Wolf 109 | | K5 V | A | | 02:02:02.81 | +03:56:20.0 | 27.827 ± 0.021 | 466.14 ± 0.03 | 10.081 | | | |
| Wolf 109 B | J02020+039 | M2.0 V | B | | 02:02:03.11 | +03:56:37.4 | 28.922 ± 0.353 | 465.77 ± 0.48 | 12.412 | 14.4 | 18.0 | |
| G3-2517912315149114240 | | … | C | | 02:02:03.16 | +03:56:36.9 | … | … | 12.832 | 16.8 | 17.739 | |
| HD 15468 | | K4.5 Vk | AB | | 02:29:02.40 | −19:58:40.8 | 51.606 ± 0.136 | 646.40 ± 0.17 | 8.394 | 145.3 | 0.182 | |
| GJ 100 C | J02285-200 | M2.5 V | C | | 02:28:32.62 | −20:02:22.5 | 50.972 ± 0.032 | 649.19 ± 0.04 | 11.784 | 242.1 | 474.569 | |
| HD 16160 | | K3 V | AB | | 02:36:06.81 | +06:53:36.0 | 138.340 ± 0.318 | 2312.10 ± 0.50 | 5.498 | 289.5 | 1.730 | |
| BXCet | J02362+068 | M4.0 V | C | | 02:36:17.20 | +06:52:41.1 | 138.437 ± 0.042 | 2312.99 ± 0.06 | 10.333 | 289.5 | 164.180 | |
| HD 16895 | | F8 V | A | | 02:44:12.53 | +49:13:41.0 | 89.685 ± 0.164 | 346.40 ± 0.24 | 3.983 | | | • |
| GJ 107 B | J02441+492 | M1.5 V | B | | 02:44:10.79 | +49:13:53.0 | 89.374 ± 0.032 | 322.95 ± 0.04 | 9.140 | 305.1 | 20.932 | |
| HD 18757 | | G1.5 V | A | | 03:04:11.26 | +61:42:09.9 | 42.459 ± 0.020 | 1000.79 ± 0.02 | 6.484 | | | |
| GJ 3195 | J03047+617 | M3.0 V | B | | 03:04:45.06 | +61:43:57.7 | 42.534 ± 0.026 | 1000.09 ± 0.03 | 11.493 | 65.8 | 263.299 | |
| V577Per | | K2 V | A | | 03:33:13.60 | +46:15:23.7 | 27.482 ± 0.031 | 188.42 ± 0.04 | 8.016 | | | |
| HD 21845B | J03332+462 | M0.0 V | B | | 03:33:14.16 | +46:15:16.2 | 27.481 ± 0.016 | 185.83 ± 0.02 | 10.497 | 142.3 | 9.502 | • |
| HD 278874 | | K2 V | Aab | | 03:39:48.92 | +33:28:24.3 | 25.434 ± 0.032 | 36.21 ± 0.04 | 8.707 | | | |
| HD 278874B | J03397+334 | M3.0 V | B | | 03:39:47.79 | +33:28:30.7 | 25.571 ± 0.041 | 38.83 ± 0.06 | 11.771 | 294.3 | 15.47 | |
| HD 23189 | | K2 V | A | | 03:48:01.40 | +68:40:26.4 | 55.825 ± 0.013 | 279.83 ± 0.02 | 8.772 | | | |
| GJ 153 C | J03480+686 | M2 V | BC | | 03:48:02.08 | +68:40:42.9 | 55.984 ± 0.105 | 288.83 ± 0.21 | 10.478 | 26.3/13.2 | 0.763/17.206 | |
| HD 275867 | | K2 V | A | | 03:52:00.35 | +39:47:43.7 | 30.985 ± 0.016 | 62.07 ± 0.02 | 9.136 | | | |
| TYC 2868-639-1 | J03519+397 | M0.0 V | B | | 03:51:58.19 | +39:46:55.8 | 31.021 ± 0.020 | 64.40 ± 0.03 | 10.393 | 207.5 | 53.919 | |
| HD 24916 | | K4 V | A | | 03:57:28.50 | −01:09:36.4 | 65.426 ± 0.023 | 234.33 ± 0.03 | 7.673 | | | |
| HD 24916B | J03574-011 | M2.5 V | Bab | | 03:57:28.68 | −01:09:25.7 | 65.492 ± 0.044 | 251.75 ± 0.06 | 10.438 | 14.1 | 11.0 | |
| HD 26965 | | K0 V | A | | 04:15:13.91 | −07:40:05.1 | 199.608 ± 0.121 | 4089.84 ± 0.16 | 4.180 | | | |
| HD 26976 | | DA2.9 | B | | 04:15:19.39 | −07:40:22.6 | 199.691 ± 0.051 | 4018.59 ± 0.06 | 9.542 | 102.2 | 83.337 | |
| HD 27848 | | … | AB | | 04:24:22.38 | +17:04:43.8 | 19.959 ± 0.026 | 103.39 ± 0.04 | 6.856 | | | • |
| V991Tau | | K4 V | C | | 04:25:00.35 | +16:59:05.2 | 18.356 ± 0.019 | 94.41 ± 0.03 | 9.970 | 121.9 | 641.2 | |

Table D.6: Multiple systems containing M dwarfs and FGK primaries (continued).

| Name | Karmn | Spectral type | Component | Class | $\alpha$ | $\delta$ | $\pi$ [mas] | $\mu_{total}$ [mas $a^{-1}$] | $G$ [mag] | $\theta$ [deg] | $\rho$ [arcsec] | Notes[a] |
|---|---|---|---|---|---|---|---|---|---|---|---|---|
| V805Tau | J04252+172 | M3.5 V | DE | | 04:25:13.67 | +17:16:05.1 | 19.313 ± 0.193 | 110.65 ± 0.26 | 11.998 | 47.1 | 1002.145 | |
| LP 415-881 | | M7.0 V | F | | 04:26:19.16 | +17:03:01.7 | 21.204 ± 0.061 | 105.70 ± 0.09 | 16.338 | 93.4 | 1677.610 | |
| HD 29391 | | F0 V | A | | 04:37:36.18 | −02:28:25.8 | 33.439 ± 0.078 | 77.72 ± 0.13 | 5.141 | 162.5 | 66.7 | |
| GJ 3305 | J04376-024 | M1.1 V | BC | | 04:37:37.51 | −02:29:29.7 | 36.009 ± 0.476 | 72.38 ± 0.63 | 9.801 | 208.2/162.6 | 0.098/66.963 | |
| V583Aur | | K5 V | A | | 04:39:25.47 | +33:32:43.9 | 11.128 ± 0.018 | 52.35 ± 0.03 | 11.002 | | | • |
| PM J04393+3331 | J04393+335 | M4.0 V | BC | | 04:39:23.22 | +33:31:48.7 | 7.425 ± 0.848 | 50.28 ± 1.83 | 13.052 | 50.6/207.1 | 0.126/61.936 | |
| HD 31412 | | F9.5 V | AB | | 04:55:56.03 | +04:40:10.5 | 27.865 ± 0.027 | 233.46 ± 0.04 | 6.891 | 17.7 | 0.443 | |
| HD 31412B | J04559+046 | M2.0 V | C | | 04:55:54.60 | +04:40:13.5 | 27.851 ± 0.026 | 234.00 ± 0.03 | 11.031 | 277.9 | 21.657 | |
| HD 35956 | | G0 V | Aab | | 05:28:51.74 | +12:32:59.6 | 33.788 ± 0.294 | 231.78 ± 0.42 | 6.597 | | | |
| HD 35956B | | M1.0 V | B | | 05:28:52.11 | +12:33:01.4 | 35.405 ± 0.020 | 237.81 ± 0.03 | 10.760 | 71.4 | 5.8 | |
| GJ 3348A | J05289+125 | M4.0 V | C | | 05:28:56.61 | +12:31:50.4 | 39.500 ± 10.200 | 230.59 ± 11.31 | 13.146 | 134.1 | 99.391 | |
| GJ 3348B | | ... | D | | 05:28:56.61 | +12:31:50.2 | ... | ... | 13.222 | 134.2 | 99.461 | |
| V538Aur | | K1 V | A | | 05:41:20.34 | +53:28:43.4 | 81.499 ± 0.025 | 523.48 ± 0.03 | 5.991 | | | |
| HD 233153 | J05415+534 | M1.0 V | B | | 05:41:30.74 | +53:29:15.0 | 81.464 ± 0.023 | 515.97 ± 0.03 | 8.919 | 71.2 | 98.035 | |
| HD 43587 | | G0 V | AB | | 06:17:15.90 | +05:06:02.6 | 51.616 ± 0.124 | 269.29 ± 0.20 | 5.565 | 70.9 | 0.840 | |
| GJ 231.1 B | J06171+051 | M3.5 V | CD | | 06:17:10.42 | +05:07:05.3 | 52.139 ± 0.083 | 261.65 ± 0.11 | 12.136 | 156.90/307.4 | 0.545/103.19 | |
| HD 263175 | | K3 V | A | | 06:46:04.47 | +32:33:22.0 | 38.828 ± 0.022 | 467.06 ± 0.03 | 8.501 | | | |
| HD 263175B | J06461+325 | M1.0 V | B | | 06:46:06.88 | +32:33:16.6 | 38.873 ± 0.017 | 472.87 ± 0.02 | 11.288 | 100.1 | 30.855 | |
| HD 50281 | | K3.5 V | AB | | 06:52:17.47 | −05:10:25.4 | 114.355 ± 0.042 | 543.70 ± 0.04 | 6.231 | 149.6 | 0.175 | |
| HD 50281B | J06523-051 | M2.0 V+ | C | | 06:52:17.42 | −05:11:24.3 | 114.291 ± 0.022 | 576.48 ± 0.03 | 9.098 | 180.6 | 58.833 | |
| GJ 3415 | | K4.5 V | A | | 06:56:28.28 | +40:04:20.6 | 39.870 ± 0.023 | 452.03 ± 0.03 | 8.687 | 6.8 | 37.808 | |
| GJ 3416 | J06564+400 | M1.0 V | B | | 06:56:28.64 | +40:04:58.3 | 39.925 ± 0.021 | 446.75 ± 0.03 | 10.299 | 6.2 | 37.9 | |
| V869Mon | | K3 V | A | | 07:39:59.40 | −03:35:55.5 | 71.032 ± 0.024 | 286.81 ± 0.03 | 6.890 | | | |
| HD 61606B | | K7 V | B | | 07:40:02.97 | −03:36:17.8 | 70.992 ± 0.025 | 294.21 ± 0.03 | 8.335 | 112.7 | 57.9 | |
| GJ 282 C | J07361-031 | M1.0 V | Cab | | 07:36:07.15 | −03:06:43.4 | 70.275 ± 0.131 | 302.35 ± 0.17 | 9.142 | 296.7 | 3894.176 | |
| HD 68146 | | F6.5 V | A | | 08:10:39.55 | −13:47:56.2 | 44.340 ± 0.053 | 257.61 ± 0.06 | 5.399 | | | |
| HD 68146B | J08105-138 | M2.5 V | B | | 08:10:34.02 | −13:48:50.1 | 44.520 ± 0.049 | 259.91 ± 0.06 | 11.008 | 236.2 | 96.934 | |
| G3-5725122965265271680 | | M4.0 V | C | | 08:10:33.96 | −13:48:49.9 | 44.150 ± 0.203 | 245.49 ± 0.32 | 12.781 | 236.6 | 97.537 | |
| V405Hya | | K2 V | A | | 09:04:20.57 | −15:54:51.8 | 36.512 ± 0.022 | 112.03 ± 0.03 | 8.464 | | | |
| 1R090406.8-155512 | J09040-159 | M2.5 V | B* | | 09:04:05.44 | −15:55:19.0 | 36.628 ± 0.021 | 113.77 ± 0.03 | 11.754 | 82.9 | 220.022 | • |

Table D.6: Multiple systems containing M dwarfs and FGK primaries (continued).

| Name | Karmn | Spectral type | Component | Class | $\alpha$ | $\delta$ | $\pi$ [mas] | $\mu_{total}$ [mas a$^{-1}$] | $G$ [mag] | $\theta$ [deg] | $\rho$ [arcsec] | Notes[a] |
|---|---|---|---|---|---|---|---|---|---|---|---|---|
| BD+22 2086A | | K5 V | A | | 09:23:06.19 | +22:18:17.6 | 29.405 ± 0.020 | 218.39 ± 0.02 | 9.274 | | | |
| BD+22 2086B | J09231+223 | M0.0 V | B | | 09:23:06.01 | +22:18:25.6 | 29.444 ± 0.022 | 223.16 ± 0.02 | 10.641 | 343.4 | 8.301 | |
| DXLeo | | G9 V(k) | A | | 09:32:43.58 | +26:59:14.8 | 55.329 ± 0.021 | 287.22 ± 0.02 | 6.813 | 64.4 | 65.558 | |
| HD 82443B | J09328+269 | M5.5 V | B | | 09:32:48.07 | +26:59:39.9 | 55.292 ± 0.071 | 284.17 ± 0.09 | 13.826 | 67.3 | 65.013 | |
| HD 82939 | | G5 V | A | | 09:36:15.78 | +37:31:44.1 | 25.740 ± 0.023 | 134.64 ± 0.02 | 10.292 | | | |
| GJ 9303 | J09362+375 | M0.0 V | Bab | | 09:36:04.14 | +37:33:08.9 | 25.836 ± 0.023 | 134.34 ± 0.02 | 8.069 | 301.5 | 162.314 | |
| GJ 397.1 | | K5 V | A | | 10:31:43.09 | +57:06:59.9 | 57.010 ± 0.014 | 185.37 ± 0.02 | 9.003 | 225.6 | 141.959 | |
| GJ 397.1 B | J10315+570 | M5.0 V | BC | | 10:31:30.64 | +57:05:20.4 | … | 161.31 ± 11.31 | 13.519 | 154.5/225.6 | 0.284/142.037 | |
| LZUMa | | G5 V | A | | 10:50:39.98 | +51:47:58.8 | 37.386 ± 0.070 | 196.65 ± 0.17 | 8.047 | | | • |
| GJ 3628 | J10506+517 | M4.1 V | B | | 10:50:37.91 | +51:45:01.6 | 37.889 ± 0.028 | 191.40 ± 0.03 | 12.763 | 186.2 | 178.3 | |
| HD 97584 | | K4 V | A | | 11:15:10.39 | +73:28:32.5 | 69.136 ± 0.015 | 418.80 ± 0.02 | 7.312 | 102.9 | 67.600 | |
| HD 97584B | J11151+734 | M2.5 V | B | | 11:15:09.56 | +73:28:38.0 | 69.134 ± 0.018 | 396.51 ± 0.03 | 10.609 | 327.1 | 6.481 | |
| GJ 426.1 A | | F1 IV | Aab | | 11:23:55.62 | +10:31:44.9 | 42.355 ± 0.398 | 169.81 ± 0.99 | 3.907 | | | |
| GJ 426.1 B | | F5 V | B | | 11:23:55.75 | +10:31:44.7 | 41.496 ± 0.351 | 185.02 ± 0.97 | 6.842 | 95.1 | 2.050 | |
| HD 108421A | | K2 V | A | | 12:27:13.83 | +27:01:24.9 | 36.504 ± 0.027 | 265.14 ± 0.03 | 8.485 | | | |
| HD 108421B | | K4 V | B | | 12:27:13.84 | +27:01:27.6 | 36.511 ± 0.031 | 256.19 ± 0.04 | 8.857 | 4.1 | 2.8 | |
| HD 115404 | | K2 V | A | | 13:16:51.76 | +17:00:57.6 | 91.018 ± 0.024 | 689.14 ± 0.04 | 6.298 | | | |
| HD 115404B | J13168+170 | M0.5 V | B | | 13:16:52.28 | +17:00:55.7 | 90.948 ± 0.023 | 701.23 ± 0.04 | 8.870 | 104.6 | 7.671 | |
| GJ 9453 | | K5 V | A | | 13:34:51.91 | +74:30:01.1 | 25.386 ± 0.026 | 439.09 ± 0.05 | 9.384 | 323.3 | 14.8 | |
| GJ 9453 B | J13348+745 | M3.5 V | B | | 13:34:49.83 | +74:30:12.6 | 25.492 ± 0.015 | 437.06 ± 0.03 | 12.363 | 324.0 | 14.192 | |
| BD+21 2602 | | K4 V | A | | 14:04:09.85 | +20:45:32.5 | 24.815 ± 0.015 | 127.65 ± 0.02 | 9.793 | | | |
| StKM 1-1119 | J14041+207 | M1.0 Ve | B | | 14:04:09.06 | +20:44:30.9 | … | 132.08 ± 3.76 | 11.499 | 190.2 | 62.7 | |
| G3-1247168140942467200 | | … | C | | 14:04:09.06 | +20:44:31.2 | … | … | 11.660 | 190.3 | 62.307 | |
| HD 126660 | | F7 V | A | | 14:25:11.39 | +51:50:56.3 | 69.069 ± 0.158 | 464.15 ± 0.23 | 3.917 | 182.5 | 70.160 | • |
| HD 126660B | J14251+518 | M2.5 V | B | | 14:25:11.17 | +51:49:46.6 | 68.812 ± 0.032 | 469.65 ± 0.05 | 10.498 | 181.7 | 69.674 | |
| KXLib | | K4 V | A | | 14:57:29.18 | −21:25:23.3 | 169.884 ± 0.065 | 2008.68 ± 0.09 | 5.364 | | | |
| HD 131976 | J14574-214 | M1.0 V | Bab | | 14:57:27.70 | −21:25:07.9 | 168.770 ± 21.540 | 1933.94 ± 26.20 | 7.249 | 306.6 | 25.762 | • |
| GJ 570 D | | T8 | C | | 15:07:55.68 | +76:12:05.4 | 169.300 ± 1.700 | 1972.88 ± 5.62 | | 317.0 | 234.000 | • |
| HD 135363 | | G5 V | AB | | 15:07:55.68 | +76:12:05.4 | 33.711 ± 0.052 | 208.70 ± 0.10 | 8.429 | 133.1 | 0.36 | |

Table D.6: Multiple systems containing M dwarfs and FGK primaries (continued).

| Name | Karmn | Spectral type | Component | Class | $\alpha$ | $\delta$ | $\pi$ [mas] | $\mu_{total}$ [mas a$^{-1}$] | $G$ [mag] | $\theta$ [deg] | $\rho$ [arcsec] | Notes[a] |
|---|---|---|---|---|---|---|---|---|---|---|---|---|
| LSPM J1507+7613 | J15079+762 | M4.5 V | C | | 15:07:56.64 | +76:14:01.8 | 33.714 ± 0.027 | 207.10 ± 0.06 | 12.379 | 1.7 | 116.445 | |
| HD 144579 | | G8 V | A | | 16:04:56.01 | +39:09:24.3 | 69.641 ± 0.014 | 573.29 ± 0.02 | 6.461 | | | |
| HD 144579B | J16048+391 | M4.0 V | B | | 16:04:50.08 | +39:09:36.6 | 69.637 ± 0.019 | 564.44 ± 0.03 | 12.855 | 280.1 | 69.998 | |
| HD 146361A | | F6 V | Aab | | 16:14:40.51 | +33:51:29.6 | 44.057 ± 0.046 | 282.06 ± 0.07 | 5.431 | | | |
| HD 146362 | | G1 V | B | | 16:14:40.01 | +33:51:25.8 | 44.134 ± 0.018 | 301.27 ± 0.03 | 6.438 | 238.5 | 7.231 | |
| TZCrB | J16139+337 | M2.5 V | CD | | 16:13:55.90 | +33:46:22.7 | 44.267 ± 0.159 | 300.69 ± 0.24 | 11.269 | 28.7/241.1 | 0.507/635.058 | |
| V1090Her | | K3 V | A | | 16:57:52.95 | +47:22:04.4 | 55.751 ± 0.016 | 308.70 ± 0.03 | 7.521 | | | |
| HD 153557B | J16578+473 | M1.5 V | B | | 16:57:53.40 | +47:22:06.7 | 55.771 ± 0.017 | 301.44 ± 0.03 | 10.419 | 63.0 | 5.091 | |
| V1089Her | | K3 V | C | | 16:57:42.01 | +47:21:47.9 | 55.718 ± 0.016 | 296.94 ± 0.03 | 7.622 | 261.6 | 112.378 | |
| HD 154363 | | K4/5 V | A | | 17:05:02.41 | −05:04:17.7 | 95.567 ± 0.024 | 1461.66 ± 0.03 | 7.265 | | | |
| HD 154363B | J17052-050 | M1.5 V | B | | 17:05:12.80 | −05:05:57.4 | 95.560 ± 0.021 | 1457.06 ± 0.03 | 9.178 | 122.7 | 184.42 | |
| HD 160269A | | G0 IV/V | AB | | 17:35:00.16 | +61:52:20.8 | 69.283 ± 0.200 | 522.56 ± 0.33 | 5.069 | 300.5 | 0.625 | |
| GJ 685 | J17355+616 | M0.5 V | C | | 17:35:35.07 | +61:40:45.4 | 69.892 ± 0.015 | 577.33 ± 0.02 | 9.175 | 160.3 | 738.164 | |
| BD+31 3330 | | K2.5 V | A | | 18:40:54.99 | +31:31:45.7 | 41.745 ± 0.016 | 841.17 ± 0.02 | 8.260 | | | |
| BD+31 3330B | J18409+315 | M1.0 V | B | | 18:40:55.33 | +31:31:37.3 | … | … | 11.037 | 152.2 | 9.45 | |
| BD+31 3330C | | … | C | | 18:40:55.31 | +31:31:37.5 | … | … | 11.139 | 153.1 | 9.210 | |
| HD 187691 | | F8 V | A | | 19:51:01.91 | +10:24:54.4 | 51.313 ± 0.090 | 277.69 ± 0.11 | 4.976 | | | |
| GJ 9671 B | J19510+104 | M4.0 V | B | | 19:51:00.97 | +10:24:38.0 | 51.373 ± 0.040 | 288.38 ± 0.05 | 11.765 | 220.2 | 21.5 | |
| HD 191785 | | K0 V | A | | 20:11:05.61 | +16:11:23.2 | 48.926 ± 0.023 | 575.39 ± 0.03 | 7.113 | | | |
| GJ 783.2 B | J20112+161 | M4.0 V | B | | 20:11:12.80 | +16:11:14.4 | 48.928 ± 0.032 | 577.77 ± 0.04 | 12.614 | 94.8 | 103.843 | |
| StKM 1-1767a | | K5 V | A | | 20:13:11.14 | +02:56:20.5 | 27.359 ± 0.018 | 108.28 ± 0.02 | 9.821 | | | |
| [R78b] 440 | J20132+029 | M1.0 V | B | | 20:13:12.94 | +02:56:02.3 | 27.359 ± 0.023 | 110.73 ± 0.03 | 10.984 | 123.9 | 32.5 | |
| HD 197076 | | G5 V | A | | 20:40:45.28 | +19:56:12.9 | 47.746 ± 0.020 | 334.03 ± 0.02 | 6.285 | | | |
| GJ 797 B | J20407+199 | M2.5 V | BC | | 20:40:44.65 | +19:54:08.2 | … | 339.09 ± 1.98 | 11.394 | 268.9/184.1 | 0.268/125.0584 | |
| V447Lac | | K1 V | A | | 22:15:54.53 | +54:40:23.5 | 45.591 ± 0.017 | 223.87 ± 0.02 | 7.283 | | | |
| GJ 4269 | J22160+546 | M4.0 V | B | | 22:16:02.99 | +54:40:00.5 | 45.625 ± 0.020 | 221.35 ± 0.03 | 12.732 | 107.4 | 76.900 | |
| HD 216385 | | F6 V | A | | 22:52:24.64 | +09:50:09.1 | 36.558 ± 0.108 | 523.35 ± 0.20 | 5.022 | 19.6 | 249.637 | |
| GJ 9801 B | J22524+099 | M3.0 V | BC | | 22:52:30.31 | +09:54:04.9 | 37.250 ± 0.760 | 530.18 ± 11.31 | 12.411 | 19.6 | 250.246 | • |
| V368Cep | | G9 V | A | | 23:19:27.78 | +79:00:13.8 | 52.784 ± 0.014 | 215.92 ± 0.02 | 7.293 | | | |

Table D.6: Multiple systems containing M dwarfs and FGK primaries (continued).

| Name | Karmn | Spectral type | Component | Class | $\alpha$ | $\delta$ | $\pi$ [mas] | $\mu_{total}$ [mas $a^{-1}$] | $G$ [mag] | $\theta$ [deg] | $\rho$ [arcsec] | Notes[a] |
|---|---|---|---|---|---|---|---|---|---|---|---|---|
| HD 220140B | J23194+790 | M3.5 V | B | | 23:19:25.64 | +79:00:04.8 | 52.840 ± 0.021 | 218.08 ± 0.04 | 10.915 | 214.1 | 10.905 | |
| LP 12-90 | J23228+787 | M5.0 V | C | | 23:22:54.97 | +78:47:39.6 | 52.834 ± 0.030 | 217.44 ± 0.05 | 14.020 | 141.1 | 962.738 | |
| StKM 2-1787 | | K4 V | A | | 23:53:35.58 | +12:06:20.4 | 26.840 ± 0.020 | 120.69 ± 0.03 | 10.426 | | | |
| PM J23535+1206S | J23535+121 | M2.5 V | B | | 23:53:35.69 | +12:06:14.8 | 26.792 ± 0.035 | 119.44 ± 0.04 | 11.297 | 165.1 | 5.785 | |

[a] HD 16895: Candidate to unresolved binary; HD 21845B: Candidate to unresolved binary; HD 27848: Hyades moving group (Röser et al., 2011; Kopytova et al., 2016; Lodieu et al., 2019; Freund et al., 2020); V583 Aur: Candidate to unresolved binary; GJ 3348A: Proper motions from Zacharias et al. (2012); 1RXS J090406.8-155512: Desidera et al. (2021) recognises the pair but it is not tabulated in WDS; GJ 397.1 B: Proper motions from Zacharias et al. (2012); LZ UMa: Candidate to unresolved binary; StKM 1-1119: Proper motions from Zacharias et al. (2012); HD 126660: Candidate to unresolved binary; HD 131976: Parallax and proper motions from van Leeuwen (2007); GJ 570 D: Parallax and proper motions from Faherty et al. (2012); Weinberger et al. (2016); GJ 9801 B: Parallax and proper motions from Dawson & De Robertis (2005); Zacharias et al. (2012).

Table D.7: Multiple systems containing M dwarfs and white dwarfs.

| Name | Karmn | Spectral type | Component | Class | $\alpha$ | $\delta$ | $\pi$ [mas] | $\mu_{total}$ [mas a$^{-1}$] | $G$ [mag] | $\theta$ [deg] | $\rho$ [arcsec] | Notes[a] |
|---|---|---|---|---|---|---|---|---|---|---|---|---|
| GJ 1015 A | J00413+558 | M4.2 V | A | A+B | 00:41:21.44 | +55:50:03.2 | 43.735 ± 0.022 | 332.50 ± 0.02 | 12.732 | | | |
| GJ 1015 B | | DBQ5 | B | | 00:41:22.64 | +55:50:07.2 | 43.659 ± 0.022 | 324.16 ± 0.03 | 13.998 | 68.4 | 10.854 | |
| GJ 3117 | J01518+644 | M2.5 V | A | AB | 01:51:51.72 | +64:26:02.8 | 57.825 ± 0.016 | 306.37 ± 0.02 | 10.409 | | | |
| GJ 3118 | | DA5.6 | B | | 01:51:51.69 | +64:25:49.3 | 57.799 ± 0.016 | 301.94 ± 0.02 | 13.947 | 180.7 | 13.503 | |
| Ross 25 | J04011+513 | M3.8 V | A | A+B | 04:01:08.18 | +51:23:06.4 | 39.816 ± 0.021 | 883.46 ± 0.03 | 12.435 | | | |
| LSPM J0401+5131 | | DC8 | B* | | 04:01:02.14 | +51:31:17.2 | 39.836 ± 0.077 | 883.24 ± 0.12 | 17.113 | 353.4 | 494.0 | |
| HD 26965 | | K0 V | A | A+B+C | 04:15:13.91 | −07:40:05.1 | 199.608 ± 0.121 | 4089.84 ± 0.16 | 4.180 | | | |
| HD 26976 | | DA2.9 | B | | 04:15:19.39 | −07:40:22.6 | 199.691 ± 0.051 | 4018.59 ± 0.06 | 9.542 | 102.2 | 83.337 | |
| DYEri | J04153-076 | M4.5 V | C | | 04:15:19.12 | −07:40:15.3 | 199.452 ± 0.069 | 4083.72 ± 0.08 | 9.775 97.5 | 78.097 | | |
| GJ 169.1 A | J04311+589 | M4.0 V | A | A+B | 04:31:14.21 | +58:58:04.7 | 181.244 ± 0.050 | 2424.35 ± 0.07 | 9.703 | | | |
| GJ 169.1 B | | DC5 | B | | 04:31:15.33 | +58:58:10.1 | 181.273 ± 0.020 | 2361.13 ± 0.03 | 12.336 | 58.2 | 10.263 | |
| GJ 3430 | J07102+376 | M4.0 V | A | A+B | 07:10:13.32 | +37:40:05.9 | 44.601 ± 0.437 | 350.22 ± 0.61 | 13.232 | | | |
| GJ 3431 | | DQ8 | B | | 07:10:14.04 | +37:40:15.0 | 41.029 ± 0.036 | 357.87 ± 0.05 | 15.512 | 43.1 | 12.5 | |
| GJ 275.2 A | J07307+481 | M4.0 V | AB | (AB)+CD | 07:30:42.46 | +48:11:38.2 | 88.723 ± 0.030 | 1287.99 ± 0.03 | 12.116 | | | • |
| GJ 275.2 B | | DA | C | | 07:30:46.99 | +48:10:05.7 | 83.484 | 1281.73 ± 3.61 | 15.065 | 153.9 | 103.053 | • |
| G 107-70B | | DA | D | | 07:30:46.96 | +48:10:06.3 | | | 15.252 | 318.1/153.9 | 0.667/102.38 | |
| GJ 283 A | | DZQA6 | A | A+B | 07:40:22.06 | −17:24:57.8 | 109.344 ± 0.018 | 1261.34 ± 0.02 | 12.970 | | | |
| GJ 283 B | J07403-174 | M6.5 Ve | B | | 07:40:20.66 | −17:24:54.5 | 109.254 ± 0.039 | 1270.98 ± 0.05 | 13.957 | 99.5 | 20.338 | |
| GJ 401 | J10456-191 | M0.5 V | A | AB | 10:45:36.98 | −19:07:01.3 | 53.172 ± 0.022 | 1963.77 ± 0.03 | 10.291 | | | |
| GJ 401 B | | DQ | B | | 10:45:36.93 | −19:06:54.7 | 53.190 ± 0.034 | 1962.56 ± 0.06 | 15.441 | 354.5 | 6.655 | |
| GJ 1142 A | J11081-052 | M3.0 V | A | A+B | 11:08:06.48 | −05:13:54.2 | 40.180 ± 0.026 | 443.17 ± 0.03 | 11.472 | | | |
| GJ 1142 B | | DA3 | B | | 11:07:59.89 | −05:09:33.1 | 40.293 ± 0.032 | 446.11 ± 0.04 | 13.091 | 339.3 | 278.996 | |
| GJ 1155 A | J12168+029 | M3.0 V | A | AB | 12:16:51.16 | +02:58:09.0 | 42.820 ± 0.035 | 700.55 ± 0.05 | 12.080 | | | |
| GJ 1155 B | | DA | B | | 12:16:51.17 | +02:58:06.9 | 42.773 ± 0.043 | 711.07 ± 0.06 | 15.330 | 177.9 | 2.126 | |
| Wolf 485 | | DA3.5 | A | A+B | 13:30:12.44 | −08:34:37.0 | 62.148 ± 0.044 | 1207.51 ± 0.06 | 12.355 | | | |
| Ross 476 | J13300-087 | M4.0 V | B | | 13:30:01.59 | −08:42:33.0 | 62.281 ± 0.042 | 1210.61 ± 0.06 | 12.706 | 198.7 | 502.447 | |
| GJ 1179 A | J13482+236 | M5.0 V | A | A+B | 13:48:11.69 | +23:36:50.7 | 84.225 ± 0.027 | 1487.63 ± 0.04 | 13.442 | | | |
| GJ 1179 B | | DC9 | B | | 13:48:01.28 | +23:34:48.4 | 84.311 ± 0.029 | 1490.91 ± 0.04 | 15.338 | 229.5 | 188.131 | |
| CMDra | J16343+571 | M4.5 V | Aab | Aab+B | 16:34:18.14 | +57:10:03.3 | 67.288 ± 0.034 | 1623.35 ± 0.06 | 11.491 | | | • |

Table D.7: Multiple systems containing M dwarfs and white dwarfs (continued).

| Name | Karmn | Spectral type | Component | Class | $\alpha$ | $\delta$ | $\pi$ [mas] | $\mu_{\rm total}$ [mas a$^{-1}$] | $G$ [mag] | $\theta$ [deg] | $\rho$ [arcsec] | Notes[a] |
|---|---|---|---|---|---|---|---|---|---|---|---|---|
| GJ 630.1 B | | DQ8 | B | | 16:34:19.37 | +57:10:28.1 | 67.354 ± 0.021 | 1634.85 ± 0.04 | 14.849 | 21.9 | 26.726 | |
| LP 387-37 | J17058+260 | M1.5 V | A | AB | 17:05:52.55 | +26:05:27.4 | 28.746 ± 0.019 | 291.03 ± 0.03 | 11.284 | | | |
| LP 387-36 | | DC7 | B | | 17:05:52.54 | +26:05:46.7 | 28.841 ± 0.054 | 285.38 ± 0.07 | 17.034 | 359.6 | 19.24 | |
| GJ 4059 | J18264+113 | M3.5 V | A | A+B | 18:26:24.58 | +11:20:52.9 | 37.056 ± 0.024 | 276.07 ± 0.03 | 11.728 | | | |
| LSPM J1826+1120S | | DA | B | | 18:26:24.42 | +11:20:45.1 | 37.055 ± 0.066 | 281.07 ± 0.09 | 16.962 | 196.8 | 8.196 | • |
| G 229-20A | J18576+535 | M3.5 V | A | AB+C | 18:57:38.42 | +53:31:14.4 | 40.348 ± 0.015 | 261.44 ± 0.03 | 12.019 | | | |
| LP 141-13 | | M3.0 V | B | | 18:57:38.34 | +53:31:12.2 | 40.365 ± 0.015 | 245.56 ± 0.03 | 12.103 | 197.0 | 2.290 | |
| LP 141-14 | | DC | C | | 18:57:39.78 | +53:30:32.5 | 40.393 ± 0.051 | 246.40 ± 0.10 | 16.942 | 163.9 | 43.653 | |
| GJ 754.1 B | J19205-076 | M2.5 V | B | A+B | 19:20:33.38 | −07:39:46.6 | 95.178 ± 0.031 | 191.46 ± 0.04 | 10.972 | | | |
| GJ 754.1 A | | DBQA5 | A | | 19:20:34.86 | −07:40:02.7 | 95.176 ± 0.029 | 173.04 ± 0.04 | 12.253 | 126.2 | 27.168 | |
| GJ 784.2 A | J20139+066 | M3.3 V | A | A+B | 20:13:58.71 | +06:41:06.8 | 43.570 ± 0.022 | 634.75 ± 0.03 | 11.951 | | | |
| V1412Aql | | DC7 | B | | 20:13:55.41 | +06:42:35.5 | 43.574 ± 0.038 | 635.70 ± 0.05 | 15.664 | 330.9 | 101.5 | |
| FRAqr | J20568-048 | M4.0 V | Aab | Aab+B | 20:56:49.39 | −04:50:52.6 | 61.824 ± 0.074 | 826.32 ± 0.09 | 10.660 | | | |
| Ross 193B | | DC10 | B | | 20:56:48.62 | −04:50:43.1 | 61.760 ± 0.052 | 807.72 ± 0.07 | 16.347 | 309.4 | 15.006 | |
| TYC 3980-1081-1 | J21516+592 | M4.0 V | A | A+B | 21:51:38.15 | +59:17:40.0 | 123.057 ± 0.594 | 79.89 ± 1.67 | 9.409 | | | |
| UCAC4 747-070768 | | DAH | B | | 21:51:39.93 | +59:17:34.5 | 118.155 ± 0.016 | 88.85 ± 0.03 | 14.373 | 111.9 | 14.642 | |
| LF 4 +54 152 | J22129+550 | M0.0 V | A | AB | 22:12:56.63 | +55:04:50.8 | 18.385 ± 0.153 | 109.13 ± 0.25 | 10.222 | | | |
| G3-2005884249925303168 | | M6.5 V | B* | | 22:13:00.46 | +55:05:48.3 | 18.550 ± 0.079 | 108.97 ± 0.12 | 17.702 | 29.8 | 66.265 | • |
| GJ 4304 | J22559+057 | M1.0 V | A | A+B | 22:55:57.20 | +05:45:14.0 | 40.347 ± 0.024 | 446.04 ± 0.03 | 10.415 | | | |
| GJ 4305 | | DA8.1 | B | | 22:55:56.09 | +05:45:18.4 | 40.325 ± 0.072 | 439.47 ± 0.07 | 16.041 | 284.6 | 17.2 | |
| G 233-42 | J23089+551 | M5.0 V | A | A+B | 23:09:58.62 | +55:06:48.1 | 60.921 ± 0.027 | 411.27 ± 0.04 | 13.986 | | | |
| LSPM J2309+5506E | | DA | B | | 23:09:59.29 | +55:06:50.2 | 60.895 ± 0.030 | 410.18 ± 0.04 | 15.604 | 69.6 | 6.150 | |
| GJ 4356 | J23389+210 | M4.3 V | A | A+B | 23:38:56.00 | +21:01:24.5 | 25.402 ± 0.028 | 328.86 ± 0.04 | 12.907 | | | |
| GJ 4357 | | DA | B | | 23:38:56.63 | +21:01:21.1 | 25.367 ± 0.100 | 325.25 ± 0.13 | 17.683 | 110.7 | 9.5 | |
| GJ 905.2 A | J23438+325 | M1.5 V | AB | (AB)+C | 23:43:52.84 | +32:35:37.8 | | 223.77 ± 13.69 | 10.611 | 308.00 | 0.109 | |
| GJ 905.2 B | | DA3.8 | C | | 23:43:50.45 | +32:32:45.8 | 53.762 ± 0.027 | 224.05 ± 0.04 | 12.967 | 190.0 | 174.652 | |

[a] GJ 275.2 A: Quadruple system, with (AB) separated ~ 0.054 arcsec (Harrington et al., 1981b); GJ 275.2 B: Parallax from Khrutskaya et al. (2010), proper motions from Monet et al. (2003); LSPM J1826+1120S: Candidate to unresolved double white dwarf based on position in HR diagram; GJ 754.1 B: White dwarf candidate from position in HR diagram (see also Jiménez-Esteban et al., 2018).

Table D.8: Stars with detected planets in our sample.

| Karmn | Name | System class | Component | Number of planets[a] | Discovery reference[b] | Flag[c] |
|---|---|---|---|---|---|---|
| J00067-075 | GJ 1002 | Single | - | 2 | Sua23 | 1 |
| J00183+440 | HD 1326 | Binary | A | 2 | How14, Pin18 | 13 |
| J00449-152 | GJ 3053 | Single | - | 2 | Dit17, Men19 | … |
| J01023-104 | GJ 3072 | Single | - | 1 | Fen20 | … |
| J01026+623 | Wolf 46 | Binary | A | 1 | Per19 | 1 |
| J01066+192 | LSPM J0106+1913 | Single | - | 2 | Cha22 | 2 |
| J01125-169 | YZ Cet | Single | - | 3 | Ast17b | 3 |
| J02002+130 | TZ Ari | Single | - | 1 | Qui22 | 1 |
| J02222+478 | GJ 96 | Single | - | 1 | Hob18 | … |
| J02489-145W | PM J02489-1432W | Binary | A | 1 | Kos21 | 2 |
| J02530+168 | Teegarden's Star | Single | - | 2 | Zec19 | 1 |
|  | HD 18143A | Triple | A | 2 | Fen22 | … |
| J02573+765 | LP 14-53 | Single | - | 1 | Sot21 | 2 |
| J03018-165S | GJ 3193 | Binary | A | 2 | Win19b, Win22 | … |
| J03133+047 | CD Cet | Single | - | 1 | Bau20 | 1 |
|  | HD 26965 | Triple | A | 1 | Dia18 | … |
| J04167-120 | LP 714-47 | Single | - | 1 | Dre20 | 2 |
| J04343+430 | PM J04343+4302 | Single | - | 1 | Blu21 | 2 |
|  | HD 29391 | Triple | A | 1 | Mac15 | … |
| J04429+189 | HD 285968 | Single | - | 1 | For09 | 13 |
| J04520+064 | Wolf 1539 | Single | - | 1 | How10 | … |
| J04538-177 | GJ 180 | Single | - | 3 | Tuo14 | … |
| J05019-069 | GJ 3323 | Single | - | 2 | Ast17a | … |
| J06105-218 | HD 42581 | Binary | AB | 2 | Tuo14, Fen20 | … |
| J06371+175 | HD 260655 | Single | - | 2 | Luq22 | 2 |
| J06548+332 | HD 265866 | Single | - | 1 | Sto20 | 1 |
| J07274+052 | Luyten's Star | Binary | AB | 2[d] | Ast17a | … |
| J07590+153 | GJ 3470 | Single | - | 1 | Bon12 | … |
| J08023+033 | GJ 3473 | Binary | A | 2 | Kem20 | 2 |
| J08409-234 | GJ 317 | Single | - | 2 | Joh07 | … |
| J08413+594 | GJ 3512 | Single | - | 2 | Mor19 | 1 |
|  | 55 Cnc | Binary | A | 5 | But97, Mar02, Mc04, Fis08 | … |
| J08551+015 | Ross 623 | Single | - | 1 | Rob13 | … |
| J08588+210 | G 41-13 | Single | - | 1 | Stef20 | … |
| J09144+526 | HD 79211 | Triple | B | 1 | Gon20 | 1 |
| J09286-121 | LP 727-31 | Single | - | 1 | Ree22 | … |
| J09360-216 | GJ 357 | Single | - | 3 | Luq19 | 12 |
| J09561+627 | GJ 373 | Binary | Aab (SB?, Rei12) | 1 | Fen20 | … |
| J10023+480 | GJ 378 | Single | - | 1 | Hob19 | … |
| J10088+692 | TYC 4384-1735-1 | Single | - | 1 | Blu20 | 2 |
| J10185-117 | LP 729-54 | Binary | A | 2 | Now20 | 2 |
| J10289+008 | Ross 446 | Single | - | 1 | Ama21 | 1 |
| J10564+070 | CN Leo | Single | - | 0[e] | Tuo19 | … |
| J11033+359 | HD 95735 | Single | - | 2 | Dia19, Ros21 | 3 |
| J11110+304E | HD 97101 | Binary | A | 2 | Ded21 | … |
| J11302+076 | K2-18 | Single | - | 2 | Mon15, Clo17 | 2 |
| J11417+427 | Ross 1003 | Single | - | 2 | Hag10, Tri18 | 13 |
| J11421+267 | Ross 905 | Single | - | 1 | But04 | 3 |
| J11477+008 | FI Vir | Single | - | 1 | Bon18 | … |
| J11509+483 | GJ 1151 | Single | - | 1[f] | Bla23 | 1 |
| J12123+544S | HD 238090 | Binary | A | 1 | Sto20 | 1 |
| J12388+116 | Wolf 433 | Single | - | 1 | Fen20 | … |

Table D.8: Planets with detected planets in our sample (continued).

| Karmn | Name | System class | Component | Number of planets[a] | Discovery reference[b] | Flag[c] |
|---|---|---|---|---|---|---|
| J12479+097 | Wolf 437 | Single | - | 1 | Tri21 | 1 |
| J13007+123 | Wolf 462 | Triple | AB | 1[g] | Bur10 | ... |
| J13119+658 | PM J13119+6550 | Single | - | 2 | Dem20 | ... |
| | HD 115404 | Binary | A | 2 | Fen22 | ... |
| J13229+244 | Ross 1020 | Single | - | 1 | Luq18 | 1 |
| J13255+688 | 2M J13253177+6850106 | Single | - | 2 | Gon22 | 1 |
| J13299+102 | Ross 490 | Single | - | 1 | Dam22 | 1 |
| J14010-026 | HD 122303 | Single | - | 1 | Sua17a | 3 |
| J15194-077 | HO Lib | Single | - | 3 | Bon05, May09, Udr07 | 3 |
| J15238+174 | Ross 508 | Single | - | 1 | Har22 | ... |
| J15583+354 | GJ 3929 | Single | - | 2 | Kem22, Bea22 | 1 |
| J16090+529 | GJ 3942 | Single | - | 1 | Per17 | ... |
| J16102-193 | K2-33 | Single | - | 1 | Dav16 | ... |
| J16126-188 | LP 804-27 | Single | - | 1 | App10 | ... |
| J16167+672S | HD 147379 | Binary | A | 1 | Rei18a | 1 |
| J16254+543 | GJ 625 | Single | - | 1 | SM17b | ... |
| J16303-126 | V2306 Oph | Single | - | 3 | Wri16 | ... |
| | V1090 Her | Triple | A | 3 | Fen22 | ... |
| J16581+257 | Ross 860 | Single | - | 1 | Joh10 | ... |
| J17153+049 | GJ 1214 | Single | - | 1 | Cha09 | ... |
| J17160+110 | GJ 3998 | Single | - | 2 | Aff16 | ... |
| J17355+616 | GJ 685 | Triple | C | 1 | Pin19 | ... |
| J17364+683 | GJ 687 | Single | - | 2 | Lur14 | ... |
| J17378+185 | GJ 686 | Single | - | 1 | Aff19 | 3 |
| J17578+046 | Barnard's Star | Single | - | 0[h] | Rib18 (Lub21) | ... |
| J18353+457 | GJ 720 A | Binary | A | 1 | Gon21 | ... |
| | LP 141-14 | Triple | C | 1 | Van20 | ... |
| J18580+059 | HD 176029 | Single | - | 1 | Tol21 | 1 |
| J19169+051N | V1428 Aql | Binary | A | 1 | Kam18 | 1 |
| J19206+731S | 2M J19204172+7311434 | Binary* | A | 1 | Cad22 | ... |
| | HD 190360 | Binary | A | 2 | Nae03, Vog05 | ... |
| J20138+133 | Ross 754 | Single | - | 1 | Mal21 | ... |
| J20451-313 | AU Mic | SKG | C | 2 | Pla20, Mar21 | 2 |
| J21164+025 | LSPM J2116+0234 | Single | - | 1 | Lal19 | 1 |
| J21221+229 | TYC 2187-512-1 | Single | - | 1 | Qui22 | 1 |
| J21466+668 | G 264-12 | Single | - | 2 | Ama21 | 1 |
| J21474+627 | TYC 4266-736-1 | Binary | A | 1 | Esp22 | 2 |
| J22096-046 | Wolf 1329 | Single | - | 2 | But06, Mon14 | ... |
| J22102+587 | UCAC4 744-073158 | Single | - | 1 | Fuk22 | ... |
| J22137-176 | GJ 1265 | Single | - | 1 | Luq18 | 1 |
| J22252+594 | GJ 4276 | Single | - | 1[i] | Nag19 | 1 |
| J22532-142 | IL Aqr | Single | - | 4[j] | Mar98/Del98, Mar01, Riv05, Riv10 | 3 |
| J23064-050 | 2MUCD 12171 | Single | - | 7[k] | Gil16, Gil17 | ... |
| J23318+199E | EQ Peg A | Quadruple | Aab (SB2, Del99b) | 1 | Cur22 | ... |

[a] In the majority of cases these planets are classified as 'Confirmed' (to the date of publication of this work) by the NASA Exoplanet Archive. A few notable stars have been included, despite of the controversial nature of the detections (e.g. Barnard's Star or CN Leo).

[b] References: Aff16: Affer et al. (2016); Aff19: Affer et al. (2019); Ama21: Amado et al. (2021); App10: Apps et al. (2010); Ast17a: Astudillo-Defru et al. (2017b); Ast17b: Astudillo-Defru et al. (2017a); Bau20: Bauer et al. (2020); Bea22: Beard et al. (2022); Bla23: Blanco-Pozo et al. (2023); Blu21: Bluhm et al. (2021); Bon05: Bonfils

et al. (2005a); Bon12: Bonfils et al. (2012); Bon18: Bonfils et al. (2018b); Burg10: Burgasser et al. (2010); But97: Butler et al. (1997); But04: Butler et al. (2004); But06: Butler et al. (2006); Cad22: Cadieux et al. (2022); Cha09: Charbonneau et al. (2009); Cha22: Chaturvedi et al. (2022); Clou17: Cloutier et al. (2017); Cur22: Curiel et al. (2022); Dam22: Damasso et al. (2022); Davi16: David et al. (2016); Ded21: Dedrick et al. (2021); Dem20: Demory et al. (2020); Dia18: Díaz et al. (2018); Dia19: Díaz et al. (2019); Dit17: Dittmann et al. (2017); Dre20: Dreizler et al. (2020); Esp22: Espinoza et al. (2022); Fen20: Feng et al. (2020); Fen22: Feng et al. (2022); Fis08: Fischer et al. (2008); For09: Forveille et al. (2009); Fuk22: Fukui et al. (2022); Gon20: González-Álvarez et al. (2020); Gon21: González-Álvarez et al. (2021); Gon22: González-Álvarez et al. (2022); Gil16: Gillon et al. (2016); Gil17: Gillon et al. (2017); Hag10: Haghighipour et al. (2010); Har22: Harakawa et al. (2022); Hob18: Hobson et al. (2018); Hob19: Hobson et al. (2019); How10: Howard et al. (2010); How14: Howell et al. (2014); Joh07: Johnson et al. (2007); Joh10: Johnson et al. (2010c); Kam18: Kaminski et al. (2018); Kem20: Kemmer et al. (2020); Kem22: Kemmer et al. (2022); Kos21: Kossakowski et al. (2021); Lal19: Lalitha et al. (2019); Lub21: Lubin et al. (2021); Luq18: Luque et al. (2018); Luq19: Luque et al. (2019b); Luq22: Luque et al. (2022b); Mac15: Macintosh et al. (2015); Mal21: Maldonado et al. (2021); Mar01: Marcy et al. (2001); Mar02: Marcy et al. (2002); Mar98: Marcy et al. (1998); Mar21: Martioli et al. (2021); May09: Mayor et al. (2009); Mc04: McArthur et al. (2004); Men19: Ment et al. (2019); Mon14: Montet et al. (2014); Mon15: Montet et al. (2015); Mor19: Morales et al. (2019); Nae03: Naef et al. (2003); Nag19: Nagel et al. (2019); Now20: Nowak et al. (2020); Per17: Perger et al. (2017); Per19: Perger et al. (2019); Pin18: Pinamonti et al. (2018); Pin19: Pinamonti et al. (2019); Pla20: Plavchan et al. (2020); Qui22: Quirrenbach et al. (2022); Ree22: Reefe et al. (2022); Rei12: Reiners et al. (2012); Rei18a: Reiners et al. (2018a); Rib18: Ribas et al. (2018); Riv05: Rivera et al. (2005); Riv10: Rivera et al. (2010); Rob13: Astropy Collaboration et al. (2013); Ros21: Rosenthal et al. (2021); Sua17b: Suárez Mascareño et al. (2017b); Sot21: Soto et al. (2021); Ste20: Stefansson et al. (2020); Sto20: Stock et al. (2020); Sua17a: Suárez Mascareño et al. (2017a); Sua23: Suárez Mascareño et al. (2023); Tol21: Toledo-Padrón et al. (2021); Tri18: Trifonov et al. (2018); Tri21: Trifonov et al. (2021); Tuo14: Tuomi et al. (2014); Tuo19: Tuomi et al. (2019); Udr07: Udry & Santos (2007); Van20: Vanderburg et al. (2020); Vog05: Vogt et al. (2005); Win19b: Winters et al. (2019b); Win22: Winters et al. (2022); Wri16: Wright et al. (2016); Zec19: Zechmeister et al. (2019).

[c] 1: Planet discovered by CARMENES; 2: Transiting planet confirmed with CARMENES follow-up observations; 3: Reanalysis with CARMENES data.

[d] Planets d and e are unconfirmed.

[e] Planets b and c are unconfirmed/controversial.

[f] Planet b (Mahadevan et al., 2021) is a false positive planet.

[g] Planet detected in Ross 458B.

[h] Unconfirmed/controversial planet.

[i] Planet c is unconfirmed.

[j] Delfosse et al. (1998a) and Marcy et al. (1998) simultaneously detected this Jovian-mass planet (published in *Astronomy & Astrophysics* and *Astrophysical Journal Letters*, respectively).

[k] Planet i is unconfirmed.

Table D.9: Eclipsing binaries in our sample.

| Karmn | Name | $P$ [d] | $\mathcal{M}_1$ [$\mathcal{M}_\odot$] | $\mathcal{M}_2$ [$\mathcal{M}_\odot$] | $\mathcal{R}_1$ [$\mathcal{R}_\odot$] | $\mathcal{R}_2$ [$\mathcal{R}_\odot$] | $a$ [au] | Reference[a] |
|---|---|---|---|---|---|---|---|---|
| J03372+691 | GJ 3236 | 0.7712600 ± 0.0000023 | 0.376 ± 0.017 | 0.281 ± 0.015 | 0.3828 ± 0.0072 | 0.2992 ± 0.0075 | 0.01430 | Irw09, Shk10 |
| J03375+178S | GJ 3240 | < 0.4 | ... | ... | ... | ... | 0.01 | Shk10 |
| J07346+318 | Castor C | 0.8142822 ± 0.0000010 | 0.5992 ± 0.0047 | 0.5992 ± 0.0047 | 0.6191 ± 0.0057 | 0.6191 ± 0.0057 | 0.01819 | Joy26, Giz02, Tor22 |
| J08316+193S | CU Cnc | 2.771468 ± 0.000004 | 0.4349 ± 0.0012 | 0.39922 ± 0.00089 | 0.4323 ± 0.0055 | 0.3916 ± 0.0094 | 0.03624 | Del99a, Del99b, Tor10 |
| J09193+620 | GJ 3547 | 20.2 | ... | ... | ... | ... | 0.16 | Shk10, Skr21 |
| J15474+451 | LP 177-102 | 3.5500184 0.0000018 | 0.2576 ± 0.0085 | 0.2585 ± 0.0080 | 0.2895 ± 0.0068 | 0.2895 ± 0.0068 | 0.03654 | Moc02, Har11, Bir12 |
| J16343+571 | CM Dra | 1.268 | 0.23102 ± 0.00089 | 0.21409 ± 0.00083 | 0.2534 ± 0.0019 | 0.2398 ± 0.0018 | 0.01750 | Mor09, Tor10, Scha19 |
| J21296+176 | Ross 775 | 53.221 ± 0.004 | 0.114 ± 0.001 | 0.114 ± 0.001 | ... | ... | ... | Mar87, Del99b |

[a] Bir12 : Birkby et al. (2012); Del99a : Delfosse et al. (1999a); Del99b : Delfosse et al. (1999b); Giz02 : Gizis et al. (2002); Har11 : Hartman et al. (2011); Irw09 : Irwin et al. (2009); Jef18 : Jeffers et al. (2018); Joy26 : Joy & Sanford (1926); Mar87 : Marcy et al. (1987); Moc02 : Mochnacki et al. (2002); Mor09 : Morales et al. (2009); Scha19 : Schanche et al. (2019); Shk10 : Shkolnik et al. (2010); Shk10 : Shkolnik et al. (2010); Skr21 : Skrzypinski (2021); Tor10 : Torres et al. (2010); Tor22 : Torres et al. (2022).

## D.10 Descriptive charts of multiple systems

# WDS 00212-4246 (KO 1)

**A:** LEHPM 494, Königstuhl 1A
**B:** 2MASS J00210589-4244433, Königstuhl 1B

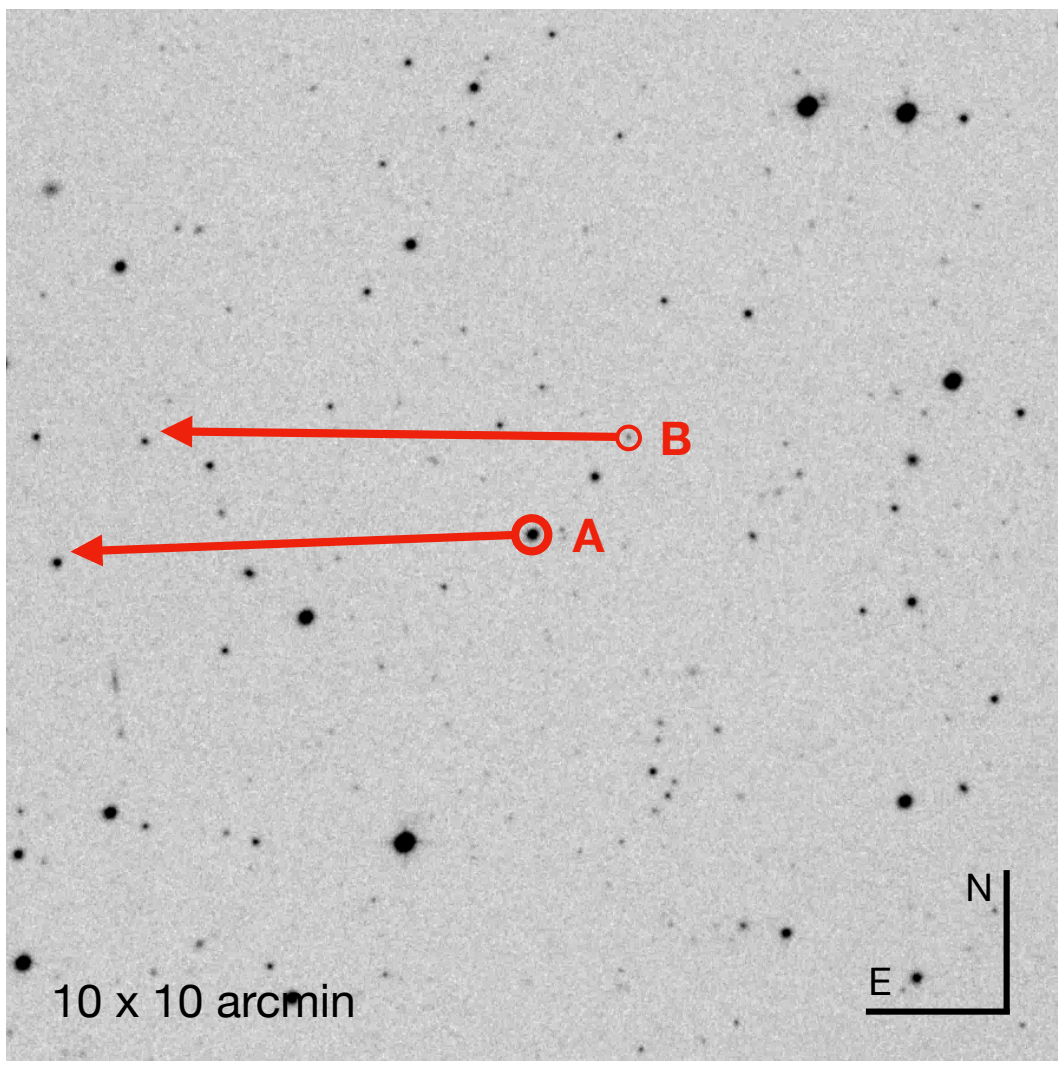

| | A-B | |
|---|---|---|
| WDS | KO 1 | |
| $\rho$ | 77.77 | arcsec |
| $\theta$ | 316.9 | deg |
| $\mu$ ratio | 0.050 | |
| $\Delta PA$ | 2.8 | deg |
| $\Delta d/d$ | 0.016 | |
| $d$ | 26.79 | pc |
| $s$ | 2083 | au |
| $P_{\rm orb}$ | 288 | $10^3$ a |
| $-U_g^*$ | 7.33 | $10^{33}$ J |

| Component | A | B | |
|---|---|---|---|
| SpT | m5.5 V | L0.6: V | |
| $\alpha$ | 00:21:11.11 | 00:21:06.29 | |
| $\delta$ | –42:45:40.4 | –42:44:43.5 | |
| $\pi$ | 37.332 ± 0.038 | 37.93 ± 0.40 | mas |
| $\mu_\alpha \cos\delta$ | 255.184 ± 0.031 | 257.82 ± 0.37 | mas a$^{-1}$ |
| $\mu_\delta$ | −12.475 ± 0.039 | −0.03 ± 0.39 | mas a$^{-1}$ |
| $\gamma$ | ... | +2 ± 1[a] | km s$^{-1}$ |
| $G$ | 15.3454 ± 0.0028 | 18.4345 ± 0.0062 | mag |
| $J$ | 12.001 ± 0.022 | 13.521 ± 0.025 | mag |
| $\mathcal{L}$ | 15.47 ± 0.17 | 2.95 ± 0.11 | $10^{-4}\mathcal{L}_\odot$ |
| $T_{\rm eff}$ | 2900 ± 50 | 2200 ± 25 | K |
| $\mathcal{M}$ | 0.109 | 0.0793 | $\mathcal{M}_\odot$ |
| ruwe | 1.074 | 2.540 | |
| Qflag 2MASS | AAA | AAA | |
| Qflag AllWISE | AAAU | AAAU | |

[a] Mohanty & Basri (2003)

WDS 00212-426 is a known physical pair of ultra-cool dwarfs (mid-M and early-L). The updated physical separation of 2083 au is wider than previously reported. This is the least massive pair in our sample, with a total mass of 0.19$\mathcal{M}_\odot$. With a separation of 1.3 arcmin, it was dubbed as 'the widest ultracool binary' by Caballero (2007). The authors also noted that confirming the binarity of the B component would help to support the hypothesis that wide triples are more prevalent than wide binaries. *Gaia* DR3 `ruwe` value indicates, in fact, a probable multiplicity of the B component.

# WDS 01568+3033 (KO 4)

**A:** NLTT 6496, Königstuhl 4A, Karmn J01567+305
**B:** NLTT 6491, Königstuhl 4B

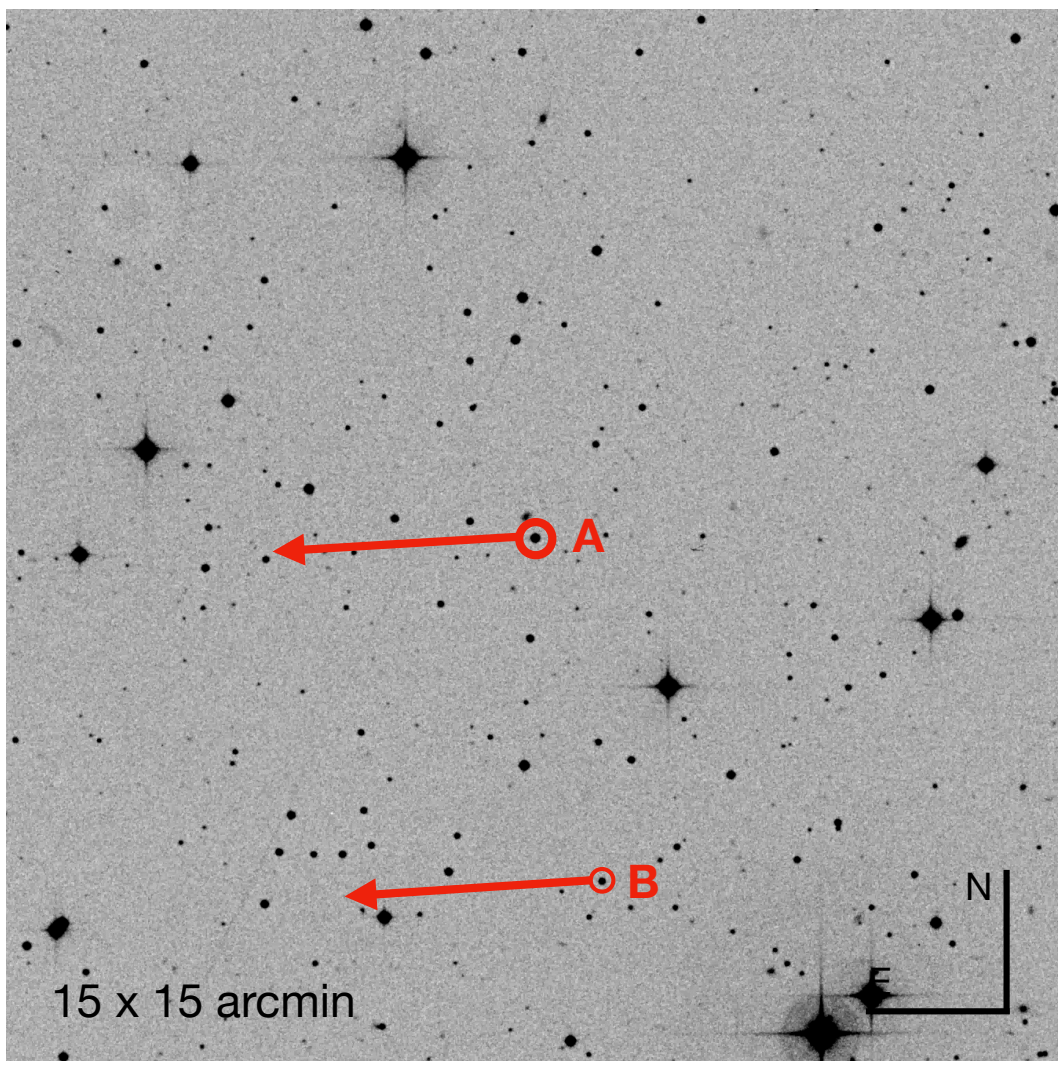

| | A-B | |
|---|---|---|
| WDS | KO 4 | |
| $\rho$ | 299.1 | arcsec |
| $\theta$ | 190.6 | deg |
| $\mu$ ratio | 0.012 | |
| $\Delta PA$ | 0.027 | deg |
| $\Delta d/d$ | 0.19 | |
| $d$ | 31.40 | pc |
| $s$ | 9392 | au |
| $P_{\rm orb}$ | 1699 | $10^3$ a |
| $-U_g^*$ | 8.27 | $10^{33}$ J |

| Component | A | B | |
|---|---|---|---|
| SpT | M4.5 V | m5 V | |
| $\alpha$ | 01:56:45.99 | 01:56:41.74 | |
| $\delta$ | +30:33:28.6 | +30:28:34.6 | |
| $\pi$ | 31.850 ± 0.076 | 26.645 ± 0.044 | mas |
| $\mu_\alpha \cos\delta$ | 210.625 ± 0.092 | 208.183 ± 0.042 | mas a$^{-1}$ |
| $\mu_\delta$ | −13.523 ± 0.090 | −13.464 ± 0.044 | mas a$^{-1}$ |
| $\gamma$ | ... | ... | km s$^{-1}$ |
| $G$ | 13.5151 ± 0.0029 | 15.1596 ± 0.0028 | mag |
| $J$ | 10.323 ± 0.023 | 11.917 ± 0.023 | mag |
| $\mathcal{L}$ | 98.6 ± 1.4 | 31.72 ± 0.46 | $10^{-4} \mathcal{L}_\odot$ |
| $T_{\rm eff}$ | 3100 ± 50 | 3100 ± 50 | K |
| $\mathcal{M}$ | 0.287 | 0.153 | $\mathcal{M}_\odot$ |
| ruwe | 4.582 | 1.109 | |
| Qflag 2MASS | AAA | AAA | |
| Qflag AllWISE | AAAB | AAAU | |

WDS 01568+3033 (Königstuhl 4A and 4B) is a wide multiple system of two intermediate M dwarfs, and one of the least bound systems found (Caballero et al., 2012). *Gaia* DR3 enlarges the distance given by the authors by more than 60%. While the pair complies with the proper motion criteria for physical parity, it exists a notable dissimilarity in distances, accompanied by an indication of poor astrometric quality. This high ruwe value of the primary is probably due to unresolved binarity, which affects the *Gaia* parallax determination.

# WDS 04309-0849 (KO 2)

## A: LP 655-23, Königstuhl 2A, Karmn J04308-088

## B: DENIS J043051.5-084900, Königstuhl 2B

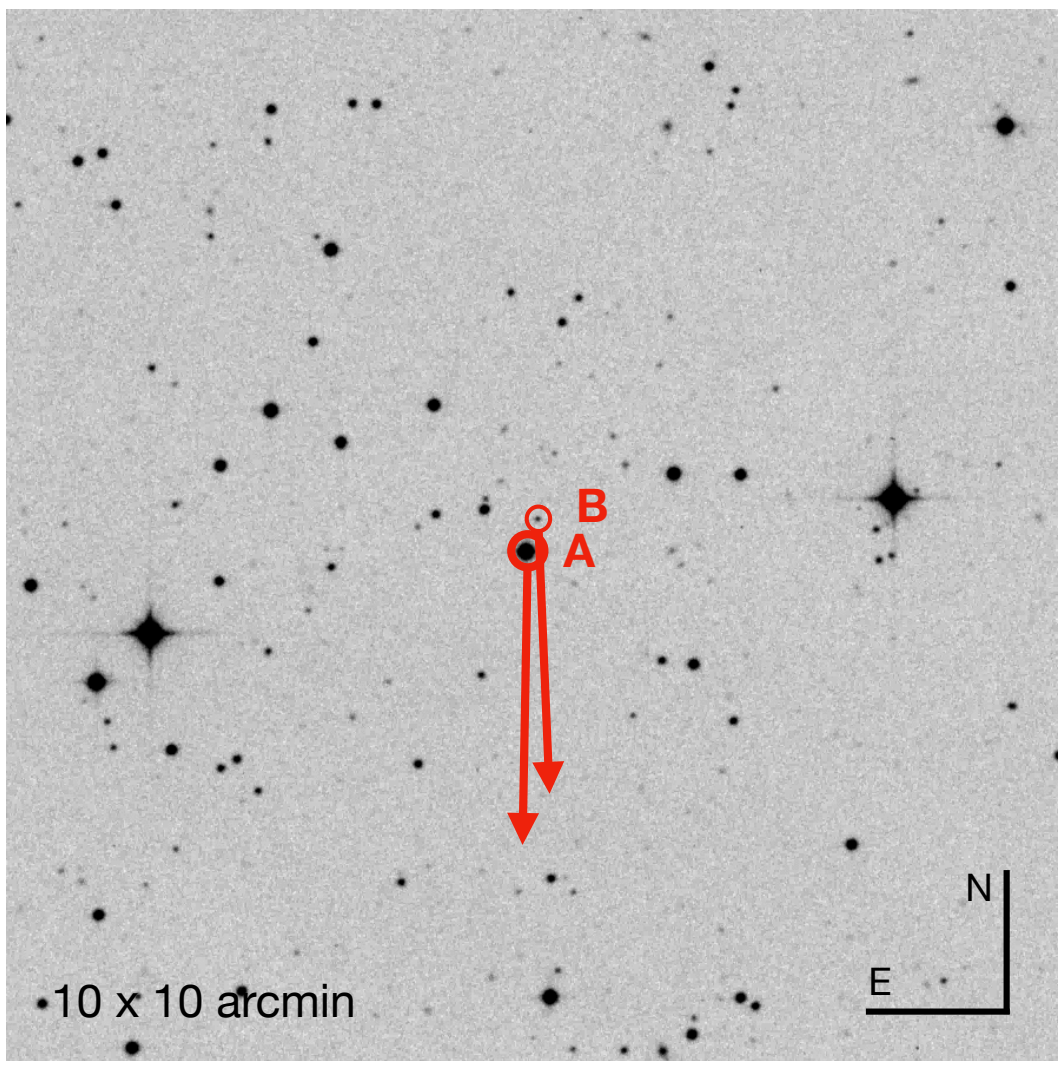

| | A-B | |
|---|---|---|
| WDS | KO 2 | |
| $\rho$ | 19.81 | arcsec |
| $\theta$ | 339.8 | deg |
| $\mu$ ratio | 0.053 | |
| $\Delta PA$ | 2.6 | deg |
| $\Delta d/d$ | 0.038 | |
| $d$ | 30.15 | pc |
| $s$ | 597 | au |
| $P_{orb}$ | 25.1 | $10^3$ a |
| $-U_g^*$ | 93.8 | $10^{33}$ J |

| Component | A | B | |
|---|---|---|---|
| SpT | M4.0 V | M8 V | |
| $\alpha$ | 04:30:52.04 | 04:30:51.58 | |
| $\delta$ | –08:49:22.0 | –08:49:03.5 | |
| $\pi$ | 33.165 ± 0.061 | 31.95 ± 0.24 | mas |
| $\mu_\alpha \cos\delta$ | 3.286 ± 0.069 | −3.97 ± 0.23 | mas a$^{-1}$ |
| $\mu_\delta$ | −161.794 ± 0.053 | −157.29 ± 0.19 | mas a$^{-1}$ |
| $\gamma$ | +1.7 ± 3.5[a] | … | km s$^{-1}$ |
| $G$ | 12.8742 ± 0.0028 | 17.4264 ± 0.0040 | mag |
| $J$ | 9.853 ± 0.024 | 12.897 ± 0.022 | mag |
| $\mathcal{L}$ | 142.2 ± 1.9 | 7.10 ± 0.19 | $10^{-4} \mathcal{L}_\odot$ |
| $T_{eff}$ | 3200 ± 50 | 2300 ± 25 | K |
| $\mathcal{M}$ | 0.339 | 0.0936 | $\mathcal{M}_\odot$ |
| ruwe | 3.370 | 1.687 | |
| Qflag 2MASS | AAA | AAA | |
| Qflag AllWISE | AAAC | AAAU | |

[a] Terrien et al. (2015)

WDS 04309-0849 is a known physical pair of spectroscopically characterised ultra-cool dwarfs (mid- and late-M). The separation of almost 20 arcsec reported in *Gaia* DR3 increases the projected separation calculated by Caballero et al. 2007. Their ruwe values suggest a possible binarity, at least for the brightest component.

# WDS 06104+2234 (LAW 14 + *New*)

**A:** 2MASS J06101775+2234199, Karmn J06102+225
**BaBb:** LP 362-121, Karmn J06103+225

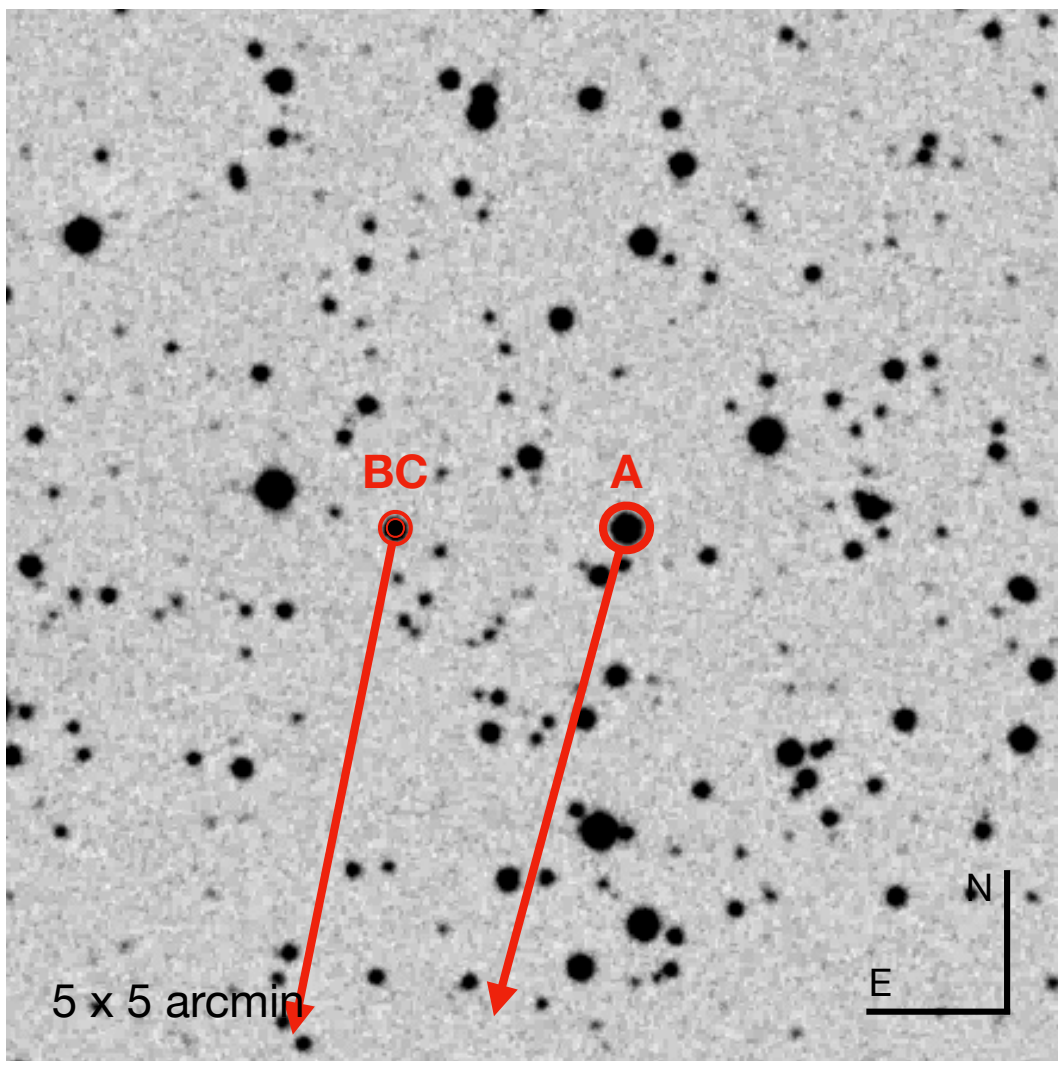

| | A-BaBb | |
|---|---|---|
| WDS | ... | |
| $\rho$ | 65.16 | arcsec |
| $\theta$ | 89.2 | deg |
| $\mu$ ratio | 0.060 | |
| $\Delta PA$ | 1.90 | deg |
| $\Delta d/d$ | 0.20 | |
| $d$ | 28.65 | pc |
| $s$ | 1867 | au |
| $P_{\rm orb}$ | 143.3 | $10^3$ a |
| $-U_g^*$ | 66.4 | $10^{33}$ J |

| Component<br>SpT | A<br>M4.0 V | BaBb<br>M6 V + m7 V[a] | |
|---|---|---|---|
| $\alpha$ | 06:10:17.81 | 06:10:22.52 | |
| $\delta$ | +22:34:17.2 | +22:34:18.1 | |
| $\pi$ | 34.900 ± 0.028 | 43.9 ± 3.7[b] | mas |
| $\mu_\alpha \cos\delta$ | 42.256 ± 0.033 | 39[c] | mas $a^{-1}$ |
| $\mu_\delta$ | −153.057 ± 0.025 | −162[c] | mas $a^{-1}$ |
| $\gamma$ | −10.390 ± 9.806 | +9 ± 4[d] | km $s^{-1}$ |
| $G$ | 12.9716 ± 0.0029 | 14.1549 ± 0.0040 | mag |
| $J$ | 9.876 ± 0.021 | 10.644 ± 0.022 | mag |
| $L$ | 121.5 ± 1.5 | ... | $10^{-4} \mathcal{L}_\odot$ |
| $T_{\rm eff}$ | 3100 ± 50 | ... | K |
| $\mathcal{M}$ | 0.289 | 0.12 + 0.10 | $\mathcal{M}_\odot$ |
| ruwe | 1.402 | ... | |
| Qflag 2MASS | AAA | AAA | |
| Qflag AllWISE | AAAC | AAAC | |

[a] From upper limits given by Law et al. (2008).
[b] Dittmann et al. (2014).
[c] Lépine & Shara (2005).
[d] Newton et al. (2014).

Karmn J06102+225 and LP 362-121 are two nearby mid-M dwarfs, the latter resolved to be a close binary (LAW 14; Law et al. 2008), and the former having a visual (non-physical) companion in the background (Janson et al. 2014). We redefine this system as triple. Next *Gaia* releases might determine more accurately the parameters for the BaBb components, for which we compiled data from several sources from the literature.

## *New*

### A: BD+37 1541

### B: Gaia DR2 943408949754423680

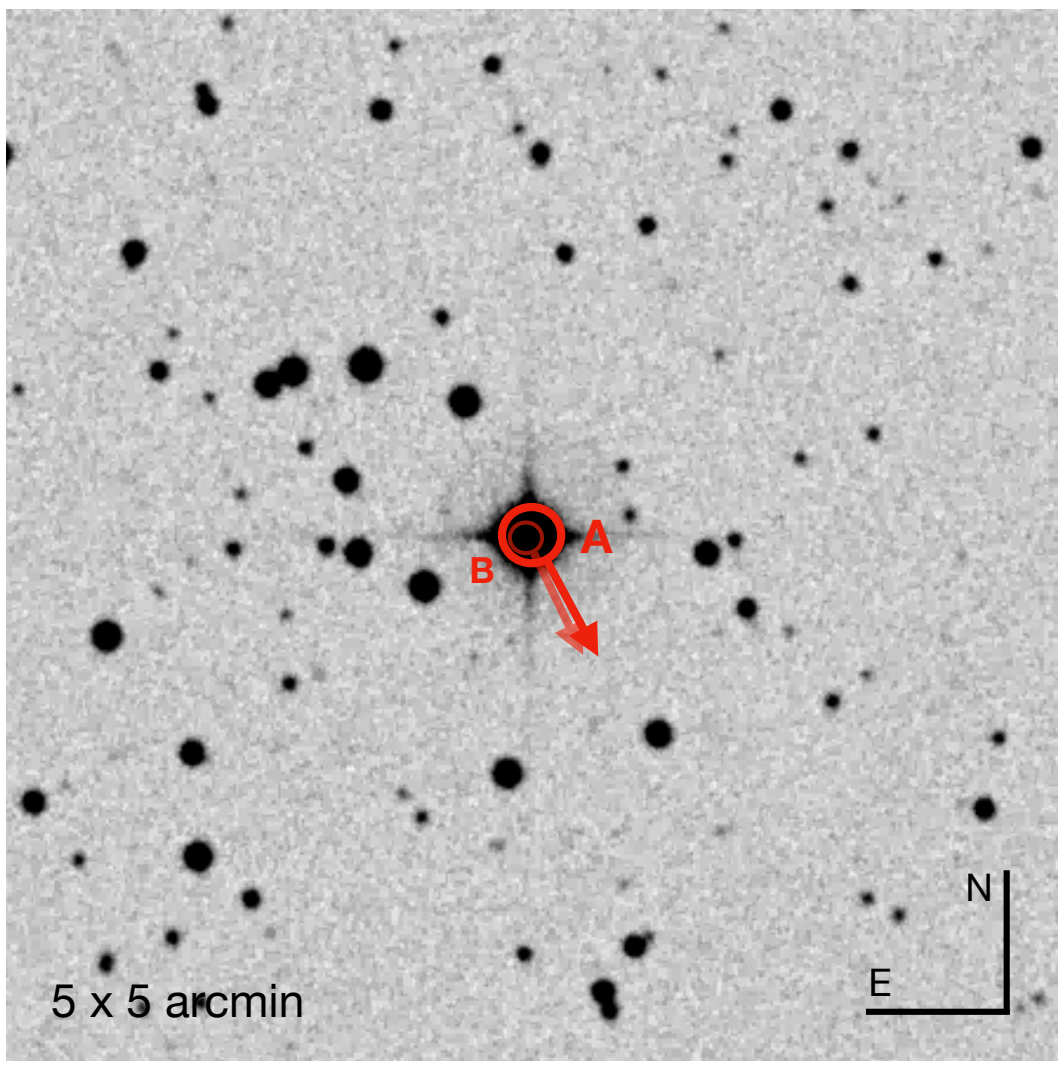

| | A-B | |
|---|---|---|
| WDS | ... | |
| $\rho$ | 3.88 | arcsec |
| $\theta$ | 201.5 | deg |
| $\mu$ ratio | 0.016 | |
| $\Delta PA$ | 0.70 | deg |
| $\Delta d/d$ | 0.0072 | |
| $d$ | 218.8 | pc |
| $s$ | 849 | au |
| $P_{\rm orb}$ | 19.5 | $10^3$ a |
| $-U_g^*$ | 1306 | $10^{33}$ J |

| Component | A | B | |
|---|---|---|---|
| SpT | f0: V | m2.5 V | |
| $\alpha$ | 06:35:38.65 | 06:35:38.53 | |
| $\delta$ | +37:51:13.6 | +37:51:10.0 | |
| $\pi$ | 4.569 ± 0.019 | 4.60 ± 0.11 | mas |
| $\mu_\alpha \cos\delta$ | −11.831 ± 0.018 | −11.69 ± 0.13 | mas a$^{-1}$ |
| $\mu_\delta$ | −19.999 ± 0.015 | −20.33 ± 0.11 | mas a$^{-1}$ |
| $\gamma$ | +0.7 ± 1.6 | ... | km s$^{-1}$ |
| $G$ | 9.1409 ± 0.0028 | 16.439 ± 0.019 | mag |
| $J$ | 8.342 ± 0.023 | ... | mag |
| $\mathcal{L}$ | 79030 ± 1750 | ... | $10^{-4}\mathcal{L}_\odot$ |
| $T_{\rm eff}$ | 6400 ± 50 | ... | K |
| $\mathcal{M}$ | 1.61 | 0.391 | $\mathcal{M}_\odot$ |
| ruwe | 0.905 | 1.099 | |
| Qflag 2MASS | AAA | ... | |
| Qflag AllWISE | AAAB | ... | |

BD+37 1541 is an early-F star (estimated by us) in which vicinity *Gaia* DR3 resolves a physical companion candidate at 3.9 arcsec that is seven magnitudes fainter in the $G$ passband, but shares similar proper motions and parallactic distances. The brightness of the very close primary strongly limits the characterisation of the early-M-dwarf secondary.

# WDS 06511+1844 (FMR 83)

## A: LSPM J0651+1843

## B: LSPM J0651+1845

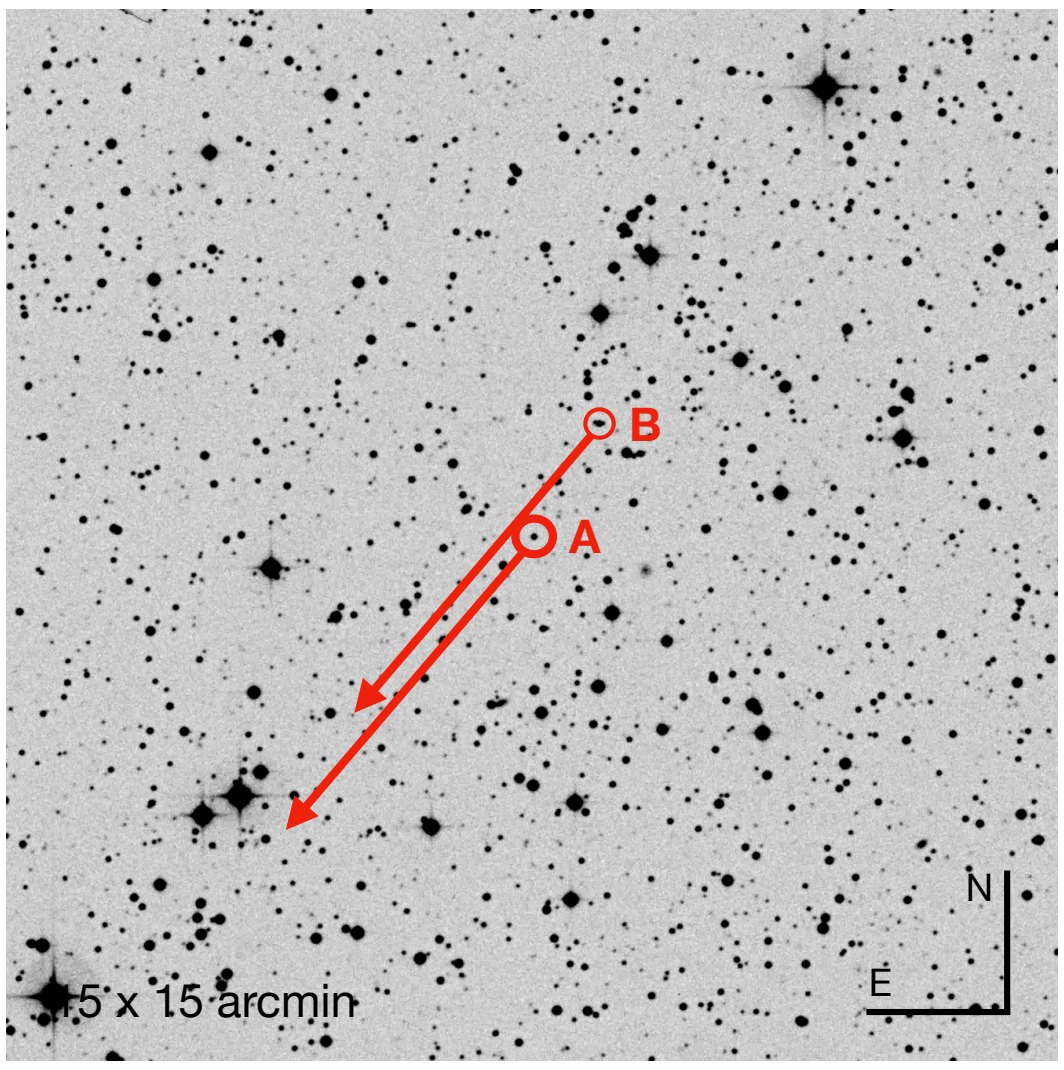

| | A-B | |
|---|---|---|
| WDS | FMR 83 | |
| $\rho$ | 111.7 | arcsec |
| $\theta$ | 330.6 | deg |
| $\mu$ ratio | 0.0020 | |
| $\Delta PA$ | 0.016 | deg |
| $\Delta d/d$ | 0.0028 | |
| $d$ | 63.85 | pc |
| $s$ | 7133 | au |
| $P_{\rm orb}$ | 1461 | $10^3$ a |
| $-U_g^*$ | 7.11 | $10^{33}$ J |

| Component | A | B | |
|---|---|---|---|
| SpT | m4.5 V | m4.5 V | |
| $\alpha$ | 06:51:00.84 | 06:51:04.70 | |
| $\delta$ | +18:45:16.1 | +18:43:38.7 | |
| $\pi$ | 15.662 ± 0.050 | 15.617 ± 0.052 | mas |
| $\mu_\alpha \cos\delta$ | 199.532 ± 0.048 | 199.078 ± 0.055 | mas a$^{-1}$ |
| $\mu_\delta$ | −244.131 ± 0.040 | −243.713 ± 0.043 | mas a$^{-1}$ |
| $\gamma$ | … | … | km s$^{-1}$ |
| $G$ | 15.9303 ± 0.0029 | 15.9455 ± 0.0029 | mag |
| $J$ | 12.981 ± 0.021 | 12.983 ± 0.023 | mag |
| $\mathcal{L}$ | 37.94 ± 0.92 | 37.65 ± 0.95 | $10^{-4} \mathcal{L}_\odot$ |
| $T_{\rm eff}$ | 3100 ± 50 | 3100 ± 50 | K |
| $\mathcal{M}$ | 0.170 | 0.169 | $\mathcal{M}_\odot$ |
| ruwe | 1.057 | 1.068 | |
| Qflag 2MASS | AAA | AAA | |
| Qflag AllWISE | AABU | AABU | |

LSPM J0651+1843 and LSPM J0651+1845 are the components of a known binary system of mid-M dwarfs. Spectral types of both stars photometrically estimated by us. Rica and Caballero (2012) characterised this pair and described it as an 'ultrafragile' system. In this work we refine their parameters with the *Gaia* DR3 solution, and also redefine the nomenclature of their components based on their bolometric luminosity (and $G$ band magnitude).

# *New*

## AaAb: 1RXS J073138.4+455718
## B: 2MASS J07310905+4556573, Karmn J07310+460
## C: PYC J07311+4556

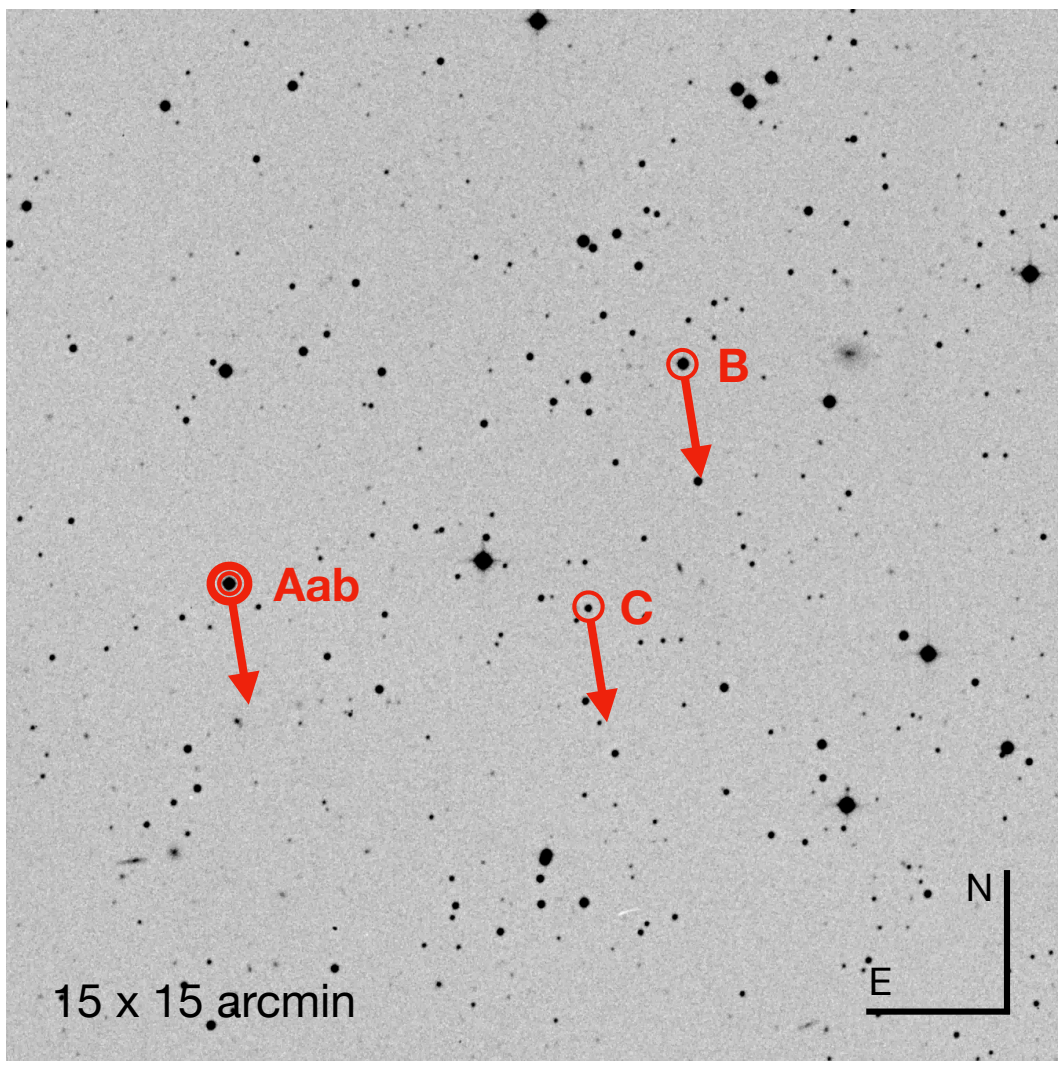

| | AaAb-B | AaAb-C | |
|---|---|---|---|
| WDS | … | … | |
| $\rho$ | 431.4 | 307.8 | arcsec |
| $\theta$ | 296.0 | 266.3 | deg |
| $\mu$ ratio | 0.086 | 0.079 | |
| $\Delta PA$ | 0.76 | 1.5 | deg |
| $\Delta d/d$ | 0.014 | 0.028 | |
| $d$ | 55.93 | | pc |
| $s$ | 24127 | 17215 | au |
| $P_{\rm orb}$ | 4217 | 2542 | $10^3$ a |
| $-U_g^*$ | 17.5 | 11.2 | $10^{33}$ J |

| Component | AaAb | B | C | |
|---|---|---|---|---|
| SpT | M3 + m4.5 V[a] | M4.0 V | m4.5 V | |
| $\alpha$ | 07:31:38.47 | 07:31:01.27 | 07:31:09.03 | |
| $\delta$ | +45:57:15.8 | +46:00:24.8 | +45:56:55.6 | |
| $\pi$ | 17.88 ± 0.42 | 18.141 ± 0.052 | 18.388 ± 0.034 | mas |
| $\mu_\alpha \cos\delta$ | −13.69 ± 0.39 | −13.516 ± 0.059 | −12.167 ± 0.034 | mas $a^{-1}$ |
| $\mu_\delta$ | −92.77 ± 0.28 | −100.846 ± 0.034 | −99.999 ± 0.024 | mas $a^{-1}$ |
| $\gamma$ | … | … | … | km $s^{-1}$ |
| $G$ | 12.7665 ± 0.0083 | 12.8963 ± 0.0029 | 15.2193 ± 0.0030 | mag |
| $J$ | 9.776 ± 0.021 | 9.948 ± 0.023 | 11.898 ± 0.022 | mag |
| $\mathcal{L}$ | … | 445.8 ± 7.1 | 65.1 ± 1.1 | $10^{-4}\mathcal{L}_\odot$ |
| $T_{\rm eff}$ | … | 3200 ± 50 | 3100 ± 50 | K |
| $\mathcal{M}$ | 0.789 | 0.504 | 0.231 | $\mathcal{M}_\odot$ |
| ruwe | 20.805 | 2.230 | 1.105 | |
| Qflag 2MASS | AAA | AAA | AAA | |
| Qflag AllWISE | AAAB | AAAB | AAAU | |

[a] From the difference in magnitudes reported in JNN 58.

Since the primary of this trio of early-to-mid M dwarfs is a 0.2-arcsec close binary (JNN 58; Janson et al., 2014b), this is a quadruple system. Components B and C are separated about 431 and 308 arcsec from the primary, respectively. Remarkably, the C component is PYC J07311+4556, a young candidate member in the $\beta$ Pictoris moving group as reported by Schlieder et al. (2012a). For this reason, for the B component we report bolometric luminosity and effective temperature values not in agreement with the expected figures for the main sequence. Additionally, its large ruwe value might be indicative of a non-resolved companion, in which case the system would turn out to be quintuple.

# WDS 07400-0336 (BGH 3 + *New*)

A: HD 61606 A

B: HD 61606 B

C: BD-02 2198, Karmn J07361-031

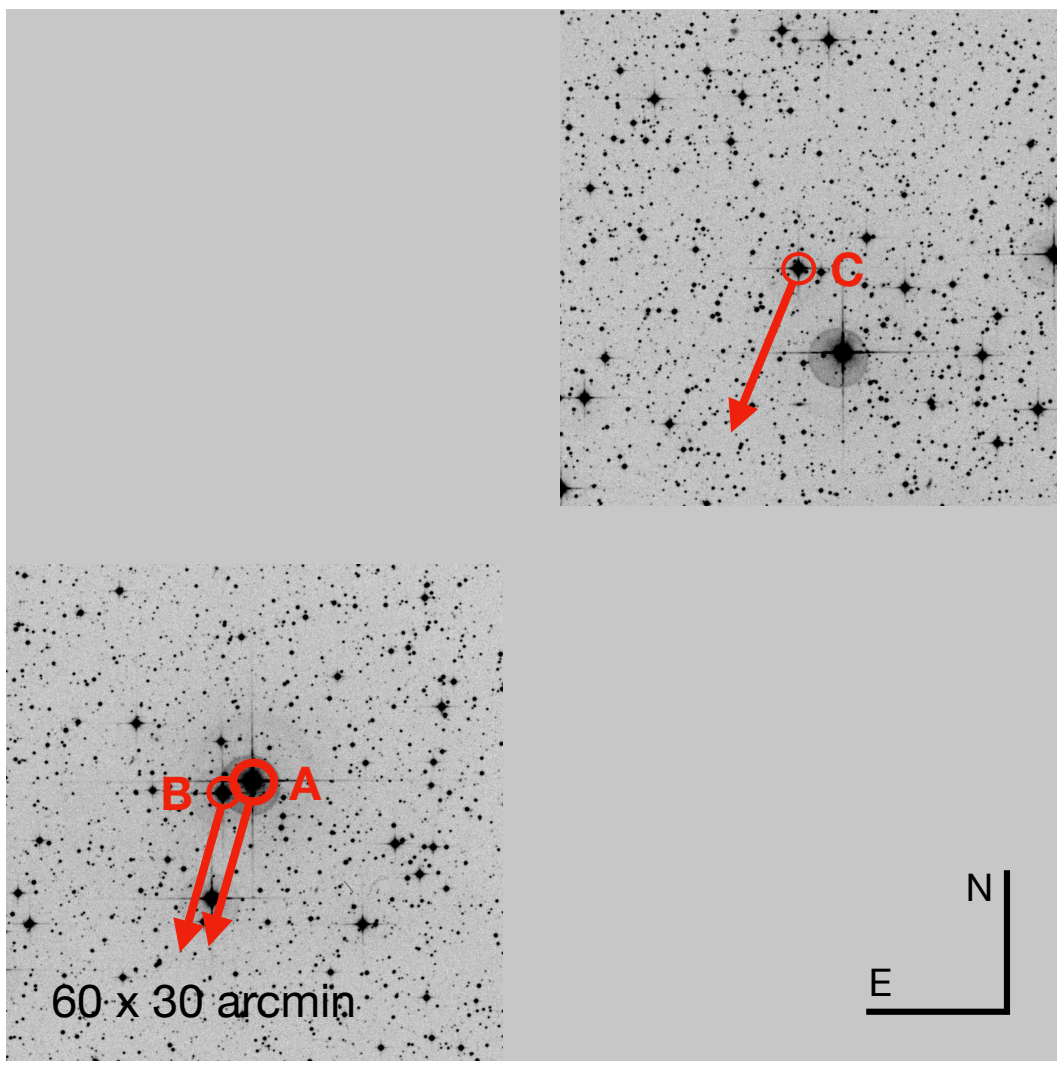

| | A-B | A-C | |
|---|---|---|---|
| WDS | BGH 3 | ... | |
| $\rho$ | 57.9 | 3894 | arcsec |
| $\theta$ | 112.7 | 296.7 | deg |
| $\mu$ ratio | 0.033 | 0.054 | |
| $\Delta PA$ | 1.18 | 0.049 | deg |
| $\Delta d/d$ | 0.00056 | 0.011 | |
| $d$ | 14.08 | | pc |
| $s$ | 815 | 54823 | au |
| $P_{\rm orb}$ | 26.6 | 10907 | $10^3$ a |
| $-U_g^*$ | 1026 | 24.4 | $10^{33}$ J |

| Component | A | B | C | |
|---|---|---|---|---|
| SpT | K3 V | K7 V | M1.0 V | |
| $\alpha$ | 07:39:59.40 | 07:40:02.97 | 07:36:07.15 | |
| $\delta$ | −03:35:55.5 | −03:36:17.8 | −03:06:43.4 | |
| $\pi$ | 71.032 ± 0.024 | 70.992 ± 0.025 | 70.27 ± 0.13 | mas |
| $\mu_\alpha \cos\delta$ | 70.078 ± 0.024 | 66.008 ± 0.024 | 74.129 ± 0.144 | mas a$^{-1}$ |
| $\mu_\delta$ | −278.117 ± 0.019 | −286.706 ± 0.020 | −293.118 ± 0.095 | mas a$^{-1}$ |
| $\gamma$ | −18.35 ± 0.18 | −19.01 ± 0.19 | −17.19 ± 0.37 | km s$^{-1}$ |
| $G$ | 6.8898 ± 0.0028 | 8.3347 ± 0.0028 | 9.1424 ± 0.0029 | mag |
| $J$ | 5.493 ± 0.027 | 6.377 ± 0.024 | 6.791 ± 0.034 | mag |
| $\mathcal{L}$ | ... | ... | 608.8 ± 9.4 | $10^{-4}\mathcal{L}_\odot$ |
| $T_{\rm eff}$ | ... | ... | 3700 ± 50 | K |
| $\mathcal{M}$ | 0.766 | 0.619 | 0.547 | $\mathcal{M}_\odot$ |
| ruwe | 1.065 | 1.022 | 6.531 | |
| Qflag 2MASS | EAA | AAA | AAA | |
| Qflag AllWISE | BBAA | BAAA | AAAA | |

WDS 07400-0336 is a known pair of K dwarfs with a projected physical separation of 815 au. Poveda et al. (2009) found that the wide early-M dwarf BD-02 2198 is an co-eval, co-moving and equi-distant companion of this pair. Ishikawa et al. (2020) concluded that this object was not a member of the system, based on the study of scape velocities and chemical abundances (from Montes et al. 2018). However, the large value of the ruwe indicator of C points towards unresolved binarity, in which case the system would be quadruple (i.e. a binary of binaries). This is the closest system of our sample (14.1 pc), but also the most separated one (55 000 au).

# *New*

## A: 1RXS J074948.5-031712
## B: 2MASS J07495087-0317194
## C: 2MASS J07494215-0320338, Karmn J07497-033

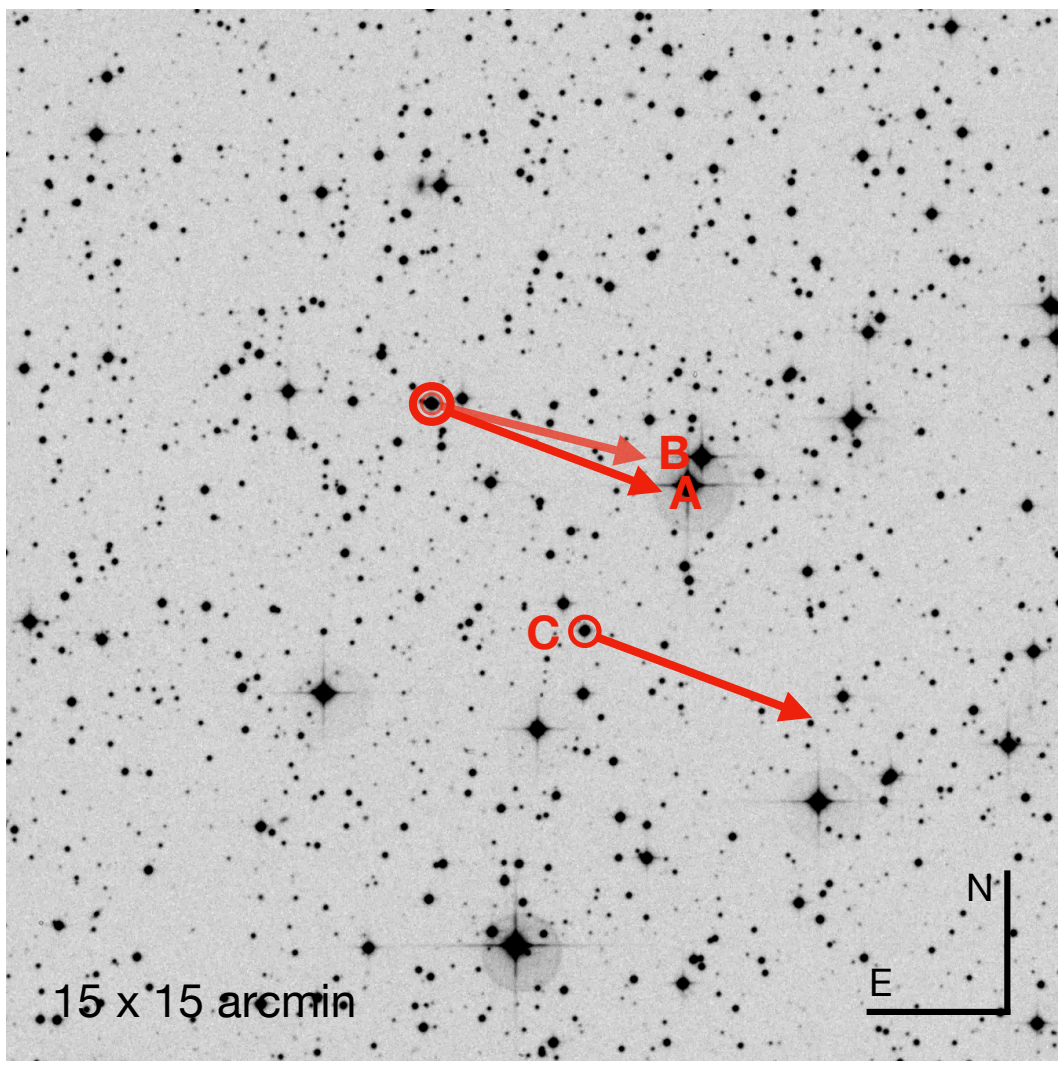

| | A-B | A-C | |
|---|---|---|---|
| WDS | … | … | |
| $\rho$ | 1.936 | 234.9 | arcsec |
| $\theta$ | 266.3 | 214.0 | deg |
| $\mu$ ratio | 0.24 | 0.084 | |
| $\Delta PA$ | 5.48 | 1.56 | deg |
| $\Delta d/d$ | 0.00043 | 0.015 | |
| $d$ | 17.04 | | pc |
| $s$ | 32.82 | 4002 | au |
| $P_{\rm orb}$ | 0.333 | 327 | $10^3$ a |
| $-U_g^*$ | 4832 | 81.4 | $10^{33}$ J |

| Component | A | B | C | |
|---|---|---|---|---|
| SpT | M3.5 V | m4: V | M3.5 V | |
| $\alpha$ | 07:49:50.75 | 07:49:50.62 | 07:49:41.97 | |
| $\delta$ | −03:17:20.3 | −03:17:20.4 | −03:20:34.9 | |
| $\pi$ | 58.683 ± 0.047 | 58.658 ± 0.045 | 57.83 ± 0.15 | mas |
| $\mu_\alpha \cos\delta$ | −174.257 ± 0.044 | −139.058 ± 0.046 | −161.90 ± 0.14 | mas a$^{-1}$ |
| $\mu_\delta$ | −65.329 ± 0.033 | −37.451 ± 0.040 | −55.71 ± 0.11 | mas a$^{-1}$ |
| $\gamma$ | … | … | −24.2 ± 5.1[a] | km s$^{-1}$ |
| $G$ | 11.5510 ± 0.0028 | 11.9199 ± 0.0028 | 11.8027 ± 0.0028 | mag |
| $J$ | 8.039 ± 0.030 | … | 8.891 ± 0.027 | mag |
| $\mathcal{L}$ | … | … | 113.8 ± 1.7 | $10^{-4}\mathcal{L}_\odot$ |
| $T_{\rm eff}$ | … | … | 3200 ± 50 | K |
| $\mathcal{M}$ | 0.3195 | 0.2814 | 0.3074 | $\mathcal{M}_\odot$ |
| ruwe | 1.575 | 1.744 | 5.936 | |
| Qflag 2MASS | AAA | … | AAA | |
| Qflag AllWISE | AAAA | … | AAAA | |

[a] Terrien et al. (2015)

At less than 2 arcsec of the M3.5-dwarf 1RXS J074948.5-031712, *Gaia* DR3 resolves a source of similar apparent brightness that is equidistant and co-moving. Although its $\mu$ ratio exceeds the limit set by our criteria for physical parity, we expect this deviation due to the closeness of the pair (less than 2 arcsec). Separated 235 arcsec from this pair is Karmn J07498-033, which astrometry is indicative of physical connection with the primary. Additionally, the large value of its `ruwe` indicator suggests an additional binarity for this star, meaning that the entire system would be quadruple.

# WDS 08371+3908 (KO 6)

### A: LP 209-28, Königstuhl 6A

### B: LP 209-27, Königstuhl 6B

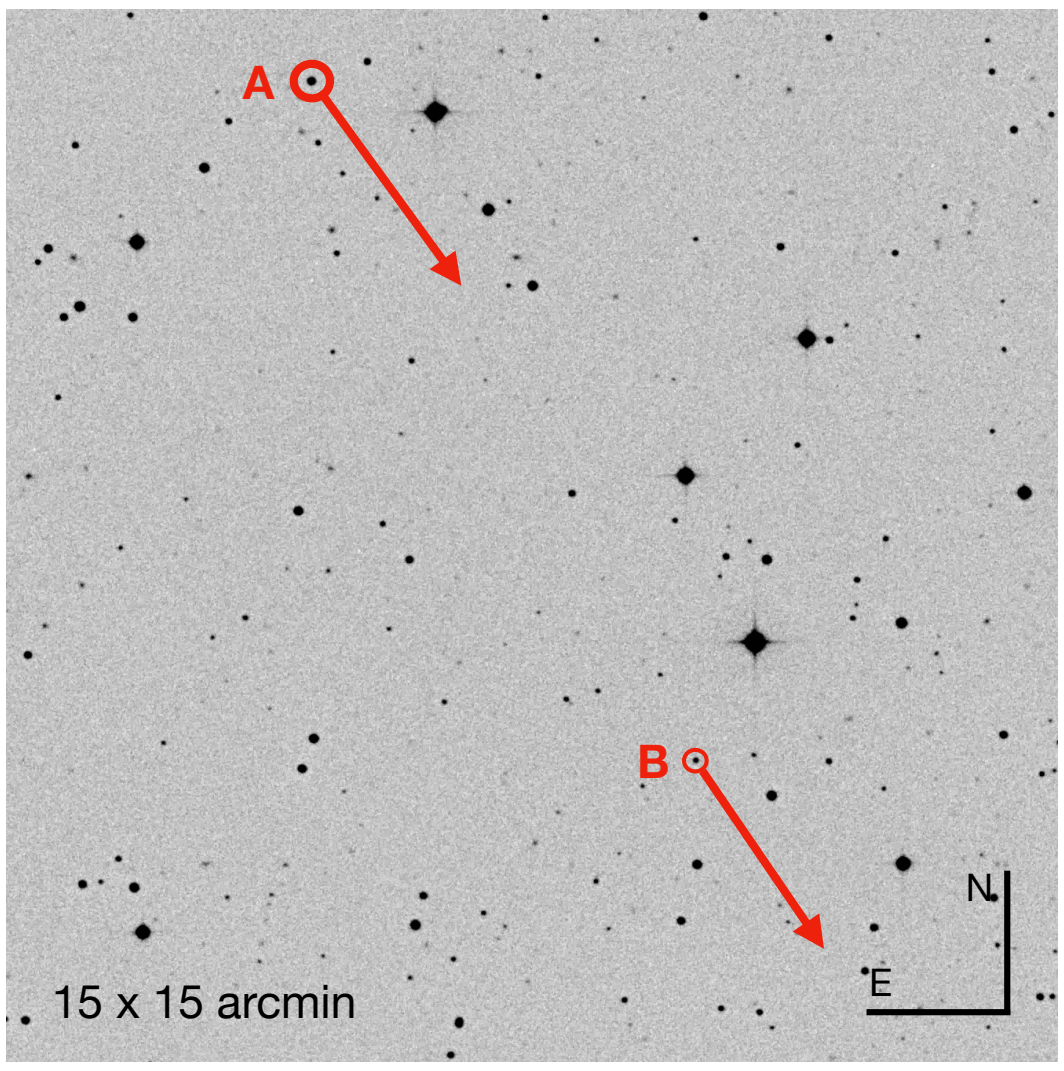

| | A-B | |
|---|---|---|
| WDS | KO 6 | |
| $\rho$ | 666.7 | arcsec |
| $\theta$ | 208.5 | deg |
| $\mu$ ratio | 0.034 | |
| $\Delta PA$ | 0.37 | deg |
| $\Delta d/d$ | 0.43 | |
| $d$ | 104.7 | pc |
| $s$ | 69837 | au |
| $P_{\rm orb}$ | 31557 | $10^3$ a |
| $-U_g^*$ | 2.26 | $10^{33}$ J |

| Component | A | B | |
|---|---|---|---|
| SpT | m3: V | m4: V | |
| $\alpha$ | 08:37:04.54 | 08:36:37.28 | |
| $\delta$ | +39:07:56.9 | +38:58:10.7 | |
| $\pi$ | 9.546 ± 0.027 | 6.688± 0.069 | mas |
| $\mu_\alpha \cos\delta$ | −119.169 ± 0.027 | −116.409 ± 0.065 | mas a$^{-1}$ |
| $\mu_\delta$ | −186.435 ± 0.023 | −179.564 ± 0.052 | mas a$^{-1}$ |
| $\gamma$ | ... | ... | km s$^{-1}$ |
| $G$ | 15.0067 ± 0.0028 | 16.8107 ± 0.0029 | mag |
| $J$ | 12.816 ± 0.022 | 14.009 ± 0.026 | mag |
| $\mathcal{L}$ | 144.8 ± 2.1 | 81.8 ± 3.2 | $10^{-4}\mathcal{L}_\odot$ |
| $T_{\rm eff}$ | 3700 ± 50 | 3200 ± 50 | K |
| $\mathcal{M}$ | 0.342 | 0.261 | $\mathcal{M}_\odot$ |
| ruwe | 1.042 | 0.964 | |
| Qflag 2MASS | AAA | AAA | |
| Qflag AllWISE | AABU | AAUU | |

LP 209-28 and LP 209-27 (KO6 AB) is a pair proposed as a binary system by Caballero et al. (2012). *Gaia* DR3 introduces a notable dissimilarity in distance between components, accompanied by a good single-star model fitting (i.e. `ruwe` < 1.4). This means that we do not expect additional multiplicity in either component, and the difference in parallactic determination would not be the cause of close, unresolved companions. We find that the separation of $\rho$ during 10 epochs of observation spanning 62 years increases by 0.248 arcsec (4 milliarcsecond per year). Therefore, we propose this system as a visual pair.

# *New*

## A: HD 77825

## B: 1RXS J090406.8-155512, Karmn J09040-159

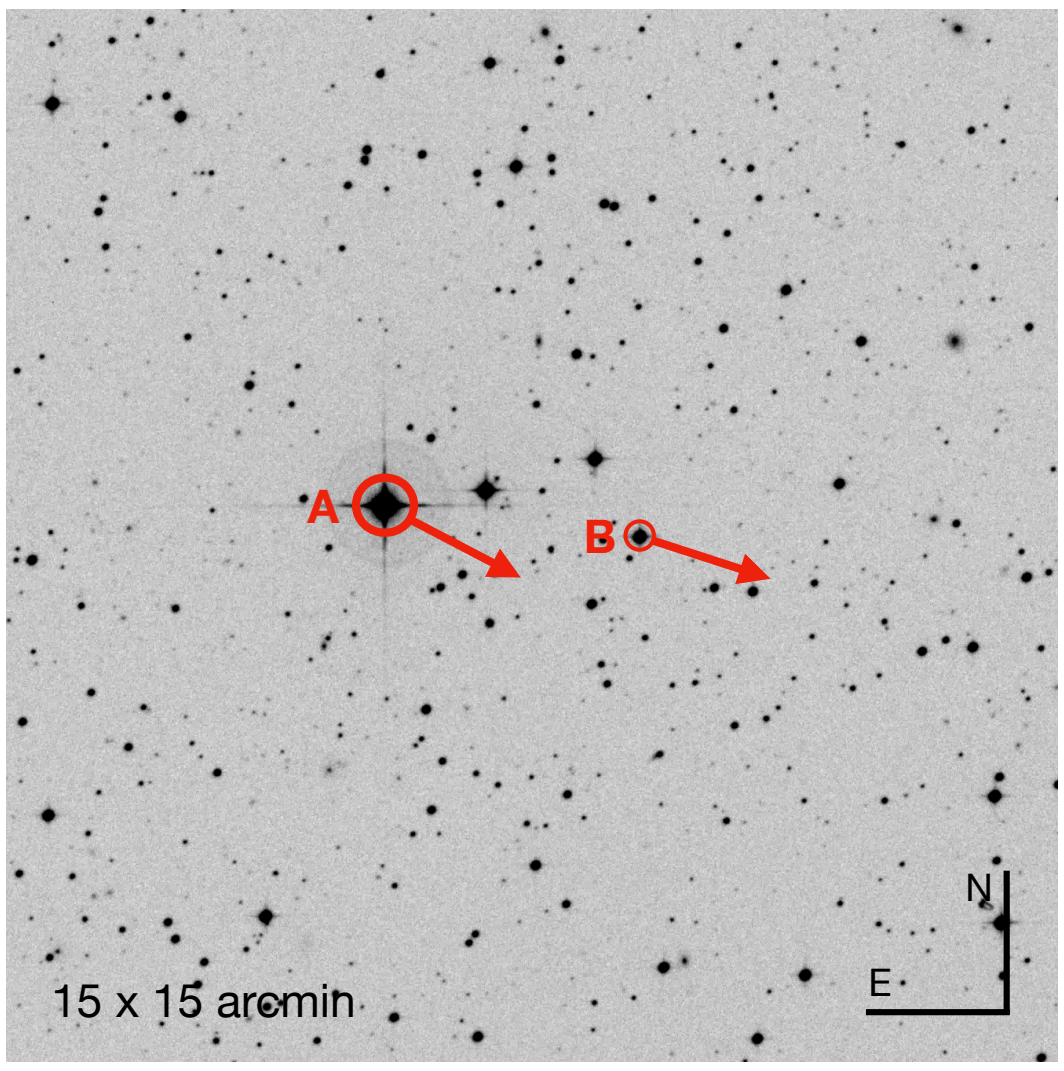

| | A-B | |
|---|---|---|
| WDS | ... | |
| $\rho$ | 220.0 | arcsec |
| $\theta$ | 262.9 | deg |
| $\mu$ ratio | 0.020 | |
| $\Delta PA$ | 0.71 | deg |
| $\Delta d/d$ | 0.0032 | |
| $d$ | 27.39 | pc |
| $s$ | 6026 | au |
| $P_{\rm orb}$ | 538 | $10^3$ a |
| $-U_g^*$ | 93.4 | $10^{33}$ J |

| Component | A | B | |
|---|---|---|---|
| SpT | K2 V | M2.5 V | |
| $\alpha$ | 09:04:20.57 | 09:04:05.44 | |
| $\delta$ | −15:54:51.8 | −15:55:19.0 | |
| $\pi$ | 36.512 ± 0.022 | 36.628 ± 0.022 | mas |
| $\mu_\alpha \cos\delta$ | −107.828 ± 0.083 | −109.034 ± 0.072 | mas $a^{-1}$ |
| $\mu_\delta$ | −30.936 ± 0.079 | −32.638 ± 0.076 | mas $a^{-1}$ |
| $\gamma$ | +4.39 ± 0.24 | +4.44 ± 0.33 | km $s^{-1}$ |
| $G$ | 8.4640 ± 0.0028 | 11.7538 ± 0.0029 | mag |
| $J$ | 7.005 ± 0.024 | 9.156 ± 0.026 | mag |
| $\mathcal{L}$ | 2601 ± 28 | 250.1 ± 3.1 | $10^{-4}\mathcal{L}_\odot$ |
| $T_{\rm eff}$ | 4800 ± 50 | 3400 ± 50 | K |
| $\mathcal{M}$ | 0.756 | 0.422 | $\mathcal{M}_\odot$ |
| ruwe | 1.002 | 1.114 | |
| Qflag 2MASS | AAA | AAA | |
| Qflag AllWISE | AAAA | AAAB | |

HD 77825 and 1RXS J090406.8-155512 (Karmn J09040-159) form a physical pair of spectroscopically-derived K2 V and M2.5 V stars located at less than 30 pc. All astrometric measurements from *Gaia* DR3 support the binarity, including their radial velocities from *Gaia* DR2.

# *New*

## A: 2MASS J13181352+7322073, Karmn J13182+733

## B: Gaia DR2 1688578285187648128

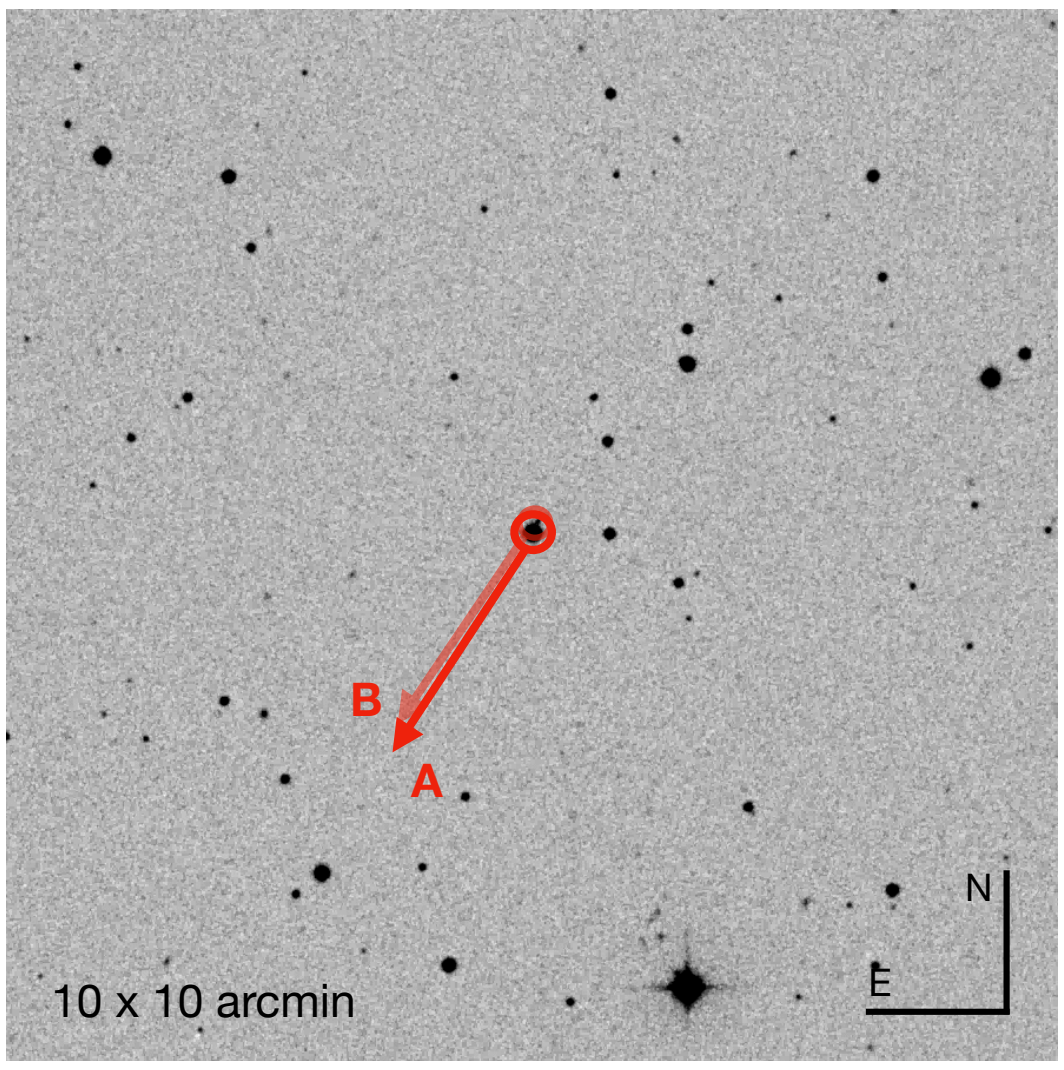

| | A-B | |
|---|---|---|
| WDS | ... | |
| $\rho$ | 7.39 | arcsec |
| $\theta$ | 155.7 | deg |
| $\mu$ ratio | 0.065 | |
| $\Delta PA$ | 1.01 | deg |
| $\Delta d/d$ | 0.007 | |
| $d$ | 25.29 | pc |
| $s$ | 187 | au |
| $P_{\rm orb}$ | 4.38 | $10^3$ a |
| $-U_g^*$ | 297 | $10^{33}$ J |

| Component | A | B | |
|---|---|---|---|
| SpT | M3.5 V | m7 V | |
| $\alpha$ | 13:18:13.82 | 13:18:13.11 | |
| $\delta$ | +73:22:05.6 | +73:22:12.4 | |
| $\pi$ | 39.534 ± 0.021 | 39.276 ± 0.098 | mas |
| $\mu_\alpha \cos\delta$ | 72.055 ± 0.027 | 70.05 ± 0.11 | mas a$^{-1}$ |
| $\mu_\delta$ | −107.549 ± 0.027 | −100.69 ± 0.12 | mas a$^{-1}$ |
| $\gamma$ | ... | ... | km s$^{-1}$ |
| $G$ | 12.3962 ± 0.0028 | 16.9130 ± 0.0030 | mag |
| $J$ | 9.541 ± 0.022 | 12.660 ± 0.025 | mag |
| $\mathcal{L}$ | 143.2 ± 1.3 | 6.20 ± 0.17 | $10^{-4}\mathcal{L}_\odot$ |
| $T_{\rm eff}$ | 3300 ± 50 | 2500 ± 50 | K |
| $\mathcal{M}$ | 0.340 | 0.092 | $\mathcal{M}_\odot$ |
| ruwe | 1.422 | 0.814 | |
| Qflag 2MASS | AAA | ABA | |
| Qflag AllWISE | AAAB | ... | |

Karmn J13182+733 is an M3.5 V star with a resolved companion at 7.4 arcsec (187 au) to the south. The astrometry of this companion indicates a physical relation to the primary, and its photometry and empirically derived parameters are compatible with a mid- to late-M dwarf.

## *New*

### A: HD 130666

### B: 2MASS J14474531+4934020

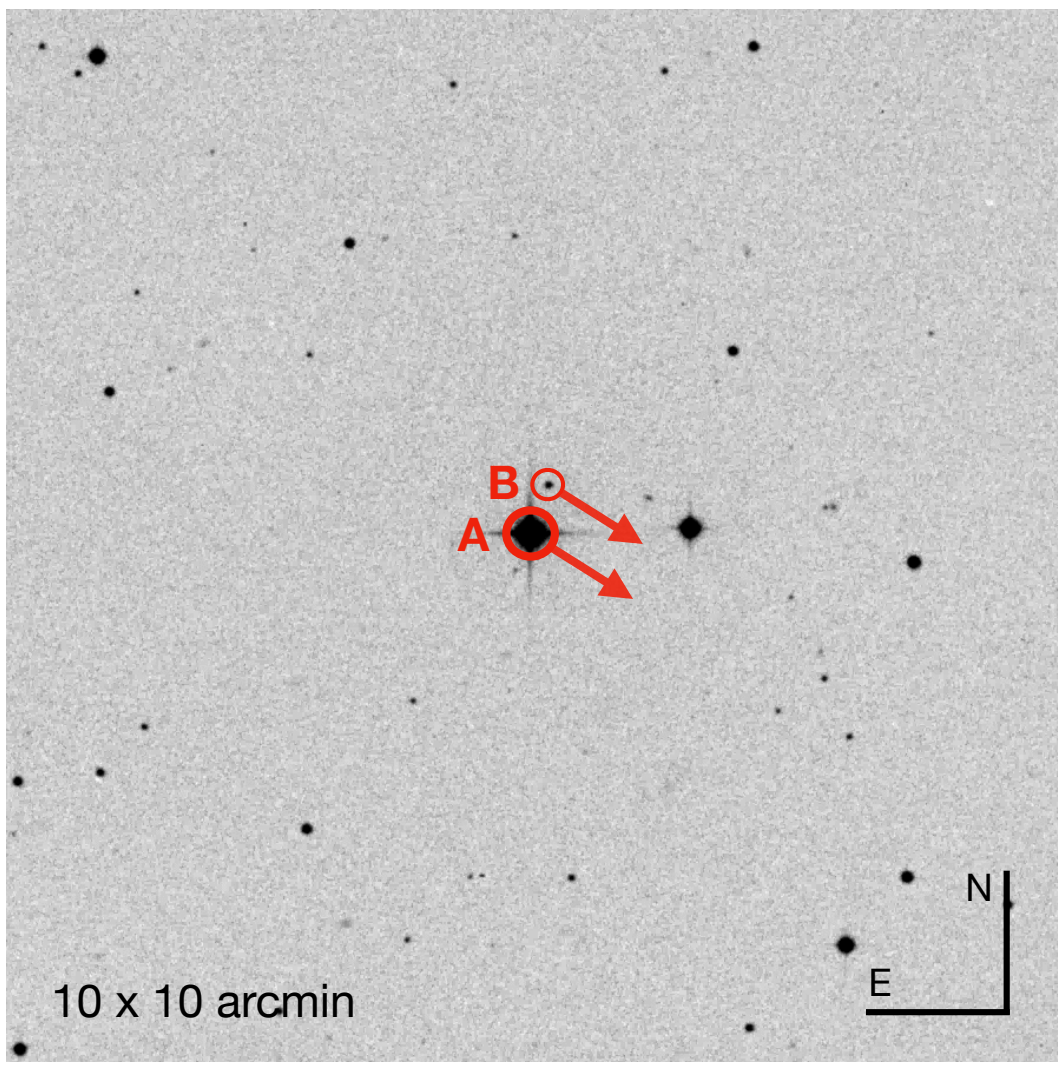

| | A-B | |
|---|---|---|
| WDS | ... | |
| $\rho$ | 29.54 | arcsec |
| $\theta$ | 336.6 | deg |
| $\mu$ ratio | 0.051 | |
| $\Delta PA$ | 21.5 | deg |
| $\Delta d/d$ | 0.0066 | |
| $d$ | 104.0 | pc |
| $s$ | 3072 | au |
| $P_{\rm orb}$ | 146 | $10^3$ a |
| $-U_g^*$ | 130 | $10^{33}$ J |

| Component | A | B | |
|---|---|---|---|
| SpT | G5 | m4.5 V | |
| $\alpha$ | 14:47:46.40 | 14:47:45.19 | |
| $\delta$ | +49:33:34.1 | +49:34:01.2 | |
| $\pi$ | 9.614 ± 0.033 | 9.551 ± 0.069 | mas |
| $\mu_\alpha \cos\delta$ | −63.589 ± 0.032 | −62.076 ± 0.068 | mas $a^{-1}$ |
| $\mu_\delta$ | −51.374 ± 0.038 | −47.446 ± 0.081 | mas $a^{-1}$ |
| $\gamma$ | −50.165 ± 0.179 | ... | km $s^{-1}$ |
| $G$ | 8.4416 ± 0.0028 | 17.2739 ± 0.0030 | mag |
| $J$ | 7.031 ± 0.030 | 14.032 ± 0.028 | mag |
| $\mathcal{L}$ | 38260 ± 9487 | 36.85 ± 1.14 | $10^{-4} \mathcal{L}_\odot$ |
| $T_{\rm eff}$ | 4900 ± 50 | 3000 ± 50 | K |
| $\mathcal{M}$ | 1.36 | 0.167 | $\mathcal{M}_\odot$ |
| ruwe | 2.094 | 1.123 | |
| Qflag 2MASS | AAF | AAA | |
| Qflag AllWISE | AAAA | AABU | |

HD 130666 is a bright Sun-like star with mid-M dwarf physical companion separated by 29.5 arsec. The components are well-characterised, both astrometric and photometrically, in *Gaia* DR3. The ruwe indicator in the primary is higher than expected for a single-star model fit.

# *New*

## A: TYC 2565-684-1

## B: 2MASS J15080798+3310222

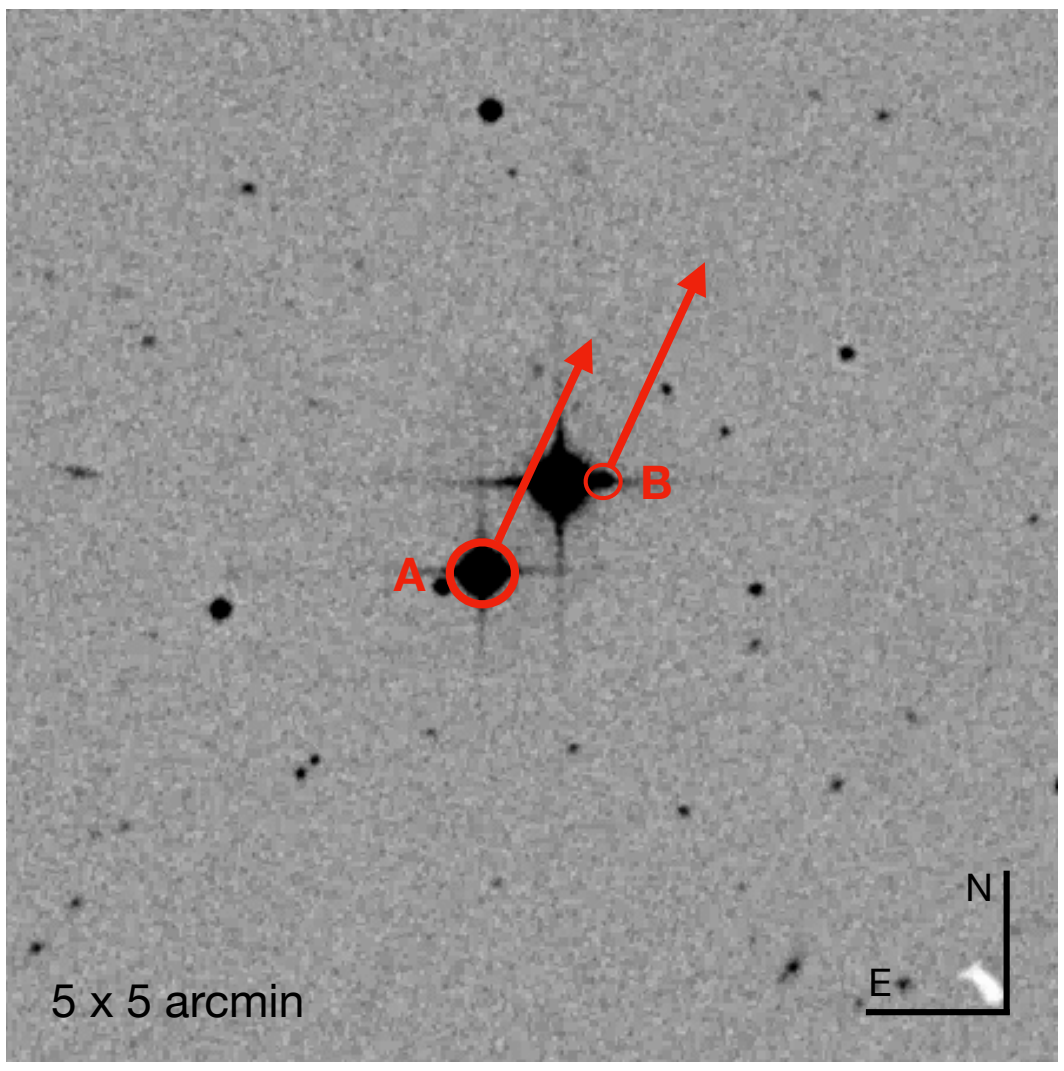

| | A-B | |
|---|---|---|
| WDS | ... | |
| $\rho$ | 43.64 | arcsec |
| $\theta$ | 306.2 | deg |
| $\mu$ ratio | 0.0041 | |
| $\Delta PA$ | 0.23 | deg |
| $\Delta d/d$ | 0.0054 | |
| $d$ | 198.1 | pc |
| $s$ | 8644 | au |
| $P_{\rm orb}$ | 787 | $10^3$ a |
| $-U_g^*$ | 79.1 | $10^{33}$ J |

| Component | A | B | |
|---|---|---|---|
| SpT | g1 V | m3 V | |
| $\alpha$ | 15:08:10.72 | 15:08:07.92 | |
| $\delta$ | +33:09:57.8 | +33:10:23.6 | |
| $\pi$ | 5.048 ± 0.019 | 5.021 ± 0.061 | mas |
| $\mu_\alpha \cos\delta$ | −39.926 ± 0.015 | −40.224 ± 0.050 | mas a$^{-1}$ |
| $\mu_\delta$ | 77.094 ± 0.018 | 76.905 ± 0.063 | mas a$^{-1}$ |
| $\gamma$ | −19.2 ± 3.4 | ... | km s$^{-1}$ |
| $G$ | 10.9314 ± 0.0028 | 16.9843 ± 0.0028 | mag |
| $J$ | 9.895 ± 0.020 | 14.081 ± 0.037 | mag |
| $\mathcal{L}$ | 11814 ± 152 | 128.2 ± 5.3 | $10^{-4}\mathcal{L}_\odot$ |
| $T_{\rm eff}$ | 5700 ± 50 | 3200 ± 50 | K |
| $\mathcal{M}$ | 1.04 | 0.324 | $\mathcal{M}_\odot$ |
| `ruwe` | 1.499 | 1.043 | |
| Qflag 2MASS | AAA | AAA | |
| Qflag AllWISE | AAAC | ... | |

TYC 2565-684-1 is a bright dwarf, estimated to be of early-G type, with a fainter companion that we classify photometrically as m3 V. The pair complies with the conditions for physical parity. It is located 13 arcsec westward from the bright background star BD+33 2544 ($G \sim 9.8$ mag).

# WDS 15092+3304 (HJ 566 + *New*)

A: HD 134494

B: BD+33 2548 B

C: Gaia DR2 1288848427727490048

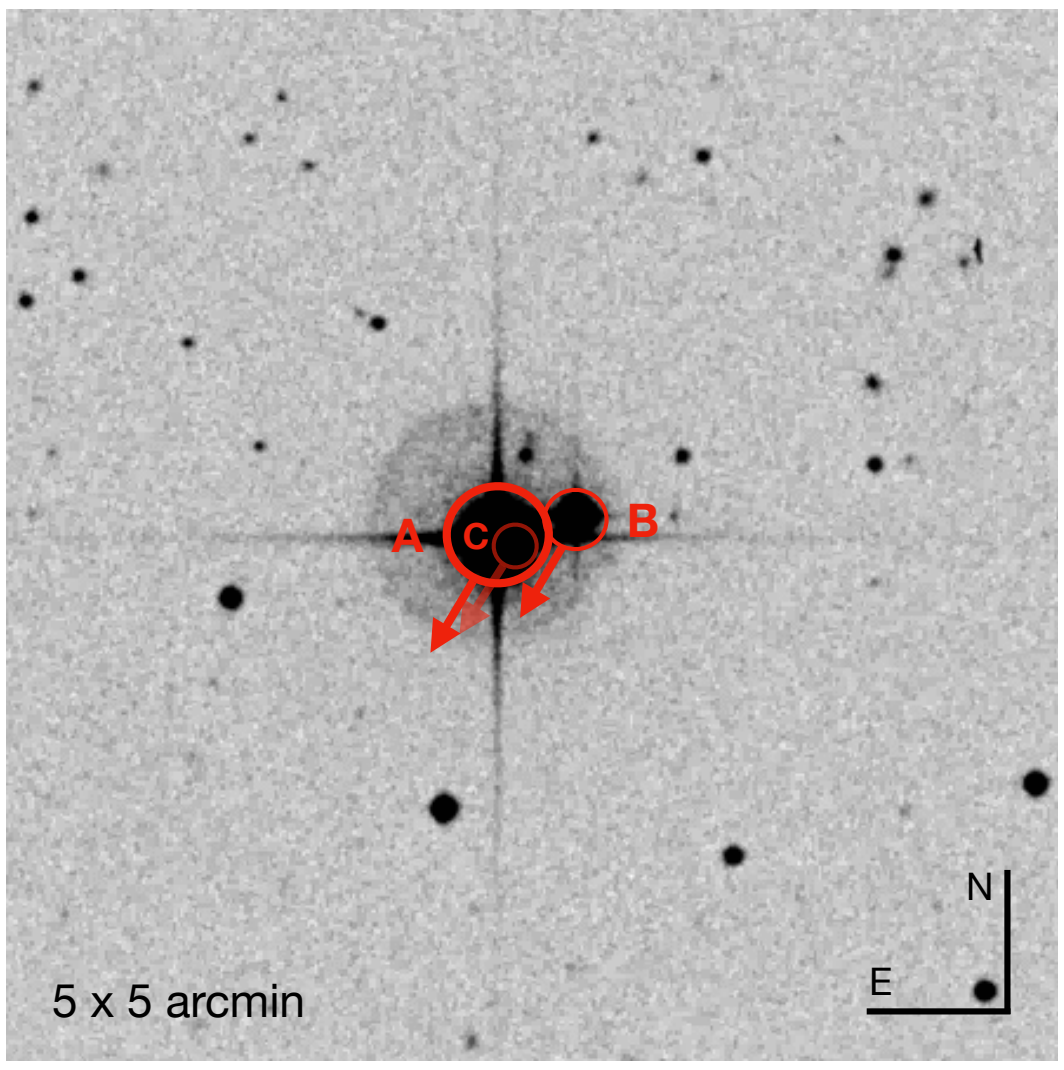

| | A-B | A-C | |
|---|---|---|---|
| WDS | HJ 566 | … | |
| $\rho$ | 23.4 | 5.85 | arcsec |
| $\theta$ | 285.0 | 180.3 | deg |
| $\mu$ ratio | 0.0089 | 0.0066 | |
| $\Delta PA$ | 0.46 | 0.21 | deg |
| $\Delta d/d$ | 0.0063 | 0.12 | |
| $d$ | 276.1 | | pc |
| $s$ | 6456 | 1615 | au |
| $P_{\rm orb}$ | 314 | 42.1 | $10^3$ a |
| $-U_g^*$ | 837 | 907 | $10^{33}$ J |

| Component | A | B | C | |
|---|---|---|---|---|
| SpT | K0 IV | f9 V | m3 V | |
| $\alpha$ | 15:09:09.91 | 15:09:08.11 | 15:09:09.91 | |
| $\delta$ | +33:03:37.4 | +33:03:43.4 | +33:03:31.5 | |
| $\pi$ | 3.622 ± 0.021 | 3.645 ± 0.019 | 4.12 ± 0.10 | mas |
| $\mu_\alpha \cos\delta$ | 13.284 ± 0.015 | 13.423 ± 0.015 | 11.27 ± 0.11 | mas a$^{-1}$ |
| $\mu_\delta$ | −23.059 ± 0.020 | −22.866 ± 0.019 | −24.23 ± 0.11 | mas a$^{-1}$ |
| $\gamma$ | +0.15 ± 0.13 | +0.02 ± 0.61 | … | km s$^{-1}$ |
| $G$ | 8.0455 ± 0.0028 | 11.2661 ± 0.0028 | 17.0507 ± 0.0055 | mag |
| $J$ | 6.511 ± 0.021 | 10.359 ± 0.021 | … | mag |
| $\mathcal{L}$ | 391600 ± 10700 | 16294 ± 307 | … | $10^{-4} \mathcal{L}_\odot$ |
| $T_{\rm eff}$ | 4800 ± 50 | 6100 ± 50 | … | K |
| $\mathcal{M}$ | 2.55 | 1.12 | 0.349 | $\mathcal{M}_\odot$ |
| ruwe | 1.011 | 1.172 | 1.343 | |
| Qflag 2MASS | AAA | AAA | … | |
| Qflag AllWISE | BAAA | AAAB | … | |

HD 134494 (K0, Cannon et al. 1993) has a known physical companion, BD+33 2548B, which we estimate to be a late-F dwarf or early-G. We propose a second candidate to physical companion resolved by *Gaia* at 5.9 arcsec from the primary and 9 mag fainter in *G*, which we estimate it to be an m3 V star. Based on the bolometric luminosity and the absolute brightnesses in *G* and *J*, we propose a new classification of the primary as subgiant.

# WDS 16329+0315 (DSG 7 + LEP 79 + DAM 649)

**Aabc:** HD 149162

**B:** G 17-23, Karmn J16330+031

**C:** LSPM J1633+0311S

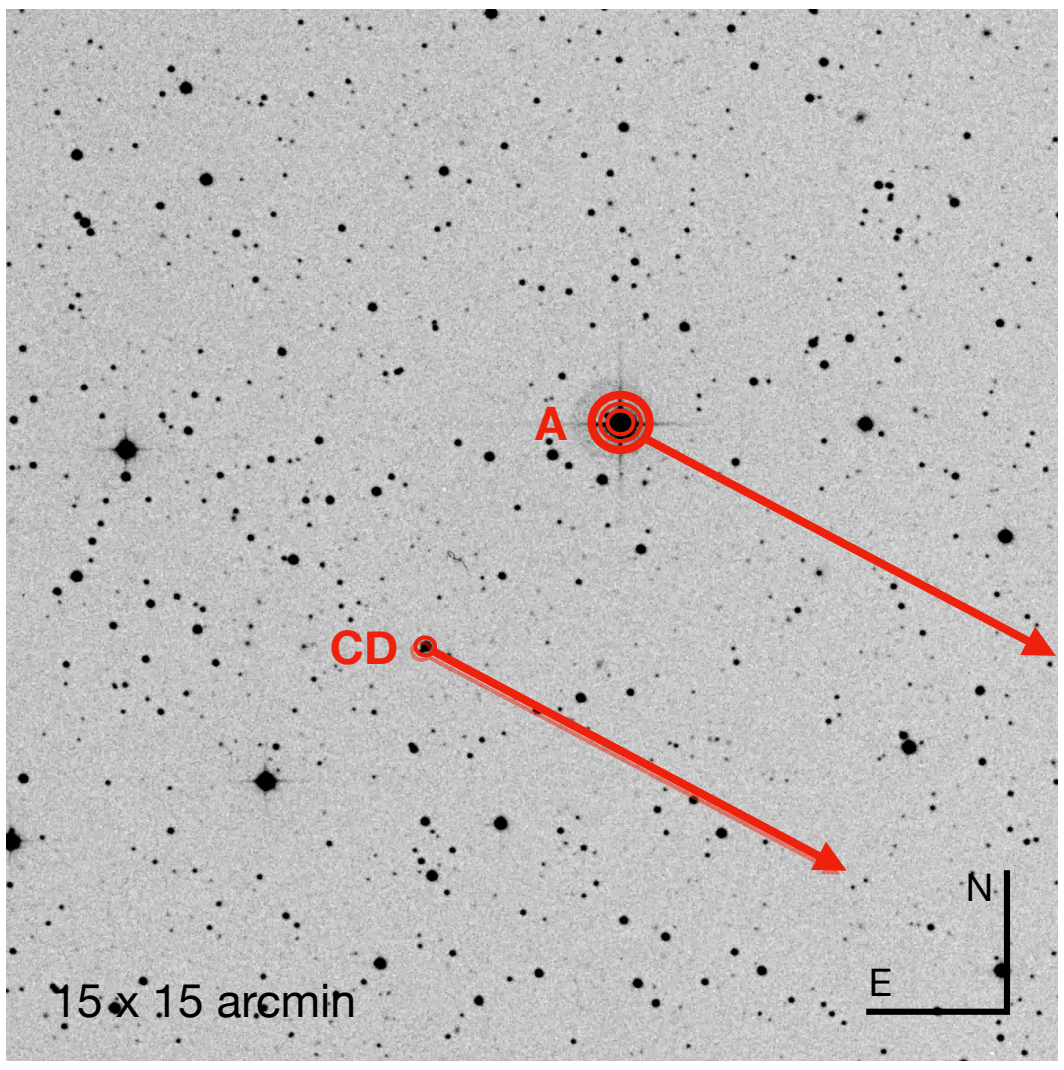

| | AaAbAc-BC | |
|---|---|---|
| WDS | LEP 79 | |
| $\rho$ | 252.0 | arcsec |
| $\theta$ | 138.4 | deg |
| $\mu$ ratio | 0.013 | |
| $\Delta PA$ | 0.67 | deg |
| $\Delta d/d$ | 0.0015 | |
| $d$ | 45.25 | pc |
| $s$ | 11406 | au |
| $P_{\rm orb}$ | 959 | $10^3$ a |
| $-U_g^*$ | 121.0 | $10^{33}$ J |

| Component | AaAbAc | B | C | |
|---|---|---|---|---|
| SpT | K0 Ve + k5 V + m5 V | M3.0 V | D: | |
| $\alpha$ | 16:32:51.24 | 16:33:02.42 | 16:33:02.71 | |
| $\delta$ | +03:14:42.8 | +03:11:34.4 | +03:11:29.7 | |
| $\pi$ | 22.09 ± 0.52 | 22.130 ± 0.016 | 22.01 ± 0.11 | mas |
| $\mu_\alpha \cos\delta$ | −373.83 ± 0.51 | −369.147 ± 0.017 | −369.41 ± 0.13 | mas $a^{-1}$ |
| $\mu_\delta$ | −183.10 ± 0.48 | −186.158 ± 0.014 | −189.69 ± 0.11 | mas $a^{-1}$ |
| $\gamma$ | −64.1 ± 7.3 | ... | ... | km $s^{-1}$ |
| $G$ | 8.5855 ± 0.0028 | 13.3872 ± 0.0028 | 17.7936 ± 0.0030 | mag |
| $J$ | 7.159 ± 0.024 | 10.625 ± 0.026 | 16.31 ± 0.28 | mag |
| $\mathcal{L}$ | ... | 167.0 ± 2.0 | ... | $10^{-4} \mathcal{L}_\odot$ |
| $T_{\rm eff}$ | ... | 3300 ± 50 | ... | K |
| $\mathcal{M}$ | 0.87 + 0.68 + 0.19 | 0.363 | 0.0985 | $\mathcal{M}_\odot$ |
| ruwe | 28.569 | 1.145 | 0.997 | |
| Qflag 2MASS | AAA | AAA | DDU | |
| Qflag AllWISE | AAAA | AAAC | ... | |

From the triple system HD 149162 (AaAbAc, DSG 7), the component Aa is spectroscopically classified as an early-K dwarf. We estimate photometrically the spectral types of Ab and Ac. At 252 arcsec from A, the M3 V G 17-23 (Karmn J16330+031) was reported to be a physical companion of the triple system (LEP 79). For this M dwarf, a close companion named LSPM J1633+0311S was additionally found at 6.4 arcsec (DAM 649), and classified as a white dwarf by Montes et al. (2018). Three of the five components are resolved by *Gaia* DR3 and 2MASS. González-Peinado et al. (2018) described this system in detail with the astrometry from *Gaia* DR1. We revisit this quintuple system with the latest astrometry from *Gaia* DR3.

# WDS 19312+3607 (GIC 158)

**Aab:** G 125-15

**B:** G 125-14, Karmn J19312+361AB

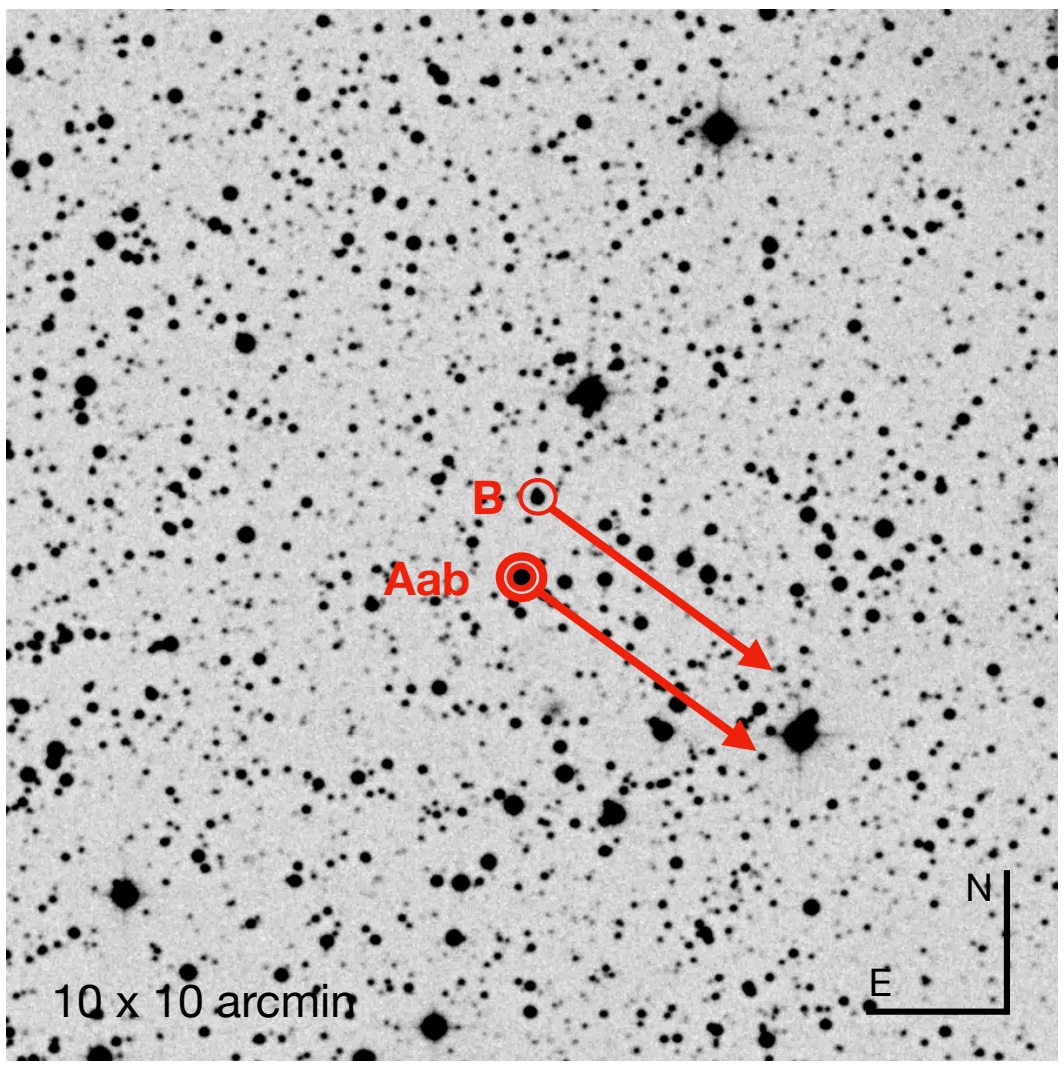

| | Aab-B | |
|---|---|---|
| WDS | GIC 158 | |
| $\rho$ | 45.78 | arcsec |
| $\theta$ | 347.4 | deg |
| $\mu$ ratio | 0.0085 | |
| $\Delta PA$ | 0.461 | deg |
| $\Delta d/d$ | 0.00072 | |
| $d$ | 40.08 | pc |
| $s$ | 1835 | au |
| $P_{orb}$ | 101.9 | $10^3$ a |
| $-U_g^*$ | 160.4 | $10^{33}$ J |

| Component | Aab | B | |
|---|---|---|---|
| SpT | M4.5 V + M5[a] | M4.5 V | |
| $\alpha$ | 19:31:12.38 | 19:31:11.56 | |
| $\delta$ | +36:07:28.2 | +36:08:12.8 | |
| $\pi$ | 24.948 ± 0.022 | 24.966 ± 0.015 | mas |
| $\mu_\alpha \cos\delta$ | −129.778 ± 0.022 | −130.266 ± 0.015 | mas $a^{-1}$ |
| $\mu_\delta$ | −106.444 ± 0.026 | −105.102 ± 0.018 | mas $a^{-1}$ |
| $\gamma$ | −22.31 ± 1.24[a] | … | km $s^{-1}$ |
| $G$ | 12.6849 ± 0.0029 | 13.9120 ± 0.0028 | mag |
| $J$ | 9.609 ± 0.022 | 10.924 ± 0.022 | mag |
| $\mathcal{L}$ | … | 93.92 ± 0.94 | $10^{-4}\mathcal{L}_\odot$ |
| $T_{eff}$ | … | 3200 ± 50 | K |
| $\mathcal{M}$ | 0.40 + 0.19 | 0.280 | $\mathcal{M}_\odot$ |
| ruwe | 1.358 | 1.203 | |
| Qflag 2MASS | AAA | AAA | |
| Qflag AllWISE | AAAB | AAAU | |

[a] Shkolnik et al. (2010)

G 125-15 is an active (Reid et al. 2004) mid-M-dwarf and double-lined spectroscopic binary (Shkolnik et al. 2010), with a known physical companion at 45.8 arcsec. The trio was investigated in detail by Caballero et al. (2010). In particular, they studied the age of the system and ruled out previous determinations of youth. Their photometrically estimated distance, as well as other previous determinations, are substantially different from the latest trigonometric value from *Gaia* DR3, which in turn enlarges the physical separation up to 1835 au.

# WDS 20198+2257 (KPP 4191 + *New*)

**Aab:** LP 395-8 A, Karmn J20198+229

**B:** LP 395-8 B

**C:** Gaia DR2 1829571684884360832

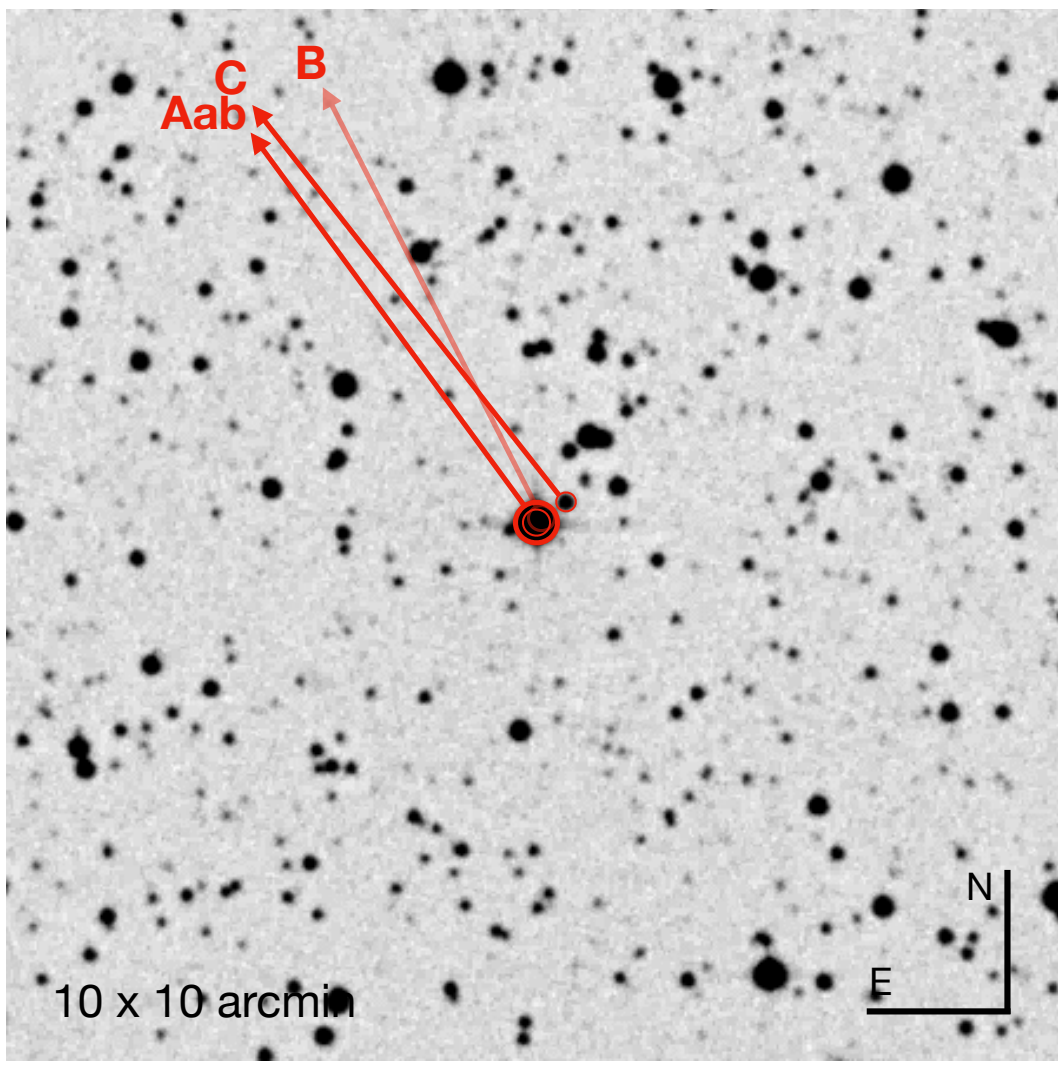

| | Aab-B | Aab-C | |
|---|---|---|---|
| WDS | KPP 4191 | ... | |
| $\rho$ | 1.92 | 11.02 | arcsec |
| $\theta$ | 355.5 | 307.4 | deg |
| $\mu$ ratio | 0.19 | 0.054 | |
| $\Delta PA$ | 10.6 | 0.0093 | deg |
| $\Delta d/d$ | 0.000044 | 0.0012 | |
| $d$ | 29.50 | | pc |
| $s$ | 56.6 | 325 | au |
| $P_{orb}$ | 0.432 | 5.95 | $10^3$ a |
| $-U_g^*$ | 9231 | 5954 | $10^{33}$ J |

| Component | Aab | B | C | |
|---|---|---|---|---|
| SpT | M3.0 V + m0 V[a] | m3.5 V | m9: V | |
| $\alpha$ | 20:19:49.36 | 20:19:49.35 | 20:19:48.72 | |
| $\delta$ | +22:56:38.1 | +22:56:40.0 | +22:56:44.8 | |
| $\pi$ | 33.897 ± 0.026 | 33.896 ± 0.053 | 33.94 ± 0.34 | mas |
| $\mu_\alpha \cos\delta$ | 83.565 ± 0.019 | 63.613 ± 0.041 | 88.09 ± 0.25 | mas $a^{-1}$ |
| $\mu_\delta$ | 106.536 ± 0.019 | 122.155 ± 0.039 | 112.33 ± 0.26 | mas $a^{-1}$ |
| $\gamma$ | ... | ... | ... | km $s^{-1}$ |
| $G$ | 11.0225 ± 0.0029 | 12.8748 ± 0.0028 | 19.4441 ± 0.0039 | mag |
| $J$ | 8.166 ± 0.021 | ... | 13.82 ± 0.11 | mag |
| $\mathcal{L}$ | ... | ... | ... | $10^{-4}\mathcal{L}_\odot$ |
| $T_{eff}$ | ... | ... | ... | K |
| $\mathcal{M}$ | 0.348 + 0.621[a] | 0.305 | 0.077 | $\mathcal{M}_\odot$ |
| ruwe | 1.484 | 1.639 | 1.046 | |
| Qflag 2MASS | AAA | ... | BBA | |
| Qflag AllWISE | AAAA | ... | ... | |

[a] Baroch et al. (2018)

LP 395-8 AB is a known binary system of M dwarfs located at less than 30 pc, which small projected separation (1.9 arcsec) affects the quality of the astrometric data. The primary (LP 395-8 A) was identified to be a spectroscopic binary by Baroch et al. (2018). To this known triple system, we add a fourth candidate to physical companion and estimated to be a very late-M dwarf from its intrinsic brightness in $G$ and $J$.

# WDS 22259-7501 (TOK 434 + DUN 238 + KO 5)

**A:** HD 212168, Königstuhl 5A
**BaBb:** CPD-75 1748B, Königstuhl 5B
**C:** DENIS J222644.3-750342, Königstuhl 5C

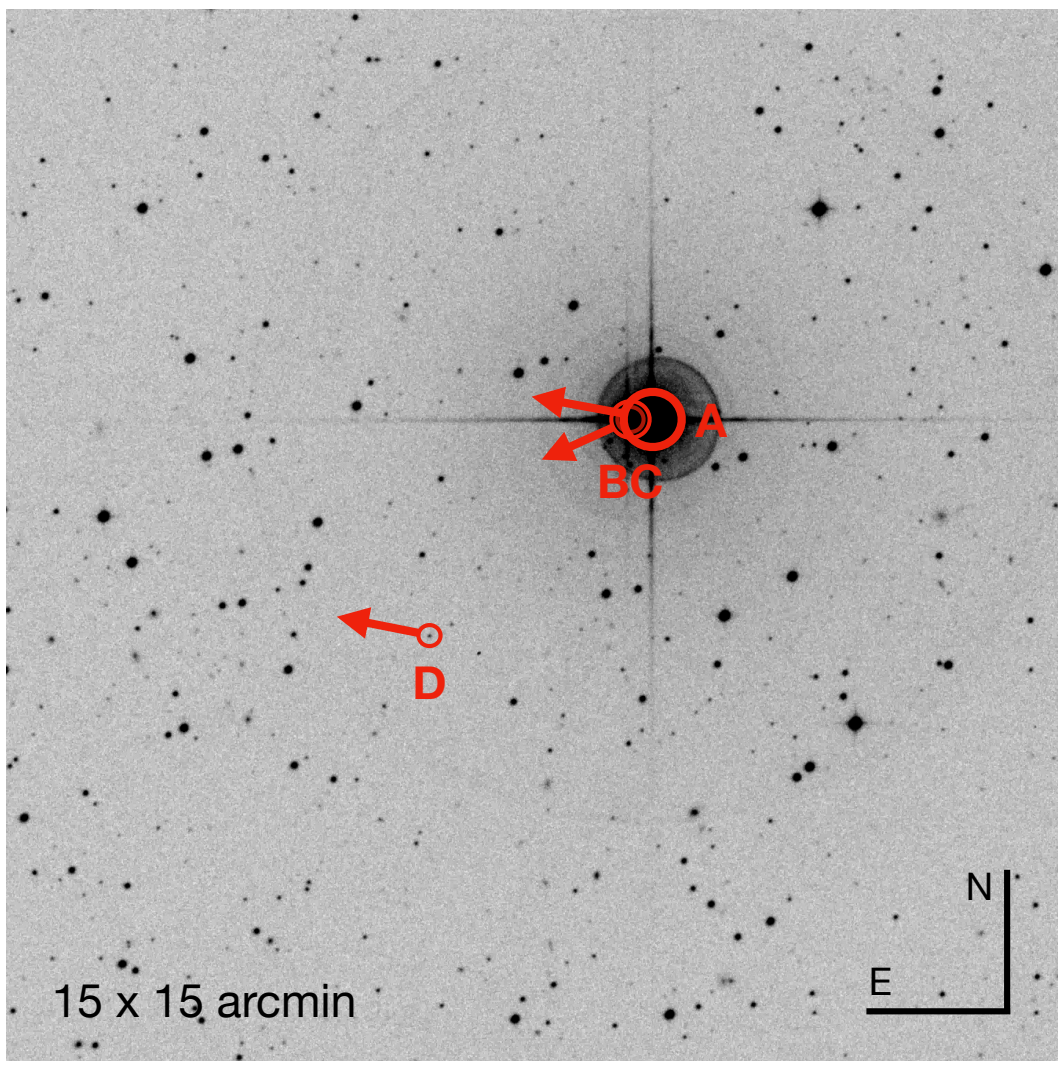

| | A-BaBb | A-C | |
|---|---|---|---|
| WDS | DUN 238 | KO 5 | |
| $\rho$ | 20.89 | 264.8 | arcsec |
| $\theta$ | 79.3 | 128.9 | deg |
| $\mu$ ratio | 0.49 | 0.029 | |
| $\Delta PA$ | 19.08 | 1.268 | deg |
| $\Delta d/d$ | 0.0007 | 0.002 | |
| $d$ | 23.41 | | pc |
| $s$ | 489.2 | 6199 | au |
| $P_{orb}$ | 10.25 | 462.3 | $10^3$ a |
| $-U_g^*$ | 2895 | 29.7 | $10^{33}$ J |

| Component | A | BaBb | C | |
|---|---|---|---|---|
| SpT | G0 V | k3 V+ | M8 | |
| $\alpha$ | 22:25:51.39 | 22:25:56.69 | 22:26:44.66 | |
| $\delta$ | –75:00:56.3 | –75:00:52.4 | –75:03:42.3 | |
| $\pi$ | 42.722 ± 0.020 | 42.69 ± 0.25 | 42.637 ± 0.078 | mas |
| $\mu_\alpha \cos\delta$ | 57.385 ± 0.021 | 33.33 ± 0.29 | 58.739 ± 0.084 | mas a$^{-1}$ |
| $\mu_\delta$ | 12.835 ± 0.023 | −3.78 ± 0.30 | 11.779± 0.092 | mas a$^{-1}$ |
| $\gamma$ | +14.51 ± 0.14 | +17.86 ± 0.59 | +17 ± 2[a] | km s$^{-1}$ |
| $G$ | 5.9771 ± 0.0028 | 8.3805 ± 0.0028 | 16.7983 ± 0.0031 | mag |
| $J$ | 5.262 ± 0.276 | 6.559 ± 0.029 | 12.353 ± 0.023 | mag |
| $\mathcal{L}$ | 15900 ± 290 | … | 7.06 ± 0.11 | $10^{-4}\mathcal{L}_\odot$ |
| $T_{\rm eff}$ | 5900 ± 50 | … | 2400 ± 50 | K |
| $\mathcal{M}$ | 1.11 | 0.720 | 0.0935 | $\mathcal{M}_\odot$ |
| ruwe | 1.017 | 16.764 | 1.189 | |
| Qflag 2MASS | DEE | AAA | AAA | |
| Qflag AllWISE | BBAA | BAAA | AAAU | |

[a] Burgasser et al. (2015)

The G0 dwarf HD 212168 and the close binary CPD −75° 1748B (TOK 434), separated 21 arcsec, form a triple system, which is additionally orbited by the M8.5 V star DENIS J222644.3-750342, separated 265 arcsec. Caballero et al. (2012) confirmed the physical binding. The Sun-lile primary and the low-mass companion share common parallax, proper motion and radial velocity. *Gaia* DR3 astrometric data for the K-dwarf secondary is, however, affected by its close binarity ($\rho$ = 0.3 arcsec).

# WDS 23059+0614 (SLW 1299 + SLW 1300)

A: SLW J2305+0613 A

B: SLW J2305+0613 B

C: SLW J2305+0613 C

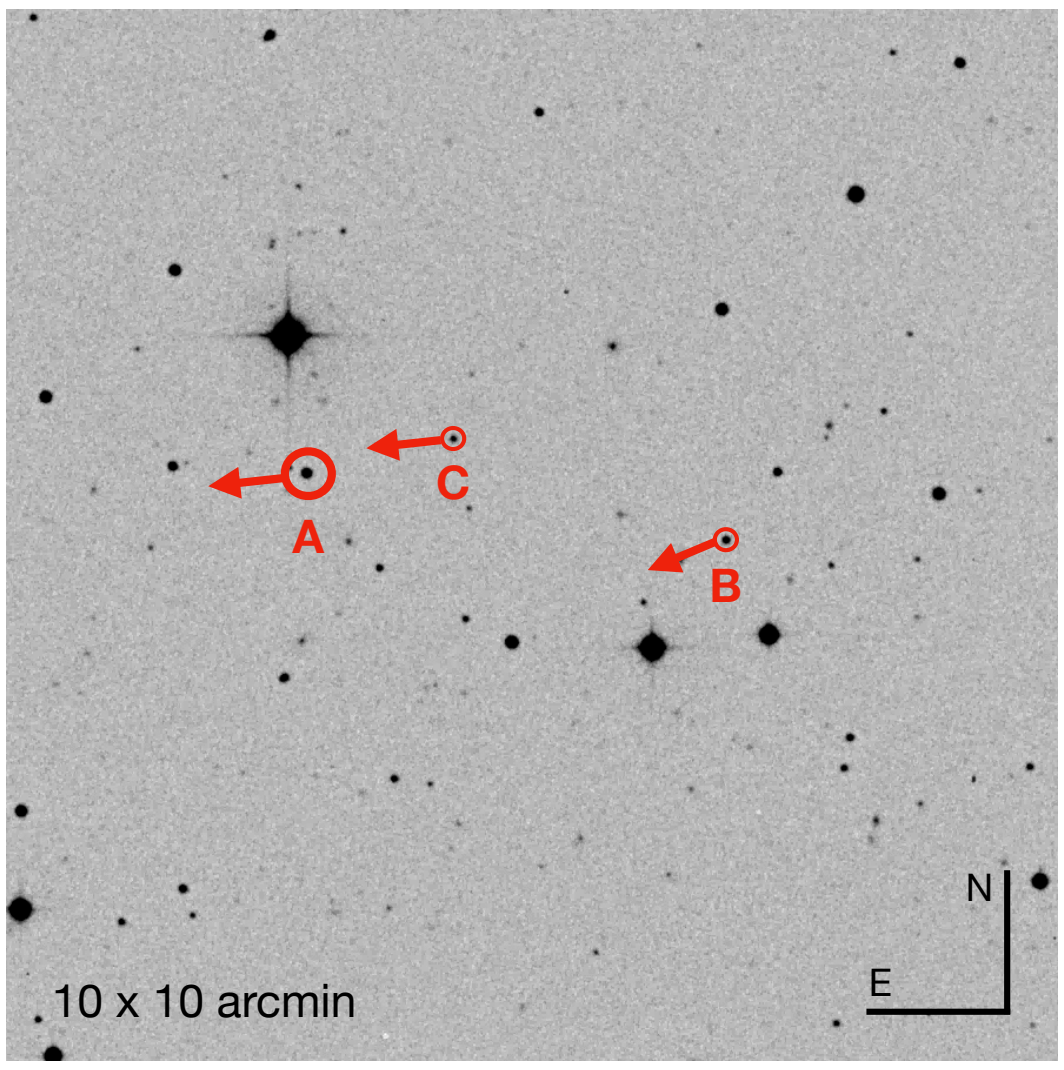

| | A-B | A-C | |
|---|---|---|---|
| WDS | SLW 1299 | SLW 1230 | |
| $\rho$ | 242.4 | 86.00 | arcsec |
| $\theta$ | 260.9 | 283.1 | deg |
| $\mu$ ratio | 0.11 | 0.029 | |
| $\Delta PA$ | 6.14 | 1.47 | deg |
| $\Delta d/d$ | 0.40 | 0.022 | |
| $d$ | 216.9 | | pc |
| $s$ | 52578 | 186579 | au |
| $P_{\rm orb}$ | 16642 | 3518 | $10^3$ a |
| $-U_g^*$ | 6.82 | 15.5 | $10^{33}$ J |

| Component | A | B | C | |
|---|---|---|---|---|
| SpT | M1.7 | M3.2 | M3.7 | |
| $\alpha$ | 23:05:51.69 | 23:05:35.65 | 23:05:46.08 | |
| $\delta$ | +06:13:34.6 | +06:12:56.1 | +06:13:54.1 | |
| $\pi$ | 4.609 ± 0.041 | 7.743 ± 0.084 | 4.71 ± 0.10 | mas |
| $\mu_\alpha \cos\delta$ | 43.584 ± 0.044 | 42.158 ± 0.092 | 42.78 ± 0.12 | mas $\rm a^{-1}$ |
| $\mu_\delta$ | −5.291 ± 0.084 | −9.784 ± 0.064 | −6.308 ± 0.084 | mas $\rm a^{-1}$ |
| $\gamma$ | … | … | … | km $\rm s^{-1}$ |
| $G$ | 16.5216 ± 0.0062 | 18.146 ± 0.036 | 18.896 ± 0.029 | mag |
| $J$ | 12.913 ± 0.025 | 13.826 ± 0.026 | 14.397 ± 0.034 | mag |
| $\mathcal{L}$ | 517 ± 22 | 196 ± 18 | 118 ± 11 | $10^{-4}\mathcal{L}_\odot$ |
| $T_{\rm eff}$ | 3500 ± 50 | 3300 ± 50 | 3200 ± 50 | K |
| $\mathcal{M}$ | 0.525 | 0.386 | 0.312 | $\mathcal{M}_\odot$ |
| ruwe | 1.081 | 1.086 | 0.937 | |
| Qflag 2MASS | AAA | AAA | AAA | |
| Qflag AllWISE | AABU | AAUU | AAUU | |

The pairs SLW J2305+0613 AB and BC are two known binary systems (SDSS SLoWPoKES Catalog; Dhital et al. 2010) of early-M dwarfs. We analyse the pairs A-B and A-C to test the candidacy for a triple system. We find that the parallactic distance of B differs by 40% to that of the primary, and also present a moderately large difference in proper motion. Given that the quality indicator ruwe does not indicate unreliable astrometry, we propose the B component to be a foreground star. Therefore, the pair A-B is visual, and the trio A-B-C does not qualify as a triple system.

# WDS 23315-0405 (KO 3 + CLO 4 + GZA 1)

## A: HD 221356, Königstuhl 3A
## BC: 2MASSW J2331016-040618, Königstuhl 3BC
## D: 2MASS J23313095-0405234

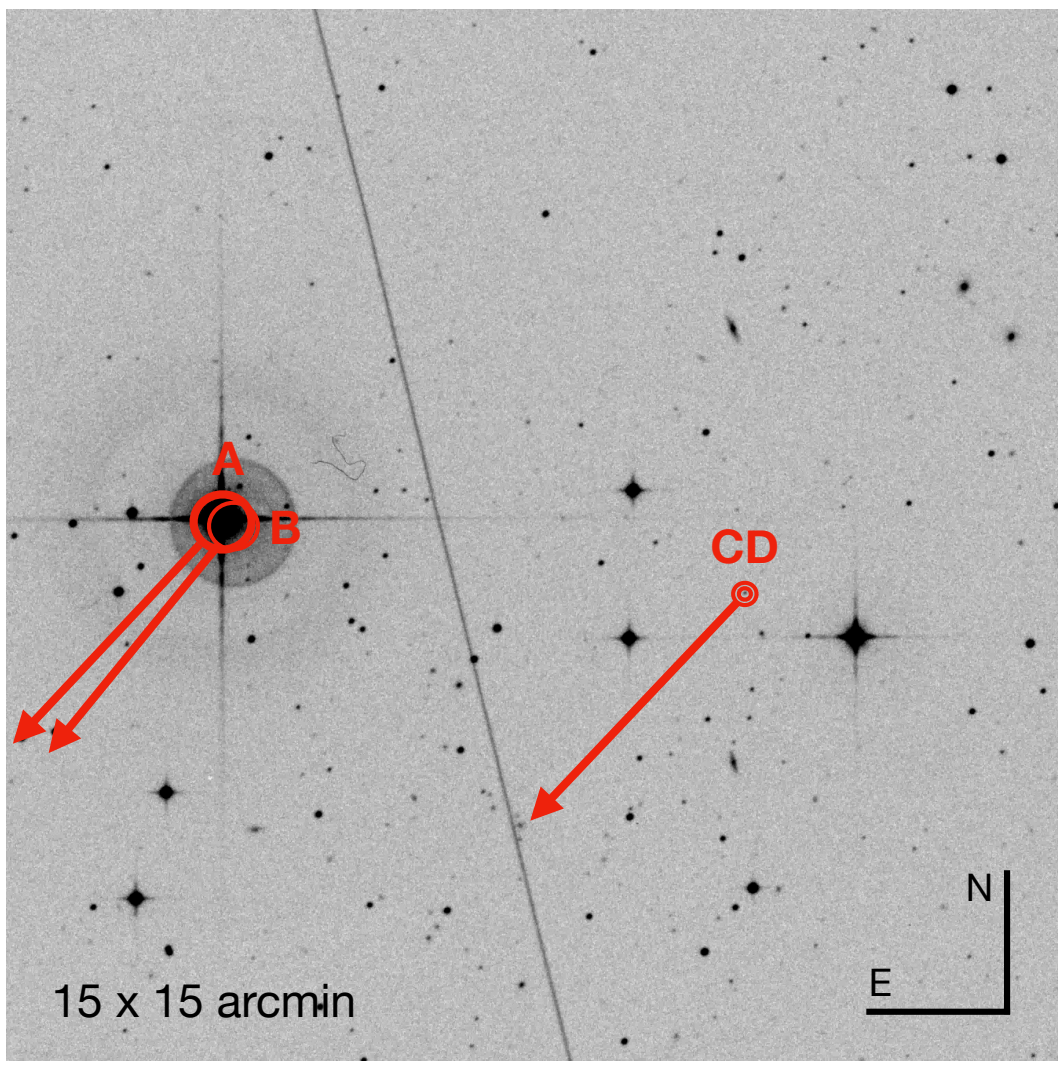

| | A-BC | A-D | |
|---|---|---|---|
| WDS | KO 3 | GZA 1 | |
| $\rho$ | 451.7 | 12.46 | arcsec |
| $\theta$ | 261.7 | 221.6 | deg |
| $\mu$ ratio | 0.012 | 0.033 | |
| $\Delta PA$ | 0.080 | 1.75 | deg |
| $\Delta d/d$ | 0.005 | 0.004 | |
| $d$ | 25.832 | | pc |
| $s$ | 11670 | 322 | au |
| $P_{orb}$ | 1205 | 5560 | $10^3$ a |
| $-U_g^*$ | 26.6 | 468.4 | $10^{33}$ J |

| Component | A | BC | D | |
|---|---|---|---|---|
| SpT | F7 V | M8.0 V + L3.0 V | L1 | |
| $\alpha$ | 23:31:31.69 | 23:31:01.82 | 23:31:31.14 | |
| $\delta$ | –04:05:17.7 | –04:06:22.5 | –04:05:27.0 | |
| $\pi$ | 38.711 ± 0.024 | 38.51 ± 0.16 | 38.54 ± 0.36 | mas |
| $\mu_\alpha \cos\delta$ | 178.130 ± 0.028 | 176.34 ± 0.17 | 169.94 ± 0.41 | mas $a^{-1}$ |
| $\mu_\delta$ | −191.845 ± 0.017 | −189.38 ± 0.13 | −194.59 ± 0.30 | mas $a^{-1}$ |
| $\gamma$ | −12.47 ± 0.18 | ... | ... | km $s^{-1}$ |
| $G$ | 6.3626 ± 0.0028 | 17.2032 ± 0.0037 | 18.527 ± 0.013 | mag |
| $J$ | 5.488 ± 0.019 | 12.938 ± 0.024 | 12.198 | mag |
| $\mathcal{L}$ | 13740 ± 150 | ... | ... | $10^{-4} \mathcal{L}_\odot$ |
| $T_{eff}$ | 6000 ± 50 | ... | ... | K |
| $\mathcal{M}$ | 1.09 | 0.161 | 0.0792 | $\mathcal{M}_\odot$ |
| ruwe | 0.857 | 1.215 | 1.198 | |
| Qflag 2MASS | AAA | AAA | UUB | |
| Qflag AllWISE | BAAA | AABU | ... | |

The triple system consisting of the F7 V HD 221356 and the 0.573 arcsec double (CLO 4; Gizis et al. 2003) 2MASSW J2331016-040618 (M8.0 V and L3.0 V), was confirmed to be a physically bound system by Caballero (2007). Gauza et al. (2012) reported a fourth component (L1 ± 1) separated by 12.5 arcsec from the primary.

# WDS 23536+1207 (VYS 11)

## A: StKM 2-1787

## B: TYC 1174-955-2, Karmn J23535+121

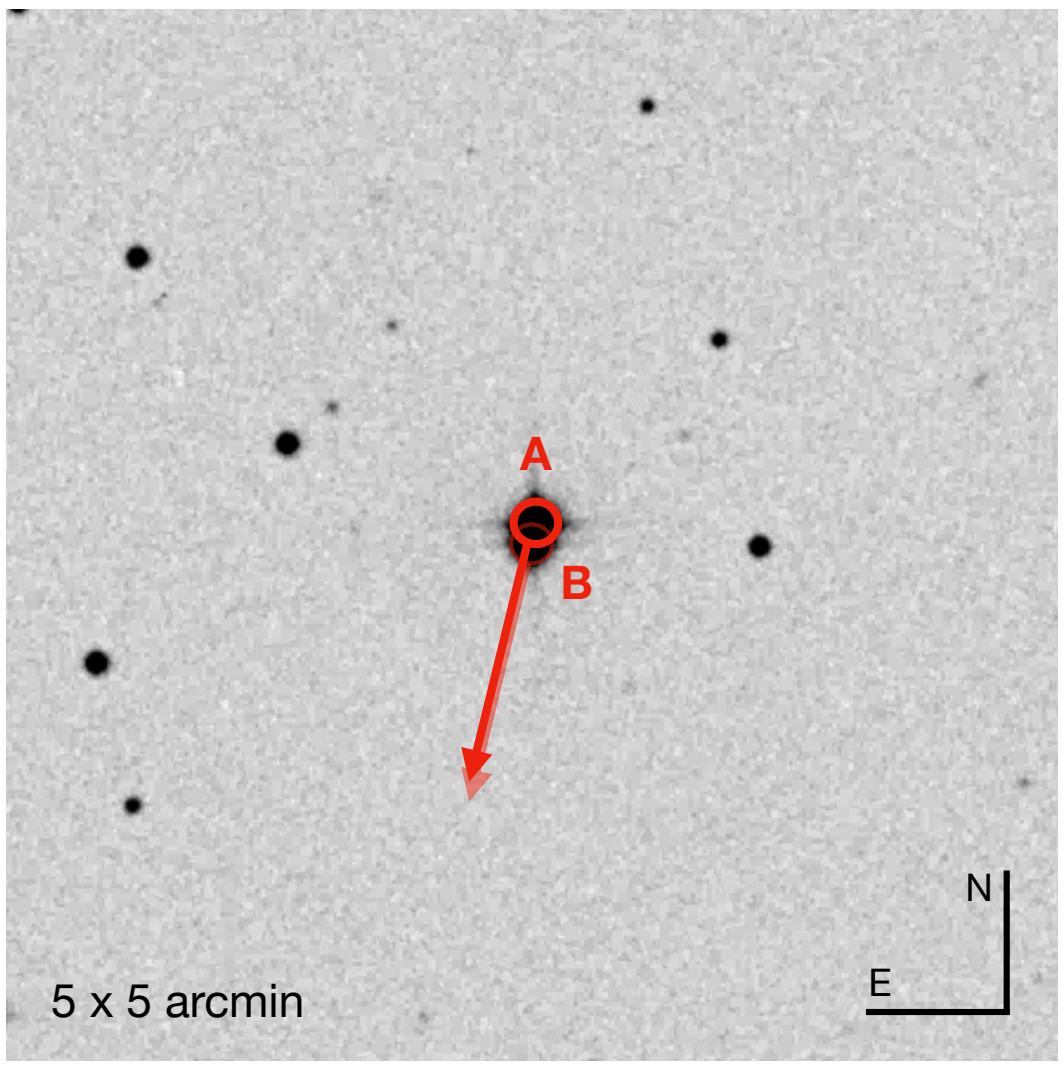

| | A-B | |
|---|---|---|
| WDS | VYS 11 | |
| $\rho$ | 5.78 | arcsec |
| $\theta$ | 165.1 | deg |
| $\mu$ ratio | 0.039 | |
| $\Delta PA$ | 2.190 | deg |
| $\Delta d/d$ | 0.0018 | |
| $d$ | 37.26 | pc |
| $s$ | 215.5 | au |
| $P_{\rm orb}$ | 3.987 | $10^3$ a |
| $-U_g^*$ | 2774 | $10^{33}$ J |

| Component | A | B | |
|---|---|---|---|
| SpT | K4 V | M2.5 V | |
| $\alpha$ | 23:53:35.58 | 23:53:35.69 | |
| $\delta$ | +12:06:20.4 | +12:06:14.8 | |
| $\pi$ | 26.839 ± 0.019 | 26.792 ± 0.035 | mas |
| $\mu_\alpha \cos\delta$ | 40.265 ± 0.023 | 44.122 ± 0.039 | mas a$^{-1}$ |
| $\mu_\delta$ | −113.773 ± 0.014 | −110.989 ± 0.022 | mas a$^{-1}$ |
| $\gamma$ | −21.69 ± 0.49 | … | km s$^{-1}$ |
| $G$ | 10.4261 ± 0.0028 | 11.2967± 0.0028 | mag |
| $J$ | 8.403 ± 0.019 | 8.670 ± 0.029 | mag |
| $\mathcal{L}$ | 1103.5 ± 9.2 | 702.7 ± 8.8 | $10^{-4} \mathcal{L}_\odot$ |
| $T_{\rm eff}$ | 4000 ± 50 | 3500 ± 50 | K |
| $\mathcal{M}$ | 0.629 | 0.538 | $\mathcal{M}_\odot$ |
| ruwe | 1.183 | 1.520 | |
| Qflag 2MASS | AAA | AAA | |
| Qflag AllWISE | AAAA | … | |

Using the latest astrometric data from *Gaia* DR3, we revisit the known binary system of StKM 2-1787 and TYC 1174-955-2, which comprises a mid-K dwarf and an early-M dwarf. We measure an orbital period of approximately 4000 years at a separation of 216 au. We confirm the binarity that Montes et al. (2018) left on hold, as new *Gaia* astrometry could come to settle the issue.

# Appendix E

## Code

In this Appendix the most relevant pieces of code used in this work can be found. Their purposes are mainly to access databases and to automate repetitive tasks. They aim for syntax simplicity and a good effort has been made to consciously abide with the style guides[1]. The full suite of code produced during this thesis can be found in the GitHub: `https://github.com/ccifuentesr/`.

[1]For instance, the `Python`'s PEP 8 (`https://peps.python.org/pep-0008/`).

## E.1 `Aladin` search

The following is a working example for a 10-arcmin automatic cone search around the source defined as \$1, loading a 15×15 arcmin$^2$ first generation DSS image, the WDS catalog, and the main astro-photometric catalogues used in this work. Use the `Tool > Macro controller` option to input this macro:

```
reset
get DSS.ESO(POSS2UKSTU_IR, 15, 15) $1 10'
pause 3  # giving time for image to load and stay on the background
get Simbad $1 10'  # Simbad
get VizieR(B/wds/wds) $1 10'  # WDS
get VizieR(II/335/galex_ais) $1 10'  # Galex DR5
get VizieR(I/259/tyc2) $1 10'  # Tycho-2
get VizieR(I/355/gaiadr3) $1 10'  # Gaia DR3
get VizieR(I/350/gaiaedr3) $1 10'  # Gaia EDR3
get VizieR(I/345/gaia2) $1 10'  # Gaia DR2
get VizieR(V/139/sdss9) $1 10'  # SDSS9
get VizieR(I/322A/out) $1 10'  # UCAC4
get VizieR(II/336/apass9) $1 10'  # APASS9
get VizieR(II/349/ps1) $1 10'  # Pan-STARRS1
get VizieR(I/327/cmc15) $1 10'  # CMC15
get VizieR(II/246/out) $1 10'  # 2MASS
get VizieR(II/328/allwise) $1 10'  # AllWISE
get VizieR(II/311/wise) $1 10'  # WISE
get VizieR(II/365/catwise ) $1 10'  # CatWISE
```

`get hips(CDS/P/SDSS9/color)`, SDSS and `get hips(CDS/P/DSS2/color)`, second generation DSS, are other surveys available for loading image adquisition. In the cases in which a close binarity is suspected, the use of several images from different epochs makes it possible to actually eyesight the change of position of the stars in the sky. Similarly, the syntax `get VizieR( ... )` accepts any catalog identifier from the VizieR database (https://vizier.cds.unistra.fr/viz-bin/VizieR).

`TOPCAT` and `Aladin` are able to exchange data, from 'Interop' menu, 'Send table to...' option. The table is then loaded as a layer in `Aladin`, with the possibility of scrolling through the objects.

## E.2 *Gaia* ADQL

The vast amount of information stored in the *Gaia* Archive (https://gea.esac.esa.int/archive/) can be retrieved using Astronomical Data Query Language (ADQL), a semantic derivative (or dialtect) of Structured Query Language (SQL). These are working examples of code snippets used in this work to obtain tailored *Gaia* data. In this work they have also been implemented programmatically within other codes using Python's astroquery.gaia package. Fields in <angle brackets> refer to user's input.

```
> Retrieval of all of the Gaia columns for a given  <gaia_id>.

 SELECT  *
 FROM gaiadr3.gaia_source
 WHERE source_id =  <gaia_id>

> Search by coordinates  <ra>,  <dec> of any source within a  <radius> in degrees,
limiting to a difference in  <parallax> of less than 10%.

 SELECT * , distance(POINT('ICRS',  <ra>,  <dec>),
 POINT('ICRS', gaia.ra, gaia.dec))  AS dist
 FROM gaiadr3.gaia_source  AS gaia
 WHERE 1=CONTAINS(POINT('ICRS',  <ra>,  <dec>),
 CIRCLE('ICRS', gaia.ra, gaia.dec,  <radius>))
 AND abs(1-(gaia.parallax/ <parallax>)) < 0.10)

> Example of double cross match with AllWISE

 SELECT  mytab.*, gaia.*
 FROM user_ccifuentes. <user_table> AS mytab
 LEFT OUTER JOIN gaiadr3.allwise_best_neighbour  AS xmatch
 ON mytab.<gaia_id> = xmatch.source_id
 LEFT OUTER JOIN gaiadr2.allwise_original_valid  AS wise
 ON xmatch.original_ext_source_id = wise.allwise_oid
```

## E.3 Software

Table E.1: Software widely used in this work.

| Software | Author(s) | Webpage |
|---|---|---|
| Aladin[a] | Bonnarel et al. (2000) | https://aladin.u-strasbg.fr |
| IRAF[b] | NOAO | http://iraf.noao.edu (404) |
| PARSEC | Bressan et al. (2012) | http://stev.oapd.inaf.it/cgi-bin/cmd |
| SteParKin[c] | Montes et al. (2001) | https://github.com/dmontesg/SteParKin |
| TOPCAT[d] | Taylor (2005) | http://www.star.bris.ac.uk/~mbt/topcat/ |

[a] Developed by the CDS.
[b] IRAF was written by the National Optical Astronomy Observatories (NOAO), but http://iraf.noao.edu does no longer exist. The latest NOAO release had a large number of problems, due to its complicated package structure, a buggy installation procedure, major bugs in the code itself (with increasing difficulty to adapt to 64-bit architectures), and (most important) major issues in the licensing and security bugs. With this, development and maintenance of IRAF is discontinued since 2013. The IRAF community distribution (https://iraf-community.github.io) notes that "[users] should be aware that IRAF is 35 years old legacy code and institutional support for IRAF and its usage is going away quickly". The same source also recommends to search for alternative solutions (e.g. in Astropy) and *not to* start new projects using IRAF. See Space Telescope Science Institute (STScI) newsletter (Vol. 2018, issue 35) "Removing the Institute's Dependence on IRAF (You can do it too!)" on the imperious necessity for de-IRAFing the community.
[c] SteParKin is based on D. Montes' and H. Tabernero's original code, originally written in FORTRAN and later in IDL, and adapted to Python by A. J. Domínguez-Fernández.

Table E.2: Python libraries most often used in this work.

| Software | Author(s) |
|---|---|
| AstroPy | Astropy Collaboration et al. (2022) |
| emcee | Foreman-Mackey et al. (2013) |
| IPython | Perez & Granger (2007) |
| Matplotlib | Hunter (2007) |
| NumPy | Harris et al. (2020) |
| Pandas | Wes McKinney (2010) |
| SciPy | Virtanen et al. (2020) |

# List of Figures

# List of Tables